# CEPC

# *Conceptual Design Report*

## Volume I - Accelerator

The CEPC Study Group

# CEPC

# *Conceptual Design Report*

## Volume I - Accelerator

The CEPC Study Group

August 2018

## Acknowledgements

The *CEPC Conceptual Design Report* (CDR) was prepared and written by the CEPC Study Group. The study was organized and led by the Institute of High Energy Physics (IHEP) of the Chinese Academy of Sciences (CAS) in collaboration with a number of institutions from various countries.

The study was supported by a large number of important entities: the National Key Programme for S&T Research and Development of the Ministry of Science and Technology (MOST), Yifang Wang's Science Studio of the Ten Thousand Talents Project, the CAS Key Foreign Cooperation Grant, the National Natural Science Foundation of China (NSFC), Beijing Municipal Science & Technology Commission, the CAS Focused Science Grant, the IHEP Innovation Grant, the CAS Lead Special Training Program, the CAS Hundred Talents Program, the CAS Center for Excellence in Particle Physics, the CAS International Partnership Program and the CAS/SAFEA International Partnership Program for Creative Research Teams.

The current volume, Volume I, is on the accelerators. A separate volume, Volume II, will be on physics and the detectors. This volume is a summary of work accomplished during the past five years by hundreds of scientists and engineers at home and abroad.

# Contents

# Executive Summary

The discovery of the Higgs boson at CERN's Large Hadron Collider (LHC) in July 2012 opened new opportunities for a large-scale accelerator. Due to the low mass of the Higgs, it is possible to produce it in the relatively clean environment of a circular electron–positron collider with reasonable luminosity, technology, cost and power consumption. The Higgs boson is a crucial cornerstone of the Standard Model (SM). It is at the center of the greatest mysteries of modern particle physics, such as the large hierarchy between the weak scale and the Planck scale, the nature of the electroweak phase transition, and many other related questions. Precise measurements of the properties of the Higgs boson serve as excellent tests of the underlying fundamental physics principles of the SM, and they are instrumental in explorations beyond the SM. In September 2012, Chinese scientists proposed a 240 GeV *Circular Electron Positron Collider* (CEPC), serving two large detectors for Higgs studies. The tunnel for such a machine could also host a *Super Proton Proton Collider* (SPPC) to reach energies beyond the LHC.

The CEPC is a large international scientific project initiated and hosted by China. It was presented for the first time to the international community at the ICFA Workshop "*Accelerators for a Higgs Factory: Linear vs. Circular*" (HF2012) in November 2012 at Fermilab. A Preliminary Conceptual Design Report (Pre-CDR, the *White Report*) was published in March 2015, followed by a Progress Report (the *Yellow Report*) in April 2017. This Conceptual Design Report (CDR, the *Blue Report*) is a summary of work during the past five years by hundreds of scientists and engineers.

The CEPC is a circular $e^+e^-$ collider located in a 100-km circumference underground tunnel. The accelerator complex consists of a linear accelerator (Linac), a damping ring (DR), the Booster, the Collider and several transport lines. In the tunnel, space is reserved for a future *pp* collider, SPPC.

The CEPC center-of-mass energy is 240 GeV, and at that collision energy will serve as a Higgs factory, generating more than one million Higgs particles. The design also allows operation at 91 GeV for a Z factory and at 160 GeV for a W factory. The number of Z particles will be close to one trillion, and $W^+W^-$ pairs about 15 million. These unprecedented large number of particles make the CEPC a powerful instrument not only for precision measurements on these important particles, but also in the search for new physics.

The heart of the CEPC is a double-ring collider. Electron and positron beams circulate in opposite directions in separate beam pipes. They collide at two interaction points (IPs) where are located large detectors described in detail in the CDR Volume II. Unlike protons, composite particles composed of quarks, electrons and positrons are fundamental leptons. Signals detected from their collision have little or no background, an important advantage over hadron colliders.

The CEPC Booster is located in the same tunnel above the Collider. It is a synchrotron with a 10 GeV injection energy and extraction energy equal to the beam collision energy. The repetition cycle is 10 seconds. Top-up injection will be used to maintain constant luminosity.

The 10 GeV Linac, injector to the Booster, built at ground level, accelerates both electrons and positrons. A 1.1 GeV damping ring reduces the positron emittance. Transport lines made of permanent magnets connect the Linac to the Booster.

The tunnel hosting the Collider and Booster will be mostly in hard rock so there is a strong and stable foundation to support the accelerators. The tunnel size is large enough to accommodate the future SPPC without removing the CEPC collider ring. This opens up the exciting possibilities of *ep* and *e*-ion physics in addition to *ee* physics (CEPC) and *pp* and ion-ion physics (SPPC).

In addition to particle physics, the Collider can operate simultaneously as a powerful synchrotron radiation (SR) light source. It will extend the usable SR spectrum into an unprecedented energy and brightness range. Two gamma-ray beamlines are included in the design.

The circulating CEPC beams radiate large amount of SR power, 30 MW per beam. Reducing power consumption is an important criterion in the design. By using superconducting radiofrequency (RF) cavities, high efficiency klystrons, 2-in-1 magnets, combined function magnets, large coil cross-section in the quadrupoles, permanent magnets in the transport lines, the total facility power consumption is kept below 300 MW. The power conversion efficiency from the grid to the beam will be more than 20%, higher than at other accelerator facilities.

Prior to the construction will be a five-year R&D period (2018-2022). During this period, prototypes of key technical components will be built and infrastructure established for industrialization for manufacturing the large number of required components.

There are numerous considerations in choosing the site. At this time six sites that all satisfy the technical requirements have been considered.

A detailed cost estimate based on a Work Breakdown Structure (WBS) has been carried out. It includes the entire accelerator complex, two detectors, the gamma-ray beamlines, conventional facilities and an adequate allowance for contingency.

Construction is expected to start in 2022 and be completed in 2030. After commissioning, a tentative operation plan will be to run 7 years for Higgs physics, followed by 2 years for operation in Z mode and 1 year for operation in W mode. This 10-year operations plan brings us to ~2040. It is expected that by then high field superconducting magnets for the SPPC will have been developed and be ready for installation, and the SPPC era will begin.

The CEPC is an important part of the world plan for high-energy physics research. It will support a comprehensive research program by scientists from throughout the world. Physicists from many countries will work together to explore the science and technology frontier, and to bring to a new level our understanding of the fundamental nature of matter, energy and the universe.

# 1 Introduction

Particle accelerators are a major invention of the 20th century. From the first linear accelerator built by Rolf Wideröe in an 88-cm long glass tube in Aachen, Germany, in the 1920s to the gigantic 27-km circumference deep-underground Large Hadron Collider (LHC) at CERN in Geneva, Switzerland, particle accelerators in the last nine decades have evolved enormously and have had a profound influence on the way we live, think and work.

Accelerators are the most powerful scientific instruments for exploring the microworld, for viewing the inner structure of materials and their constituent protons, neutrons, electrons, neutrinos, quarks, and possible still undiscovered fundamental building blocks of the universe such as dark matter and dark energy.

Colliders are a special kind of accelerator. They are often called "*engines of discovery*." The idea of colliding two particle beams to fully exploit the energy of accelerated particles was first proposed by Wideröe, who in 1943 applied for a patent on the collider concept and was awarded the patent in 1953. The first three colliders – AdA in Italy, CBX in the US, and VEP-1 in the then Soviet Union – came into operation about 50 years ago in the mid-1960s. Many other colliders followed.

Over the past decades, colliders defined the energy frontier in particle physics. Different types of colliers – proton-proton, proton-antiproton, electron-positron, electron-proton, electron-ion and ion-ion colliders – have played complementary roles in mapping out the constituents and forces in the Standard Model (SM). The latest example is the discovery of the Higgs boson at the LHC. This discovery was a showcase of what a collider can do to advance to the next level in this most basic frontier of science and technology.

The discovery of the Higgs boson, a key SM element, has placed the focus on the need to study the properties of this new particle with high precision. Because the Higgs mass is relatively low (125 GeV), a circular $e^+e^-$ collider can serve as a Higgs factory. The ring must have a large circumference in order to reduce the copious amount of synchrotron radiation from the high energy electron and positron beams. If such a large size ring were to exist, the tunnel would be ideal for housing a *pp* collider in the future with an energy much higher than that of the LHC.

The main advantage of a circular $e^+e^-$ collider of sufficiently large size is to provide a higher luminosity than a linear collider at 240 GeV and below. Also, a circular collider can accommodate more than one interaction point.

Circular $e^+e^-$ colliders have a long and glorious history. Scores of circular $e^+e^-$ colliders have been built and operated in the past five decades. The technology is mature and the experience rich. Table 1.1 lists the energy, luminosity and operation period of the $e^+e^-$ colliders (including the sole linear collider, the SLC) that have been built and operated since 1991 [1]. The highest energy collider was LEP2 with a center-of-mass (c.m.) energy of 209 GeV, whereas the collider of the highest luminosity achieved was KEKB with a luminosity of $2.1 \times 10^{34}$ cm$^{-2}$s$^{-1}$. These numbers are close to the design goal for a Higgs factory: 240 GeV and $3 \times 10^{34}$ cm$^{-2}$s$^{-1}$. However, the difficulty is to achieve both of these parameter goals in the same collider. On the other hand, the design, construction and operation of a new circular Higgs factory can benefit a great deal from the experience from other machines, especially PEP-II, LEP, KEKB and SuperKEKB.

**Table 1.1:** $e^+e^-$ colliders built and operated since 1991

| Location | Accelerator | Energy (GeV × GeV) | Luminosity ($cm^{-2}s^{-1}$) | Operation Period |
|---|---|---|---|---|
| CERN | LEP | 104.5 × 104.5 | $1 \times 10^{32}$ | 1989-2000 |
| KEK | SuperKEKB | 7 ($e^-$) × 4 ($e^+$) | $8 \times 10^{35}$ (*) | 2018-present |
| | KEKB | 8 ($e^-$) × 3.5 ($e^+$) | $2.1 \times 10^{34}$ | 1998-2010 |
| | TRISTAN | 32 × 32 | $3.7 \times 10^{31}$ | 1986-1995 |
| SLAC | PEP-II | 9 ($e^-$) × 3.1 ($e^+$) | $1.2 \times 10^{34}$ | 1999-2008 |
| | SLC | 46.2 × 46.2 | $3 \times 10^{30}$ | 1988-1998 |
| DESY | DORIS | 5.6 × 5.6 | $3.3 \times 10^{31}$ | 1974-1992 |
| Cornell | CESR | 1.8 × 1.8 to 5.5 × 5.5 | $1.3 \times 10^{33}$ | 1979-2008 |
| INFN | DAFNE | 0.51 × 0.51 | $2.4 \times 10^{32}$ | 1999-present |
| IHEP/China | BEPC & BEPC-II | 1.5 × 1.5 to 2.5 × 2.5 | $1 \times 10^{33}$ | 1988-2005, 2008-present |
| BINP | VEPP-2000 | 0.2 × 0.2 to 1 × 1 | $1.2 \times 10^{32}$ | 2010-present |
| | VEPP-4M | 1.5 × 1.5 to 5 × 5 | $5 \times 10^{30}$ | 1984 -present |

(*) Design goal

Colliders also play a critical role in advancing beam physics, accelerator research and associated technology development. In the course of building larger and more complex accelerators, technologies have been pushed to an extent never imagined before – large volume ultrahigh vacuum systems, large scale super-fluid helium applications in cryogenics, superconducting magnets, superconducting radio frequency systems, high power high efficiency microwave devices, radiation-hard materials, global control systems, advanced instrumentation and diagnostics, super-fast computing and communication networks, and giant data storage and processing systems.

In 2012 shortly after the discovery of the Higgs particle at CERN, the Institute of High Energy Physics (IHEP) in Beijing, China, in collaboration with a number of other institutions in China as well as in many other countries, launched a study of the *Circular Electron Positron Collider*, CEPC. The concept was first presented at the ICFA Workshop "*Accelerators for a Higgs Factory: Linear vs. Circular*" (HF2012) in November 2012 at Fermilab [2]. A Preliminary Conceptual Design Report (Pre-CDR, the *White Report*) was published in March 2015 [3], followed by a Progress Report (the *Yellow Report*) in April 2017 [4]. This Conceptual Design Report (CDR, the *Blue Report*) is a summary of work on CEPC during the past five years by hundreds of scientists and engineers.

The CEPC accelerator complex consists of a linear accelerator (Linac), a damping ring (DR), the Booster, the Collider and several transport lines. The Collider and the Booster are located in an underground tunnel, whereas the Linac and DR are built at ground level. As the Collider can operate simultaneously as a powerful synchrotron radiation (SR) light source, two gamma-ray beamlines are included in the design. In the tunnel, space is reserved for a future *pp* collider the *Super Proton Proton Collider*, SPPC.

A major change from the Pre-CDR is an increase of the tunnel circumference from 54 km to 100 km. Another change is in the Collider design, from single-ring (like CESR and LEP) to double-ring (like KEKB and BEPC-II). These changes greatly improve the performance.

The CEPC center-of-mass energy is 240 GeV, and at that collision energy will serve as a Higgs factory, generating more than one million Higgs particles in 7 years. The design also allows operation at 91 GeV as a Z factory and at 160 GeV as a W factory. The number of Z particles will be close to one trillion in a 2-year operation, and $W^+W^-$ pairs close to fifteen million in one year. Table 1.2 lists the CEPC top-level parameters.

**Table 1.2:** CEPC Top Level Parameters

| Parameter | Design Goal | | |
|---|---|---|---|
| | H | Z | $W^+W^-$ |
| Colliding particles | $e^+, e^-$ | | |
| Center-of-mass energy (GeV) | 240 | 91 | 160 |
| Integrated luminosity ($ab^{-1}$, per IP per year) | 0.4 | 4 | 1.3 |
| No. of IPs | 2 | | |

This report consists of 12 chapters and 8 appendices.

This chapter is an introduction.

Chapter 2 is an overview of the CEPC layout and performance goals.

Chapter 3 presents an operation plan for H, Z and W physics. It also discusses a staging scenario from the CEPC to the future SPPC and describes various collision schemes.

Chapters 4-7 are a detailed description of the CEPC accelerator complex design.

Chapter 4 is about the Collider. It discusses all the relevant collider beam physics issues, including the optics in the arcs, straight sections and interaction regions (IRs), dynamic aperture, machine errors and correction, beam-beam effects, impedance and beam instabilities, synchrotron radiation, injection and beam dump, the machine-detector interface, beam loss, background and collimation. It also describes the design of various technical systems for the Collider, including the superconducting RF (SRF) system and its power source, the cryogenic system, magnets and their power supplies, superconducting magnets in the IR, vacuum, instrumentation, control and mechanical systems.

Chapter 5 discusses the beam dynamics and technical systems of the Booster.

Chapter 6 is about the Linac, the Damping ring, and the electron and positron sources.

Chapter 7 covers three technical systems common to the CEPC accelerators: the cryogenic system (common to the Collider and the Booster), survey and alignment, and radiation protection.

Chapter 8 discusses the upgrade of the facility with the addition of the SPPC, a high-energy proton-proton collider to be installed in the same tunnel. The SPPC, has a center-of-mass energy of 75-125 TeV. It also requires an injector chain, including a proton linac, followed by a cascade of three synchrotrons to bring the proton energy up to the 2 TeV injection energy. The injection chain will not only serve as an injector to the SPPC but will also have a rich physics program of its own.

Conventional facilities are a major part of this study, not just because the tunnel circumference is large (100 km), but also because the tunnel is sufficiently wide (6 m) to accommodate three accelerators (the CEPC Collider, the CEPC Booster and the SPPC

Collider). In addition to the underground structure, there will be a large cooling water system, HVAC and other utilities. Site selection criteria are enumerated and several potential sites described. The facilities and their construction are described in Chapter 9 in the context of a possible site.

Chapter 9 is followed by a discussion of environment, safety and health considerations in Chapter 10.

Chapter 11 presents details of an extensive R&D program to be carried out prior to construction. For the CEPC, the critical path is successful implementation of the SRF R&D, including superconducting RF cavities, couplers, HOM dampers, tuners, and the associated RF power source and cryogenic system. A pre-production plan is also included. For the SPPC, the key technical component is the superconducting magnet. A 20-year road map for the magnets is outlined.

Chapter 12 discusses a possible project timeline. A detailed estimate of the construction cost is made based on a Work Breakdown Structure (WBS). The percentage of the total required for each system is shown. The operation cost (for electricity only) is also presented.

These 12 chapters constitute the main body of this report. In addition, there are eight appendices:

Appendix 1 is a detailed parameter list for the accelerators – the Collider, the Booster, the Linac, the Damping Ring, the electron and positron sources, and transport lines.

Appendix 2 is a detailed list of technical components.

Appendix 3 presents the power requirement of each technical system as well as for the entire facility.

Appendix 4 discusses an option to use an advanced partial double ring design instead of a double ring.

Appendix 5 discusses a concept to use a plasma wakefield injector instead of a conventional linac.

Appendix 6 is a discussion of using the CEPC as a high intensity $\gamma$-ray source, which would extend the usable synchrotron radiation spectrum into an unprecedented energy and brightness range.

Appendix 7 discusses the exciting possibilities of *e-p* and *e*-ion collisions in addition to *ee* collisions (CEPC) and *pp* and ion-ion collisions (SPPC).

Appendix 8 discusses polarization schemes that have the potential to be implemented in the CEPC for energy calibration and other applications.

The CEPC is a giant leap for China from the BEPC-II, the presently operating $e^+e^-$ collider at IHEP in Beijing. The challenges are big and real. But the potential payoffs are enormous. It will bring China to the forefront of world high-energy physics, open up a whole new window to fundamental research, be one of the most powerful scientific instrument ever built, and push a wide range of advanced technologies to an extent never imagined before.

The CEPC will educate and train a new generation of HEP and accelerator scientists and engineers, who will lead this field in future decades.

The CEPC is an important part of the world plan for high-energy physics research. It will support a comprehensive research program by scientists from throughout the world. Physicists from many countries will work together to explore the science and technology frontier, and bring to a new level our understanding of the fundamental nature of matter, energy and the universe.

# 2 Machine Layout and Performance

## 2.1 Machine Layout

CEPC is a double ring collider with two interaction points (IPs). The CEPC collider ring, the CEPC booster ring and the future SPPC collider ring are all housed in the same 100 km circumference tunnel. A cross section of the tunnel is shown in Fig. 2.1.

The geometry of the CEPC and the SPPC are matched as closely as possible. The large circumference is mainly determined by the SPPC high-energy goal as well as consideration of the development of bending magnets that can be designed and manufactured. The interaction region of SPPC is located in the same long straight sections where the CEPC RF cavities are placed. The 4-km long collimation regions of SPPC are located in the two interaction regions of CEPC. This poses a potential space conflict. Due to the anticipated very large detector size, bypass geometry or independent tunnel sections for SPPC or CEPC in the two regions is needed. The SPPC ring is arranged to be outside the CEPC ring. The SPPC geometry can be optimized with relatively lower magnetic field, especially in the bypass region without affecting the CEPC design. The Booster is located above the Collider and separated by 2.4 m. as shown in Fig. 2.1. This distance is sufficient to avoid magnetic interference between the Collider and the Booster.

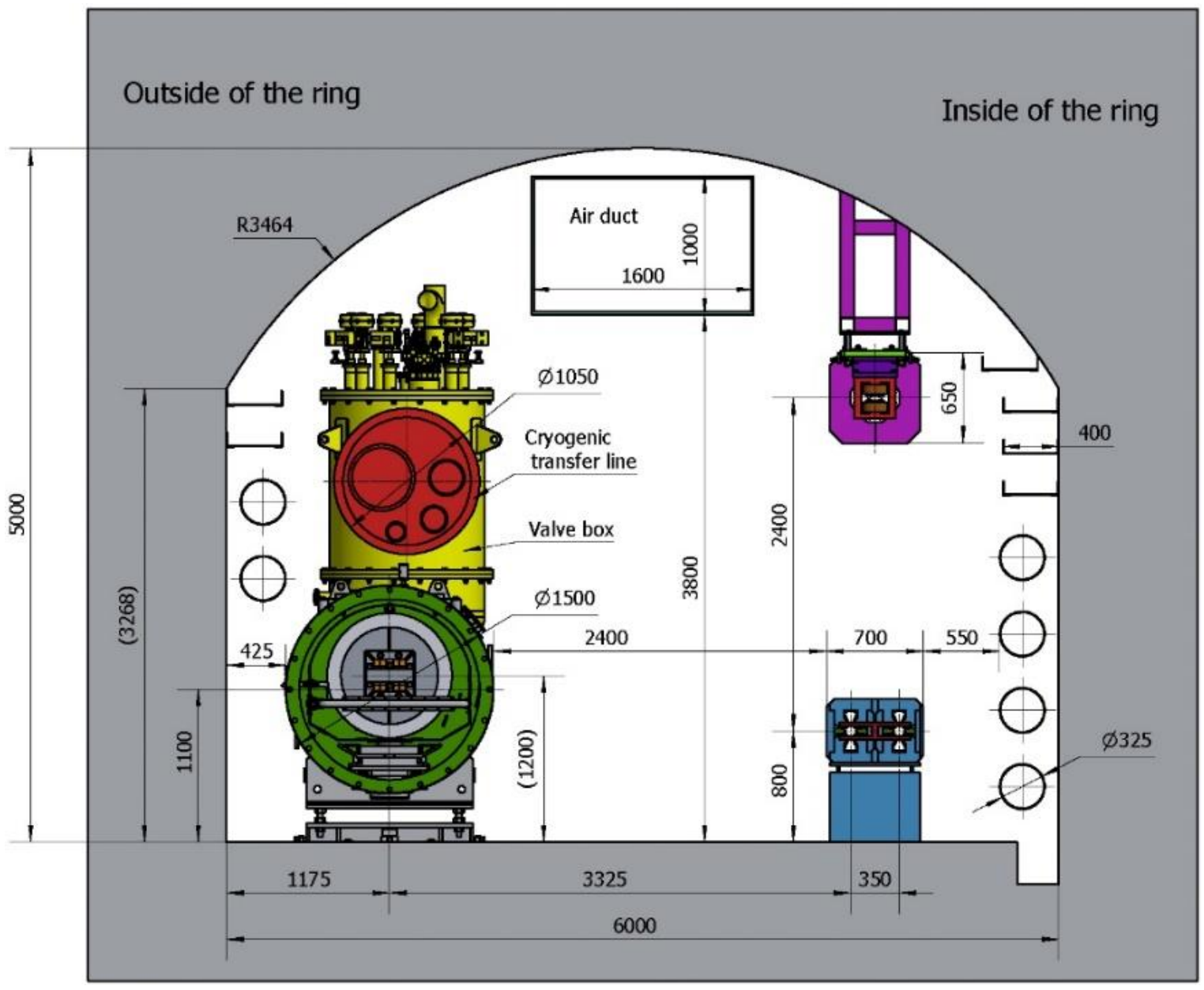

**Figure 2.1:** Tunnel cross section in the arc region

The layout of CEPC including the Linac, the transfer lines, the Booster and Collider is shown in Fig. 2.2. The Linac is at ground level and is 1.2 km. in length. The Booster is underground at a depth of approximately 100 m. The Linac and Booster are connected by two transfer lines for $e^+$ and $e^-$ respectively. These lines have a slope of 1:10. There are 8 straight sections in the Collider: 2 interaction regions, 2 RF regions and 4 injection regions. Among them, two off-axis injection regions are for operation in the Higgs, W and Z modes, whereas the two on-axis injection regions are used only during operation in the Higgs mode.

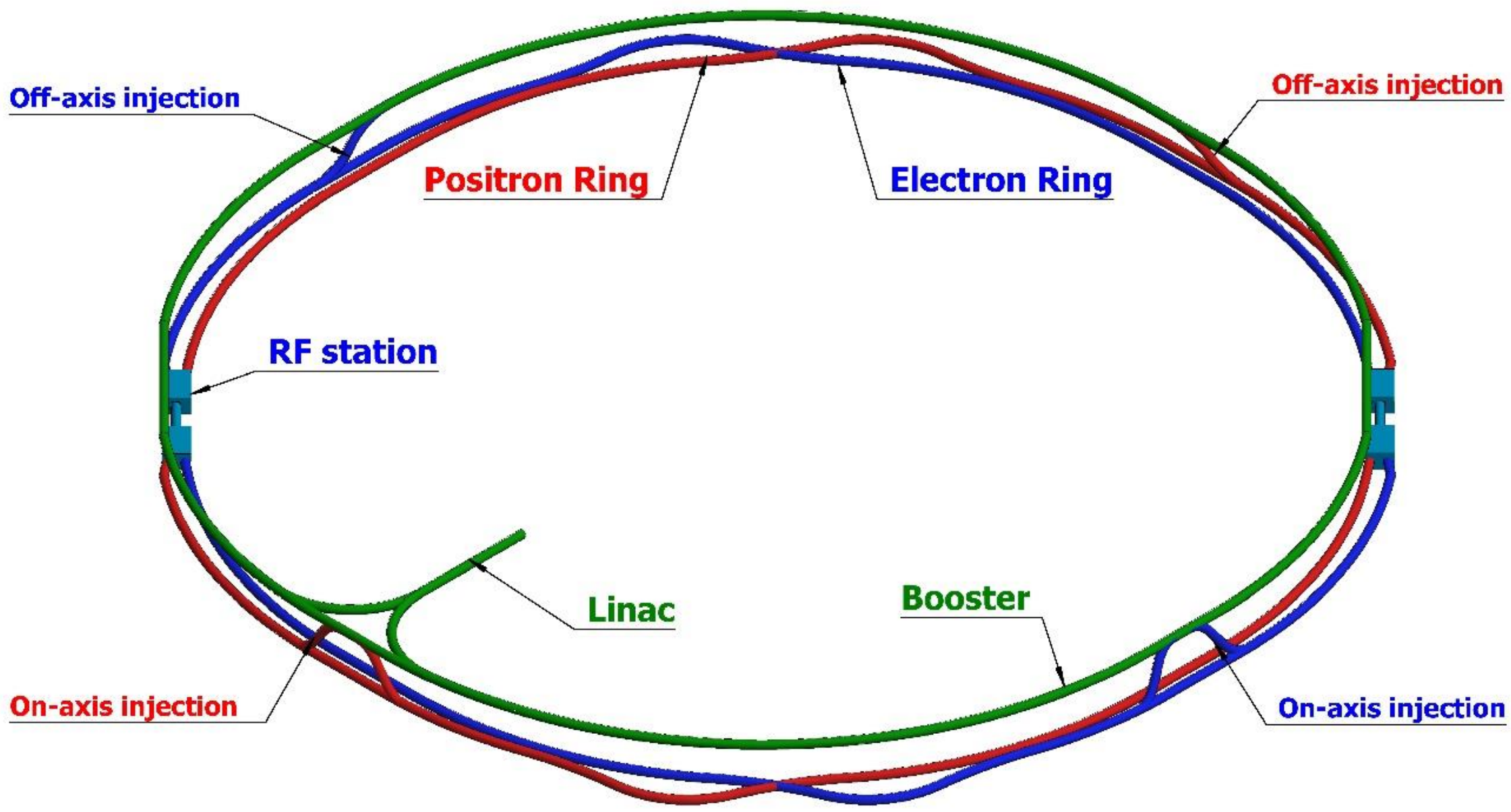

**Figure 2.2:** CEPC layout

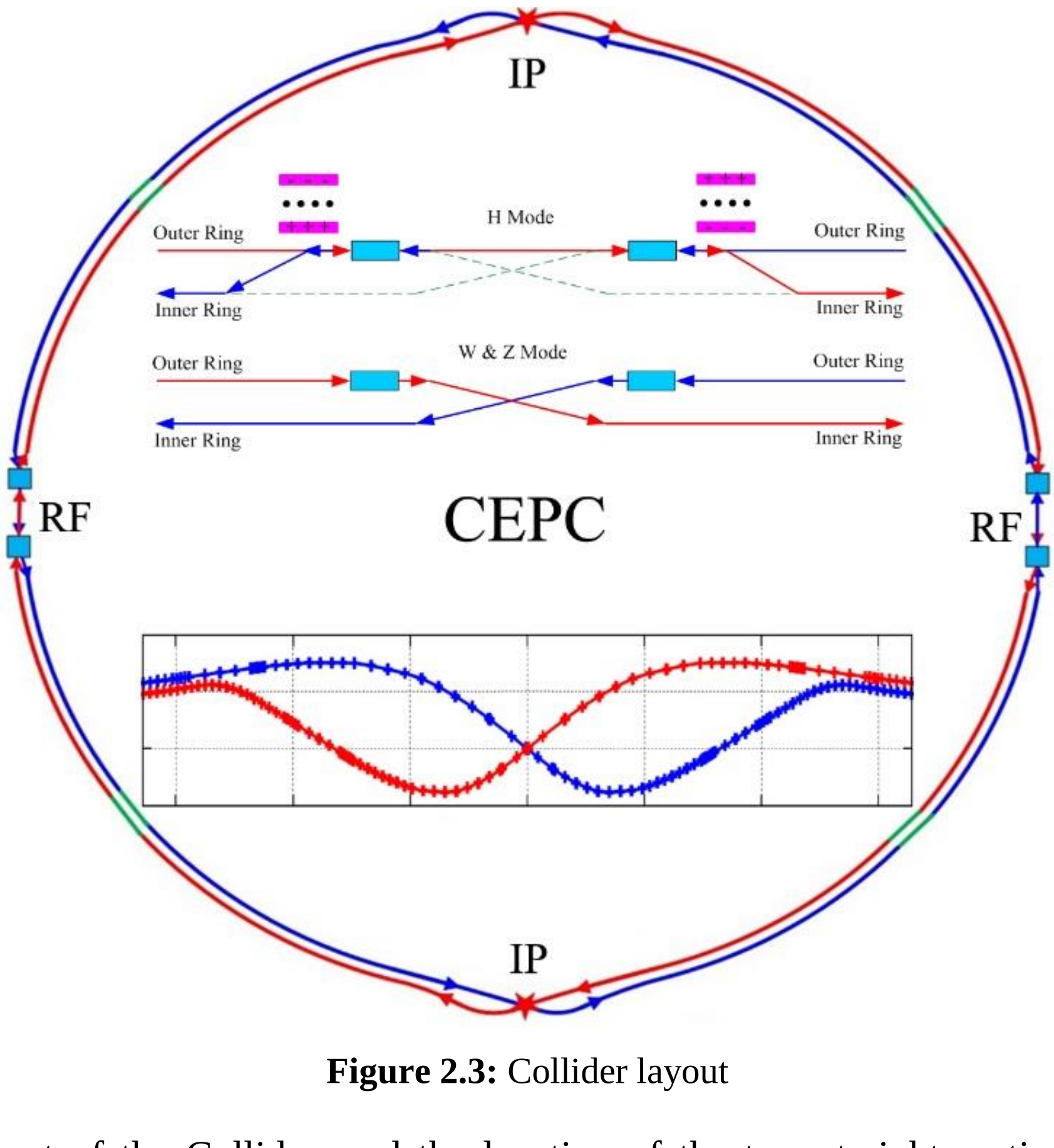

**Figure 2.3:** Collider layout

The layout of the Collider, and the location of the two straight sections used for physics, and the two straight sections used for the superconducting RF cavities is shown in more detail in Fig. 2.3. The Collider uses 650 MHz 2-cell cavities, described in more detail in Section 4.3.1.

During the operation in Higgs mode all the RF cavities are shared by both $e^+$ and $e^-$ beams using combining magnets near the cavities. Each beam fills half the ring so that all $e^+$ and $e^-$ bunches pass the RF cavities in turn. This filling scheme in Higgs mode with half the ring won't affect the luminosity because the required bunch number is relatively small and the bunch spacing is quite large.

For the W and Z modes the combining magnets in the RF region are turned off so that all bunches can be filled around the entire $e^+$ and $e^-$ rings. The beam current during operation in W and Z modes is made as high as possible to improve the luminosity. W and Z modes use the same RF cavities which are used in Higgs mode to save cost. Half the number of the cavities are used for the W and Z modes. The machine parameters for the Z mode do not increase the base budget which is based on operation as a Higgs factory.

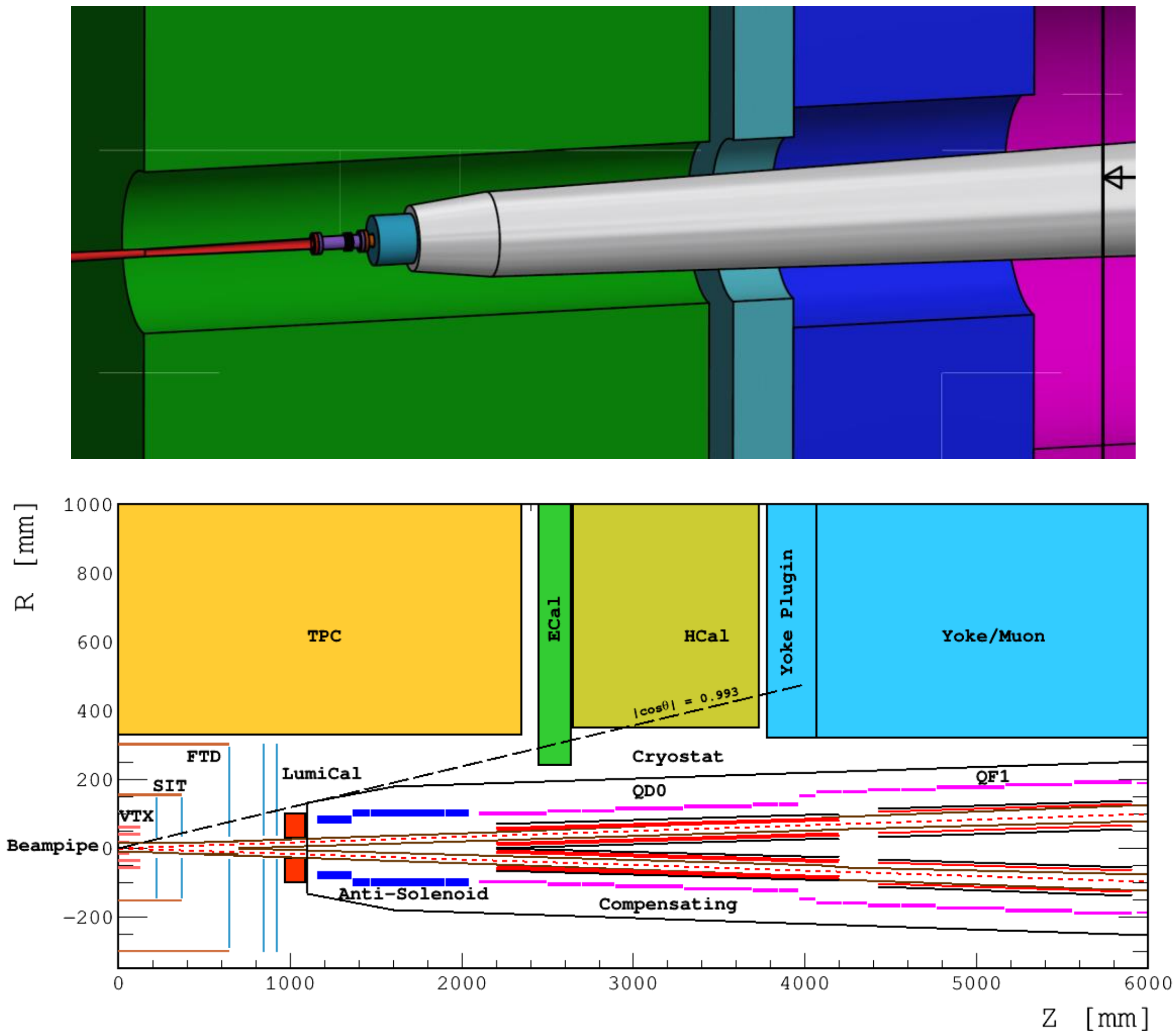

**Figure 2.4:** The central part of the interaction region

The central part of the interaction region is shown in Fig. 2.4. There is a Be pipe of length 14 cm and inner diameter 28 mm. The final focusing magnet is 2.2 m away from the IP. The horizontal crossing angle at the IP is 33 mrad to allow enough space for the superconducting quadrupole coils in a two-in-one type with space for a room temperature vacuum chamber. The detector consists of a cylindrical drift chamber surrounded by an electromagnetic calorimeter, which in turn is inside the superconducting solenoid. The accelerator components inside the detector are distributed within a conical space with an opening angle of 13.6˚. The luminosity calorimeter is located 0.95 ~ 1.11 m away from the IP and has inner radius 28.5 mm and outer radius 100 mm.

Twin-aperture dipoles and quadrupoles are in the arc region. The distance between the two beams is 0.35 m. The magnets in the other regions and all the sextupoles are independently powered for flexibility in the optics.

## 2.2 Machine Performance

The beam stay clear region is defined as ± (18 $\sigma_x$ + 3 mm.) and ± (22 $\sigma_y$ + 3 mm.) in the horizontal and vertical directions respectively. Coupling is 1%. The synchrotron radiation (SR) power per beam is limited to 30 MW. The high-energy physics goals of CEPC are to provide $e^+e^-$ collisions at a beam energy of 120 GeV and attain a luminosity of $3 \times 10^{34}$ cm$^{-2}$s$^{-1}$ at each IP for operation in the Higgs mode. Furthermore, the CEPC should be able to run at 80 GeV and 45.5 GeV for experiments running in the W and Z modes respectively. The luminosity in Z mode is $1.7 \times 10^{35}$ cm$^{-2}$s$^{-1}$ per IP, and in W mode $1 \times 10^{35}$ cm$^{-2}$s$^{-1}$ per IP

The detector solenoid is 3 T with a length of 7.6 m as the baseline design. There are 22 anti-solenoid sections with different inner diameters within the final doublet region at each side of the IP to compensate for the effects on the beam of the strong detector solenoid. For the Higgs mode, with the constraint of 30 MW, the design luminosity per IP is $3 \times 10^{34}$ cm$^{-2}$s$^{-1}$ with 242 bunches and beam current of 17.4 mA. The horizontal and vertical $\beta$ functions at the IP are 0.36 m and 1.5 mm respectively. Operation is in top-up mode. Beam lifetime is the key parameter to optimize by controlling the lost particles during data taking. The energy acceptance in Higgs mode is 1.35%. The beam lifetime with the beam-beam effect is greater than 26 minutes.

The Collider lattice in W mode is the same as in the Higgs mode. The design luminosity per IP is $1 \times 10^{35}$ cm$^{-2}$s$^{-1}$ with 1,524 bunches and beam current of 87.9 mA, again with the constraint of 30 MW beam power.

For the Z mode, the horizontal and vertical $\beta$ functions at the IP are 0.2 m and 1.5 mm respectively to avoid the strong coherent beam-beam instability with the detector solenoid at 3 T. The design luminosity per IP is $1.7 \times 10^{35}$ cm$^{-2}$s$^{-1}$ with 12,000 bunches and beam current of 461 mA. During operation in Z mode the synchrotron radiation power of each beam can only reach 16.5 MW due to the limitation of HOM heating in the RF cavities and the electron cloud instability. The coupling in Z mode is 2.2% which is much larger than expected because the solenoids' strong fringe field leads to a serious coupling growth of both beams. The coupling can be controlled and the vertical $\beta$ function at the IP can be reduced from 1.5 mm to 1.0 mm so that the luminosity in Z mode per IP can reach $3.2 \times 10^{35}$ cm$^{-2}$s$^{-1}$ with the same bunch number and beam current with the option of running the detector solenoid at 2 T.

The beam stay clear region for the Booster is defined as ± (4 $\sigma$ + 5 mm) in both horizontal and vertical directions with a round beam and emittance of 120 nm. This provides sufficient beam lifetime and transfer efficiency during injection and energy ramping. The diameter of the inner aperture of the vacuum chamber is chosen to be 55 mm from considerations of impedance.

The Booster uses 1.3 GHz 9-cell superconducting RF cavities. At the injection energy of 10 GeV from the Linac, the threshold of the single bunch current is 25.7 $\mu$A and the threshold of beam current limited by the coupled bunch instability is 127.5 mA. At the extraction energy of 120 GeV the single bunch current threshold is 300 μA and the threshold of beam current limited by RF power is 1.0 mA. The Linac bunches are injected into the Booster by horizontal on-axis injection at an energy of 10 GeV. At the extraction energies when operating in W and Z modes the circulating bunches of the Booster will be injected into the Collider by horizontal off-axis injection. However, in order to keep a sufficient margin in dynamic aperture, especially with machine errors included, at the extraction energy during Higgs mode operation a special on-axis injection scheme is used, which can significantly relax the requirements on dynamic aperture compared with conventional off-axis injection schemes. First, several circulating bunches from the Collider are extracted to the Booster while the energy is 120 GeV and the beam current limited to 1 mA. The Booster circulating bunches are then merged with the injected bunches from the Collider after 4 damping times. Then, the merged bunches in the Booster are injected back into the Collider by vertical on-axis injection. This procedure is repeated several times so that all the circulating bunches in the Booster can be accumulated into the Collider. The simulation result indicates that the collision of the stored bunches and the injected bunches is stable. The beam loading effect in the Booster RF system with the same bunch density as the Collider during the on-axis injection procedure in Higgs mode is weak. The maximum cavity voltage drop is 0.48% and the maximum phase shift is 0.63 degree. The peak HOM power per RF cavity is 62 W. which is acceptable for the Booster RF system. The dynamic aperture in the Booster is sufficient for vertical off-axis injection from the Collider. The injection duration of both beams during top-up operation are 35.4s, 45.8s and 275.2s for Higgs, W mode and Z mode respectively. The injection intervals with current decay of 3% are 47s, 153s and 504s for Higgs, W mode and Z mode based on beam lifetime. The injection duration from an empty ring are 0.17h, 0.25h and 2.2h for Higgs, W mode and Z mode respectively.

The requirements for injection efficiency are electron and positron bunch charge of 1.5 nC and repetition rate 100 Hz. The total beam transfer efficiency from transfer line to the injection point of the Collider is greater than 90% with beam emittance of 120 nm and energy spread of 0.2% at the exit of the Linac. The transfer efficiency can be made much higher with a damping ring of energy 1.1 GeV. The Linac beam emittance can be reduced to 40 nm. The Linac beam energy of 10 GeV requires the Booster injection field to be 30 Gauss. This is the minimum at which a good quality magnetic field can be obtained.

# 3 Operation Scenarios

The CEPC will operate in three different modes: H ($e^+e^- \to ZH$), Z ($e^+e^- \to Z$) and W ($e^+e^- \to W^+W^-$). The center-of-mass energies are 240, 91 and 160 GeV, and the luminosities are $3 \times 10^{34}$, $32 \times 10^{34}$ and $10 \times 10^{34}$ cm$^{-2}$s$^{-1}$, respectively, as shown in Table 3.1. The primary physics goal is to use the CEPC as a Higgs factory. Therefore, a tentative "7-2-1" operation plan is to run first as a Higgs factory for 7 years and create one million Higgs particles or more, followed by 2 years of operation as a Super Z factory and then 1 year as a W factory.

In order to make a realistic estimate of the integrated luminosity per year, we investigated the experience at 4 lepton colliders: LEP, KEKB, PEP-II and BEPC-II.

LEP (including both LEP1 and LEP2) operated from 1989 to 2000 for 12 years. From 1990-2000 the operation time ranged from 2,669 to 5,496 hours per year, and averaged 4,240 hours a year. The machine efficiency, the ratio of physics data taking time over the total operation time, ranged from 35% to 59%, with an average of 41% [1].

KEKB operated from 1998 to 2010. The statistics from 2000-2010 showed the average operation time was 5,060 hours a year and average efficiency 73% [2].

PEP-II operated from 1999 to 2008. The 2003-2008 statistics showed that in these last 6 years of running, the average operation time was 5,750 hours a year and the average efficiency 58% [3].

BEPC-II started operation in 2008 and is still running. Statistics from 3 recent years, 2015 to 2017, showed an average operation time of 7,140 hours per year and average efficiency of 67% [4]. This efficiency included BEPC-II operation for particle physics experiments as well as for synchrotron radiation light experiments.

Based on these statistics, the CEPC operation plan will be as follows:

- The CEPC will operate 8 months each year, equivalent to 250 days or 6,000 hours.
- The efficiency for physics data taking is assumed to be 60%, which reduces the above numbers to about 5 months, or 150 days, or 3,600 hours. It is equal to 1.3 Snowmass units. (Note: 1 Snowmass unit = $10^7$ seconds)

Table 3.1 lists the luminosity, integrated luminosity and total number of particles produced by the CEPC in 10 years.

**Table 3.1:** CEPC 10-year operation plan

| Particle | $E_{c.m.}$ (GeV) | $L$ per IP ($10^{34}$ cm$^{-2}$s$^{-1}$) | Integrated $L$ per year (ab$^{-1}$, 2 IPs) | Years | Total Integrated $L$ (ab$^{-1}$, 2 IPs) | Total no. of particles |
|---|---|---|---|---|---|---|
| H | 240 | 3 | 0.8 | 7 | 5.6 | $1 \times 10^6$ |
| Z | 91 | 32 (*) | 8 | 2 | 16 | $7 \times 10^{11}$ |
| $W^+W^-$ | 160 | 10 | 2.6 | 1 | 2.6 | $1.5 \times 10^7$ |

(*) Assuming detector solenoid field of 2 Tesla during Z operation

The CEPC, running at $E_{c.m.}$ = 240 GeV for 7 years, will produce one million Higgs bosons. This allows precision measurement of the Higgs couplings to the sub percent level, about an order of magnitude better than the precision achievable at the HL-LHC. Moreover, it allows for model independent determination of the Higgs width. In addition, it can also probe exotic Higgs decay branching ratios down to a level of $10^{-4}$, providing interesting probes to new physics such as Higgs portal dark matter.

The 2-year Z run at the CEPC will produce $7 \times 10^{11}$ Z, 40,000 times more than that generated during the entire LEP operation period. This is sufficient for a full compliment of electroweak precision measurements. In addition, this provides an excellent opportunity to search for new physics. It is sensitive to exotic Z decay branch ratios down to a level of $10^{-10}$, which will allow probing a variety of new physics issues including the origin of neutrino mass and the dark sector. It can produce about $10^{11}$ $b\bar{b}$ and $2 \times 10^{10}$ $\tau\bar{\tau}$, with a great potential in the study of flavor physics.

To fully realize the power of the electroweak precision measurements, it is necessary to measure the W mass to a precision of 1 MeV. In order to achieve this goal, we plan to scan the W threshold with the following runs as shown in Table 3.2:

**Table 3.2:** CEPC operation plan for WW pairs

| $E_{c.m.}$ (GeV) | Total Integrated $L$ ($ab^{-1}$, 2 IPs) | Cross section (pb) | Number of WW pairs ($\times 10^6$) |
|---|---|---|---|
| 157.5 | 0.5 | 1.25 | 0.6 |
| 161.5 | 0.2 | 3.89 | 0.8 |
| 162.5 | 1.3 | 5.02 | 6.5 |
| 172.0 | 0.6 | 12.2 | 7.3 |

The scan would also measure the W width to a precision of about 3 MeV. In addition, we plan to have a run further above the WW threshold at 172 GeV. This is important for precision QCD measurement. Taken together, 15 million WW pairs will be produced in a year, about 400 times the number produced at LEP.

This 10-year operation plan brings us to ~2040. It is expected that by then high field superconducting magnets made of iron-based high temperature superconductor (HTS) will have been developed, produced in industry and be ready for installation in the *Superconducting Proton Proton Collider*, SPPC (see Chapters 8 and 11.16).

The SPPC collider ring will be located in the same tunnel as the CEPC and placed against the outside wall. The tunnel is large enough to accommodate both the SPPC and CEPC collider rings. The two SPPC detectors will be located near the two CEPC RF straight sections. As the SPPC is on the outer side and the CEPC on the inner side, it will be easier to create the necessary SPPC bypass at the CEPC detector locations. Therefore, there will be the flexibility to operate the SPPC in *pp* or ion-ion mode with the CEPC in place, or vice versa, to operate the CEPC in *ee* mode with the SPPC in place.

Furthermore, it also opens the possibility of running both the CEPC and SPPC together for *e*-*p* and *e*-ion physics (see Appendix 7).

# 4 Collider

## 4.1 Main Parameters

### 4.1.1 Main Parameters

The CEPC parameters are listed in Table 4.1.1 based on the design method in [1].

The luminosities for Higgs and W operation are mainly limited by the SR power (30 MW). The luminosity at Higgs is $3 \times 10^{34}$ cm$^{-2}$s$^{-1}$ with 242 bunches; the luminosity at the W is $1 \times 10^{35}$ cm$^{-2}$s$^{-1}$ with 1524 bunches. At the Z pole, the luminosity for 3T detector solenoid is $1.7 \times 10^{35}$ cm$^{-2}$s$^{-1}$ and is $3.2 \times 10^{35}$ cm$^{-2}$s$^{-1}$ for 2T detector solenoid, both with 12,000 bunches. The limit of bunch number comes from the electron cloud instability of the positron beam. The minimum bunch separation for Z due to electron cloud effect is 25 ns and a 10% beam gap is left for cleaning.

The luminosities listed in Table 4.1.1 include the bunch lengthening effect determined from the impedance of whole ring and the beam-beam effect. Bunch lengthening is about 62% for Higgs, 98% for W and 250% for Z.

The beamstrahlung lifetime, which is the dominant contributor to the overall beam lifetime, is 80 minutes at Higgs energy. This is a calculation from simulation of the beam-beam effect. The quantum lifetime limited by transverse dynamic aperture including beam-beam effect and real lattice is also 80 minutes. These are calculation results from beam-beam simulation. The energy acceptance of the dynamic aperture needs to be larger than 1.35%. The energy spread due to beamstrahlung is about 34 percent of the natural energy spread.

The vertical emittance growth for Z mode is the most serious among the three operating energies due to the fringe field of the detector solenoid and anti-solenoids. A large coupling factor (2.2% (3T) / 0.9% (2T)) is chosen at Z pole; it is 0.2% for Higgs and 0.3% for W.

**Table 4.1.1:** CEPC parameters

| | Higgs | W | Z（3T） | Z（2T） |
|---|---|---|---|---|
| Number of IPs | 2 | | | |
| Beam energy (GeV) | 120 | 80 | 45.5 | |
| Circumference (km) | 100 | | | |
| Synchrotron radiation loss/turn (GeV) | 1.73 | 0.34 | 0.036 | |
| Crossing angle at IP (mrad) | 16.5 × 2 | | | |
| Piwinski angle | 3.48 | 7.0 | 23.8 | |
| Particles /bunch $N_e$ ($10^{10}$) | 15.0 | 12.0 | 8.0 | |
| Bunch number | 242 | 1524 | 12000 (10% gap) | |
| Bunch spacing (ns) | 680 | 210 | 25 | |
| Beam current (mA) | 17.4 | 87.9 | 461.0 | |
| Synch. radiation power (MW) | 30 | 30 | 16.5 | |
| Bending radius (km) | 10.7 | | | |
| Momentum compaction ($10^{-5}$) | 1.11 | | | |
| β function at IP $\beta_x^*$ / $\beta_y^*$ (m) | 0.36/0.0015 | 0.36/0.0015 | 0.2/0.0015 | 0.2/0.001 |
| Emittance x/y (nm) | 1.21/0.0024 | 0.54/0.0016 | 0.18/0.004 | 0.18/0.0016 |
| Beam size at IP $\sigma_x/\sigma_y$ (μm) | 20.9/0.06 | 13.9/0.049 | 6.0/0.078 | 6.0/0.04 |
| Beam-beam parameters $\xi_x/\xi_y$ | 0.018/0.109 | 0.013/0.123 | 0.004/0.06 | 0.004/0.079 |
| RF voltage $V_{RF}$ (GV) | 2.17 | 0.47 | 0.10 | |
| RF frequency $f_{RF}$ (MHz) | 650 | | | |
| Harmonic number | 216816 | | | |
| Natural bunch length $\sigma_z$ (mm) | 2.72 | 2.98 | 2.42 | |
| Bunch length $\sigma_z$ (mm) | 4.4 | 5.9 | 8.5 | |
| Damping time $\tau_x/\tau_y/\tau_E$ (ms) | 46.5/46.5/23.5 | 156.4/156.4/74.5 | 849.5/849.5/425.0 | |
| Natural Chromaticity | -468/-1161 | -468/-1161 | -491/-1161 | -513/-1594 |
| Betatron tune $\nu_x/\nu_y$ | 363.10 / 365.22 | | | |
| Synchrotron tune $\nu_s$ | 0.065 | 0.040 | 0.028 | |
| HOM power/cavity (2 cell) (kw) | 0.46 | 0.75 | 1.94 | |
| Natural energy spread (%) | 0.100 | 0.066 | 0.038 | |
| Energy spread (%) | 0.134 | 0.098 | 0.080 | |
| Energy acceptance requirement (%) | 1.35 | 0.90 | 0.49 | |
| Energy acceptance by RF (%) | 2.06 | 1.47 | 1.70 | |
| Photon number due to beamstrahlung | 0.082 | 0.050 | 0.023 | |
| Beamstruhlung lifetime /quantum lifetime† (min) | 80/80 | >400 | | |
| Lifetime (hour) | 0.43 | 1.4 | 4.6 | 2.5 |
| *F* (hour glass) | 0.89 | 0.94 | 0.99 | |
| Luminosity/IP ($10^{34}$ $cm^{-2}s^{-1}$) | 3 | 10 | 17 | 32 |

† include beam-beam simulation and real lattice

#### 4.1.1.1 *Constraints for Parameter Choices*

1) SR beam power
The maximum SR power for a single beam is limited to 30 MW in order to control the total AC power of the project. A future upgrade to 50 MW SR power/beam should be feasible.
2) Beam-beam tune shift limit
The beam-beam tune shift is a limit beyond which the beam emittance will blow up [2]. This is about 0.11 for Higgs, 0.13 for W and 0.096 for Z.
3) Beam lifetime due to beamstrahlung
Beam lifetime is mainly limited from the emission of single photons in the tail of the beamstrahlung spectra. At Higgs energy, it is about 80 minutes and the requirement for energy acceptance is 1.35% including errors and beam-beam nonlinearities.
4) Additional energy spread due to beamstrahlung
In order to maintain the uniformity of beam energy, the energy spread induced by beamstrahlung has to be controlled. This limit has been set to 40% of natural energy spread for Higgs.
5) HOM power per cavity
HOM power is a limit for the coaxial coupler. The bunch charge and total beam current has to be controlled. Otherwise the HOM power cannot be fully extracted and the HOM couplers will be damaged. The HOM power per 2 cell cavity has been constrained to be less than 2 KW.

#### 4.1.1.2 *Luminosity*

The luminosity for a circular e+e- collider is given by:

$$L_0[cm^{-2}s^{-1}] = 2.17\times10^{34}\left(1+r\right)\xi_y \frac{eE_0(GeV)N_bN_e}{T_0(s)\beta_y^*(cm)}F_h \tag{4.1.1}$$

where $N_e$ is the bunch population, $N_b$ is the bunch number, $T_0$ is the revolution time, $\beta_y$* is the vertical betatron function at the interaction point, *r* is the IP beam size ratio $\sigma_y^*/\sigma_x^*$ and $\xi_y$ is the vertical beam-beam tune shift. $F_h$ is the luminosity reduction factor due to the hourglass effect.

#### 4.1.1.3 *Crab Waist Scheme*

The crab waist scheme [3] increases luminosity. This scheme requires a large Piwinski angle defined as

$$\Phi = \frac{\sigma_z^*}{\sigma_x^*}tg\theta_h >> 1 \tag{4.1.2}$$

where $\theta_h$ is the half crossing angle. The Piwinski angle has to be larger than 1; a large Piwinski angle gives higher luminosity. The crossing angle is 33 mrad for CEPC. The Piwinski angle is 3.48 for Higgs, 7.0 for W and 23.8 for Z. Thus luminosity enhancement is about 100% for Higgs, 180% for W and 164% (240%) for Z.

The strong x and y betatron resonances are supressed by a pair of sextupoles placed on both sides of the IP in phase with the IP in the horizontal plane and $\pi/2$ phase shift vertically.

With large Piwinski angle the overlapping area of the colliding beams becomes much smaller than $\sigma_z$. An effective bunch length using the real overlap has been defined in order to assess the hour glass effect [1].

#### 4.1.1.4 ***Beam-beam Tune Shift***

The beam-beam tune shifts with large crossing angle are estimated by [4]

$$\xi_x = \frac{r_e N_e \beta_x^*}{2\pi\gamma\sigma_x^*\sqrt{1+\Phi^2}(\sigma_x^*\sqrt{1+\Phi^2}+\sigma_y^*)}, \quad \xi_y = \frac{r_e N_e \beta_y^*}{2\pi\gamma\sigma_y^*\left(\sigma_x^*\sqrt{1+\Phi^2}+\sigma_y^*\right)}. \tag{4.1.3}$$

The values of beam-beam tune shifts in table 4.1.1 are calculated by beam-beam simulations.

#### 4.1.1.5 ***RF Parameters***

The basic function of the RF system is to compensate for synchrotron radiation loss and provide the required bunch length and energy acceptance. The RF parameters of the collider ring are listed in Table 4.1.2. The superconducting RF cavity is chosen due to its high CW gradient, high energy efficiency and low impendence. The choice of 650 MHz RF frequency is a balance of beam stability and cost.

**Table 4.1.2:** Main RF parameters of the Collider Ring

| | **H** | **W** | **Z** |
|---|---|---|---|
| Beam Energy [GeV] | 120 | 80 | 45.5 |
| SR power / beam [MW] | 30 | 30 | 16.5 |
| Beam current / beam [mA] | 17.4 | 87.9 | 461 |
| Bunch charge [nC] | 24 | 19.2 | 12.8 |
| Bunch length [mm] | 3.26 | 5.9 | 8.5 |
| RF frequency [MHz] | 650 | | |
| RF voltage [GV] | 2.17 | 0.47 | 0.1 |
| Synchrotron phase [deg] | 37.1 | 43.7 | 68.9 |
| Cavity number | 240 | 216 | 120 |
| Cell number / cavity | 2 | | |
| Cryomodule number | 40 | 36 | 20 |
| Cavity operating gradient [MV/m] | 19.7 | 9.5 | 3.6 |
| $Q_0$ at operating gradient for long term | 1.5E10 | | |
| Total cavity wall loss @ 2 K [kW] | 6.1 | 1.3 | 0.1 |
| Input power / cavity [kW] | 250 | 278 | 275 |
| Klystron power [kW] | 800 | | |
| Klystron number | 120 | 108 | 60 |
| HOM power / cavity [kW] | 0.57 | 0.75 | 1.94 |

### 4.1.2 **Beam Lifetime**

#### 4.1.2.1 ***Beamstrahlung Lifetime***

If the beamstrahlung is so strong that particles' energy after collision is outside of the ring's energy acceptance, they may leave the beam, strike the vacuum chamber walls and hence decrease beam lifetime through the emission of single photons in the tail of the beamstrahlung spectra [5]. The analytic formulas from [6] are used to calculate the beamstrahlung lifetime and the according simulations have also been done. These two approaches give somewhat different results; from the analytical calculation about one hour for Higgs while simulation using the real lattice gives 80 minutes. For W, the beamstrahlung lifetime is longer than 400 minutes by simulation.

#### 4.1.2.2 ***Bhabha Lifetime***

The lifetime due to radiative Bhabha scattering is dominant and expressed by:

$$\tau_L = \frac{I}{eLn_{IP}\sigma_{ee} f_0} \tag{4.1.4}$$

From this, it can be seen that the lifetime is inversely proportional to the luminosity. So a balance between lifetime and luminosity needs to be made.

The cross section of the radiative Bhabha process can be calculated by both BBBrem and analytic formula [7]. In CEPC, it is $1.39\times10^{-25}$ cm$^2$ for Higgs, $1.41\times10^{-25}$ cm$^2$ for W and $1.5\times10^{-25}$ cm$^2$ (3T)/ $1.4\times10^{-25}$ cm$^2$ (2T) for Z. Therefore, the lifetime due to radiative Bhabha scattering is about 74 minutes for H, 1.8 hours for W and 5.4 (3T)/ 2.9 (2T) hours for Z.

#### 4.1.2.3 ***Touschek Lifetime***

Touschek lifetime is the result of single large-angle Coulomb scattering between particles in the beam, similar to the IBS process. Beam particles can transform their transverse momenta into longitudinal momenta randomly, which leads to a reduction of the beam lifetime when energy deviation of the particles exceeds the energy acceptance. The Touschek lifetime is about 695 hours for Higgs, 75 hours for W and about 33(3T)/21(2T) hours for Z.

#### 4.1.2.4 ***Quantum Lifetime***

Particle losses occur for Gaussian particle distributions due to finite transverse apertures or energy acceptance. Lifetimes due to these effects are

$$\tau_q = \frac{1}{2}\tau_u \frac{e^{\xi}}{\xi}, \quad \text{with} \quad \xi = \frac{A_u^2}{2\sigma_u^2} \tag{4.1.5}$$

where $u$=$x$, $y$ or $s$, $\sigma_u$ is the beam size for these three directions and $A_u$ is the limiting half apertures which are determined by the transverse dynamic aperture or the RF bucket height. Quantum lifetimes of CEPC are determined by the horizontal dynamic aperture, and are the order of $10^{30}$ hours for Higgs and $10^{340}$ hours for Z.

## 4.2 Collider Accelerator Physics

### 4.2.1 Optics

#### 4.2.1.1 *Optics Design*

Fig. 4.2.1.1 is a schematic of the 100-km circumference double-ring Collider. In the RF region, the RF cavities are shared by the two rings. Thus each beam will be only filled in half of the ring. The same RF cavities are used to realize the compatible optics for the H, W and Z modes.

The RF region layout is shown in Fig. 4.2.1.2. When operating in the W or Z modes the same cavities used in the H mode are used to save costs. Lower energy leads to a lower total RF voltage and thus only half of the cavities are used in the W and Z modes to lower the impedance. For the W and Z mode, the emittance is maintained by scaling down the magnet strength with energy. This fulfills the luminosity objective for running at the W- and Z-poles.

Twin-aperture dipoles and quadrupoles [1] are used in the arc region to reduce power. The two beams are separated by 35 cm. In other regions of the ring, independent magnets are used for flexibility. To provide enough space for nearby in two rings magnets a longitudinal separation of 30 cm is imposed.

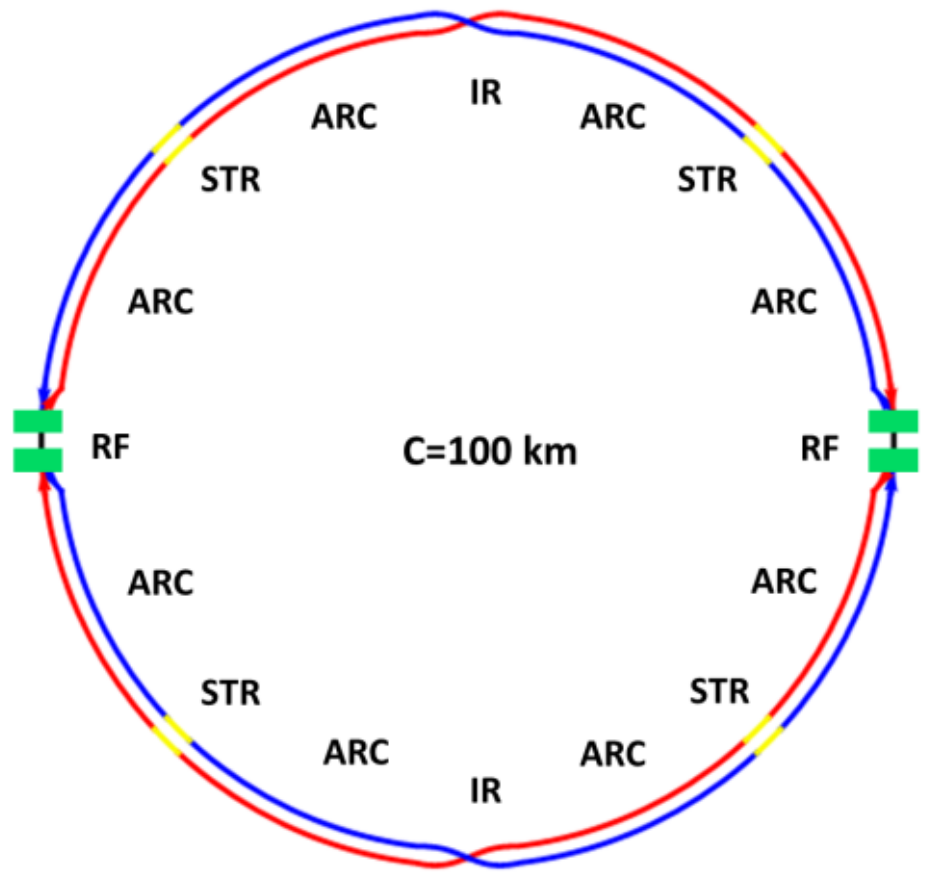

**Figure 4.2.1.1:** Collider layout.

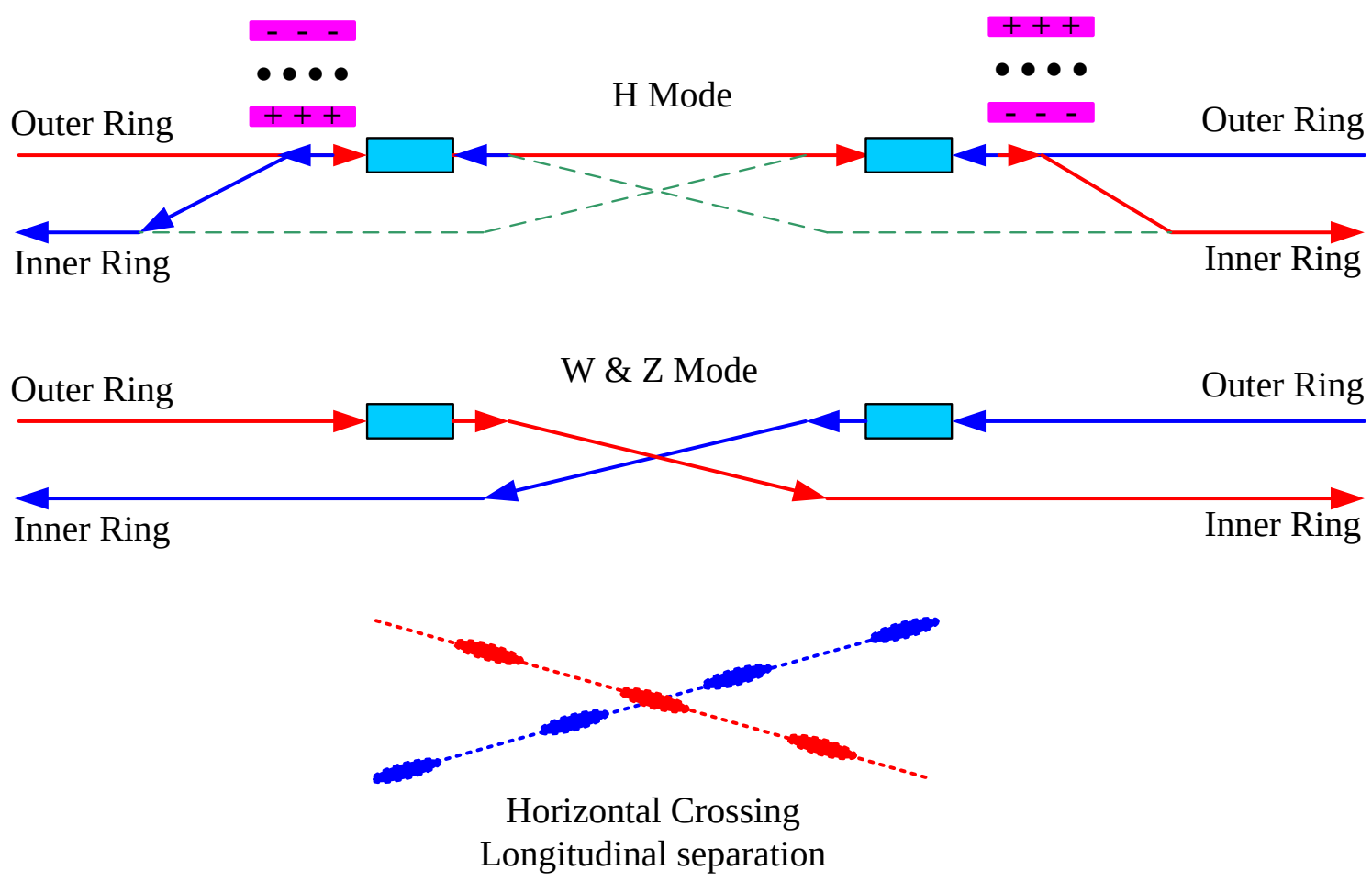

**Figure 4.2.1.2:** Layout of the RF region

#### *4.2.1.1.1 Interaction Region*

The interaction region is designed to provide local chromaticity correction generated by the final doublet magnets and the use of crab-waist [2] collisions. The region consists of modular sections including the final transformer (FT), chromaticity correction for vertical plane (CCY), chromaticity correction for horizontal plane (CCX), crab-waist section (CW) and matching transformer (MT) [3-8].

The FT consists of a quadrupole doublet, a quadrupole triplet and a weak dipole between them. The phase advance is π in the vertical plane and a bit less than π in the horizontal plane as very flat beam at IP. At the end of the FT there is the first image point and small dispersion for high order chromaticity correction.

The CCY consists of four FODO cells whose phase advances are π/2 for both planes. CCY begins with a half of a defocusing quadrupole. Four dipoles are used to make dispersion bumps. A pair of sextupoles is placed at the two beta peaks to compensate for the vertical chromaticity generated by the final defocusing quadrupole. The geometric sextupole aberrations are cancelled by the –I- transformation between the paired sextupoles. At the end of CCY, there's the second image point which is identical to the first one.

The CCX is similar to the CCY and begins with half of a focusing quadrupole.

The CW consists of two FODO cells. The crab-sextupole is placed at the beta peak. The phase advance from crab sextupole to the IP is 6π/6.5π for horizontal and vertical plane respectively.

The MT consists of 12 matching quadrupoles. With MT, The Twiss functions are matched to the arc section of the ring.

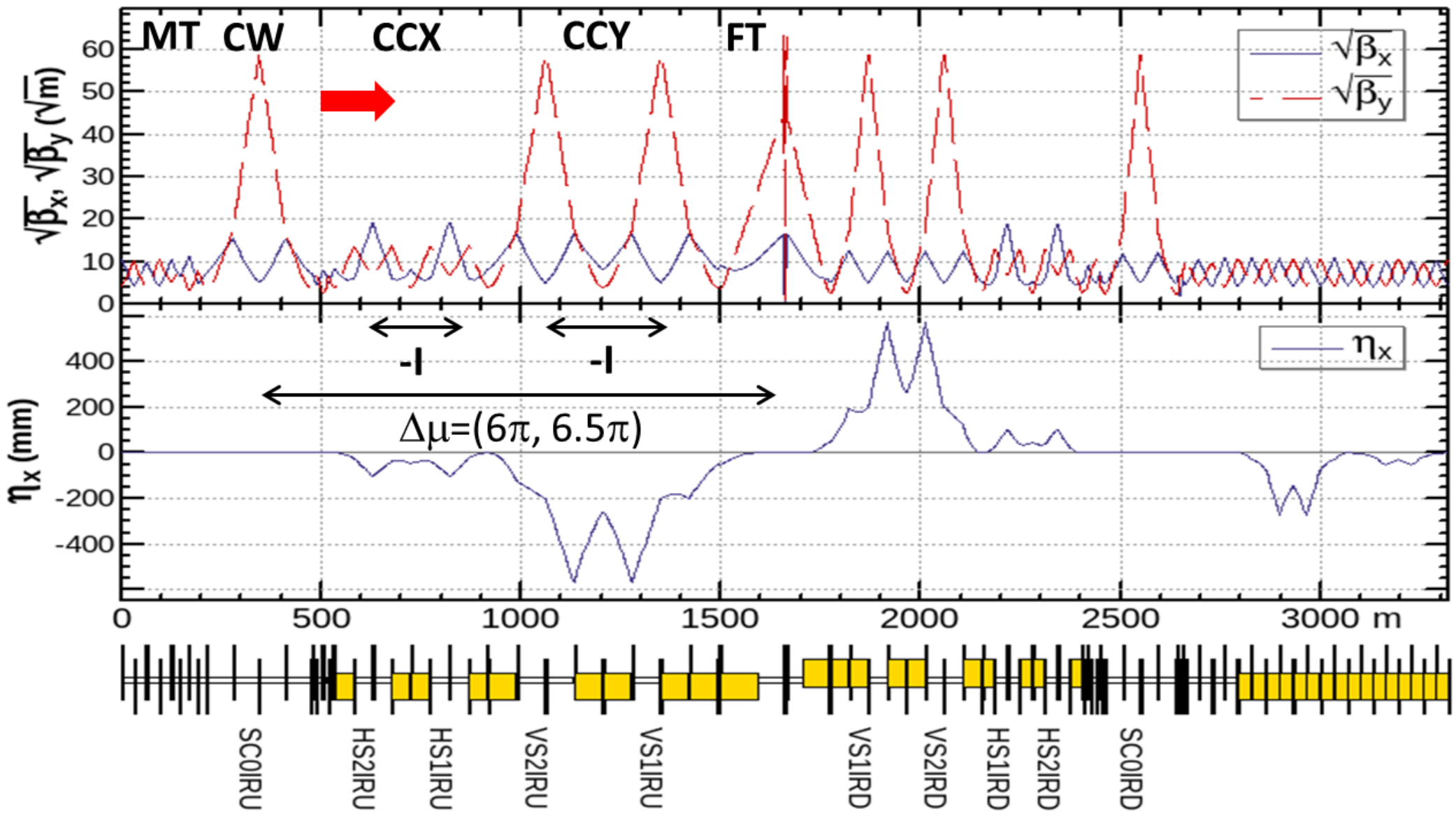

**Figure 4.2.1.3:** Optics of the interaction region for Higgs mode. ($L^*$=2.2m, $\theta_C$ =33mrad, $G_{QD0}$ =136T/m, $G_{QF1}$ =111T/m, $L_{QD0}$ =2.0m, $L_{QF1}$ =1.48m)

The second and third order chromaticity is corrected with phase tuning and additional sextupoles at the image points respectively. All the 3rd and 4th order resonance driving terms (RDT) of Lie operator due to sextupoles are almost cancelled [4, 6, 8]. The tune shift due to the finite length of the main sextupoles is corrected with additional weak sextupoles [9].

Fig. 4.2.1.3 and 4.2.1.4 show the lattice design and geometry for the interaction region, where the interaction point is located at the middle. An asymmetric lattice is adopted to allow softer bends in the upstream part of the IP [10]. Reverse bending in the last bends avoids synchrotron radiation hitting the IP. For the upstream portion of the IP, there are no bends in the last 70 m. The SR critical energy is less than 45 keV within 150 m and 120 keV within 400 m. For the downstream part of the IP, there are no bends in the last 50 m and the critical energy is less than 97 keV within 100 m and 300 keV within 250 m [11]. The maximum of the beamline distance is 9.7 m.

The IR optics for the W and Z modes is got by re-matching the IP parameters with MT.

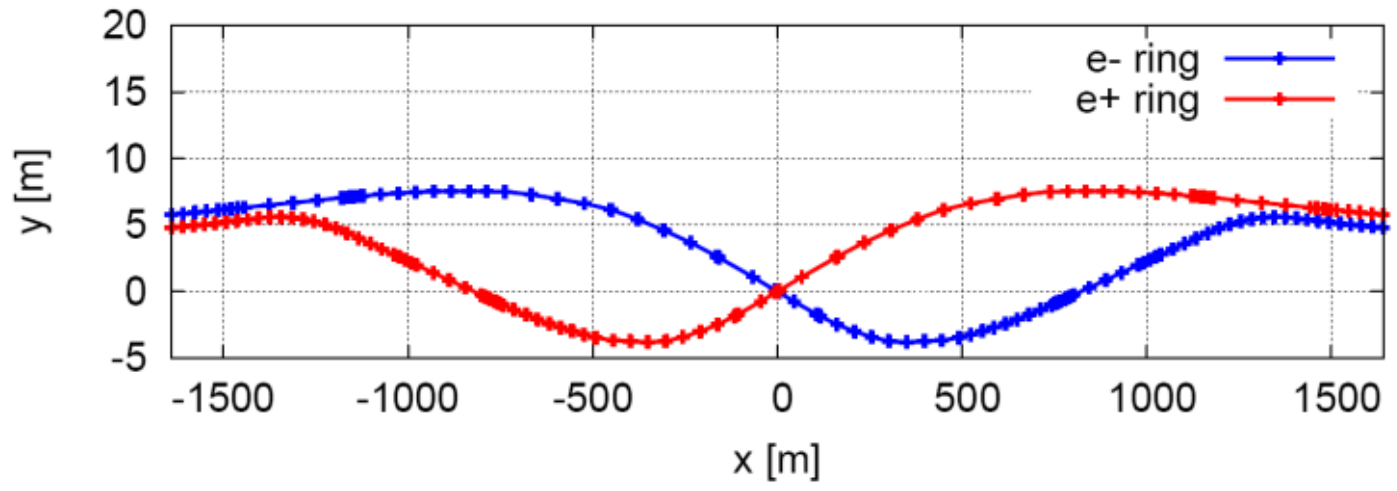

**Figure 4.2.1.4:** Geometry in the interaction region.

*4.2.1.1.2 Arc Region*

For the arc region, a FODO cell structure is chosen to provide a large filling factor of dipoles. The 90/90 degrees phase advances and non-interleaved sextupole scheme [12] is selected due to aberration cancellation. Fig. 4.2.1.5 shows the lattice optics of the arc region. The tune shift is very small even with small emittance; in each 20 cells, all the 3rd and 4th order RDT of Lie operator due to sextupoles are cancelled, except for small 4Qx, 2Qx+2Qy, 4Qy, 2Qx-2Qy [8]; the breakdown of the 3rd order RDT of Lie operator are cancelled [8] which shown in Fig. 4.2.1.6. This gives a good starting point for dynamic aperture optimization and helps reduce the number of sextupole families. The aberration that remains is mainly 2nd order chromaticity which is corrected with many families of arc sextupoles. The dispersion suppressor at the ends of the arc region is designed with a half-bending-angle FODO structure. The geometry is re-matched by adjusting the drift length.

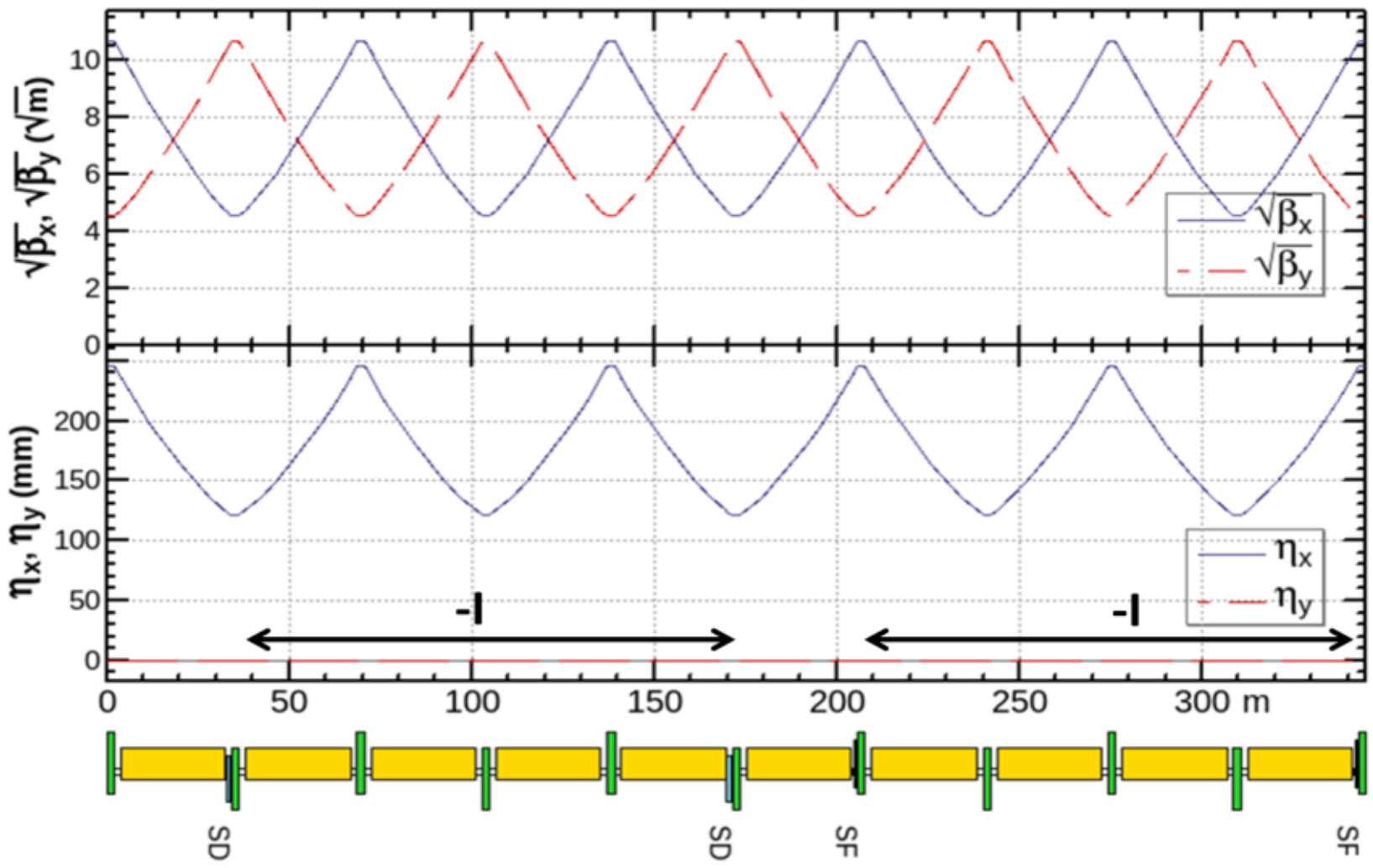

**Figure 4.2.1.5:** Optics in the arc region.

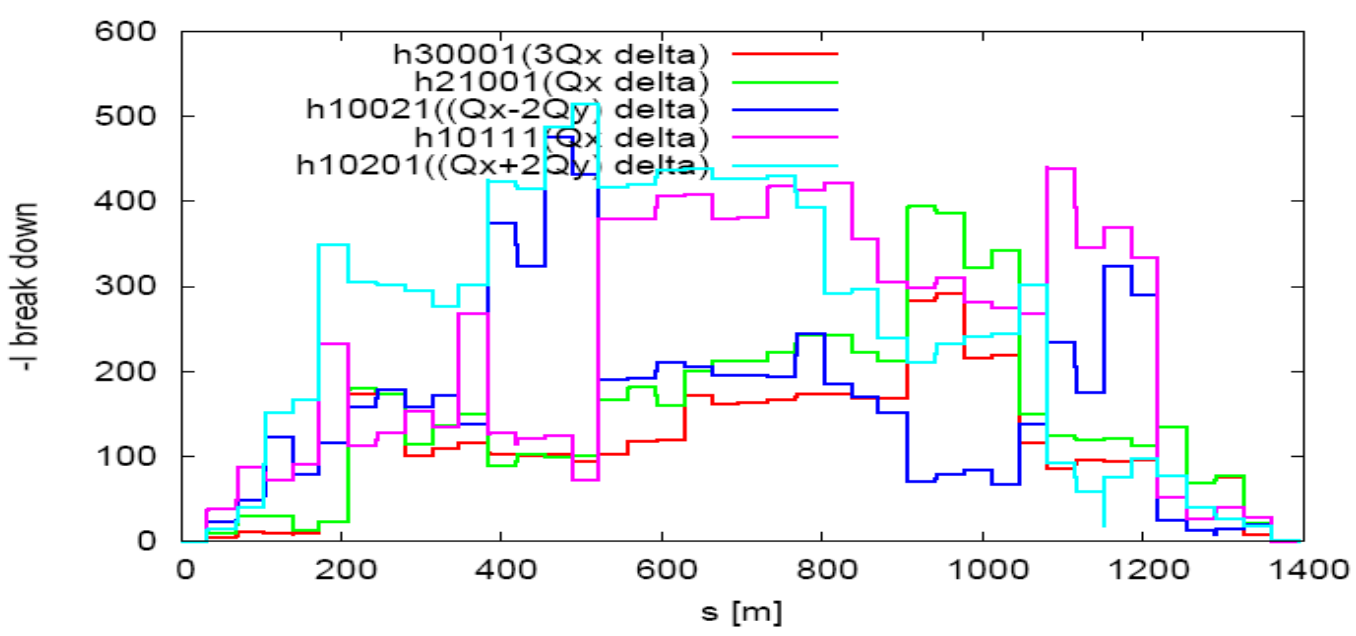

**Figure 4.2.1.6:** Optics aberrations in the arc region [8].

*4.2.1.1.3 RF Region*

In the RF region, the RF cavities are shared by the two rings. Each RF station is divided into two sections for bypassing half of the cavities when running in W or Z modes as shown above in Fig. 4.2.1.2. An electrostatic separator combined with a dipole magnet avoids bending the incoming beam [10] as shown in Fig. 4.2.1.7. The separator gradient is 2.0 MV/m and quantity is 10 with length of 4m for each one. The dipole field is 60 gauss. A 75-m drift creates a two-beam separation of 10 cm at the quadrupole entrance. In order to limit the beta functions, two triplets are used. Then the beam is further separated with dipoles. The deviation of the outgoing beam is 35 cm in the H mode and 1.0 m in the W and Z modes. This bypasses the cryo-modules whose radius is around 0.75 m. In the straight section where the cryo-modules are located, there is a small average beta function, favored for reducing the multi-bunch instability caused by the RF cavities. Thus a phase advance of 90/90 degree and a quadrupole distance of 13.7 m which accommodates a cryo-module have been chosen. The geometry of the RF region is shown in Fig. 4.2.1.8.

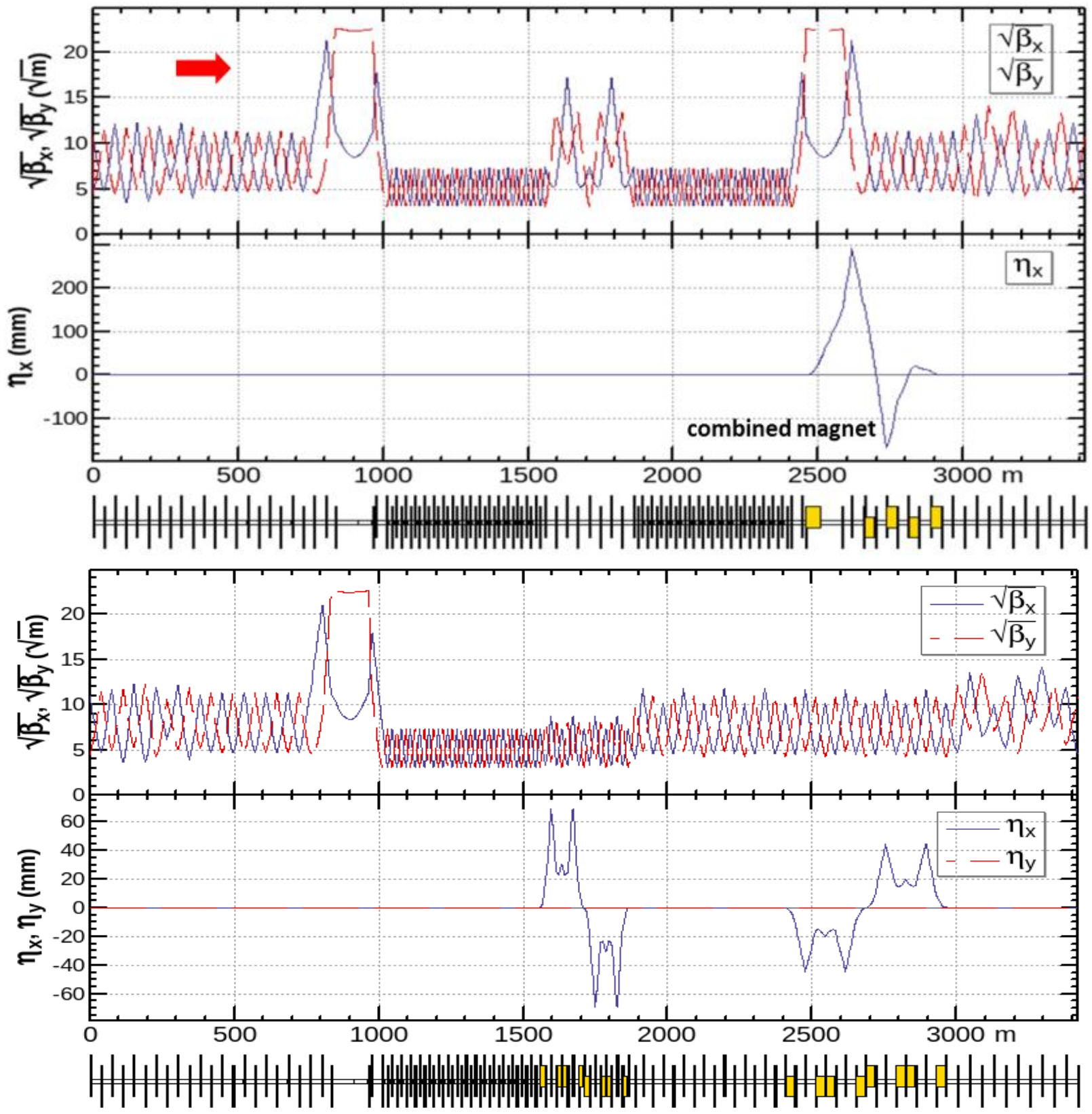

**Figure 4.2.1.7:** Optics of the RF region for Higgs (upper), W and Z (lower) modes.

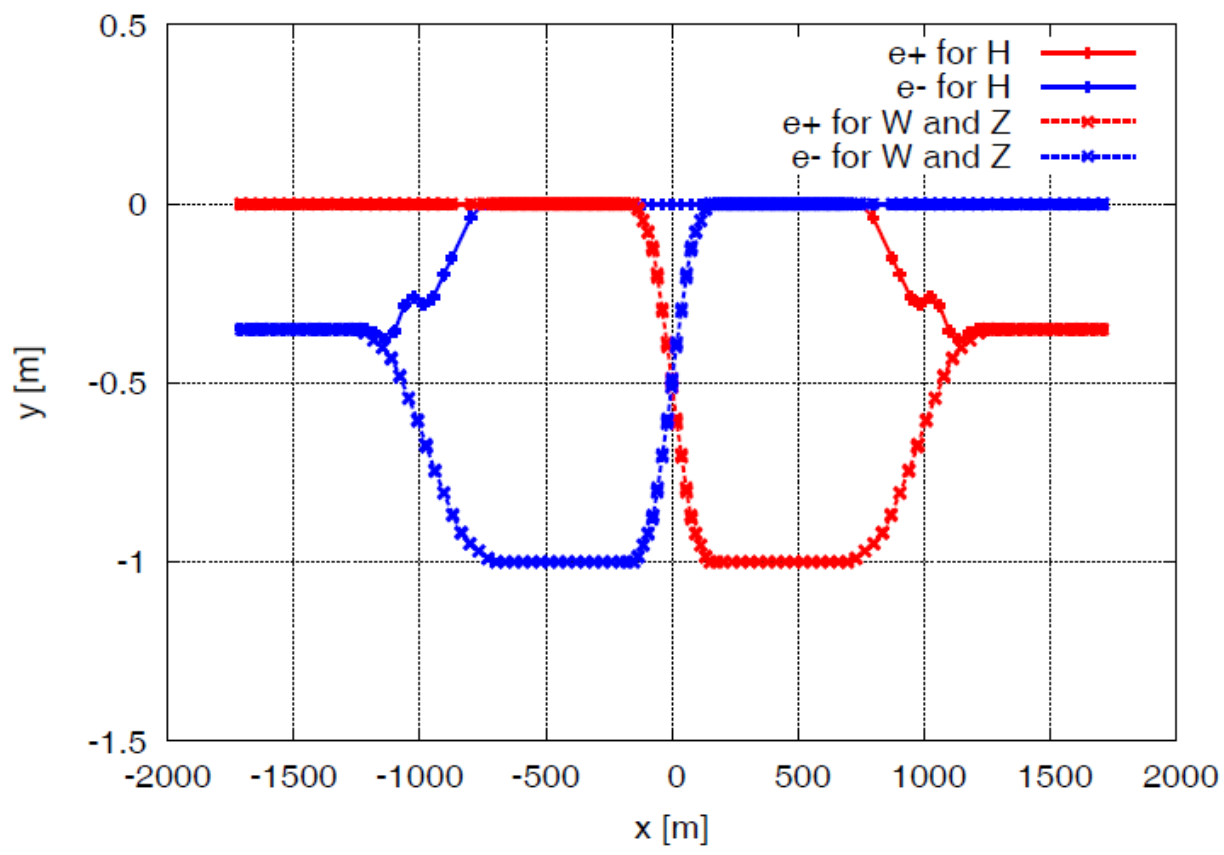

**Figure 4.2.1.8:** Geometry of the RF region.

*4.2.1.1.4 Straight Section Region*

The functions in the straight section are phase advance tuning and injection. The Fig. 4.2.1.9 shows the optics. For H mode, the on-axis injection scheme is adopted to reduce the requirement from injection while the off-axis injection scheme for W and Z modes. Independent magnets are used for the two rings and a longitudinal distance of 0.3 m between the two quadrupoles in the two rings allows a larger size of quadrupoles.

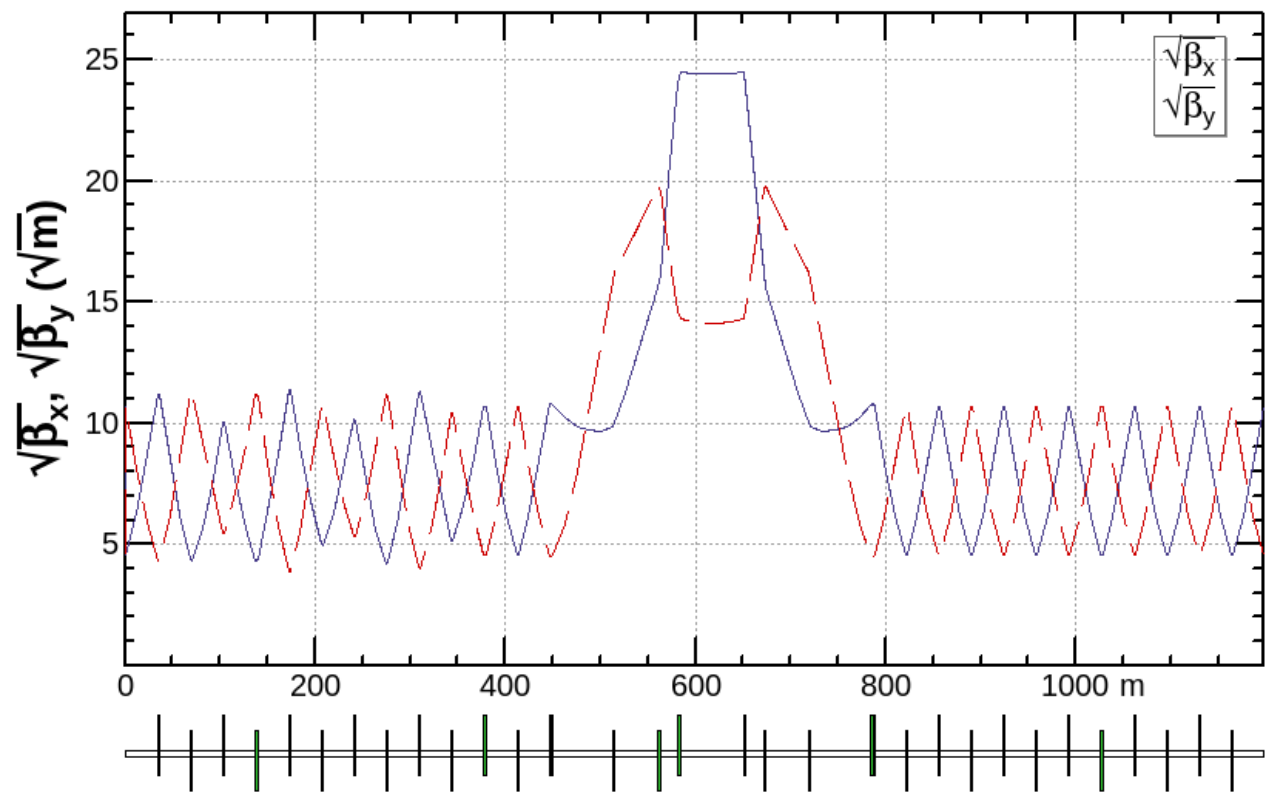

**Figure 4.2.1.9:** Optics in the straight section region for the off-axis injection.

Fig. 4.2.1.10 shows the optics of the whole collider. The lattice parameters of collider ring are listed in the Appendix 1.

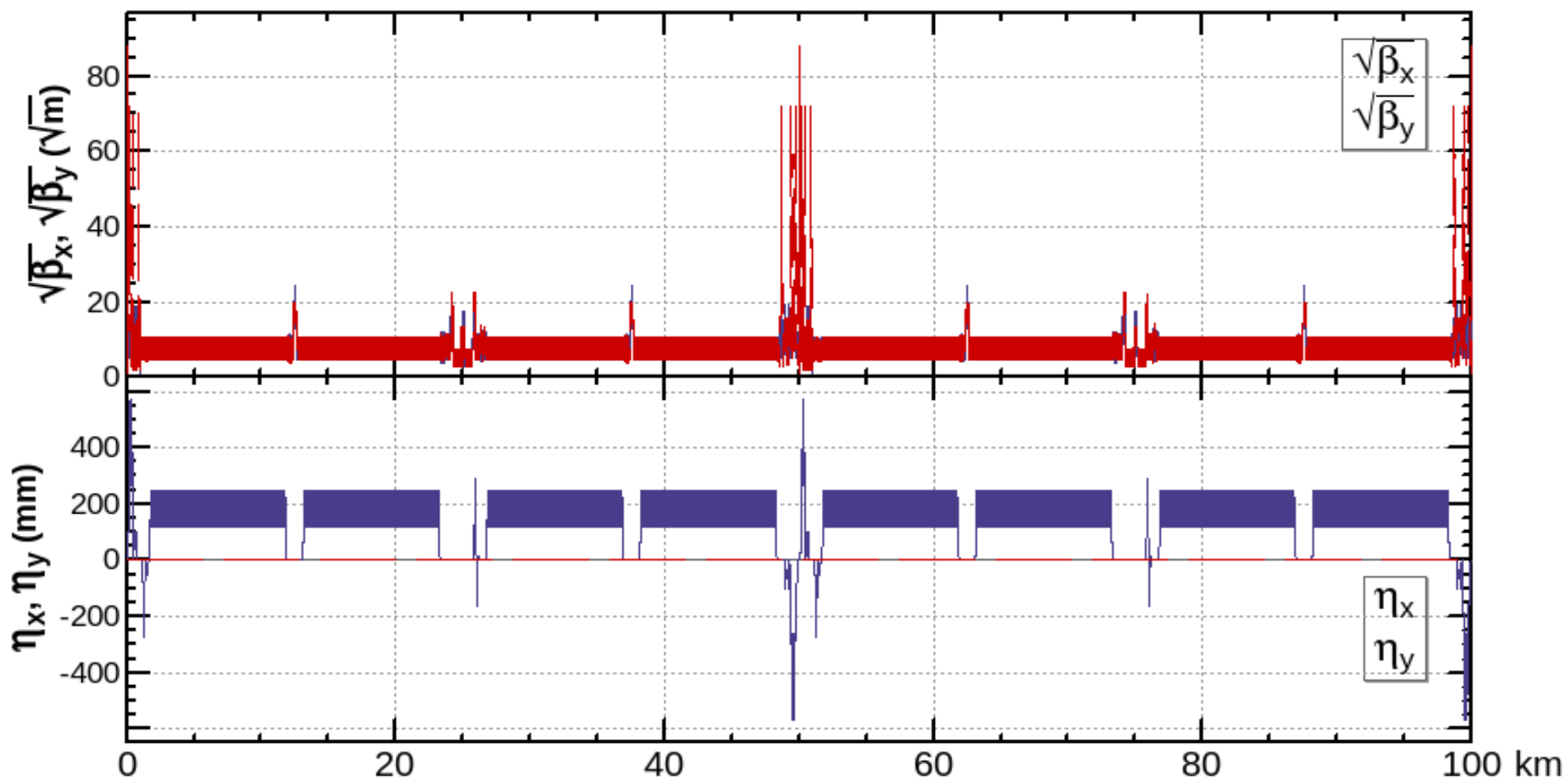

**Figure 4.2.1.10:** Collider optics for Higgs mode.

*4.2.1.1.5 Energy Sawtooth*

Synchrotron radiation has a pronounced effect on beam behaviour. With only two RF stations, the sawtooth orbit can be as large as 1 mm for H operation. Beams with this off-centre orbit will see extra fields in the magnets, which will result in ~5% distortion of the beam optics and DA reduction. The sawtooth effect is expected to be curable by tapering the magnet strength to take into account the beam energy at each magnet. This will cure beam optics problems and recover the DA. The sawtooth orbit becomes 1um after such tapering. The orbit and optics before and after tapering is shown in Fig. 4.2.1.11 (a) and (b).

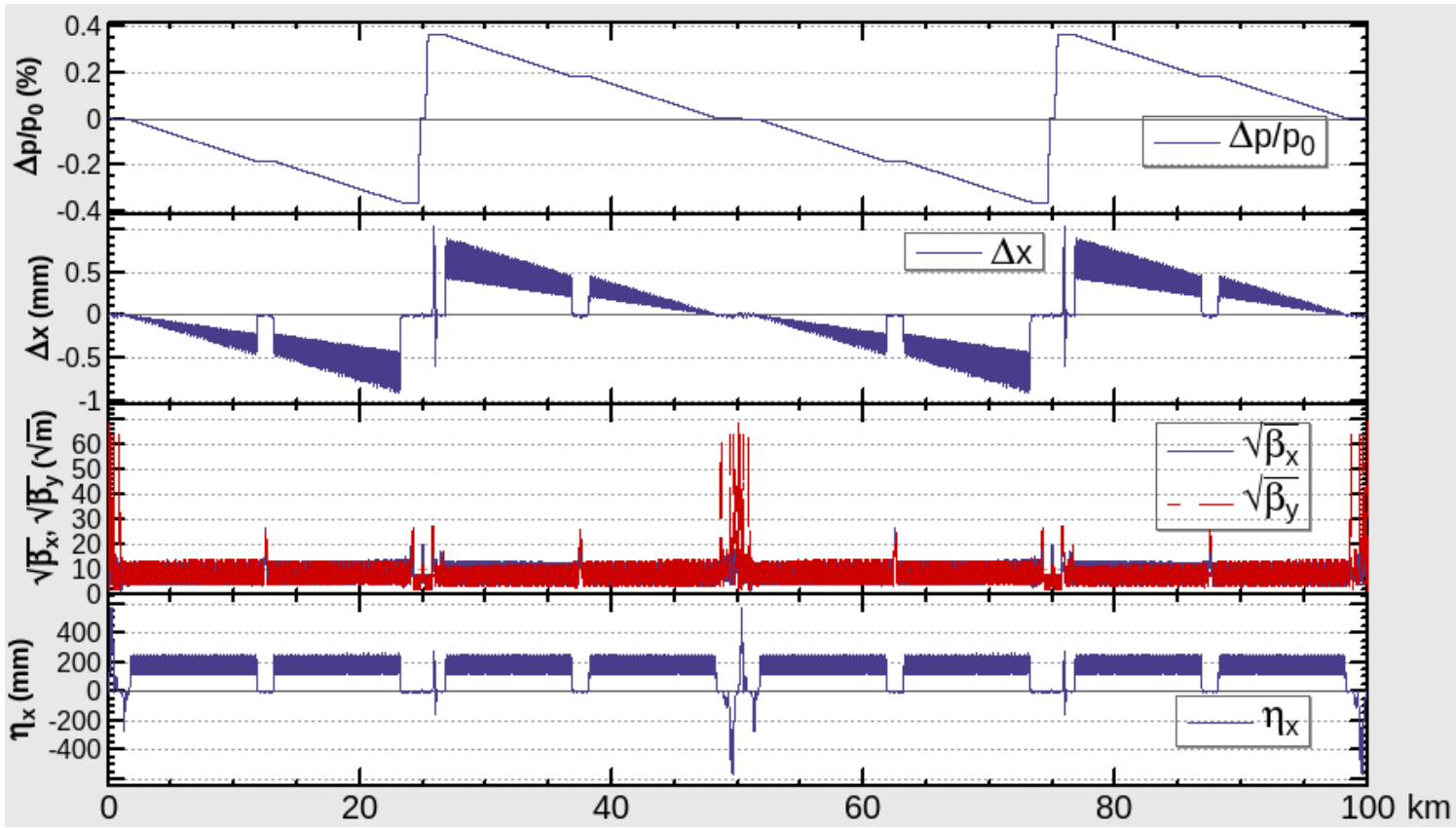

(a)

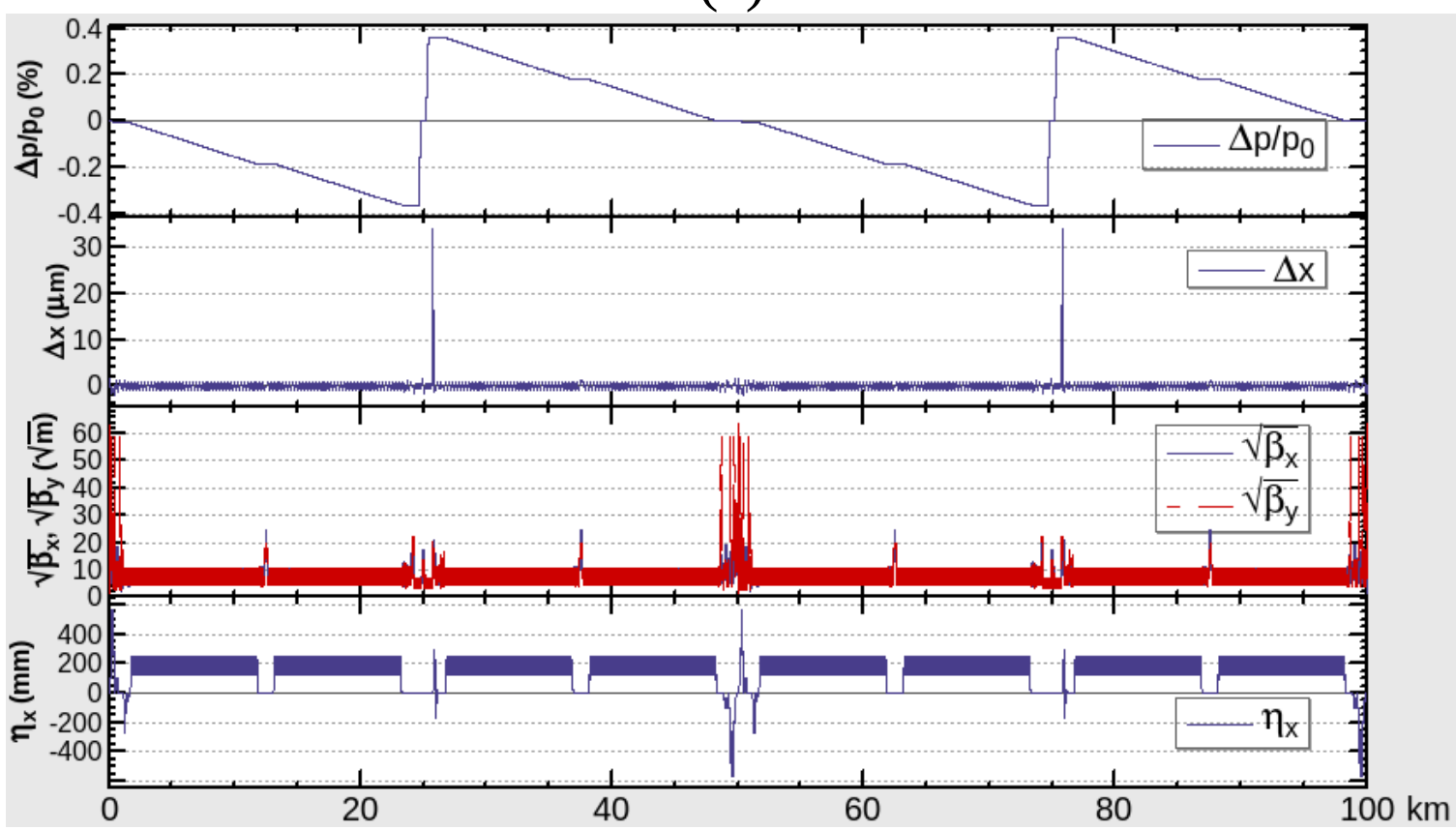

(b)

**Figure 4.2.1.11:** The orbit and optics before and after tapering the magnet strengths for the Higgs mode.

#### *4.2.1.1.6 Solenoid Coupling Effect*

Very small vertical emittance is a necessity to achieve high luminosity. The vertical emittance may come from the transverse coupling in the arc. The detector solenoid may also contribute to vertical emittance, since Bz is not constant in the z direction and there exists transverse fringe field.

The solenoid excites transverse coupling in the interaction region. The fringe field will cause a vertical closed orbit distortion and excite dispersion in the vertical direction, which contributes to an increase in the vertical emittance. The distortion in the optics for Higgs operation is shown in Fig. 4.2.1.12.

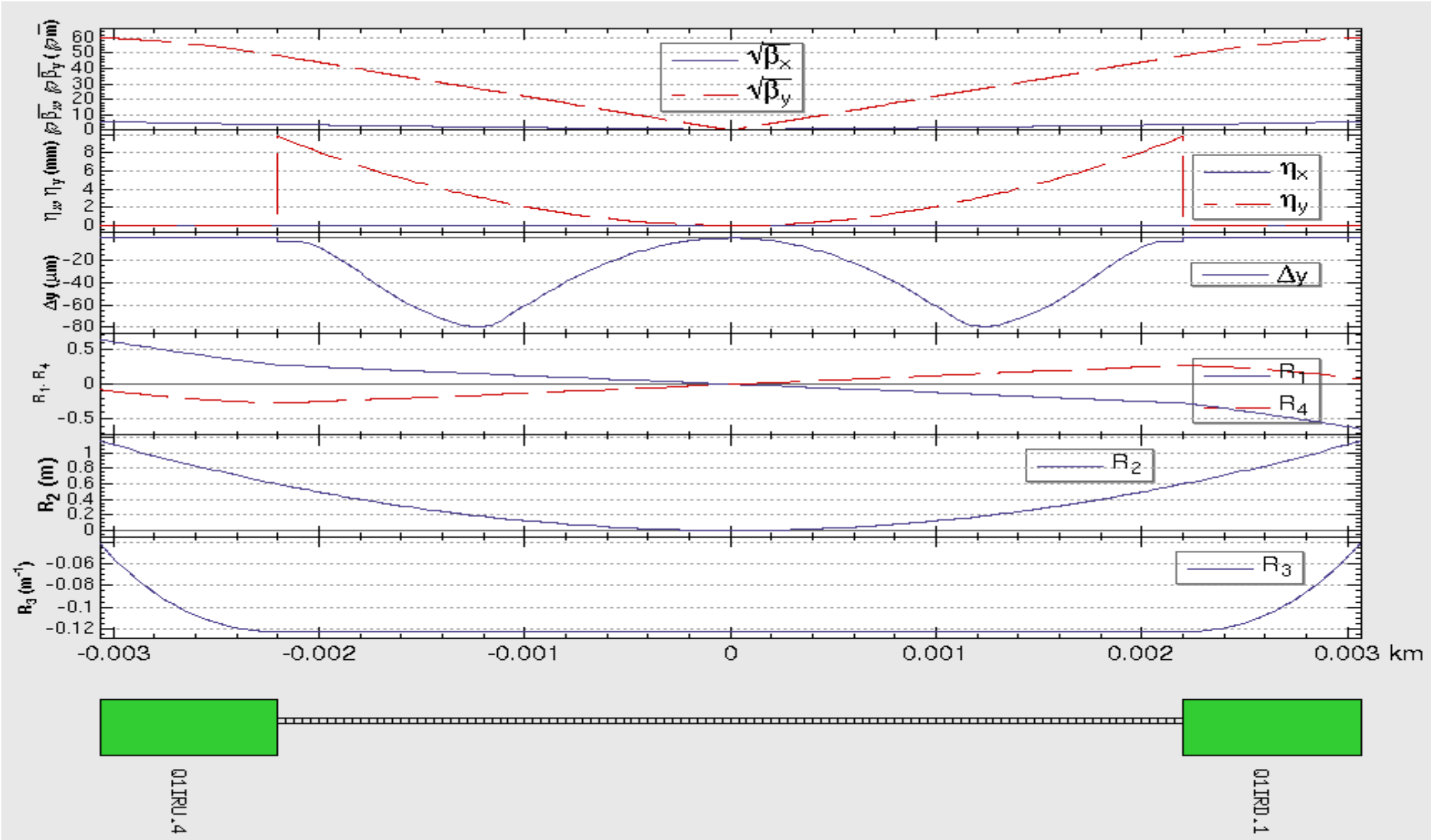

**Figure 4.2.1.12:** Optics near the IP with the detector solenoid

In order to minimize the vertical emittance, an anti-solenoid is installed as close to the IP as possible. The Bz field distribution is smooth along z which reduces the transverse field component. Sharp change in Bz should be avoided especially far from the IP, otherwise the bending will be stronger in the large beta region. The same solenoid configuration will be adopted for Higgs, W and Z operation. The vertical emittance excited by the solenoid, calculated with SAD is tabulated in Table 4.2.1.1 for the three operation modes and shows that the solenoid configuration is reasonable.

**Table 4.2.1.1:** The vertical emittance excited by solenoid

| | **Higgs** | **W** | **Z (3T)** | **Z(2T)** |
|---|---|---|---|---|
| Vertical emittance [pm·rad] | 0.16 | 0.53 | 2.9 | 0.45 |
| Normalized Vertical Emittance Budget | 6.7% | 33% | 71% | 28% |
| Emittance Coupling Budget | 0.2% | 0.3% | 2.2% | 0.89% |

### 4.2.1.2 *Dynamic Aperture*

The requirements of dynamic aperture from injection and beam-beam effect to get efficient injection and adequate beam life time are listed in the Tab. 4.2.1.2.

**Table 4.2.1.2:** The requirement on dynamic aperture

| | **Higgs** | **W** | **Z** |
|---|---|---|---|
| with on-axis injection | $8\sigma_x \times 15\sigma_y \times 1.35\%$ | - | - |
| with off-axis injection | $13\sigma_x \times 15\sigma_y \times 1.35\%$ | $15\sigma_x \times 9\sigma_y \times 0.9\%$ | $17\sigma_x \times 9\sigma_y \times 0.49\%$ |

A differential evolution algorithm based optimization code has been developed for CEPC, which is a multi-objective code called MODE [12]. The SAD code is used to do the optics calculation and dynamic aperture tracking.

Strong synchrotron radiation causes strong radiation damping which helps enlarge the dynamic aperture to some extent. The effect, shown in Fig. 4.2.1.13, is clear, especially for large off-momentum particles.

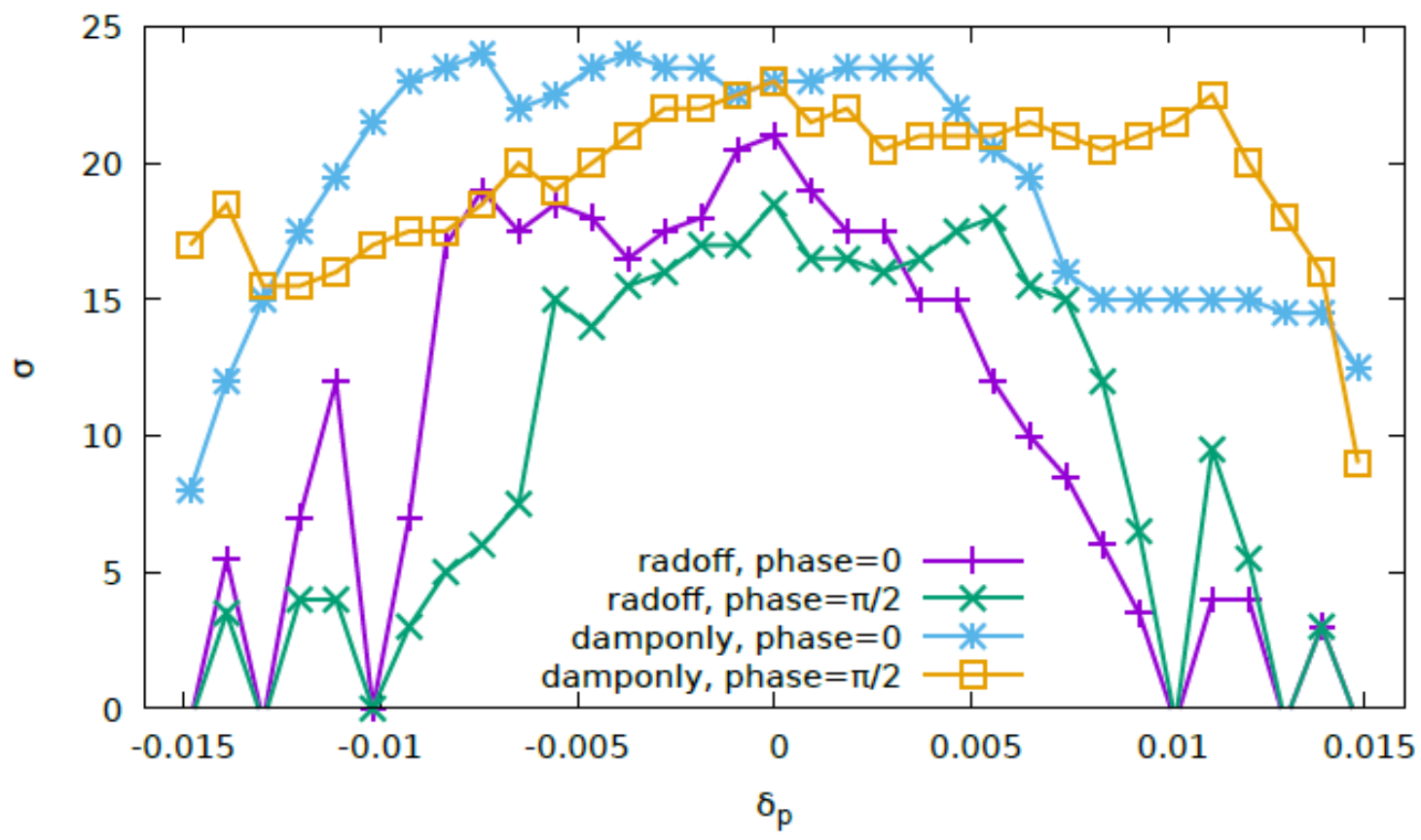

**Figure 4.2.1.13:** DA with damping at each element and with radiation off. (0.3% emittance coupling is assumed).

Quantum fluctuations in the synchrotron radiation are considered in SAD, where the random diffusion due to synchrotron radiation in the particle tracking is implemented in each magnet. The dynamic aperture for the same lattice with and without radiation fluctuations is shown in Figs. 4.2.1.14 and 4.2.1.15, which shows that the fluctuation mainly comes from the vertical direction. The difference mainly comes from the radiation in the final focus quadrupoles in the interaction region. The physics has been studied in [13].

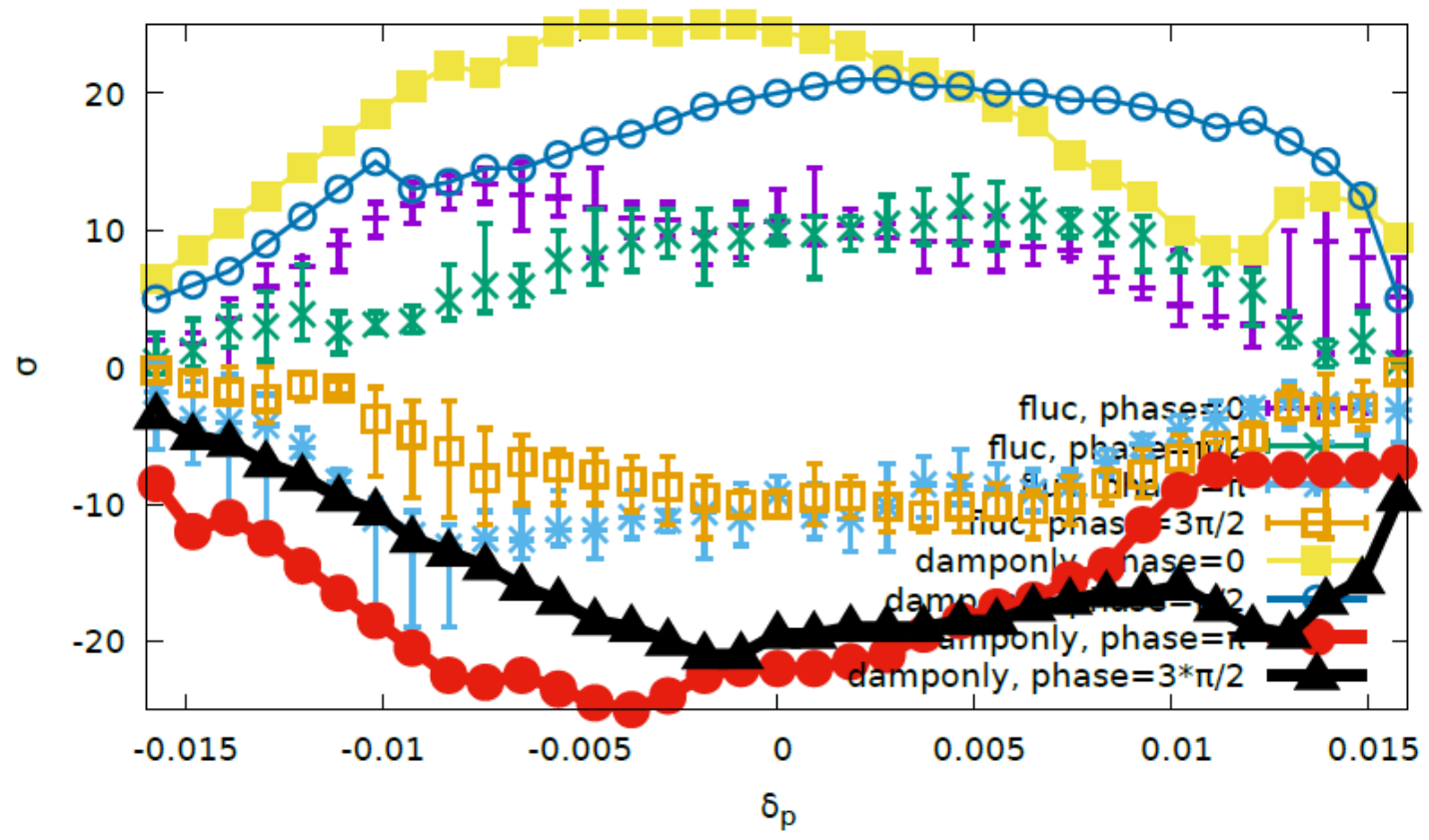

**Figure 4.2.1.14:** DA with damping at each element and radiation fluctuation at each element. The error bar shows the maximum and minimum value of 10 samples. 200 turns is tracked. 0.3% emittance coupling is assumed.

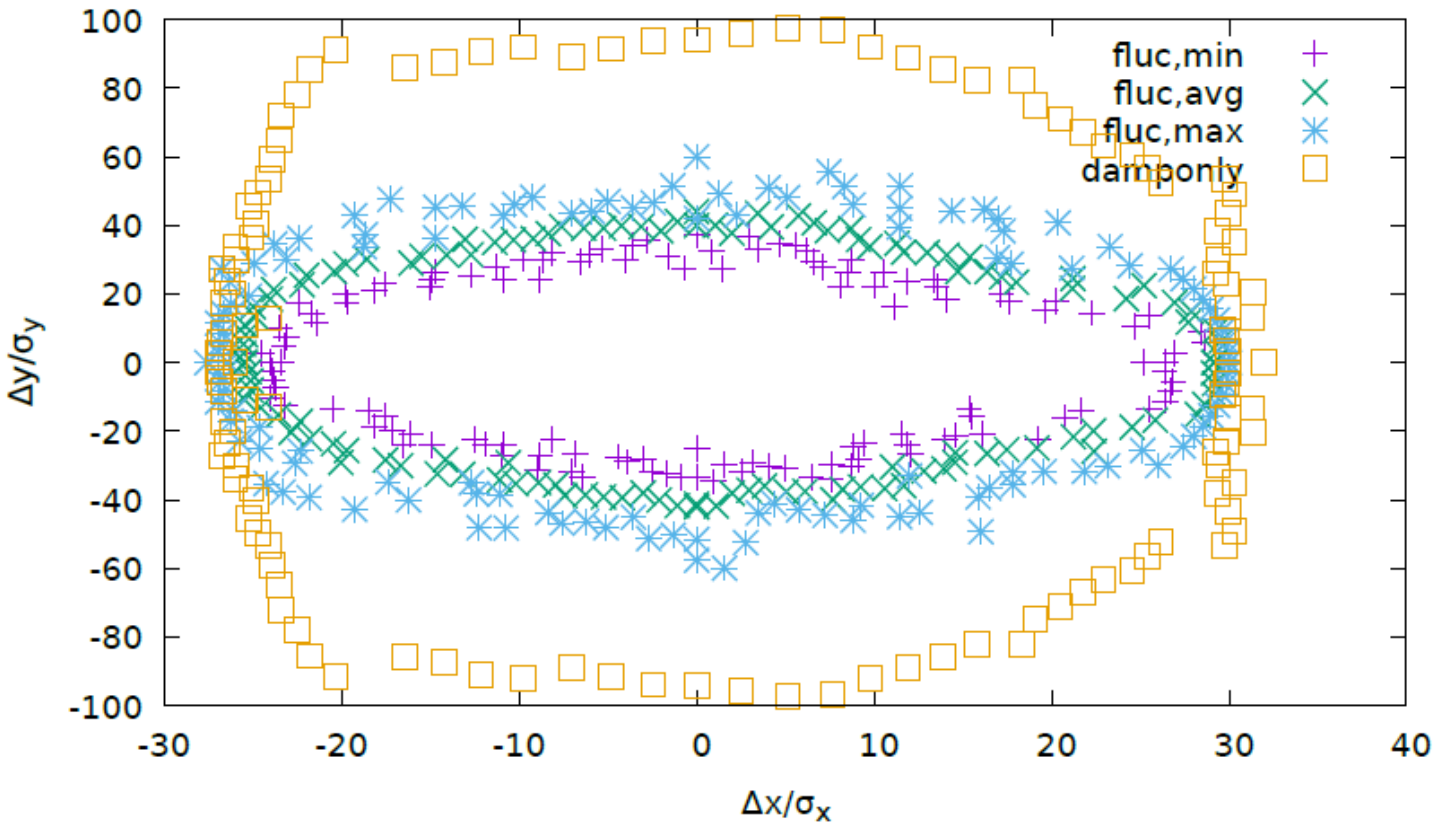

**Figure 4.2.1.15:** On-Momentum DA with damping at each element and radiation fluctuation at each element.

The dynamic aperture is tracked including radiation fluctuation during optimization. In order to reduce the random noise, the DA result is clipped to ensure the DA at large momentum deviation will be less than that at small deviation, as shown in Fig. 4.2.1.16.

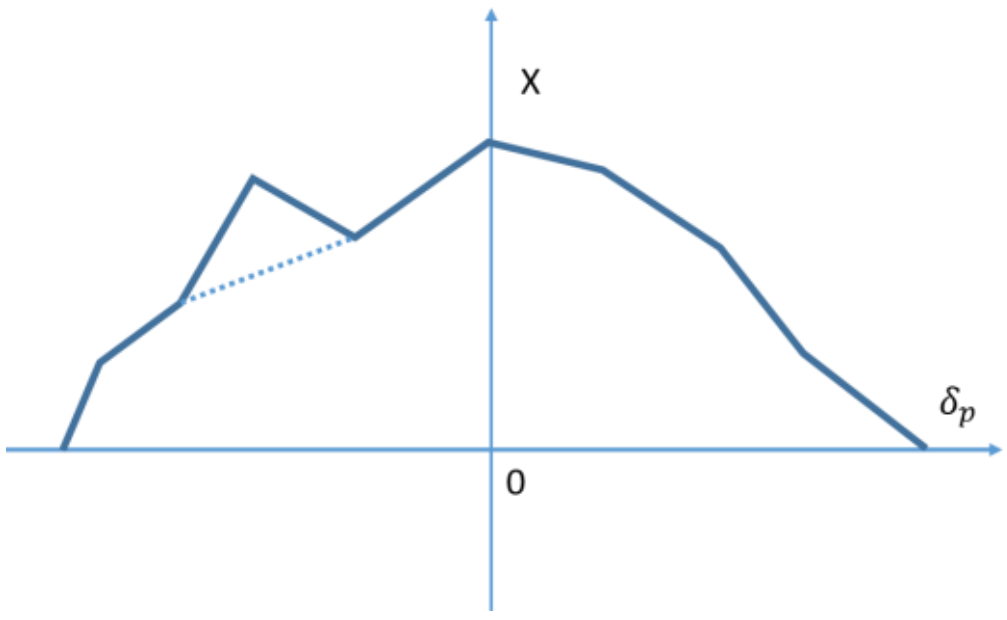

**Figure 4.2.1.16:** Clip of tracking DA

Thirty-two arc sextupole families, 10 IR sextupole families and 8 phase advance tuning knobs between different sections are used to optimize the DA. All the sextupoles (~250) could be free. There exists no clear difference with more sextupole families. The optimized DA at Higgs/W/Z energy is shown in Figs. 4.2.1.17-19. The DA could meet the requirement of injection and colliding beam lifetime.

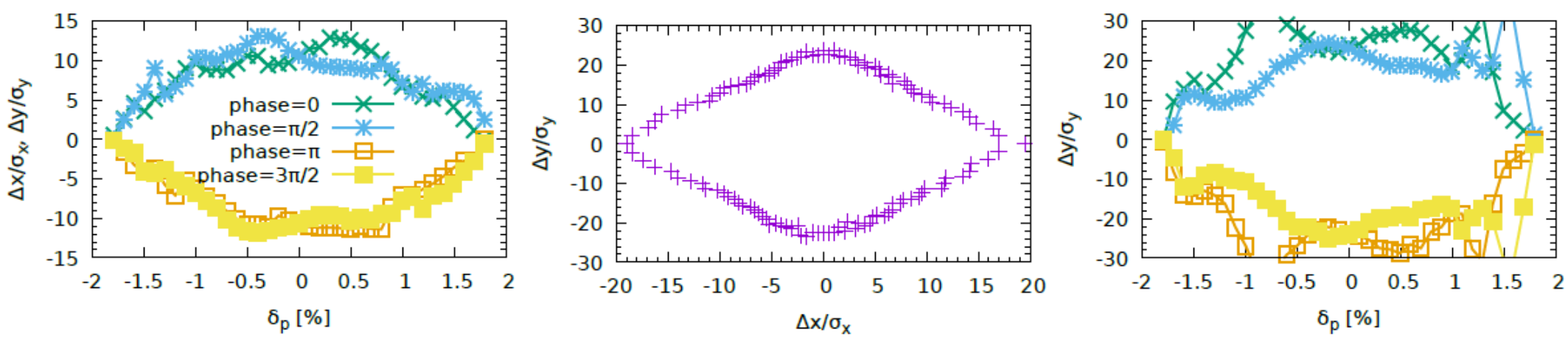

**Figure 4.2.1.17:** Higgs dynamic aperture with 90% survival of 100 samples. Radiation fluctuation is included. Design coupling is assumed in the 1st figure. On-momentum DA is shown in the 2nd one. Vertical DA is shown in the 3rd one. 145 turns are tracked.

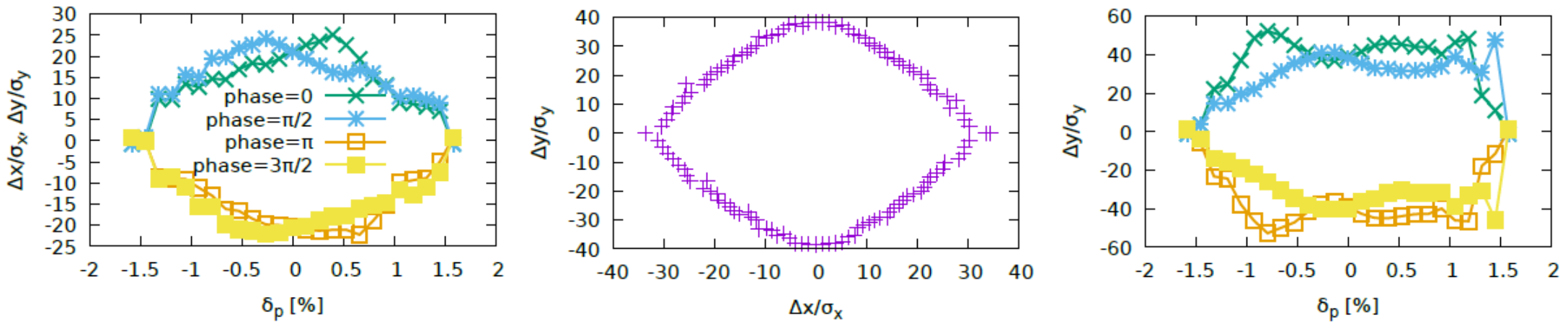

**Figure 4.2.1.18:** W dynamic aperture with 90% survival of 100 samples. Radiation fluctuation is included. Design coupling is assumed in the 1st figure. On-momentum DA is shown in the 2nd one. Vertical DA is shown in the 3rd one. 475 turns are tracked.

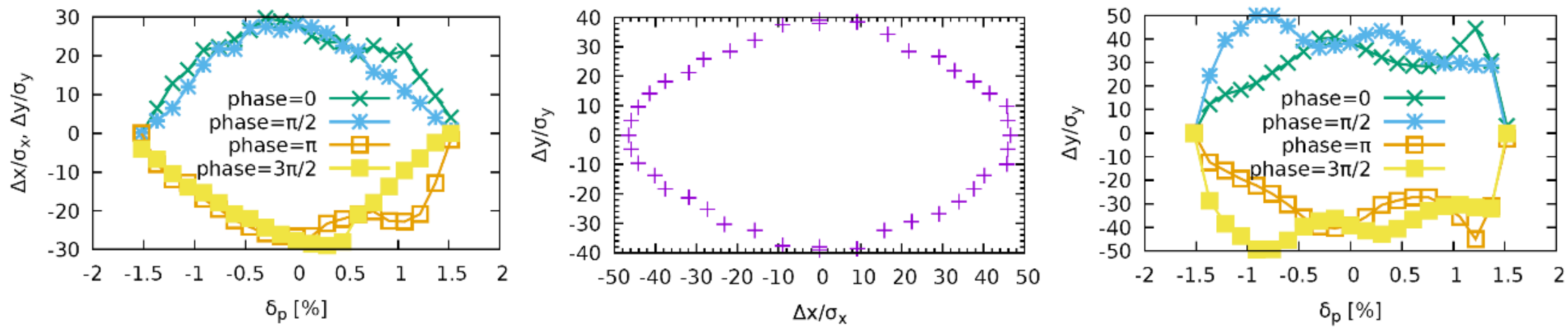

**Figure 4.2.1.19:** Z dynamic aperture with 90% survival of 100 samples (Coupling=2.2%). Radiation fluctuation is included. Design coupling is assumed in the 1st figure. On-momentum DA is shown in the 2nd one. Vertical DA is shown in the 3rd one. 2600 turns are tracked.

A lattice with combined dipole magnets in the arc region is studied as well [14]. With the help of sextupole component on the combined dipole, the power consumption of stand-alone sextupoles decrease by 75% with about half reduction of sextupole strength. 42 sextupole families are used to optimize the dynamic aperture and DA of the combined magnet scheme is as good as the normal one. The optimized DA of Higgs with the combined dipole scheme is shown in Figs. 4.2.1.20. According magnet design for this kind of special dipole is done to make sure this scheme can work.

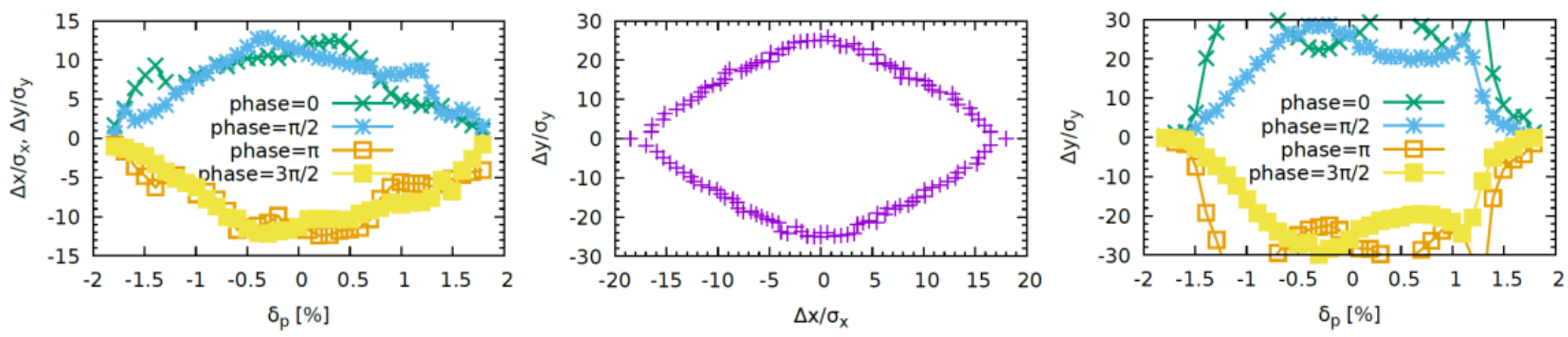

**Figure 4.2.1.20:** Higgs dynamic aperture with 90% survival of 100 samples for the combined-dipole scheme. Radiation fluctuation is included. Design coupling is assumed in the 1st figure. On-momentum DA is shown in the 2nd one. Vertical DA is shown in the 3rd one. 145 turns are tracked.

### 4.2.1.3 *Performance with Errors*

Magnet misalignments and the field errors will cause distortion of the closed orbit and optics. Because of the small beta function and emittance, the lattice is very sensitive to these errors and the correction of the optics is challenging. The correction sequence consists of closed orbit correction with sextupoles off to exclude the feed-down effect of the strong sextupoles, followed by a beta function and dispersion correction with sextupoles on, and then a betatron coupling correction. The optics corrections are performed using the response matrix fit method.

*4.2.1.3.1 Tolerance*

The misalignment errors resulting from magnet installation and field errors from manufacturing have been studied systematically. The dynamic aperture and emittance coupling after optics and closed orbit distortion corrections are considered to define the associated tolerances. Table 4.2.1.3 lists the misalignment errors and Table 4.2.1.4 lists the field errors for the simulations. Further study with larger tolerance of misalignment is undergoing and the goal for $\Delta x$ and $\Delta y$ is 100 μm.

**Table 4.2.1.3:** Misalignment RMS Error Requirements for the Collider

| Component | $\Delta x$ (mm) | $\Delta y$ (mm) | $\Delta z$ (mm) | $\Delta\theta_x$ (mrad) | $\Delta\theta_y$ (mrad) | $\Delta\theta_z$ (mrad) |
|---|---|---|---|---|---|---|
| Dipole | 0.05 | 0.05 | 0.15 | 0.2 | 0.2 | 0.1 |
| Quadrupole | 0.03 | 0.03 | 0.15 | 0.2 | 0.2 | 0.1 |

**Table 4.2.1.4:** Field Error Requirements for the Collider

| Component | Field error (RMS) |
|---|---|
| Dipole | 0.01% |
| Normal Quadrupole | 0.02% |

*4.2.1.3.2 Closed Orbit Correction*

Closed orbit distortion is caused by quadrupole misalignments and integral field errors in the dipoles. Figure 4.2.1.21 shows the closed orbit distortion around the ring in both the horizontal and vertical planes. Before orbit corrections, the rms value of orbit distortions in the arc regions are $X_{rms} \leq 6$ mm and $Y_{rms} \leq 4$ mm. The vertical orbit distortions at the IP are relatively large due to the residue vertical orbit at the strong final focus quadrupoles.

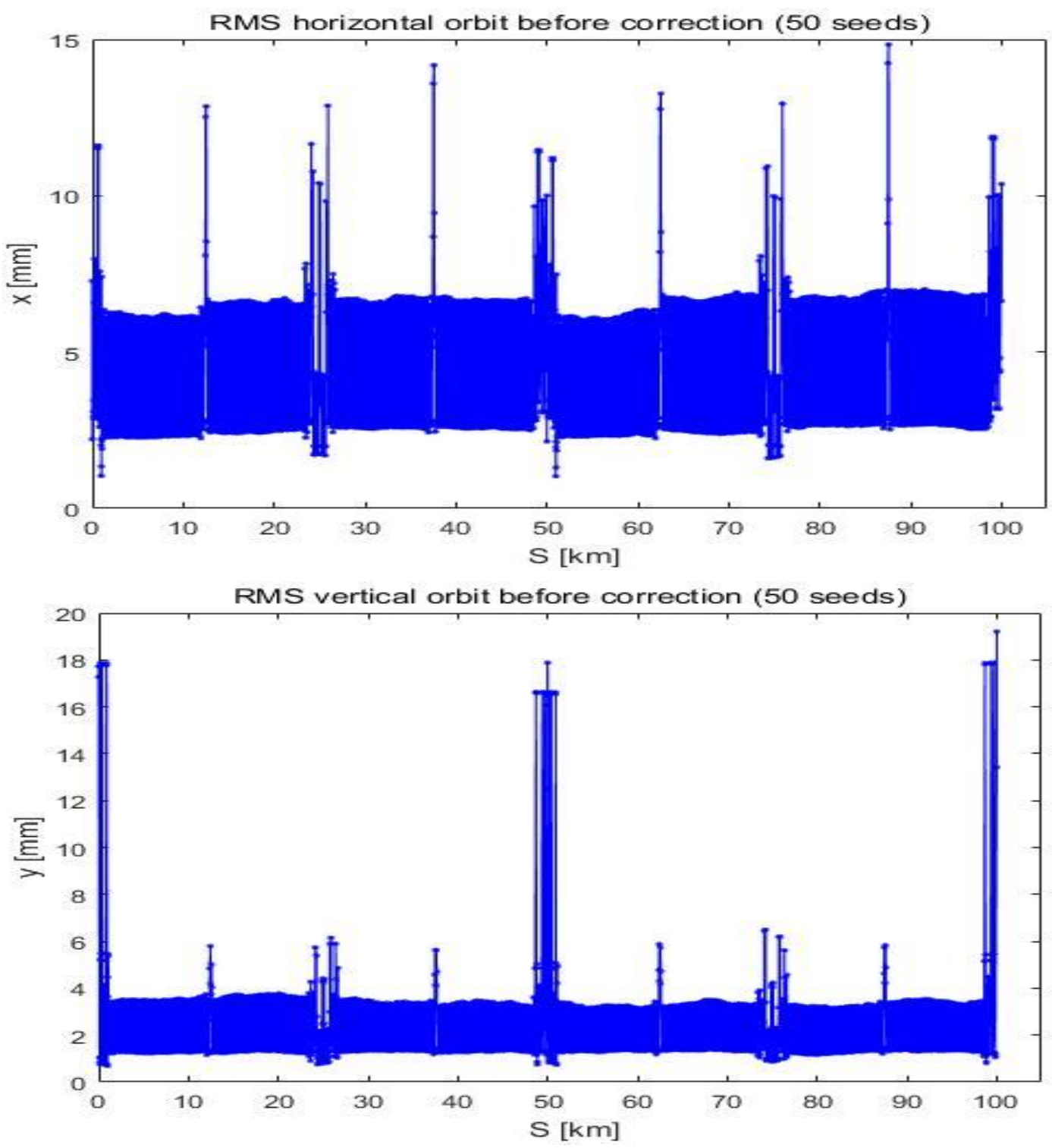

**Figure 4.2.1.21:** Closed orbit before correction

Beam position monitors (BPM) and steering magnets are arranged to correct the closed orbit. One BPM and a pair of correctors (one for horizontal and one for vertical) are installed in each cell. For the cells accommodating sextupoles, horizontal and vertical steering is produced by the sextupole trims. For other cells, individual horizontal and vertical steering magnets are located close to the focusing and defocusing quadrupoles respectively. Therefore there are four BPMs and four steering magnet pairs per betatron period. In the IR region, because of the strong final focus quadrupoles, the residual horizontal closed orbit in sextupoles will cause a large tune shift in the vertical plane. Steering magnets need to be located strategically to decrease the residual orbit.

The orbit correction algorithm is based on the response matrix. By reversing the response matrix using singular value decomposition, the steering strengths can be derived to correct the distorted orbits. The Fig. 4.2.1.22 gives the results after correction. The RMS values in the arcs are smaller than 100 μm.

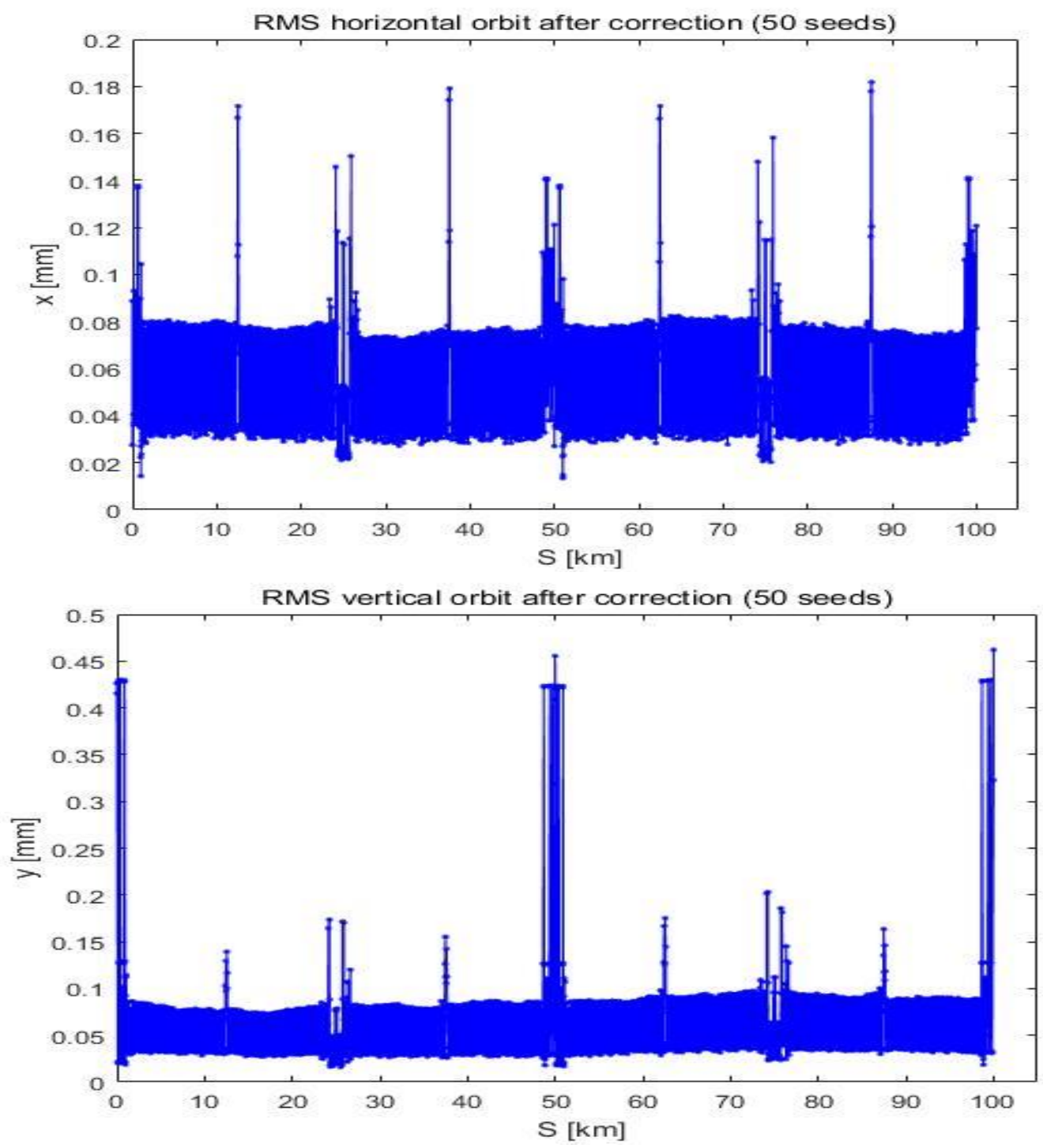

**Figure 4.2.1.22:** Closed orbit after correction

### *4.2.1.3.3 Optics Correction*

Sextupoles are switched on after the closed orbit correction. LOCO (**L**inear **o**ptics from **c**losed **o**rbits algorithm) [15] based on AT (MATLAB **A**ccelerator **T**oolbox) [16] is applied to restore the optics. By fitting the response matrix we determine the quadrupole strengths that best reduce the beta beatings and dispersion distortions. Fig. 4.2.1.23 indicates that the beta beatings amount to 0.5% after correction except for the IR region.

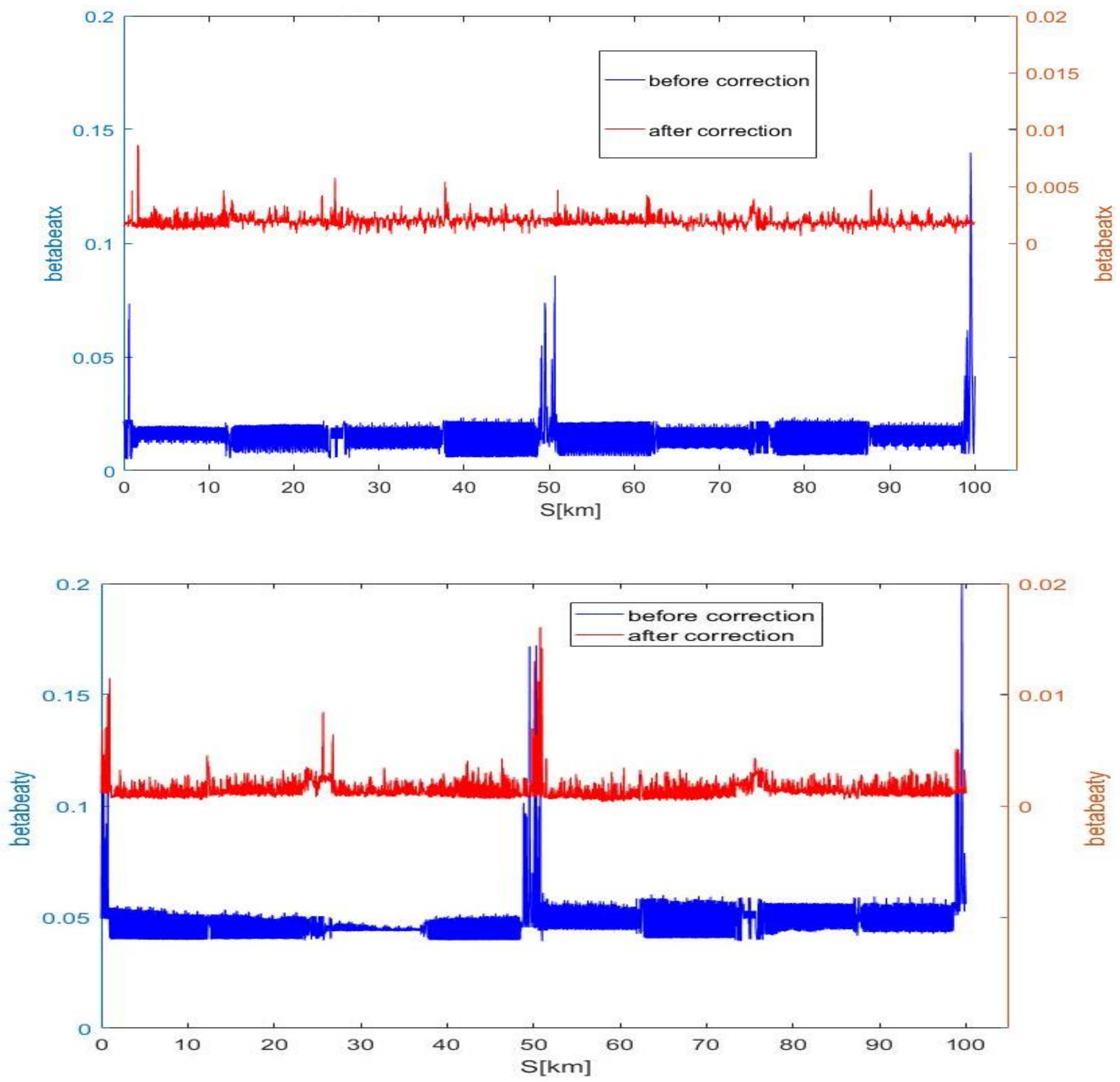

**Figure 4.2.1.23:** Beta beatings after optics correction

Quadrupole roll and the feed-down effect from sextupoles give rise to coupling. Skew coils on sextupoles and some independent skew quadrupoles are used to minimize the coupling response matrix by LOCO. Minimization limits emittance increase to less than 0.15%.

The dynamic aperture simulated with misalignment and field errors after correction for Higgs mode is shown in Fig. 4.2.1.24. It fulfils the dynamic aperture requirement.

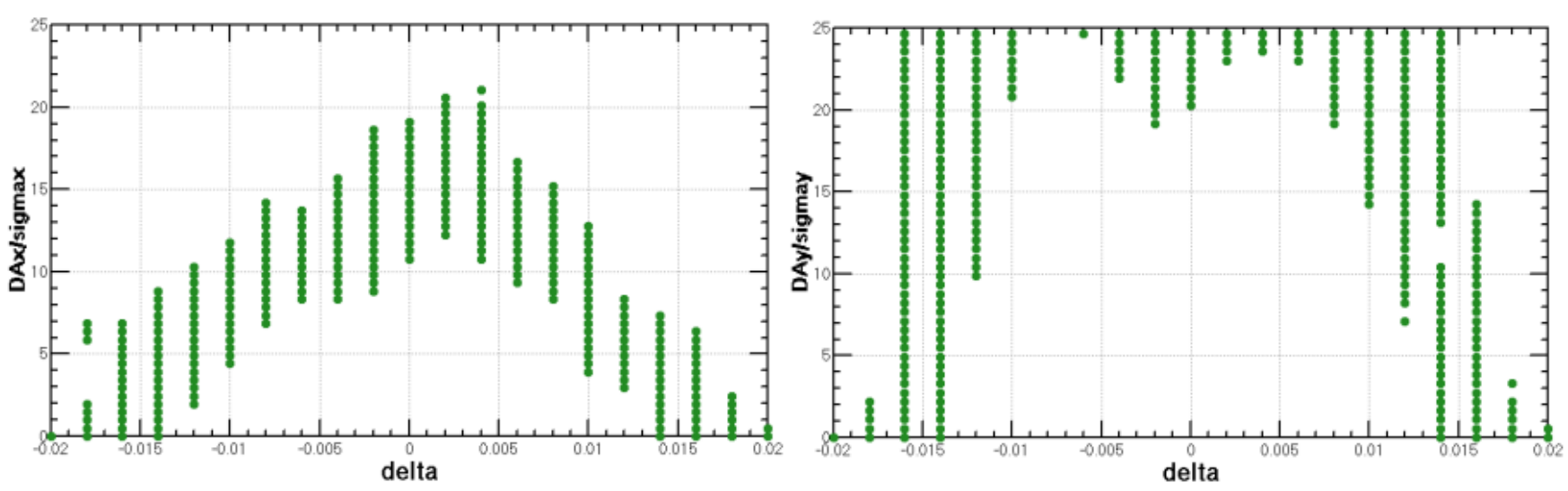

**Figure 4.2.1.24:** Higgs dynamic aperture (with errors). Radiation fluctuation is included. 145 turns are tracked.

### 4.2.2 **Beam-beam Effects**

Beam-beam interactions are the most important limitation to luminosity. They depend on a number of beam parameters and operating conditions; their impact on collider performance is most often calculated by computer simulations, also used to choose optimum operating conditions.

Beamstrahlung is synchrotron radiation excited by the beam-beam force. It will increase the energy spread, lengthen the bunch and may reduce the beam lifetime due to the long tail of the photon spectrum. The energy of the photons emitted by the beam-beam force is much harder than that from bending magnets. Fig. 4.2.2.1 is a sketch of beamstrahlung during beam-beam interactions. A beam particle (red) passes through a bunch (blue ellipse), which for carrying out simulations is sliced into several pieces. The particle trajectory is calculated slice-by-slice by a Poisson solver. The emitted photon energy is modelled using bending magnet radiation with the Monte-Carlo method.

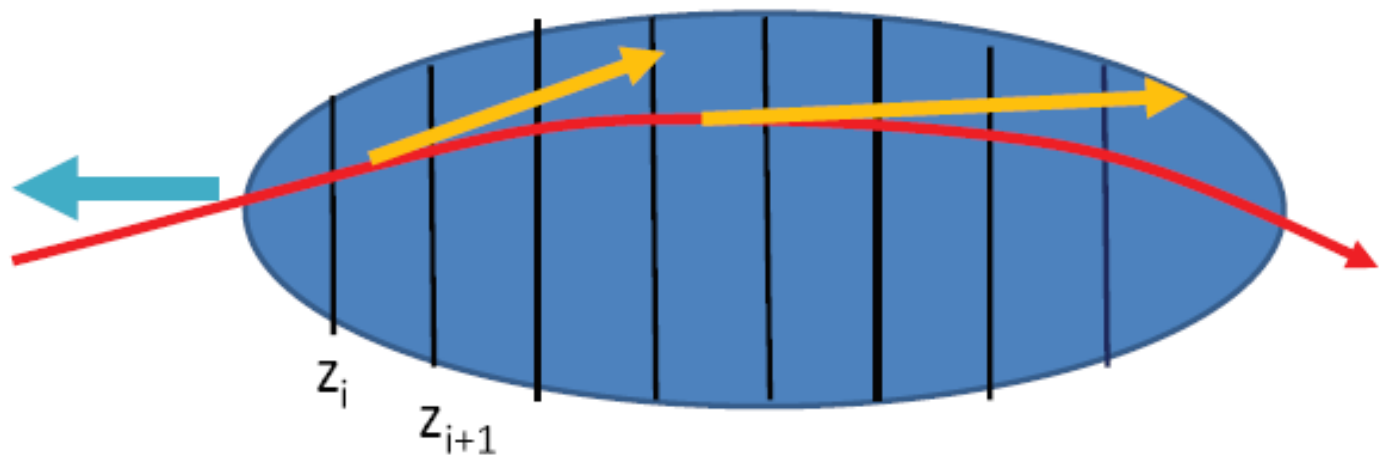

**Figure 4.2.2.1:** Energy loss of particles and photon emission.

The simulations have been carried out with LIFETRAC (weak-strong) [1], BBWS (weak-strong), BBSS (strong-strong) [2] and IBB (strong-strong) [3]. These codes has been used successfully for calculations for other colliders: DAFNE (INFN/LNF), KEKB (KEK) and BEPCII (IHEP). The weak-strong code is often used to evaluate performance, since the simulation is faster than with strong-strong codes [4,5]. In this section we only show the strong-strong simulation results, since they are more self-consistent.

We consider bunch lengthening from impedance. Fig. 4.2.2.2 shows the luminosity and RMS bunch size evolution. The simulated luminosity is $3.0 \times 10^{34} cm^{-2}s^{-1}$ with the design bunch population and bunch number. Results shown below in Figures 4.2.2.2 through 4.2.2.9 are for operation in the Higgs mode. There is about 30% bunch lengthening due to beamstrahlung. Here the ideal crab waist transformation is included before and after collision.

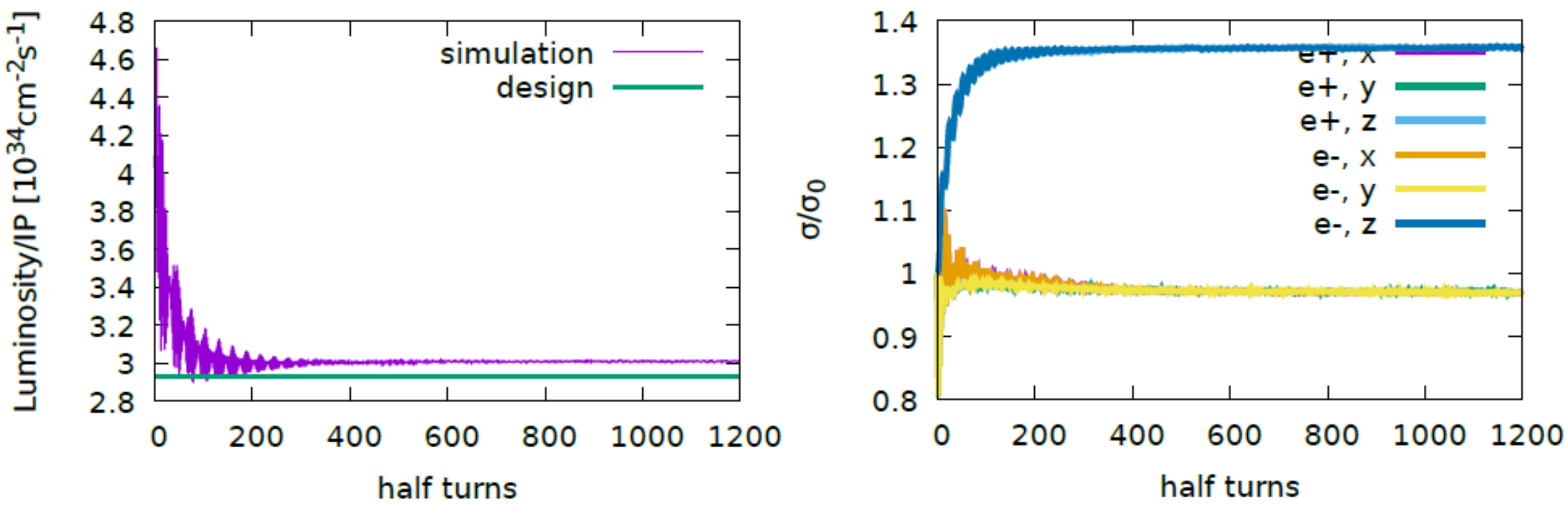

**Figure 4.2.2.2:** Luminosity and rms size evolution turn-by-turn. The half ring working point is (0.555, 0.61).

Since the working point is important for luminosity optimization, a tune scan has been done. From experience at other colliders and from weak-strong simulations, we fix the vertical tune and only vary the horizontal tune. The result is shown in Fig. 4.2.2.3. There exist two resonance lines in the range, $2\nu_x - 2\nu_s = N$ near $\nu_x = 0.535$ and $2\nu_x - 4\nu_s = N$ near $\nu_x = 0.57$. It is clear that the lowest order resonance is the more destructive. Fig. 4.2.2.4 shows how the 2nd moments change turn-by-turn at (0.535, 0.61), where the x-z coherent instability [6] is excited. The horizontal beam size blows up and there exists an in-phase x-z dipole coherent oscillation. The maximum tilt angle is about 5 mrad, which is close to 1/3 of the half crossing angle.

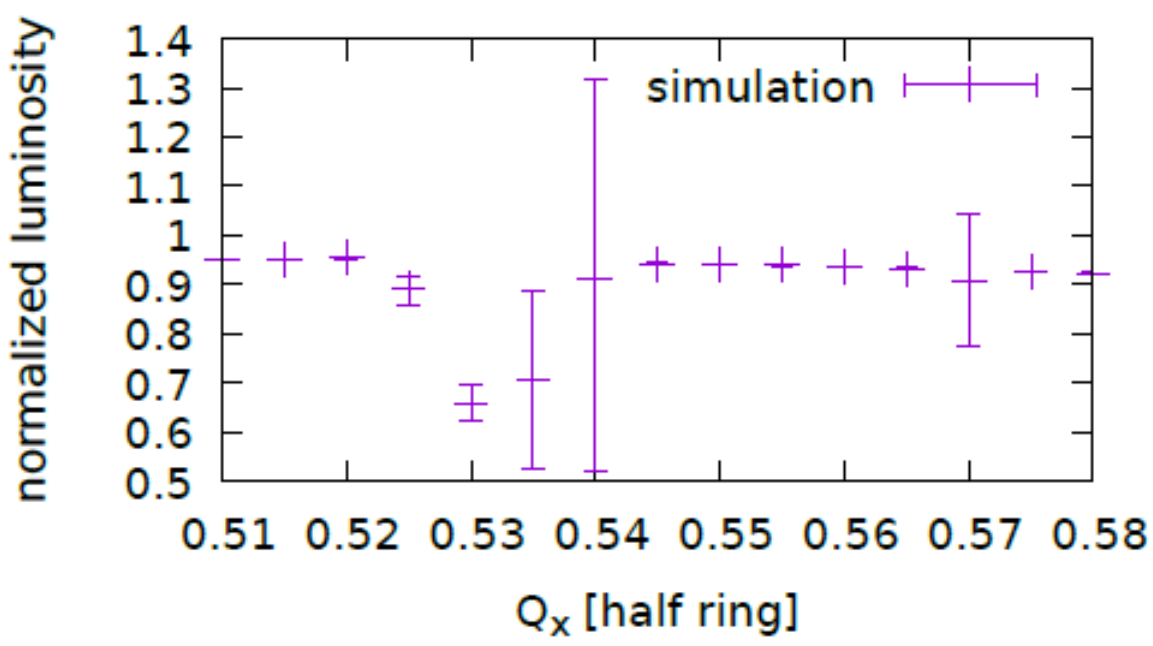

**Figure 4.2.2.3**: Luminosity versus horizontal tune. Vertical tune of the half ring is 0.61. The error bar shows the turn-by-turn luminosity difference.

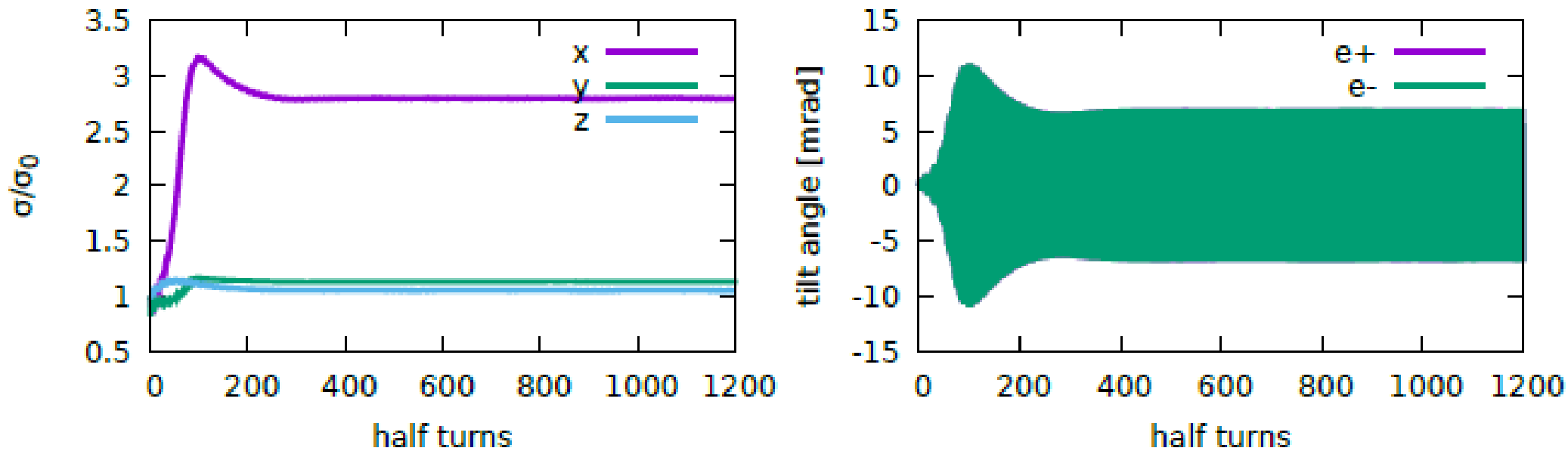

**Figure 4.2.2.4**: Evolution of the 2nd moments turn-by-turn at (0.535, 0.61).

One can obtain the equilibrium distribution with many turns of tracking data without any aperture limit. It is assumed the outgoing flow across some boundary is the same as the incident flow, which is determined by radiation damping. The lifetime limited by the boundary can be calculated by [4,5]

$$\tau = \frac{\tau_z}{2Af(A)}$$

where $f(J_z)$ gives the beam action distribution with $\int_0^\infty dJ_z f(J_z) = 1$ , $\tau_z$ is the longitudinal damping time and A is the longitudinal acceptance. It is shown in Fig. 4.2.2.5 that the lifetime estimated by damping flow agrees with that given by beam loss simulation.

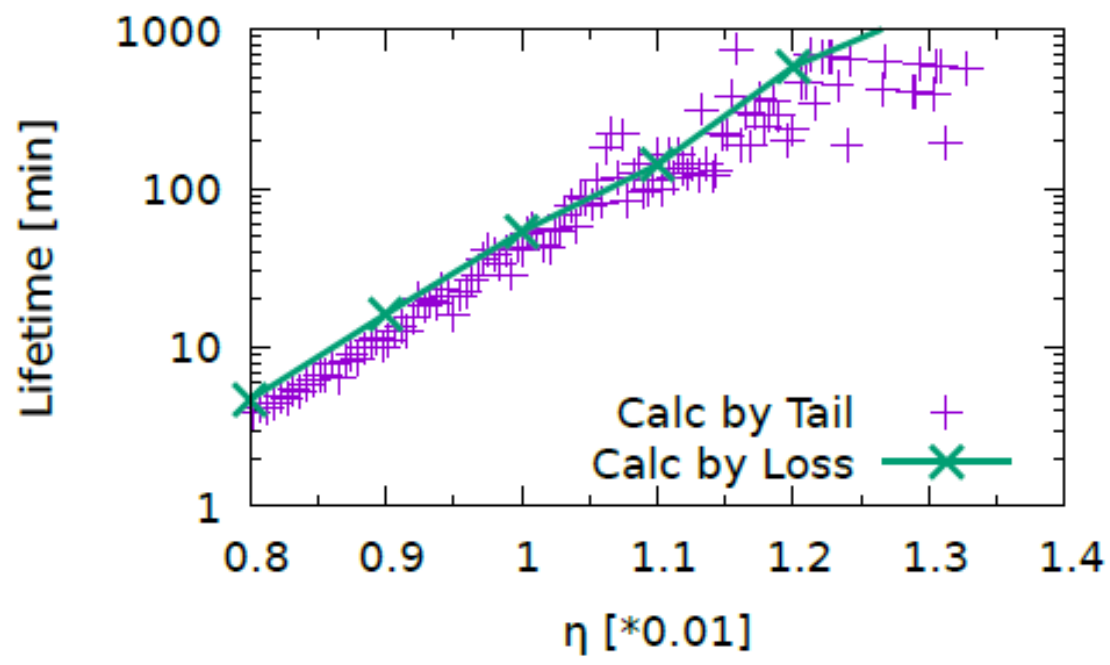

**Figure 4.2.2.5**: Comparison of beamstrahlung lifetime estimation methods.

How the luminosity, beam distribution and lifetime vary with the bunch current is important and will help us evaluate if our design goals are achievable. For a flat beam, the beam-beam parameter can be defined as

$$\xi_y = \frac{2r_e\beta_y^0}{N\gamma}\frac{L}{f_0}$$

where $r_e$ is the classical electron radius, $\beta_y^0$ is the nominal vertical beta function at the IP, $N$ is the bunch population, $f_0$ the revolution frequency and $L$ is the bunch luminosity. Fig. 4.2.2.6 shows the beam-beam parameter versus bunch current and the beamstrahlung lifetime versus momentum acceptance. The lifetime will be longer than 80 minutes with 0.013 momentum acceptance. Fig. 4.2.2.7 shows the beam tail distribution. There exists a non-Gaussian tail in the z-direction.

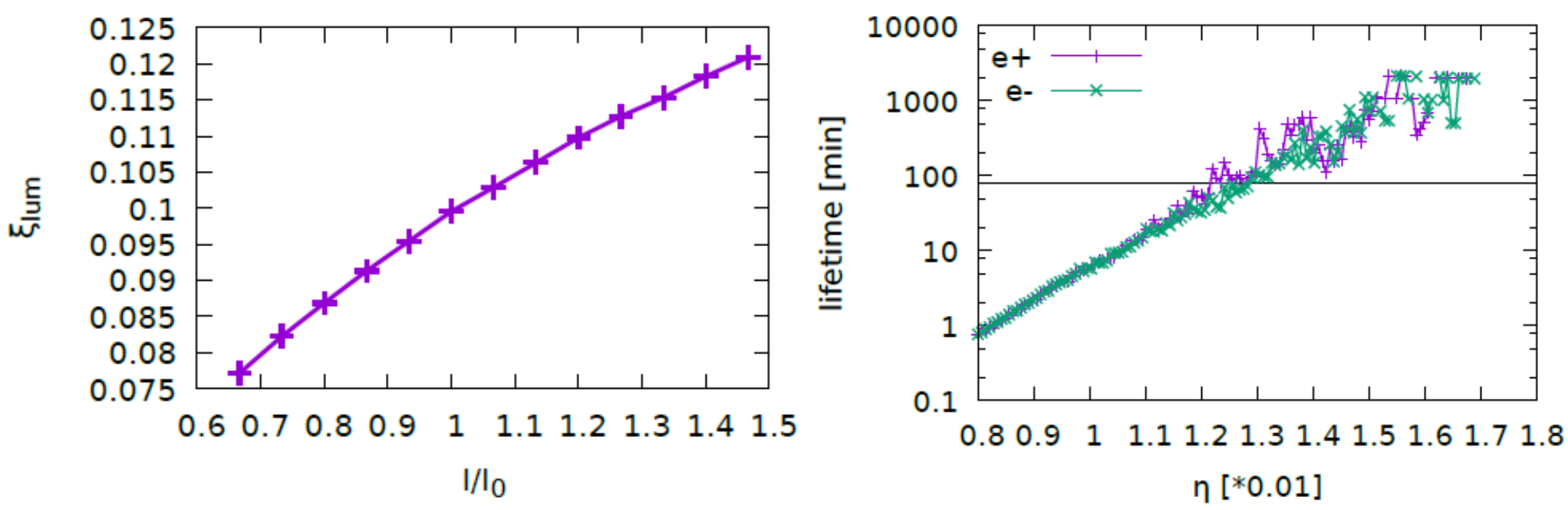

**Figure 4.2.2.6**: Beam-beam parameter and lifetime versus momenumt acceptance.

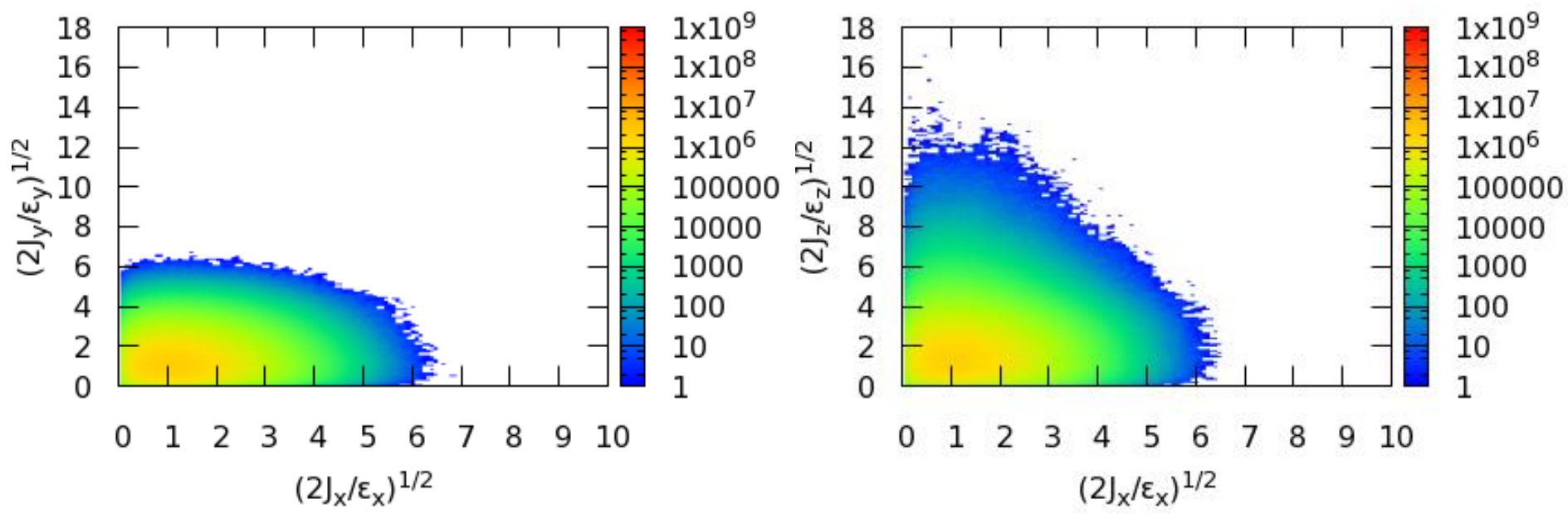

**Figure 4.2.2.7**: Beam tail distribution with crab-waist collision.

Fig. 4.2.2.8 shows luminosity versus crab-waist strength. The crab-waist transformation increases the luminosity by suppressing vertical blow up. The optimized strength is about 80% of the full crab-waist strength.

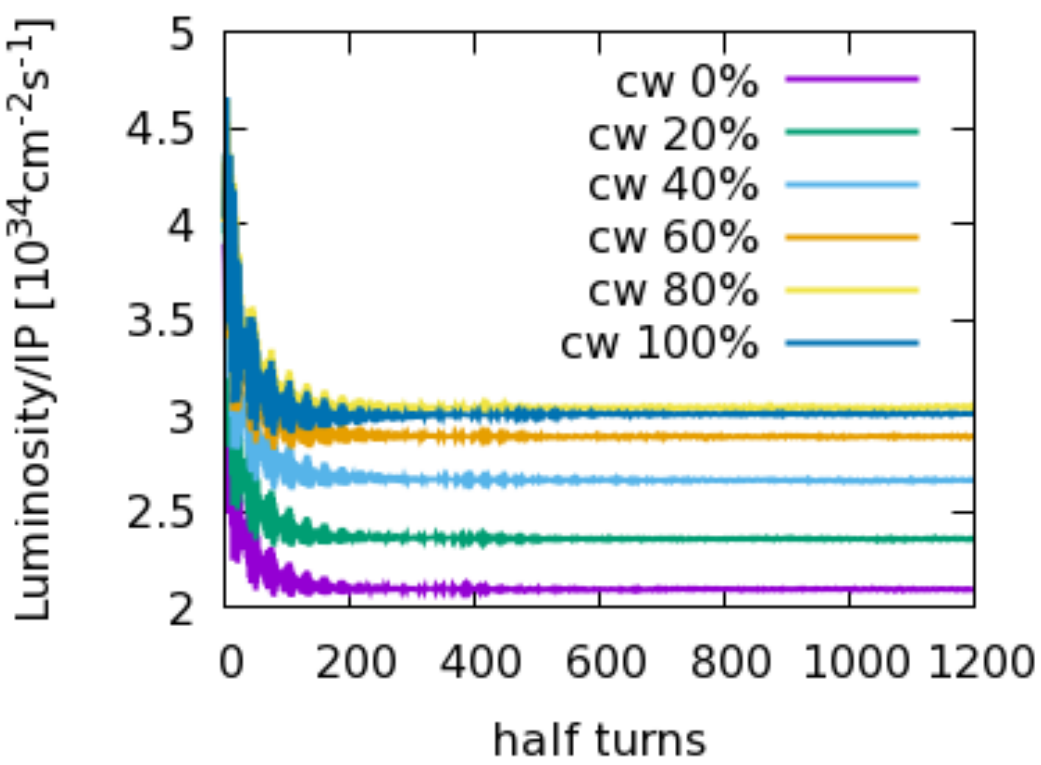

**Figure 4.2.2.8**: Luminosity versus crab-waist strength.

The beam-beam limit at the W/Z is mainly determined by the coherent x-z instability instead of the beamstrahlung lifetime as in the Higgs mode. Longer bunch length will help to suppress the coherent instability. The so-called 'bootstrapping' method [7] can help increase the colliding bunch population. That is to say we inject e+/e- alternately during collisions and the beamstrahlung effect lengthens the bunch. Longer bunch length increases the bunch current threshold and makes it possible to collide with higher bunch current. The most prominent phenomenon from the coherent instability is blow-up in the horizontal direction. The design bunch population ($8 \times 10^{10}$) is less than the stable threshold (~$12 \times 10^{10}$). The phenomenon is similar for $\nu_x = 0.562$ to 0.568; this ensures a large enough stable working point range. Simulation shows that the peak luminosity is $17.9 \times 10^{34} cm^{-2}s^{-1}(\beta_y^* = 1.5\text{mm})/35.0 \times 10^{34} cm^{-2}s^{-1}(\beta_y^* = 1\text{mm})$ with Z mode bunch population and number.

Bootstrapping injection is also used in W mode. At the design bunch population and number, the simulated luminosity is $11.6\times 10^{34} cm^{-2}s^{-1}$. The collision is stable with 10% asymmetric bunch population and also with 10% asymmetric crat-waist strength (90% vs 100%). The collision is stable at $\nu_x = 0.552$ and 0.555, which shows a narrow stable region.

With optimized lattice and with beam-beam interaction, the lifetime limited by x/y/z direction is evaluated by tracking at Higgs energy, which is shown in Fig. 4.2.2.9. To ensure 80min limited by y/z direction respectively, the dynamic aperture should be greater than $15\sigma_y \times 0.013$. It is expected the horizontal dynamic aperture is greater than $8\sigma_x$, in that case the quantum lifetime limited by x direction is long enough to be ignored.

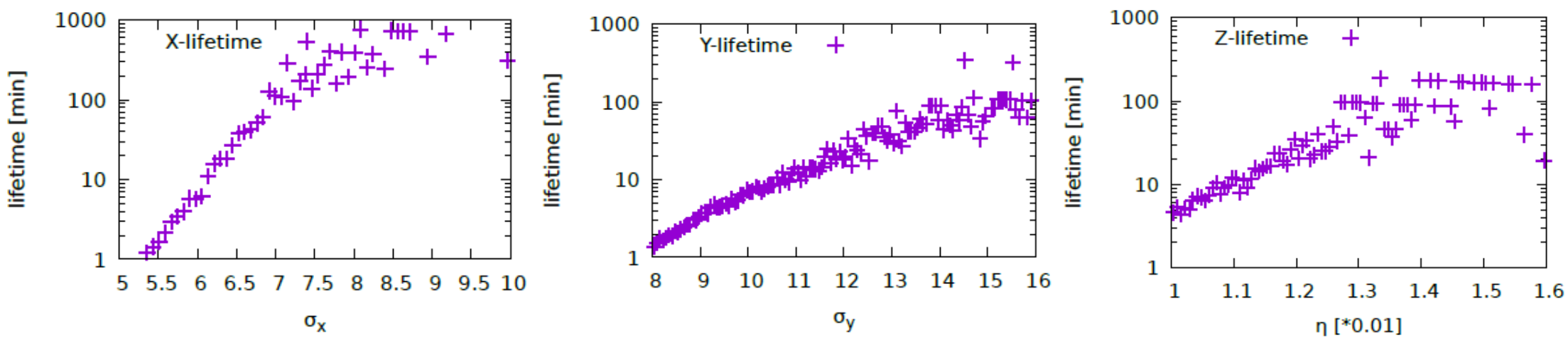

**Figure 4.2.2.9**: Lifetime with real lattice and with beam-beam interaction at Higgs.

### 4.2.3 **Beam Instability**

Interaction of an intense charged particle beam with the vacuum chamber can lead to collective instabilities. These instabilities will induce beam quality degradation or beam losses, and finally restrict the performance of the machine. In this section, the impedance budget for the Collider is given. Based on the impedance studies, beam instabilities are estimated for different operation modes.

#### 4.2.3.1 ***Impedance***

##### *4.2.3.1.1 Impedance Threshold*

The impedance thresholds are estimated analytically, which gives a rough criterion for the impedance requirements. The limitation on the longitudinal broadband impedance mainly comes from microwave instability and bunch lengthening. The threshold for microwave instability is estimated according to the Boussard or Keil-Schnell criteria. The limitation on the transverse broadband impedance mainly comes from the transverse mode coupling instability. For Gaussian bunches, the threshold current can be expressed with the transverse kick factor. The narrowband impedances are mainly contributed by cavity like structures. These impedances may induce coupled bunch instabilities in both longitudinal and transverse planes. The limitation on the shunt impedance of a HOM is evaluated in resonant condition when the resonance frequency overlaps the beam spectrum line, and the growth rate of the coupled bunch instability is less than the synchrotron radiation damping. The impedance thresholds for the different operation modes are listed in Table 4.2.3.1. The Z has the most critical restrictions for both broadband and narrowband impedances.

**Table 4.2.3.1:** Impedance thresholds

| Parameter | Symbol, unit | H | W | Z |
|---|---|---|---|---|
| Longitudinal broadband | $\|Z_\parallel/n\|_{th}$, mΩ | 8.8 | 3.2 | 0.9 |
| Transverse broadband | $k_{y,th}$, kV/pC/m | 65.6 | 32.7 | 20.1 |
| Longitudinal narrowband | $\frac{f}{\text{GHz}}\frac{\text{Re}Z_\parallel}{\text{G}\Omega}e^{-(2\pi f\sigma_t)^2}$ | 3.5 | 0.08 | 1.1E-3 |
| Transverse narrowband | $\frac{\text{Re}Z_\perp}{\text{G}\Omega/\text{m}}e^{-(2\pi f\sigma_t)^2}$ | 2.3 | 0.09 | 1.8E-3 |

##### *4.2.3.1.2 Impedance Model*

The impedance and wake are calculated both with formulas as well as simulations with ABCI and CST. The longitudinal wake contributions of different impedance objects at a bunch length of 3 mm are shown in Figure 4.2.3.1. Table 4.2.3.2 lists the impedance budget of the objects that have been considered. The calculation gives a total longitudinal effective impedance of 11.4 mΩ. The total longitudinal loss factor is 786.8 V/pC, and the total transverse kick factor is 20.2 kV/pC/m. Both longitudinal and transverse impedances are dominated by the resistive wall and elements of which there is a large quantity. The longitudinal loss factor is mainly contributed by the resistive wall and the RF cavities.

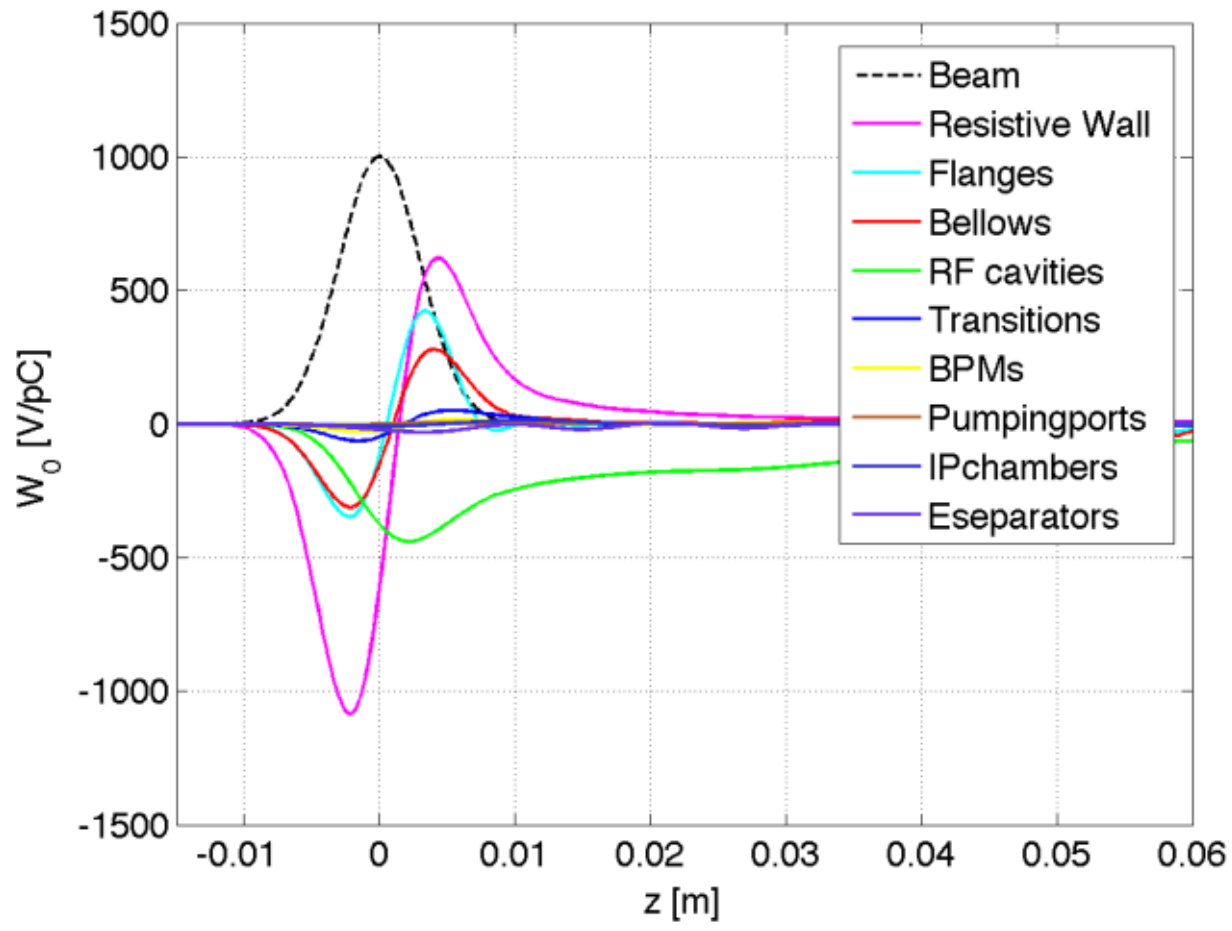

**Figure 4.2.3.1:** Longitudinal wake potential with rms bunch length of 3 mm.

**Table 4.2.3.2:** Summary of the impedance budget.

| Component | Number | $Z_{\|}/n$, mΩ | $k_{loss}$, V/pC | $k_y$, kV/pC/m |
|---|---|---|---|---|
| Resistive wall | - | 6.2 | 363.7 | 11.3 |
| RF cavities | 240 | -1.0 | 225.2 | 0.3 |
| Flanges | 20000 | 2.8 | 19.8 | 2.8 |
| BPMs | 1450 | 0.1 | 13.1 | 0.3 |
| Bellows | 12000 | 2.2 | 65.8 | 2.9 |
| Pumping ports | 5000 | 0.02 | 0.4 | 0.6 |
| IP chambers | 2 | 0.02 | 6.7 | 1.3 |
| ES separators | 22 | 0.2 | 41.2 | 0.2 |
| Transitions | 164 | 0.8 | 50.9 | 0.5 |
| Total | - | 11.4 | 786.8 | 20.2 |

### 4.2.3.2 *Impedance-driven single bunch instability*

#### *4.2.3.2.1 Microwave Instability*

The microwave instability will rarely induce beam losses, but may reduce the luminosity due to the distorted beam distribution and increasing of the beam energy spread. Figure 4.2.3.2 shows the dependency of bunch lengthening on beam intensity for Higgs and Z. The bunch lengthening and beam energy spread increase at design bunch current for different operation scenarios are shown in Table 4.2.3.3. The results show an energy spread increase for Z and W at design bunch intensity.

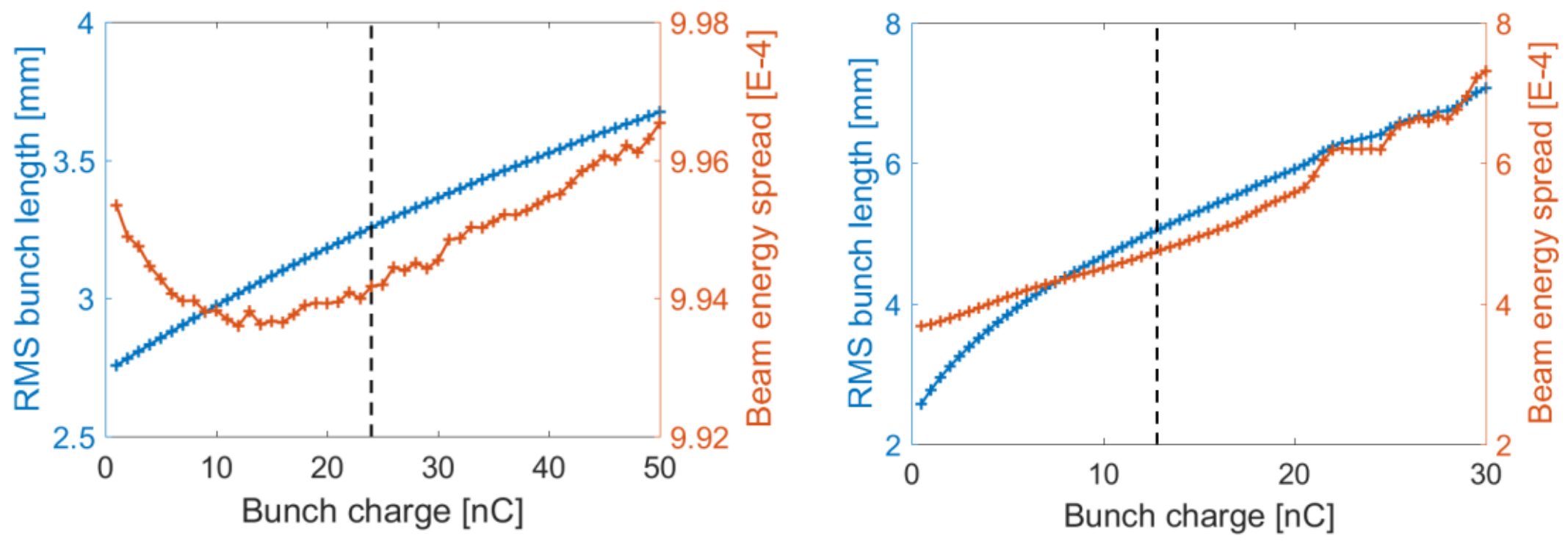

**Figure 4.2.3.2:** Dependence of bunch length and beam energy spread on the beam intensity for H (left) and Z (right).

**Table 4.2.3.3:** Bunch lengthening and beam energy spread increase at design bunch current

| Parameter | Symbol, unit | H | W | Z |
|---|---|---|---|---|
| Natural bunch length | $\sigma_{l0}$, mm | 2.7 | 3.0 | 2.4 |
| Bunch length with impedance | $\sigma_l$, mm | 3.3 | 4.2 | 5.1 |
| Beam energy spread increase | $\sigma_e/\sigma_{e0}$ | ~ 0 | 3% | 30% |

#### *4.2.3.2.2 Coherent Synchrotron Radiation*

Coherent synchrotron radiation (CSR) is generated when the beam passes through the bending magnets. Coasting beam theory with shielding has been used to estimate this threshold. The instability threshold given by the analytical theory is more than two times higher than the design bunch population. Table 4.2.3.4 shows the comparison between the design and threshold bunch population. Therefore, CSR is not a concern.

**Table 4.2.3.4:** Threshold bunch intensity for CSR

| Parameter | Symbol, unit | H | W | Z |
|---|---|---|---|---|
| Design bunch intensity | $N_e$, nC | 24.0 | 19.2 | 12.8 |
| Threshold intensity for CSR | $N_{e,\,th}$, nC | 605 | 176 | 33 |

#### *4.2.3.2.3 Transverse Mode Coupling Instability*

The threshold for the transverse mode coupling instability (TMCI) is estimated using both the analytical formula and Eigen mode analysis. The Eigen mode analysis gives the dependence of the head-tail mode frequencies on the bunch intensity, as shown in Fig. 4.2.3.3. Considering the bunch lengthening due to the longitudinal impedance, the transverse effective impedance will decrease with bunch intensity. Therefore, the threshold intensity for TMCI will increase further. The comparison between the design and threshold intensities are given in Table 4.2.3.5.

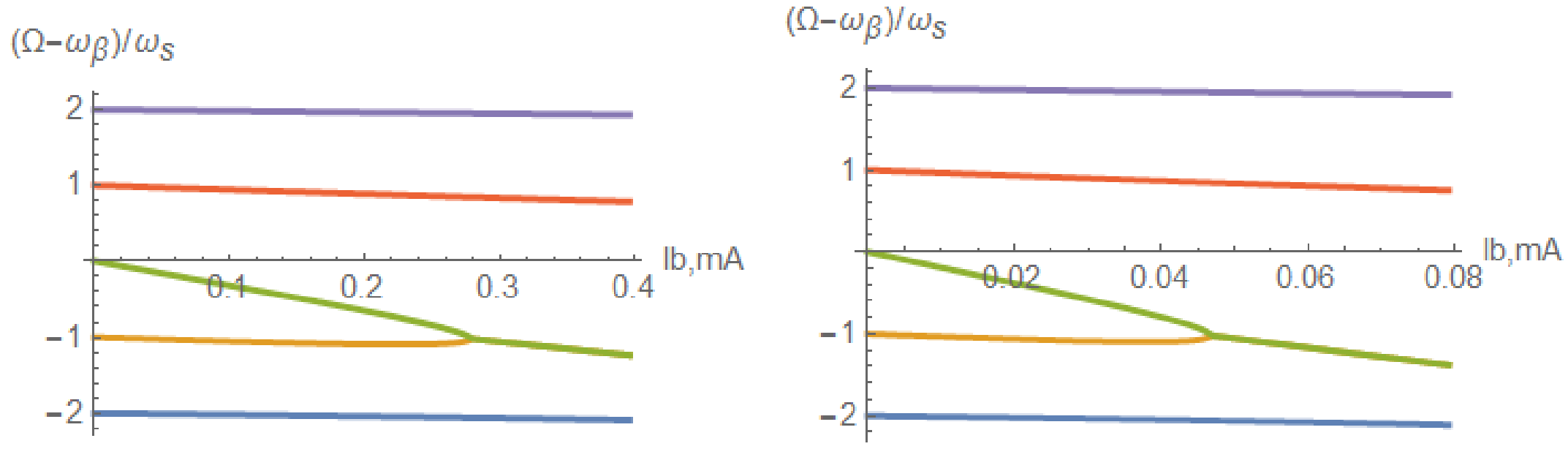

**Figure 4.2.3.3:** Dependence of the head-tail mode frequencies on the bunch intensity for H (left) and Z (right).

**Table 4.2.3.5:** Threshold bunch intensity for TMCI

| Parameter | Symbol, unit | H | W | Z |
|---|---|---|---|---|
| Design bunch intensity | $N_{e,}$ nC | 24.0 | 19.2 | 12.8 |
| TMCI threshold | $N_{e,\,th,}$ nC | >70 | ~ 40 | ~ 18 |

#### *4.2.3.2.4 Transverse Tune Shift*

The transverse operating point is (363.10, 365.22) is slightly above integer in the horizontal plane. Since the tune shift due to transverse impedance is negative, the beam could become unstable at lower beam current (or lower impedance) than that for the transverse mode coupling instability. With the estimated effective impedance around 0.7 MΩ/m, the corresponding transverse tune shift for different operation scenarios are shown in Table 4.2.3.6; they are all much lower than 0.1. So the beam should be stable from the integer resonances.

**Table 4.2.3.6:** Transverse tune shift due to impedance

| Parameter | Symbol, unit | H | W | Z |
|---|---|---|---|---|
| Transverse tune shift | $\Delta\nu_x$ | 0.014 | 0.017 | 0.020 |

### 4.2.3.3 ***Impedance-driven coupled bunch instability***

#### *4.2.3.3.1 Transverse Resistive Wall Instability*

One dominant contribution to the coupled bunch instability is the resonance at zero frequency of the transverse resistive wall impedance. The growth time for the most dangerous instability mode for Z is ~4 ms, which is much faster than the radiation damping of 843 ms. Figure 4.2.3.4 shows the growth rate of the transverse resistive wall instability with different mode numbers. An effective transverse feedback system is required to damp the instability. Considering a feedback damping time of 5 turns (transverse), we calculate the threshold beam current for different operation scenarios, as shown in Table 4.2.3.7. The threshold beam currents are higher than the design value.

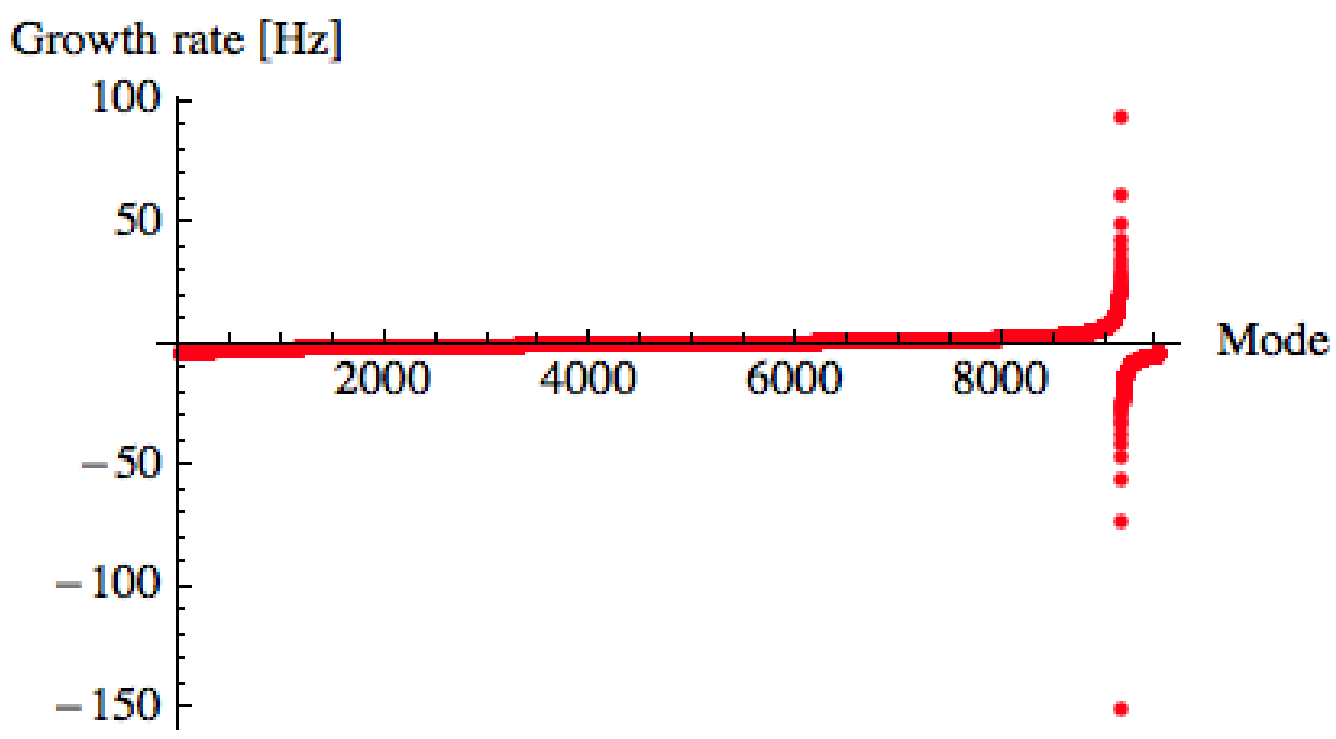

**Figure 4.2.3.4:** Growth rate of the transverse resistive wall instability with mode number for Z.

**Table 4.2.3.7:** Growth time for the transverse resistive wall instability and threshold beam current considering effective feedback damping

| Parameter | Symbol, unit | H | W | Z |
|---|---|---|---|---|
| Instability growth time | $\tau$, ms | 298 | 39 | 4.3 |
| Threshold beam current with feedback | $I_0$, A | 1.6 | 1.0 | 0.6 |

*4.2.3.3.2 Coupled Bunch Instability induced by HOMs*

Another important contribution to the coupled bunch instability is the RF HOMs. With a transverse feedback system of 5 turns and a longitudinal feedback system of one synchrotron oscillation period, all modes below cutoff frequency, except TM011 mode for Z, can be damped. However, considering the full RF system, the threshold value greatly depends on the actual tolerances of the cavity construction. With a HOM frequency scattering of 1 MHz, all the transverse and longitudinal modes below cutoff frequency can be well damped for different operation scenarios. Figure 4.2.3.5 shows the impedance spectrum of the HOMs with different frequency scattering, and compared to the threshold of the radiation damping and possible feedback damping for Z, which is the most critical case.

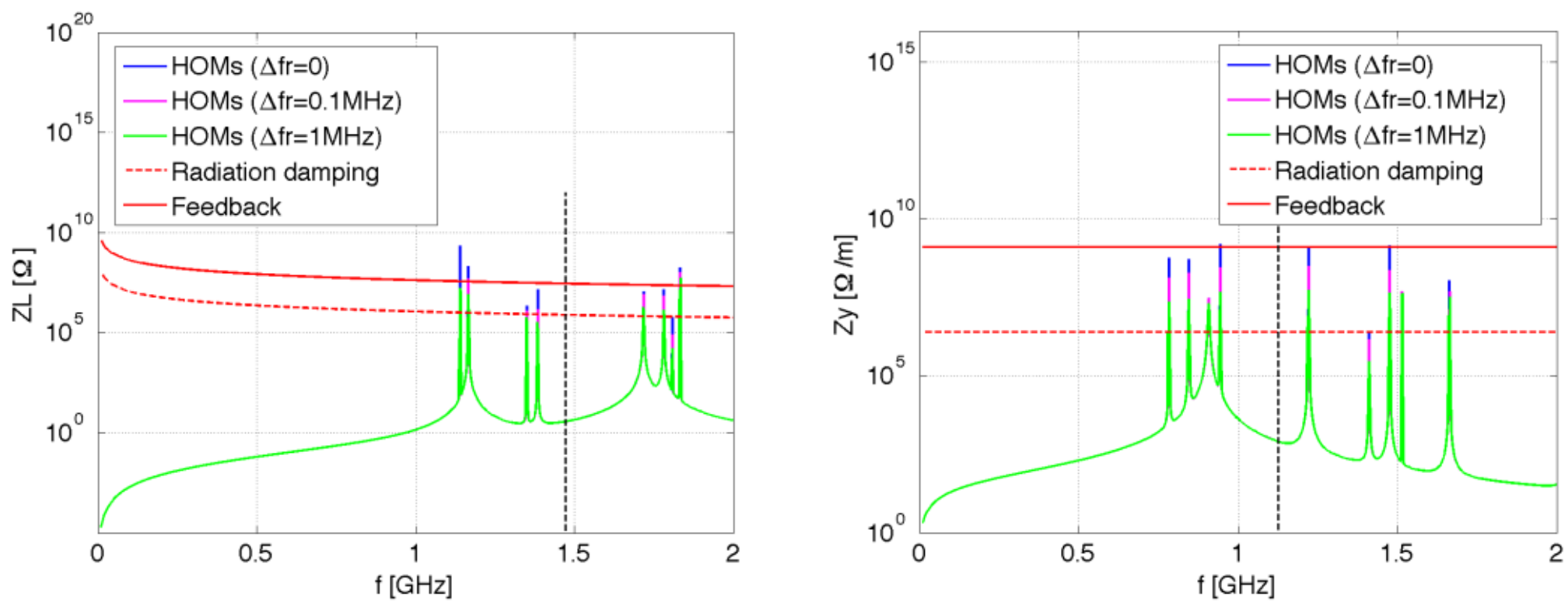

**Figure 4.2.3.5:** Impedance spectrum of HOMs compared to the threshold determined by radiation damping and feedback damping (Left: longitudinal HOMs, Right: transverse HOMs).

#### 4.2.3.4 *Electron Cloud Effect*

The build-up of accumulated photon electrons and secondary electrons has been one of the most serious restrictions on collider luminosity in PEP II, KEKB, LHC, and BEPC. The electron cloud (EC) can induce emittance blow-up and coupled bunch instabilities, and lead to luminosity degradation.

For CEPC, the photon electrons and secondary electron emission will be the main contribution to the electron cloud. The necessary condition for electron amplification is that the average secondary electron emission yield (SEY) exceeds one. The electron cloud build-up saturates when the attractive beam field at the chamber wall is compensated on the average by the electron space charge field, which corresponds to the volume density of $\rho_{e,neutr} \approx \lambda_e / \pi ab$, where $\lambda_e$ is the line density of the electron cloud, $a$ and $b$ are the half sizes of the elliptical vacuum pipe. The estimated neutralization density and simulated results for CEPC are shown in Table 4.2.3.8 and Fig. 4.2.3.6.

**Table 4.2.3.8:** Estimates on electron cloud instability for CEPC

| Components | H | Z |
|---|---|---|
| Circumference (km) | 100 | |
| Bunch population ($10^{10}$) | 15 | 8.0 |
| Bunch spacing (ns) | 10/25/50/100 | 10/25/50 |
| Beam energy (GeV) | 120 | 45.5 |
| Emittance *x*/*y* (nm) | 1.21/0.0024 | 0.18/0.004 |
| Pipe radius *H*/*V* (mm) | 37.5/28 | |
| Bunch number | 242 | 30000/12000/6000 |
| Neutralization volume density ($10^{12}$/m$^3$) | 13.0/5.2/2.6/1.3 | 8.1/3.24/1.62 |
| Simulated electron cloud density ($10^{12}$/m$^3$) | 2.1/0.52/0.11/0.018 | 0.46/0.0312/0.00227 |
| Simulated wake field *W/L* ($10^2$/m$^3$) | 2.04/0.51/0.11/0.02 | 0.72/0.049/0.0036 |
| Betatron tune | 363.10/365.22 | |
| Synchrotron tune | 0.065 | 0.028 |
| Growth time (ms) | 0.5/1.9/9.0/55.0 | 0.8/12.03/165.3 |
| Threshold electron density ($10^{11}$/m$^3$) | 7.6 | 0.324 |

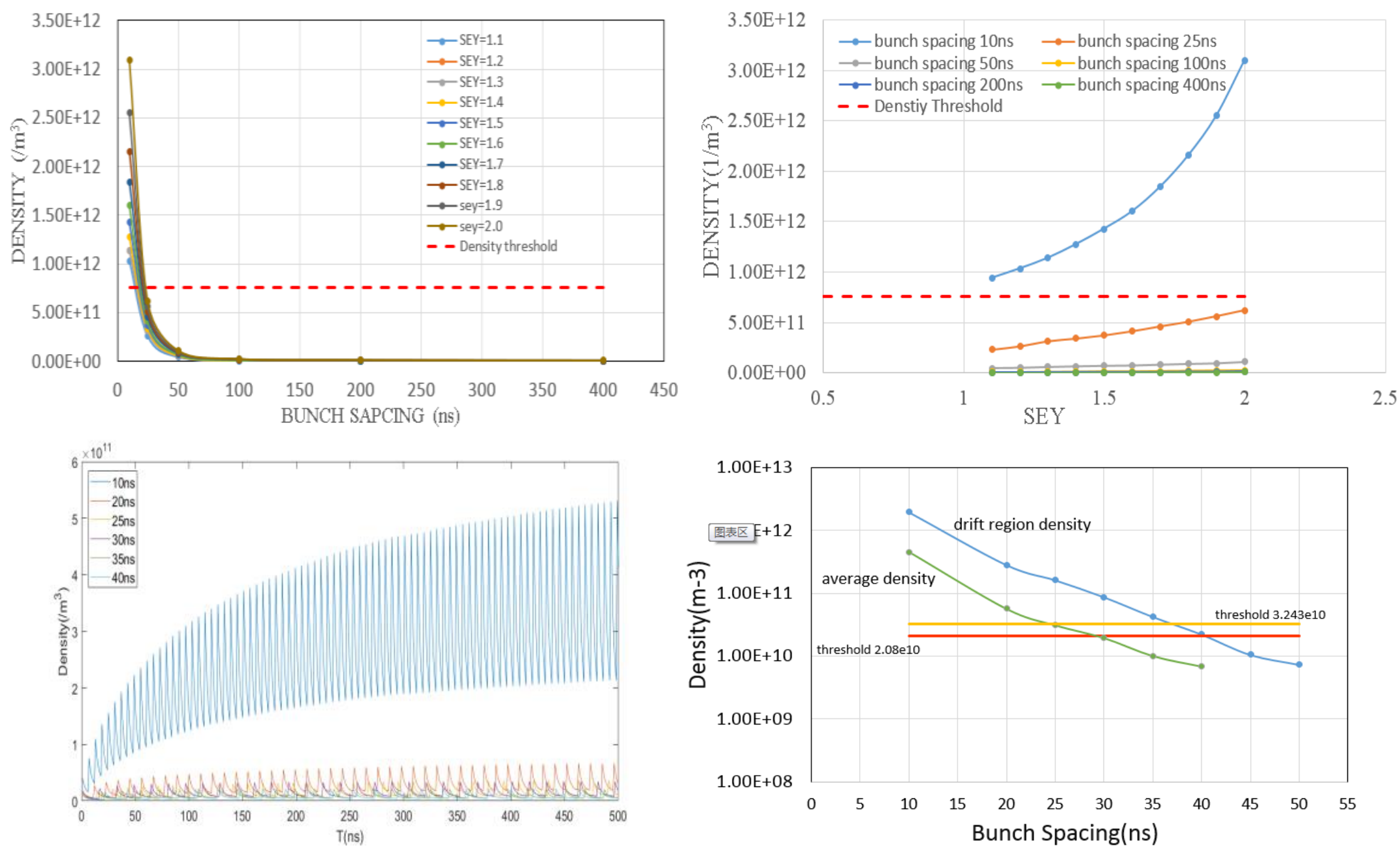

**Figure 4.2.3.6:** Electron cloud density for Higgs (Top) and Z (Bottom) with different SEY and bunch spacing.

In the strong head-tail instability model, the electron cloud threshold density for single bunch instability can be expressed as $\rho_{e,th}=(2\gamma\nu_s\omega_e\sigma_z/c)/(\sqrt{3}KQr_eC\beta)$, where $K=\omega_e\sigma_z/c$, $Q$ depends on the nonlinear interaction, and $\omega_e$ the electron oscillation frequency. Taking $Q=\pi/\sqrt{3}$, the estimate of the threshold density are summarized in Table 2.2.5-8. A SEY lower than 1.6 and bunch spacing longer than 25 ns will be enough to eliminate the electron cloud instability. Since NEG coating will be used on the inner surface of the beam pipe, it is possible to keep the maximum SEY below 1.1.

The electron cloud can also link the oscillation between subsequent bunches and may lead to coupled bunch instability. The action propagated by the EC between subsequent bunches can be presented as a wake field expressed as $W_{ec,x,y}/L=4\pi\rho_{e,neutr}/n_b$, which gives the dipole component per unit length of the wake field. Based on the wake field, the growth rate for the coupled bunch instability is calculated as $1/\tau_{e,CB}=(2r_en_bc^2)/(\gamma\omega_\beta abL_{sep})$. With a simulated EC density at SEY=1.6 and various bunch spacing, the growth times are estimated and listed in Table 2.2.5-8. A transvers feedback system with a damping time of a millisecond is required to suppress this instability.

### 4.2.3.5 *Beam-Ion Instability*

In the electron ring, instabilities can be excited by residual gas ions accumulated in the potential well of the electron beam. Fast beam ion instability is a transient beam instability excited by the beam-generated ions accumulated in a single passage of the bunch train.

The equilibrium average ion density and growth time for different beam operation scenarios are calculated and listed in Table 4.2.3.9. Here, a characteristic damping time equal to the ion oscillation frequency has been used. For the Z, the most critical case, the evolution of the beam ion density and the beam emittance with number of turns are shown in Figure 4.2.3.7. The result shows that the instability is faster than the radiation damping.

Therefore, a transverse feedback system is required for Z. Multi-bunch train filling patterns will also help to damp the instability.

**Table 4.2.3.9:** Average ion density and FBII growth time

| Parameter | Symbol, unit | H | W | Z |
|---|---|---|---|---|
| Average ion density | $\rho_{ion}$, $\times 10^{11}$m$^{-3}$ | 0.4 | 1.0 | 2.7 |
| FBII growth time | $\tau$, ms | 51.6 | 19.6 | 2.3 |

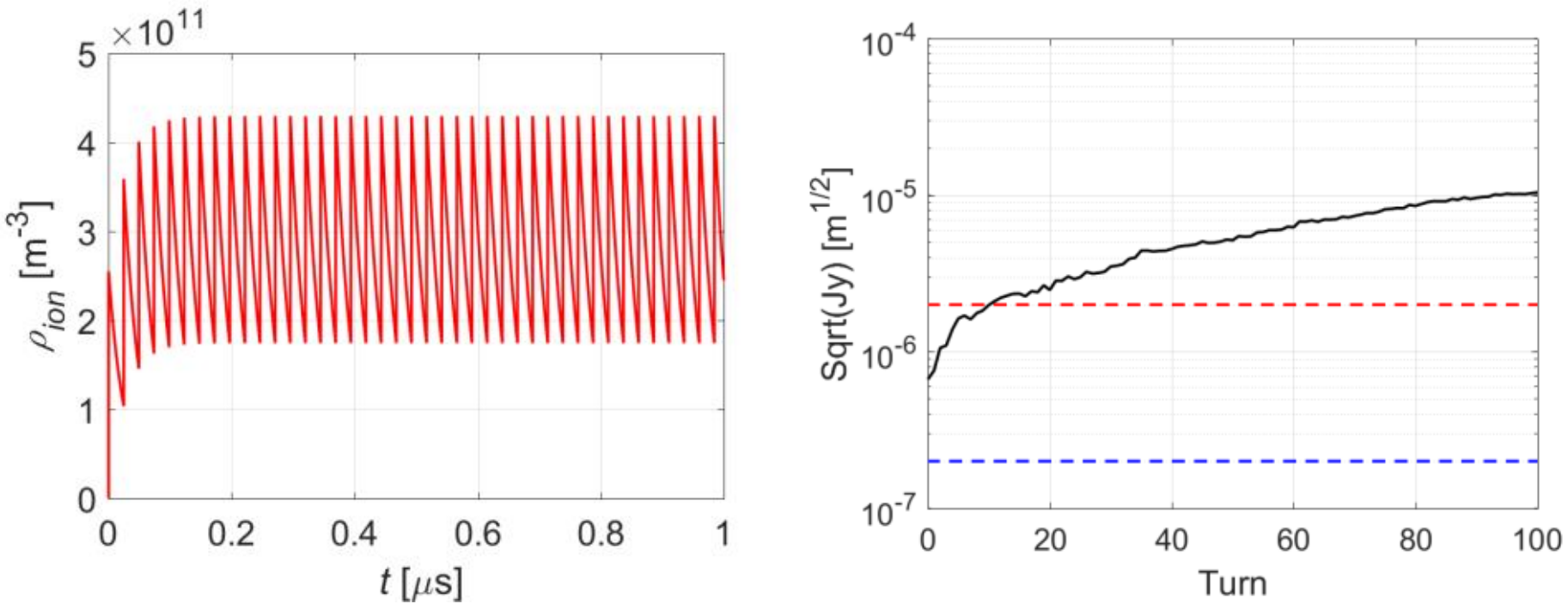

**Figure 4.2.3.7:** Evolution of the beam ion density (left) and the beam emittance (right) with number of turns for Z. The red dashed line corresponds to $1\sigma_y$ and blue dashed line corresponds to $0.1\sigma_y$.

## 4.2.4 Synchrotron Radiation

### 4.2.4.1 *Introduction*

Synchrotron radiation (SR), emitted as relatively low energy photons is the main source of undesirable energy loss and heat deposition [1]. SR in the collider has two key features. First, it has a wide energy spectrum extending from visible light through the energy range of ordinary diagnostic X-rays (hundreds of keV) up to several MeVs [2-3]. Second, it has tremendous power, usually a thousand times more than from regular beam loss [4]. Therefore, SR contributes significantly high radiation dose rates to nearby components and tunnel air [5-6]. That can cause many problems, such as heating of the vacuum chamber and air, radiation damage and formation of ozone and nitrogen oxides. To prevent damage caused by SR, the vacuum chamber must be specially designed or shielded [4].

### 4.2.4.2 *Synchrotron Radiation from Bending Magnets*

A large amount of synchrotron radiation is emitted when the electrons and positrons pass though the dipole magnets, or though the fields of the quadrupole magnets.

The total energy loss per unit length can be expressed by:

$$\Delta E = 14.08\frac{E^4}{\rho^2} \qquad (4.2.4.1)$$

where $\Delta E$ is the total energy loss per unit length in keV/m, $E$ is the energy of electrons and positrons in GeV, $\rho$ is the bending radius in meters.

The total power of the SR emitted by the electrons and positrons per turn can be calculated by:

$$P = 88.46\frac{E^4 I}{\rho} \tag{4.2.4.2}$$

where $P$ is the synchrotron power loss in watts, $I$ is the current of the circulating particles in mA.

Combing these two equations, the total SR power per unit length is:

$$P = 14.08\frac{E^4 I}{\rho^2} \tag{4.2.4.3}$$

The critical energy of the synchrotron spectrum divides the emitting radiation power into two parts and characterizes the "hardness" of the radiation. It is defined by the following expression:

$$E_c = 2.218\frac{E^3}{\rho} \tag{4.2.4.4}$$

where $E_c$ is the critical energy in keV. The photon spectrum for a single electron is given by:

$$\frac{d^2N}{d\varepsilon dt} = \left(\frac{2\alpha}{h\sqrt{3}}\right)\left(\frac{1}{\gamma^2}\right)\int_r^\infty K_{5/3}(\eta)d\eta \tag{4.2.4.5}$$

where $\alpha$ is the fine structure constant, $h$ is Planck's constant, $\gamma$ is the ratio of total energy $E$ to rest mass of the moving $e^{\pm}$, and $K$ is the modified Bessel function order 5/3.

This means that a single electron radiates at the rate of:

$$\frac{d^2N}{d\varepsilon dt} = (\frac{5.32\times 10^5}{E^2})\int_r^\infty K_{5/3}(\eta)d\eta \tag{4.2.4.6}$$

Since ds=c dt, the loss along the orbit for a single electron is:

$$\frac{d^2N}{d\varepsilon ds} = (\frac{1.775\times 10^{-3}}{E^2})\int_r^\infty K_{5/3}(\eta)d\eta \tag{4.2.4.7}$$

The total number of photons emitted by an electron per meter is:

$$\frac{dN}{ds} = \int_0^\infty E_c\left(\frac{d^2N}{d\varepsilon ds}\right)dr = 3.936\left(\frac{E}{\rho}\right)\int_0^\infty\int_r^\infty K_{5/3}(\eta)d\eta dr = 19.4\frac{E}{\rho} \tag{4.2.4.8}$$

The synchrotron radiation spectra, $R(k)$, in units of photons per MeV per electron per meter of bending length is expressed as:

$$R(k) = \frac{cS(x)}{x} \tag{4.2.4.9}$$

where $c$ is a constant, $3809.5/E^2$ and $E$ is the beam energy in GeV.

The universal function $S(x)$ can be calculated by the following formula:

$$S\left(\frac{\omega}{\omega_c}\right) = 0.4652\frac{\omega}{\omega_c}\int_{\omega/\omega_c}^\infty K_{5/3}(\eta)d\eta \tag{4.2.4.10}$$

where $\omega$ is the angular frequency of the synchrotron photon in rad/s, and $\omega_c$ is the angular frequency of the critical energy photon in rad/s.

Photons are emitted tangentially to the curved trajectory of the beam into a cone with vertical opening angle given by:

$$\varphi = \frac{1}{\gamma} = \frac{m_0 c^2}{E} \tag{4.2.4.11}$$

The parameters for SR emitted by 120 GeV and 175 GeV beams in the Collider dipoles are shown in Table 4.2.4.1.

**Table 4.2.4.1**: Parameters of Collider SR

| Parameters | Symbols | Values | | Units |
|---|---|---|---|---|
| Beam energy | $E$ | 120 | 175 | GeV |
| Beam current | $I$ | 17.4 | 3.84 | mA |
| Bending radius | $\rho$ | 10900 | | m |
| Power per unit length | $P$ | 435 | 427 | W/m |
| Critical energy | $E_c$ | 351.6 | 1090.6 | keV |
| Bending angle | $\theta$ | 2.844 | | mrad |
| Opening angle | $\varphi$ | 4.258 | 2.92 | μrad |

SR is strongly linearly polarized in the orbit plane, which means its distribution is asymmetric in the beam-vertical plane. Therefore, there is a bending angle between the direction of photons and the direction of the beam. 85% of the SR power is distributed within an opening angle.

Fig. 4.2.4.1 shows the spectra of SR photons in the dipoles at different electron energies: 45.5 GeV, 80 GeV, 120 GeV and 175 GeV. From the figure, we can see how an energy change modifies the photon spectrum.

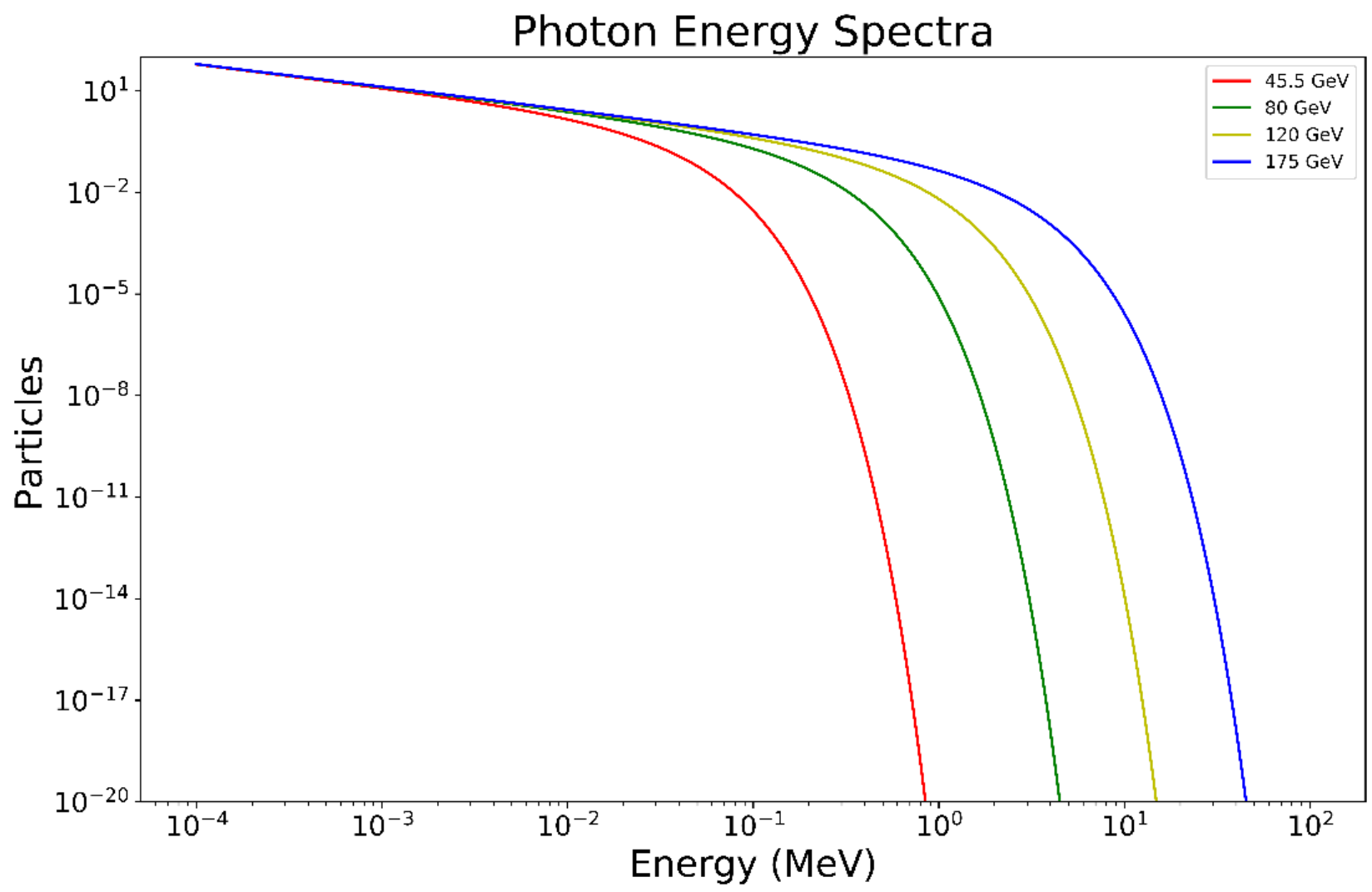

**Figure 4.2.4.1**: The photon spectrum emitted by different beam energies

We use 120 GeV to calculate the number of photons. Using Eq. 4.2.4.8:

$$N_{photon} = 6.25 \times 10^{15} \cdot I \cdot N \cdot t = 1.2125 \times 10^{17} \cdot \frac{EIt}{\rho} \tag{4.2.4.12}$$

where $I$ is the beam current in mA, $t$ is the operation time assumed to be one second. Therefore, the total number of photons is $2.363 \times 10^{16}$ per meter per second.

The average energy of SR photons is 116 keV.

### 4.2.4.3 *Monte Carlo Simulation*

We use the theoretical formulas above to calculate the energy and power distribution of the photons and use the energy spectrum, position and direction of these photons as the source in the simulation.

In Fig. 4.2.4.1, we saw that the energetic photons change rapidly with increasing energy. Though the amount is small, it gives a significant contribution to the radiation environment. We must consider this very carefully since the slope is steep. On the other hand, the share of the lowest energy component is also small. We assume that the lowest energy of incident photons is 1 keV. Therefore, all the photons with energy lower than 1 keV were set to 1 keV in the simulation and the calculation. This gives a conservative estimate of heat and dose. The spectrum used in the simulation is from 120 GeV and is calculated using a Python script.

Next, we looked at the position and direction of the photons. In a Cartesian coordinate system, the cross section of the vacuum chamber is in the x-y plane and is asymmetric in this plane. For the direction, we use the bending and opening angles from Table 4.2.4.1. Compared with the bending angle, the opening angle itself is small. Considering that the size of the vacuum chamber is several centimeters, the difference including the opening angle is several micrometers or even less. So we can assume that the opening angle was zero in the simulation.

Most SR power is converted into heat deposited in the metal. The energy deposition in the tunnel air away from the vacuum chamber and magnets, is recorded as well as the dose distribution in and around the vacuum chamber and the magnet. The cumulative dose is in unit of Gy.

### 4.2.4.4 *Energy Deposition caused by SR*

The SR power of a single beam, 435 W/m, is a significant loss and may cause damage to accelerator systems through direct radiation damage or deposition of heat. Magnets and other components must be protected from this damage.

The vacuum chamber is composed of three millimeters of aluminum or copper with a simple elliptical cross section, 75 mm×56 mm. The vacuum chamber for positrons is copper to minimize the electron-cloud effect, and the vacuum chamber for electrons is aluminum to save cost. Multiple pieces of 20 mm thick lead are placed on the outside of the vacuum chamber to absorb partial energy deposition and protect magnet coils from radiation damage. In the quadrupole, the lead is placed on the top and bottom of the vacuum chamber. Fig. 4.2.4.2 shows a dipole with lead shielding. Fig. 4.2.4.3 shows the lead shielding in quadrupole and sextupole.

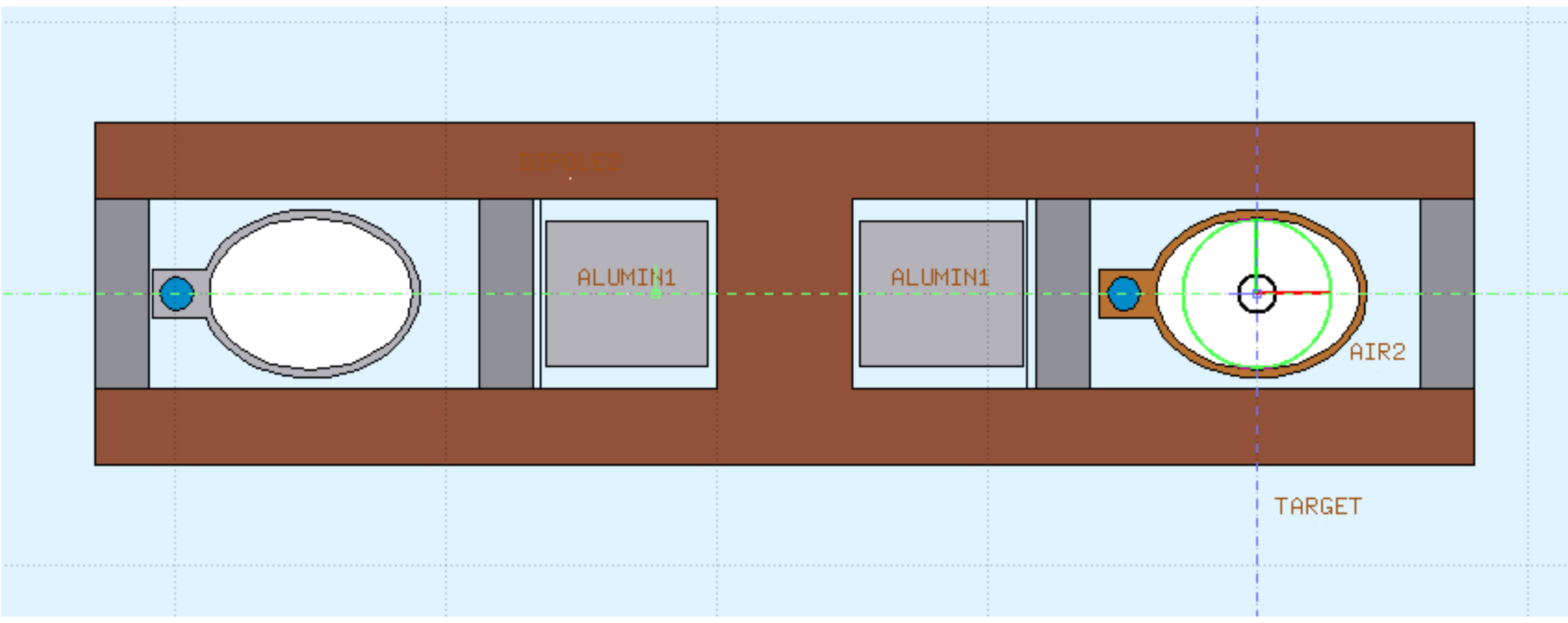

**Figure 4.2.4.2**: Dipole with lead shielding

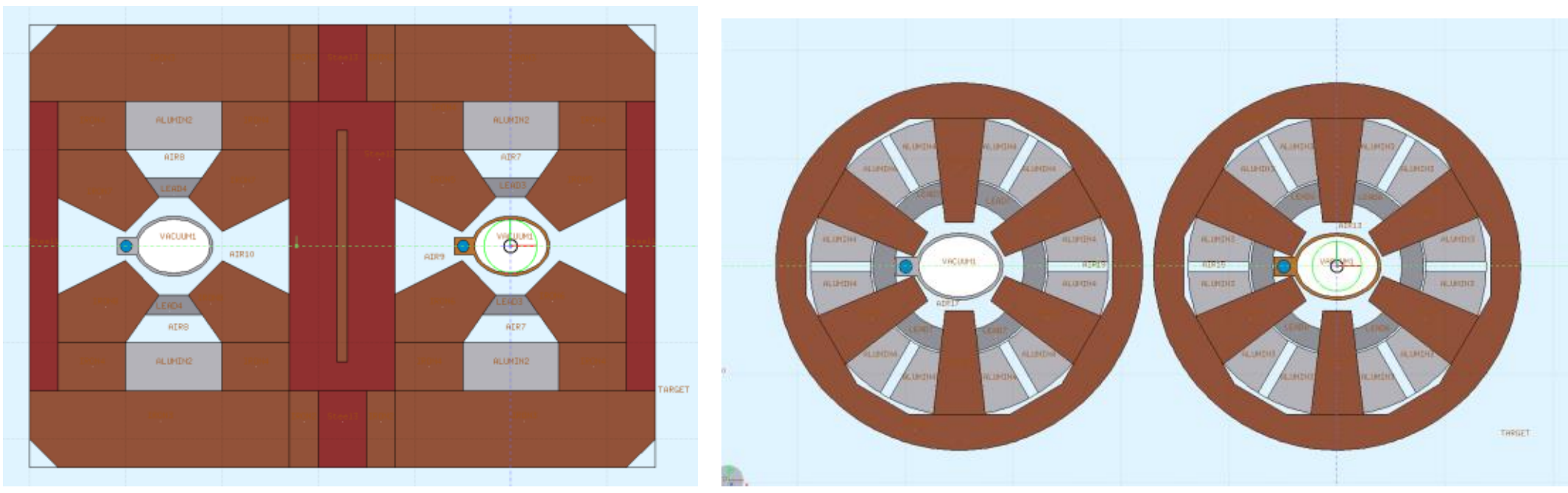

**Figure 4.2.4.3:** Left – quadrupole with lead shielding; right – sextupole with lead shielding.

FLUKA is used to simulate the energy deposition caused by photons. Without shielding, the values of energy deposition in dipole, aluminum and copper chambers and stainless steel flange are shown in Table 4.3.4.2. The flange is located among dipole every 6 meters, each flange is 4 cm long. The values of energy deposition in magnets, vacuum chamber and lead shielding are shown in Table 4.3.4.3. More than 95% of the heat is deposited in the components.

The photon energy deposited in cooling water, which will induce radionuclides. There are many reaction channels for photons interact with oxygen element. While the threshold energies is high, more than ten MeVs, the amount of radionuclides is less. We will estimate the amount of radionuclides in the future.

**Table 4.3.4.2:** Energy deposition in the dipole

| Beam direction: left W/m | | Beam direction: right W/m | |
|---|---|---|---|
| Al chamber | 199 | Al chamber | 186 |
| Cu chamber | 309 | Cu chamber | 332 |
| Dipole | 255 | Dipole | 252 |
| Stainless steel flange | 54 | Stainless steel flange | 54 |

**Table 4.3.4.3:** Energy deposition in magnets (A: dipole B: quadrupole C: sextupole)

A

| Beam direction: left W/m | | Beam direction: right W/m | |
|---|---|---|---|
| Al chamber | 199 | Al chamber | 186 |
| Cu chamber | 308 | Cu chamber | 332 |
| Dipole | 186 | Dipole | 182 |
| Lead A | 60.6 | Lead A | 29.2 |
| Lead B | 33.5 | Lead B | 80.0 |
| Lead C | 46.8 | Lead C | 18.8 |
| Lead D | 14.3 | Lead D | 20.4 |

B

| Beam direction: left W/m | | Beam direction: right W/m | |
|---|---|---|---|
| Quadrupole | 279 | Quadrupole | 268 |
| Lead A | 37.8 | Lead A | 36.4 |
| Lead B | 18.1 | Lead B | 21.7 |

C

| Beam direction: left W/m | | Beam direction: right W/m | |
|---|---|---|---|
| Sextupole | 179 | Sextupole | 174 |
| Lead A | 95.1 | Lead A | 107 |
| Lead B | 60.3 | Lead B | 43.1 |

### 4.2.4.5 *Dose Estimation for the Collider Ring*

The tunnel geometry is set up with magnets installed in the tunnel over a length of 31 meters. The 3 mm thickness pipe wall is too thin to shield effectively by itself. Therefore, a lead shield is added on both sides of the vacuum chamber. Using Fluka we study the propagation of synchrotron radiation in the magnets, the gap, and the tunnel and determine the photon spectra streaming out from the different sections. Fig. 4.2.4.4 shows the dose distributions around the coils without lead shielding. The maximum radiation dose around the coils of dipole, quadrupole and sextupole are $2.34\times10^6$ Gy/Ah, $4.69\times10^5$ Gy/Ah, and $3.25\times10^6$ Gy/Ah. The dose distributions around the coils with 20 mm lead shielding are shown in Fig. 4.2.4.5. The maximum radiation dose around the coils of dipole, quadrupole and sextupole are $1.89\times10^4$ Gy/Ah, $0.82\times10^4$ Gy/Ah, and $1.67\times10^4$ Gy/Ah.

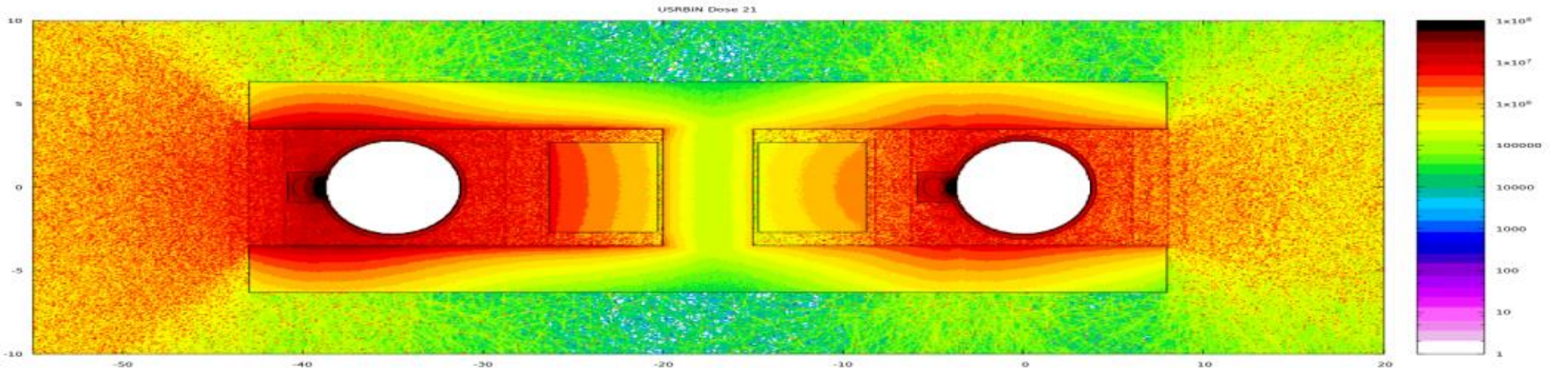

A

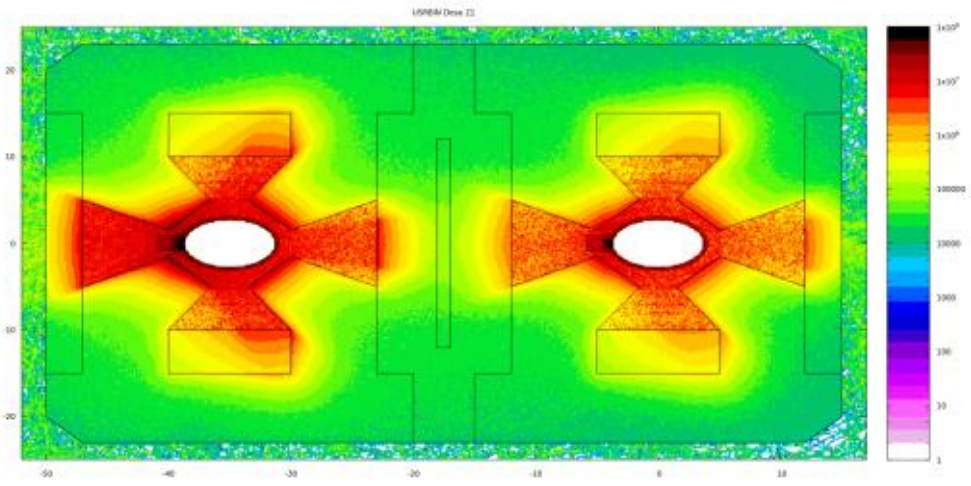

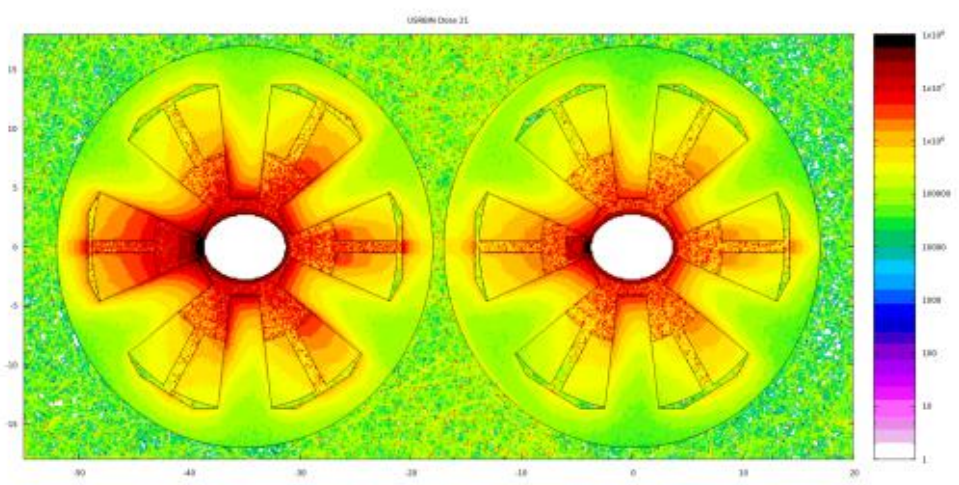

B C

**Figure 4.2.4.4**: Radiation dose around the magnets without lead shielding

(A: dipole B: quadrupole C: sextupole)

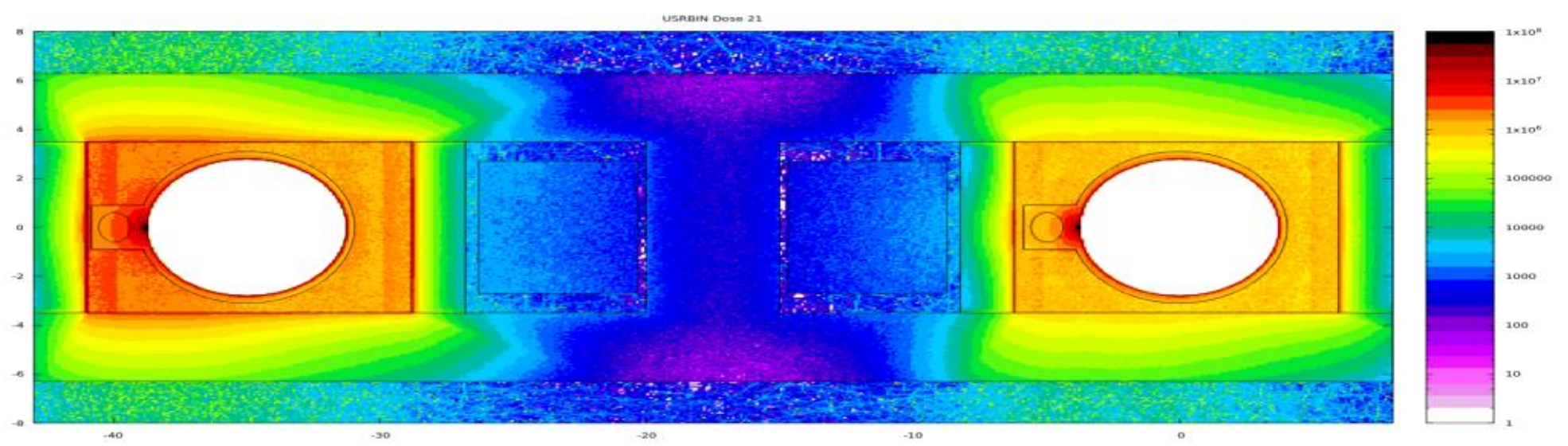

A

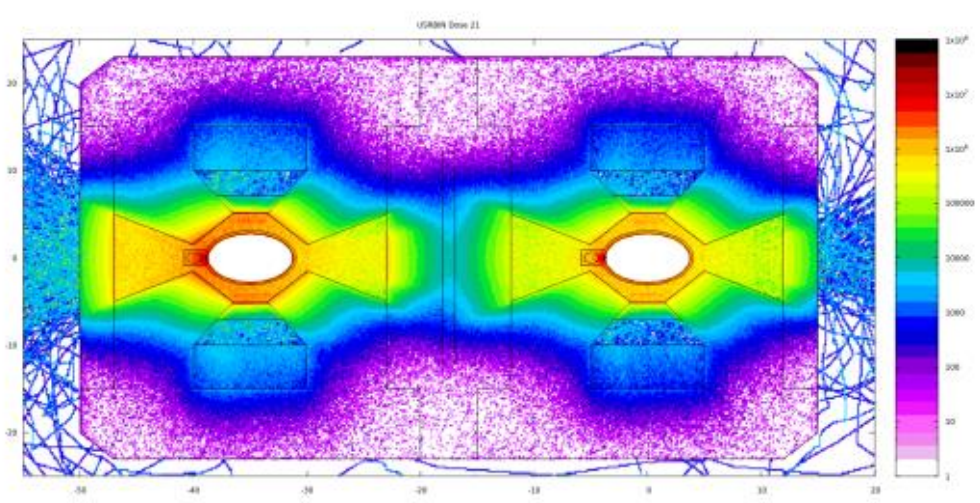

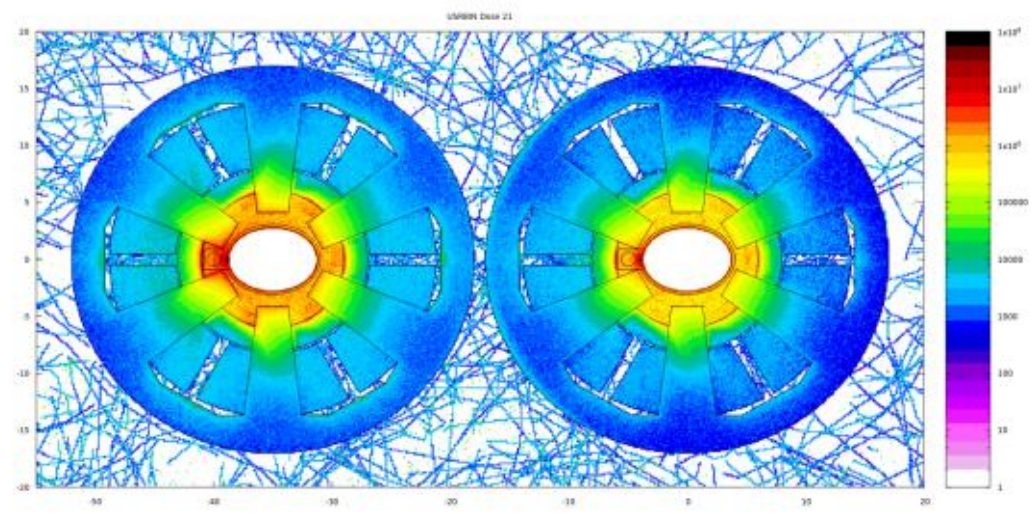

B C

**Figure 4.2.4.5**: Radiation dose around magnets with lead shielding

(A: dipole B: quadrupole C: sextupole)

Fig. 4.2.4.6 is an example at a beam energy of 120 GeV normalized to Gy/Ah. By adding 20 mm thickness lead shielding, the radiation dose around specific equipment such as control electronics, lighting, wiggler magnets, control and instrumentation cables close to the beam have been calculated.

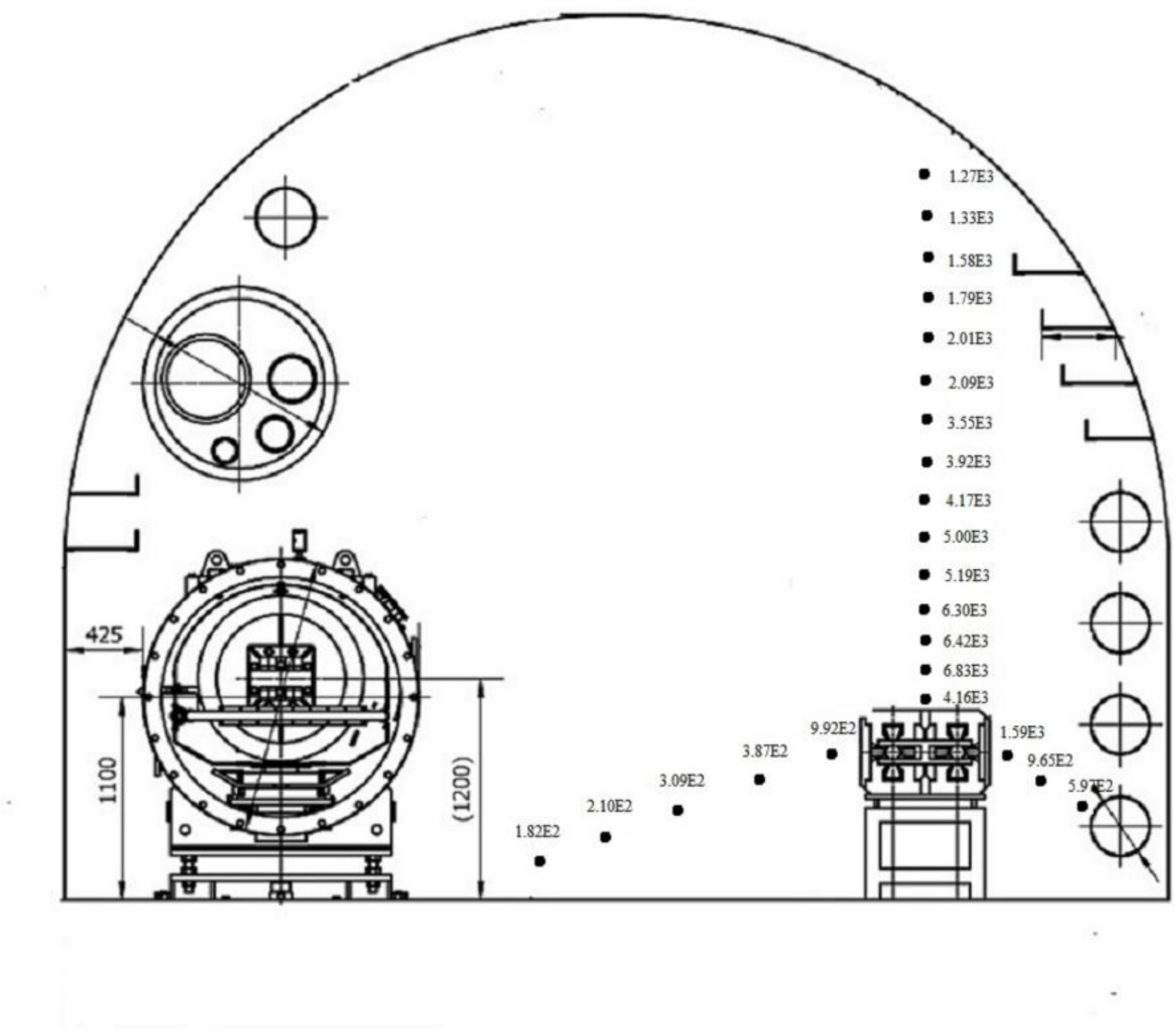

**Figure 4.2.4.6**: Dose distribution in Gy/Ah at 120 GeV in a dipole section with shielding

### 4.2.4.6 *Radiation Damage*

Radiation damage studies have been performed on organic and inorganic materials used in accelerator electronic components and metals. The organic materials are the most sensitive to radiation. For now we use the radiation damage data from CERN. At present, we are most concerned about the polymer material wrapped around the magnet coil. There are three kinds of insulation material, fiberglass, epoxy resin, semi-organic coating. The upper dose limits are assumed to be $2\times10^7$ Gy and $1\times10^8$ Gy.

**Table 4.3.4.4**: Lifetime for magnets coils

| Materials | Upper dose limit in Gy | Radiation dose in Gy/Ah | Time in h |
|---|---|---|---|
| Fiberglass | $10^8$ | $1.89\times10^4$ | $2.99\times10^5$ |
| Semi-organic coating | $10^8$ | $1.89\times10^4$ | $2.99\times10^5$ |
| Epoxy resin | $2\times10^7$ | $1.89\times10^4$ | $5.98\times10^4$ |

## 4.2.5 Injection and Beam Dump

### 4.2.5.1 *General Description*

The injection system consists of a 10 GeV Linac, followed by a full-energy Booster ring. Electron and positron beams generated and accelerated to 10 GeV in the Linac, are injected into the Booster. The beams are then accelerated to full-energy and injected into the Collider. A schematic of the injection system is in Fig. 4.2.5.1.

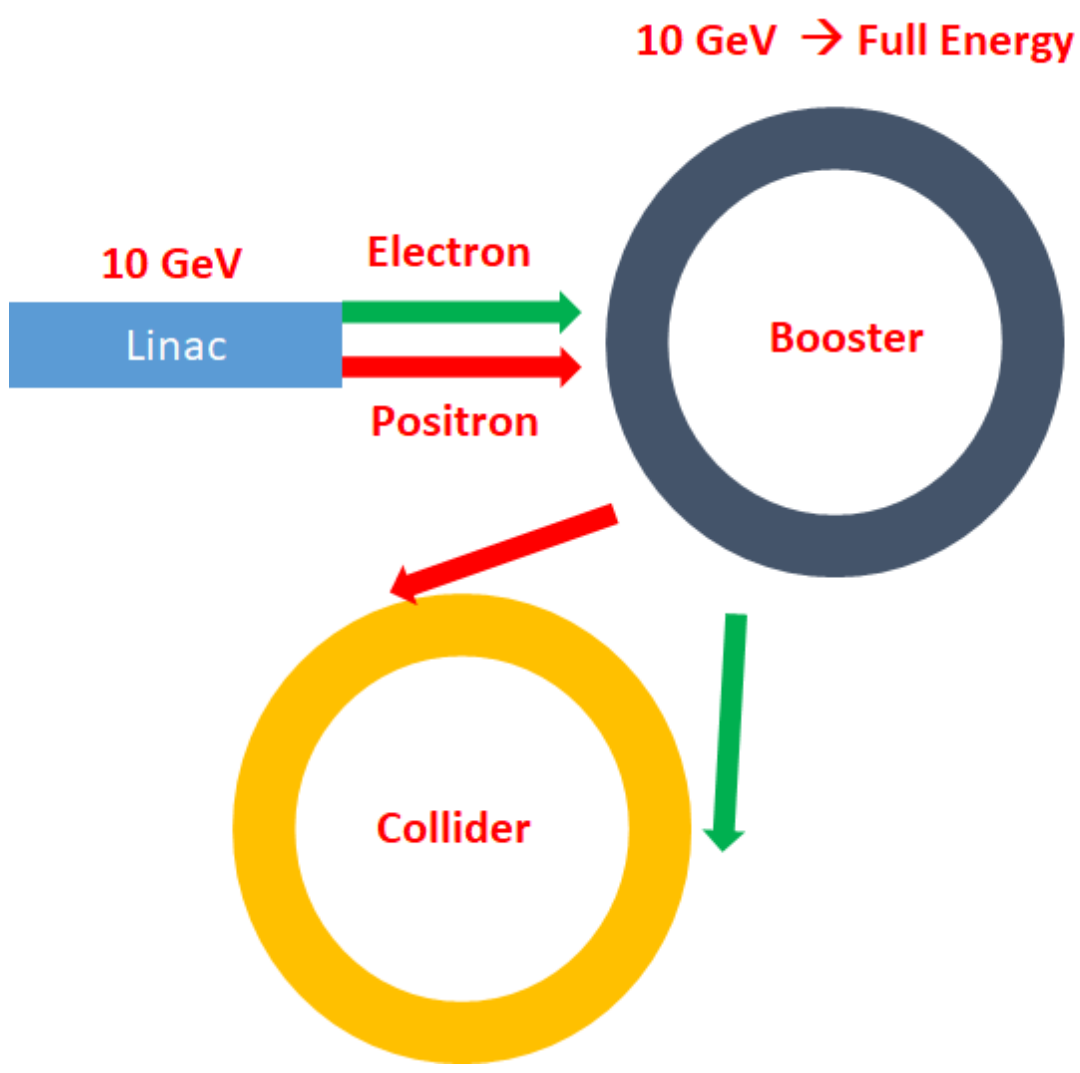

**Figure 4.2.5.1:** CEPC injection system.

For different beam energies for Higgs, W, and Z experiments, there will be different particle bunch structures in the Collider. Collider parameters important for the injection system are shown in Table 4.2.5.1. A traditional off-axis injection scheme is chosen as a baseline design of the beam injection to the main collider, and a swap-out injection is given as another choice for Higgs injection.

**Table 4.2.5.1:** Collider parameters important for the injection system.

| | Higgs | W | Z |
|---|---|---|---|
| Energy (GeV) | 120 | 80 | 45.5 |
| Bunch number | 242 | 1524 | 12000 |
| Bunch Charge (nC) | 24 | 19.2 | 12.8 |
| Bunch Current (mA) | 17.4 | 87.9 | 461 |
| Revolution Period (ms) | 0.3336 | 0.3336 | 0.3336 |
| Emittance x/y (nm) | 1.21/0.0031 | 0.54/0.0016 | 0.18/0.004 |
| Life time (Hour) | 0.43 | 1.4 | 4.6 |

### 4.2.5.2 *Off-axis Injection*

To maximize the integrated luminosity, the injection system will operate mostly in top-up mode, but also has the ability to fill the Collider from empty to full charge in a reasonable length of time. To keep the Booster current low enough, it takes several ramping cycles of the Booster to fill every bunch at W and Z mode energies. Some bunch parameters for injection have been listed in Table 4.2.5.2.

The standard top-up injection operation keeps the Collider beam current within ±3% of its nominal value. It takes the Booster 2 ramping cycles in Z mode, and one in others. An injection time structure diagram is plotted in Figure 4.2.5.2. The injection cycle frequency depends on the different beam lifetimes for different colliding modes. Filling the Collider from zero current can be done at higher injection cycle frequency.

**Table 4.2.5.2:** Injection bunch parameters for different working modes

| Mode | Higgs | W | Z |
|---|---|---|---|
| Injection Energy (GeV) | 120 | 80 | 45.5 |
| Bunch number | 242 | 1524 | 6000 |
| Bunch Charge (nC) | 0.72 | 0.576 | 0.384 |
| Beam Current (mA) | 0.523 | 2.63 | 6.9 |
| Number of Cycles | 1 | 1 | 2 |
| Current decay | 3% | 3% | 3% |
| Ramping Cycle (sec) | 10 | 6.6 | 3.8 |
| Beam damping time (sec) | 0.5 | 1 | 5 |
| Filling time (sec) | 25.84 | 45.68 | 275.2 |
| Injection frequency (sec) | 47 | 153 | 504 |

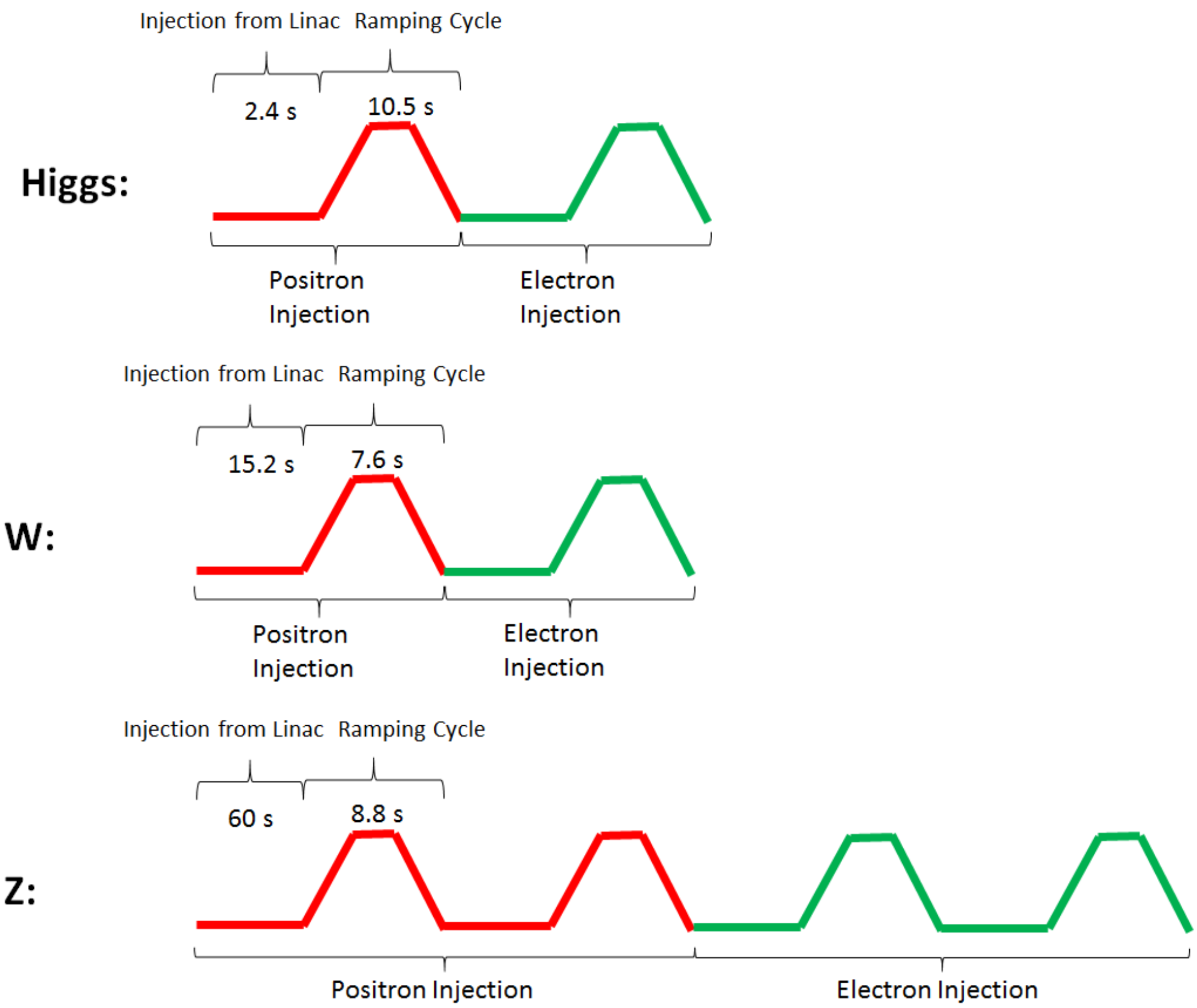

**Figure 4.5.2.2:** Injection time structure for Higgs, W and Z modes

For the off-axis injection scheme, the Twiss parameters at the exit of the transport line are optimized to minimize the required Collider acceptance. A phase-space diagram at the septum is shown in Figure 4.2.5.3. The bump height and machine aperture are determined from the septum width and beam sizes, as shown in Table 4.2.5.3.

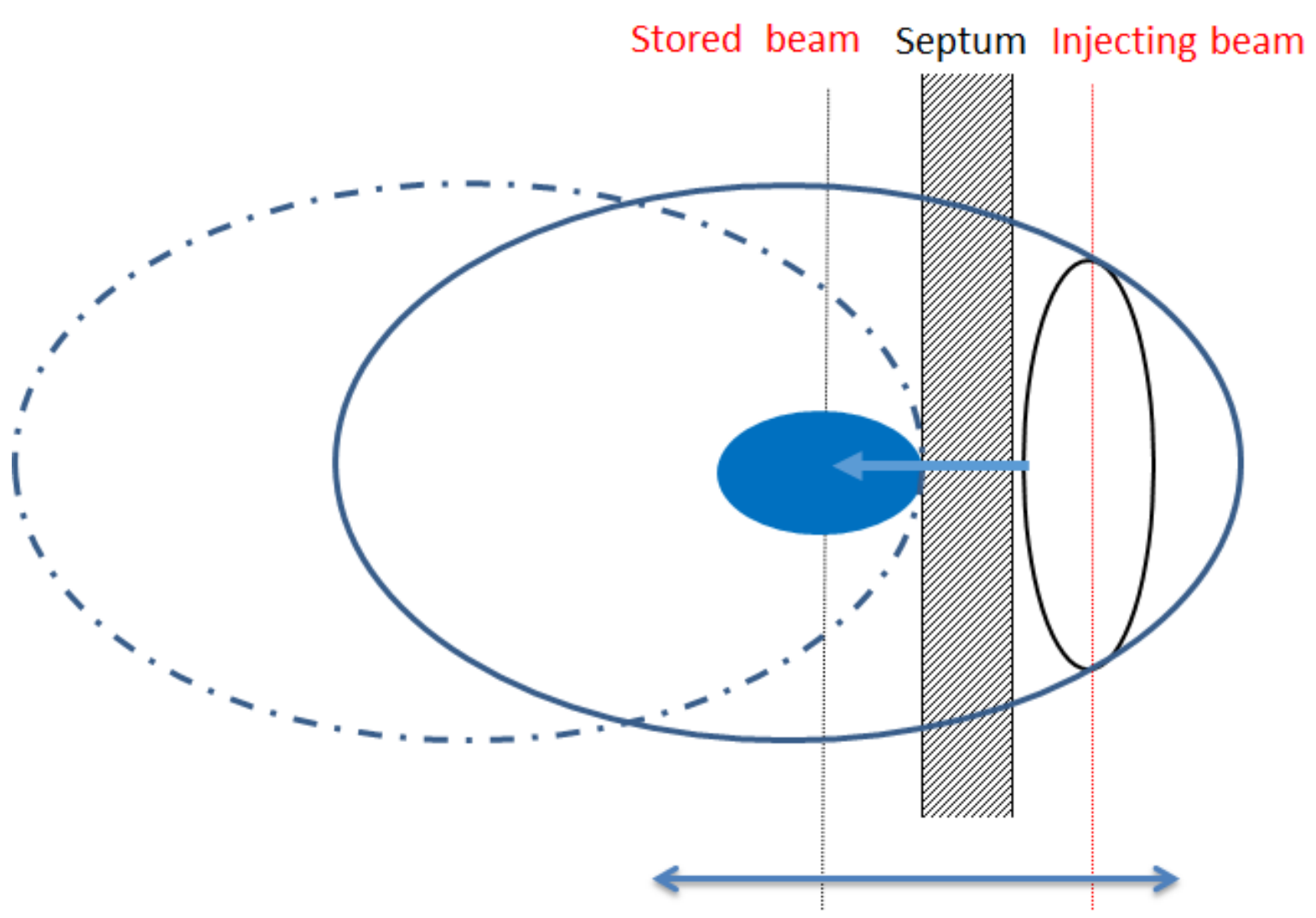

**Figure 4.2.5.3:** Phase space at the septum point

**Table 4.2.5.3:** Injection parameters of septum, beam and aperture

| Injection Energy | Higgs | W | Z |
|---|---|---|---|
| Septum width | 2 mm | 2 mm | 2 mm |
| Stored beam size | 0.85 mm | 0.57 mm | 0.32 mm |
| Injected beam size | 0.9 mm | 0.6 mm | 0.34 mm |
| Required aperture | 13.3 σ | 14.3 σ | 16.7 σ |

### 4.2.5.3 *Swap-out Injection*

At Higgs energy, the dynamic aperture in the main collider becomes much smaller than that at W and Z energies. So a swap-out on axis injection scheme is proposed to relax the requirements on DA. A diagrammatic sketch of this injection process is shown in figure 4.2.5.4. In the injection, first fill the booster with 3% bunch charge from linac, and ramp the booster up to 120 GeV, then several circulating bunches of the collider ring will be extracted to the booster while keeping the beam current in the booster under the limitation of 1mA. The circulating bunches of booster will be merged with the injected bunches from collider ring after 4 damping times, those bunches can be injected back into the same bucket in the collider. This procedure will be repeated several times so that all the circulating bunches of booster can be accumulated in the collider ring.

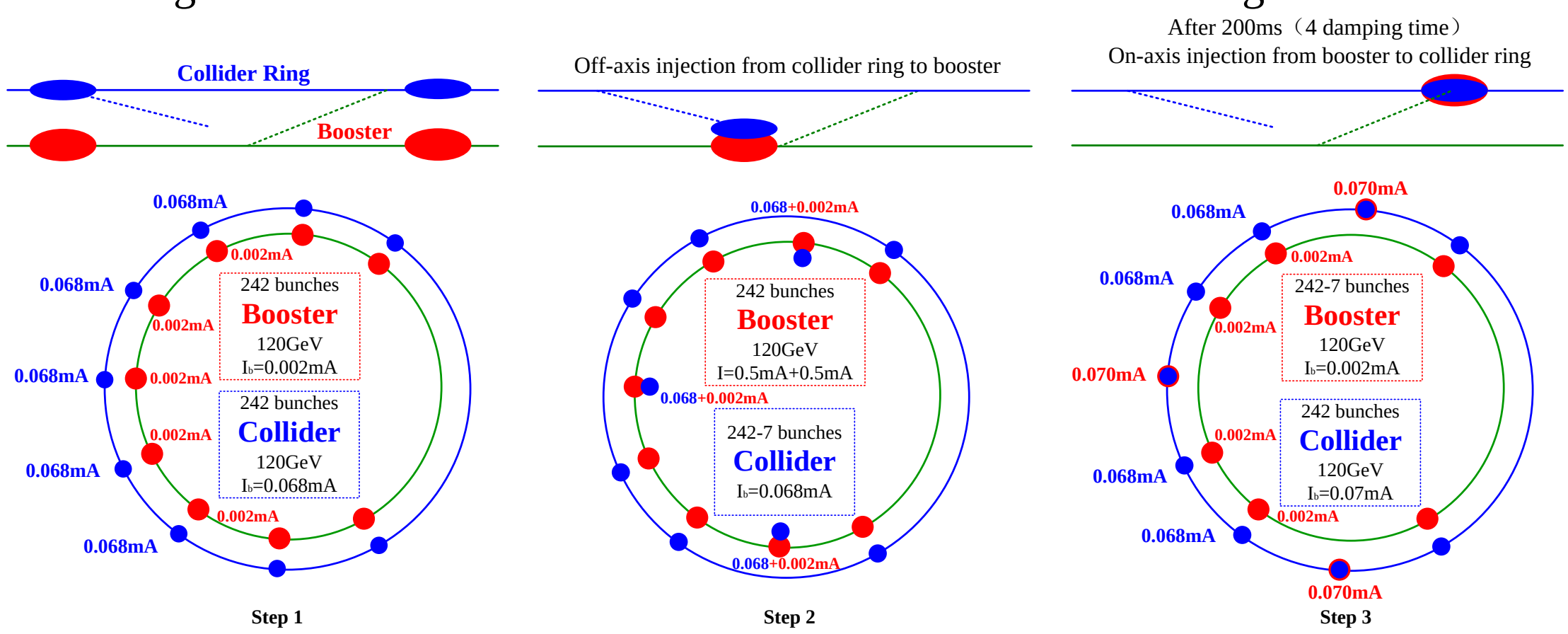

**Figure 4.2.5.4:** A diagrammatic sketch of the on axis injection process.

A strong-strong beam-beam simulation of the RMS size evolution of colliding bunches after injection has been done, as shown in figure 4.2.5.5. The simulation result shows that the asymmetry between stored beam(collider) and injected beam(booster) does not excite any flip-flop instability due to the strong radiation damping.

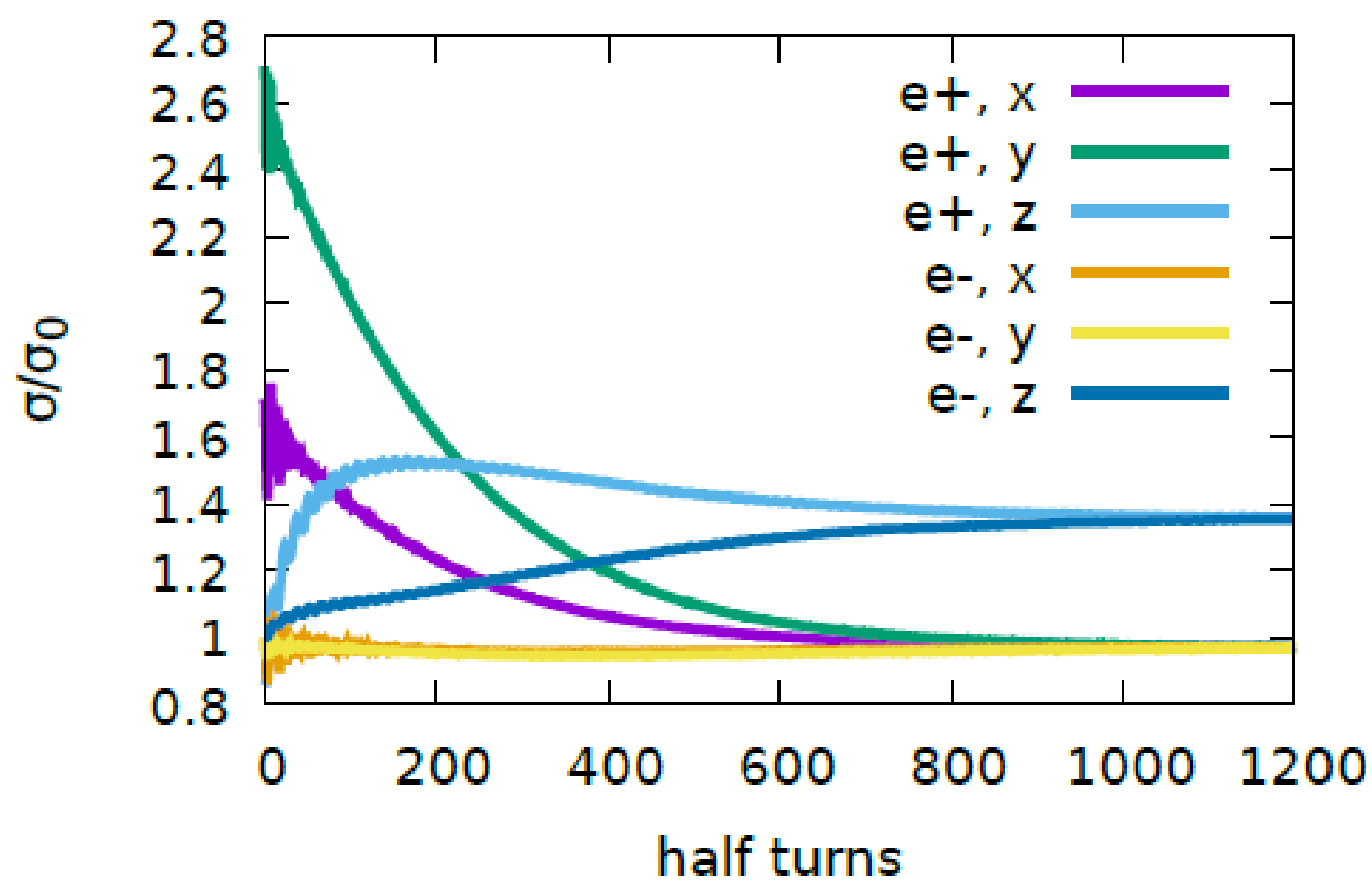

**Figure 4.2.5.5:** RMS size evolution of colliding bunches after on-axis injection.

### 4.2.5.4 *Beam Dump*

Due to the high stored energy in the Collider beams, it is important to be able to dump the beams when necessary. The beam dump system consists of a fast kicker and two beam absorbing blocks, as shown in Figure 4.5.2.4. These systems are installed in the straight section between IP3 and IP4. The kicker magnet is located right next to a defocusing quadrupole, and the two absorbing blocks are placed symmetrically on the two sides of the kicker, about 50 m from the kicker. The kicker magnet deflects both the electron and positron beams in the vertical direction; the deflection angle can change during the dumping process. In beam dumping, a vertical deflection can also be provided by off-axis passage in the horizontal focusing quadrupoles next to the kicker.

The rise time of the kicker should be fast enough so that the kicker can rise to its full strength between the two bunches passing through the kicker. This is less than 35 ns. To avoid beam energy deposition at a single point, the fall time of the kicker should be long enough so that different bunches experience gradually diminishing deflection during the 330 μs revolution time.

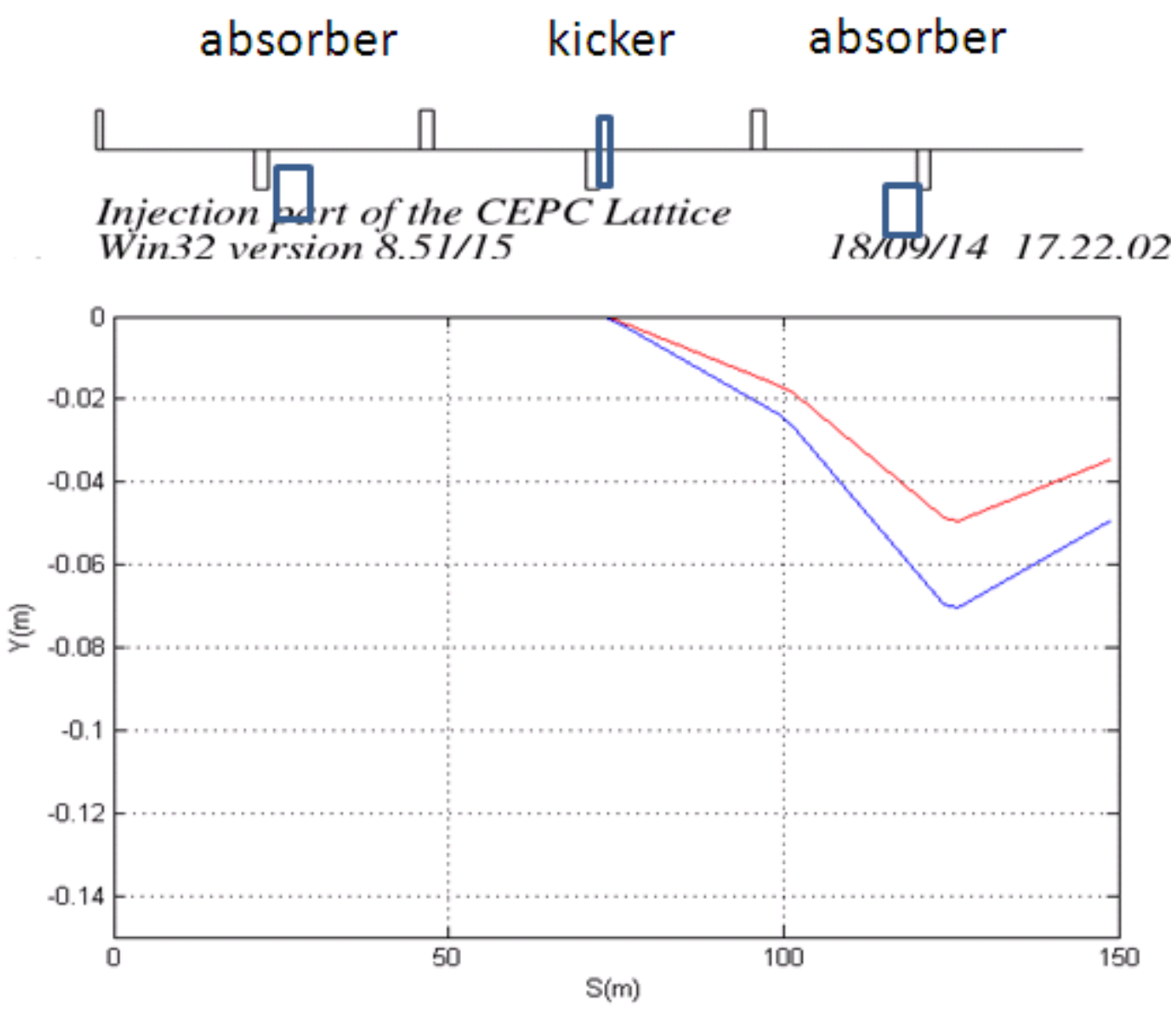

**Figure 4.2.5.6:** A schematic of the dump system and a plot of beam orbits for different deflecting angles.

### 4.2.6 **Machine-Detector Interface**

#### 4.2.6.1 ***Interaction Region Layout***

The machine-detector interface is about ±7 m in length in the IR as can be seen in figure 4.2.6.1, where many elements need to be installed, including the detector solenoid, luminosity calorimeter, interaction region beam pipe, beryllium pipe, cryostat and bellows. The cryostat includes the final doublet superconducting magnets and anti-solenoid. The CEPC detector consists of a cylindrical drift chamber surrounded by an electromagnetic calorimeter, which is immersed in a 3T superconducting solenoid of length 7.6 m. The accelerator components inside the detector should not interfere with the devices of the detector. The smaller the conical space occupied by accelerator components, the better will be the geometrical acceptance of the detector. From the requirement of detector, the conical space with an opening angle should not larger than 8.11 degrees. After optimization, the accelerator components inside the detector without shielding are within a conical space with an opening angle of 6.78 degrees. The crossing angle between electron and positron beams is 33 mrad in horizontal plane. The final focusing quadrupole is 2.2 m from the IP [1]. The luminosity calorimeter will be installed in a longitudinal location 0.95~1.11 m, with an inner radius of 28.5 mm and outer radius 100 mm. Primary results are got from the assembly, interfaces with the detector hardware, cooling channels, vibration control of the cryostats, supports and so on.

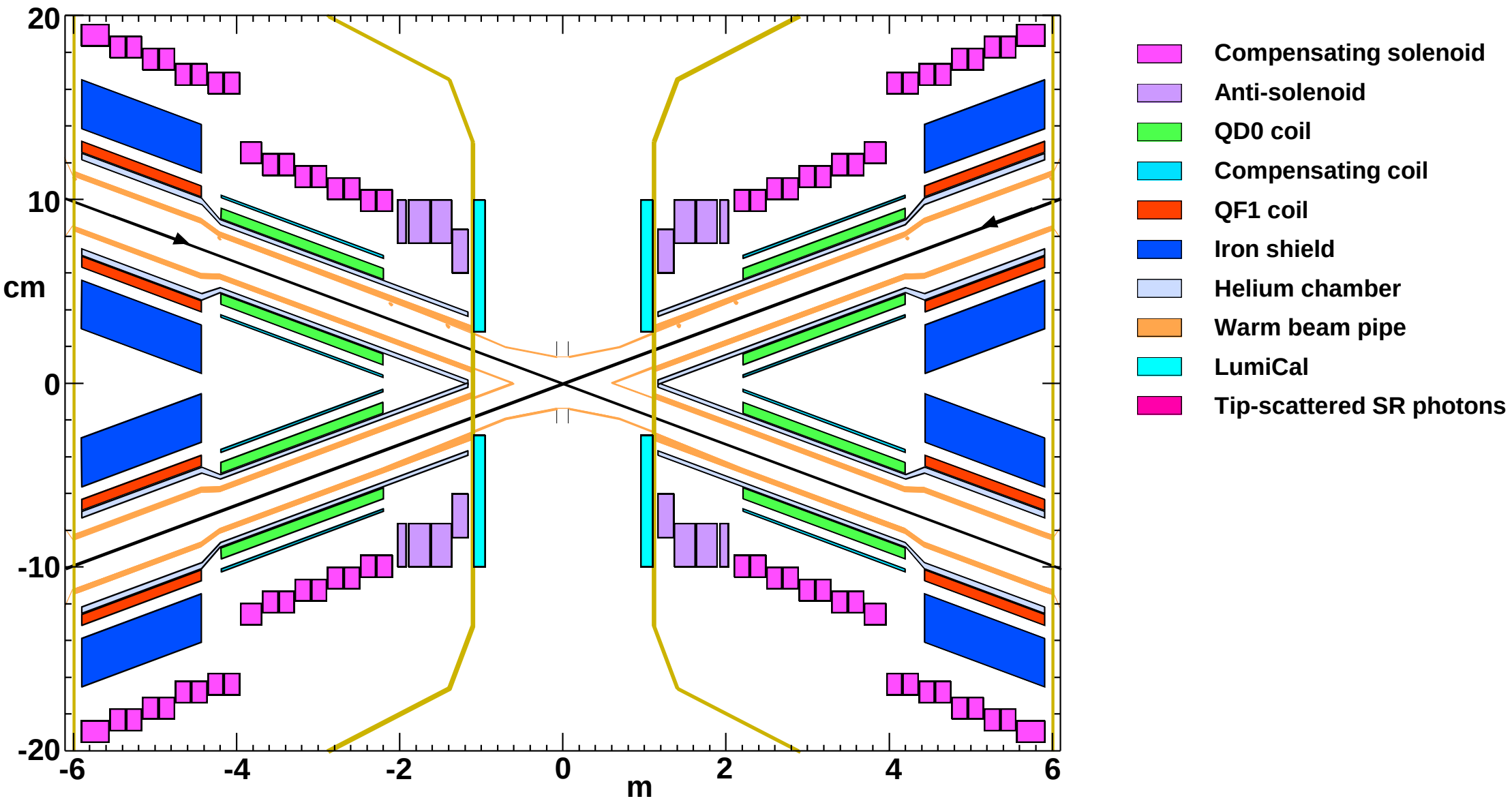

**Figure 4.2.6.1:** CPEC IR layout.

#### 4.2.6.2 ***Beam Pipe***

To reduce the detector background and radiation dose from beam loss, the vacuum chamber has to accommodate the large beam stay clear region. In order to keep precise shaping, all these chambers will be manufactured with computer controlled machining and carefully welded to avoid deformation.

The inner diameter of the beryllium pipe is chosen as 28 mm taking into account both mechanical assembly and beam background issues. The length of beryllium pipe is 14 cm in longitudinal. Due to bremsstrahlung incoherent pairs, the shape of the beam pipe between 0.2~0.5 m is selected as conic. There is a bellows for the requirements of installation in the crotch region, located about 0.7 m from the IP. A water cooling structure is required to control the heating problem of HOM. For the beam pipe within the final doublet quadrupoles, since there is a 4mm gap between the outer space of beam pipe and the inner space of Helium vessel [2], a room temperature beam pipe has been chosen.

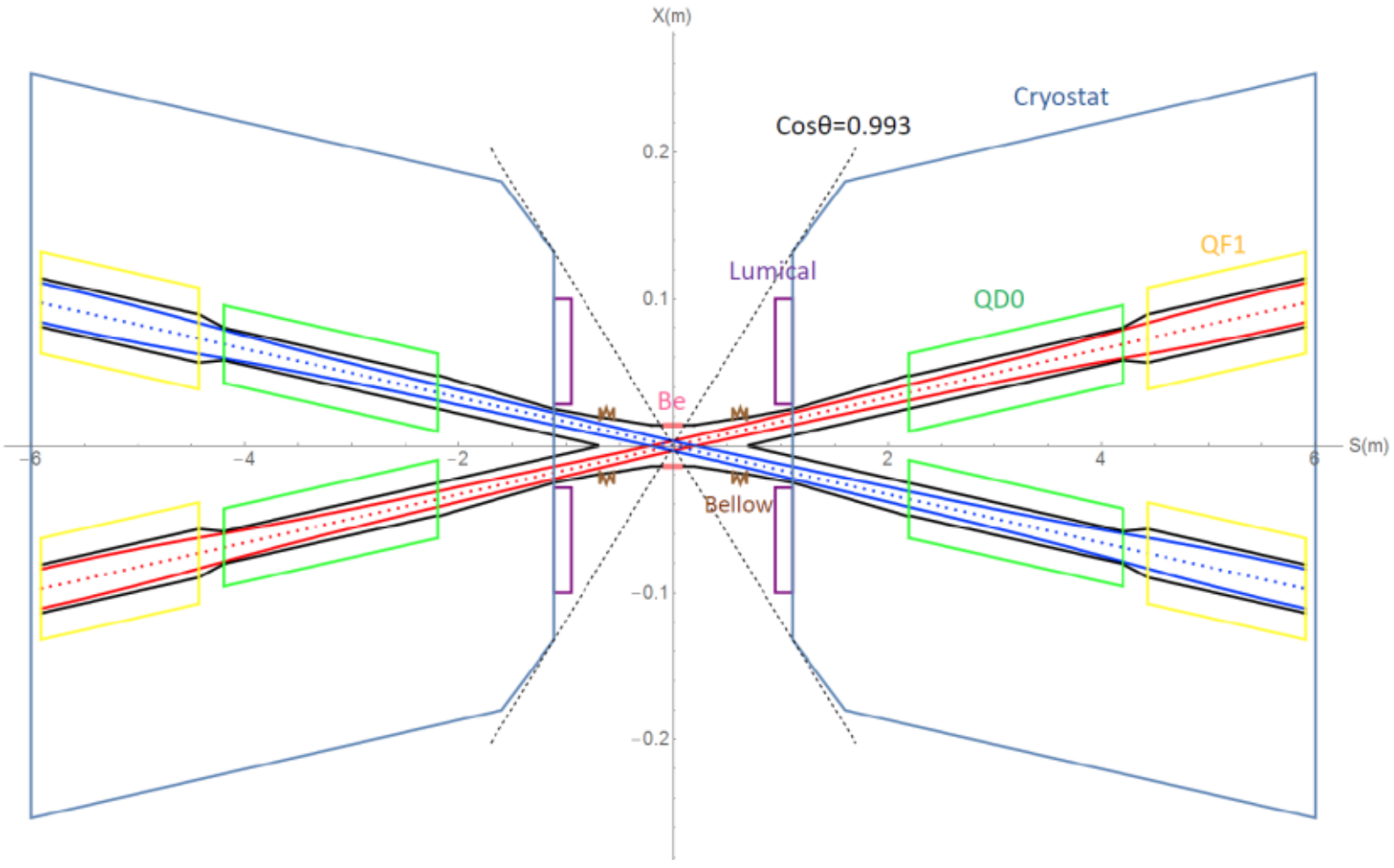

**Figure 4.2.6.2:** IR layout with beam pipes.

### 4.2.6.3 *Synchrotron Radiation*

Synchrotron radiation (SR) photons are emitted in a direction tangential to the particle trajectory [3] and contribute to the heat load of the beam pipe and can cause photon background to the experiments. Furthermore, the radiation dose can damage detector components. Therefore the beam optics should be carefully designed in order to prevent the SR photons from directly hitting or scattering into the detector beam pipe.

The maximum designed single beam current is 17.4 mA and the maximum energy is 120 GeV. The fan of SR photons in the IR are mainly generated from the final upstream bending magnet and the IR quadrupole magnets due to eccentric particles.

#### *4.2.6.3.1 SR from Bending Magnets*

An asymmetric lattice has been selected to allow softer bends in the upstream part of the IP. Reverse bending direction in the final bends avoids SR photons from hitting the IP vacuum chamber. In the upstream part of the IP the SR critical energy is less than 45 keV within 150 m and 120 keV within 400 m. For the downstream part of the IP, there are no bends in the last 50 m and the critical energy is less than 97 keV within 100 m and 300 keV within 250 m. Figure 4.2.6.3 shows the SR fans in the IR produced by the positron beam. The synchrotron radiation generated by electron beam is symmetric.

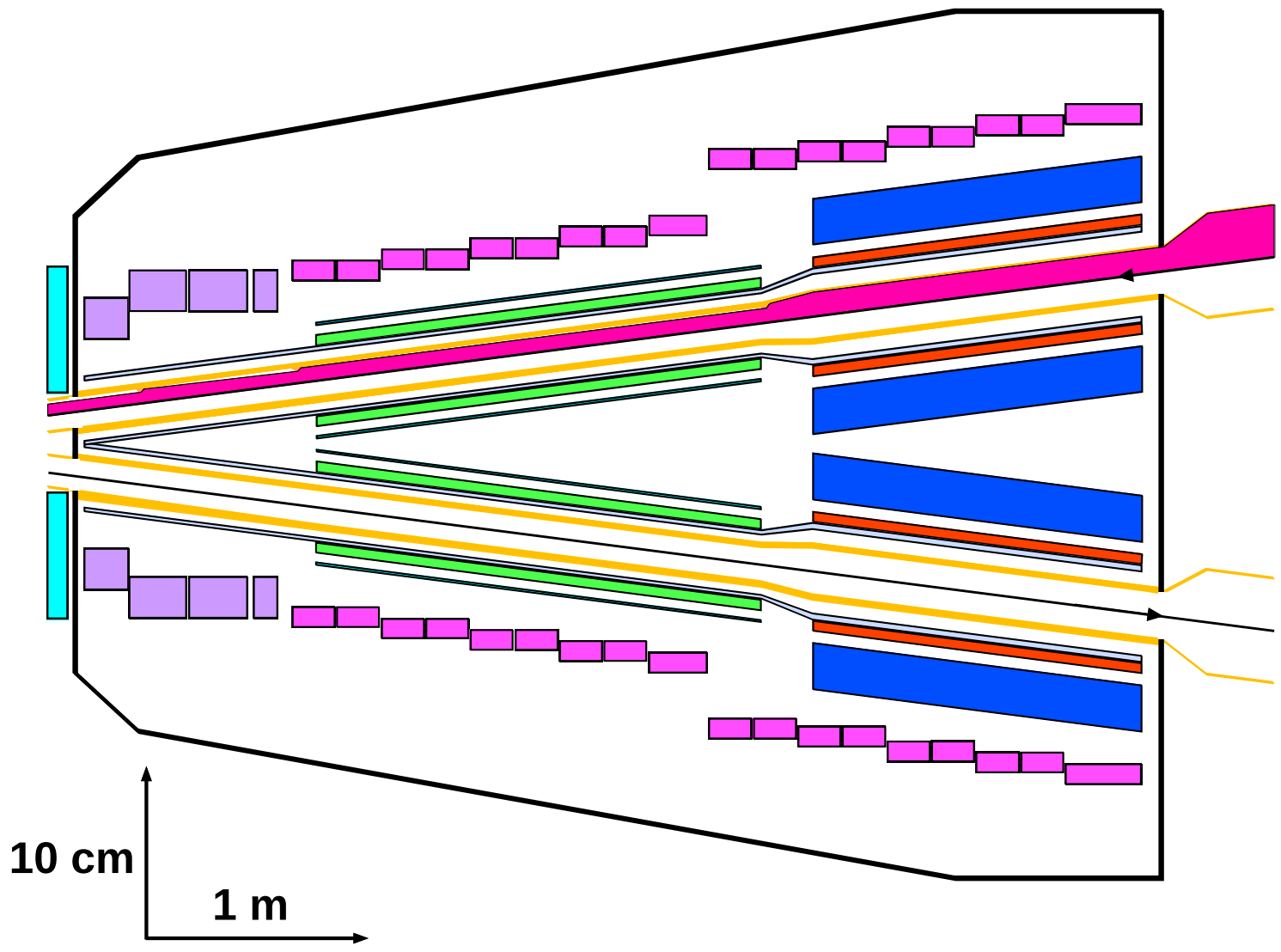

**Figure 4.2.6.3:** SR fans from the last bending magnet in the upstream part of the IR.

The incoming SR fan from the last soft bend magnet strikes the inner surface of the beam pipe inside the cryostat. The total power on the various beam pipe surfaces is listed in the following table.

**Table 4.2.6.1:** SR photons from different parts in IR

| **Surface** | **Power (W)** | **SR photons > 1 keV** |
|---|---|---|
| Under QF1 | 2.51 | $1.01\times10^{9}$ |
| Between QF1 and QD0 | 40.04 | $1.63\times10^{10}$ |
| Under QD0 | 8.08 | $3.26\times10^{9}$ |
| In front of QD0 | 4.45 | $1.80\times10^{9}$ |

A significant fraction of these incident photons will forward scatter from the beam pipe surface and hit the central Be beam pipe (a cylinder located ±7 cm around the IP with a radius of 14 mm). By installing 3 mask tips along the inside of the beam pipe to shadow the inner surface of the pipe the number of scattered photons that can hit the central beam pipe is greatly reduced to only those photons which forward scatter through the mask tips. The optimization of the mask tips (position, geometry and material) is presently under study. Figure 4.2.6.4 shows the present mask tip design in the expanded part of the picture. The current design calls for at least 3 tips for each incoming beam. Figure 4.2.6.5 shows the photon trajectories from the mask tips that can still hit the central beam pipe.

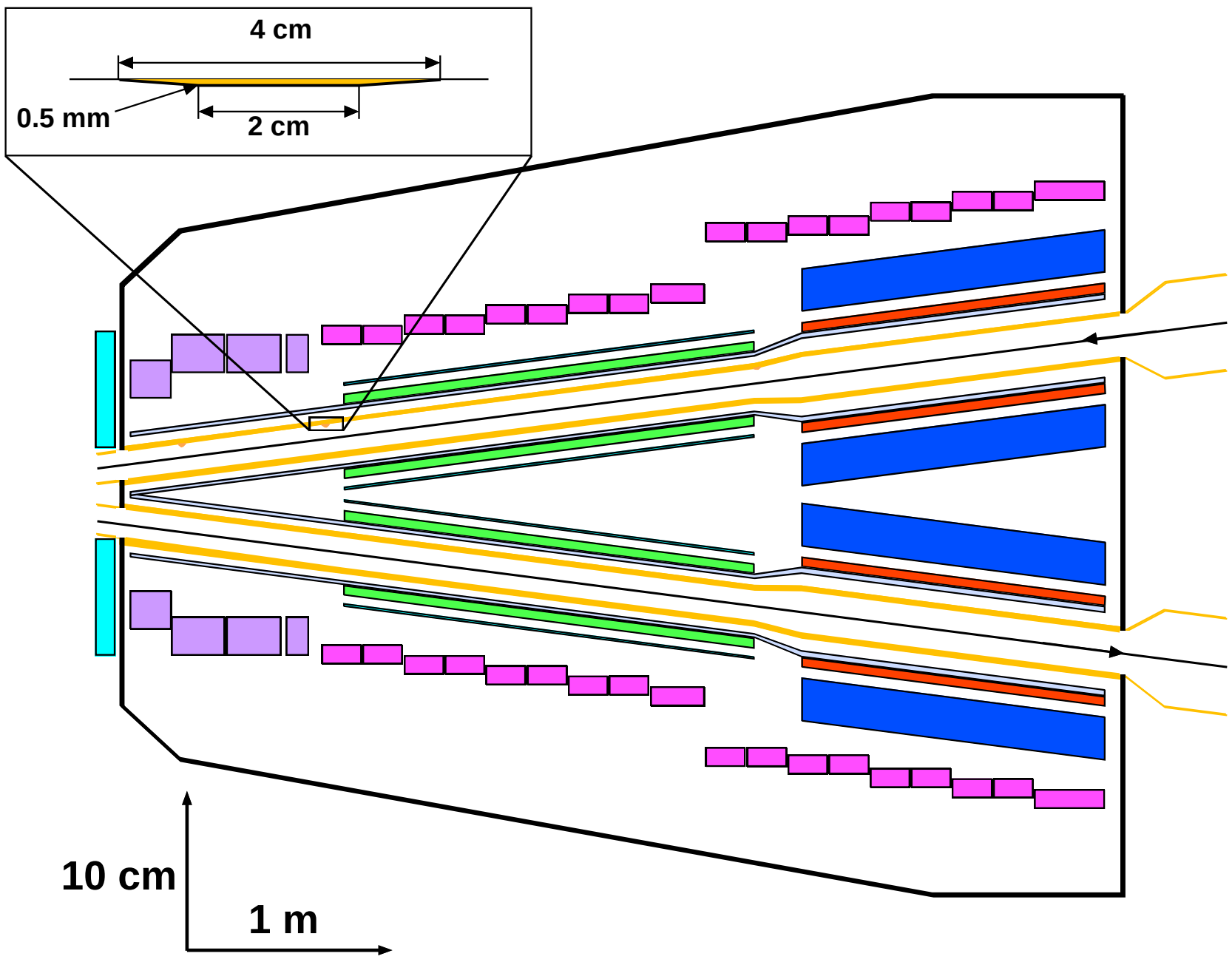

**Figure 4.2.6.4:** Mask tips design in the IR vacuum chamber.

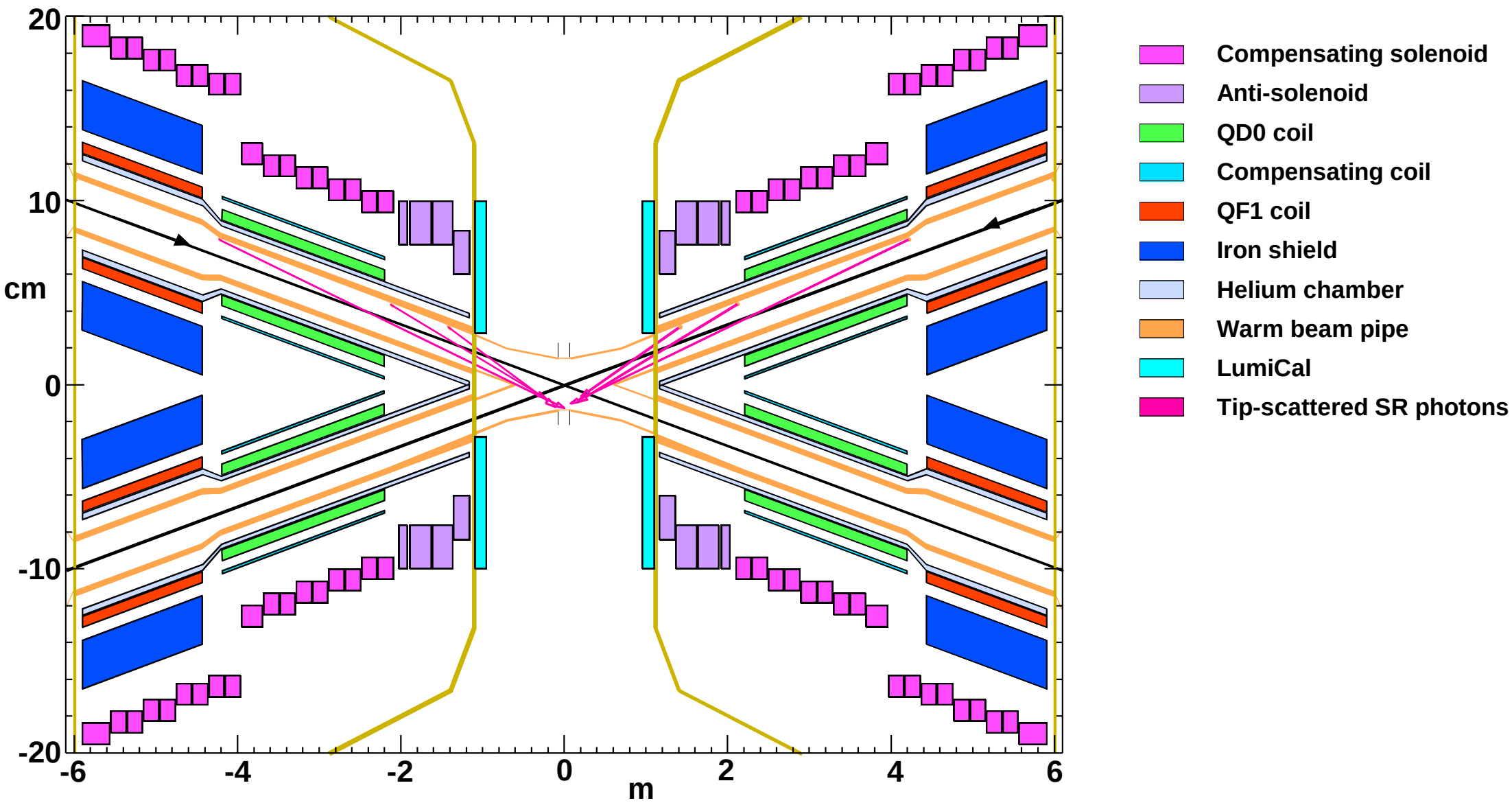

**Figure 4.2.6.5:** Photon trajectories from the mask tips that can still hit the central beam pipe.

*4.2.6.3.2 SR from the Final Doublet Quadrupoles*

The total SR power generated by the QD0 magnet is 639W horizontally and 165W vertically. The photon critical energy is about 1.3MeV horizontally and 397keV vertically. The total SR power generated by the QF1 magnet is 1567W horizontally and 42W vertically. The photon critical energy is about 1.6MeV horizontally and about 225keV vertically.

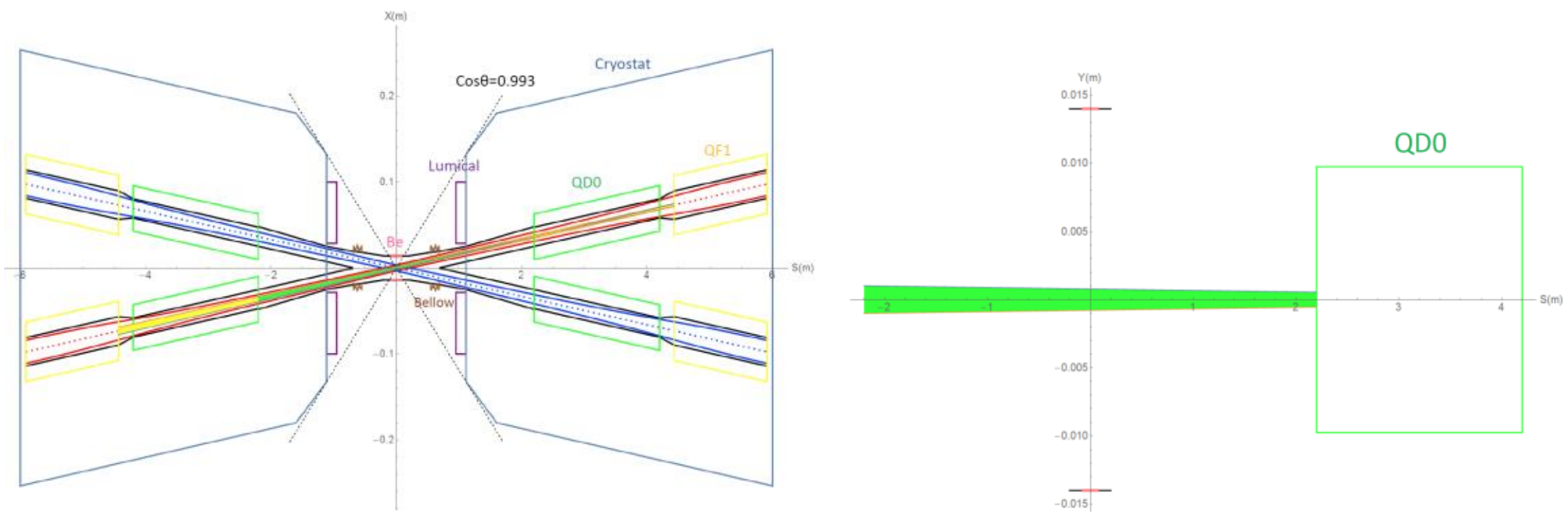

**Figure 4.2.6.6:** SR photon fans horizontally (left) from QD0 (green) and QF1 (yellow), SR photon fans vertically (right) from QD0.

There are no SR photons within $10\sigma_x$ directly hitting or once-scattering to the detector beam pipe. There is collimators for the beam loss background as can be seen in section 4.2.7.2, which will be installed in the upstream and downstream ARC far away from IP. These collimators will squeeze the beam to $13\sigma_x$. The SR photons generated from $10\sigma_x$ to $13\sigma_x$ will hit downstream of the IR beam pipe, and the once-scattering photons will not go into the detector beam pipe but goes to even far away from the IP region. Thus the SR photons from final doublet quadrupoles will not damage the detector components and cause background to experiments.

## 4.2.7 Beam Loss, Background and Collimation

The beam particles can lose a large fraction of their energy through a scattering processes such as radiative Bhabha, beamstrahlung [4], beam-gas scattering, or beam-thermal photon scattering. After optimizing the lattice, and considering the beam-beam effect and errors, the energy acceptance is about 1.5%. If the energy loss of the beam particles is larger than 1.5%, these particles will be lost from the beam and might hit the vacuum chamber. If this happens near the IR, detectors may be damaged. Beam loss production mechanisms and the associated beam lifetimes are listed in Table 4.2.7.1.

**Table 4.2.7.1:** Beam lifetime

| | Beam lifetime | others |
|---|---|---|
| Quantum effect | >1000 h | |
| Touscheck effect | >1000 h | |
| Beam-Gas elastic scattering (Coulomb scattering) | >400 h | Residual gas CO [5], $10^{-7}$ Pa |
| Beam-Gas inelastic scattering (bremsstrahlung) | 63.8 h | |
| Beam-thermal photon scattering | 50.7 h | |
| Radiative Bhabha scattering | 74 min | |
| Beamstrahlung | 80 min | |

The first three, due to the long lifetime, can safely be ignored. The next four, beamstrahlung, radiative Bhabha scattering, beam-thermal photon scattering and beam-gas inelastic scattering, especially beamstrahlung and radiative Bhabha scattering, due to shorter lifetimes, must be carefully analysed and collimated.

### 4.2.7.1 ***Beam Loss Background***

#### *4.2.7.1.1 Radiative Bhabha Scattering*

Radiative Bhabha scattering is simulated by BBBrem [6] or Py_RBB [7].

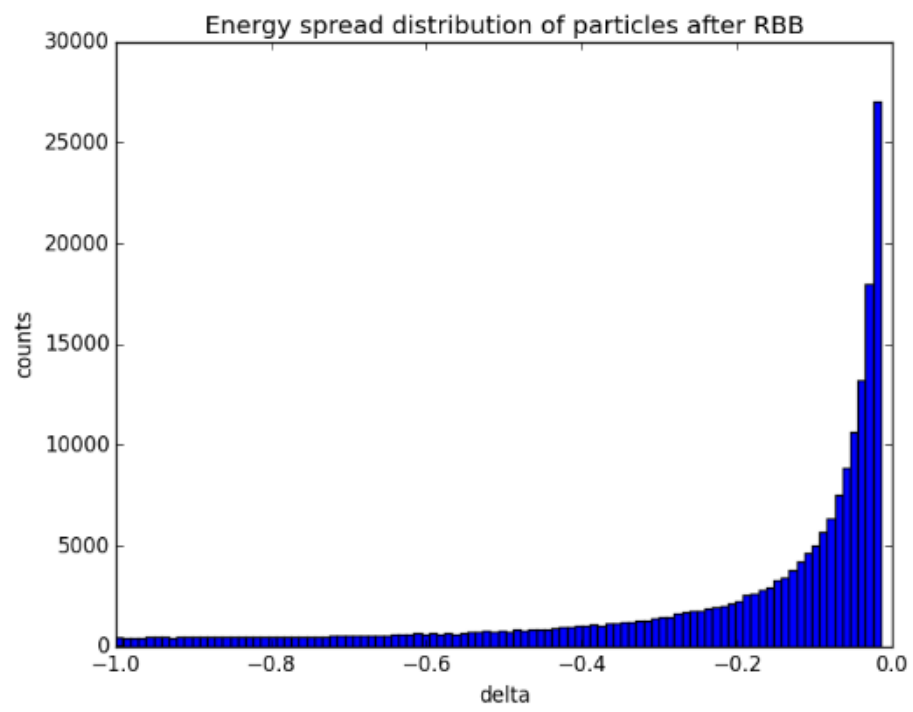

**Figure 4.2.7.1:** Radiative Bhabha Scattering events in the IR.

Figure 4.2.7.1 shows the energy spread distribution of 200,000 generated radiative Bhabha scattering events.

Due to lost particles position is unclear, scattered particles must be tracked by tools such as SAD [8] to determine the lost position. The particles will be flagged as lost if the transverse position of the particles touch the inner wall of the beam pipe. Particles might be lost at any position along the accelerator; however, only particles lost near the IP are important. The position and coordinate in phase space of lost particles are recorded, as shown in Figure 4.2.7.2 for the first two turns. This information is used as input to Geant4 [9] simulation to evaluate the radiation level at the detector.

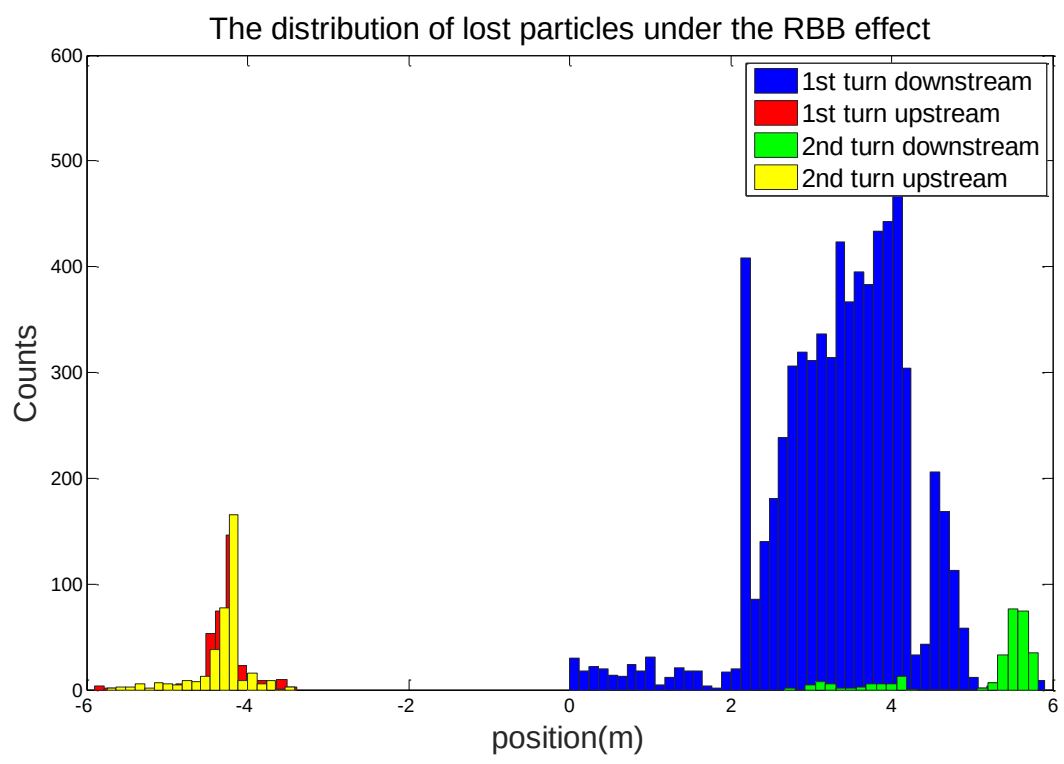

**Figure 4.2.7.2:** Distribution of lost particles positions due to radiative Bhabha scattering

As seen in Figure 4.2.7.2, most of the events are lost in the detector immediately due to their large energy loss. A few high energy particles will be lost near the IP after one revolution. Although a large fraction of events are lost in the downstream region, the radiation damage for detector components driven by these events is tolerable since most of these particles are confined within a small angle and will not strike any detector components.

Multi-turn tracking simulation results are shown as Figure 4.2.7.3.

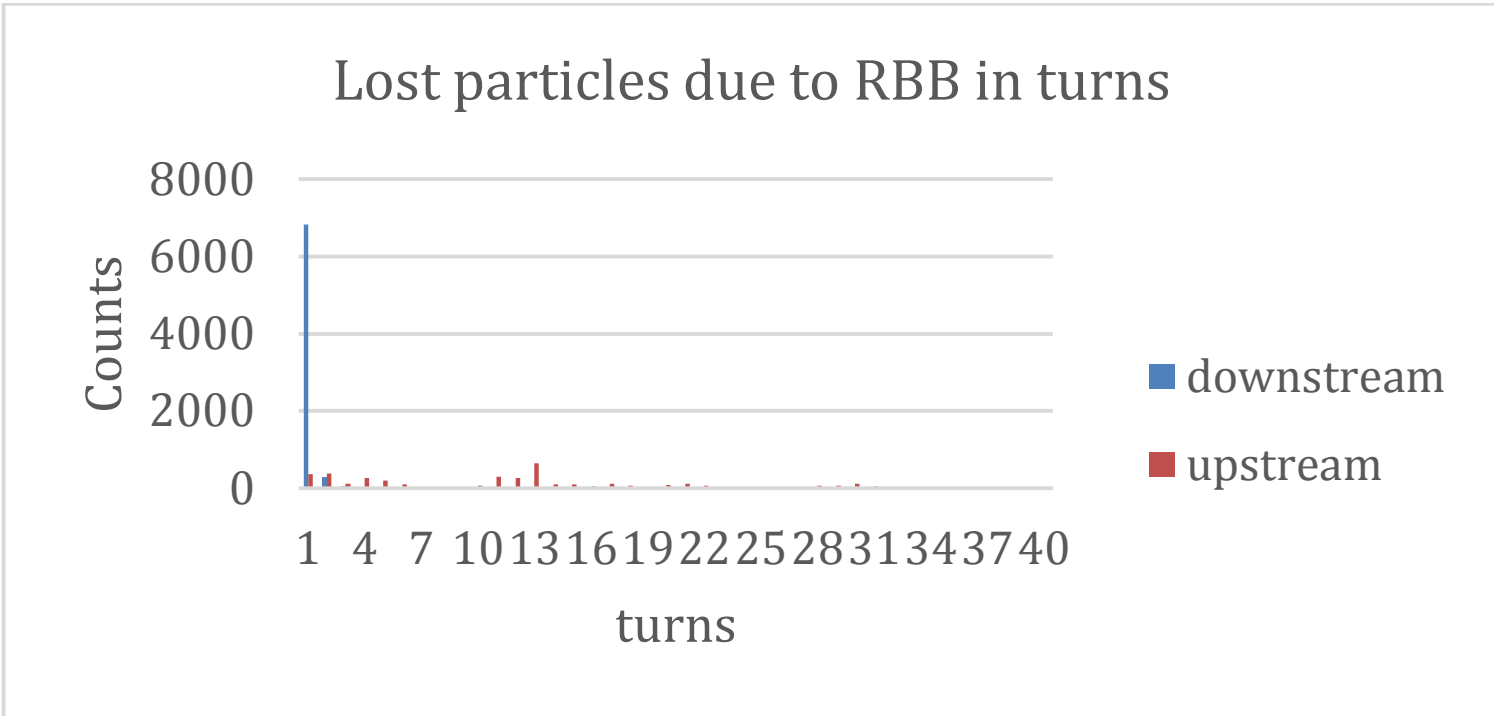

**Figure 4.2.7.3:** the statistics of lost particles in multiple turns due to RBB

Compared to the one-turn tracking, more particles get lost in the upstream region of the IR, they are potentially more harmful for they are more likely to penetrate into detector components, even with small angle trajectories with respect to the longitudinal direction.

*4.2.7.1.2 Beamstrahlung*

High energy photons from beamstrahlung can interact with each other and induce electron-positron pair production or hadronic background [10]. Also, if the energy spread is larger than the energy acceptance, particles might be lost.

Beamstrahlung events have been generated with Guinea-Pig++ [11] or Py_BS [7]. Below in Figure 4.2.7.4, is the energy spread of 200,000 beamstrahlung events.

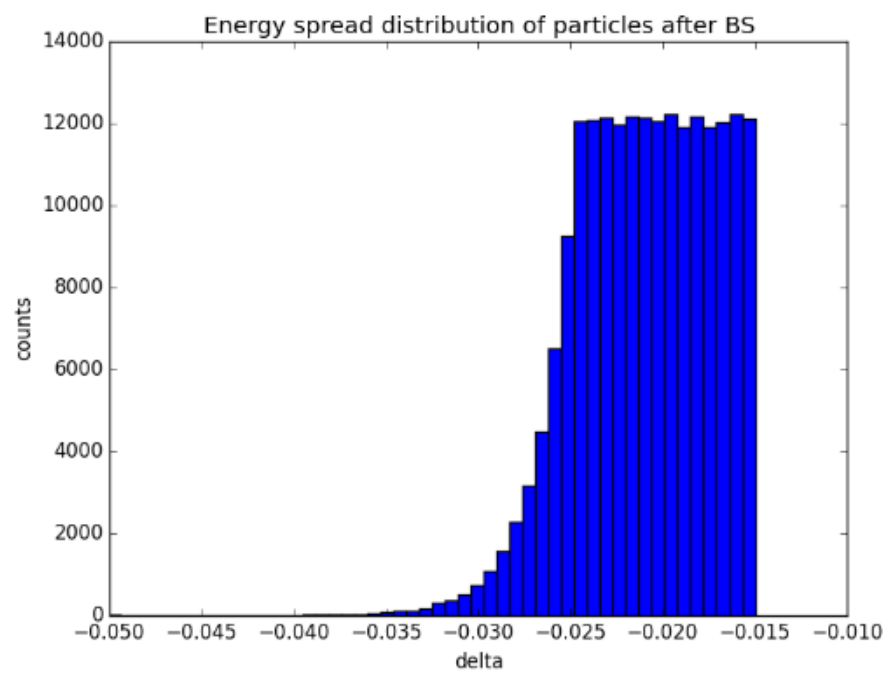

**Figure 4.2.7.4:** Beamstrahlung events in the IR.

Due to the energy spread distribution close to the energy acceptance and not having a large tail, the beam loss particles do not appear in the downstream of the first-turn tracking in SAD but do appear in multi-turn tracking, as can be seen in Figures 4.2.7.5 and 4.2.7.6.

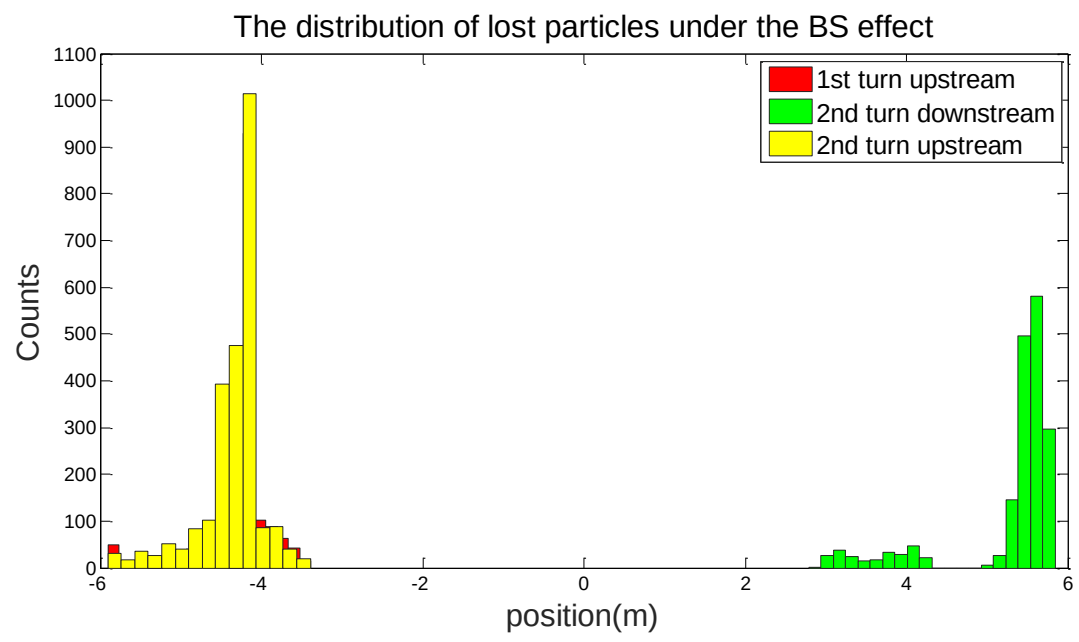

**Figure 4.2.7.5:** The distribution of lost particles positions due to beamstrahlung

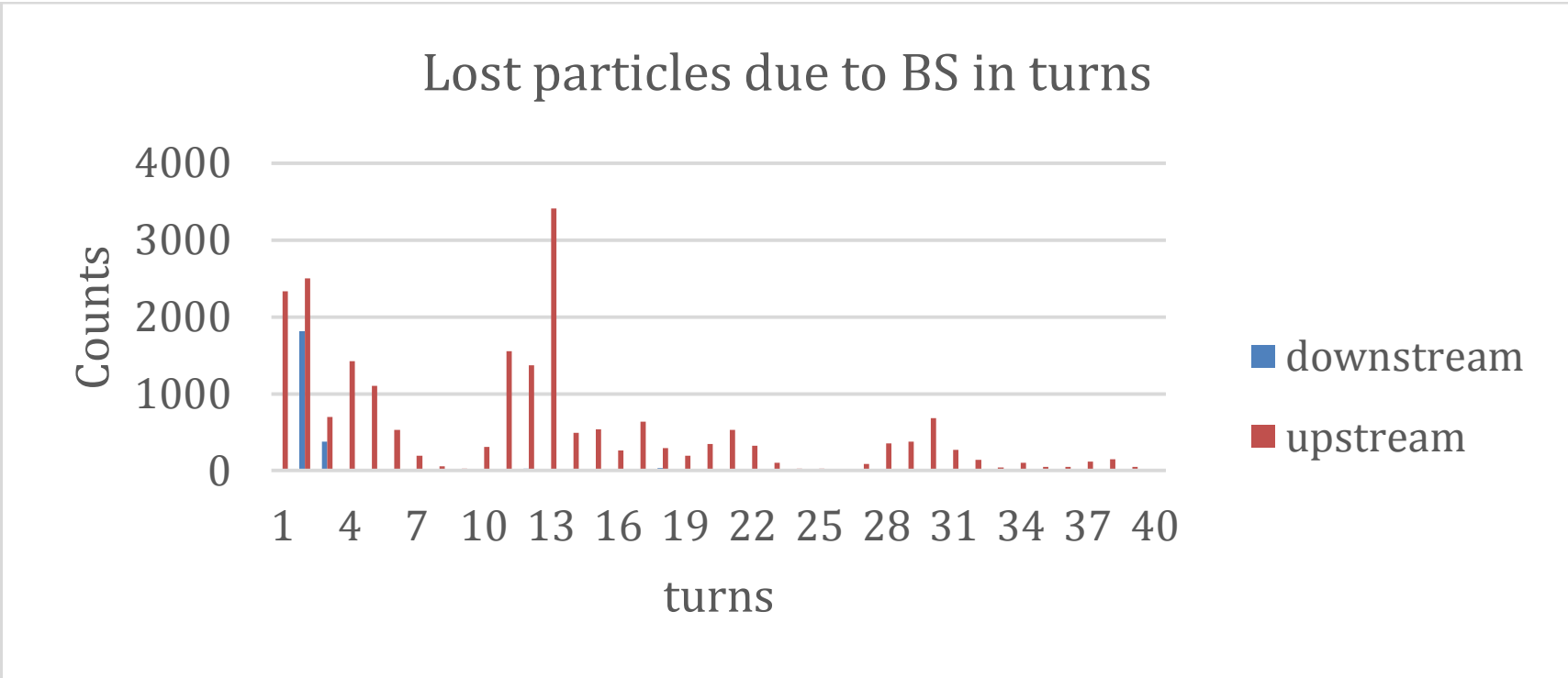

**Figure 4.2.7.6:** The distribution of lost particles vs. turns due to beamstrahlung

#### *4.2.7.1.3 Beam-Gas Inelastic Scattering*

Beam particles are lost by inelastic scattering from the residual gas in the vacuum chamber. During this process, the particle emits a photon and loses energy. If the energy loss is large enough beyond the energy acceptance, particles are lost from the beam and cause background in the detector.

Beam-gas bremsstrahlung simulation is similar to radiative Bhabha scattering and beamstrahlung, but the latter two are only generated during collisions at the IP, whereas beam-gas bremsstrahlung can occur at any position around the ring. For the MDI, background caused by beam-gas bremsstrahlung in the last 200 m upstream of the IP can be serious. We have sliced up the 200m upstream region of the IP, and calculated the number of events based on the beam lifetime. The events are generated from a Monte-Carlo generator and are placed at the entrance of the slice region. The tracking method in SAD is the same as for RBB and BS. Below are the lost particles statistics after tracking for 40 turns.

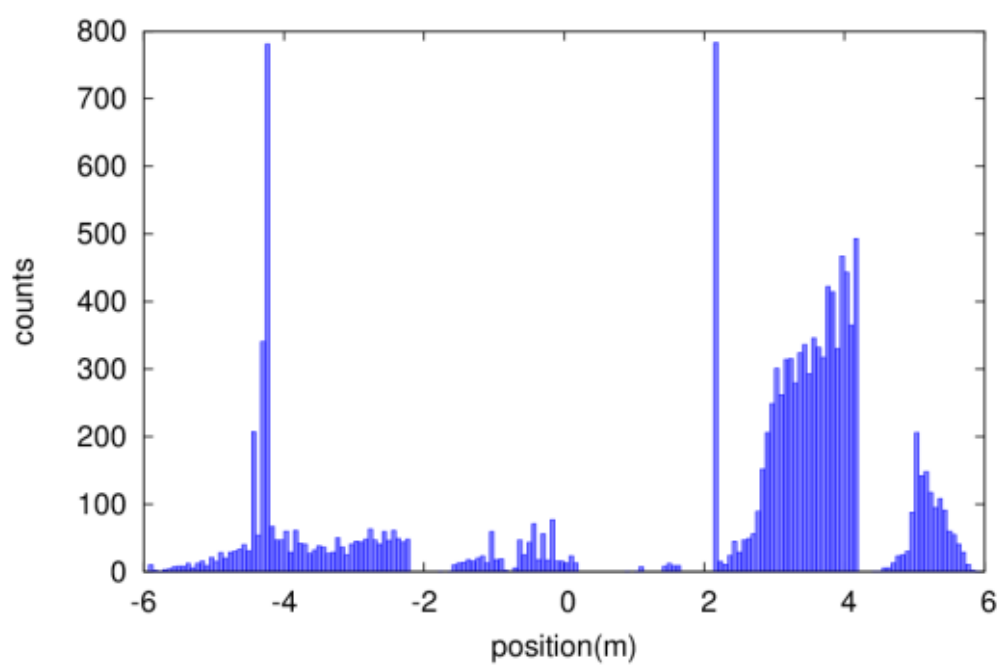

**Figure 4.2.7.7:** The distribution of lost particles from beam-gas bremsstrahlung

#### *4.2.7.1.4 Beam-Thermal Photon Scattering*

Accelerator components such as beam pipe, will emit a large number of thermal photons with different energies and directions due to thermal radiation. Beam particles will lose energy from scattering of thermal photons on electrons which is known as Compton Effect. The resulting off-momentum particles constitute a potential source of background in the detectors.

The simulation method used is the same as for beam-gas bremsstrahlung. The distribution of lost particles after tracking for 40 turns is shown in Figure 4.2.7.8.

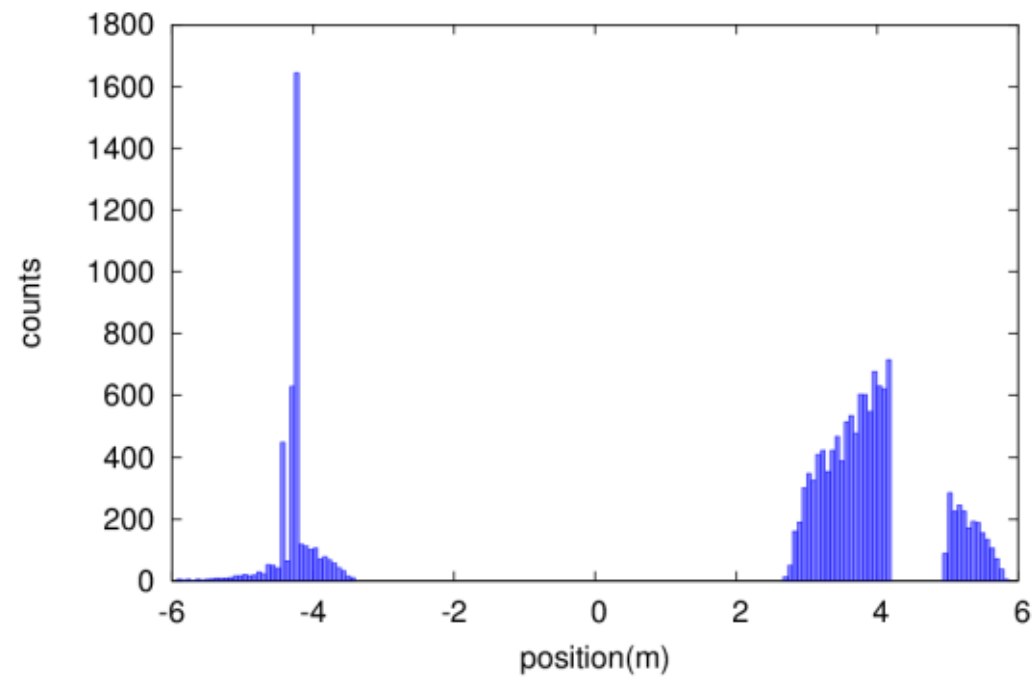

**Figure 4.2.7.8:** The distribution of lost particles positions due to beam-thermal photon scattering

Compared with RBB and BS, the lost particles due to beam-gas bremsstrahlung and beam-thermal photon scattering are at a much lower level.

### 4.2.7.2 *Collimators*

Collimators inserted into the beam line to reduce the number of particles lost in the IR. The aperture of the collimator should be as small as possible to absorb lost particles without however, affecting the beam core.

There are several requirements that must be satisfied in the collimator design:

- Aperture of collimator should be smaller than beam stay clear region: BSC_x=± (18 $\sigma_x$+3 mm), BSC_y=± (22 $\sigma_y$+3 mm)
- Impedance requirement: slope angle of collimator < 0.1 rad
- To shield against large energy spread particles, the phase between a pair collimators: $\pi/2+n*\pi$
- Collimators should be located in a large dispersion region to shield from large energy spread particles:

$$\sigma = \sqrt{\varepsilon\beta + (D_x\sigma_e)^2}$$

Four collimators are used in this design, only for horizontal plane (APTX1, APTX2, APTX3 and APTX4). Two of them (APTX1 and APTX2) are located in the upstream of the IP, and the others (APTX3 and APTX4) are located downstream of the IP. The distance to the IP range from 1800 meters to 2300 meters. Table 4.2.7.2 shows the design parameters, and the stopping efficiency as a function of collimator aperture is shown in Figure 4.2.7.9 which can be seen that vertical collimators are not needed.

**Table 4.2.7.2:** Collimator Design Parameters

| Name | Position | Distance to IP (m) | Beta function (m) | D(x) (m) | Phase | BSC/2 (m) | Range of half width allowed (mm) |
|---|---|---|---|---|---|---|---|
| APTX1 | D1I.1897 | 2139.06 | 113.83 | 0.24 | 356.87 | 0.00968 | 2.2~9.68 |
| APTX2 | D1I.1894 | 2207.63 | 113.83 | 0.24 | 356.62 | 0.00968 | 2.2~9.68 |
| APTX3 | D1O.10 | 1832.52 | 113.83 | 0.12 | 6.65 | 0.00382 | 2.2~9.68 |
| APTX4 | D1O.14 | 1901.09 | 113.83 | 0.12 | 6.90 | 0.00382 | 2.2~9.68 |

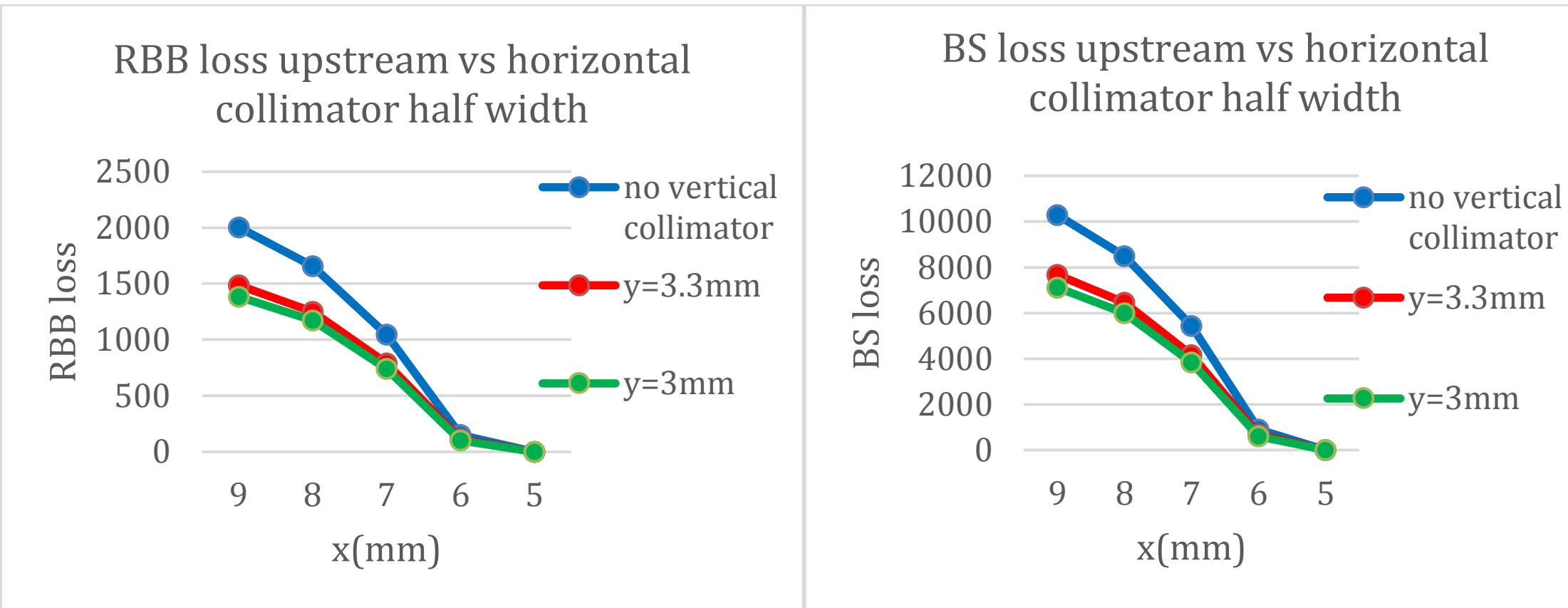

**Figure 4.2.7.9:** RBB and BS loss as the function of the collimator half width.

The collimator half widths are set as 5 mm (~13$\sigma_x$) in the horizontal plane. The collimators do not affect the beam quantum lifetime. The statistic for lost particles in 40 turns is illustrated in Figures 4.2.7.10.

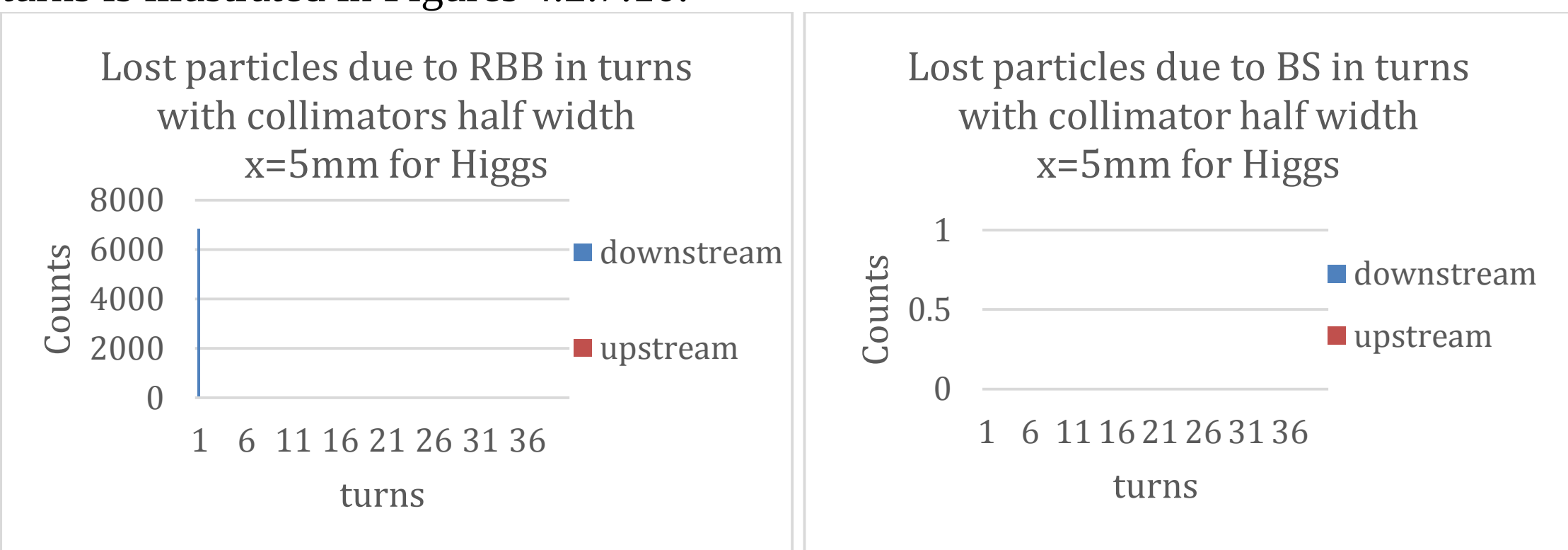

**Figure 4.2.7.10:** Distribution of lost particles due to RBB (left) and BS (right) vs. turns with collimators set at half width x=5 mm for Higgs mode operation

Compared with the results shown in Figures 4.2.7.3 and 4.2.7.6, the lost particles upstream of the IP have been reduced to zero with this system of collimators. Although the beam loss downstream part of the IP is still fairly large in the first turn, the radiation damage and the detector background are not as serious as the loss rate for the relatively small angle to the ideal orbit for these particles and the direction is away from detector.

With the collimators designed to supress background from radiative Bhabha scattering and beamstrahlung, the lost particles from beam-gas bremsstrahlung and beam-thermal photon scattering are supressed to an even lower level as shown in Figures 4.2.7.11. Although the beam loss in the downstream of the IP is still remained, the radiation damage and the detector background are not serious, since the direction is leaving the detector.

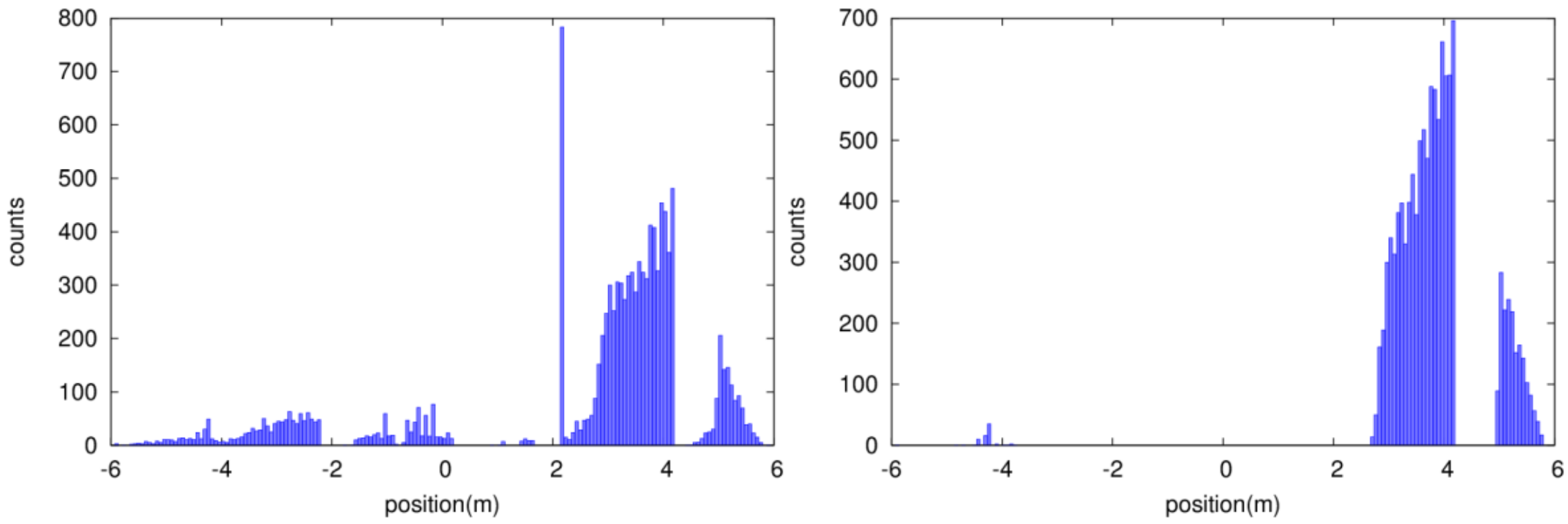

**Figure 4.2.7.11:** Lost particles due to beam-gas bremsstrahlung (left) and beam-thermal photon scattering (right) for 40 turns with collimators at half width x=5 mm for Higgs mode

For the 45.5 GeV Z mode, according to the off-momentum dynamic aperture after optimizing the CEPC lattice, and considering the beam-beam effect and errors, the energy acceptance is about 1.0%. Although Z lattice is the same as the one in Higgs, the emittance is about 7 times lower. Thus the beam stay clear region will be small. The collimators design width will be accordingly smaller than in Higgs case. After optimization the collimators with half width x=2.5mm (~17$\sigma_x$) can be accepted, results are shown below in Figure 4.2.7.12.

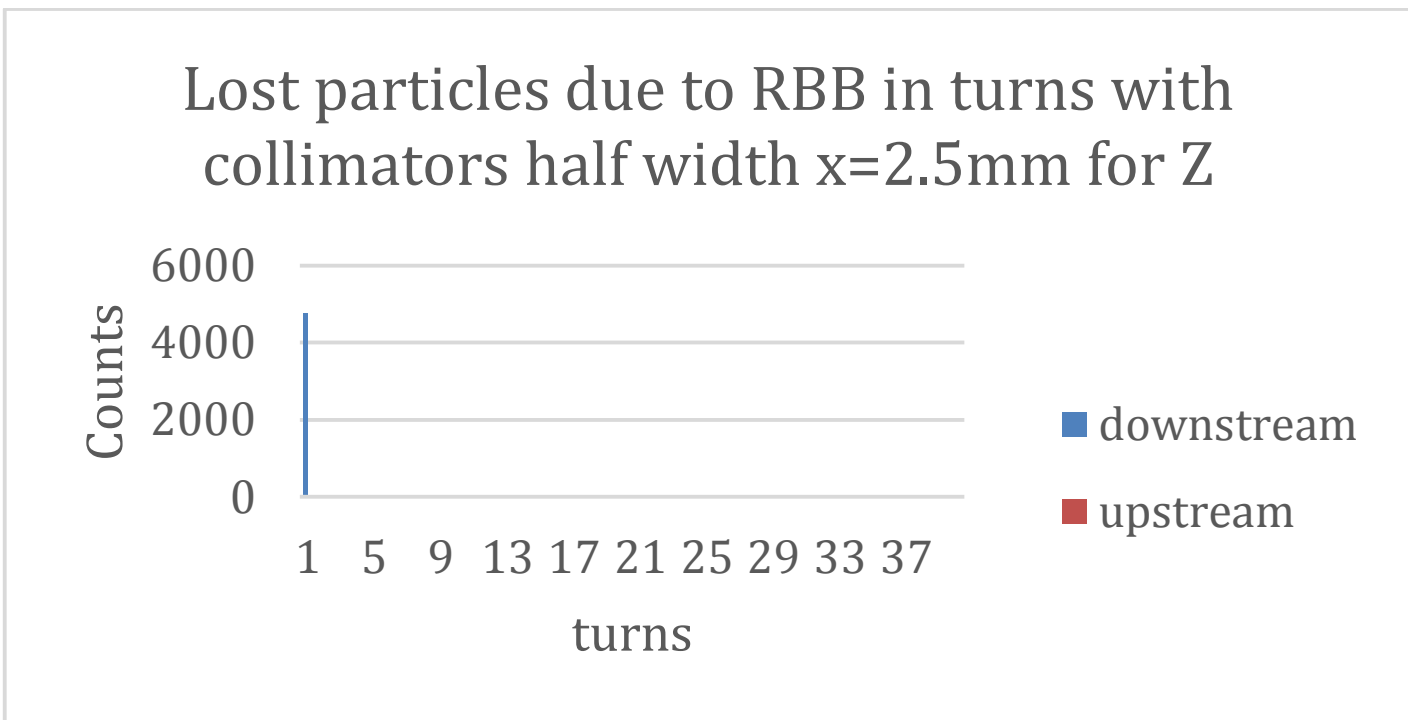

**Figure 4.2.7.12:** Lost particles due to RBB vs. turns with collimators at x=2.5 mm for Z mode

Since in the 45.5GeV Z factory, beamstrahlung effect is not present, therefore we have no background from this source. However, the Beam-Gas bremsstrahlung and Beam-Thermal photon scattering are still two important processes that will affect the beam lifetime at a level of ~ 57.26 and 70.17 hours respectively. The beam loss results with and without collimators are shown below:

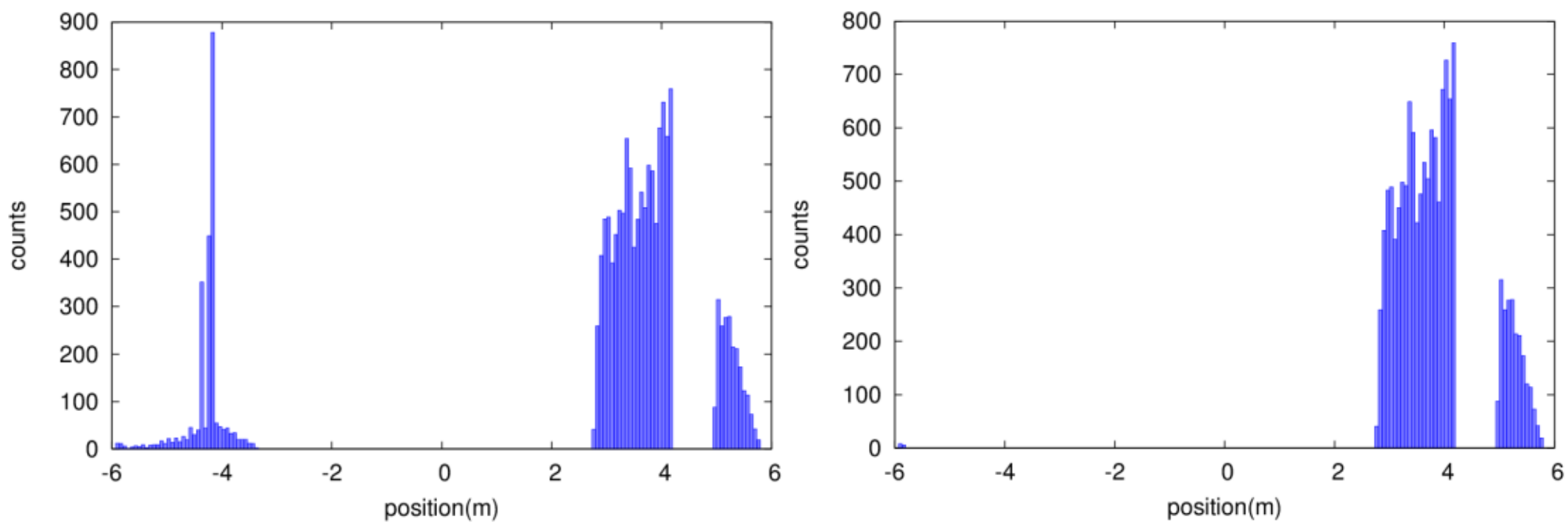

**Figure 4.2.7.13:** Lost particles statistics due to Beam-Gas bremsstrahlung for 40 turns without (left) and with (right) collimators half width x=2.5mm for Z mode

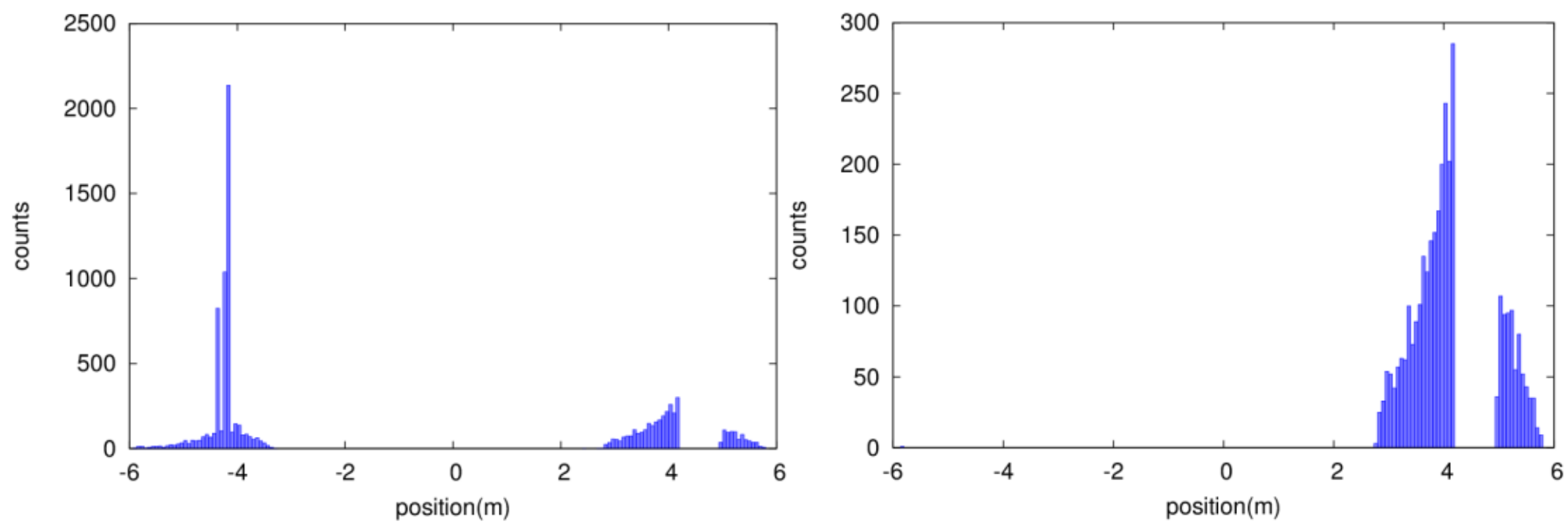

**Figure 4.2.7.14:** Lost particles statistics due to Beam-Thermal photon scattering for 40 turns without (left) and with (right) collimators half width x=2.5mm for Z mode

As shown in Figures 4.2.7.13 and 4.2.7.14, with collimators half width settings of x=2.5mm, the beam loss caused by beam-gas bremsstrahlung and beam-thermal photon scattering has almost been eliminated upstream of the IP. Although the beam loss in the downstream of the IP is still fairly large in the first turn, the radiation damage and the detector background are not as serious since the direction of the background particles is away from the detector.

## 4.3 Collider Technical Systems

### 4.3.1 **Superconducting RF System**

#### 4.3.1.1 ***CEPC SRF System Overview***

The RF system accelerates the electron and positron beams, compensates for synchrotron radiation loss and provides sufficient RF voltage for energy acceptance and the required bunch length in the Booster and Collider. Superconducting radio frequency (SRF) cavities are used because they have much higher continuous wave (CW) gradient and energy efficiency as well as larger beam aperture compared to normal conducting cavities.

CEPC will use a 650 MHz RF system with 240 cavities for the Collider and a 1.3 GHz RF system with 96 cavities for the Booster. Since the Booster has a few percent of the beam current of the Collider and low duty cycle (less than 10 %), 1.3 GHz is chosen as the frequency. The Collider RF is the second sub-harmonic of the Booster frequency. This choice of frequencies minimizes the construction and operating cost, fulfills the beam dynamics and luminosity requirements and uses mature technology developed by TESLA and adopted for Euro-XFEL and LCLS-II. These frequencies have the most synergy with other ongoing SRF accelerator projects in China and abroad.

There are two RF sections located at two long straight sections respectively. Each RF section contains two Collider RF stations and one Booster RF station between the two Collider RF stations. Each of the 11 m-long Collider cryomodules contains six 650 MHz 2-cell cavities, and each of the 12 m-long Booster cryomodules contains eight 1.3 GHz 9-cell cavities. The Collider cryomodules will be mounted on the tunnel floor, and the Booster cryomodules will be hung from the ceiling at a different beamline height. The Collider cavities operate in CW mode, and the Booster cavities operate in fast voltage ramp mode.

The Collider is double-ring with shared cavities for Higgs operation and separate cavities for W and Z operations. High RF voltage and low current is required for Higgs and low voltage and high current for W and Z [1]. This common cavity scheme will reduce the total cavity and cryomodule number as well as the cryogenics by half compared to the usual double ring with separate cavities for the two rings. The electron or positron beams will go through the two RF stations in each RF section for Higgs operation. Less than half of the buckets will be filled to avoid collisions in the RF section. When operating for W and Z, part of the Higgs cavities will be used in each RF station, and the electron or positron beams will go through only one of the two RF stations of a RF section. The W and Z-pole bunches will be quasi-uniformly distributed in the two rings. This configuration reduces by half the beam current in the W and Z cavities and the cavity impedance seen by the beam. Since the idle cavities have large impedance, they will be moved off the beam-line or detuned (parked) in the W and Z operation.

Full installation of the same cavities and cryomodules for Higgs, W, and Z-pole runs is the baseline for the Collider SRF system design. The maximum synchrotron radiation (SR) power limit per beam is 30 MW. As a result of these conditions and the higher physics priority and limited construction cost for Higgs runs, the Collider SRF system is optimized for the Higgs mode of 30 MW SR power per beam, with enough tunnel space and operating margin to allow higher RF voltage and/or SR power (50 MW SR power per beam) by adding cavities.

From LEP2 and LHC experience of handling large HOM power in a multi-cavity cryomodule with coaxial HOM couplers, the upper-limit of average HOM power produced in each 2-cell 650 MHz cavity is set to be 2 kW. This HOM power limit and the fast-growing longitudinal coupled-bunch instabilities (CBI) driven by both the fundamental and higher order modes impedance of the RF cavities determine to a large extent the highest beam current and luminosity obtainable in the Z mode. It is impossible to have a single common SRF system for the highest possible luminosity in each mode (Higgs, W, and Z) with up to 50 MW SR power per beam, due to the wide range of SRF parameters in terms of RF voltage and beam current. A staged SRF complex is inevitable. For a higher luminosity Z upgrade, because of the high HOM power and the need to have the smallest number of cavities, KEKB / BEPCII type cryomodules with very high input coupler power will be needed.

The Booster performance will not be limited by the SRF system in the luminosity range foreseen provided the beam current is small enough and the beam feedback system works well for CBI during the energy ramp, especially in the low energy region. If the Booster current is increased for Z-pole, SSA power should be increased accordingly and the HOM power will finally reach the max HOM coupler capacity of the TESLA cavity.

The SRF technical challenges that require R&D include: achieving the cavity gradient and high quality factor in the real cryomodule environment, robust and variable high power input couplers that are design compatible with cavity clean assembly and low heat load, efficient and economical damping of the HOM power with minimum dynamic cryogenic heat load, and fast RF ramp and control of the Booster.

In parallel with design and key R&D, extensive development of SRF personnel, infrastructure and industrialization is essential for the successful realization of CEPC. This will have synergy with ongoing SRF-based accelerator projects in China, such as SHINE (Shanghai HIgh repetition rate XFEL aNd Extreme light facility) and ADANES (Accelerator Driven Advanced Nuclear Energy System) in Huizhou, Guangdong. For this purpose, a large SRF infrastructure facility (PAPS-RF) is being built at the Huairou Science City near Beijing.

#### 4.3.1.2 *Collider RF Parameters*

The Collider RF parameters (Table 4.3.1.1) are chosen to meet the baseline luminosity requirement for each operating energy. Given the total synchrotron radiation power, parasitic loss and RF voltage, the cavity number and voltage for the Collider are mainly determined by the input coupler power handling capability, chosen to be 300 kW. This is a balance of SRF system capital cost, coupler operating risk and cavity gradient. The cavity gradient is determined by the cell numbers when the cavity voltage is fixed. More cells are better for low gradient, but will increase the cavity HOM power and impedance, also lower the coupling and increase the trapping probability for HOMs. We have chosen 2-cell and 20 MV/m CW operation for the 650 MHz cavity. Due to RF mismatch at different beam energy and current, the input coupler should have variable coupling to avoid extra power and match for higher beam current. Each cavity has two detachable coaxial HOM couplers mounted on the cavity beam pipe with HOM power handling capacity of 1 kW. Each 11 m-long cryomodule consists of six cavities. (Figure 4.3.1.1) Each cryomodule has two beamline HOM absorbers at room temperature outside the vacuum vessel with HOM power handling capacity of 5 kW each.

Each two cavities will share a klystron with 800 kW maximum output power. The beam power in the two cavity is less than 70 % of the klystron maximum output power for all the operation modes, the extra power will be enough for the waveguide loss, cavity detuning, parasitic loss and LLRF headroom. Some of the Higgs cavities will be used for W and Z operation (108 out of 120 cavities per ring for W, 60 out of 120 cavities per ring for Z), which has the same RF power source and distribution as the Higgs. The unused idle cavities will be detuned and kept at 2 K to extract fundamental mode and HOM power.

Beam loading effects define the RF system design and configuration of the W and Z mode. Transient beam loading of beam gaps, large cavity detuning with small revolution frequency, and cavity fundamental mode and higher order mode induced coupled bunch instabilities with small radiation damping are of the most concern. Less cavities and cell numbers are preferred to have high stored energy and low impedance.

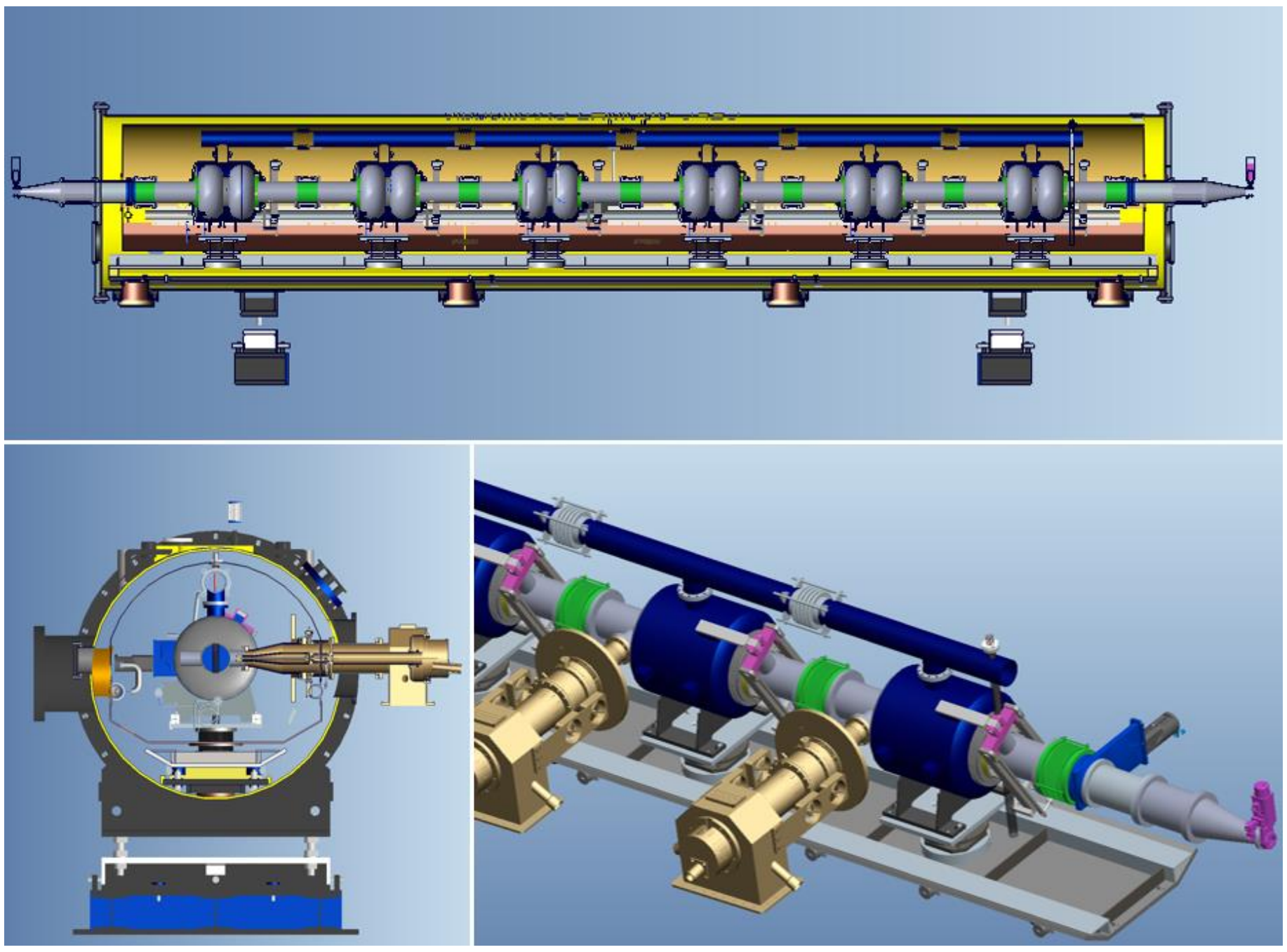

**Figure 4.3.1.1:** CEPC Collider Ring 650 MHz 2-cell cavity and cryomodule

**Table 4.3.1.1:** CEPC Collider RF parameters

| | H | W | Z |
|---|---|---|---|
| Beam Energy [GeV] | 120 | 80 | 45.5 |
| Luminosity / IP [$10^{34}$ $cm^{-2}s^{-1}$] | 2.93 | 10.1 | 16.6/32.1 |
| SR power / beam [MW] | 30 | 30 | 16.5 |
| Circumference [km] | 100 | 100 | 100 |
| RF frequency [MHz] | 650 | 650 | 650 |
| Harmonic number | 216816 | 216816 | 216816 |

| | H | W | Z |
|---|---|---|---|
| Revolution time [μs] | 333.563 | 333.563 | 333.563 |
| SR loss / turn [GV] | 1.73 | 0.34 | 0.036 |
| RF voltage [GV] | 2.17 | 0.47 | 0.1 |
| Synchrotron phase from crest [deg] | 37.1 | 43.7 | 68.9 |
| Beam current / beam [mA] | 17.4 | 87.7 | 460 |
| Bunch charge [nC] | 24 | 19.2 | 12.8 |
| Bunch length [mm] | 3.26 | 5.9 | 8.5 |
| Bunches / beam | 242 | 1524 | 12000 |
| Synchrotron oscillation period [ms] | 5.2 | 8.5 | 11.9 |
| Longitudinal damping time [ms] | 23.1 | 78.5 | 421.6 |
| RF station number | 4 | 4 | 4 |
| Total RF section length [m] | 640 | 576 | 320 |
| Cavity number | 240 | 216 | 120 |
| Cavity number / cryomodule | 6 | 6 | 6 |
| Cryomodule number | 40 | 36 | 20 |
| Cryomodule length [m] | 11 | 11 | 11 |
| Cell number / cavity | 2 | 2 | 2 |
| Cavity effective length [m] | 0.46 | 0.46 | 0.46 |
| R/Q [Ω] | 213 | 213 | 213 |
| Cavity operating gradient [MV/m] | 19.7 | 9.5 | 3.6 |
| $Q_0$ at operating gradient for long term | 1.5E10 | 1.5E10 | 1.5E10 |
| Acceptance gradient in vertical test [MV/m] | 22 | 22 | 22 |
| $Q_0$ at acceptance gradient in vertical test | 4E10 | 4E10 | 4E10 |
| Optimal $Q_L$ | 1.5E6 | 3.2E5 | 4.7E4 |
| Cavity bandwidth at optimal $Q_L$ [kHz] | 0.4 | 2.0 | 13.7 |
| Optimal detuning [kHz] | -0.2 | -1.0 | -17.8 |
| Input power / cavity [kW] | 250 | 278 | 275 |
| Input coupler power capacity [kW] | 300 | 300 | 300 |
| External Q adjusting range | 2E5 ~ 2E6 | 2E5 ~ 2E6 | 4E4 ~ 2E6 |
| Cavity number / klystron | 2 | 2 | 2 |
| Klystron max output power [kW] | 800 | 800 | 800 |
| Klystron number | 120 | 108 | 60 |
| HOM power / cavity [kW] | 0.57 | 0.75 | 1.94 |
| HOM coupler power capacity [kW] | 1.0 | 1.0 | 1.0 |
| HOM absorber power capacity [kW] | 5 | 5 | 5 |
| Wall loss / cavity @ 2 K [W] | 25.6 | 5.9 | 0.9 |
| Total cavity wall loss @ 2 K [kW] | 6.1 | 1.3 | 0.1 |

#### 4.3.1.3 ***Beam Cavity Interaction***

*(1) Transient Beam Loading*

Transient beam loading is of major concern in a large ring with uneven fill patterns. Uneven fills with beam gaps are necessary for beam abort and ion-clearing or in bunch train operation. This results in a bunch phase shift, less longitudinal focusing, smaller energy acceptance, and possible lifetime and luminosity degradation. The bunch phase shift can be estimated by analytical calculation or more accurately by transfer function simulation and time domain simulation [2-4].

Even a 1% beam gap to mitigate ion-trapping and fast beam ion instability (FBII) will create a large bunch phase shift in CEPC Z mode. If the fill pattern is changed from one long gap to many small gaps and short bunch trains, the phase shift will be reduced to a negligible level [5]. For one of the fill patterns of the CEPC Z mode, 48 trains of 250 bunches locate symmetrically in the ring with bunch spacing of 24.6 ns (16 buckets) and gaps of 820 ns (533 buckets). The calculated largest phase shift in a train is 6.1 degree, which is not trivial. A ~ 5 μs long gap is needed for the Higgs mode to avoid collision of the two half-ring-filled bunch trains in the RF sections. This gap will result in a 0.5 degree phase shift for the cavity in the end of the RF section. Bunch swapping during the on-axis injection of the Higgs mode will cause transient beam loading in both Collider and Booster. Total phase shift of electron and positron may be different due to different fill patterns of the two beams during bunch swapping.

There are several methods for transient beam loading compensation: (1) increase cavity stored energy; (2) change the fill pattern and RF distribution (spread as uniformly as possible): (3) increase synchrotron phase (change beam parameters). Correction can be done by RF generator modulation by providing an additional current to fully cancel out beam current variations in each cavity. But this method needs a special RF source with high peak power and high repetition rate. Special techniques are needed to reduce the filling power and average power due to low RF-to-beam power efficiency [6,7]. Phase modulation of the generator current without additional demands on the available RF power could work [8]. A second correction method is to add a travelling wave cavity or (detuned) beat cavity [9,10].

*(2) Cavity Fundamental Mode Instability*

Due to the large circumference, low RF voltage and high current, fundamental mode longitudinal coupled-bunch-instability (CBI) will occur for CEPC W and Z. There are more than 20 unstable modes (parked cavities not considered) for Z (Figure 4.3.1.2). It's necessary to use direct and comb RF feedback loops (as in PEP-II [11] and KEKB [12]) to decrease the effective fundamental-mode impedance and the growth rates to a manageable level for the longitudinal bunch-by-bunch feedback system to suppress the instability. Less powered cavities will reduce the detuning thus alleviate the instability, but will increase the input coupler power. Higher total RF voltage and thus larger cavity stored energy will alleviate the problem because of less detuning, but will affect bunch length and luminosity optimization.

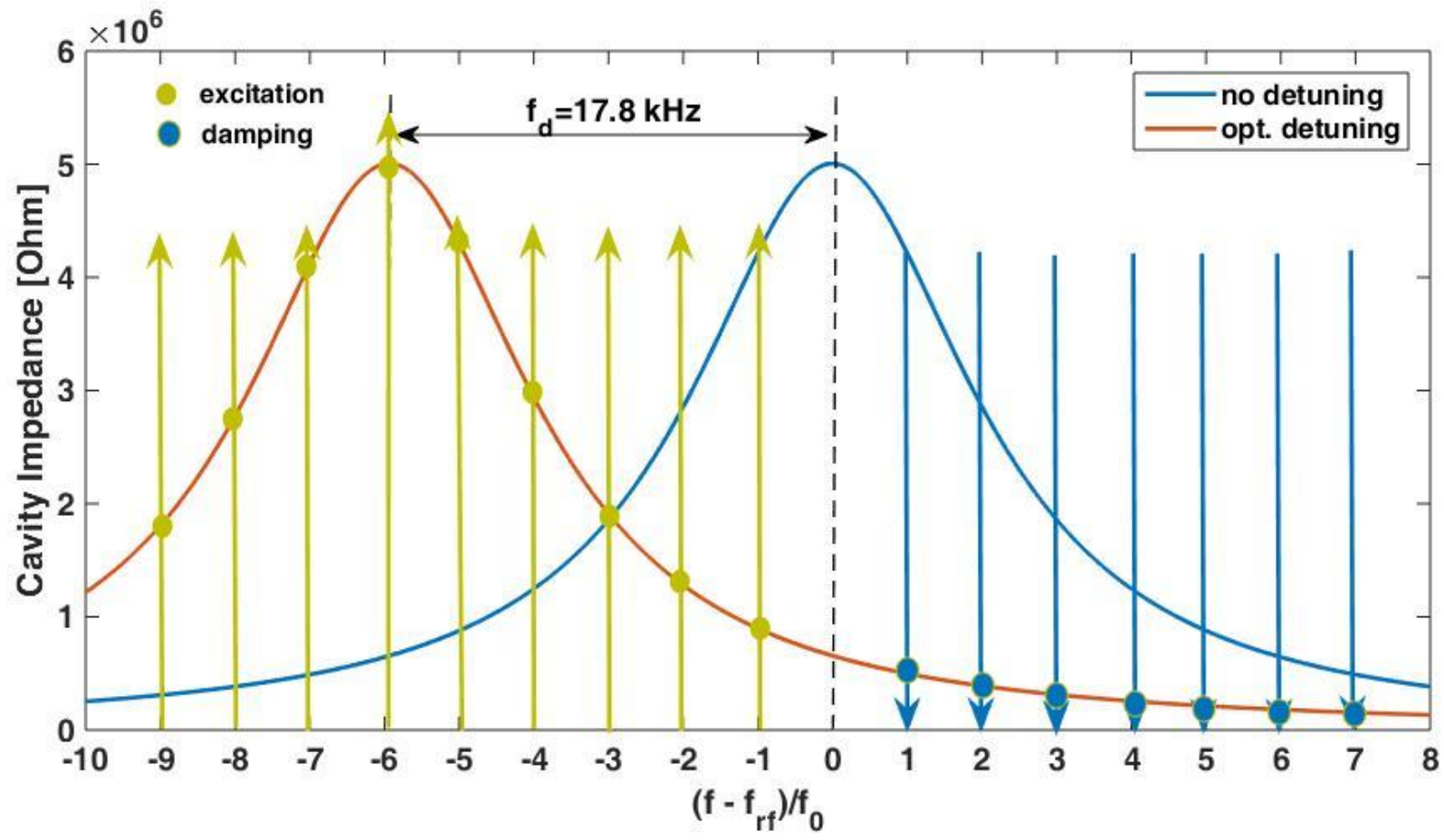

**Figure 4.3.1.2**: Fundamental mode impedance and beam spectrum at the Z-pole

For the large number of parked cavities in the Z mode, the longitudinal coupled-bunch instabilities excited by the fundamental impedance are also important issues. To solve this problem, symmetry detuning of the parked cavities is proposed, i.e. positive detuning of half of the parked cavities and negative detuning of the other half.

A vector sum direct feedback of each pair of cavities will be used to cure the second Robison instability by decreasing the effective impedance seen by the beam [13]. Enough klystron RF power margin is necessary for control.

*(3) Cavity HOM CBI*

Table 4.3.1.2 gives the threshold for the external quality factor of the HOMs with high $R/Q$. The comparison of the CBI impedance threshold and cavity impedance including the HOM frequency spread and feedback is discussed in 4.2.3.3.2.

**Table 4.3.1.2:** Damping requirements of prominent HOMs of the 650 MHz 2-cell cavity

| Mode | $f$ (MHz) | $R/Q^*$ (monopole Ω, dipole Ω/m) | $Q_e$ (H) | $Q_e$ (W) | $Q_e$ (Z) |
|---|---|---|---|---|---|
| TM011 | 1165.574 | 65.2 | $1.9\times10^5$ | $1.8\times10^4$ | $2.6\times10^2$ |
| TM020 | 1383.898 | 1.3 | $8.2\times10^6$ | $7.6\times10^5$ | $1.1\times10^4$ |
| TM021 | 1717.475 | 19.9 | $4.3\times10^5$ | $4.0\times10^4$ | $5.8\times10^2$ |
| TM012 | 1832.801 | 17.26 | $4.6\times10^5$ | $4.3\times10^4$ | $6.2\times10^2$ |
| TE111 | 844.738 | 279.8 | $4.9\times10^4$ | $7.7\times10^3$ | $1.6\times10^2$ |
| TM110 | 907.592 | 420.1 | $3.3\times10^4$ | $5.1\times10^3$ | $1.0\times10^2$ |
| TE121 | 1475.553 | 125.8 | $1.1\times10^5$ | $1.7\times10^4$ | $3.5\times10^2$ |
| TM120 | 1662.599 | 18.8 | $7.4\times10^5$ | $1.2\times10^5$ | $2.3\times10^3$ |

$^*$ Longitudinal $R/Q$ with the accelerator definition and $k_{//\,\mathrm{mode}} = 2\pi f\cdot(R/Q)/4$ [V/pC]. Transverse $R/Q$: $k_{\perp\,\mathrm{mode}} = 2\pi f\cdot(R/Q)/4$ [V/(pC·m)]

#### 4.3.1.4 *Cavity*

The 650 MHz 2-cell cavity (Figure 4.3.1.3) is made of bulk niobium and operates at 2 K with $Q_0 > 4\times10^{10}$ at 22 MV/m for the vertical acceptance test, $Q_0 > 2\times10^{10}$ at 20 MV/m for the horizontal test. The normal operation gradient is below 20 MV/m, and the lower limit of $Q_0$ is $1.5\times10^{10}$ for long term operation. The main RF parameters are listed in Table 4.3.1.3. The cavity mechanical structure is optimized with the helium vessel to minimize pressure sensitivity (df/dp) and mechanical stress [14]. The cavity wall thickness is 4 mm. The length of the cavity beam pipes, HOM coupler ports and input coupler port should be long enough to ensure negligible power dissipation in the gaskets and flange surfaces compared to the wall loss of the high Q cavity, but cannot be too long to go above the Nb critical temperature. Special gapless gaskets will be used to avoid additional dissipation at different joints. Cooling of cavity ports by an extended helium vessel could be considered, especially for the power coupler and HOM coupler. Copper plating is necessary for the bellows between cavities. RF shielded bellow might be needed. It is necessary to achieve compliance with Chinese pressure codes and permission from authorities to operate low-temperature containers made from niobium and titanium. CEPC will gain experience from SHINE in Shanghai, which will be the first accelerator project with SRF cavity to have compliance with pressure codes in China.

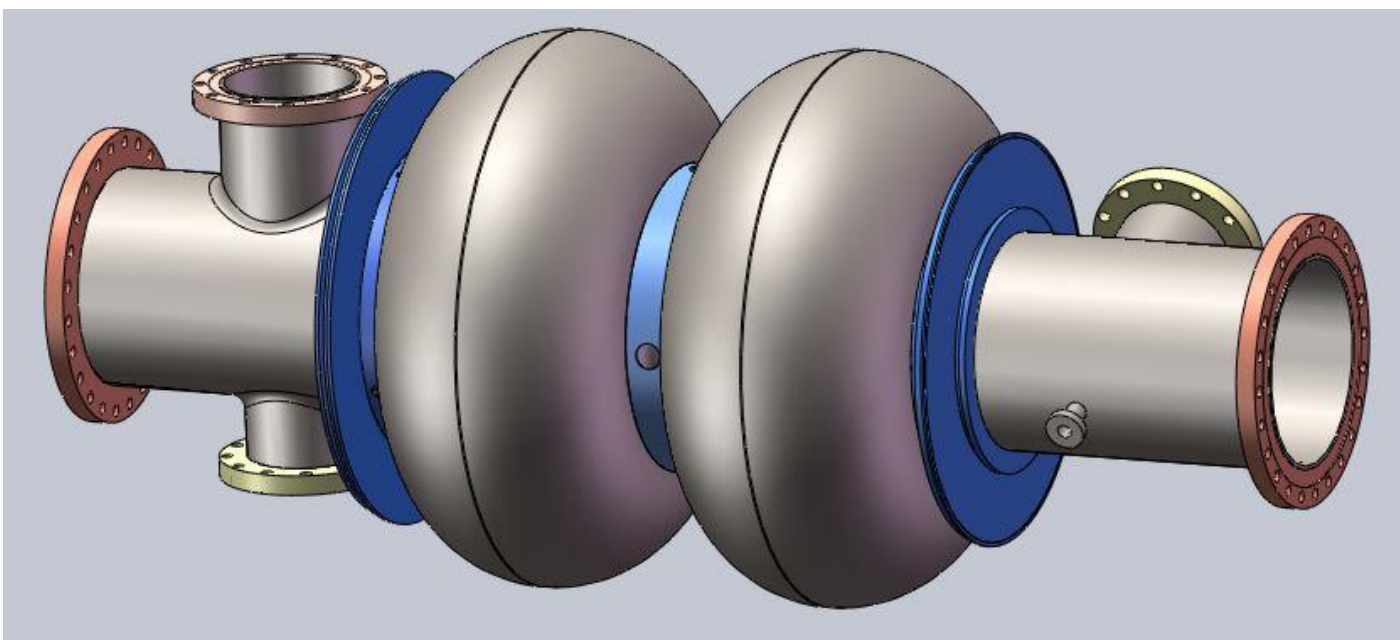

**Figure 4.3.1.3:** 3-D model of the 650 MHz 2-cell cavity

**Table 4.3.1.3:** Main parameters of the CEPC 650 MHz 2-cell cavity

| Parameter | Unit | Value |
|---|---|---|
| Cavity frequency | MHz | 650 |
| Number of cells | - | 2 |
| Cavity effective length | m | 0.46 |
| Cavity iris diameter | mm | 156 |
| Beam tube diameter | mm | 156 |
| Cell-to-cell coupling | - | 3 % |
| R/Q | Ω | 213 |
| Geometry factor | Ω | 284 |
| $E_{peak}/E_{acc}$ | - | 2.4 |
| $B_{peak}/E_{acc}$ | mT/(MV/m) | 4.2 |

Most SRF cavities are made of bulk niobium. New cavity material and techniques are being studied worldwide, and Nitrogen doping (N-doping) and $Nb_3Sn$ are the most promising. N-doping can double $Q_0$ for the 1.3 GHz cavity as well as the 650 MHz cavity [15]. N-doping has been adopted for the Linac Coherent Light Source II (LCLS-II) at SLAC [16].

In order to achieve high Q in the horizontal test and operation, many issues should be considered. Ultra-clean cavity surface processing is required to avoid field emission. Cavity string assembly with input couplers and HOM couplers in a Class 10 room is necessary. Magnetic field at the cavity position inside the cryomodule should be reduced to 5 mGs. Fast cool down at the transition temperature is also key to flux expulsion.

Particle-free vacuum near the SRF cryomodules is critical to maintain the cavity performance during long-term operation. For the vacuum installation near cavity, it is suggested to clean and pre-assemble the vacuum parts in cleanroom and assemble the remaining connections in a portable clean enclosure onsite. To reduce the particle transport, components will be transported under vacuum rather than in dry nitrogen. Slow pumping and slow venting procedures are important to reduce particle transport. NEG coated vacuum chamber will be helpful to reduce outgassing beside the cryomodule, but it is necessary to verify the dust production during activation of NEG coated chamber by using vacuum particle sensor.

A 4500 $m^2$ SRF lab is being built in Huairou Science City in the north of Beijing [18]. The SRF facility is aimed at cavity R&D as well as testing of several hundreds of SRF cavities and couplers, and assembly and testing of about 20 cryomodules per year for different users including CEPC.

#### 4.3.1.5 *Power Coupler*

One of the key technology challenges is achieving the very high power handling capability of the input power coupler for the Collider SRF cavity. Both the $Q_0$ and the accelerating gradient for SRF cavities are high, which requires that the coupler be assembled with the cavity in a Class 10 cleanroom. In addition, considering the large number of couplers, heat load (both dynamic and static) is another important issue to be solved. The main challenges are: variable coupling with wide range; high power handling capability (CW 300 kW); single window coupler and cavity clean assembly; very small heat load; simple structure for cost saving; and high yield and high reliability.

Considering the excellent performance, close frequency and experiences obtained at IHEP, the BEPC-II 500 MHz coupler design (Tristan Type window) is taken as the baseline. Several modifications are considered: reduce the distance between the window and the coupling port, put the window into the cryostat profile and thus have the window and cavity assembled in a Class 10 cleanroom; and redesign the mechanical structure for higher power capacity and lower heat load.

**Table 4.3.1.4:** Parameters of the power coupler for the Collider.

| Parameters | Collider |
|---|---|
| Frequency | 650 MHz |
| Maximum power | CW, 300 kW |
| $Q_{ext}$ | 1E5~2E6 |
| Coupling type | Antenna |
| Coupler type | Coaxial |
| Number of windows | 1 |
| Window type | Coaxial Tristan type warm window |

To reduce the loss on the surface of the coaxial line, a 75 Ω coaxial line is chosen. The RF design is shown in Figure 4.3.1.4. The power loss in the ceramic and on the coaxial line surface is shown in Table 4.3.1.5.

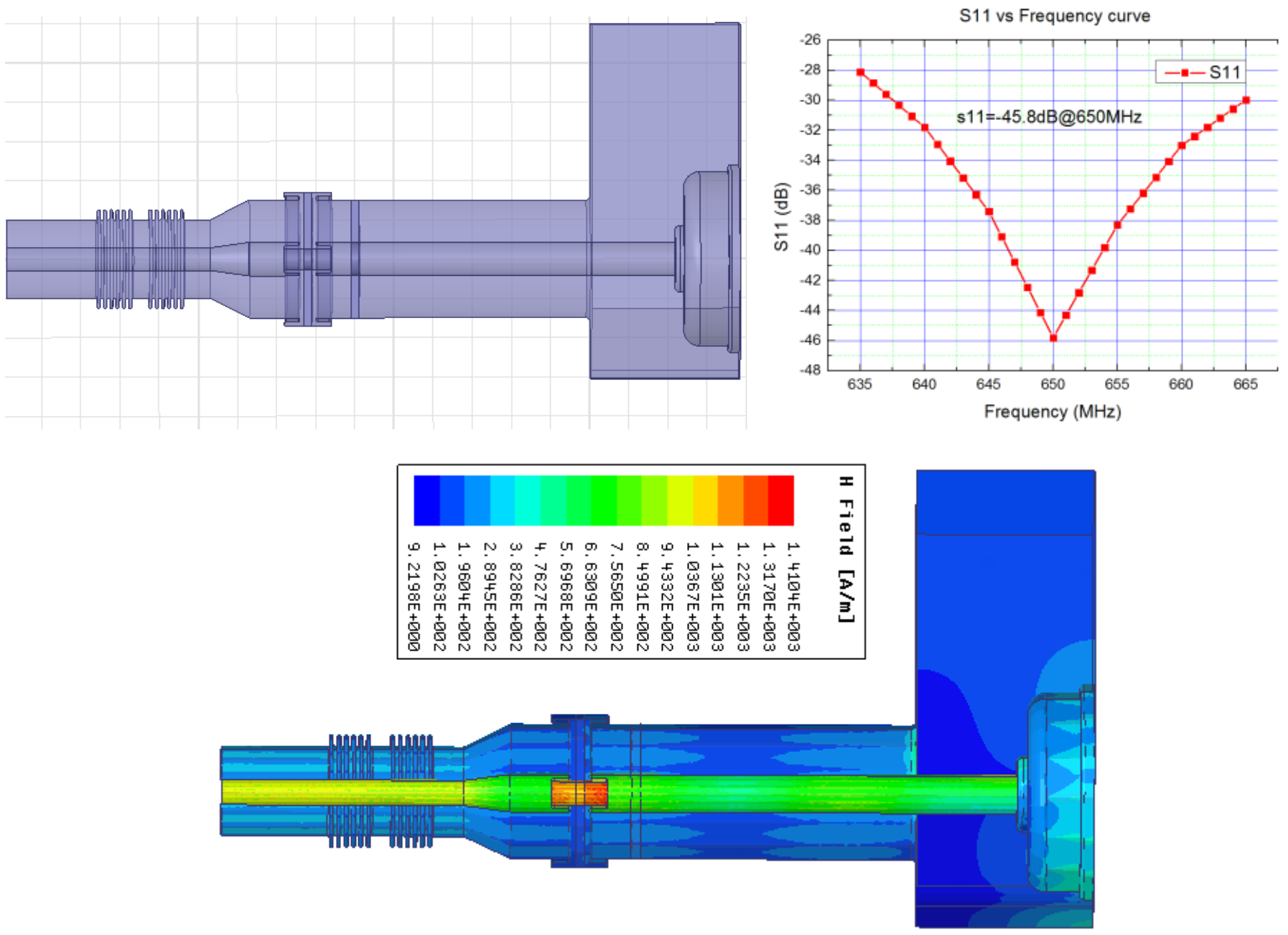

**Figure 4.3.1.4:** 650 MHz coupler RF design

**Table 4.3.1.5:** RF loss in 650 MHz coupler (at 300 kW CW travelling wave)

| | Power loss (W) |
|---|---|
| Ceramic | 113 |
| Inner conductor of window | 70.4 |
| Outer conductor of window | 15.8 |
| Inner coaxial of vacuum part | 103.4 |
| Outer coaxial of vacuum part | 50.0 |
| Inner coaxial of air part | 102.6 |
| Outer coaxial of air part | 23.1 |
| Doorknob | 77.0 |
| Total | 555 |

A disk window is used for the coupler. The coupler has a water cooled inner conductor and a low thermal conduction outer conductor. The length of the window and the vacuum part of the coaxial line was reduced to fit the cryomodule inner diameter. A doorknob structure is used for the waveguide-coax transition.

For the fixed-coupling coupler, the outer conductor is cooled by He-gas. The mechanical design is shown in Figure 4.3.1.5.

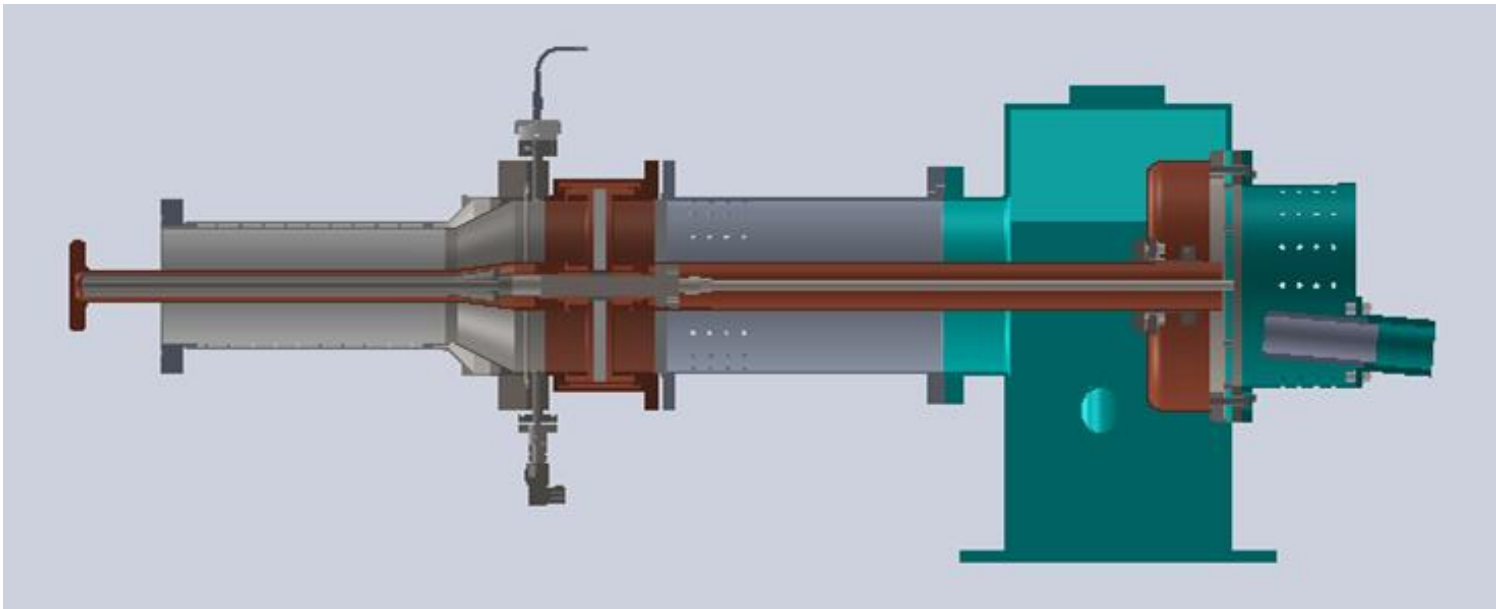

**Figure 4.3.1.5:** 650 MHz fixed-coupling coupler design

For the variable coupler, we have two preliminary designs (Figure 4.3.1.6): one is to add the bellows to the inner conductor, and use a He-gas-cooled outer conductor; another is to add the bellows to the outer conductor, and use a thermal-anchor-cooled outer conductor.

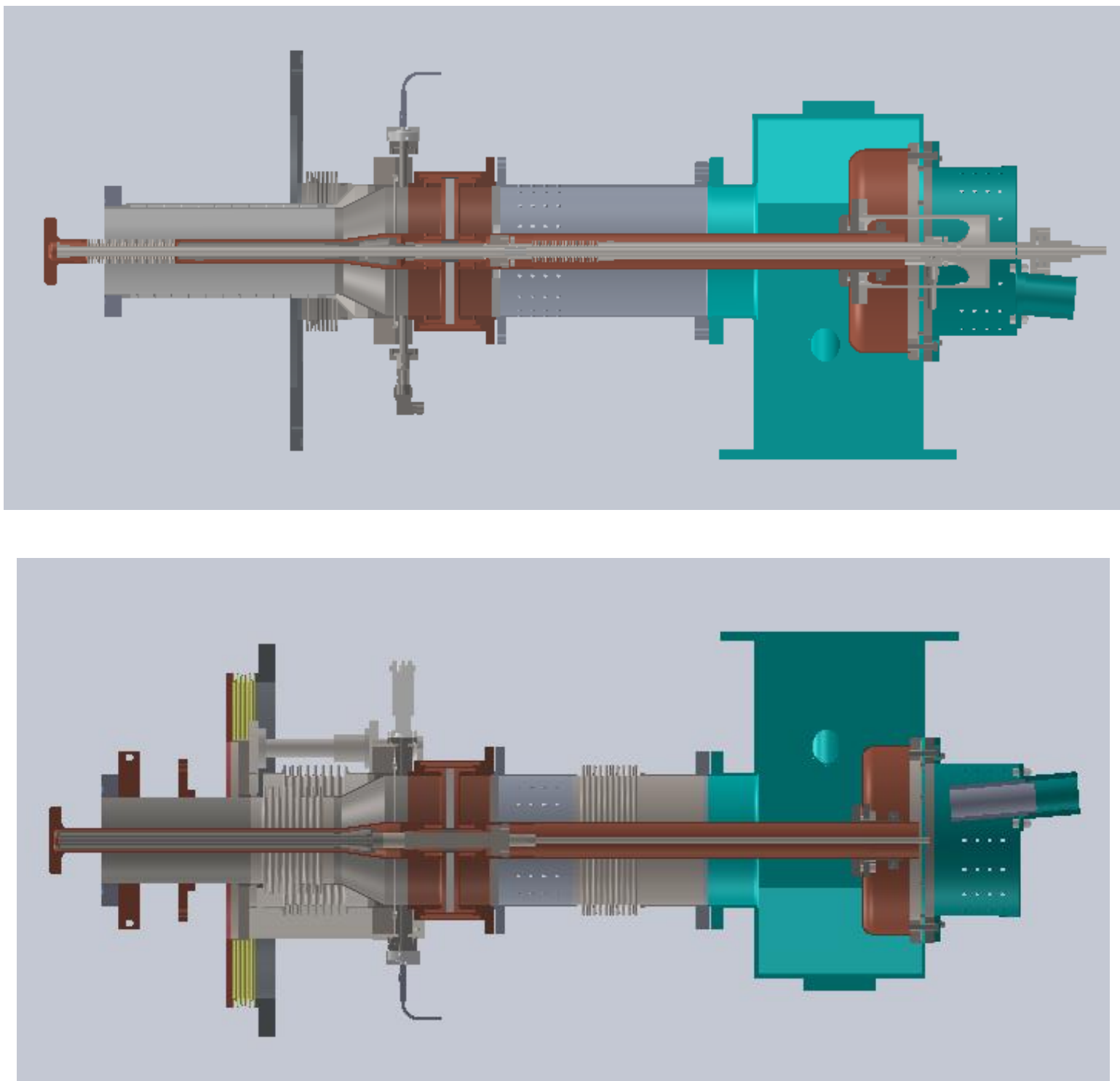

**Figure 4.3.1.6:** 650 MHz variable coupler designs

The heat load limit of the power coupler with 300 kW CW travelling wave power is: dynamic 1 W / 6 W / 10 W @ 2 K / 5 K / 80 K, static 0.3 W / 3 W / 6 W @ 2 K / 5 K / 80 K. Thermal performance will be optimized.

$Al_2O_3$ ceramic will be used for the window and the whole window will be brazed. The

inner conductor will be welded to the window by EBW. The outer conductor will be welded by TIG. The bellow will be welded by laser or TIG. Copper plating will be made by electroplating.

### 4.3.1.6 ***HOM Damper***

Higher-order-modes excited by the intense beam bunches must be damped to avoid additional cryogenic loss and beam instabilities. This is accomplished by extracting the stored energy via coaxial HOM couplers mounted on both sides of the cavity beam pipe. The HOM absorbers are outside the cryomodule. The kW level coaxial HOM couplers were used for LEP2 and LHC with good performance [19,20].

The average power losses can be calculated as a single pass excitation. As shown in Fig. 4.3.1.7, HOM power damping of 0.48 kW for each 650 MHz 2-cell cavity is required for the Collider Higgs mode.

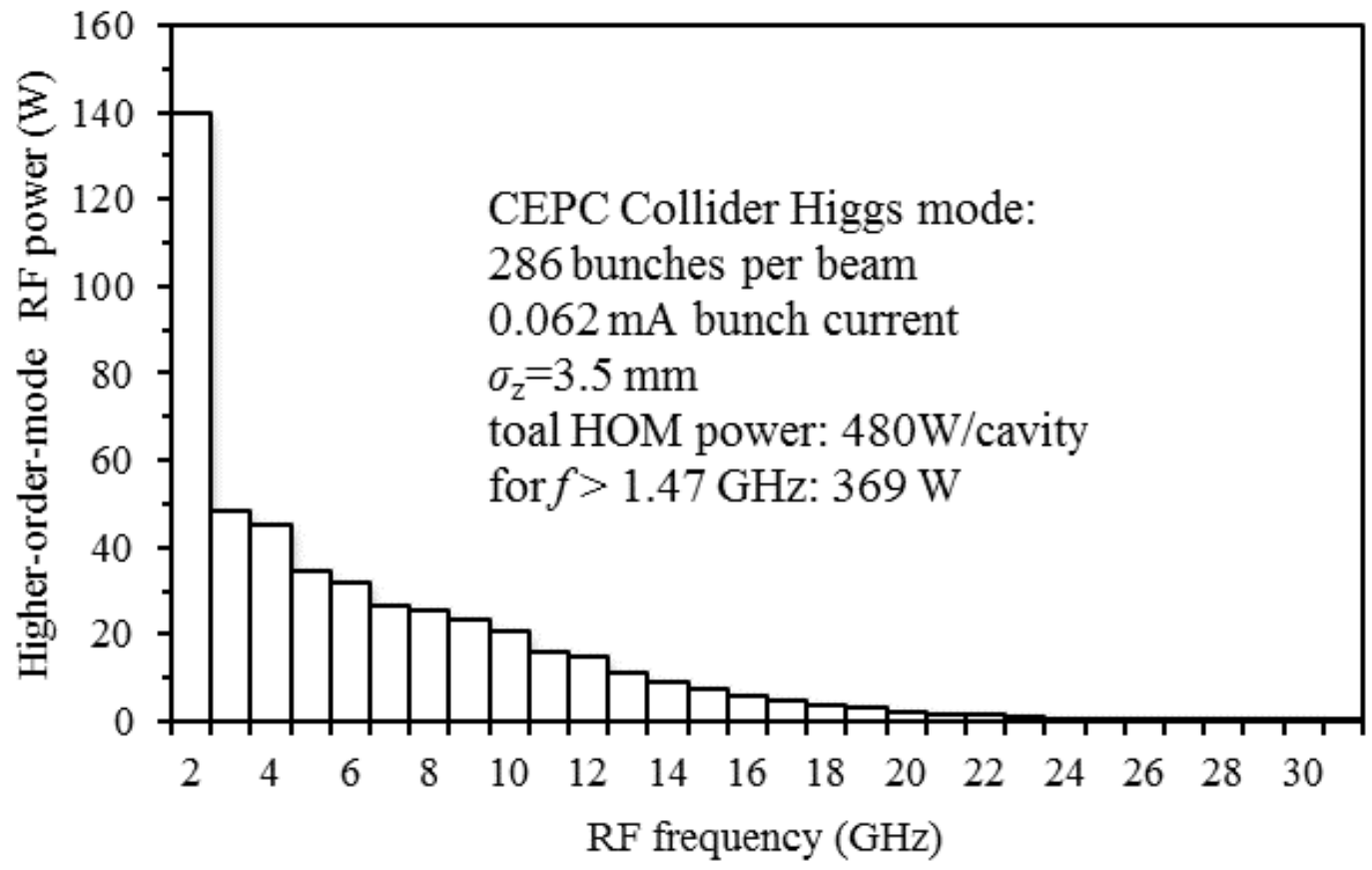

**Figure 4.3.1.7:** Frequency distribution of HOM power

Resonant excitation should be considered especially for the low frequency modes below cut-off. Different fill patterns will move the beam spectrum peaks and may generate smaller or larger power. By slightly changing the detuning of each cavity or fill pattern, resonance can be avoided. The cut-off frequencies of the beam pipe are 1.471 GHz (TM01) and 1.126 GHz (TE11). All the HOM power below the cut-off frequency is coupled by the HOM coupler mounted on the beam pipe. The propagating modes will be absorbed by the two HOM absorbers at room temperature outside the cryomodule. Each absorber has to damp several kWs of HOM power; thus the absorber can't be placed in the cryogenic region. Another concern related to the HOMs is that some modes far above the cut-off frequency may become trapped among cavities in the cryomodule due to the large frequency spread. HOM propagating properties of the entire cavity string will be an important issue for further study [21].

#### *4.3.1.6.1 HOM Coupler*

To damp different polarized HOMs, two HOM couplers per cavity with 110 degree angle between them are used. The HOM coupler design must be optimized for the operating frequency (high damping) and the HOM spectrum (low damping) of the cavities. A loop type HOM coupler is designed with the transmission line models (Figure 4.3.1.8) [22-24]. The coupler is designed to transfer 1 kW HOM power (expected in the worst case) and operate from 780 MHz to 1471 MHz. A double notch coupler is chosen due to its large bandwidth for the fundamental mode (Figure 4.3.1.9). To prevent the leakage of fundamental power from the HOM coupler, the $Q_e$ for the fundamental mode is designed to be larger than $10^{12}$. With good control of fabrication tolerance, the notch filter may not need to be tuned because of its large bandwidth (~ 100 MHz). A specially designed rigid coaxial line will be used to extract the HOM power.

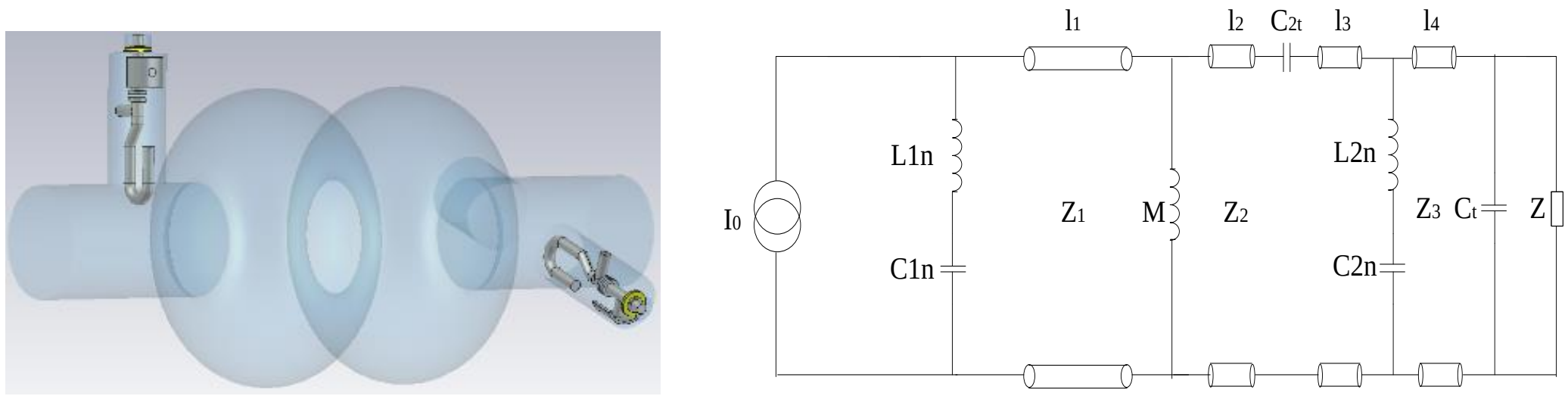

**Figure 1.3.1.8:** RF model and transmission line equivalent circuit of the HOM coupler

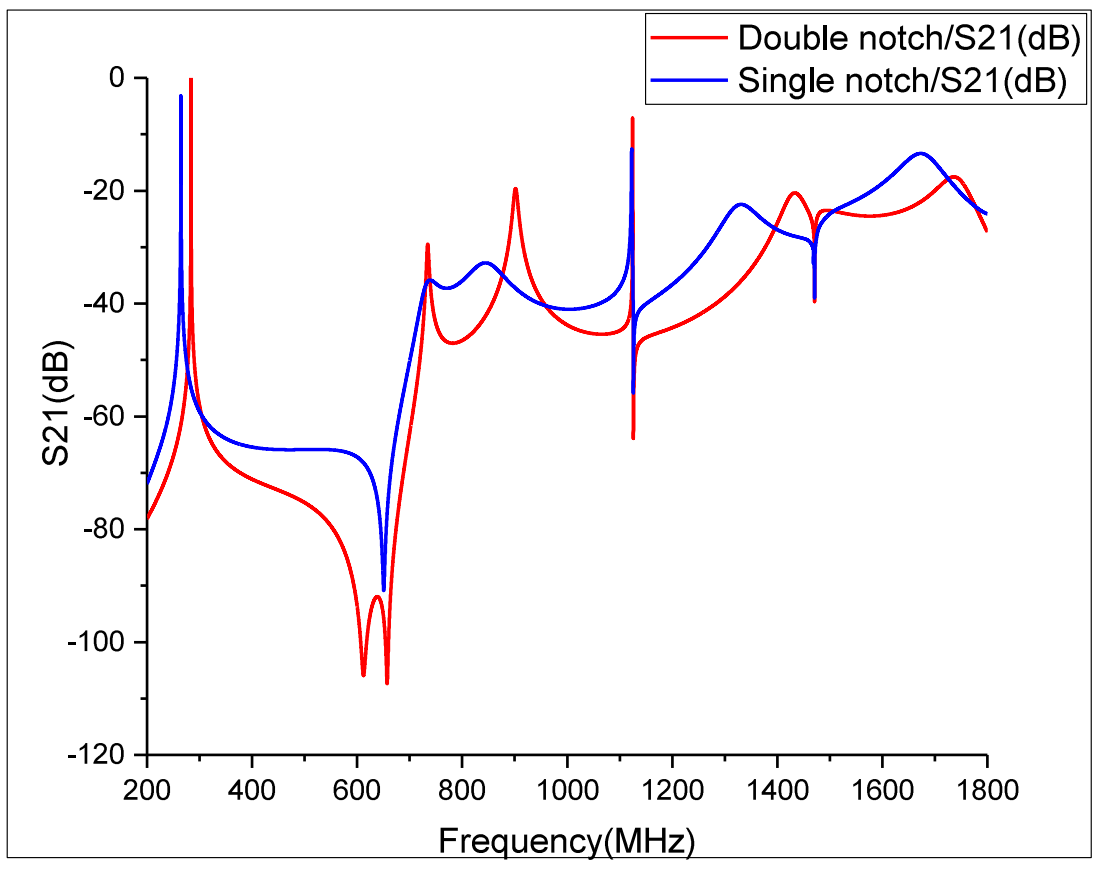

**Figure 4.3.1.9:** Transmission property of single notch and double notch couplers.

The external *Q* values are simulated for a 650 MHz 2-cell cavity equipped with two loop HOM couplers (Table 4.3.1.6). All modes under the cut off frequency are below the threshold except the TM011 mode of Z. The external Q of the TM011 mode could be reduced to 1E5 by optimising the loop shape and angle. Taking into account the frequency spread, the effective external Q will be under the threshold with beam feedback (9.2E3).

**Table 4.3.1.6:** External Q of the HOM coupler for 650 MHz 2-cell cavity

| Modes | *f* (GHz) | *R/Q* (monopole Ω, dipole Ω/m) | $Q_e$ simulation |
|---|---|---|---|
| TM011 | 1165.574 | 65.2 | 2.46E+05 |
| TM020 | 1383.898 | 1.3 | 1.55E+05 |
| TE111 | 844.738 | 279.8 | 1.50E+04 |
| TM110 | 907.592 | 420.1 | 4.58E+02 |

The TM010 0 mode of the 2-cell cavity (the Same Order Mode, SOM) may also drive instabilities or extract RF power from the beam.  Since the SOM is so close in frequency to the operating mode, it can’t be damped in the same way as HOMs using HOM couplers or beam tubes. The input coupler can be used as the SOM coupler.

The loop part of the HOM coupler is made with niobium. The 2 W coaxial line heat load, the 1 kW HOM power as well as the power dissipation under the operational condition of $E_{acc}$=20 MV/m caused by the fundamental mode are considered in the thermal analysis. Simulation results show that a helium tank outside the HOM coupler is needed in order to keep the loop part in a superconducting state. The HOM coupler structure includes two parts. One part consists of a niobium loop, a niobium shell, two flanges and helium jacket made of Nb55Ti. The other consists of a ceramic window, a copper probe and a stainless-steel flange. The mechanical structure and assembly steps are shown in Fig. 4.3.1.10.

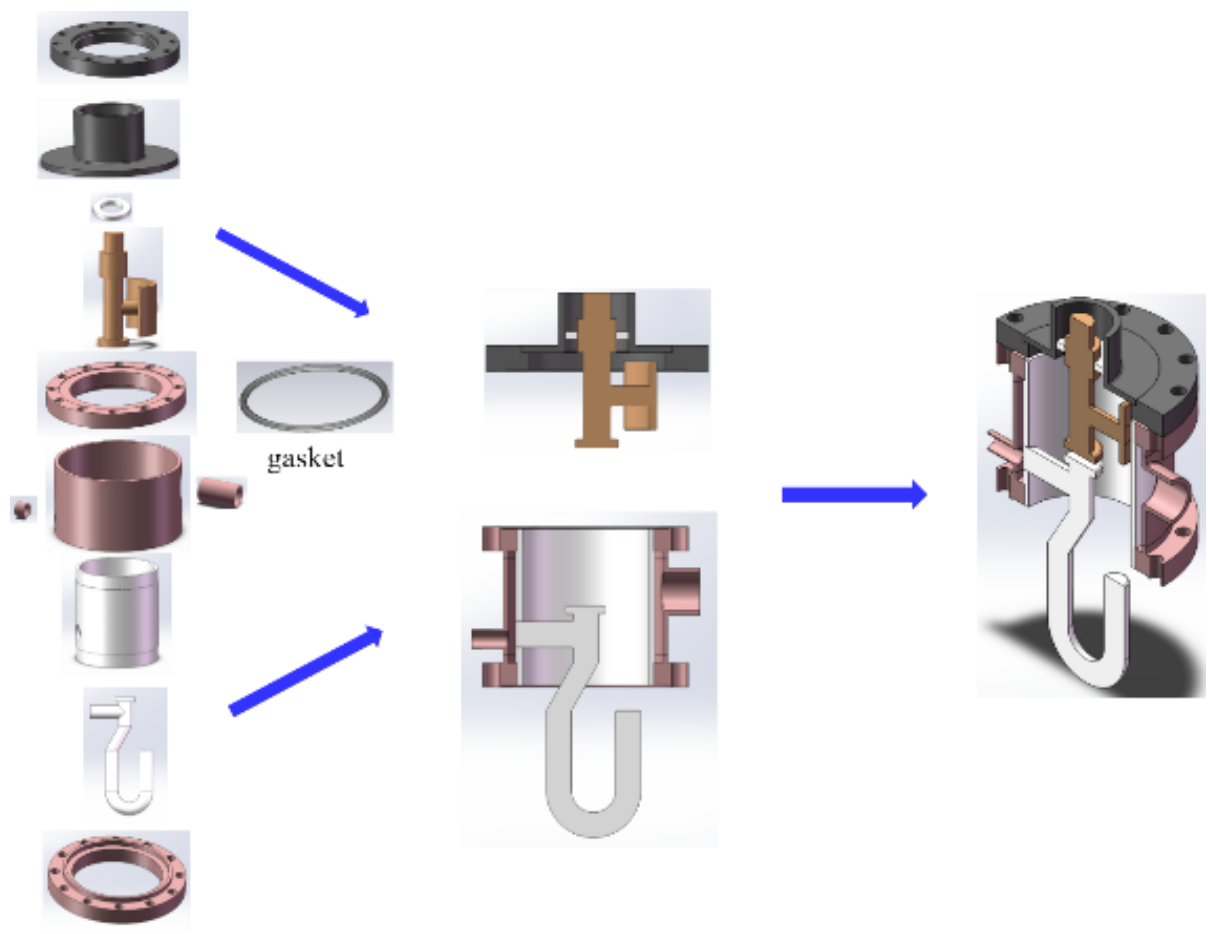

**Figure 4.3.1.10:** Mechanical structure of the HOM coupler.

#### *4.3.1.6.2 HOM Absorber*

The HOM absorber is mainly used to damp the HOM power above 1.4 GHz. The structure of the HOM absorber is similar to a circular waveguide. The microwave absorbing material will be brazed onto the inner surface of the waveguide. One or two types of absorbing material will be needed for the wide frequency range due to the short bunch length.

The size of the waveguide of the absorber is the same as the beam pipe of the 650 MHz cavity. In the initial design, ferrite is used to absorb the HOM power. The ferrite is cut into rectangular bricks. This will reduce the fabrication cost dramatically and will also reduce the difficulty in ferrite machining and brazing. For further higher damping requirements, this structure makes it feasible to mount different kinds of absorbing material to fulfil broad band operation. Effective cooling for each ferrite brick is accomplished with the cooling structure shown in Figure 4.3.1.11. There is a balance between structure complexity and cooling efficiency.

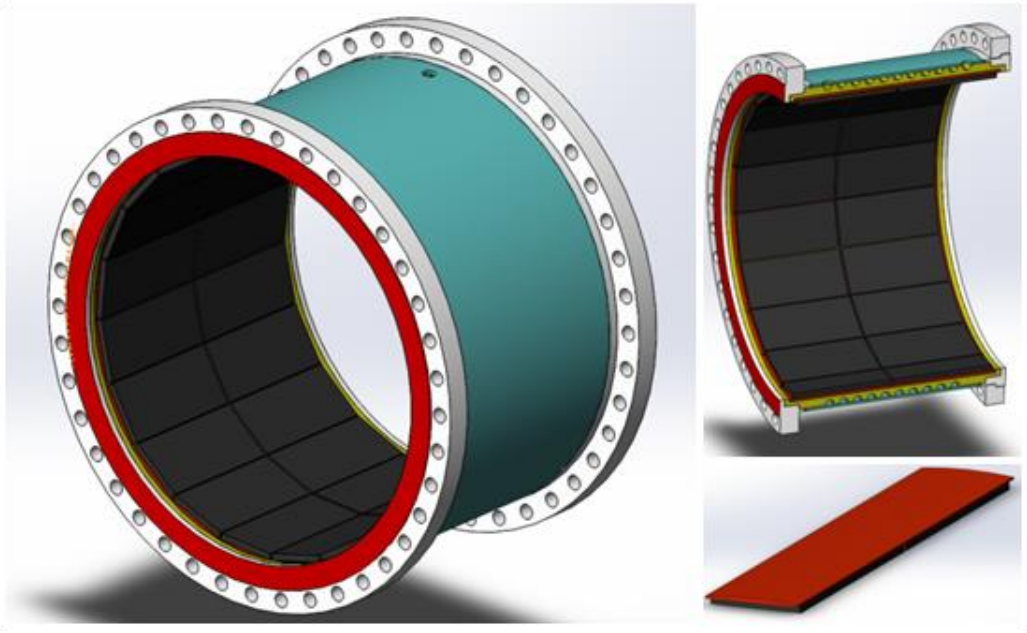

**Figure 4.3.1.11:** Mechanical and cooling structure of the HOM absorber

### 4.3.1.7 *Frequency Tuner*

The cavity tuner is used to compensate or damp frequency changes due to Lorenz force, beam loading effects and microphonics. Development of the frequency tuning system is challenging in terms of material shrinkage, large tuning range, high resolution, extreme working conditions at low temperature and the ultra-high vacuum and radiation environment. Highly reliable and maintainable tuners are required. The main parameters of the tuning system are listed in Table 4.3.1.7. Access ports will be necessary for easy maintenance.

**Table 4.3.1.7:** Main parameters for the Collider 650 MHz 2-cell cavity tuner

| Parameters | Unit | Value |
|---|---|---|
| Cavity tuning sensitivity | kHz / mm | 310 |
| Cavity spring constant | kN / mm | 16 |
| Coarse (slow) tuner frequency range | kHz | 340 |
| Coarse tuner frequency resolution | Hz | < 20 |
| Fine (fast) tuner frequency range | kHz | > 1.5 |
| Fine tuner frequency resolution | Hz | 3 |
| Motor and piezo temperature | K | 5~10 |
| Motor number | | 1 |
| Piezo number | | 2 |
| Operating pressure | Torr | < 5E-5 |
| Operating lifetime | year | 20 |

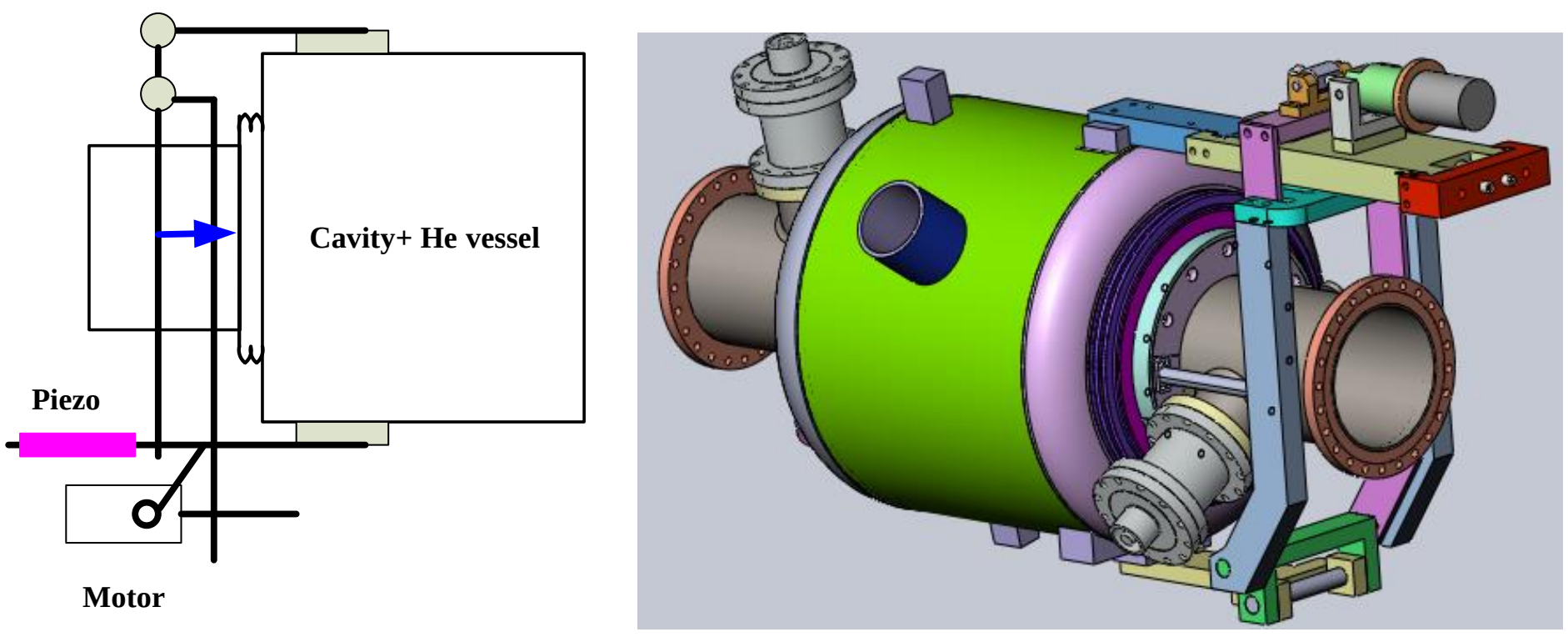

**Figure 4.3.1.12:** Working principle and 3D model of tuner for 650 MHz 2- cell cavity

A double-lever tuner is designed for the Collider Ring cavity (Figure 4.3.1.12) [25]. A main arm is hinged at one end and connects to the actuation system at the other end, and pushes the flange on the cavity end plate through two rods in the middle. The actuation system consists of a low temperature stepper motor and two piezoelectric ceramic actuators with large tuning stroke. The motor is held by a bracket and connected to a secondary arm.

### 4.3.1.8 *Cryomodule*

The 650 MHz cavity cryomodule operates at 2 K in superfluid helium. It houses six 2-cell 650 MHz superconducting cavities, six power couplers, six tuners and two HOM absorbers. High $Q_0$ requirement drives new design features (fast cool down and magnetic hygiene). The static heat load of the whole cryomodule will be 5 W at 2 K.

The cryomodule consists of the outer vacuum vessel, the cavity string of six 650 MHz cavities and their auxiliary components, cryogenic pips, support fixtures and thermal shields (Figure 4.3.1.13).

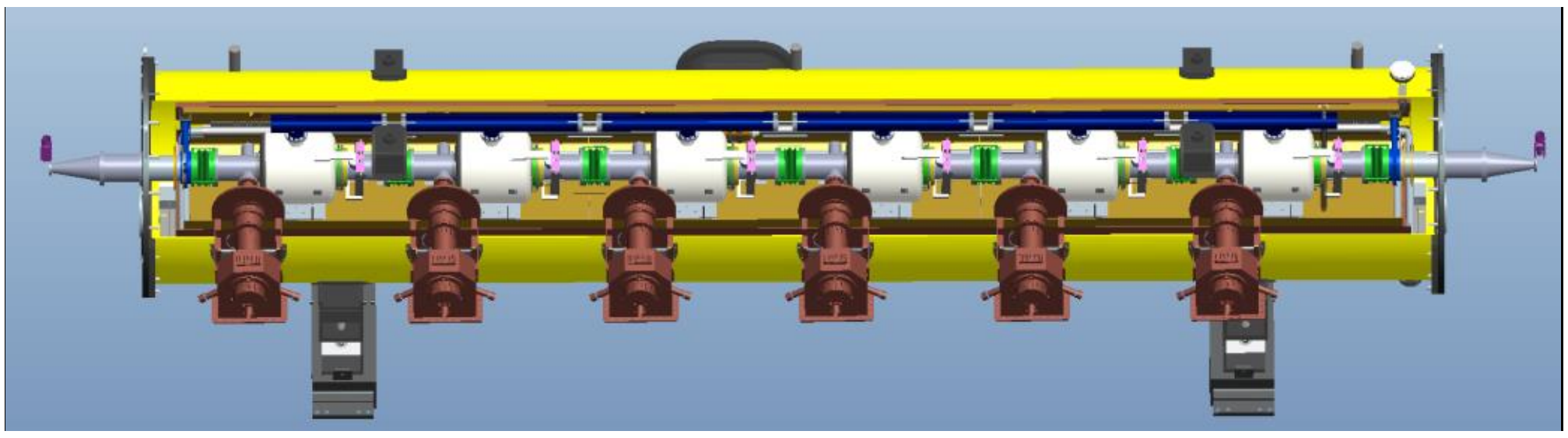

Overall view of the cryomodule

**Figure 4.3.1.13:** Conceptual model of the Collider Ring cryomodule

The structure of the cryomodule is as follows: a stainless steel vacuum vessel with a diameter about 1400 mm, strongback at room temperature acting as a support structure, together with 6 posts on top of the strongback, a 2 K two-phase pipe connected to the cavity Helium vessels, a 5 K forward and return line, an 80 K forward and return line, a warm-up/cool-down line with capillaries to the bottom of each cavity vessel, and aluminium thermal shields with stiff upper parts for 5 K and 80 K with 10 layers of super-insulation (MLI) for 5 K and 30 layers for 80 K, attached to the support structure. Due to large dynamic heat load at 2 K, 5 K shield is not necessary while only keeping the intercepts to cool individual components.

To maintain the high $Q_0$ performance from vertical tests to cryomodules, fast cool down and flux expulsion should be used [26]. A fast cool down rate is from 10 K/min from 45 K to 4.5 K. The goal is to achieve more than a 4 K cavity top to bottom temperature difference at the beginning of superconducting transition. Large helium mass flow is needed for this cool down rate. A closed-end warm-up/cool-down manifold will be created for each cryomodule by providing a cool-down/warm-up valve on each cryomodule.

Nitrogen-doped cavities are more sensitive to surface magnetic field which causes surface resistance increase. The ambient magnetic field at the cavity surface should be less than 5 mG. Magnetic shielding and demagnetization of parts and the whole module should be implemented for the magnetic hygiene control [27].

## 4.3.2 RF Power Source

### 4.3.2.1 *Introduction*

High power radio frequency sources are required to provide the energy needed to accelerate particles. The RF power needs to be stable such that any variation in the supplied RF power has a limited and acceptable impact on the Collider beam quality.

The RF power source delivers energy to the electrons to compensate for their energy loss from synchrotron radiation and from interactions with the beam chamber impedance. The RF power source also delivers energy to the beam when ramping to higher energy and captures and focuses the electrons into bunches. The beam and the RF stations are two strongly interacting dynamic systems and this complicates stability considerations for the combined system.

The Collider beam power is more than 60 MW; high power klystrons are the more attractive choice because of their high efficiency, more than the solid state amplifier and other power sources.

### 4.3.2.2 *Collider RF Transmission System*

The Collider SRF system consists of 240 2-cell cavities. A minimum of 300 kW transmitted to the cavity is required to meet the sum of the radiated, HOM and reflected power demands [1]. Table 4.3.2.1 summarizes the RF power demands.

**Table 4.3.2.1:** SRF system parameters

| Parameters | Higgs |
|---|---|
| Operation frequency(MHz) | 650+/-0.5 |
| Cavity number | 240 |
| Coupler input power（kW） | 300 |

A single 800 kW klystron amplifier will drive two 2-cell cavities through a magic tee and one circulator and load. The choice of one klystron for two cavities is technically justified by better control of microphonic noise and minimum perturbation in the case of a klystron trip [2].

The linear operation of the klystron as well as the transmission losses is taken into account. The RF power produced by the klystron will be delivered by a standard WR1500 aluminium waveguide. A schematic of the RF Transmission System (RFTS) and the placement of the klystron and waveguide are shown in Figures 4.3.2.1 and 4.3.2.2.

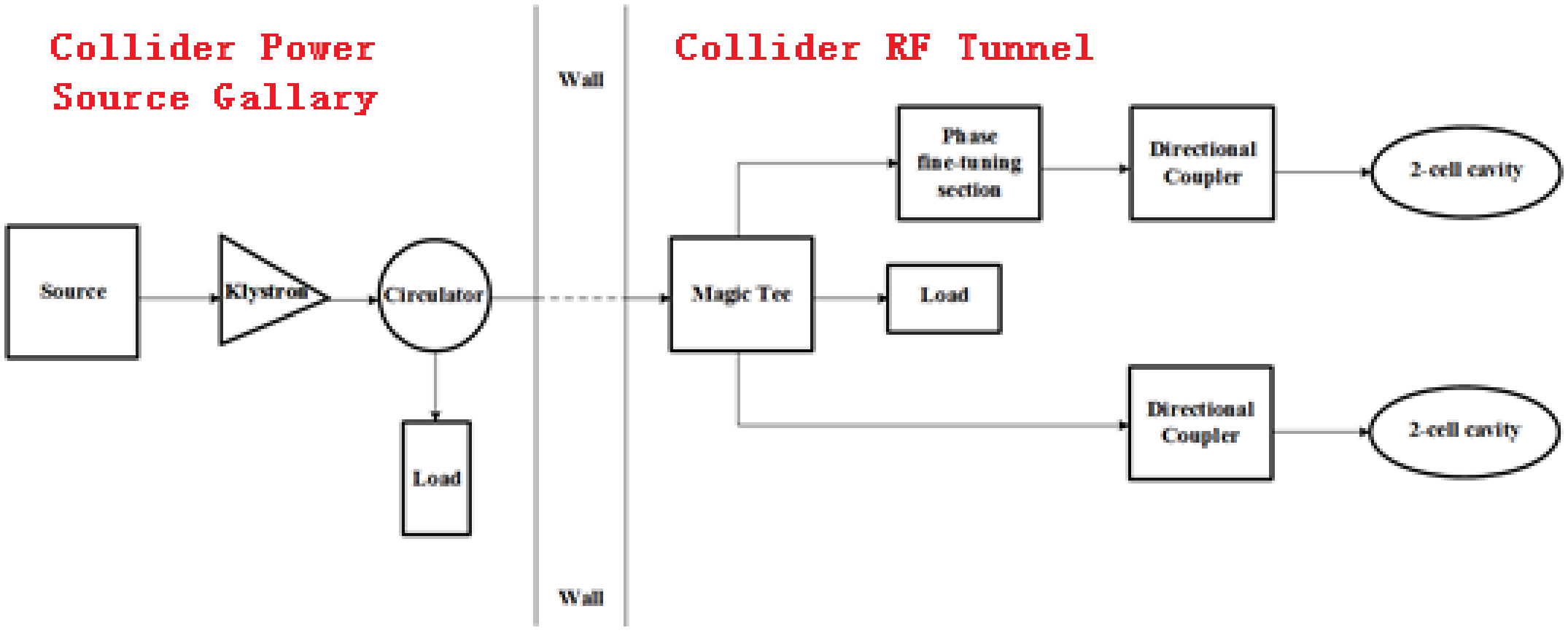

**Figure 4.3.2.1:** Schematic of the Collider RFTS

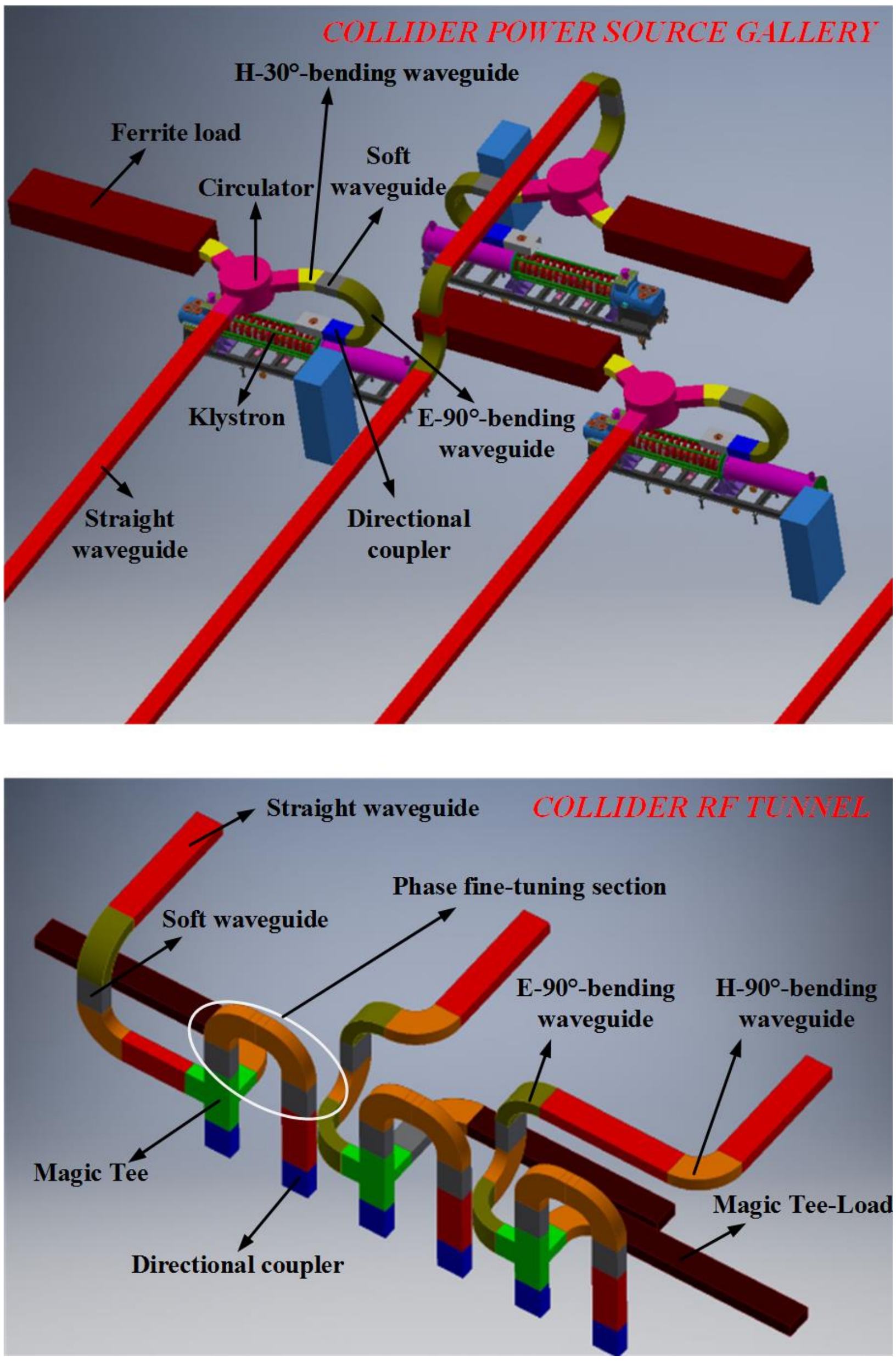

**Figure 4.3.2.2:** Placement of the klystron and waveguide

In order to protect the klystron, a CW 800 kW circulator is installed in each feed line to block power either reflected or discharged from the cavity. The directional coupler will provide both forward and backward (reflected) pick up signals with directivity better than 30 dB. In order to compensate any phase differences introduced in the system, a phase fine-tuning section is inserted into one of the paths to further adjust the phase accurately, which is done by slightly changing the length of the two soft waveguides in the section.

#### 4.3.2.3 *High Efficiency Klystron*

The acquisition of a high efficiency RF power source for CEPC is a key issue. It is well known that klystron efficiency is strongly dependent on beam perveance of the tube. For pulsed klystrons which have relatively high perveance, the efficiency ranges between 40% and 45%. For klystrons that operate in CW or long pulsed mode, the perveance is relatively low and efficiency can reach 65%. Multi-beam klystrons are favored for their high efficiency of more than 65% [3]. In a recent theoretical calculation [4] 90% RF power conversion efficiency is achieved. Considering this recent high efficiency approach, our design goal is to achieve around 80%. It should be noted that the higher the efficiency, the greater the probability for unstable oscillations due to back scattering of electrons generated at the output cavity gap. Another issue to consider is how to increase the efficiency at the operating point (linear region of the power transfer curve). It is also necessary to consider the focusing electromagnet power consumption. The design parameters for a 650 MHz klystron for the case of conventional single beam, high efficiency and MBK approach are listed in Table 4.3.2.2.

**Table 4.3.2.2:** CEPC Klystron Key Design Parameters

| Parameters | Units | Values |
|---|---|---|
| Centre frequency | MHz | 650±0.5 |
| Output power | kW | 800 |
| Efficiency(Goal) | % | 80 |

#### 4.3.2.4 *PSM Power Supply*

The RF amplifier-klystron power supply is PSM type HVDCPS (PSM: pulse step modulation, HVDC: high voltage direct current，PS：power supply). The performance of the PSM power supply will determine the beam quality. Other auxiliary power supplies are integrated into the PSM power supply. These include anode DC power, filament power, focus power and ion pump power. The power supplies and the control system are housed in three different cabinets which together comprise a single unit.

The DC power supply for the cathode is a PSM supply currently used in broadcast transmitters. PSM switching power supplies have the benefit of low energy storage and fast turn-off capability of the IGBT (Insulated Gate Bipolar Transistor). This eliminates the need for a protection crowbar circuit. Designed for 120 kV 15 A, this PSM essentially consists of 168 power modules connected in series and supplied through their own secondary windings from four transformers. The four transformers are shifted in phase, resulting in 24-pulse loading of the mains with 6-pulse rectification in the module chain. Every HV transformer distributes power to 42 DC modules. Each module can provide 800-volt power, which can be switched on/off individually by fast IGBT switches operating at 1 kHz. Figure 4.3.2.3 is a schematic of the PSM power supply.

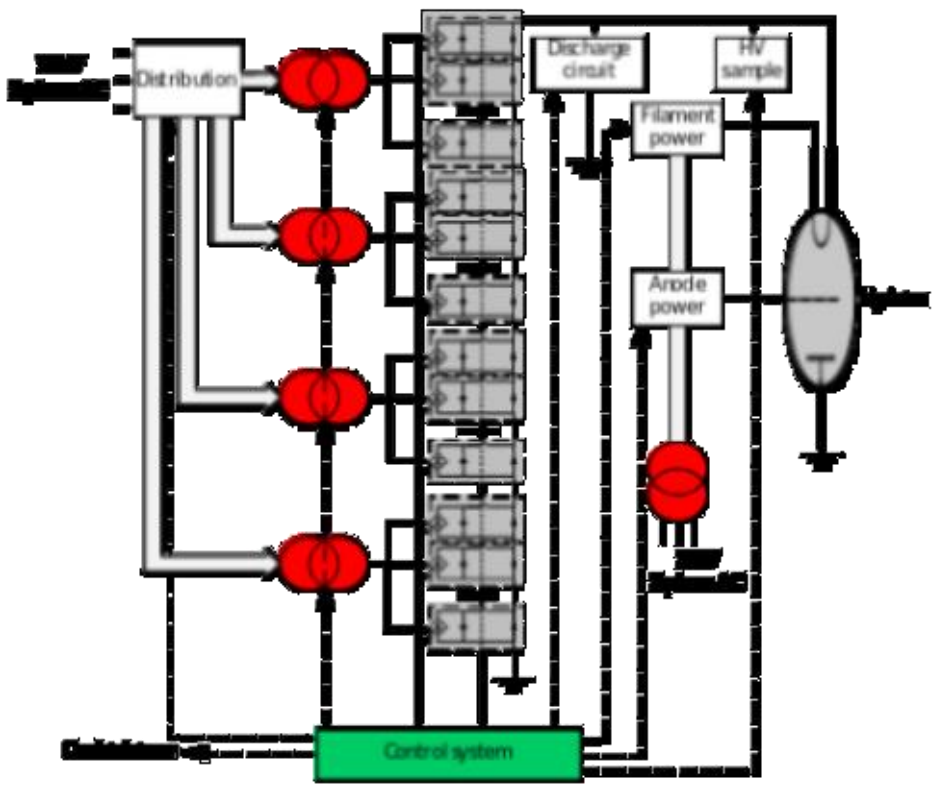

**Figure 4.3.2.3:** PSM power supply schematic

The main PSM features are efficiency, regulation speed, accuracy and compatibility to large variations in load impedance and are a good fit to the performance specifications in Table 4.3.3. For better RF source quality, the PSM can operate in PWM (pulse width modulation) mode [5] with switch frequency 1 kHz. Output voltage is smoother and ripple stability can easily be controlled to better than 0.2% [6].

**Table 4.3.2.3:** PSM performance specification

| Parameters | Units | Values |
|---|---|---|
| High voltage | kV | 120 |
| Current | A | 15 |
| Module quantity | | 168 |
| Module voltage | V | 800 |
| Module switch frequency | Hz | 1k |
| Module number of redundancy | | 9 |
| Voltage stability | % | < 0.2 |
| Efficiency | % | >95 |
| Turn-off time | us | <5 |
| Stored energy | J | <15 |

### 4.3.2.5 *Low Level RF System*

The successful experience of allocating phase reference and maintaining coherence in large facilities like PEPII and LHC [7-9] helps greatly in the design. We will take the combined advantage of vector sum and signal cavity control methods to meet user and system requirements. The LLRF system controls that the klystrons work effectively and compensates for the beam loading, Lorentz Force detuning and slow drifts or other interference caused by environmental dependency of components. Figure 4.3.2.4 shows the LLRF layout.

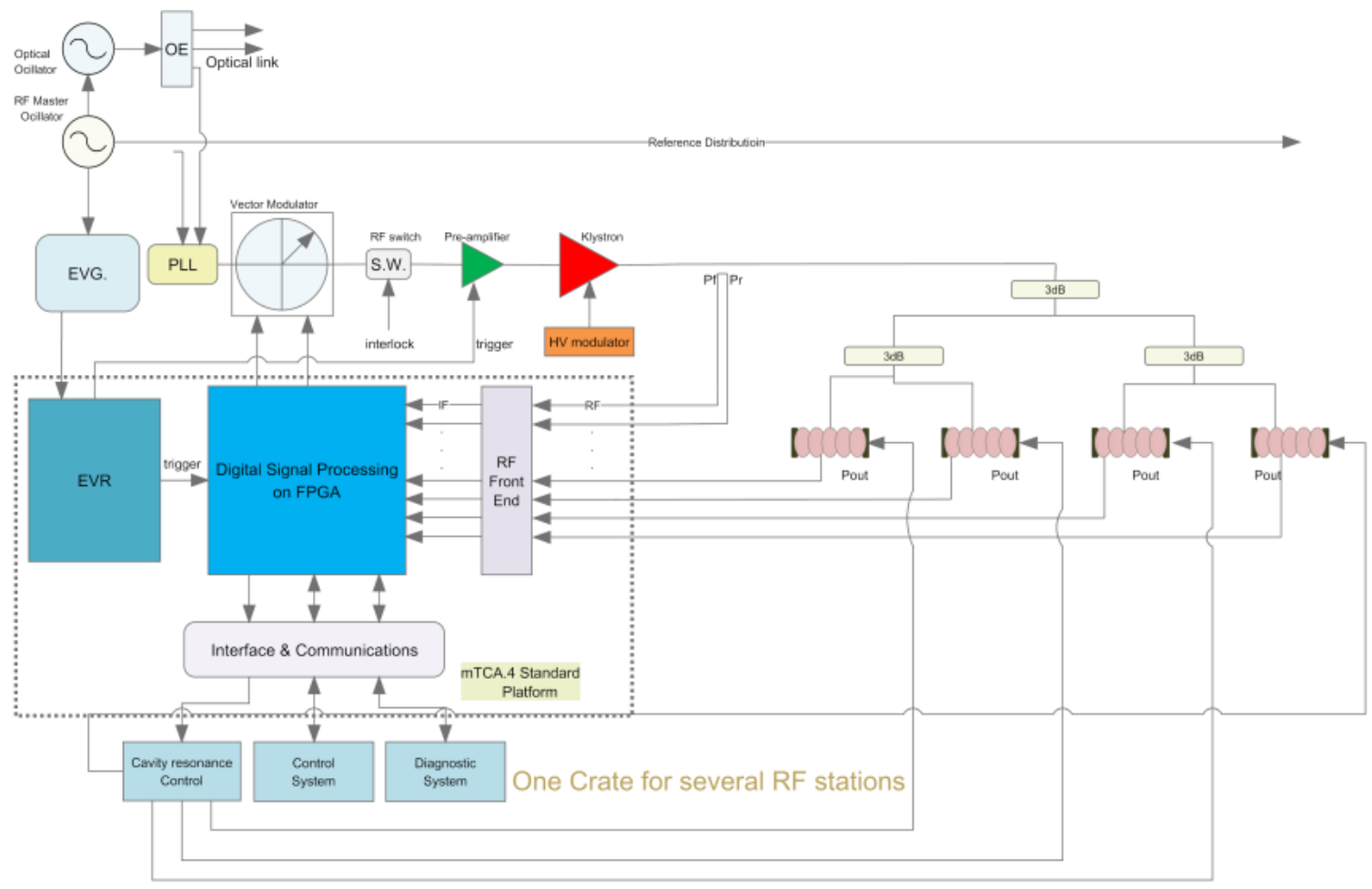

**Figure 4.3.2.4:** The LLRF system layout

The LLRF specifications are listed in Table 4.3.2.5. The signals from the DAC output, the pre-amplifier output, the klystron output forward & reverse, the 3 dB power splitter output forward & reverse, and the pickup of the cavities are sampled simultaneously to monitor the full system status.

**Table 4.3.2.5:** LLRF specification

| Parameters | Value/Unit |
|---|---|
| Phase stabilization | 0.1deg (rms) |
| Amplitude stabilization | 0.1% (rms) |
| Tuner range | >200kHz |
| Piezo range | >500Hz |
| Signal number | 20 channel |

### 4.3.3 **Magnets**

#### 4.3.3.1 ***Overview of the Collider Magnets***

For each ring, there are 2466 dipoles, 3052 quadrupoles, 948 sextupoles and 2904 correctors. Most of the magnets are conventional except 8 quadrupoles and 16 sextupoles are superconducting. There are 9370 magnets in one Collider ring and the magnets occupy over 80% of the circumference. The cost and power consumption are two of the most important issues for the magnet design. To reduce the cost and power consumption, 2384 dipoles and 2392 quadrupoles are designed to be dual aperture magnets to provide magnetic field for both beams. Besides the dual aperture magnet design, several special technologies are used to reduce the cost of the magnets, including core steel dilution for dipoles and aluminum coils instead of copper. To reduce the power consumption of the magnets, the current density of the coils is designed to be a relatively low value. In addition, the magnets are designed for low-current high-voltage operating mode as much as possible to reduce the power consumption in the power cables. Dual aperture dipoles, dual aperture quadrupoles and the sextupoles are designed and studied with 2D software.

From LEP experience, aging of the magnet coil from radiation will be a serious problem. Radiation shielding is considered in the magnet design. The details are discussed in Section 4.2.4.

#### 4.3.3.2 ***Dual Aperture Dipole***

The total length of the dual aperture dipoles is 68% of the length of the ring, so the design should be compact, simple and power saving. The ‘I’ shaped core sharing one coil and providing two identical fields is chosen for the dual aperture dipole. This can save about 50% power consumption compared to two separate dipoles. To make the dual aperture dipole fabrication easy, the magnet core is divided into 5 segments of about 5.7 m length each. For the first and the last segments, sextupole field is combined with the dipole field. This configuration is used to reduce the required strength of the individual sextupoles.

To facilitate the installation of the vacuum chamber, the pole gap of the dipole is increased to 70 mm. To shield from radiation from the high energy e+/e- beams, shielding blocks made of 30 mm thick lead will be added over the vacuum chamber. Considering the beam energy saw tooth effect, trim coils are used for both apertures to adjust the field by ±1.5% independently. 2D simulation results show that adjusting the trim coil current in one aperture has no obvious effect on the field in the other aperture. Subsequently, 3D model will be created to simulate the effect of fringe field. Figure 4.3.3.1 shows the cross section of the dual aperture dipole and Figure 4.3.3.2 gives the horizontal field error distribution at the mid plane of the dipole only segments. After pole shimming, the field uniformity meets the design requirement.

The cross section of the main coils are designed as large as possible to decrease the current density of the coil. The coils are made of aluminum to reduce the cost as well as the magnet weight.

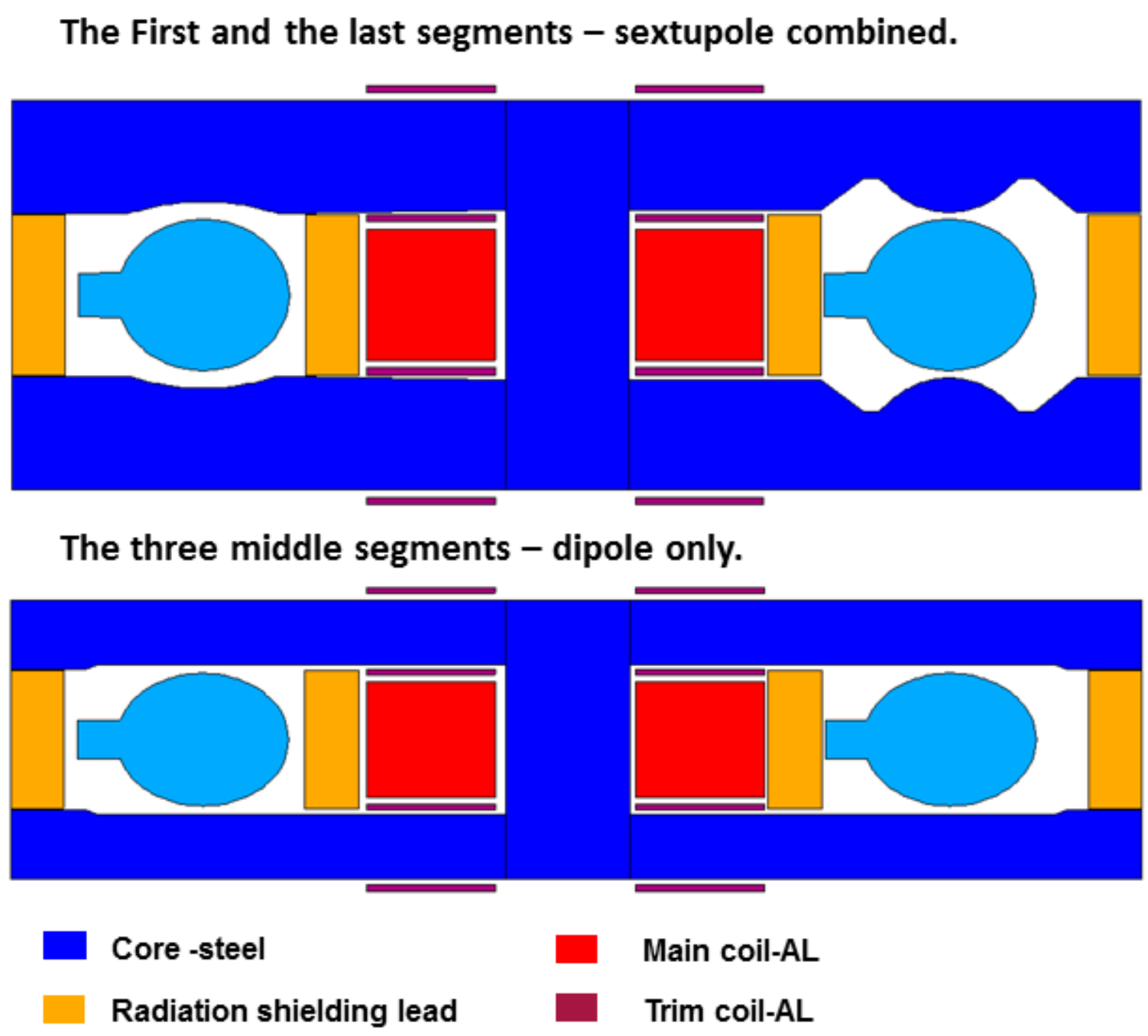

**Figure 4.3.3.1:** Cross section of the dual aperture dipole.

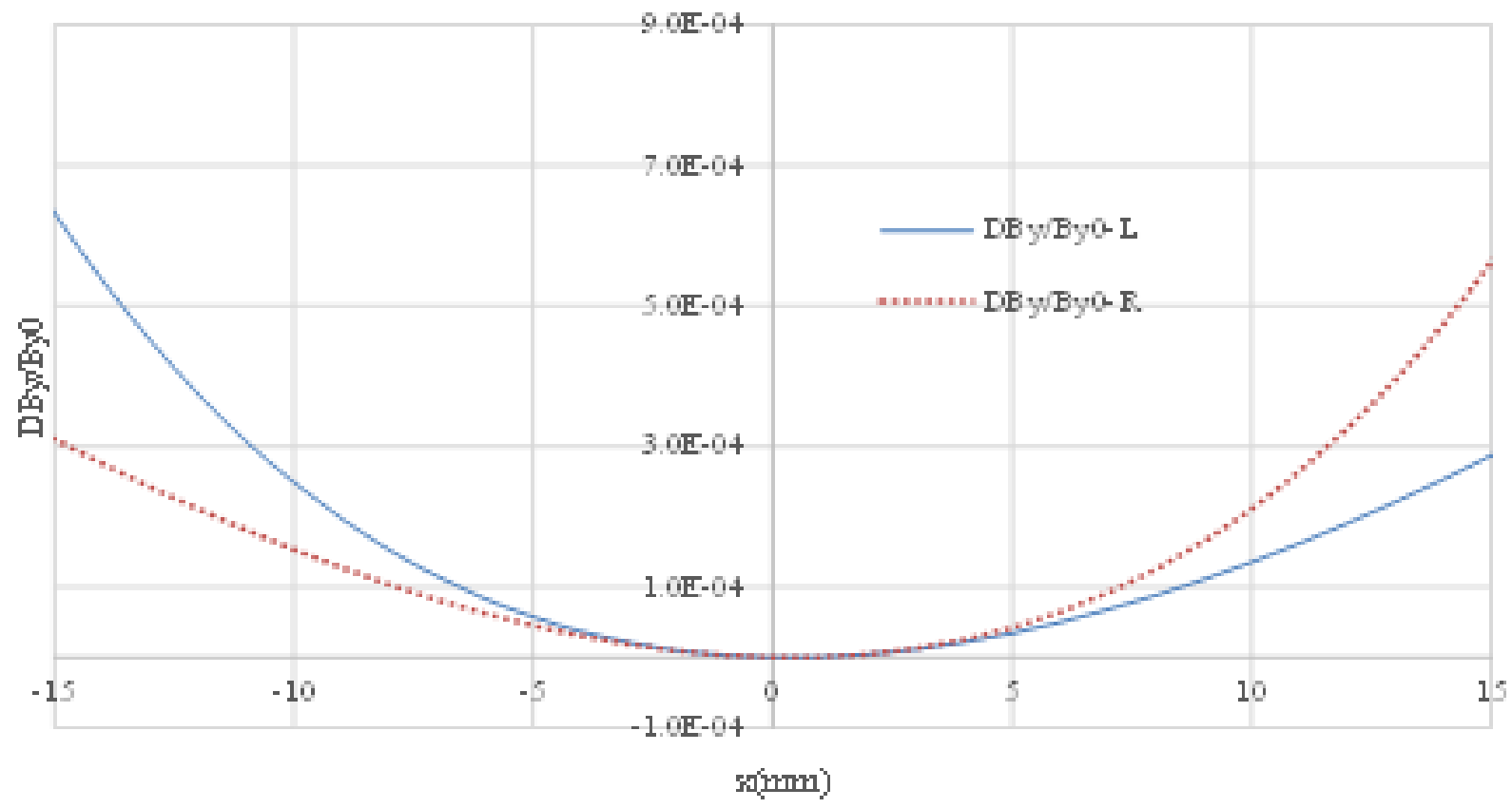

**Figure 4.3.3.2:** The normalized field error distribution of the dual aperture dipole.

The design parameters of dual aperture dipole are listed in Table 4.3.3.5.

**Table 4.3.3.5:** Parameters of the dual aperture dipole.

| | | |
|---|---|---|
| Beam center separation [mm] | | 350 |
| Magnetic length [m] | | 28.686 |
| Magnetic strength [Gs] | | 373.4 |
| Gap [mm] | | 70 |
| Coil | Number | 2 |
| | Shape | Racetrack |
| | Material | Aluminum |
| | Conductor specs. [mm] | 30×54 |
| Current [A] | | 1058 |
| Current density [A/mm$^2$] | | 0.67 |
| Resistance [mΩ] | | 2.44 |
| Voltage [V] | | 2.58 |
| Power consumption [kW] | | 2.73 |
| Cooling water | Loop number | 1 |
| | Pressure drop [kg/cm$^2$] | 6 |
| | Velocity [m/s] | 1.75 |
| | Flux [l/s] | 0.138 |
| | Temperature rise [°C] | 4.7 |

#### 4.3.3.3 *Dual Aperture Quadrupole*

The dual aperture quadrupole has different polarities in the two apertures. There are two racetrack main coils made of aluminum. Compared with two independent single aperture quadrupoles, the dual aperture quadrupole saves nearly 50% in energy. Trim coils in both apertures have ±1.5% adjustment capability to compensate for the sawtooth in beam energy. To reduce the field coupling between the two apertures, a 50 mm gap filled with stainless steel is inserted into the yoke and separates the yoke into two parts. Figure 4.3.3.3 gives the cross section of the dual aperture quadrupole and Figure 4.3.3.4 shows the magnetic flux.

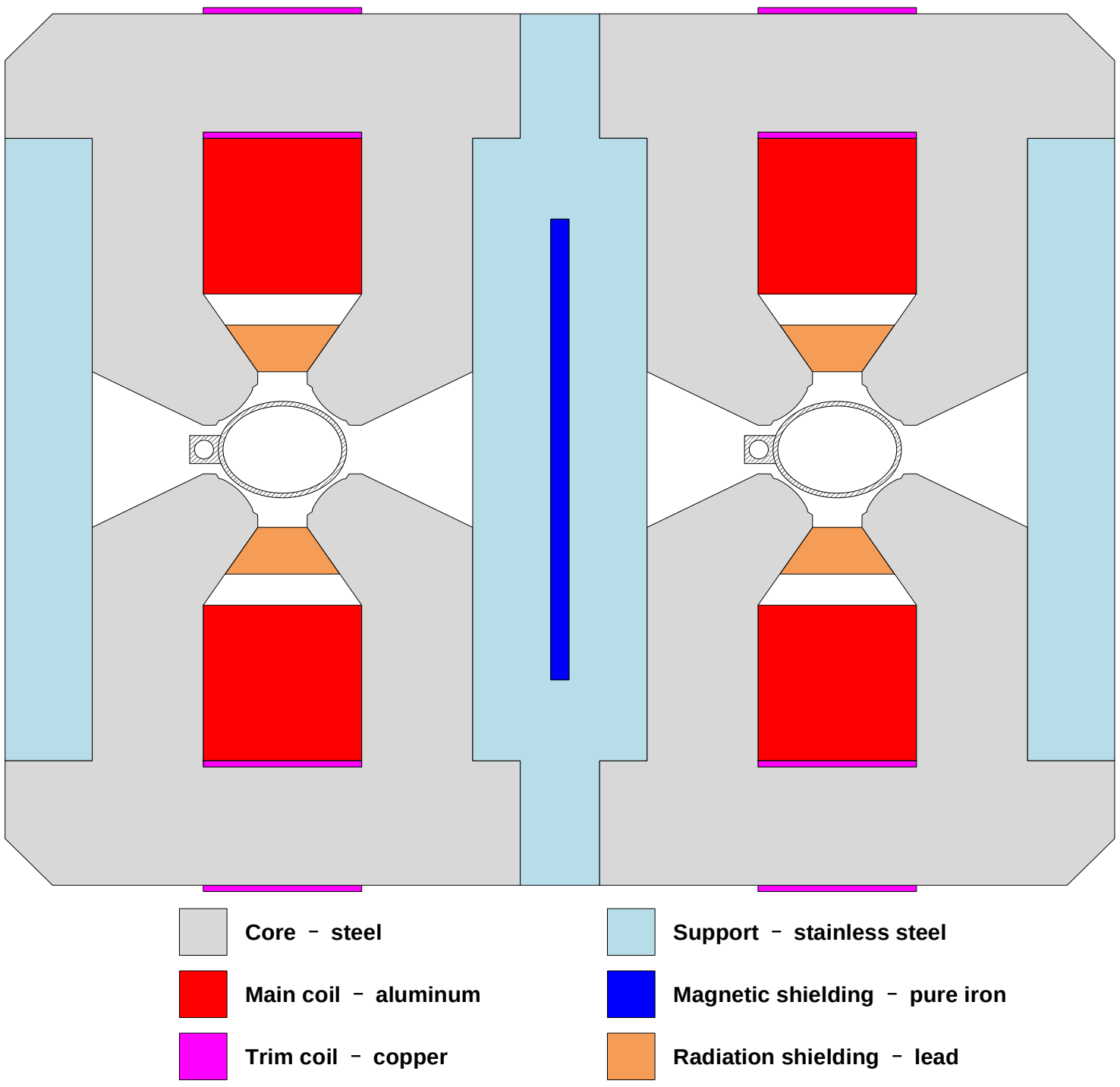

**Figure 4.3.3.3:** Cross section of the dual aperture quadrupole.

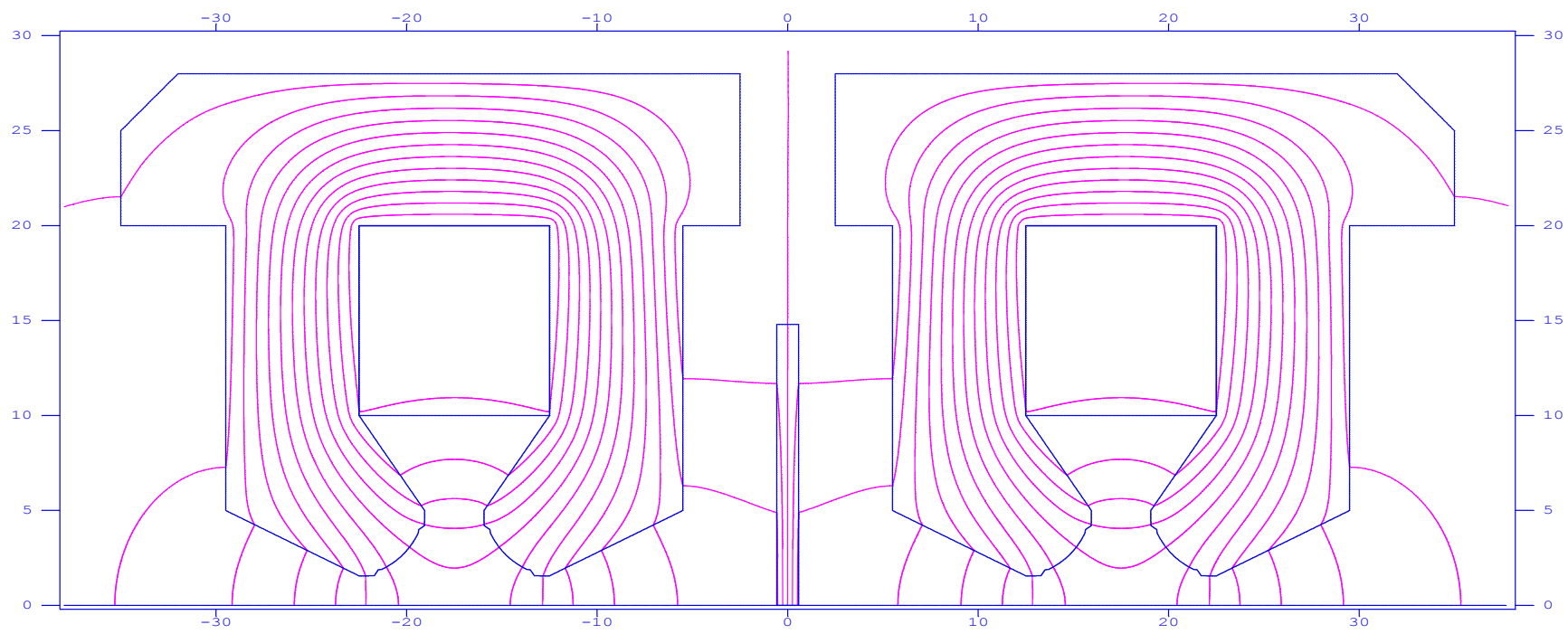

**Figure 4.3.3.4:** The magnetic flux in the dual aperture quadrupole.

By optimizing the profile of the pole, the systematic harmonics are suppressed to be less than $3\times10^{-4}$. However, field coupling introduces non-systematic harmonics even if there is a 50 mm gap between the two yokes. To compensate these non-systematic harmonics, a pure iron shielding plate is inserted between the two apertures. Simulation shows that the non-systematic harmonics in both apertures are sensitive to the thickness of the shielding plate. An optimal shielding thickness of 11.52 mm will compensate the non-systematic harmonics and reduce them to close to zero. Table 4.3.3.6 shows the harmonics before and after shielding. It is seen from the table that the shielding does not affect the systematic harmonics in both apertures.

**Table 4.3.3.6:** The harmonics before and after shielding. [$10^{-4}$]

| $B_n/B_2$ | Before shielding | After shielding |
|---|---|---|
| 1 | -2132.7 | -0.2 |
| 3 | -169.9 | 0.0 |
| 4 | 0.4 | 0.5 |
| 5 | -2.3 | 0.0 |
| 6 | -0.1 | 0.0 |
| 7 | 0.2 | 0.0 |
| 10 | 0.0 | 0.0 |
| 14 | 0.0 | 0.0 |

The trim coils effect the non-systematic harmonics in both apertures. Table 4.3.3.7 shows how the harmonics drift in the right aperture while adjusting the field by ±1.5% in the left aperture. This drift can be compensated by tuning the thickness of the shielding plate.

**Table 4.3.3.7:** The harmonics drift with adjusting the field by ±1.5%. [$10^{-4}$]

| $B_n/B_2$ | No variation | +1.5% | -1.5% |
|---|---|---|---|
| 1 | -1.3 | -51.4 | 52.3 |
| 3 | -0.1 | -4.1 | 4.1 |
| 4 | 0.5 | 0.5 | 0.5 |
| 5 | 0.0 | 0.0 | 0.0 |
| 6 | 0.0 | 0.0 | 0.0 |
| 7 | 0.0 | 0.0 | 0.0 |
| 10 | 0.0 | 0.0 | 0.0 |
| 14 | 0.0 | 0.0 | 0.0 |

The cross talk between the two apertures will be further studied with 3D simulation. The design parameters of the dual aperture quadrupole are listed in Table 4.3.3.8.

**Table 4.3.3.8:** Design parameters of the dual aperture quadrupole.

| | | |
|---|---|---|
| Beam center separation [mm] | | 350 |
| Magnetic length [m] | | 2 |
| Gradient [T/m] | | 8.42 |
| Aperture [mm] | | 76 |
| Coil | Number | 2 |
| | Shape | Racetrack |
| | Material | Aluminum |
| | Turns | 64 |
| | Conductor specs. [mm] | 11×11, φ7, R1 |
| Current [A] | | 154 |
| Current density [A/mm$^2$] | | 1.89 |
| Resistance [mΩ] | | 221.2 |
| Voltage [V] | | 34.1 |
| Power consumption [kW] | | 5.3 |
| Cooling water | Loop number | 4 |
| | Pressure drop [kg/cm$^2$] | 6 |
| | Velocity [m/s] | 1.3 |
| | Flux [l/s] | 0.201 |
| | Temperature rise [°C] | 6.3 |

#### 4.3.3.4 *Sextupole SD/SF*

The sextupoles are two individual parallel magnets instead of a dual aperture component. To use dual aperture dipoles and quadrupoles, the distance between the e+ and e- beam is quite close; therefore, the sextupole size is limited and the space between two neighboring sextupoles is restricted. Figure 4.3.3.5 shows the cross sections and positions in two neighboring sextupoles in the two rings.

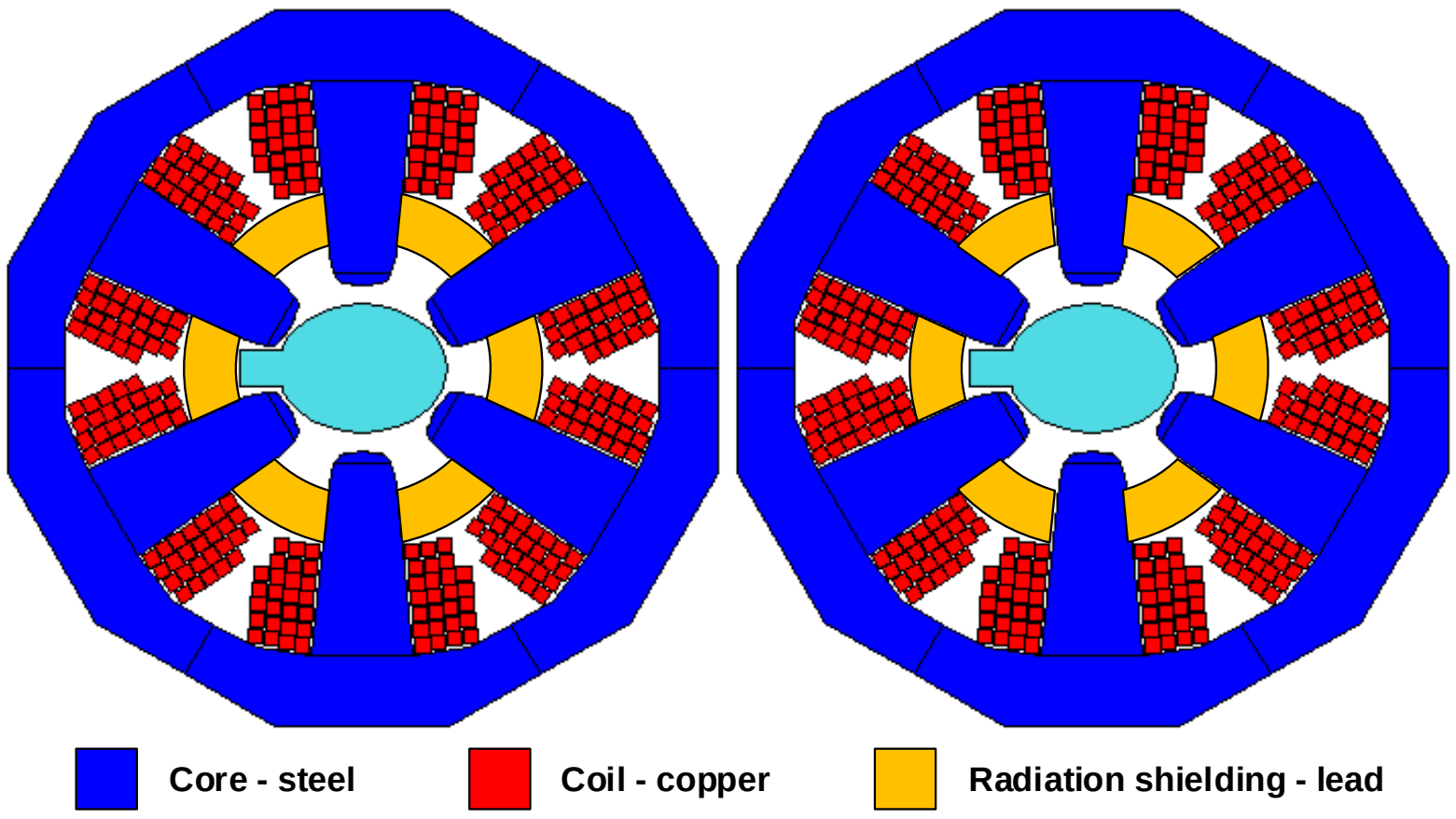

**Figure 4.3.3.5:** Cross sections of two neighboring sextupoles in the two rings.

Although the two sextupoles are close to each other, the field interference between them is negligible. The pole surface profile is optimized to compensate the harmonics. Table 4.3.3.9 gives the systematic harmonics of the sextupole; the design parameters are listed in Table 4.3.3.10.

**Table 4.3.3.9:** Systematic harmonics of the sextupole [$10^{-4}$]

| n | $B_n/B_3$ |
|---|---|
| 3 | 10000 |
| 9 | 1.34 |
| 15 | 0.74 |
| 21 | -0.31 |

**Table 4.3.3.10:** Sextupole design parameters

| | | | |
|---|---|---|---|
| Magnet type | | SF | SD |
| Magnetic length [m] | | 0.7 | 1.4 |
| Gradient [T/m$^2$] | | 506.22 | |
| Aperture [mm] | | 80 | |
| Coil | Number | 6 | |
| | Shape | Racetrack | |
| | Material | Copper | |
| | Turns | 26 | |
| | Conductor specs. [mm] | 7×7, φ3, R1 | 7×7, φ4, R1 |
| Current [A] | | 168.4 | |
| Current density [A/mm$^2$] | | 4.10 | 4.73 |
| Resistance [mΩ] | | 115.7 | 245.7 |
| Voltage [V] | | 19.5 | 41.4 |
| Power consumption [kW] | | 3.28 | 6.97 |
| Cooling water | Loop number | 12 | |
| | Pressure drop [kg/cm$^2$] | 6 | |
| | Velocity [m/s] | 2.06 | 1.79 |
| | Flux [l/s] | 0.175 | 0.269 |
| | Temperature rise [°C] | 4.5 | 6.2 |

### 4.3.4 Superconducting Magnets in the Interaction Region

Compact high gradient quadrupole doublet QD0 and QF1 are required on both sides of the collision points. The requirements for QD0 and QF1 are based on a circumference of 100 km, L* of 2.2 m and a beam crossing angle of 33 mrad.

QD0 and QF1 are twin aperture quadrupoles and are operated fully inside the solenoid field of the detector magnet which has a central field of 3.0 T. To minimize the effect of the longitudinal solenoid field on the accelerator beam, anti-solenoids before QD0 and outside QD0 and QF1 are needed. Their magnetic field direction is opposite to the detector solenoid, and the total integral longitudinal field generated by the detector solenoid and anti-solenoid coils is zero. It is also required that the total solenoid field inside the QD0 and QF1 magnet be close to zero.

The Machine Detector Interface (MDI) imposes the condition that accelerator devices can only start after 1.1 m along the longitudinal axis, so the available space for the anti-solenoid before QD0 is limited. In addition, the angle of the accelerator magnet seen from the IP point must be small and satisfy the detector requirements. Taking into account the high field strength of twin aperture quadrupole magnets, the high central field of the anti-solenoid, and the limited space, superconducting technology based on NbTi conductor will be used for these interaction region superconducting quadrupole magnets and anti-solenoids. Furthermore, some superconducting sextupole magnets are required.

#### 4.3.4.1 *Superconducting Quadrupole Magnet QD0*

##### *4.3.4.1.1 Overall Design*

The final focus QD0 is a double aperture superconducting magnet. The distance from QD0 to the IP point is 2.2 m, and the minimum distance between two aperture center lines is only 72.61 mm, so a very limited radial space is available for QD0. The outer diameter of a single aperture of QD0 is determined by the separation of two beams at the IP side.

QD0 has a two layer cos2$\theta$ coil using Rutherford cable without an iron yoke. The four coils are clamped with stainless steel collars. The beam pipe at room temperature is held inside the helium vessel with a clearance gap of 4 mm.

The magnetic field harmonics in the good field region are required to be less than $3\times10^{-4}$. The field cross talk of the two apertures in QD0 with such a small aperture separation distance is serious, and each multipole field inside one aperture is affected by the field from the other aperture. So a shield coil is introduced outside the quadrupole coil to improve the field quality in each aperture.

*4.3.4.1.2 2D Field Calculation*

QD0 is an iron-free small-aperture long magnet. Its coils will be made of Rutherford cable with a width of 3 mm, a mid-thickness of 0.94 mm and a keystone angle of 1.8 degrees. The QD0 coil cross section is optimized with four coil blocks in two layers; there are 23 turns for each pole.

2D field calculations are performed using OPERA [1]. First one aperture of QD0 is included in the calculation, and only one quarter is modelled. After optimization, good field quality in the good field region is obtained. The magnetic flux lines is shown in Figures 4.3.4.1.

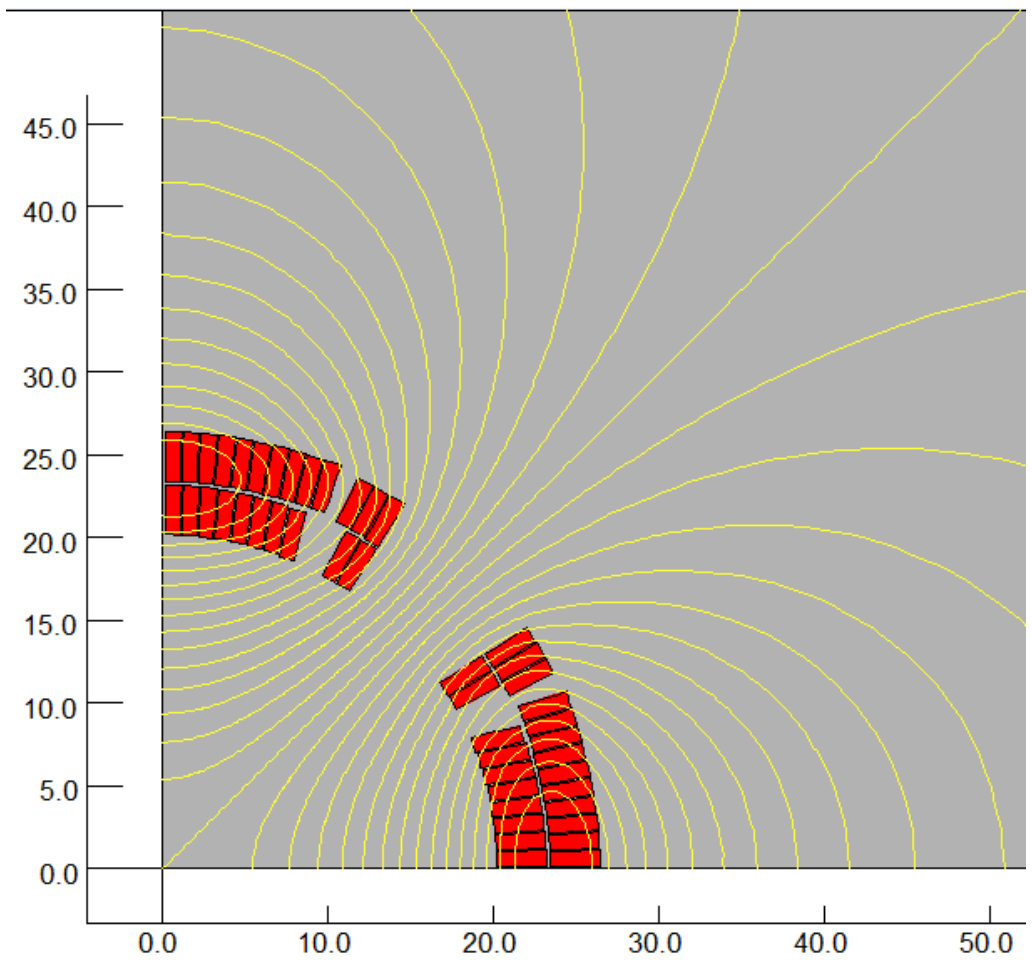

**Figure 4.3.4.1:** 2D flux lines (One quarter cross section)

The calculated relative multipole field components normalized to the main quadrupole field are listed in Table 4.3.4.1.

**Table 4.3.4.1:** 2D field harmonics（unit, $1\times10^{-4}$）

| n | $B_n/B_2$ @ R=9.8mm |
|---|---|
| 2 | 10000 |
| 6 | -0.77 |
| 10 | -0.45 |
| 14 | -0.098 |

The field in one aperture is affected due to the field generated by the coil in the other aperture. Field cross talk of the two apertures is modelled using OPERA-2D and shown in Figure 4.3.4.2.

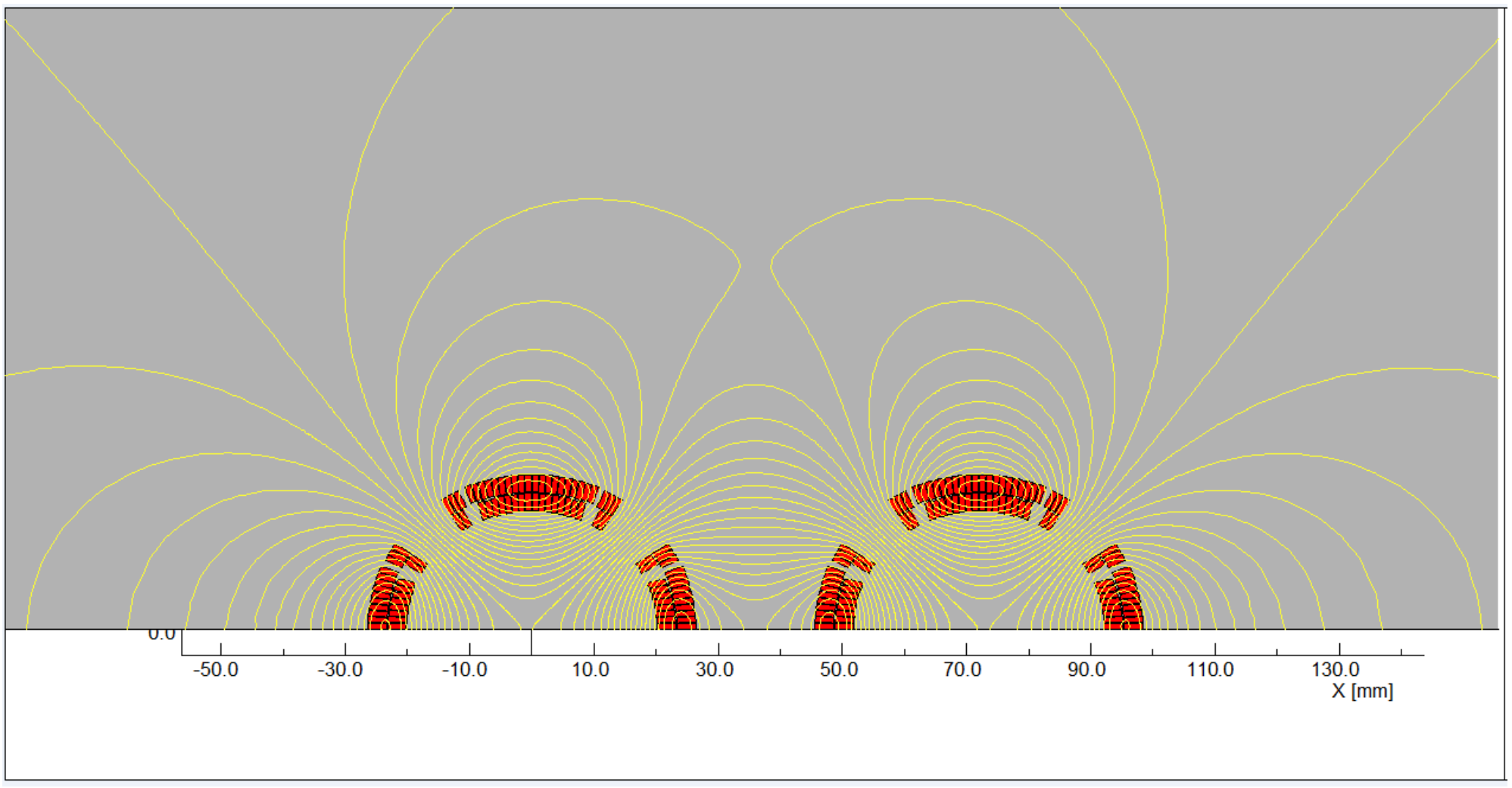

**Figure 4.3.4.2:** Flux lines of two aperture coils

Multipole fields in one aperture as a function of aperture central distance is presented in Fig. 4.3.4.3 (unit, $1{\times}10^{-4}$)

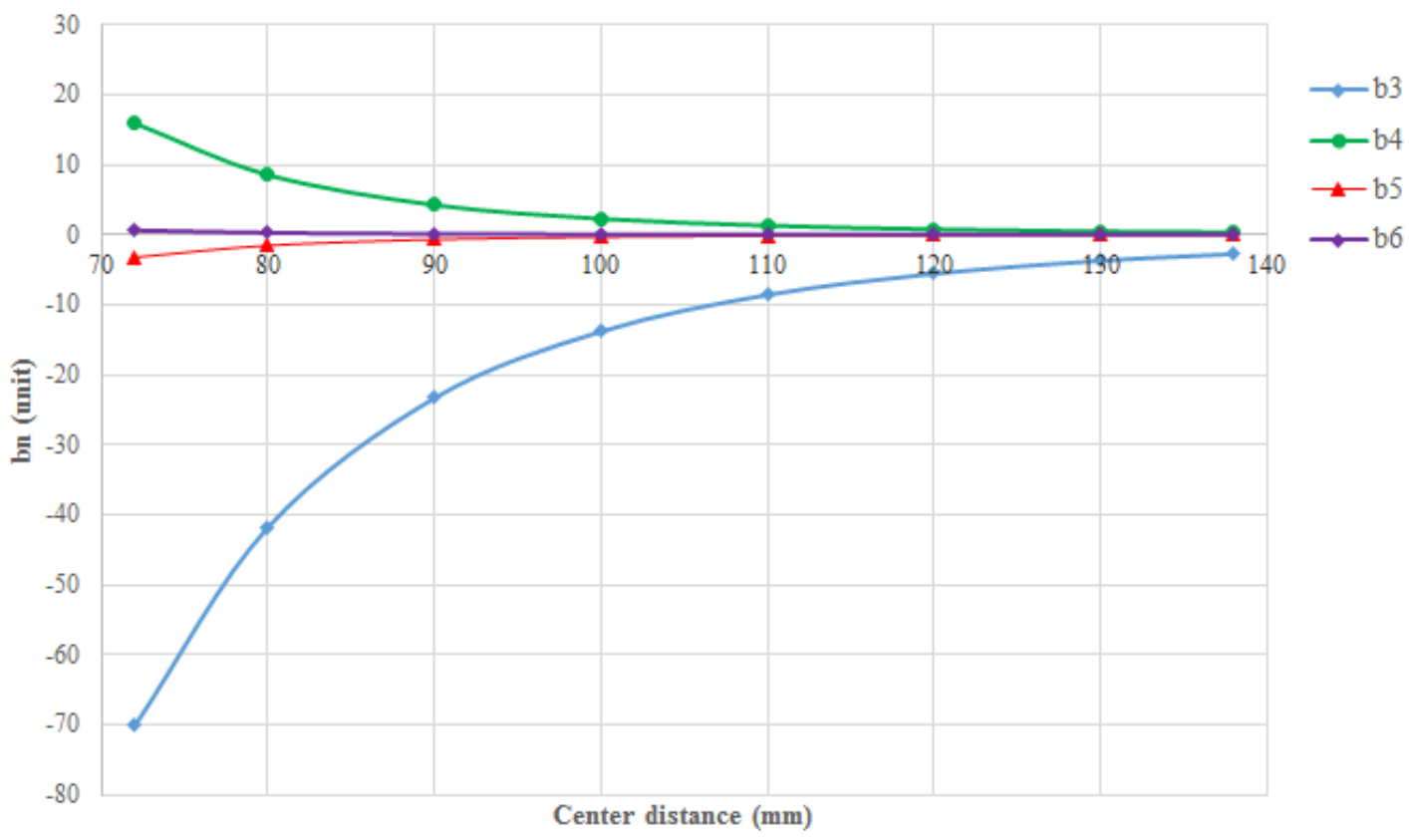

**Figure 4.3.4.3:** Multipole field in each aperture as a result of field cross-talk

Due to the small distance between the two QD0 apertures, the field cross talk is serious. The most serious multipole field is the sextupole field [2].

*4.3.4.1.3 3D Field Calculation*

QD0 coils are simplified and modelled in OPERA-3D. First, the field quality in the single aperture is calculated, and then the multipole fields induced by the field cross talk of the two apertures are obtained.

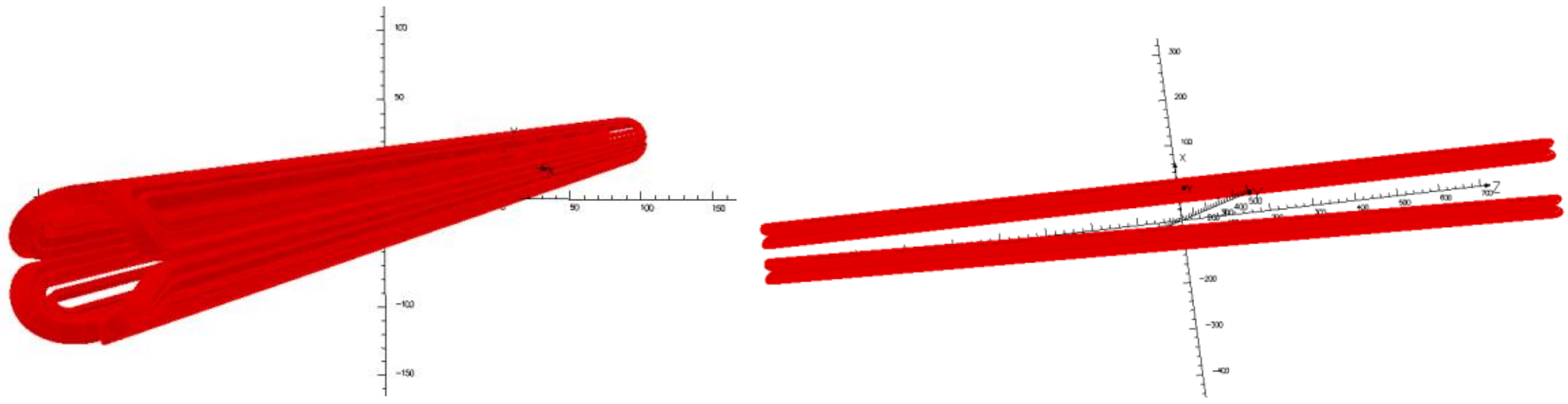

(a) Single aperture coil (b) Two aperture coils

**Figure 4.3.4.4:** OPERA-3D model of QD0 coil

The calculated integrated multipole field in one aperture as a result of field cross talk are quite large, especially the sextupole and octupole field components [2].

*4.3.4.1.4 Shield Coil Design*

A two-layer shield coil is placed just outside the quadrupole coil to improve field quality. The shield coil is not symmetric within each aperture, but the shield coils for the two apertures are symmetric. The conductor for the shield coil is round NbTi 0.5 mm diameter wire, and there are 44 turns for each pole. After optimization, the shield coil in the left aperture is shown in Fig. 4.3.4.5.

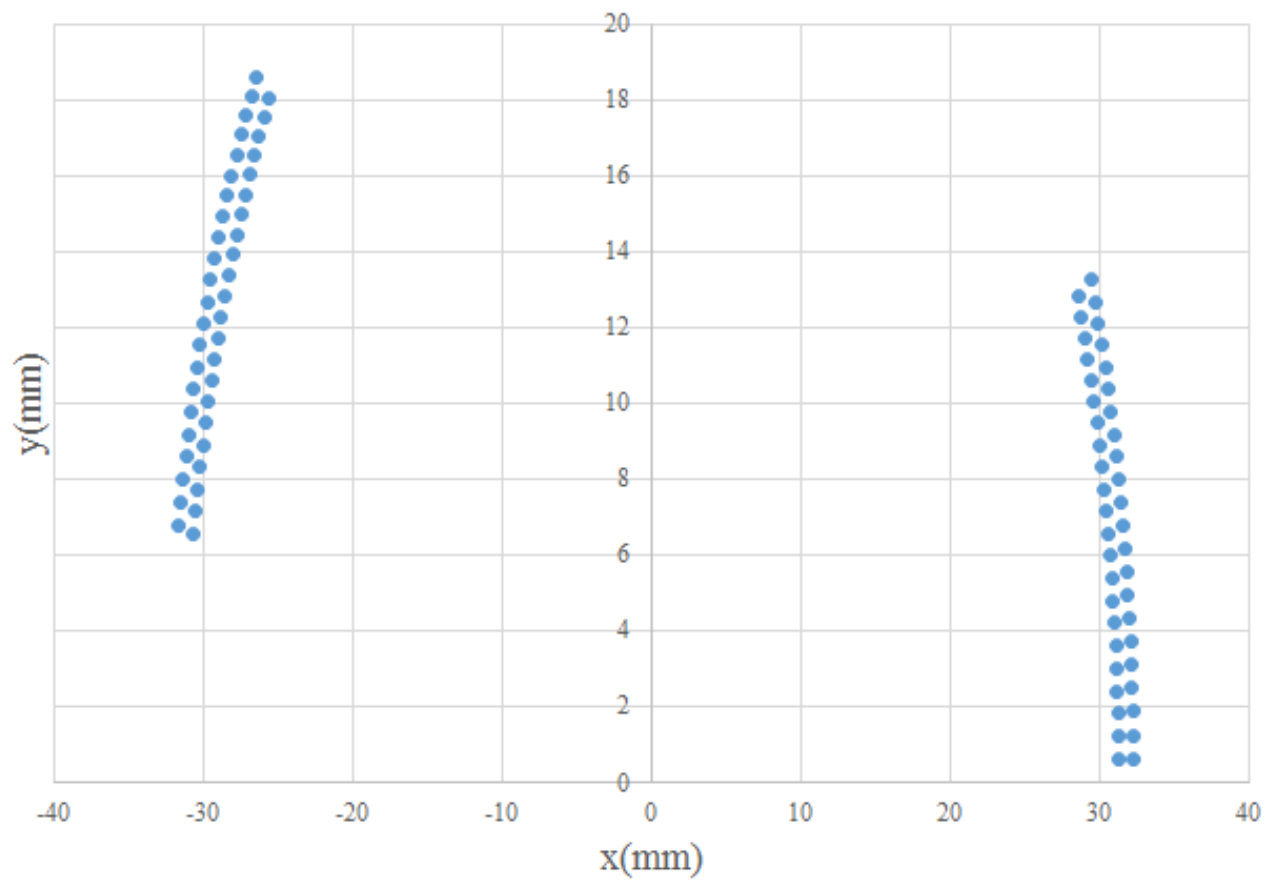

**Figure 4.3.4.5:** Shield coil in one aperture (half)

The calculated field quality in each aperture with a shield coil is very good. Each integrated multipole field is smaller than $2\times10^{-4}$. To match the fall off of field harmonics caused by the field cross talk when the distance between the two beam lines increases, the conductor lengths of the shield coil at each angular position are different. Therefore, each multipole field is optimized to be smaller than $3\times10^{-4}$ at different longitudinal positions in each aperture. In addition, the integrated dipole field in each aperture is very small.

The schematic cross-section of a single aperture QD0 magnet is shown in Ref. [2]. Design parameters and forces are listed in Table 4.3.4.3.

### 4.3.4.2 ***Superconducting Quadrupole Magnet QF1***

#### *4.3.4.2.1 Overall Design*

The design of QF1 is similar to QD0, except that there is an iron yoke around the coil. The Rutherford cable used is similar to that of QD0. The QF1 coil consists of four coil blocks in two layers separated by wedges, with 29 turns for each pole. Since the distance between the two apertures is much larger and there is an iron yoke, the field cross talk between the two apertures of QF1 is not the issue as it is for QD0.

#### *4.3.4.2.2 2D Field Calculation*

The QF1 cross section is optimized using OPERA-2D. One quarter of a single aperture of QF1 is modelled. After optimization, the field quality in each aperture is good. The magnetic flux lines is show in Figures 4.3.4.6.

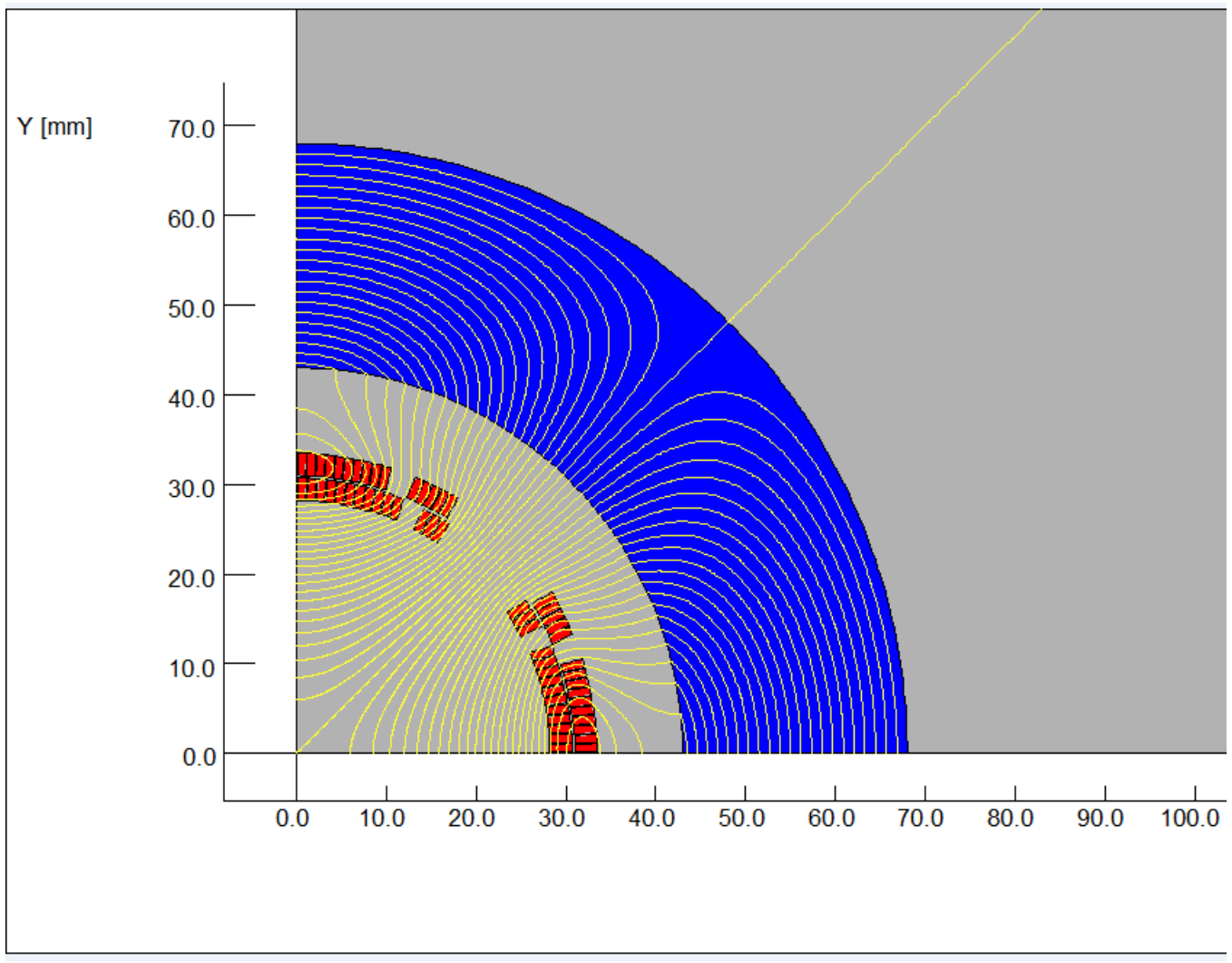

**Figure 4.3.4.6:** 2D flux lines of single aperture QF1 (one quadrant)

The calculated relative multipole field contents are listed in Table 4.3.4.2.

**Table 4.3.4.2:** 2D field harmonics of QF1 (unit, $1\times10^{-4}$)

| n | $B_n/B_2$ @ R=13.5 mm |
|---|---|
| 2 | 10000 |
| 6 | 1.08 |
| 10 | -0.34 |
| 14 | 0.002 |

#### *4.3.4.2.3 Field Cross Talk*

Two aperture cross talk in QF1 is modelled and studied with OPERA-2D. Figure 4.3.4.7 shows flux lines in the two coils. The results show that the iron yoke can shield the leakage field of each aperture. The field harmonics from field cross talk is negligible.

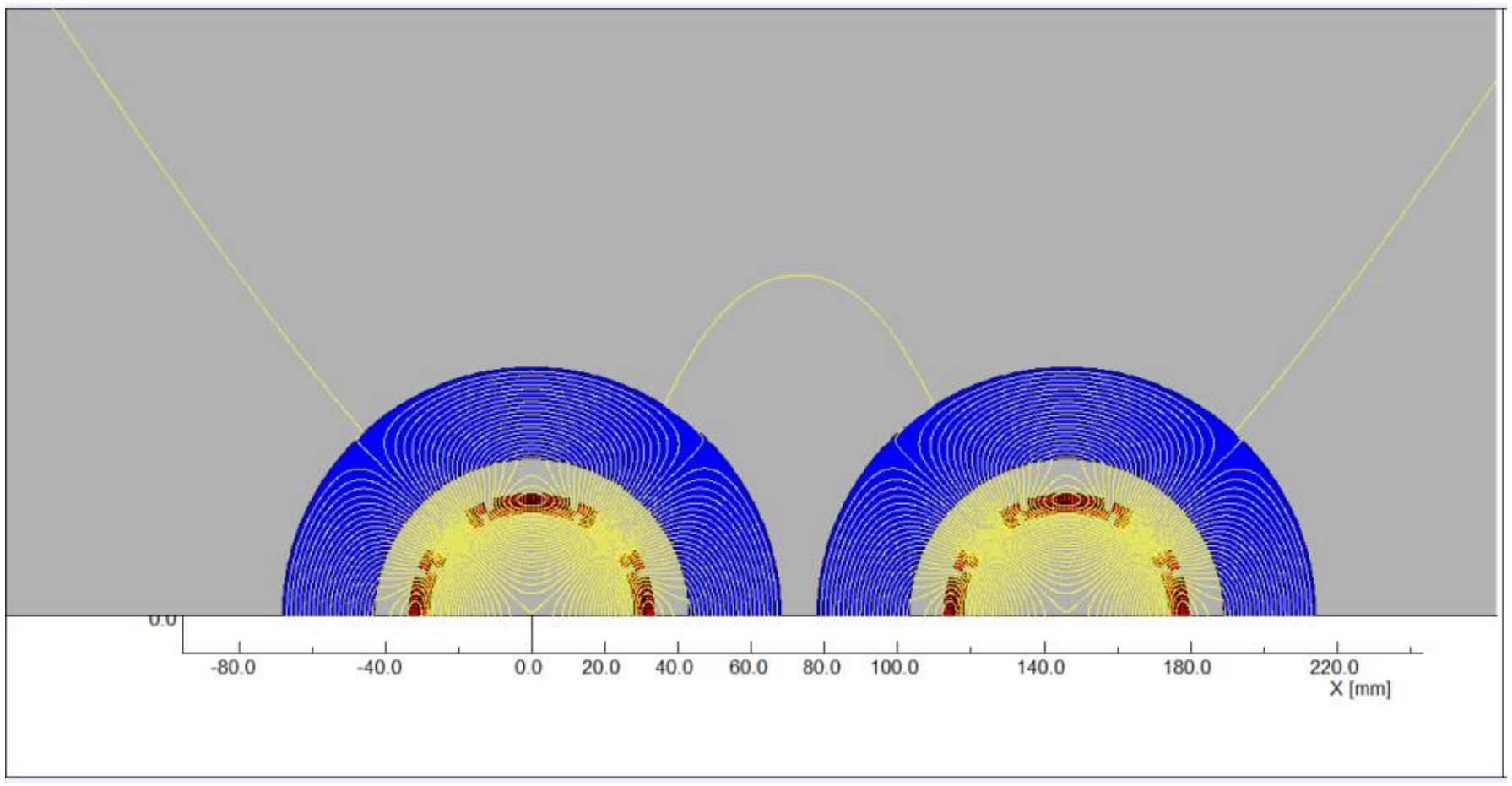

**Figure 4.3.4.7:** Flux lines for two QF1 aperture coils

*4.3.4.2.4 Design Parameters, Force and Magnet Layout*

The schematic cross-section of a single aperture of QF1 can bee found in Ref. [2]. Parameters and forces of QD0 and QF1 are listed in Table 4.3.4.3.

**Table 4.3.4.3:** Design parameters of quadrupoles QD0 and QF1

| Magnet name | QD0 | QF1 |
|---|---|---|
| Field gradient (T/m) | 136 | 110 |
| Magnetic length (m) | 2.0 | 1.48 |
| Coil turns per pole | 23 | 29 |
| Excitation current (A) | 2510 | 2250 |
| Coil layers | 2 | 2 |
| Conductor size (mm) | Rutherford NbTi-Cu Cable, width 3 mm, mid thickness 0.94 mm, keystone angle 1.8 deg | Rutherford NbTi-Cu Cable, width 3 mm, mid thickness 0.95 mm, keystone angle 1.6 deg |
| Stored energy (KJ) | 25.0 | 30.5 |
| Inductance (H) | 0.008 | 0.012 |
| Peak field in coil (T) | 3.3 | 3.8 |
| Coil inner diameter (mm) | 40 | 56 |
| Coil outer diameter (mm) | 53 | 69 |
| X direction Lorentz force/octant (kN) | 68 | 110 |
| Y direction Lorentz force/octant (kN) | -140 | -120 |

### 4.3.4.3 ***Superconducting Anti-Solenoid***

*4.3.4.3.1 Overall Design*

The requirements are summarized below:

1) The total integral longitudinal field generated by the detector solenoid and anti-solenoid coils is zero.

2) The longitudinal field inside QD0 and QF1 should be smaller than a few hundred Gauss at each longitudinal position.
3) The distribution of the solenoid field along the longitudinal direction should meet the requirement of the beam optics for vertical emittance.
4) The angle of the anti-solenoid seen at the collision point satisfies detector requirements.

The anti-solenoid design fully takes into account these requirements. The anti-solenoid will be wound with rectangular NbTi-Cu conductor. Since the magnetic field of the detector solenoid is not constant, and decreases slowly along the longitudinal direction, and also in order to reduce the magnet size, energy and cost, the anti-solenoid is divided into a total of 22 sections with different inner coil diameters. These sections are connected in series, but the current in some sections of the anti-solenoid can be adjusted using auxiliary power supplies if necessary.

#### *4.3.4.3.2 2D Field Calculation*

The calculation is performed using an axial-symmetric model in OPERA-2D. Figures 4.3.4.8 shows the anti-solenoid flux lines.

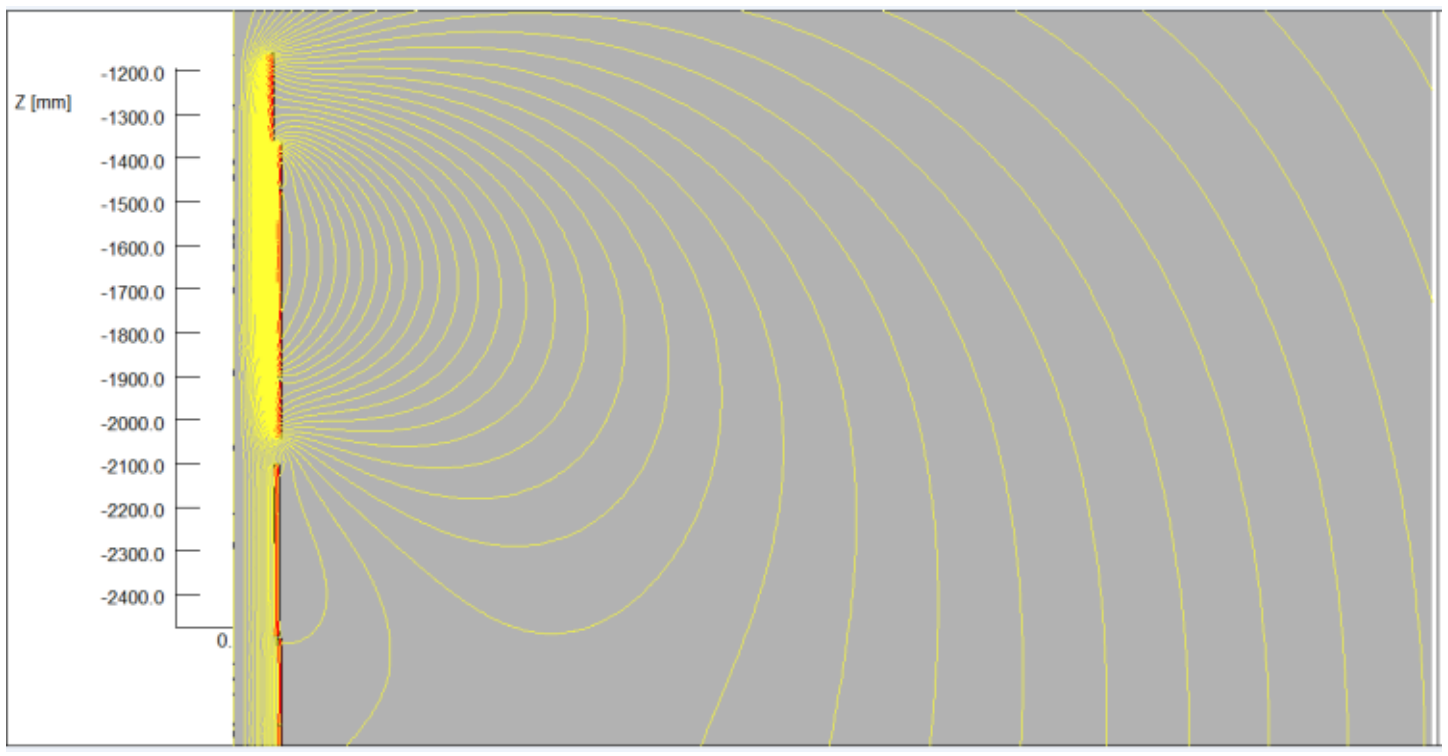

**Figure 4.3.4.8:** Anti-solenoid flux lines

The distribution of the combined field along the longitudinal direction can be found in Ref. [2].

The central field in the first section in the anti-solenoid is the strongest, with a peak value of 7.2T. The combined field distribution of anti-solenoid and detector solenoid magnet meets the design requirements.

#### *4.3.4.3.3 Design Parameters, Force and Magnet Layout*

Anti-solenoid parameters and forces are listed in Table 4.3.4.4.

**Table 4.3.4.4:** Parameters of interaction region anti-solenoids

| Item | Anti-solenoid before QD0 | Anti-solenoid at QD0 | Anti-solenoid after QD0 |
|---|---|---|---|
| Central field (T) | 7.2 | 2.8 | 1.8 |
| Magnetic length（m） | 1.1 | 2.0 | 1.7 |
| Conductor (NbTi-Cu, mm) | 2.5×1.5 | | |
| Coil layers | 16 | 8 | 4/2 |
| Excitation current（kA） | 1.0 | | |
| Stored energy (KJ) | 715 | | |
| Inductance (H) | 1.4 | | |
| Peak field in coil (T) | 7.7 | 3.0 | 1.9 |
| Number of sections | 4 | 11 | 7 |
| Solenoid coil inner diameter (mm) | 120 | | |
| Solenoid coil outer diameter (mm) | 390 | | |
| Total Lorentz force $F_z$ (kN) | -75 | -13 | 88 |
| Lorentz force $F_z$ with detector solenoid operation (kN) | -95 | -81 | -92 |
| Cryostat diameter (mm) | 500 | | |
| Total number | 4 | | |

Since the field in the last section of the anti-solenoid is very low and to reduce the length of the cryostat, the last section of the anti-solenoid will be operated at room-temperature.

The superconducting QD0, QF1, and anti-solenoid coils (except the last section) are in the same cryostat; the layout is shown in Figure 4.2.6.5.

### 4.3.4.4 *Superconducting Sextupole Magnet*

The requirements of superconducting sextupole magnets for Higgs operation are listed in Table 4.3.4.5.

**Table 4.3.4.5:** Requirements of interaction region sextupoles

| Magnet | Number | Central field strength ($T/m^2$, for Higgs) | Magnetic length (m) | Aperture diameter (mm) | Reference radius (mm) |
|---|---|---|---|---|---|
| VSIRD | 8 | 1635 | 0.6 | 66 | 8.5 |
| HSIRD | 8 | 1882 | 0.8 | 66 | 15.0 |
| VSIRU | 8 | 1562 | 0.6 | 66 | 8.5 |
| HSIRU | 8 | 1999 | 0.6 | 66 | 15.5 |

The superconducting sextupole magnets have an iron yoke around the coils to enhance the field strength and reduce the operating current. The four types of sextupole magnets are designed to have the same cross section. They use Rutherford cable similar to QD0.

The cross section is optimized using OPERA-2D. One quarter of the cross section is modelled. The sextupole coil consists of two coil blocks in two layers; there are 33 turns in each pole. The field quality in the aperture is good. The magnetic flux density distribution for type HSIRU sextupole magnet at Higgs operation is shown in Figure 4.3.4.9.

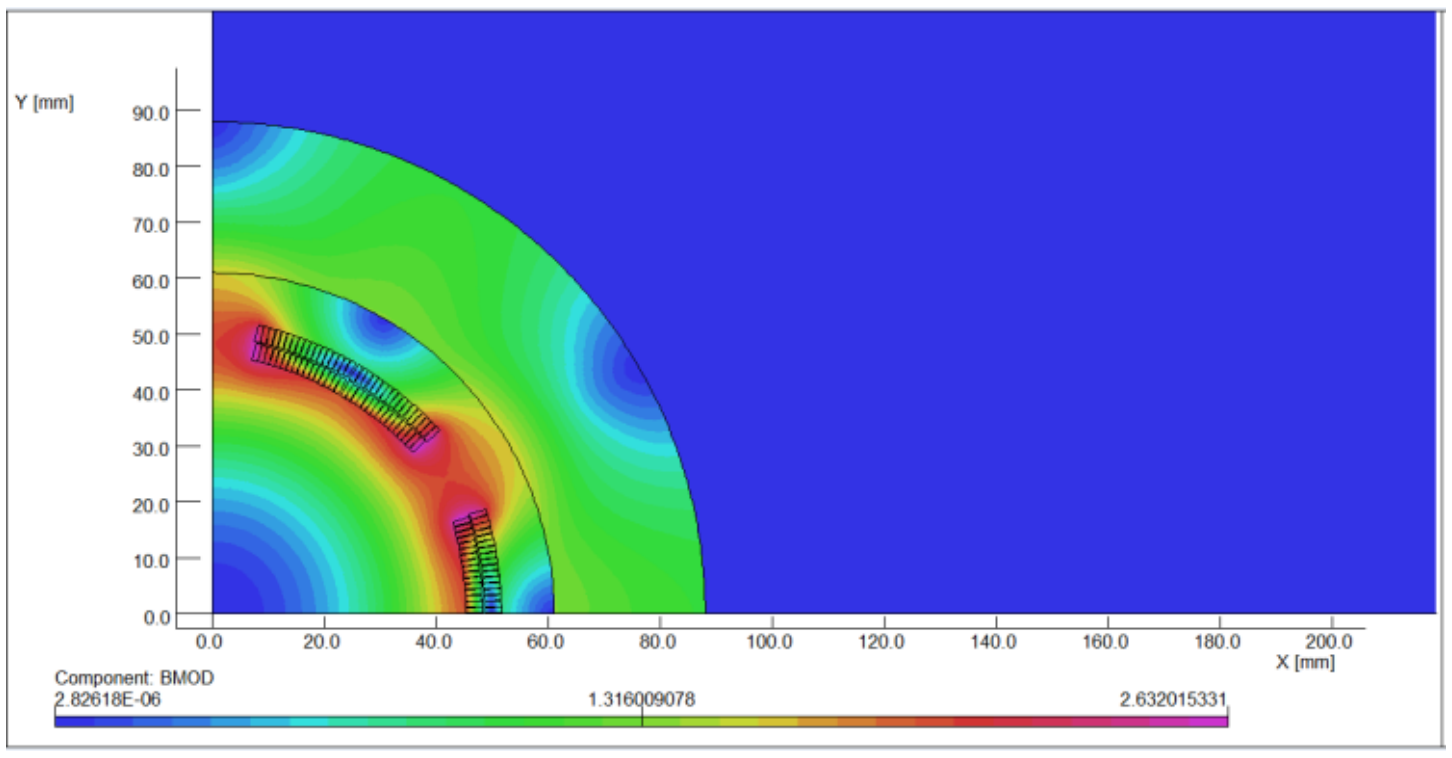

**Figure 4.3.4.9:** Magnetic flux density distribution

Parameters for the four superconducting sextupoles are listed in Table 4.3.4.6.

**Table 4.3.4.6:** Sextupole parameters

| Magnet name | VSIRD | HSIRD | VSIRU | HSIRU |
|---|---|---|---|---|
| Field strength (T/m$^2$) | 1635 | 1882 | 1562 | 1999 |
| Magnetic length (m) | 0.6 | 0.8 | 0.6 | 0.6 |
| Coil turns per pole | 33 | | | |
| Excitation current (A) | 1200 | 1380 | 1150 | 1450 |
| Coil layers | 2 | | | |
| Conductor size (mm) | Rutherford NbTi-Cu Cable, width 3 mm, mid thickness 0.95 mm | | | |
| Stored energy (KJ) | 3.4 | 6.0 | 3.1 | 5.1 |
| Inductance (H) | 0.005 | 0.006 | 0.005 | 0.005 |
| Peak field in coil (T) | 2.2 | 2.55 | 2.1 | 2.7 |
| Coil inner diameter (mm) | 90 | | | |
| Coil outer diameter (mm) | 104 | | | |
| Cryostat diameter (mm) | 300 | | | |
| Total magnet number | 8 | 8 | 8 | 8 |

### 4.3.5 **Magnet Power Supplies**

A large number of power supplies are required for powering the magnets of the Collider ring, the Booster ring, the low energy beam transport (LEBT) and the Linac.

The Collider power supplies are DC supplies. All the power supplies are rated for 120 GeV operation, and in addition have 10 ~ 15% safety margin in both current and voltage. All the dipole, quadrupole and sextupole power supplies are unipolar, and all correction power supplies are bipolar to allow current reversal.

Following are the basic design criteria for the power supplies:

- Meet the accelerator physics requirements.
- Collaborate and communicate with the magnet designers so as to choose the most suitable circuits.
- Use a modular design for most of the power supplies.
- High reliability, better EMC and convenient structure for maintenance.
- Full digital design for all supplies.
- Switching mode as the main topology.

The parameters for the magnet power supply system are as follow [1]:

- The magnet parameters and connection modes, including cable losses, determine the power supplies current and voltage ratings.
- Cable current density less than 2A/mm$^2$
- Water cooling for power supplies with power greater than 1 kW, and forced air cooling for the others.
- Power factor of the main network $\cos\phi = 0.9$.
- Efficiency of power supplies: $\eta = 0.87$.
- Computation of cable resistance: $R = \rho \times \frac{l}{s}(1 + \Delta T\alpha)$

where ρ is the resistivity of copper (= 0.0182 at 20°C), α is the temperature coefficient (= 0.00393/°C), l is the cable length and s is the total cross section of the cable in mm$^2$.

#### 4.3.5.1 ***Collider Power Supplies***

The design of the power supply system and the power supply halls is based on the CEPC layout, which has 8 arcs and 8 straight sections. The power supplies for dual aperture dipole are housed in the surface halls. The others are housed underground.

Power supply halls will be built on the surface and located between two arcs. In order to save power cables, two adjacent half-arcs will use one or two or four power supplies. There are 8 power supply halls corresponding to the 8 arcs in the tunnel as in Figures 4.3.5.1a and 4.3.5.1b.

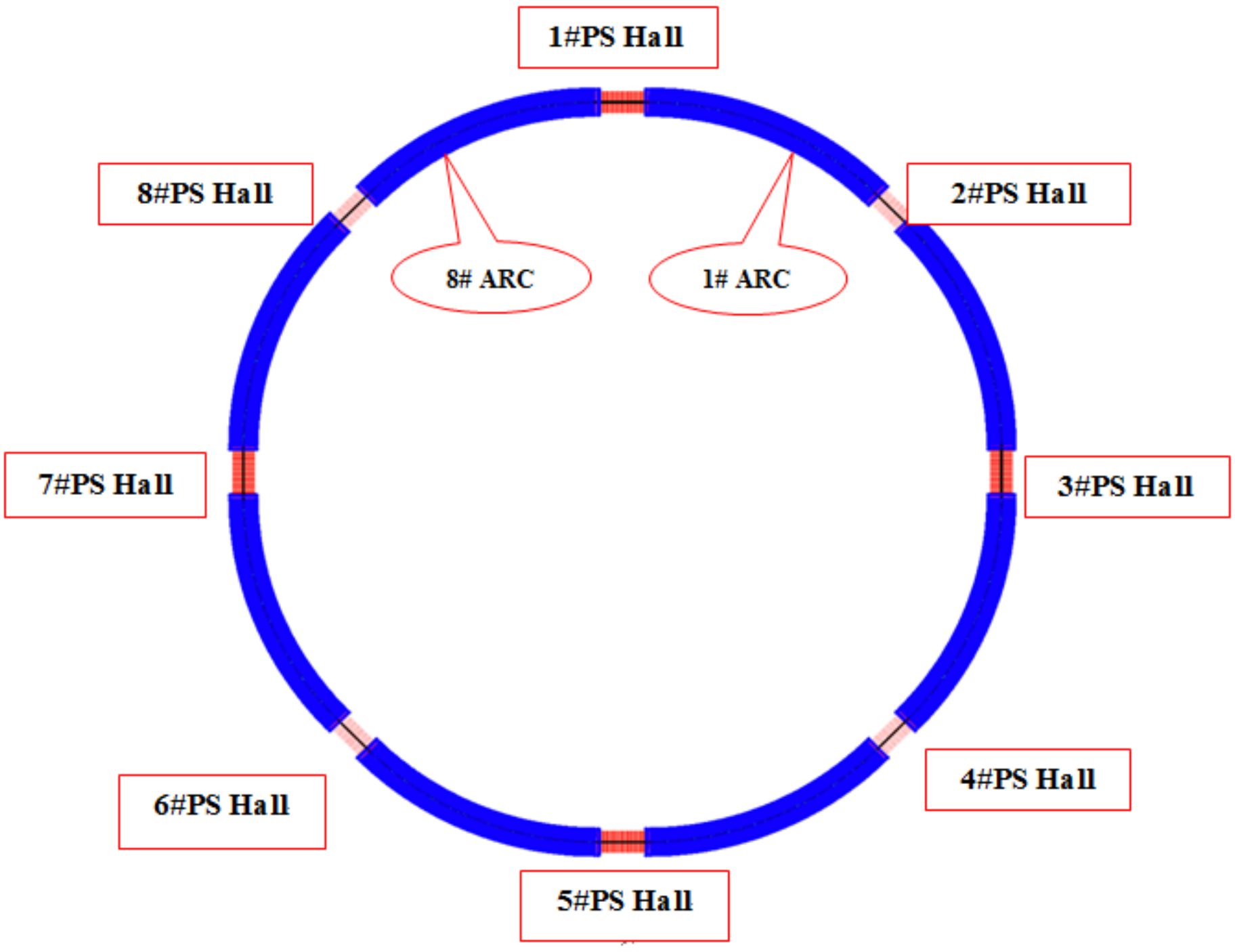

**Figure 4.3.5.1a:** PS Hall layout

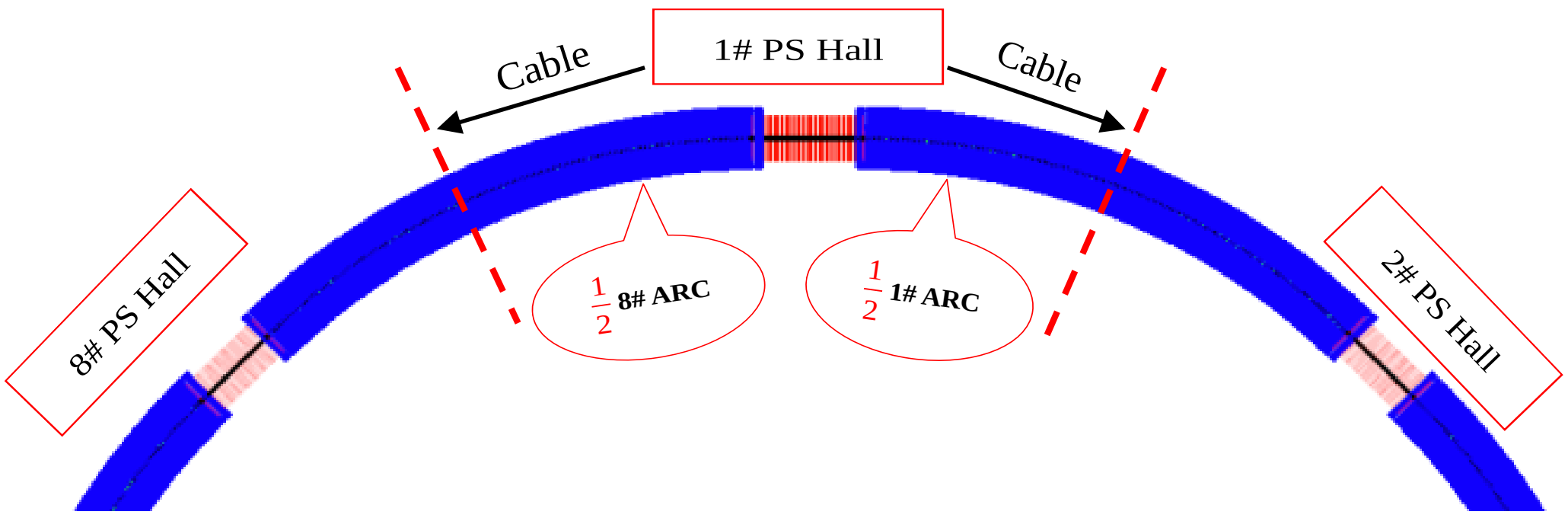

**Figure 4.3.5.1b:** PS Hall layout

System reliability is required to be very high and the system should operate with a mean-time-between-failures (MTBF) of several months. The power supply is designed with built-in redundancy by using a modular approach. The power part is divided into n + 1 modules, n supplying nominal current, and one module in reserve in case of a trip. Efficient monitoring and diagnosis methods will be adopted to anticipate faults.

### 4.3.5.2 *Dipole Magnet Power Supplies*

The Collider has 2384 dual-aperture dipole magnets and 162 single-aperture dipole magnets. Dual-aperture dipoles in two adjacent half-arcs are connected in series and powered by one power supply. Thus there are 8 dipole power supplies, each 1.40 MW (including an allowance for cable losses). The single-aperture dipole magnets are powered independently. The rate power of the supplies is about 9 kW. These power supplies will be installed in an auxiliary underground tunnel, to be close to the magnet load.

Underground installation is the main factor for reduced volume and high efficiency. The power supply manufacturer ratings include 10 ~ 15% safety margins in both current and voltage.

#### 4.3.5.3 *Quadrupole Magnet Power Supplies*

The Collider quadrupoles include 1196 dual-aperture QI, 1196 dual-aperture QO and 1132 other single-aperture quadrupoles. The dual-aperture quadrupoles are divided into 96 focusing families and 96 defocusing families. Each family consists of 12 or13 series-connected magnets powered by one separate power supply. The current in each quadrupole can be separately adjusted with a shunt up to ±2%. The single-aperture quadrupoles are powered independently, and the supplies will be installed in the auxiliary stub tunnel around the main tunnel.

#### 4.3.5.4 *Sextupole Magnet Power Supplies*

There are 1,792 sextupoles in the arc region. According to the accelerator physics requirements, each 2 adjacent magnets are connected in series and powered by one power supply. All other 72 magnets in RF region are powered independently, and the supplies will be installed in the auxiliary tunnel around the main tunnel too.

#### 4.3.5.5 *Corrector Power Supplies*

The total number of correction BH and BV magnets is about 5,808, each powered by a separate supply. For convenient maintenance and repair, the ratings for all correction power supplies is the same. They are a module-based design.

**Table 4.3.5.1:** Power supply requirements

| Power Supply | Quantity | Stability /8hours | Output Rating |
|---|---|---|---|
| D-aperture Dipole | 8 | 100ppm | 1170A/1200V |
| S-aperture Dipole | 162 | 100ppm | 180A/50V |
| D-aperture Quadrupole | 192 | 100ppm | 180A/750V |
| S-aperture Quadrupole | 1022 | 100ppm | 180A/70V,240,850V |
| Sext.D | 448 | 100ppm | 180A/140V |
| Sext.F | 448 | 100ppm | 180A/140V |
| Sext | 72 | 100ppm | 40A/75V |
| Corrector | 5808 | 500ppm | 40A /75V, 110V |
| **Total system power** | | | **48MW** |

#### 4.3.5.6 *Topology of the Unipolar Power Supplies*

All power supplies use switched-mode as the main topology, because the switching mode is easy to control with a digital controller.

A typical single quadrant switched-mode power converter structure is shown in Figure 4.3.5.2 [2].

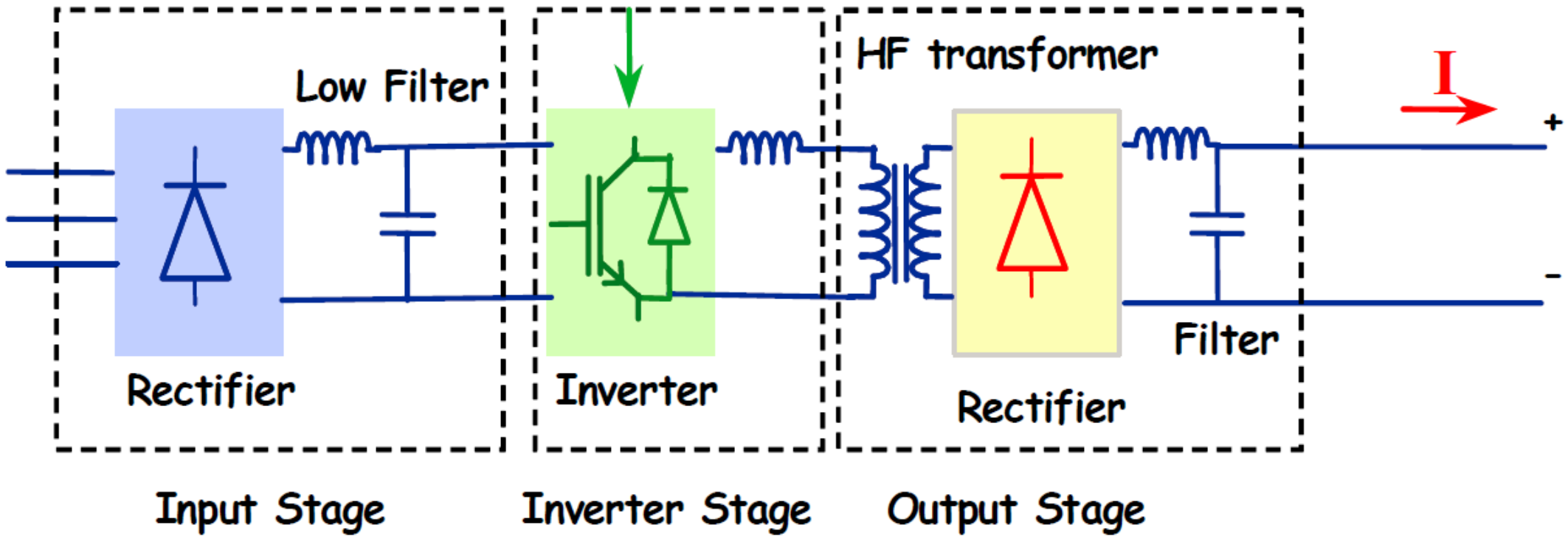

**Figure 4.3.5.2:** Diagram of a unipolar power supply.

The switched-mode power converter can be separated into three parts:

1. Input stage – includes magnetic and thermal protection, an AC contactor, a three-phase six-pulse diode rectifier, the necessary filtering on the ac and dc sides and a soft-start circuit to limit the inrush current.
2. Inverter stage – includes a Full-Bridge Zero-Voltage Zero-Current Switching Phase-Shift inverter (FB-ZVZCS-PS) with a switching frequency around 20 kHz.
3. Output stage – includes a high-frequency (HF) transformers for insulation and adaptation, a rectifier stage and an output filter.

Input Stage: there are a variety of input stages that could be considered. The easiest would be a six-pulse diode rectifier connected to the mains and feeding all the units. An extension to this, which is more suitable for high power applications, is a 12-pulse diode rectifier.

Inverter stage: the DC bus voltage generated by the input stage maintains voltage to the high frequency switching unit. This unit converts the DC supply to an alternating sou2]rce that drives the high-frequency transformer. Switching at high frequency reduces the size and cost of the components.

Output Stage: the output stage performs the rectification, filtering as well as impedance matching to the load. The size of the transformer and the output filter is inversely proportional to the frequency of operation. The size difference between the line transformer and the switched-mode transformer is approximately 20:1.

### 4.3.5.7 *Topology of the Bipolar (Corrector) Power Supplies*

The switched-mode power converter structure is shown in Figure 4.3.5.3 [2]. It includes:

- A mains rectifier stage with magnetic and thermal protection, an AC contactor, a three-phase six-pulse diode rectifier, the necessary filtering on the AC and DC sides and a soft-start circuit to limit the inrush current.
- An inverter stage using a soft-commutated bridge with IGBT switching at high frequencies >20 kHz.
- A high frequency transformer and a bipolar output stage: this part comprises a high-frequency transformer for insulation and adaptation. A bipolar output stage provides reversal of the polarity. The magnet energy, during the ramp-down of the current, is dissipated by the converter or sent back to the mains.

- An output circuit with a free-wheel safety and discharge circuit (also called crowbar), both DCCT transducer heads and the earthing circuit.

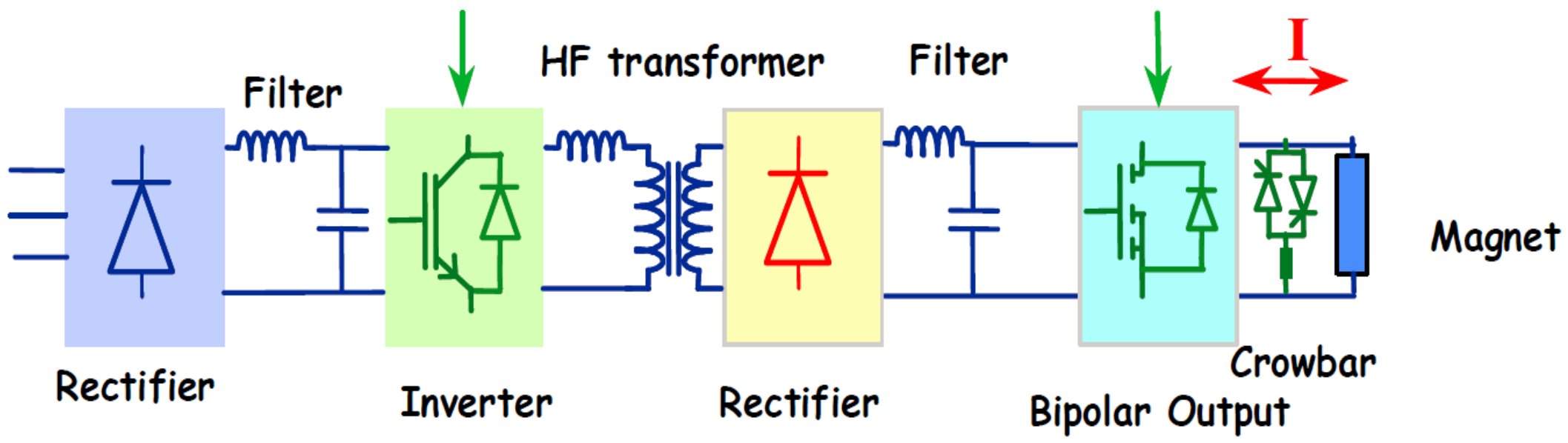

**Figure 4.3.5.3:** Topology diagram of the corrector power supply.

### 4.3.5.8 *Electronics of the Power Supplies*

All CEPC power supplies are digitally controlled.

MCU, DSP and FPGA has made it possible to replace analog regulation functions like PID controllers with digital algorithms implemented in these intelligent processors.

The advantages of digital control over analog include:

- A complex fast control algorithms can be implemented and remain stable in relation to the process dynamics;
- Flexible for different projects;
- No extra offset or drift and better noise immunity;
- Parameter optimization and changes of the control system can be done by software; no hardware redesign is required;
- Friendly for debugging and diagnostics;
- Easy to extend functionality.

Fig. 4.3.5.4. shows the digital controller for CEPC power supplies. For a digital control system, the performance of the converter is mainly determined by the ADC (Analog-to-Digital Converter) and the DCCT (Direct Current Current Transducer). Based on the digital controller there is only one DCCT used for both feedback and display, compared to the analog system where two DCCTs are necessary for feedback and display separately.

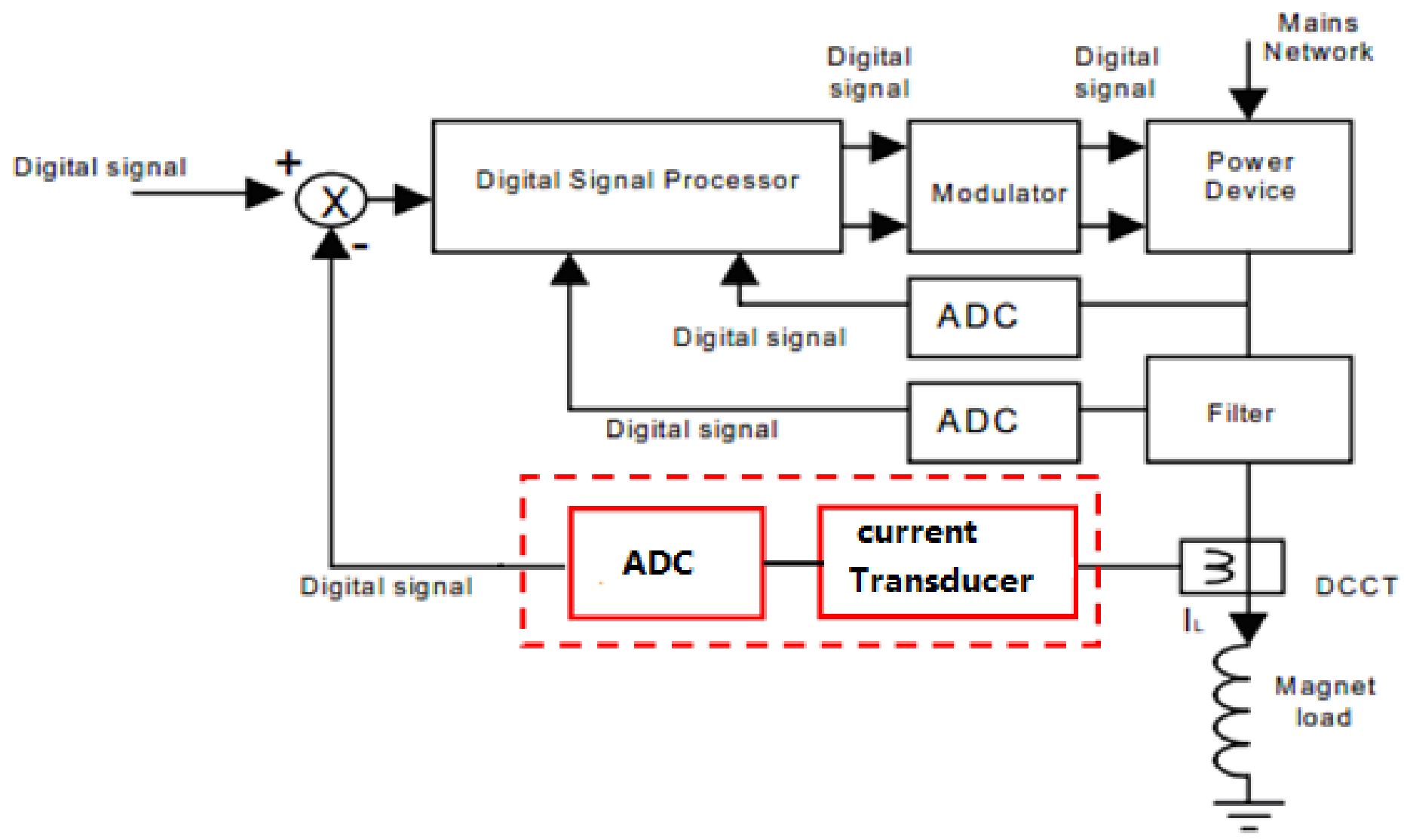

**Figure 4.3.5.4:** Digital controller for CEPC magnet power supplies.

### 4.3.5.9 *Digital Controller*

A typical structure diagram of a digital controller called a DPSCM (Digital Power Supply Control Module) is shown in Fig. 4.3.5.5. It uses FPGA as the core device for data processing. Included are digital PID regulation of the current and voltage control loops; control of various types of ADCs and DACs; control of the external interface, such as remote control systems, timing system and synchronization with other power supplies, local display and monitoring.

To meet the high performance of magnet power supplies, the following issues must be considered during the DPSCM design.

- Chip choice of digital signal processing: based on the system-on-chip of FPGA (Altera)
- ADC design: Low noise design on PCB; Constant temperature protection for ADC; anti-dithering circuit design
- Implementation of the digital control algorithm on FPGA: Embedded fuzzy logic and expert system into the digital control platform (DCP) for better diagnostics, faults analysis , auto-detection and self-calibration

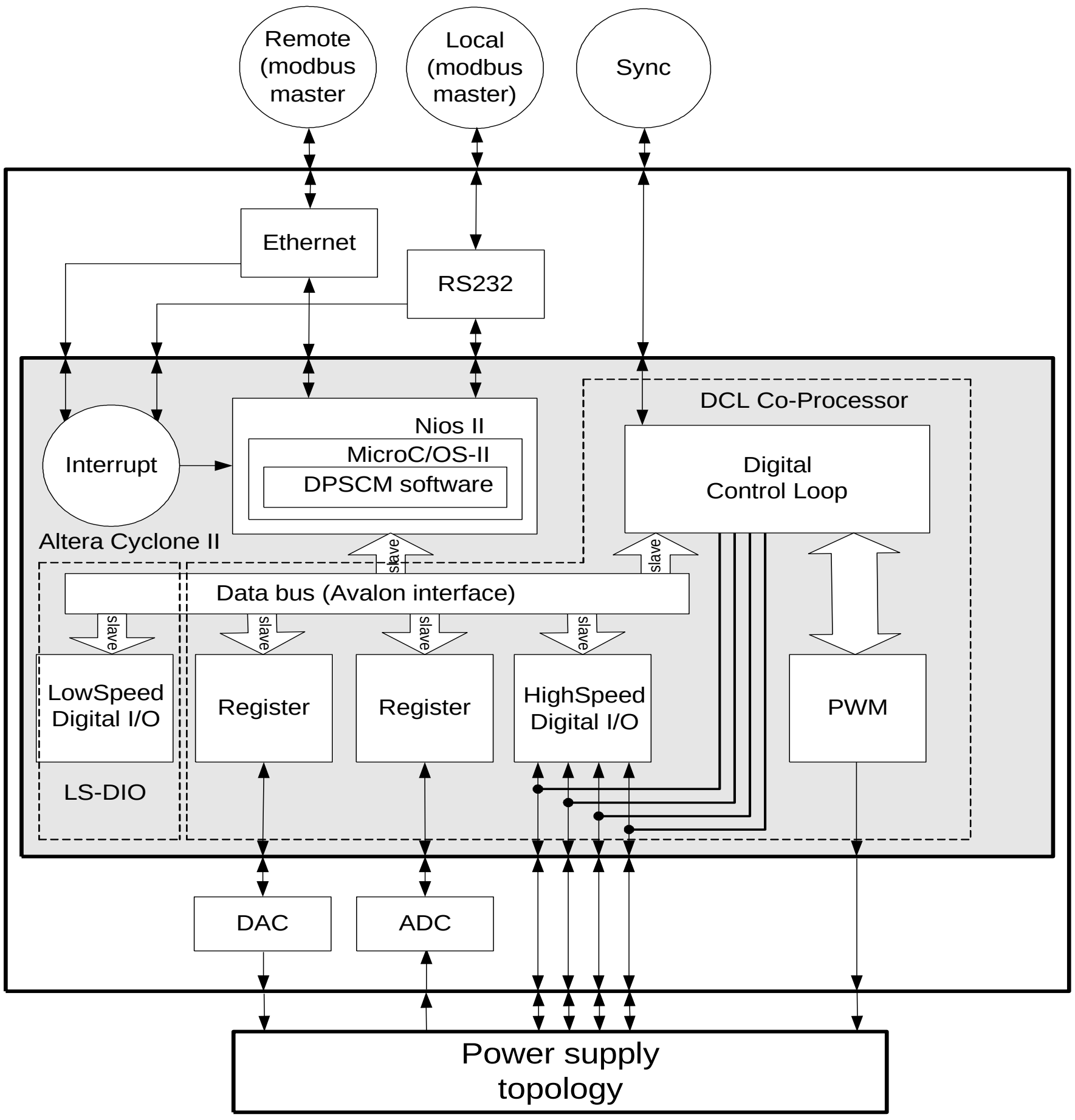

**Figure 4.3.5.5:** Structure diagram of embedded DPSCM power supply

## 4.3.6 Vacuum System

Beam lifetime and stability are of major importance in a storage ring. The interaction of the stored particles with the molecules of the residual gas leads to particle losses and gives rise to background in the detector. There are two 120 GeV circulating beams, each 17.4 mA. These beams emit intense synchrotron radiation in a forward-directed narrow cone. This energetic photon flux produces strong outgassing from the vacuum chamber and a large dynamic pressure increase, which limits the beam lifetime and may cause increased background in the experiments. Therefore, the pumping must maintain the specified operating pressure under the condition of a large dynamic photo-desorption gas load.

To estimate the beam-gas lifetime, knowledge of the residual gas composition is required. Generally, the dynamic pressure is dominated by desorbed $H_2$ (> 60%) and $CO+CO_2$ (< 40%). Heavy molecules such as Ar are particularly bad whereas light molecules such as $H_2$ are less critical.

The basic requirements for the ultra-high vacuum system are:

- A vacuum lower than $3\times10^{-9}$ Torr. It can be shown that the beam lifetime would exceed 20 h if there were only beam gas interactions.
- Good lifetime must be achieved soon after the initial startup with a stored beam.
- The system must be capable of quick recovery after sections are let up to air for maintenance or repairs.
- The chamber wall must be as smooth as possible to minimize electromagnetic fields induced by the beam.
- Very low pressure must be achieved in the interaction regions to minimize detector backgrounds from beam-gas scattering, ideally $3\times10^{-10}$ Torr or lower outside of the Q1 magnet.
- Sufficient cooling is required to safely dissipate the heat load associated with both synchrotron radiation and higher-order-mode (HOM) losses.

Several colliders with double storage rings for both electrons and positrons have been constructed to date. The vacuum system parameters of these different colliders are compared in Table 4.3.6.1.

**Table4.3.6.1:** Comparison of vacuum-related parameters in several storage rings

| Parameter | PEPII (USA) | | KEKB (Japan) | | LEP2 (CERN) | CEPC (China) | |
|---|---|---|---|---|---|---|---|
| | $e^+$ | $e^-$ | $e^+$ | $e^-$ | $e^+$ $e^-$ | $e^+$ | $e^-$ |
| Energy [GeV] | 3.11 | 9.00 | 3.5 | 8.0 | 96 | 120 | |
| Beam current [A] | 2.14 | 0.95 | 2.6 | 1.1 | 2×0.007 | 0.0174 | |
| Circumference [m] | 2199.32 | | 3016.26 | | 26700 | 100000 | |
| Bending radius [m] | 13.75 | 165 | 16.31 | 104.46 | 3096.18 | 10700 | |
| Arc beam pipe material | Extruded aluminum, TiN coating | Extruded copper | Extruded copper | | Extruded aluminum, Lead shielding | Extruded copper, NEG coating ($e^+$); Extruded aluminum ($e^-$) | |
| Arc beam pipe shape | Ellipse with antechamber 95×55 | Octagon 90×50 | Circle 94 | Racetrack 104×50 | Ellipse 131×70 | Ellipse 75×56 | |
| Pump type in arcs | TSP, IP | IP | NEGs, IP | NEGs, IP | NEGs, TSP, IP | NEGs, IP | |

### 4.3.6.1 *Synchrotron Radiation Power and Gas Load*

In the vacuum system design two issues related to the synchrotron radiation must be considered. One is heating of the vacuum chamber walls from the high thermal flux and the other is the strong gas desorption from both photon-desorption and thermal desorption. The dynamic pressure induced by synchrotron radiation can rise by several orders of magnitude once a beam starts circulating. In this section, we quantify the effects and evaluate their impact.

*4.3.6.1.1 Synchrotron Radiation Power*

To estimate the heat load, we start from the well-known expression [Sands, 1970] for the synchrotron radiation power (in kW) emitted by an electron beam in uniform circular motion [1]:

$$P_{SR} = \frac{88.5E^4 I}{\rho} \quad (4.3.6.1)$$

where $E$ is the beam energy (in GeV), $I$ is the total beam current (in A), and ρ is the bending radius of the dipole (in meters). The linear power density (in kW/m) along the circumference is given by

$$P_L = \frac{P_{SR}}{2\pi\rho} = \frac{88.5E^4 I}{2\pi\rho^2} \quad (4.3.6.2)$$

For CEPC, $E$ = 120 GeV, $I$ = 0.0174 A, $\rho$ = 10700 m, and we find from Eqs. (4.3.6.1) and (4.3.6.2) the total synchrotron radiation power $P_{SR}$=29.8 MW and a linear power density of $P_L$ =444 W/m.

*4.3.6.1.2 Gas Load*

The gas load arises from two processes: thermal outgassing and synchrotron-radiation-induced photo-desorption. To estimate the desorption rate, we follow the approach of Grobner et al. [1983]. The effective gas load due to photo-desorption is [1]

$$Q_{gas} = 24.2EI\eta \text{ [Torr·L/s]}, \quad (4.3.6.3)$$

where $E$ is the beam energy in GeV, $I$ the beam current in A, and $\eta$ the photo-desorption coefficient in molecules/photon. The photo-desorption coefficient $\eta$ is a property of the chamber that depends on several factors:

- Chamber material
- Material fabrication and preparation
- Amount of prior exposure to radiation
- Photon angle of incidence
- Photon energy

Experimental measurements indicate that a copper (or aluminum) chamber may eventually develop an effective $\eta \approx 10^{-6}$ [1]. For a vacuum chamber with a desorption coefficient of $\eta = 2 \times 10^{-5}$, the dynamic gas load is

$$Q_{gas} = 4.84 \times 10^{-4} EI \text{ [Torr·L/s]}, \quad (4.3.6.4)$$

and the linear gas load is

$$Q_L = \frac{Q_{gas}}{2\pi\rho} \text{ [Torr·L/s·m]}. \quad (4.3.6.5)$$

We obtain the total dynamic gas load of $Q_{gas} = 1.0 \times 10^{-3}$ Torr·L/s, and a linear SR gas load of $Q_{LSR} = 1.5 \times 10^{-8}$ Torr·L/s/m. Assuming the thermal outgassing rate of the vacuum chambers is $1 \times 10^{-11}$ Torr·L/s·cm$^2$，for an elliptical cross section of the vacuum chamber (H×V = 75 mm × 56 mm), a linear thermal gas load $Q_{LT} = 2.1 \times 10^{-8}$ Torr·L/s/m. The total linear gas load will be $3.6 \times 10^{-8}$ Torr·L/s/m.

### 4.3.6.2 ***Vacuum Chamber***

#### *4.3.6.2.1 Vacuum Chamber Material*

The synchrotron radiation power deposited requires a water-cooled high electrical conductivity chamber (aluminum or copper). Extruded aluminum chambers have been used in LEP [2]; they were water-cooled and covered with lead cladding to prevent other components from radiation damage.

Copper is preferred because of its naturally lower molecular yields, lower electrical resistance, and its smaller radiation length, giving more efficiency in preventing photons from escaping through the vacuum chamber wall and damaging the magnets and other components. Also since the chamber walls in the arcs are subjected to very high thermal loads, copper with its excellent thermal conductivity is preferred. Vacuum chambers in the straight sections will be fabricated from stainless steel.

Copper has been extensively used for B-factory vacuum chambers [3], and it has been found that its initial molecular yields were lower than aluminum by nearly 1~2 orders of magnitude [4-5]. The PSD (Photon Stimulated Desorption) tests on copper at DCI have shown that a photo-desorption coefficient of $10^{-6}$ can be achieved in a reasonable time at high current. [1] Such a low photo-desorption coefficient allows us to design the vacuum chamber with a conventional elliptical or octagonal shape, instead of being driven to adopt an antechamber design that is more difficult and expensive to fabricate. The apparent cost disadvantage of copper is offset by the relative simplicity of the copper shape, by the reduction in the amount of pumping needed, and by shortening the vacuum system commissioning time.

Considering the cost of fabrication, the vacuum chambers of the electron ring will be made of aluminum alloy, and that of the positron ring will be produced from copper in order to reduce secondary electron yields and avoid e-cloud instability. Radiation shielding will be done with lead blocks put inside the magnets.

#### *4.3.6.2.2 Vacuum Chamber Shape*

The cross-section of the dipole vacuum chamber is elliptical, 75 mm wide by 56 mm high (Fig. 4.3.6.1). The length of this dipole chamber is 6 m, and the chamber wall thickness is 3 mm. A cooling channel attached to the outer wall of the beam duct carries away the heat produced by synchrotron radiation hitting the chamber wall. The beam duct of the positron ring will be extruded from full lengths of UNS C10100, high-purity, oxygen-free, high-conductivity copper, and the cooling channel will be fabricated from USN C10300, an oxygen-free copper alloy. The vacuum flanges are made of stainless steel.

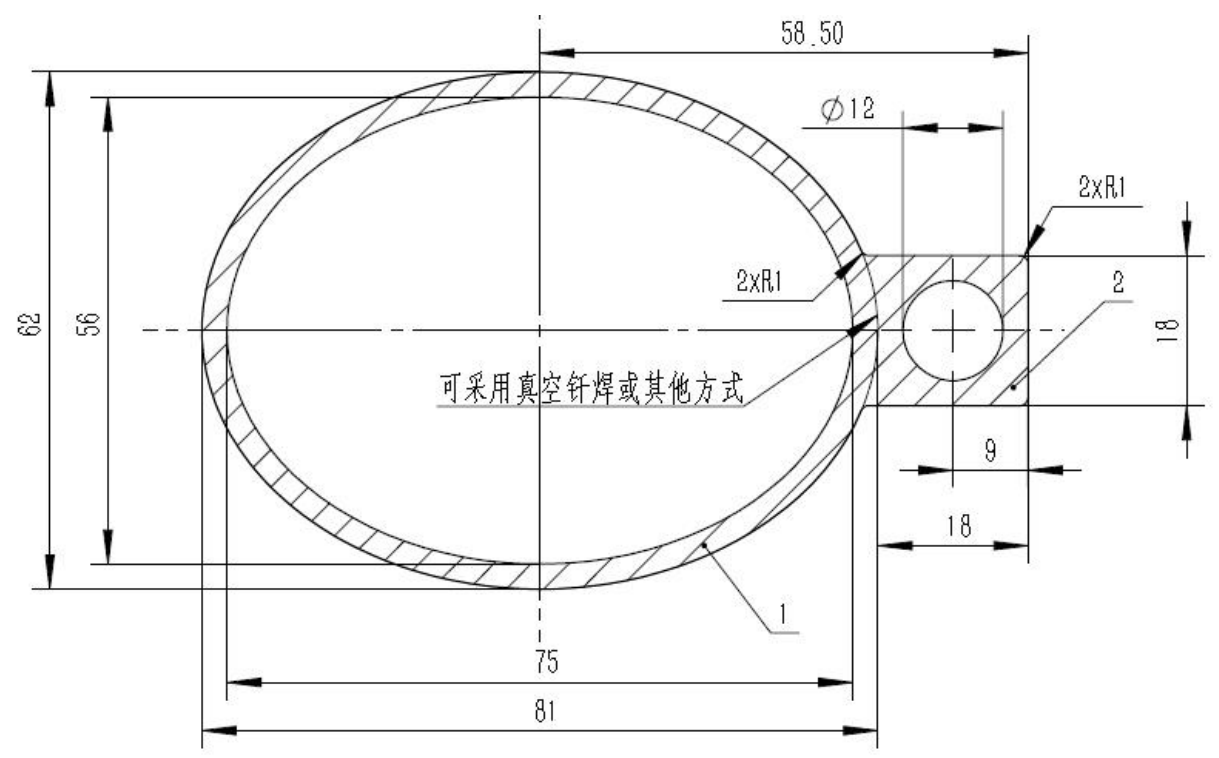

**Figure 4.3.6.1:** Copper dipole vacuum chamber

One of the main challenges in designing the vacuum chamber is to adequately handle the high thermal synchrotron radiation power incident on the vacuum chamber wall. The linear power density reaches 444 W/m. Finite-element analysis of a dipole chamber subjected to this power shows that the highest temperature reaches 37.4°C when a convective heat transfer coefficient of $1\times10^{-3}$ W/mm²·℃ is chosen. The maximum stress is 12 MPa, and the maximum deformations of x, y, z directions are 0.033 mm, 0.016 mm and 0.25 mm/m which are in the safety range. Figure 4.3.6.2 shows the results of the finite-element analysis.

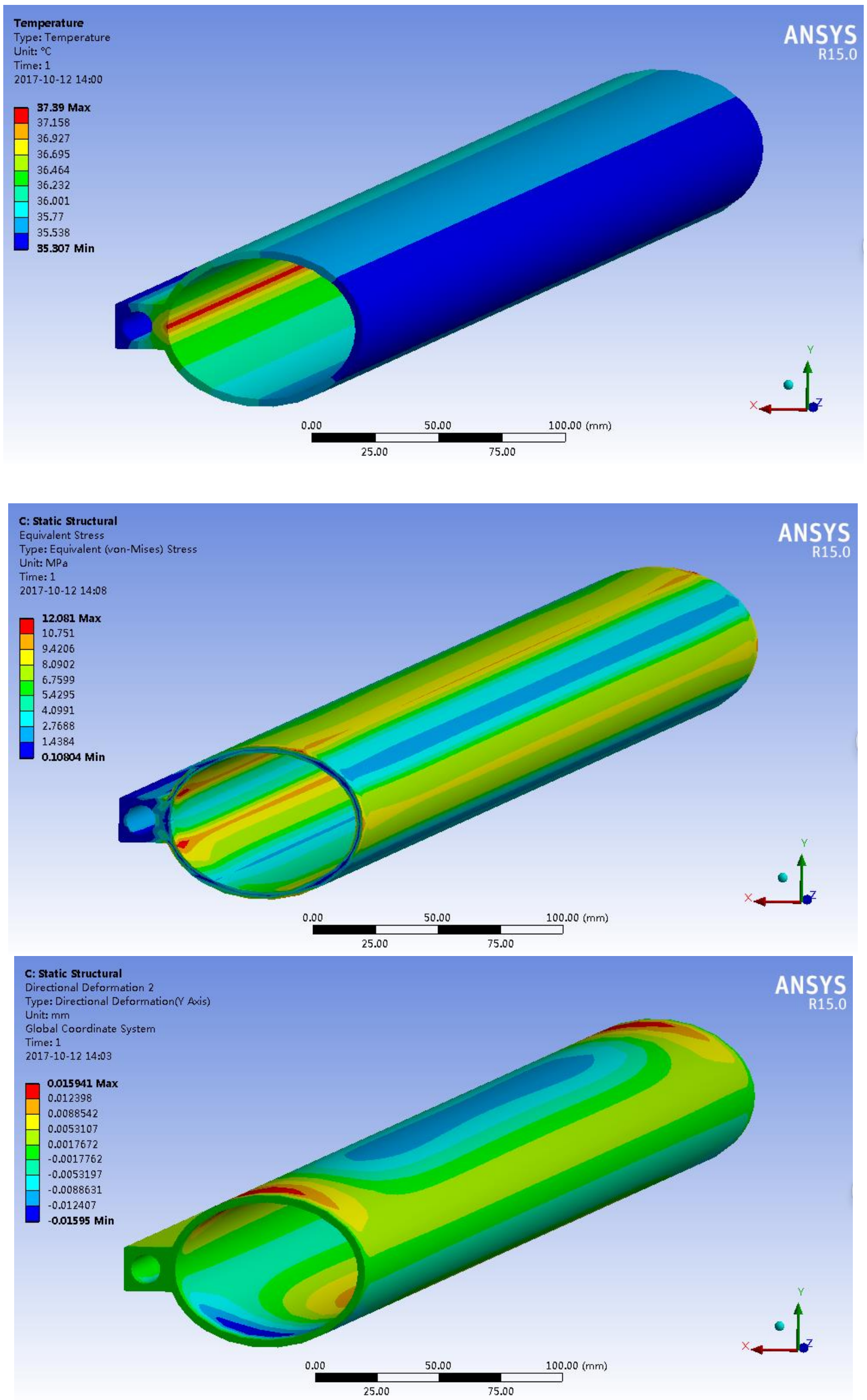

**Figure 4.3.6.2:** Results of the finite-element analysis of a copper dipole vacuum chamber. Top – the temperature rise; middle – the stress; bottom – the deformation.

The chamber consists of an extruded copper chamber and cooling channel with two conflate-type end flanges. These are both drawn to their final shape and to produce a minimum half-hard temper. The pieces are then cleaned and joined by electron-beam welding. After welding, the subassembly is stretch-formed to its correct radius, then the ends are machined and the part cleaned. Finally, the end flanges are TIG-brazed onto the ends of the chamber [1]. A one-piece chamber extrusion eliminates all longitudinal vacuum welds, which affords a more accurate and dependable chamber.

The aluminum chamber of the electron ring will be extruded from Al-6061, and stainless steel conflat flanges are welded onto the ends of the chambers though transition material. The results of the finite element analysis indicate that the highest temperature is 39.9°C; the maximum stress is 15.7 MPa; the maximum deformations of x, y, z directions are 0.06 mm, 0.029 mm and 0.53 mm/m, respectively, which are in the safe region.

The synchrotron radiation will deposit 5 W heating continually on the stainless steel flange, the thermal analysis indicate that the maximum temperature is 58 °C with 5 W power uniformly distributed in the flange. While the powers of 2.5 W and 5 W are concentrated in a small spot with 1.2 mm in diameter, the maximum temperature are 56 °C and 90 °C, respectively, which are in a safe region for stainless steel material. Using the coefficient of expansion for stainless steel which is $1.7\times10^{-5}$ m/(m·K), the temperature difference between 90 °C and 56 °C for a flange pair will result in an expansion of $5.78\times10^{-4}$ m/m. If we assume the flange is 10 cm in diameter then we have an expansion of $5.78\times10^{-5}$ m or 57.8 microns, which will still be tight for vacuum.

An oxide layer on the chamber inner surface contains a large amount of carbon which would be released as CO and $CO_2$ in photo-desorption. To remove this first oxide layer and to produce a new oxide layer that is free of carbon, a commercially available chemical cleaner containing $H_2O_2$ and $H_2SO_4$ will be used, or a standard acid etch with $H_2SO_4$, $HNO_3$, HCI and water will be applied.

### 4.3.6.3 *Bellows Module with RF Shielding*

The primary function of the bellows module is to allow for thermal expansion of the chambers and for lateral, longitudinal and angular offsets due to tolerances and alignment, while providing a uniform chamber cross section to reduce the impedance seen by the beam. Figure 4.3.6.3 is a drawing of the RF shielding bellows module.

The usual RF-shield has many narrow Be-Cu fingers that slide along the inside of the beam passage as the bellows is compressed. One of the key issues for this finger-type RF-shield is the strength of the contact force. Each contact-finger should touch a beam tube with an appropriate contact force to maintain sufficient electrical contact against a high frequency current. The larger the force, of course, the better the electrical contact, but the more abrasion (dust generation) during mechanical flexing [3]. It is important to have a minimum contact force to avoid excess heating and arcing at the contact point. The leakage of HOM RF from the slits between contact fingers into the inside of bellows is another important issue.

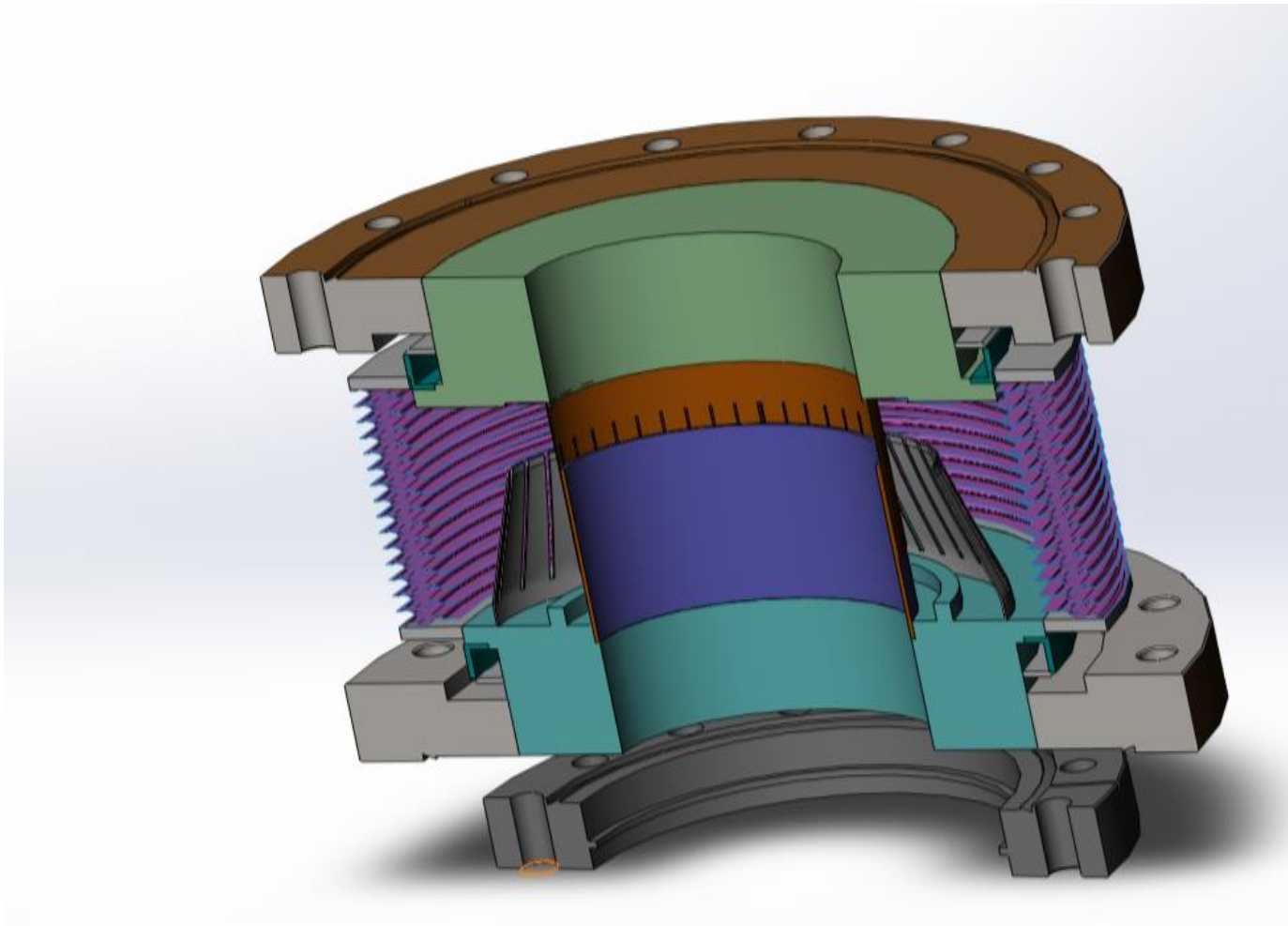

**Figure 4.3.6.3:** RF shielding bellows module

The fingers maintain a relatively high contact pressure of 125±25 g/finger. The slit length between fingers is 20 mm. The RF-shield can accommodate a maximum expansion of 10 mm and contraction of 20 mm, allowing for a 2 mm offset. The step at the contact point is limited to less than 1 mm. The cooling water channel takes care of synchrotron radiation power, Joule loss and HOM heat load on the inner surface, and leaked HOM power inside the bellows. The RF fingers will be protected by the mask located in the upstream vacuum chamber from synchrotron radiation irradiation.

### 4.3.6.4 *Pumping System*

The 100 km circumference of the ring will be subdivided into 520 sectors by all metal gate valves. These allow pump down from atmospheric pressure, leak detecting, bake-out, and vacuum interlock protection to be done in sections of manageable length and volume. Roughing down to approximately $10^{-7}$ Torr will be achieved by an oil free turbo-molecular pump group. The main pumping is achieved with Non Evaporable Getter (NEG)-coated copper chambers in the positron ring, sputter ion pumps will be used to maintain pressure and pump off $CH_4$ and noble gases that can't be pumped off by the NEG pump. The aluminum chambers in the electron ring will be evacuated by both ion pumps and NEG pumps spaced about 6 m apart. For the pumping system in the interaction regions where the detectors are located, depending on the space available, NEG pumps, sublimation pumps and sputter ion pumps will be used.

#### *4.3.6.4.1 NEG Coating*

The NEG coating is a titanium, zirconium, vanadium alloy, deposited on the inner surface of the chamber through sputtering. The NEG-coated chamber is first inspected for gross contamination or surface defects, which could cause poor film adhesion [6]. Each dipole chamber will be fitted with three cathodes (made of twisting together Ti, Zr and V metal 0.5 mm wires) mounted along the chamber axis to achieve uniform thickness distribution along the perimeter. To keep the cathode close to the chamber's axis, several ceramic spacers are spaced along the chamber length, plus two adaptors at the extremities. Chambers are then evacuated to the $10^{-9}$ mbar range by a turbo-molecular pump group and before coating baked overnight and leak tested with helium. A Residual Gas Analyzer (RGA) is also used to monitor partial pressure. The process gas and pressure were krypton at ~0.1 mbar, and the chamber temperature around 200 °C. [7]

The vacuum in a NEG-coated chamber is improved by both reduced desorption yield and direct pumping by the NEG alloy. When exposed to air, the NEG surface is saturated and loses its pumping activity. An essential operation is activation, which produces diffusion of the saturated surface layer into the bulk of NEG material by heating the NEG-coated chamber. NEG films can be fully activated at relatively low temperature, like 250°C for 2 hours. Even lower activation temperature for longer times (e.g. 180°C for 24 hours) has been successfully applied in the case of aluminum chambers which cannot withstand high temperature bake-out.

##### *4.3.6.4.2 Sputter Ion Pumps*

Sputter ion pumps are required to pump Ar, He and $CH_4$, which are not absorbed by the NEG. During the period of beam cleaning, $CH_4$ will be an important component of the residual gas, and this determines the number of sputter ion pumps to be installed. A sputter ion pump will be mounted at intervals of 6 to 18 m. The number of pumps can easily be doubled if necessary. The sputter ion pumps are started only after the NEG has been activated, i.e. at a pressure of $10^{-7}$ Torr or lower. This allows several pumps to be connected in parallel to one common power supply.

Sputter ion pumps have high reliability, no moving parts, long life and high radiation resistance. In addition, the ion pump current is proportional to vacuum pressure and can provide a detailed pressure profile around the ring. The power supplies of ion pumps will trip to protect the ion pumps from damage if the ion current rises above a pre-set value. The leakage current of the pumps is less than 10% of the current drawn at $1\times10^{-9}$ Torr, which make the pump suitable for use as a pressure monitor. Ion pump currents can be stored in a databank, enabling the operators to find problems conveniently.

### 4.3.6.5 ***Vacuum Measurement and Control***

The size of CEPC excludes the installation of vacuum gauges at short intervals. Only some special sections such as the injection regions, RF cavities and interaction regions are equipped with cold cathode gauges and residual gas analyzers. For the remainder of the ring only the current of the sputter ion pumps will be monitored continuously and should provide adequate pressure measurements down to $10^{-9}$ Torr. Mobile diagnosis equipment can be brought to places of interest during pump down, leak detection and bake-out when the machine is accessible.

The control of the vacuum system will be part of the general computer control system and includes the control of the sputter ion pumps, vacuum gauges, sector valves, and the monitoring of the water cooling of the vacuum chambers. The vital interlocks (sectors valve, RF cavities, water cooling) will be hard-wired. Other controls will only be needed locally and temporarily, and therefore will be handled by mobile terminals.

Due to the high radiation levels in the tunnel, all the vacuum electronic devices will be located at the service buildings.

### 4.3.7 **Instrumentation**

#### 4.3.7.1 ***Introduction***

The beam instrumentation system consists of various beam monitors and signal processing electronics, and must provide precise and sufficient information so that accelerator physicists and machine operators can improve the injection efficiency, optimize the lattice parameters, monitor the beam behaviour and increase the luminosity. Good instrumentation is also crucial for efficient commissioning.

There are unique problems specific to the large size of the ring. Considering the long distances, it is not a good choice to use copper cables to send signals; we should digitize the analog signals in the tunnel and use optical fibers to send data from electronics near the monitors to local stations in an auxiliary tunnel. The positrons and electrons pass through the same monitors, and we distinguish them by polarity. We summarize our design philosophy:

- Satisfy the requirements for long-term stable operation;
- Appropriate precision and speed for parameter measurements;
- Large dynamic range under different conditions;
- Coupling impedance of the devices must be as small as possible;
- In house construction of components should be used as much as possible to save money.

We need to monitor beam status quickly and accurately, measure and control the bunch current efficiently, and cure beam instabilities. The beam orbit measurement is important, especially in the interaction region. It can help us know the beam position, offset and crossing angle and it is advantageous for increasing the luminosity. There are several subsystems, including BPMs for beam position, the DCCT for average beam current measurement, the tune measurement system, the photon monitoring system which includes a CCD camera for monitoring the beam profile, and a streak camera for measurement of bunch length measurement. These systems are summarized in Table 4.3.7.1.

**Table 4.3.7.1:** Main technical parameters of the Collider Beam Instrumentation Systems

| Subsystems | | Parameters | Quantity |
|---|---|---|---|
| BPM | Bunch by Bunch | Measurement area (x × y)：±40 mm×±20 mm<br>Accuracy：1 mm<br>Resolution：0.1 mm | 2900 |
| | Closed orbit | Measurement area (x × y)：±20 mm×±10 mm<br>Accuracy：0.1 mm<br>Resolution：<0.001 mm<br>Measurement time of COD：< 4 s | |
| BLM | | Dynamic range:$10^6$-$10^8$<br>Counting rates: ≤10 MHz<br>Radiation environment: <$10^8$ Rad<br>Response time: ~ns | 5800 |
| Tune | | Resolution：0.0001 (0.1 kHz)<br>Accuracy：0.0005 (0.5 kHz) | 2 |
| DCCT | | Dynamic measurement range：0.0~1.5 A<br>Linearity：0.1 %<br>Zero drift: <0.05 mA<br>Remarks: shielding needed | 2 |
| BCM | | Measurement range：10 mA / per bunch<br>Relative precision：1/4095<br>Smallest bunch spacing：0.5 m | 2 |
| Feedback system | Transverse | Damping rate > 20 $ms^{-1}$<br>Oscillation amplitude < 1 mm | 2 |
| | Longitudinal | Damping rate > 0.5 $s^{-1}$<br>Energy error < 0.6% | 2 |
| Synchrotron light monitor | Beam size measurement | Resolution：10% beam size | 4 |
| | Bunch length measurement | Resolution：0.5 ps (using streak camera)<br>Measurement time：1s | 2 |
| Vacuum chamber displacement measurement | | Resolution:0.001mm | 500 |

### 4.3.7.2 ***Beam Position Measurement***

With one Beam Position Monitor (BPM) near each quadruple, there will be 2,324 BPMs; this also includes some additional ones at specific locations. We will set up 32 local stations for BPMs and other instruments in an auxiliary tunnel, each controlling 72 BPMs. Frontend electronics and digital electronics of the BPMs are in the tunnel. Considering the limited space in the tunnel, it is a good option to place them under the magnet girder.

The data from the digital electronics will be sent by optical fiber to local stations where there will be a multiplex system for communicating with other local stations and the central control room.

High resolution is necessary for the BPM at the special regions near the IP or the local chromaticity correction sextupoles. We also will consider implementing non-linear mapping for the off-axis position of the pretzel beams, an orbit slow feedback system and IP point orbit feedback system based on the BPM signals.

#### *4.3.7.2.1 Mechanical Construction*

The design criteria are the following:

1. Short length to save space;
2. Skewed sensor positions to avoid direct impact from synchrotron radiation;
3. Minimum RF loading and higher modes coupling to the beam;
4. High precision for interchangeability;
5. Flanges for replacement in case of a leak;
6. Resistance to corrosion and to baking up to 300°C.

The best solution is a capacitive monitor with a button-like electrode, as used in most other electron machines. In order to attain the required geometrical accuracy, the four buttons will be mounted on a machined block of aluminum welded into the copper vacuum chamber. A matched feedthrough needs to be designed in order to avoid reflections and endure high temperature up to 200℃. The feedthrough is made of titanium and the outer conductor as well as the flange with its plug is made of stainless steel. The pick-up output voltage can be expressed by

$$V_{\text{pickup}}(\text{t})=I_{\text{image}}(\text{t}) \times \text{Z}_{\text{t}}$$

In the time domain, where the $I_{\text{image}}$(t) is proportional to the bunch current $I_{\text{bunch}}$(t), $\text{Z}_{\text{t}}$ is the longitudinal transfer impedance. [1,2,3] The transfer impedance depends on the capacitance *C* of the electrodes and an external input resistor *R*. Figure 4.3.7.1(a) shows the absolute value and phase of the transfer impedance for $l$ = 10 cm length cylindrical pick-up with a capacitance of $C$ = 100 pF and an ion velocity of $\beta$= 50% for high (1MΩ) and low (50 Ω) input impedances of the amplifier [2]. Figure 4.3.7.1(b) shows the frequency spectrum of the pickup electrode voltage, which is high-pass, the same as the transfer impedance. A Gaussian function in time domain of width of $\sigma_t$ has a Fourier transformation described by a Gaussian function of width $\sigma_f = 1/(2\pi\sigma_t)$ centered at $f = 0$. The $\sigma_f$ of the beam is 16 GHz. The cut-off frequency of the BPM $f_{\text{cut}} = \omega_{\text{cut}}/2\pi = (2\pi\text{RC})^{-1}$ is about 3 GHz, where the $\sigma_t$ is the bunch length, *R* and *C* is the input impedance and capacitance of the electrode.

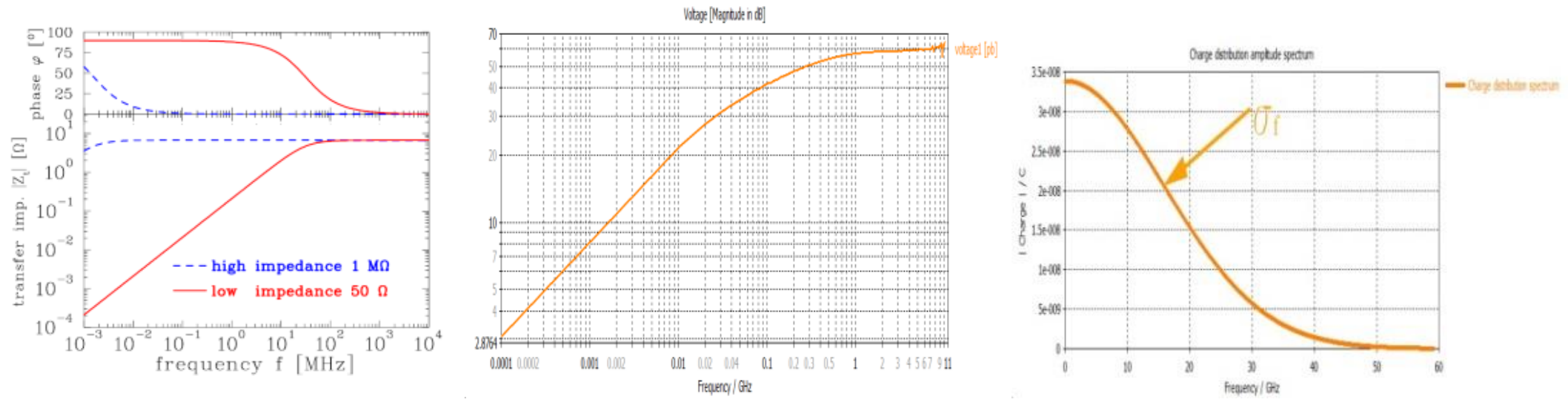

**Figure 4.3.7.1:** (a) Absolute value and phase of the transfer impedance (b) the frequency spectrum of the pickup electrode voltage signal obtained by CST simulation (c) the frequency spectrum of a Gaussian bunch.

In order to study the response of a pick-up electrode to the beam, CST particle studio wake-field simulations are done on the beam parameters and vacuum pipe. [4,5] Figure 4.3.7.2 shows the button BPMs design. As shown in Figure 4.3.7.2, the radius and height of the electrode is 3 mm and the center of the pickup electrode is located in α= 45°where the pipe size is 37.5 mm × 28 mm.

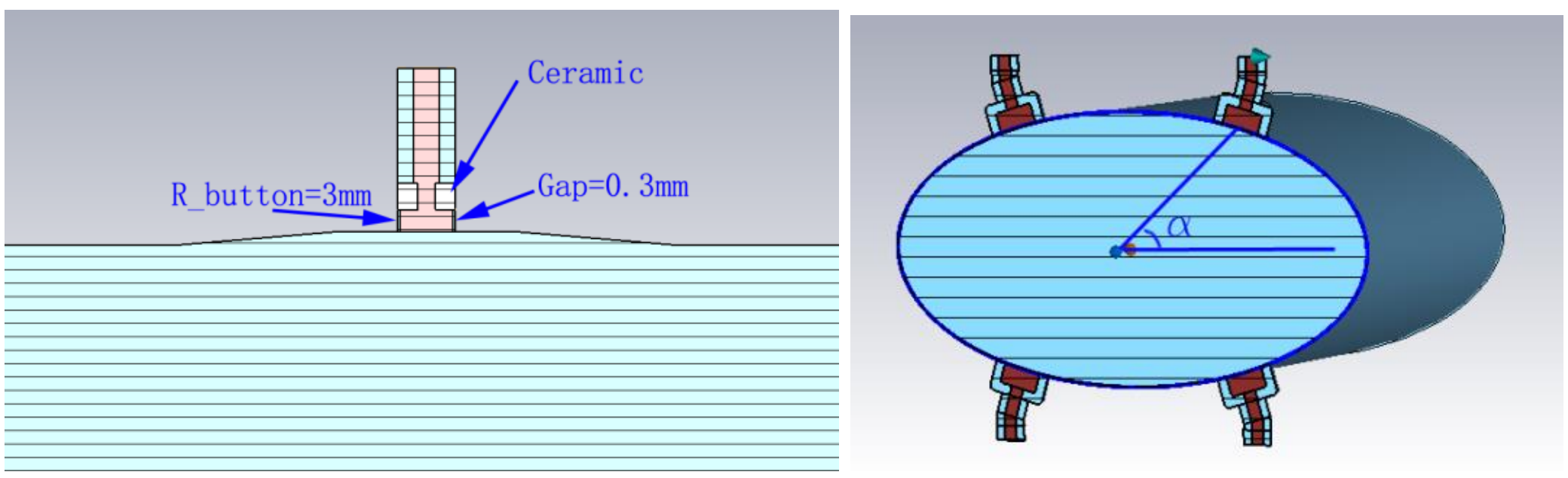

**Figure 4.3.7.2:** (a) button pick up detail (b) approximate model with a port aligned with Cartesian coordinate planes.

Figure 4.3.7.3(a) shows the signal of a pickup. The simulation parameters and design parameters are shown in the table 4.3.7/2. The $I_{\mathrm{bunch}}$ is over 2000 A as shown in 4.3.7.3(b), which means that a transfer impedance of 0.1Ω can induce a voltage over 100 volts. As shown in 5.7.3(a), the amplitude of the signal is hundreds of volts though a small size electrode with a radius of 3 mm. 4.3.7.3(c) is the sensitivity mapping of the simulation in a range of 60 mm × 40 mm. The scan range is shown in the inset figure of 4.3.7.3(c), where U= Δx/Σy, V=Δy/Σy. The transverse response of the signal is non-linear for large amplitudes, the horizontal and vertical sensitivities near the center of the pipe is 4.17 %/mm and 2.99 %/mm as shown in 4.3.7.3(d).

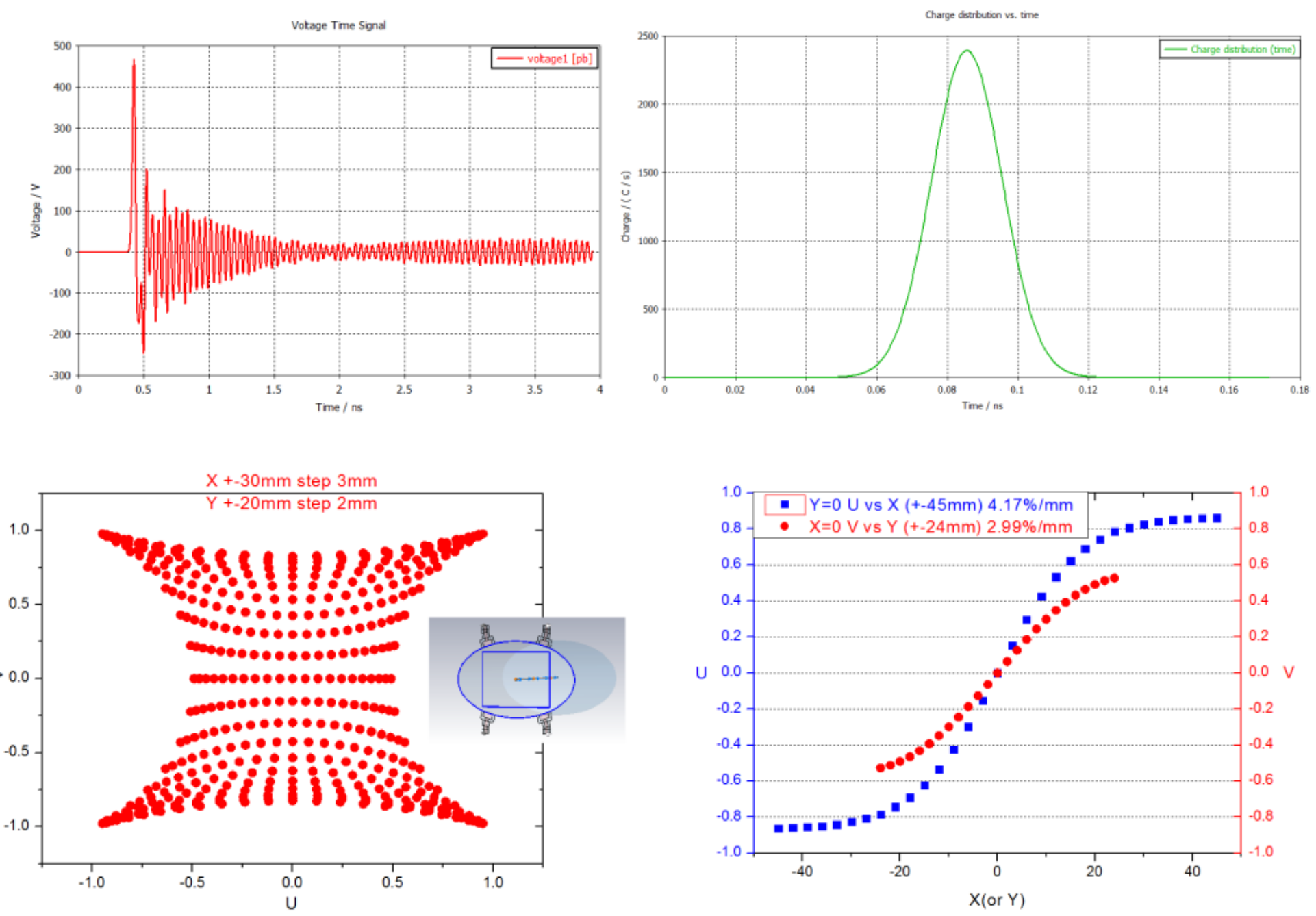

**Figure 4.3.7.3:** (a) the signal of an electrode (b) bunch charge distribution in the time domain (c) sensitivity mapping of the button BPMs on elliptical pipes (d) sensitivity near the center of the pipe.

**Table 4.3.7.2:** Main parameters used for CST simulation

| Parameters | | Design report | CST simulation |
|---|---|---|---|
| Bunch | Length (mm) | 2.72/2.98/3.67 | 3 |
| | Population | $1.7\times10^7$ | |
| Vacuum pipe | Elliptic cylinder<br>a/b (mm) | Elliptic cylinder<br>37.5/28 | |

As shown in the Figure 4.3.7.3 (d), the horizontal sensitivity is higher than the vertical when the electrode located at the 45°. The sensitivity is determined by the azimuthal angle in an elliptical pipe or the distances to the x and y axes. At an azimuthal angle of 50 there is equal horizontal and vertical sensitivity.

We define the coupling factor between the beam and the pickups as the voltage at the middle of the vacuum pipe when applying 1V at one of the pickups. This is also suitable for the coupling factor between the electrodes. [6] We can also get the electrode-to-ground capacity using the CST electrostatic solver. [7,8,9] The capacity of one button electrode C=Q/U=2.93 pF, where Q is the charge on the electrode when we applied a 1V potential to it. Because of the small size of the pickups and long distance between the pickups and beam, the coupling factor between the beam and pickups is 4%, the coupling factor between the pickups is much smaller. Figure 4.3.7.4 (a) is the isoline of potential when 1V potential applied on top right electrode.

The BPM thermal analysis has been done using the CST multi-physics studio because of peak power is very high. The peak power of the pickups is $P_{peak}=(U_{peak})^2/R$=3.2kW and the average power

$$P_{average} = \frac{c\int \frac{V^2}{R}}{\Delta s} = \frac{\sqrt{\pi}\sigma_s P_{peak}}{2\Delta s} = 0.0086 \text{ w},$$

where the $\sigma_s$=3 mm and $\Delta s$=1.07 km is the length and the bunch spacing. The pipe and pickups are aluminum; the loss factor of the beam in a 1000 mm distance is 45.5V/nC. Fig. 4.3.7.4 shows the wake potential and impedance.

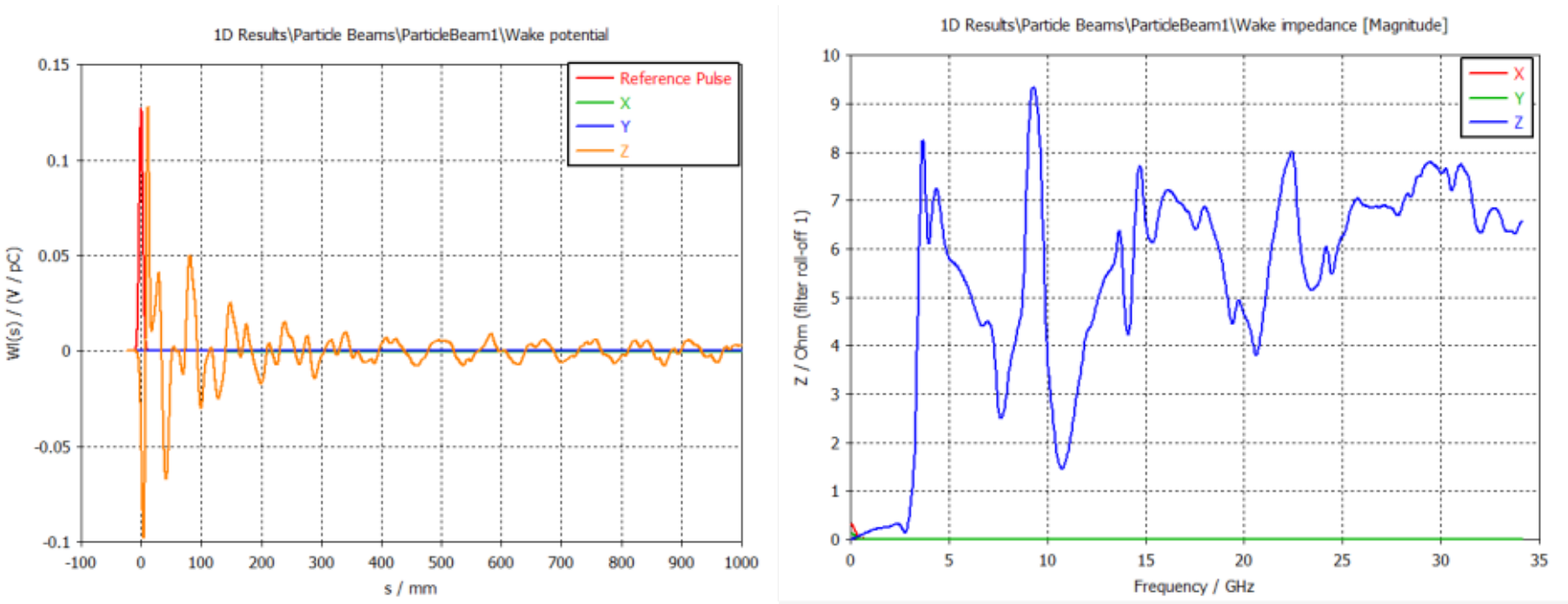

**Figure 4.3.7.4:** (a) the wake potential of the BPM (b) wake impedance of the BPM

### *4.3.7.2.2 BPM Signal Processing*

With high speed ADCs and high resolution we can acquire bunch by bunch positions. The entire system will use microTCA.4 standard structure. This includes RFFE (radio frequency front-end electronics), high speed ADCs, digital electronics and clock signals. 4 channel ADCs will be adopted. The beam position monitor electronics has three different modes: first turn, the FA mode is for fast data acquisition of individual turns and SA data for close orbit measurements. The schematic diagram of the bunch by bunch BPM electronics is shown in Fig. 4.3.7.5.

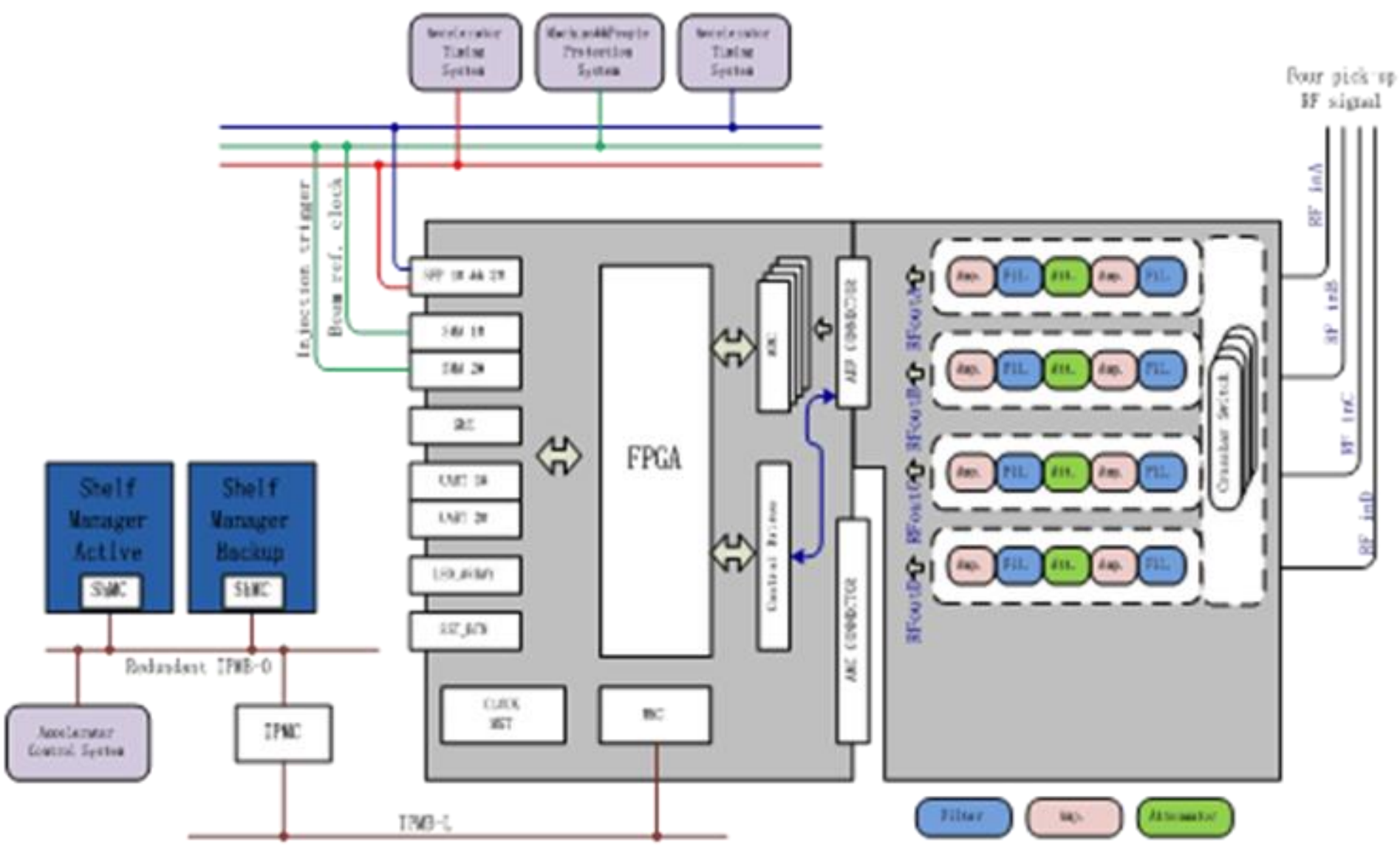

**Figure 4.3.7.5:** Schematic of the bunch by bunch BPM system

BPM electronics will be installed under the magnet girder in the tunnel with two layers of shielding, one polyethylene, and the other lead as shown in the Fig. 4.3.7.6.

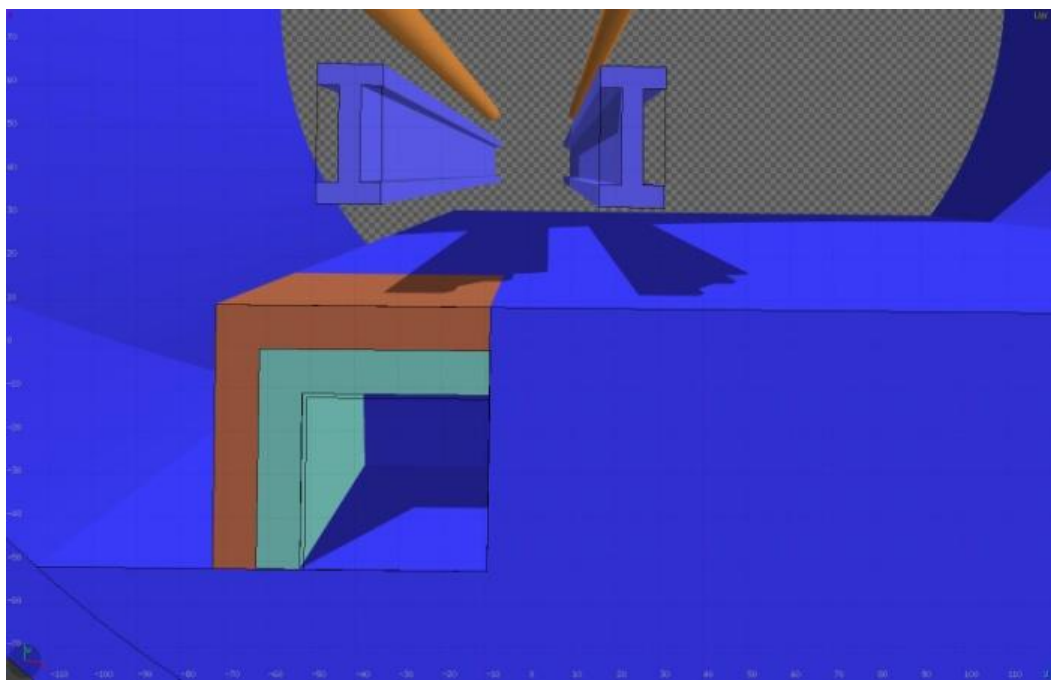

**Figure 4.3.7.6:** Schematic of electronics shielding

### 4.3.7.3 *DC Beam Current Measurement*

The average current and life time are important parameters. DCCT (Direct-current current transformer) are used to measure average beam current. DCCT's are essential for machine commissioning, operation and machine/personnel safety. They have large dynamic range, wide bandwidth and high resolution and often are the only truly calibrated beam instrument in an accelerator and serve as a reference to calibrate other beam diagnostics.

The DCCT includes three parts: magnetic coils, control electronics and a data acquisition system. The DCCT needs to be shielded from stray electromagnetic fields in order to have sufficient resolution. The ceramic pipe must be carefully designed to avoid a discontinuous structure and thus decrease beam impedance. The DCCT is always installed in a straight section but must be placed distant from superconducting cavities, quadruple and corrector magnets and power supply cables. The design of the magnetic coils is the core technology of the DCCT. There are three magnetic coils, one is a fast response transformer, and the other two are second harmonic magnetic modulators. The fast response transformer senses rapid beam changes for coarse adjustments; the magnetic modulator senses slow beam drifts for precision adjustments.

We will select amorphous and nanocrystalline materials because of their high permeability. We could of course select a mature commercial product like the Bergoz NPCT shown is in Figure 4.3.7.7. The modulator, second harmonic detection, PI adjustment, signal amplifier and I/V converting circuits also need to be designed.

A special vacuum chamber with ceramic gap is needed for isolation from the wall current. When a charged particle passes through, it loses energy due to the structure.

For high precision measurement, a Keithley DMM7500 7½-Digit multimeter is chosen for the A/D; the resolution for DC voltage is 10 nV, the noise and one year stability can reach 14 ppm.

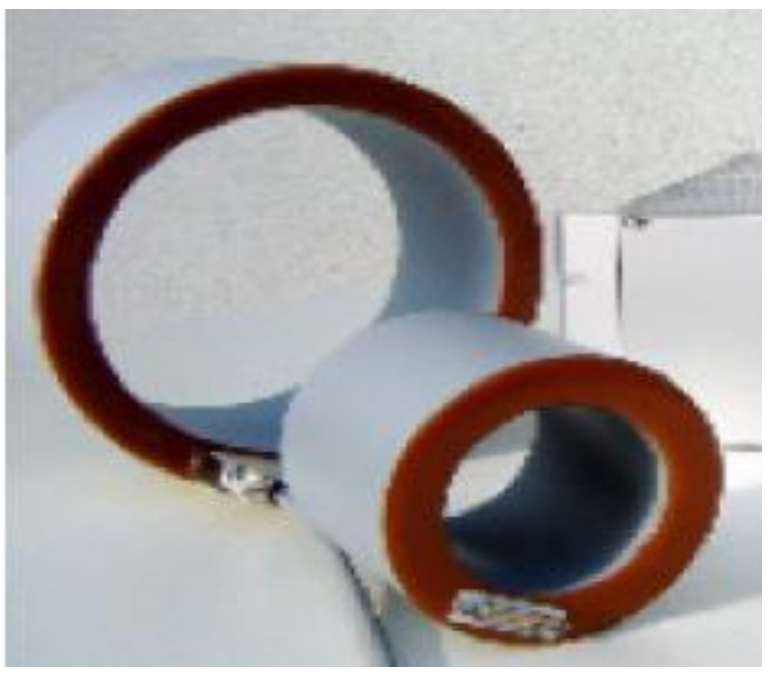

**Figure 4.3.7.7:** Bergoz DCCT

GSI lab has proposed a new design DCCT using the Tunneling Magneto Resonance (TMR) effect. But this new type DCCT cannot develop further because of the lower sensitivity and resolution of the MR device. The resolution of the newest TMR chip can reach nano-tesla and make a new type DCCT possible.

The basic design of the TMR current is something like clamp ammeter. Beam passes through a magnetic ring with a gap, a TMR sensor is put in a gap of the ring. The output of the sensor changes with the beam current.

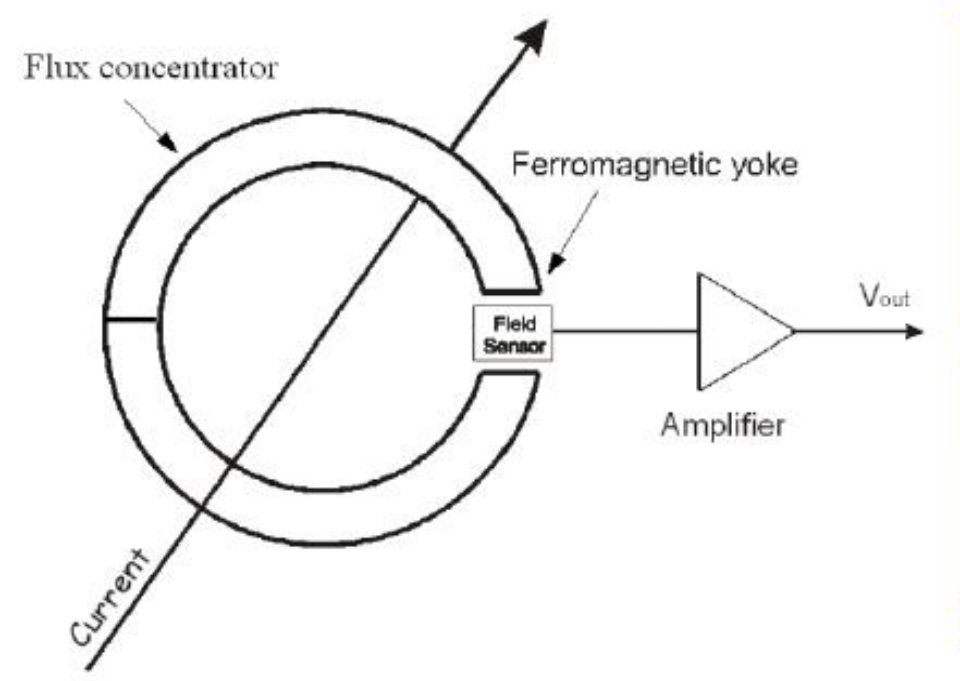

**Figure 4.3.7.8 (a):** The principle of a new type DCCT

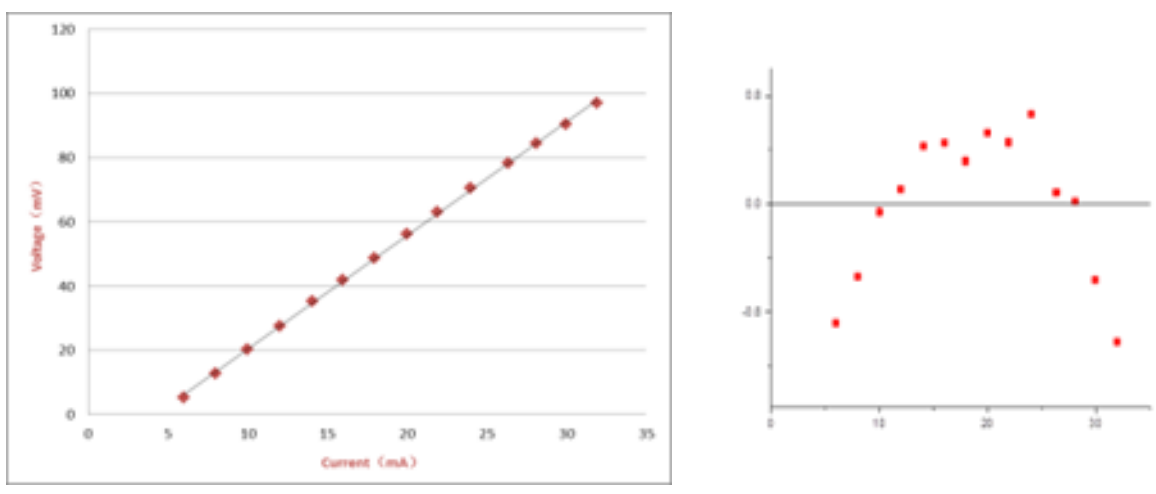

**Figure 4.3.7.8 (b):** Linearity test result (left) and standard deviation (right)

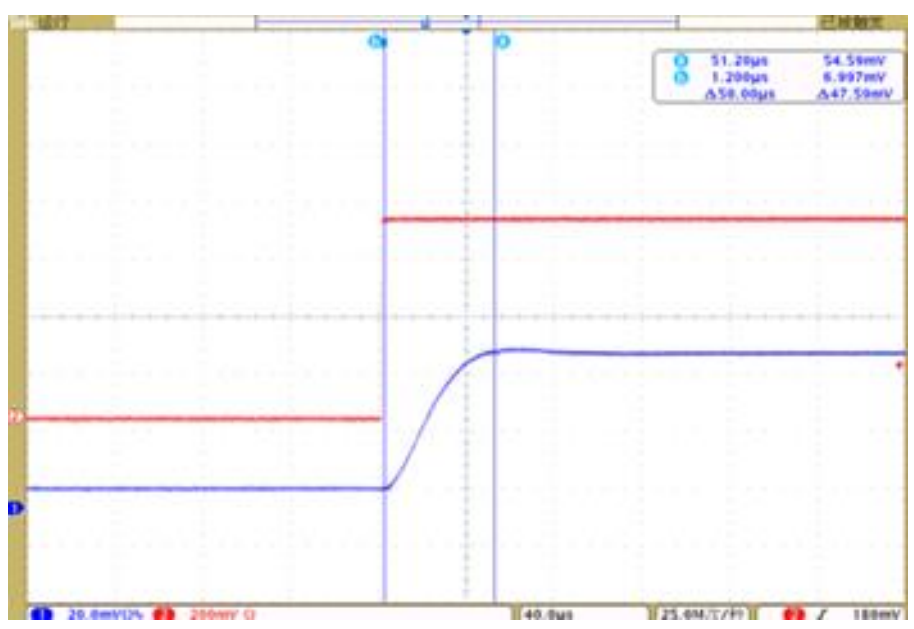

**Figure 4.3.7.8 (c):** TMR current sensor’s response (blue) to a standard square wave (red)

### 4.3.7.4 *Bunch Current Measurement*

The Bunch Current Monitor (BCM) needs to measure the relative charge distribution among the 100 bunches ($50e^+$ and $50e^-$) circulating in the ring. Through absolute calibration of the DCCT these BCM measurements will be continuously transformed into individual bunch intensities, the results stored in the accelerator's database for display and control of the current of every bunch.

The bunch current measurement system includes picking up the signal from the FCT, high speed digital signal acquisition and processing it locally at the beam instrumentation station. To achieve the feedback time requirements (less than the Booster injection period), the signal processing will occupy a very short time synchronous with the injection frequency, and the signal transmission is by fibre. The schematic of the bunch current measurement system is shown as the Fig. 4.3.7.9. The hardware of this system will be purchased commercially and software developed in house.

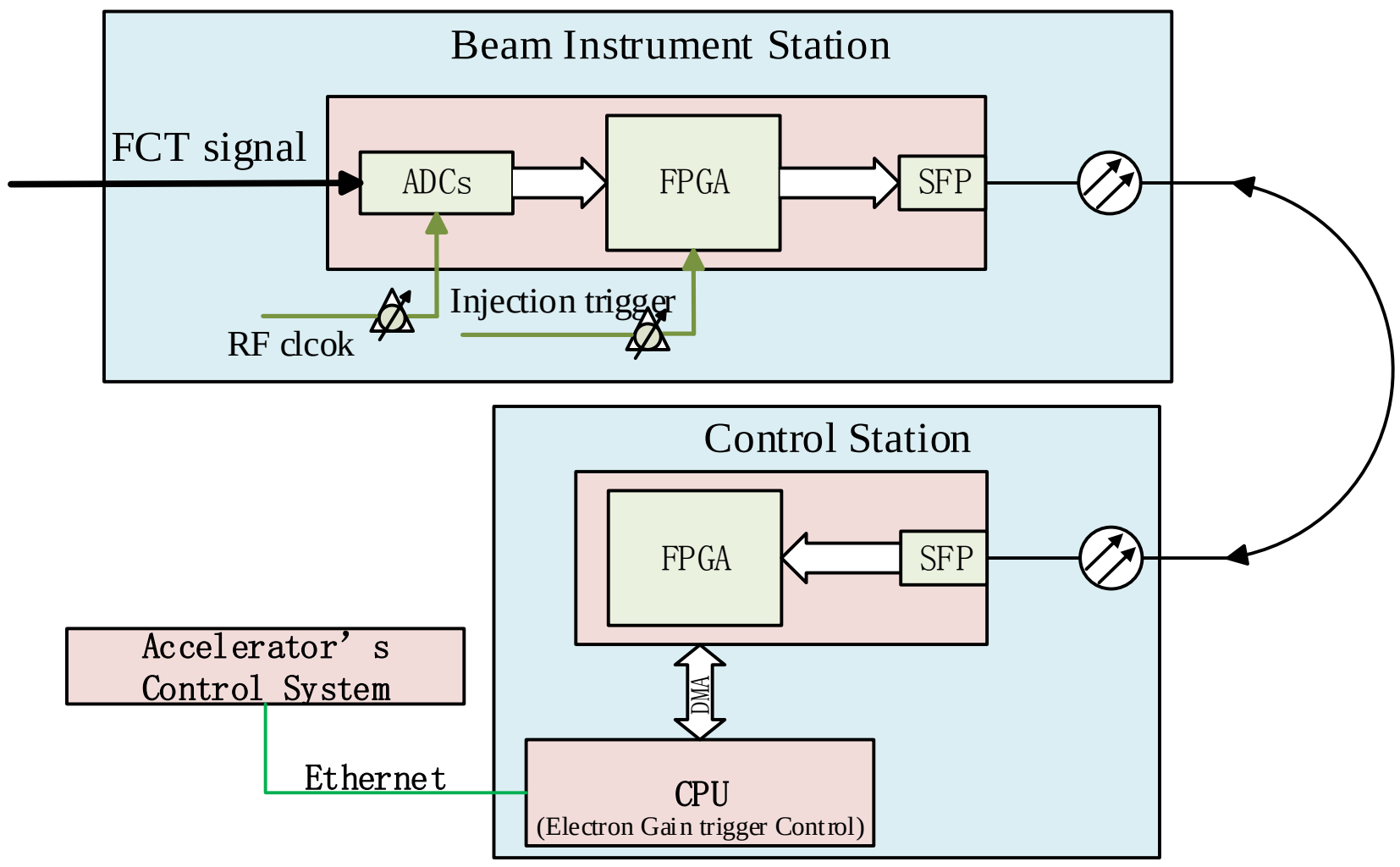

**Figure 4.3.7.9:** Schematic of the bunch current measurement system

### 4.3.7.5 *Synchrotron Light Monitor*

Synchrotron light including visible light, UV, X-rays and gamma rays, is a powerful non-destructive diagnostic tool for beam profile or bunch length measurements. The visible light beam line, which is extracted by a water-cooled Be mirror with a thick "cold finger" absorber to block the X ray and gamma ray fan, will be used for beam size measurement with direct imaging method or a double slit interferometer.

Another beam line is the X-ray beam line; hard X-rays and gamma rays pass through the front end, shielding wall. An Al window isolates the vacuum from air, which acts also as a filter. An X-ray pinhole imaging system will be set up and an X-ray camera placed in the X-ray optical laboratory would be used to record the transverse profile.

#### *4.3.7.5.1 Visible Light Beam Line*

The visible light and UV from the bending magnet is extracted by the first extraction mirror (Be). To avoid mirror deformation by heat load, a thick "cold finger" absorber is placed horizontally in front of the Be mirror to block hard X-rays and gamma rays. Visible light and UV will be reflected to the visible light lab outside the tunnel. For measuring beam size around 40 μm×1000 μm, the direct imaging method and the double slit interferometer methods are good choices. A streak camera will be used for bunch length measurement.

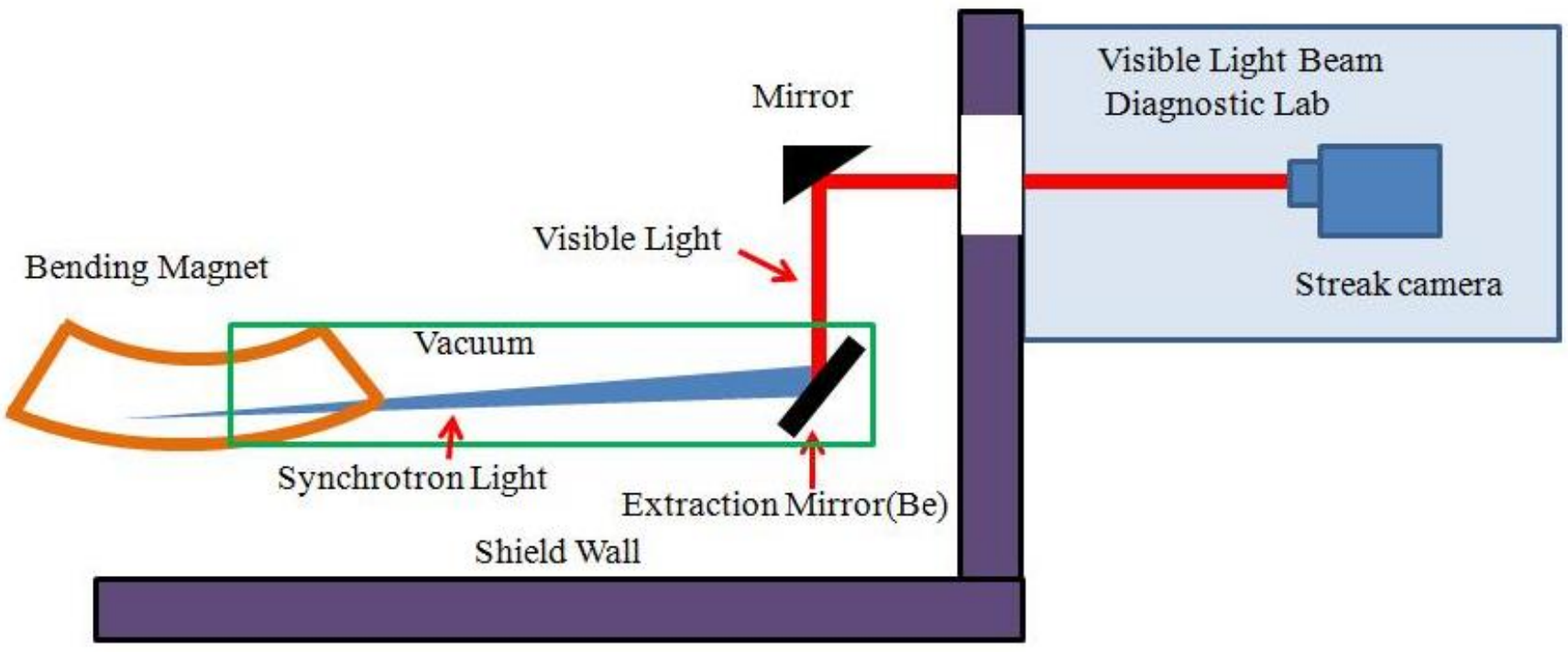

**Figure 4.3.7.10:** Schematic of the visible light diagnostic beam line

The extraction Be mirror should be specifically designed similar to the one in KEK. (Fig.4.3.7.11). The middle of mirror in the backside is thin enough to let X-rays pass through as much as possible. However, there are still residual X-ray absorbed by the mirror and cause deformation. A thick absorber will be a good choice to place in front of the extraction mirror. Both ends of the mirror have water-cooled holes to decrease the temperature.

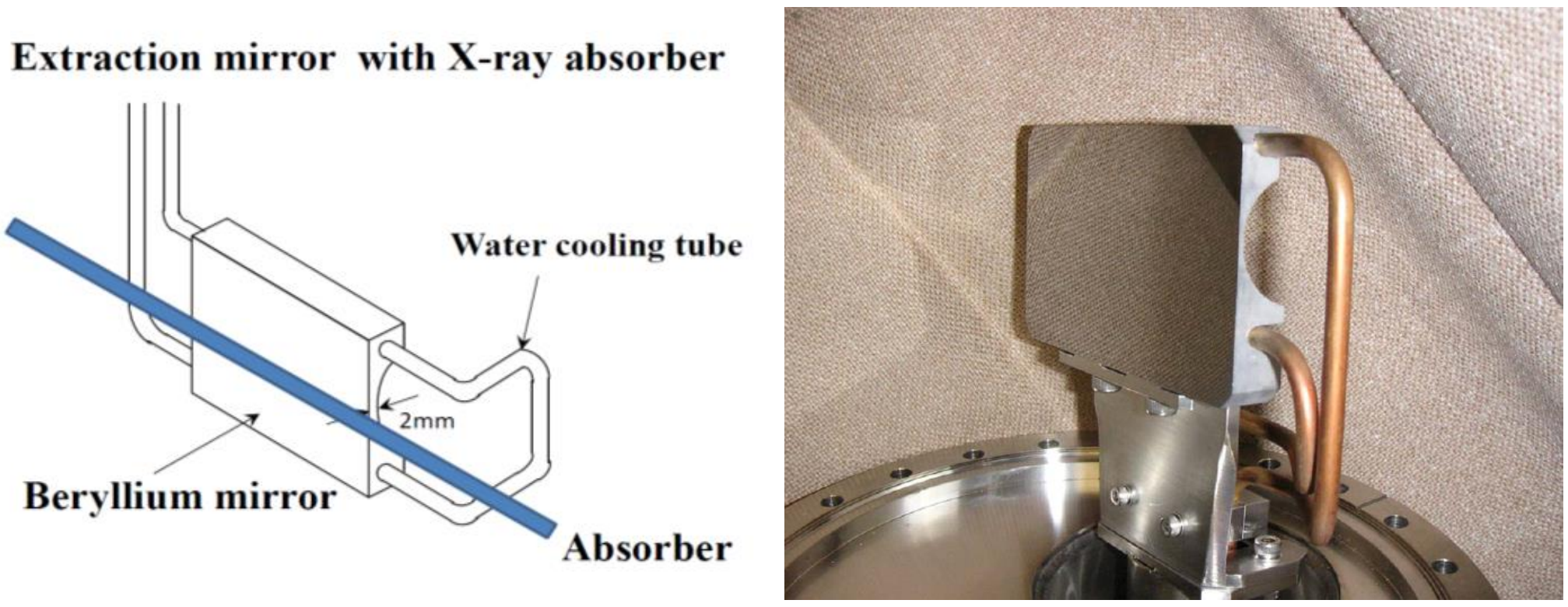

**Figure 4.3.7.11:** Design of the extracted Be mirror at KEK

#### *4.3.7.5.2 X-ray Beam Line*

The resolution of the X-ray pinhole imaging system can reach 10 μm. and is widely used for its simple setup and high reliability. The resolution of pinhole optics is a balance between the diffraction limit (hole too small) and geometric blurring (hole too large). The beam size is about 40 μm×1000 μm, so an X-ray pinhole is a good choice for real-time measurement. The schematic is shown in Fig.4.3.7.12.

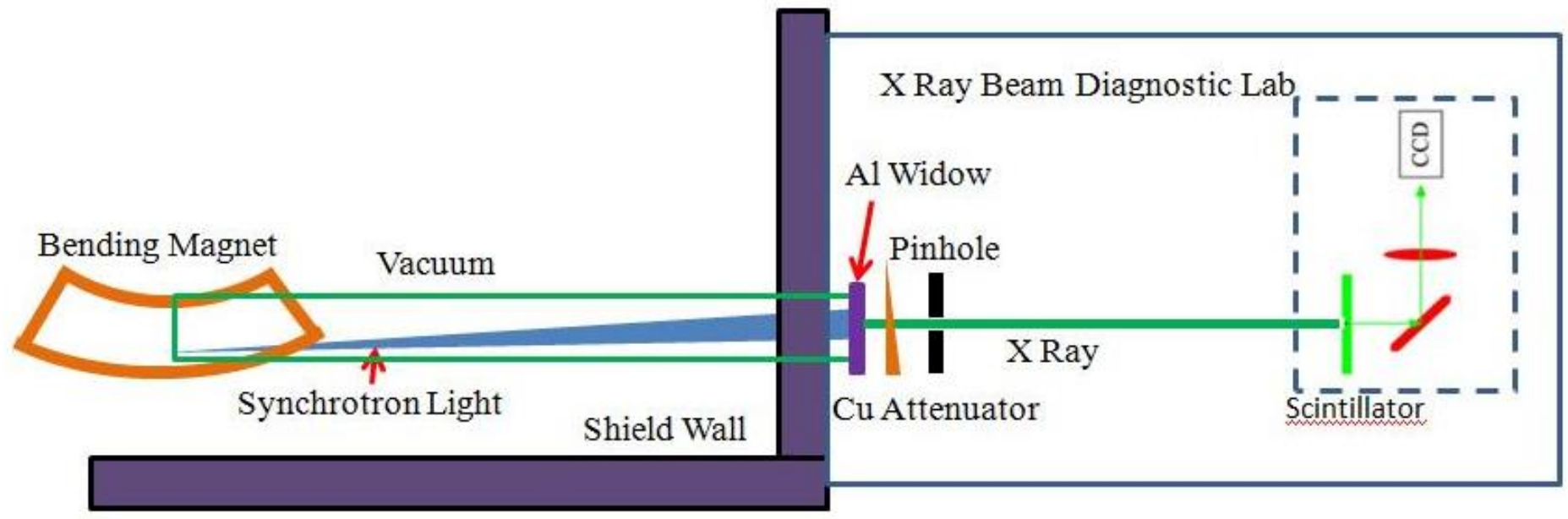

**Figure 4.3.7.12:** X ray beam line

X-rays from the bending magnet goes through an aluminum window which transmits only the high-energy photons from vacuum to air. A pinhole array combined with horizontal and vertical tungsten slits is placed after the window, as close as possible to the source. To obtain the two-dimensional beam profile, a scintillator screen based X-ray camera is placed at the end of the beam line.

*4.3.7.5.3 Bunch Length Measurement*

Synchrotron radiation can also provide bunch length measurements using a steak camera. Hamamatsu and Optronis both manufacture precision streak cameras with resolution that can reach sub-picoseconds.

On the other hand, a two photons interferometer can be used to measure the very short beam length. Fig.4.3.7.13 shows the calculation and the experiment result at KEK PF. [10]

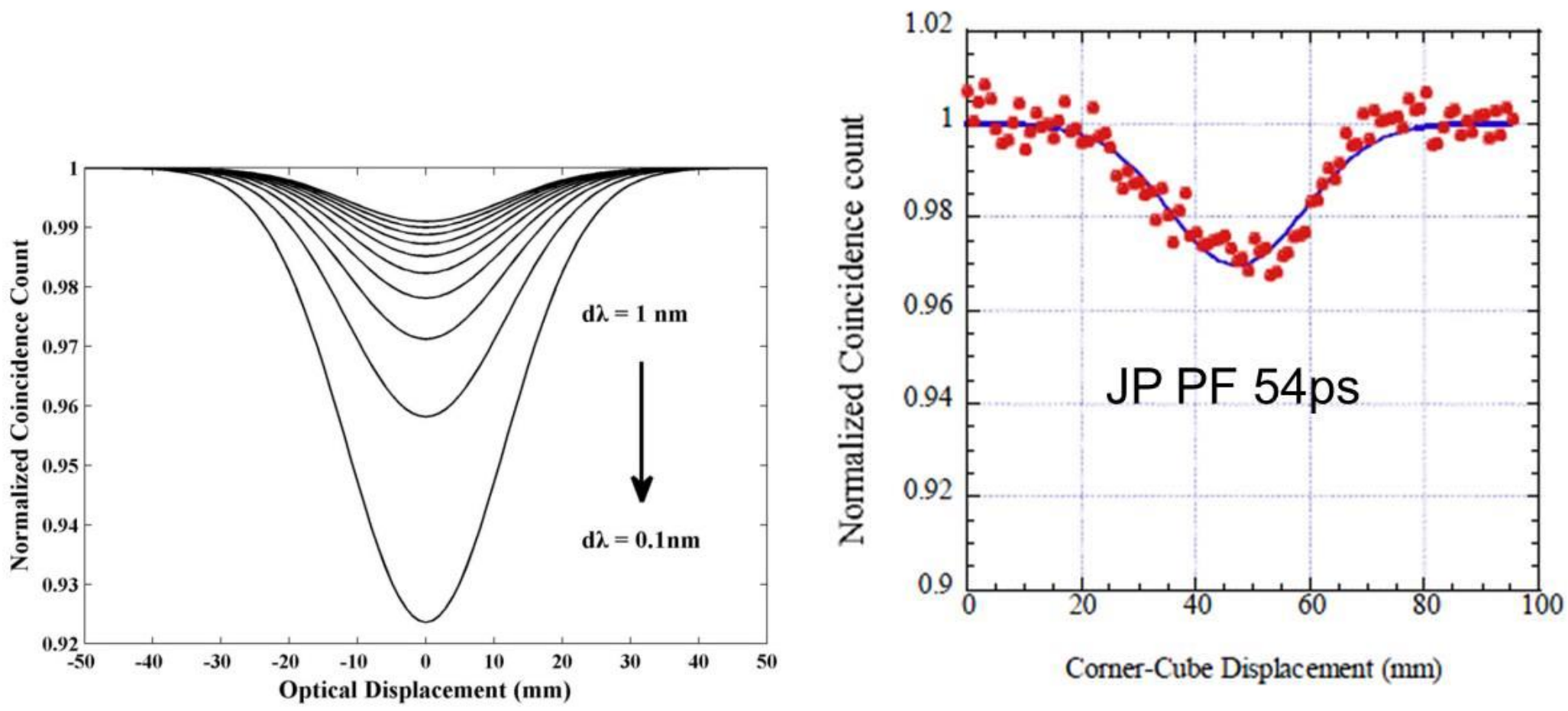

**Figure 4.3.7.13:** The calculation and experiment result

### 4.3.7.6 ***Beam Loss Measurement***

Radiation losses can be monitored with Beam Loss Monitors (BLM). These are commercially available radiation detectors, mounted on the outside of the vacuum chamber and indicate whether, when and where the beam is lost.

The following factors are considered for selecting the right type of BLM: intrinsic sensitivity, dynamic range, radiation hardness, response time and sensitivity to synchrotron radiation (SR). Table 4.3.7.4 summarizes the performance parameters of the four type of BLM.

**Table 4.3.7.4:** Parameters of four BLM types

| Type of BLM | Dynamic range | Response time | Sensitivity (for MIPs) | Radiation resistance | Sensitivity to SR |
|---|---|---|---|---|---|
| Ionization chamber | $10^8$ | 89μs | 600($Elec_{gain}$) (1L) | >100Mrad | Sensitive |
| PIN-photodiode | $10^8$ | 5ns[11] | 100($Elec_{gain}$) (1$cm^2$) | >100Mrad | Insensitive |
| Cherenkov counters | $10^5$~$10^6$ | 10ns | 270 ($PMT_{gain}$)(1L) | 100Mrad | Insensitive |
| Scintillators+PMT | $10^6$ | 20ns | ≈$18\cdot10^3$ ($PMT_{Gain}$) | ≈20Mrad | Sensitive |

An ionization chamber in its simplest form consists of two parallel metallic electrodes separated by a gap. The gap is filled with compressed air, argon or helium to improve the linearity and dynamic range. Its advantages are wide dynamic range and high irradiation capability. At the same time, they have the disadvantages of being slower and being sensitive to synchrotron radiation [12].

The scintillators+PMT are very fast and bunch to bunch measurements can be achieved. This detector is expensive and also sensitive to synchrotron radiation.

Cherenkov based fibers are much more radiation hard but much less sensitive to beam loss than scintillators. However, with the additional gain of a PMT their sensitivity exceeds the ionization chamber. Cerenkov light is instantaneous, unlike scintillators, and the threshold for light output is several hundred keV, making Cerenkov detectors insensitive to the background radiation from synchrotron radiation. For example, electrons below about 150 keV will not produce any light, while 1 GeV protons or 0.5 MeV electrons produce about 169 photons/cm. [13]

The PIN-photodiode is very fast, not very expensive, and the radiation resistance is rather good. They have a large dynamic range and a high sensitivity but they exist only in small sizes. In Fig.4.3.7.14, the PIN-photodiodes detector consists of two PIN-photodiodes mounted face-to-face. In contrast to a high energy charged particle which produces signals in both diodes, a photon interacts in one diode only. Although the PIN-photodiode is relatively insensitive to synchrotron radiation background, in the electron beam energy which is more than 45 GeV, the synchrotron radiation photon undergoes a photo effect or a Compton Effect; the emitted electron may reach the second diode, resulting in coincident signals [14]. However, this problem got resolved elegantly at both HERA and LEP. [15, 16] As shown in the Fig.4.3.7.14, a thin copper (or lead) layer between the two diodes can reduce the probability for the emitted electron to reach the second diode. In this way the background counts due to synchrotron radiation can be reduced. In LEP, the copper layer further reduces the background rate by a factor of 10[15]. The optimal thickness of the layer can be calculated from the range of electrons in matter. The penetration depth R in which 90~95% of the incident electrons are stopped is given by (for Al), [16]

$$\mathrm{R(Al)} = \mathrm{A}\cdot\mathrm{E}\cdot[1-\frac{B}{(1+C\cdot E)}]mg/cm^{-2}$$

where A=0.55 • $10^{-3}$ $gcm^{-2}$ $keV^{-1}$, B=0.984, C=3 • $10^{-3}$ $keV^{-1}$, E = energy of the electron. For energies above 100 keV and for materials with higher Z (e.g. copper) the range is approximately:

$$0.6 \leqslant R/R(Al) \leqslant 1$$

The synchrotron radiation photons can generate electrons by photoelectron effect. The range of the emitted electrons in copper is calculated to be ($\rho_{Cu}$ =8.96 g/$cm^3$)

$$R\ (Cu) \approx\ 0.114\ mm$$

Therefore, a thin layer of 120 μm copper between the diodes is sufficient to stop most of the Compton- and photoelectrons. This layer will not have an influence on the MIPs produced by beam losses. So, the detection efficiency of the BLM from beam losses is unchanged.

A thin layer of a high Z material like lead or copper between the two photodiodes inside the BLM leads to a decrease of background counts due to SR. but a fraction of the coincidence rate at high dose rates comes from multiple photons which interact at the same time in the two diodes. To reduce these coincidence rates, an additional lead shield will be needed around the BLMs. [15,16]

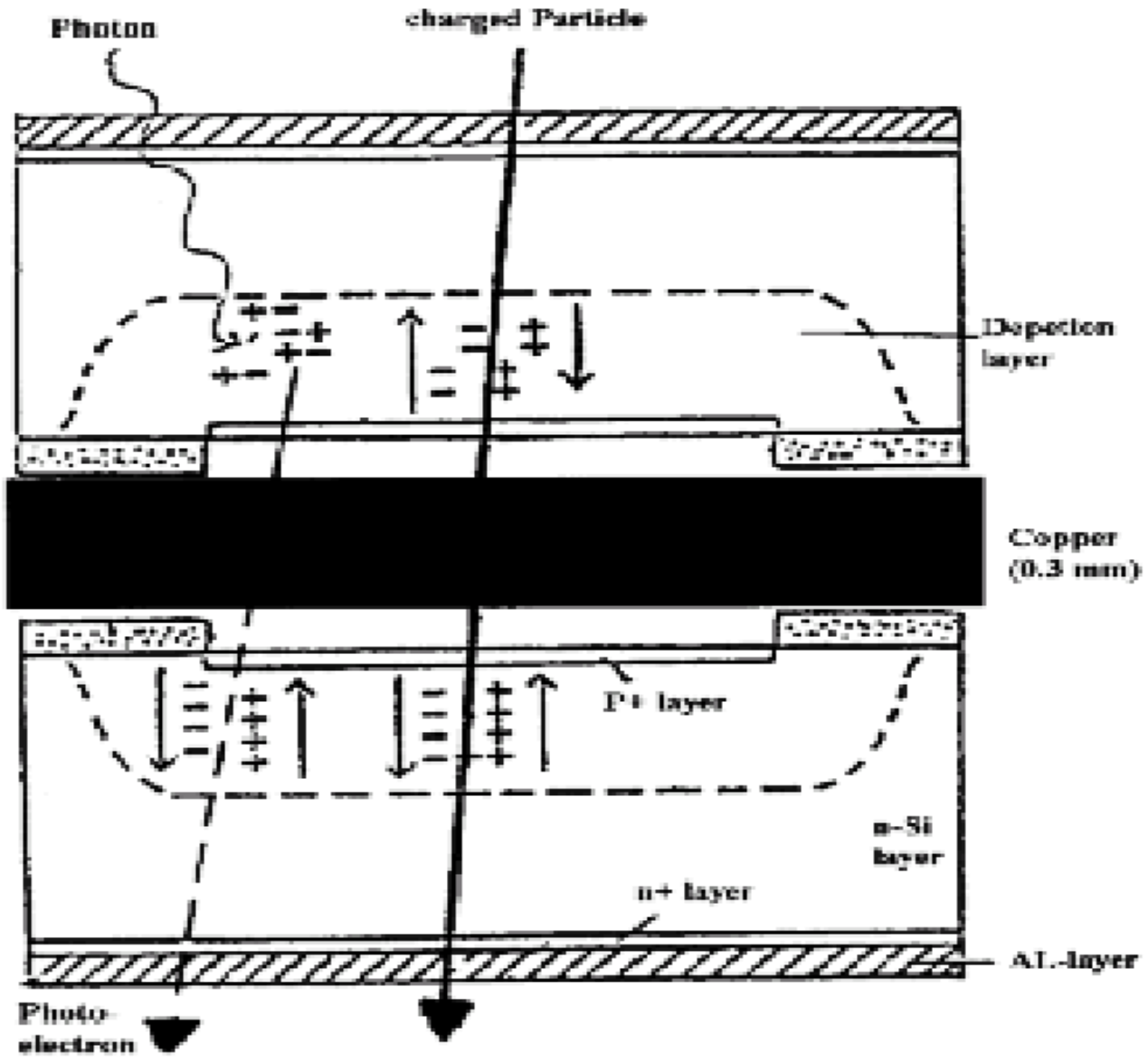

**Figure 4.3.7.14:** Two PIN-photodiodes with a copper layer between them

According to these comparisons between the four types of BLMs and experience with the electron BLM system in HERA and LEP, the PIN-Photodiodes detector is chosen. The detectors will be placed around the machine, at locations where the betatron amplitude functions reach a maximum i.e. in the arcs near each quadrupole. The efficiency of a BLM will be highest if it is located at the maximum of the shower. Monte Carlo simulations are needed to find the exact optimum locations for the monitors, as well as to calibrate the BLMs in terms of lost particles/signal.

The structure of the Beam Loss Monitor System is shown in the Fig. 4.3.7.15. The pulse signals from detectors are fed to electronics. The network card and Ethernet are used to connect all electronics to the PC. The beam loss distribution will be seen in the control room. Every vacuum chamber near a quadrupole magnet needs 2 BLMs, for a total of 10800.

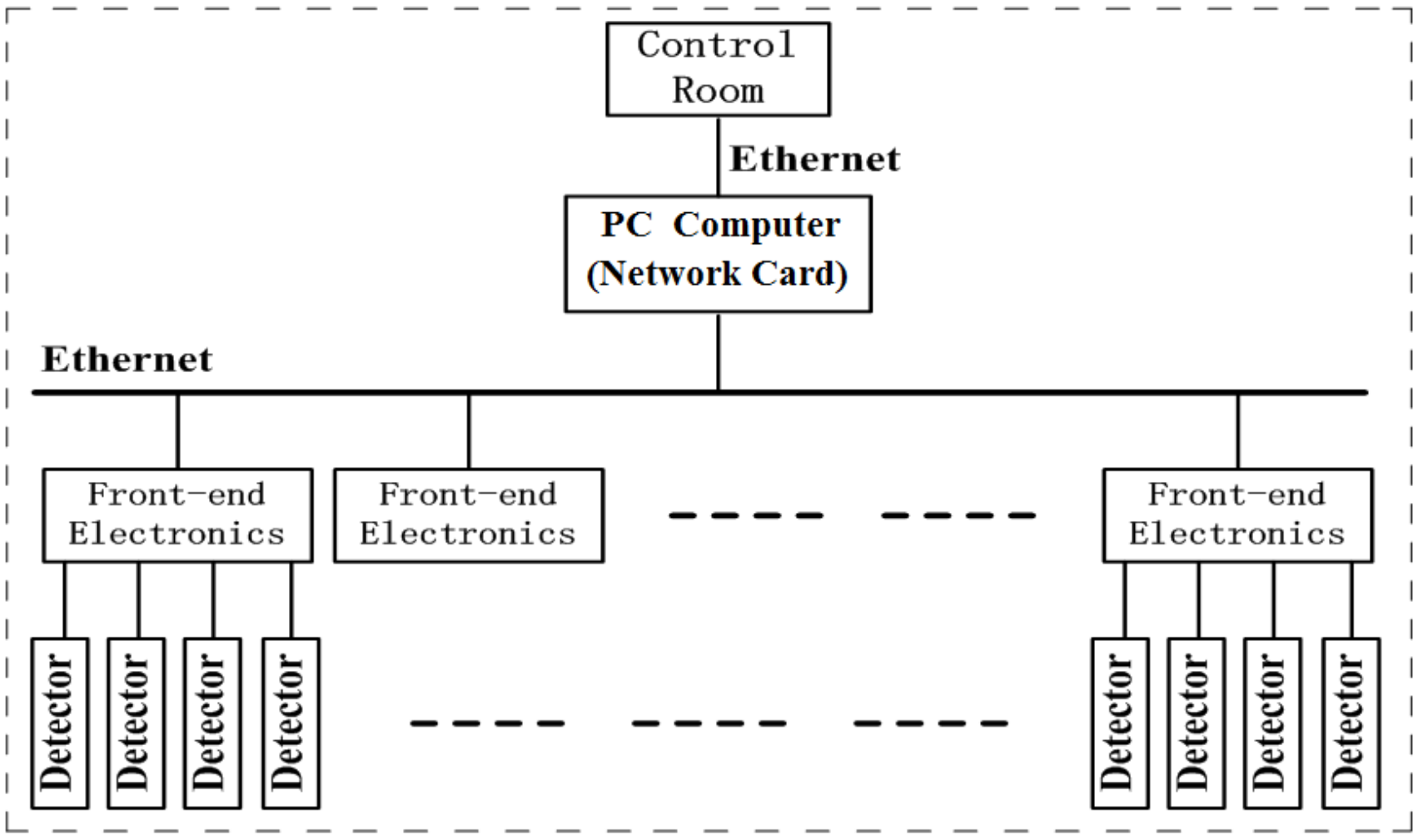

**Figure 4.3.7.15:** The structure of the Beam Loss Monitor System

### 4.3.7.7 *Tune Measurement*

The system to measure betatron tunes is a must. For example, betatron tunes may differ for positrons and electrons because of the large orbit separation resulting from the momentum saw tooth. Individual bunch measurement is useful to identify intensity related tune shifts. Furthermore, beam-beam effects result in various coherent oscillation modes.

The system foreseen will consist of a magnetic shaker and an electrostatic pick-up for each plane. The H and V systems will both be installed in a straight section, at a sufficient distance from the IP so that gating on individual bunches is possible.

Apart from the traditional swept-frequency excitation for the tune measurement, newer technology like direct diode detection will be considered. Direct Diode Detection (3D) is a technique developed at CERN initially for the LHC tune measurement system [17] and recently for observing beam motion of very small amplitude. [18] The basic idea is to time stretch the beam pulse from the pick-up in order to increase the betatron frequency content in the baseband. This can be accomplished by a simple diode detector followed by an RC low pass filter. [19]

This 3D method has many advantages: simplicity and low cost, revolution frequency suppression, robustness against saturation, flattening out the beam dynamic range, independent to filling pattern. At the same time, it also has a disadvantage: operation in the low frequency range, response is dominated by the largest bunches; cannot measure bunch by bunch tune. [19]

### 4.3.7.8 *Vacuum Chamber Displacement Measurement*

Due to heat effects caused by synchrotron radiation and beam loss, the vacuum chamber will be displaced. Then it can cause a decrease in BPM resolution. So in order to calibrate the BPMs, the displacement need to be measured.

The entire system includes Linear Variable Differential Transformer (LVDT), signal processing unit, computer and network and can be mounted near the BPM; the sense signal is sent by the 600 m cable to the local station.

#### 4.3.7.9 *Feedback System*

There are 27083 bunches in the storage ring; the bunch current is 298.5 mA, and due to HOMs and resistive wall instability, multi-bunch instability may occur. Because the longitudinal fundamental mode instability of the cavity is very fast, synchrotron damping time is far slower than rise time of the instability, In order to cure these instabilities, feedback systems in all three dimensions are necessary. A short filter means less delay and fast damping, so we can design a good filter to apply the kick signal to the beam in several turns. Or we can use a multi-pick to process the oscillation signal [20], or use multi-feedback systems in one ring. [21]

The feedback system must sense bunch motion and deliver either deflection or acceleration independently to each bunch in order to damp all of the possible dipole multi-bunch instabilities.

In the last 10-15 years, digital bunch-by-bunch feedback systems with single pickup and single kicker topology have become popular.

The transverse feedback system consists of front end electronics, digital electronics, amplifier and a kicker. Front end electronics convert the BPM oscillation signal and process it digitally, where 90 degrees phase shift, closed orbit component is removed and the single turn delay has been done. The analog signal will be sent to the amplifier and kicker to give the beam an angular kick; a power amplifier drives a 50 Ω stripline kicker shorted at one end. [20,21] The plates of the kicker are powered differentially using a hybrid power divider driven by the combined amplifier output.

Longitudinal feedback is more difficult than transverse feedback; there is back end electronics for a longitudinal feedback system; digital signal processing is required to convert to carrier frequency. For the longitudinal feedback system, a pillbox cavity kicker is necessary.

#### 4.3.7.10 *Other Systems*

Beam polarization measurement system and energy and an energy spread measurement system are additional instrumentation systems for CEPC.

Transvers beam polarization could be measured by a laser Compton polarimeter which is based on spin-dependent Compton scattering of circularly polarized photons from polarized electrons and positrons.

## 4.3.8 **Control System**

### 4.3.8.1 ***Control System Overview***

The control system covers the entire100 km ring, housing the Collider and Booster rings as well as a Linac injection system about 1.2 km long.

To build so large a control system, the more commercial/industrial products and techniques that are used, the better quality the whole project will be. Using commercial products will also benefit maintenance and upgrades. Distribution of large volume control messages and collection of an even larger volume of monitoring data, together with the different level system alarms and data archiving from such a large area, present many design challenges.

For the whole system, the time relationship among the widely distributed devices related to beam source, injection, accumulation, acceleration, extraction, diagnostics and post-mortem analysis must be defined and maintained. For the Booster, a synchronization accuracy of several micro-seconds among the hundreds of related power supplies is necessary.

With the evolution of electronic techniques, hardware prices decrease rapidly, at the same time with better performance. This argues for delaying purchase and final mass-production as late as possible. On the other hand, technical studies and interfaces between different systems should be made as early as possible to ease the system development, integration and commissioning. A full-scale prototype system should be set up first for development and function tests.

The CEPC control system consists of a global control system, including timing system, MPS, Network, and local control systems such as power supply control, vacuum control, RF control, Injection/Ejection control, and temperature monitoring. This chapter covers the global control system and local system (sub-system) for the Collider ring. The other sub-system controls will be descripted in the Booster and Linac chapters.

The whole control system will be divided into 3 layers (Figure 4.3.8.1): presentation tier, middle ties and front end tier, based on Ethernet. Ethernet will continue to be the backbone of the control system. For those systems with high intensity real-time computing, FPGA plus MTCA/XTCA will be better for ease of scalability and possible reliability. For most of the slow speed applications, PLC (Programmable Logic Controller) should be used as much as possible to improve the overall reliability.

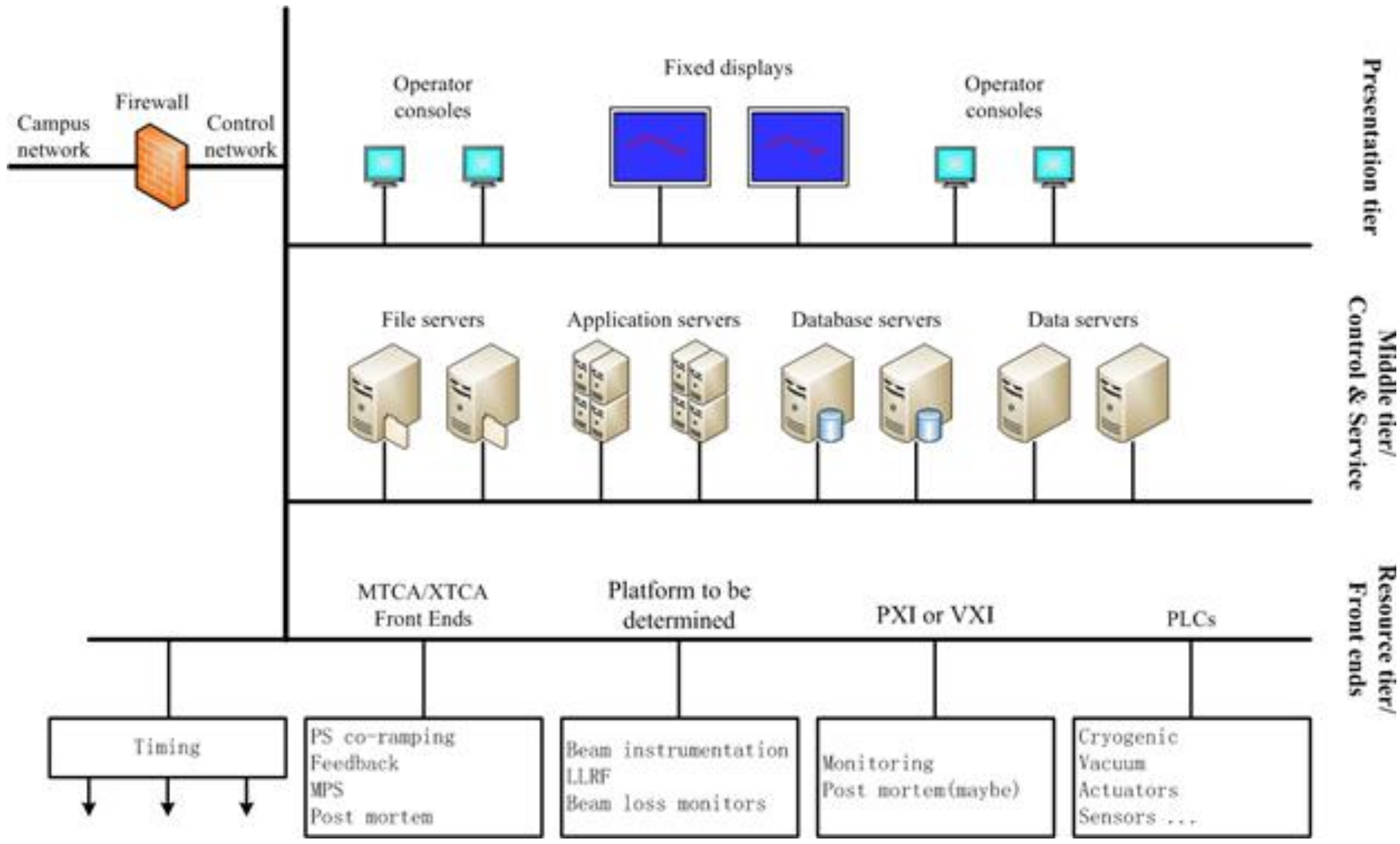

**Figure 4.3.8.1:** Control system architecture

### 4.3.8.2 *Control Software Platform*

EPICS (Experimental Physics and Industrial Control System) has been widely applied in large experimental facilities around the world. Originally developed by Los Alamos and Argonne National Laboratories, it has been continuously improved in the past years with many available application tools. Many device drivers have been implemented, which eases overall system integration. EPICS is chosen as our control software platform since its successful application in the accelerator control system of the BEPCII (Beijing Electron Positron Collider II) and CSNS (China Spallation Neutron Source).

EPICS is based on a client-server model with communications through Ethernet. Its main components are OPI (Operator Interface), IOC (Input Output Controller) and CA (Channel Access). OPI is the client side module for operators, IOC is the server side, and CA is the communication module. Fig. 4.8-2 shows the EPICS structure.

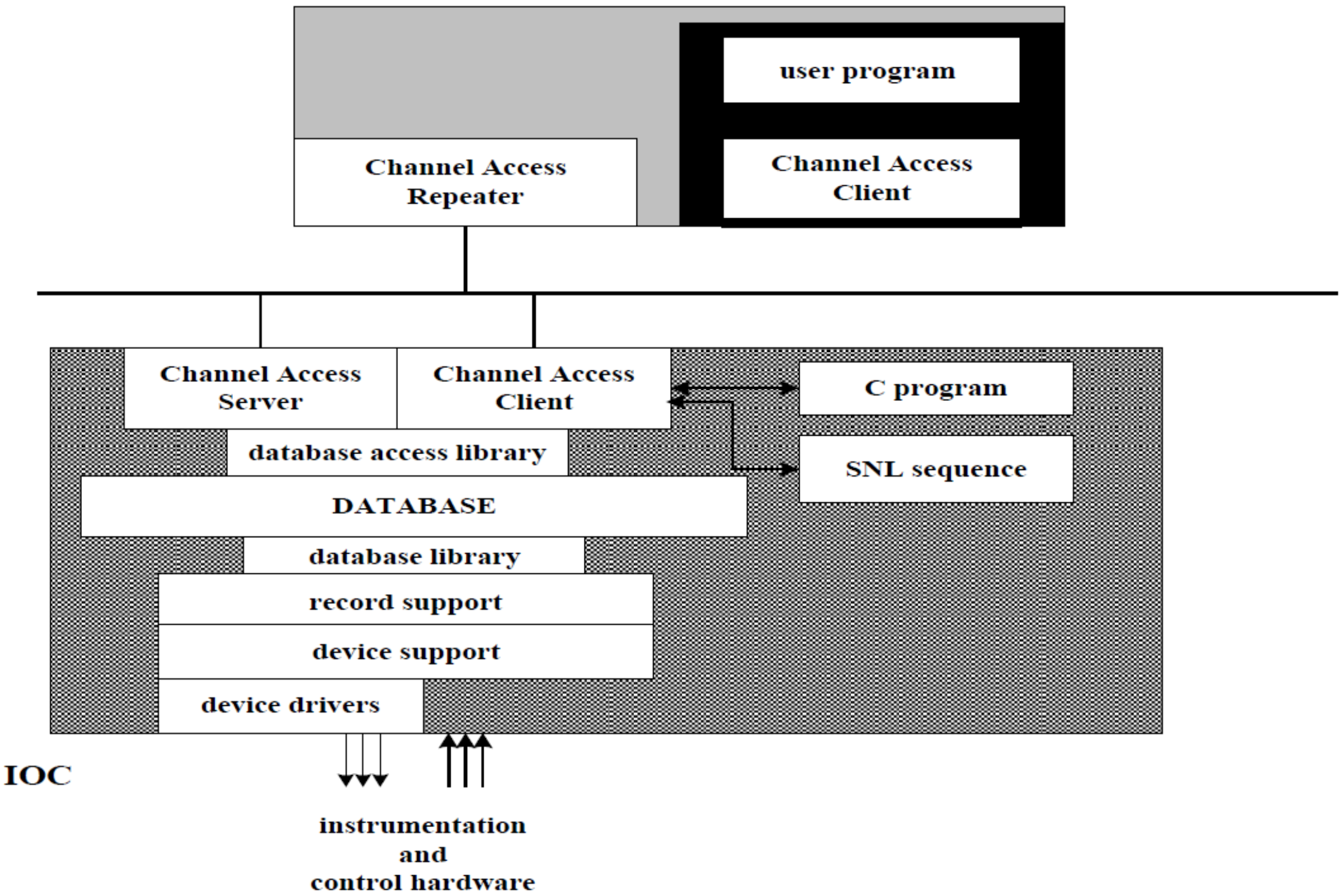

**Figure 4.3.8.2:** Structure of EPICS

### 4.3.8.3 *Global Control System*

#### *4.3.8.3.1 Computers and Servers*

Servers, workstations and PCs will be used as operator consoles at the presentation layer with a friendly graphical man-machine interface. The operator can monitor and control the accelerator equipment from the consoles. They can store and recall the machine parameters, integrate the device data into office products, and define a sequencer for machine operators. The consoles also provide tools to display the beam status and equipment alarms, to access all system data and plot real-time and historical data.

#### *4.3.8.3.2 Software Development Environment*

The software development environment should be defined initially to ease future compatibility during system integration and to ease maintenance. This includes mainly the operation systems, the development tools, the software upgrade strategy and the hardware platform. Standardization should be done first; version control is a must. Also, a global software and hardware platform should be provided for co-development.

Upgradability must be considered first during the setup of the environment and the software development. Careful study is needed and rules should be made as early as possible.

#### *4.3.8.3.3 Control Network*

The control system network will be a core redundant design with a 40 Gb/s or 100 GB/s interface to the different aggregation switches. Edge switches are connected to the aggregation switches to provide communication to the devices. Links of 10 Gb/s will be provided to the different devices by the aggregation switches. Links of 1 Gb/s and 10 Mb/s will be provided to the different devices by the edge switches.

Aggregation and edge switches will be provided at each local control station and the CCR (Central Control Room). A three-layer structure is preferred from cost-saving considerations.

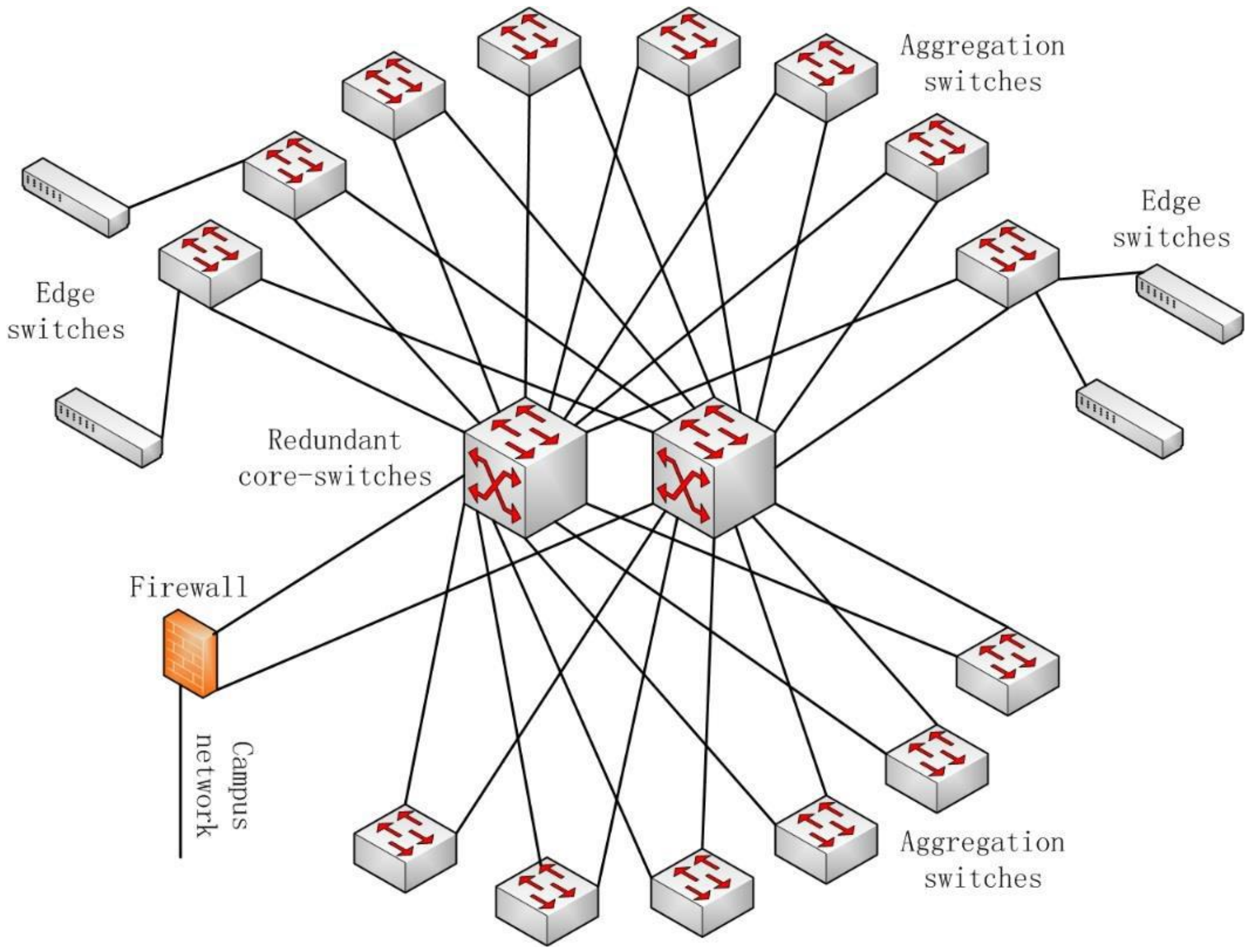

**Figure 4.3.8.3:** The accelerator control network

#### *4.3.8.3.4 Timing System*

Operation frequencies of the Linac, the Booster and the Collider are 2856 MHz, 1300 MHz and 650 MHz respectively.

The timing system synchronizes all the relevant components in the CEPC complex. This includes generating a trigger sequence to synchronize the electron gun, modulators, pulsed power supplies, injection/ejection kickers. It matches the delays between the gun triggering and the pulse of the injection kicker so that the bunch could be injected into the appropriate bucket. Meanwhile, the timing system also sends timing pulses as the reference time triggers to the related beam diagnostic devices for bucket selection and beam parameter measurement.

#### *4.3.8.3.5 Machine Protection System*

Energy stored in the Collider ring is about 690 kJ for both beams. Energy injected into the Collider from the Booster is about 34.5KJ per pulse train. To prevent equipment damage, beam in the Collider must be ejected into the beam dump and injection stopped whenever there are problems with main components or the dose rates due to the lost particles are too high.

Redundant controllers will be used to ease maintenance and to increase reliability. IO connections will not be redundant in order to save cost.

The response time of the MPS would be comparable to the 0.17 ms Collider revolution time, so a low level FPS (Fast Protection System) may be needed. Fig. 4.3.8-4 shows the architecture of one station of the MPS.

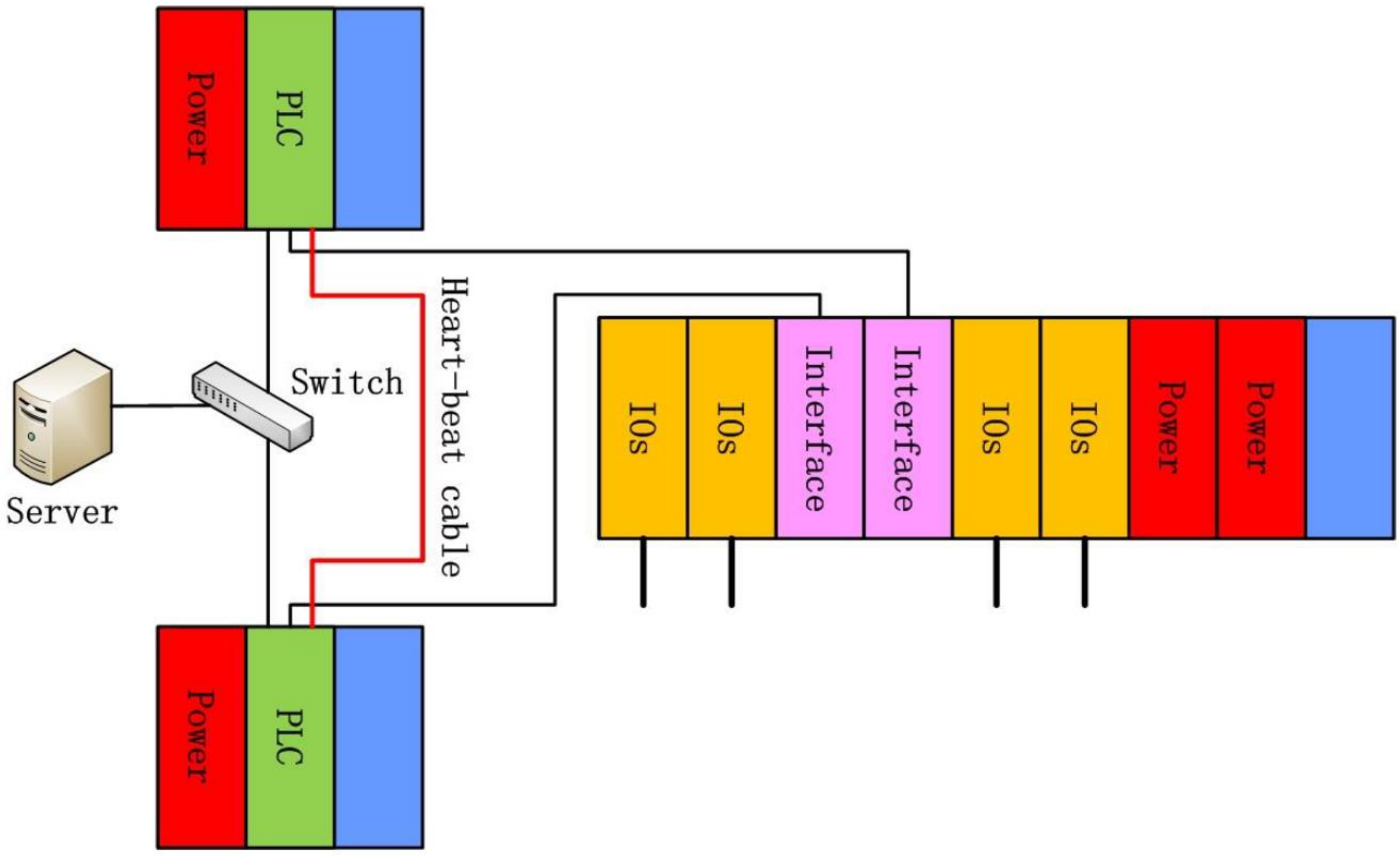

**Figure 4.3.8.4:** Structure of the MPS of one station

*4.3.8.3.6 Data Archiving and Retrieval*

The number of signals that need to be archived is in the tens of thousands. Several data archivers are needed with data recording and retrieval separated to improve overall efficiency.

An individual retrieval server is used for fast data retrieval. This server and the OPIs can also be deployed in the facility campus network to unburden the control network.

The overall system will be developed with JAVA and a relational database.

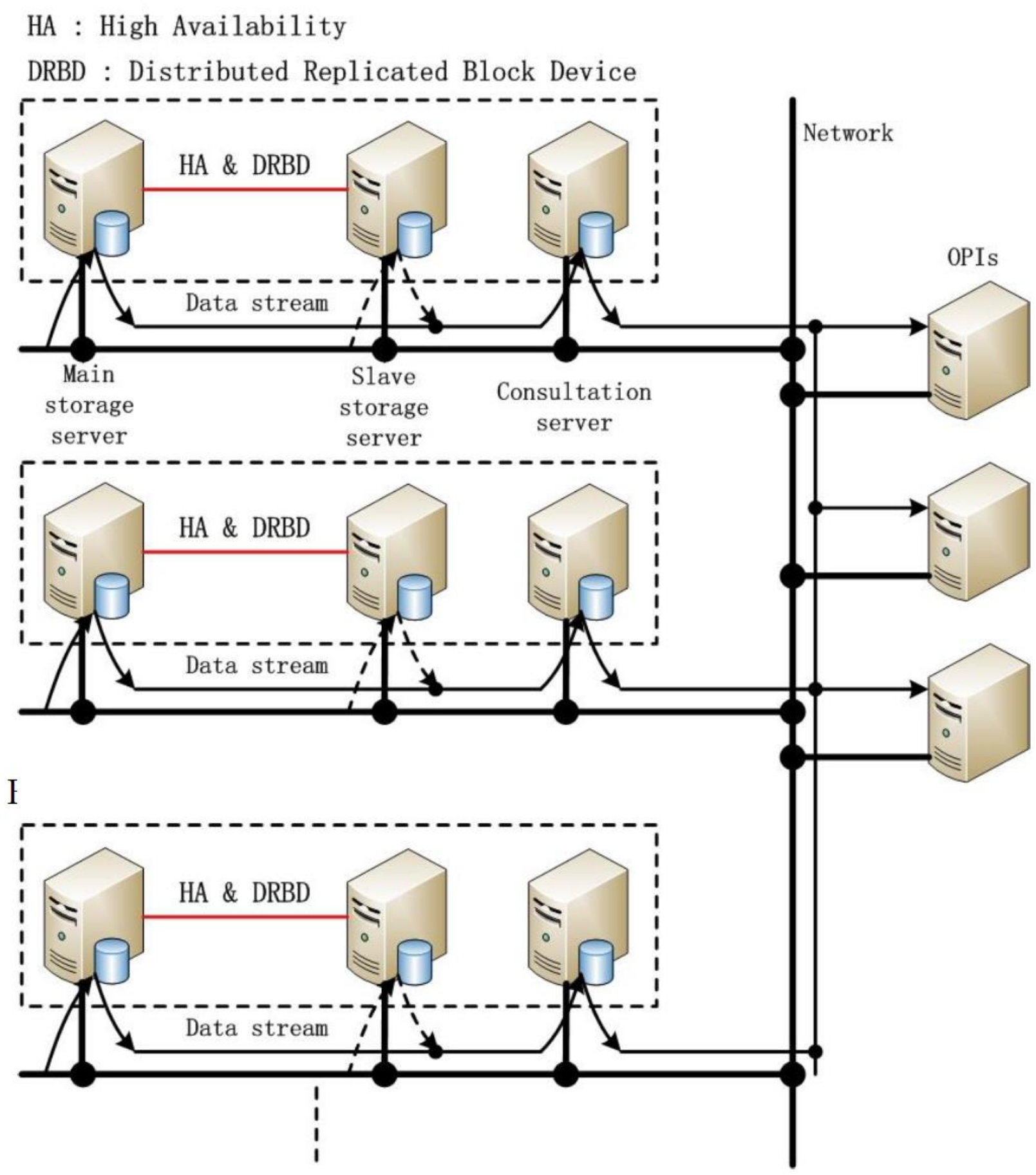

**Figure 4.3.8.5:** Hardware configuration and data flow of data archiving

*4.3.8.3.7 System Alarms*

Data on fault conditions need to be collected, stored and reported. Thousands of alarm signals are expected from the many different systems. To deal with the alarms properly and to provide as meaningful information as possible to the operators, the alarm system should be designed with several levels.

The fundamental level will be the equipment level. Data with an accurate timestamp will be recorded at this level and overall alarm signals provided to the network for server level storage and treatment. OPIs should be provided to the operators for them to deal with the abnormal situations. Fig. 4.3.8-6 shows the architecture of the alarm system.

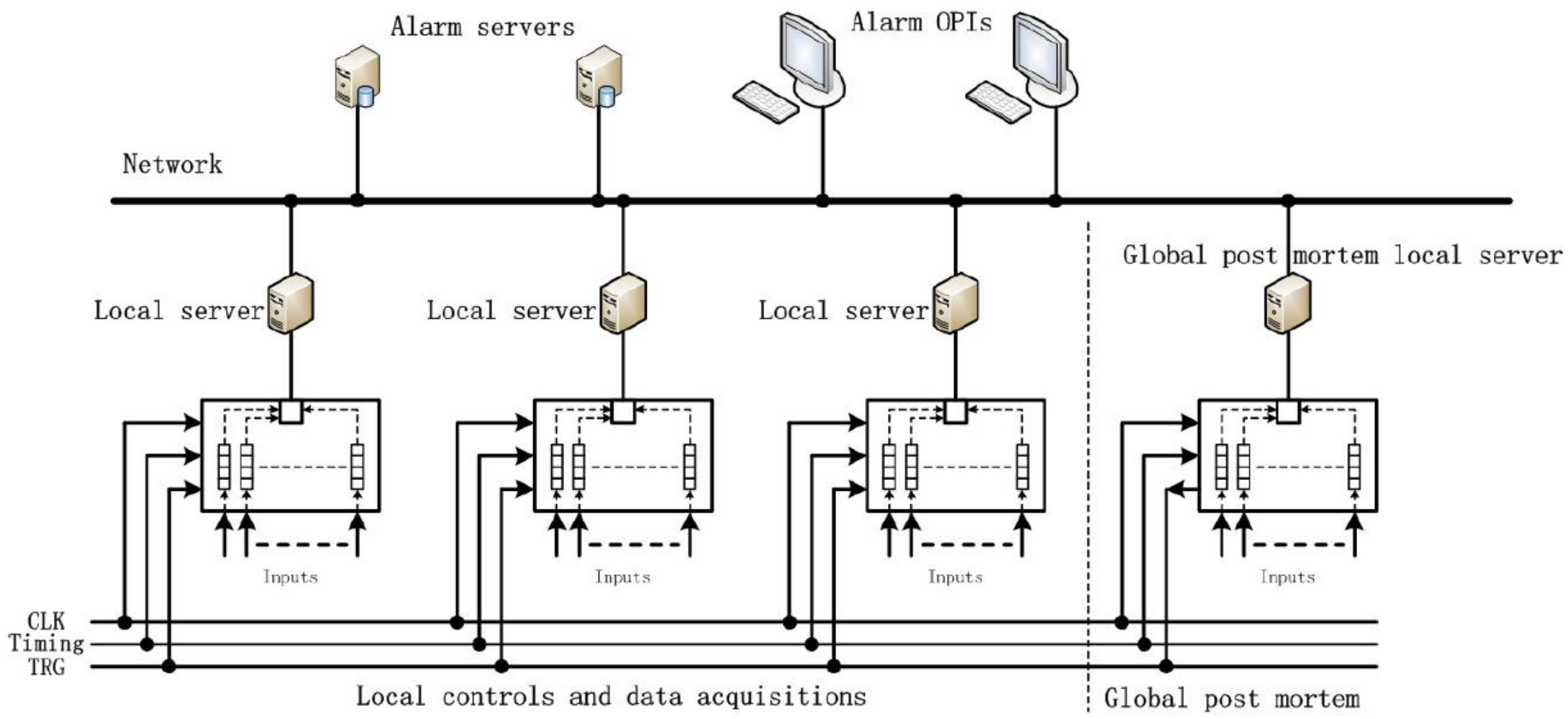

**Figure 4.3.8.6:** Alarm system architecture

### *4.3.8.3.8 Post Mortem, Large Data Storage and Analysis*

Every control or data acquisition system should record a period of data with accurate timestamp whenever there is a problem. The required time accuracy time window are defined differently for each systems. Some small systems can provide analog and/or digital signals directly to the global post mortem system. The global post mortem system then will record the data with the required time accuracy and window.

Fig. 4.3.8.7 shows the structure of the post mortem system. Local controls and data acquisition are the same as in the alarm system. The triggered data in the local devices will be transmitted into the post mortem servers whenever there is a problem. The software analysis interface should be defined and provided by the post mortem servers.

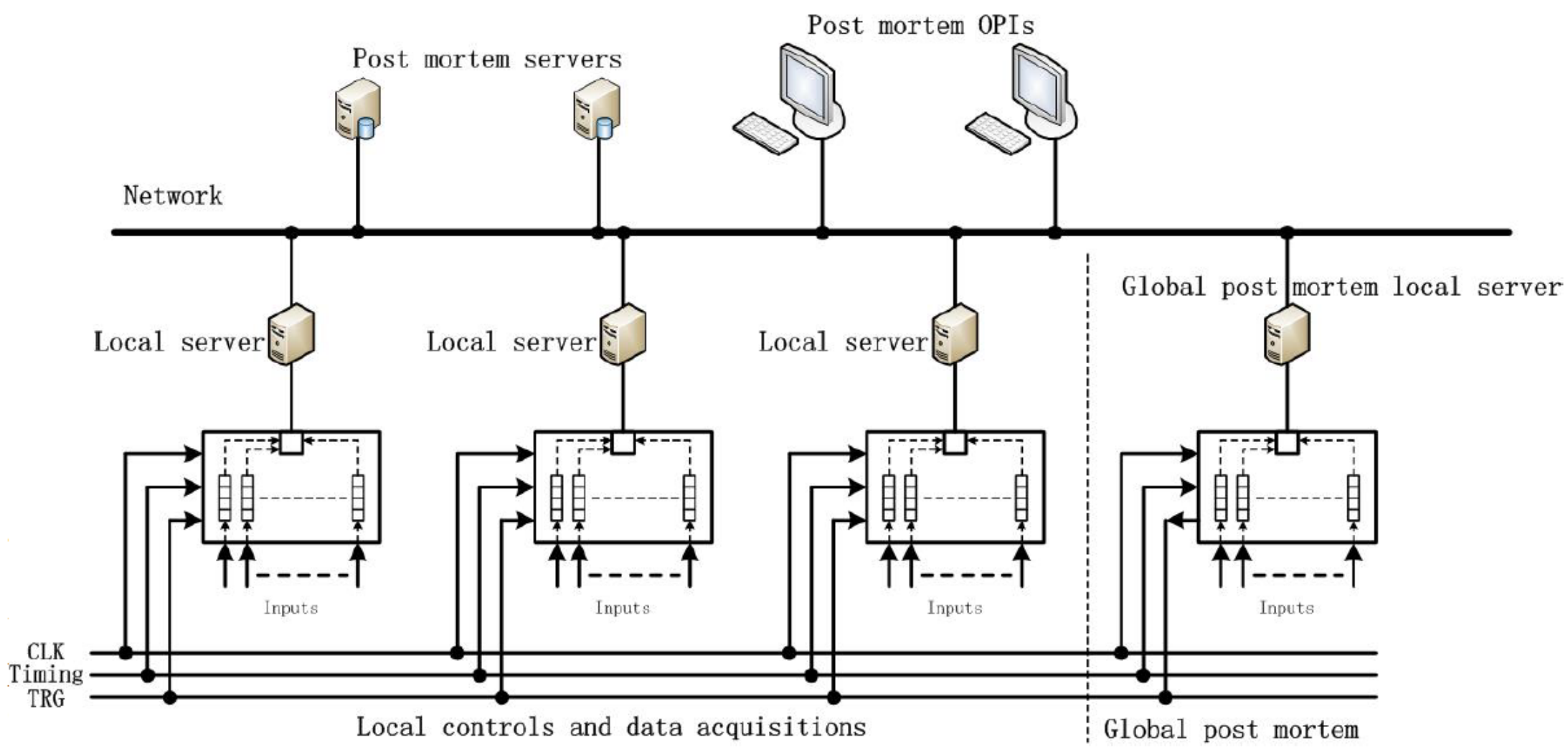

**Figure 4.3.8.7:** Post mortem system architecture

### *4.3.8.3.9 Event Log*

An automatic way to store status display a running report should be developed. This helps the operators and the experts to learn the health of the whole complex and to find minor problems before they can become serious. A tree structure distinguishes the information from different systems and different levels. Also, logs for comments should be provided.

### 4.3.8.4 ***Front-end Devices Control***

The front-end control system of the Collider ring includes power supply control, vacuum control, temperature monitoring, RF control, cryogenic control.

MTCA crates will probably be used for the LLRF (Low Level Radio Frequency) control for the high bandwidth real-time data exchange between different RF cavities.

Self-designed modules will be used for the approximate 4000 power supplies and co-ramping functions will be designed for all the power supply control modules.

PLCs will be used for the control of the cryogenic system, the vacuum system, the movable collimators and the vacuum chamber temperature monitoring system.

Beam instrumentation devices will be integrated into the whole system via Ethernet directly.

Commercial PXI crates and modules or self-designed modules will be used for post-mortem data acquisition and analysis.

Industrial computers will be used as IOCs for some of the device level controls.

#### *4.3.8.4.1 Power Supply Control*

There are a total of about 4000 power supplies of which more than 3000 are for correctors. Power supplies in the Booster need to be co-ramped for acceleration, while the others just need to be powered to the correct currents. The power supply remote control is designed to fulfil both requirements and to ease maintenance with only the differences being in the physical connection and software configuration.

Since the CEPC complex is so large it is not easy to repair or replace a broken device within a short time, so redundant design will be used wherever possible. Two redundant controllers will be installed in a self-designed control crate. Each interface card has two completely isolated connectors. All the connections will be done through a self-designed passive backplane. Two isolated crate power supplies will provide power to the two control routes respectively. Global redundancy will be done through the network packets exchange or heart-beat cable connections between the two controllers. Fig. 4.8.3-8 shows the redundant connections of the power supply remote control. Fig.4.8.3-9 shows the preliminary remote control crate arrangement.

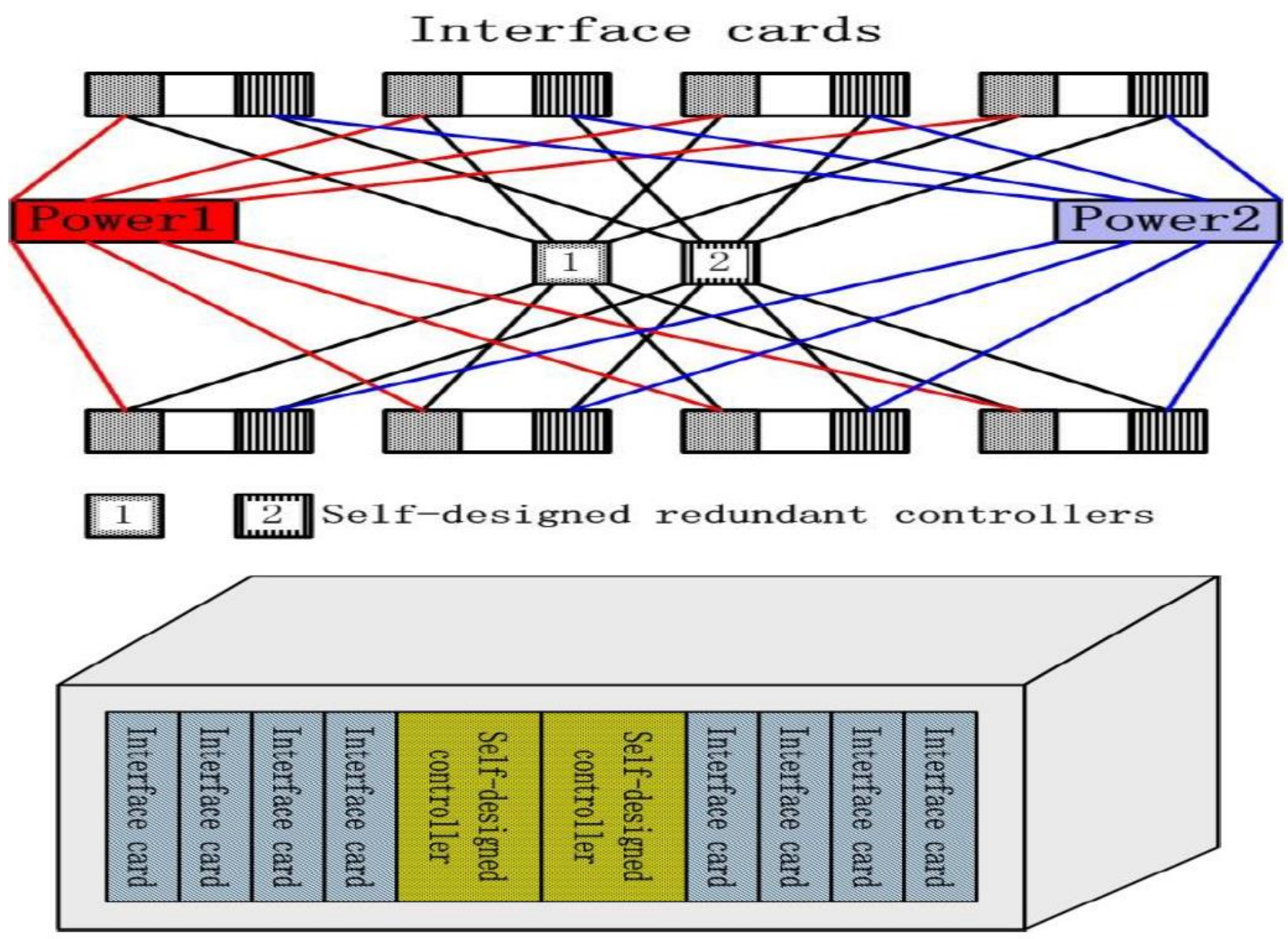

**Figure 4.3.8.9:** Preliminary power supply remote control crate arrangement

#### *4.3.8.4.2 Vacuum Control*

Normally, interfaces to the vacuum devices are relay contacts and RS232/485 ports, where, relay contacts are for on/off control, protections and monitoring. Similar to the design of MPS, redundant PLCs will be used for the relay contacts with a similar structure as shown in Fig.4.3.8.4. Integration of specific RS232/485 devices will be determined later.

#### *4.3.8.4.3 Vacuum Chamber Temperature Monitoring*

Heat on the vacuum chambers from synchrotron photons can increase greatly due to miss steering of the beam. Too much heat will raise the temperature, worsen the vacuum and possibly damage the chambers. Temperature monitoring of the vacuum chambers is essential, especially in the Collider arcs. Two temperature sensors are planned for each dipole joint. The total is about 8000 sensors in the Collider.

The relationship between the vacuum chamber temperature monitoring system and the MPS need to be made clear later, since there are so many sensors. Sensors in critical areas can be monitored separately and used in protection systems.

#### *4.3.8.4.4 Integration of Other Subsystems Control*

The Collider ring RF system consists of 336 superconducting accelerating cavities, high power sources to drive them and the low-level control system. In general, a short bunch length requires a high RF voltage and high RF frequency.

The interlock system switches off the cavity tuning in the event of a fault or unsafe condition. Faults might occur in the cooling water system, the vacuum and temperature of a cavity, or in the cryogenic system. The local interlock system sends a warning message and failure signal to the MPS when a fault has been detected

## 4.3.9 **Mechanical Systems**

### 4.3.9.1 *Introduction*

Most of the dipole and quadrupole magnets in the Collider are twin-aperture magnets, while the sextupole magnets and correctors are single aperture. The two sextupoles at the same location are supported together. The quantities of the magnets and their supports are listed in Table 4.3.9.1. The supports for the superconducting quadrupoles are more complex. But as their quantity is less, the details will not be described here. Also not described are the numerous supports for the vacuum system and instrumentation.

**Table 4.3.9.1:** Quantities of magnets and their supports in the Collider

| Magnet type | Quantity | Magnet length (mm) | Core number per magnet | No. of supports per core |
|---|---|---|---|---|
| Dipole | 2384 | 28686 (twin-aperture) | 5 | 4 |
| | 162 | 9667~93378 (single-aperture) | 2~17 | 4 |
| Quadrupole | 2384 | 2000 (twin-aperture) | 1 | 2 |
| | 8 | 1000 (twin-aperture) | 1 | 1 |
| | 1132 | 500~3500 (single-aperture) | 1 | 1~3 |
| | 8 | 1480/2000 (superconducting) | 1 | 1 |
| Sextupole | 996 | 700/1400 | 1 | 0.5 |
| | 72 | 300/1000 | 1 | 1 |
| | 32 | 300 (superconducting) | 1 | 1 |
| Corrector | 5808 | 875 | 1 | 1 |

For the dipoles, five magnet cores of length of 5,670 mm each are connected to become a magnet unit of total length 28618 mm. The number and location of support points minimizes deformation. The goal of the magnet supports are to achieve a simple and flexible structure, minimize magnet deformation, have good stability and low cost.

### 4.3.9.2 *Magnet Support System*

#### *4.3.9.2.1 Requirements*

Each magnet support contains the pedestal, magnet mounting plate and adjusting mechanism. The pedestal is made of concrete poured during construction. The magnet mounting plate is between magnet and adjusting mechanism, reducing stress and magnet displacement by increasing the contact area. The adjusting mechanism has 6 DOFs (degrees of freedom).

Assume the +Z axis of the support system is along the girder, the +Y axis is up, and the coordinate system is a right-hand one. Design goals are the following:

- Range and accuracy of adjustment shown in Table 4.3.9.2.
- Stability with large time constants, avoiding creep and fatigue deformation.
- Simple and reliable mechanics for safe mounting and easy alignment.
- Good vibration performance.

**Table 4.3.9.2:** Adjustment range and accuracy of Collider magnet supports

| Direction | Range of adjustment | Direction | Range of adjustment |
|---|---|---|---|
| X | ≥±20 mm | Δθx | ≥±10 mrad |
| Y | ≥±30 mm | Δθy | ≥±10 mrad |
| Z | ≥±20 mm | Δθz | ≥±10 mrad |

#### *4.3.9.2.2 Structure of Dipole Magnet Support*

In the Collider there are 2,546 dipole magnets. 2,384 of them have the same length of 28,618 mm.

Five magnet cores of length 5,670 mm each are connected to become a magnet unit, shown in Figure 4.3.9.1. For each core, there are two main supports and two auxiliary supports, as shown in Figure 4.3.9.2. The main support is for support and adjustment (6 DOFs), while the auxiliary supports are only for support (1 DOF).

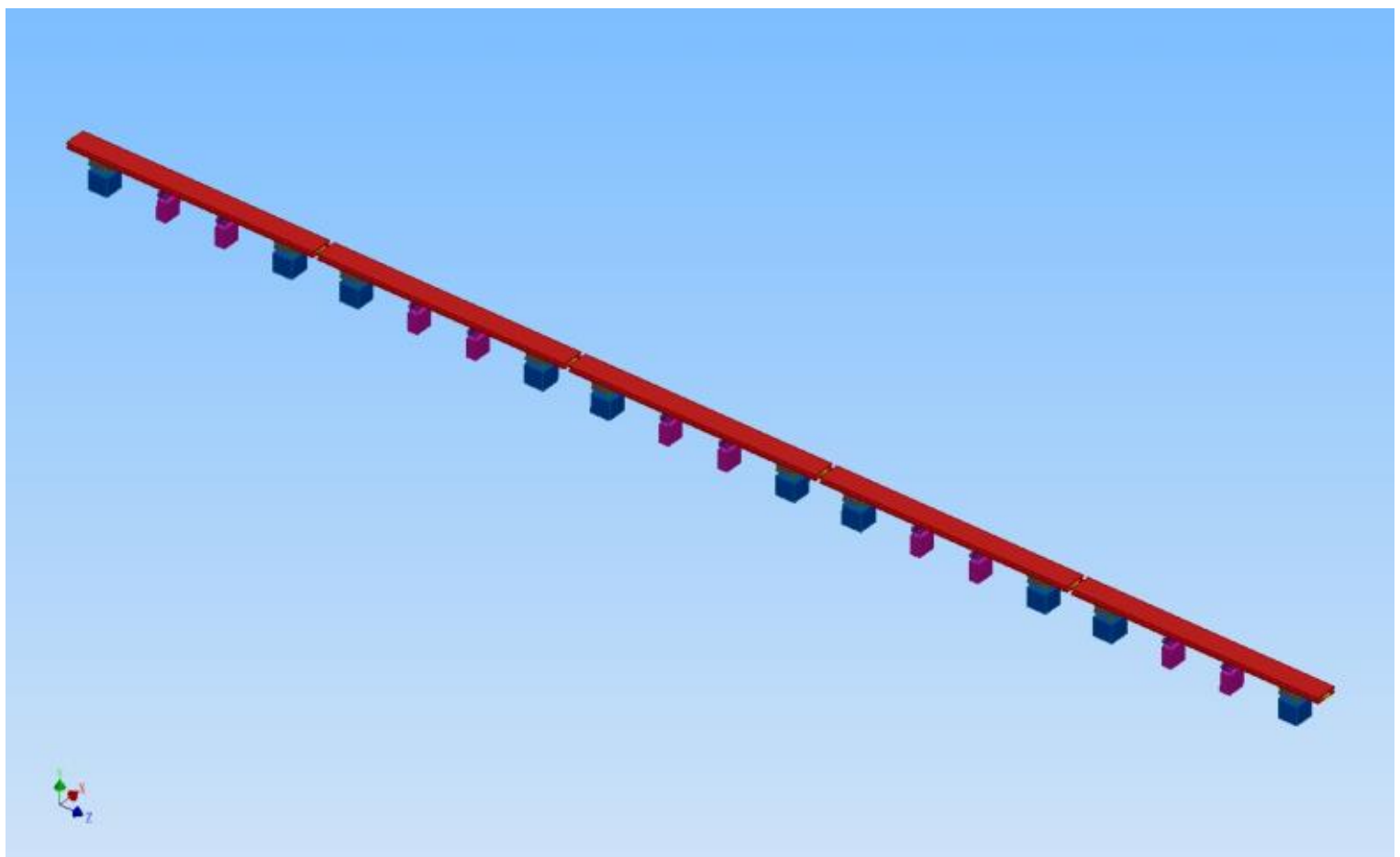

**Figure 4.3.9.1:** One dipole magnet unit and its support

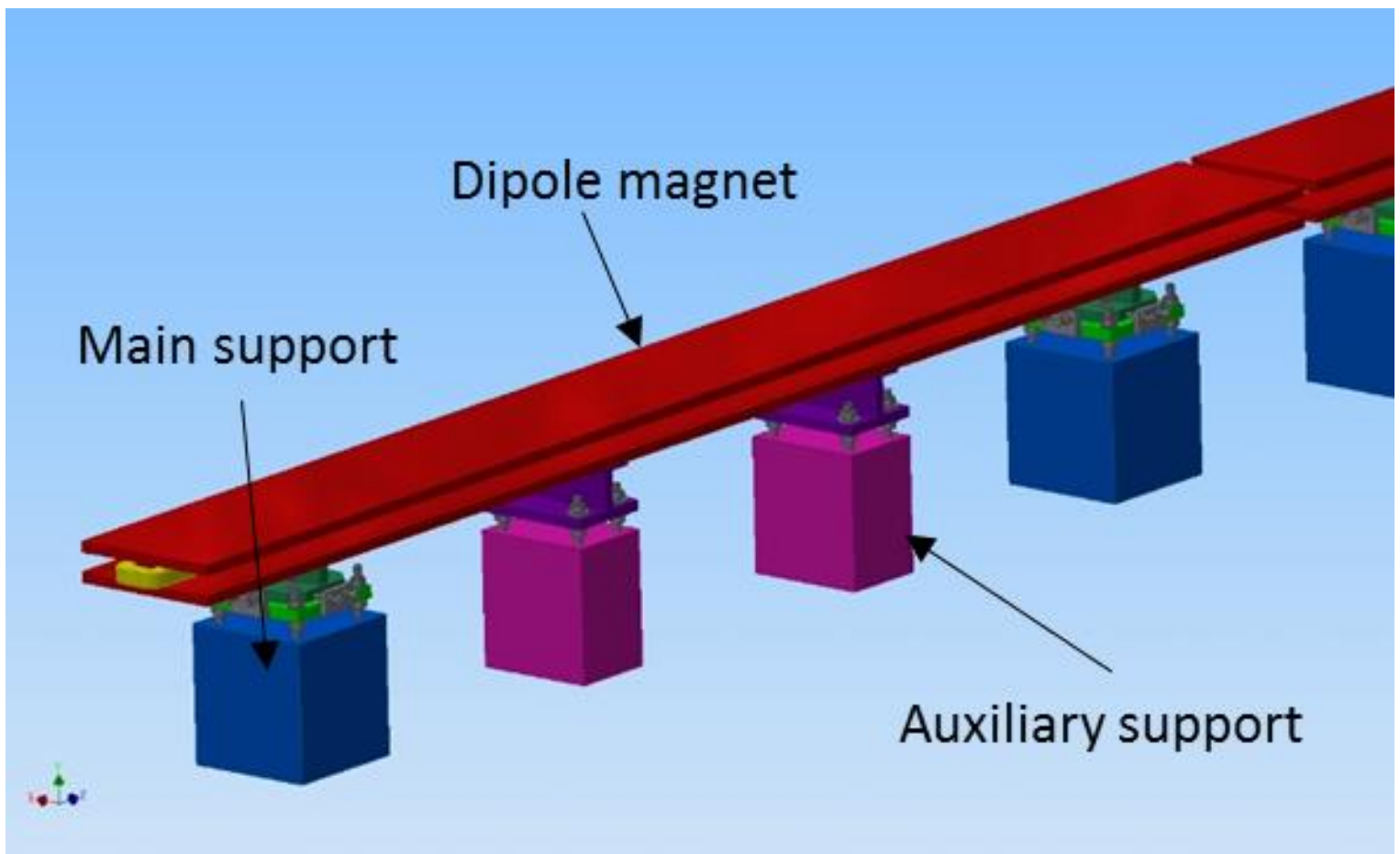

**Figure 4.3.9.2:** Dipole magnet and its supports for each module

The main support can adjust in 6 DOFs. To avoid coupling horizontal and vertical adjustments, the adjusting mechanism is a separated type and consists of 2 layers, a top layer and a bottom layer. The magnet is supported by the top layer. When horizontal alignment is done, the magnet mounting plate is fixed to the bottom layer with bolts. The vertical position is adjusted by screw nuts. The horizontal position is adjusted by push-pull bolts.

The auxiliary support has only a Y-axis DOF. Together with the main supports, they support the magnet module and avoid deformation. Similar to the vertical adjustment mechanism of the main support, the adjusting mechanism is a screw bolt.

#### *4.3.9.2.3 Distribution of Support Points for Dipole Magnets*

Figure 4.3.9.3 shows the force diagram for the long dipole module, assuming it is a slender beam. Theoretical calculation or Response surface design in ANSYS can be used to minimize the deflection of the magnet [1].To obtain a common result, the beam length is uniformed to 1 m. When *l1* equals 0.132 m and *l2* equals 0.263 m, the maximum deflection of the calculated three points is minimum.

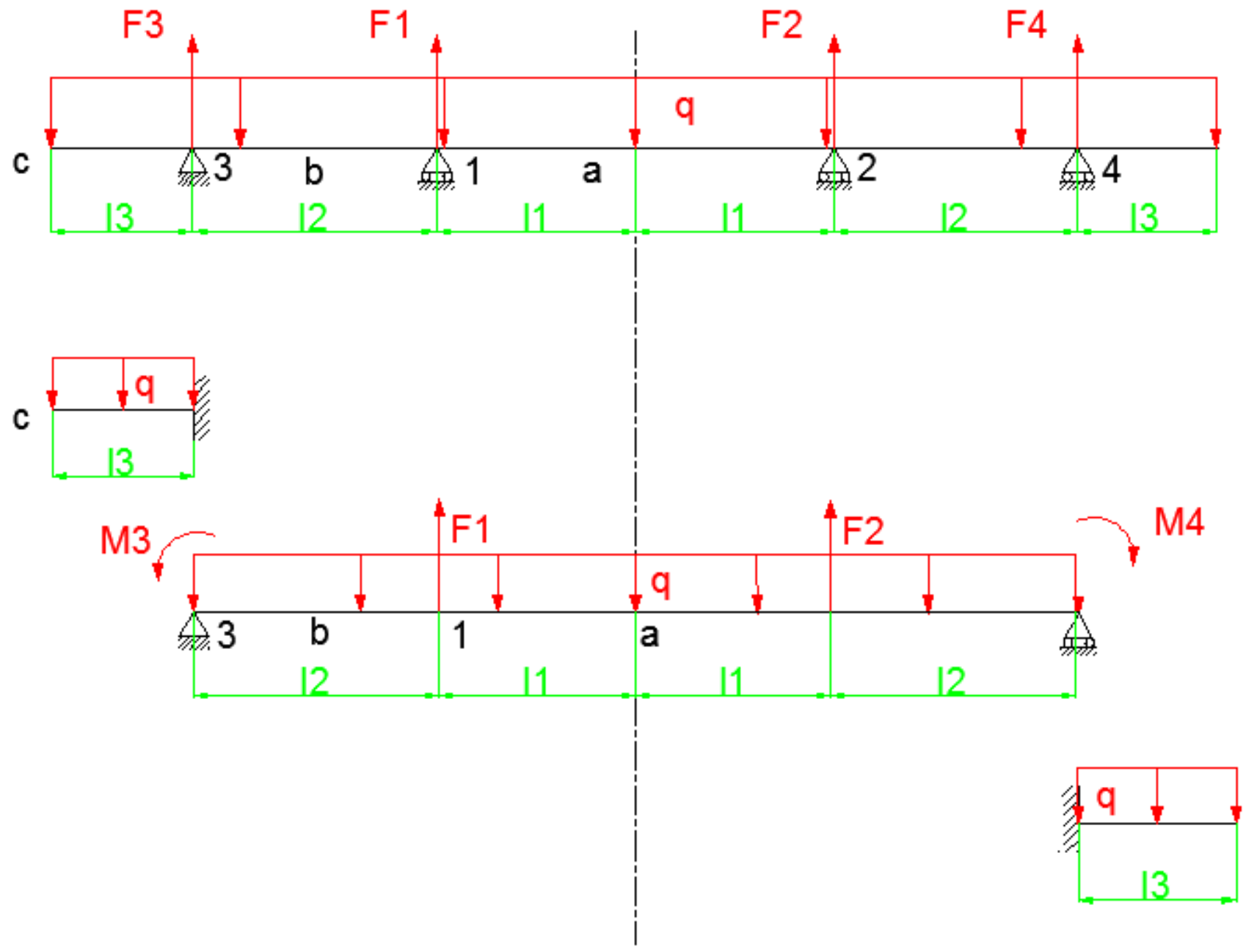

**Figure 4.3.9.3:** Force diagram of 4 support points

The method is independent to the length or the cross section, as long as the magnet core can be as assumed to be a slender beam. If the length changes, the supporting locations can change easily according to the calculation above.

#### *4.3.9.2.4 Quadrupole Support System*

The quadrupole magnet type of the largest quantity is 2000 mm long. It is supported by two supports, as shown in Figure 4.3.9.4. The supports are similar to the main supports of the dipole magnet.

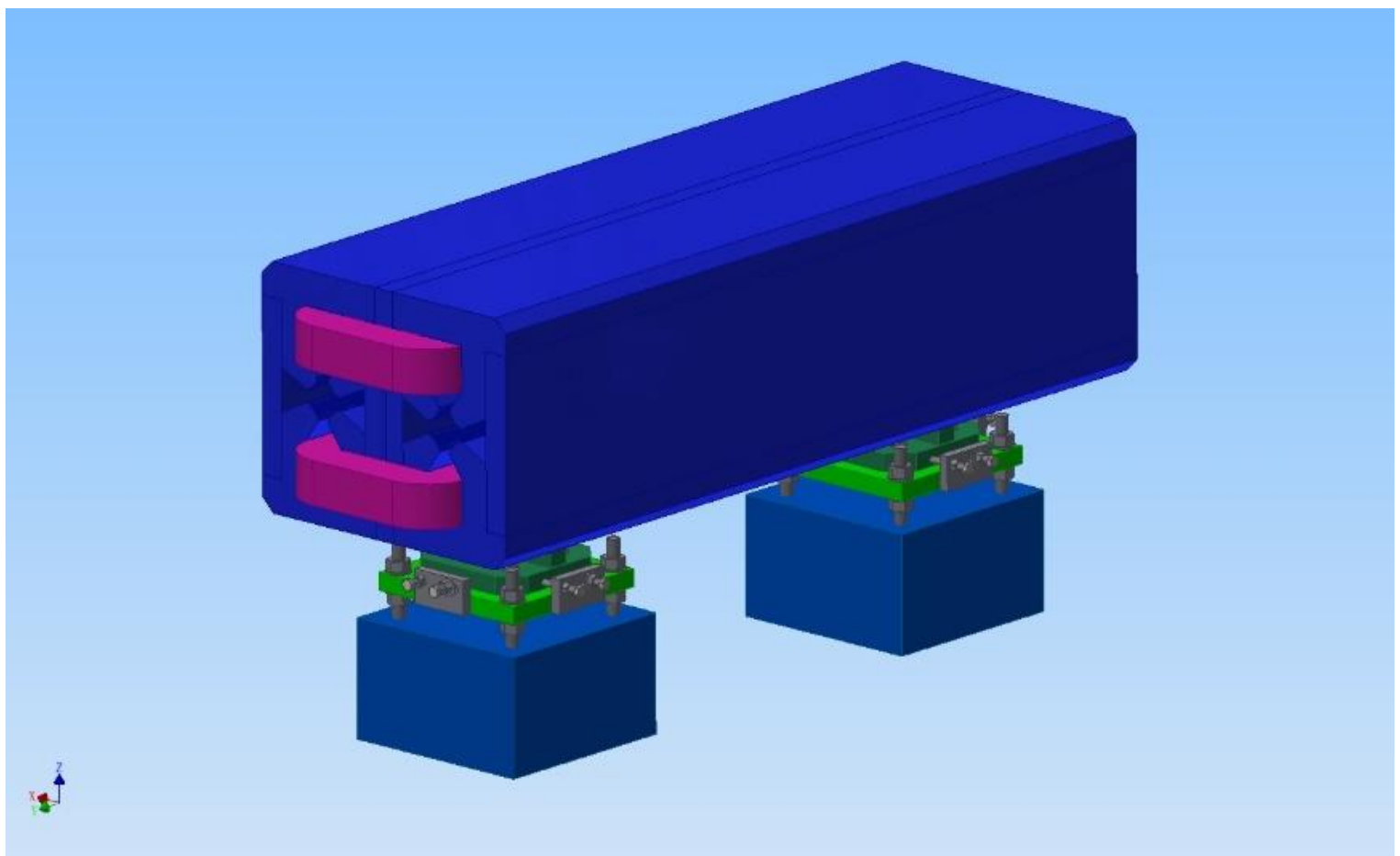

**Figure 4.3.9.4:** Quadrupole and its supports

*4.3.9.2.5 Sextupole Support System*

There are 448 sextupole magnets of length 700 mm and 448 of length 1400 mm. The two sextupole magnets at the same location are supported together, by one support because the length is relatively short, as shown in Figure 4.3.9.5. The structure of the supports are also similar to the dipole main supports.

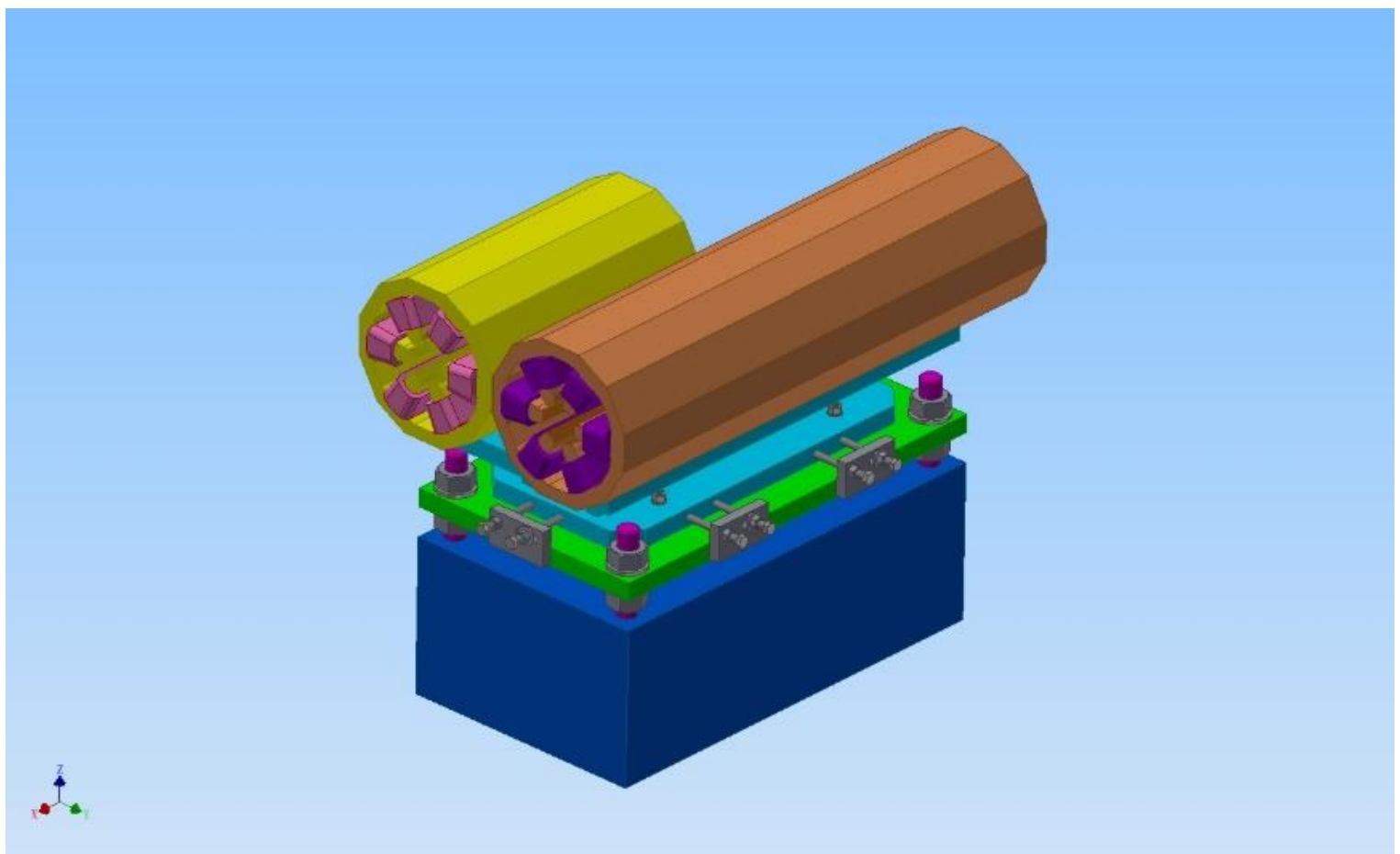

**Figure 4.3.9.5:** Sextupole magnets and their support

*4.3.9.2.6 Corrector Support System*

There are 5808 correctors of length 875 mm in the Collider; half are horizontal correctors and half are vertical. The supports of the correctors are similar to the dipole main supports, one support for each magnet.

### 4.3.9.3 *Layout of the Tunnel Cross Section*

The tunnel is 6,000 mm wide and 5,000 mm high, both in the arc section and the RF section.

In the tunnel, the Collider ring is near the inner wall, and the SPPC near the outer wall. The layout of the arc section is shown in Figure 4.3.9.6. The operation space near the two walls is about half a meter, enough for one person. The aisle between the CEPC and SPPC is 2,400 mm, for transportation, installation, alignment and other operations.

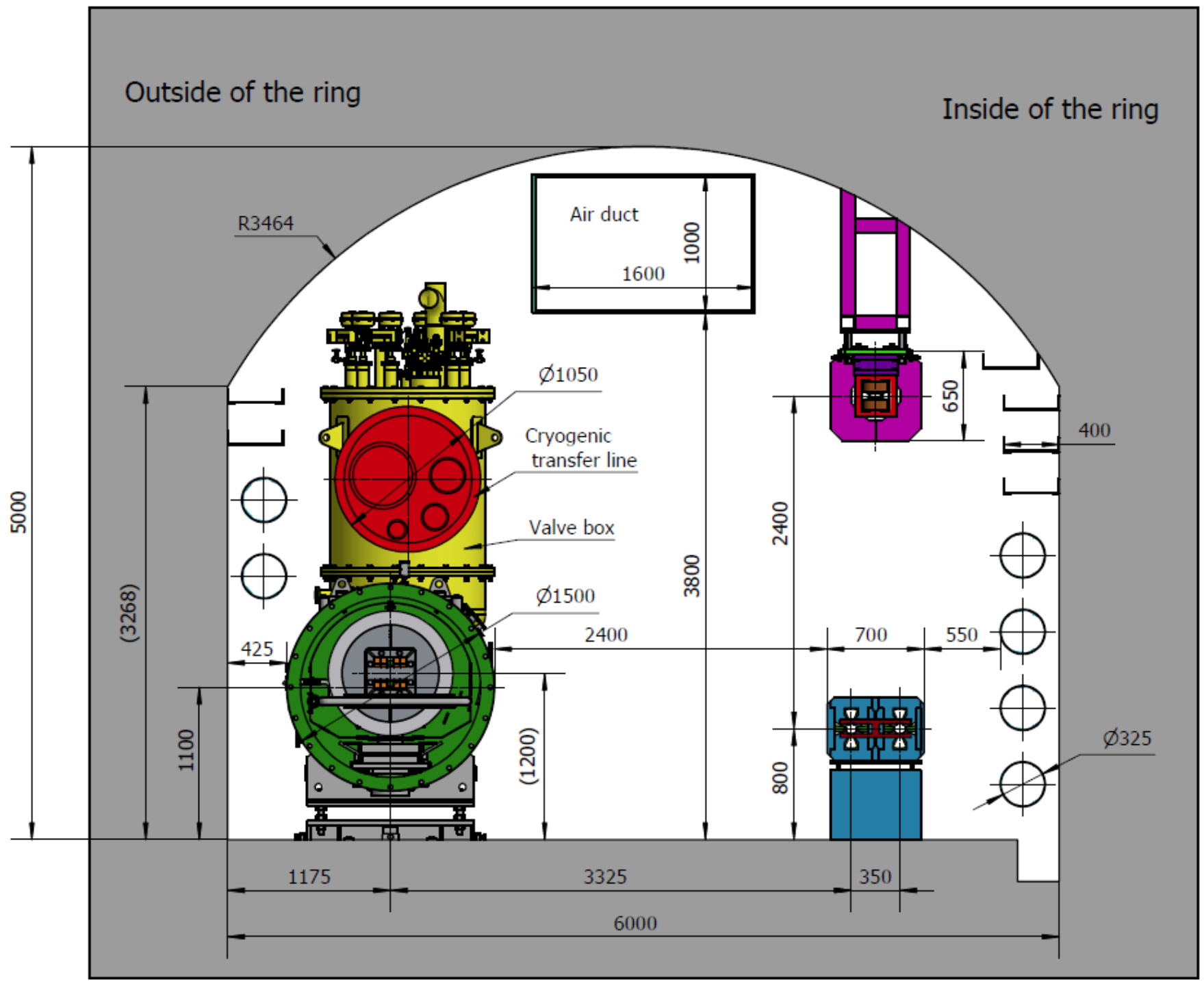

**Figure 4.3.9.6:** Tunnel cross section in the arc-section

CEPC has two RF sections. The RF tunnel can be divided into two parts, one where there are only Collider cryomodules but no Booster cryomodules; the other is the reverse. At the RF section, there are no SPPC devices because colliding points of SPPC are near the RF sections of CEPC. In the CEPC Collider, the cryomodules are in the "bigger ring" due to their physical design. The cross sections of the RF tunnel and gallery are shown in Figs. 4.3.9.7 and 4.3.9.8.

At the RF section, there are many auxiliary devices such as the power sources, cryogenic equipment and cooling water. They are all in the two auxiliary tunnels (galleries) parallel to the two RF main tunnels. Each auxiliary tunnel is 818.6 meters long, 8 meters wide and 7 meters high. The layout of the auxiliary tunnel is symmetrical to the symmetry point of the Collider and Booster. In an auxiliary tunnel, there are two Collider RF power source galleries, 235 meters each, two Cryogenic system galleries 37 meters each, two utilities galleries 70 meters each and one Booster RF power source gallery of 134.6 meters, as shown in Fig. 4.3.9.9. In order to accommodate new devices that will be required for future upgrade for $t\bar{t}$ operation, an additional space of 1,130 meters is added to each auxiliary tunnel for a total length of 1,948.6 meters.

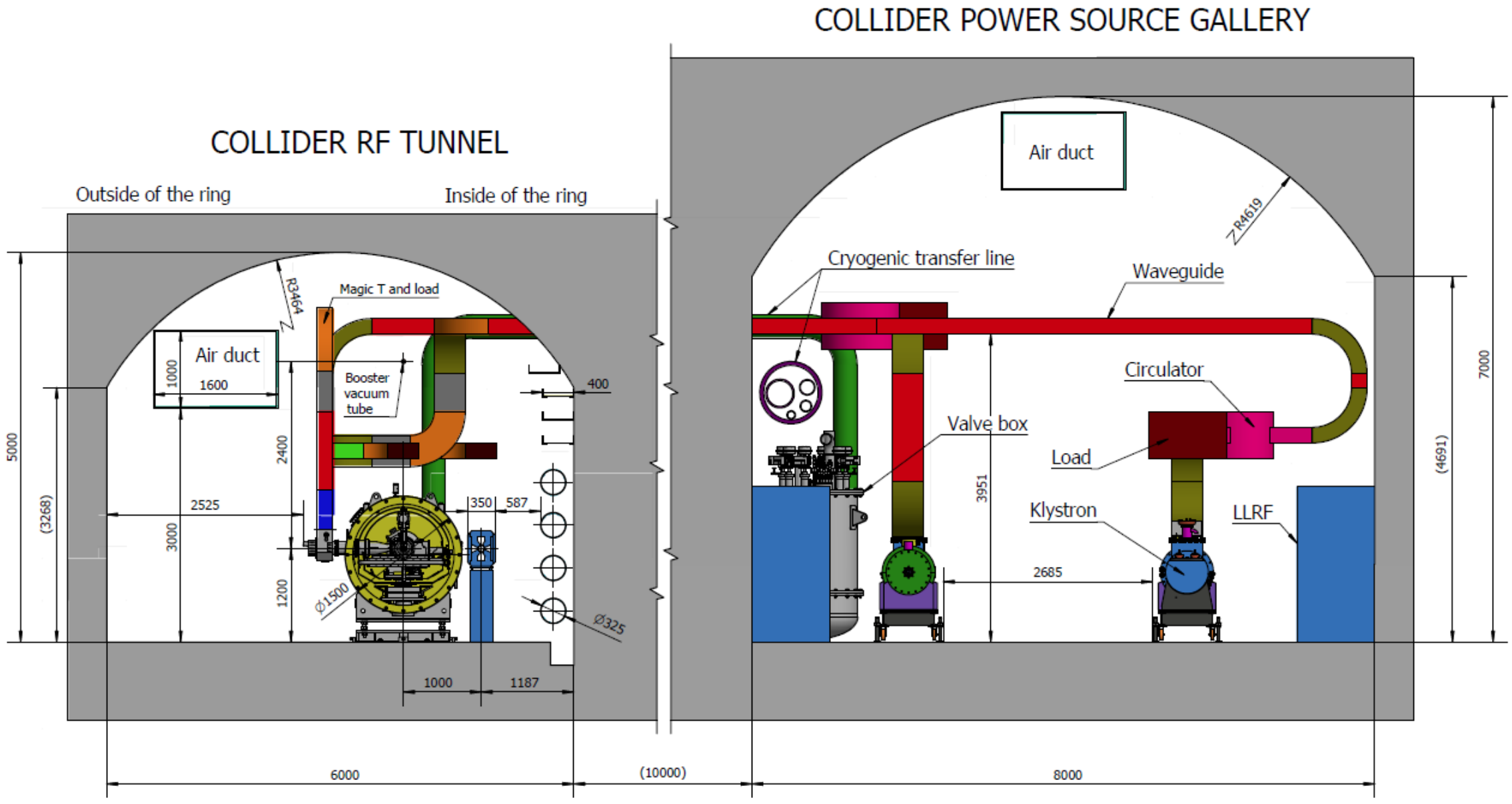

**Figure 4.3.9.7:** Tunnel cross section at the RF-section (Collider RF main tunnel and gallery)

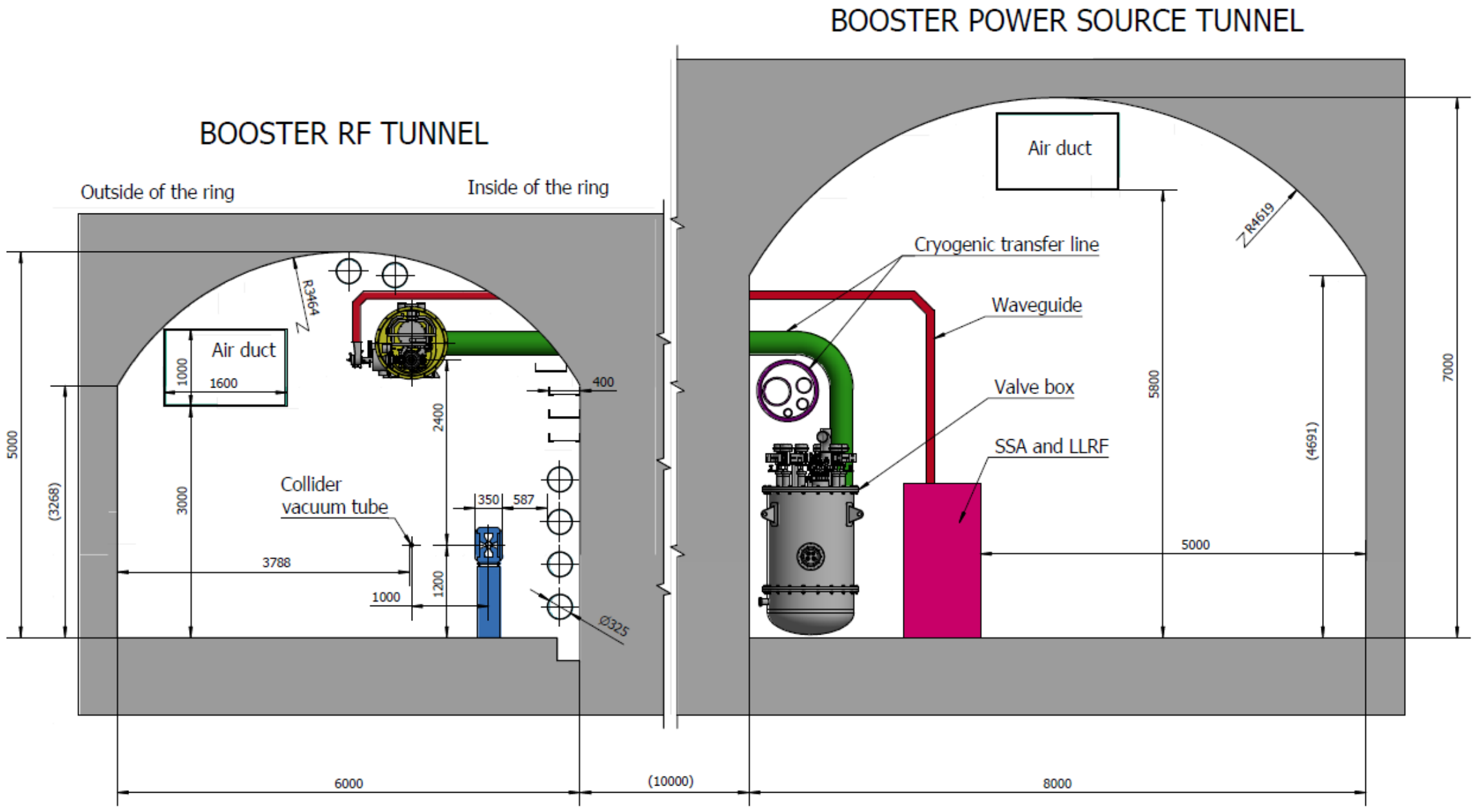

**Figure 4.3.9.8:** Tunnel cross section at RF-section (Booster RF main tunnel and gallery)

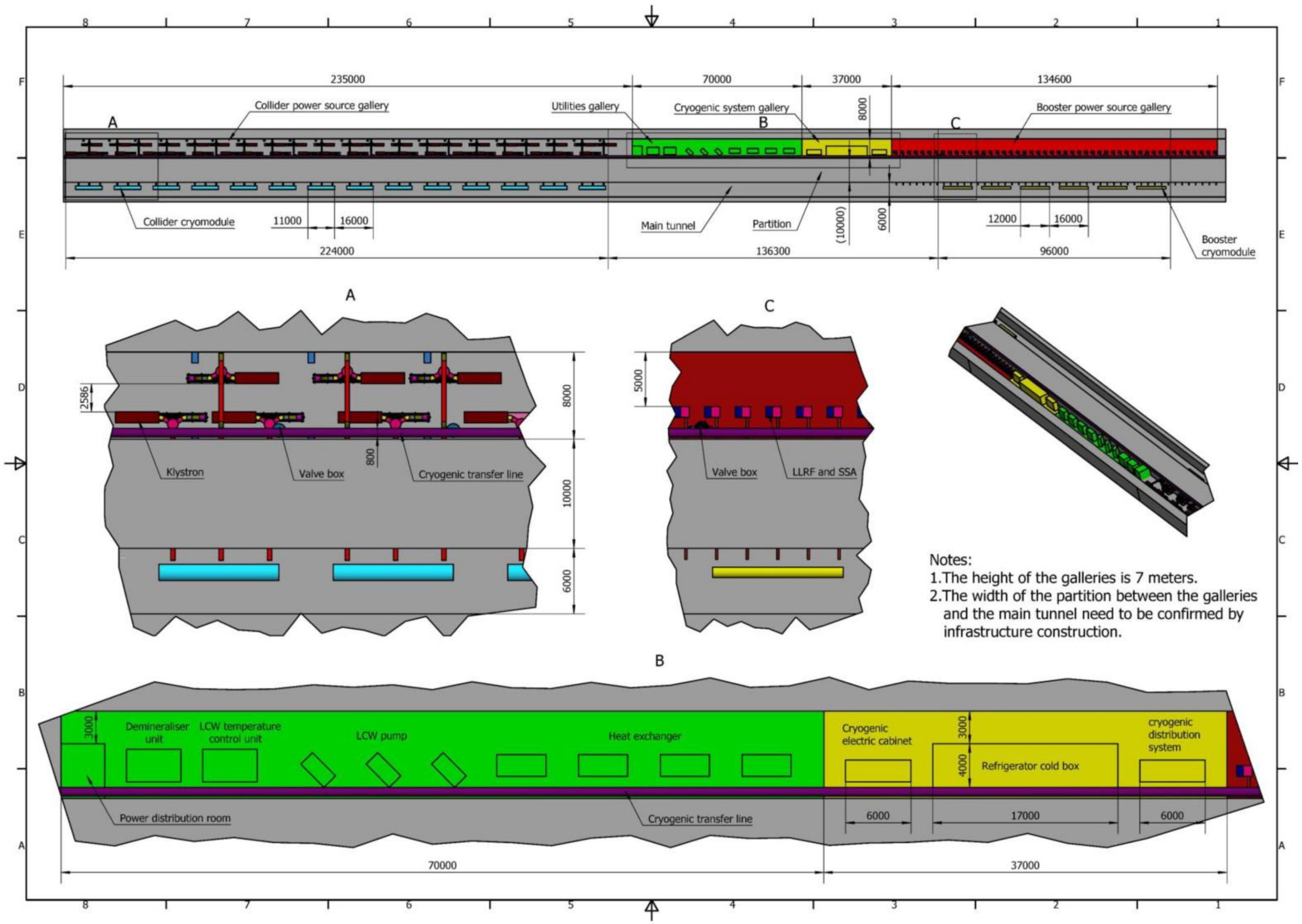

**Figure 4.3.9.9:** Cross section of the auxiliary tunnel at the RF straight section

Lots of installation and adjustment should be done in the tunnel, so there must be enough space for the major components, the services and the civil engineering components. Careful utilization of available space is important. 3D models of key areas like the arc-section, the RF section and the shaft section will be built. All the elements of these sections integrated into a computer aided design model ensures that there is no interference between components or tunnel boundaries and there is enough space reserved for transportation and installation.

### 4.3.9.4 *Collimator Mechanical System*

A collimator is a device that captures spent electrons/positrons near the beam orbit and reduces background in the particle detector [2]. There are two types of collimators for CEPC, movable and fixed.

There are dozens of movable collimators and the goals in their design is to achieve a simple structure with good heat dissipation.

The movable collimators can be adjusted the remotely, as shown schematically in Figure 4.3.9.10. The movable head is driven by a stepping motor and ball screw. Two methods for the movable head have been considered. One is to make the collimator chamber to a certain profile with constant cross-section with bellows on both sides, similar to the design at KEKB [2]. The other method is the movable head is separated from the chamber and can be moved, similar to the designs at Super KEKB and LHC [3-4].

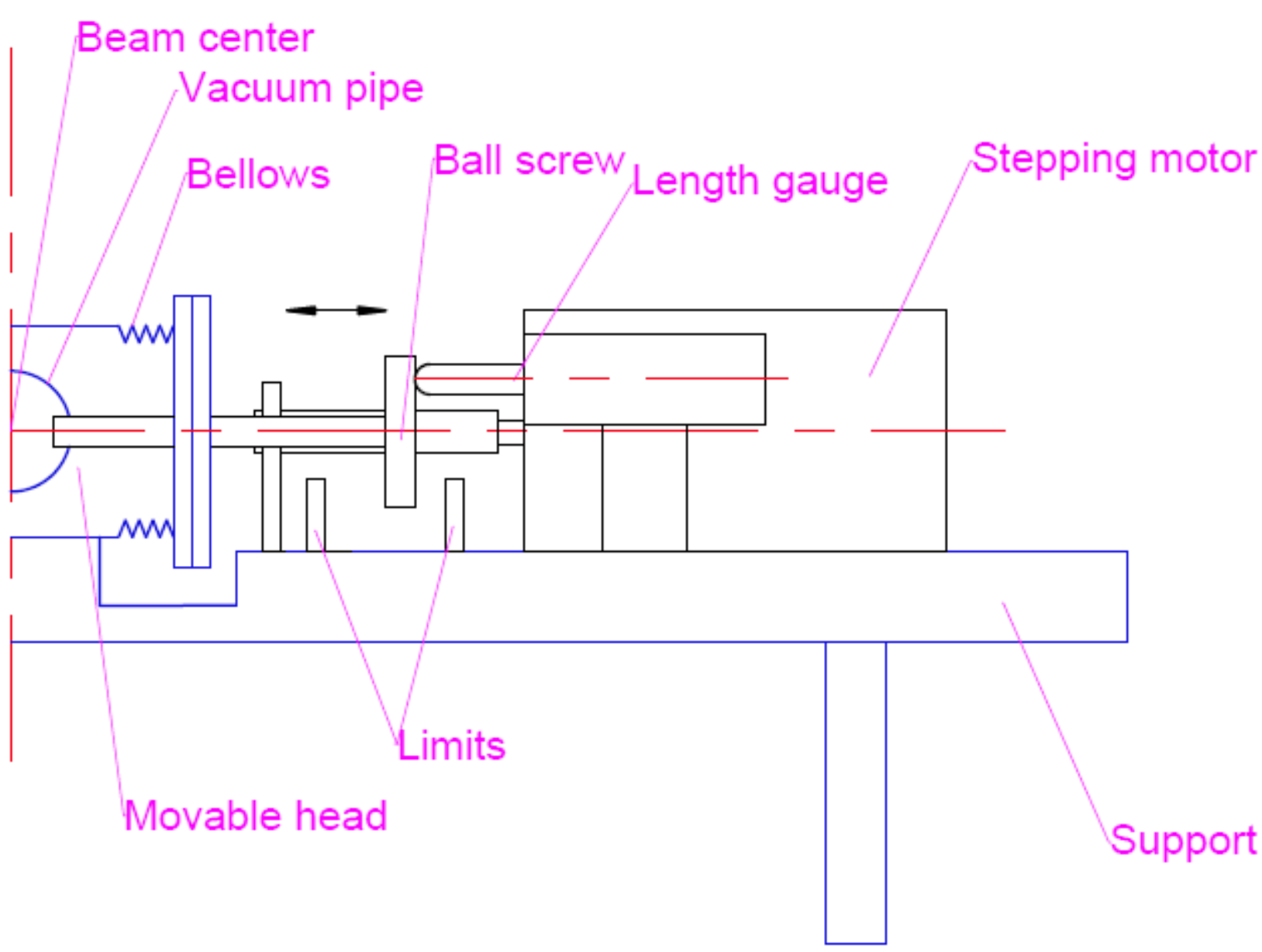

**Figure 4.3.9.10:** Schematic of movable collimators

The fixed collimator is relatively simple. The method is to make the fixed profile of the collimator chamber (or to fix blocks to the chamber) during fabrication.

The technical design details and goals are summarized:

- Design the collimators with simple and flexible structure, both for the movable collimators and fixed collimators.
- Calculate the thermal and mechanical stress and optimize the structure and materials to improve performance.

- Design the inner section of the collimator according to the physical requirements. Optimize the profile to obtain low impedance.

# 5 Booster

## 5.1 Main Parameters

### 5.1.1 Main Parameters of Booster

The booster provides electron and positron beams to the collider at different energies. Both the initial injection from zero current and the top-up injection should be fulfilled. Fig. 5.1.1 shows the overall layout. The beam is accelerated in a 10 GeV Linac and then injected and accelerated in the booster to the specific energy required by the three collider operating modes (Higgs, W and Z). The booster is in the same tunnel as the collider, placed above the collider ring except in the interaction region where there are bypasses to avoid the detectors at IP1 and IP3.

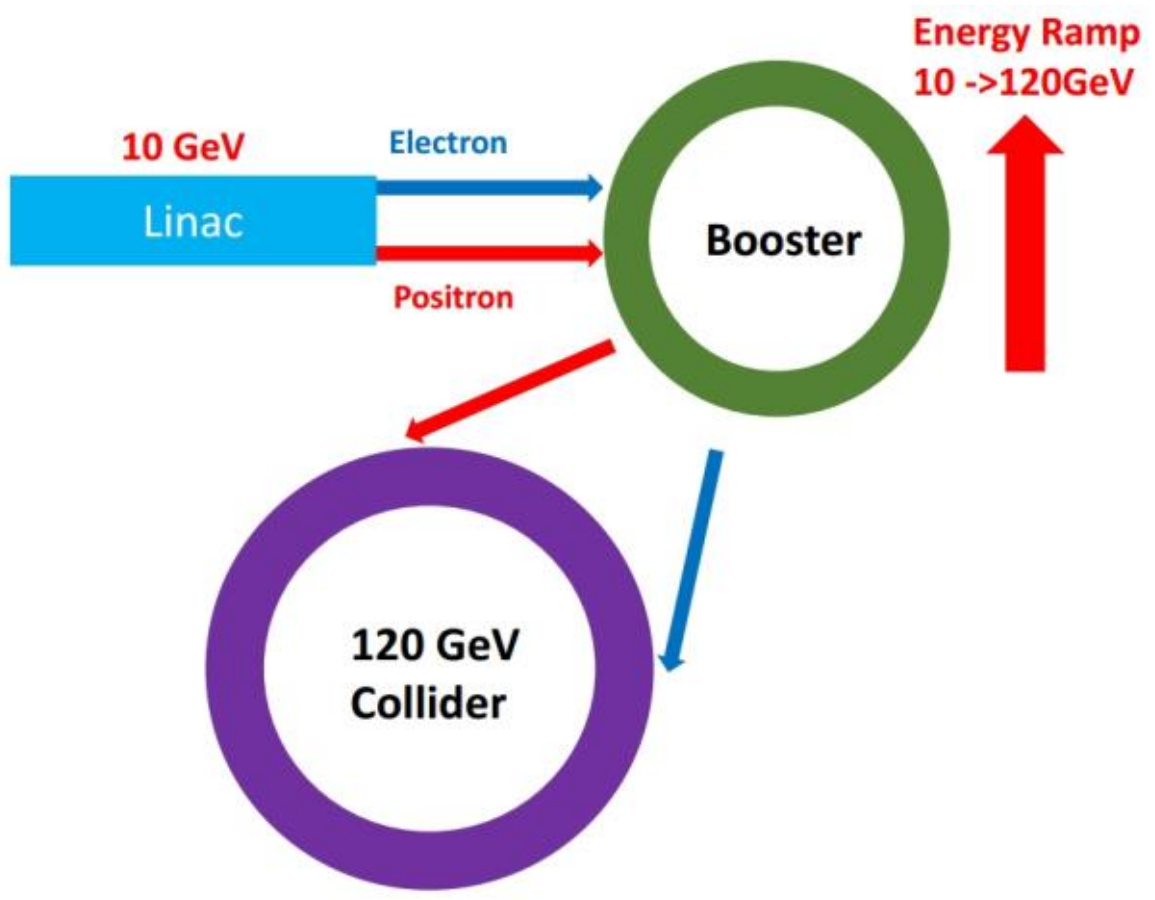

**Figure 5.1.1:** Overall layout of the CEPC injection chain.

The main booster parameters at injection and extraction energies are listed in Tables 5.1.1 and 5.1.2. Top up injection is required. Assumptions are 3% current decay, 92% booster transfer efficiency including 3% beam loss due to the quantum lifetime and 5% beam loss during ramping. The total beam current in the booster is less than 1 mA for running in Higgs mode, 4 mA for W mode and 10 mA for Z. These limits are set by the RF system. The beam is injected from the linac to the booster by on-axis scheme and is injected from booster to collider by off-axis scheme at three different energies for Higgs, W and Z. Also the on axis injection from booster to collider has been designed for Higgs in case the dynamic aperture of collider ring at Higgs energy is not good enough for the off axis injection. The threshold of single bunch current due to TMCI at 120GeV is 300μA which is higher than the maximum single bunch current in the on-axis scheme. The top up injection time is 25.8 seconds for Higgs off-axis mode, 35.4 seconds for Higgs on-axis mode, 45.8 seconds for W and 4.6 minutes for Z. The full injection time from 0 current for both beams is 10 minutes for Higgs, 15 minutes for W and 2.2 hours for Z (bootstrapping start from half of the design current).

After energy ramping, the booster emittance for Higgs and W approaches the value small enough to inject into the Collider. The beam emittance for Z mode after energy ramping still cannot fulfil the collider injection requirement and further damping (5s) is needed before extraction from the booster.

The RF voltage and longitudinal tune of the booster during ramping for the three energy modes are shown in Fig. 5.1.2. The longitudinal tune is constant (0.1) during ramping for Z and W. The longitudinal tune is 0.13 for Higgs to get larger energy acceptance for the on axis injection scheme. The beam lifetime at 10 GeV is $4.0\times10^9$ hours, dominated by the transverse quantum lifetime, and is $2.3\times10^{16}$ hours at 120 GeV, dominated by the longitudinal quantum lifetime.

**Table 5.1.1:** Main parameters for the booster at injection energy

| | | ***H*** | ***W*** | ***Z*** |
|---|---|---|---|---|
| Beam energy | GeV | 10 | | |
| Bunch number | | 242 | 1524 | 6000 |
| Threshold of single bunch current | μA | 25.7 | | |
| Threshold of beam current (limited by coupled bunch instability) | mA | 100 | | |
| Bunch charge | nC | 0.78 | 0.63 | 0.45 |
| Single bunch current | μA | 2.3 | 1.8 | 1.3 |
| Beam current | mA | 0.57 | 2.86 | 7.51 |
| Energy spread | % | 0.0078 | | |
| Synchrotron radiation loss/turn | keV | 73.5 | | |
| Momentum compaction factor | $10^{-5}$ | 2.44 | | |
| Emittance | nm | 0.025 | | |
| Natural chromaticity | H/V | -336/-333 | | |
| RF voltage | MV | 62.7 | | |
| Betatron tune $\nu_x/\nu_y/\nu_s$ | | 263.2/261.2/0.1 | | |
| RF energy acceptance | % | 1.9 | | |
| Damping time | s | 90.7 | | |
| Bunch length of linac beam | mm | 1.0 | | |
| Energy spread of linac beam | % | 0.16 | | |
| Emittance of linac beam | nm | 40~120 | | |

**Table 5.1.2:** Main parameters for the booster at extraction energy

| | | ***H*** | | ***W*** | ***Z*** |
|---|---|---|---|---|---|
| | | Off axis injection | On axis injection | Off axis injection | Off axis injection |
| Beam energy | GeV | 120 | | 80 | 45.5 |
| Bunch number | | 242 | 235+7 | 1524 | 6000 |
| Maximum bunch charge | nC | 0.72 | 24.0 | 0.58 | 0.41 |
| Maximum single bunch current | μA | 2.1 | 70 | 1.7 | 1.2 |
| Threshold of single bunch current | μA | 300 | | | |
| Threshold of beam current (limited by RF power) | mA | 1.0 | | 4.0 | 10.0 |
| Beam current | mA | 0.52 | 1.0 | 2.63 | 6.91 |
| Injection duration for top-up (Both beams) | s | 25.8 | 35.4 | 45.8 | 275.2 |
| Injection interval for top-up | s | 47.0 | | 153.0 | 504.0 |
| Current decay during injection interval | | 3% | | | |
| Energy spread | % | 0.094 | | 0.062 | 0.036 |
| Synchrotron radiation loss/turn | GeV | 1.52 | | 0.3 | 0.032 |
| Momentum compaction factor | $10^{-5}$ | 2.44 | | | |
| Emittance | nm | 3.57 | | 1.59 | 0.51 |
| Natural chromaticity | H/V | -336/-333 | | | |
| Betatron tune $\nu_x/\nu_y$ | | 263.2/261.2 | | | |
| RF voltage | GV | 1.97 | | 0.585 | 0.287 |
| Longitudinal tune | | 0.13 | | 0.10 | 0.10 |
| RF energy acceptance | % | 1.0 | | 1.2 | 1.8 |
| Damping time | ms | 52 | | 177 | 963 |
| Natural bunch length | mm | 2.8 | | 2.4 | 1.3 |
| Injection duration from empty ring | h | 0.17 | | 0.25 | 2.2 |

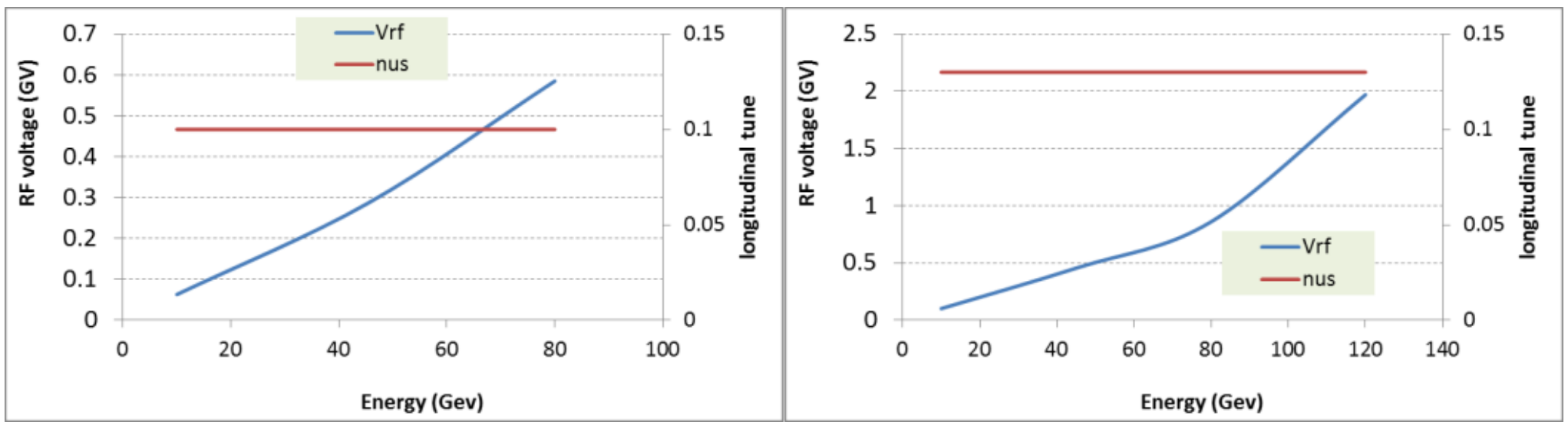

**Figure 5.1.2:** Booster RF ramping curve (left: Z & W, right: H)

### 5.1.2 RF Parameters

The RF parameters are listed in Table 5.1.3. The superconducting RF cavity is chosen due to its high CW gradient, high energy efficiency and low impendence. The choice of 1.3 GHz RF frequency is a balance of beam stability and cost.

**Table 5.1.3:** Main RF parameters of the booster Ring

| | H | W | Z |
|---|---|---|---|
| | off-axis injection | | |
| Extraction beam energy [GeV] | 120 | 80 | 45.5 |
| Bunch number | 242 | 1524 | 6000 |
| Bunch charge [nC] | 0.72 | 0.576 | 0.384 |
| Beam current [mA] | 0.52 | 2.63 | 6.91 |
| Extraction RF voltage [GV] | 1.97 | 0.585 | 0.287 |
| Extraction bunch length [mm] | 2.7 | 2.4 | 1.3 |
| Cavity number (1.3 GHz TESLA 9-cell) | 96 | 64 | 32 |
| Cryomodule number (8 cavities / cryomodule) | 12 | 8 | 4 |
| Cavity operating gradient [MV/m] | 19.8 | 8.8 | 8.6 |
| $Q_0$ @ 2 K at operating gradient (long term) | 1E10 | 1E10 | 1E10 |
| $Q_L$ | 1E7 | 6.5E6 | 1E7 |
| Cavity bandwidth [Hz] | 130 | 200 | 130 |
| Input peak power per cavity [kW] | 18.2 | 12.4 | 7.1 |
| Input average power per cavity [kW] | 0.7 | 0.3 | 0.5 |
| SSA peak power [kW] (one cavity per SSA) | 25 | 25 | 25 |
| HOM average power per cavity [W] | 0.2 | 0.7 | 4.1 |
| Total average cavity wall loss @ 2 K eq. [kW] | 0.2 | 0.01 | 0.02 |

## 5.2 Booster Accelerator Physics

### 5.2.1 Optics

#### 5.2.1.1 *Optics and Survey Design*

The design goal for the booster optics is to make sure the geometry is the same as the collider and satisfy the requirements of beam dynamics. The total number of magnets and sextupole families is minimized taking into account capital and operating costs. The maximum cell length and hence the maximum emittance in the booster is limited by the collider injection requirements.

##### *5.2.1.1.1 Survey Design*

Both CEPC collider and booster are in the same tunnel, and booster is on the top of collider. The horizontal position of the booster has been designed in the center of collider two beams. The horizontal position error of booster is controlled under ±0.17m. CEPC booster has the same circumference as the collider (100016.4m). The geometry of booster compared with the collider and the cross section of the tunnel is shown in Fig. 5.2.1.1.

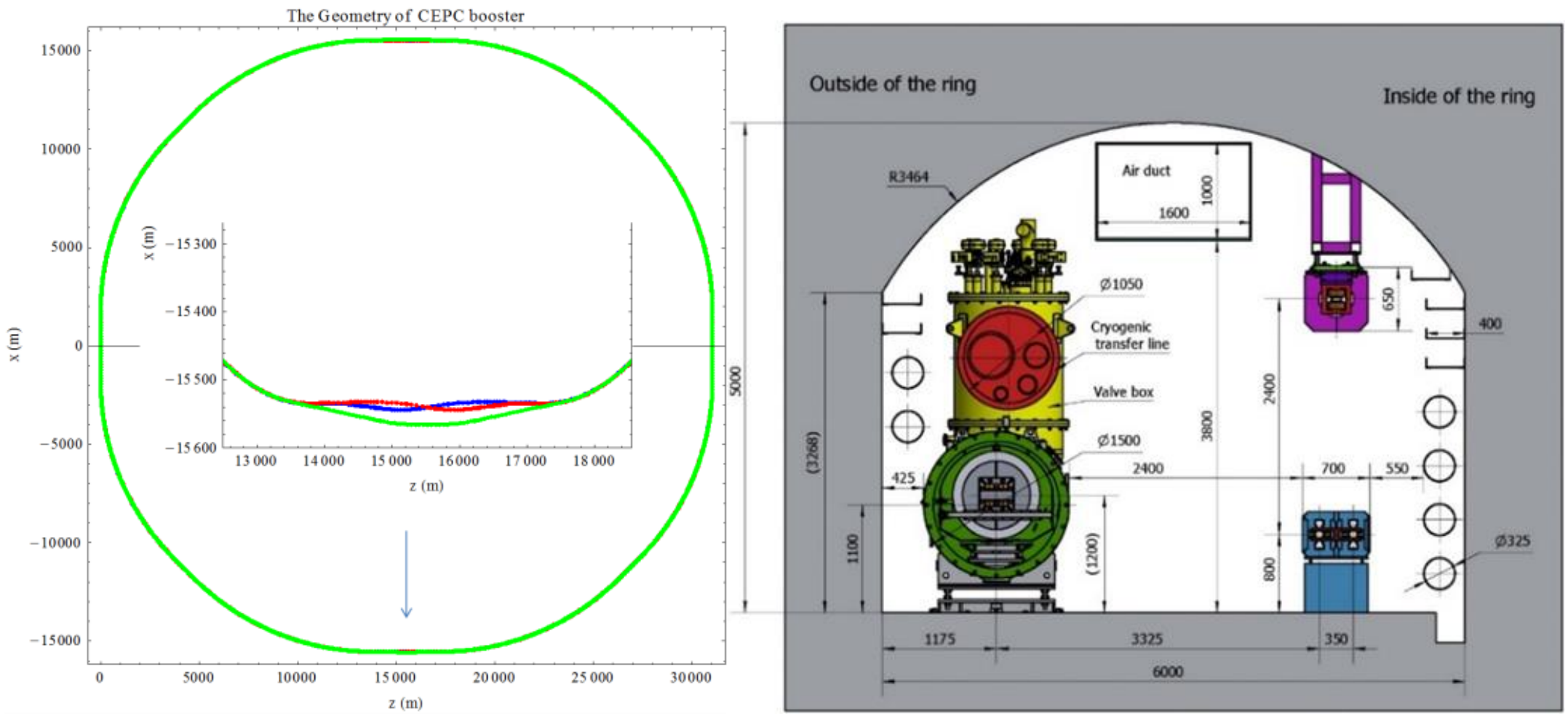

**Figure 5.2.1.1:** Booster vs. collider layouts and the cross section of tunnel.

##### *5.2.1.1.2 Arc Region*

Standard FODO cells have been chosen for the booster lattice [1]. The length of two FODO cells in the booster corresponds to three FODO cells in the collider. The phase advance of each cell is 90/90 degrees in the horizontal and vertical planes. The length of each bend is 46.4 m including ten short dipole magnets. The length of each quadrupole is 1.0 m, while the distance between each quadrupole and the adjacent bending magnet is 1.6 m. Thus the total length of each FODO structure is 101 m. 97 FODO structures make up an octant. At the two ends of each octant, there are dispersion suppressors and straight sections. Fig. 5.2.1.2 shows the twiss functions of the FODO cell and the dispersion suppressor.

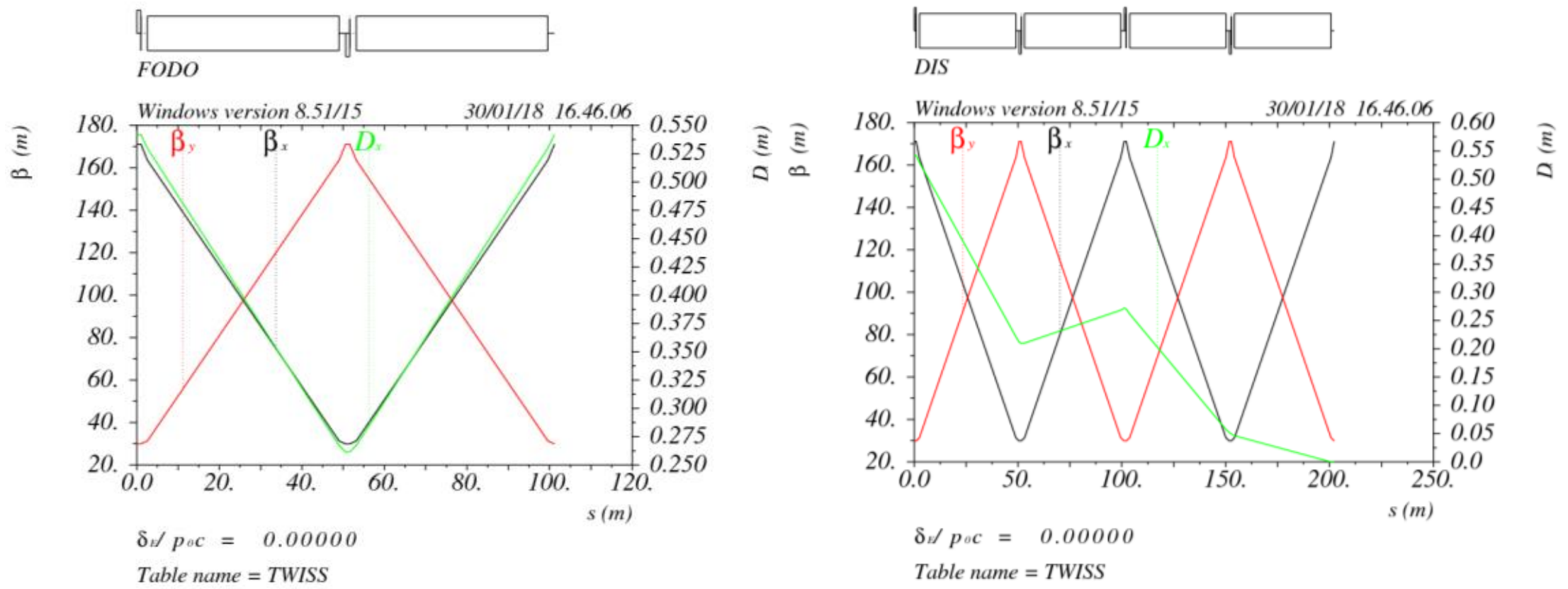

**Figure 5.2.1.2:** The Twiss functions of the FODO cell and the dispersion suppressor.

#### *5.2.1.1.3 Injection Region*

The length of the straight sections for injection/extraction is exactly the same as in the collider. Fig. 5.2.1.3 shows the lattice functions in the injection/extraction region. The phase advance in the injection straight section is tunable for adjusting the working point of the entire ring and also optimizing the off-momentum DA.

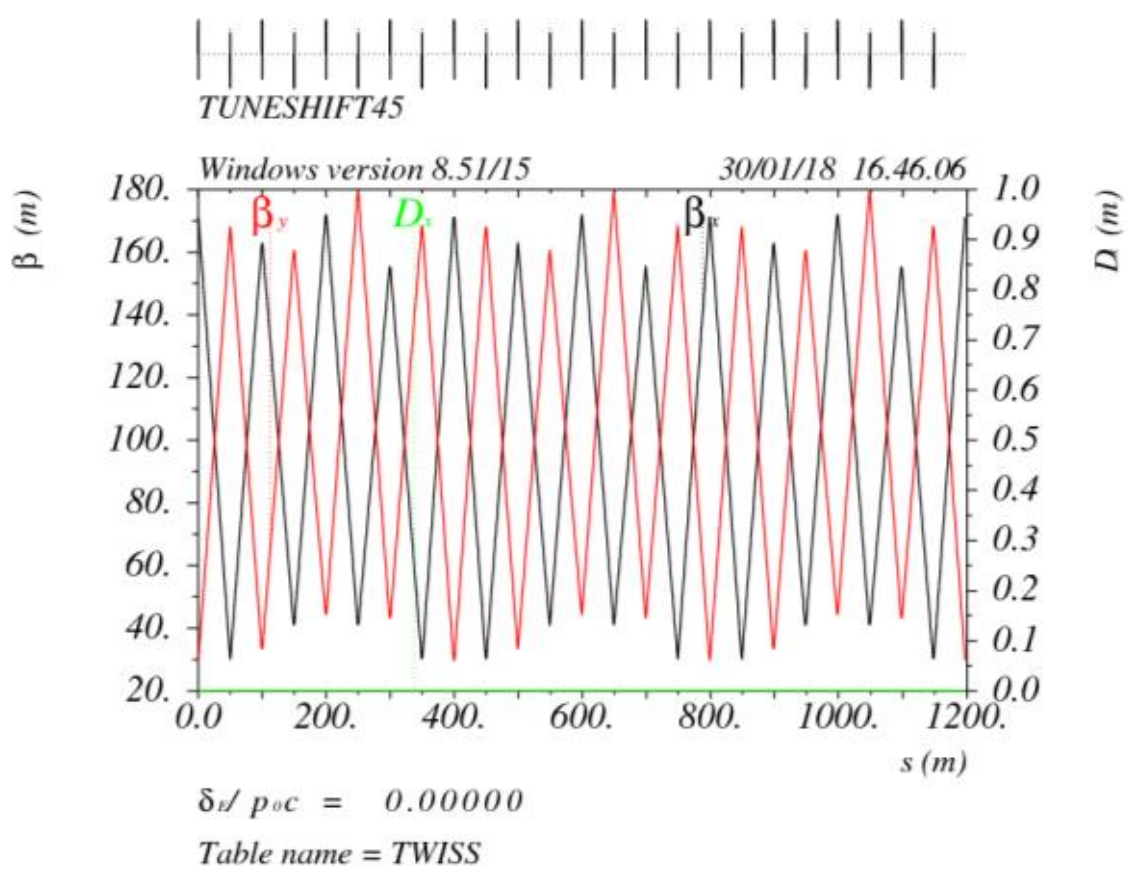

**Figure 5.2.1.3:** The Twiss functions of injection straight section.

#### *5.2.1.1.4 RF Region*

In the RF section, dedicated optics with a lower beta function is designed to reduce the multi-bunch instability due to the RF cavities. Two matching sections whose phase advances are tunable at the two ends of the RF straight section transfer the beta function from the standard arc to the low beta section. Fig. 5.2.1.4 shows the lattice functions in the RF region. The length of the low beta section is 1.6 km and the total length of the RF straight section is 3.4 km which is exactly same as the collider ring.

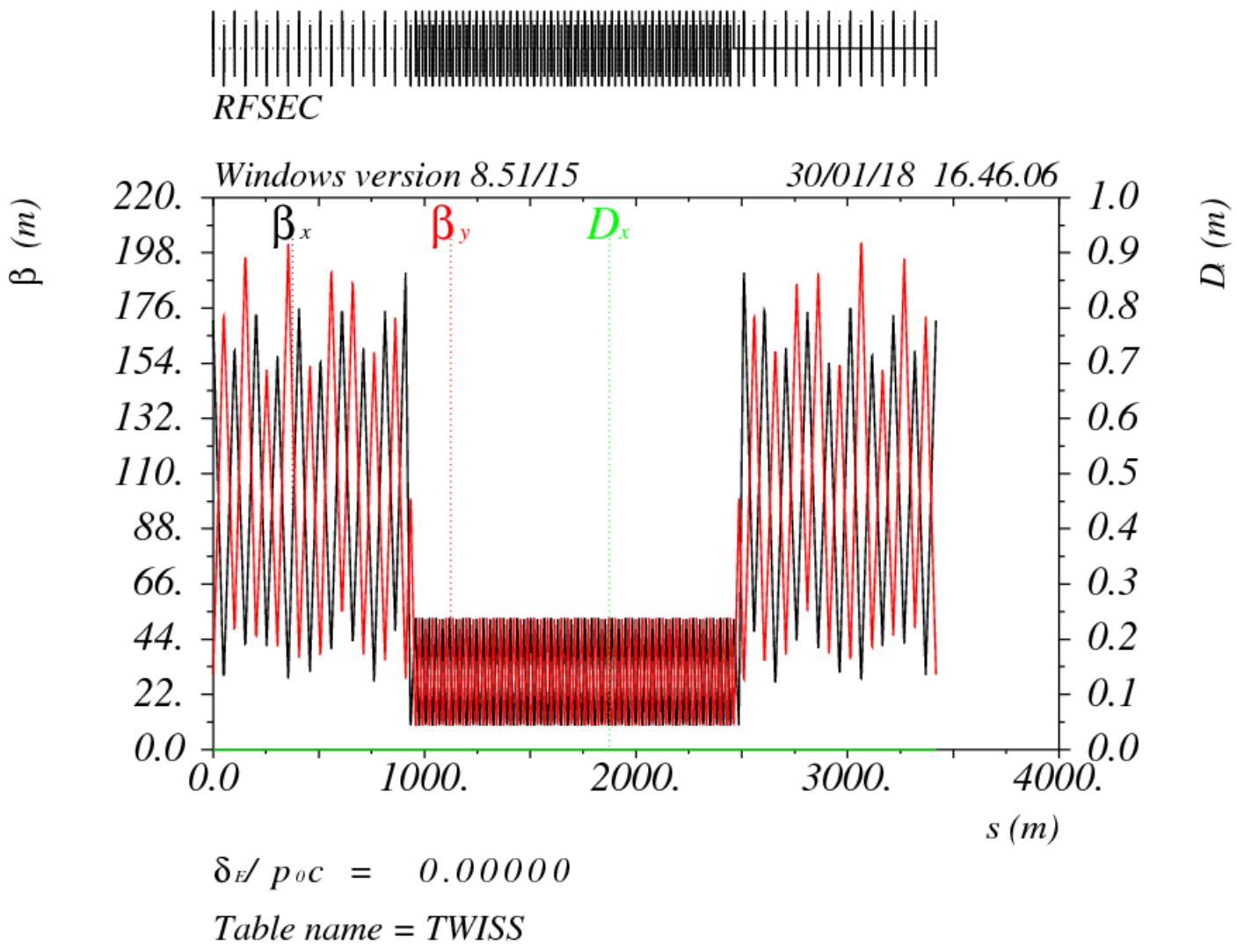

**Fig. 5.2.1.4:** RF straight section.

### *5.2.1.1.5 IR Region*

The geometry of booster is same as collider ring except for the IR. In the IR region, the booster is bypassed from the outer side to avoid a conflict with the CEPC detectors. The separation between the detector centre and booster is 25 m. Fig. 5.2.1.5 shows the lattice functions in the IR.

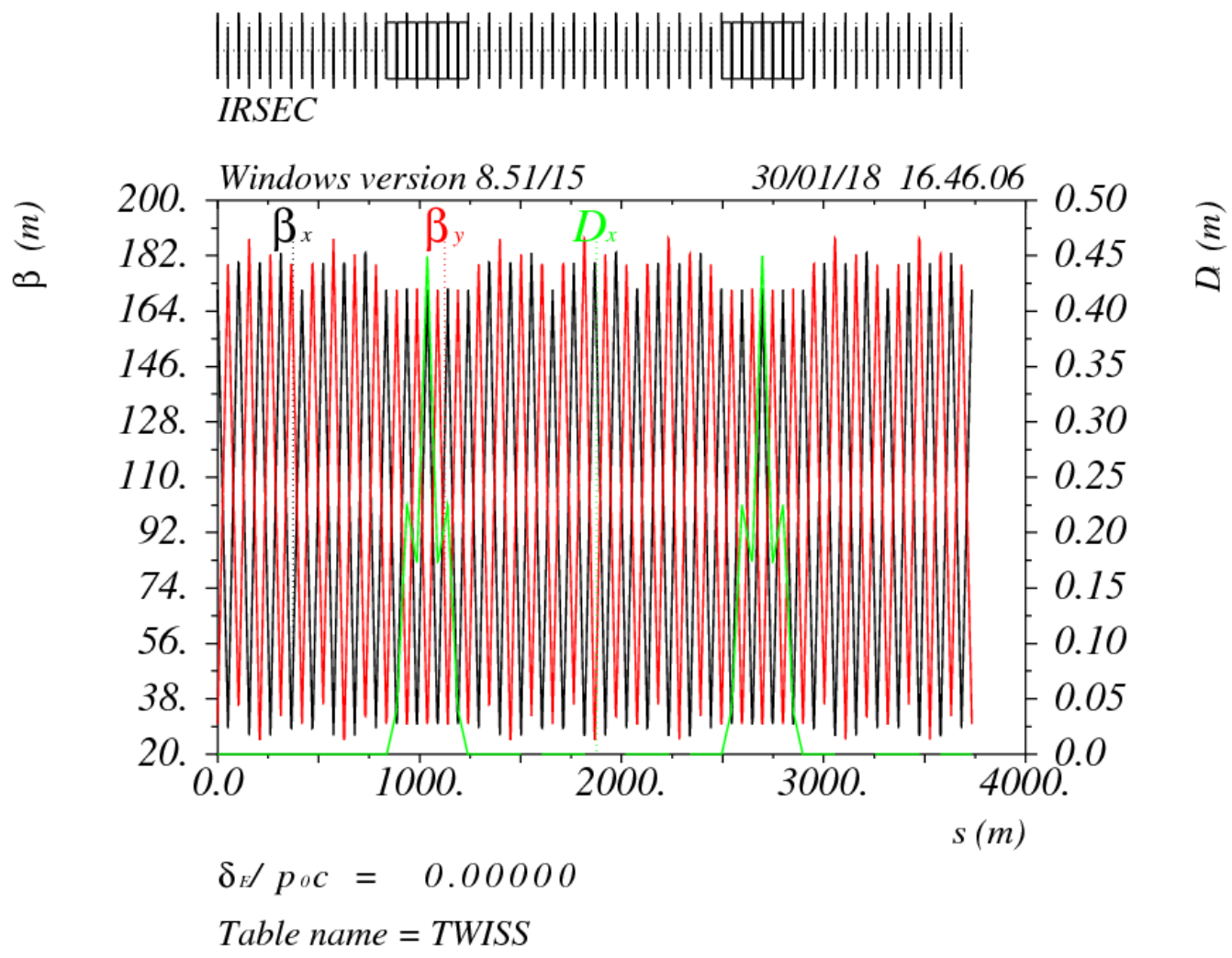

**Figure 5.2.1.5:** IR bypass.

### *5.2.1.1.6 Sawtooth Effect*

With only two RF stations, the maximum sawtooth orbit is 1.7 mm at 120GeV. The off-centre orbits in sextupoles result in extra quadrupole fields and hence result in ~6% distortion of optics. The orbit and optics of booster with sawtooth effect at 120GeV is shown in Fig. 5.2.1.6. No DA reduction due to sawtooth effect is seen in booster. So magnets energy tapering is unnecessary in booster.

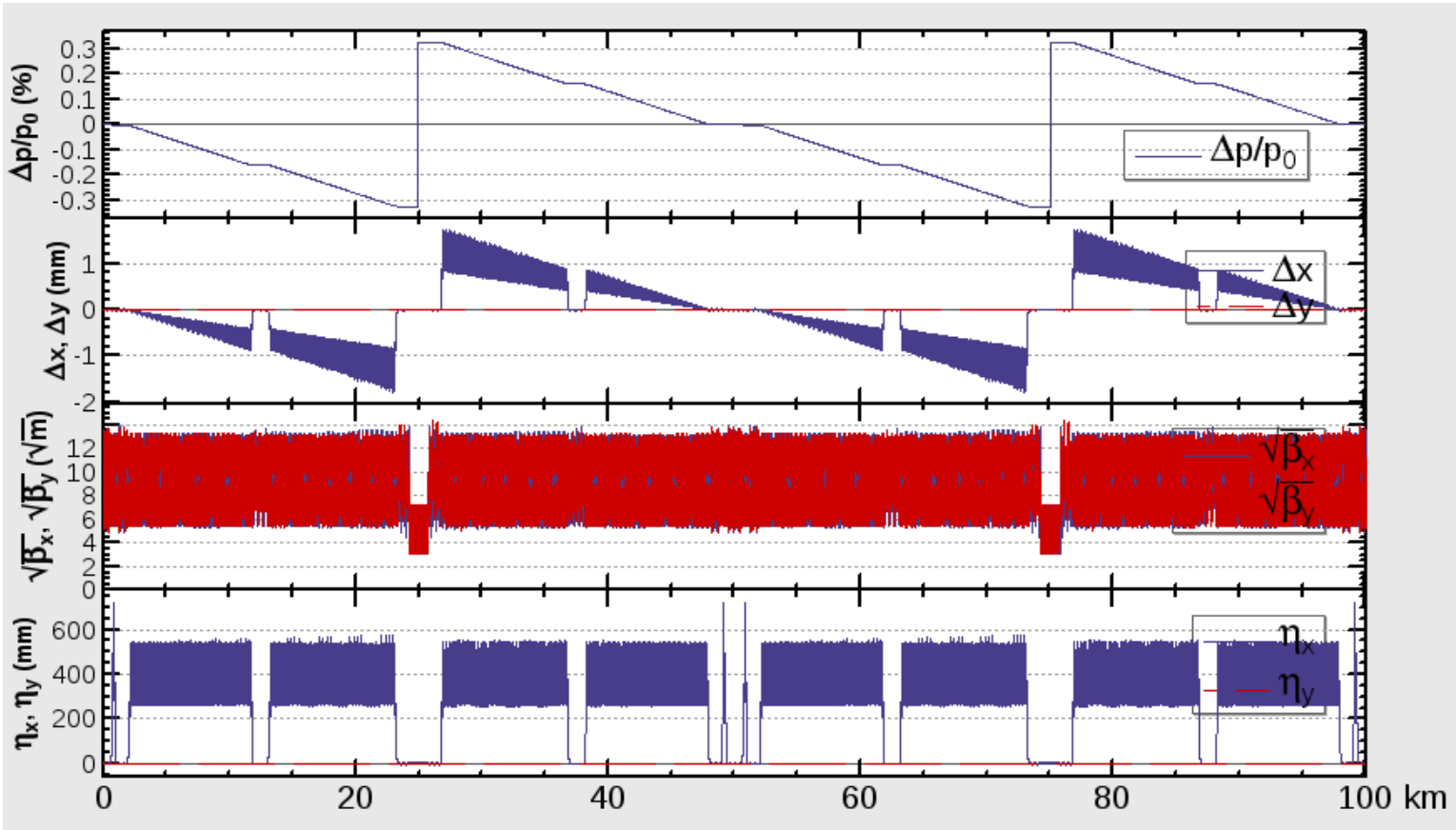

**Figure 5.2.1.6:** Booster orbit and optics with sawtooth effect at 120GeV.

### *5.2.1.1.7 Off-momentum DA Optimization*

The phase of the injection/extraction straight section between two octants is optimized automatically by downhill method [2]. The DA result at 120GeV after optimization without errors is shown in Fig. 5.2.1.7. The design goal is to reach 1% energy acceptance at 120GeV including errors which is the requirement of re-injection process for the on-axis injection scheme.

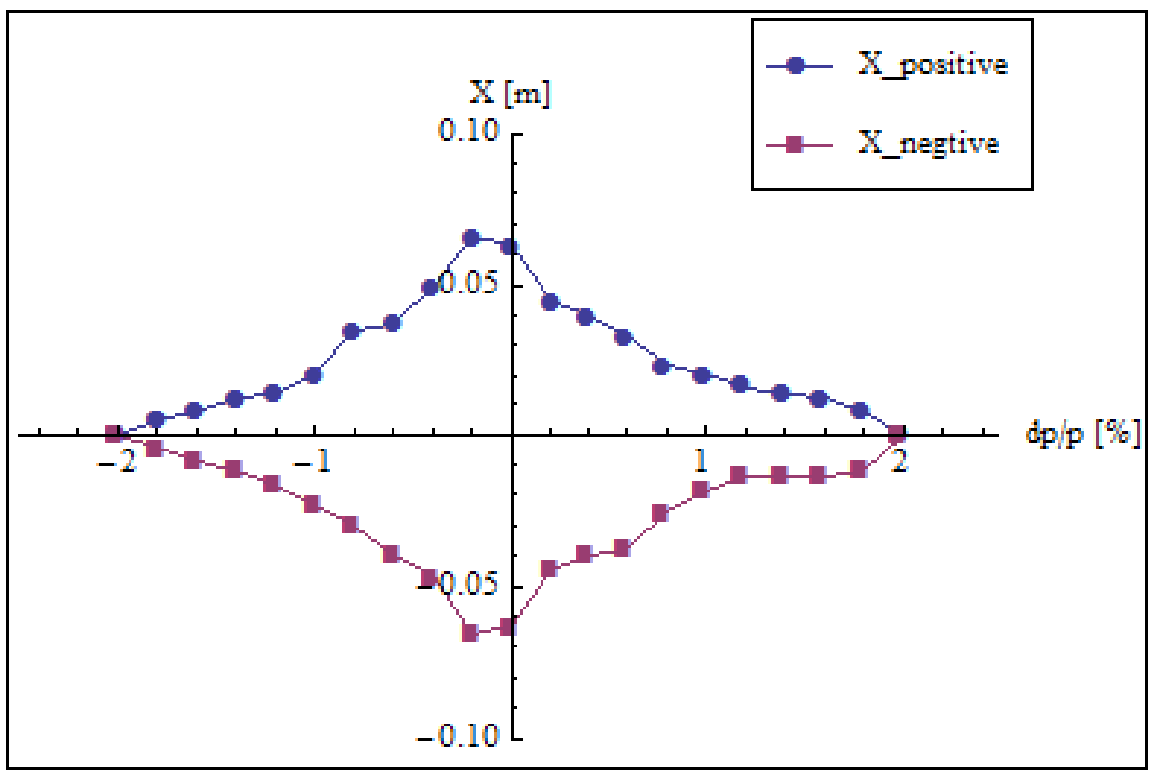

**Figure 5.2.1.7:** Off-momentum DA at 120 GeV (without errors) after optimization with phase tuning in injection/extraction section.

## 5.2.1.2 ***Performance with Errors***

### *5.2.1.2.1 Error Analysis*

Table 5.2.1.1 lists the details of the error settings. Gaussian distribution for the errors is used and is cut off at 3σ. With these errors, the closed orbits (Fig. 5.2.1.8) are smaller than the beam pipe whose diameter is 55mm so first turn trajectory correction is unnecessary. Four horizontal/vertical correctors and 8 BPMs are inserted every 2π phase advance, so totally 1054 horizontal correctors and 1054 vertical correctors are used to correct orbit distortions. Fig. 5.2.1.9 shows the closed orbit after correction. The maximum orbit is smaller than 1mm with orbit correction.

**Table 5.2.1.1:** Error analysis settings.

| Parameters | Dipole | Quadrupole | Sextupole | Parameters | BPM (10Hz) |
|---|---|---|---|---|---|
| Transverse shift x/y (μm) | 50 | 70 | 70 | Accuracy (m) | $1\times10^{-7}$ |
| Longitudinal shift z (μm) | 100 | 150 | 100 | Tilt (mrad) | 10 |
| Tilt about x/y (mrad) | 0.2 | 0.2 | 0.2 | Gain | 5% |
| Tilt about z (mrad) | 0.1 | 0.2 | 0.2 | Offset after BBA(mm) | $30\times10^{-3}$ |
| Nominal field | $3\times10^{-4}$ | $2\times10^{-4}$ | $3\times10^{-4}$ | | |

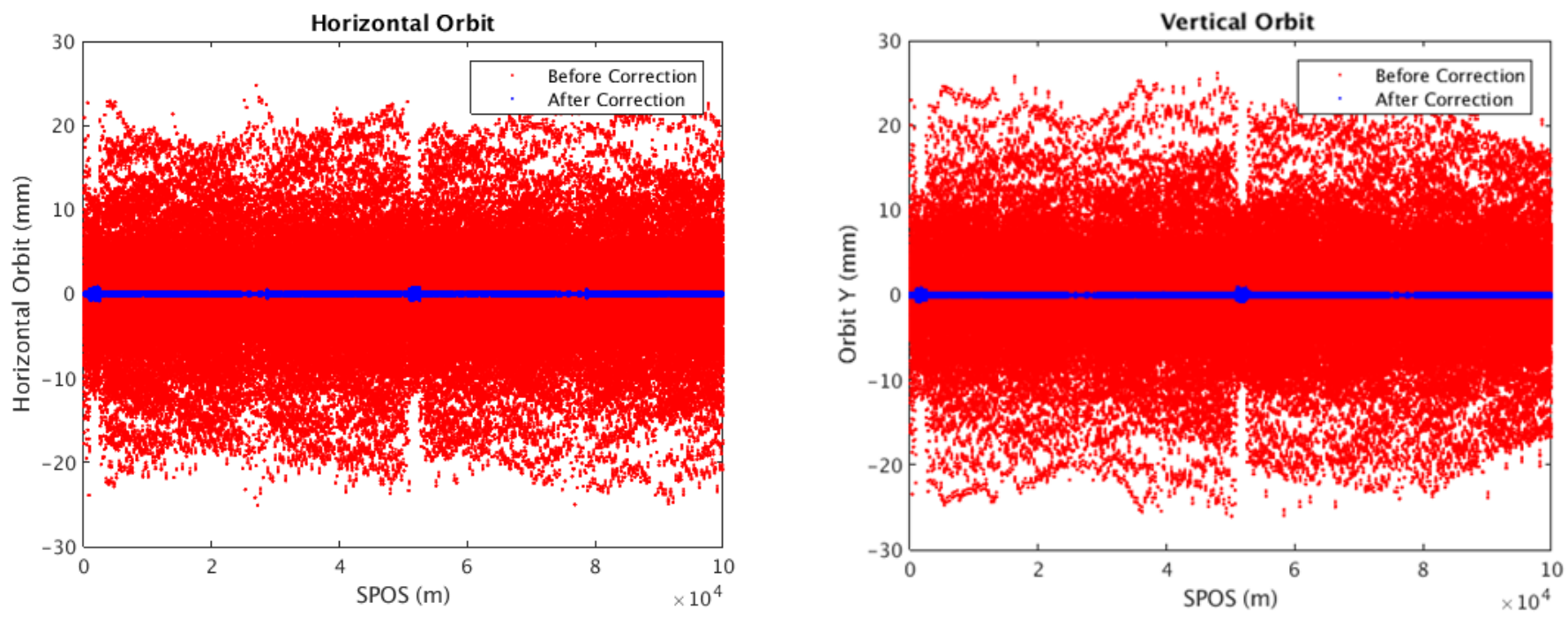

**Figure 5.2.1.8:** Closed orbit with errors (left: horizontal, right: vertical).

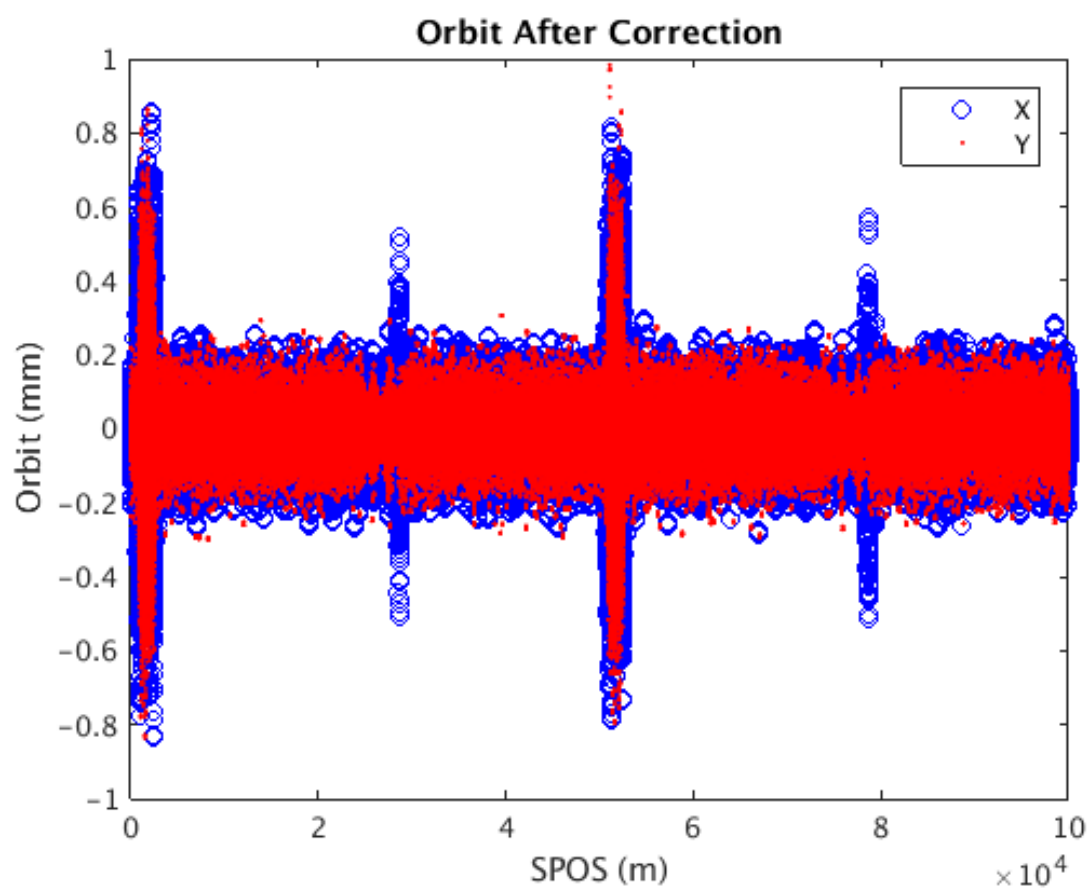

**Figure 5.2.1.9:** Closed orbit after correction

Fig.5.2.1.10 shows the distribution of orbit, dispersion and beta-beat after the orbit correction. 11000 locations in the ring and 10 random seeds are chosen to generate these distributions for a total of 110000 samples. From those results, the rms orbit after COD correction is 80 μm; the rms dispersion is 14 mm and the rms beta-beat is 3.5%.

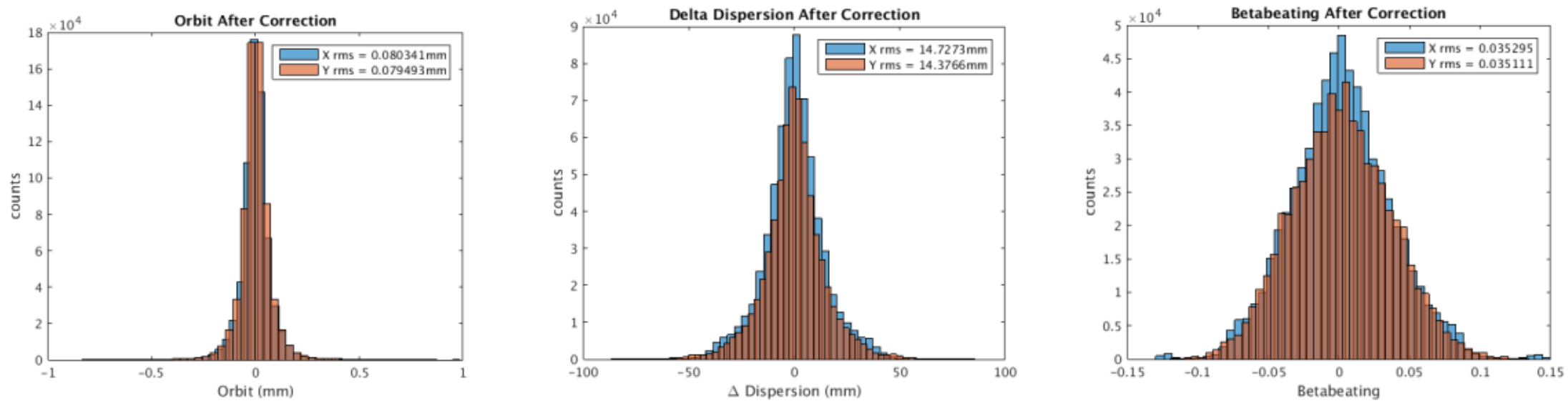

**Figure 5.2.1.10:** Distribution for orbit, dispersion and beta-beating after COD correction.

Fig.5.2.1.11 shows the distribution of emittance and coupling for the 10 random seeds. The right plot shows that the relative vertical emittance due to coupling is less than 10% for all the simulation seeds. 512 Sextupoles are used to correct the coupling and residual coupling after correction is controlled under 0.5% which is shown in Fig. 5.2.1.12.

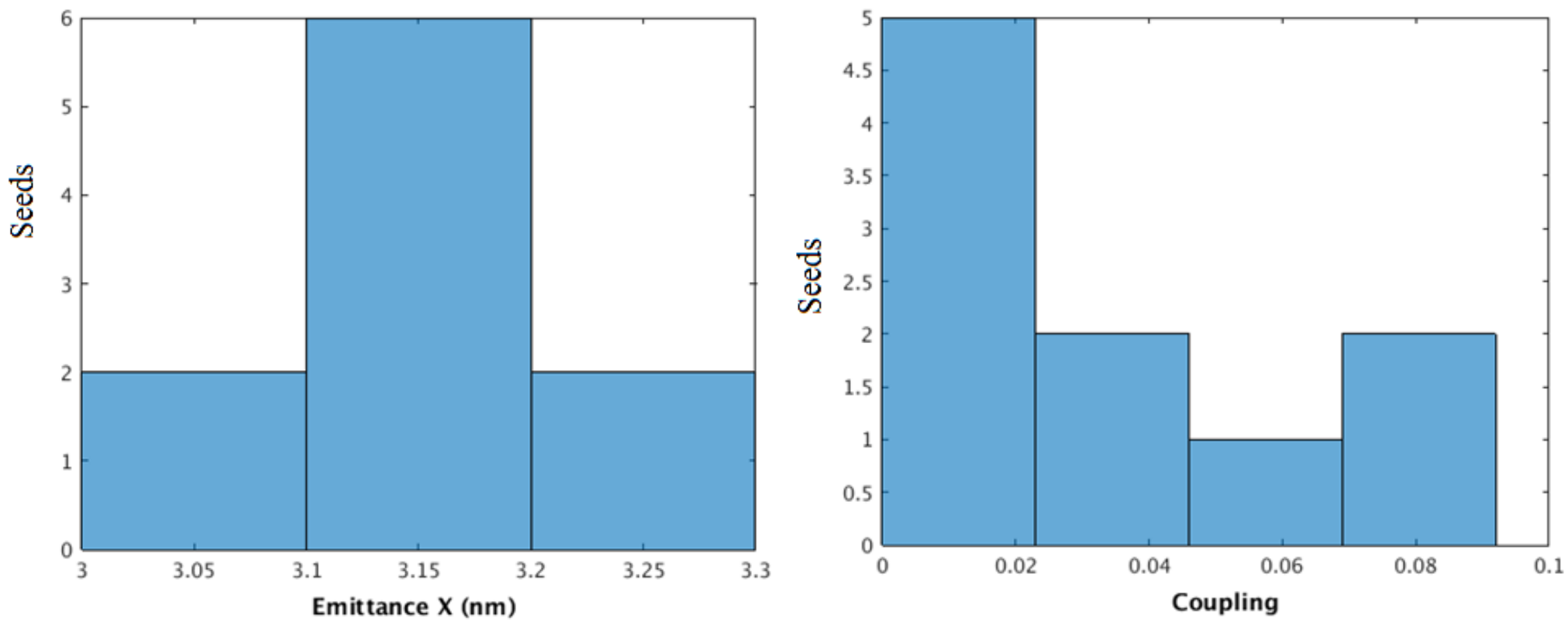

**Figure 5.2.1.11:** Emittance distribution & coupling distribution after COD correction.

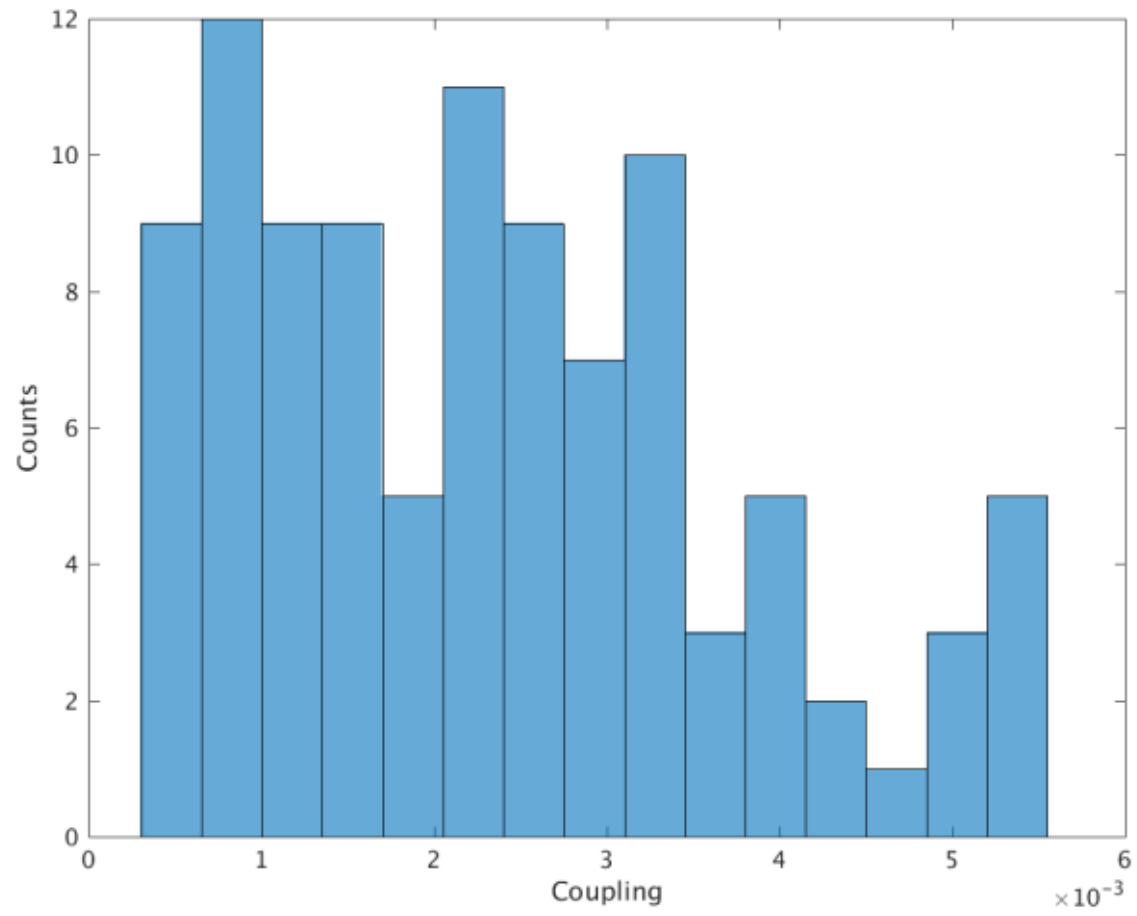

**Figure 5.2.1.12:** Coupling distribution after coupling correction.

### *5.2.1.2.2 Dynamic Aperture*

A non-interleave scheme and two sextupole families are adopted for linear chromaticity correction. With dedicated sextupole arrangement, the nonlinearity of the booster is corrected to fifth order [1, 3]. Both the phase advances between sextupole pairs and the ones between octants are optimized carefully in order to achieve larger dynamic aperture for both on-momentum and off-momentum particles. The thick black line in Fig. 5.2.1.10 shows the dynamic aperture of bare lattice; the purple line is the DA with errors and orbit corrections; the red line is the DA with errors; the thin black line in the left plot is the beam stay clear region at 10GeV and the blue line in the right plot is the DA with orbit corrections and radiative damping effect at 120GeV.

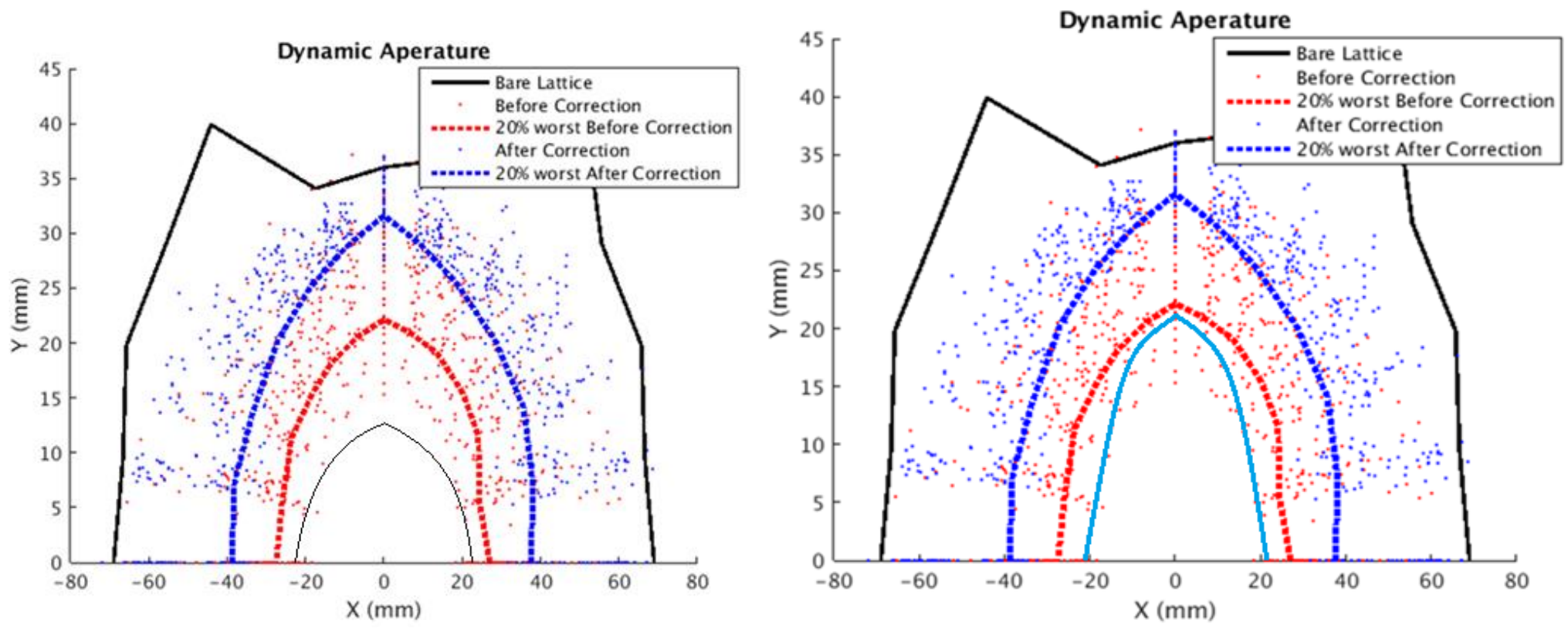

**Figure 5.2.1.13:** Dynamic aperture of booster (left: 10GeV; right: 120GeV).

With errors and orbit corrections, the dynamic aperture of the booster is nearly two thirds of that for the bare lattice as shown in Fig. 5.2.1.13. At 10GeV, the DA with errors should be larger than the beam stay clear region which is shown in the left plot of Fig. 5.2.1.13. At 120GeV, the radiative damping effect and sawtooth effect is also considered except for the error effect, and the according DA result including damping and sawtooth is shown as the blue line in the right plot of Fig. 5.2.1.13. The DA requirement and the real DA results which have been realized are listed in table 5.2.1.2. Where the DA requirement at 10GeV is determined by the beam stay clear region and is determined by the re-injection process from the collider in the on-axis injection scheme at 120GeV.

In conclusion the errors in the booster are tolerable.

**Table 5.2.1.2:** Summary of booster DA results.

| | DA requirement | | DA results | |
|---|---|---|---|---|
| | H | V | H | V |
| 10GeV ($\varepsilon_x$= $\varepsilon_y$ =120nm) | 4$\sigma_x$ +5mm | 4$\sigma_y$ +5mm | 7.7$\sigma_x$ +5mm | 14.3$\sigma_y$ +5mm |
| 120GeV ($\varepsilon_x$=3.57nm, $\varepsilon_y$= $\varepsilon_x$*0.005) | 6$\sigma_x$ +3mm | 49$\sigma_y$ +3mm | 21.8$\sigma_x$ +3mm | 779$\sigma_y$ +3mm |

### 5.2.2 **Beam Instability**

Because of the lower energy during injection, beam instability caused by impedance can be more serious than that in the Collider. Resistive wall impedance varies with different vacuum pipe materials. After comparing the impedance with stainless steel, copper and aluminum, aluminum has been chosen for the Booster vacuum chamber. The longitudinal and transverse resistive wall impedance is shown in Figure 5.2.2.1.

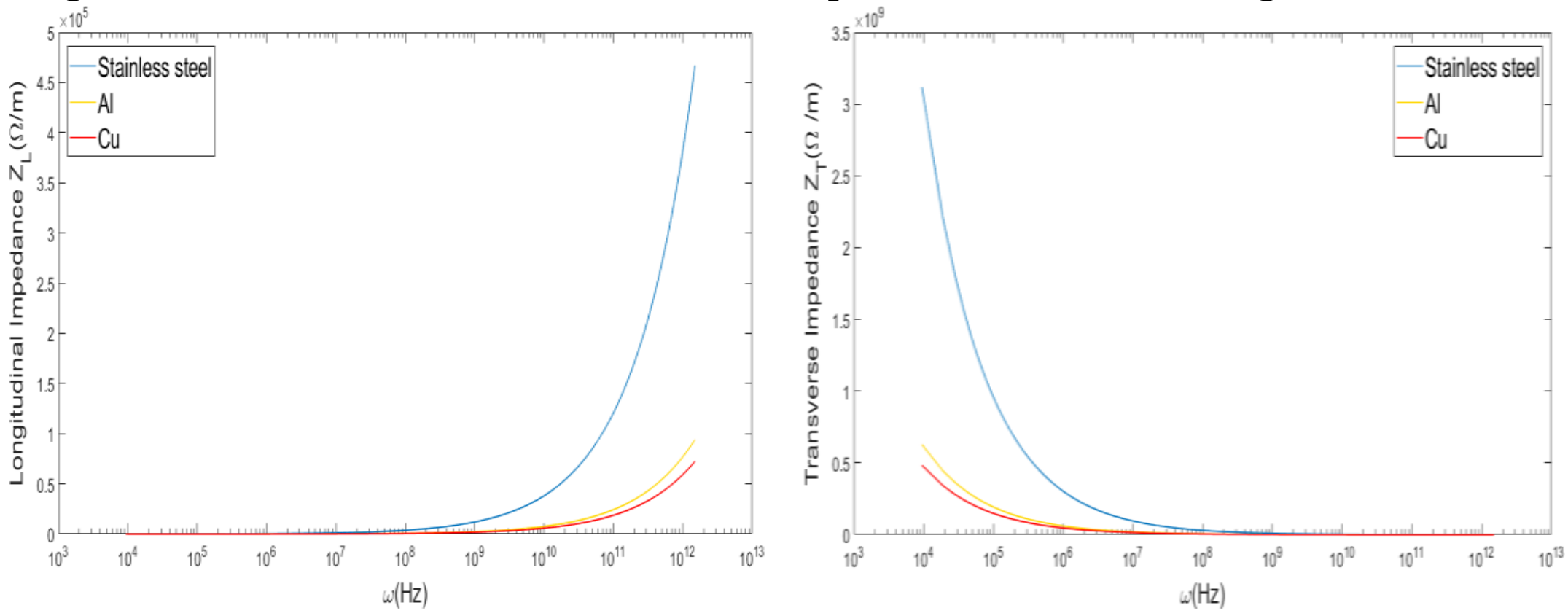

**Figure 5.2.2.1:** Booster resistive wall impedance

The transverse mode coupling instability (TMCI) caused by the transverse resistive wall impedance is the dominant factor in limiting the bunch current. The formula,

$$I_b^{th} \approx 0.7 \frac{4\pi c \nu_s (E/e)}{C} \frac{1}{\sum_i \beta_{t,i} \kappa_{t,i}}$$

gives a bunch current threshold of 25.1 μA, higher than the design current 3.3 μA. In the above formula $\nu_s$ is the synchrotron tune, $\beta_{t,j}$ is the average beta function in the $j^{th}$ element, $\kappa_{t,j}$ is the transverse loss factor and $E$ is the beam injection energy. The dependence of bunch current threshold on the pipe radius is shown in Figure 5.2.2.2.

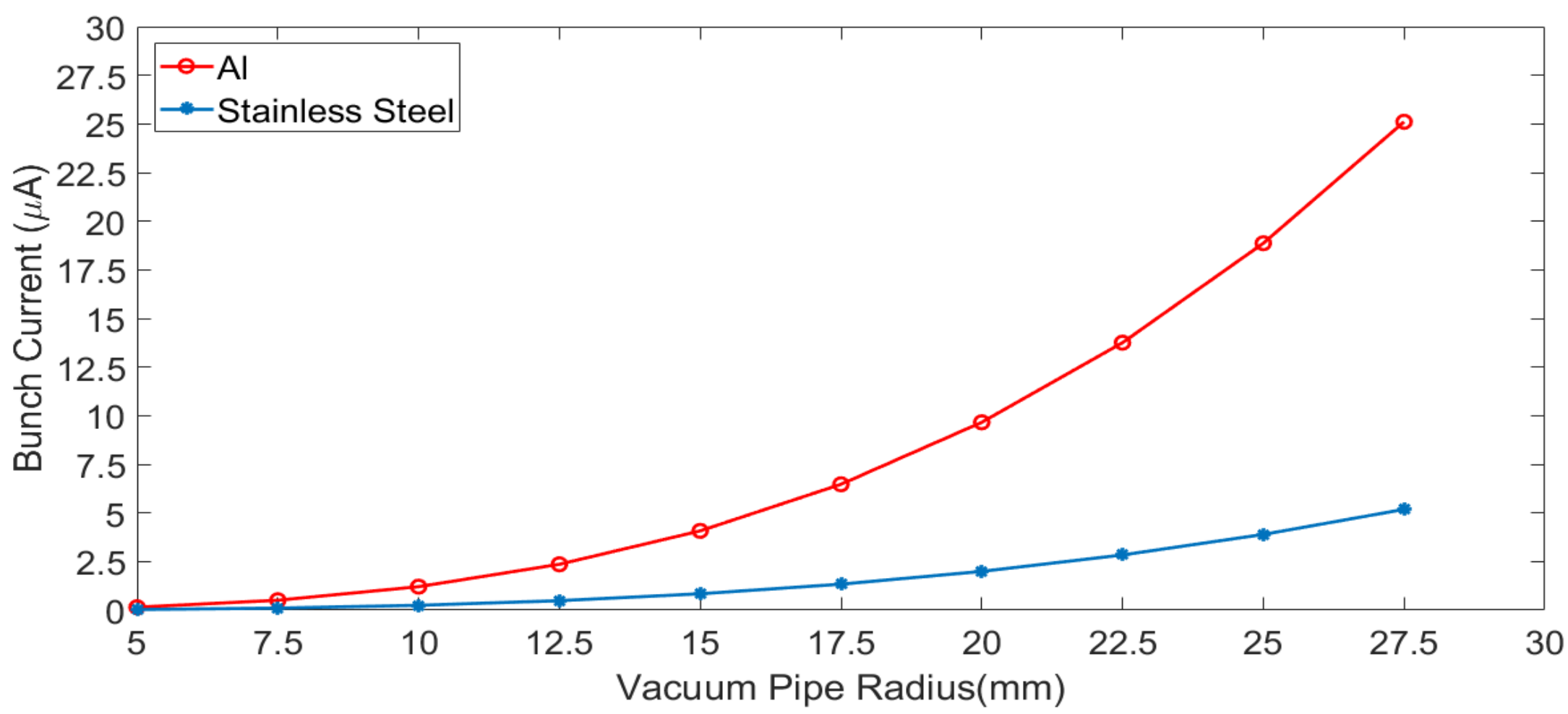

**Figure 5.2.2.2:** Bunch current threshold & the pipe radius

With multi-bunch operation, the growth rate in the transverse resistive wall instability was estimated by

$$\frac{1}{\tau_t} = -\frac{I_b c}{4\pi(E/e)\nu_{x,y}} \sum_{p=-\infty}^{\infty} \mathrm{Re}\, Z_t(\omega_{pn}),$$

where $\omega_{pn} = \omega_0\,(pn_b + n + \nu_{x,y})$ and $\mathrm{Re}Z_t(\omega_{pn})$ is the real part of transverse impedance. At injection with low and high beam current (2 mA and 10 mA, respectively), the dependence of instability growth rate on the pipe radius is shown in Figs. 5.2.2.3 and 5.2.2.4.

The aluminum pipe with radius 27.5 mm is chosen to accommodate high current injection. The mode distribution of instability for Higgs and Z are shown in figure 5.2.2.3 and 5.2.2.4, respectively. The most dangerous modes number is 224 for Higgs and 340 for Z. A transverse feedback system is necessary to stabilize the bunch oscillation.

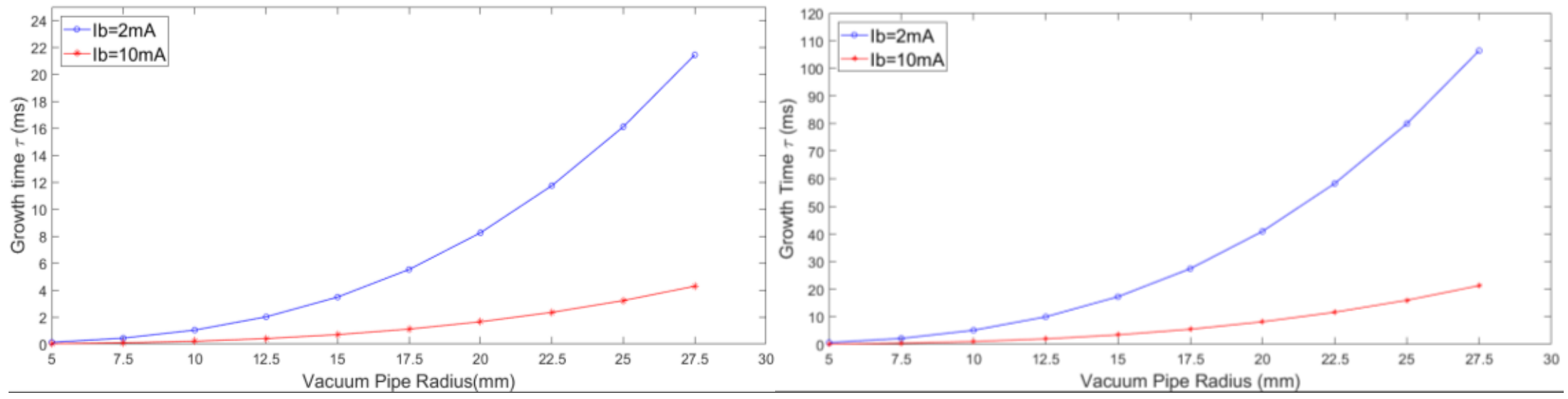

**Figure 5.2.2.3:** The instability growth rate vs. vacuum pipe radius (Left: stainless steel; right: Al)

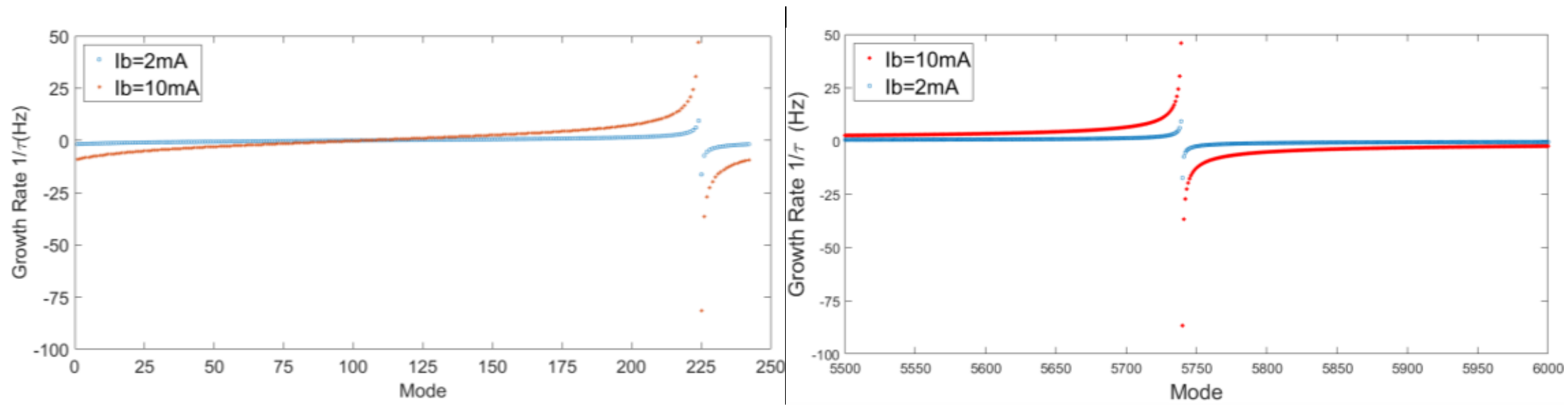

**Figure 5.2.2.4:** The mode distribution of instability for Higgs and Z (Left: Higgs; right: Z)

For Higgs commissioning, an alternative beam injection scheme is on-axis swap-out method, that is, some high current bunches from the collider (7-14 bunches with 70 μA per bunch) are injected into the booster and combined with existing bunches and then re-injected into the collider. With this scheme, the bunch current threshold is 300 μA and multi-bunch instability growth time is about 154 ms. The instability threshold and mode distribution of this scheme are shown in Figure 5.2.2.5.

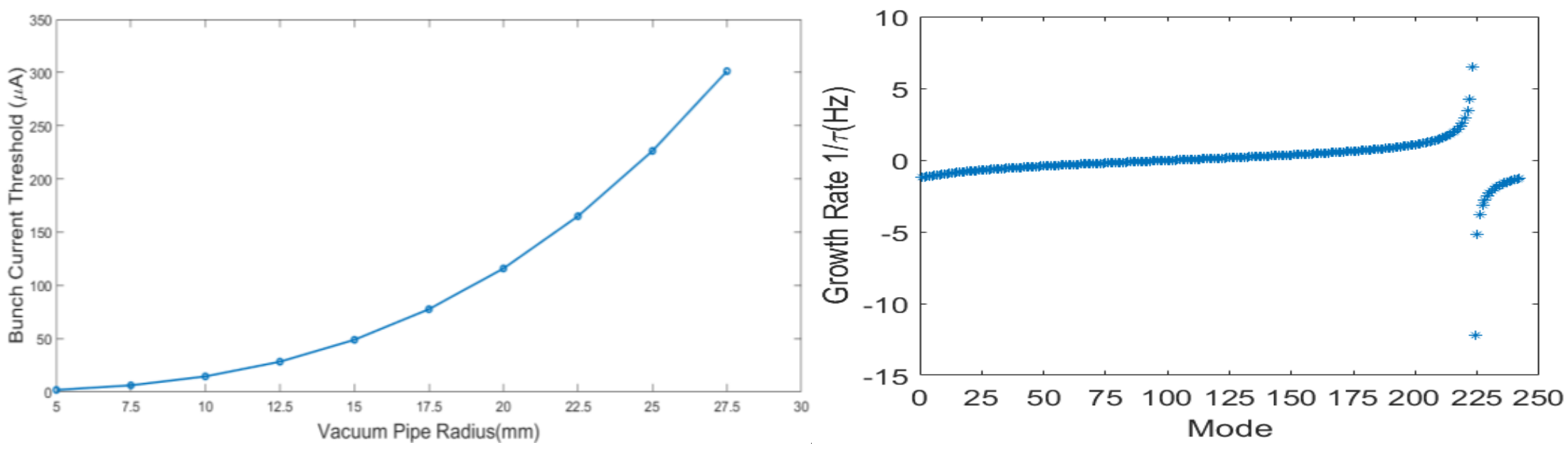

**Figure 5.2.2.5:** The instability threshold and mode distribution for on-axis injection scheme

### 5.2.3 Injection and Extraction

On-axis injection is used for the Booster. The system consists of a horizontal septum and a single kicker in a Booster long straight section as shown in Fig. 5.2.3.1 and summarized in Table 5.2.3.1. Booster extraction is similar to injection. It is located near IP1; magnet parameters are shown in Table 5.2.3.2.

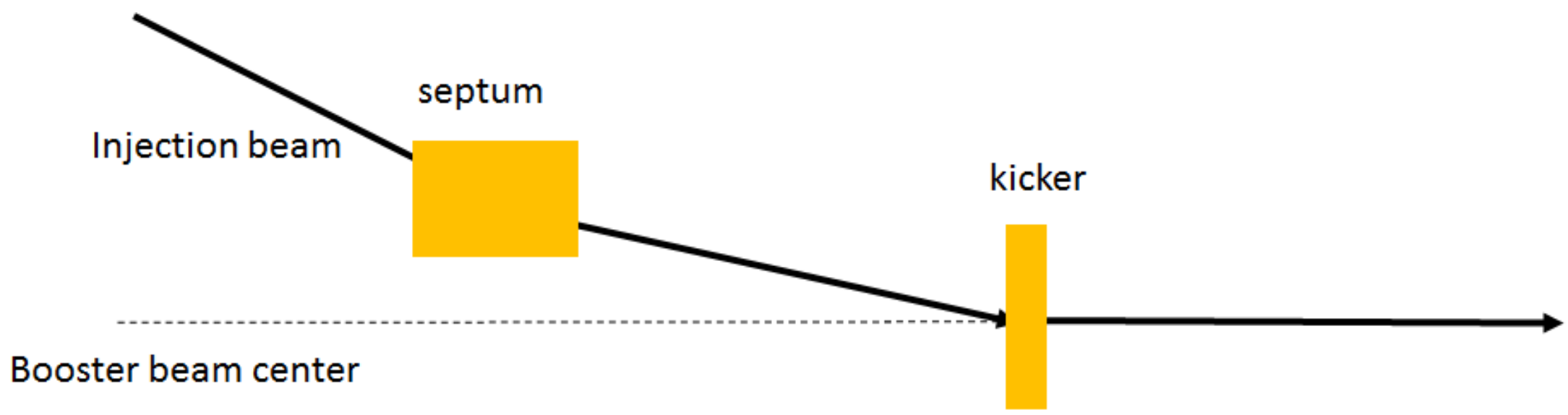

**Figure 5.2.3.1:** The Booster on-axis injection scheme

**Table 5.2.3.1:** Parameters of the injection septum and kickers

| Component | Length (m) | Waveform | Deflection angle (mrad) | Field (T) | Beam-Stay-clear | |
|---|---|---|---|---|---|---|
| | | | | | H(mm) | V(mm) |
| Septum | 2 | DC | 9.1 | 0.152 | 88 | 88 |
| Kicker | 0.5 | Half_sine | 0.5 | 0.034 | 88 | 88 |

**Table 5.2.3.2:** Parameters of the extraction septum and kickers

| Component | Length (m) | Waveform | Deflection angle (mrad) | Field (T) | Beam-Stay-clear | |
|---|---|---|---|---|---|---|
| | | | | | H(mm) | V(mm) |
| Septum | 10 | DC | 10.4 | 0.58 | 20 | 20 |
| Kicker | 2 | Half_sine | 0.2 | 0.056 | 20 | 20 |

### 5.2.4 Transport Lines

The Booster-to-Collider transport line consists of three parts. The first and third parts are horizontal bends, and the second part is a vertical bend. The Twiss parameters are shown in Fig. 5.2.4.1.

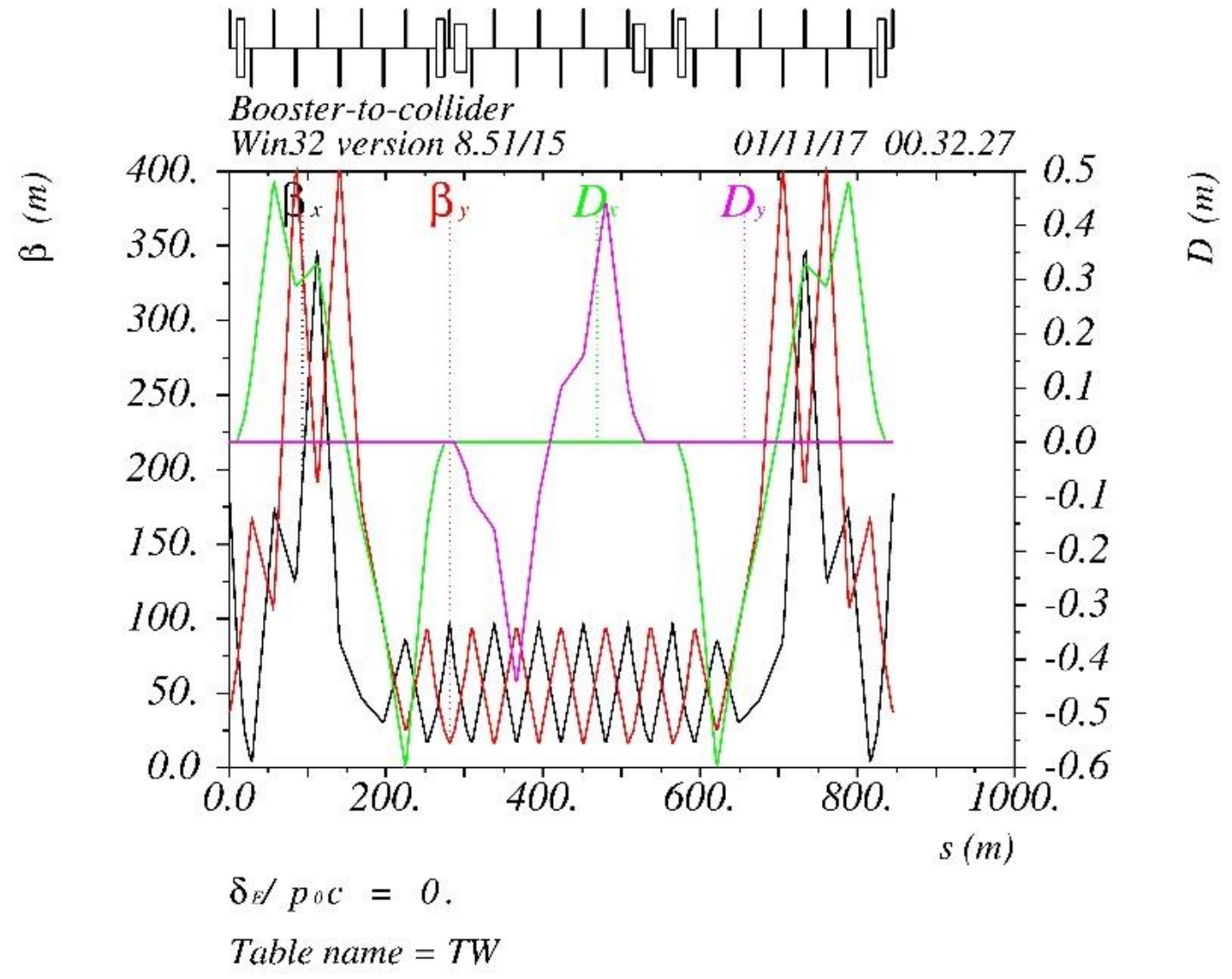

**Figure 5.2.4.1:** Twiss parameters of the Booster to Collider transport line

## 5.2.5 Synchrotron Radiation

### 5.2.5.1 *Synchrotron Radiation from Bending Magnets*

The energy and bending circumference of the Booster are the same as the Collider; however, the currents are different.

The total SR power per unit length can be calculated by:

$$P = 14.08 \frac{E^4 I}{\rho^2} \tag{5.2.5.1}$$

where *P* is the synchrotron power loss in watts, *I* is the current of the circulating particles in mA, *E* is in GeV and $\rho$ is in meter as before.

The photon spectrum can be calculated by formulas 4.2.4.5 through 4.2.4.12. The parameters of the SR emitted in the dipoles are shown in Table 5.2.5.1.

**Table 5.2.5.1:** Parameters of Booster SR

| Parameter | Symbol | Value | Unit |
|---|---|---|---|
| Beam energy | $E$ | 120 | GeV |
| Beam current | $I$ | 0.5227 | mA |
| Bending radius | $\rho$ | 10900 | m |
| Power per unit length | $P$ | 12.85 | W/m |
| Critical energy | $E_c$ | 351.6 | keV |
| Bending angle | $\theta$ | 2.844 | mrad |
| Opening angle | $\varphi$ | 4.258 | μrad |

Fig. 5.2.5.1 shows the spectra of SR photons produced in the Booster dipoles at different electron energies.

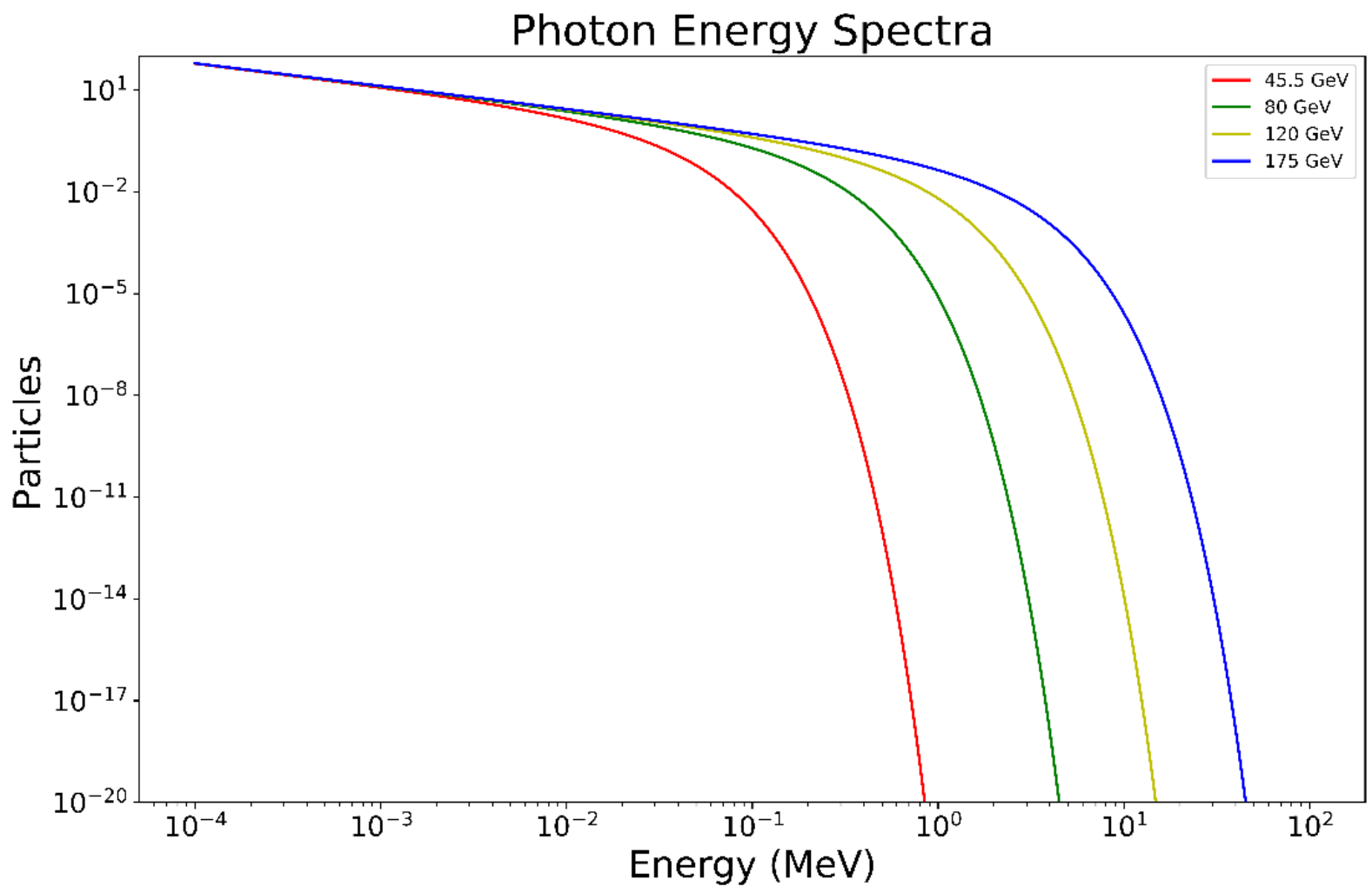

**Fig. 5.2.5.1:** The SR photon spectrum at different beam energies

#### 5.2.5.2 *Dose Estimation for the Booster*

The SR power of a single beam is 12.8 W/m, which is three percent of the Collider SR power so energy deposition is not a problem. However, since the magnets are not shielded, radiation dose should be considered. A Monte Carlo simulation is used to scale dose rates for the Booster from results obtained for the Collider. Besides the current, the Booster has a lower duty cycle. The actual dose is lower than $1.5\times10^4$ Gy/Ah at 120 GeV compared to $5.14 \times 10^5$ for the Collider.

## 5.3 Booster Technical Systems

### 5.3.1 Superconducting RF System

#### 5.3.1.1 *Booster RF Parameters*

The Booster RF system parameters are summarized in Table 5.3.1.1. Booster cavities operate in fast ramp mode. The maximum energy ramp rate is 22 GeV/s. The maximum total RF voltage ramp rate is 0.4 GV/s. The maximum cavity voltage ramp rate is 3.8 MV/s. The major challenge is the narrow bandwidth operation with microphonics and Lorentz detuning and higher order mode instabilities during energy and voltage ramp in a short time.

The Higgs parameters in Table 5.3.1.1 are for the off-axis injection. The average Higgs beam current will increase to 1 mA for the on-axis injection. Up to 13 bunches with 33 times more charge will be injected back from the Collider Ring and result in much higher peak and average cavity HOM power and moderate transient beam loading.

**Table 5.3.1.1:** CEPC Booster RF parameters

| | H | W | Z |
|---|---|---|---|
| Injection beam energy [GeV] | 10 | | |
| Extraction beam energy [GeV] | 120 | 80 | 45.5 |
| Extraction average SR power [MW] | 0.03 | 0.02 | 0.016 |
| Total RF input power (peak) [MW] | 1.75 | 0.8 | 0.23 |
| RF frequency [MHz] | 1300 | | |
| Beam current [mA] | 0.52 | 2.63 | 6.91 |
| Bunch charge [nC] | 0.72 | 0.58 | 0.38 |
| Bunches / beam | 242 | 1524 | 6000 |
| Injection RF voltage [GV] | 0.0627 | | |
| Extraction RF voltage [GV] | 1.97 | 0.585 | 0.287 |
| One injection duration for top-up (e+ and e-) [s] | 25.8 | 45.7 | 137.6 |
| Injection from linac [s] | 2.42 | 15.24 | 60 |
| Ramp-up or ramp-down duration [s] | 5.0 | 3.3 | 1.9 |
| Damping duration [s] | 0.5 | 1.0 | 5.0 |
| Extraction duration [ms] | 0.3 | 0.3 | 0.3 |
| Number of injection times | 1 | 1 | 2 |
| Total injection duration | 25.8 | 45.7 | 275.2 |
| Injection interval for top-up [s] | 73 | 153 | 438 |
| Injection synchrotron phase from crest [deg] | 89.93 | | |
| Extraction synchrotron phase from crest [deg] | 39.5 | 59.1 | 83.6 |
| Bunch length injected from linac [mm] | 1 | | |
| Extraction equilibrium bunch length [mm] | 2.7 | 2.4 | 1.3 |
| Injection longitudinal damping time [s] | 45.2 | | |
| Extraction longitudinal damping time [ms] | 26 | 89 | 474 |
| RF station number | 2 | | |
| Cavity number | 96 | 64 | 32 |
| Cavity number / cryomodule | 8 | | |
| Cryomodule number | 12 | 8 | 4 |
| Cell number / cavity | 9 | | |
| Effective length [m] | 1.038 | | |
| R/Q [Ω] | 1036 | | |
| Injection cavity gradient [MV/m] | 0.6 | 0.9 | 1.9 |
| Extraction cavity gradient [MV/m] | 19.8 | 8.8 | 8.6 |
| $Q_0$ at operating gradient for long term | 1E10 | | |
| Acceptance gradient in vertical test [MV/m] | 24 | | |
| $Q_0$ at acceptance gradient in vertical test | 3E10 | | |

| | H | W | Z |
|---|---|---|---|
| $Q_L$ | 1.0E7 | 6.5E6 | 1.0E7 |
| Cavity bandwidth [Hz] | 130 | 200 | 130 |
| Beam power (peak) / cavity [kW] | 8.3 | 12.3 | 6.9 |
| Input power (peak) / cavity [kW] | 18.2 | 12.4 | 7.1 |
| RF power duty factor | 3.8 % | 2.4 % | 7.4 % |
| Input power (average) / cavity [kW] | 0.7 | 0.3 | 0.5 |
| Input coupler power capacity (peak) [kW] | 20.0 | | |
| Input coupler power capacity (average) [kW] | 4.0 | | |
| External Q adjusting range | 4E6 ~ 2E7 | | |
| Cavity number / SSA | 1 | | |
| SSA power [kW] | 25 | | |
| SSA number | 96 | 64 | 32 |
| HOM power (peak) / cavity [W] | 1 | 4.4 | 12.1 |
| HOM power (average) / cavity [W] | 0.18 | 0.69 | 4.1 |
| HOM coupler power capacity [W] | 5 | | |
| HOM absorber power capacity @ 80 K [W] | 10 | | |
| Cryogenic dynamic heat load duty factor | 4.3 % | 2.9 % | 7.9 % |
| Total cavity wall loss (average) @ 2 K [kW] | 0.17 | 0.01 | 0.02 |

The Booster cavity HOM CBI growth rates of the dangerous modes have been calculated with the measured external Q of the TESLA cavity (Table 5.3.1.2) assuming all the idle cavities are in the beamline for W and Z. In the calculation, the average $\beta_{x,y}$ in the RF cavity is 30 m.

**Table 5.3.1.2:** CEPC Booster TESLA 9-cavity HOM CBI growth time

| Modes | $f$ (GHz) | $R/Q$ (monopole Ω, dipole Ω/m) | $Q_e$ measured | CBI Growth Time (ms) | | |
|---|---|---|---|---|---|---|
| | | | | H-extraction | W-extraction | Z-extraction |
| TM011 | 2.45 | 156 | 5.9E4 | 2307 | 236.7 | 51.3 |
| TM012 | 3.845 | 44 | 2.4E5 | 1281.3 | 131.5 | 28.5 |
| TE111 | 1.739 | 4283 | 3.4E3 | 7308.6 | 967.1 | 209.6 |
| TM110 | 1.874 | 2293 | 5.0E4 | 928.3 | 122.8 | 26.6 |
| TM111 | 2.577 | 4336 | 5.0E4 | 490.9 | 65 | 14.1 |
| TE121 | 3.087 | 196 | 4.4E4 | 12341 | 1633 | 353.9 |
| | | | | H-injection | W-injection | Z-injection |
| TM011 | 2.45 | 156 | 5.9E4 | 149 | 29.6 | 11.3 |
| TM012 | 3.845 | 44 | 2.4E5 | 82.7 | 16.4 | 6.3 |
| TE111 | 1.739 | 4283 | 3.4E3 | 609 | 120.9 | 46.1 |
| TM110 | 1.874 | 2293 | 5.0E4 | 77.4 | 15.4 | 5.9 |
| TM111 | 2.577 | 4336 | 5.0E4 | 40.9 | 8.1 | 3.1 |
| TE121 | 3.087 | 196 | 4.4E4 | 1028.4 | 204.1 | 77.8 |

Due to the low energy injection to the Booster, the growth time of all the modes are much shorter than the radiation damping time during injection and while the Booster is in the low energy region. With the assumed bunch-by-bunch transverse feedback time of 3.3 ms and longitudinal feedback time of 2.5~3.2 ms (synchrotron oscillation period), nearly all the modes are safe with enough margin except TM111 (above beam tube cut off). There will be more margin (usually more than one to two orders of magnitude) if the HOM frequency spreads among the cavities are included.

The fundamental mode RF power and longitudinal coupled-bunch instabilities excited by the detuned fundamental mode of the parked TESLA cavities for W and Z operation, as well as the effect of the other eight passband modes of TM010 is to be studied.

#### 5.3.1.2 ***RF Voltage Ramp***

The Booster beam energy increases nearly linearly with time. During the energy ramp the total RF voltage must be increased to avoid synchro-betatron oscillations, in other words to keep the synchrotron tune constant. At higher energy, the bunch length (should not be too short) and quantum lifetime are the dominating factors. Figure 5.3.1.1 shows the voltage ramp and the associated beam dynamic parameters change with constant synchrotron tune. Figure 5.3.1.2 shows the RF parameters change during the injection and ramp up. Both figures are for Higgs operation.

The LLRF tuning system must compensate for the relatively large Lorentz force detuning and microphonics compared with the narrow cavity bandwidth (130 Hz) during the voltage ramp. The Lorentz force detuning coefficient is assumed to be 1 Hz / $(MV/m)^2$, which is typical for a TESLA 9-cell cavity with the helium vessel and end lever tuner. Ten percent of the Lorentz force detuning is assumed to remain after the piezo compensation. The microphonics is assumed to be 10 Hz peak to peak. First and second

Robison instability should be considered during injection and ramping. Counter-phasing is a possible way to ramp the effective RF voltage.

During the voltage ramp, multipacting may happen in the input coupler and cavity. Counter-phasing ramp of the total effective RF voltage could be used to avoid cavity trip.

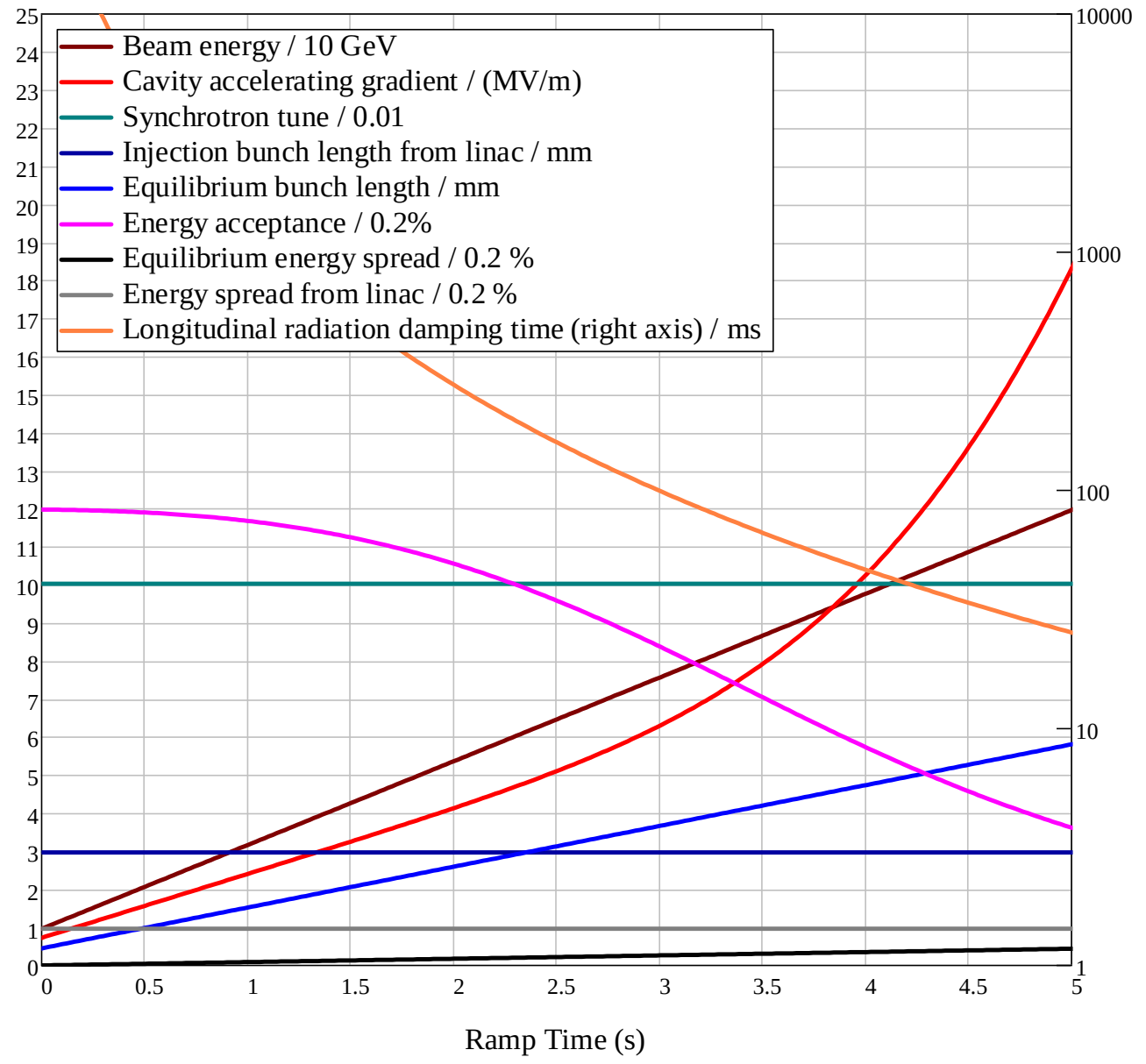

**Figure 5.3.1.1:** Booster RF voltage ramp with constant synchrotron tune (Higgs mode)

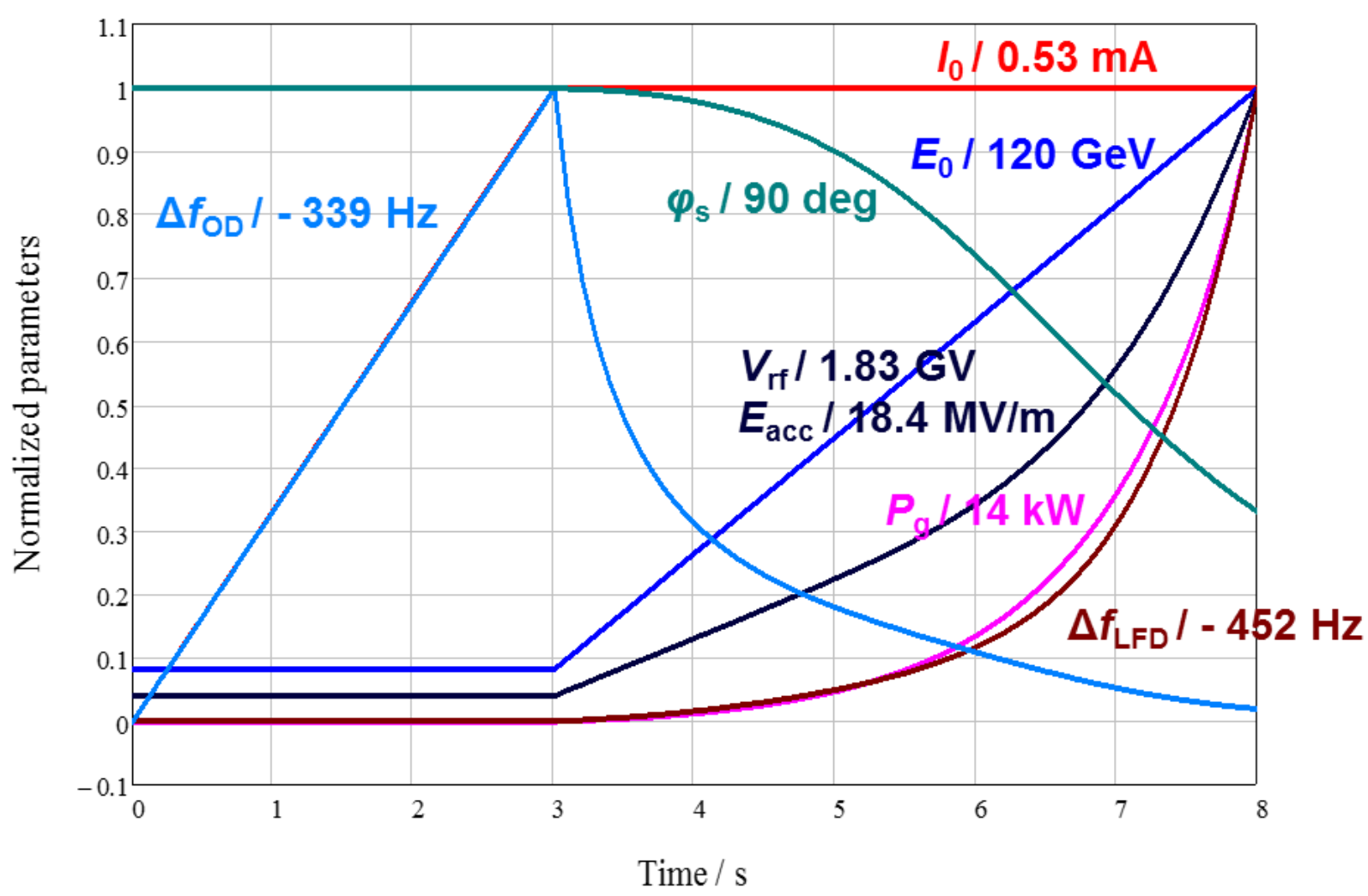

**Figure 5.3.1.2:** Booster RF parameters during injection and ramp up (Higgs mode)

### 5.3.1.3 ***1.3 GHz SRF Technology***

TESLA 1.3 GHz 9-cell cavities will be used for the Booster. The cavity is made of bulk niobium and operates at 2 K with $Q_0 = 1 \times 10^{10}$ below 20 MV/m for long term and should reach $Q_0 = 3 \times 10^{10}$ at 24 MV/m by nitrogen-doping for the vertical acceptance test.

The TTF III coupler used for Euro-XFEL and modified for LCLS-II could be used with peak power of 20 kW and average power up to 4 kW. Due to the fast voltage ramp, beam loading variation and narrow bandwidth of the Booster 1.3 GHz cavity, it is critical for the tuner to compensate the dynamic Lorenz force detuning and micro-phonics, and keep the beam stable with proper detuning.

## 5.3.2 RF Power Source

### 5.3.2.1 *Introduction*

The Booster RF system consists of 1300 MHz superconducting RF cavities. There are 12 cryo-modules for Higgs operation, each containing eight 9-cell superconducting cavities. These cavities need 96 1300 MHz power sources. Solid state amplifiers (SSAs) provide the required power. Booster SRF system parameters are listed in Table 5.3.2.1.

**Table 5.3.2.1:** Booster SRF system parameters

| | |
|---|---|
| Operation frequency (MHz) | 1300+/-0.5 |
| Cavity number | 96 |
| RF source number | 96 (25 kW SSA) |

### 5.3.2.2 *RF Power Source Choice*

Different possible alternatives for the Booster power sources were considered in terms of modularity and technology: vacuum tubes (Klystron, IOT – Inductive Output Tube, Diacrode) and solid state [1]. With the progress of transistor technology, especially the emergence of the sixth generation LDMOSFET, the output power and efficiency of a single transistor has been greatly improved. High power can be obtained by a combination of numerous transistors. SSAs are being considered for an increasing number of accelerators, both circular and linear. Their capabilities extend from a few kW to several hundred kW, and from less than 100 MHz to above 1 GHz. Reasonable efficiency (~50%), high gain, and modular design provide high reliability.

The SSA has many important advantages: high reliability for redundancy design, high flexibility for module design, high stability, low maintenance requirements, absence of warm-up time and low voltage operation and reasonable efficiency. So the SSA has been chosen for the Booster RF power source system.

### 5.3.2.3 *Solid State Amplifier*

Each cavity requires 14.1 kW peak power to accelerate 0.53 mA beam current at a gradient of 18.4 MV/m for operation in the Higgs mode. Taking into account LLRF control reserve, transmission losses, reflection and power redundancy, the total power requirement of each power source is 25 kW. One cavity per source is chosen to simplify the LLRF control and waveguide distribution.

The 1 dB bandwidth of the SSA needs more than 1 MHz for the LLRF control requirement. The nominal maximum power is achieved with less than 1 dB compression, and the harmonic power output is less than -30 dBc. The AC to RF efficiency goal for the SSAs at the rated power is at least 45%. The mean time between failures (MTBF) should be larger than 30,000 hours, with less than 5% of the power modules failing per year. The Booster can still run in the event of the failure of one module. The output port is standard WR650 waveguide. The Booster SSA specifications are listed in Table 5.3.2.2.

**Table 5.3.2.2:** 25 kW/1.3 GHz SSA Specifications

| Parameters | Values |
|---|---|
| Frequency | 1.3 GHz |
| Power (< 1 dB Compression) | 25 kW |
| Gain | ≥65 dB |
| Bandwidth (1 dB) | ≥ 1 MHz |
| Amplitude stability | ≤0.1% RMS |
| Phase stability | ≤0.1°RMS |
| Phase Variation(1kW25kW) | ≤10° |
| Harmonic | < -30 dBc |
| Spurious | < -60 dBc |
| Efficiency at 25kW | ≥45% |
| MTBF | ≥30000 h |
| Redundancy | Booster can still run with 1 power module failure |

### 5.3.2.4 *RF Transmission System*

The purpose of the RF transmission system (RFTS) is to deliver the power from the RF power sources to the RF cavities. The RFTS design must have high transmission efficiency, low reflection, and be flexible, reliable as well as cost-effective.

Each cavity will be fed by a 25 kW solid state amplifier (SSA) operating at 1.3 GHz. The RF power produced by the SSA will be delivered by standard WR650 waveguides. The schematic of the Booster RFTS and the placement of the SSA+LLRF are shown in Figures 5.3.2.1 and 5.3.2.2.

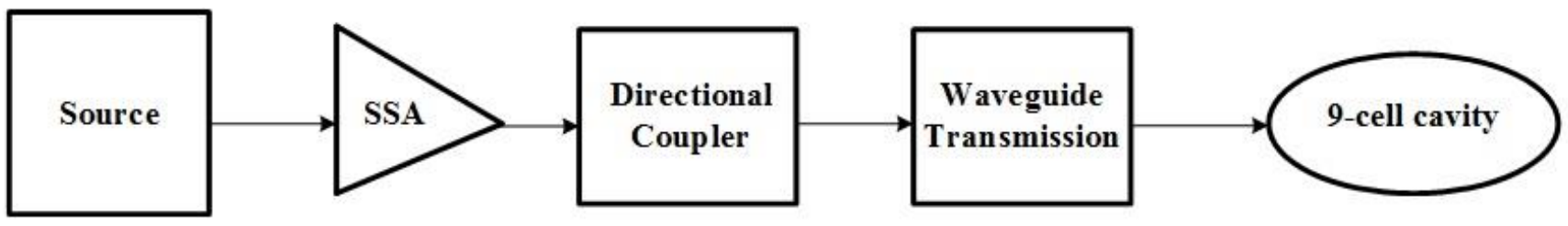

**Figure 5.3.2.1:** Schematic of the Booster RFTS

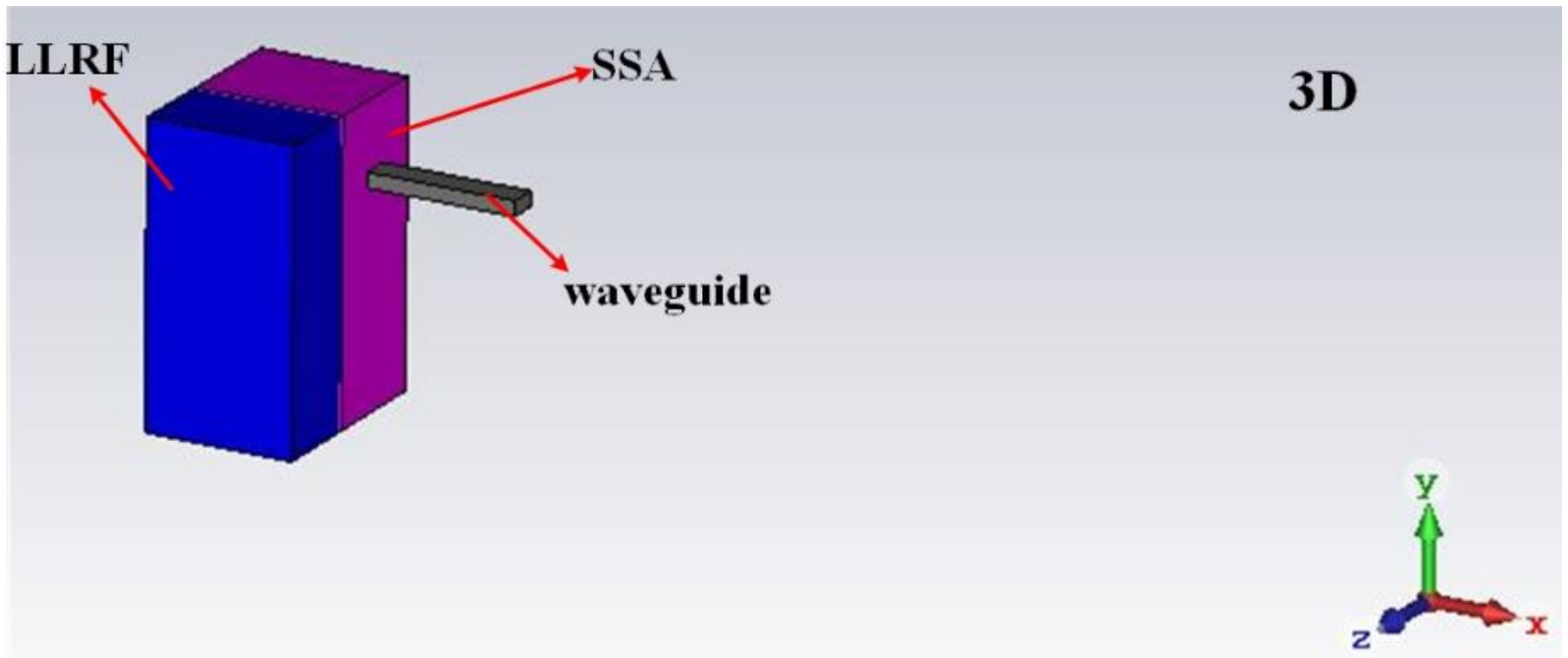

**Figure 5.3.2.2:** Placement of the SSA and LLRF

One source per cavity simplifies the RF power delivery. There will be no power splitters or phase shifters in the Booster RFTS system, and the length differences as well as the thermal phase drifts will be easily tracked and compensated by the individual LLRF controls. Besides the waveguides, each power transmission line will have a directional coupler and a waveguide-coaxial transition [2]. The waveguide-coaxial transition is used to connect the waveguide system with the coaxial coupler of the 9-cell superconducting cavity.

### 5.3.2.5 *LLRF System*

One Booster cycle is divided into four periods: beam injection, energy ramp, a flat top for beam extraction and ramp down. RF voltages in the cavities are modulated and controlled by the LLRF. The LLRF system layout for the Booster is shown in Figure 5.3.2.3. Operation and cavity parameters are listed in Table 5.3.2.3. The amplitude and phase stabilization requirements for these cavities are 0.1% (rms) and 0.1 deg (rms).

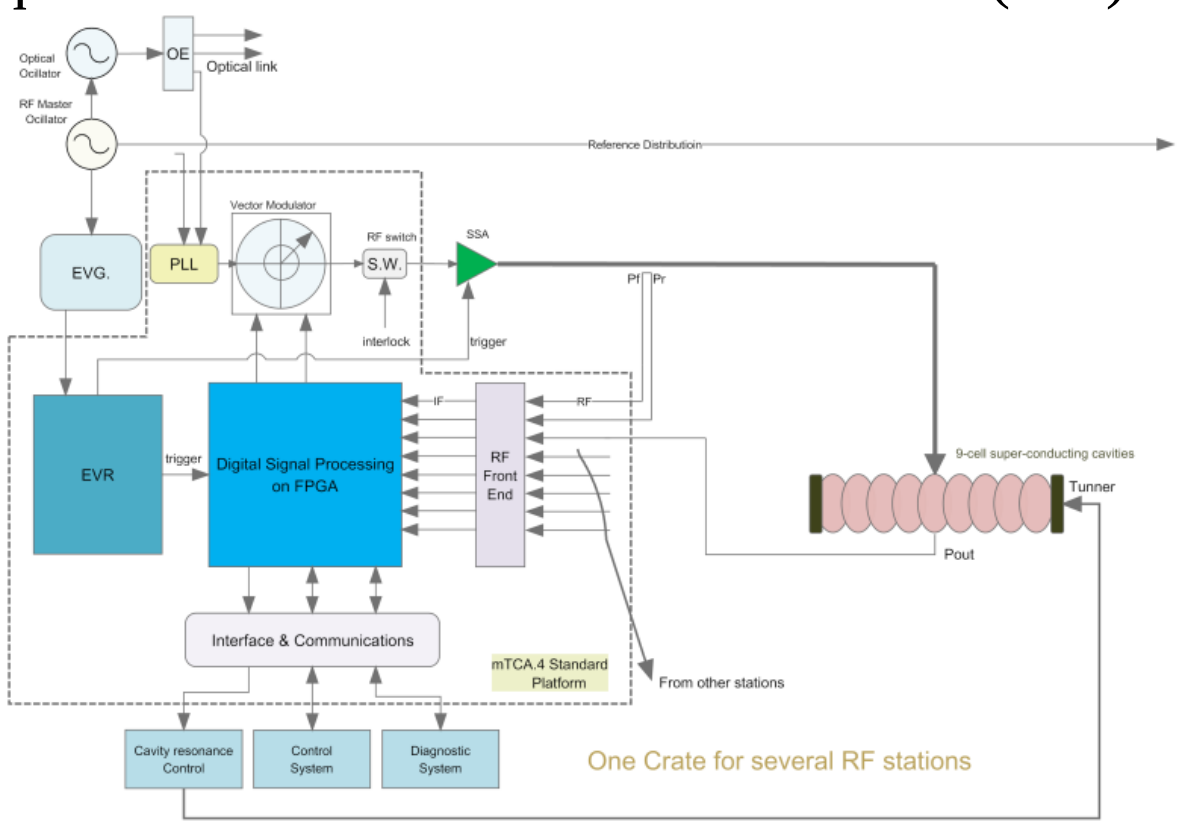

**Figure 5.3.2.3:** The LLRF system layout for the Booster

The LLRF system for the superconducting cavities includes the cavity resonance controller, which is placed in the same crate as the motion control for the fast and slow tuners. Those fast motion controllers drive the piezo actuators. It may become necessary to adopt iterative learning algorithms corresponding to timing triggers for the operation mode.

**Table 5.3.2.3.** Parameters of the Booster

| | H | W | Z |
|---|---|---|---|
| RF frequency [MHz] | 1300 | 1300 | 1300 |
| Injection and extraction cycle [s] | 13.5 | 16.5 | 26.5 |
| Injection duration [s] | 3 | 10 | 24 |
| Ramping duration [s] | 5.0 | 3.0 | 0.5 |
| Extraction duration [s] | 0.5 | 0.5 | 0.5 |
| Cavity number | 96 | 64 | 32 |
| Effective length [m] | 1.038 | 1.038 | 1.038 |
| R/Q [Ω] | 1036 | 1036 | 1036 |
| Injection cavity voltage [MV] | 0.9 | 1.4 | 2.8 |
| Injection cavity gradient [MV/m] | 0.9 | 1.4 | 2.7 |
| Extraction cavity voltage [MV] | 19.1 | 10.9 | 11.3 |
| Extraction cavity gradient [MV/m] | 18.4 | 10.5 | 10.8 |
| $Q_0$ at operating gradient | 1E+10 | 1E+10 | 1E+10 |
| Optimal $Q_L$ | 4E+07 | 4.4E+07 | 2.3E+08 |
| Optimal detuning [Hz] | -9.3 | -29.2 | -30.4 |
| Microphonics detuning [Hz] | 20.0 | 20.0 | 20.0 |
| Lorentz force detuning [Hz] | -337.3 | -111.0 | -117.5 |
| $Q_L$ | 1.0E+07 | 1.0E+07 | 1.0E+07 |
| Cavity bandwidth [Hz] | 130 | 130 | 130 |
| Cavity time constant [μs] | 2449 | 2449 | 2449 |
| Cavity number / SSA | 1 | 1 | 1 |
| SSA power [kW] | 25 | 25 | 25 |
| SSA number | 96 | 64 | 32 |

### 5.3.3 Magnets

#### 5.3.3.1 *Introduction*

The circumferences of the Booster and the Collider are nearly similar, about 100 km. The Booster has 16320 dipoles, 2036 quadrupoles and 448 sextupoles. The length of most of the Booster dipoles is 4.7 m, which means that more than 74% of the Booster circumference will be filled with dipoles. Cost is an important issue in the design, especially for the dipoles. Since the field at injection from the Linac is very low, a new type of dipole magnet that has a core made by steel-aluminium laminations with thin return yokes and several holes in the pole area, is proposed. There are two advantages: one is the reduction of the magnet weight and cost; another is an increase of the working magnetic field in the iron cores, which makes the field less sensitive to differences in iron quality and in particular to the coercive force. To further reduce the cost of the magnets, aluminium conductors instead of copper conductors will be used for the coils.

For the Booster, the field of the dipole, quadrupole and sextupole magnets will change when the particle beams are accelerated from 10 GeV to 120 GeV. The ratio of the max. field to min. field for the Booster magnets is 12, and a typical cycle is shown in Fig. 5.3.3.1

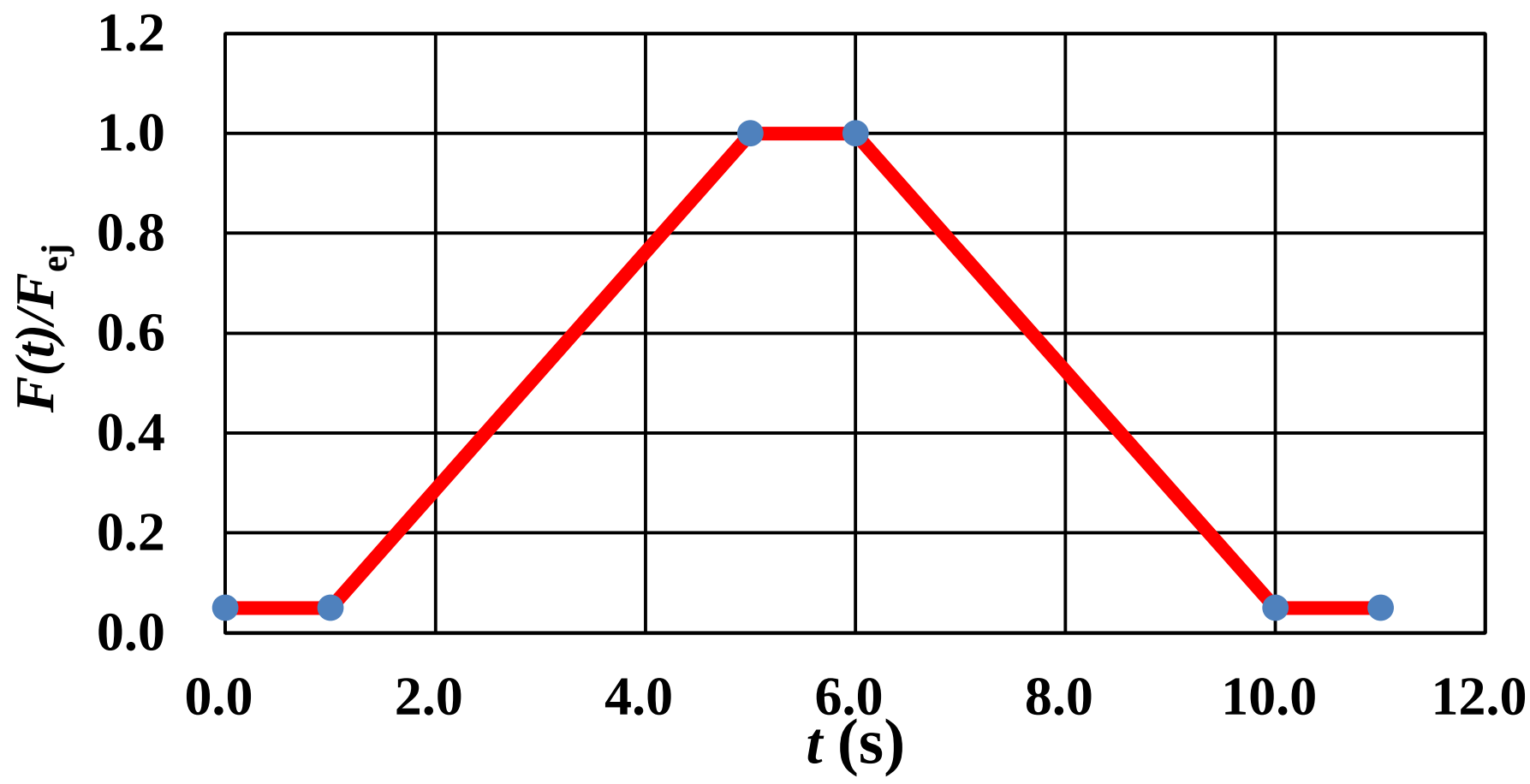

**Figure 5.3.3.1:** The magnetic field cycle of the Booster

### 5.3.3.2 *Dipole Magnets*

Most magnets are 4.7 m long, the others are 2.4 m and 1.7 m long. The field will change from 29 Gauss to 392 Gauss during acceleration. Due to this very low injection field level, the cores are composed of stacks of 1 mm thick low carbon steel laminations spaced by 1 mm thick aluminium laminations. Because magnetic force on the poles is very small, the return yoke of the core can be made as thin as possible. In the pole areas of the laminations, some holes will be stamped to further reduce the weight of the cores as well as to increase the field in the laminations. The considerations of steel-aluminium core, the thin return yoke and the holes in pole areas can improve the performance of the iron core and considerably reduces the weight and the cost.

Also for economic reasons, the excitation bars are made from 99.5% pure aluminum of cross section 30×40 mm$^2$. Thanks to low Joule loss in the bars, the magnets are cooled by air, not water.

The uniformity of the integral field of the 4.7 m long dipole cores can be optimized within $5\times10^{-4}$ by pole shimming in 2D or end chamfering in 3D. The cross section and magnetic flux of the dipole magnet is shown in Fig. 5.3.3,2, and its main parameters are listed in Table 5.3.3.1.

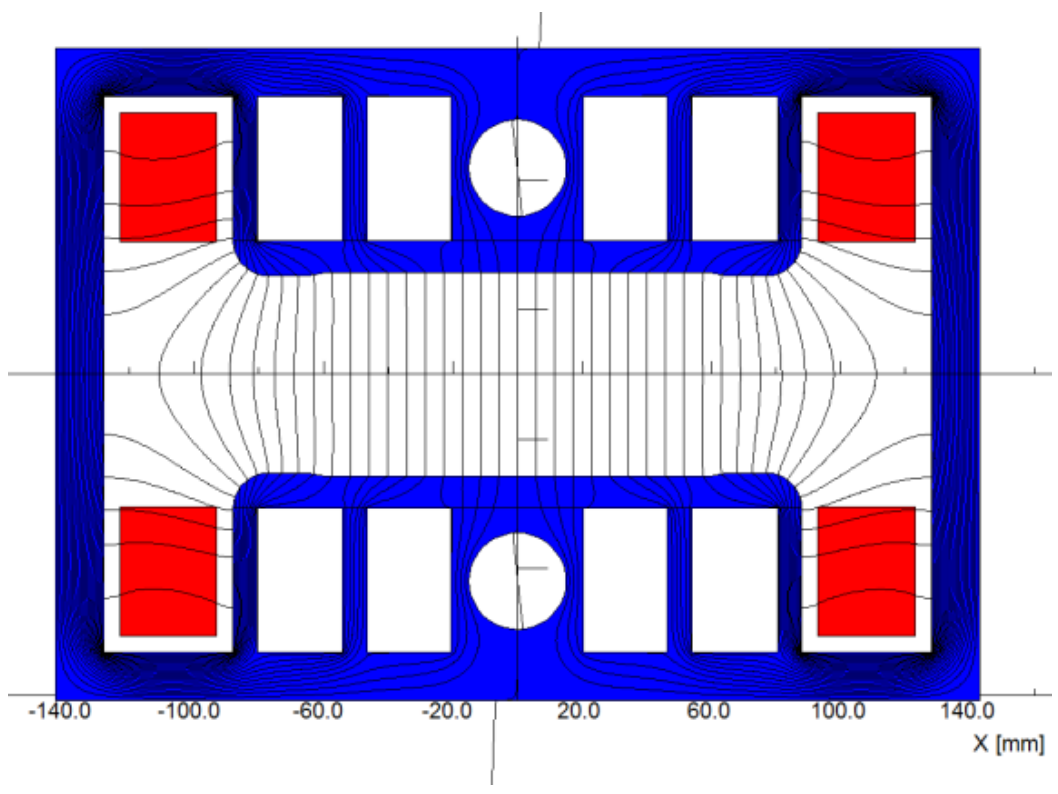

**Figure 5.3.3.2:** The magnetic flux distribution of the Booster dipole magnet

**Table 5.3.3.1:** The main parameters of the Booster dipole magnets

| | BST-63B-Arc | BST-63B-Arc-Dis | BST-74B-IR |
|---|---|---|---|
| Quantity | 15360 | 640 | 320 |
| Max. field [Gs] | 338 | 338 | 392 |
| Min. field [Gs] | 29 | 29 | 33 |
| Gap (mm) | 63 | 63 | 63 |
| Magnetic Length (mm) | 4711 | 2355.5 | 1682.5 |
| Good field region (mm) | 55 | 55 | 55 |
| Field uniformity | 0.1% | 0.1% | 0.1% |
| Turns per pole | 1 | 1 | 1 |
| Max. current[A] | 856 | 856 | 992 |
| Min. current[A] | 71 | 71 | 84 |
| Conductor size (mm) | 30×40(Al) | 30×40(Al) | 30×40(Al) |
| Max. current density(A/mm$^2$) | 1.04 | 1.04 | 1.21 |
| Resistance of coil(mΩ) | 0.56 | 0.29 | 0.21 |
| Max. power loss (W) | 407.7 | 212.5 | 210.8 |
| Avg. power loss [W] | 163.1 | 85.0 | 84.3 |
| Inductance (mH) | 0.09 | 0.04 | 0.05 |
| Core width/height（mm） | 330/220 | 330/220 | 330/220 |
| Core length (mm) | 4650 | 2295 | 1620 |
| Core weight (ton) | 1.22 | 0.61 | 0.66 |
| Conductor weight (ton) | 0.18 | 0.09 | 0.1 |
| Magnet weight (ton) | 1.4 | 0.7 | 0.76 |

### 5.3.3.3 *Quadrupole Magnets*

Because of the large number of quadrupole magnets, reducing the cost is important. Hollow aluminum conductor is selected for the coil instead of conventional copper, because of lower price and reduced weight. The iron core is made of 0.5 mm thick laminated low carbon silicon steel sheets. The magnet will be assembled from four identical quadrants, and can also be split into two halves for installation of the vacuum chamber. The coil windows leave a certain amount of space to place radiation shielding blocks.

Due to the long length, 2D magnetic field analysis is sufficient. The pole profiles are designed to introduce positive 12-pole and 20-pole multipole fields to compensate for end field effects.

The cross sections for the quadrupole magnets have been designed and optimized using OPERA-2D. In the simulation, only one quarter of the magnet is modelled. The magnetic flux lines are show in Fig. 5.3.3.3.

The main design parameters are listed in Table 5.3.3.2.

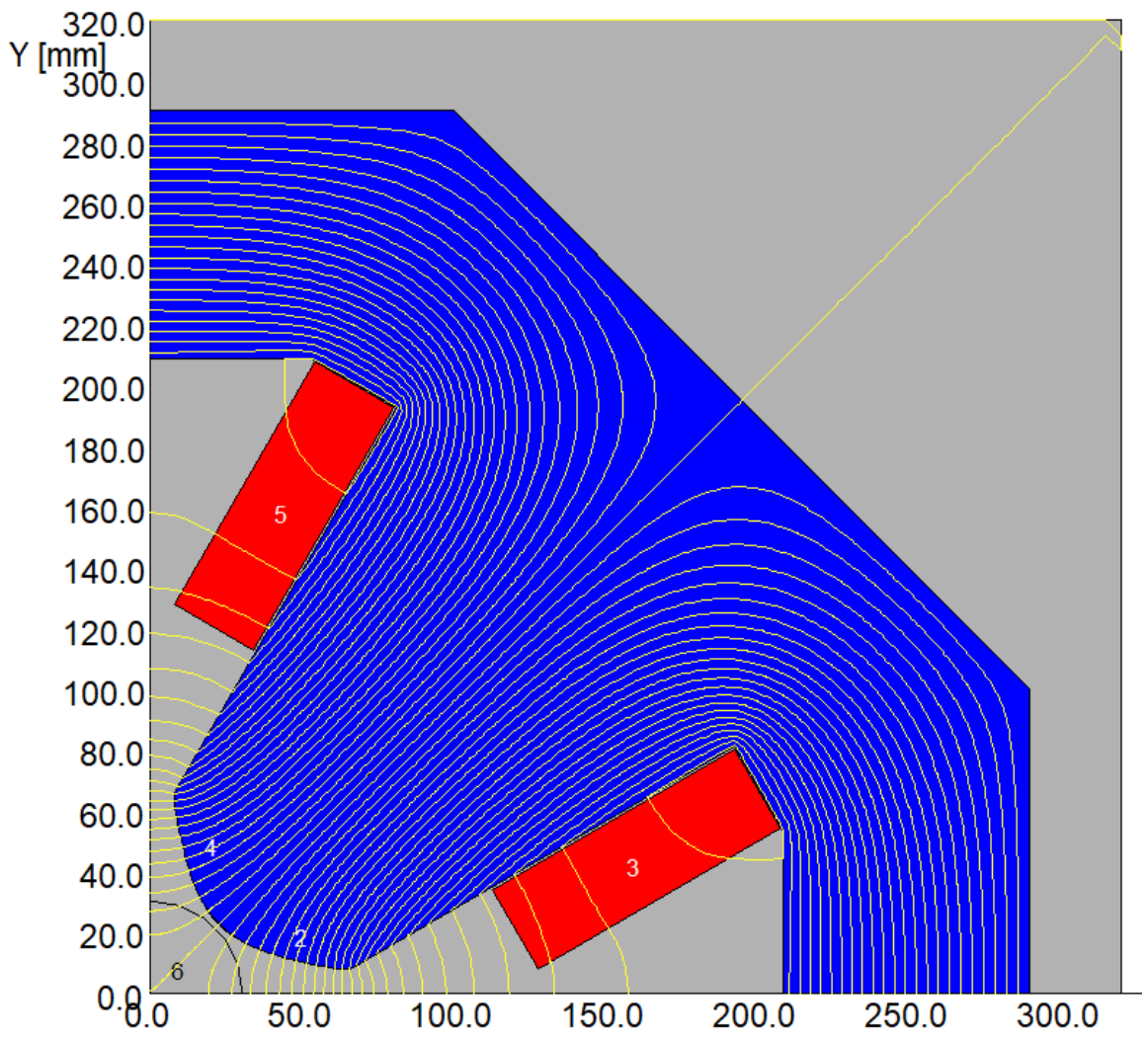

**Figure 5.3.3.3:** Magnetic flux lines of the Booster quadrupole magnet (one quarter)

**Table 5.3.3.2:** Main design parameters of CEPC Booster quadrupole magnets

| Magnet name | BST-QM | BST-QARC | BST-QMRF | BST-QRF |
|---|---|---|---|---|
| Quantity | 246 | 1664 | 8 | 118 |
| Aperture diameter(mm) | 64 | 64 | 64 | 64 |
| Magnetic length（mm） | 1000 | 1000 | 1500 | 2200 |
| Max. Field gradient（T/m） | 12.18 | 11.07 | 15.29 | 16.61 |
| Min. Field gradient（T/m） | 0.99 | 0.92 | 1.03 | 1.38 |
| GFR radius (mm) | 28 | 28 | 28 | 28 |
| Harmonic errors | 5.0E-4 | 5.0E-4 | 5.0E-4 | 5.0E-4 |
| Coil turns per pole | 16 | 16 | 20 | 20 |
| Max. current（A） | 313 | 285 | 316 | 344 |
| Al conductor size（mm） | 10×10D6 | 10×10D6 | 10×10D6 | 10×10D6 |
| Max. current density（$A/mm^2$） | 4.42 | 4.02 | 4.46 | 4.86 |
| Resistance（Ω） | 0.063 | 0.063 | 0.111 | 0.158 |
| Max Power loss（kW） | 6.14 | 5.08 | 11.11 | 18.74 |
| Avg. Power loss (kW) | 2.46 | 2.03 | 4.44 | 7.50 |
| Inductance（mH） | 19.9 | 19.9 | 46.8 | 68.1 |
| Conductor weight（kg） | 29 | 29 | 51 | 73 |
| Core length（mm） | 940 | 940 | 1440 | 2140 |
| Core width & height（mm） | 600 | 600 | 622 | 622 |
| Core weight（kg） | 1478 | 1478 | 2594 | 3855 |
| Number of water circuits | 4 | 4 | 8 | 8 |
| Water pressure drop（kg/cm$^2$） | 6 | 6 | 6 | 6 |
| Flow velocity (m/s) | 2.53 | 2.53 | 2.71 | 2.22 |
| Water flux (l/s) | 0.29 | 0.29 | 0.61 | 0.50 |
| Temperature rise(℃) | 4.4 | 3.6 | 3.7 | 7.6 |

### 5.3.3.4 *Sextupole Magnets*

The sextupole magnets are divided into two families, focusing or defocusing (horizontally), both of which have the same aperture diameter and magnetic length but different working magnetic field. The field of the magnets will change with beam energy. The minimum sextupole field is 1/12 of the maximum.

The cores of the magnets have a two-in-one structure, made of low-carbon silicon steel sheets and end plates. By using end chamfering, the field errors can be reduced to meet the strict field requirements. The coils of the magnets have a simple racetrack-shaped structure. The coils are wound from solid copper conductors. The windows of the coils leave a certain amount of space for placing radiation shielding blocks.

The cross sections for the sextupole magnets have been designed and optimized by using OPERA-2D. Thanks to symmetry, only one quarter of the magnet is modelled. The magnetic flux lines are show in Fig. 5.3.3.4.

The main design parameters are listed in Table 5.3.3.3.

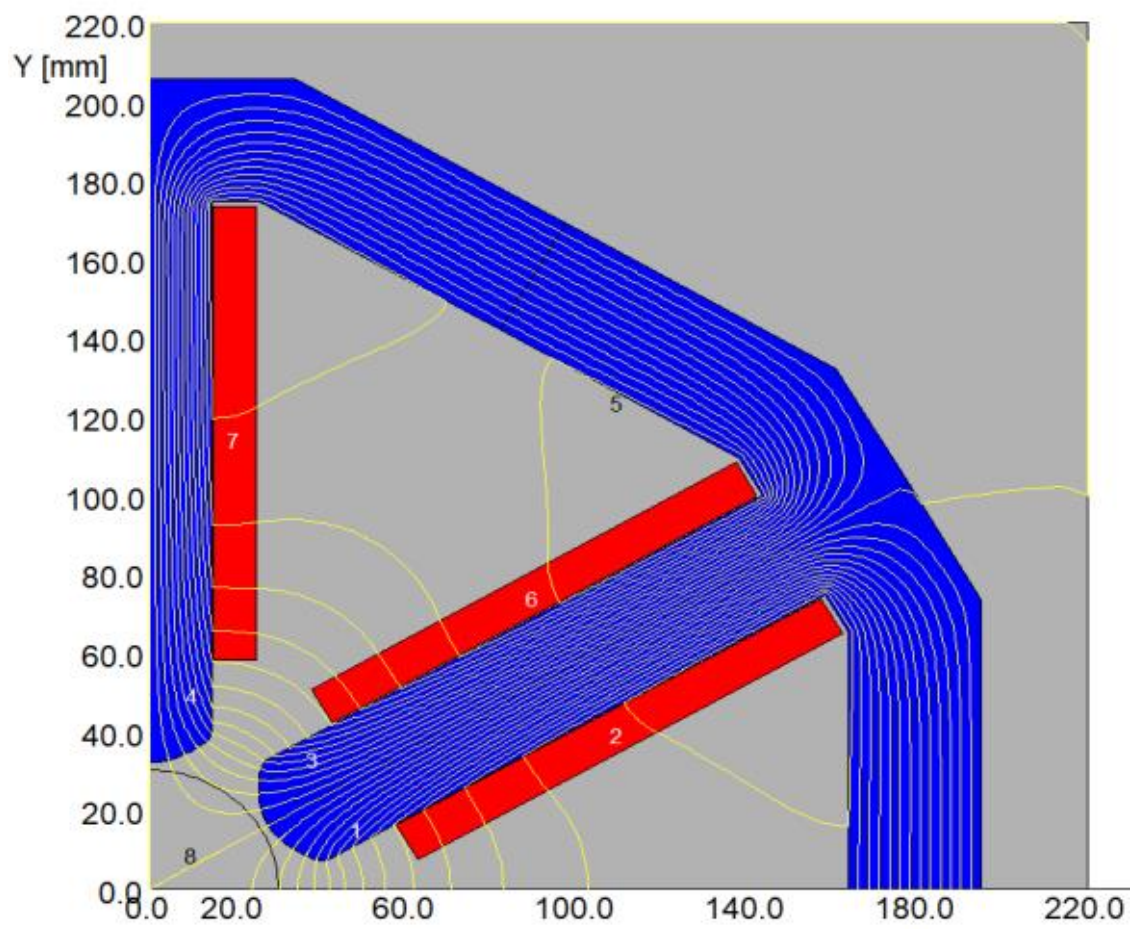

**Figure 5.3.3.4:** Magnetic flux lines of the Booster sextupole (one quarter)

**Table 5.3.3.3:** Main parameters of the sextupole magnets

| Magnet name | BST-64SF | BST-64SD |
|---|---|---|
| Quantity | 224 | 224 |
| Aperture diameter(mm) | 64 | 64 |
| Magnetic length（mm） | 400 | 400 |
| Max. sextupole field (T/m$^2$) | 216.9 | 437.1 |
| Min. sextupole field (T/m$^2$) | 18.1 | 36.4 |
| GFR radius (mm) | 28 | 28 |
| Harmonic errors | 1.0E-3 | 1.0E-3 |
| Coil turns per pole | 8 | 16 |
| Max. current（A） | 120.4 | 122.2 |
| Cu conductor size（mm） | 6×6D3 | 6×6D3 |
| Max. current density（A/mm$^2$） | 4.19 | 4.25 |
| Resistance（mΩ） | 27.2 | 54.5 |
| Max Power loss（kW） | 0.40 | 0.81 |
| Avg. Power loss (kW) | 0.16 | 0.33 |
| Inductance（mH） | 1.76 | 7.59 |
| Core length（mm） | 360 | 360 |
| Core width & height（mm） | 300 | 400 |
| Magnet weight（kg） | 120 | 200 |
| Number of water circuits | 2 | 6 |
| Water pressure drop（kg/cm$^2$） | 6 | 6 |
| Flow velocity (m/s) | 2.12 | 2.67 |
| Water flux (l/s) | 0.03 | 0.11 |
| Temperature rise(℃) | 6.68 | 3.72 |

### 5.3.3.5 *Correction Magnets*

Two types of correctors in the Booster are used for vertical and horizontal correction of the closed-orbit. Since two kinds of magnets have the same gap, same maximum field and same effective length, the design of them is similar.

To meet the field quality requirements, the correctors have H-type structure cores so the pole surfaces can be shimmed to optimize the field. The cores are stacked from 0.5 mm thick laminations. The racetrack shaped coils are wound from solid copper conductor 5.5 mm by 5 mm. Each coil has 24 turns, which are formed from 4 layers; no water cooling is required.

The OPERA software is used to simulate the field. Magnetic flux distributions are shown in figure 5.3.3.5.

The main parameters of the correctors are listed in Table 5.3.3.4.

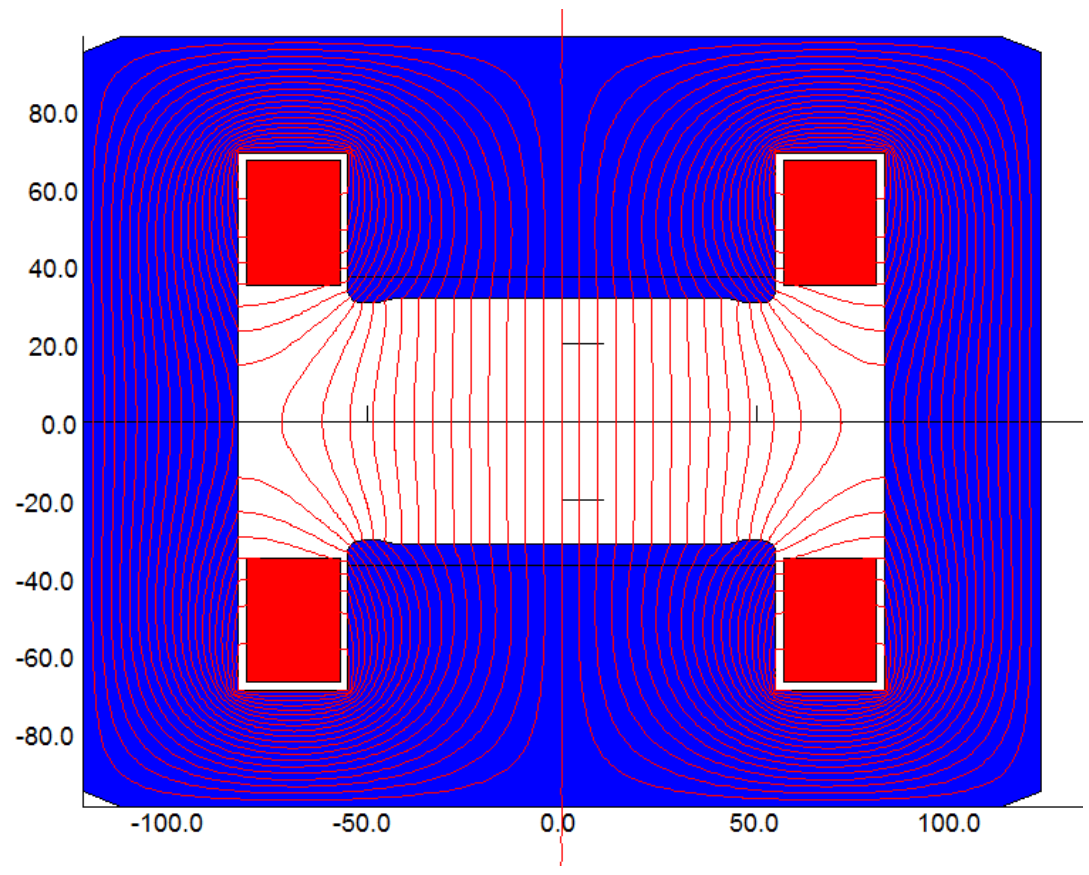

**Fig.5.3.3.5:** Magnetic flux distribution of the Booster corrector

**Table 5.3.3.4:** Design parameters of Booster correctors

| Magnet name | BST-63C |
|---|---|
| Quantity | 350 |
| Magnet gap (mm) | 63 |
| Magnetic field (T) | 0.02 |
| Effective length (mm) | 583 |
| Good field region (mm) | 55 |
| Field errors | 0.1% |
| Ampere turns per pole (AT) | 540 |
| Maximum current (A) | 22.5 |
| Coil turns per pole | 24 |
| Conductor size （mm） | 5.5×5 |
| Current density（A/mm$^2$） | 0.82 |
| Resistance (Ω) | 0.05 |
| Voltage drop (V) | 1.12 |
| Power loss（W） | 25 |
| Core length (mm) | 550 |
| Core width/height(mm) | 246/198 |
| Magnet weight（kg） | 240 |

### 5.3.3.6 *Septum Magnets for Beam Injection and Extraction*

There are three types of septum magnets for beam injection and extraction, Septum1, Septum2 and Septum3. Septum3 is divided into Septum3-1, Septum3-2, Septum3-3 and Septum3-4 because of the magnetic strength and the septum thickness.

Because of limitation of septum thickness and horizontal injection of the beam, eddy-current septum magnets are used. Since each type of magnet has a long effective length, several “short” magnets are installed in groups to obtain a “long” magnet. The eddy-current septum magnet has a C shaped core and is excited by a half sine wave pulsed current with a width of 60 μs. The iron core is made of 0.15 mm thick laminated low carbon silicon steel sheets and an insulating coating exists between two sheets. The coil and septum plate are both made of copper and cooled with air due to the low power. The outer cover of the iron core is made from stainless steel and is welded with the septum plate to form a cavity. The cavity provides the vacuum environment and produces magnetic field shielding by the eddy current effect. The circulating beam vacuum chamber is made of high permeability material DT4, which further shields the stray field.

The cross section of the septum magnet is designed by OPERA-2D. According to the simulation results, the magnetic field uniformity is better than $3\times10^{-3}$ and the stray field is less than $5\times10^{-4}$. The distribution of the magnetic flux is shown in Fig. 5.3.3.6.

The main parameters are listed in Table 5.3.3.5.

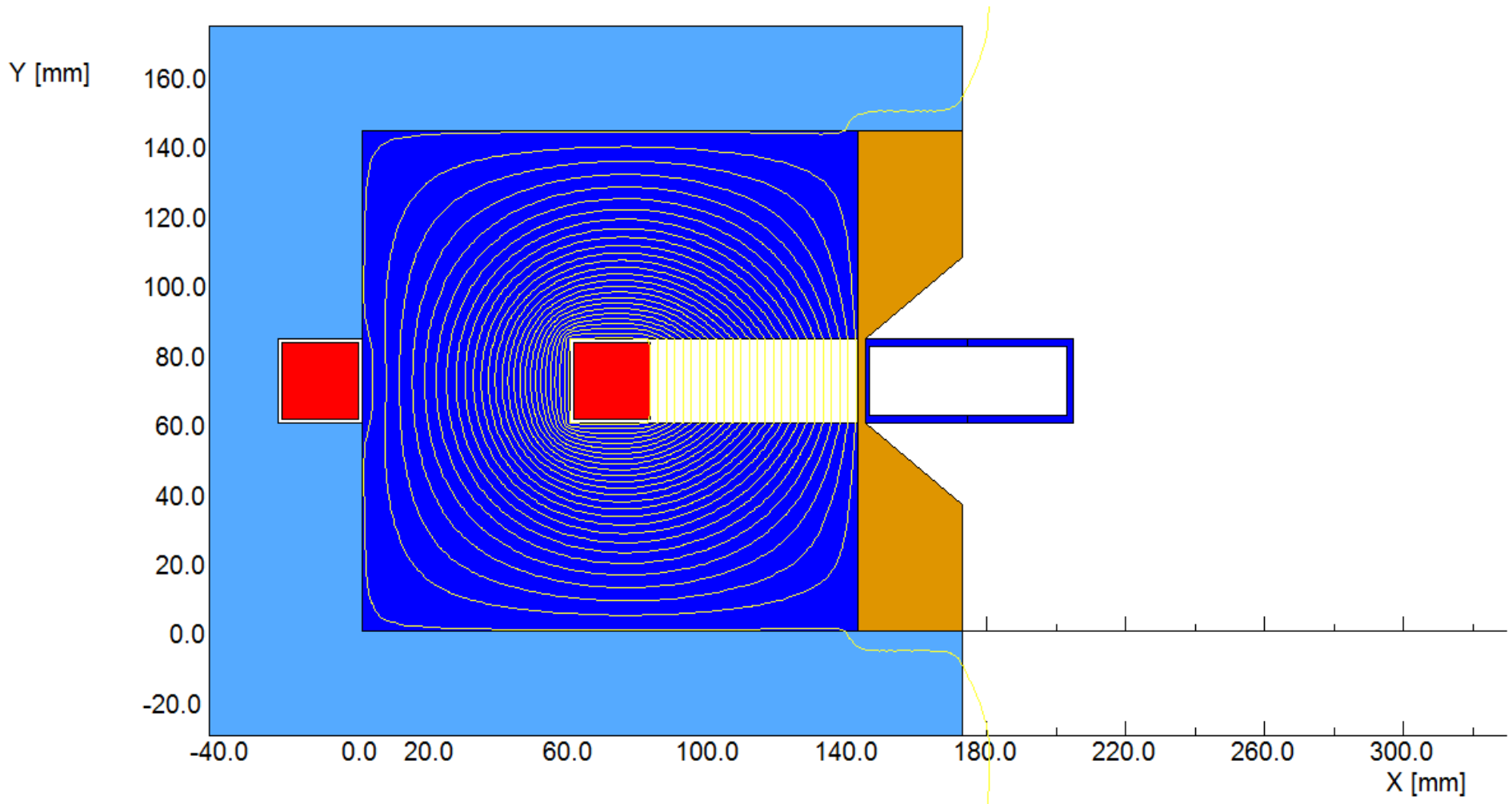

**Figure 5.3.3.6:** The distribution of the magnetic flux in the septum.

**Table 5.3.3.5:** The main parameters of the septum magnets for beam injection and extraction

| Magnet name | Septum1 | Septum2 | Septum3-1 | Septum3-2 | Septum3-3 | Septum3-4 |
|---|---|---|---|---|---|---|
| Quantity | 4 | 20 | 30 | 30 | 30 | 30 |
| Magnetic length (m) | 1 | 1 | 1 | 1 | 1 | 1 |
| Magnetic field (T) | 0.152 | 0.58 | 0.54 | 0.272 | 0.136 | 0.068 |
| Gap (mm) | 66 | 24 | 24 | 24 | 24 | 24 |
| Good field region (mm) | 88 | 20 | 20 | 20 | 20 | 20 |
| Field uniformity | $3\times10^{-3}$ | $3\times10^{-3}$ | $3\times10^{-3}$ | $3\times10^{-3}$ | $3\times10^{-3}$ | $3\times10^{-3}$ |
| Septum thickness (mm) | 3 | 3 | 8 | 6 | 4 | 2 |
| Stray field ($\Delta B/B$) | $5\times10^{-4}$ | $5\times10^{-4}$ | $5\times10^{-4}$ | $5\times10^{-4}$ | $5\times10^{-4}$ | $5\times10^{-4}$ |
| Coil turns | 1 | 1 | 1 | 1 | 1 | 1 |
| Conductor size (mm) | 64×64 | 22×22 | 22×22 | 22×22 | 22×22 | 22×22 |
| Field waveform | Half sine | Half sine | Half sine | Half sine | Half sine | Half sine |
| Max current (A) | 8223 | 11409 | 10623 | 5351 | 2675 | 1338 |
| Max voltage drop (V) | 1324 | 1529 | 1424 | 717 | 359 | 179 |
| Pulse width (μs) | 60 | 60 | 60 | 60 | 60 | 60 |
| Inductance (μH) | 3.07 | 2.56 | 2.56 | 2.56 | 2.56 | 2.56 |
| Effective resistance (mΩ) | 0.23 | 0.59 | 0.59 | 0.59 | 0.59 | 0.59 |
| Avg. power loss (W) | 0.92 | 4.59 | 3.98 | 1.01 | 0.25 | 0.06 |

### 5.3.3.7 *Kicker Magnets for Beam Transport Lines*

For injection and extraction of electron and positron beams, there are two similar injection and extraction systems for the Booster and two injection systems into the Collider Ring. For these systems, the kicker magnets are in two families according to their apertures. The field waveform is half-sine wave and the pulse width is 300 ns. Because the field changes very fast, all the kicker magnets are put inside a vacuum tank. The cores of the magnets have a window frame structure, which consists of two C-shaped Ni-Zn type ferrite blocks. The ferrite developed in a Chinese ferrite company has the desired properties of high frequency response, low loss and low outgassing rate.

The coils of the magnets are made of a single-turn copper conductor, which is welded to copper bars. Since the average power loss of the coils is very low, the coil is not cooled by water. The pulsed current is fed from one end of the ferrite core by high voltage feedthroughs. The inner surfaces of the magnet will be coated with TiN films to reduce the secondary electron yield.

The OPERA-2D/TRANSIENT program is used to simulate the pulsed field of the kicker magnets. The flux flow at the peak field is shown in figure 5.3.3.7. The field ha satisfactory uniformity in the good field region. The main parameters of the magnets are listed in table 5.3.3.6.

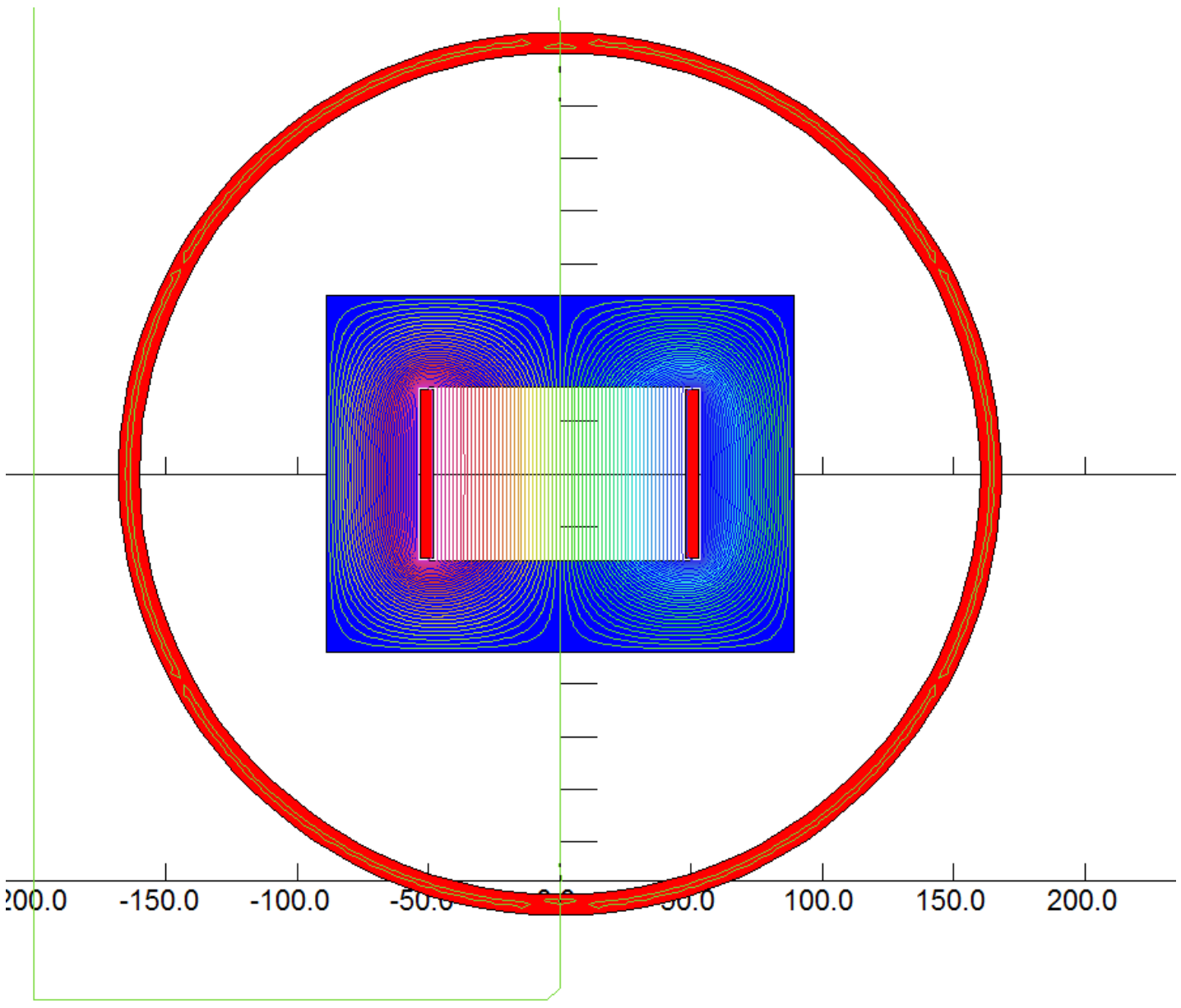

**Figure 5.3.3.7:** The distribution of the magnetic flux at peak field

**Table 5.3.3.6:** The main parameters of the kicker magnets for beam injection and extraction

| | Kicker-Inj-BST | Kicker-Ext-BST/Inj-MR |
|---|---|---|
| Quantity | 2 | 20 |
| Field amplitude(Gs) | 340 | 560 |
| Magnetic length (mm) | 500 | 1000 |
| Gap(mm) | 66 | 24 |
| Field waveform | Half-sine | Half-sine |
| Pulse width(ns) | 300 | 300 |
| Repetitive rate (Hz) | 100 | 100 |
| GFR (mm) | 88 | 20 |
| Field uniformity | 1% | 1% |
| Turns of coil | 1 | 1 |
| Current amplitude (A) | 1803 | 1080 |
| Conductor size (mm) | 5*65 | 5*23 |
| Core material | Ni-Zn ferrite | Ni-Zn ferrite |
| Inductance (uH) | 0.92 | 1.26 |
| Max. voltage (kV) | 17.2 | 14.1 |
| Core width/height (mm) | 180/140 | 90/75 |
| Core weight (kg) | 67.3 | 28.5 |
| Radius of vacuum tank(mm) | 160 | 100 |

### 5.3.3.8 *Dipole for Beam Transport Lines*

There are two types of transport lines, one for electrons, the other for positrons. The structures are symmetric; only the directions of magnetic fields are opposite. Every transport line employs four types of dipole magnets, named BT0&BT1, Btv, BT2 and BT3 respectively (BT0 and BT1 are similar type magnets but with different names); their lengths are 5 m, 28 m, 35 m and 15 m. The highest field among them is 0.5 T. All of these magnets are operated with DC excitation, which means solid iron can be used for the yoke. It is difficult to fabricate very long yokes: so the dipoles of 15 m and 35 m length need to be divided into 3 and 7 sections. The length of each of these sections is 5 m, so they are 3in1 and 7in1 dipoles. The dipoles of 28 m can be divided into 7 parts, each of them are 4 m long, so this type of dipole is 7in1. The choice of coils is a balance between material costs, machining costs, cooling water flow rate and temperature rise. Based on these considerations, it is better to adopt TU2 copper as the coil material; the coil cross-sections for these four types of dipoles have the same dimension, 12 mm $\times$ 12 mm $\Phi$ 8.

The optimizations of the magnetic field are performed by using OPERA/TOSCA, the cross-section shapes of three types of magnets are illustrated in Figures 5.3.3.8 through 5.3.3.10. The parameters of three types of dipoles are summarized in Table 5.3.3.7.

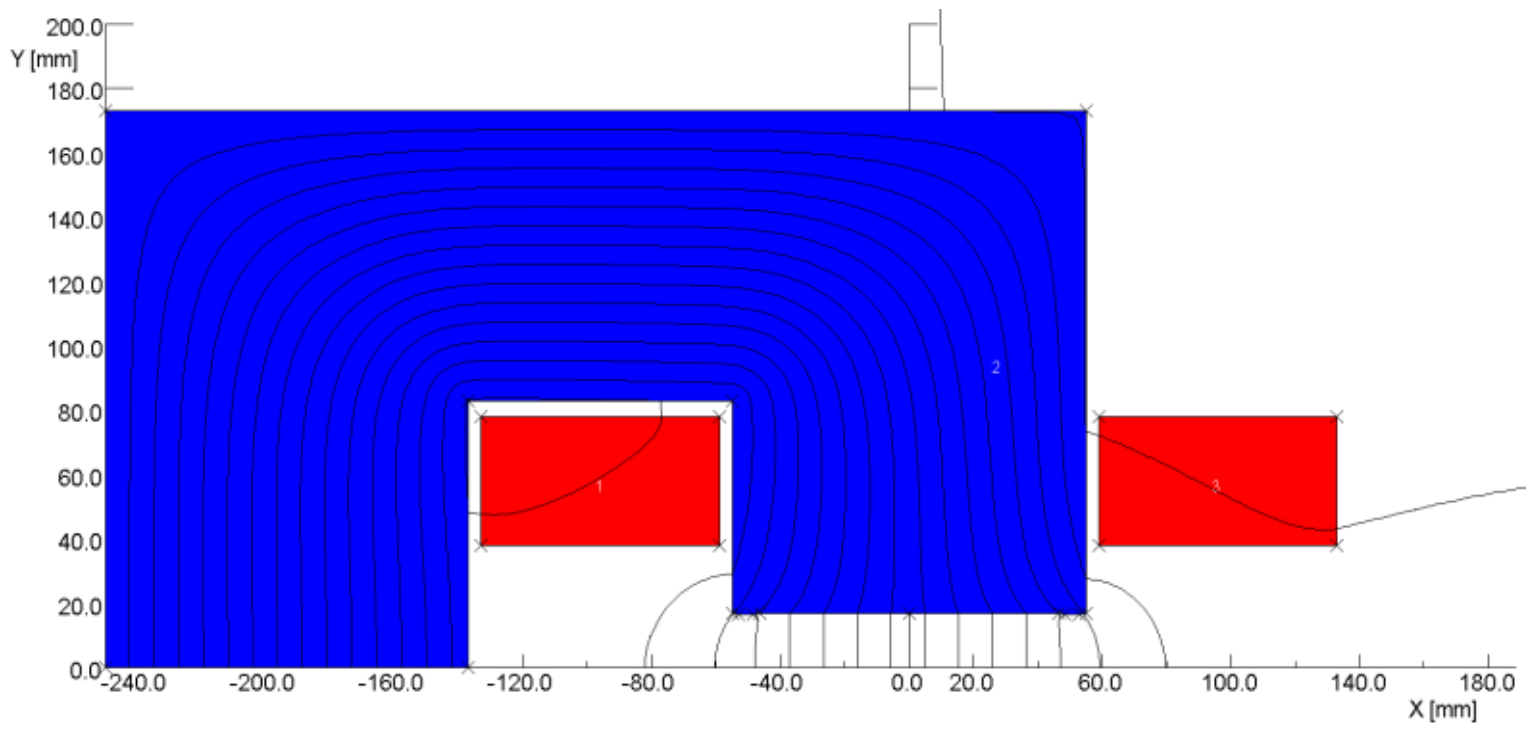

**Fig. 5.3.3.8:** The cross-section of the BT1 dipole

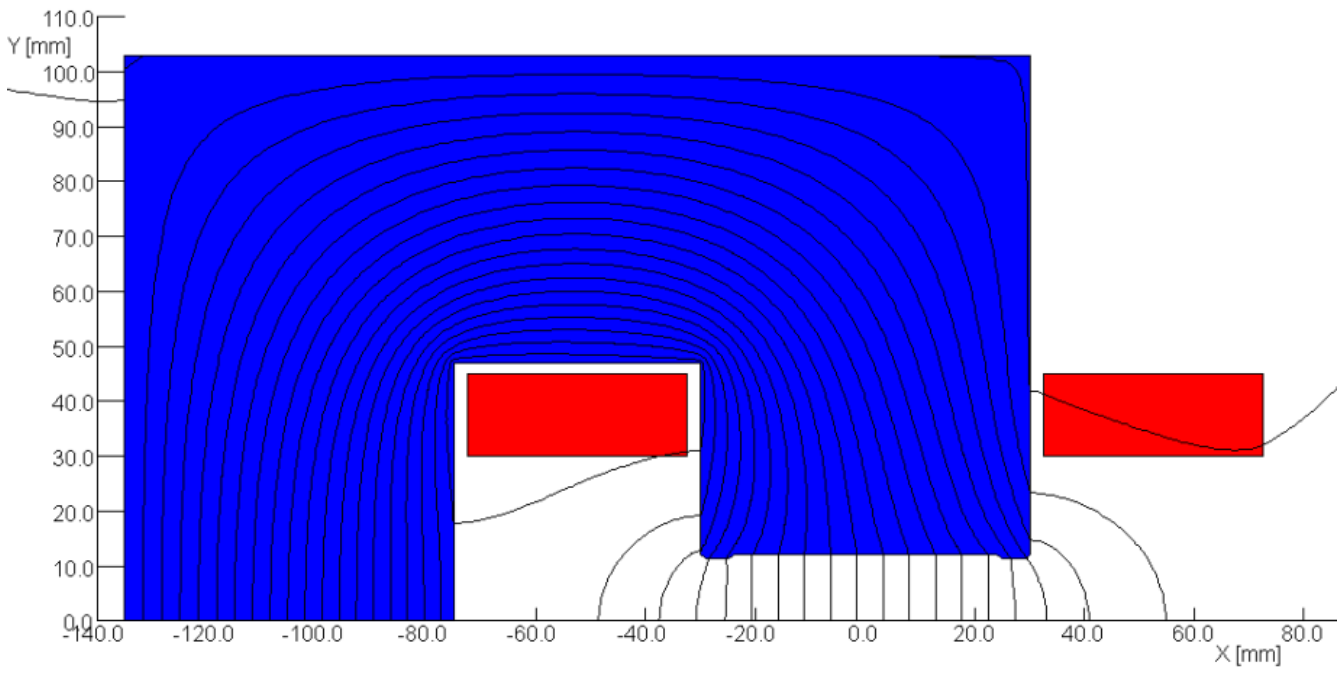

**Fig. 5.3.3.9:** The cross-section of the BT2 dipole

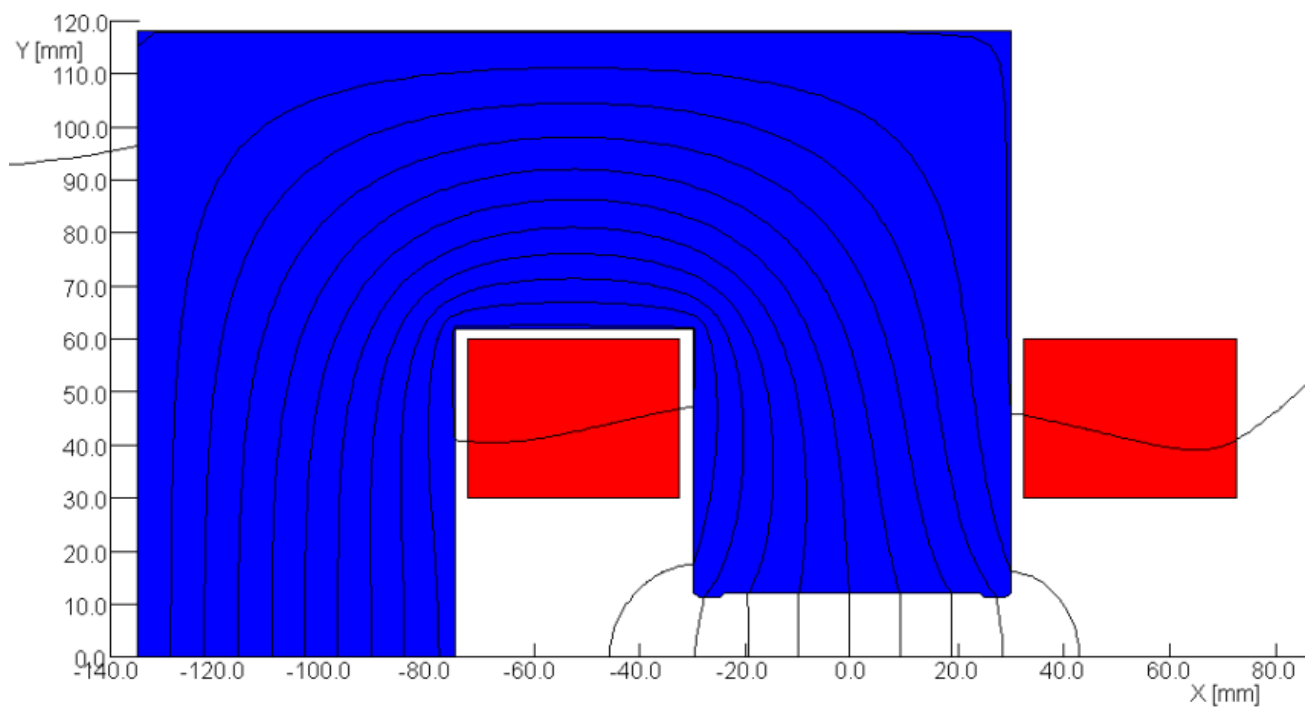

**Fig. 5.3.3.10:** The cross-section of the BT3 dipole

**Table 5.3.3.7:** The parameters of dipoles for Booster to Collider transport lines

| Magnet name | BT0&BT1 | Btv | BT2 | BT3 |
|---|---|---|---|---|
| Magnet type | whole | 7in1 | 7in1 | 3in1 |
| Quantity | 48 | 4 | 8 | 4 |
| Magnetic Length (m) | 5 | 4*7=28 | 5*7=35 | 5*3=15 |
| Strength of field (T) | 0.5 | 0.5 | 0.142 | 0.27 |
| Gap (mm) | 37 | 37 | 24 | 24 |
| Good field region (mm) | 33 | 33 | 20 | 20 |
| Field uniformity | 0.1% | 0.1% | 0.1% | 0.1% |
| Excitation amp-turns（At） | 7421 | 7421 | 1418 | 2840 |
| Size of conductor（mm） | 12*12Φ8 | 12*12Φ8 | 12*12Φ8 | 12*12Φ8 |
| Coils turns on each pole | 24 | 24 | 4 | 8 |
| Current（A） | 310 | 310 | 355 | 355 |
| Current density（A/mm$^2$) | 3.33 | 3.33 | 3.82 | 3.82 |
| Weight of conductor (kg) | 436.6 | 2445 | 510 | 436.4 |
| Resistance (Ohm） | 0.1 | 0.56 | 0.114 | 0.1 |
| Voltage drop (V) | 30.3 | 170 | 40.5 | 34.8 |
| Power loss (kW) | 9.37 | 52.5 | 14.4 | 12.4 |
| Inductance (mH) | 5 | 28 | 9.2 | 11 |
| Width/Height of core (mm) | 310/360 | 310/360 | 165/206 | 165/236 |
| Length of core (m) | 5 | 4*7=28 | 5*7=35 | 5*3=15 |
| Weight of core (ton) | 3.5 | 19.6 | 3.86 | 1.87 |
| Total weight of magnet (ton) | 3.93 | 22 | 4.37 | 2.4 |
| Number of cooling circuits | 4 | 4*6=24 | 2*7=14 | 2*3=6 |
| Water pressure drop (kg/cm$^2$) | 6 | 6 | 6 | 6 |
| Water flow per magnet（L/s） | 0.313 | 1.837 | 2.06 | 0.594 |
| Temperature rise($^o$C) | 7.1 | 7.1 | 1.66 | 5.1 |

### 5.3.3.9 *Quadrupole for Beam Transport Lines*

There are two kinds of quadrupole magnets with different apertures for the transport lines. One kind of quadrupole has a large aperture, 33 mm; the other has a smaller aperture, 24 mm. Figure 5.3.3.11 shows the 2D magnetic flux distribution for the larger aperture quadrupole. With the help of pole face optimization, all the high harmonic errors at the reference radius of 13 mm and 10 mm for the quadrupole magnets can meet the field requirements. The design parameters for the two kinds of quadrupole magnets are shown in Table 5.3.3.8.

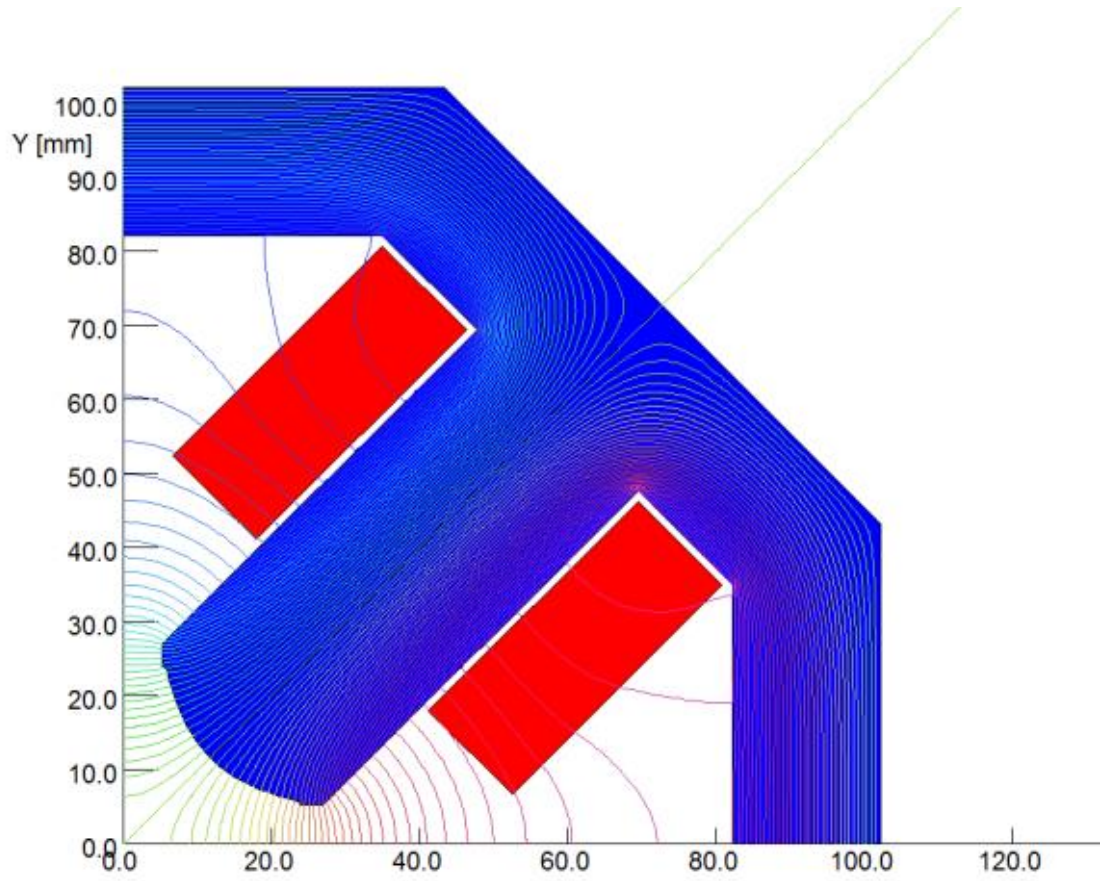

**Figure 5.3.3.11:** The magnetic flux distribution in the quadrupole magnets

**Table 5.3.3.8:** Main parameters of the quadrupole magnets for the transport lines

| Magnet name | TL-33Q | TL-24Q |
|---|---|---|
| Quantity | 80 | 40 |
| Bore diameter（mm） | 33 | 24 |
| Field gradient（T/m） | 4.9 | 9 |
| Magnetic length（m） | 0.9 | 2.0 |
| Ampere-turns per pole（AT） | 541 | 526 |
| Coil turns per pole | 32 | 28 |
| Excitation current（A） | 17.2 | 19.1 |
| Conductor size（mm） | 5×4 | 5×4 |
| Current density（$A/mm^2$） | 0.86 | 0.95 |
| Resistance（Ω） | 0.24 | 0.44 |
| Inductance（mH） | 5.2 | 8.8 |
| Voltage drop（V） | 4.2 | 8.5 |
| Power loss（W） | 71.5 | 161.3 |
| Core width/height（mm） | 210/210 | 170/170 |
| Core length（mm） | 884 | 1988 |
| Core weight（kg） | 240 | 354 |
| Conductor weight（kg） | 67.5 | 123.5 |
| Magnet weight (kg) | 308 | 478 |

### 5.3.3.10 *Correction Magnets for the Beam Transport Lines*

There are two kinds of correctors with different gaps, 37 mm and 24 mm. Each kind of magnet is independently used for vertical and horizontal correction of the closed-orbit.

The correctors have an H-type structure because their pole shapes can be shimmed to optimize the field. The cores are stacks of 0.5 mm thick laminations. Racetrack-shaped coils are wound from solid copper conductor of 5.5 mm by 4 mm. Since the current density is lower than $1A/mm^2$, the coils have no water cooling.

The OPERA software is used to simulate the field of the correctors. Magnetic flux distribution for one kind of corrector is shown in figure 5.3.3.12. The main parameters of the two kinds of correctors are listed in Table 5.3.3.9.

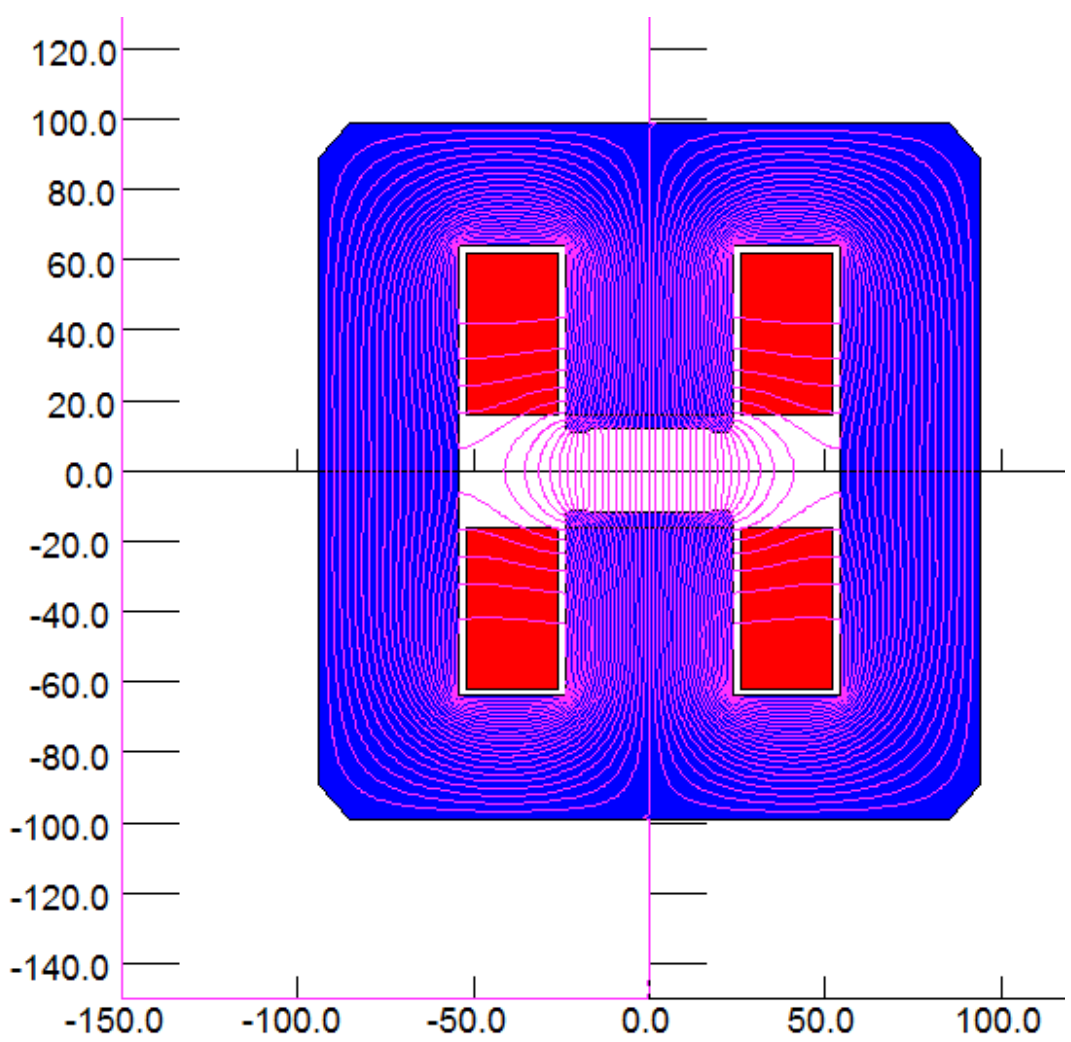

**Fig. 5.3.3.12:** Magnetic flux distribution in the correctors for transport lines

**Table 5.3.3.9:** Parameters of correctors for the beam transport lines

| Magnet name | T-37C | T-24C |
|---|---|---|
| Quantity | 24 | 30 |
| Gap [mm] | 37 | 24 |
| Max. Field [Gs] | 560 | 1000 |
| Magnetic Length [mm] | 200 | 300 |
| Good Field Region [mm] | 33 | 20 |
| Field Uniformity | 0.1% | 0.1% |
| Ampere turns per pole[At] | 850 | 1000 |
| Turns per pole | 40 | 48 |
| Max. current[A] | 21.25 | 20.8 |
| Size of conductor [mm*mm] | 5.5*4 | 5.5*4 |
| Current density[A/mm$^2$] | 0.97 | 0.95 |
| Resistance(Ohm) | 0.047 | 0.072 |
| Power loss (W) | 21 | 31 |
| Voltage[V] | 1.003 | 1.5 |
| Height of core [mm] | 232 | 198 |
| Width of core [mm] | 200 | 188 |
| Core Length [mm] | 180 | 280 |
| Magnet weight [kg] | 85 | 110 |

## 5.3.4 **Magnet Power Supplies**

For the Booster power supplies, the design principles are the same as for the Collider power supplies. There is however, a difference since the Collider supplies are DC and the Booster supplies ramp at a repetition rate of 0.1 Hz. Fig. 5.3.4.1 shows the Booster magnetic field cycle. All the unipolar supplies are housed in the surface halls. The bipolar supplies (correctors) are housed underground.

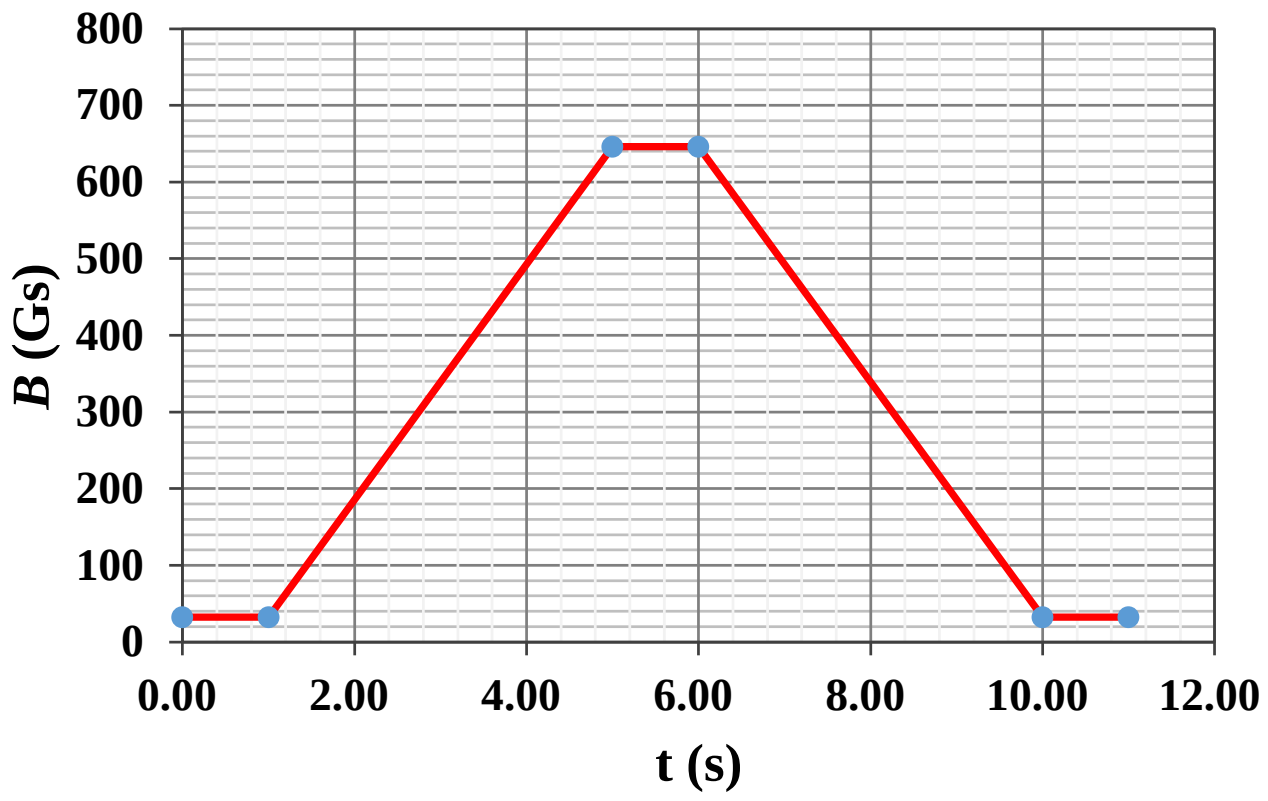

**Figure 5.3.4.1:** The magnetic field cycle of the Booster

### 5.3.4.1 ***Dipole Magnet Power Supplies***

There are 8 arc units for the Booster, the same as for the Collider. There are 16,320 dipoles. For each half arc, all the dipoles are connected in series and powered by one power supply. Thus there are 16 dipole supplies, each with a load of 1020 dipoles. The supply power is 0.77 MW, including an allowance for cable losses. There are 10 ~ 15% safety margins in both current and voltage as part of the manufacturer's ratings.

#### 5.3.4.2 *Quadrupole Magnet Power Supplies*

The quadrupoles are divided into 16 focusing families and 16 defocusing families; each family corresponds to half of an arc. Each family consists of 63 or 64 series-connected magnets powered by one supply.

#### 5.3.4.3 *Sextupole Magnet Power Supplies*

For each arc, there are four families of focusing sextupoles and four families of defocusing sextupoles. Each family consists of 14 series-connected magnets and is powered by one power supply.

#### 5.3.4.4 *Corrector Magnet Power Supplies*

The total number of corrector magnets, BH and BV, is 350, each independently powered by its own bipolar supply. For convenient maintenance and repair, the rating for all corrector power supplies is the same and they use a module-based design.

**Table 5.3.4.1:** Booster power supplies requirements

| Power Supply | Quantity | Stability / /tracking error | Output Rating |
|---|---|---|---|
| Dipole | 16 | 500ppm / 0.1% | 940A/820V |
| Quadrupole.D | 16 | 500ppm / 0.1% | 320A/2100V |
| Quadrupole.F | 16 | 500ppm / 0.1% | 320A/2100V |
| Sext.D | 16 | 1000ppm / 0.1% | 140A/650V |
| Sext.F | 16 | 1000ppm / 0.1% | 140A/650V |
| Corrector | 350 | 1000ppm | ±25A /±20V |
| **Average system power** | | | **11.62MW** |

Booster power supplies are cycled at a repetition of 0.1 Hz. In order to reduce the tracking error, a digital controller shown in Figure 5.3.4.2 is used. Modern control algorithms will be implemented for decreasing the tracking error in output current to obtain better performance.

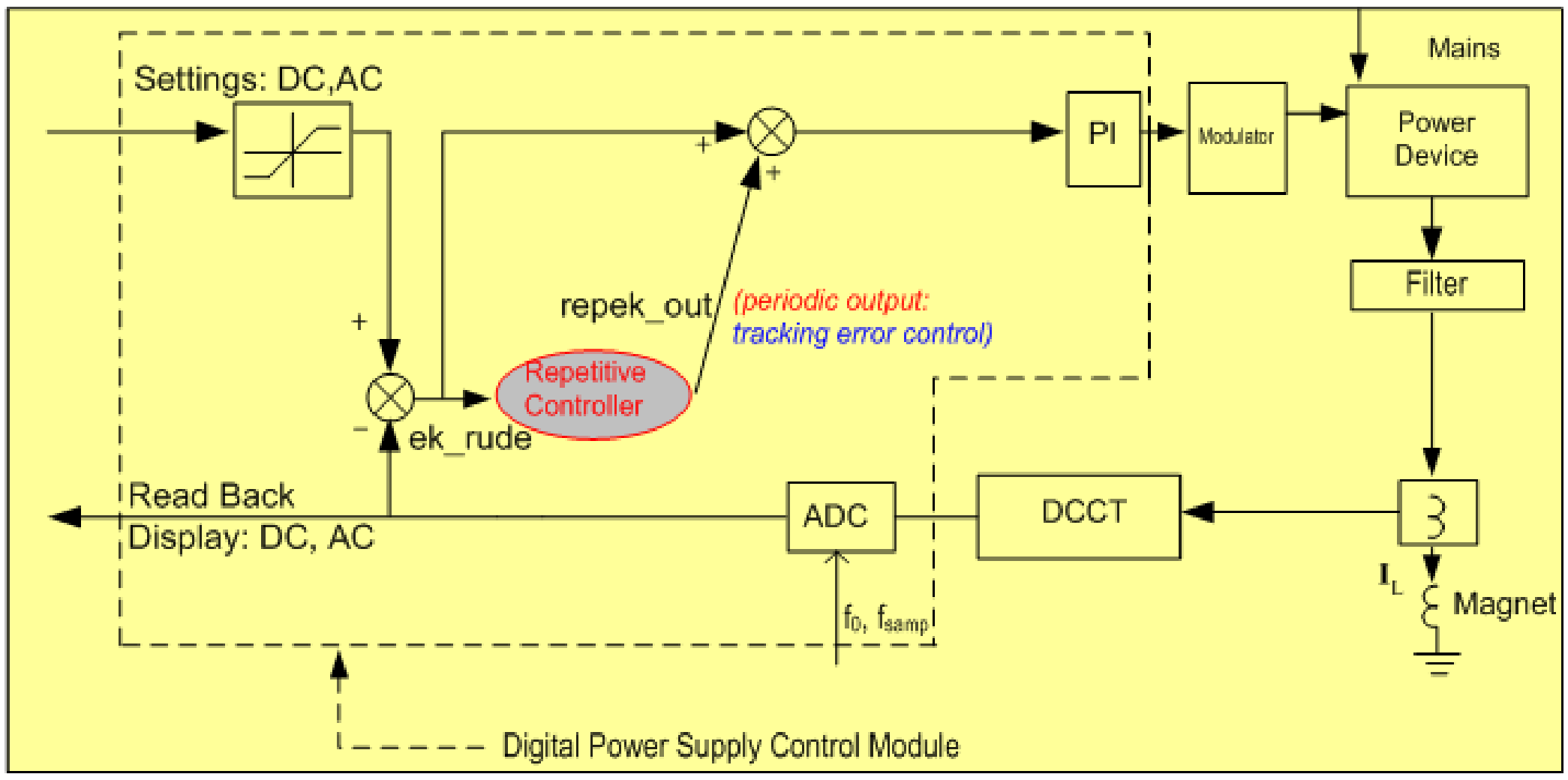

**Figure 5.3.4.2:** Digital controller for the Booster power supplies

## 5.3.5 Vacuum System

The Booster vacuum system includes chambers, bellows, pumps, gauges, valves and other components. An average pressure of less than 3×10$^{-8}$ Torr is required to minimize beam loss and bremsstrahlung from residual gas. Stainless steel has low electrical conductivity which limits the beam current due to instability. Therefore, aluminum with its higher conductivity has been chosen for the Booster chambers. Conventional high vacuum technology will be implemented and high vacuum will be achieved with small ion pumps distributed around the circumference.

### 5.3.5.1 *Heat Load and Gas Load*

To estimate the heat load, we start from formulas 4.3.6.1 and 4.3.6.2 for the synchrotron radiation power. For the Booster, $E$ = 120 GeV, $I$ = 0.00053 A, $\rho$ = 11380.8 m, giving a total SR power of $P_{SR}$ = 854.6 kW and a linear power density of $P_L$ =12 W/m. These cause a negligible heating load. Since the magnetic field changes very slowly, the heat load induced by the eddy current is almost zero.

The gas load includes thermal outgassing and synchrotron-radiation-induced photo-desorption. Thermal outgassing contributes mainly to the base pressure in the absence of a circulating beam. To estimate the desorption rate induced by SR we use formulas 4.3.6.4 and 4.3.6.5. The effective gas load due to photo-desorption is 3.1×10$^{-5}$ Torr·L/s. The linear SR gas load is 4.3×10$^{-10}$ Torr·L/s/m. Assuming the thermal outgassing rate of the vacuum chambers is 1×10$^{-11}$ Torr·L/s·cm$^2$ for a circular cross section of 5.5 cm in diameter, the linear thermal gas load $Q_{LT}$ = 1.7×10$^{-8}$ Torr·L/s/m. Compared to the thermal outgassing, the photo-desorption gas load is negligible.

### 5.3.5.2 *Vacuum Chamber*

Aluminum alloys are widely used in electron/positron storage rings, due to their high electric and thermal conductivity, easy extrusion and welding. Aluminum is cheaper than stainless steel or copper. Most aluminum alloys will not become magnetized from machining and welding and they do not form long lifetime radioactivity. However, the relatively low strength and hardness prevent aluminum alloys from being used for all-metal sealing flanges.

The cross-section of the dipole vacuum chamber is a circle with a diameter of 55 mm, length about 6 m and wall thickness 2 mm. The chambers are extruded from Al 6061 and stainless steel conflat flanges are welded onto the ends though transition material. Finite element analysis indicates that the highest temperature is 28.9°C in an ambient temperature of 25°C when a convective heat transfer coefficient of $2\times10^{-5}$ W/mm$^{2}\cdot$℃ is used. The maximum stress and deformation are 1.78 MPa and 0.0045 mm respectively, which are in the safe region. Figure 5.3.5.1 shows the results of the finite-element analysis.

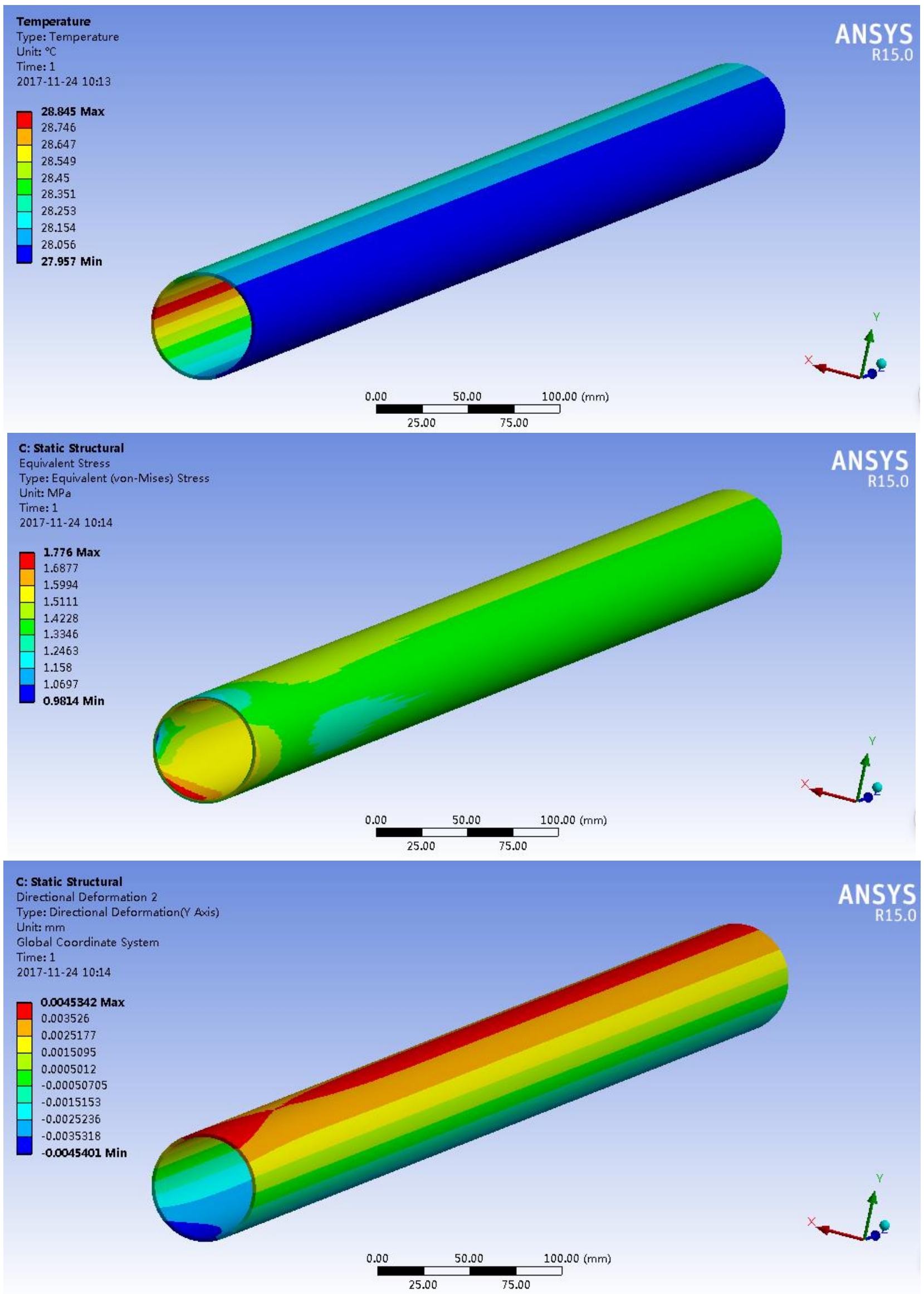

**Figure 5.3.5.1:** Results of the finite-element analysis on the aluminum vacuum chamber. Top – the temperature; middle – the stress; bottom – the deformation.

### 5.3.5.3 *Vacuum Pumping and Measurement*

The Booster circumference will be subdivided into 520 sectors with all metal gate valves, thus permitting pumping down, leak detection, bake-out, and vacuum interlock protection to be done in sections of manageable length and volume. Roughing down to approximately $10^{-7}$ Torr will be achieved by an oil free turbo-molecular pump group; the main pumping will be followed by ion pumps spaced about 6 m apart.

The size of the Booster excludes the installation of vacuum gauges at sufficiently short intervals. Some special sections such as injection regions, RF cavities and extraction regions are equipped with cold cathode gauges and residual gas analyzers, but the bulk of the pressure monitoring is done with the current of the sputter ion pumps. They will be monitored continuously and will provide adequate pressure measurements down to $10^{-9}$ Torr. Mobile diagnosis equipment can be brought to a place of particular interest during pump down, leak detection and bake-out when the machine is accessible.

## 5.3.6 **Instrumentation**

### 5.3.6.1 *Introduction*

The requirements of the Booster instrumentation system are to monitor beam status quickly and accurately, to measure and control the bunch current efficiently, and to cure beam instabilities. The beam orbit measurement is particularly important. Much of the instrumentation is the same as for the Collider and is described in Chapter 4. In this Chapter we will repeat a few of those details and also point out differences in the two systems.

### 5.3.6.2 *Beam Position Measurement*

BPMs in the Booster are spaced approximately every 60 m. The total number is 1,808 which also includes additional ones at specific locations for special purposes. Front end electronics and digital electronics are in the tunnel. Considering the limited space they are placed under the magnet girders. Radiation shielding of these electronics needs to be carefully designed.

Fig.5.3.6.1 shows the button BPM; the electrode radius is 3 mm and the gap between the button and the pipe is 0.3 mm. The difference from Fig. 4.3.7.2 (Collider) is in the shape of the vacuum chamber.

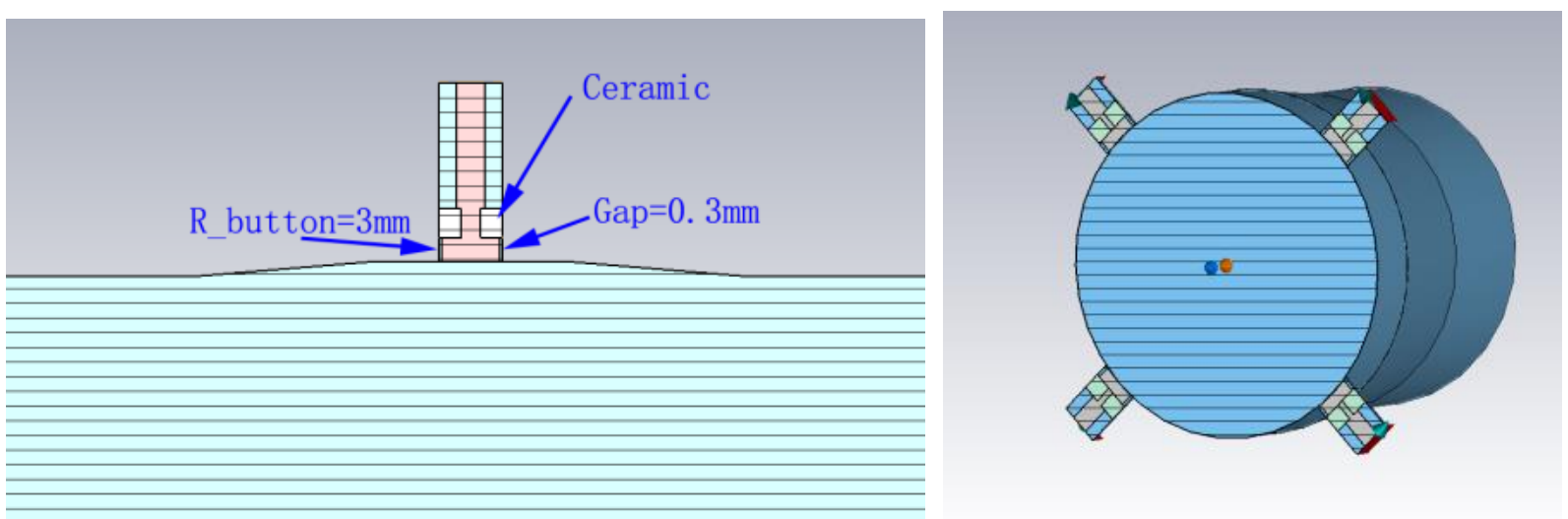

**Figure 5.3.6.1:** Left – Button pick up detail; Right – Model of the Booster BPM.

In order to study the pickup response to the beam, simulations are done using the CST particle studio. Figures 5.3.6.2 (left) and (middle) show the signal. Figure 5.3.6.2 (Right) shows how a current bunch of 45.8 A can induce a voltage of over 5 volts. The transfer impedance is about 0.37 Ω.

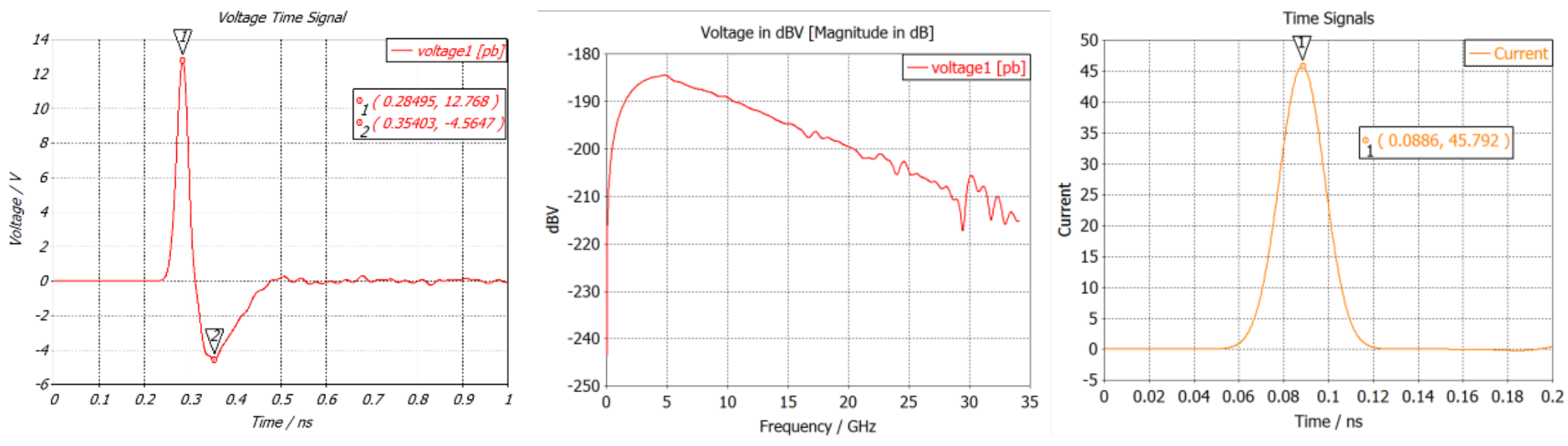

**Figure 5.3.6.2:** Left – signal of an electrode in the time domain; Middile – signal of an electrode in the frequency domain; Right – bunch current distribution in the time domain

**Table 5.3.6.1:** Parameters used for Booster BPM CST simulation

| | Bunch_charge (nC) | Bunch_length (mm) | Current_peak (A) | V_pp (V) |
|---|---|---|---|---|
| tt | 1.16 | 3 | 45.79 | 17.32 |
| Higgs | 0.648 | | 25.82 | 9.68 |
| W | 0.173 | | 6.89 | 2.55 |
| Z | 0.23 | | 9.08 | 3.41 |

The horizontal and vertical sensitivities near the center of the pipe are 7.49 %/mm and 7.48 %/mm as shown in Fig.5.3.6.3 (left) is the sensitivity mapping of the simulation in an area of 24 mm × 24 mm. $U = \Delta x/\Sigma x = \Delta y/\Sigma y$. The transverse response of the signal is non-linear for the beam off center, especially if the distance between the beam and pipe center is larger than 10 mm. Fig. 5.3.6.3 (right) is the response over the area of 16 mm × 16 mm.

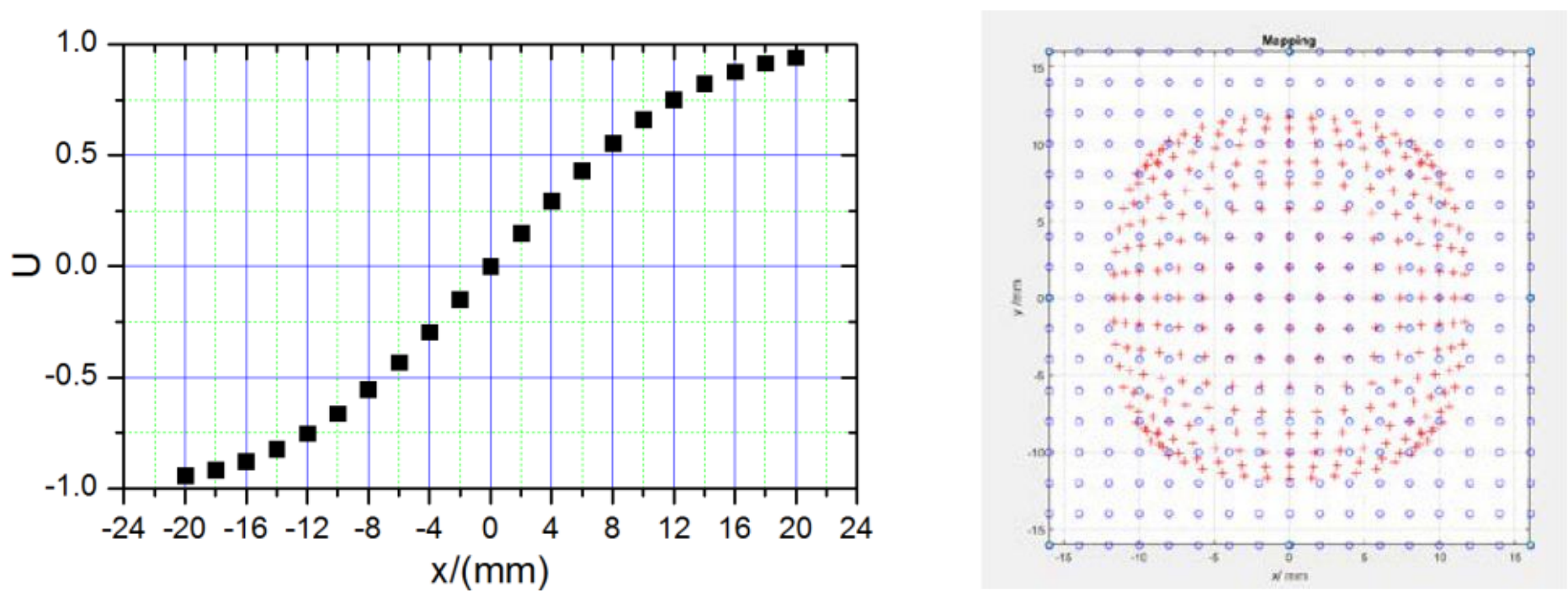

**Figure 5.3.6.3:** Left – sensibility mapping over a range of + 24 mm; Right – response in the area.

With high speed ADCs and high resolution we can acquire bunch by bunch positions. The entire system will use microTCA.4 standard structure. This includes RFFE (**r**adio **f**requency **f**ront-end **e**lectronics), high speed 4-channel ADCs, digital electronics and clock signals. Each pickup has a set of electronics. The BPM electronics have three different modes: (1) first turn (especially important during commissioning); (2) FA mode for fast data acquisition; (3) SA data for closed orbit measurement. The schematic of the bunch by bunch BPM electronics is the same as for the Collider, Fig. 4.3.7.5.

### 5.3.6.3 *Beam Current Measurement*

Beam current measurements include average current measurement and the bunch current monitor (BCM). The requirements and the kinds of instrumentation for these two systems are the same as discussed in detail in Chapter 4, Section 4.3.7.2. Also discussed in that section is the interesting possibility of a new type of beam current monitor based on the Tunnelling Magnet Resonance effect.

### 5.3.6.4 *Synchrotron Light Monitor*

The synchrotron light monitor (SLM) in the Booster is simpler than in the Collider because the emittance is larger and beam sizes in both x and y directions are from several hundreds of microns up to millimetres. A visible light beam line will be built. The design of the extraction mirror is the same as in the Collider. Visible light imaging will obtain the beam profile by use of a telescope to image the source point in the bending magnet on to a CCD camera. A neutral density filter will be placed in front to attenuate the visible light, especially during energy ramping. The emittance will be calculated with the observed beam sizes and lattice parameters of the source point. The beam profile can be observed at various electron energies.

### 5.3.6.5 *Beam Loss Monitor*

Beam loss monitors (BLMs) are an important part of the machine protection systems of particle accelerators and are used to minimize losses to protect equipment. Furthermore, they are a sensitive tool to localize losses and determine the time in the acceleration cycle when they occur. BLMs are mounted outside of the accelerator vacuum chamber. The signal from the BLM is proportional to the number of lost particles. Since the Booster ramps from 6 GeV to 120 GeV, the Booster BLM needs to have a wide dynamic range.

The BLM chosen for the Booster uses Cherenkov light generated in a long quartz fiber. As shown in figure 5.3.6.4 two PMT's set at both ends of the fiber, read out with flash ADCs, can be used to determine the beam loss position because the time for the Cherenkov light to reach the two PMTs is different.

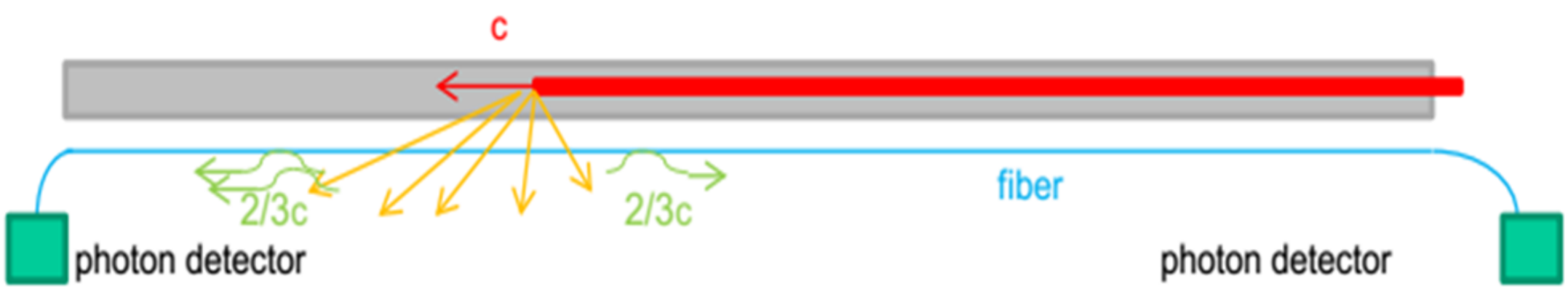

**Figure.5.3.6.4.** Principles of loss detection using optical fibers

Based on a careful study [6] of different fibers for BLM systems, SC400 fibers have been chosen for the Booster system. The fiber length is 150 m. The photomultiplier is a Hamamatsu H6780-02 equipped with a FC adapter, which has a typical insertion loss below 0.3 dB. Since the Booster circumference is 100 km, about 670 fibers will be installed along the beam pipe.

The signals from the PMTs will be transmitted by coaxial cables to a flash ADC (CAEN V1729A, 4channel, 14 bits 2GS/s) located outside the shield wall. A trigger signal from the accelerator master oscillator will be used as time reference to determine the position of the beam loss. The flash ADC starts data acquisition when the external trigger arrives. If there is a beam loss in the number N+1 bunch after the arrival of the trigger, the beam loss position will be calculated as shown in Fig 5.6.3.5. The upstream and downstream PMTs detect the beam loss pulse signal at the time of $T_a$ and $T_b$, respectively. L is length of the fiber which is parallel to the beam pipe, l is the distance from the upstream PMT to the beam loss position. $T_1$ = l/c is the time that the bunch travels from the upstream of the fiber to the beam loss location. $T_2$ and $T_3$ are the time that the Cherenkov light signal travels from the beam loss position to the upstream and downstream PMT respectively. △t is the interval time between two bunches. We can obtain formulas (5.3.6.1) and (5.3.6.2). Then the beam loss position l and the bunch number N can be calculated with formulas (5.3.6/3) and (5.3.6.4).

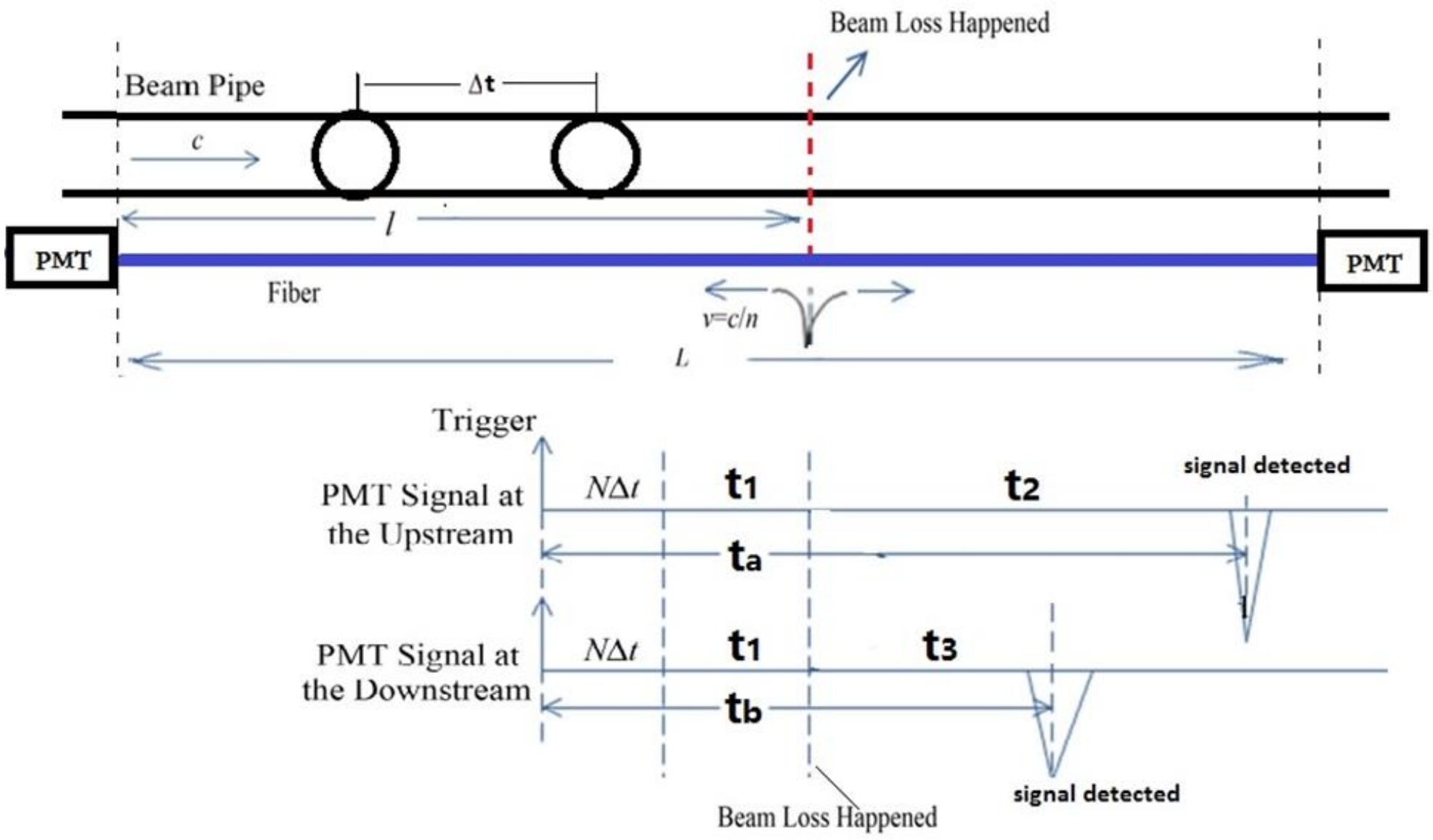

**Figure 5.3.6.5.** Detection principle for beam loss position and bunch number.

$$t_a = t_1 + t_2 + N\Delta t = \frac{l}{c} + \frac{l}{v} + N\Delta t \quad (5.3.6.1)$$

$$t_b = t_1 + t_3 + N\Delta t = \frac{l}{c} + \frac{L-l}{v} + N\Delta t \quad (5.3.6.2)$$

$$l = \frac{L - v(t_b - t_a)}{2} \quad (5.3.6.3)$$

$$N = \frac{2t_a cv - (L + t_a v - t_b v)(v + c)}{2cv\Delta t} \quad (5.3.6.4)$$

### 5.3.6.6 *Feedback System*

Since the Booster ramps from 6 GeV to 120 GeV in a very short time and the tunes will change during the ramp and the FB system must take this into consideration. One option is to design the ramp in a way that maintains transverse and longitudinal tunes in a relatively small range of values that can be controlled by a single filter. Another option is to have multiple feedback filters designed for different parts of the tune range (with overlap) and to switch from one to another during the ramp.

#### 5.3.6.7 *Other Beam Instrumentation in the Booster*

The tune measurement system and the vacuum chamber displacement measurement system are all similar to those in the Collider and are described in Chapter 4.

### 5.3.7 **Control System**

#### 5.3.7.1 *Introduction*

The Booster control system controls and monitors all of the functions in the Booster. There are many elements in common between the control of the three accelerators (Collider, Booster, Linac) that are described in considerably more detail in Chapter 4.

#### 5.3.7.2 *Power Supply Control*

There are magnet power supplies for bending magnets, quadruple and sextuple magnets, and correctors. These power supplies are distributed in several buildings. A major difference between the Booster and Collider supplies is that the Booster supplies need to be co-ramped for beam acceleration. The power supply control systems for Booster and Collider are similar. Two redundant controllers will be installed in a control crate and two isolated crate power supplies will provide power to the two control routes.

#### 5.3.7.3 *Vacuum System Control*

There are total 520 vacuum valves, and 2160 gauges and other devices such as pump controllers distributed around the Booster. Programmable logic controllers (PLCs), the heart of the vacuum protection interlock system, will be used to monitor gauge set-point outputs and to provide control of the sector gate valves. The PLCs will also output interlock signals to the RF and other subsystems and receive interlock signals from other subsystems. A ladder logic program will reside and run in the PLC processor module to control the gate valves on a fail-safe basis. A sector valve can be opened only if the adjacent vacuum conditions are satisfied. A vote-to-close algorithm will be adopted to close a sector valve when the vacuum pressure is above the gauge set-point on both sides of the valve. Besides, a sector valve will also be closed by the PLC under a number of conditions, such as power loss, controller fail and operator input.

#### 5.3.7.4 *RF System Control*

The Booster RF system controls include an interlock system to switch off cavity tuning, when the RF high voltage power supplies are in a fault or unsafe condition. A fault might happen in the cooling water system, the vacuum and temperature of a cavity, or the liquid helium in the cryogenic system. The system sends a warning message and failure signal to the MPS when a fault has been detected

### 5.3.8 Mechanical Systems

#### 5.3.8.1 *Introduction*

The Booster is mounted above the Collider. The number of magnets and their supports in the Booster and the Transport lines are listed in Table 5.3.8.1. In addition there are supports for vacuum and instrumentations components. The Booster magnet support structure has three parts; the steel frame is mounted to the embedded plate in the top of the tunnel wall; the adjusting mechanism and magnet mounting plate are similar to those in the Collider.

There are two transport lines. One connects the Linac and the Booster (LTB), while the other connects the Booster and the Collider (BTC). The magnets in LTB are supported from the ground, similar to the supports in the Collider, while the magnets in BTC are hung from the tunnel wall, similar to the supports in the Booster.

**Table 5.3.8.1:** Quantities of magnets and their supports in the Booster

| | Magnet type | Quantity | Magnet (core) length (mm) | No. of supports per magnet |
|---|---|---|---|---|
| Magnets in Booster | Dipole magnet | 15360 | 5445 | 4 |
| | | 640 | 2645 | 3 |
| | | 320 | 2945 | 3 |
| | Quadrupole magnet | 1910 | 940 | 1 |
| | | 8 | 1440 | 2 |
| | | 118 | 2140 | 2 |
| | Sextupole magnet | 448 | 360 | 1 |
| | Correctors | 350 | 550 | 1 |
| Magnets in Transport line BTC | Dipole magnet | 68 | 5000 | 4 |
| | Quadrupole magnet | 40 | 1988 | 2 |
| | Corrector | 30 | 300 | 1 |
| | Kicker | 20 | 1000 | 1 |
| | Septum | 140 | 1000 | 1 |
| Magnets in Transport line LTB | Dipole magnet | 48 | 5000 | 4 |
| | | 28 | 4000 | 4 |
| | Quadrupole magnet | 80 | 884 | 1 |
| | Corrector | 24 | 200 | 1 |
| | Kicker | 2 | 500 | 1 |
| | Septum | 4 | 1000 | 1 |

#### 5.3.8.2 *Requirements and Key Technologies for Magnet Supports*

Every Booster magnet is hung and supported independently. Requirements are similar to those for the Collider.

- Range and accuracy of adjustment are the same as the supports in Collider.
- Simple and reliable mechanics for safe mounting and easy alignment.
- Stability over a large time constant, avoiding creep and fatigue deformation.
- Good vibration performance.
- The steel frame must have sufficient strength to achieve these goals.

To install the magnet a transport system will be used. The transport system can move vertically and horizontally to the designated locations of the magnets. The mounting and adjusting structure of the magnet should be simple and flexible in operation.

#### 5.3.8.3 *Topology Optimization of Booster Magnet Supports*

The dipole magnets are very long and hanged to the top of the tunnel, so the deformation and stability of the magnet must be considered. Topology optimization has been used for the Booster magnet supports, including the 2D and 3D optimization. The steel frames are then designed according to the optimization results.

The 2D optimization shows that when the frames is almost vertical, it has the minimum degree of flexibility (the best static strength). The details of the topology optimization processes are described in Reference 1 [1]. And the structure of the steel frame can be re-calculated once the location of Booster magnets are changed.

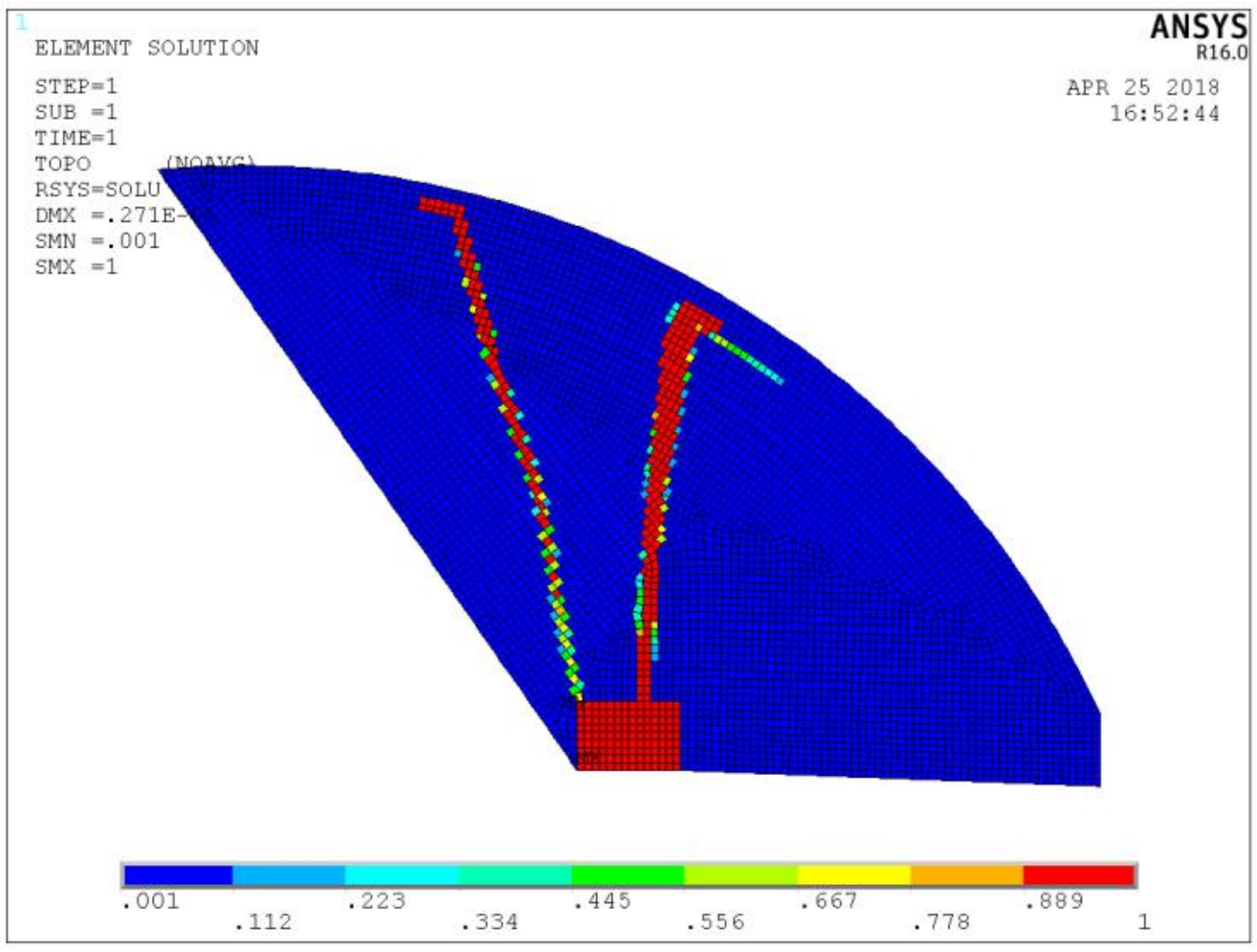

**Figure 5.3.8.1:** Mesh density of the steel frame in the plane perpendicular to the beam

#### 5.3.8.4 *Structure Design of Magnet Supports*

The dipole is the component with the largest quantity. It is about 5,445 mm long. Each magnet is hung from four supports, two main supports and two auxiliary supports, as shown in Fig. 5.3.8.2. The two main supports are for support and adjustment (6 DOFs), while the two auxiliary supports are only for support (1 DOF).

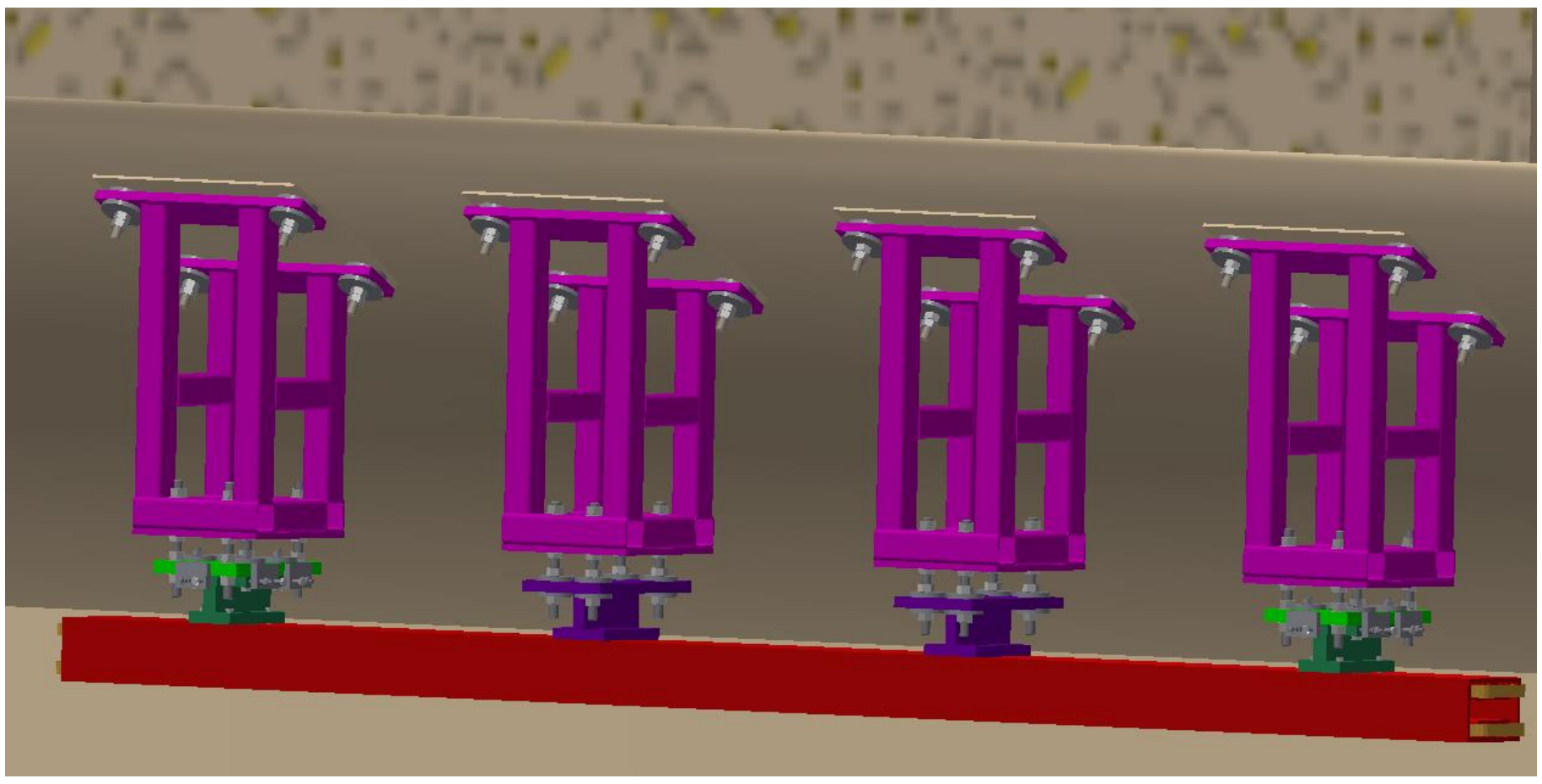

**Figure 5.3.8.2:** Booster dipole magnet and its supports for each module

Other magnets are supported in a similar way. The magnet supports for LTB are similar to those in the Collider, and the magnet supports for the BTC are similar to those in Booster. Only the height changes.

# 6 Linac, Damping Ring and Sources

## 6.1 Main Parameters

The CEPC injector consists of a Linac and a Booster. A normal conducting S-band linac with frequency 2860 MHz provides electron and positron beams at an energy of up to 10 GeV at a repetition rate of 100 Hz and one-bunch-per-pulse. The Linac parameters are shown in Table 6.1.1.

**Table 6.1.1:** Linac parameters

| Parameter | Symbol | Unit | Value |
|---|---|---|---|
| $e^-$ /$e^+$ beam energy | $E_{e^-}/E_{e^+}$ | GeV | 10 |
| Repetition rate | $f$ | Hz | 100 |
| $e^-$ /$e^+$ bunch population | $N_{e^-}/N_{e^+}$ | | >$9.4\times10^9$ |
| | $N_{e^-}/N_{e^+}$ | nC | >1.5 |
| Energy spread ($e^-$ /$e^+$ ) | $\sigma_E$ | | <$2\times10^{-3}$ |
| Emittance ($e^-$ /$e^+$ ) | | nm | <120 |
| $e^-$ beam energy on Target | | GeV | 4 |
| $e^-$ bunch charge on Target | | nC | 10 |
| Length | $L$ | m | 1200 |

The Linac is comprised of both an electron linac and a positron linac as shown in Fig. 6.1.1. There is an electron source and bunching system (ESBS or pre-injector), the first accelerating section (FAS) where the electron beam is accelerated to 4 GeV, a positron source and pre-accelerating section (PSPAS) where the positron beam is produced and accelerated to more than 200 MeV, the second accelerating section (SAS) where the positron beam is accelerated to 4 GeV and the third accelerating section (TAS) where both beams are accelerated to 10 GeV. There is also an electron bypass transport line (EBTL) and a damping ring with energy in 1.1 GeV. The damping ring reduces injection difficulties, and provides higher injection efficiency. This saves damping time in the Booster where the emittance is damped to the required value for the Collider.

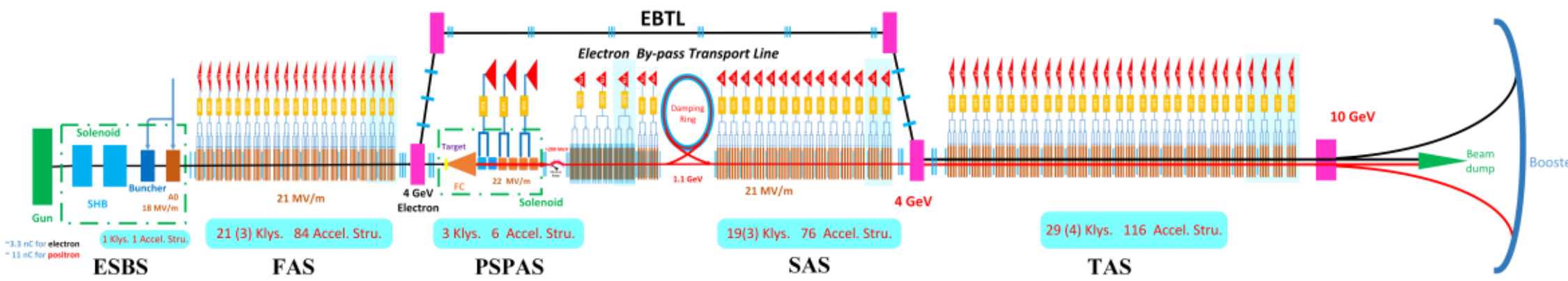

**Figure 6.1.1:** Linac layout.

The Linac as the first injector part, it's high availability and simplicity are the design principles and very important. So there are 15% backups for klystrons and accelerating structures. The Linac should be have potential to meet higher requirements and update in the future, so a damping ring is introduced to decrease emittance, the bunch charge both for electron beam and positron beam should be larger than 3 nC and the two-bunch-per-pulse also is possible.

The S-band accelerating structure is adopt in the whole Linac. Compared with S-band accelerating structure, the C-band accelerating structure have higher accelerating gradient and can reduce tunnel length. However the C-band accelerating structure have smaller aperture and higher short-range Wakefield, so the required beam emittance is samller, the sustainable bunch charge is lower and beam orbit control is more strict; the C-band accelerating structure have higher accelerating gradient and higher frequency, so the phase errors and accelerating gradient errors should be controled more strictly with same exit energy and energy jitter requirements and the bunch length should be smaller with same energy spread requirement. The C-band accelerating structure is a good choice for high energy linac with not very high bunch charge. Considering that the linac energy only is 10 GeV and the S-band accelerating structure have larger error tolerance, the S-band accelerating stucture is adopt even in high energy part in the CEPC linac.

## 6.2 Linac and Damping Ring Accelerator Physics

### 6.2.1 Linac and Damping Ring Design – Optics and Beam Dynamics

#### 6.2.1.1 *Pre-injector*

The pre-injector contains the electron source and a bunching system. The required bunch charge for both electrons and positrons at the Linac exit is larger than 1.5 nC. We have designed for 3 nC. Two operation modes of the electron source are required; one is to provide a 3.3 nC bunch charge for electron injection and the other is to provide a 11 nC bunch charge as the primary electron beam for positron production. The bunching system consists of two sub-harmonic bunching cavities, an S-band buncher and a normal S-band accelerating structure, as shown in Fig. 6.2.1.1 and further details are in section 6.5.1 (RF system).

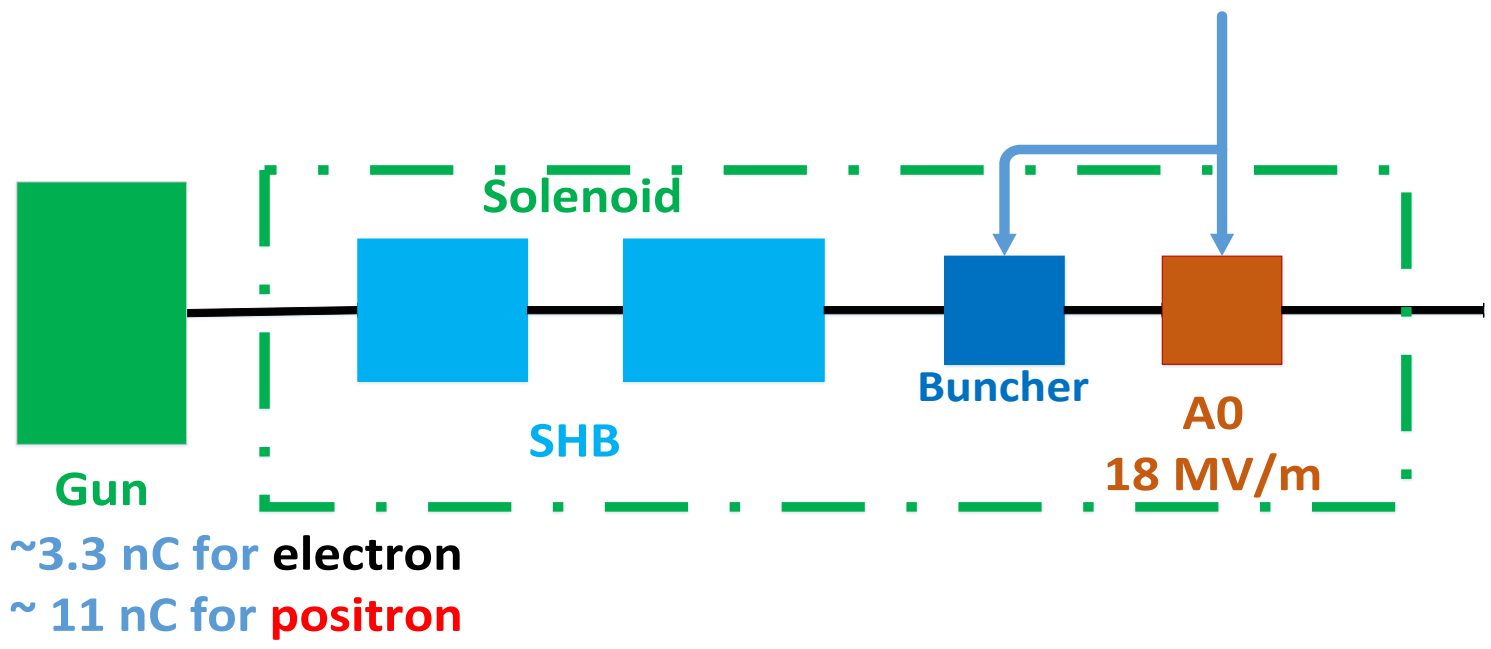

**Figure 6.2.1.1:** Layout of the bunching system and pre-accelerating section.

The two sub-harmonic pre-bunchers and the one S-band buncher act to velocity modulate the non-relativistic electron beam emerging from the gun, and compress the pulse length before it passes into the Linac. A pulsed beam with 1.0 ns FWHM length from the gun is compressed into a single bunch of about 10 ps (FWHM) by the bunching system.

After bunching the electron beams are accelerated to 50 MeV with one S-band accelerating structure. Here one klystron provides power to the buncher and the first accelerating structure. The transverse focussing element in the bunching system is solenoid; the magnetic field is shown in Fig. 6.2.1.2.

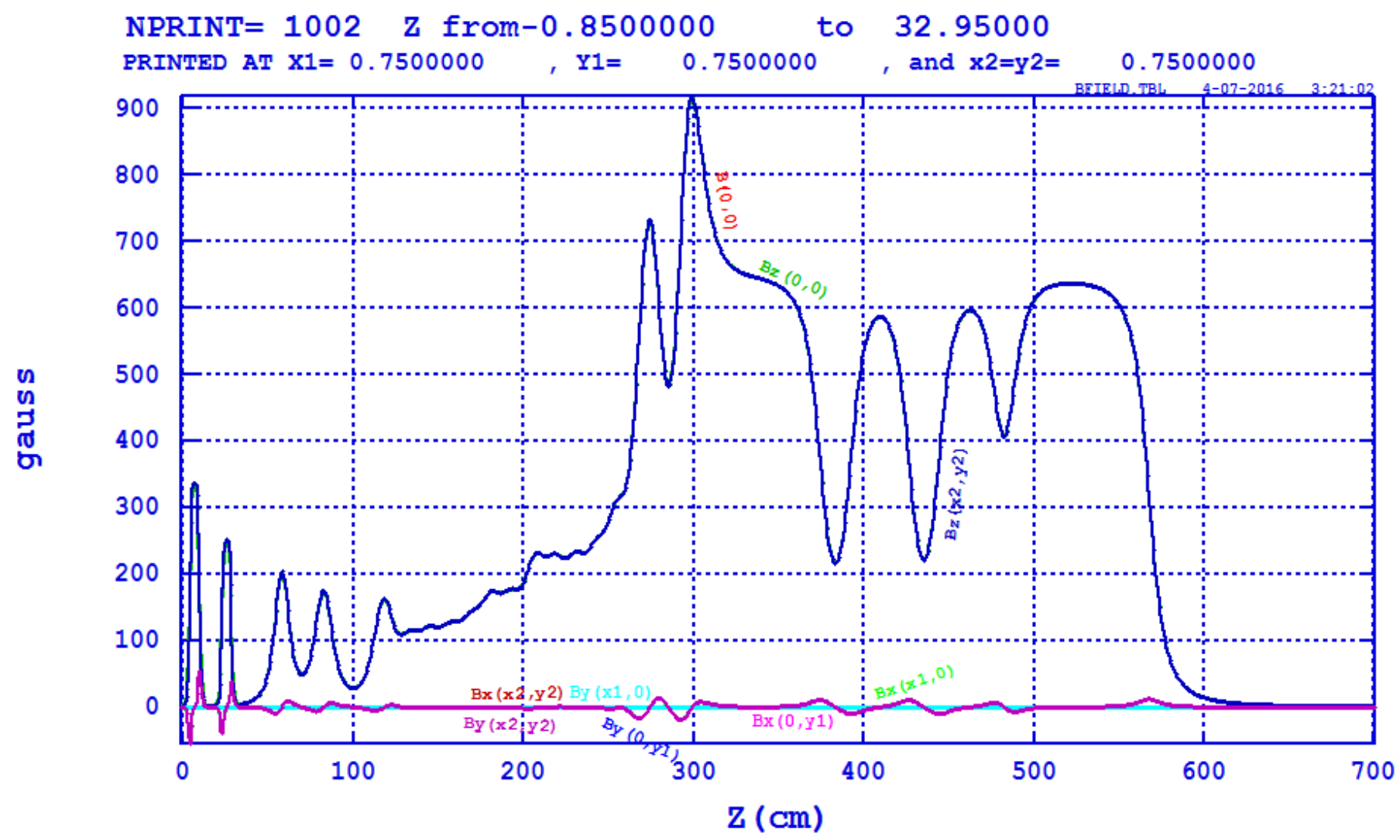

**Figure 6.2.1.2:** Magnetic field in the pre-injector.

Parmela is used to simulate the beam dynamics of the pre-injector including space charge effect [1]. The beam envelopes in the transverse and longitudinal planes are shown in Fig. 6.2.1.3; the beam sizes are controlled carefully to reduce beam loss. Beam distribution at exit of the pre-injector and normalized rms emittance along the beam direction is shown in Fig. 6.2.1.4. The normalized rms emittance at the pre-injector exit is 80 mm-mrad and the transmission efficiency of the pre-injector is about 90%.

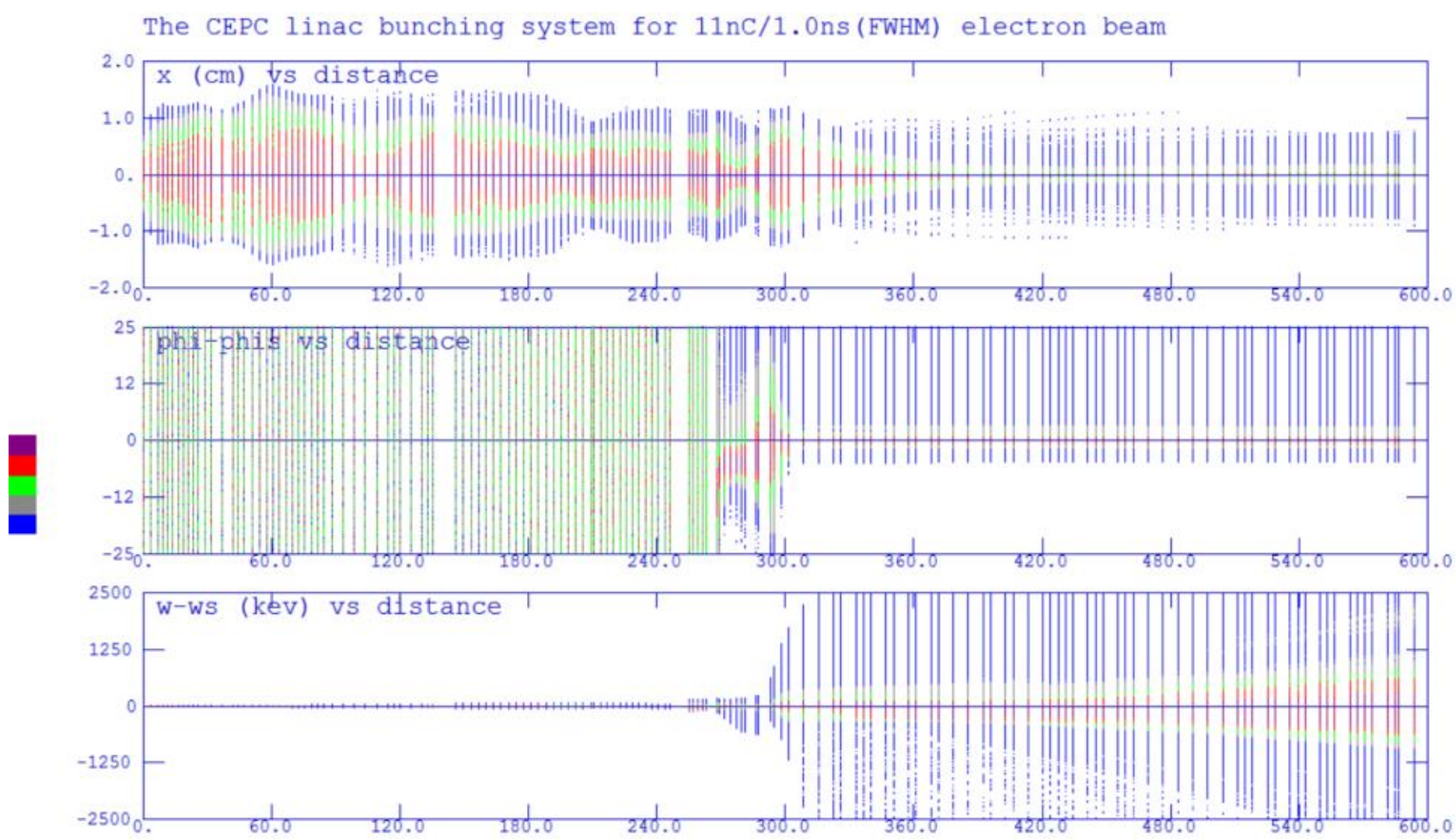

**Figure 6.2.1.3:** Beam envelopes in the bunching system and the pre-accelerating section.

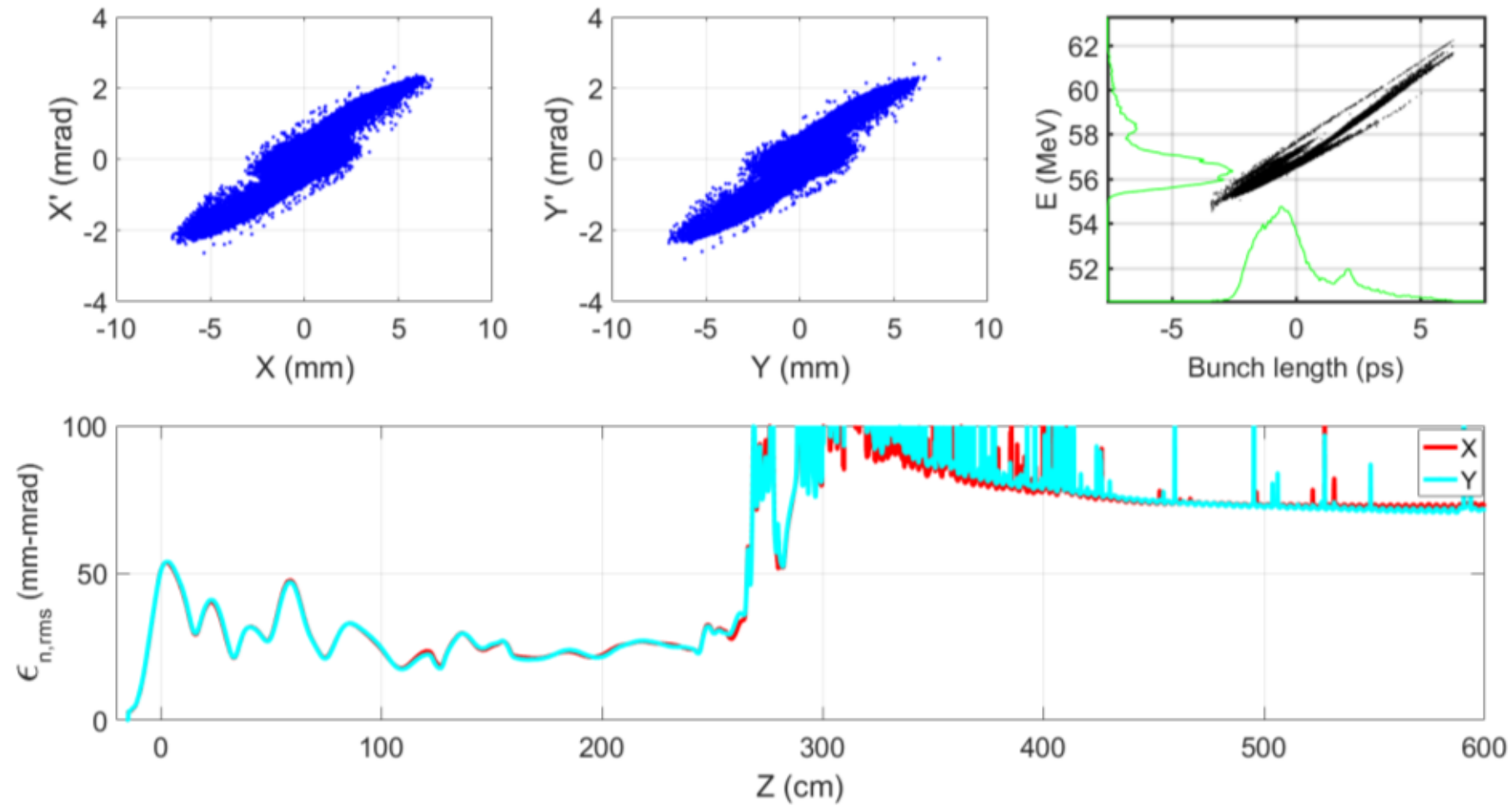

**Figure 6.2.1.4:** Beam distribution at the exit of the pre-injector (top) and normalized rms emittance (below) along the beam direction.

### 6.2.1.2 *High Bunch Charge Electron Linac*

The high bunch charge electron Linac is the first accelerating section (FAS) for positron beam production. FAS accelerates the electron beam from 50 MeV to 4 GeV with a maximum bunch charge of 10 nC. In this section one klystron with SLED drives 4 accelerating structures; further details are in section 6.5.1 (RF system). There are two transverse focucsing structures placed in different energy sections, one-triplet-four-accelerating-structures in one period and one-triplet-eight-accelerating-structures in one period. These are shown in Fig. 6.2.1.5.

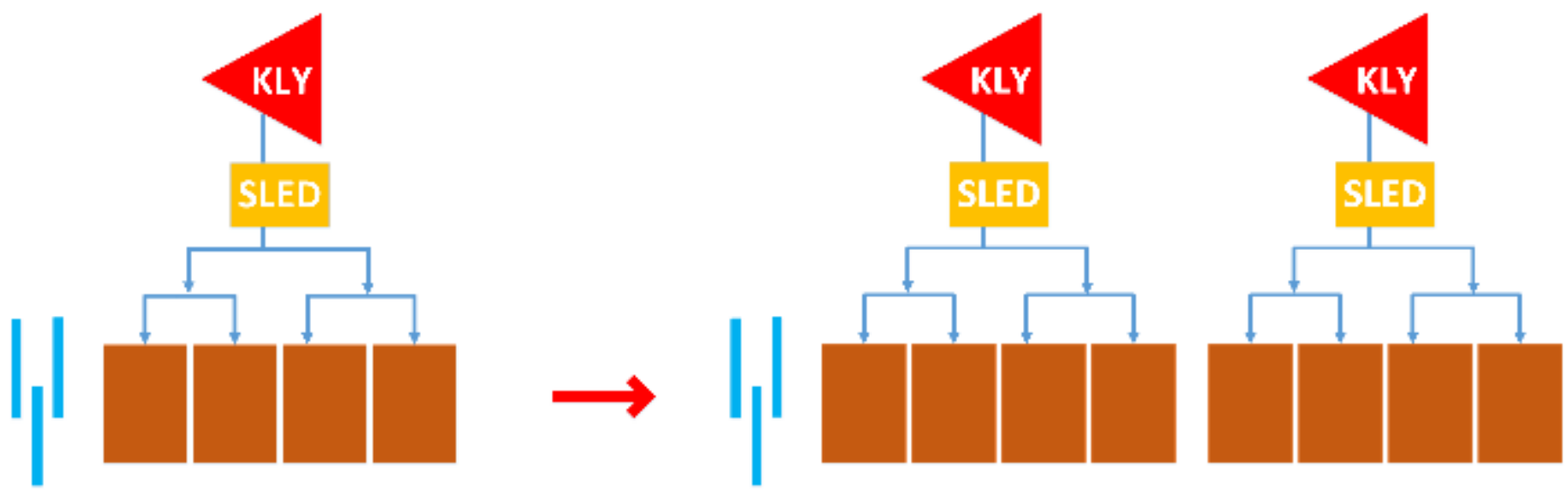

**Figure 6.2.1.5:** Transverse focusing structures in FAS.

The short-range longitudinal and transverse wakefields are simulated with a short-range wakefield model for periodic linac structures [2]. Results are shown in Fig. 6.2.1.6. To overcome the effect of the longitudinal wakefield the accelerating phase must be carefully designed. The beam distribution at the FAS exit is shown in Fig. 6.2.1.7 and the beam envelope is shown in Fig.6.2.1.8. The energy spread is large but can meet the requirements for positron production; the rms beam size can be controlled within 0.5 mm.

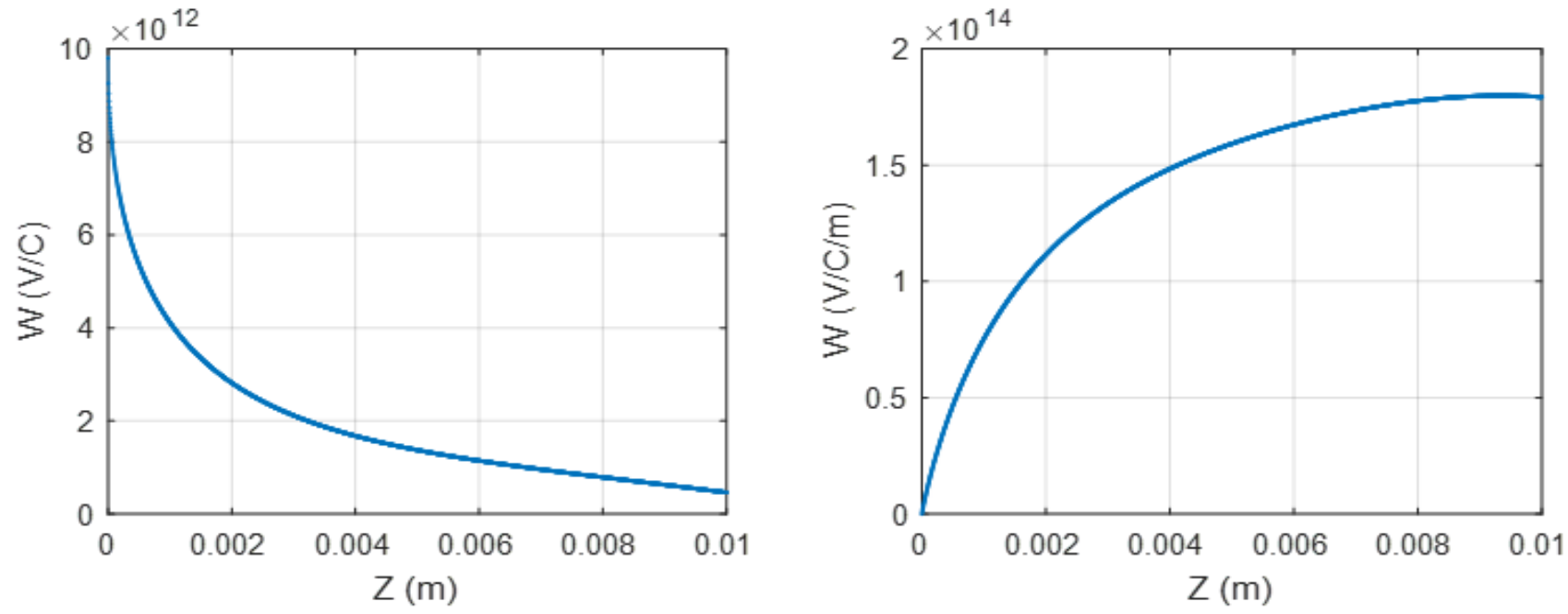

**Figure 6.2.1.6:** The short-range wakefields of the S-band accelerating structure. Left: longitudinal; right: transverse.

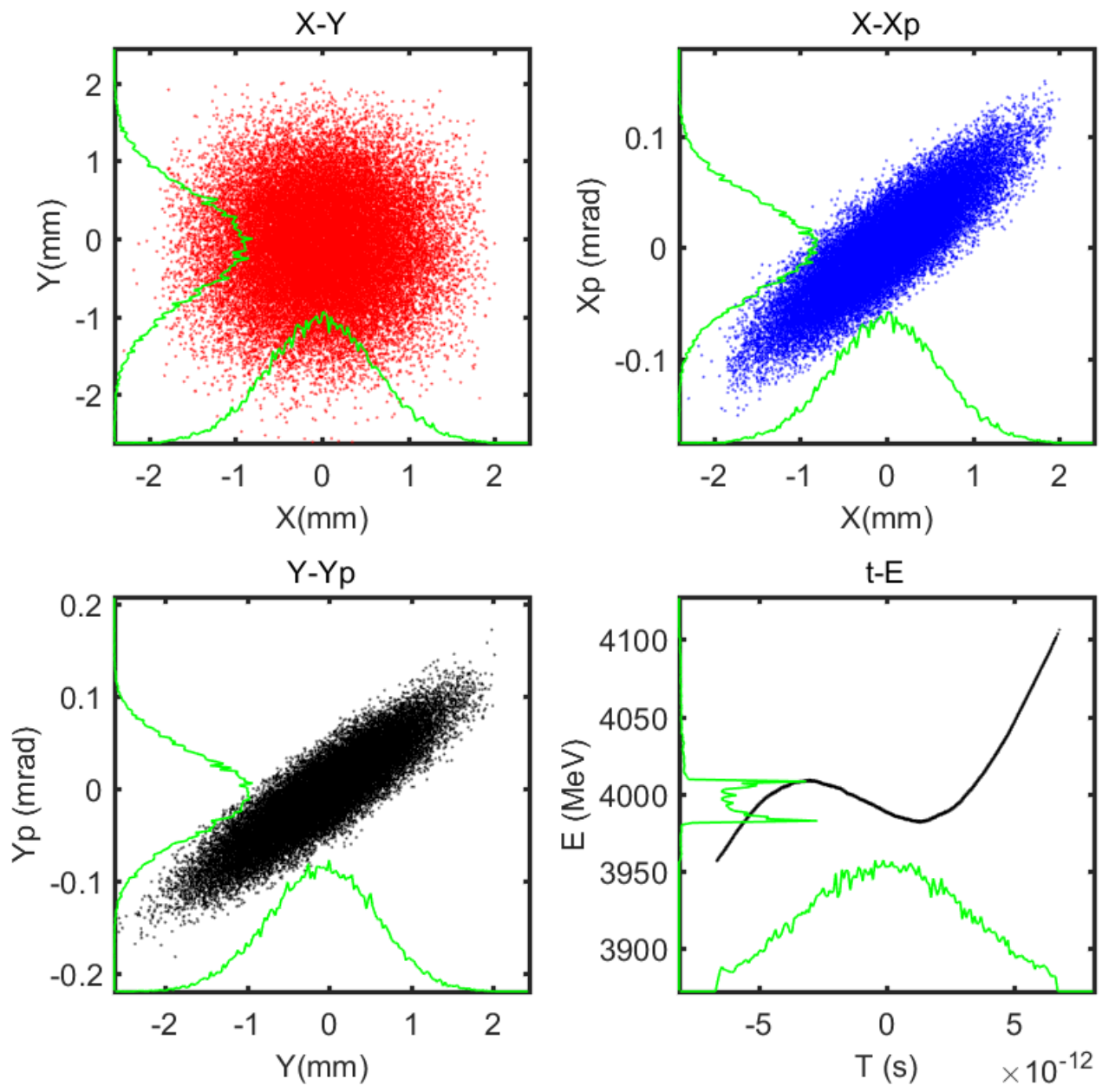

**Figure 6.2.1.7:** Beam distributions at the FAS exit in the high bunch charge mode.

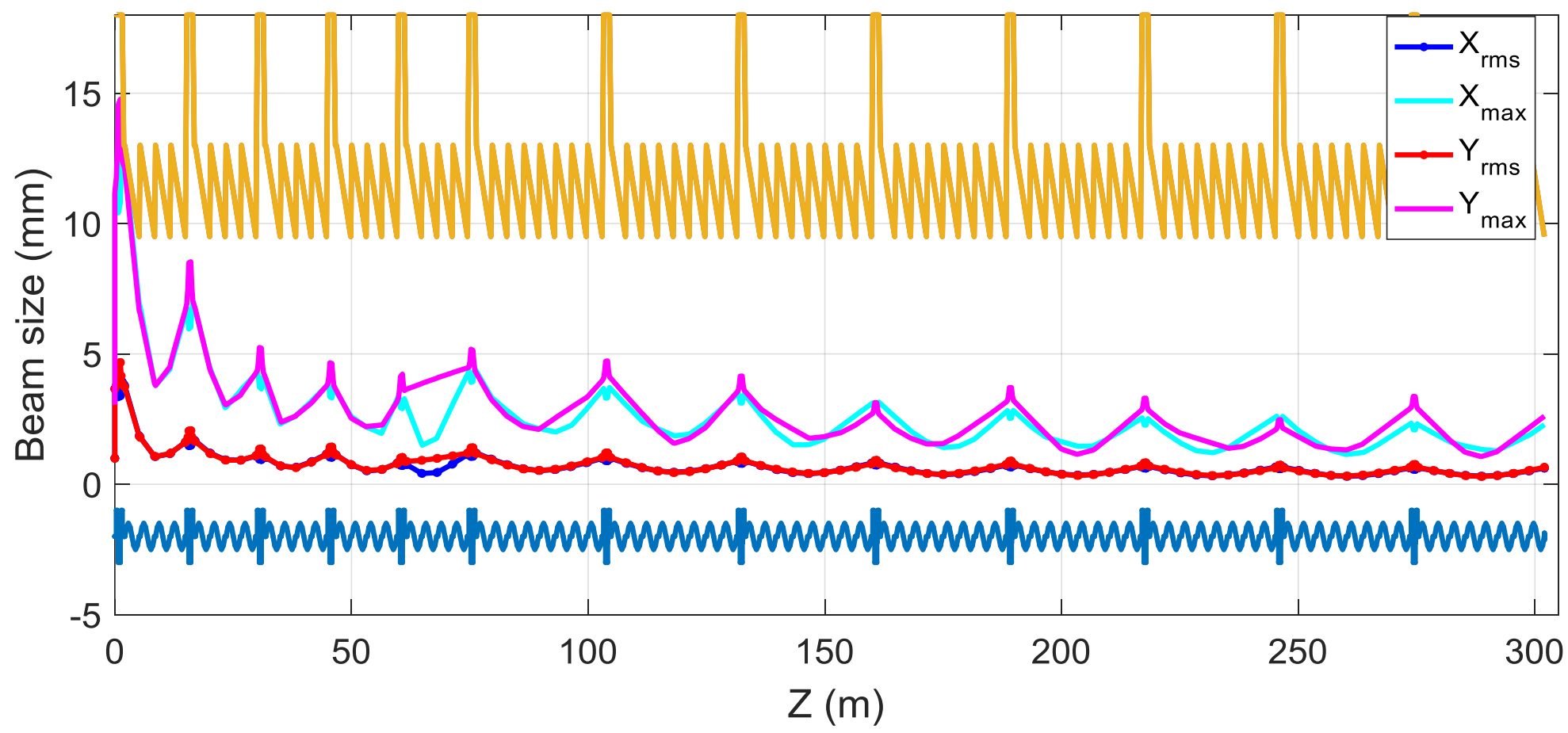

**Figure 6.2.1.8:** Beam envelope along the beam direction in FAS.

### 6.2.1.3 *Positron Capture and Pre-Accelerating Section*

A schematic of the positron source and pre-accelerating section (PSPAS) is shown in Fig. 6.2.1.9, composed of the target, flux concentrator (FC) which is an adiabatic matching device (AMD), caputre accelerating structures (blue), pre-acclerating structures (orange) and a chicane system. To achieve a 3 nC bunch charge positron beam, a 4 GeV primary electron beam with an maximum intensity of 10 nC/bunch is designed. The maximum average beam power is 4 kW at a repetition rate of 100 Hz.

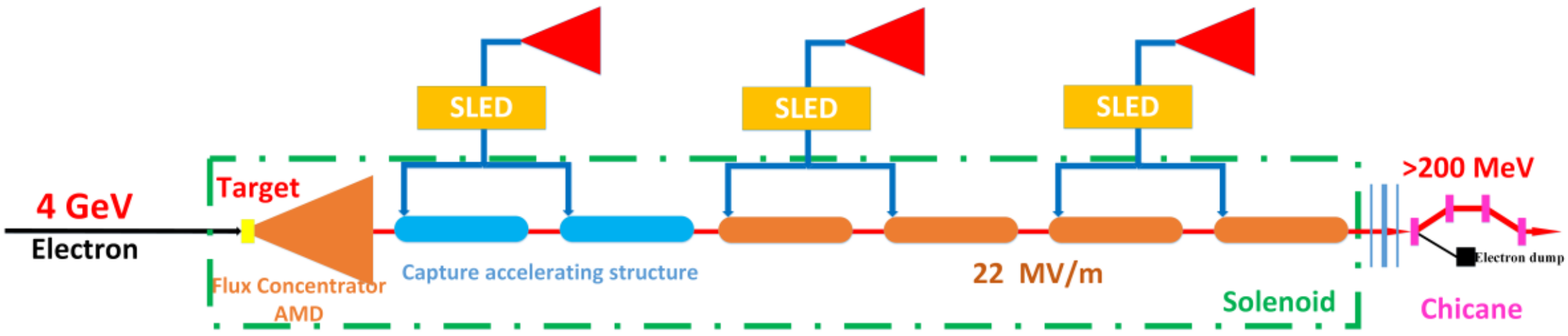

**Figure 6.2.1.9:** PSPAS layout.

The layout of target and AMD is shown in Fig. 6.2.1.10 and the magnetic field of the PSPAS is shown in Fig. 6.2.1.11. Immediately following the target there are six 2-m long high-gradient constant-impedance S-band (2860 MHz) accelerating structures with large aperture. The PSPAS parameters are listed in Table 6.2.1.1. The detailed design of the positron source target is dicussed in Section 6.4. The chicane system is designed to separate the electron beam into the beam dump.

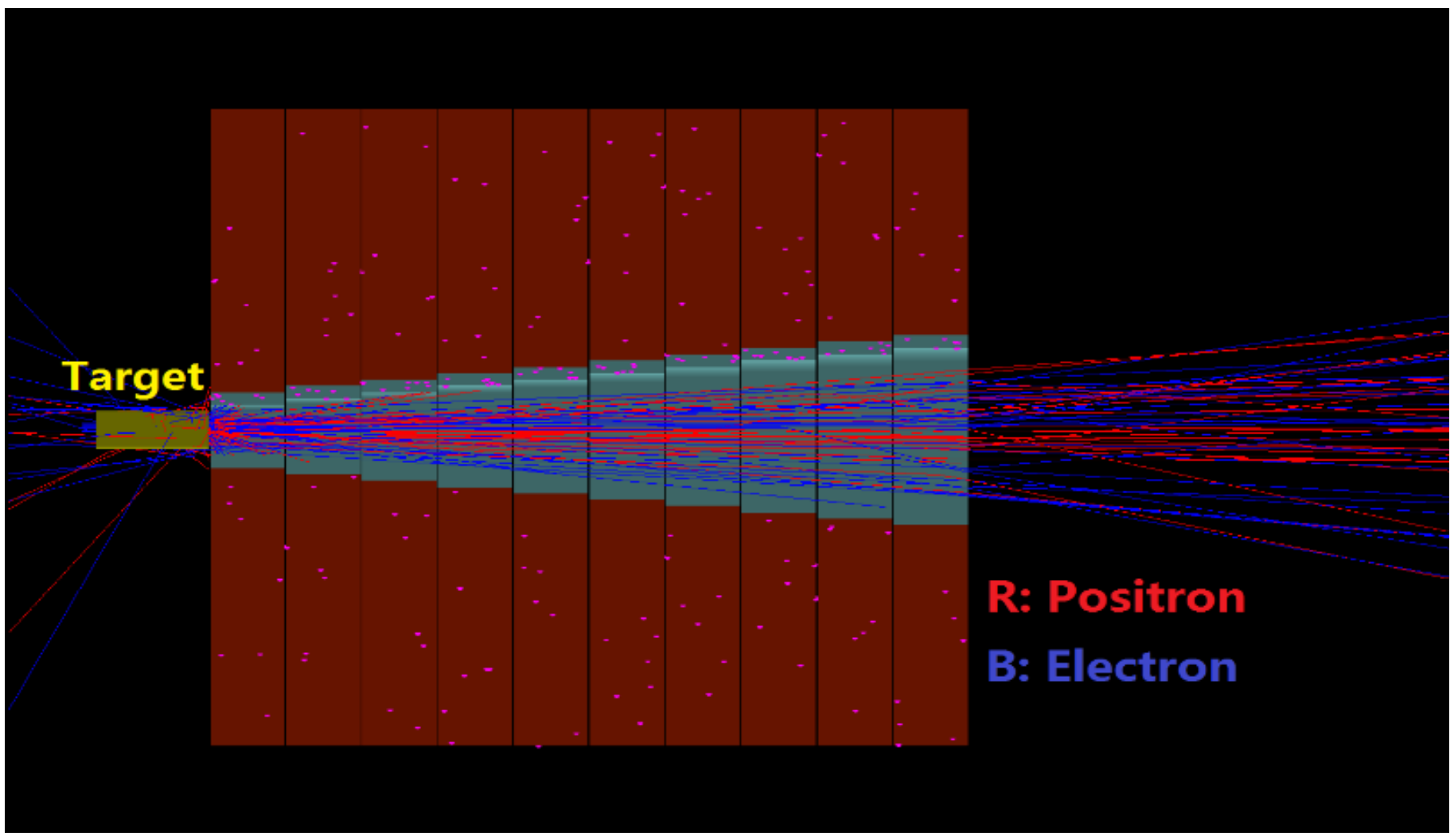

**Figure 6.2.1.10:** Layout of the target and the AMD.

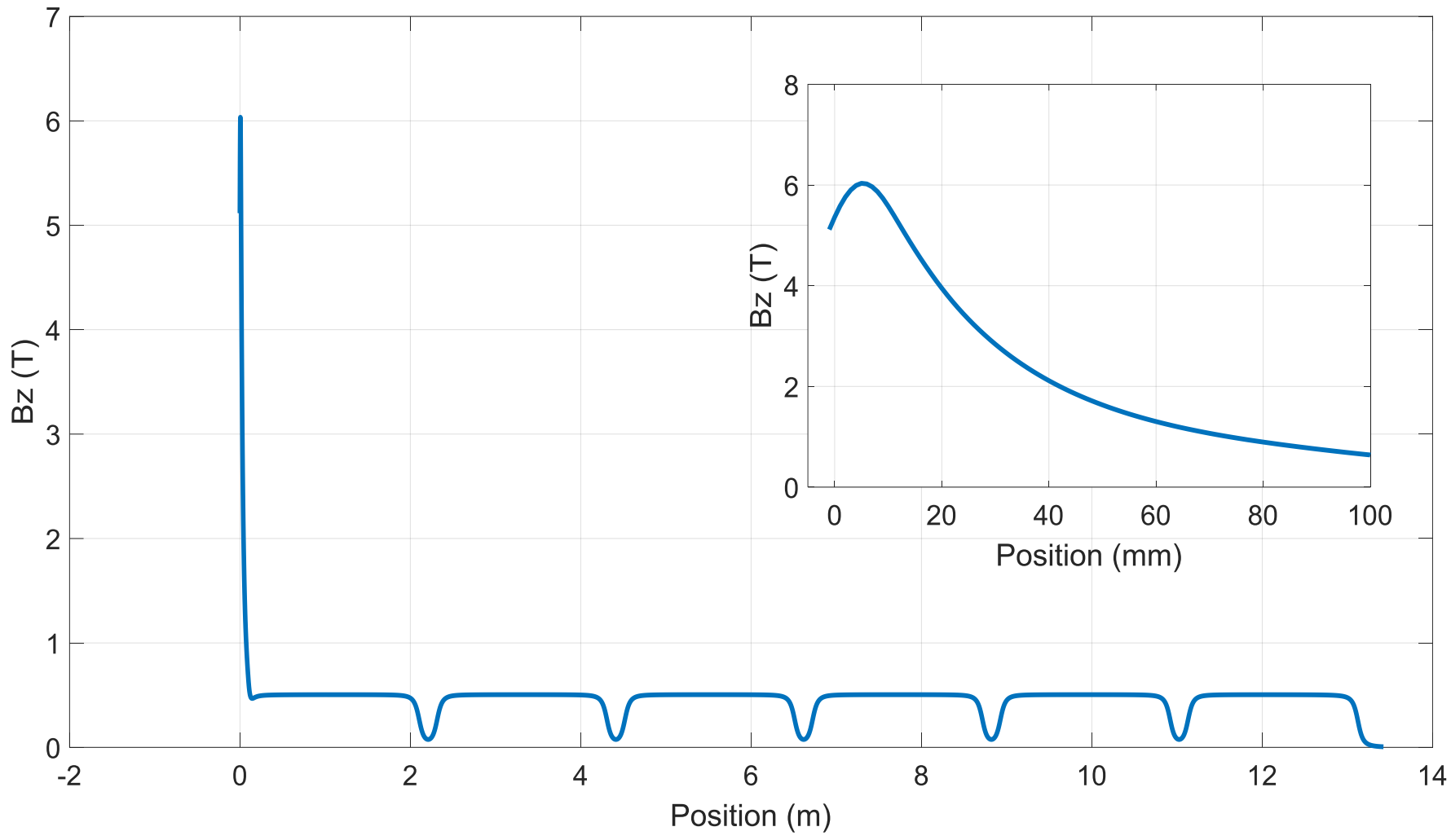

**Figure 6.2.1.11:** PSPAS magnetic field.

**Table 6.2.1.1:** PSPAS parameters

| Positron source | Unit | Value |
|---|---|---|
| $e^-$ beam energy on the target | GeV | 4 |
| $e^-$ bunch charge on the target | nC | 10 |
| Target material | | W |
| Target thickness | mm | 15 |
| Focus device (Flux Concentrator) peak magnet field | T | 5.5 |
| $e^+$ bunch charge after capture | nC | >3 |
| $e^+$ Energy after capture section | MeV | >200 |

In the pre-accelerating section one klystron drives two accelerating structures. The accelerating structure aperture is chosen as 25 mm. This choice is based on considerations of capture efficiency, emittance and accelerating structure design. Figure 6.2.1.12 shows the positron yield at the second capture accelerating structure exit with different accelerating gradients and with different input phase matching the RF phase. There are two phase ranges where there is higher positron yield. These are called deceleration mode and acceleration mode. Based on positron yield and considering the beam energy, the accelerating gradient chosen is 22 MV/m.

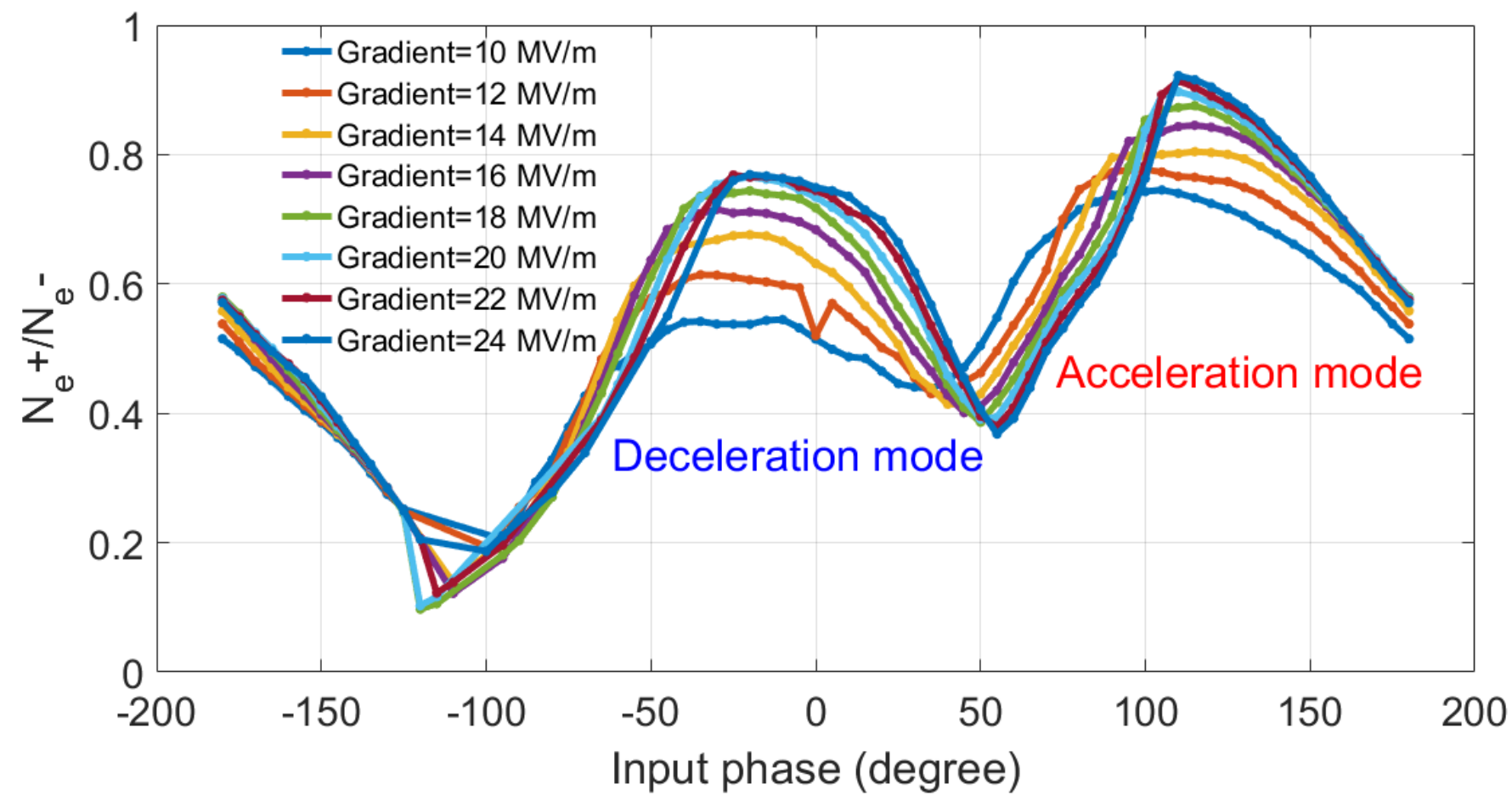

**Figure 6.2.1.12:** Positron yield at the second capture accelerating structure exit with different accelerating gradients and input phases.

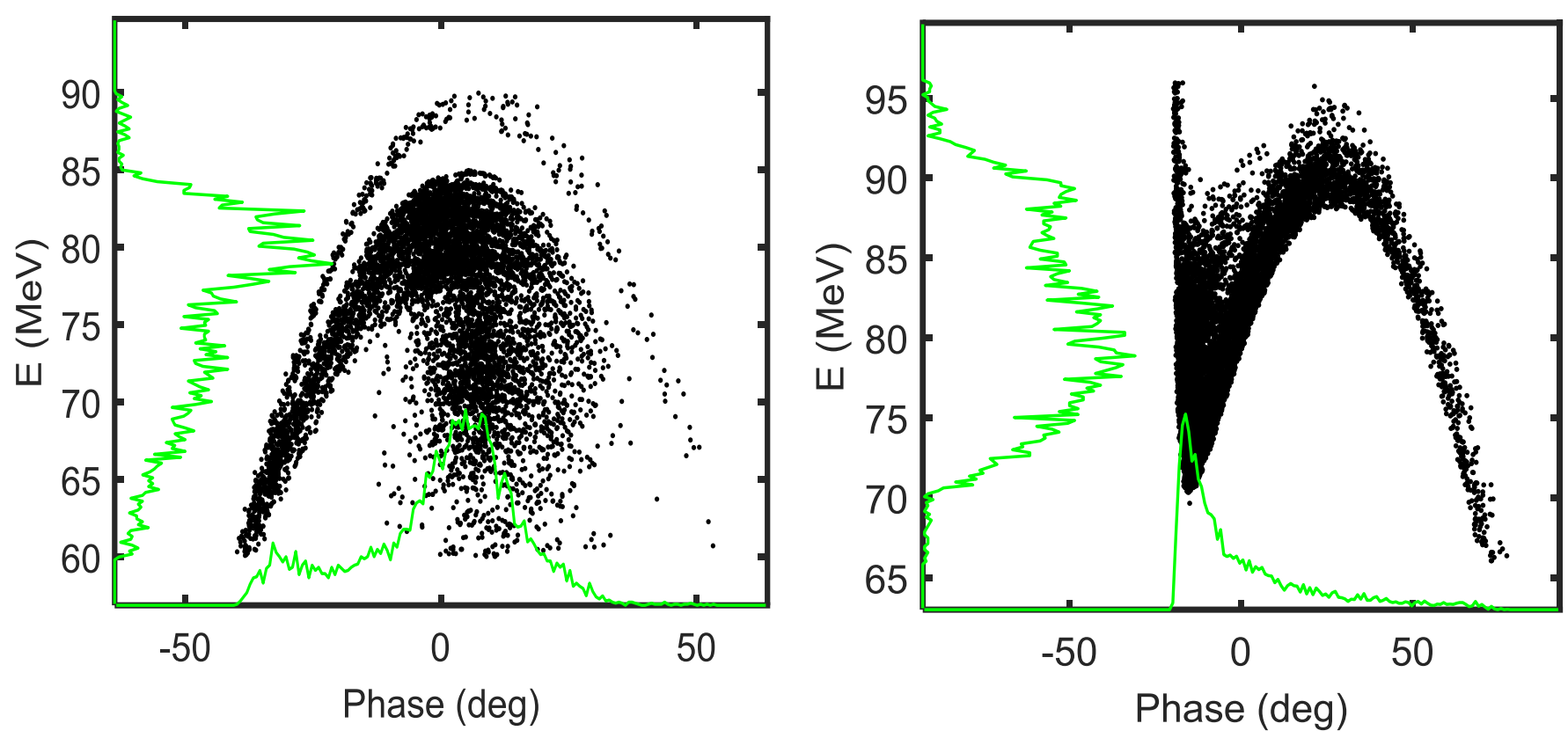

**Figure 6.2.1.13:** Beam distribution at the capture accelerating structure exit. Left deceleration mode. Right: acceleration mode.

The beam distributions in deceleration and acceleration modes at the capture accelerating structure exit are shown in Fig. 6.2.1.13. From the simulation results acceleration mode has a more compact phase spectrum. The acceleration mode is used in the simulation, although the deceleration mode is also possible in the operation.

After detailed optimization of the RF phase of the pre-accelerating structure, beam envelopes are calculated and shown in Fig. 6.2.1.14. The beam distributions at the PSPAS exit are shown in Fig. 6.2.1.15, where the energy cut off condition is [235 MeV, 265 MeV] and the phase cut off condition is [-8 degree, 12 degree]. The positron yield ($N_e^+/N_e^-$) is larger than 0.5 which means a 10 nC electron beam can produce a 5 nC positron beam at the PSPAS exit with the cut off condition. So it is possible that to meet the requirements the electron beam for positron production could be smaller than 10 nC.

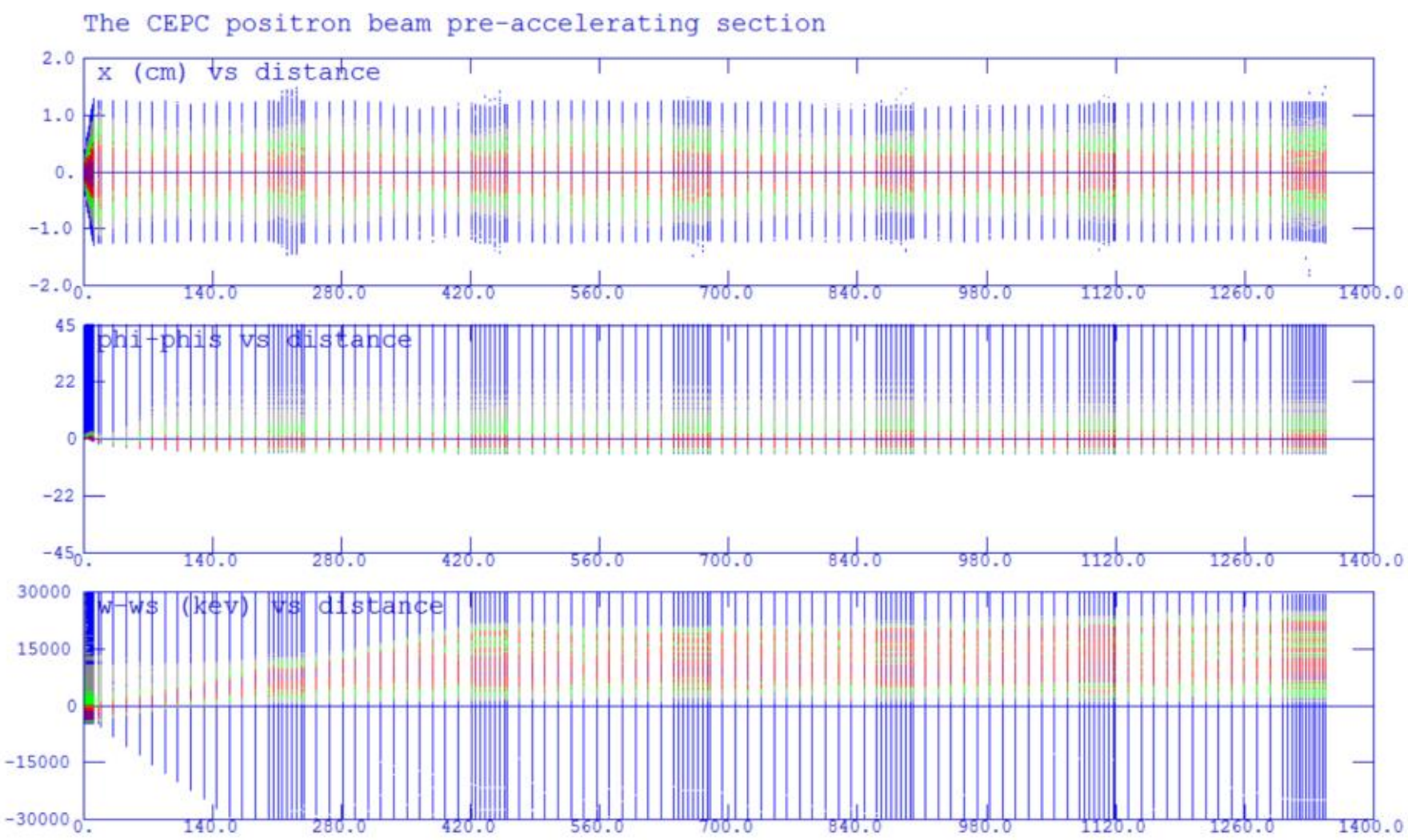

**Figure 6.2.1.14:** Beam envelopes in the pre-accelerating section.

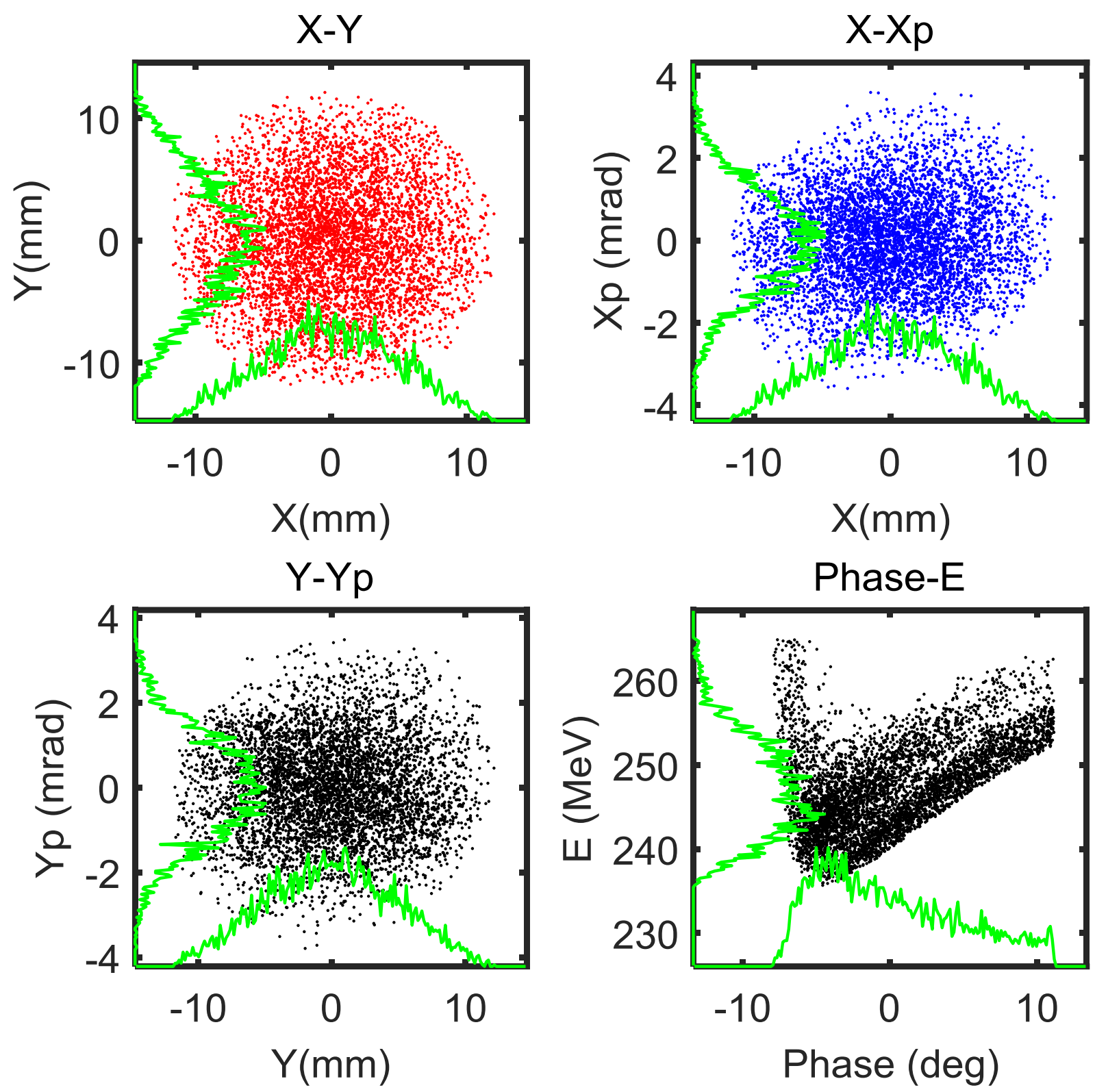

**Figure 6.2.1.15:** Beam distribution at the PSPAS exit.

### 6.2.1.4 *Positron Linac*

The positron Linac is composed of the SAS and the TAS with bunch charge 3 nC. Simulations are performed over an energy range from 200 MeV to 10 GeV. The third accelerating section is shared and accelerates both prositron and electron beams from 4 GeV to 10 GeV. Because the emittance of the positron beam is larger than the electron beam, the lattice of the TAS is based on the positron beam requirements. In the low-energy part of the SAS, the focusing structure is FODO and the quadrupoles nest on the accelerating structure. As the emittance decreases with energy rises and the damping ring the focusing structures are varied to decrease the number of required quadrupoles: one-triplet-one-accelerating-structure, one-triplet-four-accelerating-structures and one-triplet-eight-accelerating-structures. Four focusing structures are shown schematically in Fig. 6.2.1.16. The maximum magnetic field on the quadrupole pole tip should be smaller than 0.6 T.

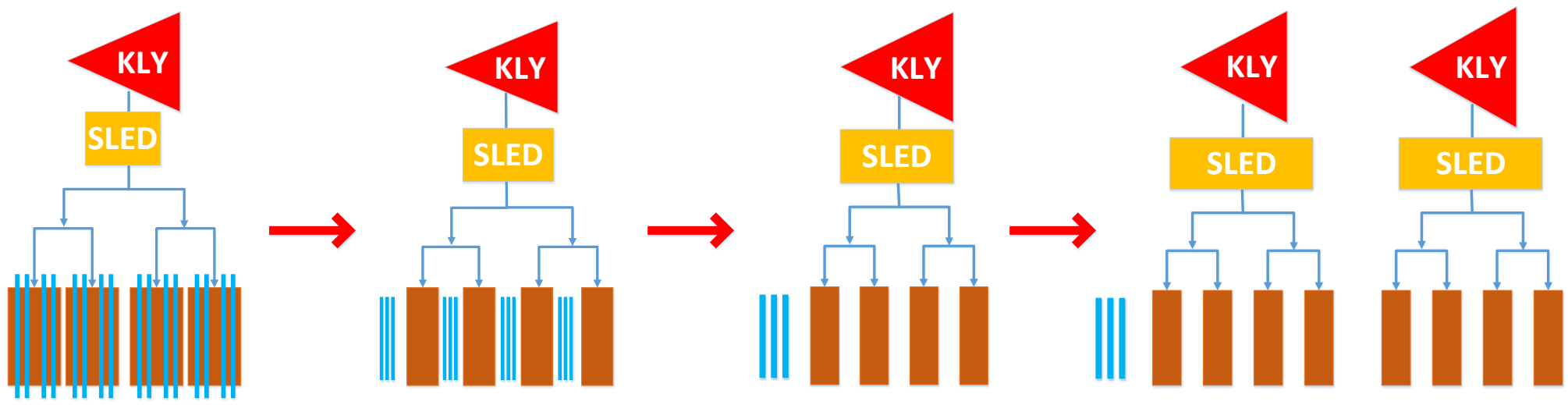

**Figure 6.2.1.16:** The focusing structures of positron Linac.

Beam simulation results are shown in Fig. 6.2.1.17. These calculations take into account the short-range wakefield with a bunch charge 3 nC. At linac exit the energy spread is 0.16% and rms emittance is 40 nm with the damping ring, which all meet the Booster requirements. The break in the plots is at the position of the DR.

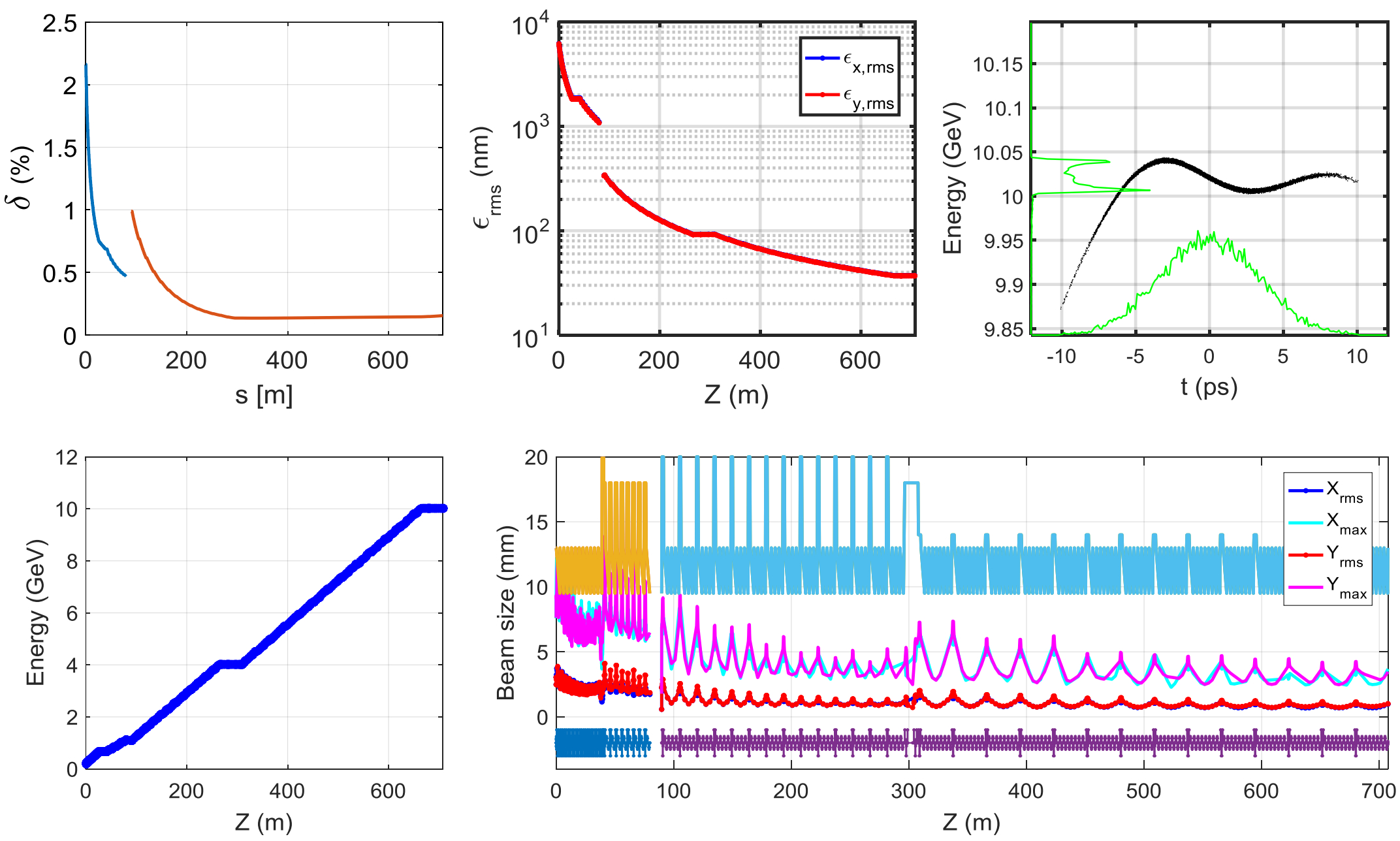

**Figure 6.2.1.17:** Simulation results along the positron Linac (the break in the graph is at the DR position). Shown are rms energy spread (top left), rms emittance (top middle), longitudinal phase space distribution (top right), energy (down left) and beam sizes (down right).

### 6.2.1.5 *Electron Linac*

The electron Linac is composed of FAS, EBTL and TAS with bunch charge 3 nC. The optical functions are shown in Fig.6.2.1.18. The horizontal distance between EBTL and the Linac is 2 m. Dynamics results with the bypass section are shown in Fig. 6.2.1.19; the rms energy spread is 0.11% and the rms emittance is about 5 nm at the electron Linac exit. This meets the Booster requirements.

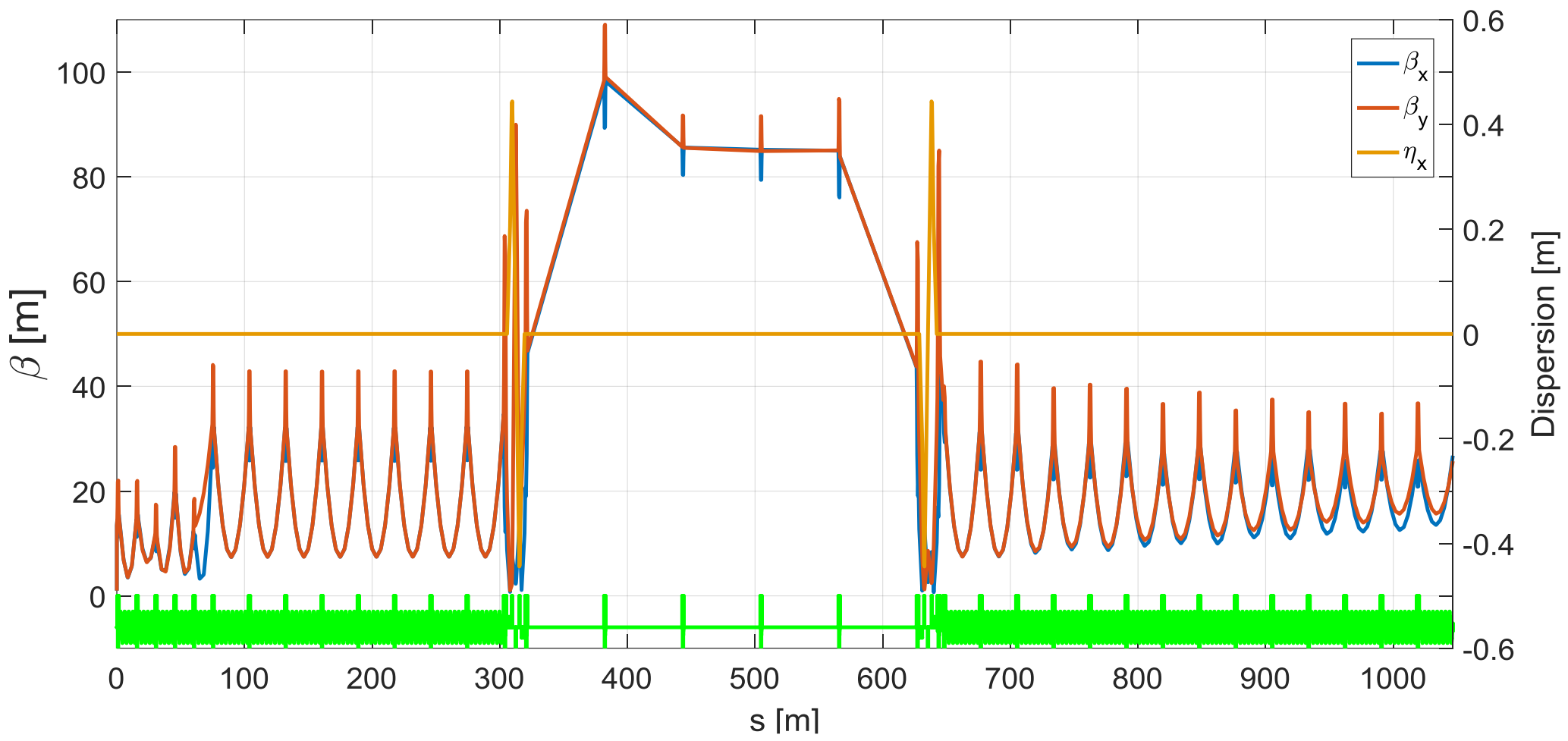

**Figure 6.2.1.18:** Optical functions of the electron Linac.

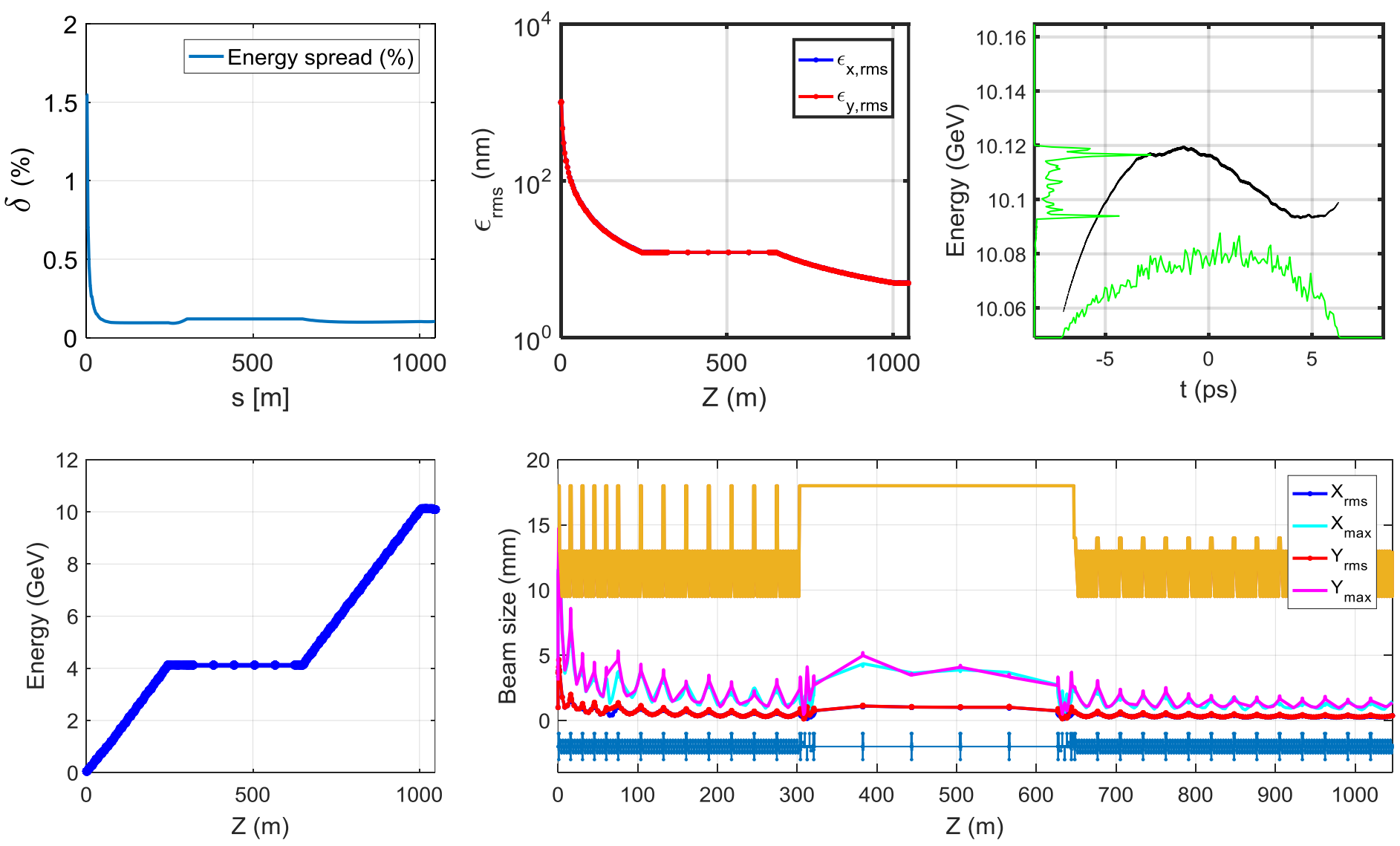

**Figure 6.2.1.19:** Beam dynamic simulation results for the electron Linac.

## 6.2.1.6 *Damping Ring*

The damping ring (DR) reduces the emittance. It has an energy of 1.1 GeV and circumference 58.5 m [3]. It is racetrack shaped as shown in Fig. 6.2.1.20. The DR parameters are shown in Table 6.2.1.2. The lattice uses 60°/60° FODO cells and interleaved sextupoles. The optical functions are shown in Fig.6.2.1.21. The chromaticity of the damping ring is corrected by two sextupole families. The DA is calculaed using SAD and the results of the simulation are shown in Fig. 6.2.1.22, The DA is 7 times larger than the rms size of the injected beam. Considering the coherent synchrotron radiation (CSR) and the DR momentum acceptance, a longer bunch length and smaller energy spread is desirable. So there is an energy-spread compression system (ECS) to compress energy spread and lengthen the bunch before the injection of DR. To meet the requirement of energy spread at the Linac exit, a bunch compression system (BCS) compresses the bunch length to 1 mm after DR [3].

**Table 6.2.1.2:** Main parameters of the Damping Ring

| DR V1.0 | **Unit** | **Value** |
|---|---|---|
| Energy | GeV | 1.1 |
| Circumference | M | 58.5 |
| Repetition frequency | Hz | 100 |
| Bending radius | m | 3.62 |
| Dipole strength $B_0$ | T | 1.01 |
| Momentum Compaction Factor $\alpha_c$ | | 0.076 |
| $U_0$ | keV | 35.8 |
| Damping time x/y/z | ms | 12/12/6 |
| $\delta_0$ | % | 0.05 |
| $\varepsilon_0$ | mm.mrad | 287.4 |
| Bunch length $\sigma_z$ | mm | 7 (23ps) |
| $\varepsilon_{inj}$ | mm.mrad | 2500 |
| $\varepsilon_{ext\ x/y}$ | mm.mrad | 704/471 |
| $\delta_{inj}$ /$\delta_{ext}$ | % | 0.3/0.06 |
| Energy acceptance by RF | % | 1.0 |
| $f_{RF}$ | MHz | 650 |
| $V_{RF}$ | MV | 1.8 |

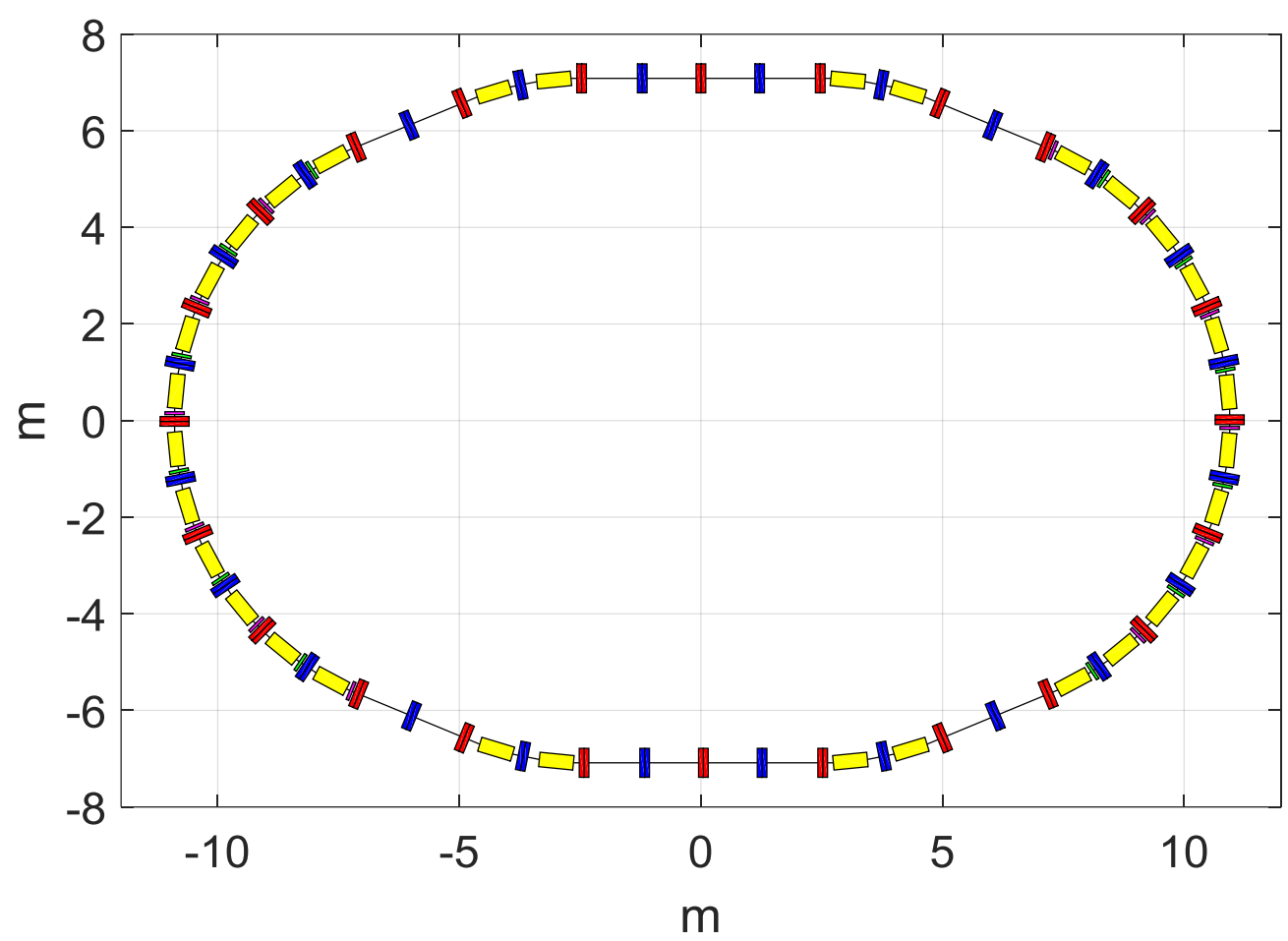

**Figure 6.2.1.20:** The damping ring layout.

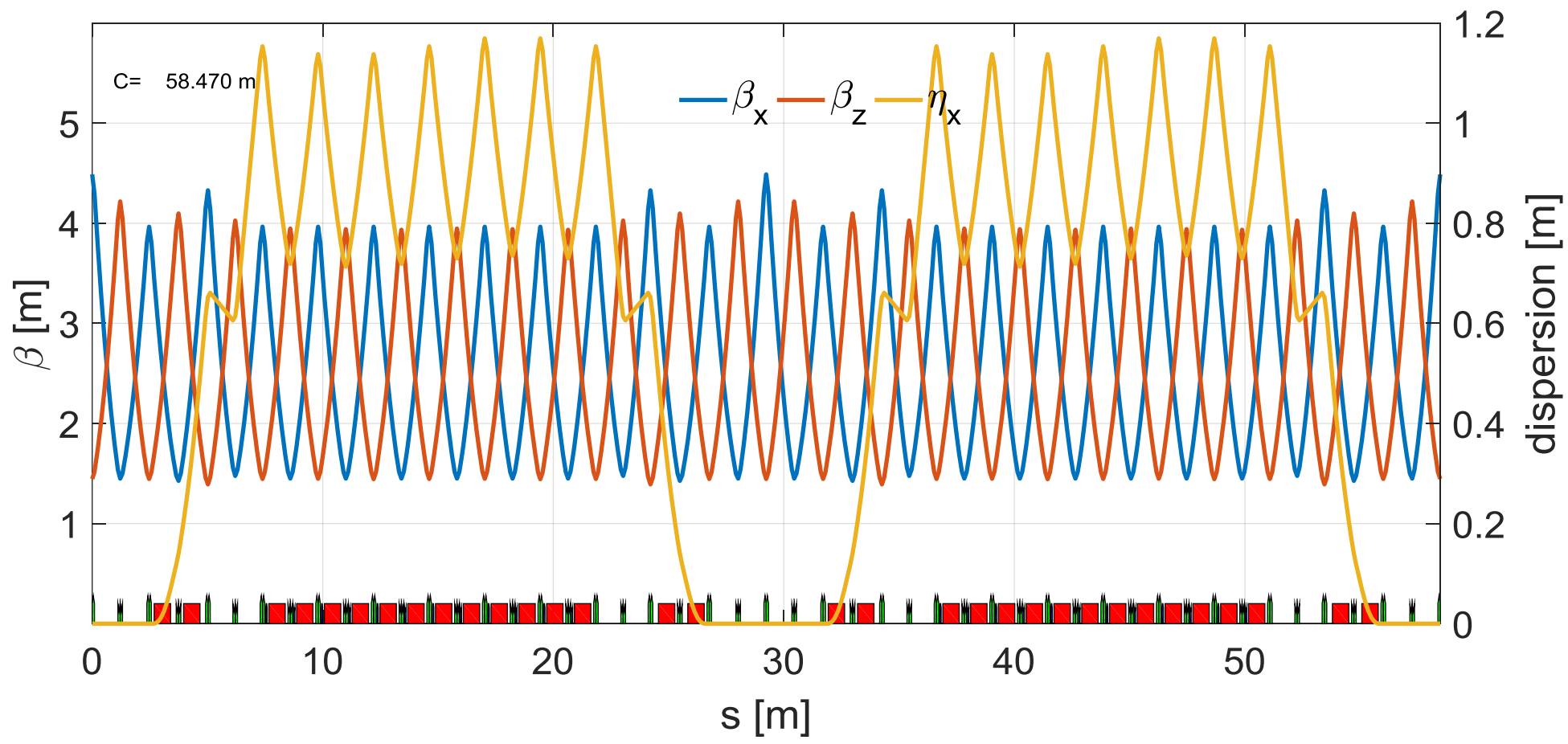

**Figure 6.2.1.21:** Optical functions of the entire DR.

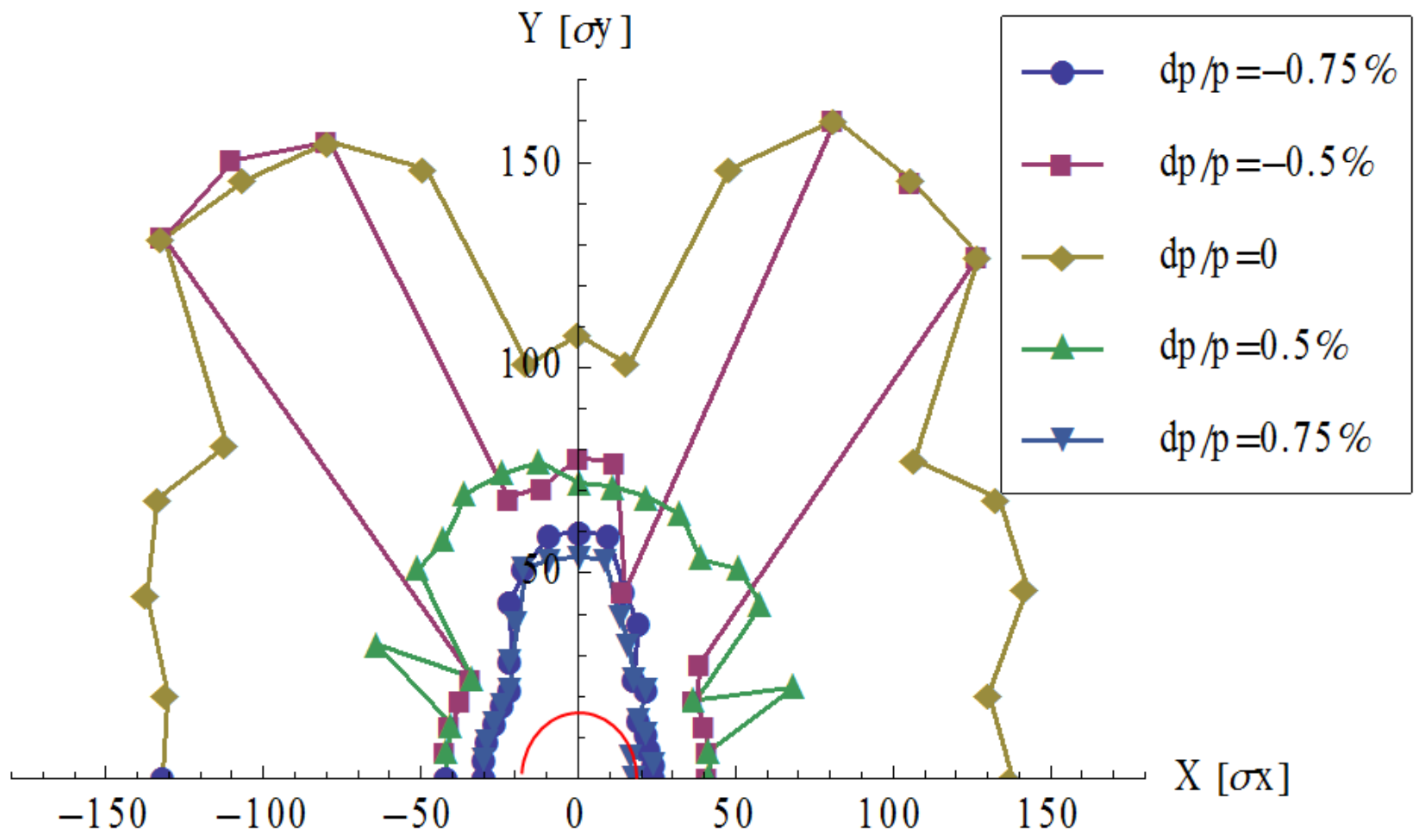

**Figure 6.2.1.22:** DA of the DR; Red line is 7 times the injection beam size.

### 6.2.1.7 *Error Study*

We classify the error sources into three groups:

1. Misalignment errors (translation and rotation) affects all devices with e.m. fields: solenoids, dipoles, quadrupoles, accelerating cavities.
2. Field errors affect the fields as well as the phases of accelerating structures and the fields of magnets.
3. BPM uncertainty errors affects the orbit correction.

All these error sources can be static or dynamic. Beam orbit jitter caused by magnet vibration from ground vibration or other mechanical vibrations cannot be corrected. We need to control the beam orbit jitter carefully to meet the Booster injection requirements. Fig. 6.2.1.23 shows the rms beam orbit jitter for different quadrupole vibration amplitudes. The rms orbit jitter should be smaller than 0.2 mm which means that the maximum orbit jitter is about 0.6 mm. Quadrupole vibration amplitude need be controlled to within 5 μm.

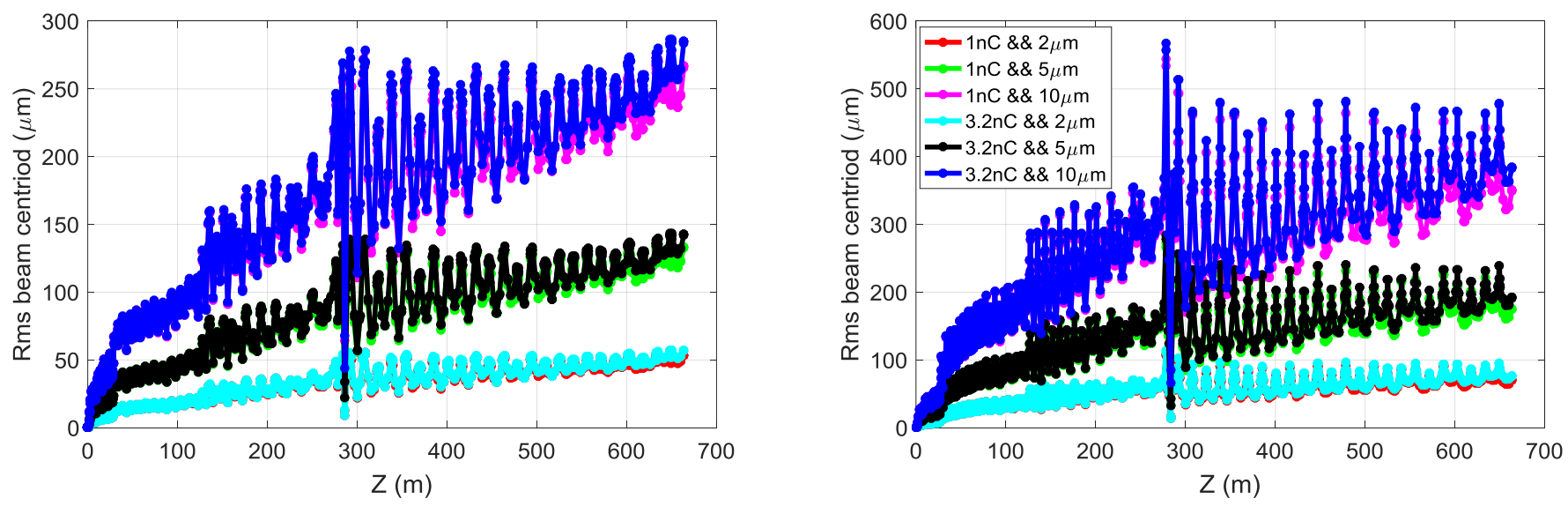

**Figure 6.2.1.23:** The rms beam orbit jitter with different quadrupole vibration amplitudes along the Linac.

The errors settings for the error study are shown in Table 6.2.1.3. They are based on actual engineering experience. The error distribution is a 3σ truncated Gaussian distribution. The rms beam orbits with error are shown in Fig.6.2.1.24. From simulation one can observe that the beam centriods with errors are too large and correction is necessary. A one-to-one correction scheme is used and each period has one pair of correctors and one BPM. The simulation with correction are also shown in Fig.6.2.1.24. The maximum beam centriod is smaller than 0.6 mm in the low-energy section and 0.3 mm in the high-energy section.

**Table 6.2.1.3:** Error settings for the error study

| Error description | Unit | Value |
|---|---|---|
| Translational error | mm | 0.1 |
| Rotation error | mrad | 0.2 |
| Magnetic element field error | % | 0.1 |
| BPM uncertainty | mm | 0.1 |

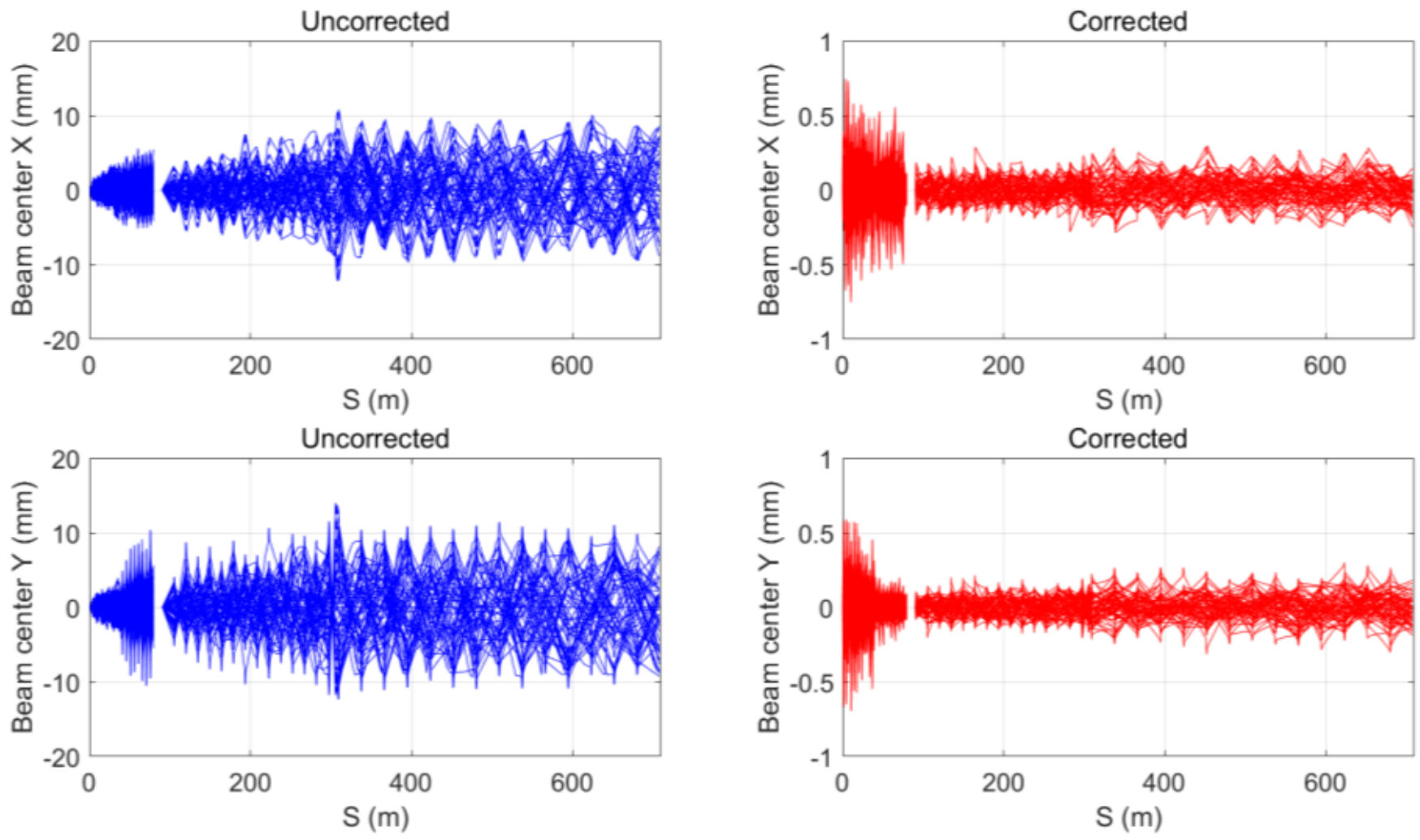

**Figure 6.2.1.24:** The rms beam orbit with errors without correction (left) and with correction (right) in the positron Linac.

The phase error and accelerating gradient error in the accelerating structures can cause energy jitter and an energy spread increase. The energy jitter reqirement for the Booster is within ±0.2% and the energy spread is less than 0.2%. Figure 6.2.1.25 shows the energy jitter and energy spread with different phase and accelerating gradient errors. This shows that the phase error should be controlled within 0.5 degree and the accelerating gradient errors should be controlled within 0.5%.

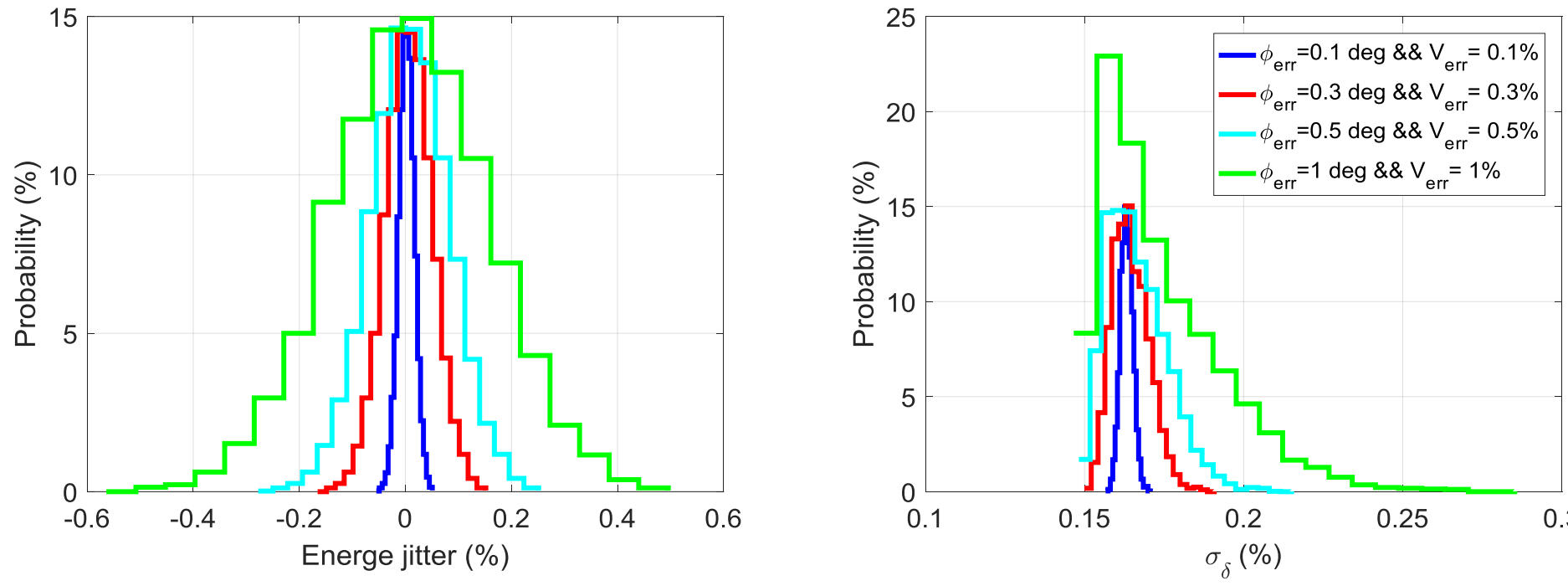

**Figure 6.2.1.25:** Energy jitter with different phase and accelerating gradient errors.

## 6.2.2 Transport Lines

To reduce the project cost, the Linac is at ground level while the Booster is in a tunnel about 100 m underground. As a result, the Linac to Booster transport line consists of two parts, a vertical sloping line and a horizontal bending line as shown in Fig. 6.2.2.1; the Twiss parameters are in Fig. 6.2.2.2.

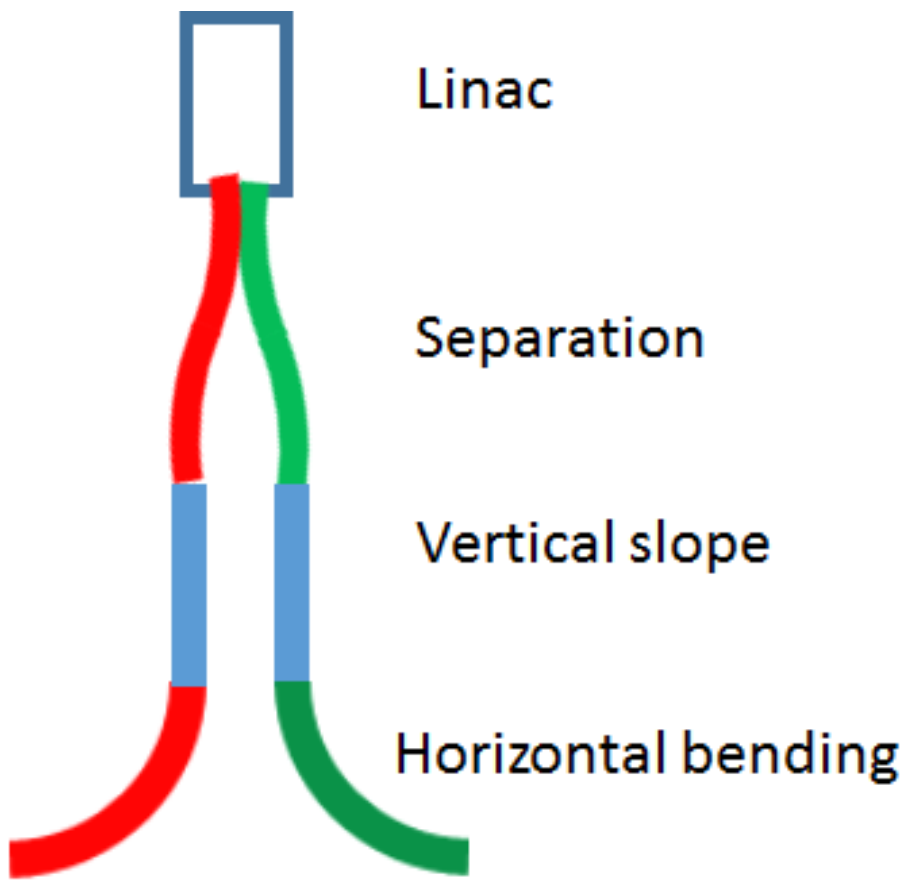

**Figure 6.2.2.1:** Layout of the transport line from the Linac to the Booster

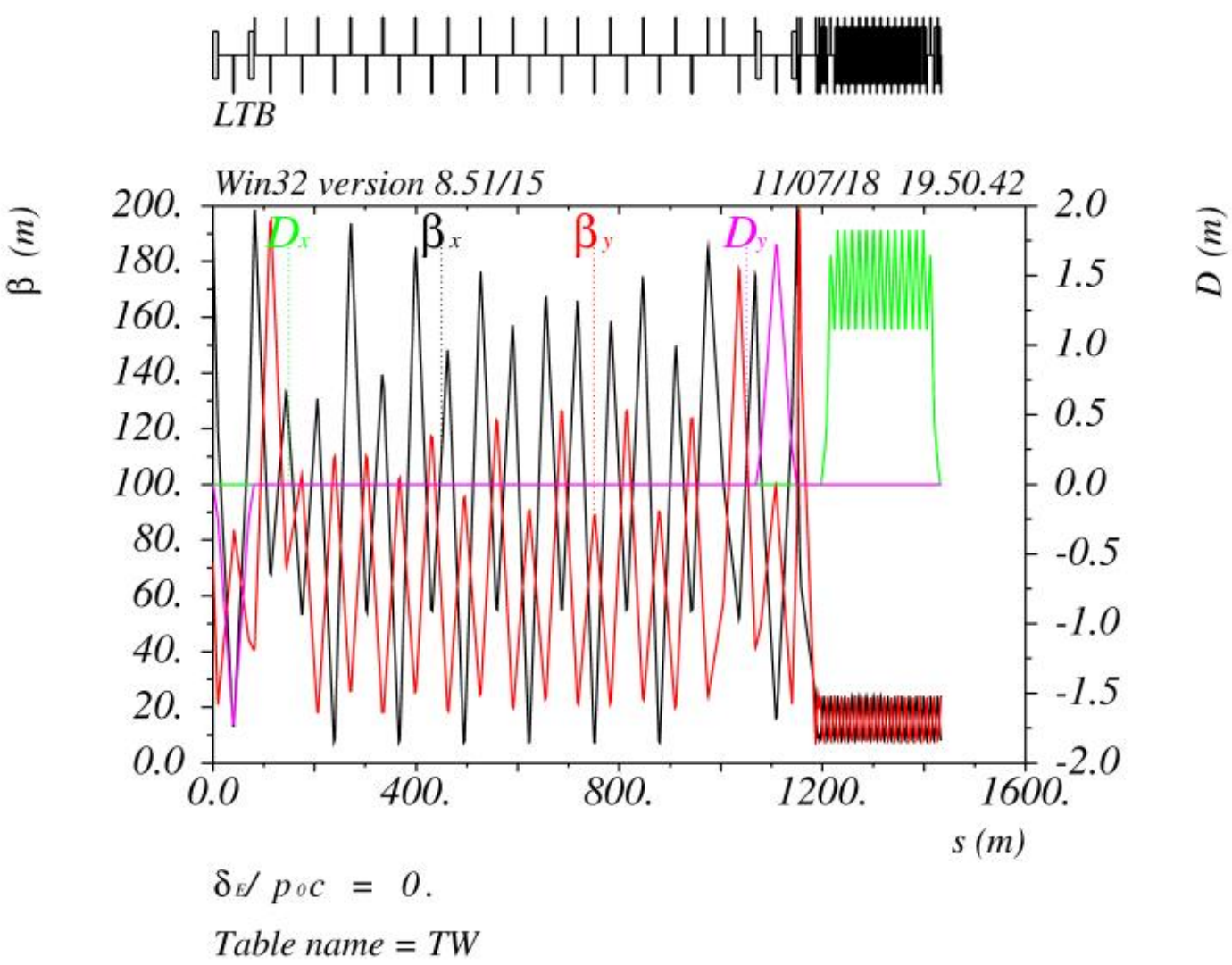

**Figure 6.2.2.2:** Twiss parameters of the Linac to Booster transport line

## 6.3 Electron Source

### 6.3.1 Source Design

A conventional thermionic electron gun is chosen. It is similar to those used at BEPC-Ⅱ and KEKB and consists of a flat surface cathode-grid assembly, a focusing electrode and an anode. The widely used EIMAC-Y796 cathode-grid assembly, which has a cathode area of 2 $cm^2$, will be the dispenser cathode. It can provide a current density as high as 12 $A/cm^2$ and has a long lifetime. The EGUN [1] code is used for the beam optics simulations and geometry optimization. The goal is to obtain minimum emittance at the end port with voltage of 160 kV and a current of 10A. The electron trajectories are shown in Fig. 6.3.1. Fig. 6.3.2 shows the phase space in the x and y planes and Fig. 6.3.3 shows the current density on the cathode surface.

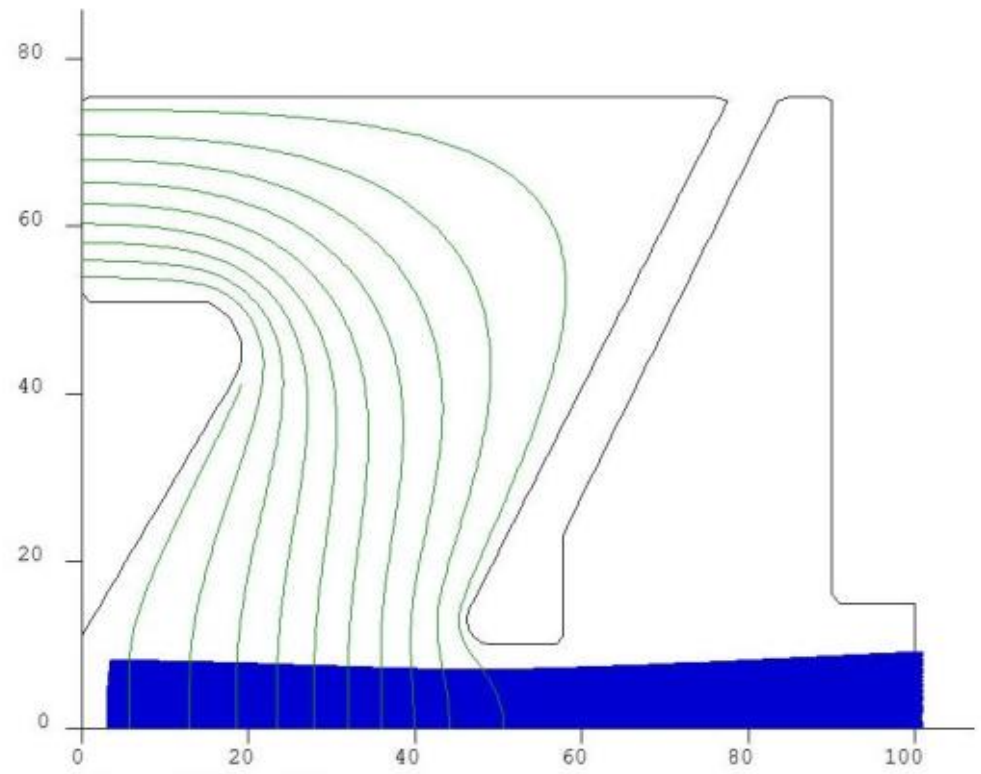

**Figure 6.3.1:** Beam trajectory of the injector electron gun.

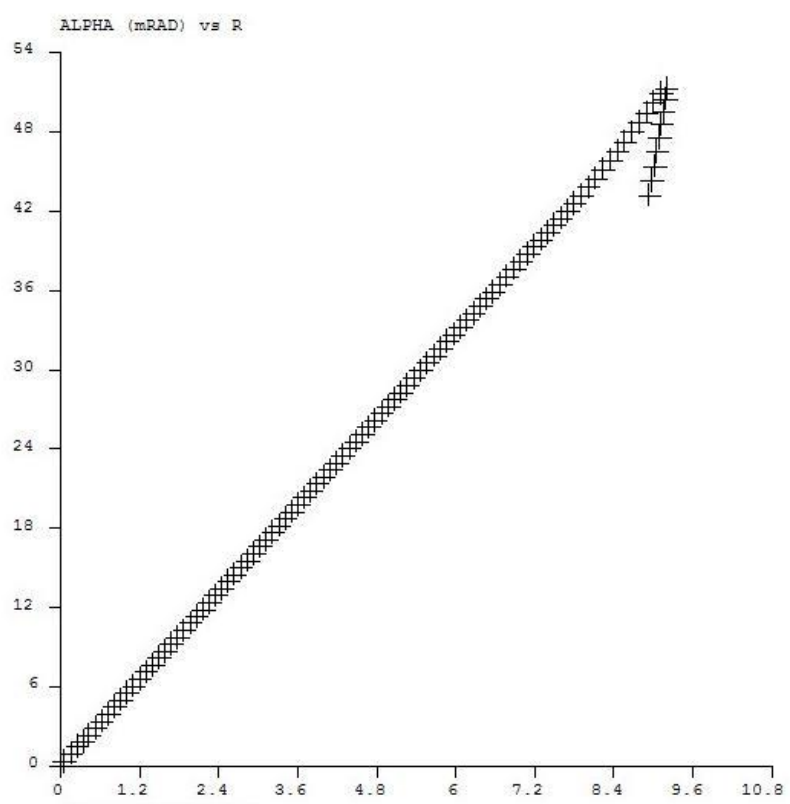

**Figure 6.3.2:** Phase plane at the exit port

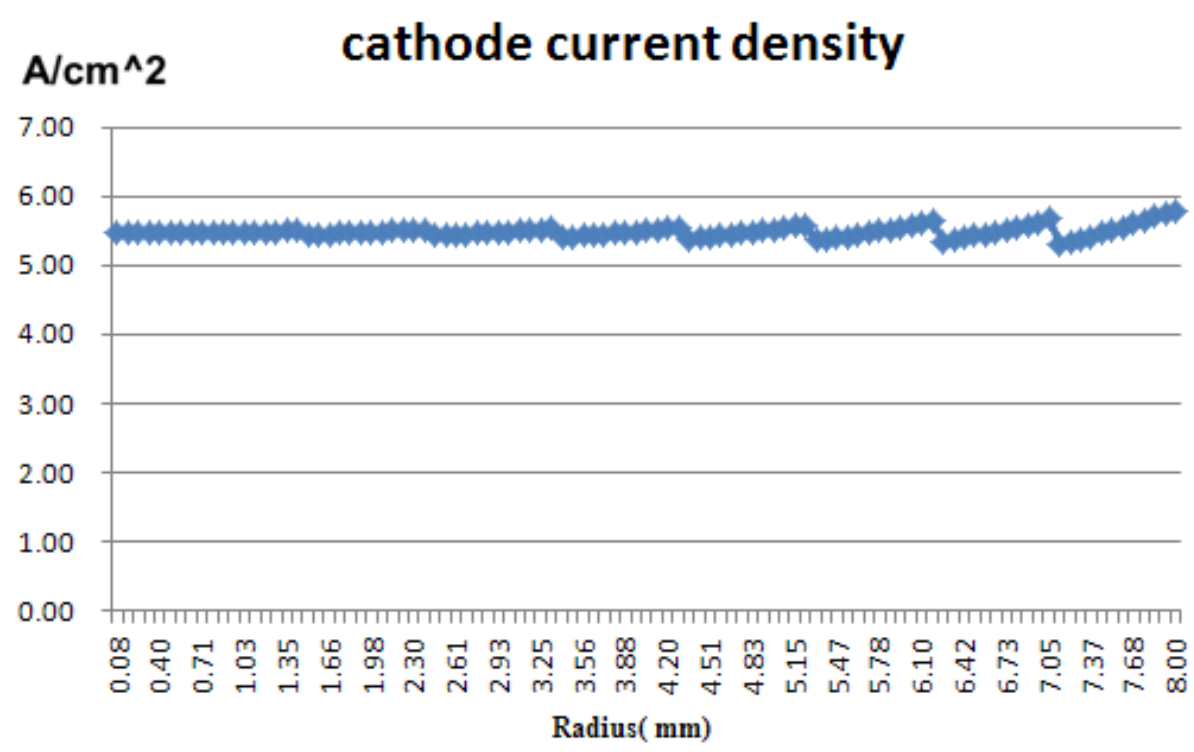

**Figure 6.3.3:** Current density at the cathode.

The beam trajectory is almost parallel to obtain low emittance. At the end port, which is 100 mm from the cathode surface, the x and y emittancs are 17.84 π (mm·mrad). EGUN predicts a beam perveance of 0.169 μP. A uniform current density on the cathode leads to a long lifetime.

### 6.3.2 **Pulser System**

The electron gun is a triode gun. The pulsed system includes a DC power supply, control box and pulser. Figure 6.2.4 is the schematic. This pulser, used in BEPC-II [2], is an already proven technique. The DC power supply varies from zero to 1 kV and stores and discharges the energy with a switch. The control box controls the trigger pulse amplitude, provides a trigger pulses, monitors temperatures, processes the trigger signal and converts between the optical and electrical signals. Specifications are listed in Table 6.3.1.

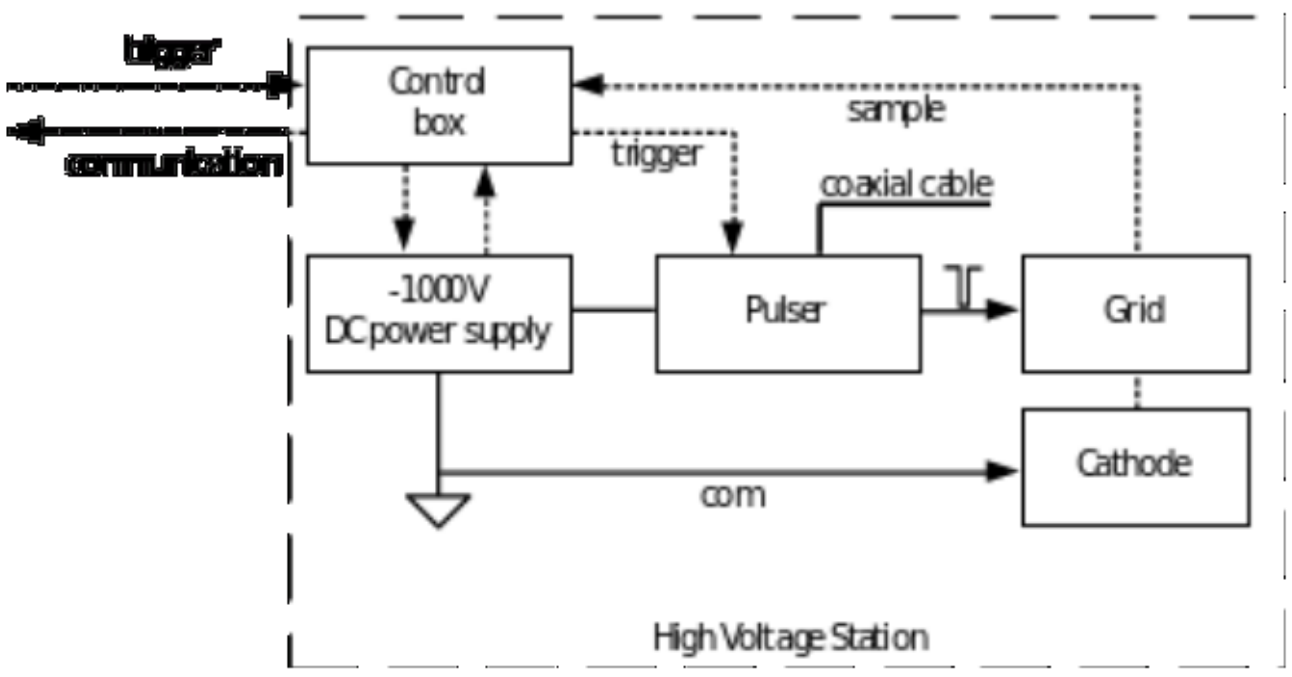

**Figure 6.3.4:** Pulsed system schematic

The pulser is a critical device. The discharge switch is a series-stacked avalanche transistor. Coaxial cable is a pulse forming line and energy storage device connected to the end of the switch. The pulse width is determined by the length of the coaxial cable which is matched to the pulse width of 1 ns to 10 ns. The polarity of the pulse is negative.

**Table 6.3.1:** Pulser system specifications.

| Parameter | Unit | Values |
|---|---|---|
| Pulse voltage | V | 1000 |
| Pulse width | ns | 1-10 |
| Rise time | ns | 0.8 |
| Polarity | | Negative |
| Jitter | ps (RMS) | 20 |
| AC power supply | V | 220 (-10%~+10%) |

## 6.3.3 **High Voltage System**

The electron gun system consists of an electron gun body, a high voltage power supply, a high voltage deck, a pulser and a control unit. The gun should be able to operate in a 1 ns single beam pulse mode to generate the electron beam. The gun parameters are listed in Table 6.3.2 and Fig. 6.3.5 is a schematic diagram. The high voltage system is shown in Fig. 6.3.5.

**Table 6.3.2:** Electron gun specifications.

| Parameters | Values |
|---|---|
| Type | Triode |
| Maximum Beam Current(A) | 10 |
| Anode High Voltage(kV) | 120~200 |
| Filament voltage(V) | 6~8 |
| Filament current(A) | 5~7.5 |
| Grid bias voltage(V) | 0~200 |
| Pulse width (ns) | 1 |
| Pulse rate (pps) | 100 |

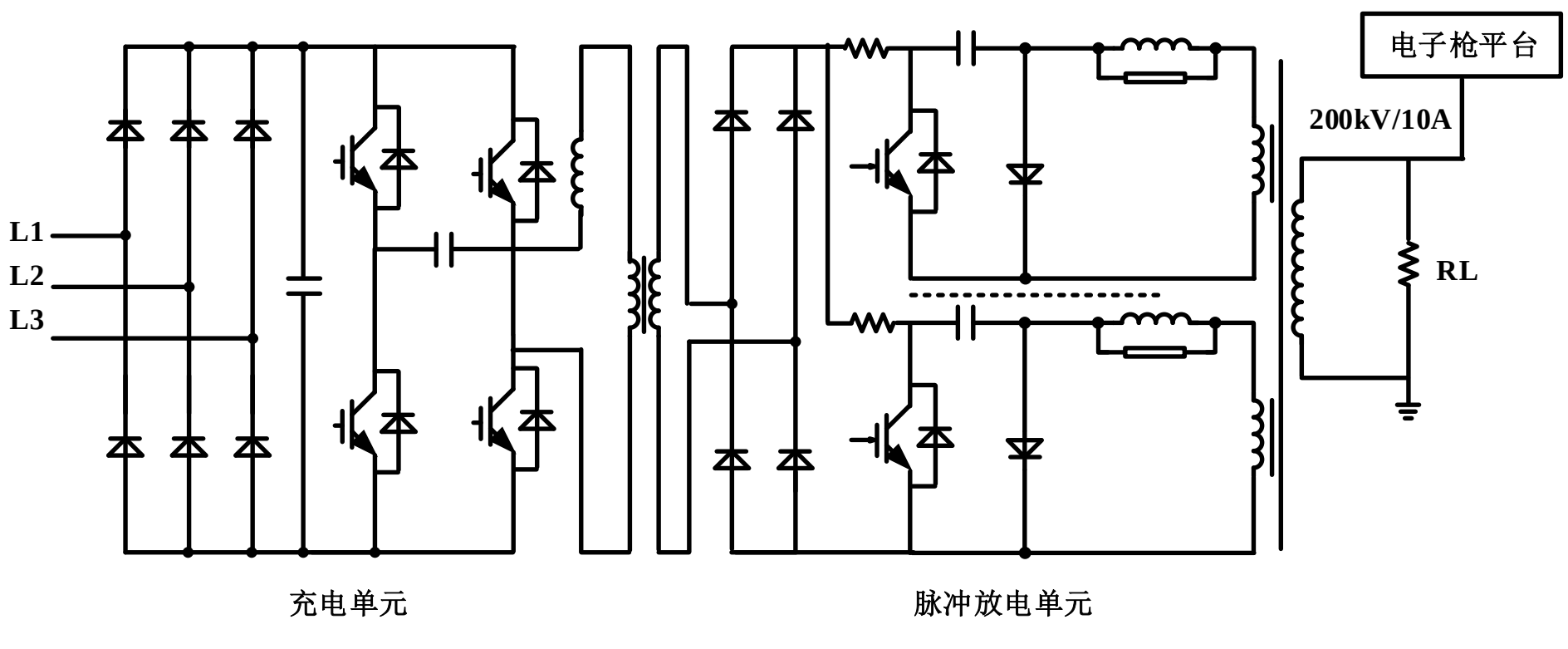

**Figure 6.3.5:** Electron gun schematic

## 6.4 Positron Source

### 6.4.1 Target

The simulation study on positron source target design has been done using the G4beamline [1] and FLUKA codes [2]. The initial electron beam energy is 4 GeV and beam size is small, rms 0.5 mm. The positron yield at the target exit was optimized by scanning the W target thickness at different electron beam energies, as shown in Fig.6.4.1.1. From these results and energy deposition the length chosen is 15 mm.

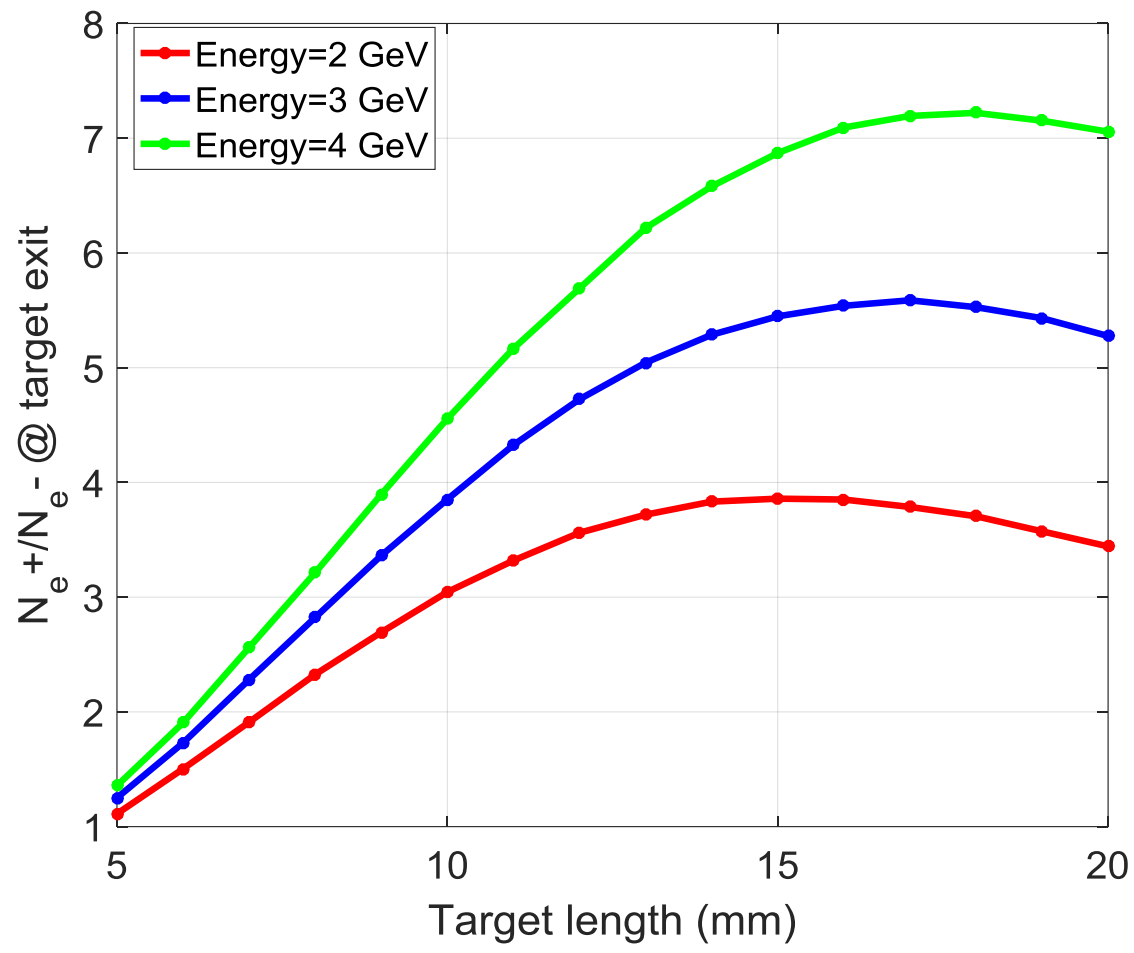

**Figure 6.4.1.1:** Positron yield with different target lengths and electron energies.

FLUKA code is used to calculate the energy deposition in the structure. As shown in Fig. 6.4.1.2, the total energy deposition is 0.784 GeV per electron particle for a 4 GeV electron beam. This means that the power deposition is about 784 W and water cooling is necessary. The cylindrical W target is embedded in a cuboid copper block for supporting and cooling.

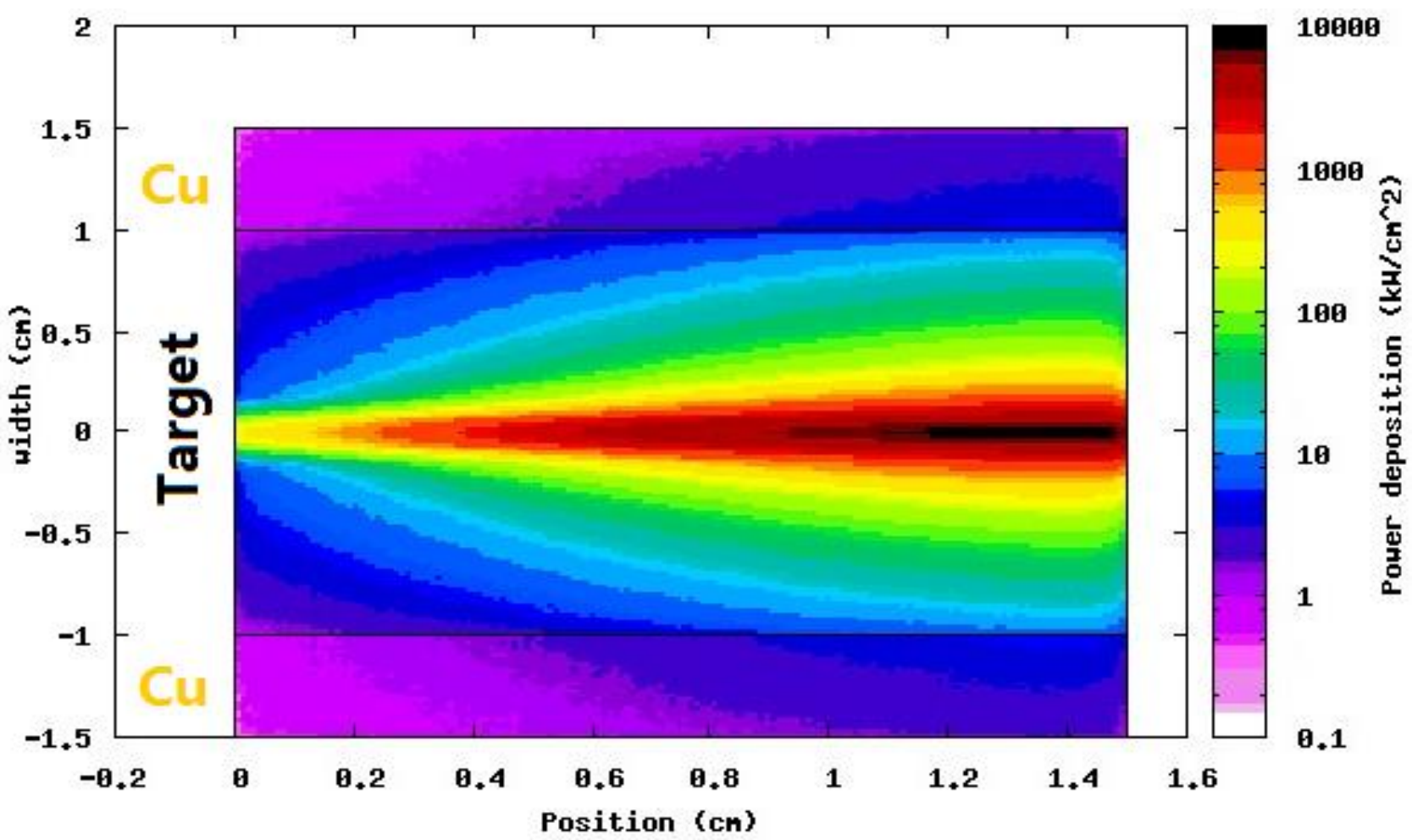

**Figure 6.4.1.2:** Energy deposition in the target.

### 6.4.2 **Flux Concentrator**

The large transverse emittance of the positron beam emerging from the target is transformed to match the pre-accelerating section with an AMD flux concentrator in the capture section. Using the AMD the beam with large divergence and small beam size is transformed to small divergence and large beam size; the simulation results are shown in Fig.6.4.2.1. Following the AMD the positrons are accelerated to 200 MeV in the pre-accelerating section. The magnetic field is a pseudo-adiabatically changing solenoid field from peak 6-T to 0.5 T. This is a flux concentrator superimposed on a 0.5-T DC solenoid field as shown in Fig.6.4.2.2.

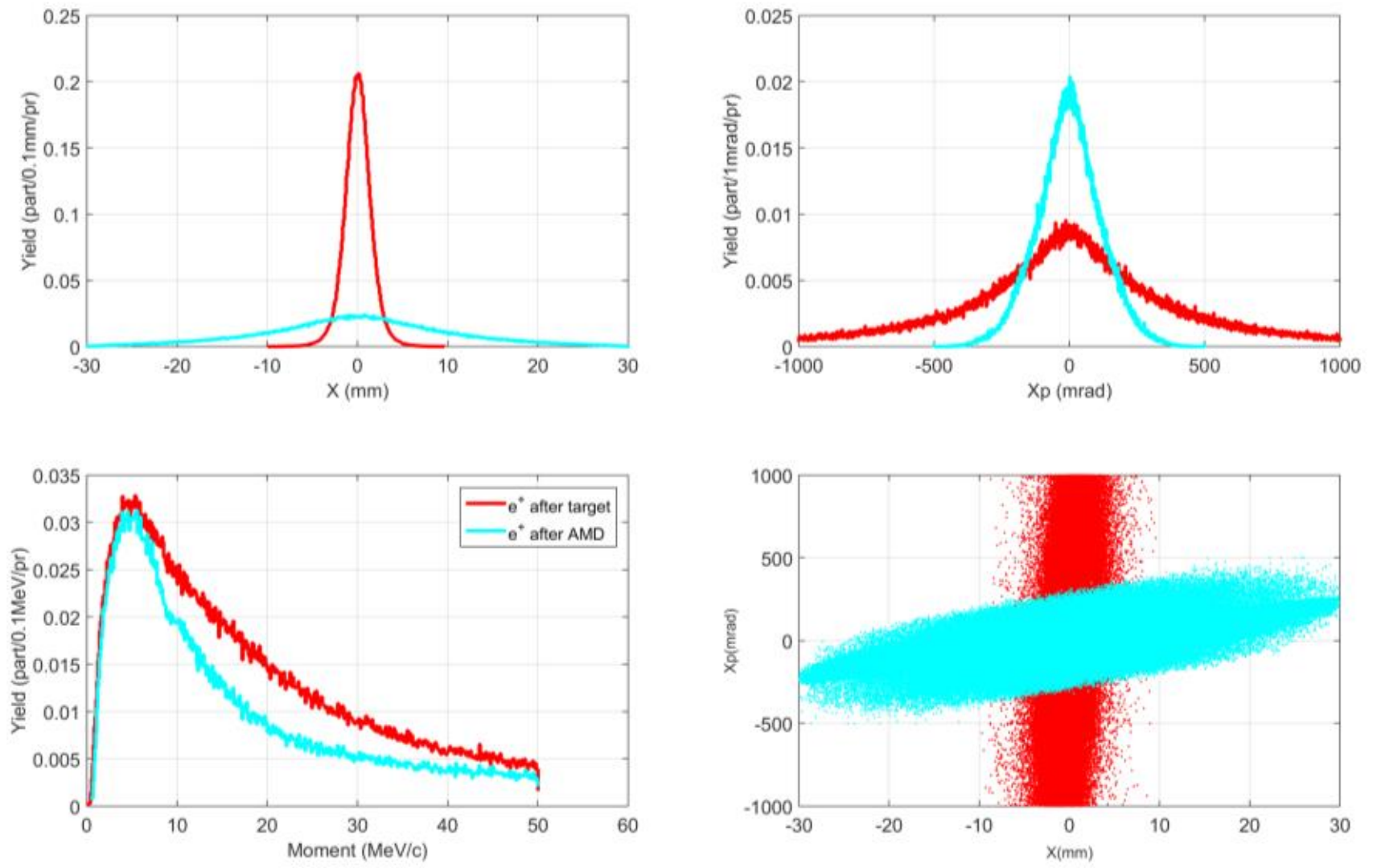

**Figure 6.4.2.1:** Beam transformation by the AMD section.

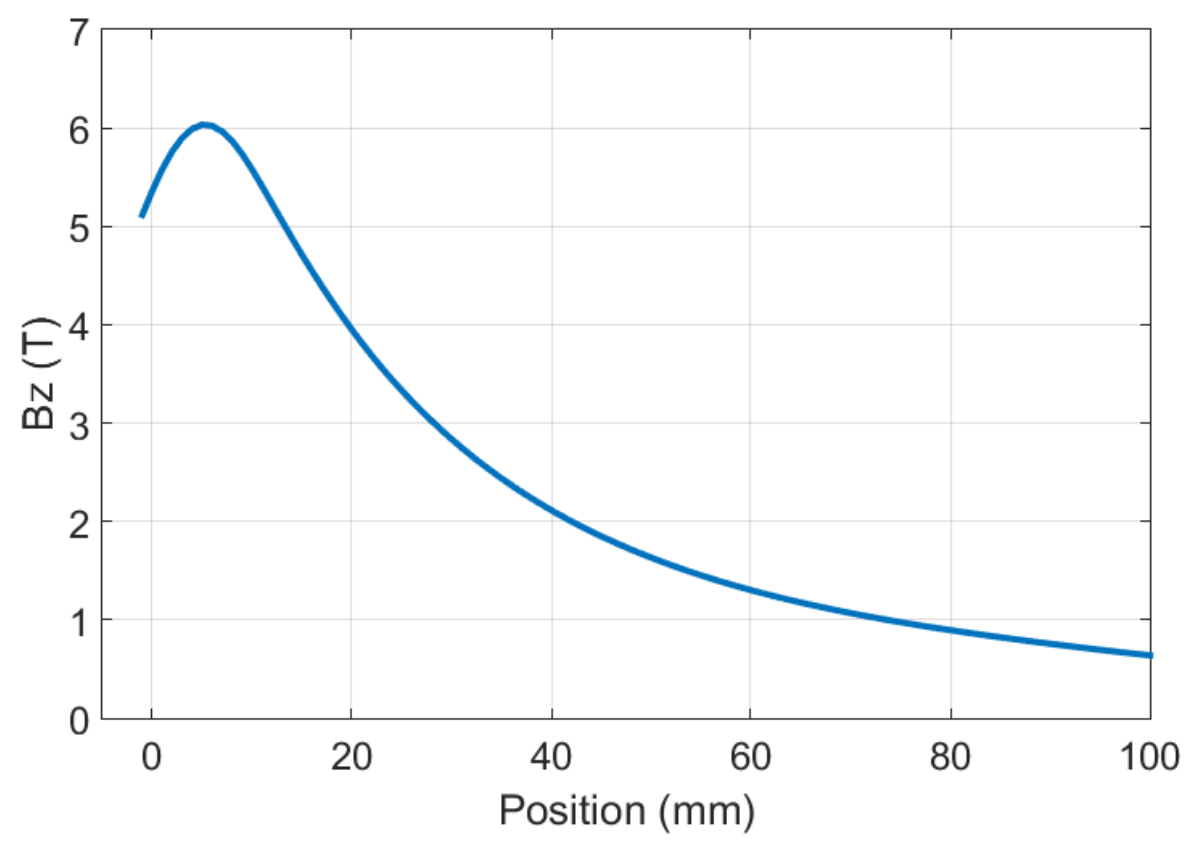

**Figure 6.4.2.2:** Magnetic field of flux concentrator.

The flux concentrator is an adiabatic matching device placed between the target and accelerating structure. It produces a magnetic field with a sharp rise over less than 5 mm to its peak value, and then falls off adiabatically over 10 cm. Simulation of the magnetic field from the flux concentrator has been calculated by the Opera code, as shown in Fig.6.4.2.3. The peak magnetic field reaches 5.5 T with a 12 kA current drive.

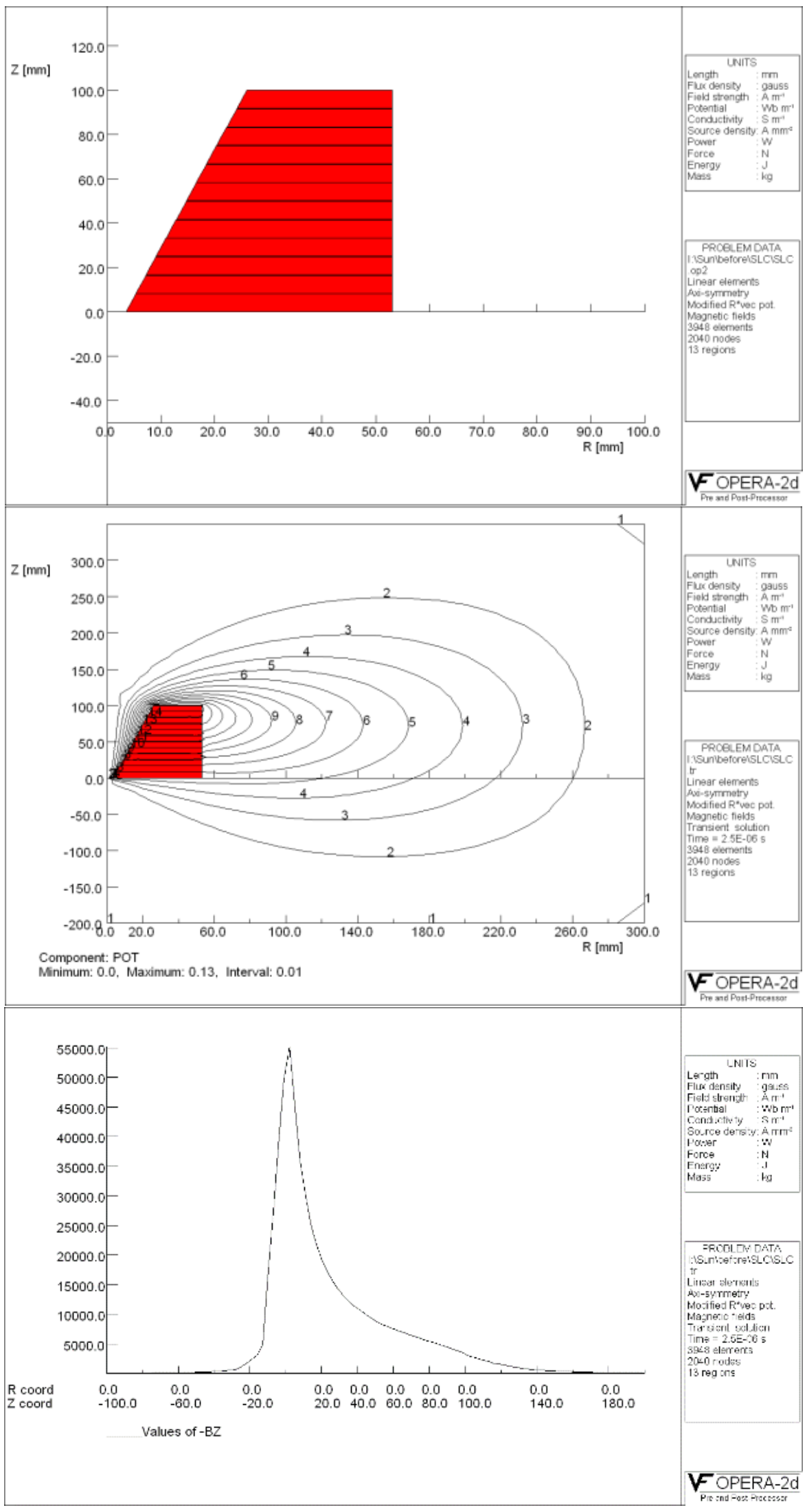

**Figure 6.4.2.3:** Simulation of the magnetic field generation from a flux concentrator

Mechanical design of the flux concentrator is shown in Fig.6.4.2.4, a design initially developed by SLAC. The copper coil is 100 mm, 12 turns and a current of 16 kA. The copper core has an outer radius of 40 mm and a conical inner radius increasing from 3.5 mm to 26 mm. Excitation current and water cooling is provided by a hollow circular copper conductor brazed to the outside of the coil. The flux concentrator is a rather complicated device because it is difficult to machine.

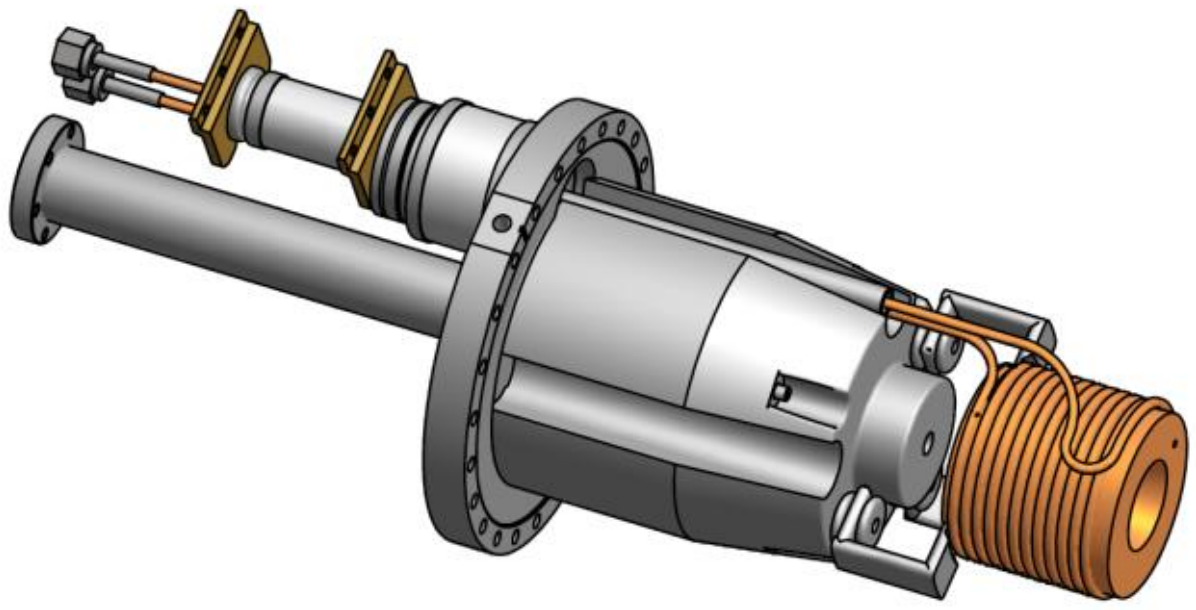

**Figure 6.4.2.4:** Mechanical design of the flux concentrator

## 6.5 Linac Technical Systems

### 6.5.1 RF System

The 10 GeV Linac operates at 2860 MHz. The pulse length is 4 μs and the repetition frequency is 100 Hz. The RF system includes the bunching system, the main RF system and the positron pre-accelerating section. The main Linac and the positron pre-accelerating section are powered by 80 MW klystrons.

#### 6.5.1.1 *Bunching System*

After leaving the electron guns, the electron bunches go into the bunching system, which consists of the following components: the first subharmonic buncher (SHB1) operating at 143 MHz ($20^{th}$ subharmonic), the second subharmonic buncher (SHB2) operating at 572 MHz ($5^{th}$ subharmonic), and a constant-impedance travelling-wave buncher operating in $2\pi/3$ mode at 2860 MHz. Fig. 6.5.1.1 shows the layout of the bunching system.

The two subharmonic pre-bunchers and the one S-band buncher act to velocity modulate the non-relativistic electron beam emerging from the gun, and compress the pulse before it passes into the buncher.

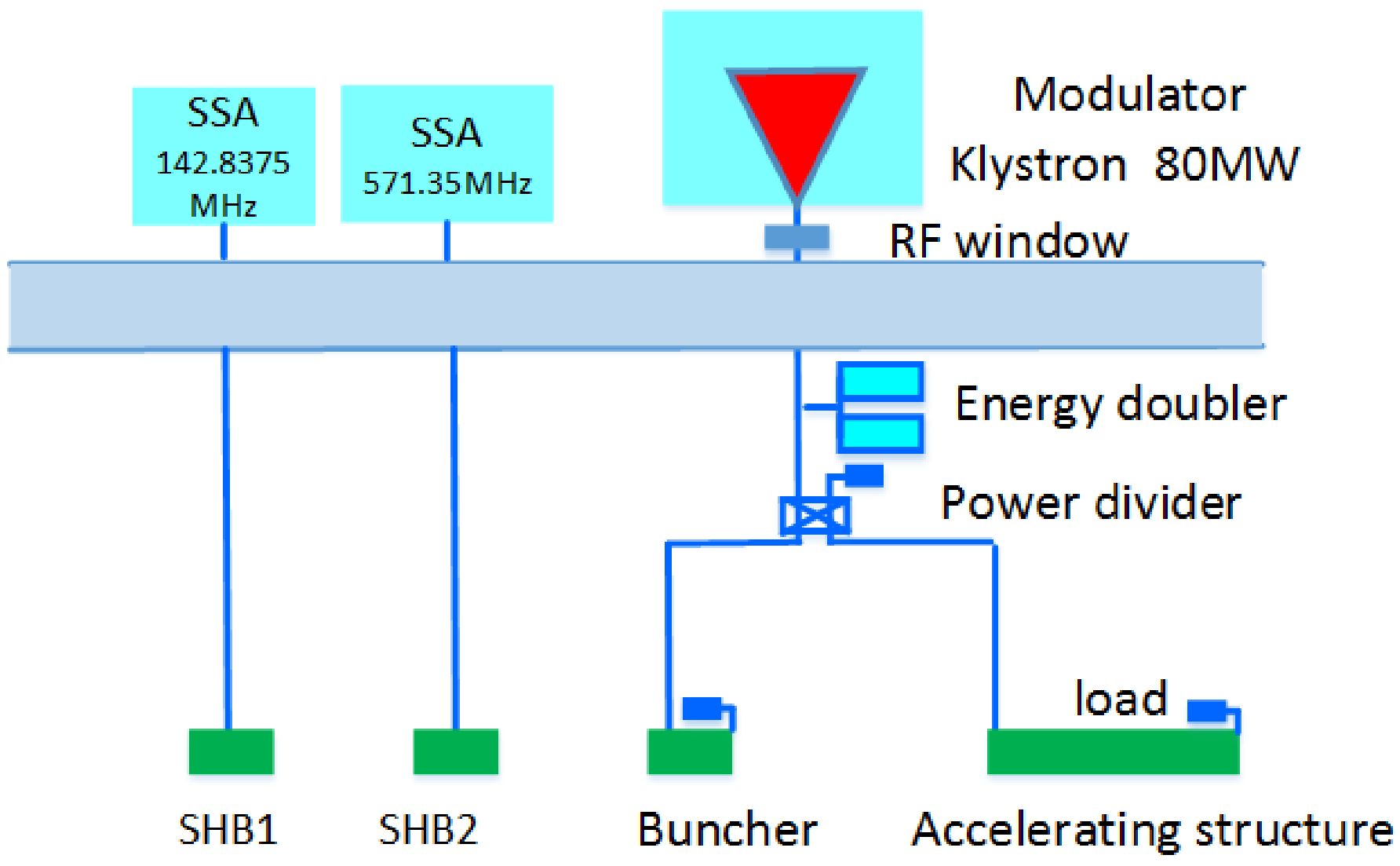

**Figure 6.5.1.1:** Bunching system layout

The resonant frequency of SHB1 is 143 MHz; the material used is OFC (oxygen free copper) and the Q value is about 8200. The bunching voltage, chosen to be 105 kV, provides an operating margin. The shunt impedance is 1.4 MΩ assuming that the power from the power supply system is about 10 kW. The frequency tuning range is 400 kHz, where we use the KEKB linear accelerator as a reference. [2]

The resonant frequency of SHB2 is 572 MHz. The material is OFC and the Q value is about 13,000. To retain a margin, the designed bunching voltage is chosen as 145 kV. The shunt impedance is 3.7 MΩ and the power from the power supply system at 572 MHz is about 7 kW.

The buncher is a 6-cavity (including 2 coupling cavities) traveling-wave structure operating in 2π/3 mode at 2860 MHz; the relative phase velocity is 0.75. The beam bunches from the two SHBs go into the input coupler of the buncher at the correct phase. They are then focused by the microwave field and accelerated at the same time. The input power of the buncher is about 3 MW.

The main parameters of the subharmonic pre-bunchers and S-band buncher are shown in Table 6.5.1.1.

**Table 6.5.1.1:** The main parameters of the sub-harmonic pre-buncher and S-band buncher.

| **First sub-harmonic pre-buncher** | | |
|---|---|---|
| Type | Re-entrant | |
| Frequency | MHz | 143 |
| Unloaded Q | | 8200 |
| Shunt impedance | MΩ | 1.4 |
| $E_{surface, max}/E_{gap, max}$ | | 2.53 |
| **Second sub-harmonic pre-buncher** | | |
| Frequency | MHz | 572 |
| Unloaded Q | | 13000 |
| Shunt impedance | MΩ | 3.7 |
| $E_{surface, max}/E_{gap, max}$ | | 2.44 |
| **S-Band Buncher** | | |
| Type | Constant impedance,TW, 2π/3-mode | |
| Frequency | MHz | 2860 |
| Input and output VSWR | | ≤1.2 |
| Bandwidth (VSWR ≤ 1.2) | MHz | ≥4.0 |
| Peak RF input power | MW | 3 |
| Phase velocity ($V_p$/c)/group velocity ($V_g$/c ) | | 0.75 / 0.0119 |
| Shunt impedance | MΩ/m | 36 |
| Unloaded Q | | 11000 |
| RF attenuation parameter | Neper/m | 0.228 |
| Number of cavities | | 4＋2×0.5 |

### 6.5.1.2 *Main Linac RF System*

The power is then evenly divided among four 3-m constant gradient accelerating sections on a support girder. At 21 MV/m accelerating gradient, each klystron is thus capable of providing 252 MeV to each particle at a repetition rate of 100 Hz. Fig. 6.5.1.14 shows one unit power supply system for the accelerating structures.

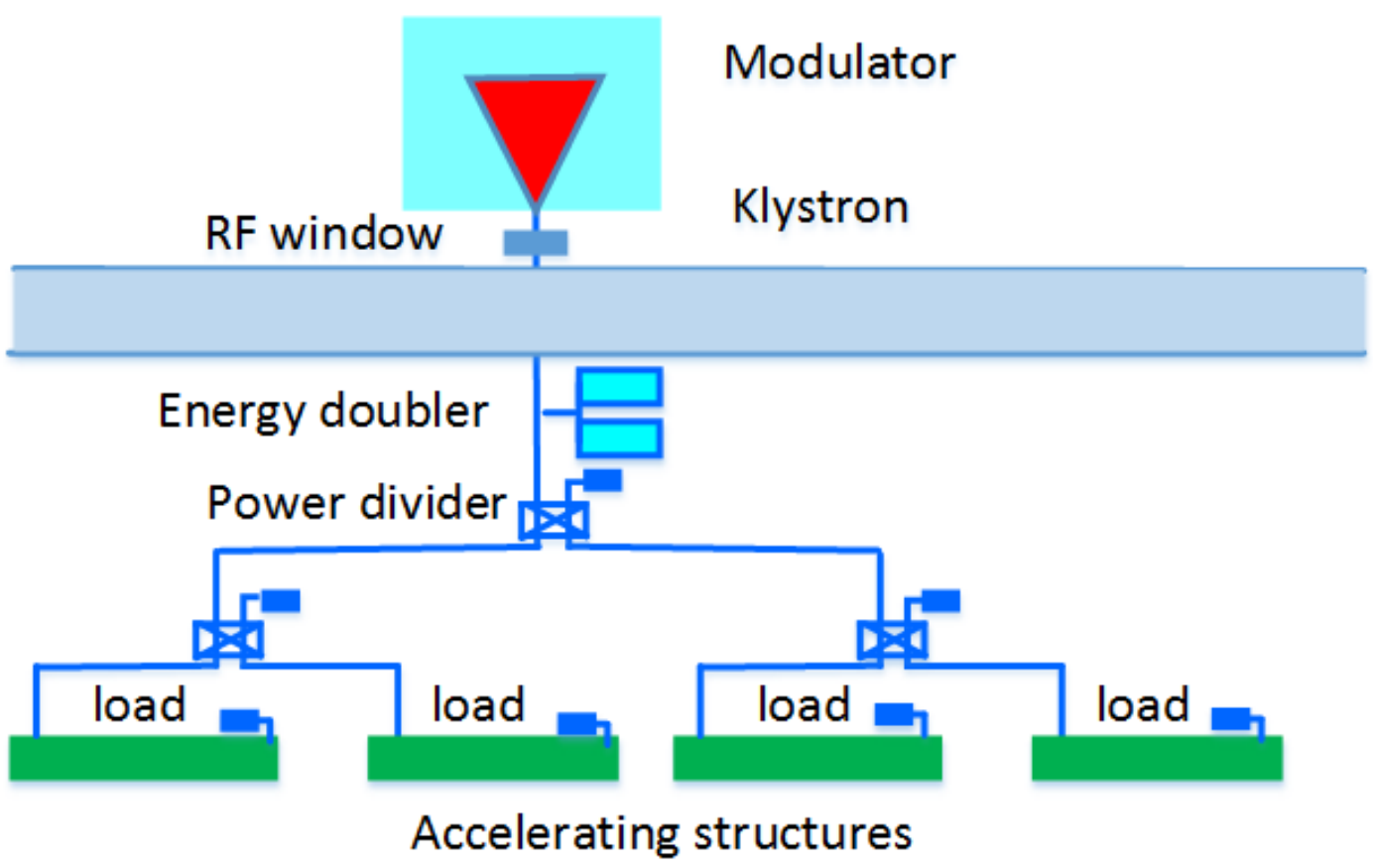

**Figure 6.5.1.14**: The layout of one unit of main linac RF system.

#### *6.5.1.2.1 RF Transmission Design*

The RF transmission and measurement system is a microwave path between the klystron and the accelerating sections composed of a waveguide and other microwave components. These include straight waveguide, E-band waveguide, H-band waveguide, directional coupler, power divider, high power waveguide phase shifter and attenuator, dry load, microwave monitoring unit, and peak power meter.

The waveguide model is WR-284, the section size is 72.14 mm by 34.04 mm. The material is OFC and the flange is made of stainless steel. The microwave signal can be phase-shifted or attenuated by a high power waveguide phase shifter and attenuator. The microwave signal coupled from the directional coupler can be tested not only to observe the waveform, but also for power measurement. In order to absorb the excess microwave power from the accelerating structure, a high power SiC dry load is required. It uses brazing rod β-phase SiC ceramics as a peak microwave energy absorber and incorporates indirect water cooling. The SiC dry load withstands peak power up to 60 MW in high-power testing. Its performance has reached the international advanced level for similar products. Parameters are in Table 6.5.1.2.

**Table 6.5.1.2:** High power SiC dry load parameters.

| ***Parameters*** | | ***Unit*** |
|---|---|---|
| Frequency | 2860 | MHz |
| VSWR | <1.1 | |
| Maximum peak power | 30 (with SLED)<br>10 (without SLED) | MW |
| Repetition frequency | 100 | Hz |
| Pulse width | 4 | μs |

The modular microwave monitoring unit will contain several independent parts, such as the attenuation unit, filter unit, detection unit and virtual oscilloscope unit. The modular multi-unit design not only facilitates maintenance, but also reduces the space required to avoid electromagnetic interference. The microwave monitoring unit realizes three functions at the same time and provides three kinds of signals: signals for power measurement, signals for observing the waveform on a oscilloscope, and virtual oscilloscope signals for input into the local computer.

#### *6.5.1.2.2 RF Pulse Compressor*

We draw on the experience from operation of the Stanford Linear Accelerator Energy Doubler (SLED) at the maximum klystron output peak power of 80 MW, with a pulse length of 4 μs. The S-band SLED, which is an RF pulse compression system using high-Q resonant cavities, is one of the most important RF components in the S-band high-power RF station. The S-band SLED consists of a 3-dB power hybrid and two identical over-coupled cylindrical cavities resonant at the 2860 MHz. A fast-acting triggered phase-shift-keying (PSK) π-phase-shifter, which reverses the RF phase of the klystron output power, is inserted into the klystron drive line. The cavities begin by storing klystron output power during a large fraction of the time duration of each pulse. Then the phase of the klystron output is reversed, and the cavities emit the stored power rapidly into the accelerating section, adding to the klystron output power during the remaining pulse length. This means that the peak power is enhanced at the expense of the pulse length without increasing the average input power consumption.

The specifications of the S-band SLED are listed in Table 6.5.1.3. Two coupling slots are located between the waveguide and the cavity to decrease the peak surface field, and thus to increase the operating stability in high power conditions. The input pulse length is 4 μs with a 180° phase reversal at time 3.17 μs. The energy multiplication factor can be larger than 1.6 from the operations experience of the BEPC-II Linac.

**Table 6.5.1.3:** The main parameters of the pulse compressor.

| **Parameters** | | **Unit** |
|---|---|---|
| Operating frequency | 2860 | MHz |
| Resonant mode | $TE_{0,1,5}$ | |
| Coupling coefficient | 5 | |
| Peak power gain | > 5 | |
| Unload Q factor | ~100,000 | |
| Energy multiplication factor | ~1.6 | |
| Max. input peak power | 80 | MW |
| Input pulse length | 4 | us |
| Output pulse length | 1 | us |
| Repetition rate | 100 | Hz |

#### *6.5.1.2.3 Accelerating Structure*

S-band constant-gradient copper accelerating structures operating in 2π/3 mode at 2860 MHz will be used to accelerate the bunched electron and positron beams up to the final energy. The accelerating structure parameters are shown in Table 6.5.1.4. A dual-feed racetrack symmetry coupler design will be used to reduce emittance growth from the asymmetry coupler. The accelerating structure operation temperature is 30℃, which is maintained within 0.1˚ so the phase shift along the entire length of an accelerator section is kept within 2˚.

**Table 6.5.1.4:** Accelerating structure parameters.

| **Parameters** | | **Unit** |
|---|---|---|
| Operating frequency | 2860 | MHz |
| Operating temperature | 30.0 ± 0.1 | ºC |
| Number of cells | 84 +2 coupler cells | |
| Section length | 3048 | mm |
| Phase advance per cell | 2π/3 - mode | |
| Cell length | 34.966 | mm |
| Disk thickness (t) | 5.5 | mm |
| Iris diameter (2a) | 26.231～19.243 | mm |
| Cell diameter (2b) | 83.460～81.781 | mm |
| Shunt impedance ($r_0$) | 60～69 | MΩ/m |
| Q factor | 15465～15370 | |
| Group velocity ($v_g$/c) | 0.020～0.0080 | |
| Filling time | 850 | ns |
| Attenuation factor | 0.50 | Neper |

### 6.5.1.3 *Positron Pre-accelerating Section*

The positron pre-accelerating section includes 6 accelerating structures. In order to obtain high capture efficiency, the accelerating structures are big-hole structures. Constant impedance structures are used and the frequency is 2860 MHz. For klystron power 80 MW, the gradient of the accelerating structure is about 22 MV/m. The layout is in Fig. 6.5.1.15 and the parameters are shown in Table 6.5.1.5.

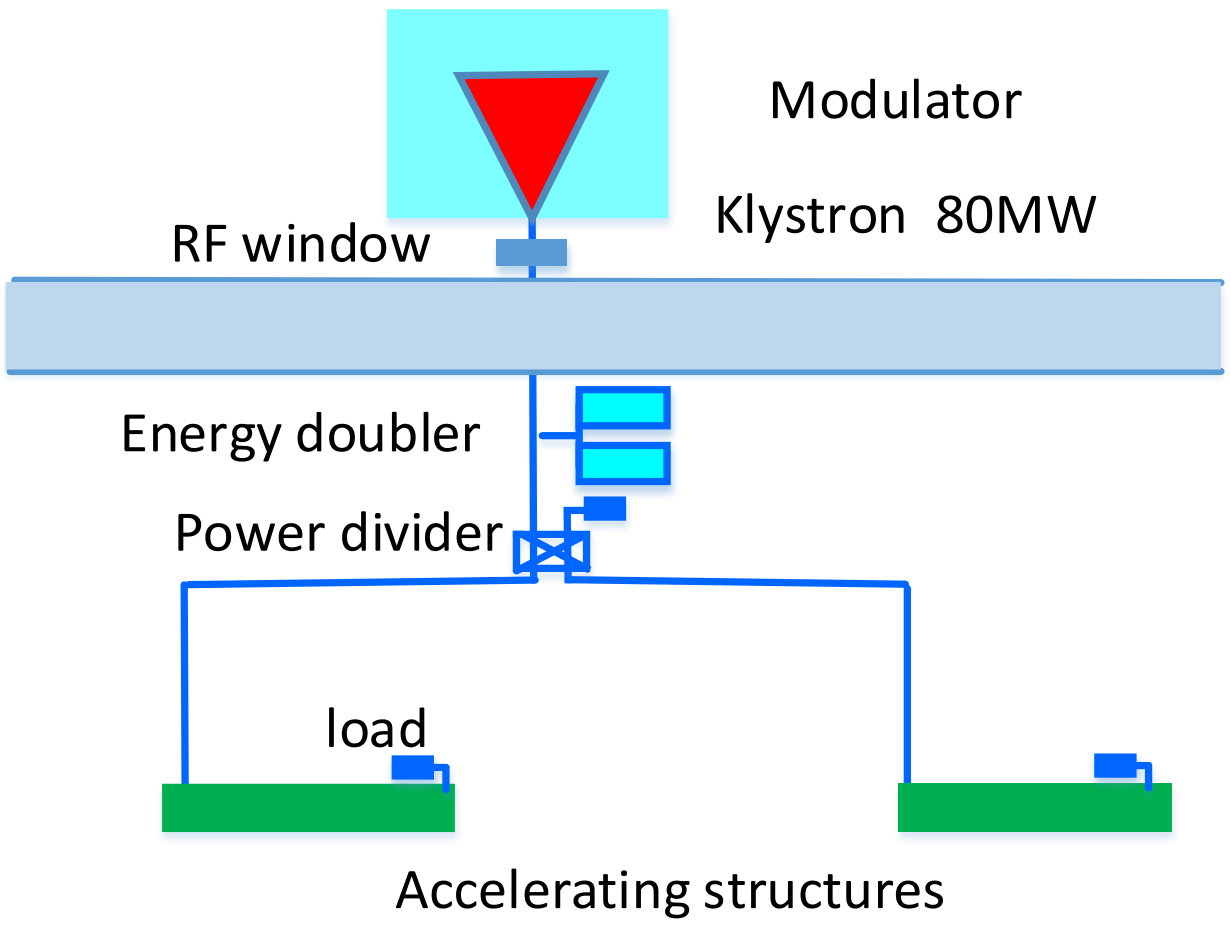

**Figure 6.5.1.15:** The layout of positron pre-accelerating section RF system.

**Table 6.5.1.5:** Big-hole accelerating structure parameters.

| **Parameters** | | **Unit** |
|---|---|---|
| Operating frequency | 2860 | MHz |
| Operating temperature | 30.0 ± 0.1 | ºC |
| Number of cells | 55 +2 coupler cells | |
| Section length | 2000 | mm |
| Phase advance per cell | 2π/3 - mode | |
| Cell length | 34.966 | mm |
| Disk thickness (t) | 5.5 | mm |
| Iris diameter (2a) | 25 | mm |
| Cell diameter (2b) | 89.475 | mm |
| Shunt impedance ($r_0$) | 60.8 | MΩ/m |
| Q factor | 15448 | |
| Group velocity ($v_g/c$) | 0.019 | |
| Filling time | 336 | ns |
| Attenuation factor | 0.195 | Neper |

## 6.5.2 RF Power Source

### 6.5.2.1 *Introduction*

To have a reasonable length linac, operation at high accelerating gradient is required. For copper (non-superconducting) accelerator structures, this implies a high peak power per unit length and a high peak power per RF source assuming a limited number of discrete sources. To enhance the peak power produced by an RF source, the SLED RF pulse compression scheme is used.

The main high power RF components are 75 units of 80 MW S-band klystrons and conventional solid state modulators. A waveguide system is used for power transmission from the klystrons to the accelerating structures, 75 klystrons are used to provide power for 288 accelerating structures, An RF window is used for vacuum isolation between the klystron and the waveguide transmission system.

### 6.5.2.2 *S Band Klystron*

The RF power source system includes 75 sets of 80 MW pulsed klystrons operating at a frequency of 2860 MHz. Based on the existing S band 65 MW klystron applied in the BEPCII linac injector, the BAC method will be adopted to increase the klystron efficiency from 40% to 55% to meet the CEPC power requirement. Cavities are inserted between the $4^{th}$ and $5^{th}$ cavities but the same layout and total length is retained [1-3]. The other parts such as gun, coil and collector will be re-used to save R&D effort and reduce fabrication cost. The klystron specification is shown in Table 6.5.2.1.

**Table 6.5.2.1:** 2856 MHz/80 MW klystron specification

| Parameters | Values |
|---|---|
| Frequency | 2860 MHz |
| Output power | 80 MW |
| Pulse width | 4μs |
| Voltage | 350 kV |
| Current | 416 A |
| Perveance | 2μP |
| Gain | >50 dB |
| Efficiency | 55% |

### 6.5.2.3 *Solid State Modulator*

The klystrons are powered by pulsed solid state modulators, a well-established high-reliability technology.

To accelerate an electron beam with a pulse width of 1 μsec, the flat-top of the klystron beam voltage must be more than 2 μsec long. Long-term regulation and pulse flatness of the klystron beam voltage must be less than ±0.15% to prevent RF phase modulation and microwave power fluctuations. Modulator specifications are shown in Table 6.5.2.2.

**Table 6.5.2.2:** Modulator specifications

| Parameters | Values |
|---|---|
| Peak output power (MW) | 200/150 |
| Average output power (kW) | 80 |
| Pulse width (μs) | ＞4 (flat top) |
| Pulse rate (pps) | 100 |
| Pulse Flatness | ＜0.5% peak-to-peak |
| Pulse-Pulse Regulation | ＜0.3% |

The modulator can be divided into four major sections: a charging section, a discharging section, a pulse transformer tank and a klystron load. In the charging section, there is a series resonant type high-voltage charging power supply (HVDC) detector. Fig. 6.5.2.3 is a simplified modulator circuit diagram.

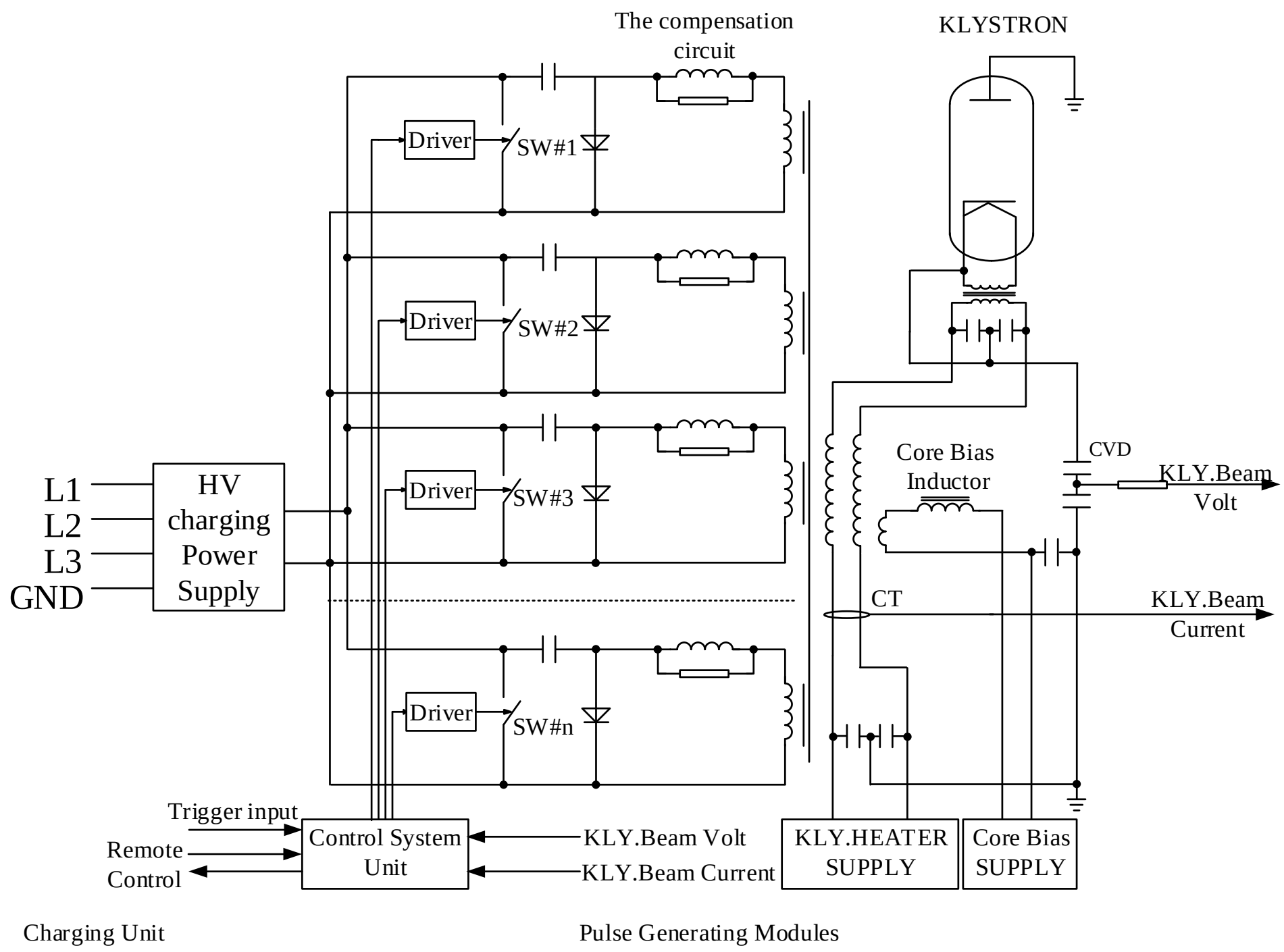

**Figure 6.5.2.3**: Simplified modulator circuit diagram

For system and personnel safety, the interlock has static and dynamic modes. The static mode includes door interlocks, ground hooks, heater PS trips, cooling water flow and temperature status, and over voltage and current trips. The dynamic mode uses an analog signal from the vacuum system.

### 6.5.2.4 *LLRF System*

The Linac is comprised of room temperature bunching systems and hundreds of S-band accelerating structures with the high level RF S-band signals modulated by 50 Hz short pulses. There are 74 power source units and the pulse length is about 4 μs. The pulse-to-pulse amplitude fluctuation and phase drift should be compensated by LLRF system.

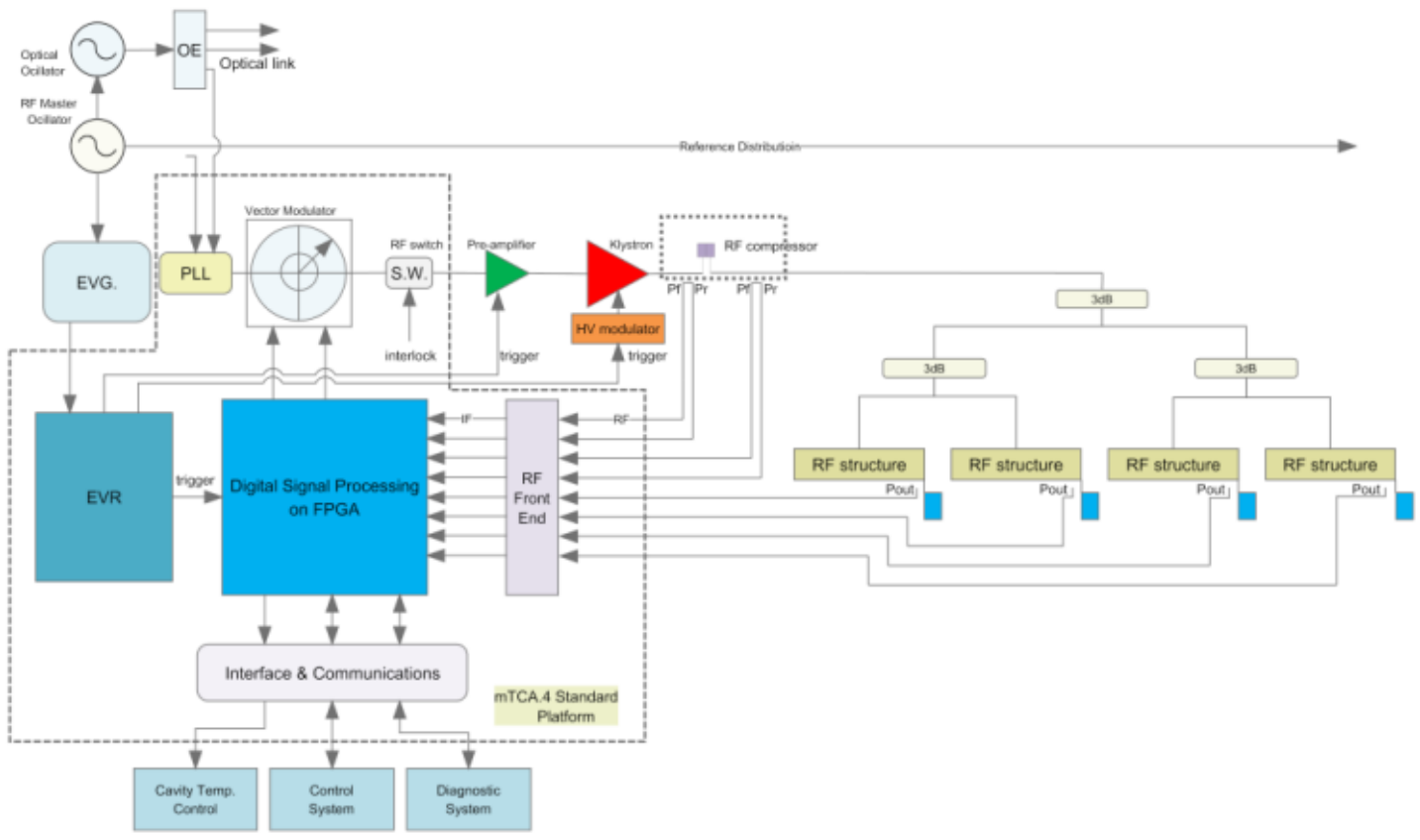

**Figure 6.5.2.4:** The LLRF system layout for the injector accelerating structures

The LLRF systems are synchronized to a phase reference system. They control the RF fields in the accelerating structures to meet the beam dynamics tolerance requirements of energy stability, luminosity loss and emittance growth. One LLRF unit controls one klystron and each single klystron feeds four accelerating tubes. A layout of one RF station is shown in Fig. 6.5.2.4. The controller will regulate amplitude and phase of the vector sum of a string of those accelerating structures.

The sub-harmonic bunchers are standing wave structures and the S-band buncher and accelerating tubes are travelling wave structures. Since the RF pulse duration is short, we adopt pulse-to-pulse feedback and adaptive feed-forward techniques to correct slow drifts and repetitive distortions. We also provide an interface for the implementation of beam-based feedback. The phase reference distribution, the phase-locked loops, the vector modulator, the pre-amplifier and the high-voltage modulator together with the klystron determine the pulse-to-pulse stability. Each of these components must meet the required short-term stability. The amplitude and phase stabilization requirements for these cavities are 0.2% (rms) and 0.2 deg (rms), respectively.

## 6.5.3 **Magnets**

### 6.5.3.1 *Dipole Magnets*

There are 5 types of dipole magnets in the Linac: 4 dipole magnets are 2.35 m long, 4 dipole magnets are 0.279 m long, 1 dipole magnet is 0.262 m long, 1 dipole magnet is 5.236 m long and 1 dipole magnet is 5.847 m long. The total number is 12. All the dipole magnets are excited by DC current. The iron core is made of 0.5 mm thick laminated low carbon silicon steel sheets.

The cores of the magnets have a curved structure so that racetrack-shaped coils can be used. The magnet is split into two halves for vacuum chamber installation.

The cross sections for the dipole magnets have been designed and optimized using OPERA-2D, sufficient in the conceptual design. Half of the magnet is modelled. Magnetic flux lines for one of the dipole magnets are show in Fig. 6.5.3.1. All 19 dipoles use the same H type structure. The parameters are listed in Table 6.5.3.1.

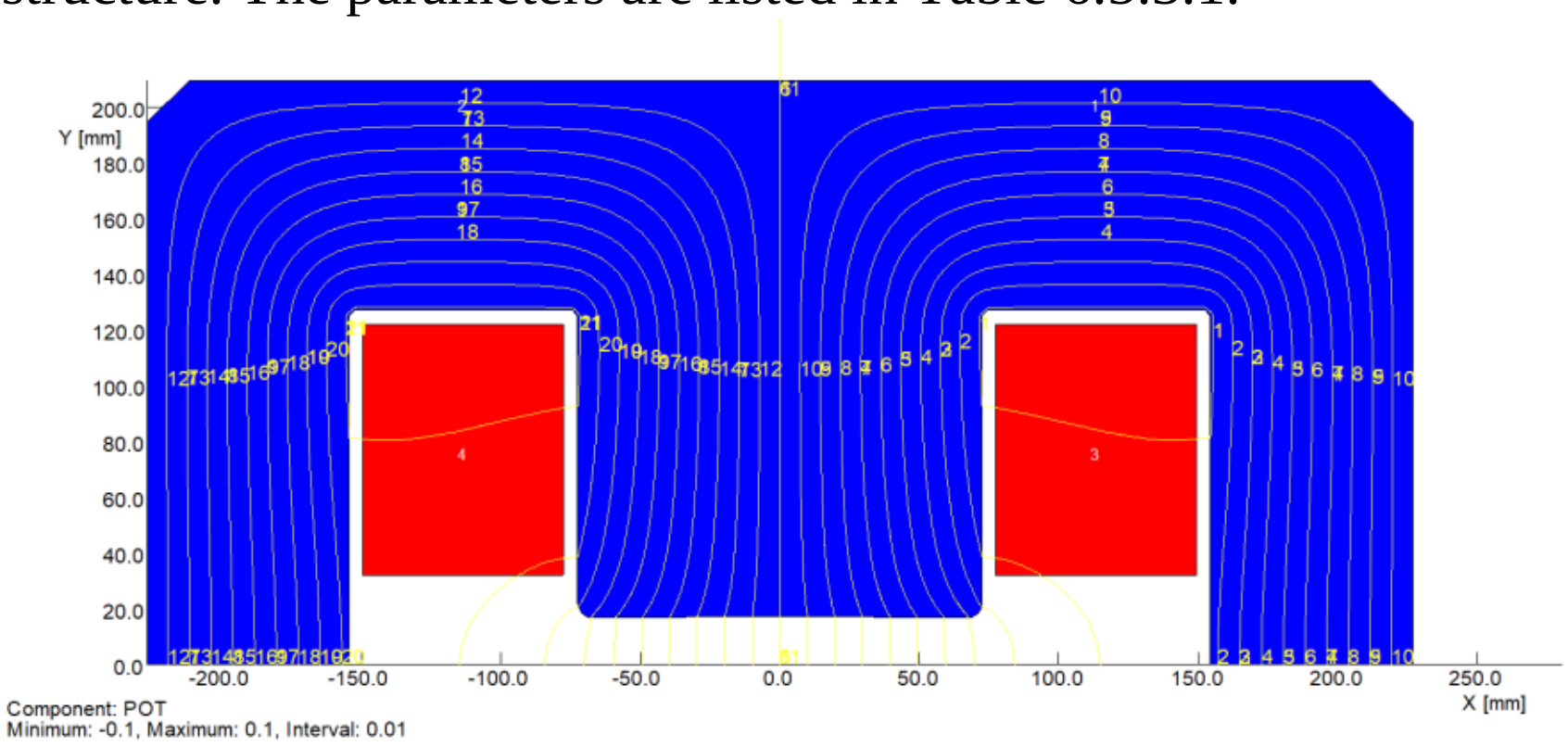

**Figure 6.5.3.1:** 2D flux lines of the B dipole magnet (Half cross section)

**Table 6.5.3.1:** Parameters of the Linac dipole magnets

| **Magnet name** | **B** | **CB1** | **AM1** | **AM2** | **AM3** |
|---|---|---|---|---|---|
| Quantity | 4 | 4 | 2 | 1 | 1 |
| Gap (mm) | 34 | 54 | 34 | 44 | 44 |
| Max. Field (T) | 1 | 0.50 | 0.3 | 0.8 | 1 |
| Deflection Angle | 10 | 8 | 20 | 15 | 10 |
| Deflection Radius (m) | 13.5 | 2.0 | 0.75 | 20.0 | 33.5 |
| Magnetic Length (mm) | 2356 | 279 | 262 | 5236 | 5847 |
| Good Field Region (mm) | 60×30 | 100×50 | 60×30 | 100×40 | 100×40 |
| Field Uniformity | 0.1% | 0.1% | 0.1% | 0.1% | 0.1% |
| Ampere-turns per pole (A-T) | 13900 | 10900 | 4100 | 14200 | 17700 |
| Turns per pole | 80 | 80 | 32 | 64 | 80 |
| Max. current (A) | 174 | 136.3 | 128.2 | 221.9 | 221.3 |
| Conductor size (mm) | 9×9Φ6 | 6.5×6.5Φ4 | 6.5×6.5Φ4 | 9×9Φ6 | 9×9Φ6 |
| Current density (A/mm$^2$) | 3.35 | 4.73 | 4.45 | 4.28 | 4.27 |
| Resistance （Ω） | 0.3 | 0.11 | 0.035 | 0.52 | 0.73 |
| Inductance （mH） | 430 | 51.1 | 8.3 | 651 | 1152 |
| Voltage drop (V) | 52 | 15.1 | 4.5 | 115.5 | 160 |
| Power loss (kW) | 9.1 | 2.05 | 0.58 | 25.6 | 35.4 |
| Core length (mm) | 2326 | 230 | 230 | 5200 | 5805 |
| Core width/height (mm) | 460/420 | 600/440 | 365/305 | 600/420 | 600/500 |
| Core weight (t) | 3.04 | 0.53 | 0.21 | 8.8 | 9.1 |
| Water pressure (kg/cm$^2$) | 6 | 6 | 6 | 6 | 6 |
| Cooling circuits | 4 | 4 | 2 | 8 | 10 |
| Water flow velocity (m/s) | 0.95 | 1.75 | 2.28 | 1.03 | 0.97 |
| Total water flow (l/s) | 0.21 | 0.176 | 0.057 | 0.93 | 1.37 |
| Temperature increase (°C) | 10 | 2.8 | 2.5 | 6.6 | 6.2 |

### 6.5.3.2 *Quadrupole Magnets*

The quadrupole magnets are divided into two types: one is a conventional quadrupole magnet with aperture 100 mm; the other is a small aperture triple quadrupole magnet, in which the quadrupole is divided into three apertures, 32 mm, 40 mm and 60 mm.

All magnets are DC powered. The core material is silicon steel sheet or DT4 solid iron; the coils are hollow copper conductors. Although the maximum field on the pole tips is around 5 to 6 kGs, local saturation is reduced by optimizing the pole surface design. The temperature rise is controlled to 10 ℃ by water cooling.

The magnetic flux in one of the quadrupole magnets is shown in Fig. 6.5.3.2. The parameters are listed in Table 6.5.3.2.

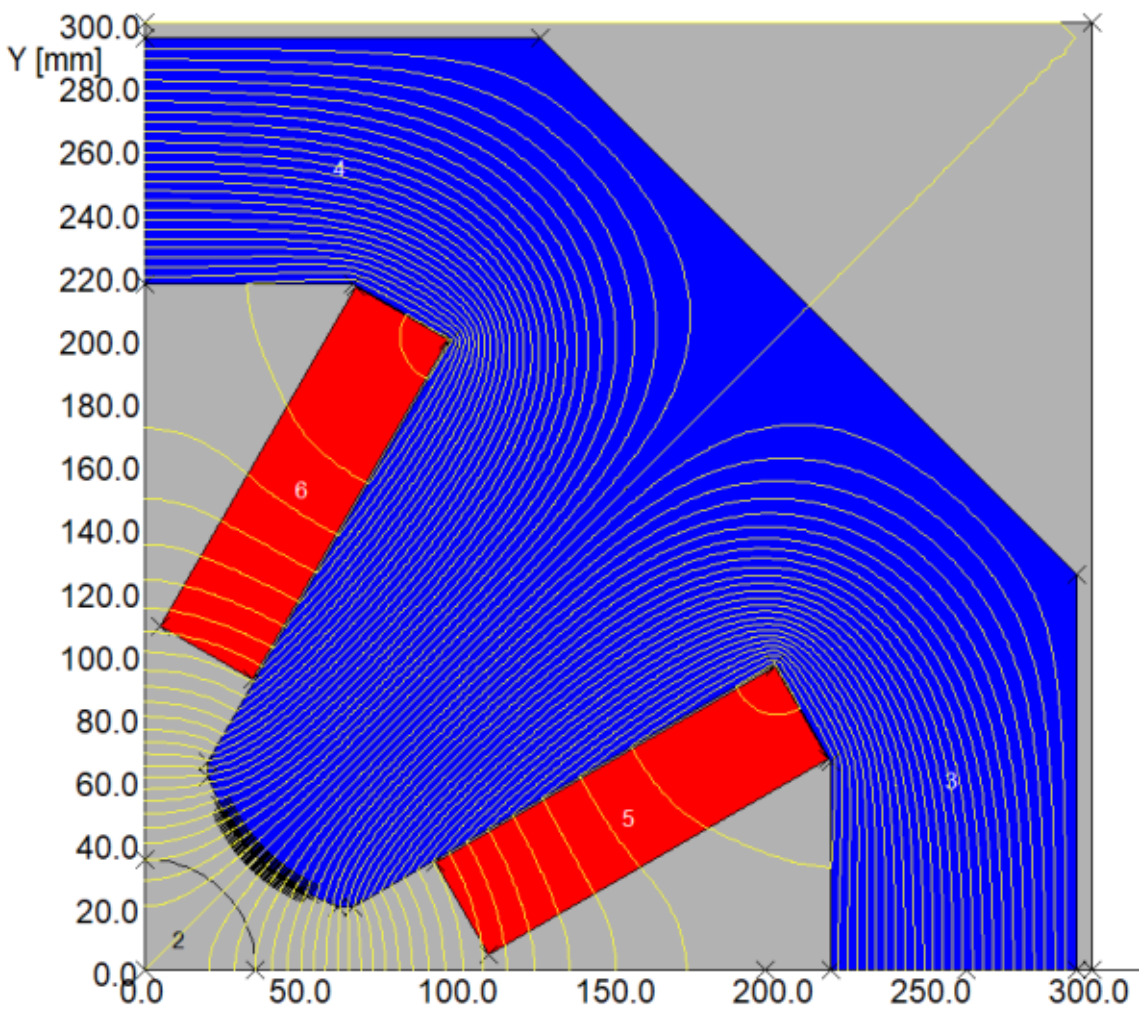

**Figure 6.5.3.2:** The magnetic flux for one of the aperture magnets

**Table 6.5.3.2:** Linac quadrupole magnet parameters

| **Magnet name** | **100Q** | **60SQ** | **60LQ** | **40SQ** | **40LQ** | **32SQ** | **32LQ** |
|---|---|---|---|---|---|---|---|
| Quantity | 48 | 6 | 3 | 88 | 50 | 38 | 19 |
| Aperture (mm) | 100 | 60 | 60 | 40 | 40 | 32 | 32 |
| Magnetic length (mm) | 300 | 100 | 200 | 200 | 400 | 300 | 600 |
| Field gradient (T/m) | 10 | 15 | 15 | 28 | 28 | 36 | 36 |
| GFR-radius (mm) | 25 | 25 | 25 | 16 | 16 | 13 | 13 |
| Field errors | 0.001 | 0.001 | 0.001 | 0.001 | 0.001 | 0.001 | 0.001 |
| AT per pole | 9947 | 5371 | 5371 | 4456 | 4456 | 3667 | 3667 |
| Turns per pole | 64 | 40 | 32 | 32 | 28 | 28 | 40 |
| Current (A) | 160.4 | 135.6 | 141.2 | 141.2 | 132.8 | 132.8 | 135.6 |
| Conductor size (mm) | 6.5×6.5 D4 | 6.5×6.5 D4 | 6.5×6.5 D4 | 6.5×6.5 D4 | 6.5×6.5 D4 | 6.5×6.5 D4 | 6.5×6.5 D4 |
| Current density (A/mm$^2$) | 5.44 | 4.60 | 4.60 | 4.79 | 4.79 | 4.51 | 4.51 |
| Resistance (mΩ) | 114 | 51 | 71 | 57 | 90 | 64 | 108 |
| Voltage drop (V) | 23 | 9.7 | 8.1 | 12.7 | 8.5 | 14.3 | 9.7 |
| Power loss（kW） | 3.7 | 1.3 | 1.2 | 1.8 | 1.2 | 1.9 | 1.3 |
| Inductance (mH) | 44.1 | 18.3 | 17.7 | 35.4 | 18.0 | 36.1 | 18.3 |
| Core length (mm) | 270 | 80 | 180 | 180 | 380 | 280 | 580 |
| Core width & height（mm） | 600 | 500 | 500 | 470 | 470 | 600 | 600 |
| Magnet weight（kg） | 500 | 120 | 240 | 210 | 420 | 240 | 480 |
| Water pressure (kg/cm$^2$) | 6 | 3 | 3 | 3 | 3 | 3 | 3 |
| Cooling circuits | 8 | 4 | 4 | 4 | 4 | 4 | 4 |
| Water flow velocity (m/s) | 2.19 | 1.80 | 1.48 | 1.68 | 1.30 | 1.57 | 1.17 |
| Total water flow (l/s) | 0.22 | 0.09 | 0.07 | 0.08 | 0.07 | 0.08 | 0.06 |
| Temperature increase (°C) | 4.0 | 2.5 | 4.2 | 3.2 | 6.5 | 3.4 | 7.7 |

### 6.5.3.3 *Solenoids*

There are 5 types of solenoid in the Linac: FS and S1 solenoid is 80 mm long; S2 solenoid is 120 mm long; S3 solenoids are 50 mm long and the number of them is 20; S4 solenoids are 1000 mm long and the number of them is 15. The maximum magnetic fields of FS, S1 to S3 are 0.06 T and 0.1 T and the maximum magnetic fields of S4 is 0.5 T.

The FS, S1 to S3 solenoids operate without water cooling. The S4 solenoids are water cooled because of the high magnetic fields. All the solenoids are excited by direct-current. The parameters are listed in Table 6.5.3.3.

**Table 6.5.3.3:** Parameters of Linac solenoids

| **Magnet name** | FS | **S1** | **S2** | **S3** | **S4** |
|---|---|---|---|---|---|
| Quantity | 4 | 1 | 1 | 20 | 15 |
| Aperture [mm] | 90 | 100 | 100 | 100 | 400 |
| Max. Field [T] | 0.06 | 0.1 | 0.1 | 0.1 | 0.5 |
| Magnetic Length [mm] | 80 | 80 | 120 | 50 | 1000 |
| Ampere-turns [AT] | 5250 | 9300 | 15500 | 4900 | 440000 |
| Turns | 525 | 930 | 1550 | 490 | 1200 |
| Current [A] | 10 | 10 | 10 | 10 | 366.7 |
| Current density [A/mm$^2$] | 1.38 | 1.38 | 1.38 | 1.38 | 4.4 |
| Conductor size [mm] | 1.95×3.72 | | | | 10×10Φ5 |
| Resistance [Ω] | 0.55 | 3.01 | 5.49 | 1.74 | 0.38 |
| Voltage drop [V] | 5.5 | 30.1 | 54.9 | 17.4 | 139 |
| Power loss [W] | 55 | 300 | 550 | 170 | 51 |
| Coil weight [kg] | 40 | 75.01 | 136.66 | 43.20 | 1500 |
| Max diameter [mm] | 200 | 395 | 432 | 432 | 640 |
| Water pressure [kg/cm$^2$] | No water cooling | | | | 6 |
| Cooling circuits | | | | | 10 |
| Water flow velocity [m/s] | | | | | 0.94 |
| Total water flow [l/s] | | | | | 1.84 |
| Temperature increase [°C] | | | | | 6.6 |

### 6.5.3.4 *Correctors*

There are four kinds of correctors in the Linac. Two of them have a low field of 150 Gs; the others have a high field of 850 Gs. For the low field correctors, the magnets can be designed to combine horizontal and vertical corrections. For the high field correctors, the magnets provide separate horizontal and vertical correction.

The correctors have a window frame core, formed by solid iron bars. The coils are wound with solid copper conductors, installed on the yokes of the cores. Since the correctors work in DC mode and the current density is lower than 1A/mm$^2$, the coils have no water cooling.

The OPERA software is used to simulate the field of the correctors. Magnetic flux distributions of the low field and high field correctors are shown in Figs. 6.5.3.4 and 6.5.3.5. The main parameters of the four types of correctors are listed in Table 6.5.3.4.

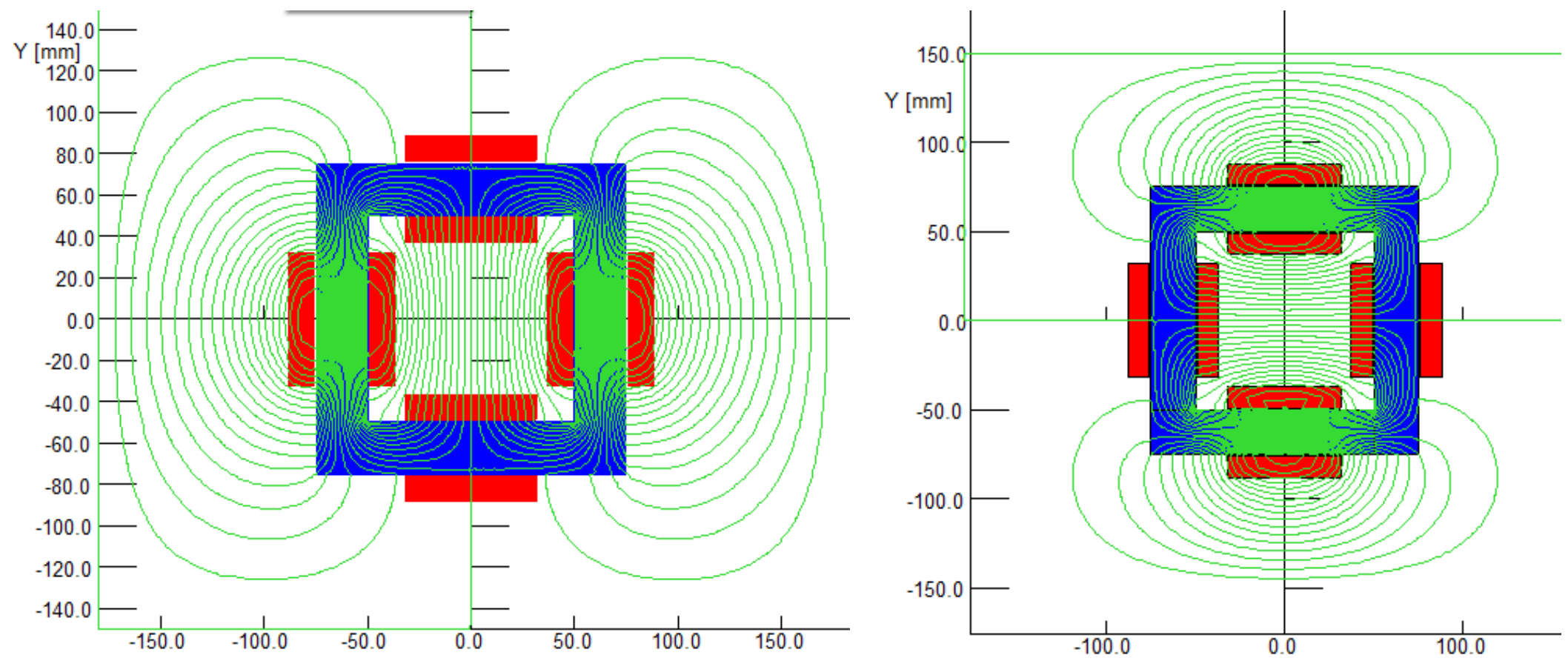

**Figure 6.5.3.3:** Magnetic flux distribution for the low field correctors

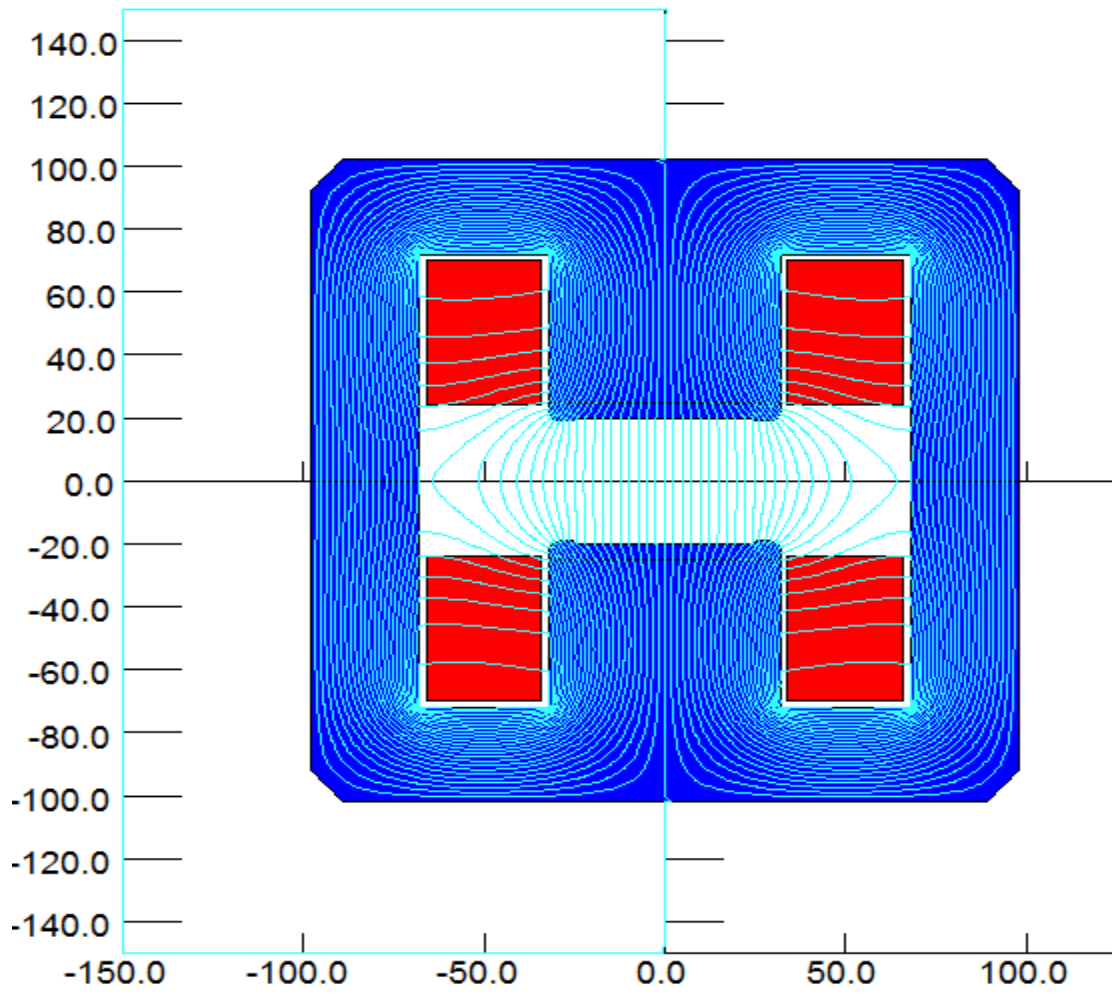

**Figure 6.5.3.4:** Magnetic flux distribution for the high field correctors

**Table 6.5.3.4:** The Main parameters of the Linac correctors

| **Magnet name** | **L-100C** | **L-60C** | **L-40C** | **L-34C** |
|---|---|---|---|---|
| Quantity | 17 | 3 | 46 | 19 |
| Gap [mm] | 100 | 60 | 40 | 34 |
| Max. Field [Gs] | 150 | 150 | 800 | 850 |
| Magnetic Length [mm] | 100 | 100 | 100 | 200 |
| Good Field Region [mm] | 30 | 25 | 30 | 26 |
| Field Uniformity | 1% | 1% | 1% | 1% |
| Ampere turns per pole[At] | 1670 | 1200 | 1300 | 1215 |
| Turns per pole | 48 | 40 | 48 | 40 |
| Max. current [A] | 34.8 | 30 | 27.08 | 30 |
| Size of conductor [mm*mm] | 5.5×5 | 5.5×5 | 5.5×5 | 5.5×5 |
| Current density [A/mm$^2$] | 1.265 | 1.091 | 0.98 | 0.92 |
| Resistance [Ω ] | 0.017 | 0.013 | 0.037 | 0.037 |
| Power loss [W] | 20 | 12 | 27 | 24 |
| Voltage [V] | 0.58 | 0.4 | 0.99 | 0.93 |
| Height of core [mm] | 200 | 150 | 204 | 216 |
| Width of core [mm] | 200 | 150 | 196 | 166 |
| Core Length [mm] | 80 | 80 | 80 | 180 |
| Total weight of magnet [kg] | 40 | 30 | 45 | 70 |

### 6.5.4 **Magnet Power Supplies**

The Linac power supplies follow the same design principles as the Booster and the Collider supplies.

The power supplies are DC supplies, used switched-mode as the main topology. All the power supplies are being designed in accordance to the parameters and ratings of the magnet, and in addition have 10 ~ 15% safety margin in both current and voltage. All the dipole, quadrupole and solenoid power supplies are unipolar, and all correction power supplies are bipolar to allow current reversal. All power supplies are housed along the Linac hall, to be close to the magnet load.

For convenient maintenance and repair, all power supplies are module-based design and digitally controlled.

According to the requirements for the accelerator physics, the power supplies in Linac include 11 diode power supplies, 177 quadrupole power supplies, 27 solenoid power supplies and 110 correctors. The total power consumption of converters for magnets is 1.45 MW, the detail parameters as shown in Table 6.5.4.1

**Table 6.5.4.1:** Power supply requirements

| Power Supply | Number | Stability 8hours | Output Ratings |
|---|---|---|---|
| Dipole | 11 | 500 ppm | 240A/200V |
| Quadrupole | 177 | 500 ppm | 150A/40V |
| Solenoids-1 | 5 | 500 ppm | 400A/460V |
| Solenoids-1 | 22 | 500 ppm | 11A/30V |
| Corrector | 110 | 500 ppm | 40A/7V |
| Total system power | | | 1.45MW |

### 6.5.5 **Vacuum System**

#### 6.5.5.1 ***Vacuum Requirements***

The Linac vacuum system will provide a stable and acceptable pressure for beam transfer efficiency and protection of the waveguides, accelerating tubes and electron gun from damage induced by high-voltage arcing. A dynamic pressure of less than $2\times10^{-7}$ Torr is required in the Linac and less than $2\times10^{-8}$ Torr is necessary in the electron gun in order to avoid the e-gun cathodes from being contaminated. Table 6.5.5.1 shows the design specifications of the Linac vacuum system.

**Table 6.5.5.1:** Design specifications of the Linac vacuum system

| Equipment and section | Static pressure (Torr) | Dynamic Pressure (Torr) |
|---|---|---|
| E-gun | $<1\times10^{-9}$ | $<2\times10^{-8}$ |
| ESBS | $<5\times10^{-8}$ | $<2\times10^{-7}$ |
| Accelerator section | $<5\times10^{-8}$ | $<2\times10^{-7}$ |
| Waveguide section | $<5\times10^{-8}$ | $<2\times10^{-7}$ |

#### 6.5.5.2 ***Vacuum Equipment***

The vacuum chambers will be fabricated from low magnetic permeability stainless steel and utilize conflat flanges. Pumping is done with 1310 ion pumps. There are 611 cold cathode gauges to measure pressure. The Linac vacuum system is divided into 29 sectors with metal gate valves. Vacuum sectors will be roughed down from atmosphere with portable turbo-molecular pumps (TMP) backed with dry scroll pumps. When a pressure of less than $1\times10^{-6}$ Torr is obtained, the sputter ion pumps will be turned on. The TMP will be manually isolated with all metal valves while the vacuum sector is at high vacuum, in order to prevent the vacuum sector from being exposed to atmosphere due to pump or power failure.

Power supplies and controllers for the Linac vacuum system will be located in the service area due to the high radiation level in the tunnel. The vacuum devices such as gauge controllers, pump controllers, gate valves, residual gas analyzers with local and remote capability, will be interfaced to the machine control system for remote monitoring, operation and control.

### 6.5.6 **Instrumentation**

#### 6.5.6.1 ***Introduction***

The types of instrumentation required include beam position monitors, beam profile monitors, beam current monitors and beam loss monitoring. There are large differences in the electron and positron diagnostic signals. Those that work well for positrons are likely to be saturated with electrons. The stripline BPM will be used to measure beam positions and angles and determine the beam trajectory. The beam shape measurement can provide emittance, energy and energy spread. The integrated charge detectors (ICT) will be used to measure the bunch charge. Faraday cups are used to measure beam current. The beam diagnostics system will have sufficient dynamic range from the minimum to the maximum single-shot bunch charge. The type, quantity and function of linear beam detectors are listed in Table 6.5.4.1.

**Table 6.5.4.1:** Linac instrumentation

| Type | | Quantity | Function |
|---|---|---|---|
| Beam profile monitor | | 80 | Beam transverse size |
| ICT | | 42 | Bunch charge and transmission efficiency |
| Faraday cup | | 2 | Beam current |
| Beam position monitor | button | 1 | Beam position and trajectory |
| | Stripline | 110 | Beam position and trajectory |
| Energy analysis station（YAG /OTR） | | 3 | Beam energy and energy spread |
| Emittance measurement（YAG /OTR） | | 4 | Beam emittance |

#### 6.5.6.2 ***Beam Position Monitor***

The stripline BPM has relatively high sensitivity and a simple structure, and measures beam positions and intensities.  At the exit from the electron gun, only one button type pick-up is used because of the limited space. All BPM electronics will be based on Micro TCA architecture. SMA-type feedthroughs will be used. An image of the electronics is shown in Figure 6.4.6.1. The stripline BPM operating frequency is 500 MHz and the ADC sampling rate is about 170 MHz.

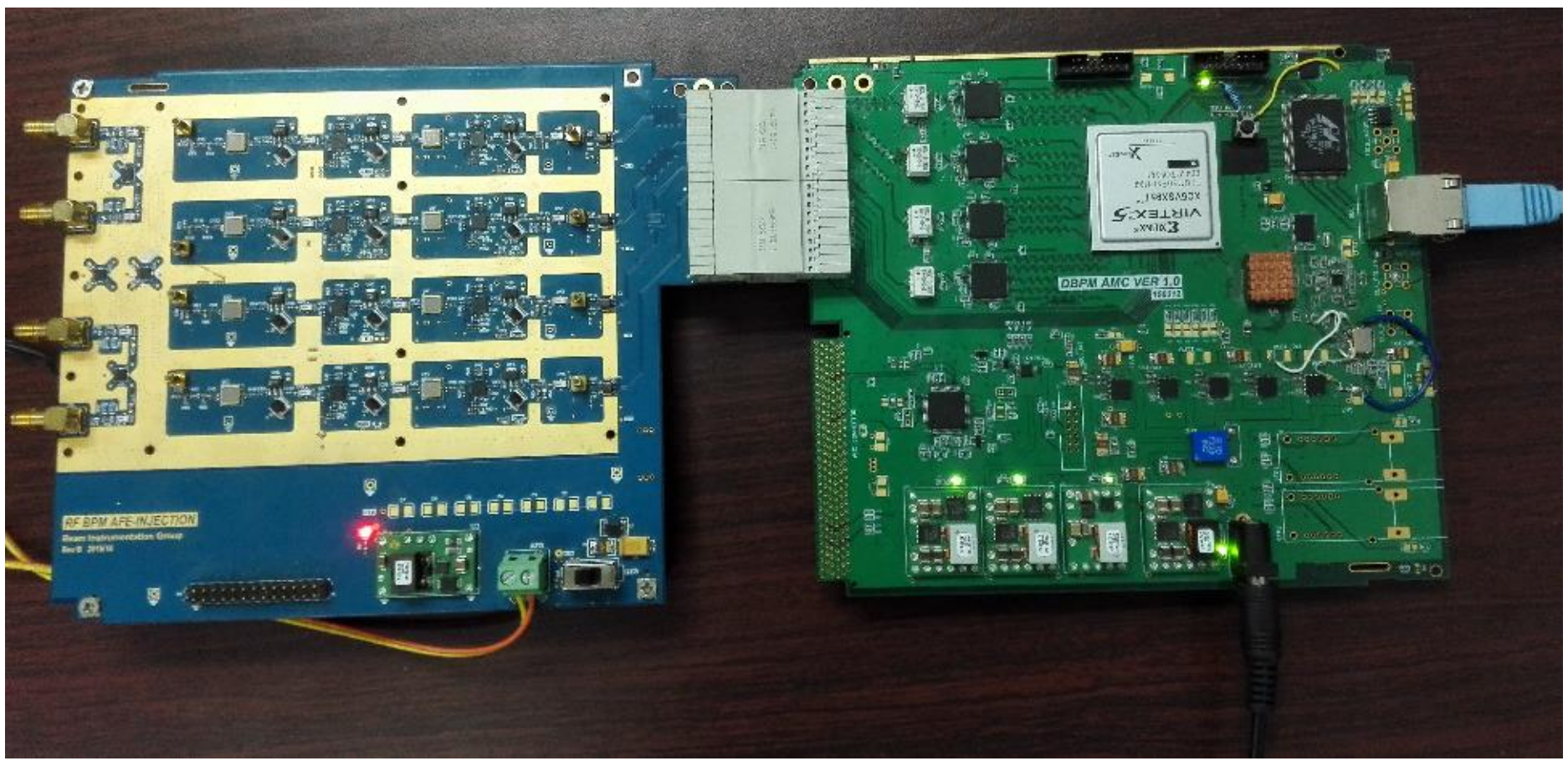

**Figure 6.4.6.1:** BPM electronics

### 6.5.6.3 ***Beam Profile Monitor***

The YAG / OTR target is used to measure the beam profile. The YAG / OTR system can be divided into a vacuum mechanical part and an optical path imaging acquisition part. The vacuum mechanical component is used for positioning the target slice, motion control and vacuum sealing, and the optical path component is used for the transmission of the outgoing light and the image acquisition.

The YAG / OTR is typically placed at 45$^{o}$ to the beam. The target is a YAG crystal doped with cesium (Ce) and the OTR is an aluminized film silicon wafer. The target is outside the beam pipe and is moved to the middle of the beam pipe (and cuts off the beam) during measurement. The beam hits the target, generates light, and the light is led out to the CCD camera. The beam profile can be determined by digitizing the camera image.

### 6.5.6.4 ***Beam Current Monitor***

The Linac is equipped with 42 integrated charge detectors (ICT) to monitor beam intensity and transmission efficiency. Compared with the Faraday cup, ICT has an unobstructed feature that allows on-line measurements without the need for radiation protection. Through the post-level electronics system processing, one can calculate the bunch charge. Bergoz company produces beam charge detectors and electronics. The whole system is shown in Figure 6.4.6.2.

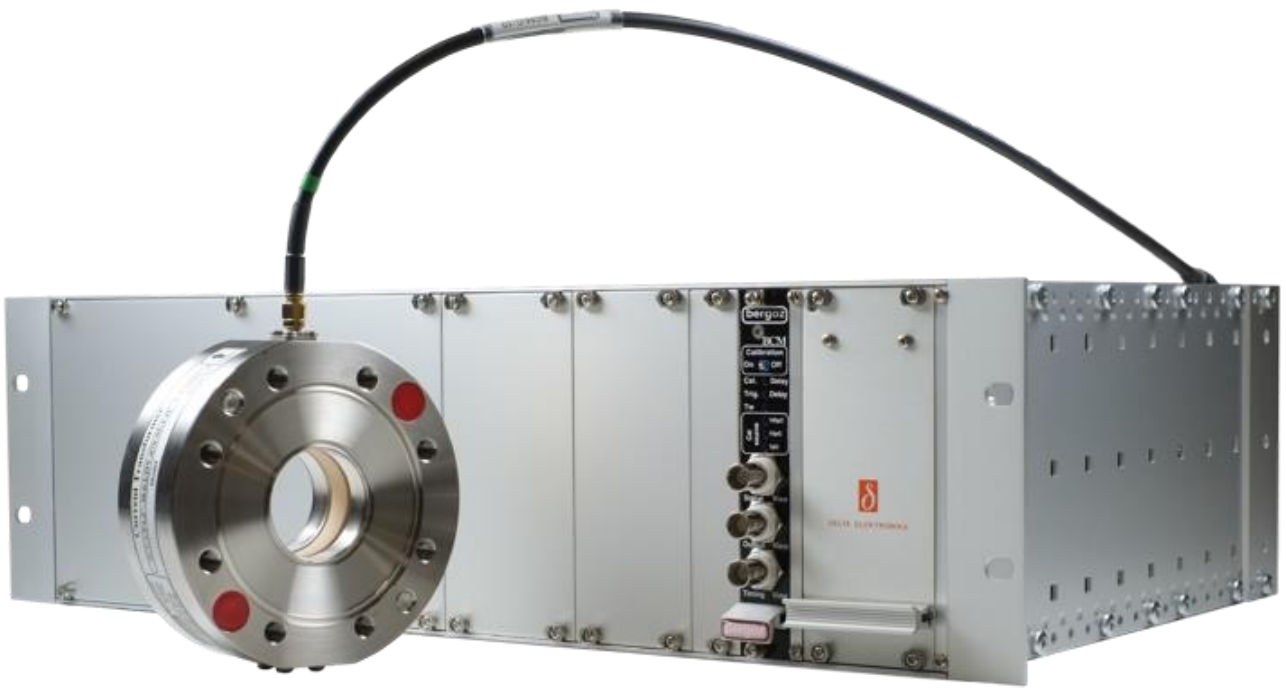

**Figure 6.4.6.2:** ICT system produced by Bergoz

There are 2 Faraday cups for beam current measurement and ICT calibration. The Faraday cup is copper and the output signal is transmitted to the electrometer through a coaxial cable. The electrometer can be a Keithley-6514 and uses an IEEE-488 computer interface. Due to the high X-ray intensity generated in the Faraday cups, lead shields should be used outside the Faraday barrels for radiation protection.

#### 6.5.6.5 *Beam Energy and Energy Spread*

Beam energy and energy spread can be measured by deflecting the beam with a dipole and using the spot position and size on a fluorescent screen.

#### 6.5.6.6 *Beam Emittance*

Beam Emittance is measured using an upstream quadrupole and observing the changes in the beam profile as a function of quadrupole strength.

### 6.5.7 Control System

The control system allows the operators to monitor and control equipment distributed along the 1.2 km Linac gallery. This can be done both from local and central control rooms. There is a machine protection system to keep the devices in a safe condition. All of the useful parameters are stored in a database for later retrieval.

The devices to be controlled include magnet power supplies, klystrons and modulators, vacuum valves, pumps and gauges, electron gun, positron target, microwave system and beam instruments.

Operators will be able to adjust the current and choose the operating mode of the electron/positron gun.

Parameters of klystrons and modulators will be monitored and displayed. These include the high voltage, the output power, the RF phase and the amplitude of the output envelope. There are interlock loops for klystrons and modulators. In case the pressure outside a vacuum klystron window exceeds a specified limit, the HV of corresponding modulator will be turned off.

### 6.5.8 Mechanical Systems

The mechanical system supports the accelerator tubes, dipole and quadrupole magnets, solenoids, correctors, as well as vacuum system components and instrumentation. Table 6.5.8.1 lists the number of components.

There are two types of quadrupoles. One type is supported separately. The other is triplet quadrupoles which have a common support system.

**Table 6.5.8.1**: Quantities of accelerator tubes, magnets and their supports in the Linac

| Magnet type | Quantity | Device length (mm) | No. of supports per device |
|---|---|---|---|
| Accelerator tubes | 277 | 3000 | 3 |
| | 6 | 2000 | 2 |
| Dipole | 4 | 2356 | 2 |
| | 6 | 262/279 | 1 |
| | 2 | 5236/5847 | 4 |
| Quadrupole | 54 | 300/400 | 1 |
| | 3 | ~600 (triplet) | 1 |
| | 63 | ~1200/1800 (triplet) | 2 |
| Solenoids | 1 | 80-1000 | 1 |
| Corrector | 85 | 100-250 | 1 |

The Linac is near the ground of the tunnel. Each support structure contains the pedestal, mounting plate and adjusting mechanism, similar to that in the Collider. The requirements are also the same, which are described in Section 4.3.9.

Different from the Collider and Booster which are in the tunnel deep underground, the Linac tunnel is near the ground. The gallery of Linac is above the Linac tunnel for preliminary design, which is shown in Figure 6.5.8.1. The tunnel is 3.5 meters wide, and at the bypass location, the width is changed to 5.5 meters.

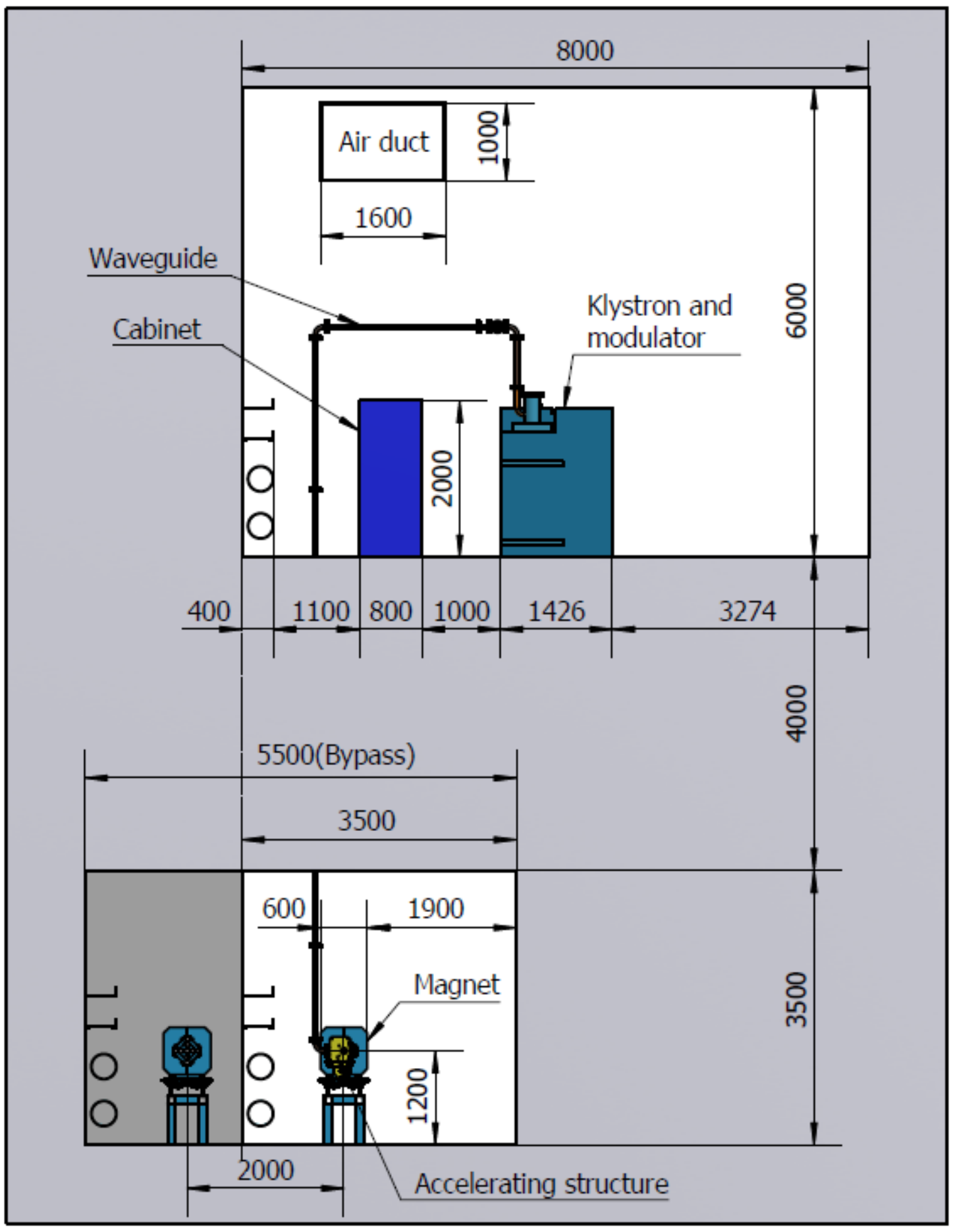

**Figure 6.5.8.1:** Cross section of Linac and its gallery

## 6.6 Damping Ring Technical Systems

### 6.6.1 RF System

#### 6.6.1.1 *RF System Design*

Two 650 MHz normal conducting 5-cell cavities will be used for the Damping Ring RF system. The RF parameters are listed in Table 6.6.1.1. Each cavity can provide up to 1.3 MV cavity voltage. Two 50 kW solid state amplifiers (SSA) will be used to feed RF power to the two cavities separately. This provides an operation margin. The cavity structure is based on the 500 MHz normal conducting 5-cell cavity used for HERA/PETRA. [1] The cavity includes five magnetically slot-coupled re-entrant cells, one coaxial input coupler and two plunger tuners.

**Table 6.6.1.1:** Damping Ring RF parameters

| | |
|---|---|
| Circumference [m] | 58.5 |
| Beam energy [GeV] | 1.1 |
| SR loss / turn [keV] | 35.8 |
| Beam current [mA] | 15.4 |
| Bunch charge [nC] | 3 |
| RF frequency [MHz] | 650 |
| Harmonic number | 127 |
| RF voltage [MV] | 1.8 |
| Number of cavities | 2 |
| Number of cells / cavity | 5 |
| Cavity effective length [m] | 1.15 |
| Cavity total length [m] | 1.5 |
| Shunt impedance [MΩ] (accelerator definition) | 37 |
| Cavity operating voltage [MV] | 0.9 |
| Cavity operating gradient [MV/m] | 0.78 |
| $Q_0$ | 33600 |
| Input power / cavity [kW] | 22.2 |
| Coupling coefficient | 1.01 |
| Max output power / SSA [kW] | 50 |

## 6.6.2 **Magnets**

### 6.6.2.1 ***Dipole Magnets***

There are 32 dipole magnets of length of 0.71 m for the Damping Ring (DR), 24 dipole magnets of length 0.646 m and 8 dipole magnets of length 1.614 m for its transport line. They are DC magnets designed with conventional technology. The dipole parameters are listed in Table 6.6.2.1.

**Table 6.6.2.1:** DR dipole magnets

| Magnet name | DR-36B | LTD-44B-I | LTD-44B-II |
|---|---|---|---|
| Quantity | 32 | 24 | 8 |
| Gap [mm] | 36 | 44 | 44 |
| Max. Field [T] | 1.015 | 1 | 1 |
| Magnetic Length [mm] | 710 | 646 | 1614 |
| Good Field Region [mm] | 50*30 | 100*40 | 100*40 |
| Field Uniformity | 0.1% | 0.1% | 0.1% |
| Ampere-turns per pole（AT） | 14730 | 17700 | 17700 |
| Turns per pole | 64 | 64 | 64 |
| Max. current（A） | 230 | 276.6 | 276.6 |
| Conductor size（mm） | 9×9Φ6 | 9×9Φ6 | 9×9Φ6 |
| Current density（$A/mm^2$） | 4.44 | 5.34 | 5.34 |
| Resistance（Ω） | 0.081 | 0.082 | 0.205 |
| Inductance（mH） | 78.5 | 80.1 | 200 |
| Voltage drop (V) | 18.5 | 22.6 | 54.5 |
| Power loss (kW) | 4.25 | 6.3 | 15.7 |
| Core length(mm) | 710 | 646 | 1614 |
| Core width/height（mm） | 440/380 | 600/465 | 600/465 |
| Magnet weight（t） | 1 | 1.5 | 4.5 |
| Water pressure（$kg/cm^2$） | 6 | 6 | 6 |
| Cooling circuits | 4 | 4 | 4 |
| Water flow velocity（m/s） | 2.01 | 2 | 1.5 |
| Total water flow（l/s） | 0.45 | 0.45 | 0.3 |
| Temperature increase（°C） | 3 | 3.5 | 13 |

### 6.6.2.2 ***Quadrupole Magnets***

The DR quadrupoles are DC and their design and manufacturing requirements are the same as the Linac conventional quadrupoles. Parameters are listed in Table 6.6.2.2.

**Table 6.6.2.2:** DR quadrupole parameters

| Magnet name | DR-36Q | LTD-54Q |
|---|---|---|
| Quantity | 48 | 44 |
| Aperture (mm) | 36 | 54 |
| Magnetic length（mm） | 200 | 200 |
| Field gradient（T/m） | 20 | 20 |
| GFR-radius (mm) | 15 | 23 |
| Field errors | 0.1% | 0.1% |
| AT per pole | 2578 | 5801 |
| Turns per pole | 20 | 40 |
| Current（A） | 130.2 | 147.9 |
| Conductor size（mm） | 6.5×6.5D4 | 6.5×6.5D4 |
| Current density（$A/mm^2$） | 4.42 | 5.02 |
| Resistance（mΩ） | 36 | 71 |
| Voltage drop（V） | 4.7 | 10.7 |
| Power loss（kW） | 0.60 | 1.56 |
| Inductance（mH） | 6.2 | 20.7 |
| Core length（mm） | 180 | 180 |
| Core width & height（mm） | 410 | 500 |
| Magnet weight（kg） | 170 | 240 |
| Water pressure（$kg/cm^2$） | 4 | 4 |
| Cooling circuits | 3 | 3 |
| Water flow velocity（m/s） | 2.20 | 1.48 |
| Total water flow（l/s） | 0.11 | 0.07 |
| Temperature increase（°C） | 1.30 | 5.00 |

### 6.6.2.3 *Sextupole Magnets*

The sextupole magnets for the Linac Damping Ring are divided into two types: focusing and defocusing. Because of the low working magnetic field, the magnet design and parameters are calculated according to the maximum magnetic field requirements. Magnet are a two-in-one structure; the core material is DT4 solid iron, and the coils are solid copper conductors. The magnetic flux distribution is is shown in Figure 6.6.2.1 and parameters are in Table 6.6.2.3.

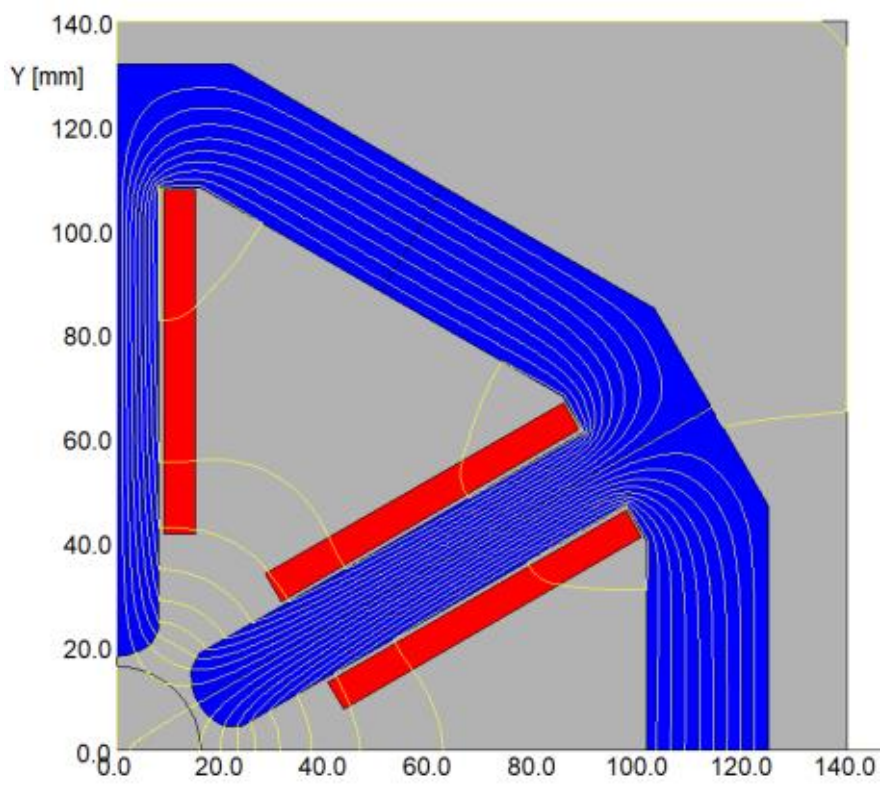

**Figure 6.6.2.1:** DR sextupole magnet flux distribution

**Table 6.6.2.3:** DR sextupole magnet parameters

| | |
|---|---|
| Magnet name | CEPC-DR-36S |
| Quantity | 24 |
| Aperture diameter (mm) | 36 |
| Magnetic length（mm） | 60 |
| Max. sextupole field (T/m$^2$) | 160 |
| GFR radius (mm) | 15 |
| Harmonic errors across GFR | 0.1% |
| Ampere-turns per pole（AT） | 124 |
| Coil turns per pole | 14 |
| Conductor size（mm） | 2×4 |
| Excitation current（A） | 8.9 |
| Current density（A/mm$^2$） | 1.12 |
| Resistance（mΩ）@35° | 41.7 |
| Voltage drop（V） | 0.37 |
| Max Power loss（W） | 3.32 |
| Inductance（mH） | 1 |
| Core length（mm） | 54 |
| Core width & height（mm） | 220 |
| Net core weight（kg） | 20 |

### 6.6.2.4 *Septum and Kicker Magnets*

Beam injection and extraction for the Damping Ring requires 2 Lambertson septum magnets and 2 fast kicker magnets.

The Lambertson magnet core is made from solid DT4 iron. The coil is a conventional racetrack consisting of 4 loops, 2 layers per loop, 8 turns per layer, making a total of 64 turns. The excitation current is 564 A. Parameters are in Table 6.6.2.4.

**Table 6.6.2.4:** DR Lambertson magnet parameters

| | |
|---|---|
| Magnet name | DR-LAM |
| Quantity | 2 |
| Effective magnetic length（mm） | 646 |
| Max. field（T） | 1 |
| Gap（mm） | 44 |
| Good field region（mm） | 100*40 |
| Field uniformity | 0.1% |
| Coil turns | 64 |
| Excitation current（A） | 564 |
| Conductor size（mm） | 15*15D10 |
| Current density（A/mm$^2$） | 3.85 |
| Resistance（mΩ） | 14.77 |
| Inductance（mH） | 4.17 |
| Voltage drop（V） | 8.32 |
| Power loss（kW） | 4.69 |
| Core length （mm） | 600 |
| Core width/height（mm） | 660/460 |
| Magnet weight（t） | 1.36 |
| Water pressure（$kg/cm^2$） | 6 |
| Cooling circuits | 4 |
| Water velocity（m/s） | 4.17 |
| Water flow（l/s） | 1.31 |
| Temperature rise（℃） | 0.85 |

Because the field changes rapidly, the kicker magnets for injection and extraction are inside a vacuum tank. The cores of the magnets have a window frame structure, consisting of two C-shaped Ni-Zn type ferrite blocks. The coils are made of a single-turn of copper conductor welded by copper bars. The pulsed current is fed from one end of the ferrite core by a high-voltage feedthrough. Kicker magnet parameters are listed in Table 6.6.2.5.

**Table 6.6.2.5:** DR kicker magnet parameters

| | DR-Kicker |
|---|---|
| Quantity | 2 |
| Field amplitude(Gs) | 600 |
| Magnetic length (mm) | 500 |
| Gap(mm) | 44 |
| Field waveform | Half-sine |
| Pulse width(ns) | 300 |
| Repetitive rate (Hz) | 100 |
| GFR (mm) | 40 |
| Field uniformity | 1% |
| Turns of coil | 1 |
| Current amplitude (A) | 2121 |
| Conductor size (mm) | 5*42 |
| Core material | Ni-Zn ferrite |
| Inductance (uH) | 0.63 |
| Max. voltage (kV) | 13.9 |
| Core width/height (mm) | 150/100 |
| Core weight (kg) | 50 |
| Radius of vacuum tank(mm) | 140 |

### 6.6.3 **Magnet Power Supplies**

The Damping Ring power supplies are DC, used in switched-mode. All the supplies have a 10 ~ 15% safety margin in both current and voltage. All the dipole, quadrupole and sextupole power supplies are unipolar, and are housed along the Linac hall, to be close to the magnet loads.

For convenient maintenance and repair, all power supplies are module-based and digitally controlled.

The design criteria are the same as for the Collider supplies. There are 38 diode power supplies, 78 quadrupole power supplies and 24 sextupole power supplies. The total power for the DR power supply system is 0.3 MW. Parameters are in Table 6.6.3.1.

**Table 6.6.3.1:** Power supply requirements

| Power Supply | Number | Stability 8 hours | Output Ratings |
|---|---|---|---|
| BARC | 32 | 500 ppm | 240A/25V |
| LTD-B | 6 | 500 ppm | 300A/35V |
| DR-Q | 52 | 500 ppm | 140A/11V |
| LTD-Q | 13 | 500 ppm | 160A/30V |
| DR-S | 2 | 500 ppm | 10A/7V |
| Total system power | | | 0.3MW |

### 6.6.4 Vacuum System

The Damping Ring is a 58.47 m storage ring containing vacuum chambers, pumps, gauges, and valves. An average pressure of less than $5\times10^{-8}$ Torr is required to minimize beam loss and bremsstrahlung radiation due to beam residual gas scattering. The vacuum chambers are elliptical tubes made of stainless steel. The inner cross section of the vacuum chamber is 33 mm × 30 mm. Conventional high vacuum technologies will be implemented and high vacuum will be achieved with small ion pumps distributed around the damping ring.

To estimate the heat load, we start from formulas (4.3.6.1) and (4.3.6.2) for the synchrotron radiation power emitted by an electron beam in uniform circular motion. For our damping ring, $E$ = 1.1 GeV, $I$ = 0.016 A, $\rho$ = 3.616 m, which from these equations gives a total synchrotron radiation power, $P_{SR}$=573 W, and a linear power density of $P_L$ =25 W/m, which are negligible for the heating load.

The gas load includes thermal outgassing and synchrotron-radiation-induced photo-desorption. Thermal outgassing contributes mainly to the base pressure in the absence of circulating beam. To estimate the desorption rate induced by synchrotron radiation, we use formulas (4.3.6.4) and (4.3.6.5). The effective gas load due to photo-desorption is found to be $8.5\times10^{-6}$ Torr·L/s and the linear SR gas load will be $3.8\times10^{-7}$ Torr·L/s/m. The thermal outgassing rate for stainless steel is taken as $1\times10^{-11}$ Torr·L/s·cm$^2$, assuming a good bake-out and careful handing. For a circular cross section of diameter of 30 mm, a linear thermal gas load is $Q_{LT}$ =$9.4\times10^{-9}$ Torr·L/s/m. Therefore, compared to the photo-desorption gas load, the thermal outgassing load is negligible.

There are a total of 40 ion pumps (30L/s each) distributed around the DR circumference. The pumps are connected to the vacuum chamber through a 50 mm-long $\phi$ 63 pipe, which limits the pumping speed to ~20 L/s. There are RF screens on the pump ports, which further reduce the effective pumping speed of each ion pump to ~15 L/s. For the thermal outgassing rate of the cavity, we use a conservative value of $5\times10^{-11}$ Torr·L/s·cm$^2$ because of the difficulty of an in-situ bake-out of the cavity. In anticipation of a large gas load, two 400 L/s ion pumps are deployed in the RF sectors. The effective pumping speed is ~360 L/s due to the orifice conductance limitation. Because of the stronger photo-desorption in the damping ring, we assume the residual gas composition to be closer to that of the Collider (i.e., 75% hydrogen and 25% CO), and the calculated conductance is thus 3.4 times higher than that of air.

The vacuum chamber are made from 316L stainless steel tubing cold drawn to an elliptical shape with inside major and minor axes of 33 mm and 30 mm, respectively. The vacuum chambers are connected by bellows with conflat-type flanges.

For the vacuum design of the transport lines, conventional vacuum technologies will be used.

### 6.6.5 **Instrumentation**

There are 40 BPMs in the damping ring with resolution 0.1 mm. The BPM is button type, similar to those in the Booster and the Collider. The diameter of the button is 6 mm with a 0.3 mm gap between the button and vacuum pipe. The signal and transfer impedance is calculated and shown in Fig. 6.6.5.1. The peak to peak amplitude is 14.4 V and the transfer impedance is about 0.15 Ω at the 500 MHz. The spectral intensity of the signal processing electronics is 4.68 $\times 10^{-10}$ V/Hz at 500 MHz, and the maximum value is 1.14 $\times 10^{-9}$ V/Hz at 2.18 GHz

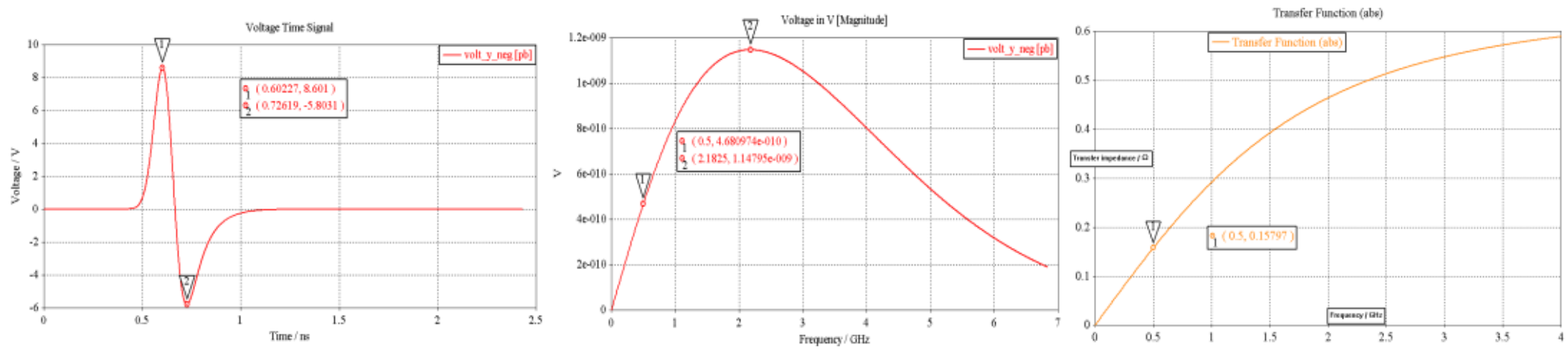

**Figure 6.6.5.1:** Left – signal of an electrode in the time domain; Middle – signal of an electrode in the frequency domain for a bunch charge 3 nC and bunch length 15 mm; Right – Transfer impedance.

The horizontal and vertical sensitivities near the center of the pipe are 12.25 %/mm and 12.26 %/mm as shown in Fig. 6.6.5.2(a). Fig. 6.6.5.2(b) is the sensitivity mapping of the simulation in an area of 10 mm × 10 mm. The scan range is shown in the inset figure of Fig. 6.6.5.2(b), where U= Δx/Σx, V=Δy/Σy. The transverse response of the signal is non-linear for the beam off center, especially if the distance between the beam and pipe center is greater than 4 mm.

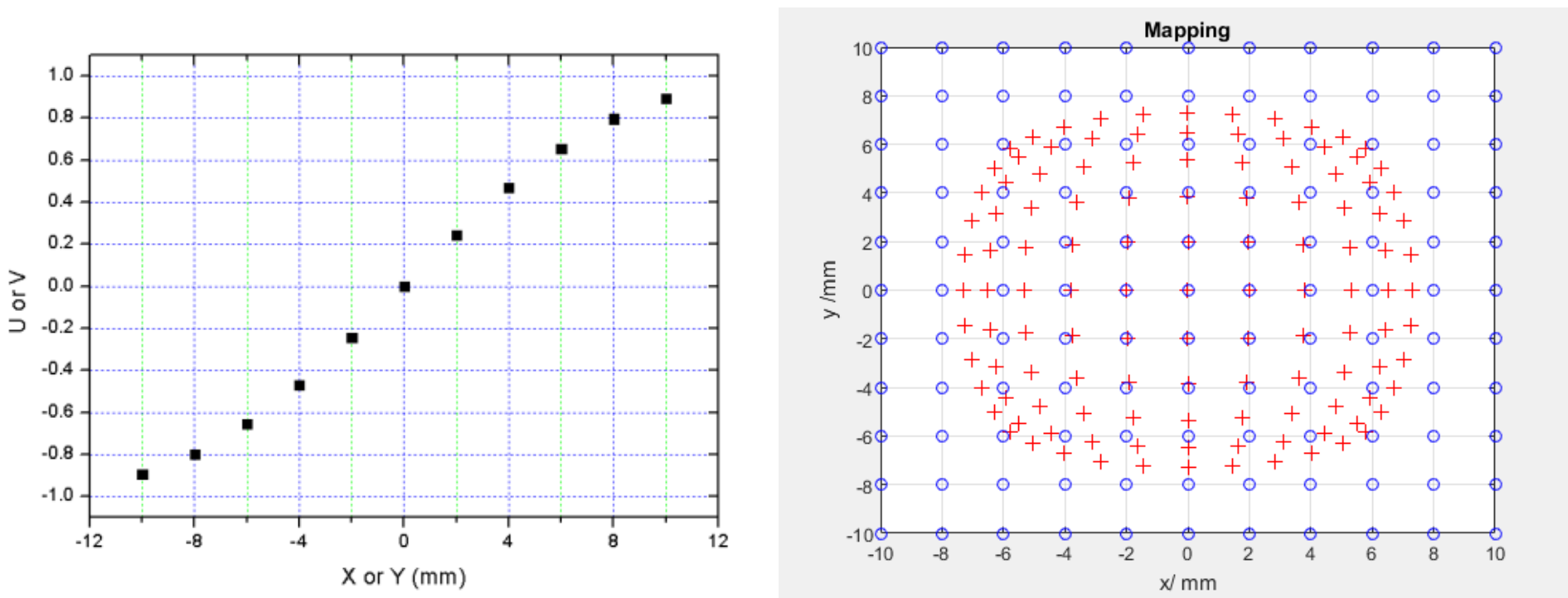

**Figure 6.6.5.2:** (a) sensitivity scan in the range ±10 mm (b) sensitivity mapping of the button BPMs in an area 10 mm × 10 mm.

### 6.6.6 **Mechanical Systems**

The DR is 58.47 meters in circumference. Together with a 60-meter long transport line there are about 120 m of components to support. Table 6.6.6.1 lists the number of magnets and their supports. In addition the support system also supports accelerator tubes as well as the vacuum system and instrumentation.

**Table 6.6.6.1:** Quantities and their supports in the DR

| Magnet type | Quantity | Device length (mm) | No. of supports per device |
|---|---|---|---|
| Dipole | 32 | 710 | 1 |
| | 26 | 646 | 1 |
| | 8 | 1614 | 2 |
| | 2 | 500 | 1 |
| Quadrupole | 92 | 200 | 1 |
| Sextuple | 24 | 60 | 1 |
| Corrector | 64 | 100 | 1 |

The design and requirements are similar to those in the Collider and the Linac, which are described in Sections 4.3.9. and 6.5.8, respectively.

# 7 Systems Common to CEPC Accelerators

## 7.1 Cryogenic System

### 7.1.1 Overview

All the cavities in the Booster and Collider rings will be cooled in a 2 K liquid-helium bath to achieve a good cavity quality factor. The cooling benefits from helium II properties of large effective thermal conductivity and heat capacity as well as low viscosity. It is a technically safe and economically reasonable choice. The 2 K cryostat will be protected against heat radiation by means of two thermal shields cooled with 5-8 K helium as well as 40-80K helium from a refrigerator.

There are 4 cryo-stations around the 100 km circumference as shown in Fig 7.1.1. Generally, each cryo-station, supplied from a common cryogenic plant, has one refrigerator and one distribution box.

The cryogenic system is designed for fully automatic operation, stable and reliable.

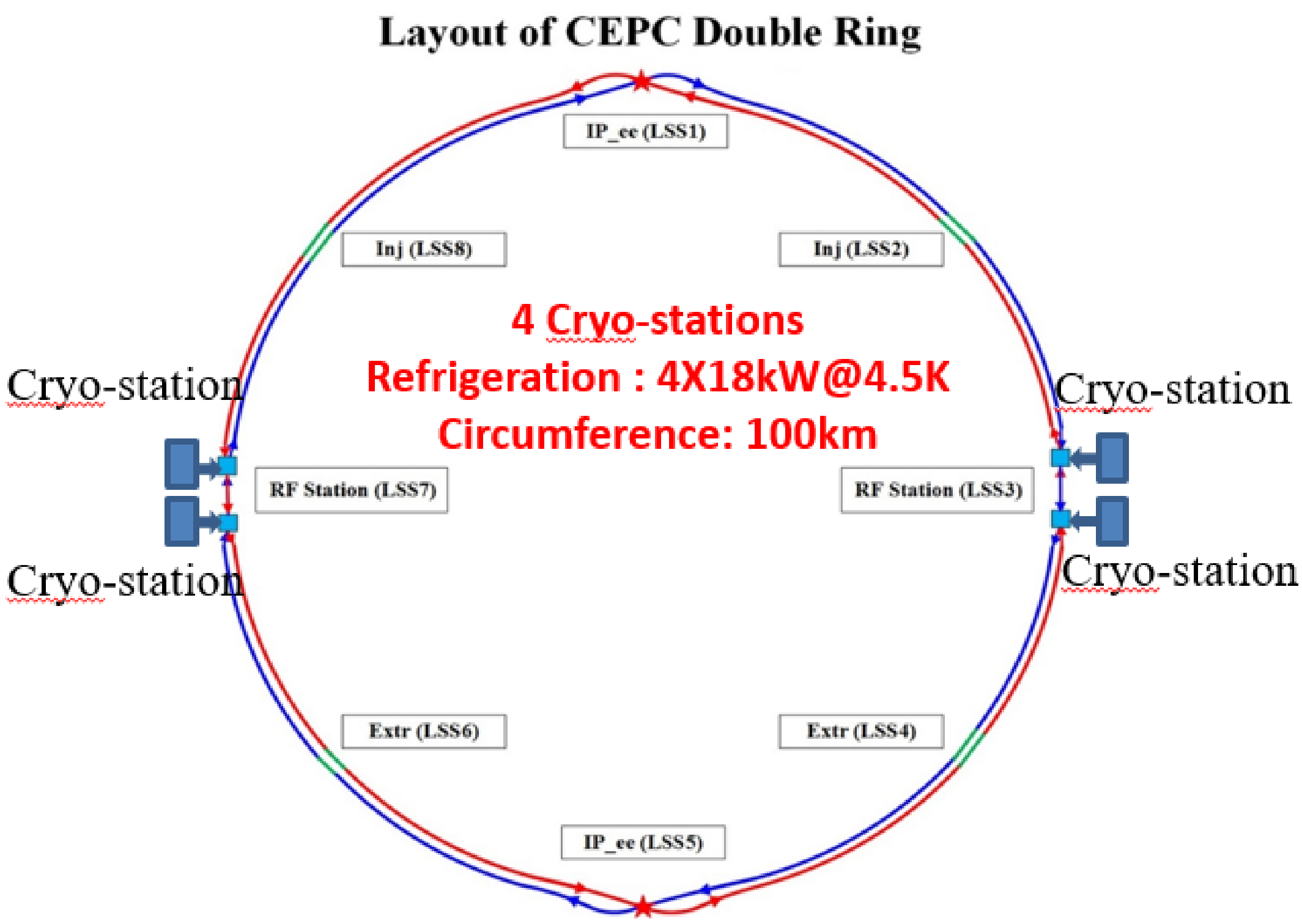

**Figure 7.1.1:** Layout of the CEPC cryogenic system

### 7.1.2 Layout of Cryo-Unit and Cryo-Strings

Each cryo-station (Fig 7.1.2) includes two strings; one string groups 3 modules from the Booster and the other groups 10 modules from the Collider.

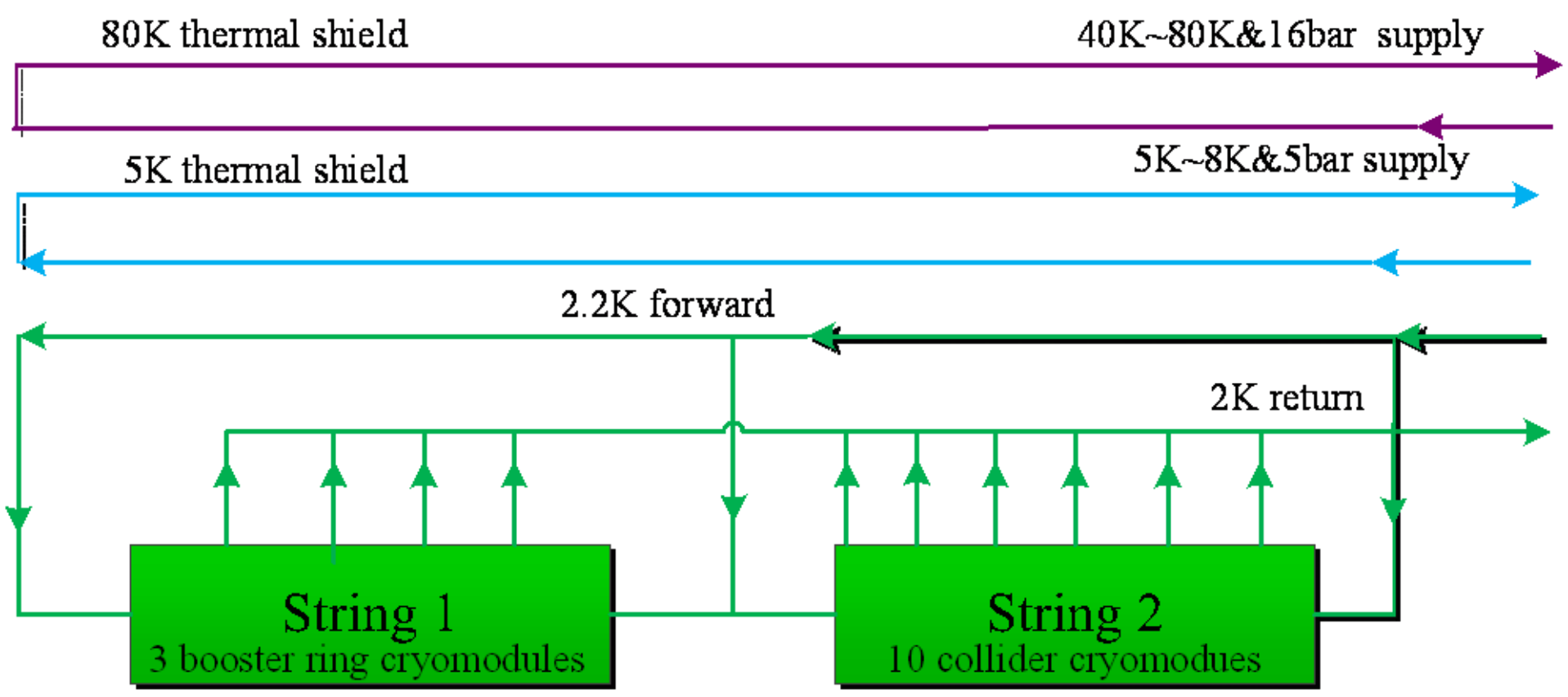

**Figure 7.1.2:** Cryogenic station

Saturated He II cools the RF cavities at 2 K. In view of the high thermodynamic cost of refrigeration at 2 K, the design of the cryogenic components aims at intercepting heat loads at higher temperatures. Hence helium-gas-cooled shields intercept both radiation and conduction at two temperatures: 40 ~80 K and 5~8 K. The 40~80 K thermal shield is the first major heat intercept, sheltering the cold mass from the bulk of heat leaks from ambient temperature. The 5~8 K shield provides lower temperature heat interception.

During operation, one-phase helium of 2.2 K and 1.2 bar is provided by the refrigerator to all cryomodules. Each cryomodule has one valve box with two valves. The JT-valve is used to expand helium to a liquid helium separator. A two-phase line (liquid-helium supply and concurrent vapor return) connects each helium vessel and connects to the major gas return header once per-module. A small diameter warm-up/cool-down line connects the bottom of the helium vessels at both ends. The cavities are immersed in baths of saturated superfluid helium, gravity filled from a 2 K two-phase header. Saturated superfluid helium flows along the two-phase header which is connected to the pumping return line and then to the refrigerator.

Details of Strings 1 (Booster) and 2 (Collider) are in Figs. 7.1.3 and 7.1.4.

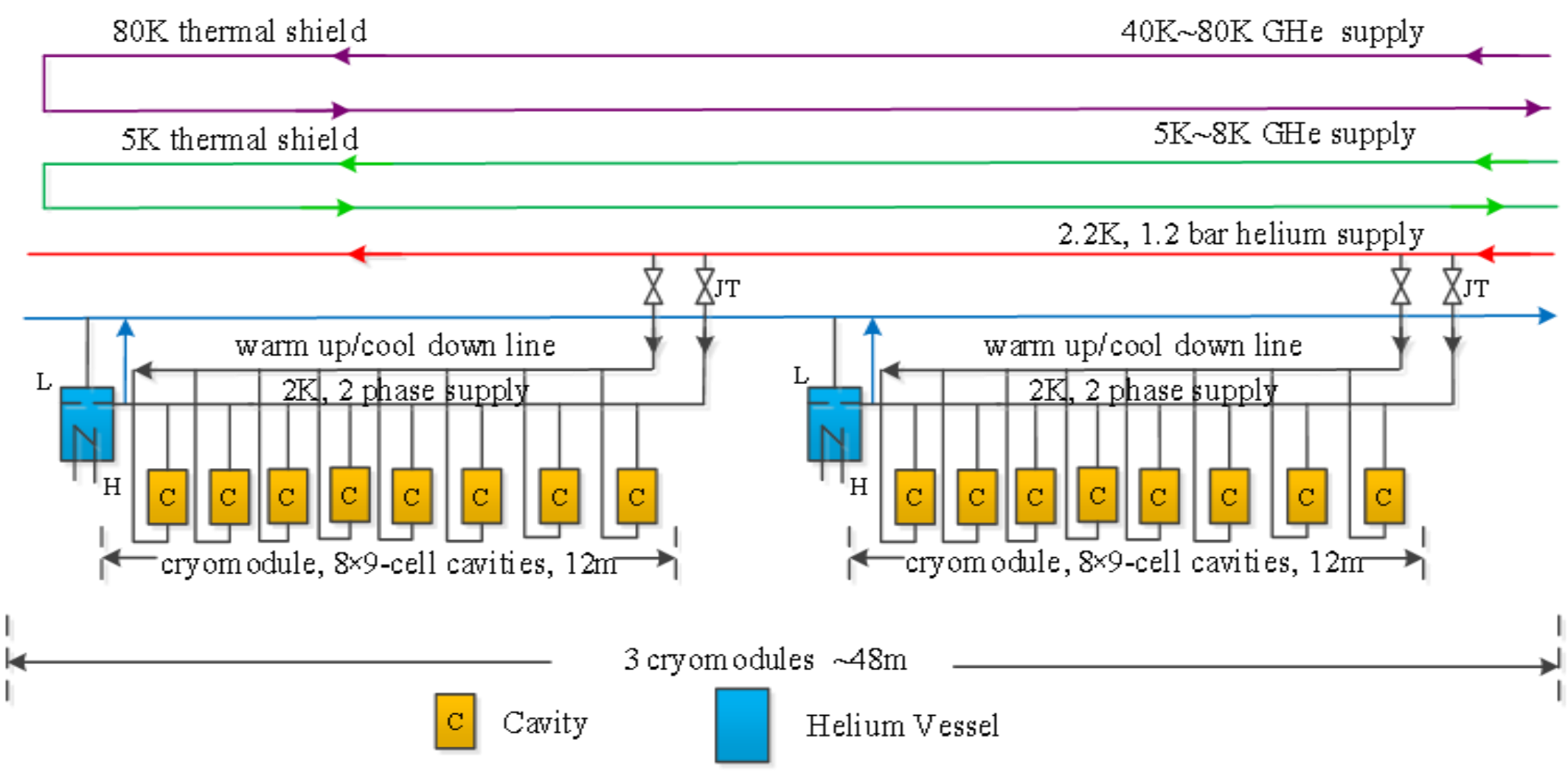

**Figure 7.1.3:** Booster cryomodules for string 1

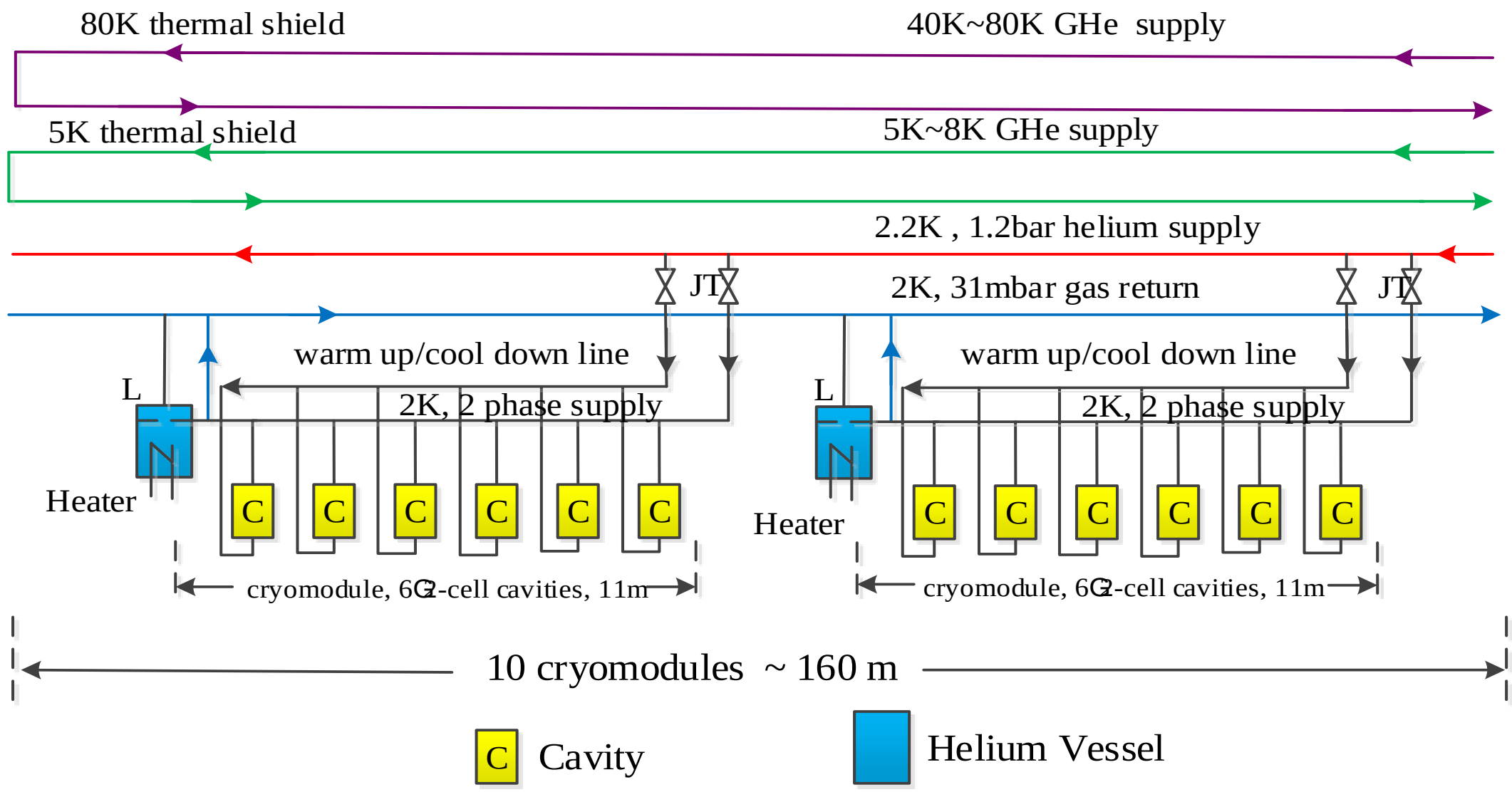

**Figure 7.1.4:** Collider cryomodules for string 2

### 7.1.3 Heat Load

The heat load is mainly from the superconducting cavities. The 1.3 GHz TESLA like 9-cell cavities with quality factor $1 \times 10^{10}$ @ 19.8 MV/m are for the Booster and the 650 MHz 2-cell cavities with quality factor $1.5 \times 10^{10}$ @19.7 MV/m are for the Collider. Table 7.1.1 lists parameters, as well as the module dynamic heat load from the cavity walls.

**Table 7.1.1:** Parameters of the Booster and Collider cavities

| 100km RF parameters | Collider | | | Booster | | |
|---|---|---|---|---|---|---|
| | H | W | Z | H | W | Z |
| Frequency(MHz) | 650 | 650 | 650 | 1300 | 1300 | 1300 |
| Cavity operating voltage (MV) | 9.04 | 4.35 | 1.67 | 19.17 | 11.72 | 12.19 |
| Duty factor | CW | CW | CW | 0.043 | 0.029 | 0.079 |
| HOM power/cavity (KW) | 0.57 | 0.75 | 1.94 | 0.0002 | 0.0007 | 0.0041 |
| Number of cells per cavity | 2 | 2 | 2 | 9 | 9 | 9 |
| Total number of cavities | 240 | 216 | 120 | 96 | 64 | 32 |
| Total number of modules | 40 | 36 | 20 | 12 | 8 | 4 |
| Eacc (MV/m) | 19.66 | 9.46 | 3.62 | 19.77 | 8.81 | 8.64 |
| R/Q | 213 | 213 | 213 | 1036 | 1036 | 1036 |
| Q0 | 1.50E+10 | 1.50E+10 | 1.50E+10 | 1.00E+10 | 1.00E+10 | 1.00E+10 |
| Operation temperature (K) | 2 | 2 | 2 | 2 | 2 | 2 |
| Cavity dynamic heat load@ 2 K (W/cavity) | 25.59 | 5.93 | 0.87 | 1.75 | 0.23 | 0.61 |
| Module dynamic heat load@ 2 K (W/module) | 153.52 | 35.57 | 5.22 | 13.98 | 1.87 | 4.91 |

As shown in the Table 7.1.1, the module dynamic heat load with Higgs mode is the largest in the Collider and Booster. So the cryogenic system will be designed with Higgs mode.

Table 7.1.2 summarizes the static and dynamic heat during Higgs mode at different temperature levels. This amounts to a total equivalent entropic capacity of 47.53 kW at 4.5 K.

**Table 7.1.2:** Heat loads for SC cavities

| Higgs Mode | Unit | Collider | | | Booster | | |
|---|---|---|---|---|---|---|---|
| | | 40-80K | 5-8K | 2K | 40-80K | 5-8K | 2K |
| Predicted static heat load per cryomodule | W | 300 | 60 | 12 | 140 | 20 | 3 |
| Cavity dynamic heat load per cryomodule | W | 0 | 0 | 153.59 | 0 | 0 | 13.98 |
| HOM dynamic heat load per cryomodule | W | 20 | 12 | 2 | 2 | 1 | 1 |
| Input coupler dynamic heat load per cryomodule | W | 60 | 40 | 6 | 40 | 3 | 0.4 |
| Module dynamic heat load | W | 80 | 52 | 161.59 | 42 | 4 | 15.38 |
| Connection boxes | W | 50 | 10 | 10 | 50 | 10 | 10 |
| Cryomodule number | | 40 | | | 12 | | |
| Total heat load | kW | 17.20 | 4.88 | 7.34 | 2.78 | 0.41 | 0.34 |
| Total predicted mass flow | g/s | 82.42 | 152.26 | 346.58 | 13.34 | 12.73 | 16.07 |
| Overall net cryogenic capacity multiplier | | 1.54 | 1.54 | 1.54 | 1.54 | 1.54 | 1.54 |
| 4.5K equiv. heat load with multiplier | kW | 1.99 | 6.80 | 36.18 | 0.32 | 0.57 | 1.68 |
| Total 4.5K equiv. heat load with multiplier | kW | 44.96 | | | 2.57 | | |
| Total 4.5K equiv. heat load of booster and collider | kW | 47.53 | | | | | |
| Installed Power | MW | 9.84 | | | 0.56 | | |
| | | 10.4 | | | | | |

The figures in table 7.1.2 include an “overall net cryogenic capacity multiplier” in general use in the cryogenic community for estimating heat loads. This factor includes a margin for plant regulation, and buffers for transient operating conditions, performance decreases during operation and for general design risks. This multiplier parameter is from the ILC Design report [1].

In the ILC design, the real COP at 40~80 K, 5~8 K and 2 K are 16.4, 197.9 and 700.2 respectively. The estimated total installed power for both Booster and Collider is 10.4 MW.

### 7.1.4 Refrigerator

The heat loads shown in Table 7.1.2 require the helium refrigerator plants to have a total capacity over 47.53 kW at 4.5 K. Four individual refrigerators will be employed. Including an operating margin, the cryogenic plant capacities are 18 kW at 4.5 K for each cryogenic station. The total cryogenic capacities are equivalent to 72 kW at 4.5 K.

Many aspects must be taken into account during refrigerator design, including cost, reliability, efficiency, maintenance, appearance, flexibility and convenience of use. The initial capital cost of the cryogenic system as well as the high energy costs of its operation over the life of the facility represent a significant fraction of the total project budget, so reducing these costs has been the primary focus of our design. Reliability is also a major concern, as the experimental schedule is intolerant of unscheduled down time.

The refrigerator main components include a compressor station with oil removal system, vacuum pumps and the cold box which is vacuum insulated and houses the aluminum plate-fin heat exchangers and several stages of turbo-expanders.

The fundamental cooling process expanding compressed helium gas to do work against low-temperature expansion engines, then recycling the lower pressure exhaust gas through a series of heat exchangers and subsequent compression is a variant of the Carnot process that has been in use for many decades.

There are five pressure levels in the cryoplant: 20 bara, 4 bara, 1.05 bara, 0.4 bara and 3 kPa. These are obtained with the high pressure screw compressor group, middle pressure screw compressor group, warm compressors and cold compressors. There are 6 temperature levels in the system with 5-class turbine expansions and 11 heat exchangers. At the 40 K and 5 K temperature levels helium flows are directed to the thermal shields of the cryomodules. The corresponding return flows are fed back to the refrigerator at suitable temperature levels. Inside the refrigerator cold-box the helium is purified of residual air, neon and hydrogen by switchable absorbers at the 80 K and 20 K temperature levels. Fig. 7.1.5 is the refrigerator flow diagram.

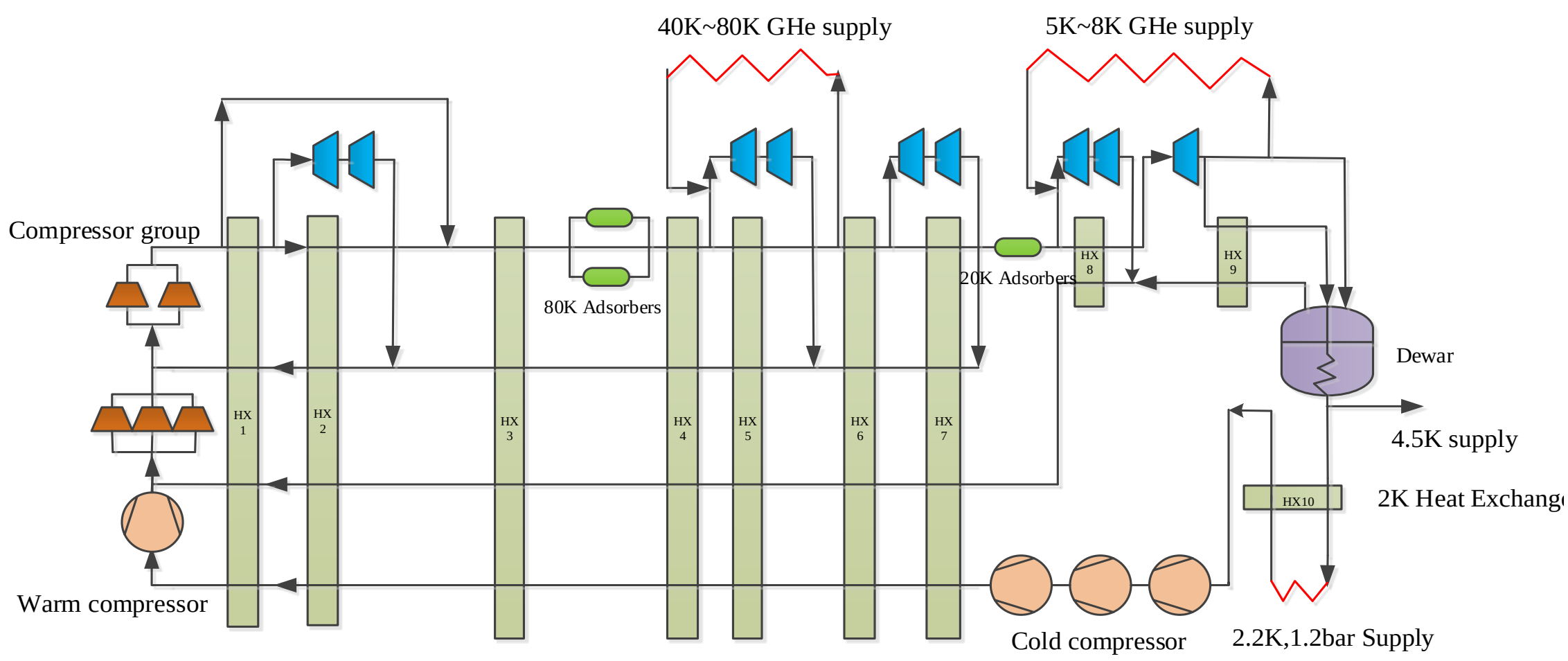

**Figure 7.1.5:** 2 K refrigerator flow diagram

The cryoplant will supply 4.5 K and 2.2 K helium to the cryomodules. At each cryomodule the helium goes through a phase separator and a 2K counter flow heat exchanger to recover the cooling power, then expanded to 31 mbar via a JT-valve, resulting in liquid He II at 2K. The low pressure helium vapor from the 2K saturated baths surrounding the cavities returns to the refrigerator through the gas return pipe. The vapor is pumped away and returned to the cryoplant.

There are two options for such a pumping system. One relies solely on cold compressors; the other employs a set of cold compressors followed by a final stage of warm compression. After superheating in the counter flow heat exchanger, the gas is compressed in the multiple-stage cold compressors to a pressure in the 0.5 to 0.9 bar range. This stream is separately warmed up to ambient in exchangers and goes back to the warm compressors. The choice of a warm vacuum compressor makes it easier to adjust for the heat load variations. This approach, which CERN uses in the LHC [3], also allows for an easier restart of the 2 K system after a system stoppage.

### 7.1.5 **Infrastructure**

The 2 K cryogenic system consists of oil lubricated screw compressors, a liquefied-helium storage vessel, a 2 K refrigerator cold box, cryomodules, a helium-gas pumping system and high-performance transfer lines. The cryogenic station is located near the RF station. The cooling power required at each RF station will be produced by a refrigerator with a capacity of 18 kW at 4.5 K, installed at four cryogenic stations, and distributed to the adjacent superconducting cavities [2, 3].

For reasons of simplicity, reliability and maintenance, the number of active cryogenic components distributed around the ring is minimized and the equipment locations chosen following these principles:

1) Equipment is installed as much as possible above ground to avoid excavation. Normal temperature equipment will be installed at ground level.
2) To decrease heat loss, low-temperature equipment will be installed nearby the cryomodules [4].

Equipment at ground level includes the electric substation, the warm compressor station, helium storage tanks, cooling towers and helium purification. Underground are the cold-boxes, cold compressor, 2 K cryomodules, cryogenic transfer lines and distribution valve boxes. Fig. 7.1.6 shows the general architecture of the cryogenic system. Fig. 7.1.7 shows the overall schematic. .

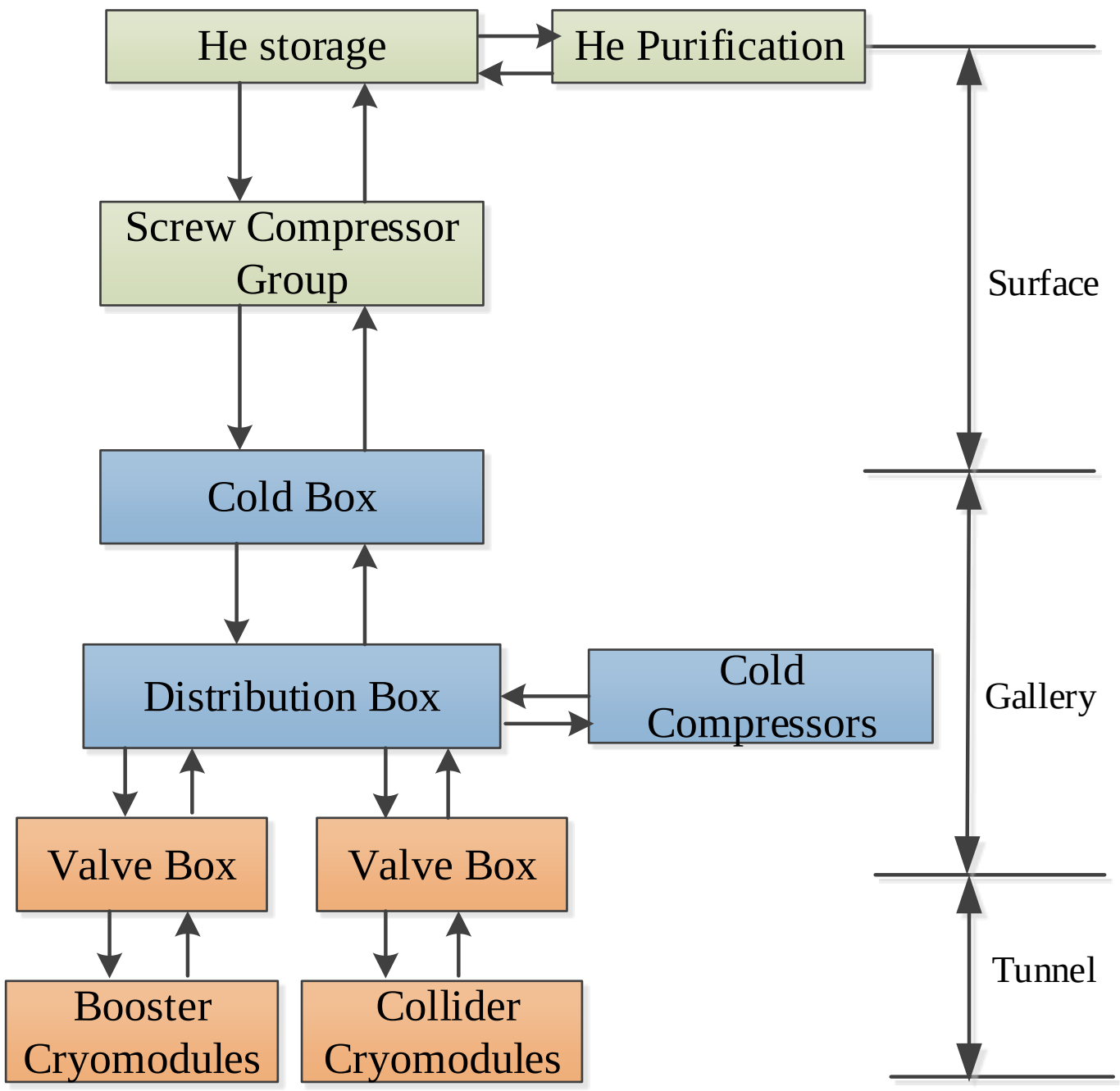

**Figure 7.1.6:** General architecture of the cryogenic system

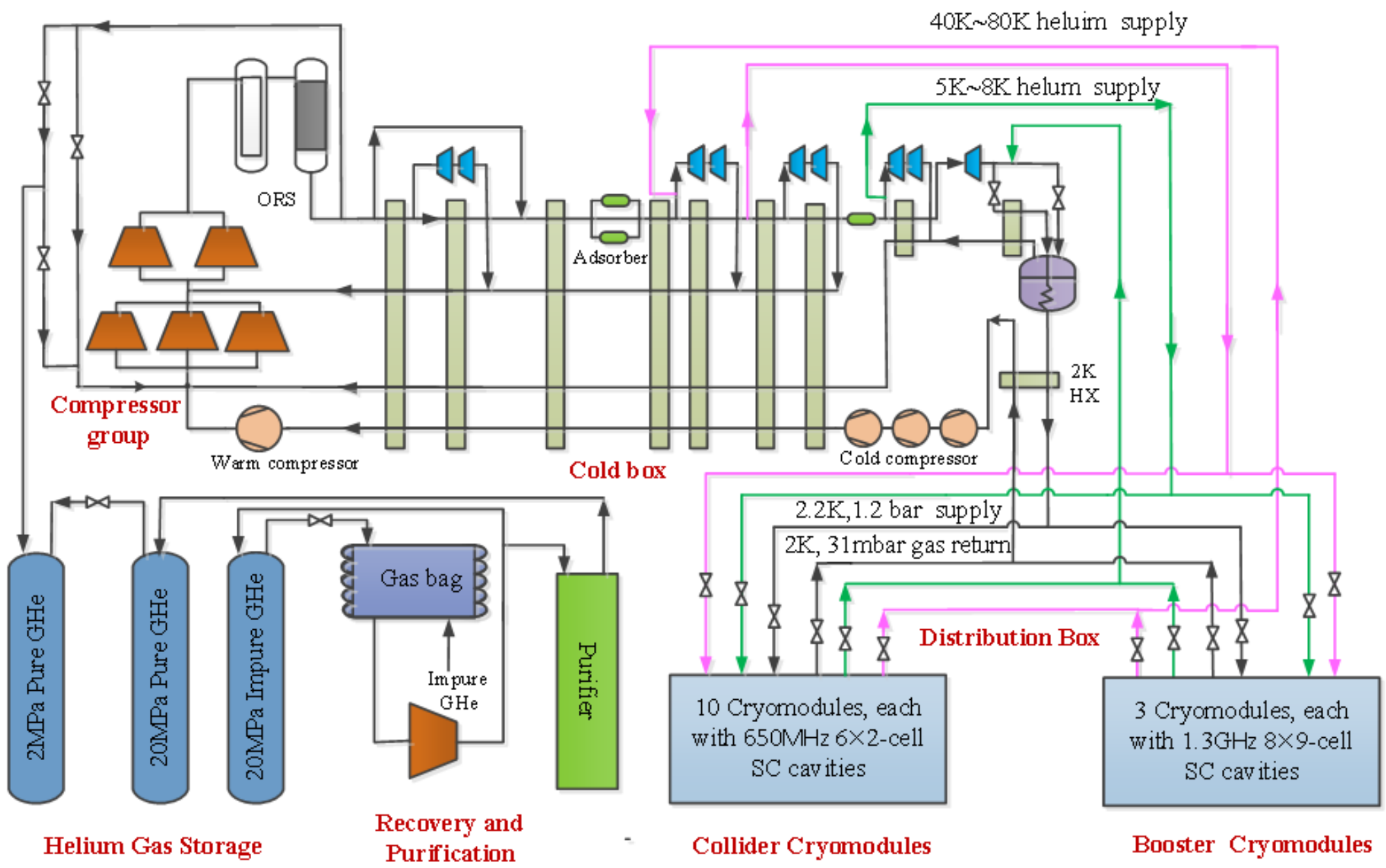

**Figure 7.1.7:** Cryogenic system schematic

### 7.1.6 Helium Inventory

Most of the helium inventory is liquid which bathes the RF cavities and is roughly 70% of the whole system. The volumes of one 1.3 GHz and one 650 MHz module is about 320 liters and 346 liters, respectively. The total liquid helium volume in the system is 17,680 liters.

Accounting for the liquid in the Dewar and in the transfer lines, and using the 70% factor mentioned above, the liquid volume in the system is about 25,257 liters, or about 3,679 kg [5].

Assuming that all the helium is returned to the helium tanks after machine shutdown, the inventory will be $2.3 \times 10^4$ m$^3$. To safely operate the cryogenic system, a factor of 60% is added, so $3.8 \times 10^4$ m$^3$ is required helium inventory system. The total helium inventory of the whole machine is about 6,131 kg.

### 7.1.7 Cryogenics for IR Superconducting Magnets

In CEPC Interaction Region (IR), there are 4 QD0 magnets, 4 QF1 magnets, 4 anti-solenoids, 32 sexupole magnets and some correctors. All the superconducting magnets work at the temperature of 4.5K. The IR SC magnet cryostat is grouped by 1 QD0 magnet, 1 QF1 magnet, 1 anti-solenoid and several correctors. There are 4 IR SC magnet cryostats. Since the distance between two sextupole magnets is far, each sextupole magnet has a cryostat. There are 32 IR SC sextupole magnet cryostats. Every cryostat has a valve box. The 4.5 equivalent heat load with multiplier is 5.73 kW.

There are 2 cryo-stations, each cryo-station has a 3kW@4.5K refrigerator. A roughly estimated heat load of SC magnets is shown in table 7.1.3. Every cryo-station includes one IR SC magnet cryostat and eight IR SC sextupole magnet cryostats. The corresponding installed power is 1.8 MW with the COP of 300 W/1W.

**Table 7.1.3:** Heat loads for SC magnets

| Name | Unit | No. | Heat load for each | Heat load |
|---|---|---|---|---|
| IR SC sextupole magnet | W | 32 | 10 | 320 |
| Valve Box of IR SC sextupole magnet | W | 32 | 20 | 640 |
| Current lead of IR SC sextupole magnet | g/s | 32 | 0.1 | 3.2 |
| IR SC magnet | W | 4 | 30 | 120 |
| Valve Box of IR SC magnet | W | 4 | 30 | 120 |
| Current lead of IR SC magnet | g/s | 4 | 0.5 | 2 |
| Main distribution valve box | W | 2 | 50 | 100 |
| Cryogenic transfer-line | W | 4000 | 0.5 | 2000 |
| Total equiv. heat load @4.5K | W | / | / | 3820 |
| Total equiv. heat load @4.5K with multiplier 1.5 | W | / | / | 5730 |
| Cooling capacity of refrigerator@4.5K | W | 2 | 3000 | 6000 |
| Installed power (COP(300W/1W)) | MW | / | / | 1.8 |

## 7.2 Survey and Alignment

### 7.2.1 Overview

The main goal is to provide a relatively smooth path between adjacent components, based on the absolute position accuracy provided by the beam orbit. The survey and alignment work will be carried out in two stages: (1) geometric straightening during the construction period; (2) beam based alignment. During construction, all of the components will be aligned to their designated geometric positions to a defined precision. This provides an initial orbit for commissioning. During operation the beam will be the benchmark and will be used for final alignment.

Because of the large circumference and the large number of components, error accumulation is an important issue as shown in Table 7.2.1.1. Survey and alignment during construction and installation will require a great deal of time. Therefore it is incumbent to improve efficiency by developing efficient accelerator survey and alignment technologies.

**Table 7.2.1.1:** Error accumulation in the accelerator tunnel control network

| **Name** | **Value/mm** |
|---|---|
| Error in one station of the laser tracker | 0.1 |
| Interval between two stations | 7000 |
| Circumference of RCS tunnel in CSNS | 228000 |
| Error accumulation of RCS in CSNS | 3.3 |
| Error accumulation in the 100km tunnel | 1428 |

Alignment work can be divided as follows:

1. Accelerator alignment control network. The main purpose of the alignment control network is to unify all the accelerator components into an alignment frame and control error accumulation. The control network provides a unified reference frame and a baseline.
   The alignment control network will be a three-level hierarchy: primary control network, backbone control network and tunnel control network. Control network precision is controlled by a variety of measurement methods that are used to check each other and maximize accuracy
2. Accelerator component alignment includes component fiducialization, pre-alignment, and tunnel installation alignment.
3. Alignment during running periods. During operations, uneven settlement of the foundation will cause a position change in key components such as magnets.

These will have a prominent effect on the beam orbit, especially if the geology of the site is complicated. Long-term real-time monitoring of the settlement of the Collider foundation will be done with the HLS (Hydrostatic Levelling System).

4. Unconventional alignment technology for the special accelerator components. The vision instrument is composed of the fuselage, horizontal dial, vertical dial, lens, image sensor, distance sensor, horizontal sensor, horizontal adjustment knob, vertical adjustment knob, display screen, and flash. It integrates an angle measuring system, laser ranging system, digital photogrammetry system and is characterized by non-contact, high precision and high efficiency.

## 7.2.2 Control Network

### 7.2.2.1 *The Primary Control Network*

The primary control network, shown in Fig. 7.2.2.1, is used to control the mutual locations between the Linac, transport lines, Booster and Collider with3mm relative accuracy. It provides high-precision location reference data for the tunnel control network.

This network will also be used to monitor long-term deformations. The network consists of 4 external permanent points and 12 device area permanent points. The location of the latter are two in the Linac, two in the transport lines and 8 in the Collider.

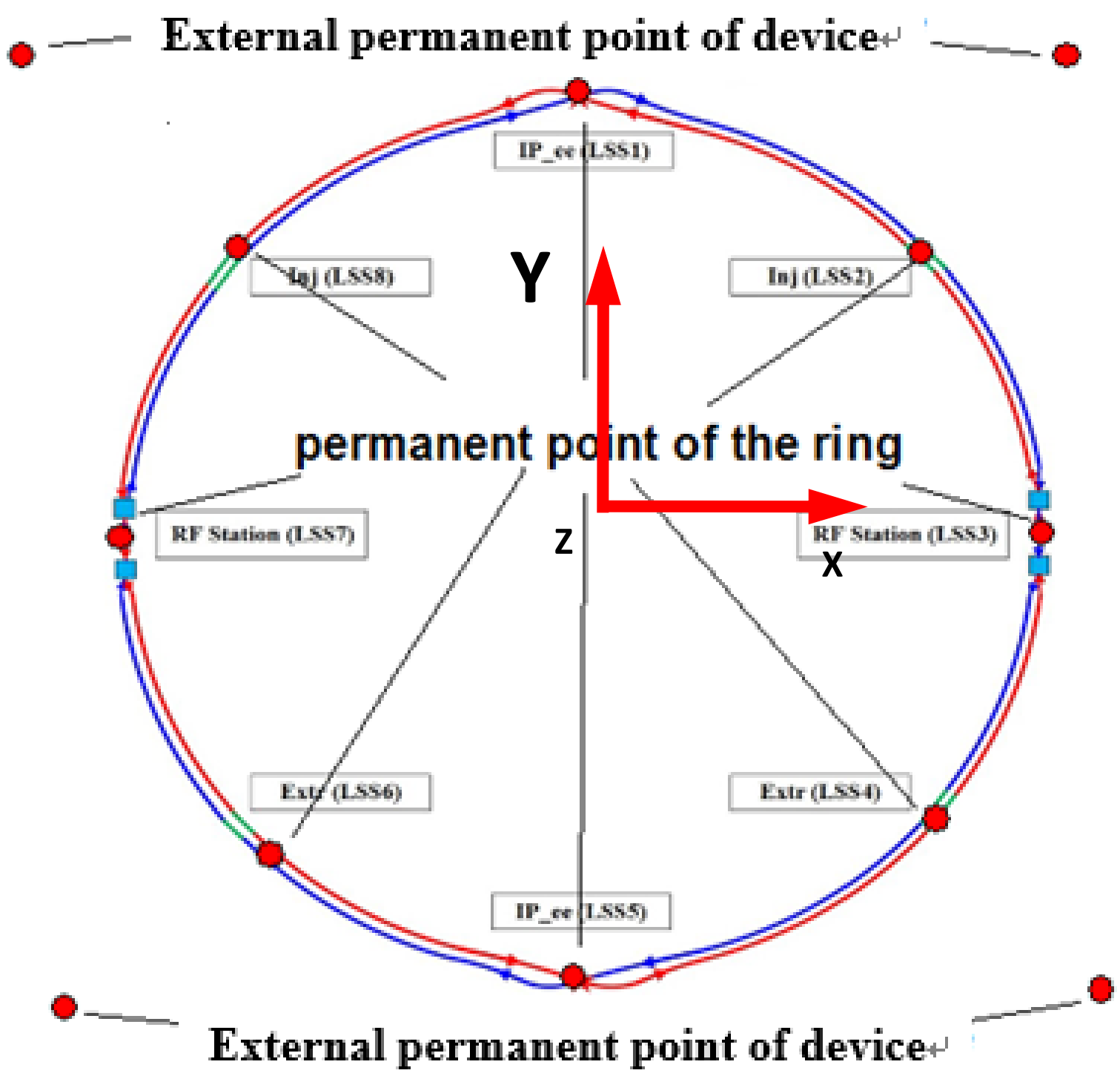

**Figure 7.2.2.1:** Primary control network layout

The external points are at ground level supported by steel bars and poured concrete. These points need to be located on stable ground with minimal disturbance from road traffic. A photograph of an example is in Fig. 7.2.2.2. Also here should be no high-voltage cables within 200 meters, there should be no strong reflection of satellite signals in the vicinity and no large water area to avoid multi-path effects and there should be no obstruction within 10 degrees above the point [1].

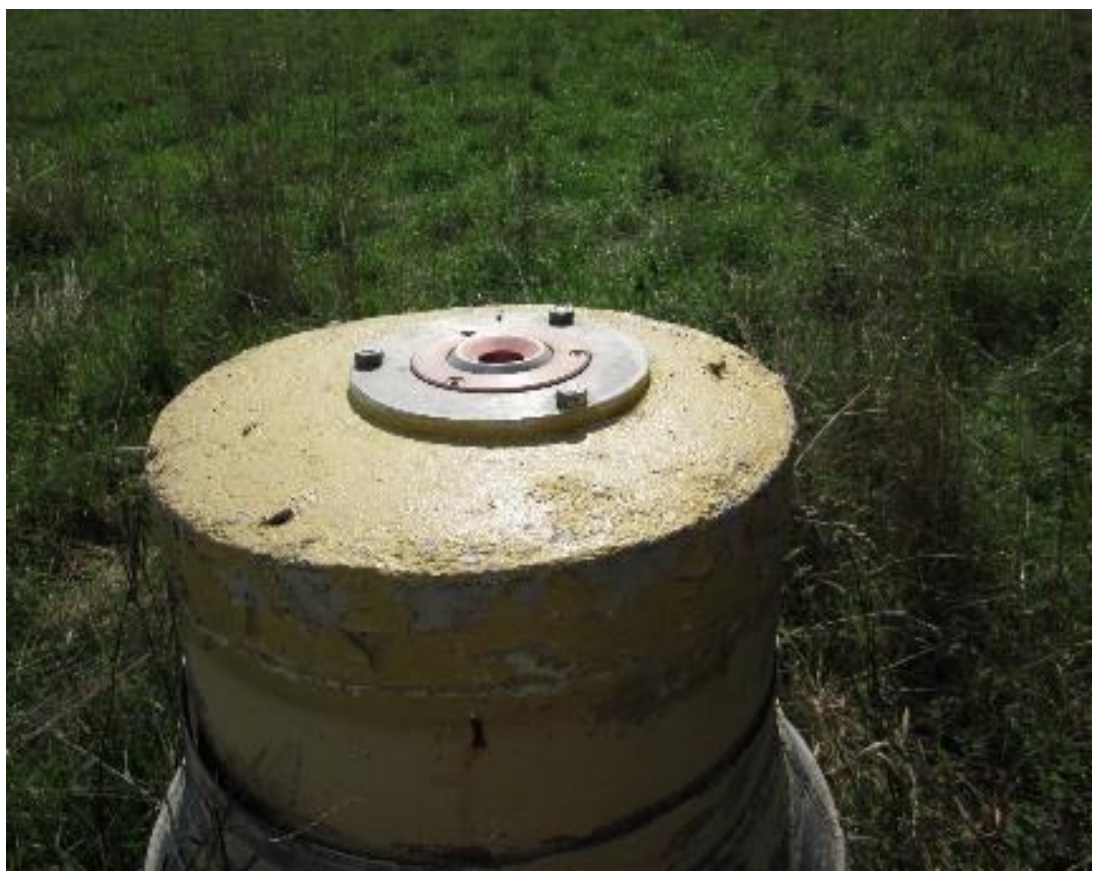

**Figure 7.2.2.2:** External permanent point mark

The 12device area permanent points belong both to the primary control network and the tunnel control network. They are in the tunnel about 100 meters underground. Above each permanent point there is an inter-visible hole leading up to the surface, so through these holes the location of these permanent points can be measured from ground [2].

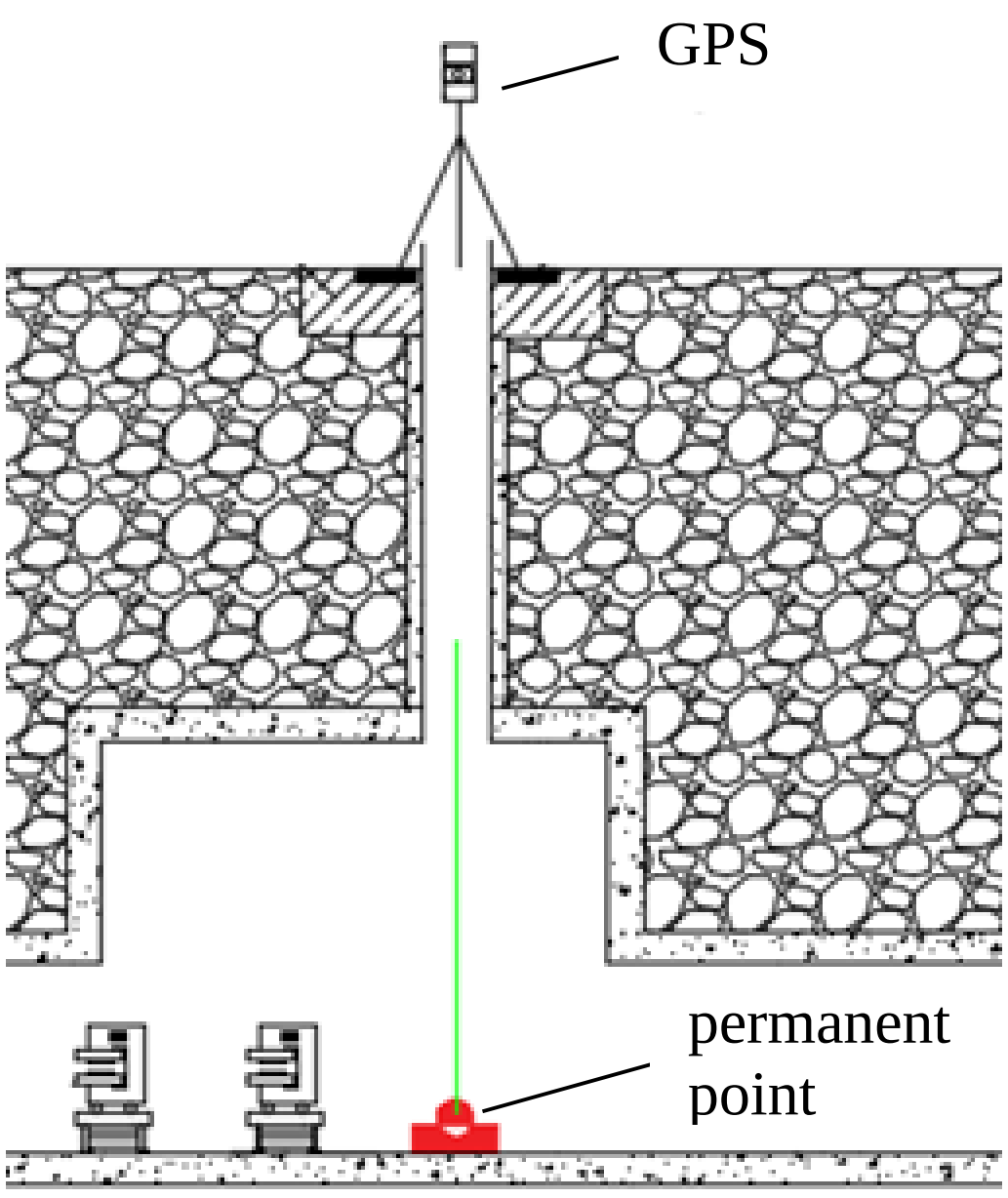

**Figure 7.2.2.3:** Measure permanent point from ground

During construction, the primary control network will be measured twice. The first GNSS measurement will be carried out after construction of the inter-visible holes and the permanent points are completed. The second measurement will be carried out before installation of accelerator components in the tunnel.

Survey observations are done with the LEICA GS10 receiver and the AR20 choke antenna shown in Fig. 7.2.2.4.

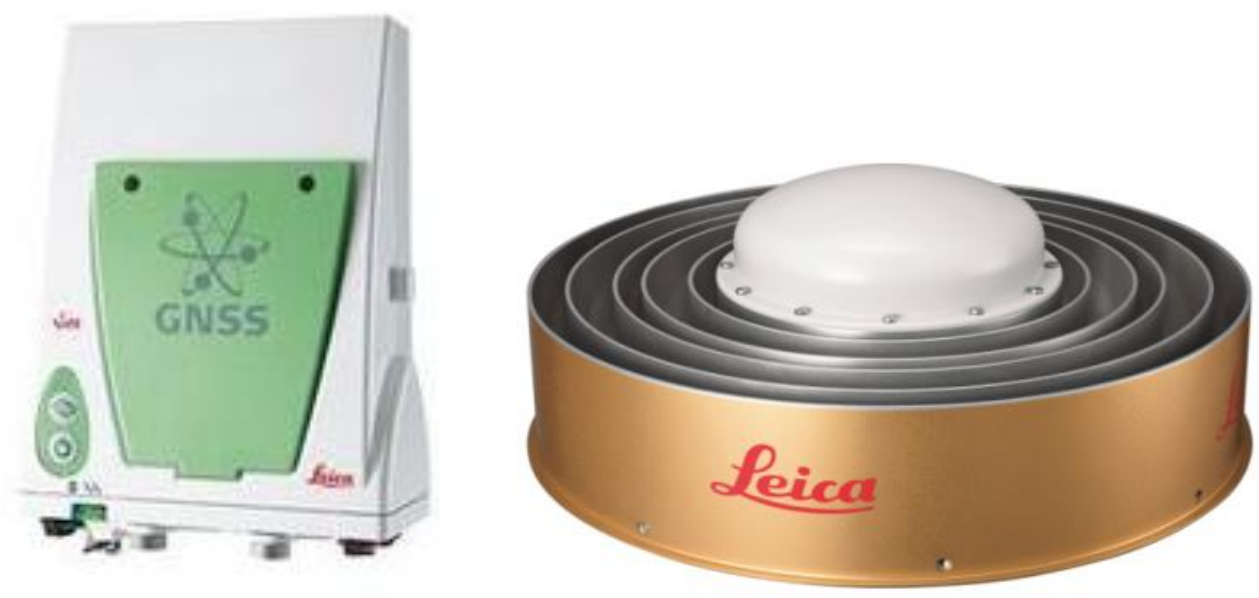

**Figure 7.2.2.4:** GS10 receiver and AR20 choke antenna

The accuracy of a projected point is 0.5mm; the accuracy of the height measurement between the tunnel permanent point and the projected point is 0.2mm. The final horizontal accuracy of a permanent point is about 10mm.

The level observation of the primary control network is made with a Leica DNA03 electronic level measurement system. Observation requirements [3] will be second-class leveling. Back and forth observation distances are about 50 meters. The observation leveling route is divided into the Linac leveling route, the Linac to ring leveling route and the ring leveling route.

The ring leveling route is divided into 4 sections and each quadrant measurement is a closed route. Fig. 7.2.2.5 shows the leveling route of the fourth quadrant R07P-R08P-R01P-R08P-R07P.

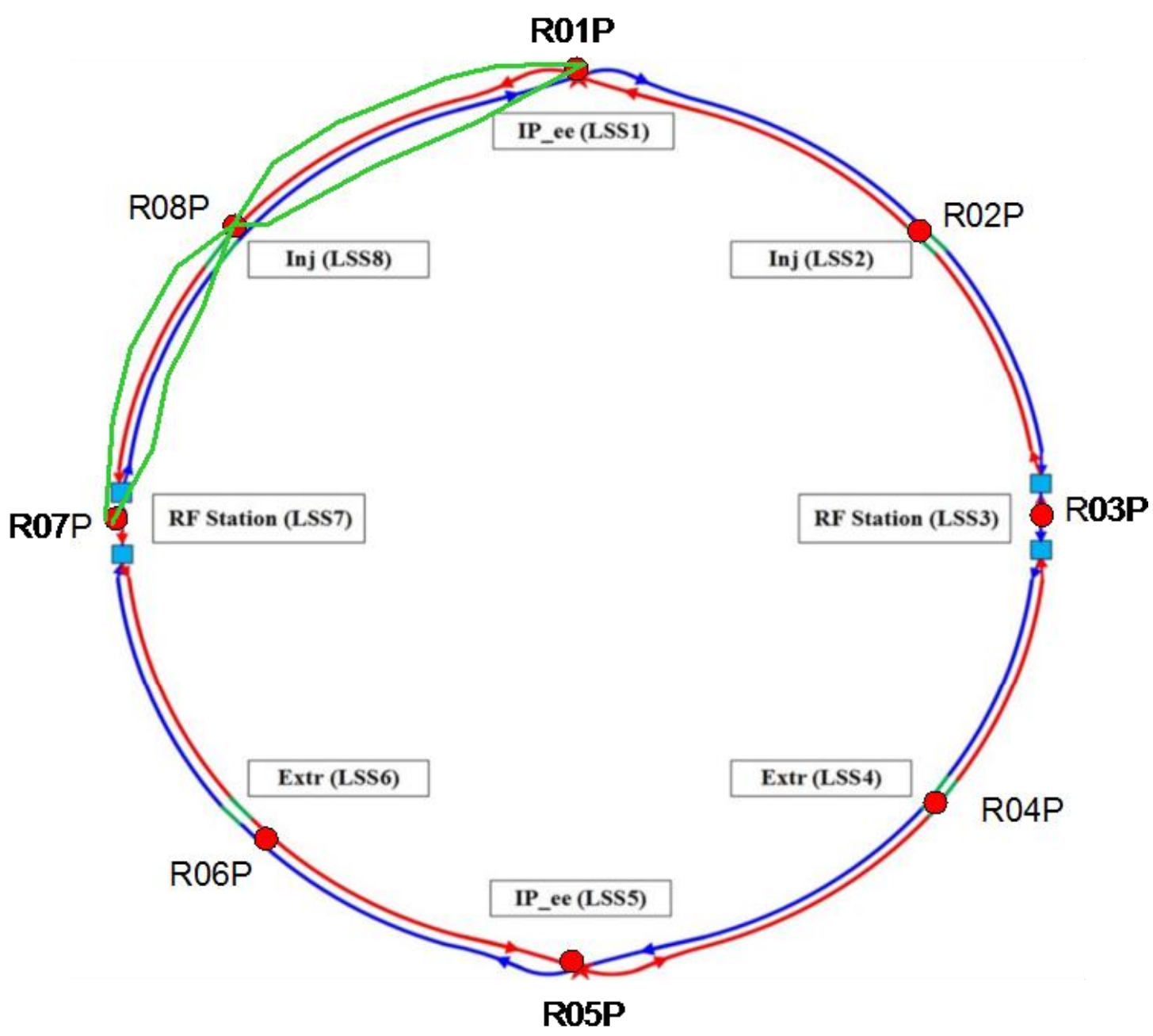

**Figure 7.2.2.5:** Leveling route of one quadrant

The final level measurement accuracy of a permanent point is about7 mm.

### 7.2.2.2 *Tunnel Backbone Control Network*

The tunnel backbone network connects the primary network and the tunnel network, Utilizing the large span and high accuracy measurement result to constrain the tunnel network error accumulation, it is a good extension of the primary network. Backbone network survey is carried out using total stations and high-precision gyroscopes.

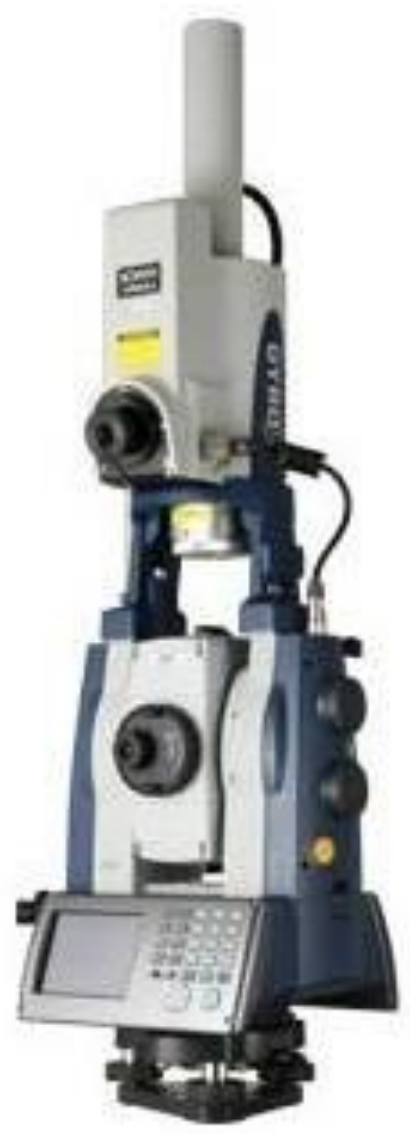

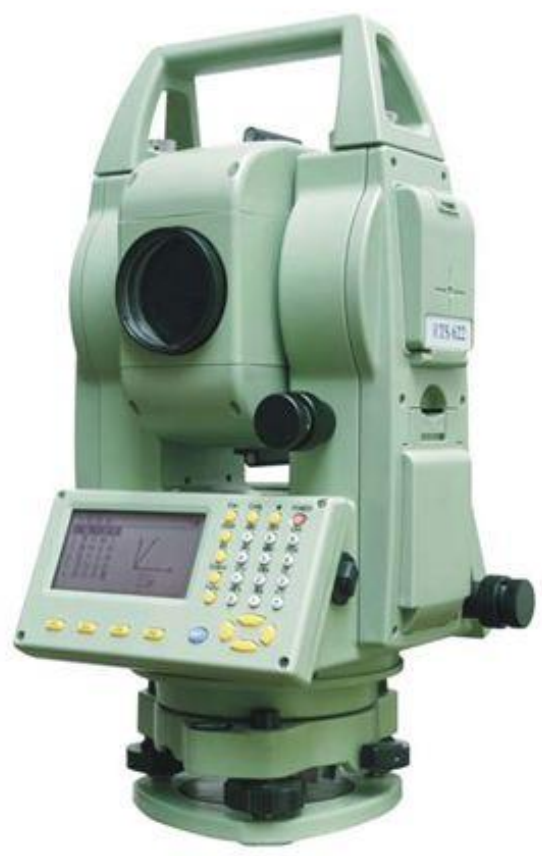

**Figure 7.2.2.6:** High precision gyroscope **Figure7.2.2.7:** Total Station

The tunnel backbone network is composed of many sections. In each section there are two control points, one near the inner wall and the other near the outer wall. The control points of adjacent sections must be inter-visible and the distances between adjacent sections should be as long as possible, as shown in Fig. 7.2.2.8.

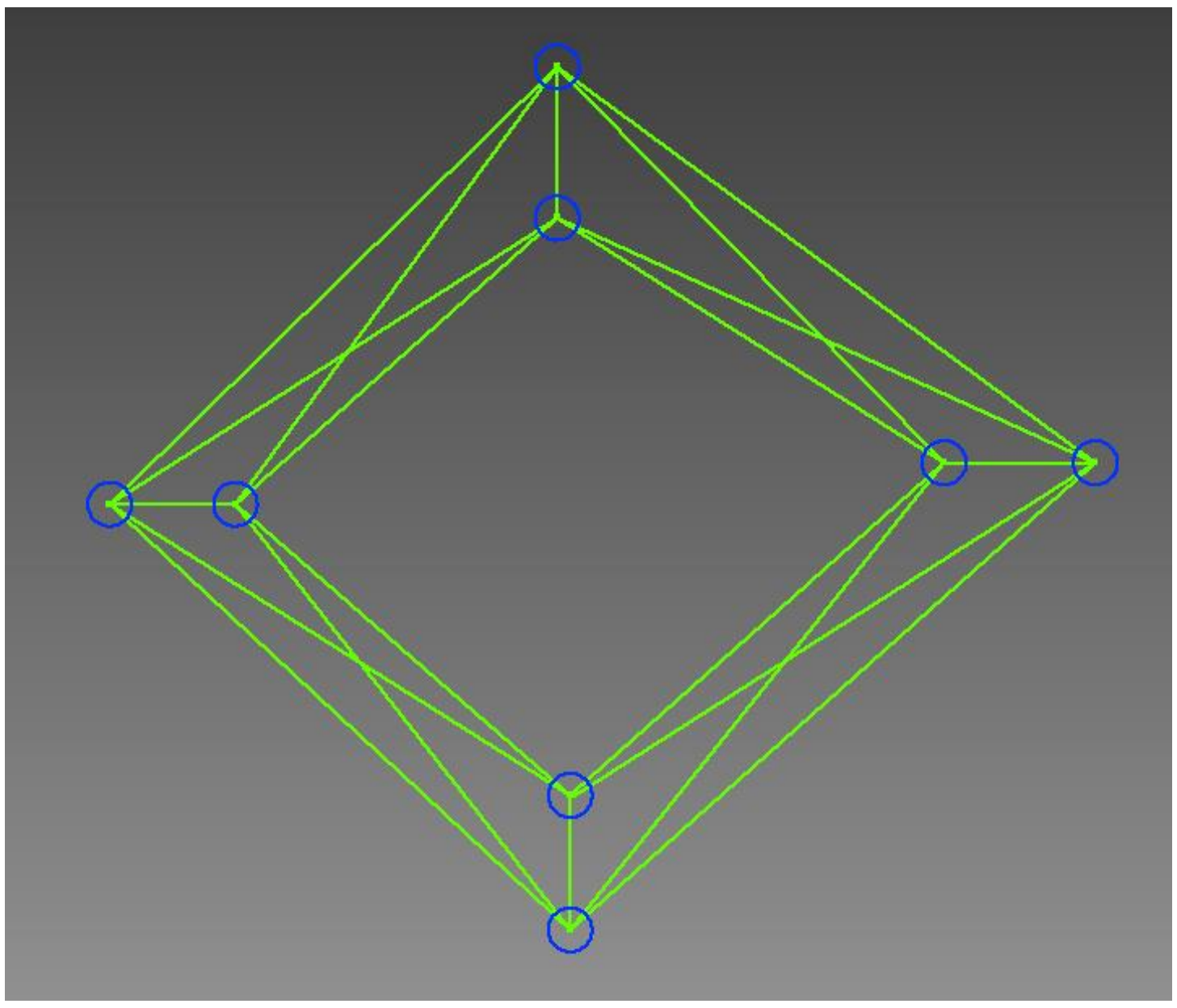

**Figure 7.2.2.8:** The closed double-traverse control network

### 7.2.2.3 *Tunnel Control Network*

Tunnel control network is a coordinate reference system for accelerator component installation, adjustment and for monitoring position changes. The tunnel control network is a three-dimensional control network [4,5].

1. To improve the accuracy of component alignment, the tunnel control network should fully cover the entire component area and with a dense number of points. The layout is shown in Fig. 7.2.2.9. Points are evenly distributed at intervals of 6 meters. Each section consists of 4control points, 2 on the ground, 2 on the two sides of the tunnel wall. The 2 ground points are placed on both sides of the pedestrian passage way; points on the walls are installed 2 meters above ground.

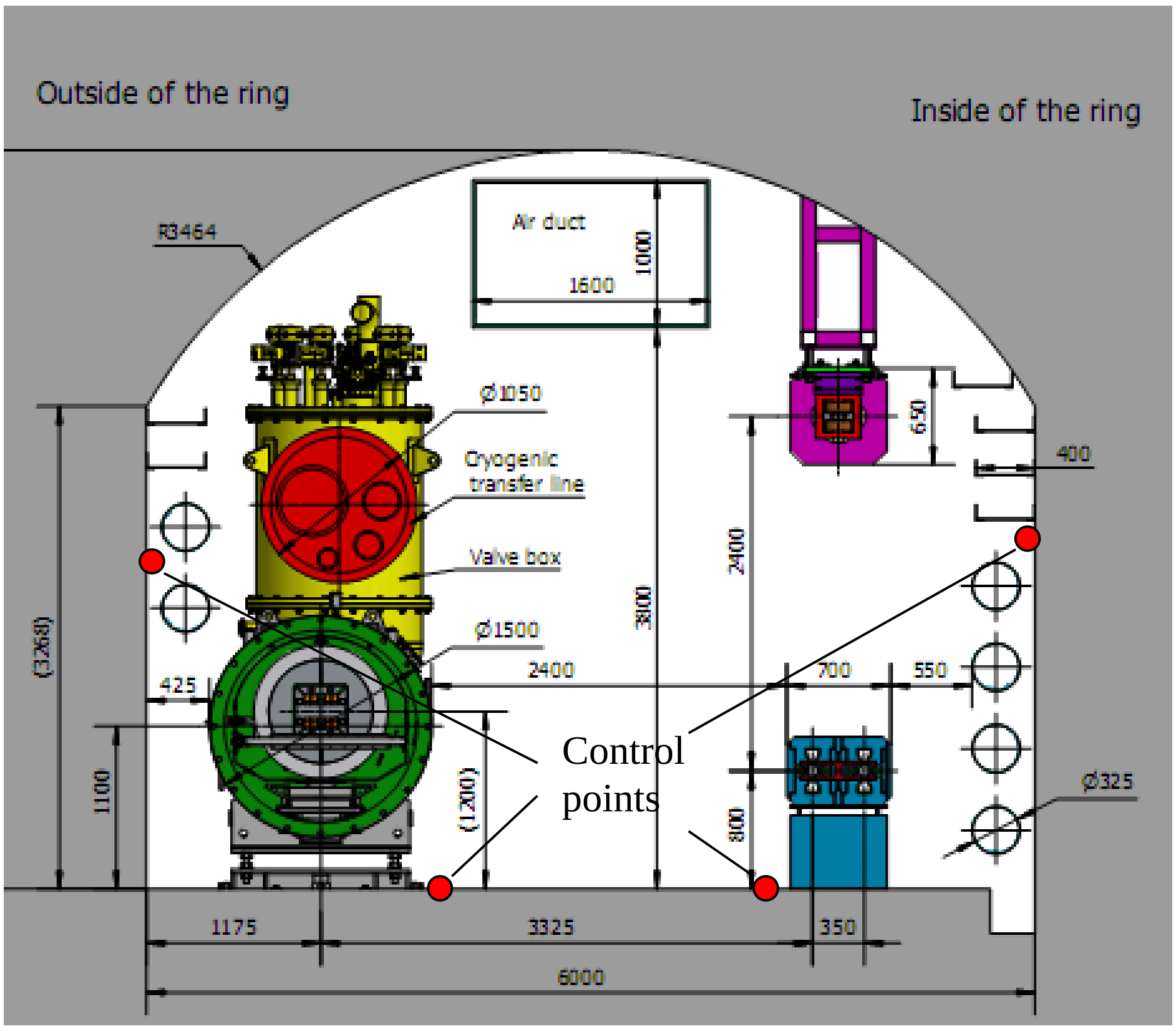

**Figure 7.2.2.9:** Layout of tunnel control network

The number of control network sections is about 16667 and the number of control points is about 66668.

The tunnel control network survey, shown diagrammatically in Fig. 7.2.2.10, is generally carried out by a laser tracker. In order to improve the measurement accuracy of the control network, there should be enough redundant measurements between adjacent measurement stations. A laser tracker station will be placed in the middle of every adjacent control network section. In each station the level will be measured to establish a level coordinate system and 6 network sections will be measured, the measurement range is about 30 meters. There will be 16,667 measurement stations for the tunnel control network, which can be measured by multiple work teams working in different measurement regions simultaneously.

The tunnel network measurement data are processed by horizontal and vertical adjustment independently. The horizontal relative accuracy is less than 0.12 mm, and the elevation relative accuracy is less than 0.1 mm.

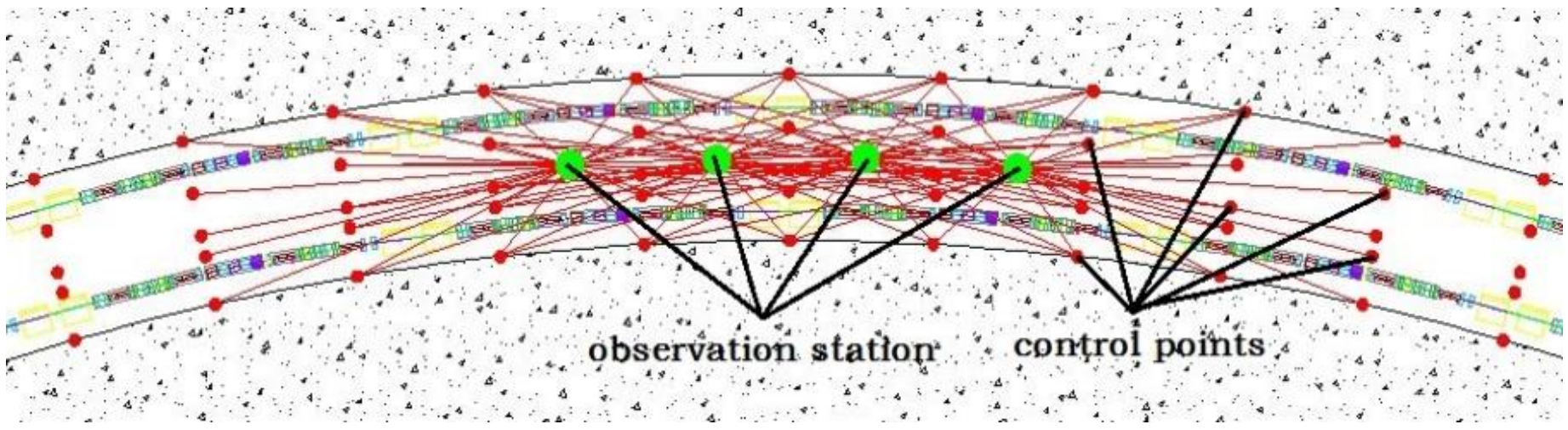

**Figure 7.2.2.10:** Tunnel control network survey

2. An automatic observation system is a new type instrument that integrates an automatic mobile platform, photogrammetry, distance, angle measurement functions and has high precision and high efficiency characteristics. We plan to use it and automate the tunnel control network survey.  This system, illustrated in Fig. 7.2.2.11uses an infrared sensing tracking system. A strong reflective stripe placed on the tunnel passage way ground guides it.

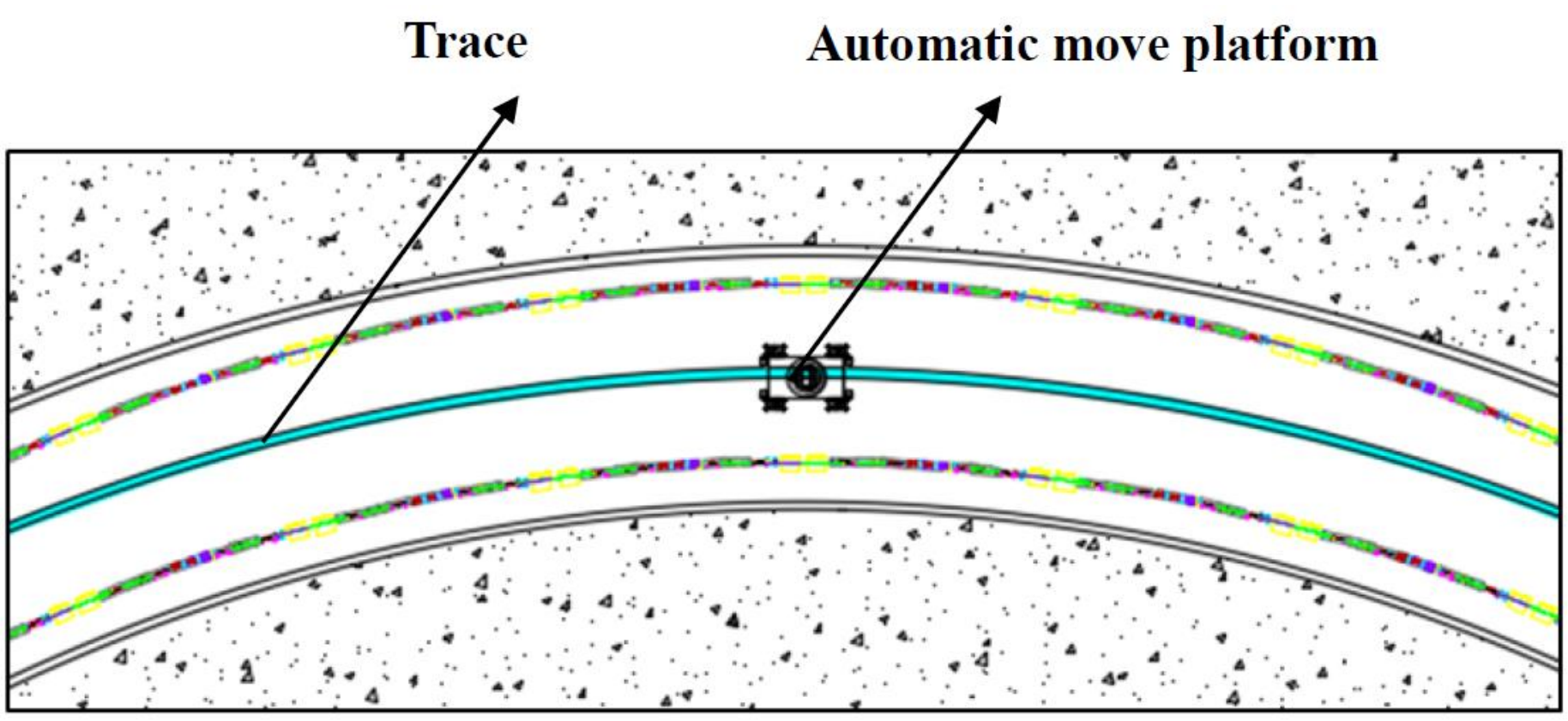

**Figure 7.2.2.11:** Layout of track

A measurement station will be set in the middle of every two adjacent sections. When the platform moves to a predetermined station location, the supporting legs will automatically hold up the platform and adjust the level, after which it begins to measure the control network.

As illustrated in Figs. 7.2.2.12 and 7.2.2.13, tunnel observation is divided into horizontal and vertical plane observations. To provide sufficient redundancy and overlap to improve measurement accuracy multiple photos are taken in each plane and at each station. [6-9]. Point accuracy will reach 0.2 mm.

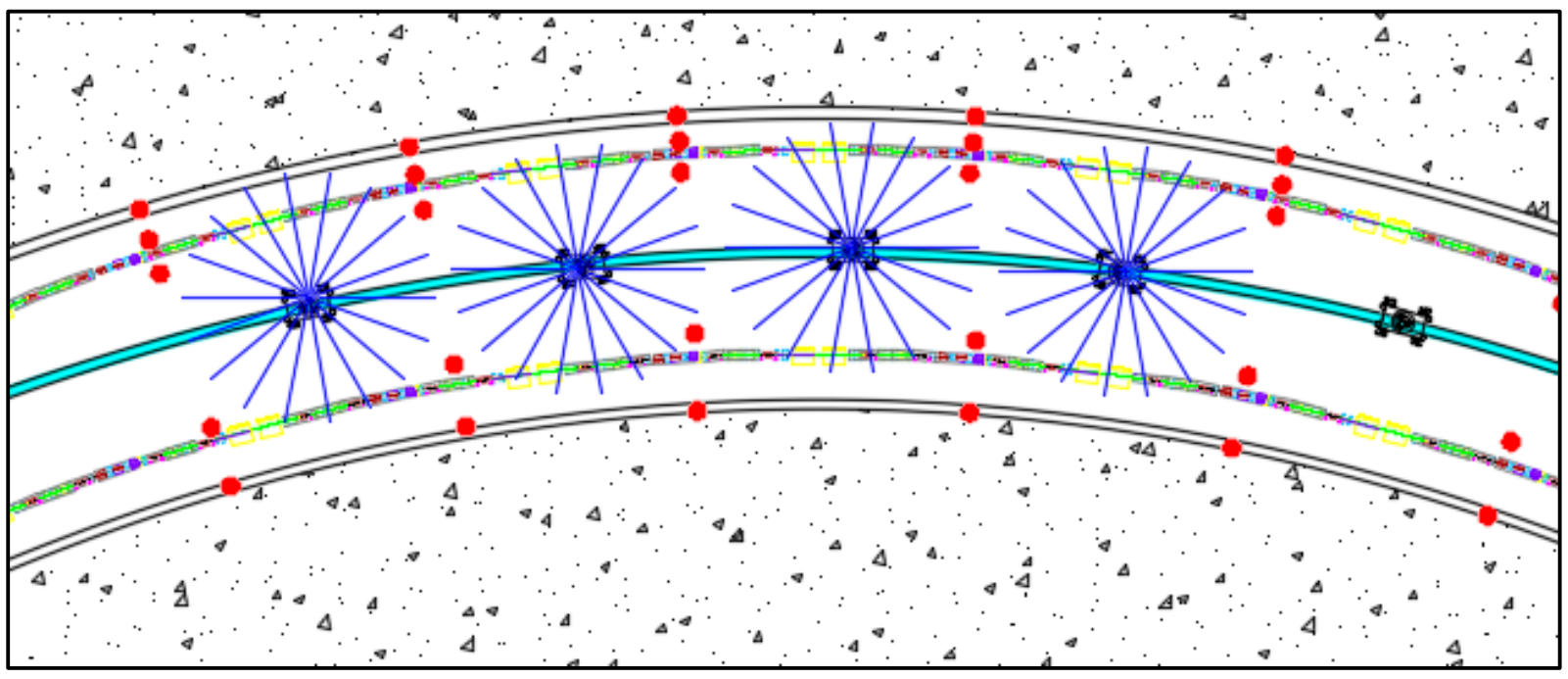

**Figure 7.2.2.12:** Horizontal plane observation

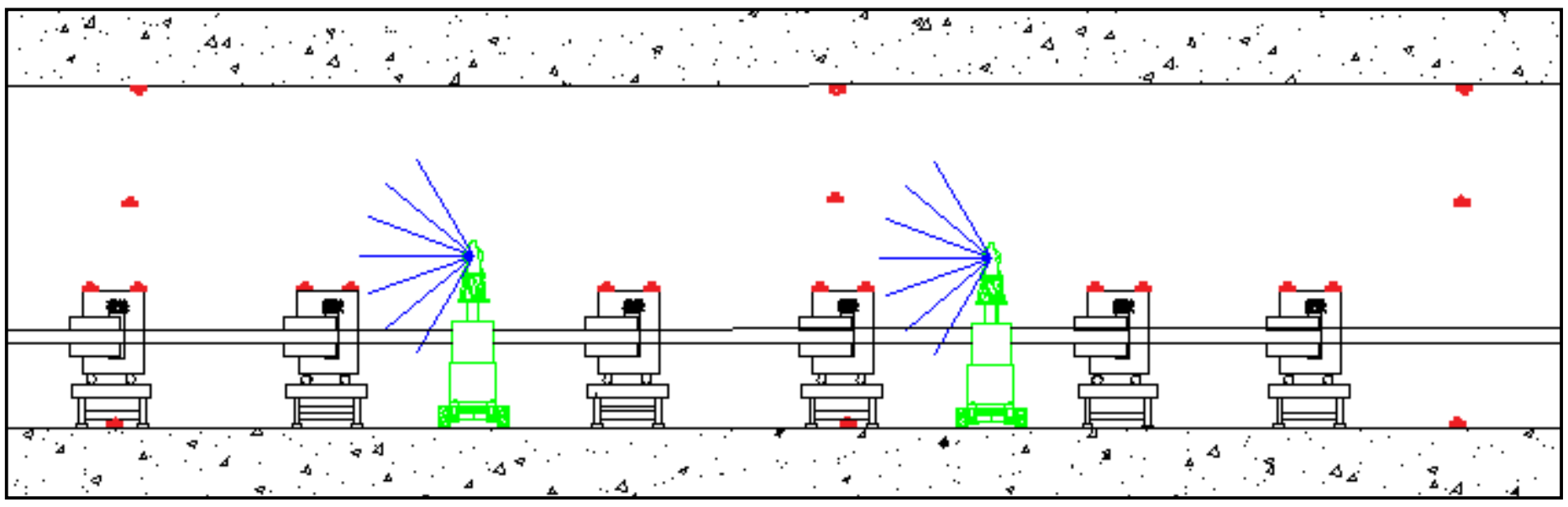

**Figure 7.2.2.13:** Vertical plane observation

### 7.2.3 Component Fiducialization

All the components requiring precision alignment need to be done fiducialization in advance. The process will relate the mechanical center (beam center) to its fiducial points. Then the fiducial point coordinates can be transferred to the global coordinate system and used for component alignment in the tunnel.

#### 7.2.3.1 *Component Fiducialization by Laser Tracker*

Using a laser tracker to measure the fiducial planes and fiducial points of the component, we can establish a coordinate system for the component. In this coordinate system we can obtain the position relation between the fiducial points and the beam center. Most of the components which need fiducialization are magnets. Below we use a quadrupole as an example.

1. Establish magnet coordinate system:
   (1) The fiducial points on the top surface can be marked F1-F4;
   (2) Using a rotating magnetic measurement coil to detect the magnetic center of the quadruple, a laser tracker can measure the rotating center of the coil to get a line named the Z-line;
   (3) Measure the beam enter and exit end planes of the quadrupole to create the E-plane and F-plane;
   (4) Insert a fiducial plate on the lower two poles of the quadrupole and measure it to get the T-plane;
   (5) Establish two intersection points P-A and P-B of the Z-line with the E- and F-planes;
   (6) Create the midpoint O of points P-A and P-B;

(7) Establish the magnet coordinate system: O is the origin; z-line is the Z axis (first direction); the normal of T-plane is the Y axis (second direction).
(8) F1-F4 in the magnet coordinate system are the fiducial coordinates of the magnet.

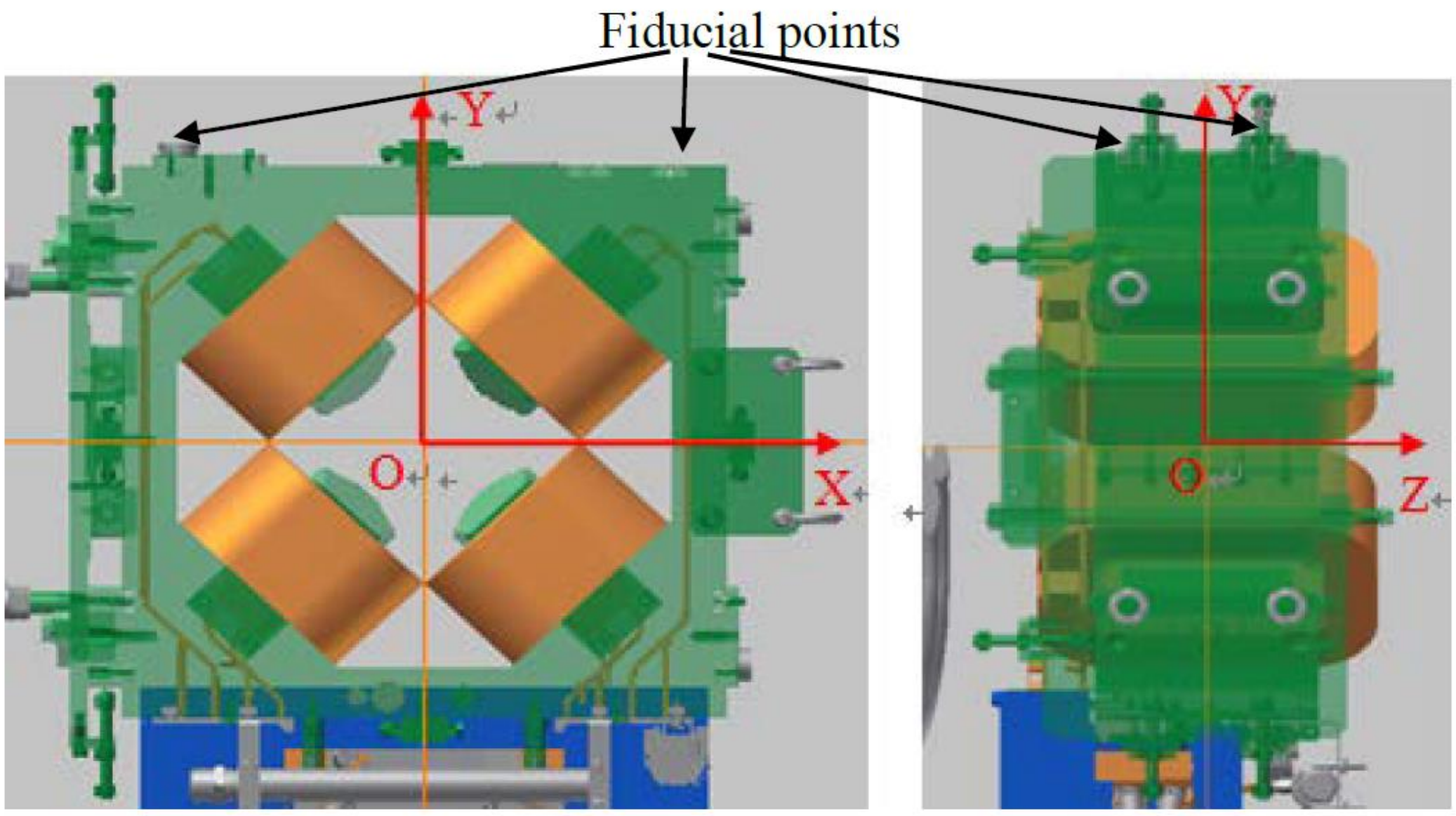

**Figure 7.2.3.1:** Fiducialization coordinate system of a quadrupole magnet

2. Fiducialization certification requires measuring at least twice and comparing the results. Furthermore, other tools such as rulers, levels, transit squares are also needed to examine relative positions. Errors for the example of a quadrupole magnet are summarized in Table 7.2.3.1.

**Table7.2.3.1:** Fiducialization errors of a quadrupole magnet

| Component | Transverse (mm) | Elevation (mm) | Longitudinal (mm) | Pitching angle (mard) | Swinging angle (mard) | Rolling angle (mard) |
|---|---|---|---|---|---|---|
| Quadrupole magnet | 0.04 | 0.04 | 0.01 | 0.25 | 0. 2 | 0.2 |

#### 7.2.3.2 *Component Fiducialization Using the Vision Instrument*

The laser tracker work described above will be augmented by photogrammetry to better cope with the large number of components. Measuring multiple points in each picture can improve fiducialization efficiency [10]. This photogrammetry based on vision is very suitable for measuring large number of components and more efficient than a laser tracker. One example of its use is for the yoke of the detector, which is described in the detector volume of the CEPC CDR.

### 7.2.4 Alignment of Components in the Tunnel

The tunnel control network final coordinates, obtained after adjustment calculations will be used as a position reference for components alignment. Components tunnel alignment will include three phases, they are initial installation alignment, over all survey and smooth alignment.

### 7.2.4.1 *Installation Alignment of Components in Tunnel*

As illustrated in Fig. 7.2.4.1 the tunnel control network and laser trackers will be used to align components. First, set a laser tracker station near the component to be aligned. Then measure the control network and through a best-fit we can get the position of the laser tracker relative to the control network. Since the adjustment direction of a component is generally not parallel to the axis of the global coordinate system, it is more convenient to convert the coordinate of the component from the global coordinate system to the component coordinate system. Using the laser tracker to measure the fiducial points on the component, we can obtain the offset and adjust it to the required tolerance of ±0.05mm.

In order to carry out component alignment to the desired precision there should be no obstruction to block the laser beam, nor air currents to distort it. There should be no large heat sources and light sources near the instrument and also no large vibrations during the adjusting process.

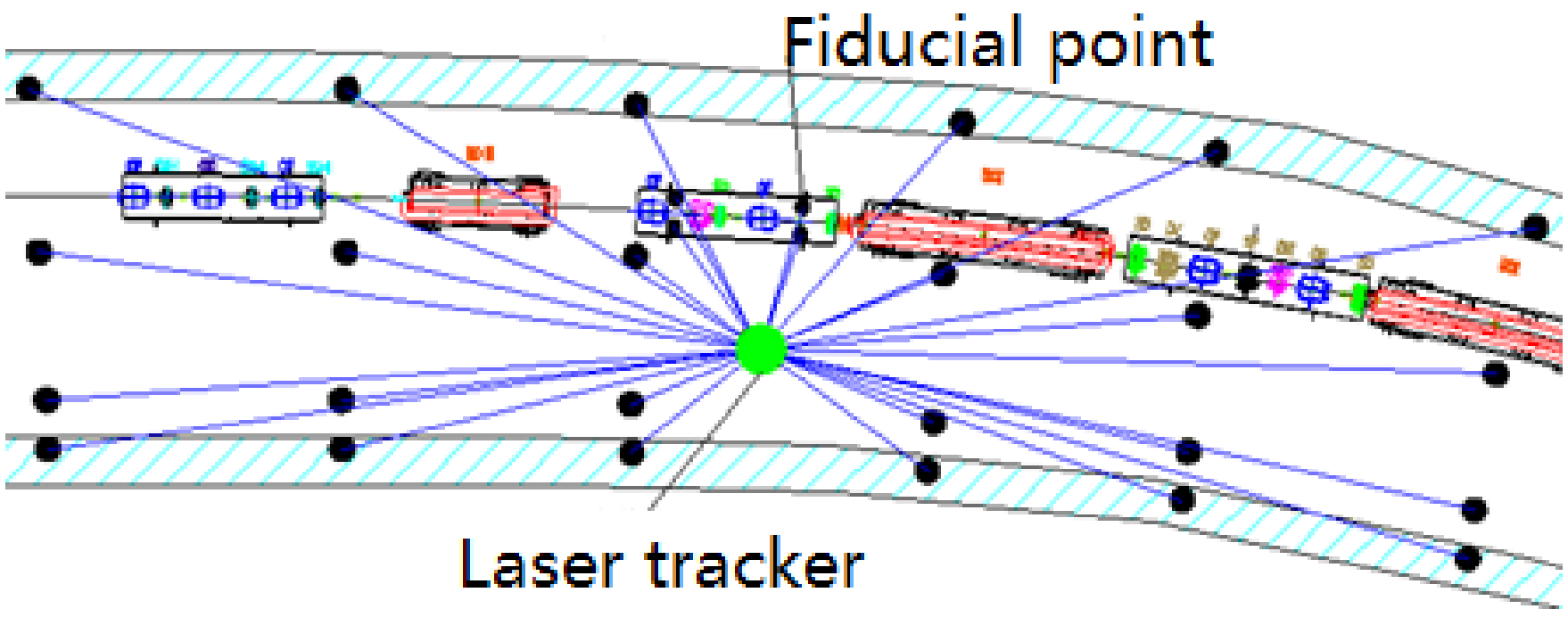

**Figure 7.2.4.1:** Components tunnel alignment scheme

### 7.2.4.2 *Overall Survey*

After the initial alignment following installation an overall survey is required to judge whether all the components are within the required tolerance and whether the beam orbit is smooth. We will set the laser tracker stations near every BPM to insure all the components can be measured. Using single station round-trip measurement scheme, the overall survey will be completed by the transfer station method. The best-fit accuracy of adjacent stations should be less than 0.15mm.Points measured at each station include the control network points in the upstream three sections and in the downstream three sections, the fiducial points of the component, two checking points horizontally and two more vertically.

Each station should establish a level coordinate system to distinguish the horizontal coordinates and the elevation coordinates. Measurement result will be processed by horizontal and vertical adjustment independently to avoid twist.

### 7.2.4.3 *Smooth Alignment of Tunnel Components*

Based on the overall survey, we first examine the offsets between actual positions and their theoretical positions. If there are no critical errors then an orbit smoothing calculation can be performed. We can best-fit the theoretical values of the components to their actual coordinates segmentally and calculate the errors between them. After that, the errors of adjacent components will be examined. If we set a 0.05mm tolerance, the accelerator components exceeding this will be identified and adjusted to within that tolerance. This smooth alignment should be repeated once or twice to insure the final alignment accuracy meet the physics requirement.

The alignment tolerances of the components provided by the physics group are in Table 7.2.4.1 (also see Table 4.2.1.3 in Sec.4.2.1.3)

**Table 7.2.4.1:** Requirement of some components alignment accuracy

| Component | Position (mm) | | | Angle (mrad) | | |
|---|---|---|---|---|---|---|
| | Transverse $\Delta X$ | Vertical $\Delta Y$ | Longitudinal $\Delta Z$ | Pitch $\Delta\theta_X$ | Yaw $\Delta\theta_Y$ | Roll $\Delta\theta_Z$ |
| Dipole | 0.05 | 0.05 | 0.15 | 0.2 | 0.2 | 0.1 |
| Quadrupole | 0.03 | 0.03 | 0.15 | 0.2 | 0.2 | 0.1 |

The X,Y,Z directions in the table refer to a component coordinate system where X is the transverse direction, Y is the vertical direction and Z is the beam direction, $\Delta\theta_X$, $\Delta\theta_Y$, $\Delta\theta_Z$ are angles of rotation around the corresponding axis.

This requirement is too tight for the alignment to realize at present. Alignment errors can be divided into four parts according to the previous introduction. Their error estimates are in Table 7.2.4.2.

**Table 7.2.4.2:** Alignment error analysis

| Fiducialization /mm | Control network /mm | Measurement /mm | Adjustment /mm | Totall /mm |
|---|---|---|---|---|
| 0.05~0.1 | 0.12 | 0.05 | 0.05~0.1 | 0.15~0.2 |

From this analysis we can see that at present, the final alignment accuracy of a component is about 0.15~0.2 mm, which does not meet the current physics requirement. If we want to improve the accuracy we need to use higher precision instruments to carry out the component fiducialization and develop new methods to do the smooth alignment. By doing these we hope the final alignment accuracy can reach 0.1 mm. The physics group has also set 0.1 mm as the goal of tolerances in their study.

#### 7.2.4.4 *Interaction Region Alignment*

Alignment in the interaction region will be more demanding than in the arcs. Fig. 7.2.4.2 will guide you through the following text. The superconducting quadruples (SCQ) on both sides of the detector should be precisely aligned relative to the detector. Laser trackers will initially be used to do a rough alignment. Setting a laser tracker station on each side of the detector and using the tunnel control network as a reference, we can roughly align the SCQs on both sides. Then, using a laser collimator system, a refined alignment will be used to attain high precision.

We will install two laser collimators on one side of an IP and two targets on the other side. Using laser trackers and levels the laser collimators and targets will be aligned to the nominal positions and then the visual lines between the targets and the collimators will become a positional reference system and we can use this system to align all the components between them to the beam orbit. The accuracy goal is 0.1mm for the relative positional accuracy of SCQ and other adjacent components.

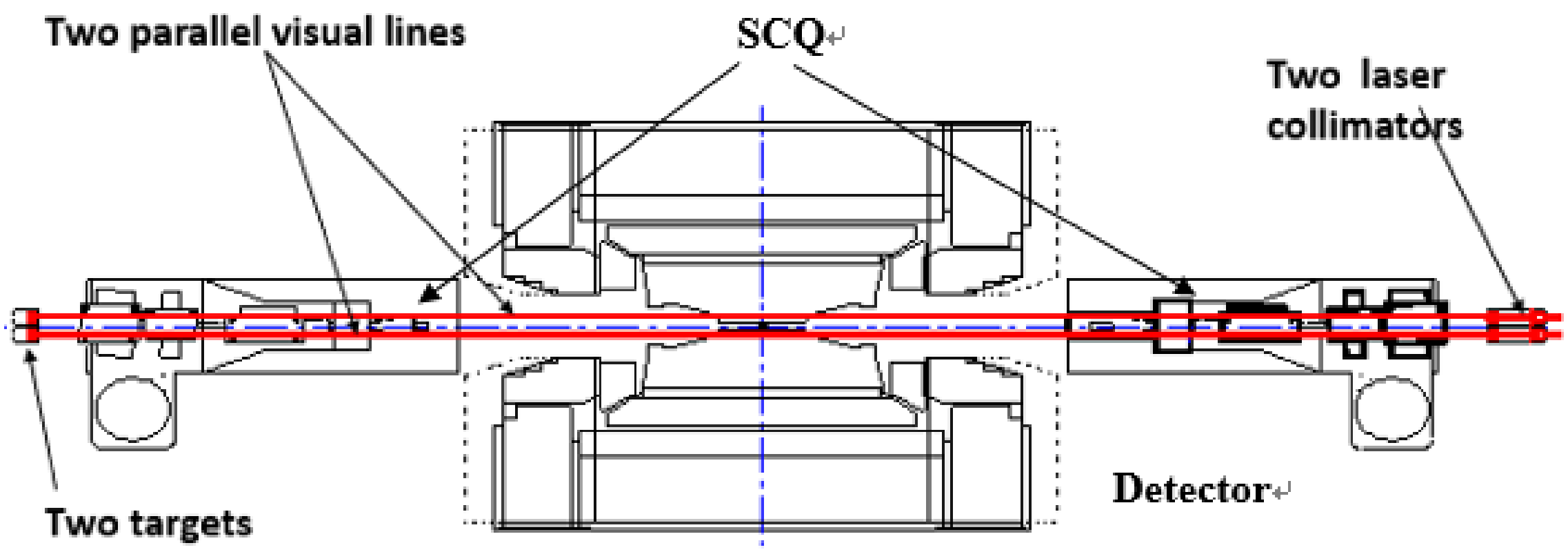

**Figure 7.2.4.2:** Interaction region alignment

### 7.2.5 Geoid Refinement of the Tunnel

All of the accelerator components are installed on a plane but all measurements carried out on the earth are based on the geoid, so a transformation is required from the geoid to the CEPC plane.

Affected by the density of the earth and the ground's undulation, the real geoid is an irregular surface, and does not completely coincide with the earth ellipsoid as shown in Fig. 7.2.5.1. For small accelerators of less than 1 km circumference, the geoid can be approximated as a sphere; for a medium accelerator (less than 10 km), the geoid can be approximated as an ellipsoid. But for our large-scale installation the irregular degree of the Earth's internal density and terrain undulation, especially close to mountains, results in large differences between the geoid and an ellipsoid. So irregular curvature variations cannot be ignored. It is necessary to establish an ultra-precision geoid model. [11,12]

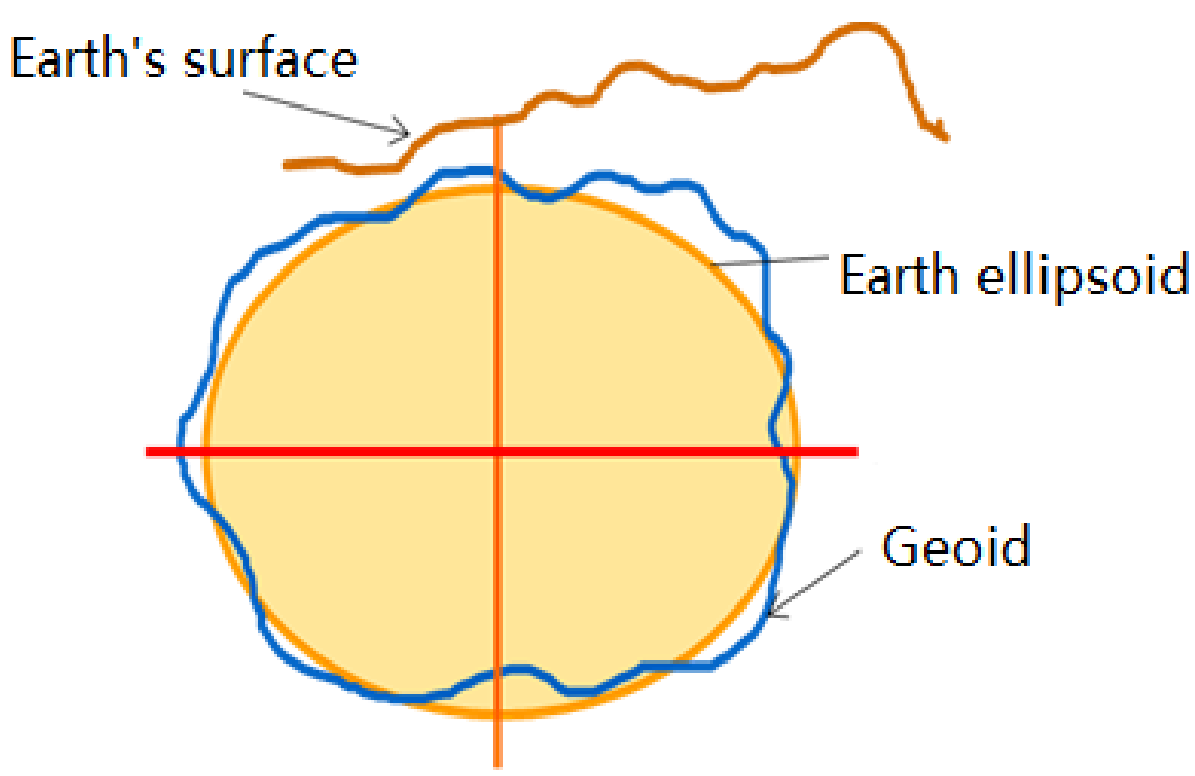

**Figure 7.2.5.1:** The ellipsoid and the true geoid

LHC carried out precise local gravity measurements with a zenith camera, and absolute and relative gravimeters. They established a precise geoid model and used it to correct the alignment data.

Our proposed plan is to measure gravity data and vertical deviation observations every 100 meters along the accelerator area.

From these data a gravitational field model can be obtained as shown below:

$$\mathbf{W}(x,y,z)=\mathbf{U}(x,y,z)+\mathbf{T}(x,y,z) \tag{7.2.5.1}$$

where U (x,y,z) is the regular gravity term, which can be derived from theory, and T (x,y,z) is the irregular term, known as the disturbance field. Based on the tunnel horizontal plane, the geometric variation between points A and B generated by the disturbance field can be shown as follows with vertical deviation $\varepsilon$ along the path from A to B and with positive correction $E_{AB}^{H_{tnl}}$ :

$$\Delta N_{AB}^{H_{tnl}}=-\int_A^B \varepsilon\cdot ds-E_{AB}^{H_{tnl}} \tag{7.2.5.2}$$

The positive correction $E_{AB}^{H_{tnl}}$ can be obtained by measuring the gravity value, and ε can be obtained by measuring the vertical deviation value. Therefore, the geoid difference between the accelerator's points can be obtained by solving the above equation. Combining this with the gravity geiod model, the real geiod model of the grid-based accelerator tunnel can be established.

Non- parallel correction of geoid:

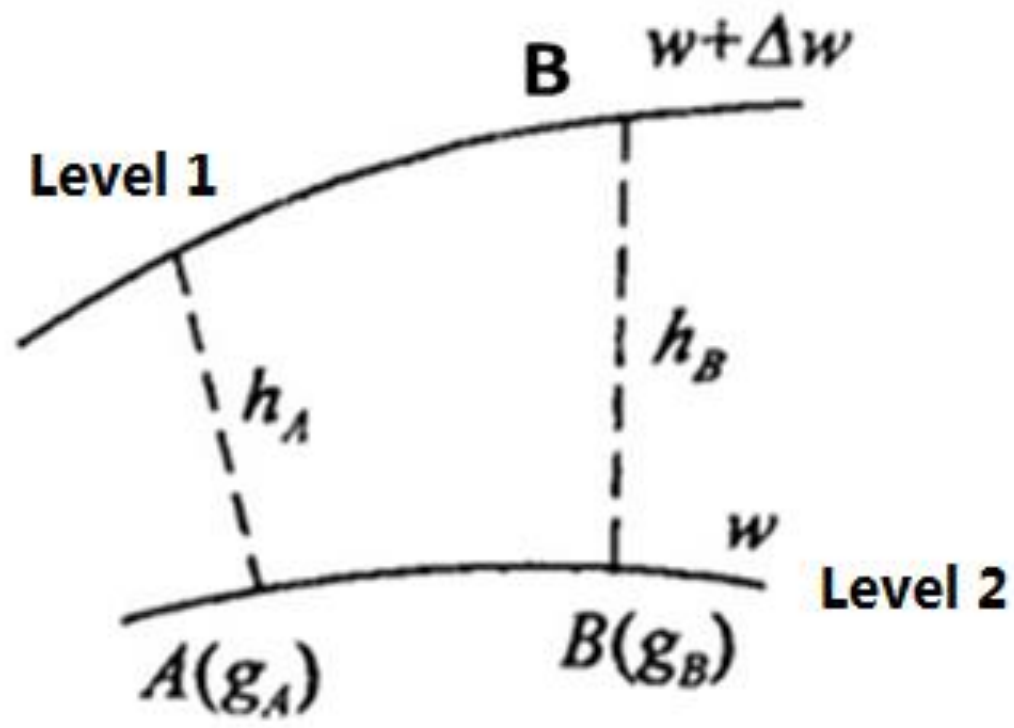

**Figure 7.2.5.2:** Non-parallelismof the geiod

The gravity difference between two points of ground result in the unparallelism of the geoid is illustrated in Fig. 7.2.5.2. In actual alignment, the height difference between the two points consists of three parts: the measured height difference, the non-parallel correction of the normal geoid and the gravity anomaly correction. In CEPC, it is necessary to measure the gravity at the control point, correct the unparallelism of the geoid and gravity abnormality to achieve the high-precision elevation value.

The effect of the geoid curvature on the level elevation, as shown in Fig. 7.2.5.3 can be corrected by the equation:

$$\Delta h=S^2/2R \tag{7.2.5.3}$$

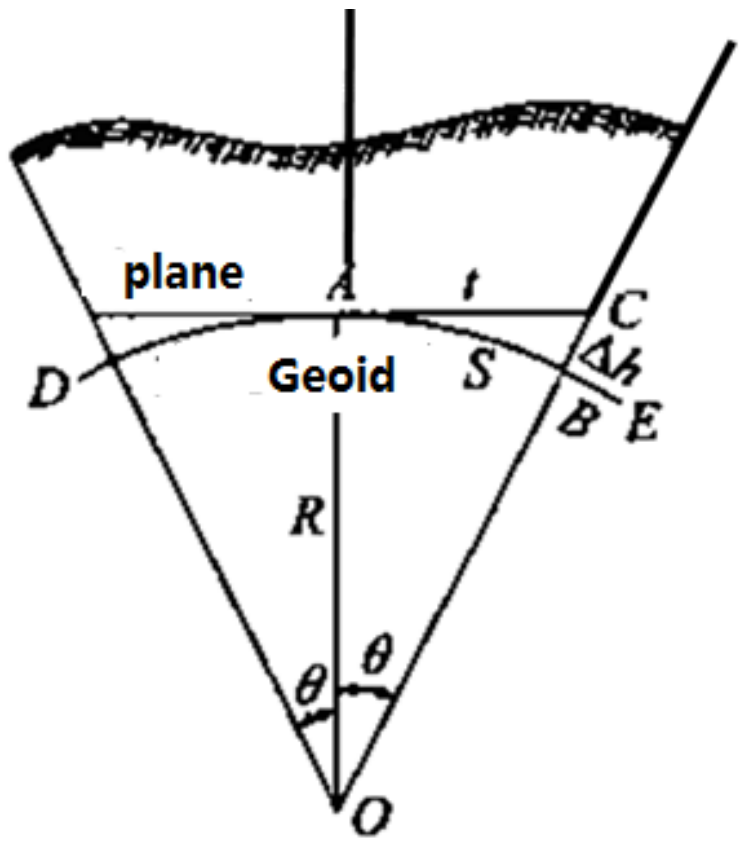

**Figure 7.2.5.3:** Water-level, geoid and elevation corrections

S is the distance between two points, R is the radius of the earth, where the R is no longer a sphere radius or an ellipsoid radius but now is based on our precise geoid model.

### 7.2.6 **Monitoring of Tunnel Settlement**

Uneven settlement of the foundation has a large effect on the beam orbit, especially when the site geologic conditions are complicated. It is necessary to monitor the foundation settlement by the HLS (Hydrostatic Levelling System).

The HLS has the following design parameters: sensor sensitivity 0.005 mm, measurement standard errors of height difference between two points: <± 0.05mm, accuracy instability (three months) <± 0.05mm, range ±10mm. Environmental conditions for using the HLS are temperature 5-40 ° C and humidity 100 %.

The HLS system is composed of some sensors, water pipes, support system, injecting and pumping system, signal collector and power cables. A schematic diagram is in Fig. 7.2.6.1.

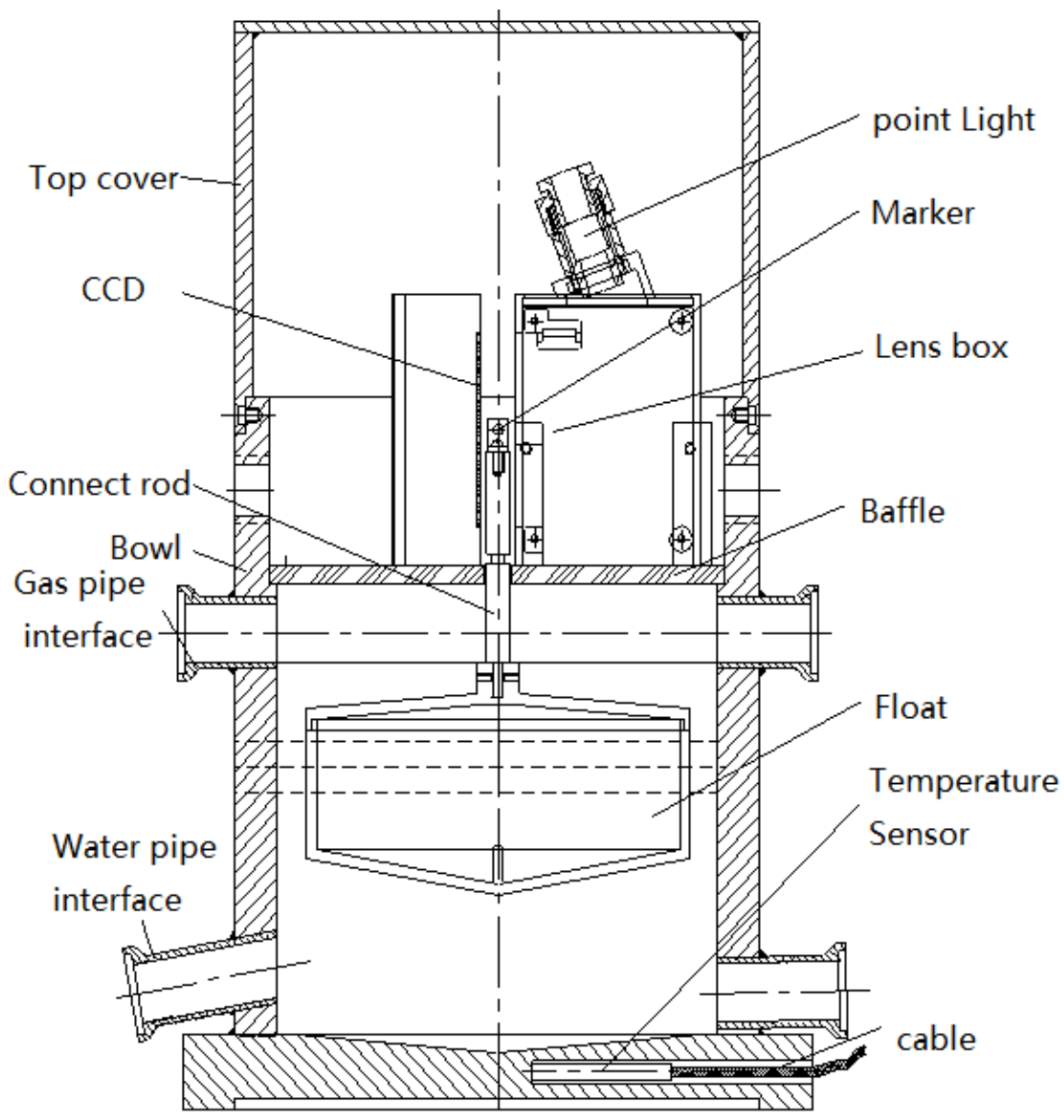

**Figure 7.2.6.1:** Schematic diagram of sensor

To have convenient maintenance of the HLS, each quadrant of the CEPC ring will have an HLS. The end sensors of adjacent HLS will be adjusted to the same height, by doing this, both independence and correlation of each HLS can be realized.

The layout is shown in Fig. 7.3.6.2. In the ring tunnel, along the pedestrian passageway, about 100 mm from the component girder, HLS sensors will be installed at an interval of 500 m. Each quadrant is 25km, so each quadrant needs 50 sensors or a ring total of 200 sensors. In the IP areas due to the very heavy detectors and ultra-high alignment precision requirement, the ground settlement need to be pay more attention. HLS sensors will be densified in the IP areas.

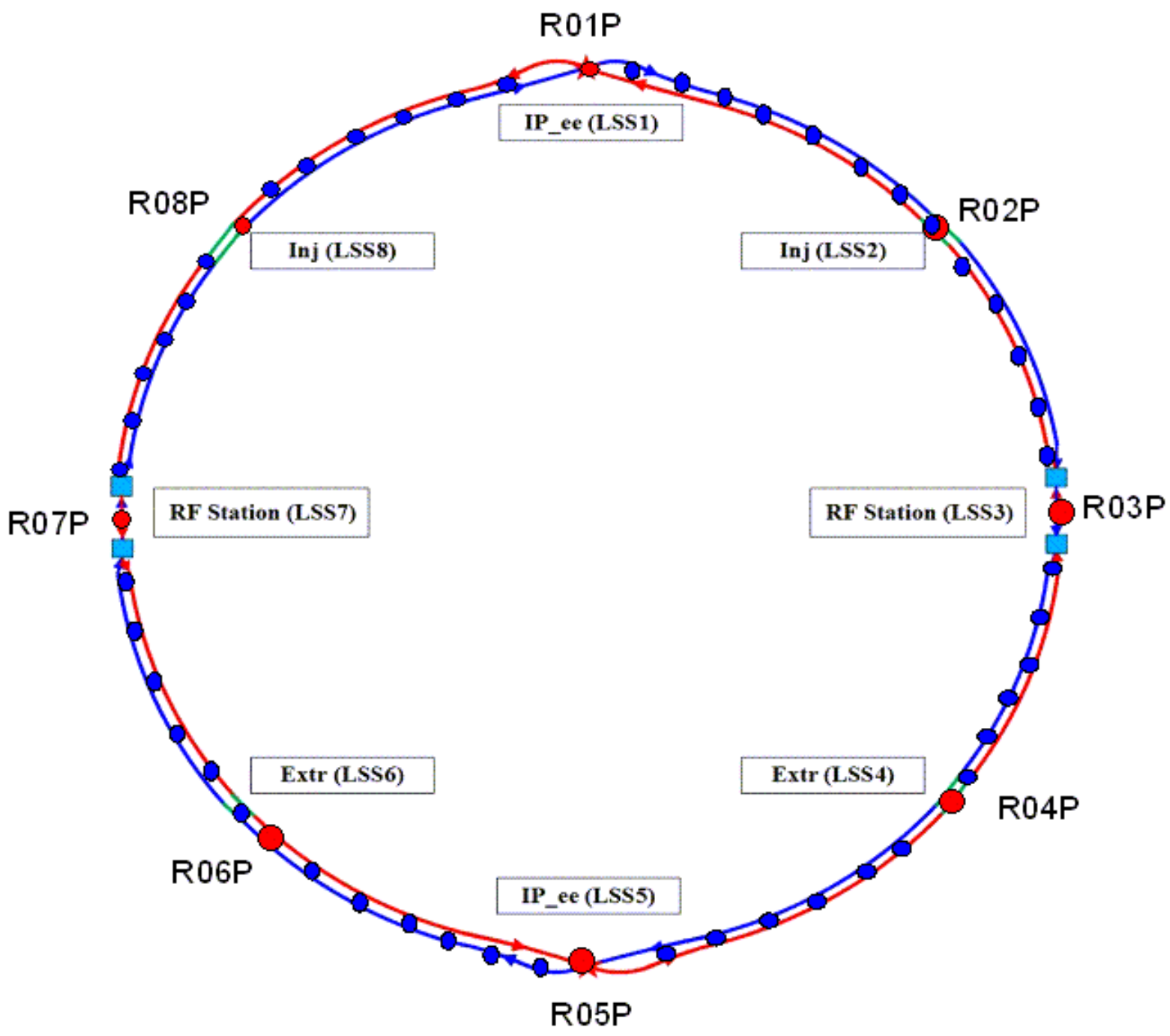

**Figure 7.2.6.2:** Layout of the HLS in the CEPC ring

HLS is vulnerable to temperature, air pressure, tides and earth motion. Specific designs need to be considered. In each HLS sensor a temperature sensor will be integrated, and the temperature influence can be calibrated. HLS sensor will be designed to a half-filled type, adjacent sensors will be connected by a tube. This can decrease the affection of uneven air pressure. Tides and earth motion will bring a regular influence to HLS. Its model need to be researched [13,14].

## 7.3 Radiation Protection

Described in this section is the expected radiological situation. We enumerate the provisions made to minimize the consequences for workers as well as to the general public living in the vicinity or visiting the site. Included is adequate shielding, a state-of-the-art radiation monitoring and alarm system, as well as a rigorous access control system.

### 7.3.1 Introduction

#### 7.3.1.1 *Workplace Classification*

Radiation areas are classified as follows [1]:

a) **Radiation monitored area**. Registered radiation workers can enter freely any time. This includes facilities, halls and areas and surfaces outside concrete shielding;
b) **Radiation controlled area**. Access to this area is limited and permission and required access procedures are required. An example is the auxiliary tunnel;

These occupancy factors will be clearly defined for each structure and area.

#### 7.3.1.2 *Design Criteria*

Standards and rules are listed as follows. Table 7.3.1 lists the dose limits from the national standards [2].

1. The national standard of the People's Republic of China, "Basic standards for protection against ionizing radiation and for the safety of radiation sources," GB 18871-2002.

2. The national standard of the People's Republic of China, "The rule for radiation protection of particle accelerators," GB 5172-1985.
3. ICRP publication 103 "The 2007 Recommendations of the International Commission on Radiological Protection."

**Table 7.3.1:** Dose limits in the national standards GB18871-2002.

| Item | | **Worker** | **Public** |
|---|---|---|---|
| Effective Dose | Average in 5 years | 20 mSv/year | 1 mSv/year |
| | Maximum in a single year of the 5 year period | 50 mSv/year | 5 mSv/year |
| Equivalent Dose | Lens of the eye | 150 mSv/year | 15 mSv/year |
| | Skin | 500 mSv/year | 50 mSv/year |

According to these standards, the maximum occupational exposure limit is 50 mSv per year. However, in applying the ALARA ("as low as reasonably achievable") philosophy, the goal is to maintain exposures well below this limit, for occupational exposure, the annual dose limit should below 5 mSv and for the public the annual dose limit should below 0.1 mSv (compare with CERN: the dose limit for designed area should below 6mSv/y and for off-site area should below 0.3mSv/y, in addition, the optimization goal is 100μSv/y and 10μSv/y respectively). The deduced dose rate limits for shielding design are listed in Table 7.3.2.

**Table 7.3.2:** Prompt dose rate limits for different areas

| **Area** | **Design Value** | **Example** |
|---|---|---|
| Radiation monitored area | < 2.5 μSv/h | Outside the tunnels, where a worker can stay longer |
| Radiation controlled area | < 25 μSv/h | Outside the tunnels, where a worker can stay occasionally, also including some areas that is forbidden for works, such as the dose rate higher than 1 mSv/h |
| Site boundary | 0.1 mSv/year | |

The dose limits for soil, ground water and air activation are based on GB18871-2002 (the radionuclide and its exempted activity or specific activity are detailed descripted in Appendix A of GB18871-2002): "If there is more than one kind of radionuclide, only if the ratio of activity (or specific activity) to its exempt value of each kind of the radionuclide was less than 1, it is exemptible." This is expressed as:

$$\sum_{i=1}^{n} \frac{S_{i_saturation}}{S_{i_exempt}} < 1 \tag{7.3.1}$$

This is convenient to evaluate soil and ground water activation. The prompt dose rate is ~5.5 mSv/h in a thickness of 1 m of soil to ensure it is below the exempt value above [3].

## 7.3.2 Radiation Sources and Shielding Design

### 7.3.2.1 *Interaction of High Energy Electrons with Matter*

The particles of a high-energy electron beam interact with matter via various processes such as beam-gas, beam-collimator, beam-target or beam-dump interactions. Electromagnetic cascades and nuclear reactions dominate and will result in the production of ionizing radiation fields (prompt, mixed radiation fields) and the production of radioactive nuclei inside the target material (induced activity).

### 7.3.2.2 *Radiation Sources*

In a high energy accelerator, there are two main radiation sources: the prompt and the residual radiation fields, as depicted in Fig. 7.3.1.

The so-called prompt, mixed radiation fields are mainly composed of neutrons and photons; there are also some charged hadrons (protons, pions, kaons) and leptons (e.g. muons). The composition of the fields at a given point in or outside the tunnel strongly depends on position with respect to the beam loss and the kind of shielding in between.

Radioactive isotopes are produced in the accelerator components and tunnel in nuclear reactions caused by a high energy primary or secondary particle. The isotopes decay, mainly by beta and gamma emission until the “Valley of Stability” is reached. Since the half-lives of the radioactive isotopes range from fractions of seconds to years and more, radiation fields will always be present in the machine once it becomes operational.

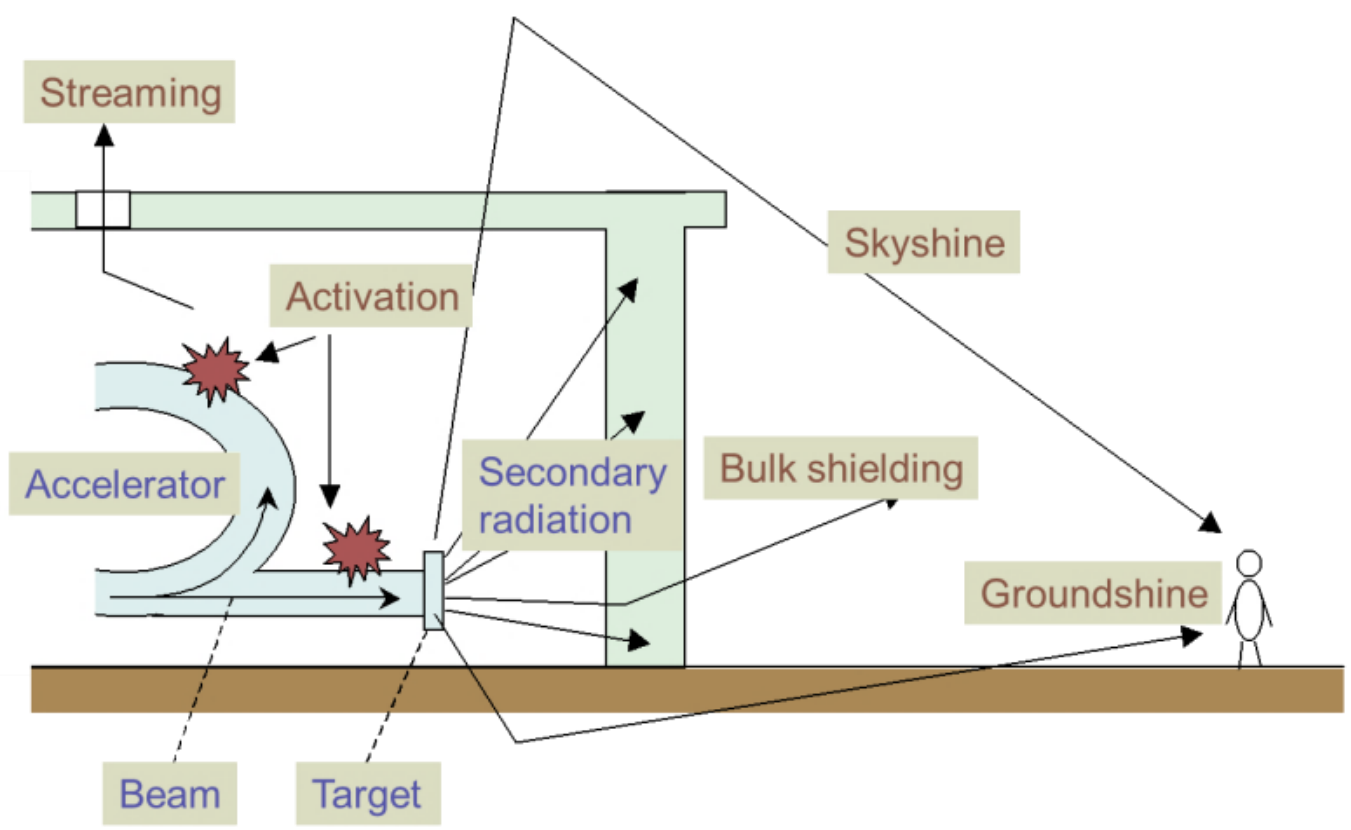

**Figure 7.3.1:** Sketch map of the radiation sources around an accelerator.

### 7.3.2.3 *Shielding Calculation Methods*

Radiation shielding design is based on Monte-Carlo (MC) simulations using the FLUKA and MCNP codes and the results checked with empirical formulas.

### 7.3.2.4 *Radiation Shielding Design for the Project*

The radiation shielding design philosophy adopted is that the shielding thickness of the main tunnel be determined by the radiation level caused by average beam loss along the tunnel. Hot spots, such as places for collisions, injection, collimation and beam dumps need to have additional local shielding to reduce the radiation level to be the same as for the main tunnel.

Dose rate simulation with 120 GeV electrons at the loss rate of one W/m and shielded with a 1 cm thick Fe beam pipe are shown in Fig. 7.3.2. The radiation levels for other beam loss parameters can be deduced with conversion coefficients.

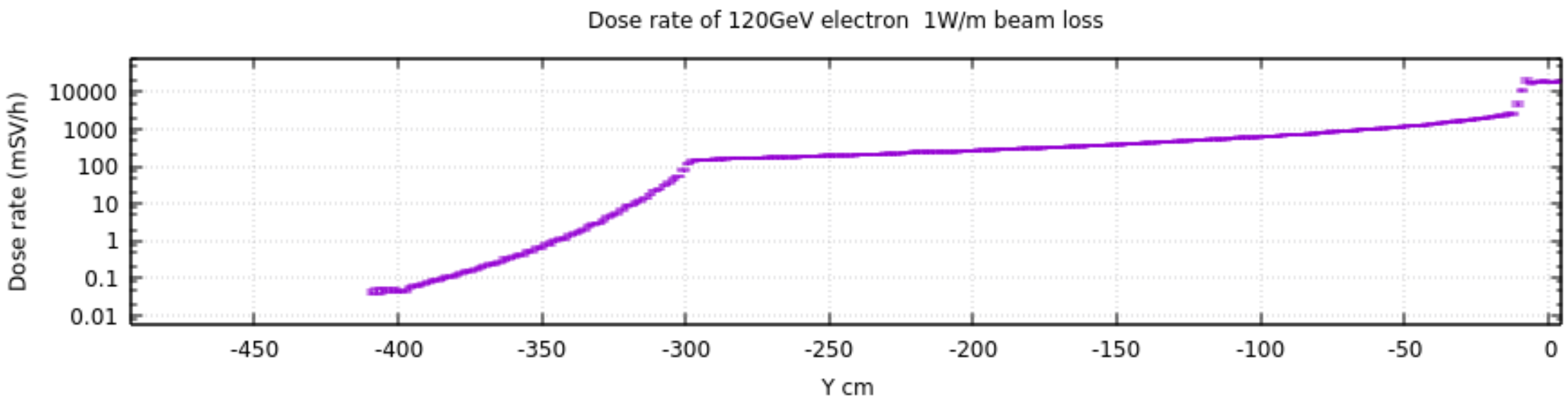

**Figure 7.3.2:** The distribution of prompt radiation dose rate (300 to 400 cm is concrete).

The critical components of the dump are the absorbers. Several materials (aluminum, iron, copper, graphite) have been investigated, taking into account thermal conductivity and melting temperature. For an absorber for the Linac beam we chose iron, and for the high energy absorber in the Collider we chose graphite. The iron should be chemically, thermally and mechanically processed and forged. A water cooling system is incorporated to prevent excess heating.

Monte Carlo simulations were done to find the optimal absorber dimensions. The energy deposition and ambient dose equivalent distribution by the primary beam was calculated using FLUKA. The beam parameters used in the FLUKA simulations are given in Table 7.3.1. Preliminary dimensions of the absorber were chosen to be 200 cm in radius and 500 cm in length. The limitation of average ambient dose in the range of 1 meter outside the absorber is 5.5 mSv/h.

**Table 7.3.1:** Parameters used in the FLUKA simulations for the dumps

| | Linac | Collider |
|---|---|---|
| Energy(GeV) | 10 | 120 |
| Current(uA) | 0.16 | 62.7 |
| No. of Electrons | $1 \times 10^{12}$ | $3.92 \times 10^{14}$ |

Preliminary dimensions of the Linac dump were chosen to be 10 cm in radius and 30 cm in length. Since photons dominate, iron can be used as shielding material. Figs. 7.3.3 and 7.3.4 are the dose curves of longitudinal and transverse (respectively) ambient dose equivalent distribution with different thicknesses of iron. If we chose the minimum iron thickness, the concrete shielding would be 140 cm transverse and 240 cm longitudinal and if we chose the minimum total thickness, the iron would be 70 cm transverse and 130 cm longitudinal.

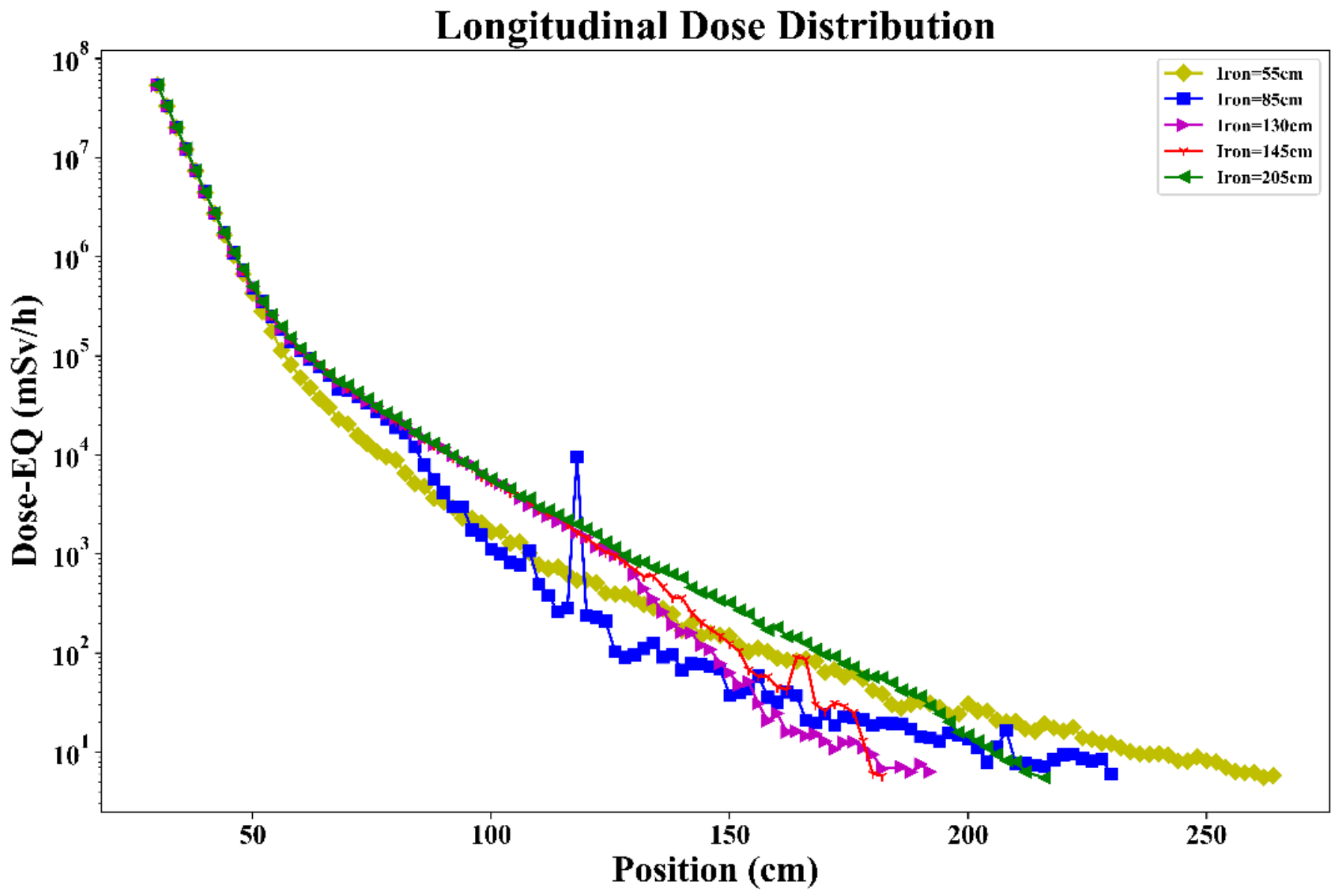

**Figure 7.3.3:** Longitudinal ambient dose equivalent distribution of the Linac absorber

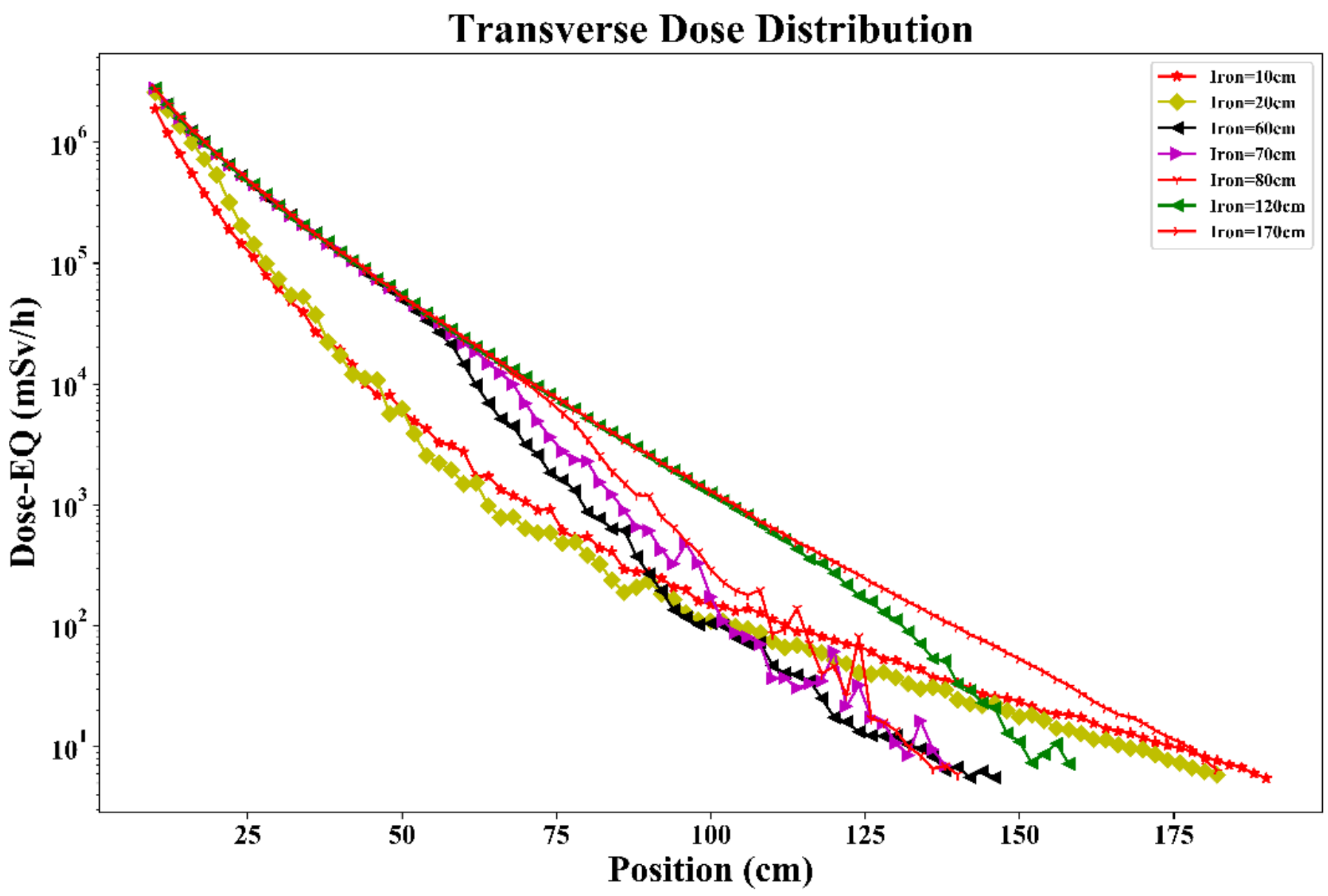

**Figure 7.3.4:** Transverse ambient dose equivalent distribution of Linac absorber

## 7.3.3 **Induced Radioactivity**

### 7.3.3.1 ***Specific Activity and Calculation Methods***

Lost beam interacts with surrounding components and induces radionuclides. In addition, while the air absorbs the secondary γ–rays, $O_3$ and $NO_X$ and other harmful gases are generated.

The main activation isotopes are:

- Concrete shielding: $^{24}Na$
- Air activation: $^{11}C$, $^{13}N$, $^{15}O$
- Cooling water activation: $^{11}C$, $^{13}N$, $^{15}O$, $^{7}Be$, $^{3}H$
- Soil activation: $^{7}Be$, $^{3}H$

FLUKA is used to calculate the specific activity induced by electron interactions in the beam line, shielding components and in the environment Also the traditional method of folding particle fluence with energy-dependent cross sections for the production of specific isotopes can be used for some of the simulations.

#### 7.3.3.2 *Estimation of the Amount of Nitrogen Oxides*

Gamma rays decompose atmospheric oxygen into free radicals. These combine with $O_2$ to form $O_3$. which combines with NO in the air to form NOx. $NO_2$ combines with $H_2O$ in the air to form $HNO_3$. In this process, the production (defined as the number of molecules produced per absorption of 100 eV energy of the γ ray) of $O_3$, $NO_X$, HNO3 is 10, 4.8, 1.5 respectively. For simplicity, we only calculate the amount of $O_3$. The amount of $NO_X$ can be obtained from the above proportions.

In an irradiation space of volume V, chemical decomposition and ventilation for the removal of ozone are considered. The amount of ozone molecules, N, obeys the following equation:

$$\frac{dN}{dt} = PG - \left(\alpha' + \frac{KF}{V}\right)N \tag{7.3.2}$$

Solving this differential equation, gives the following:

$$\mathrm{N} = \frac{\mathrm{PG}}{\alpha' + \frac{\mathrm{KF}}{\mathrm{V}}}\left[1 - \mathrm{e}^{-(\alpha' + \frac{\mathrm{KF}}{\mathrm{V}})\mathrm{t}}\right] \tag{7.3.3}$$

In this formula:

P – the power absorbed by the air in eV/s;

G – the production of $O_3$, usually the G value is between 0.03 to 0.09 molecules per eV; G = 0.06 molecules per eV is used in the calculation;

F – the ventilation rate of the irradiated area in $cm^3/s$;

V – the volume of the irradiated area in $cm^3$;

K – the mixing uniformity coefficient, K=1/3;

α' – chemical decay constant of O3, chemical half-life of O3 is about 50;

t – the irradiation time in seconds.

One ppm of $O_3$ in the air is equivalent to $2.463\times10^{13}$ $O_3$ molecules in one $cm^3$ of air:

$$\mathrm{C_p} = \frac{\mathrm{N}}{2.463\times10^{13}\mathrm{V}} \tag{7.3.4}$$

According to the basic concept of energy transmission when γ rays pass through the air, it is easy to deduce the power absorbed by the air with the following formula:

$$\mathrm{P} = 6.25\times10^{18}\sum_i\left[\mathrm{E_{\gamma i}}\Psi_{\gamma i}(\mathrm{K}/\Phi)_i\right]\rho_{air}\mathrm{V_{air}} \tag{7.3.5}$$

In this formula:

$E_{\gamma i}$ – energy interval of the γ ray, eV;

$\Psi_{\gamma i}$ – average flux in the energy interval of the γ ray, $cm^{-2}s^{-1}eV^{-1}$

$(K/\Phi)_I$ – the conversion coefficients from mono-energetic photon flux to air kerma, $Gy{\bullet}cm^2$

$\rho_{air}$ – air density in standard conditions, $kg/m^3$

$V_{air}$ – the volume of air, $m^3$

$6.25\times10^{18}$ – conversion coefficient, eV/J

The physical meaning of $E_{\gamma i}\Psi_{\gamma i}$ is fluence rate of specific energy photons. The physical meaning of $\rho_{air}V_{air}$ is the mass of air. $E_{\gamma i}$ and $\Psi_{\gamma i}$ could be simulated by Monte-Carlo; $(K/\Phi)_i$ can be interpolated from the ICRU74 or ICRP74 reports.

Hence, the $O_3$ concentration can be expressed by:

$$C_p = \frac{PG}{2.463\times10^{13}V(\alpha'+\frac{KF}{V})}\left[1-e^{-(\alpha'+\frac{KF}{V})t}\right] \quad (7.3.6)$$

The density of $O_3$ is $1.964\times10^{-3}g/cm^3$, so the concentration of $O_3$ in the air can be expressed in $g/cm^3$:

$$C_g(t) = 7.97\times10^{-17}\frac{PG}{V(\alpha'+\frac{KF}{V})}\left[1-e^{-(\alpha'+\frac{KF}{V})t}\right] \quad (7.3.7)$$

Because the chemical half-life of $O_3$ is only 50 minutes, the concentration of $O_3$ in the tunnel could become easily saturated. The saturated concentration is Cg ($g/cm^3$):

$$C_g(t) = 7.97\times10^{-17}\frac{PG}{V(\alpha'+\frac{KF}{V})} \quad (7.3.8)$$

### 7.3.4 **Personal Safety Interlock System (PSIS)**

#### 7.3.4.1 ***System Design Criteria***

The PSIS is designed following these criteria:

1. Hardware is reliable so all critical device interlock signals are generated by hardware.
2. At the highest interlock level the PSIS has top priority to shut off the beam in the CEPC Central Control System.
3. Fail safe: the beam will be shut off when a critical device has a breakdown.
4. Redundancy ensures reliability, reduces fault time and preserves upgrade possibilities.
5. Multilayer protection: interlock key, emergency shut-off button, emergency door-open button, acousto-optic alarm, patrol search and secure before beam start-up and surveillance cameras will ensure multilayer personal safety.
6. People oriented: the primary purpose is personal safety, but in addition it should be convenient to operate and maintain the PSIS with a good human-computer interface.

### 7.3.4.2 *PSIS Design*

The PSIS consists of a Programmable Logic Controller (PLC) and Access Control System (ACS). PLC monitors interlocked equipment, ACS administers interlock information. The access conditions for the interlocked areas stipulate that the names and identification numbers of all persons who enter must be known/shown and recorded. Fig. 7.3.5 shows the layout of the PSIS.

The PLC system: consists of the access controller, interlock key, emergency/patrol button， emergency door open button, acousto-optic alarm and interlock equipment. Multilayer personnel protection is guaranteed by programming interlock signals from this equipment. Meanwhile, host double backup and double lines for signal transmission will guarantee reliability of the system.

The ACS includes a camera, LED display and data server. The PSIS can monitor the interlock areas, and display and store the interlock information.

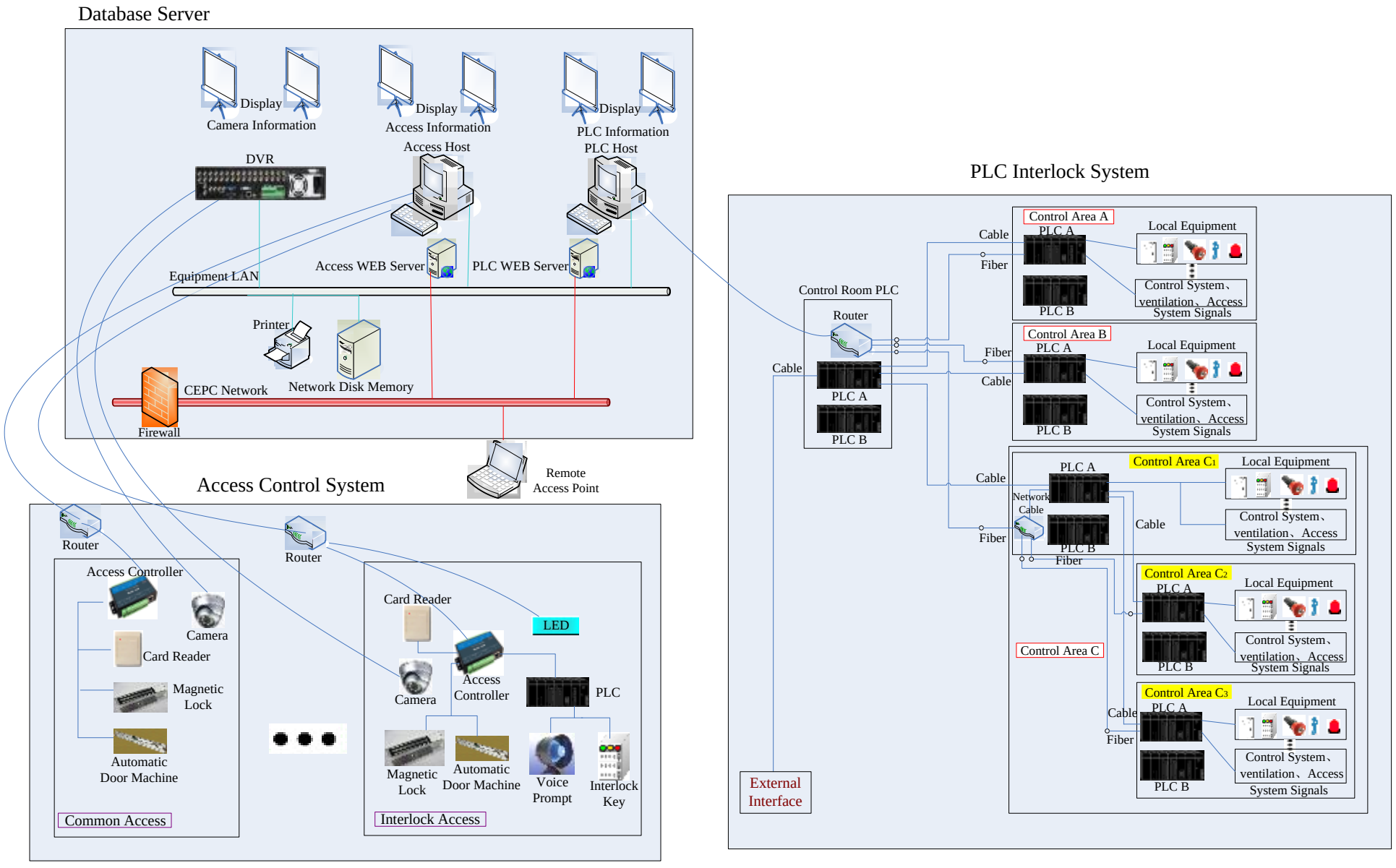

**Figure 7.3.5:** PSIS layout

A patrol search and secure must be done in every interlock area before startup to be absolutely certain that no person is left in that area. An acousto-optic alarm will signal start-up from the “ready” signal from Central Control System (CCS) and warn anyone in the area to quickly leave. The access control system will still be working in a “shutdown” phase. The beam will be shut off immediately by an emergency button in case of an accident. Fig. 7.3.6 shows the PSIS operation flow chart.

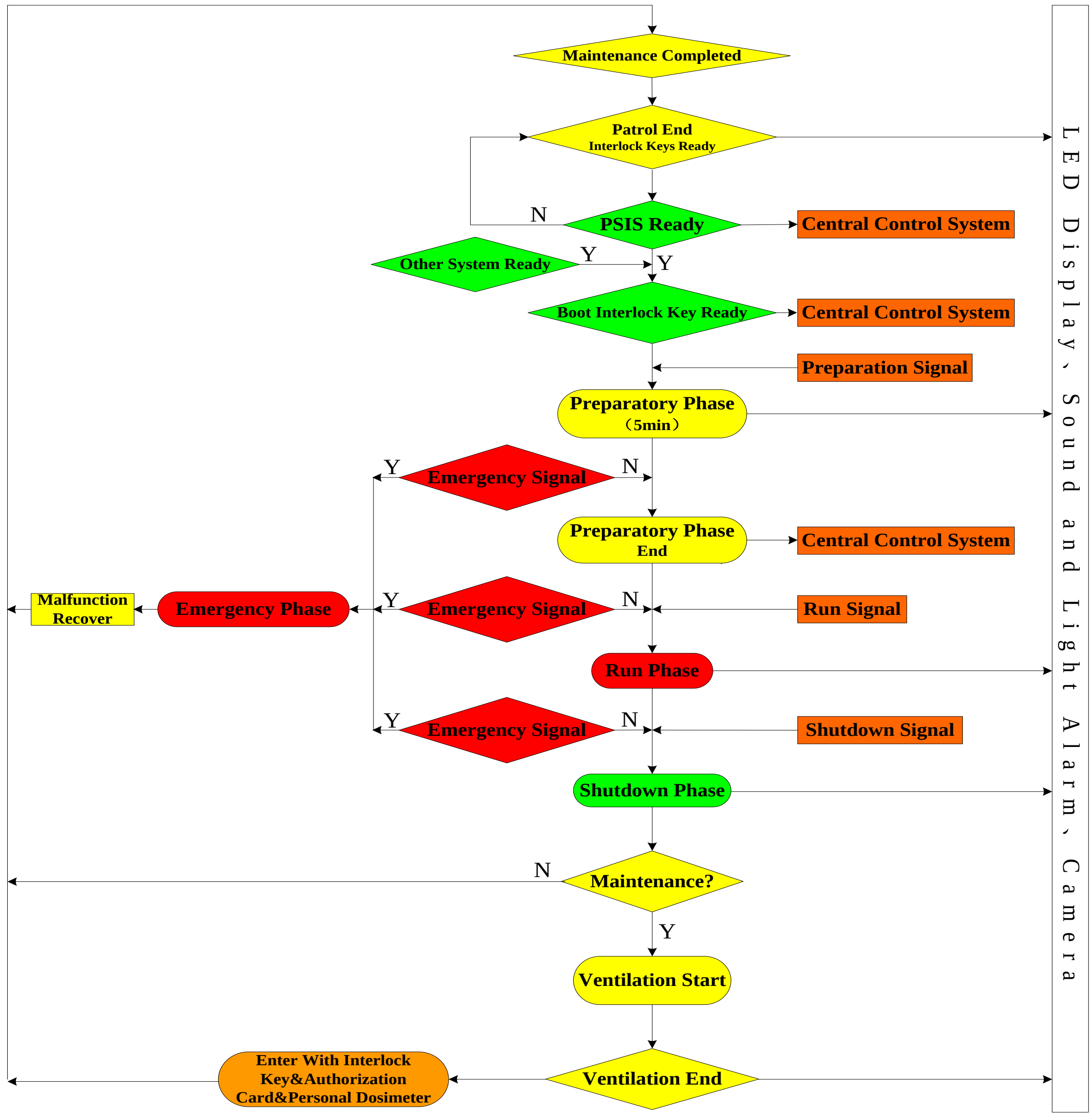

**Figure 7.3.6:** PSIS operation flow chart

## 7.3.5 Radiation Dose Monitoring Program

### 7.3.5.1 *Radiation Monitoring System*

The radiation monitoring system will be a new, state-of-the-art system. It will conform to the latest legal requirements and international standards. It will be based on the results of the preliminary hazard analysis, the latest technical developments and specific requirements such as the time structure and composition of the radiation fields.

The radiation monitoring system will provide continuous measurements of the ambient dose equivalent and the ambient dose rate equivalent in the underground areas together with the surface areas inside and outside the project perimeter. If preset radiation levels are exceeded within radiation controlled areas, an alarm will be triggered and transmitted. Remote alarms will sound in the control rooms. It will permanently monitor the level of radioactivity in water and air released from the project. iIt will also include hand-foot monitors, site gate monitors, tools and material monitors.

The radiation monitoring system provides remote supervision, long term database storage and off-line data analysis. A typical frame diagram is given in Fig. 7.3.7. The data will be accessible via the internet.

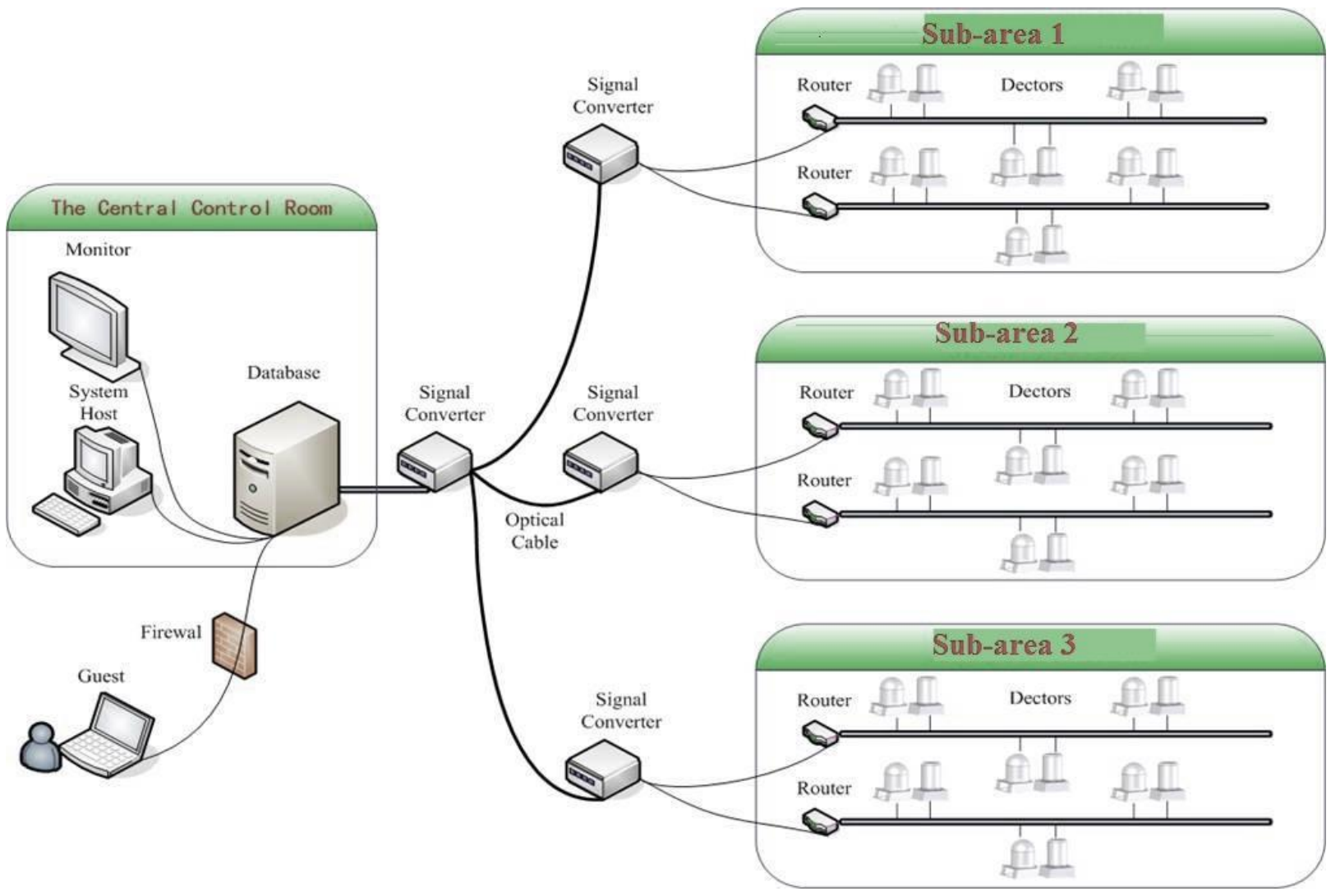

**Figure 7.3.7:** Frame diagram of the radiation monitor system.

To achieve control and data communication in different conditions, the communication system has four communication paths: Ethernet, wireless, GPRS (General Packet Radio Service) and the offline record.

### 7.3.5.2 *Workplace Monitoring Program*

The workplace monitoring program is organized so as to guarantee that the workplace radiation levels comply with relevant regulations. Monitoring sites are established at all the main entrances to high radiation level areas and in the workplaces near the accelerator or radioactive sources. Once a radiation level exceeds the set value, monitors will sound the alarm to inform people to evacuate. Each site has one gamma detector and one neutron detector.

### 7.3.5.3 *Environmental Monitoring Program*

The objective of the environmental monitoring program is to prove that the facility complies with the regulatory limits and to provide early warning if violation of these limits is imminent. The program includes monitoring and measurements of dose rates in the environment: air, water, plants and soil. In order to evaluate the impact to the environment a base line will be established with a complete background survey of 2-3 years before equipment is installed and continue 3-5 years after operation begins.

The dose rate and integrated dose will be monitored at stations located at critical or representative places. Each station will consist of a pressurized ionization chamber for gamma monitoring and a rem-counter for neutrons. These are on-line and generate alarms when dose-rate thresholds are exceeded.

Each fluid or gas extraction duct likely to contain radioactivity produced in the facility will be equipped with a monitoring station. Each station consists of an on-line real-time monitor for short-lived radioactive substances together with a sampler. Whilst the readings of the monitor will be stored in a database, the filters will be replaced periodically and analyzed in an off-line laboratory for longer-lived beta and gamma activity. These measurements will be carried out especially after upgrades to the facility.

#### 7.3.5.4 *Personal Dose Monitoring Program*

All staff will participate in the personal dose monitoring program. OSL (Optically Stimulated Luminescence) was chosen for gamma dose monitoring, and the CR-39 solid track dosimeter was chosen for neutron dose monitoring. An electronic personal dose alarm should be used by persons entering the tunnel for maintenance.

### 7.3.6 **Management of Radioactive Components**

Accelerator components and maintenance tools and fixtures will become radioactive from accelerator operation. The specific activity depends on the material composition, the location of the material, the irradiation history and on the elapsed decay time. All radioactive items will be transferred to temporary storage and then sent to long-term repositories or disposed of according the legislation.

# 8 SPPC

## 8.1 Introduction

### 8.1.1 Science Reach of the SPPC

The SPPC (Super Proton-Proton Collider) is envisioned to be an extremely powerful machine, far beyond the scope of the existing LHC, with center of mass energy 75 TeV, a nominal luminosity of 1.0 x $10^{35}$ $cm^{-2}s^{-1}$ per IP, and an integrated luminosity of 30 $ab^{-1}$ assuming 2 interaction points and ten years of running. A later upgrade to even higher luminosities is possible. Luminosity has a modest effect on energy reach, in comparison with higher beam energy [1], but raising the luminosity will likely be much cheaper than increasing the energy. The ultimate upgrade phase for SPPC is to explore physics at the center of mass energy of 125-150 TeV.

The CEPC (Circular Electron-Positron Collider) and the SPPC together will have the capability to precisely probe Higgs physics. [2] However, what the high energy physics community expects more is that SPPC will explore directly a much larger region of the landscape of new physics models, and make a huge leap in our understanding of the physical world. There are many issues in energy-frontier physics that SPPC will explore, including the mechanism of Electroweak Symmetry Breaking (EWSB) and the nature of the electroweak phase transition, the naturalness problem, and the understanding of dark matter. While these three questions may be correlated, they also point to different exploration directions leading to more fundamental physics principles. SPPC will explore new ground and have a great potential for making profound breakthroughs.

As a "Higgs factory", the CEPC can measure with high precision the properties of the Higgs boson. The total Higgs width can be measured to a relative precision of 2.9%. Using the recoil mass method, CEPC can precisely measure the absolute Higgs couplings to the Z bosons g(HZZ) and the invisible decay branching fraction to gluons, W bosons and heavy fermions [g(Hgg), g(HWW), g(Hbb), g(Hcc), and g(H$\tau\tau$)] at the few percentage level. In addition, it can measure the rare decay couplings [g(H$\gamma\gamma$) and g(H$\mu\mu$)] to the 10% level. However, limited by its center of mass energy, CEPC cannot directly measure g(Htt) and g(HHH). These two couplings are extremely important for understanding EWSB and naturalness. [3]

Extending the CEPC Higgs factory program, billions of Higgs bosons will be produced at the SPPC. This huge yield will provide important physics opportunities, especially for the rare but relatively clean channels. For example, SPPC can improve the measurement of Higgs-photon coupling, observe the coupling g(H$\mu\mu$), and test other rare decays such as t $\rightarrow$ Hc, H $\rightarrow \mu\tau$. Reaching a higher energy threshold than CEPC, SPPC could measure g(HHH) to the 10% level [4], and directly determine the coupling g(Htt) to the sub-percentage level [5]. The Higgs self-coupling is regarded as the holy grail of experimental particle physics, not only because of the experimental challenges, but also because this coupling is a key probe to the form of the Higgs potential. By measuring g(HHH), SPPC can help to answer the question whether the electroweak phase transition is of the $1^{st}$ or $2^{nd}$ order, crucially connected to the idea of electroweak baryogenesis.

As an energy frontier machine, the SPPC could discover an entirely new set of particles in the *O* (10 TeV) regime, and unveil new fundamental physics principles. One of the most exciting opportunities is to address the naturalness problem. This problem stems from the vast difference between two energy scales: the currently probed electroweak scale and a new fundamental scale, such as the Planck scale. Solutions to the naturalness problem almost inevitably predict the existence of a plethora of new physics particles not far from the electroweak scale. Discovery of such new particles will be a stunning success for an understanding of the naturalness principle. Searching for these possible new particles at the LHC can probe the level of fine-tuning down to $10^{-2}$, while SPPC would push this down to the unprecedented level of $10^{-4}$, beyond the common concept of the naturalness principle.

Dark matter remains one of the most puzzling issues in particle physics and cosmology. Weakly interacting massive particles (WIMPs) are still the most plausible dark matter candidates. If dark matter interacts with Standard Model particles with coupling strength similar to that of the weak interaction, the mass of a WIMP particle could easily be in the TeV range, and likely to be covered at SPPC energy. Combining the relevant bounds on the mass and coupling from the direct (underground) and the indirect (astroparticle) dark matter searches, SPPC would allow us to substantially extend the coverage of the WIMP parameter space for large classes of models.

At the SPPC energy regime, all the SM particles are essentially “massless”, and electroweak symmetry and flavor symmetry will be restored. The top quark and electroweak gauge bosons should behave like partons in the initial state, and like narrow jets in the final state. Understanding SM processes in such an unprecedented environment poses new challenges and offers unique opportunities for sharpening our tools in the search for new physics at higher energy scales.

### 8.1.2 **The SPPC Complex and Design Goals**

SPPC is a complex accelerator facility and will be able to support research in different fields of physics, similar to the multi-use accelerator complex at CERN. Besides the energy frontier physics program in the collider, the beams from each of the four accelerators in the injector chain can also support their own physics programs. The four stages, shown in Fig. 8.1.1 and described in more detail in Section 8.4, are a proton linac (p-Linac), a rapid cycling synchrotron (p-RCS), a medium-stage synchrotron (MSS) and the final stage synchrotron (SS). This research can occur during periods when beam is not required by the next-stage accelerator. The option of heavy ion collisions also expands the SPPC program into a deeper level of nuclear matter studies. There would also be the possibility of electron-proton and electron ion interactions. In summary, SPPC will play a central role in experimental particle physics in this post-Higgs discovery world. It is the natural next stage of the circular collider physics program after CEPC. Combining these two world class machines will be a significant milestone in our pursuit of the fundamental laws of nature.

Given the 100 km circumference tunnel, we will try to achieve 75 GeV center of mass energy in p-p collisions with the anticipated accelerator technology in the 2030’s, but a more ambitious goal to go higher energy is possible. This, of course, depends on the magnetic field that bends the protons around the ring, 12 T using magnets with iron-based high-temperature superconductors (Fe-HTS) for the initial SPPC and 20-24 T also using Fe-HTS magnets in an upgrade. Taking into account the expected evolution in detector technology we can expect that the nominal luminosity of $1.0 \times 10^{35}$ $cm^{-2}s^{-1}$ will be usable at the early phase of running but a high-luminosity upgrade is also considered. At least two IPs will be made available. Some key parameters are shown in Table 8.1.1.

**Table 8.1.1:** Key parameters of the SPPC baseline design

| Parameter | Value | | Unit |
|---|---|---|---|
| | Initial | Ultimate | |
| Center of mass energy | 75 | 125-150 | TeV |
| Nominal luminosity | $1.0\times10^{35}$ | - | $cm^{-2}s^{-1}$ |
| Number of IPs | 2 | 2 | |
| Circumference | 100 | 100 | Km |
| Injection energy | 2.1 | 4.2 | TeV |
| Overall cycle time | 9-14 | - | Hours |
| Dipole field | 12 | 20-24 | T |

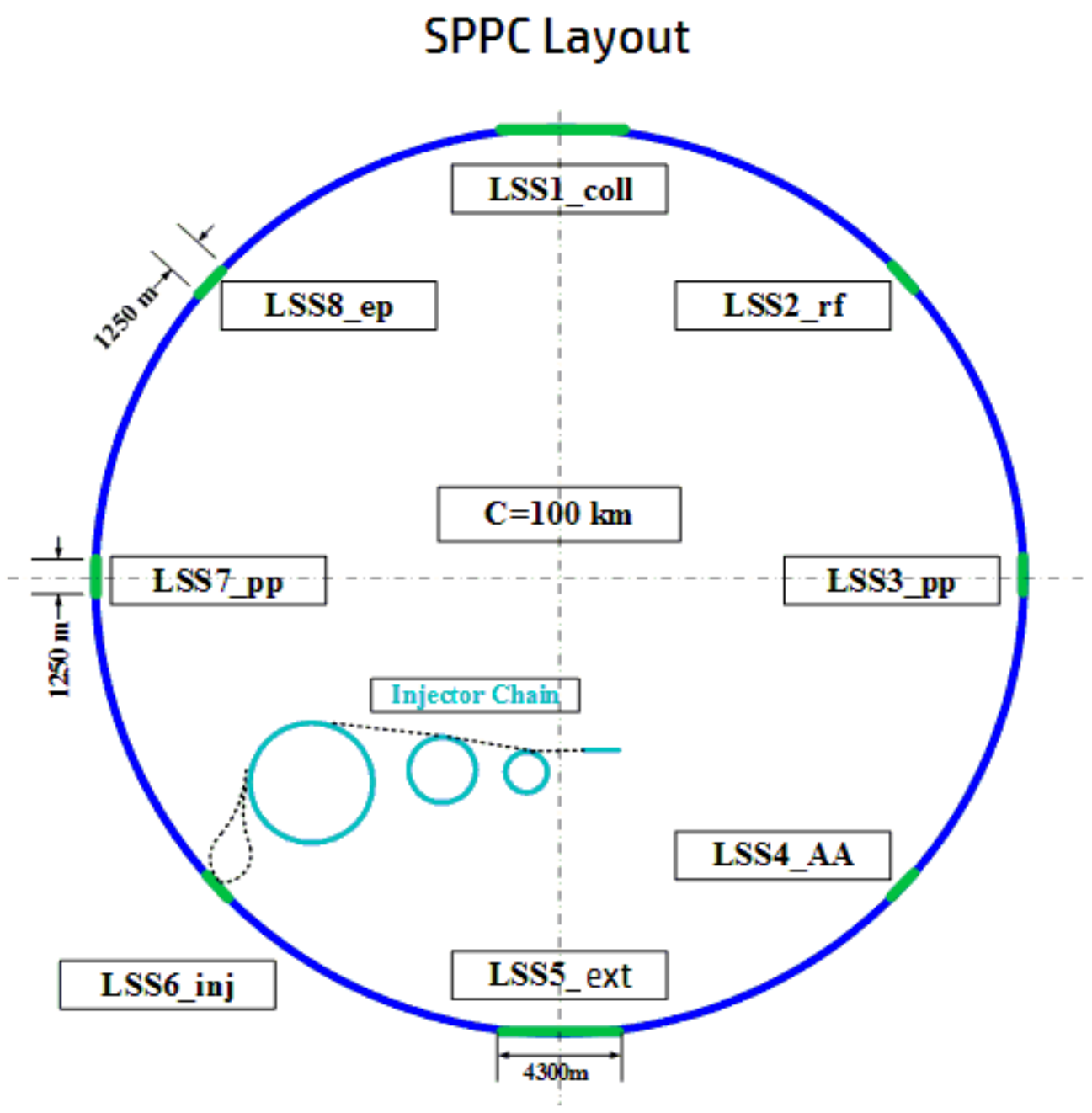

**Figure 8.1.1:** SPPC accelerator complex

## 8.1.3 **Overview of the SPPC Design**

The collider will coexist with the previously built CEPC, housed in the same tunnel, of circumference 100 km. The shape and symmetry of the tunnel is a compromise between the two colliders. The SPPC requires relatively longer straight sections which will be described below. This means eight identical arcs and eight long straight sections for two large detectors, injection and extraction, RF stations and a complicated collimator system. Based on expected progress in HTS technology, especially Fe-HTS technology and also high-field magnet technology in the next fifteen to twenty years, we expect that a field of 12 T will be attainable for the main dipole magnets with reasonable cost and cheaper than that based on $Nb_3Sn$ superconductors. Twin-aperture magnets will be used for the two-ring collider. A filling factor of 78% in the arcs (similar to LHC) is assumed.

With a circulating beam current of about 0.73 A and small beta functions ($\beta$*) of 0.75 m at the collision points, the nominal luminosity can reach $1.0 \times 10^{35}$ $cm^{-2}s^{-1}$ per IP. The high beam energy, high beam current and high magnetic field will produce strong synchrotron radiation which will impose critical requirements on the vacuum system. We expect that this technical challenge will be solved in the next two decades by developing efficient beam screens to extract the heavy heat load from the synchrotron radiation and reduce the electron cloud density. If forced to lower the synchrotron radiation power, we would have to reduce the bunch population or the number of bunches and try to achieve a smaller $\beta^*$ to retain the luminosity goal.

As in other proton colliders using superconducting magnets, the injection energy is mainly defined by the field quality of the magnets at the bottom of their range. Persistent currents in the coils (magnetization) distort the field distribution at injection energy. Other factors favouring relatively higher injection energy are the coupling impedance, which is important to collective beam instabilities, the smaller geometrical emittance required to reduce apertures of beam screen and magnet, and the requirement on the good-field-region of the magnets. If we use the LHC ratio of 15 for top to bottom fields the injection energy would be 2.5 TeV. A larger ratio of 20 could be considered, which would mean an injection energy of 1.875 TeV. This would make the injector chain cheaper. We have adopted a compromise with an injection energy of 2.1 TeV.

The injector chain pre-accelerates the beam to injection energy with the required bunch current, bunch structure, emittance and beam fill period. To reach 2.1 TeV, a four-stage injector chain is proposed: the p-Linac to 1.2 GeV, the p-RCS to 10 GeV, the MSS to 180 GeV and the SS to 2.1 TeV. High repetition rates for the lower energy stages help reduce the SS cycling period. This is important because the SS uses superconducting magnets. The beams of high repetition rates can also be used for other research applications when the accelerators are not preparing beam for injection into the SPPC.

If not controlled, synchrotron cooling would rapidly reduce the beam emittances and cause excessive beam-beam tune shifts. Noise in transverse deflecting cavities must be used to limit the minimum transverse emittances (emittance heating), and thus tune shifts. Without leveling, and with constant beam-beam tune shift, the luminosity decays exponentially from its initial peak with a lifetime of approximately 14.2 hours. To maximize the integrated luminosity, the turnaround time (defined as the period in a machine cycle excluding the collision period) should be made as short as possible, preferably short compared to the beam decay time. The initially assumed average 2.4-hour is acceptable, giving an optimized complete cycle time of about 17 hours, but a turnaround time of as little as 0.8 hours would certainly be preferred.

The peak and average luminosities could be raised by allowing the synchrotron damping to lower the transverse emittance and allowing higher but acceptable tune shifts (0.02-0.03). But, if not leveled, the peak luminosities and thus the numbers of interactions per beam crossing could become excessive. Limiting the peak luminosity (leveling) would limit this number, yet still allow an increase in the average luminosity. Using more and closer spaced bunches could reduce the number of interactions per bunch crossing, without lowering the peak luminosities. However, if the beam current is not to be raised, the numbers of protons per bunch must be proportionally reduced, and, if luminosity is to be preserved, the synchrotron damping must be allowed to further lower the emittances, while not increasing the tune shifts.

There are many technical challenges in designing and building the collider, including its injector chain. The two most difficult ones are the development and production of 12-T magnets with Fe-HTS technology, and the beam screen associated with very strong synchrotron radiation. Significant R&D efforts in the coming decade are needed to solve these problems.

### 8.1.4 Compatibility with CEPC and Other Physics Prospects

The present proposal calls for retaining the CEPC collider rings and its full energy booster in the tunnel after SPPC is constructed and brought into operation. This arrangement will enable a future *e-p* collision program. Thus, even when the *e+e-* program is ended, the CEPC may still be operated to provide an electron beam for the *e-p* program. In this case, both CEPC and SPPC colliders will be operated simultaneously.

Housing both CEPC and SPPC, the two largest and most complex particle accelerators in the world, in a common underground tunnel and potentially operating them simultaneously are unprecedented. Thus, there is no prior experience we can learn from. While in principle it is plausible, there are high technical and operational risks associated to such a plan in addition to the complications of machine operation, maintenance and protection. Therefore, we must apply careful consideration and planning at an early stage of the CEPC-SPPC project. First of all, the layouts for the two colliders and CEPC injector should be compatible. This is not easy. One needs to produce several beam bypasses to detour around the large detectors in the two colliders

Heavy ion collisions will be also foreseen in the SPPC, so that a dedicated IP and ion beam acceleration from the injector chain will be planned.

## 8.2 Key Accelerator Issues and Design

### 8.2.1 Main Parameters

#### 8.2.1.1 *Collision Energy and Layout*

To reach the design goal for 75-TeV center of mass energy with a circumference of 100 km, a magnetic field of 12 T is required. This is not far from current state-of-the-art magnet technology using $Nb_3Sn$ superconductors. However, Fe-HTS technology optimistically has raised an expectation of becoming available and much cheaper in 10-15 years, and able to generate a field higher than 20 T in the future. Thus Fe-HTS magnet technology is chosen for SPPC. Even with the large circumference, the arc sections should be designed to be as compact as possible to provide the necessary long straight sections. Although the lattice study is still under way, it is assumed that traditional FODO focusing will be used everywhere, except at the IPs where triplets are used to produce the very small $\beta^*$. The arcs represent most of the circumference, and the arc filling factor is taken as 0.78, similar to LHC [1]. A key issue here is to define the number of long straight sections and their lengths. They are needed to produce those very small beta functions where the large physics detectors are placed, and for hosting the beam injection and extraction systems (abort), collimation systems and RF stations. A total length of about 16.1 km is reserved for the long straight sections, with eight long straight sections of which 2 are 4.3 km long and the 6 others are 1.25 km long. With this configuration, the top beam energy is 37.5 TeV which provides 75 TeV in collision energy. The main parameters are listed in Table 8.2.1.

**Table 8.2.1:** Main SPPC parameters

| Parameter | Value | Unit |
|---|---|---|
| **General design parameters** | | |
| Circumference | 100 | km |
| Beam energy | 37.5 | TeV |
| Lorentz gamma | 39979 | |
| Dipole field | 12 | T |
| Dipole curvature radius | 10415.4 | m |
| Arc filling factor | 0.78 | |
| Total dipole magnet length | 65.442 | km |
| Arc length | 83.9 | km |
| Number of long straight sections | 8 | |
| Total straight section length | 16.1 | km |
| Energy gain factor in collider rings | 17.86 | |
| Injection energy | 2.1 | TeV |
| Number of IPs | 2 | |
| Revolution frequency | 3.00 | kHz |
| **Physics performance and beam parameters** | | |
| Nominal luminosity per IP | $1.0\times10^{35}$ | $cm^{-2}s^{-1}$ |
| Beta function at collision | 0.75 | m |
| Circulating beam current | 0.73 | A |
| Nominal beam-beam tune shift limit per IP | 0.0075 | |
| Bunch separation | 25 | ns |
| Number of bunches | 10080 | |
| Bunch population | $1.5\times10^{11}$ | |
| Accumulated particles per beam | $1.5\times10^{15}$ | |
| Normalized rms transverse emittance | 2.4 | μm |
| Beam life time due to burn-off | 14.2 | hours |
| Total inelastic cross section | 147 | mb |
| Reduction factor in luminosity | 0.85 | |
| Full crossing angle | 110 | μrad |
| rms bunch length | 75.5 | mm |
| rms IP spot size | 6.8 | μm |
| Beta at the first parasitic encounter | 19.5 | m |
| rms spot size at the first parasitic encounter | 34.5 | μm |
| Stored energy per beam | 9.1 | GJ |
| SR power per beam | 1.1 | MW |
| SR heat load at arc per aperture | 12.8 | W/m |
| Energy loss per turn | 1.48 | MeV |

#### 8.2.1.2 *Luminosity*

The initial luminosity (or nominal luminosity) of $1.0\times10^{35}$ $cm^{-2}s^{-1}$ is much higher than in previously built machines such as the Tevatron [2] and LHC [1] and in designs such as SSC [3], HE-LHC [4], and comparable to FCC-hh [5]. While using the same bunch spacing, the number of interactions per bunch crossing is higher than present-day detectors could handle. It is believed, however, that ongoing detector R&D efforts and general technical evolution will be able to solve this problem.

Another important parameter is the average, and thus integrated luminosity. One must consider the loss of stored protons from collisions, the cycle turnaround time, and the shrinking of the transverse emittance due to synchrotron radiation. Emittance shrinkage from synchrotron radiation could maintain or even raise the peak luminosity after the start of collisions, but would eventually also increase the beam-beam tune shift to an unacceptable level. An emittance blow-up system is thus used to counteract the emittance shrinkage, and can be used to limit the tune shift to an acceptable level. Another method to increase the luminosity is to adjust $\beta^*$ during the collisions by taking advantage of emittance shrinking while keeping the beam-beam tune shift constant.

#### 8.2.1.3 *Bunch Structure and Population*

Many bunches with relatively small bunch spacing are desirable for achieving high luminosity operation. However, the bunch spacing can be limited both by parasitic collisions in the proximity of the IPs, and by the electron cloud instability. One also needs to consider the ability of the detector trigger systems to cope with short bunch spacing. We have adopted 25 ns for the nominal bunch spacing at SPPC, just like LHC. The bunch spacing of 25 ns is defined by the RF system in the MSS of the injector chain and preserved from there on. The possibility of shorter bunch spacing will be investigated. Time gaps between bunch trains are needed for beam injection and extraction in both SPPC and the injector chain. Their lengths depend on the practical designs of the injection and extraction (abort) systems, and the rise time of the kickers for beam extraction from SPPC, assumed now to be a few microseconds. The bunch filling is taken to be about 76% of the ring circumference, which is smaller than in LHC and is due to requiring more injection gaps.

Bunch population is first defined in the p-RCS of the injector chain, where the beam from the p-Linac fills the RF buckets using both transverse and longitudinal paintings. With the nominal bunch number and bunch population, the circulation current will be about 0.73 A in the collider rings.

#### 8.2.1.4 *Beam Size at the IPs*

The beam size is determined by the $\beta^*$ of the insertion lattice and the beam emittance. The initial normalized emittance is predefined in the p-RCS of the injector chain and preserved with a slight increase in the course of reaching the top energy of the SPPC due to many different factors such as nonlinear resonance crossings. However, at the top energy of 37.5 TeV and in the later part of the acceleration stage, synchrotron radiation will take effect, with damping times about 2.35 hours and 1.17 hours for the transverse and longitudinal emittances, respectively. This will allow emittances after the collision start, to become significantly smaller than the initial values at the time when the beams reach the top energy. However, the emittances cannot be allowed to fall without limit because of the beam-beam tune shift and luminosity considerations. Thus a stochastic emittance heating system is required to limit the synchrotron radiation cooling and control the emittance level during the collision process.

#### 8.2.1.5 *Crossing Angle*

To avoid parasitic collisions near the IPs producing background for the experiments, it is important to separate the two beams, except at the IPs, using a crossing angle. The crossing angle is chosen to avoid the beams overlapping at the first parasitic encounters at 3.75 m from the IPs when the bunch spacing is 25 ns. At these locations the separation is greater than 12 times the rms beam size. At the SPPC, this crossing angle at the collision energy is about 110 μrad. Compared to head-on collisions, this bunch crossing angle will result in a luminosity reduction of about 15%. With a smaller bunch separation, the crossing angle must be larger, and the luminosity reduction would be greater, but luminosity loss with crossing angles can be recovered when crab cavities are used. The crossing angle may be different at injection due to different lattice settings and larger emittance.

#### 8.2.1.6 *Turnaround Time*

Turnaround time is the total time period in a machine cycle when the beams are out of collision, including the programmed count down checking time before injection, the final check with a pilot shot, the beam filling time with SS beam pulses, the ramping up (or acceleration) time, and the ramping down time. Filling one SPPC ring requires 10 SS beam pulses, which means a minimum filling time of about 14 minutes including pilot pulses. The ramping up and down times are each about 12.4 minutes. Altogether, the minimum turnaround time is 48 minutes. However, the experience at LHC and other proton colliders [6] shows that only about one third of the operations from injection to the top energy are successful and the average turnaround time is closer to 2.4 hours. This is considered acceptable, and with a luminosity run time of 4-8 hours, during which the beams are in collision, it gives a total cycle time of about 7-11 hours.

### 8.2.2 **Key Accelerator Physics Issues**

#### 8.2.2.1 *Synchrotron Radiation*

Synchrotron radiation (SR) power is proportional to the fourth power of the Lorentz factor and the inverse of the radius of curvature in the dipoles, and becomes an important effect for protons at multi-TeV energies using high field superconducting dipoles. With the beam current of 0.73 A and magnetic field of 12 T, the synchrotron radiation power reaches about 12.8 W/m per aperture in the arc sections, about sixty times LHC. The average critical photon energy is about 1.8 keV. There is also synchrotron radiation as the beam passes through the high-gradient superconducting quadrupole magnets.

At SPPC, synchrotron radiation imposes severe technical challenges to the vacuum system and a probable limit on the circulating current. If absorbed at the liquid helium temperature of the magnet bores, the synchrotron radiation's heat load would be excessive, so it must be absorbed at a higher temperature. A beam screen, or other capture system, must be situated between the beam and the vacuum chamber. This limits the beam tube aperture, raising the beam impedance, and/or increases the required superconducting magnet bore radius. The working temperature at the beam screen is a key parameter in the design. The beam screen is also important in controlling the coupling impedance and reducing the electron cloud effect.

Synchrotron radiation also has an important impact on the beam dynamics at, and approaching, the top energy. Without intervention, both the longitudinal and transverse emittances will shrink with lifetimes of about 2.35 and 1.17 hours, respectively. The short damping times may help eliminate collective beam instabilities. One may exploit this feature to enhance the machine performance by allowing the transverse emittances to fall and to increase the luminosity. But nevertheless, to avoid excessive beam-beam tune shift, a stochastic transverse field noise systems will have to be installed to control the emittance reduction.

#### 8.2.2.2 *Intra-beam Scattering*

Intra-beam scattering (IBS) within bunches can couple longitudinal momentum into transverse motion, and it will increase the transverse emittances, or in our case, slow the emittance cooling from synchrotron radiation. With the initial SPPC parameters at the collision energy, IBS has a negligible effect and the lifetime is more than a hundred hours. But as the emittance shrinks from the synchrotron cooling, it becomes significant, and eventually limits the emittance reduction.

#### 8.2.2.3 *Beam-beam Effects*

Beam-beam effects, which could lead to emittance growth, lifetime reduction, and instabilities, have an important effect on the luminosity. These effects come from both head-on interactions and long-range or parasitic interactions. There are several different beam-beam effects affecting the performance: the incoherent beam-beam effects which influence beam lifetime and dynamic aperture; the PACMAN effects which will cause bunch to bunch variation; and coherent effects which will lead to beam oscillations and instabilities.

It is reasonable to choose a conservative beam-beam parameter of 0.0075 for one IP or 0.015 for two IPs, though LHC has reported stable operation with a total value of $\Delta Q_{tot}$~0.03 with 3 interaction points. [7]

#### 8.2.2.4 *Electron Cloud Effect*

The electron cloud (EC) can cause beam instability. The build-up of accumulated photon electrons and secondary electrons has proved to be one of the most serious restrictions on collider luminosity in PEP II, KEKB, LHC, and BEPC. The EC links together the motion of subsequent bunches and induces coupled bunch instability. It also leads to emittance blow-up and luminosity degradation [8]. For next-generation super proton colliders such as SPPC, a bunch population higher than $10^{11}$ and a bunch spacing less than or equal to 25 ns, the EC effect will be critical for reaching the luminosity of $1\times10^{35}$ $cm^{-2}s^{-1}$.

Because of the low-temperature beam pipes for the superconducting magnets at SPPC, the deposited power on the beam screen from the secondary electron multipacting may be a serious issue. The measured deposited power in the dipole magnets of LHC has proven to increase exponentially to about 10 W/m, when the secondary emission yield (SEY) is larger than 1.4. Therefore, SEY at SPPC should be controlled to stay below 1.4 or even 1.2 by coating TiN or NEG on the internal walls of the vacuum chamber and devices inside the vacuum.

### 8.2.2.5 *Beam Loss and Collimation*

Beam losses will be extremely important for safe operation in a machine like SPPC where the stored beam energy will be 9.1 GJ per beam. Beam losses can be divided into two classes, irregular and regular. Irregular beam losses are avoidable losses and are often the result of a misaligned beam or a fault in an accelerator element. A typical example is a RF trip, which causes loss of synchronization during acceleration and collisions. Vacuum problems also fall into this category. Such losses can be distributed around the machine. A well-designed collimator system might collect most of the lost particles, but even a fraction of the lost particles may cause problems at other locations. Regular losses are non-avoidable and localized in the collimator system or on other aperture limits. They will occur continuously during operation and correspond to the lifetime and transport efficiency of the beam in the accelerator. The lowest possible losses are set by various effects, e.g. Touschek effect, beam-beam interactions, collisions, transverse and longitudinal diffusion, residual gas scattering, halo scraping and instabilities.

Halo particles might potentially impinge on the vacuum chambers and get lost. The radiation from the lost particles will trigger quenching of the superconducting magnets, generate unacceptable background in detectors, damage radiation-sensitive devices, and cause residual radioactivity that limits hands-on maintenance. These problems can be addressed by collimation systems which confine the particle losses to specified locations where better shielding and heat-load transfer are provided. For the SPPC, the situation is even more complicated, mainly because extremely high collimation efficiency is required. The cleaning inefficiency at SPPC should be lower than $3.0\times10^{-6}$. To achieve this goal, a more efficient collimation method than the one at LHC is needed.

## 8.2.3 **Preliminary Lattice Design**

### 8.2.3.1 *General Layout and Lattice Consideration*

Different lattice designs are under study, especially for the arcs. Typically, there are three kinds of arc designs, with difference in the dispersion suppression methods. One method uses full bend suppressors which have same number of dipole magnets per cell as the regular arc cells; one method uses half bend suppressors which have half dipole strength per cell as the regular cells; the third uses LHC-like suppressors which have less dipole strength but shorter cell length. The third design is more compatible with an electron-positron collider in the same tunnel and thus is chosen for the baseline design.

### 8.2.3.2 *Arcs*

The arcs are composed of standard FODO cells. The basic phase advance chosen for one FODO cell of SPPC is 90 degrees in both transverse planes. Larger beta functions and dispersions will lead to larger magnet apertures, which critically affects magnet cost. What's more, the dispersion function also influences the momentum collimation design.

The magnet arrangement and beam optics of one standard FODO cell in the present design are shown in Fig. 8.2.2. The horizontal beta function and dispersion function have their maximum values at the middle quadrupole. There are 12 bending magnets in the cell, each with length 14.45 m and strength 11.8 T. The quadrupole length is 6 m. The length of a standard FODO cell is 213.4 m, and there are 44 such cells in each arc.

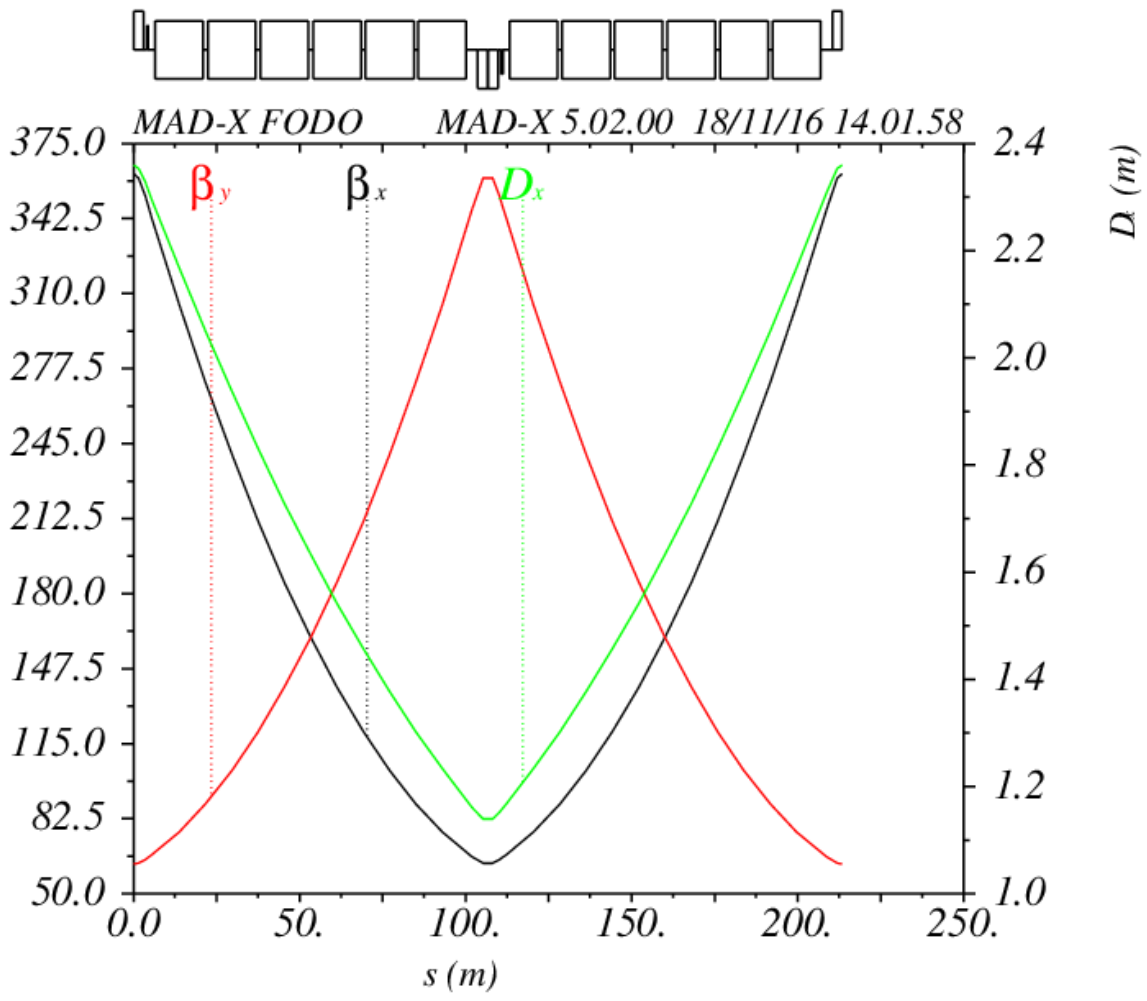

**Figure 8.2.2:** Lattice functions of a standard arc FODO cell.

### 8.2.3.3 *Dispersion Suppressor*

The dispersion suppressor (DS) matches the dispersion functions and beta function between the arc and adjacent long straight section. Besides matching the optics, the dispersion suppressor of SPPC is designed to have the ability to slightly adjust the layout of SPPC to meet the compatibility requirement between CEPC and SPPC. The left part of the dispersion suppressor is shown in Fig. 8.2.3, and the right part is similar. It is composed of two FODO cells, which are shorter than a standard arc cell, with only 4 dipoles or empty in each half cell. There is a drift space in the first half DS cell which is designed to be flexible, from 30 m to 80 m. The last half cell is not filled with dipoles to make betatron function matching easier.

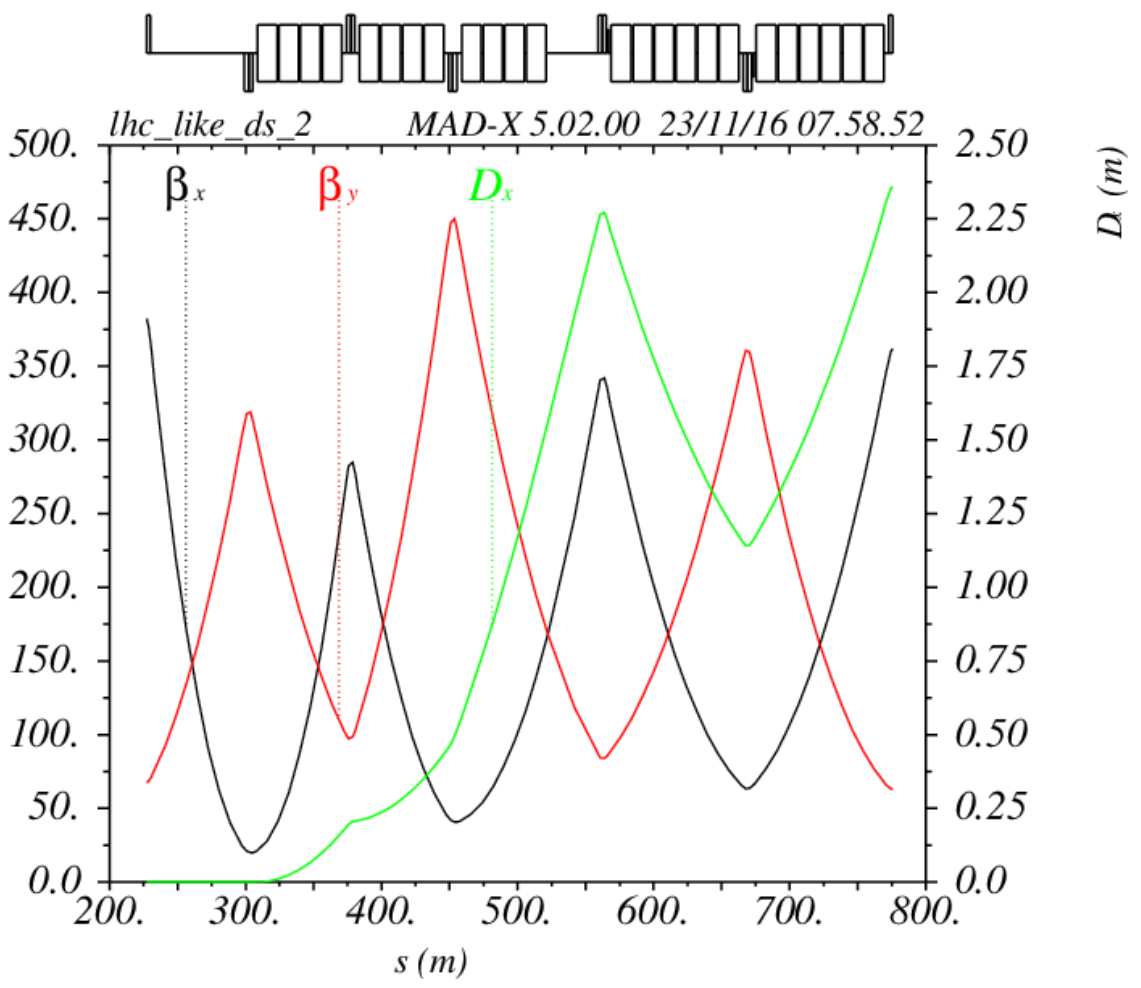

**Figure 8.2.3:** Left part of the dispersion suppressor.

### 8.2.3.4 *High Luminosity Insertions*

The long straight sections LSS3 and LSS7 are for p-p collisions. The identical lattice structures of LSS3 and LSS7 are anti-symmetrical, so a crossing angle at the IPs can be produced and the beams go from outer ring to inner ring or reverse. Preliminary lattice designs for LSS3 and LSS7 are shown in Fig. 8.2.4.

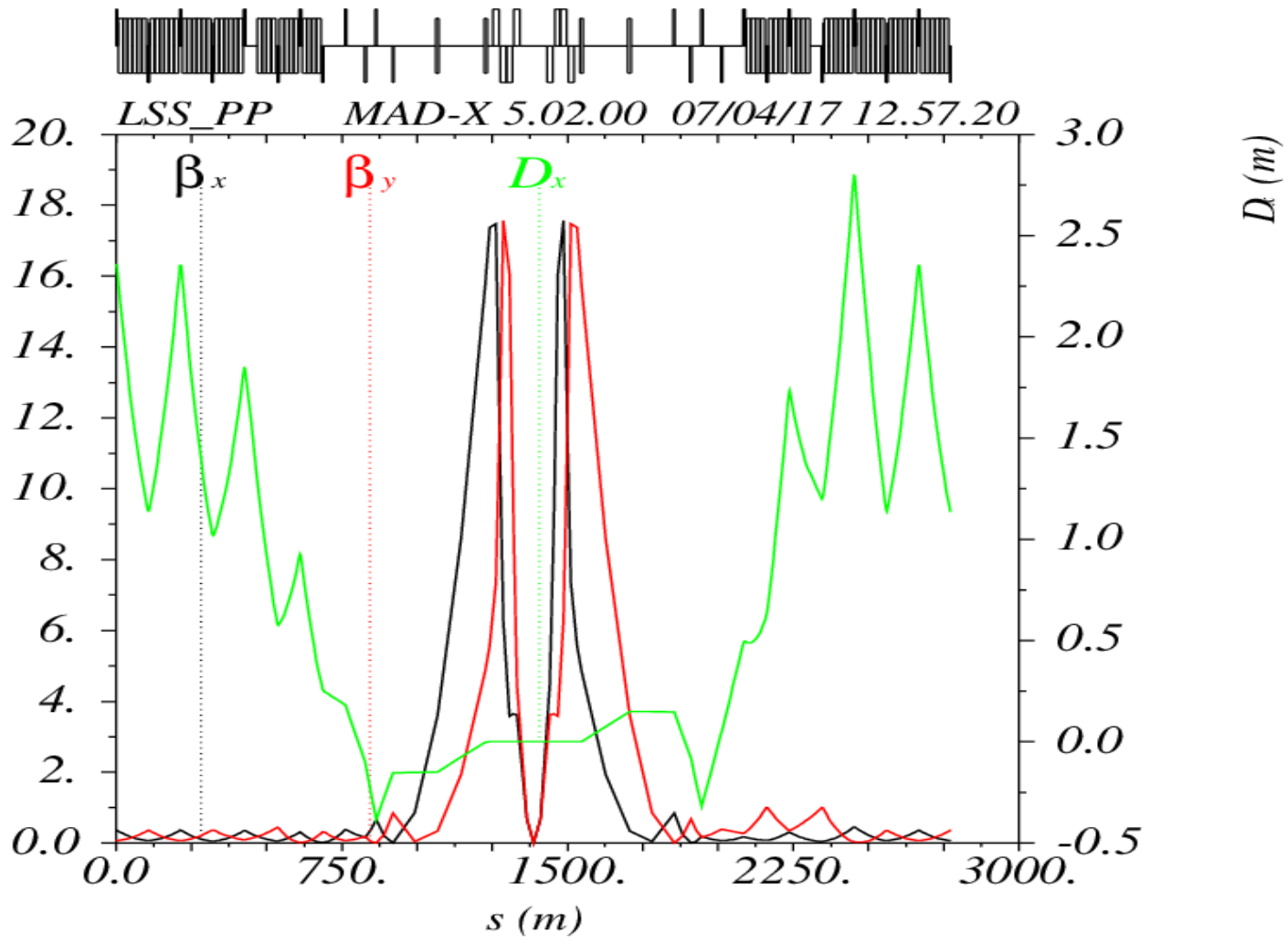

**Figure 8.2.4:** Lattice function in the LSS3 and LSS7

### 8.2.3.5 *Dynamic Aperture*

To carry out a preliminary study of dynamic aperture, a full lattice is required, as in Fig. 8.2.5 where periodic cells are used for six other long straight sections except the IPs. A preliminary dynamic aperture study was performed using SIXTRACK. Chromaticity correction magnets and dipole field errors were used in the tracking. Both the dynamic apertures at the collision energy and injection energy seem to be acceptable, with the former shown in Fig. 8.2.5, about 10 $\sigma$ in both the horizontal and vertical directions.

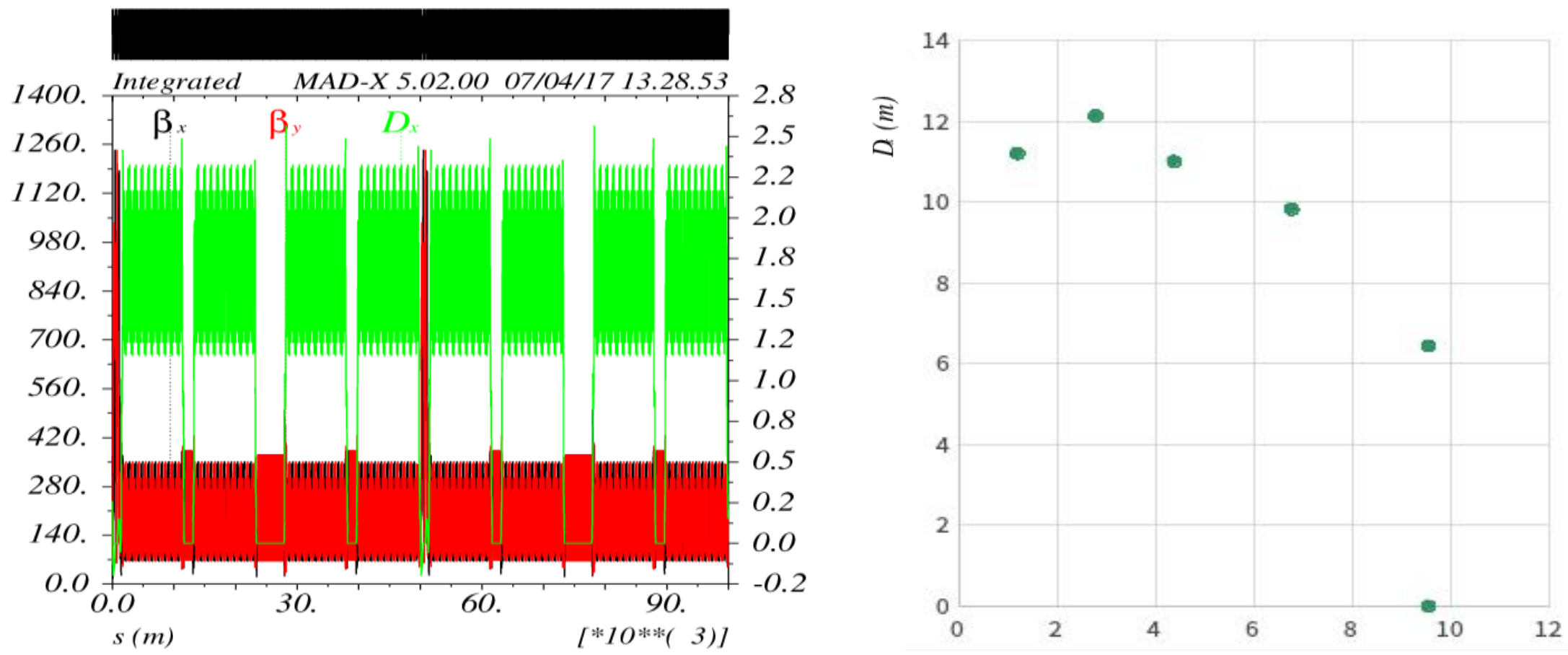

**Figure 8.2.5:** Lattice functions of a full lattice (left) and dynamic aperture (right) at top-energy

## 8.2.4 Luminosity and Leveling

The initial (or nominal) luminosity of $1.0\times10^{35}$ $cm^{-2}s^{-1}$ is modest for a next-generation proton-proton collider and comparable to FCC-hh [5, 9-10] and lower than in the HL-LHC [11]. This design also allows for a future luminosity upgrade.

Besides the synchrotron radiation power limits on the circulation current and luminosity, the number of interactions per bunch crossing is also a limit to the luminosity. It is believed that ongoing R&D efforts on detectors and general technical evolution will be able to solve the data pile-up problem. On the other hand, it is important to increase the average, and thus integrated luminosity while maintaining the maximum instantaneous luminosity [12]. Thus a luminosity leveling scheme should be used. By taking into account the loss of stored protons from collisions, cycle turnaround time, shrinking of the transverse emittance due to synchrotron radiation, and the beam-beam shift, one can design different leveling schemes, as shown in Fig. 8.2.6. An emittance blow-up system is needed to control the emittance shrinkage. Another method to increase the luminosity is to adjust *β** during the collisions by taking advantage of emittance shrinking while keeping the beam-beam tune shift constant.

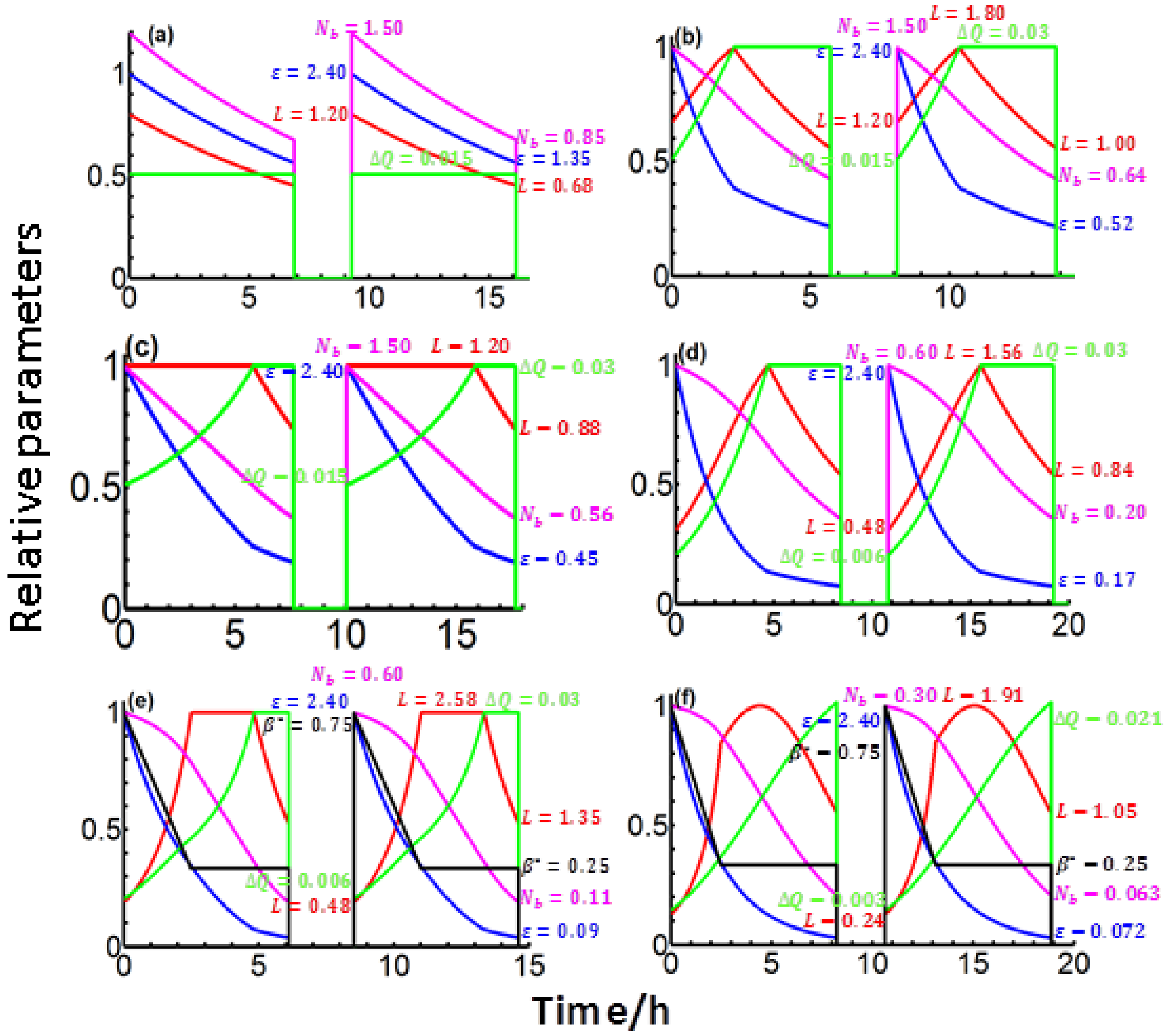

**Figure 8.2.6:** Evolution of parameters vs time with a turnaround time of 2.4 hours and bunch spacing of 25 ns. Red: luminosity, magenta: number of protons per bunch, blue: transverse emittance, green: beam-beam tune shift, black: β* at the IP. (a) with fixed tune shift; (b) allowing the tune shift to rise to 0.03; (c) as in (b) but with the luminosity "leveled" at its initial value; (d) as in (c) but bunch spacing of 10 ns; (e) as for (d) but reducing β* in proportion to emittance down to 25 cm; (f) as for (e) but with bunch spacing of 5 ns. In plots a), b), c) and d), β* is kept constant at the nominal 0.75 m.

### 8.2.5 Collimation Design

The proposed collimation method at SPPC is to have both the betatron and momentum collimation systems in the same insertion. In this way the momentum collimation system cleans the particles with significant energy loss in the transverse collimators due to the Single Diffractive effect (SDE) [13] Otherwise those particles would be lost in the downstream cold region. Our studies show that this method is very efficient to improve the cleaning efficiency. This is different than the momentum collimation section at the LHC [14-15] where dispersion is intentionally designed to be non-zero between the two adjacent DS sections; here it is required to have an achromatic end between the momentum collimation and the transverse collimation sections. Besides, it provides the indispensible protection from beam loss and radiation for the cold dipole and quadrupole magnets. Fig. 8.2.7 illustrates the betatron and dispersion functions of the full cleaning insertion. The related parameters are listed in Table 8.2.3.

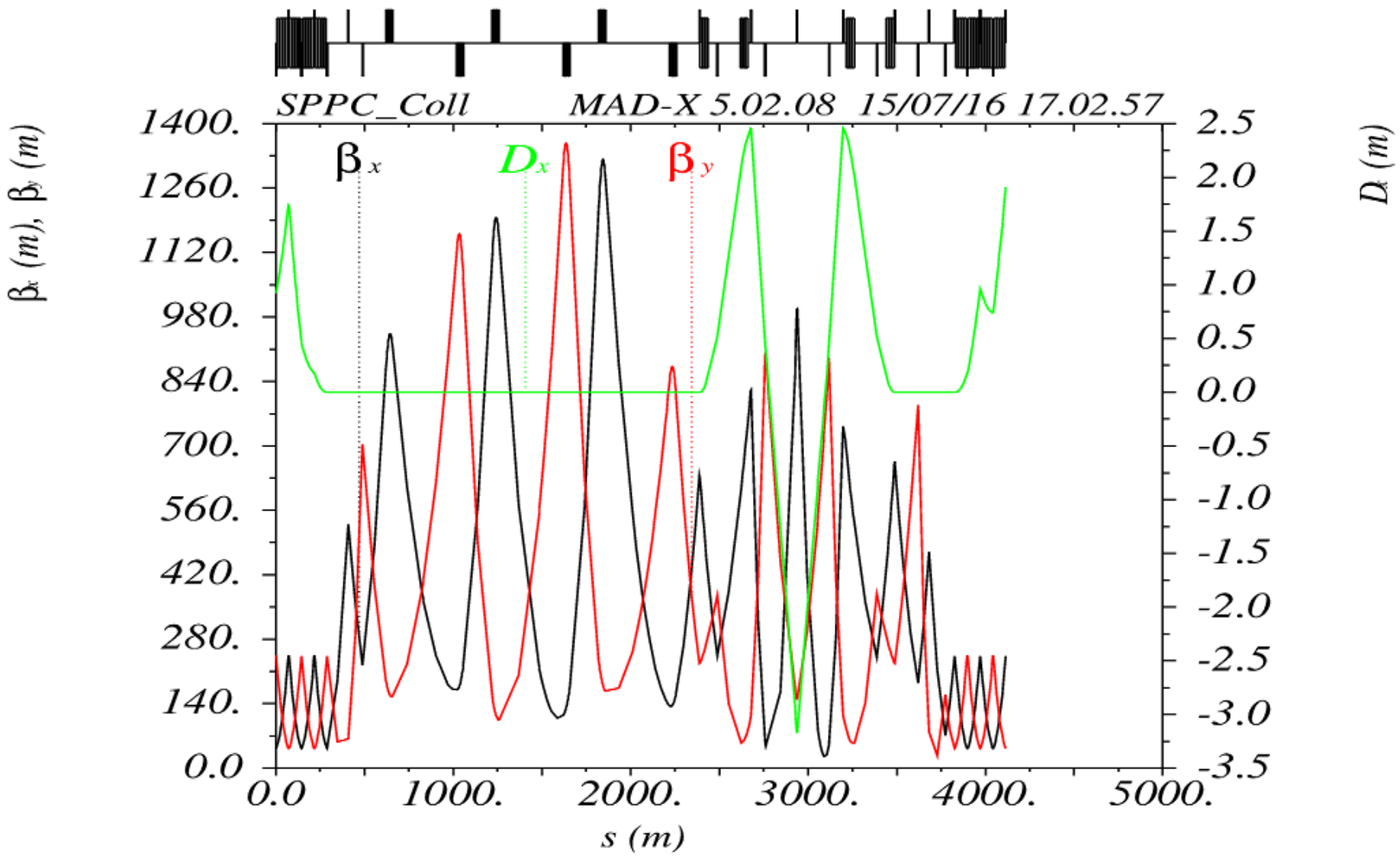

**Figure 8.2.7:** The betatron and dispersion functions of the cleaning insertion

**Table 8.2.3:** Basic parameters of the collimation insertion

| Parameter | value | Unit |
|---|---|---|
| Total cleaning insertion length | 4.2 | km |
| Length of betatron / momentum collimation section | 2.6 / 1.6 | km |
| Horizontal phase advance of betatron / momentum collimation section | 1.87π /2.25π | rad |
| Warm/cold quadrupole length | 3.3/4.62 | m |
| Warm quadrupole strength | 0.00014 | $m^{-2}$ |
| Dipole length in the momentum cleaning section | 14.62 | m |
| Number of dipoles per group | 4 | |
| Number of dipole groups | 4 | |
| Dipole field | 16 | T |
| Maximum beta function ($\beta_x$ / $\beta_y$) | 1819/1896 | m |

Multi-particle simulations using the lattice parameters and collimator settings have been carried out with the code Merlin. The initial settings for the betatron collimator are similar physical gaps and phase advances as the LHC, chosen to verify the effectiveness of the novel collimation method. To increase the simulation efficiency with a huge number of particles ($10^8$), he initial beam distributions are chosen as a ring type distribution in the horizontal plane and a Gaussian distribution in the vertical plane. According to the positions of the lost particles, three protective collimators in Tungsten with the aperture just the same as the primary momentum collimator intercept the particles. In this study, the maximum energy spread of particles which can pass through the primary collimator is about 0.1 %. As shown in Fig. 8.2.8, one can see that the collimation method has extremely low cleaning inefficiency in the downstream DS regions, at least better than $5\times10^{-7}$, which can meet the SPPC requirement.

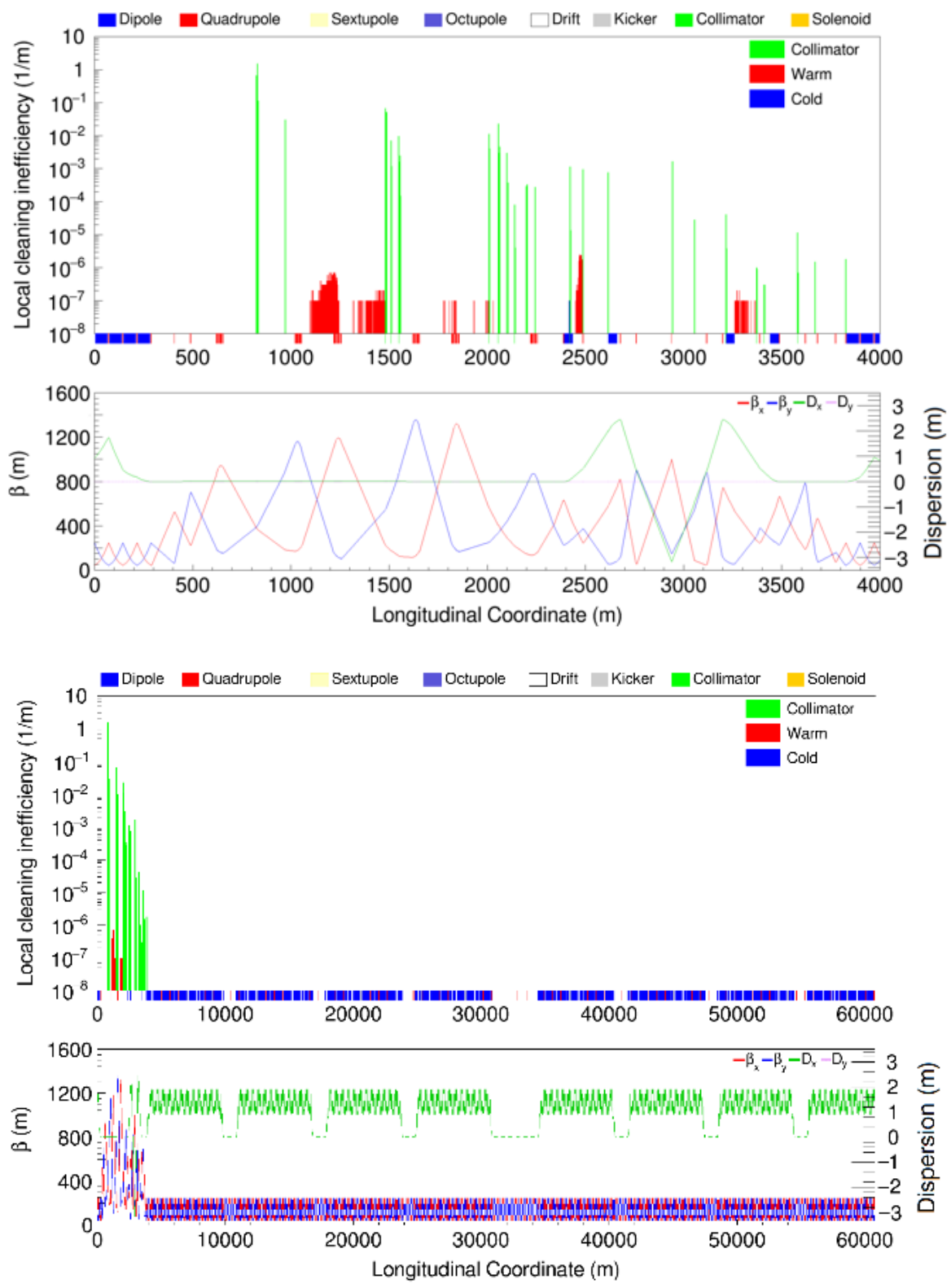

**Figure 8.2.8:** Beam loss distribution with energy spread $\delta$ = 0.1 % (Left: the cleaning insertion; right: the whole ring)

In order to improve the transverse collimation efficiency, superconducting quadrupoles with relatively low fields and large apertures are used to provide more phase advance in the section so that one more collimation stage can be added. Some protective collimators will have to be used to protect the quadrupoles from high radiation dose in the region. The first study shows promising results, both in collimation efficiency and radiation dose reduction in the quadrupoles.

### 8.2.6 **Cryogenic Vacuum and Beam Screen**

#### 8.2.6.1 ***Vacuum***

SPPC has three vacuum systems: insulation vacuum for the cryogenic system; beam vacuum for the low-temperature sections; and beam vacuum for the chambers in the room-temperature sections.

#### *8.2.6.1.1 Insulation Vacuum*

The aim here is to avoid convective heat transfer and there is no need for high vacuum. The room-temperature pressure in the cryostats before cool-down does not have to be better than 10 Pa. Then, so long as there is no significant leak, the pressure will stabilize around $10^{-4}$ Pa, when cold. As a huge volume of insulation vacuum is required, careful design is needed to reduce the cost.

#### *8.2.6.1.2 Vacuum in Cold Sections*

In principle, with HTS magnets the cold bore temperature can go higher than 4 K. Compared to LTS magnets this saves the cost of a huge cryogenics system. However, very high vacuum is required to limit beam loss or beam quality deterioration. This imposes a critical limitation in selecting the temperature, as the $H_2$ pumping speed is strongly related to the temperature. Currently we are considering either to adopt the conventional temperature of 1.9 K as used at LHC or to exploit a more aggressive solution with 3.8 K. In the latter case, an auxiliary pumping system such as cryo-absorbers used at LHC is required [16-17]. Further study is under way.

In interaction regions or around experiments where superconducting quadrupoles are used, the vacuum has to be very good (less than $10^{13}$ $H_2$ per $m^3$) to avoid creating background in the detectors. But the beams are straight here and there is relatively little synchrotron radiation.

In the arcs, the requirement is based on the beam lifetime, which depends on the nuclear scattering of protons on the residual gas. To ensure a beam lifetime of about 100 hours, the equivalent hydrogen gas density should be below $10^{15}$ $H_2$ per $m^3$. The problem here is the huge synchrotron radiation power. If allowed to fall directly on the magnet bore at the magnet temperature of 3.8 K, the wall power needed to remove it would be grossly too high. It has to be intercepted on a beam screen, which works at a higher temperature, e.g. 40-60 K and is located between the beam and cold bore (see below). This screen, at such a temperature, will desorb hydrogen gas, particularly if it is directly exposed to synchrotron radiation. The space outside the screen will be cryopumped by the low temperature of the bore. Slots must be introduced in the shield to pump the beam space.

#### *8.2.6.1.3 Vacuum in Warm Sections*

The warm regions are used to house the beam collimation, injection, and extraction systems. They use warm magnets to avoid quenching from the inevitable beam losses in these locations. They have difficult vacuum pumping requirements due to desorption from the beam losses. Non-Evaporable Getter (NEG) is probably required. These sections are of limited overall length or much shorter than the cold sections.

### 8.2.6.2 ***Beam Screen***

The main function of a beam screen is to shield the cold bore of the superconducting magnets from Synchrotron Radiation (SR) [18]. The estimated SR power is about 12.8 W/m per aperture in the arc dipoles. This is much higher than the 0.22 W/m at LHC [19], and greatly increases the difficulty of the beam screen design. The beam screen design is a compromise to extract the heat load, minimize the occupation of the bore aperture, provide vacuum pumping, reduce coupling impedance and mitigate the electron cloud effect. An ideal design separately addresses these functions. . The screen itself which encircles the beam, with the slot on the outer side would be run at a relatively lower temperature to control the impedance, while the absorption structures which synchrotron radiation penetrates through the slot would be at a higher temperature to minimize the wall power needed to extract the synchrotron radiation power.

The operating temperature of the screen must be high enough to avoid excessive wall power needed to remove the heat, but not too high to avoid excessive resistivity of the high-temperature superconducting material or copper coating on its inside surfaces, leading to excessive impedance, and to avoid radiating too much power to the 3.8 K bore at.

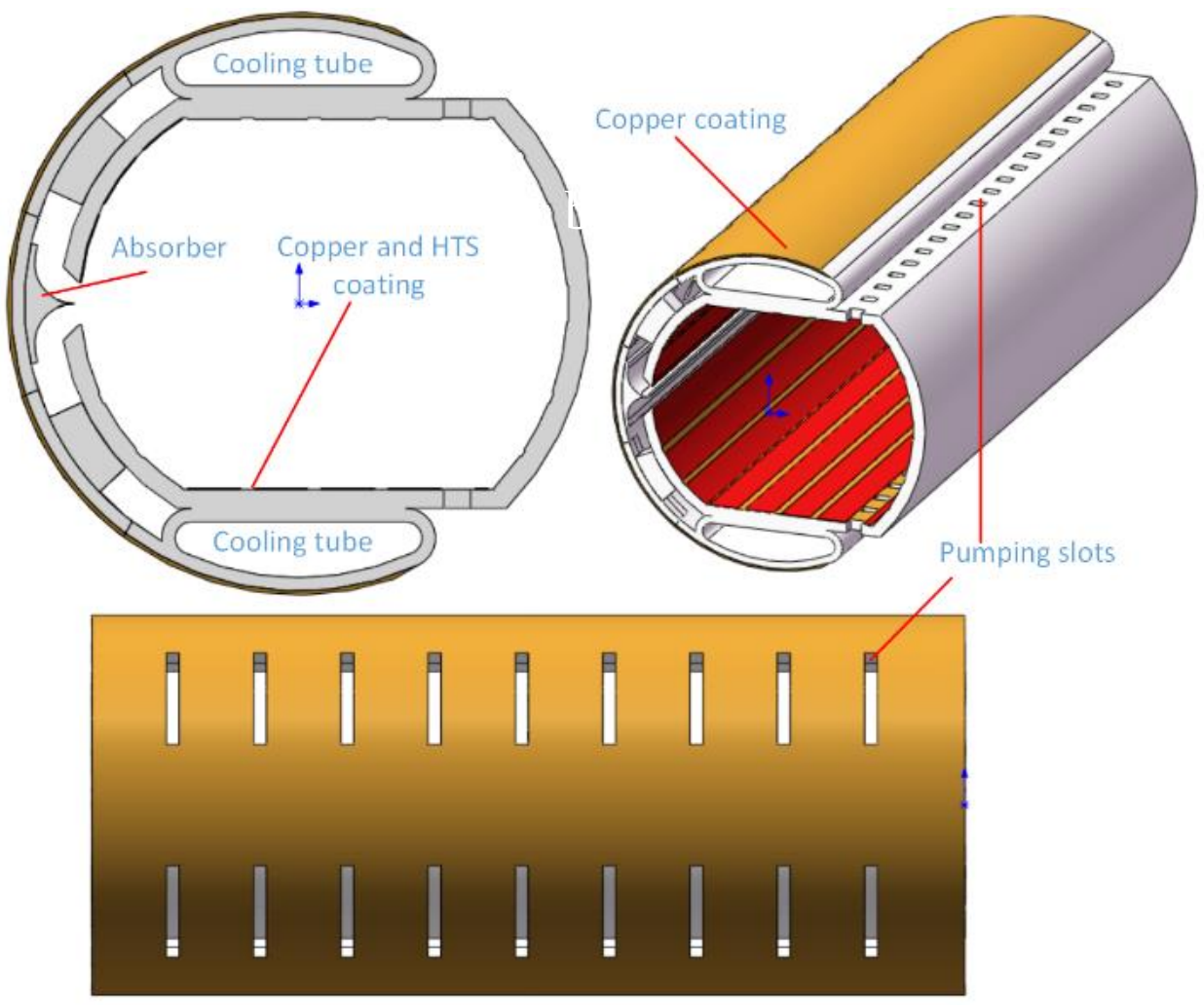

**Figure 8.2.9:** Beam screen schematic with inner HTS coating

The design must satisfy requirements of vacuum stability, mechanical support, influence on beam dynamics and refrigeration power. Fig. 8.2.9 is a schematic for a basic beam screen under consideration at SPPC, similar to the idea developed at FCC-hh,

The main challenges for the beam screen  are:

### *I. Synchrotron Radiation*

The inside of the screen is coated with a layer of high-temperature superconducting material (e.g. Fe-HTS or YBCO) or copper to reduce the resistive impedance. At higher temperatures this impedance, from its higher electrical resistivity, will be increased, leading to worse collective beam instability. The operating temperature is also constrained to limit heat radiation and conduction to the cold bore, and by considerations of desorption. The operating temperature should be chosen carefully. Different refrigerants can be considered, such as liquid neon or liquid oxygen.

### *II. Electron cloud*

A proper beam screen structure can restrain the generation of photo-electrons feeding an electron cloud. The proposed FCC-hh design has a slit in the outer mid-plane of the screen, and curved interception surfaces that reflect the synchrotron radiation up or down into confined absorption structures where photo-desorption is not a problem. The disadvantages of the design are that it occupies more aperture and increases the difficulty of the mechanical structure. Even without direct synchrotron radiation, the inner screen’s surface should be coated by a thin film of low secondary electron yield to reduce electron production.

***III. Vacuum***

Vacuum in the beam screen will depend on several factors: the beam structure, the beam energy, the beam population, the critical photon energy and synchrotron radiation power. Beam structure has an important effect on the buildup of the electron and ion clouds which may lead to vacuum instability. Pumping speed is the dominant factor for vacuum stability. The beam screen must be designed with sufficient transparency to the cold vacuum duct to retain an effective pumping speed. However, good transparency obtained by adding more slots will increase the resistive impedance which may cause beam instabilities.

***IV. Magnet quenches***

The beam screen should have sufficient strength to resist the pulsed electromagnetic forces generated by a magnet quench [20]. Stainless steel can also be used as the base structure material, reducing such forces, but a thick copper film of 75 μm coated on the base to decrease the wall impedance produces a strong source of electromagnetic force. The thinner the film, the smaller the force, but the higher the resistive impedance. Coating of a high-temperature superconducting film looks like a good solution, though it increases the technical difficulty.

***V. Impedance***

Together with the inner surface coating, the shape and size of the beam screen structure needs to be optimized in order to decrease the transverse wall impedance.

### 8.2.7 **Other Technical Challenges**

Besides the two key technologies, high-field magnets and vacuum/beam screens, there are other important technologies requiring development in the coming decade in order to build SPPC. Among them are the machine protection system that requires extremely high efficiency collimation, and a very reliable beam abort system. These are important for dumping the huge energy stored in the circulating beams, when a magnet quenches, or another abnormal operating condition occurs. If the extraction system has to be installed in a relatively short straight section, one has to develop more powerful kickers.

A complicated feedback system is required to maintain beam stability. The beam control system also controls emittance blow-up which is important for controlling beam-beam induced instabilities and for leveling the integrated luminosity.

Beam loss control and collimation in the high-power accelerators of the injector chain pose additional challenges. A proton RCS of 10 GeV and a few MW are still new to the community, and needs special care. The gigantic cryogenic system for magnets, beam screens and RF cavities also needs serious consideration.

## 8.3 High-field Superconducting Magnet

SPPC needs thousands of 12~20 T (upgrade phase) dipole and quadrupole magnets to bend and focus the beams. The nominal aperture in these magnets is 40~50 mm with field uniformity of $10^{-4}$ attained in at least 2/3 of the aperture radius. The magnets will have two beam apertures of opposite magnetic polarity within the same yoke to save space and cost. The currently assumed distance between the two apertures in the main dipoles is 200~300 mm, but this could be changed based on detailed design optimization to control cross-talk and considering the overall magnet size. The outer diameter of the main dipole and quadrupole magnets should not be larger than 900 mm, so that they can be placed inside cryostats having an outer diameter of 1500 mm. The total magnetic length of the main dipole magnets is about 65.4 km out of the total circumference of 100 km. If the length of each dipole magnet is about 15 m, then about 4360 dipole magnets are required [1, 2].

High gradient quadrupoles for SPPC are divided into three groups:

1) at the IPs with single aperture, diameter D = 60 mm, and $B_{pole}$ = 12 T;
2) in the matching section, D = 60 mm, $B_{pole}$ = 12 T;

3) in the arcs, D = 45 mm, $B_{pole}$=12 T.

The ones in the matching sections and arcs are 2-in-1 yoke-sharing magnets.

All the superconducting magnets used in existing accelerators are based on NbTi technology. These magnets work at significantly lower field than the required 12~20 T, e.g., 3.5 T at 4.2 K at RHIC and 8.3 T at 1.9 K at LHC [3, 4]. There are a total of 4 coil configurations which can provide a dipole field: cos-theta type [5], common coil type [6], block type [7] and canted cos-theta type [8]. Among these the common coil type is the simplest structure. The coils have larger bending radius and there is less strain. The critical current density $J_c$ of both $Nb_3Sn$ and HTS will be greatly reduced by high strain level. Thus, the common coil configuration has been chosen for the SPPC dipoles.

### 8.3.1 Conceptual Design Study of 12-T Iron-based HTS Dipole Magnet

A conceptual design study of 12-T 2-in-1 dipole magnets has been carried out with the assumption that the $J_e$ level of IBS in 10 years will increase about 10 times higher than the present level. This is shown in Fig. 8.3.1. Besides significant improvement in $J_e$, we are also expecting that the IBS superconductor will have better mechanical performance than present high field conductors like $Nb_3Sn$ and Bi-2212, and also significantly lower cost.

The magnet aperture is 45 mm diameter. The main field is 12 T in the two apertures with $10^{-4}$ field uniformity. The common-coil configuration is adopted because of its simple structure and easy fabrication. Two types of coil ends are considered and compared for the field quality and structure optimization: soft-way bending and hard-way bending. For hard-way bending the coil is wound with flared ends and in a way to minimize the amount of superconductors. The main parameters, coil layouts and the field quality optimization of this design is summarized here.

A study of two coil layouts, #1 and #2, have been completed, as shown in Fig. 8.3.2. The main parameters are listed in Tables 8.3.1 and 8.3.2. The minimum bending radius of the cables is around 80 mm. The outer diameter of the magnet is 620 mm and the inter-aperture spacing is 236-258 mm. For #1, we put 4 coil blocks with 8 turns per block in the inner two layers, 4 coil blocks with 21 turns per block in the middle and outside. With a current of 9400 A, we obtain 12 T main field in the aperture and 12.78 T peak field in the coils. For #2 design, there are 4 coil blocks with 4 turns per block in the inner two layers, 2 coil blocks with 33(16+17 for gap) turns per block in the middle and 2 coil blocks with 28(14+14) turns per block on the outside. We can get a 12 T main field and 12.85 peak field with a current of 8100 A. Field distributions of the two designs are shown in Fig. 8.3.3. The operating margin is 21% at 4.2 K for the two designs.

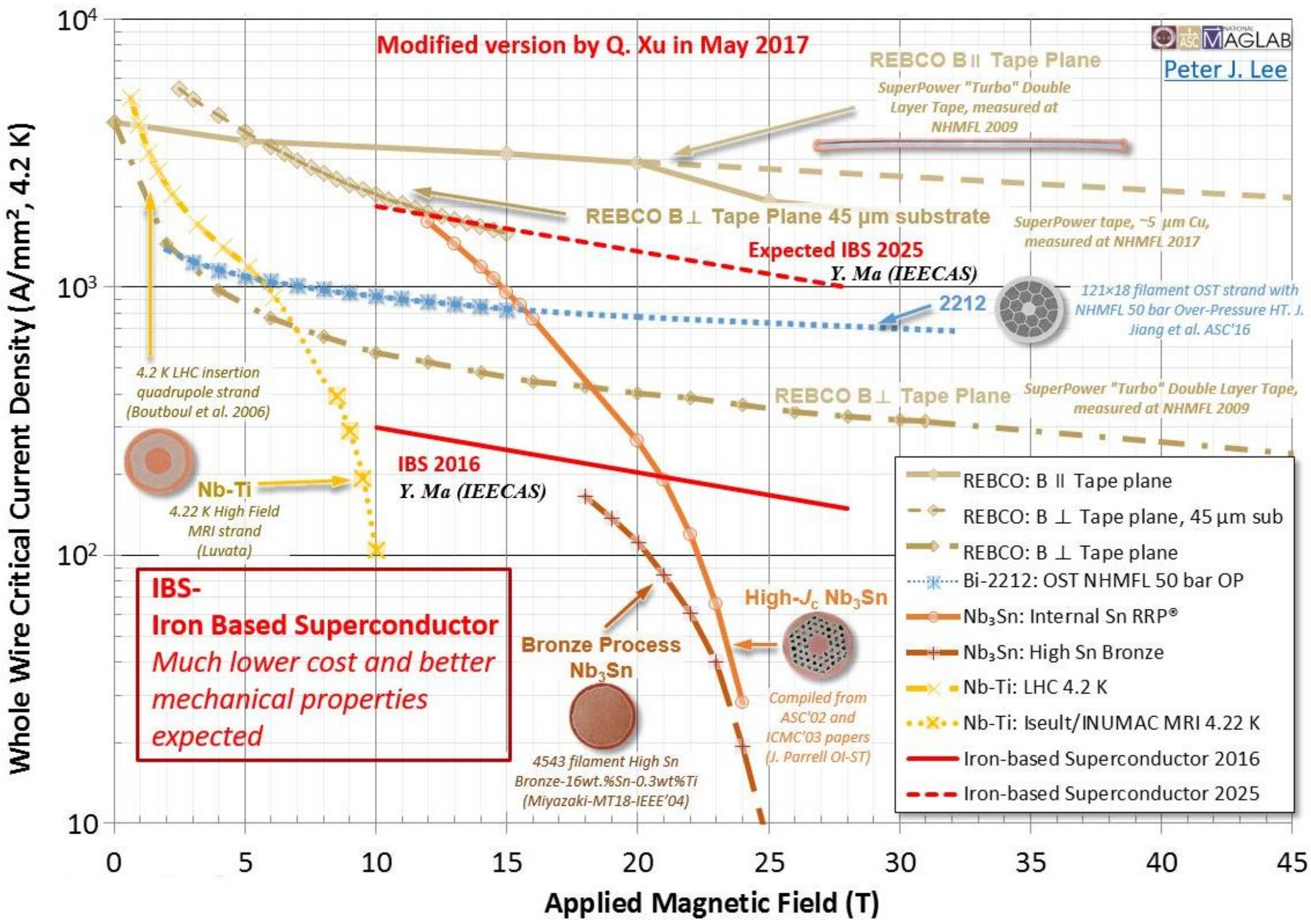

**Figure 8.3.1.** $J_e$ of IBS in 10 years compared to other practical materials

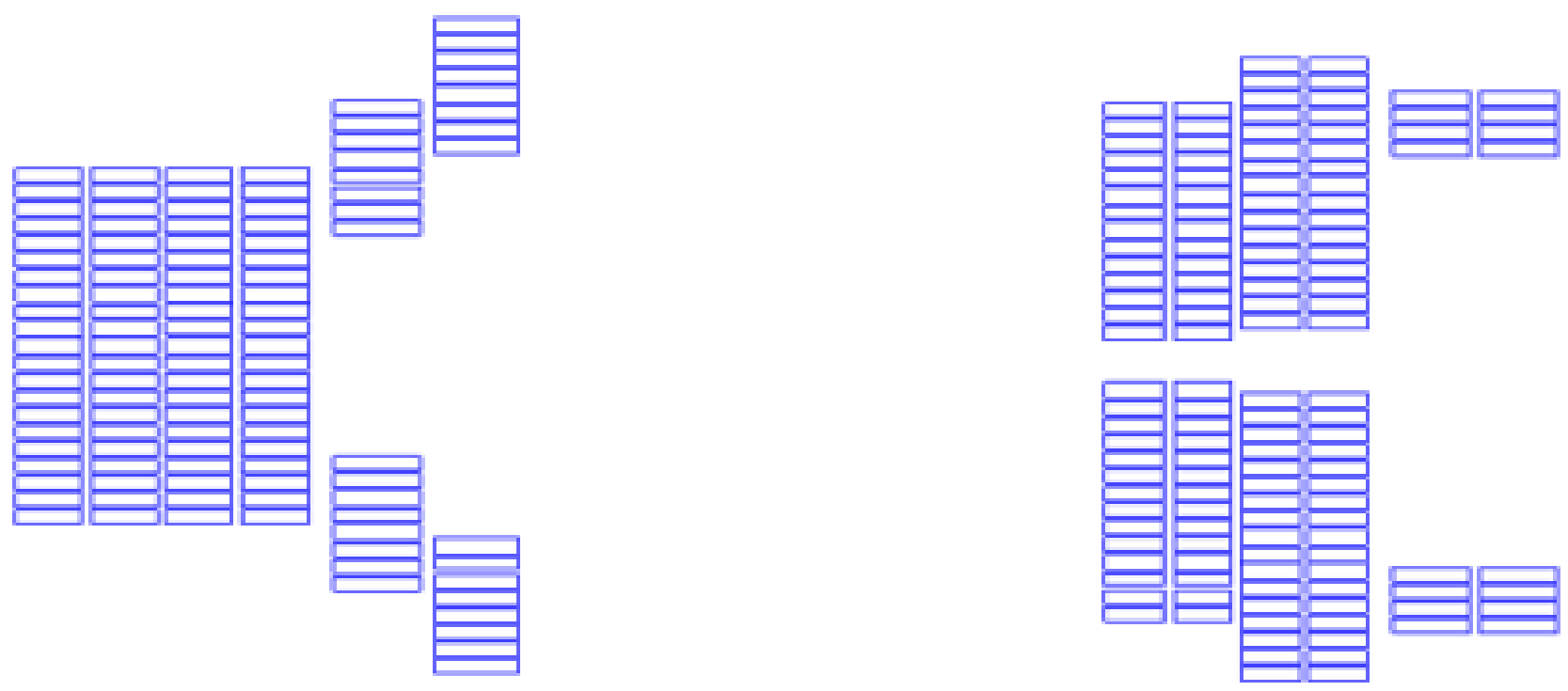

**Figure 8.3.2:** Left: the #1 coil cross section in the first quadrant. Right: the #2 coil cross section in the first quadrant.

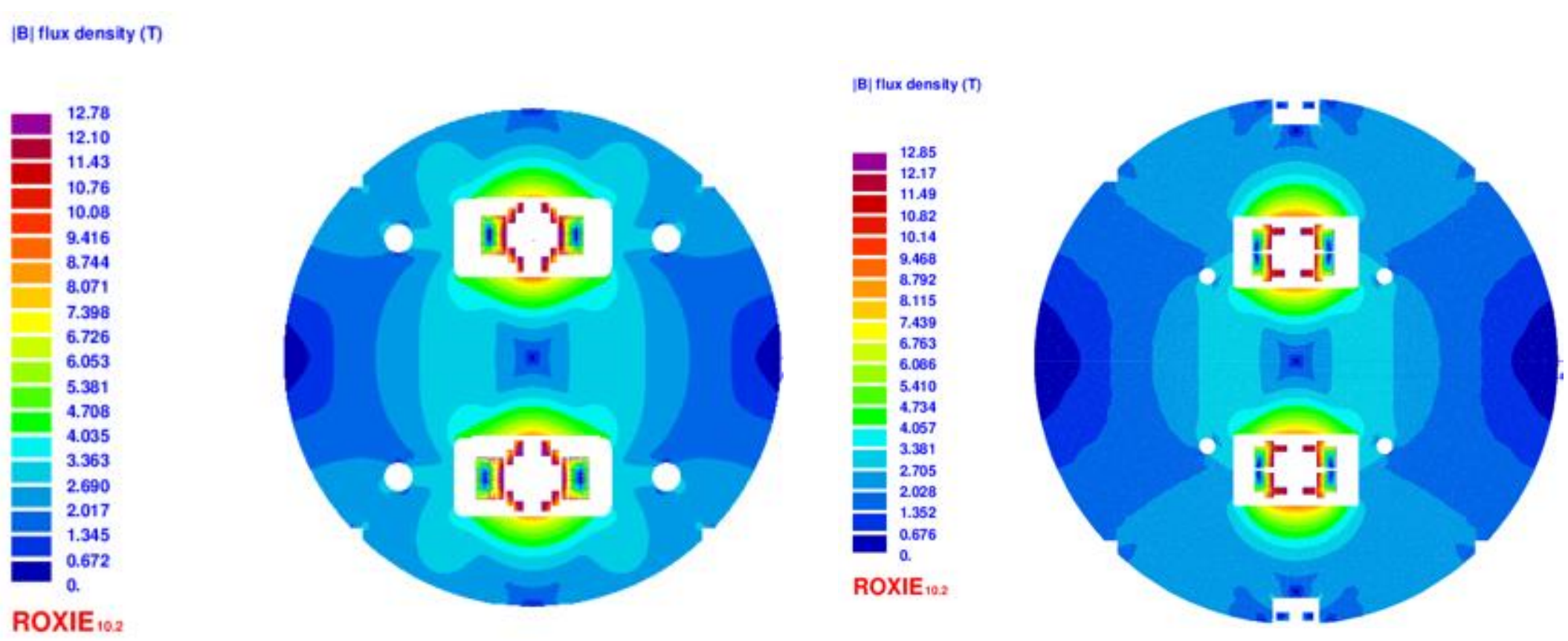

**Figure 8.3.3:** Left: field distribution of the #1 design. Right: Field distribution of the #2 design.

**Table 8.3.1:** Main parameters of the 12-T iron-based dipole magnet

| Parameter | Unit | Value |
|---|---|---|
| Number of apertures | - | 2 |
| Aperture diameter | mm | 45 |
| Inter-aperture spacing | mm | 236/258 |
| Operating current | A | 9400/8100 |
| Operating temperature | K | 4.2 |
| Load line ratio | / | 79% |
| Main field in the aperture | T | 12 |
| Coil peak field | T | 12.78/12.85 |
| Number of iron-based coils | - | 6 |
| Outer diameter of the magnet | mm | 620 |
| Minimum bending radius | mm | 84.97/77.27 |

**Table 8.3.2:** Cables and strands parameters for #1 and #2 dipole

| Parameter | Unit | Inner | Middle | outer |
|---|---|---|---|---|
| Number of layers | - | 2 | 2 | 2 |
| Number of turns per layer | - | 8/4 | 21/16+17 | 21/14+14 |
| Cable width | mm | 7.2/6.6 | 5.6/5 | 5.6/5 |
| Cable inner height | mm | 1.5 | 1.5 | 1.5 |
| Cable outer height | mm | 1.5 | 1.5 | 1.5 |
| Number of strands | - | 18/16 | 14/12 | 14/12 |
| Insulation thickness | mm | 0.15 | 0.15 | 0.15 |
| Strand diameter | mm | 0.802 | 0.802 | 0.802 |
| Copper to superconductor ratio | - | 1 | 1 | 1 |
| Reference field | T | 10 | 10 | 10 |
| Jc @ reference field, 4.2K | $A/mm^2$ | 4000 | 4000 | 4000 |
| dJc/dB | $A/mm^2$ | 111 | 111 | 111 |

We bend the #1 design in the hard-way and #2 design in the soft-way, as shown in Fig 8.3.4. For design #1, we choose to bend the upper two blocks in hard-way to save conductors and make space for the beam pipe. Hard-way bending parts are on an ellipsoid with a 5 degree inclination angle to decrease the influence to field quality. For design #2, we bend the inner two blocks upward and all the other coil blocks in the soft-way. By optimizing lengths of coil straight sections one can achieve a $10^{-4}$ integrated field quality along the axis. Fig. 8.3.5 shows field harmonic variation along the axis for the design #2.

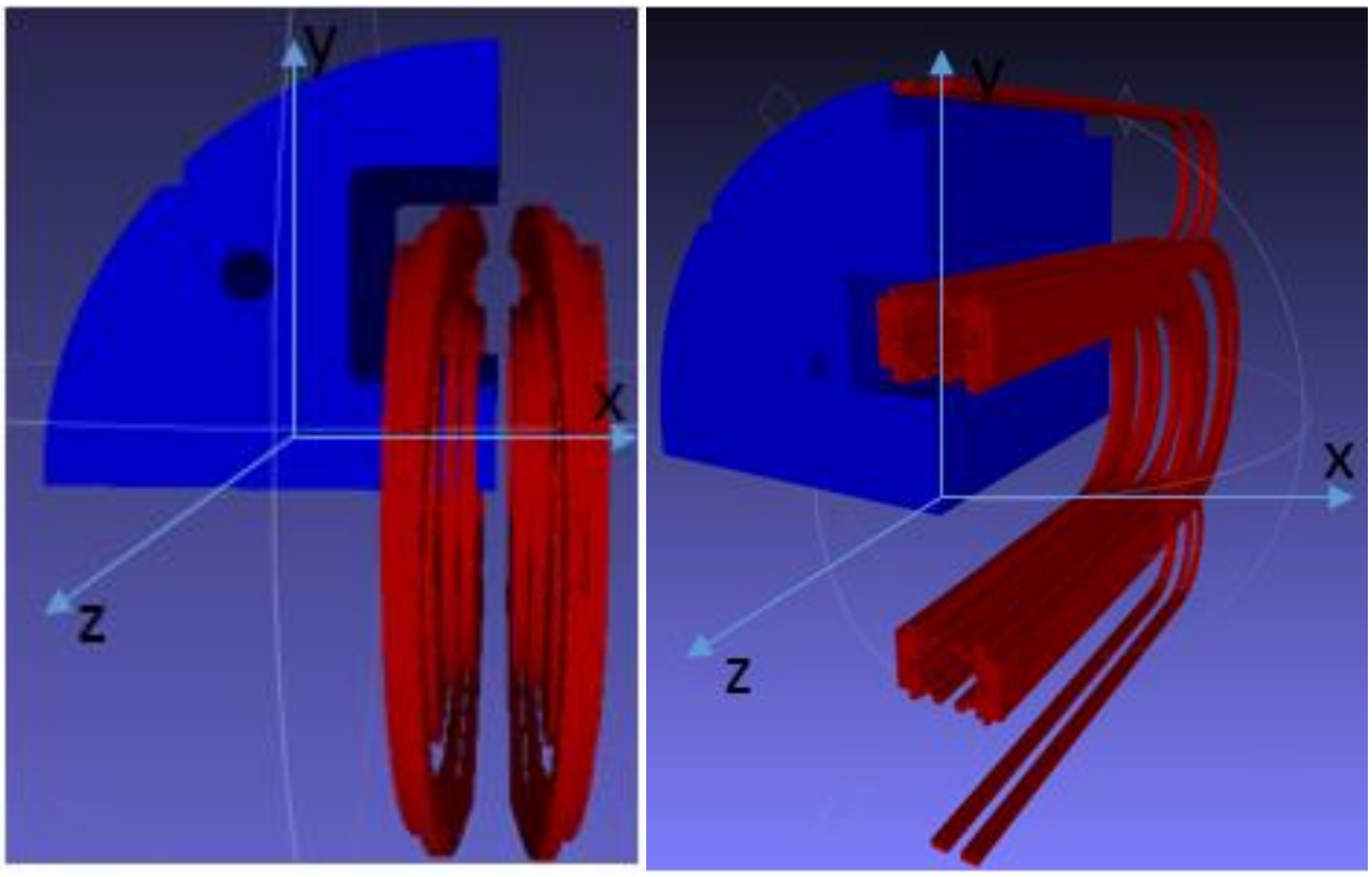

**Figure 8.3.4:** Left: the layout of hard-way coil ends. Right: the layout of soft-way coil ends

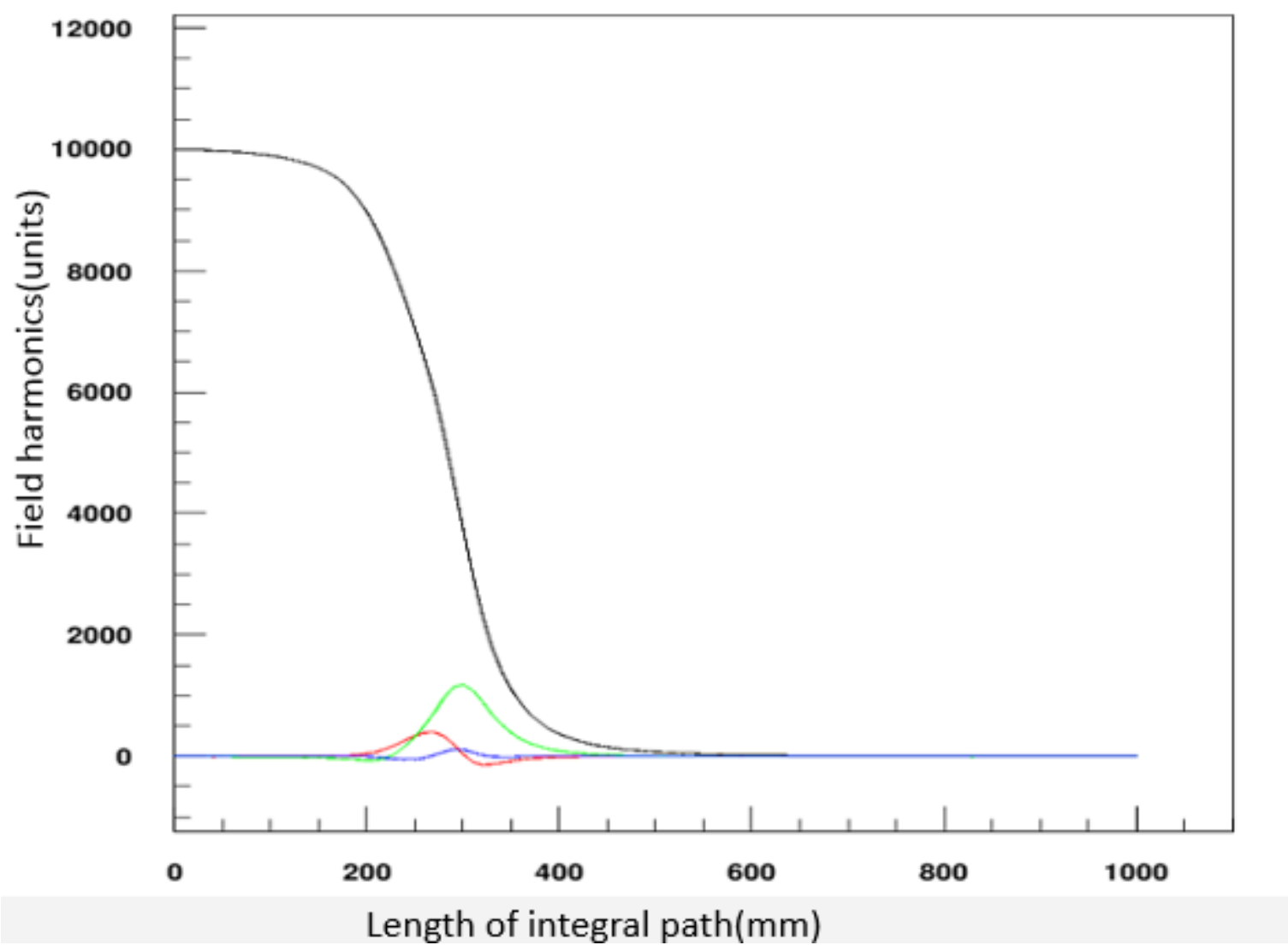

**Figure 8.3.5:** Field harmonic variations along the axis for the design #2 (black $b_1$, green $b_3$, red $b_5$, blue $a_2$).

### 8.3.2 Conceptual Design Study of 20-T Hybrid Dipole Magnet

Instead of using flared ends to make space for beam pipes, all the coils in this design are flat race-track configuration as shown in Fig. 8.3.6 left. One of the most inner coil block in each quadrant bends to the top (or bottom) of the iron yoke to make space for beam pipes. This design will simplify the coil fabrication procedures and lower the strain level in coils. A trade-off is that a little more superconductor is needed to go through the top and bottom of the magnet without any positive contribution to the main field in the aperture;

The coil layout in the straight section is re-optimized to further increase the main field and decrease the stress level, i.e., the ratio of the coil height to the coil width is increased, in other words, the coil blocks become thinner to make them more efficient to generate high field with a fixed quantity of superconductor, as shown in Fig. 8.3.6 right. One more benefit of this "thinner" design is that the stress level is reduced due to the increased area perpendicular to the Lorentz force direction;

By carefully re-optimizing the position and size of each coil block, the ratio of the peak field in the coil to the main field in the aperture is reduced from 20.71 T / 20.06 T = 1.032 to 20.42 T / 20.09 T = 1.016. The operating load line is reduced from 90% to 89% at 4.2 K.

The main design parameters of the magnet are listed in Table 8.3.3. The final diameter should be around 40~50 mm, determined by results from SPPC accelerator physics. The operating margin is 11% at 4.2 K, ~ 20% at 1.9 K. The outer diameter of the iron yoke is assumed to be 800 mm, subject to possible reduction to make space for a thicker mechanical support structure. We are trying to limit the size of the 20-T magnet to 900 mm, to reduce the tunnel size and the cost of civil construction. The magnet is made of 12 superconducting coils: 2 $Nb_3Sn$ outer coils, 2 $Nb_3Sn$ inner coils, 2 HTS outer coils, 4 HTS inner-b coils and 2 HTS inner-c coils. The 12 coils are wound with 3 different types of superconducting cables: 15 mm width $Nb_3Sn$ outer cable (38 strands), 22 mm width $Nb_3Sn$ inner cable (56 strands) and 20 mm width HTS cable (50 strands). The 3 superconducting cables are fabricated with 2 types of strands: high Jc $Nb_3Sn$ strand and HTS strand. The Jc of these strands is assumed to be the same as the current level [10]; possibly the $J_c$ of these conductors will be greatly increase in the next several years [11, 12].

The coil layout in the ends has been optimized to reduce the integrated high order multiples to less than $10^{-4}$ of the main field. To make the simplest structure, we assume each coil block bends with the same radius at the ends, in other words, the shape of each coil is always a "semicircle" at the ends. The length of straight section of each coil block is the main variable for optimizing the integrated field quality. Fig. 8.3.7 shows the distribution of high order multipoles along the axis for optimized coil ends shown in Fig. 8.3.6 left, with the reference radius of 15 mm. The integrated multipoles (from 0 mm to 1500 mm in Fig. 8.3.6) are listed in Table 8.3.4. All of them are less than 1 unit. The bending radius of each coil varies from 100.9 mm to 135.7 mm.

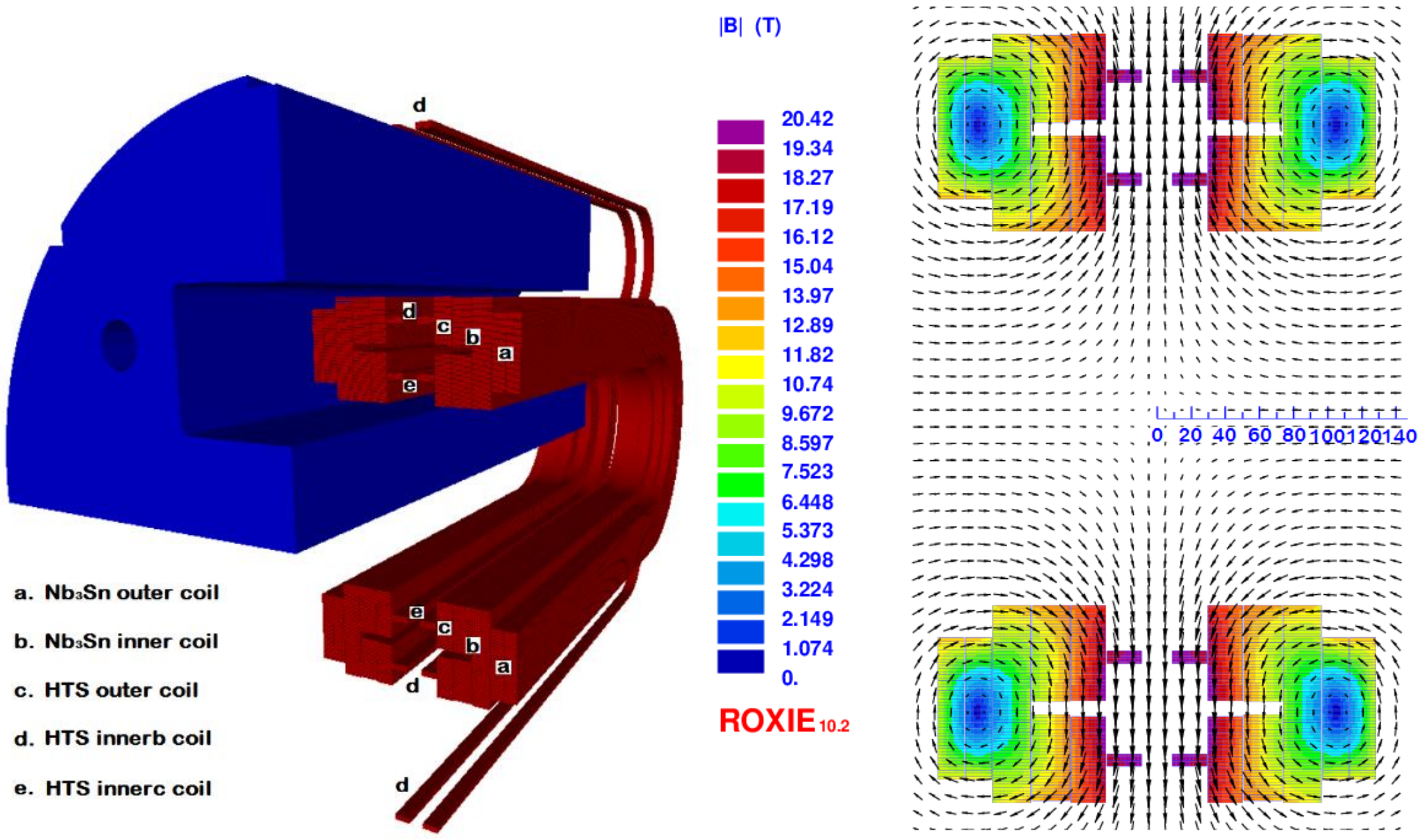

**Figure 8.3.6:** Design of the 20-T dipole magnet with common coil configuration. Left: coil layout and iron; Right: Magnetic flux in the straight section.

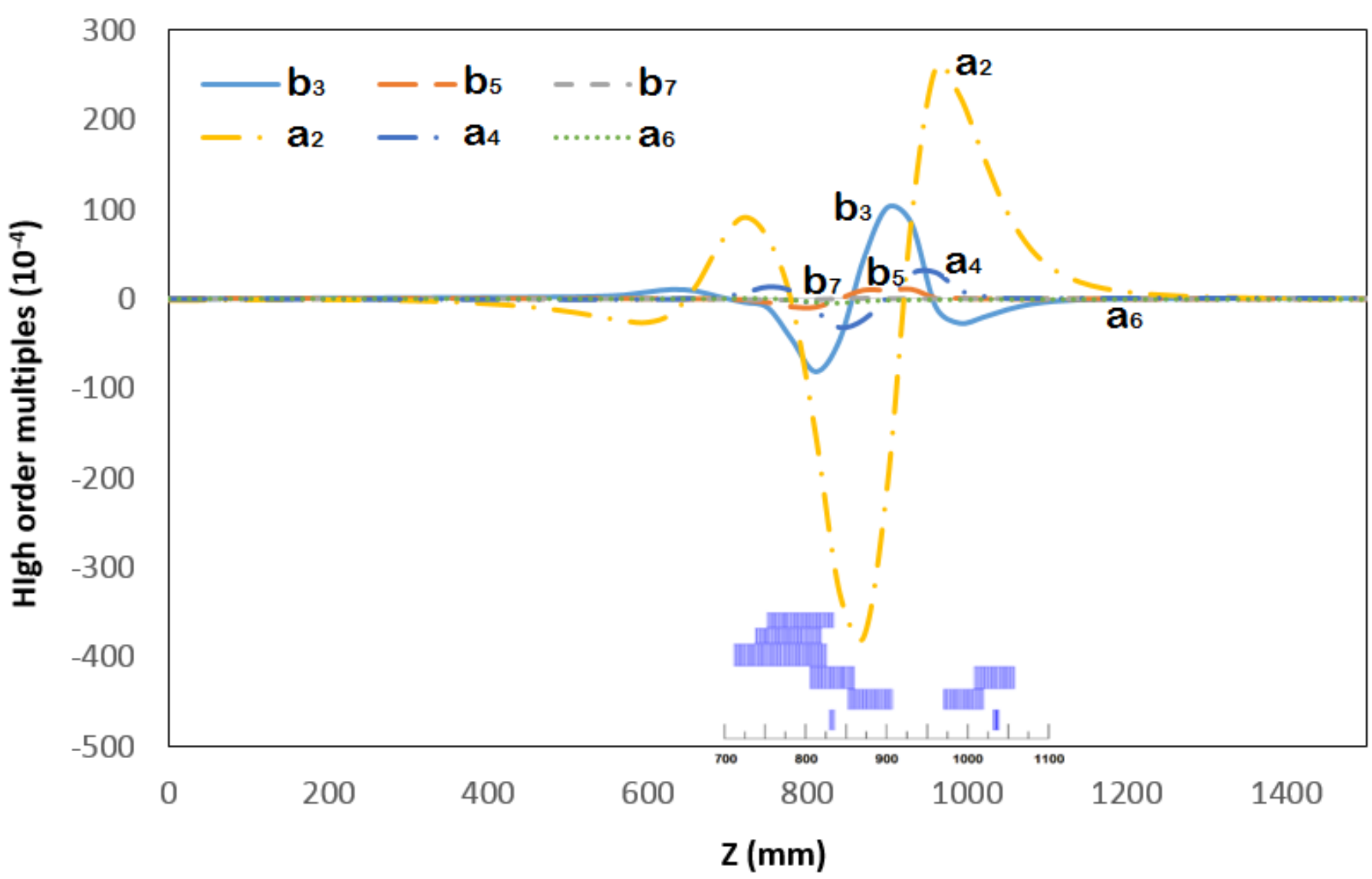

**Figure 8.3.7:** High order multiples along the axis for an optimized coil ends shown in Fig. 1 left (with the reference radius of 15 mm).

**Table 8.3.3:** Main parameters of a 20-T dipole magnet

| Item | | | Value |
|---|---|---|---|
| Magnet | | Number of apertures | 2 |
| | | Aperture diameter (mm) | 50 |
| | | Inter-aperture spacing (mm) | 333 |
| | | Operating current (A) | 14700 |
| | | Operating temperature (K) | 4.2 |
| | | Operating field (T) | 20 |
| | | Peak field (T) | 20.4 |
| | | Margin along the load line (%) | 11 |
| | | Stored magnetic energy (MJ/m) | 7.8 |
| | | Inductance (mH/m) | 72.1 |
| | | Yoke ID (mm) | 260 |
| | | Yoke OD (mm) | 800 |
| | | Weight per unit length (kg/m) | 3200 |
| | | Energy density (coil volume) ($MJ/m^3$) | 738 |
| | | Force per aperture – X / Y (MN/m) | 23.5/4.4 |
| | | Peak stress in coil (MPa) | 240 |
| | | Fringe Field @ r = 750 mm (T) | 0.02 |
| Coil | $Nb_3Sn$ outer | Number of layers | 2 |
| | | Number of turns per layer | 46 |
| (2 $Nb_3Sn$ outer + 2 | | Bending radius (mm) | 127.8 |
| $Nb_3Sn$ inner + | $Nb_3Sn$ inner | Number of layers | 2 |
| 2 HTS outer +4 HTS | | Number of turns per layer | 59/64 |
| inner-b + | | Bending radius (mm) | 109.1 |
| 2 HTS inner-c) | HTS outer | Number of layers | 1 |
| | | Number of turns per layer | 59 |
| | | Bending radius (mm) | 109.0 |
| | HTS inner-b | Number of layers | 1 |
| | | Number of turns per layer | 4 |
| | | Bending radius (mm) | 135.7 |
| | HTS inner-c | Number of layers | 1 |
| | | Number of turns per layer | 4 |
| | | Bending radius (mm) | 100.9 |
| Cable | $Nb_3Sn$ outer | Number of strands | 38 |
| | | Cable dimension ($mm^2$) | 15*1.5 |
| ($Nb_3Sn$ outer+ $Nb_3Sn$ | | Insulation thickness (mm) | 0.15 |
| inner | $Nb_3Sn$ inner | Number of strands | 56 |
| + HTS cable) | | Cable dimension ($mm^2$) | 22*1.5 |
| | | Insulation thickness (mm) | 0.15 |
| | HTS | Number of strands | 50 |
| | | Cable dimension ($mm^2$) | 20*1.5 |
| | | Insulation thickness (mm) | 0.15 |
| Strand | $Nb_3Sn$ | Diameter (mm) | 0.82 |
| | | Copper/Superconductor ratio | 1 |
| ($Nb_3Sn$ + HTS) | | Non-Cu $J_c$ ($A/mm^2$ @15 T, 4.2 K) | 1500 |
| | | $dJ_c/dB$ (A/T) | 350 |
| | HTS | Diameter (mm) | 0.82 |
| | | Copper/Superconductor ratio | 1 |
| | | Non-Cu $J_c$ ($A/mm^2$ @20 T, 4.2 K) | 1300 |
| | | $dJ_c/dB$ (A/T) | 13 |

**Table 8.3.4:** Integrated field quality along the axis with reference radius of 15 mm

| Integrated $b_n/a_n$ along axis | Value ($10^{-4}$) |
|---|---|
| b3 | 0.24 |
| b5 | 0.78 |
| b7 | -0.48 |
| b9 | -0.70 |
| a2 | 0.37 |
| a4 | 0.00 |
| a6 | 0.17 |

Fig. 8.3.8 shows the direction of magnetic force in the coils. The total force per aperture is 23.5 MN/m in the horizontal direction and 4.4 MN/m in the vertical direction. The coils tend to move outward in both directions after excitation. If we divide the horizontal magnetic force by the area perpendicular to its direction, we can get a rough estimation of the stress in the coils: 23.5/0.121=194 MPa. FEM simulation results show that the peak stress in the coil is around 249 MPa. Such a stress level may reduce the critical current density of $Nb_3Sn$ and Bi-2212 superconductors [13, 14]; Effective methods for stress management need to be investigated. The other type of HTS superconductor, ReBCO, can tolerate much higher stress and strain without showing any degradation, but the magnetization effect in the tape conductor is more severe than multifilament round wires like Bi-2212 and $Nb_3Sn$, which will make it difficult to obtain a $10^{-4}$ field quality.

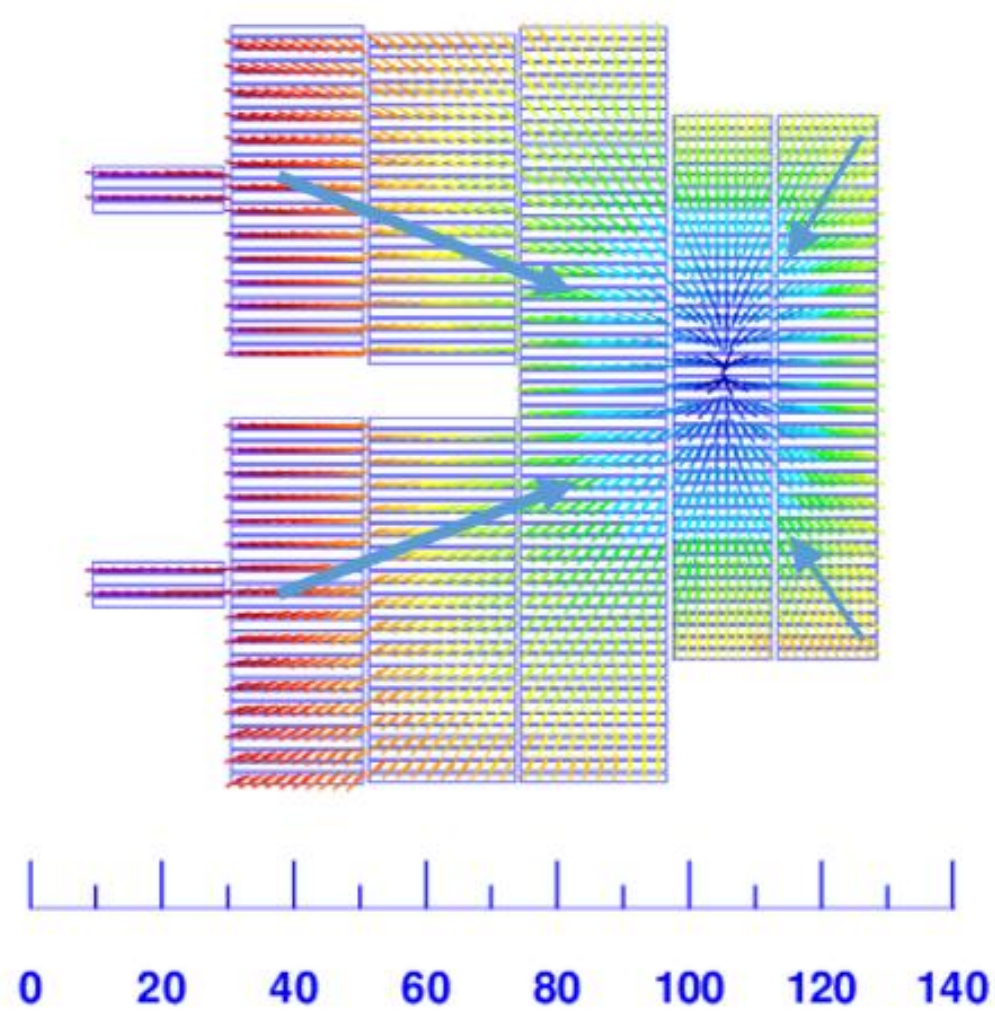

**Figure 8.3.8:** Direction of magnetic force in coils.

Different than in other coil configurations such ascos-theta or block, the two apertures of the common coil configuration are located in the vertical direction, which doubles the requirement of mechanical support strength horizontally, i.e., for each aperture the magnetic force is 23.5 MN per meter; for the whole magnet it becomes 47 MN per meter. If a shell based structure is adopted to provide support for the magnet [15], to constrain such a stress level, the thickness of the aluminum shell would be 75 mm, assuming the stress level in the shell is limited to less than 300 MPa [16]. To reduce the shell thickness and magnet size, aluminum alloys with higher yield strength will be tested in future.

### 8.3.3 Challenges for Fabrication and R&D Steps

Several main challenges are listed below for mass production of the 12~20 T magnets:

a) *$J_c$ and cost of superconductors:* Thousands of tons of superconductor are needed to fabricate the high field magnets. Significant increase of $J_c$ and decrease of cost of the superconductors are expected to reduce the cost of the project.

b) *Stress management in coils*: The stress in superconducting coils is above 200 MPa at 20 T operation. As both $Nb_3Sn$ and HTS superconducting materials are strain sensitive, effective methods need to be investigated to reduce the stress to a more acceptable level.

c) *Achieving $10^{-4}$ field quality with HTS coils*: The current distribution within a conductor is related to the field history. This magnetization depends on the dimensions of the conductor. Finer strands give much less magnetization. LTS (Low Temperature Superconductor) conductors such as NbTi are made of thousands of small filaments with diameter of only a few microns. The filaments in current Bi-2212 conductors are larger than those in NbTi, and the ReBCO tape is a single 'filament' and is orders of magnitude larger. This will make it difficult for the magnets with the HTS coils to reach a field uniformity of $10^{-4}$.

d) *Achieving quench protection of HTS coils*: The quench propagation speed in HTS coils is hundreds of times lower than in LTS coils. This makes the present quench detection and protection methods unsuitable for HTS coils.

## 8.4 Injector Chain

### 8.4.1 General Design Considerations

The injector chain by itself is a large accelerator complex. To reach 2.1 TeV required for injection into the SPPC, we require a four-stage acceleration system, with energy gain per stage between 8 and 18. It accelerates the beam and prepares it with the required properties of bunch current, bunch structure, and emittance, as well as the beam fill period.

The four stages are shown in Fig. 8.4.1, with additional parameters given in Table 8.4.1 The p-Linac is a superconducting linac with a repetition rate of 50 Hz. The p-RCS is a rapid cycling synchrotron with a repetition rate of 25 Hz. The MSS has a lower repetition rate of 0.5 Hz. The SS is based on superconducting magnets with maximum dipole field of about 8 Tesla. The higher repetition rates for the earlier stages help reduce the SS cycling period and thus the overall SPPC beam fill time. For easier maintenance and cost efficiency, as well as the physics programs, the first three stages will be built relatively shallow underground, e.g., -15 m, whereas the SS with a much larger circumference will be built at the same level as the SPPC, -50 m to -100 m.

As shown in Table 8.4.1, the different stages are needed for only fractions of the time. They could operate with longer duty cycles, or continuously, to provide high-power beams for other research applications, when they do not serve the SPPC. As the present bunch population at the SPPC is limited mainly by the SR power, the accelerators of the injector chain have the potential to load more accumulated particles in a pulse or deliver higher beam power for their own diverse applications. This capability is also very useful for future SPPC upgrades.

For such a complex injector system, it will take about 10 years to build and commission it stage-by-stage. Thus, hopefully the construction of the injector accelerators can begin several years earlier than the SPPC, and overlap in time with the CEPC physics operation.

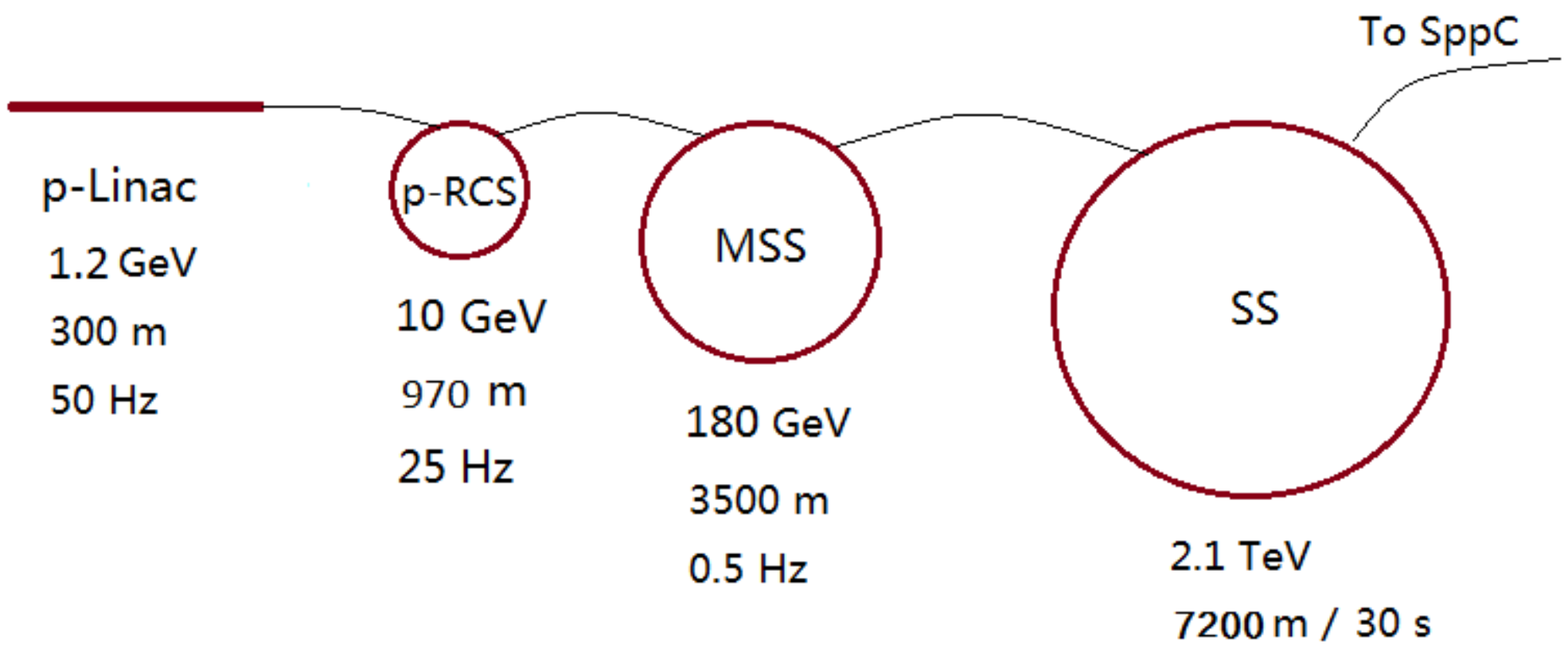

**Figure 8.4.1:** Injector chain for the SPPC

## 8.4.2 **Preliminary Design Concepts**

### 8.4.2.1 ***Linac (p-Linac)***

Superconducting linacs have undergone tremendous development in the last twenty years [1] and will presumably make even more progress in the next decade. Hence we have chosen 1.2 GeV in energy and 50 Hz in repetition rate for the p-Linac. The continuous beam power is 1.63 MW. At least half of this could be available for other applications.

### 8.4.2.2 ***Rapid Cycling Synchrotron (p-RCS)***

The continuous beam power from p-RCS is 3.4 MW. Only one other proton driver (study for a future Neutrino Factory) has performance close to this [2]. The high repetition rate of 25 Hz will shorten the beam filling time in the MSS. Only a fraction of this power is needed to fill the MSS. Thus most of the beam pulses from the p-RCS could be used for other physics programs. The p-RCS will use mature accelerator technology but be on a larger scale than existing rapid-cycling proton synchrotrons. High-Q ferrite loaded RF cavities provide RF voltage of about 3 MV, RF frequency swing of 36-40 MHz and bunch spacing of 25 ns.

### 8.4.2.3 ***Medium Stage Synchrotron (MSS)***

The MSS has beam power similar to the p-RCS but with much higher beam energy and much lower repetition rate. The SPS at CERN and the Main Injector at Fermilab are two good examples for its design. But due to much higher beam power, the beam loss rate must be more strictly controlled. The same RF system as in the p-RCS is planned, but a 200-MHz RF system could be used in the future for bunch splitting to provide 5-ns bunch spacing. The beam from the MSS will find additional physics programs other than only being the injector for the SS.

#### 8.4.2.4 *Super Synchrotron (SS)*

The SS will use superconducting magnets similar to those used at the LHC, but with a higher ramp rate. We do not need to consider synchrotron radiation because of the much lower energy. There are no apparent critical technical risks in building the SS. It is still unclear if the beam from the SS can find its own physics programs besides being the SPPC injector.

**Table 8.4.1:** Main parameters for the injector chain at SPPC

| | Energy | Average current | Length/ Circum. | Repetition Rate | Max. beam power or energy | Dipole field | Duty factor for next stage |
|---|---|---|---|---|---|---|---|
| | GeV | mA | km | Hz | MW/MJ | T | % |
| p-Linac | 1.2 | 1.4 | ~0.3 | 50 | 1.6/ | - | 50 |
| p-RCS | 10 | 0.34 | 0.97 | 25 | 3.4/ | 1.0 | 6 |
| MSS | 180 | 0.02 | 3.5 | 0.5 | 3.7/ | 1.7 | 13.3 |
| SS | 2100 | - | 7.2 | 1/30 | /34 | 8.3 | 1.3 |

A dedicated heavy-ion linac (i-Linac) together with a new heavy-ion synchrotron (i-RCS), in parallel to the proton linac/RCS, is needed to provide heavy-ion beams at the injection energy of the MSS, with a beam rigidity of about 36 Tm which is the same as the 10 GeV proton beam.

# 9 Conventional Facilities

## 9.1 Introduction

CEPC consists of a Collider, the injection system into the Collider whose main components are a Linac, a Booster, and transport lines, and two large physics detectors. Civil construction, sometimes called conventional facilities, house all of the components of the CEPC and reserve space for SPPC, as illustrated in Fig. 9.1.1. The layout and construction of each part is determined by their geometric relationships, environmental conditions and safety considerations. Practicality, adaptability and operating efficiency are criteria to be carefully considered in the design of the civil construction.

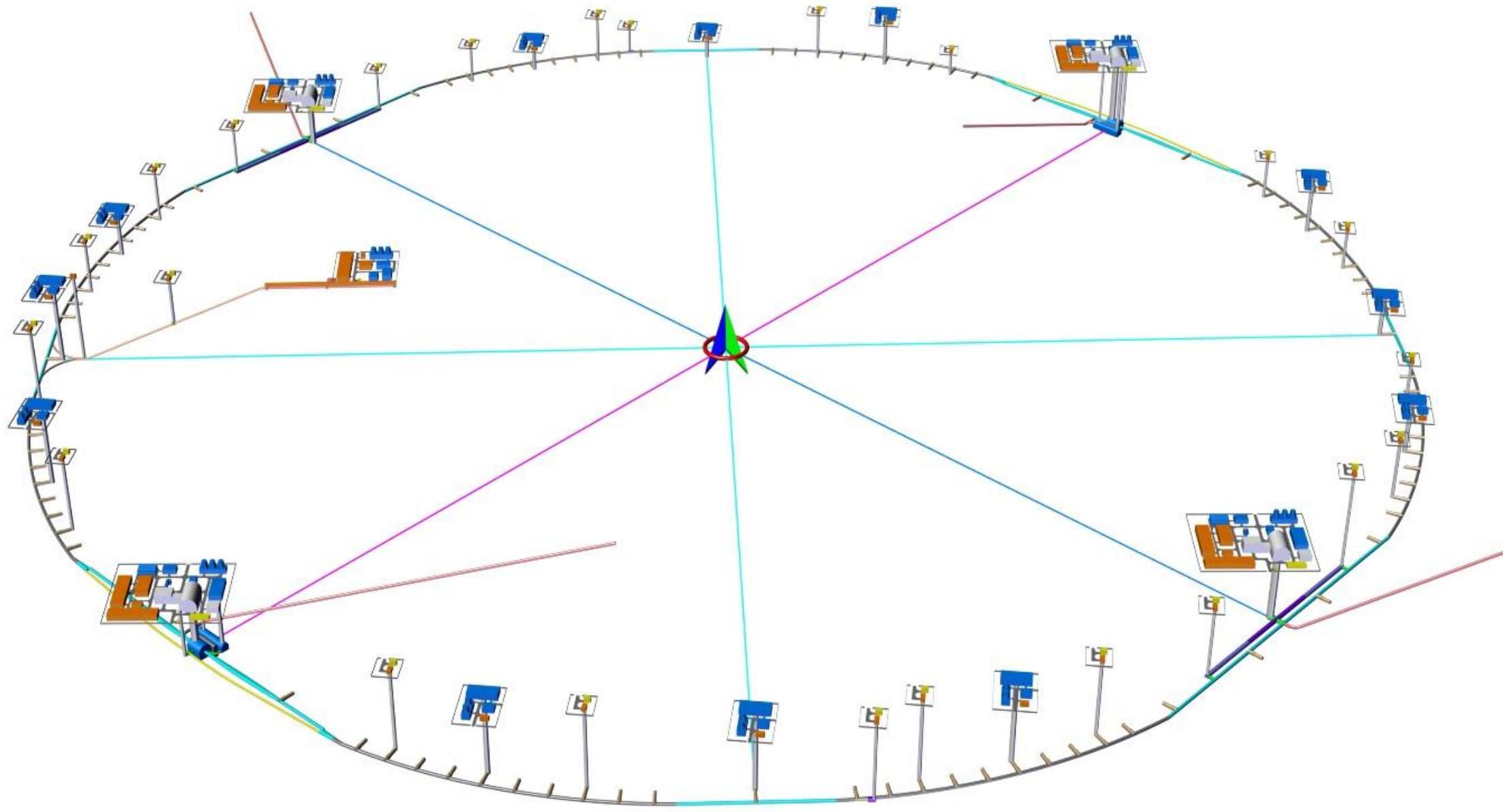

**Figure 9.1.1:** Layout of surface and underground CEPC structures

The following defines the scope of work and the requirements to be met.

1) The main tunnel to house the Collider and Booster synchrotrons, auxiliary tunnels for the Booster bypass and RF equipment, the Linac tunnel and equipment gallery and transport line tunnels. The main tunnel is 100 km in circumference and 100 m below ground.
2) The experiment halls are 100 m below ground and span 30~40 m. There are additional chambers such as power source halls, cryogenics halls and spaces for the water cooling system, etc.
3) There are accesses to the experiment halls, such as access tunnels, transport shafts, and emergency exits.
4) At ground level there are ancillary structures with a total area of 140,450 m$^2$. These include structures near the shaft openings, structures to house substations and electric distribution, cryogenics rooms, and ventilation fan rooms.
5) Space for staging the construction equipment and materials and dumping sites.
6) Included in the project scope are related lifting equipment, conveyance, systems for electric supply, drainage, ventilation and air conditioning, communication, controls and monitoring, safety escape, and firefighting. The firefighting system includes fire alarms, hydrants, gas fire-extinguishing system, and a smoke exhaust system. Maintenance of these systems as well as their potential for future upgrades is fully considered in their design.

## 9.2 Site and Structure

### 9.2.1 **Preliminary Site Selections**

#### 9.2.1.1 ***Basic Principles of Site Selection***

In the selection of the CEPC site, besides engineering technology conditions such as topography and geology, the construction conditions that need to be considered include location, local government support, social and cultural environment, regional development and environmental impact. These external construction conditions may sometimes be the decisive factor in site selection.

Following are factors that should be considered in the site selection:

1) Geography
   The site should be sufficiently large and appropriately located to accommodate the future development of the Institute of High Energy Physics of the Chinese Academy of Sciences. The site should promote the CEPC project and the construction of an international science city.
2) Natural conditions
   a) The structural stability conditions are good and avoid deep faults and motions and deformations that are recent in geologic time. Seismic peak acceleration is generally less than 0.10g.
   b) Good rock conditions. Large area hard rock with stable lithology are suitable for construction of underground caverns.
   c) No large height differences, mostly low mountains and hilly areas.
   d) The quaternary overburden is not thick.
   e) The permeability of rock is relatively low.
   f) External dynamic geological phenomena are relatively small...
3) Access conditions
   In order to minimize capital costs and accelerate the progress in the initial stages of the construction, the site should be located where access is convenient.
4) Environmental factors
   Few environmental impact problems and no environment sensitive zones should be involved, such as nature preserves, parks, military areas, or other environmental constraints such as wetlands.
5) The site should be where construction conditions and related economic factors are good.

#### 9.2.1.2 ***Brief Introduction of Each Potential Site***

A preliminary study of geological conditions for CEPC's potential site location was carried out in Hebei, Guangdong, Shaanxi, Jiangsu, and Zhejiang provinces. The geological survey of site selection in this conceptual design stage was carried out in the Funing site (Hebei Province), in the site of the Shen-Shan Special Cooperation zone (Guangdong Province), and in the site of Huangling area (Shaanxi province).

I. The Funing site is located in the Funing District, Qinhuangdao City of Hebei Province, Beidaihe District, Changli and Lulong Counties. This is a hilly area, with elevations of 0 to 600 m. The main strata are Archaean gneiss, Mesozoic magmatic rocks, volcanism from the Yanshan period, and some Mesozoic sand shale. The rock is mainly hard without thick overburden, and has a basic seismic intensity of degree VII. The site conditions are suitable for construction of large underground caverns and tunnels. The depths of the underground caverns do not vary a great deal.

II. The Shen-Shan site is located in the Shen-Shan Special Cooperation zone, Haifeng and Huidong Counties of Guangdong Province. The landform is dominated by low mountain areas with elevations of 20 to 800 m. The main strata consist of Mesozoic volcano rock and sand-mudstone, granite of the Yanshan period, and a small amount of Cenozoic shaly glutenite. These rocks are mainly hard with fracture structure, no thick overburden, and the basic seismic intensity degree VI~VII. Some of the caverns will be quite deep and require a long shaft. The layout is relatively complex and difficult to construct.

III. The Huangling area site is located in Huangling County and Luochuan County (Yan'an City, Shanxi Province), Yijun County (Tongchuan City), and Baishui County (Weinan City). The landform belongs to a plateau gully region with elevations of 600 to 1600 m. The stratum on the horizontal layer and its lithology is Mesozoic Triassic terrestrial clastic rock, with 100 to 150 m of overlying loess. The rocks are generally of moderate hardness with simple structure, and the basic seismic intensity is mainly of degree VI. The buried depth of underground caverns and the shaft depths vary considerably. The layout for construction is relatively easy, and the construction work of moderate difficulty.

At this stage in the conceptual design and planning we use the Funing site as representative.

### 9.2.2 **Site Construction Condition**

#### 9.2.2.1 ***Geographical Position***

Funing is located in the northeast part of Hebei province and northwest of Qinhuangdao city, 478 km north of Shijiazhuang (provincial capital), 240 km east of Beijing, and 23 km west of Qinhuangdao.

#### 9.2.2.2 ***Transportation Condition***

Funing District has several convenient transportation assets. There are the Tianjin-Qinhuangdao passenger dedicated line, the main national railway and local railway trunk lines such as the Beijing-Qinhuangdao high-speed railway, and the Beijing-Harbin, Tianjin-Shanhaiguan, Datong-Qinhuangdao and Qinhuangdao-Shanhaiguan railways, as well as the Beijing-Shenyang, Coastal, and Qinhuangdao-Chengde expressways. Two national roads 102 and 205 as well as five provincial roads are transportation hubs in the Qinhuangdao area. Funing is 35 km from Qinhuangdao port, 45 km from Shanhaiguan airport and 25 km from Qinhuangdao's new airport.

### 9.2.2.3 *Hydrology and Meteorology*

The climate is semi humid continental monsoon and in the warm temperate zone. It is also affected by the marine climate with four distinct seasons, adequate illumination and abundant rainfall. The mean air temperature is 10.2℃. There is a frost-free period of 177 days, 2745 hours of sunshine and average rainfall of 730.7 mm yearly.

The main rivers in Funing include the Luanhe, Yinma, Yanghe, Daihe, Tanghe. They are all perennial rivers, running from north to south and emptying into the Bohai Sea. The Yanghe Reservoir is a large one with a total capacity of $3.86 \times 10^9$ $m^3$ and there are a large number of small reservoirs in the Funing area.

### 9.2.2.4 *Economics*

As of 2016, Funing District with 1100 square kilometers, had a total population of 334,123.

In 2016, the total GDP of the Funing area was 11.05226 billion yuan, an increase of 6.1% compared with the same period last year. The per capita disposable income of urban residents in the region was 29,842 yuan, rising 8% yearly. The per capita disposable income of rural residents was 12,839 yuan, an increase of 7.8% over the same period last year.

### 9.2.2.5 *Engineering Geology*

The selected site is located in a low mountain and hilly area, high in the west and low in the east. The river systems include the Yanghe River and the Yinma River systems. The formation lithology mainly includes the Archaeozoic metamorphic rocks dominated by gneiss and schist, the Mesozoic magmatic rocks dominated by granite and the Mesozoic volcanic rocks dominated by tuff, as well as a small amount of the Mesozoic sandstone and sandstone intercalated with mudstone. Rocks are mainly hard. The surface overburden is relatively thin and the thickness of river alluvium is 15 to 20 m. No deep regional fractures are distributed in the selected site area which has a ground motion peak acceleration of 0.10~0.15 g and a basic seismic intensity of degree VII, so the area is basically stable. There are two types of groundwater: pore water in loose rocks and fissure water in the bedrock weathered zone, with a relatively low abundance of the latter. Exogenic geological processes are not developed within the selected site area, and the thickness of the weathered zone is 20 to 30 m, so there are no major engineering geological restrictions, and the site area is suitable for the construction of a large-sale underground project.

Main engineering geological considerations:

1) Water gushing into the tunnel: The zones where this is a possibility include the section passing through the Yanghe River downstream of the Yanghe Reservoir, the section passing through the Yanghe River in the southeast of the Yanghe Reservoir, and corresponding alluvial plains. Water bursting may also be found near local fault fractured zones, especially in the area with a thick partially weathered zone.
2) Stability of surrounding rocks: Most of the tunnel sections are composed of slightly weathered to fresh rock mass, so the surrounding rocks are relatively stable. A few of the tunnel sections are composed of moderately weathered rock mass, so there may exist problems concerning the stability of surrounding rocks. When the tunnel passes through fault fractured zones, the stability of the surrounding rocks is poor and alleviating measures are needed... The inlet section of a vertical shaft is composed of a moderately weathered rock mass with poor stability, so corrective measures will be required. The experiment halls have a large span and high side walls. Whether underground excavation or an open excavation method is chosen, there exist problems concerning the block stability of the side walls. If the open excavation method is used, the surrounding rocks above the upper moderately weathered zone are of poor stability.

### 9.2.3 **Project Layout and Main Structure**

#### 9.2.3.1 ***General Layout of the Tunnel and Surface Structures***

##### *9.2.3.1.1 General Layout Principles and Requirements*

1) The layout, length and buried depth of the tunnel meet the needs of the accelerator and the detectors.
2) The operation needs to be secure, with easy management and convenient traffic flow.
3) The geologic structure around the circumference of the tunnel is simple and the hydrogeological conditions are suitable for construction.
4) There is easy access to hydroelectricity.
5) Shafts and adits provide entrance to the tunnel.
6) Shafts will avoid densely populated areas. Auxiliary facilities, such as cooling towers and substations are close to the access shafts.
7) The layout must meet the requirements for transportation and installation of experimental equipment.
8) The number and length of construction adits should be determined by the terrain and geologic conditions, the construction methods and the external traffic situation. It will help to balance and optimize the required person-hours and time requirements among tunnel sections.
9) Minimize the impact on the local ecology. The surface facilities should avoid existing buildings.
10) Meet the requirements of government regulations and norms.

##### *9.2.3.1.2 General Layout*

Figures 9.2.1 and 9.2.2 show the 100-km circumference tunnel in plan and profile. The tunnel has an inverted U-shape, of 6.00 m. width 5.00 m. height. Considering the relatively thick overburden of the Yanghe River alluvial plain in the southeast part of the site, point B that is the midpoint of LSS2, through which the tunnel passes the Yinma River, is designated as the lowest tunnel point, and point A (midpoint of LSS4), opposite to point B along the diameter, is designated as the highest tunnel point. The longitudinal slope of the tunnel is 0.3% from topology as well as drainage requirements during construction and operation. Surrounding rocks of the tunnel consist of granite, gneiss, schist and tuff and are mainly of Class II ~ III.

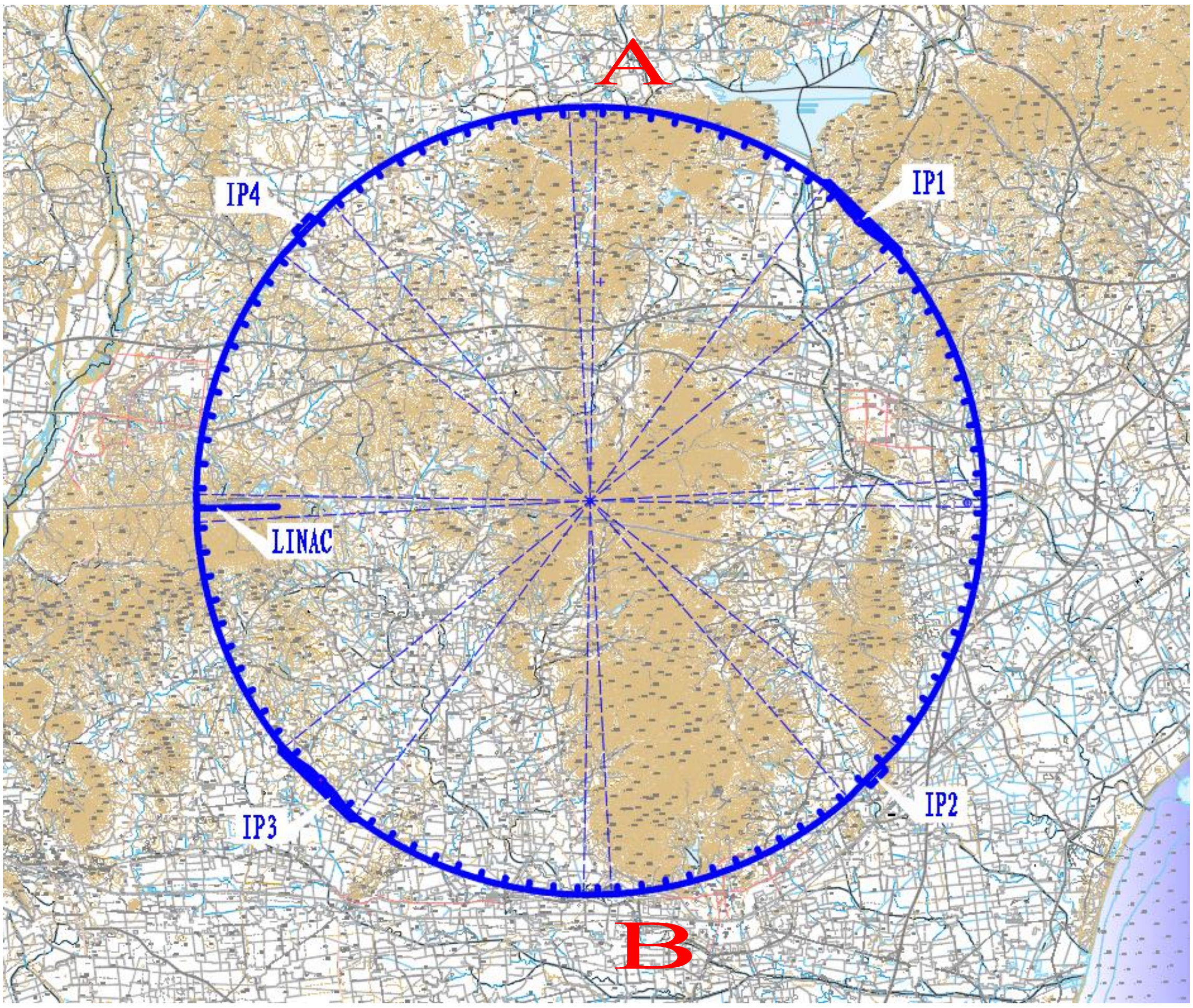

**Figure 9.2.1:** Layout plan of the CEPC tunnel

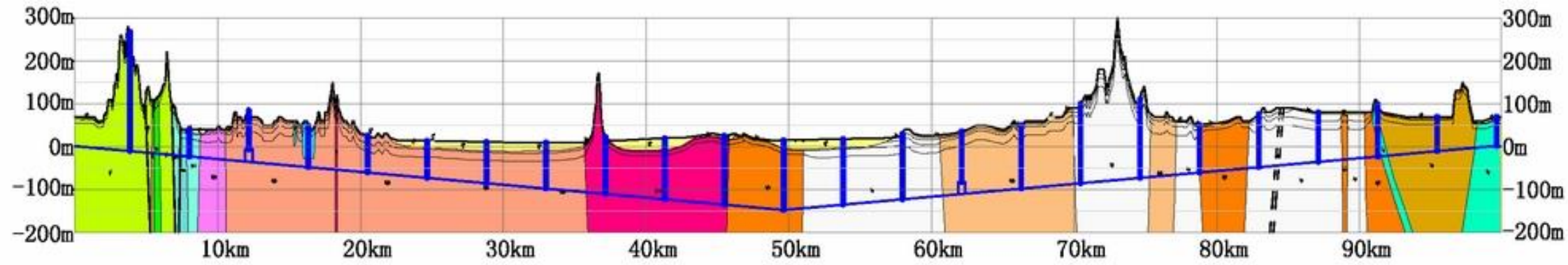

**Figure 9.2.2:** Longitudinal profile of the CEPC tunnel

Underground structures consist of the following as shown in Fig. 9.2.3:

- Collider ring tunnel
- Experiment halls (includes main and service caverns): IP1 and IP3 are experiment halls for CEPC, and IP2 and IP4 are future experiment halls for SPPC
- Linac and BTL tunnels: Linac tunnel, klystron gallery, hall for the damping ring, BTL tunnel and its branch tunnels
- Auxiliary tunnels: RF auxiliary tunnels, Booster bypass tunnels in the IR and many short auxiliary tunnels
- Vertical shafts in experiment halls and RF zones and along the ring tunnel for personnel and delivery of equipment to tunnels and halls, and for providing channels for ventilation, refrigeration, cooling and control and monitoring lines.

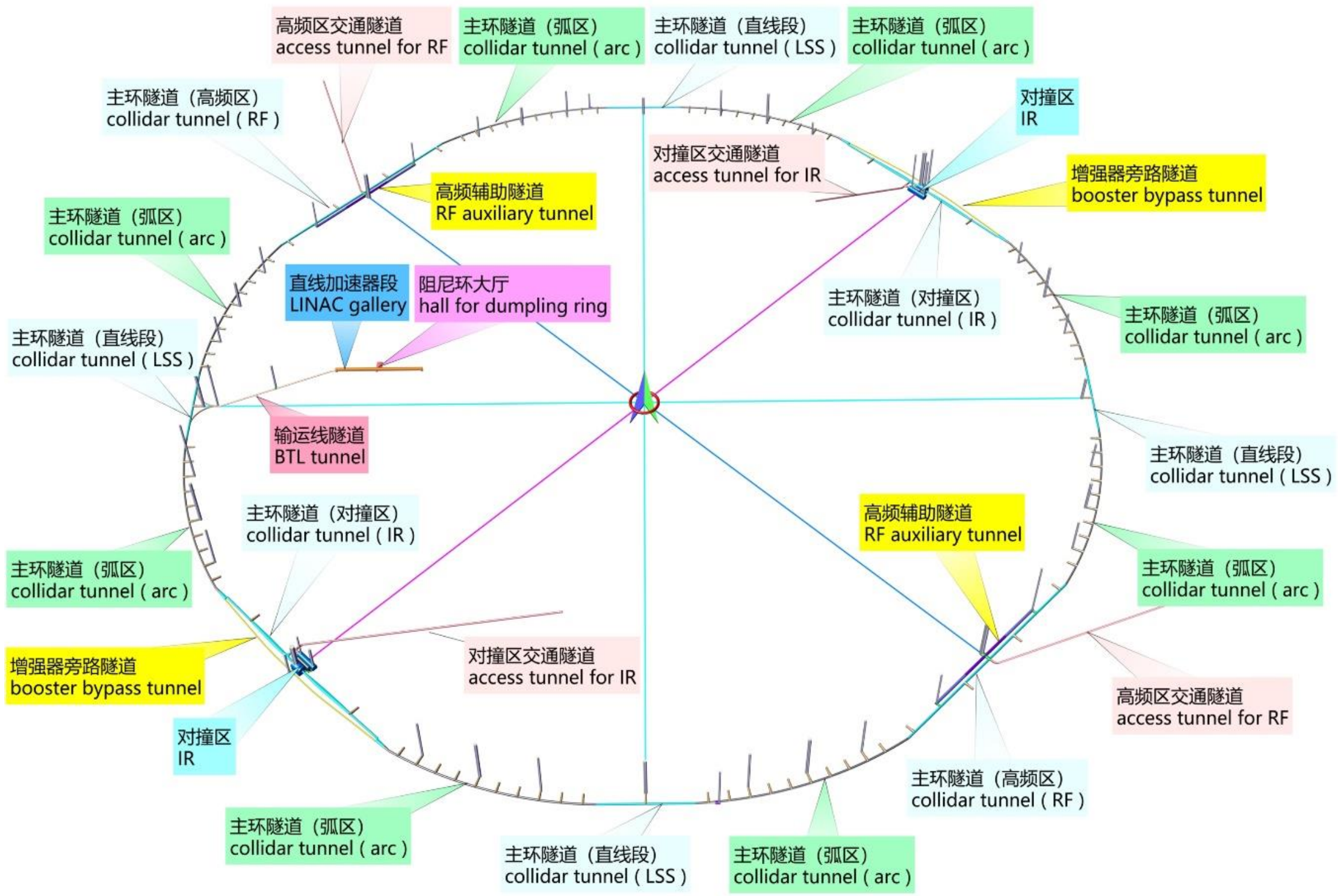

**Figure 9.2.3:** Underground structure layout

Surface structures within the main ring area, such as auxiliary equipment structures, cooling towers, substations and ventilation systems, are located close to the vertical shafts which provide access to the underground network.

### 9.2.3.2 ***Underground Structures***

#### *9.2.3.2.1 Collider Tunnel*

The Collider tunnel consists of:

- 4 curved arc sections L= 10246.70 m
- 4 curved arc sections, L = 10192.25 m
- 2 IRs, IP1/IP3, L = 3320.00 m
- 2 linear sections for RF IP2/IP4, L = 3240.00 m
- 4 linear sections (LSS), L = 1197.50 m

The total tunnel length is 99.67 km. These are shown in Fig. 9.2.4 and itemized in Table 9.2.3.1.

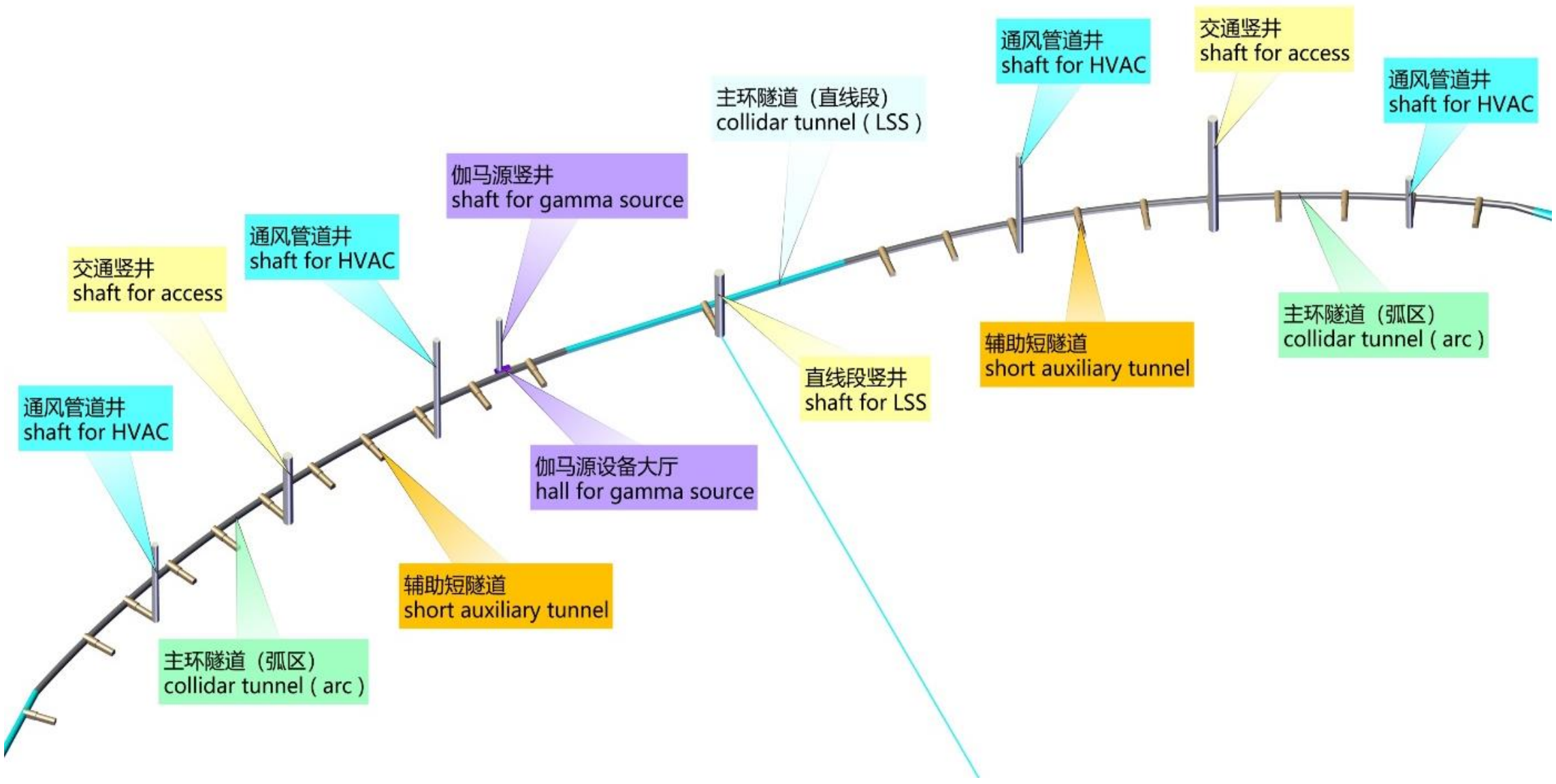

**Figure 9.2.4:** Layout of the Collider ring tunnel

**Table 9.2.3.1:** Characteristics of Collider tunnel structures

<table>
<tr><th colspan="2">Item</th><th>Unit</th><th colspan="3">Qty</th><th>Remarks</th></tr>
<tr><td rowspan="3">Linear section</td><td>Qty</td><td>Section</td><td>2</td><td>2</td><td>4</td><td rowspan="2">IR=3320.00 m<br>RF=3240.00 m<br>LSS=1197.50 m<br>Width varies from 6.00 to 11.40 m. H is 4.5 m</td></tr>
<tr><td>Length</td><td>m</td><td>3320.00</td><td>3240.00</td><td>1197.50</td></tr>
<tr><td>Dimension (∩-shaped)</td><td>m</td><td colspan="3">6.00 × 5.00</td><td></td></tr>
<tr><td rowspan="3">Arc section</td><td>Qty.</td><td>Section</td><td colspan="2">4</td><td>4</td><td rowspan="2">IR-LSS = 10246.70m<br>LSS-RF = 10192.25m</td></tr>
<tr><td>Length</td><td>m</td><td colspan="2">10246.70</td><td>10192.25</td></tr>
<tr><td>Dimension (∩-shaped)</td><td>m</td><td colspan="3">6.00 × 5.00</td><td></td></tr>
<tr><td colspan="2">Collider ring diameter</td><td>m</td><td colspan="3">12822</td><td></td></tr>
<tr><td colspan="2">Total length of tunnel</td><td>km</td><td colspan="3">99.666</td><td></td></tr>
<tr><td colspan="2">Longitudinal gradient of tunnel</td><td></td><td colspan="3">0.30%</td><td></td></tr>
</table>

The cross section of the tunnel is divided into three parts shown in Fig. 9.2.5.:

- Outer side, reserved for SPPC;
- Inner side, on which CEPC equipment and components will be installed;
- Central part of the tunnel for equipment operation and transport.

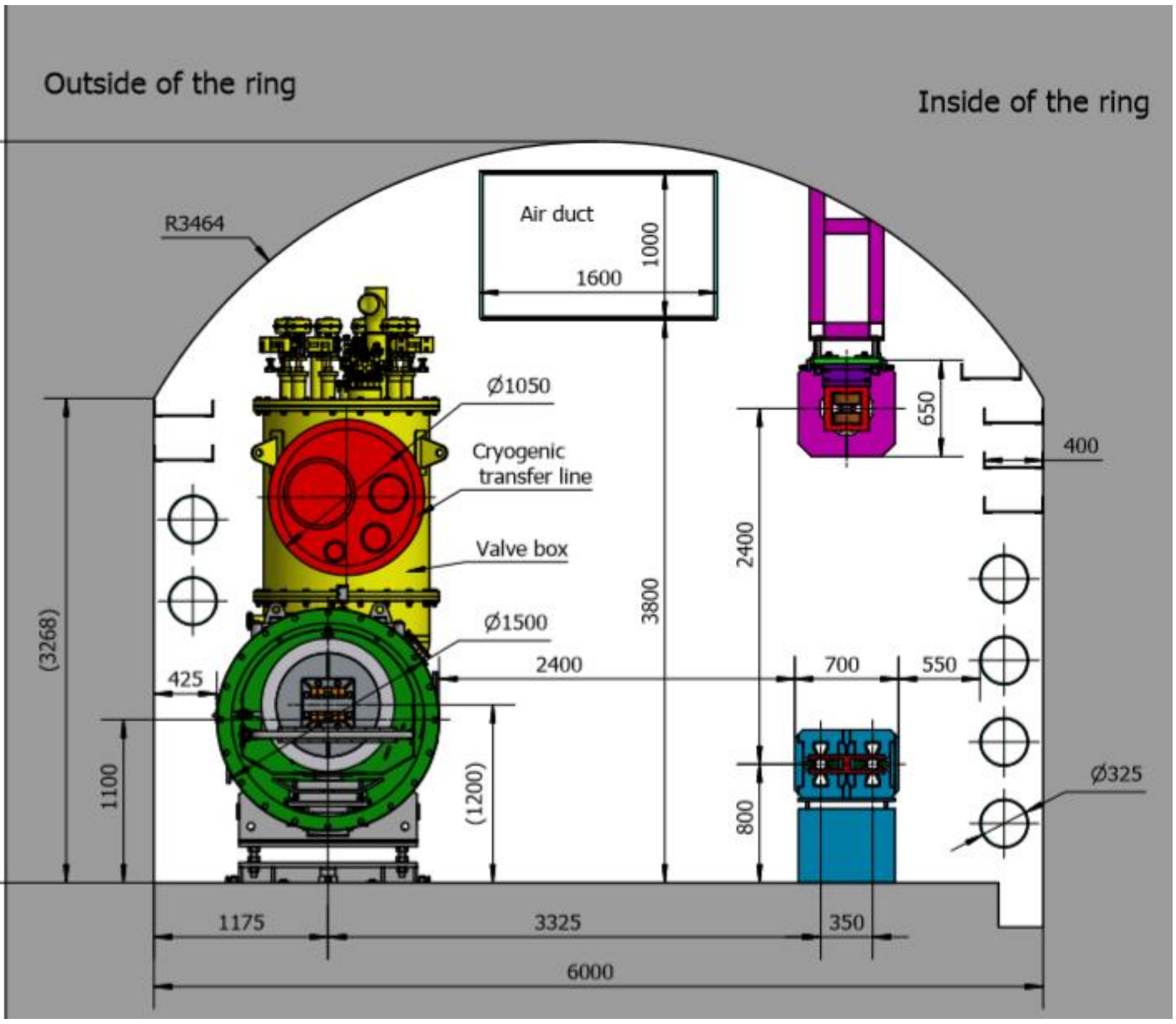

**Figure 9.2.5:** Tunnel cross section in the arc sections

*9.2.3.2.2 IR Sections IP1 and IP3*

IP1 and IP3 are where the large detectors will be located. Therefore, in these straight sections the Booster beam must bypass around the detectors. Zones within 1532.00 m. of both ends are divided into two tunnels, one for the e+/e- colliding beams 6.00 m to 11.40 m. wide, 4.50 m. high and 1509.30 m. long in each direction... The tunnel for the Booster bypass is 3.50 m. wide, 3.50 m. high, and 3018.60 m. long. Figs. 9.2.6 and 9.2.7 illustrate the arrangement with additional details in Table 9.2.3.2. The dimensions are determined from the requirements of local control, electrical systems, beam instrumentation, vacuum and cooling equipment as well as space for the passage of personnel and maintenance equipment.

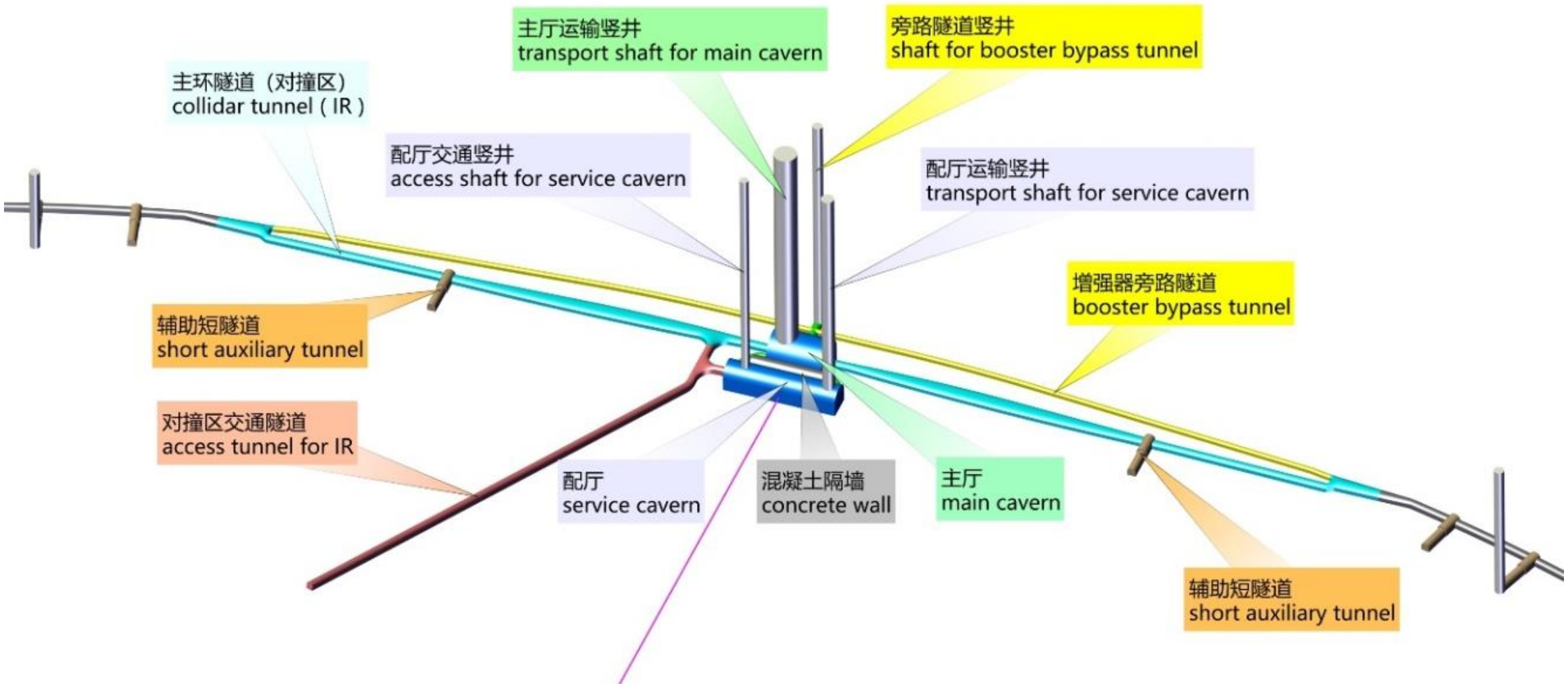

**Figure 9.2.6:** Tunnel layout in the IR section

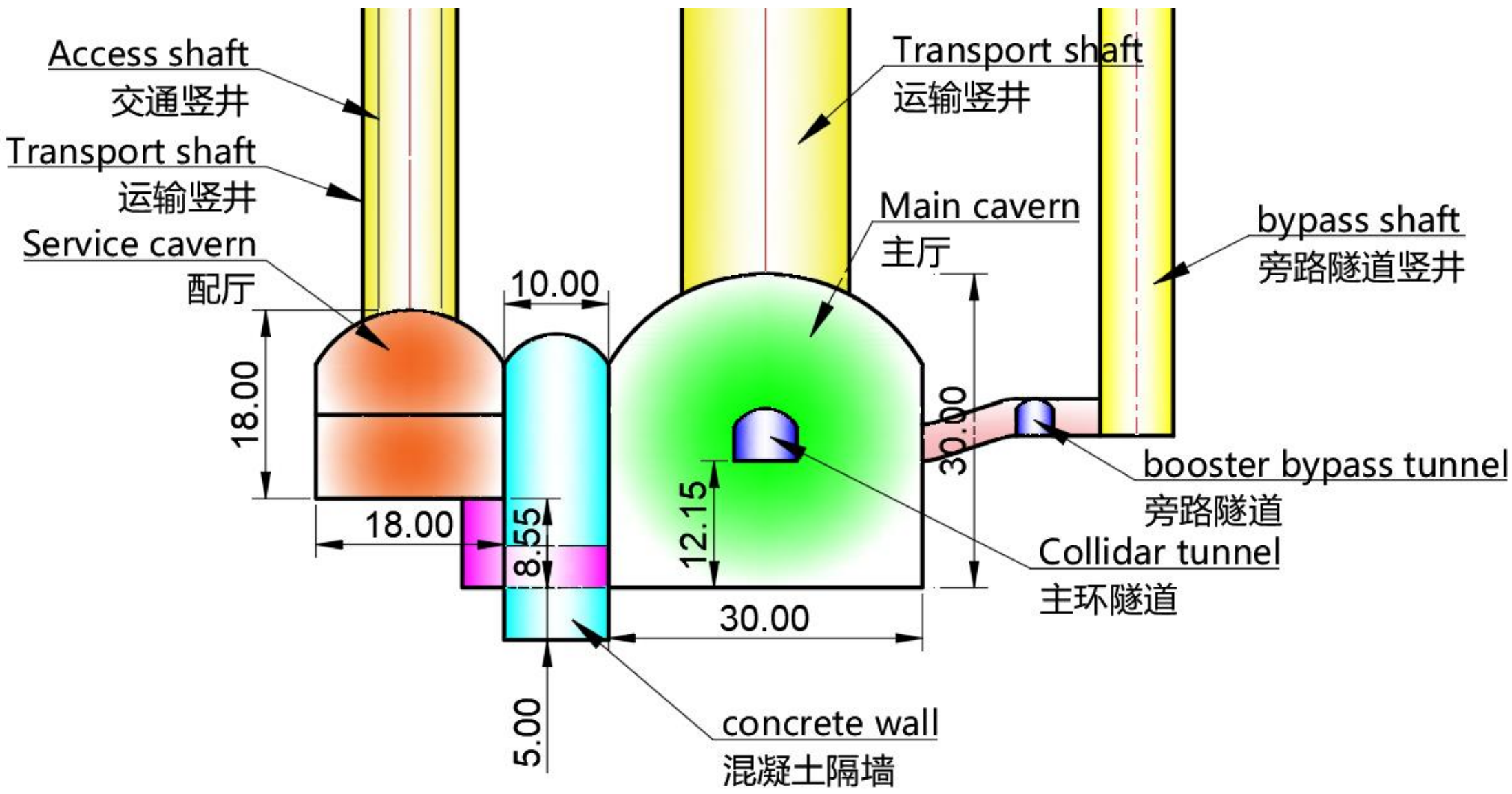

**Figure 9.2.7:** IR tunnel cross section

**Table 9.2.3.2:** Structures of the bypass tunnel in the collision area

| Item | | Unit | Qty |
|---|---|---|---|
| **Bypass tunnels in the collision area** | Qty | Nos. | 2 |
| | Length of individual tunnel | m | 3018.60 |
| | Dimension (∩-shaped) | m | 3.50 × 3.50 |

*9.2.3.2.3 IR Sections IP2 and IP4*

The RF cavities are in IP2 and IP4. These straight sections will also house the future SPPC detectors. The tunnel is 6.00 m. wide, 5.00 m. high and 3240.00 m. long. An auxiliary tunnel of length 2000.00 m is symmetric around the RF system midpoint. It is 8.00 m. wide and 7.00 m. high. This RF tunnel, illustrated in Figs. 9.2.8 and 9.2.9 with additional details in Table 9.2.3.3, is used to house RF power sources, water cooling equipment, vacuum and cooling equipment and has sufficient room for personnel and maintenance equipment.

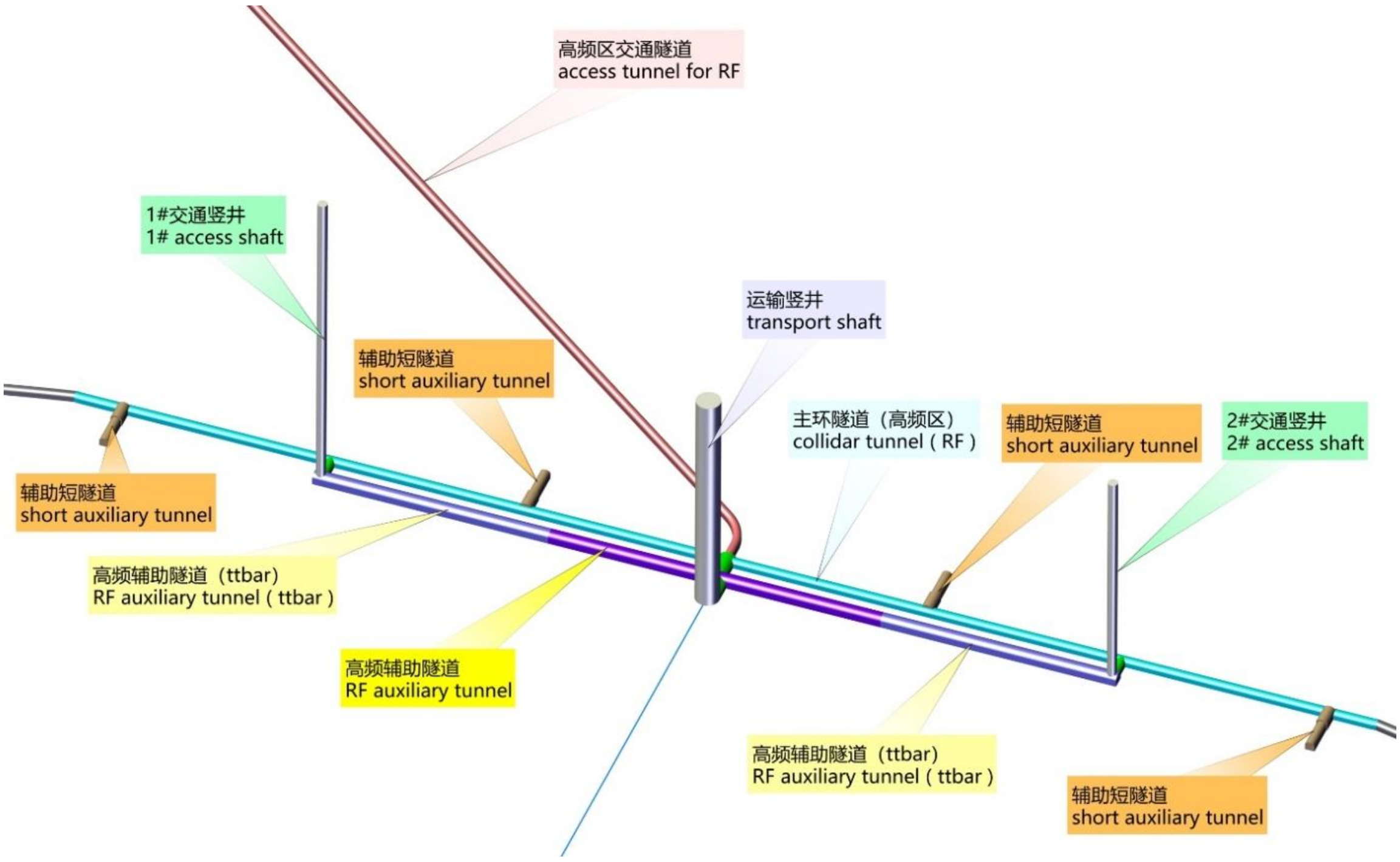

**Figure 9.2.8:** Layout of the RF Zone

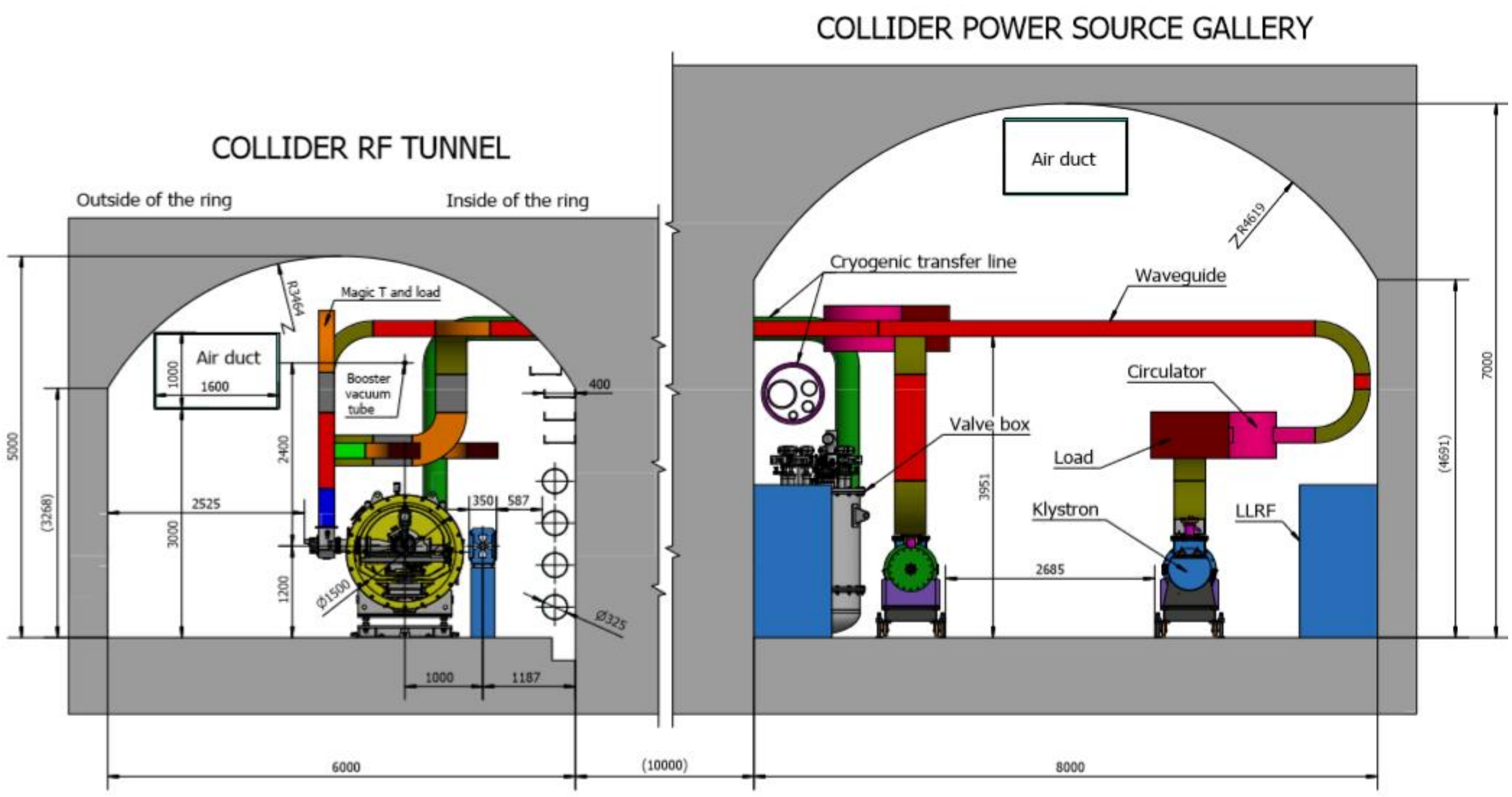

**Figure 9.2.9:** Tunnel cross section at the RF Zone

**Table 9.2.3.3:** Structure of the service tunnel in the RF zone

| Item | | Unit | Qty |
|---|---|---|---|
| **Service tunnels in the RF zone** | Qty | Nos. | 2 |
| | length of individual tunnel | m | 2000 |
| | Dimension | m | 8.00 × 7.00 |

*9.2.3.2.4 Liner Section of the Collider Tunnel*

A linear section for the Collider and Booster, of length of 1197.50 m. has the same cross section as in the arc sections, 6.00 m. wide and 5.00 m. high.

*9.2.3.2.5 Auxiliary Tunnels*

96 auxiliary tunnels (see Table 9.2.3.4) are uniformly distributed around the ring for the underground power transmission system, electronic equipment and other auxiliary components. Each auxiliary tunnel is 30.00 m long, and is 7.00 m × 7.00 m for the first 10.00 m, and 6.00 m × 6.00 m for the remaining 20.00 m.

**Table 9.2.3.4:** Auxiliary tunnels

| Item | | Unit | Qty |
|---|---|---|---|
| **Auxiliary tunnels** | Qty | Nos. | 96 |
| | length of individual tunnel | m | 30 |
| | Dimension (∩-shaped) | m | 7.00 × 7.00 (Front 20 m) |
| | | m | 6.00 × 6.00 (Rear 10 m) |

*9.2.3.2.6 Collider Experiment Hall*

The experimental halls are large, 40 m × 30 m × 30 m (L×W×H) and the service caverns for the detectors are 80 m × 18 m × 18 m. The main cavern floor level is 12.15 m lower than the collider ring tunnel and 8.55 m lower than the service cavern base as shown in Fig. 9.2.7 and itemized in Table 9.2.3.5.

**Table 9.2.3.5:** Experimental halls

| Item | | Unit | Qty |
|---|---|---|---|
| Cavern Qty | | Nos. | 2 |
| Main cavern | IP1、IP3 Dimension (L×W×H) | m | 40 × 30 × 30 |
| Service cavern | IP1、IP3 Dimension (L×W×H) | m | 80 × 18 × 18 |

*9.2.3.2.7 Linac to Booster Transfer Tunnel*

The Linac to Booster transfer tunnel connects to the inner side of the linear section between IP2 and IP3. This BTL tunnel is 1200.00 m. long. The Linac tunnel is 3.50 m. wide and 3.50 m. high with local width expanded to 5.50 m. The klystron gallery is 6.00 m. high and 8.00 m. wide. A damping ring hall is at the middle of the Linac. Its dimensions are 25 m wide, 3.50 m. high and of length of 40.00 m, The Linac tunnel section is near the surface and is inclined downward along 1825.00 m and then levels off to be parallel with the main ring. There it divides into two branch tunnels to transport electrons and positrons in clockwise and counterclockwise directions. These are 459.00 m in length. The transfer tunnel cross section width is 4.5 m and height 3.50 m. As with the other tunnels there is room for cooling, electrical and local control equipment, with clear space for pedestrians and room for moving and using maintenance equipment. Details are in Fig. 9.2.10 and Table 9.2.3.6.

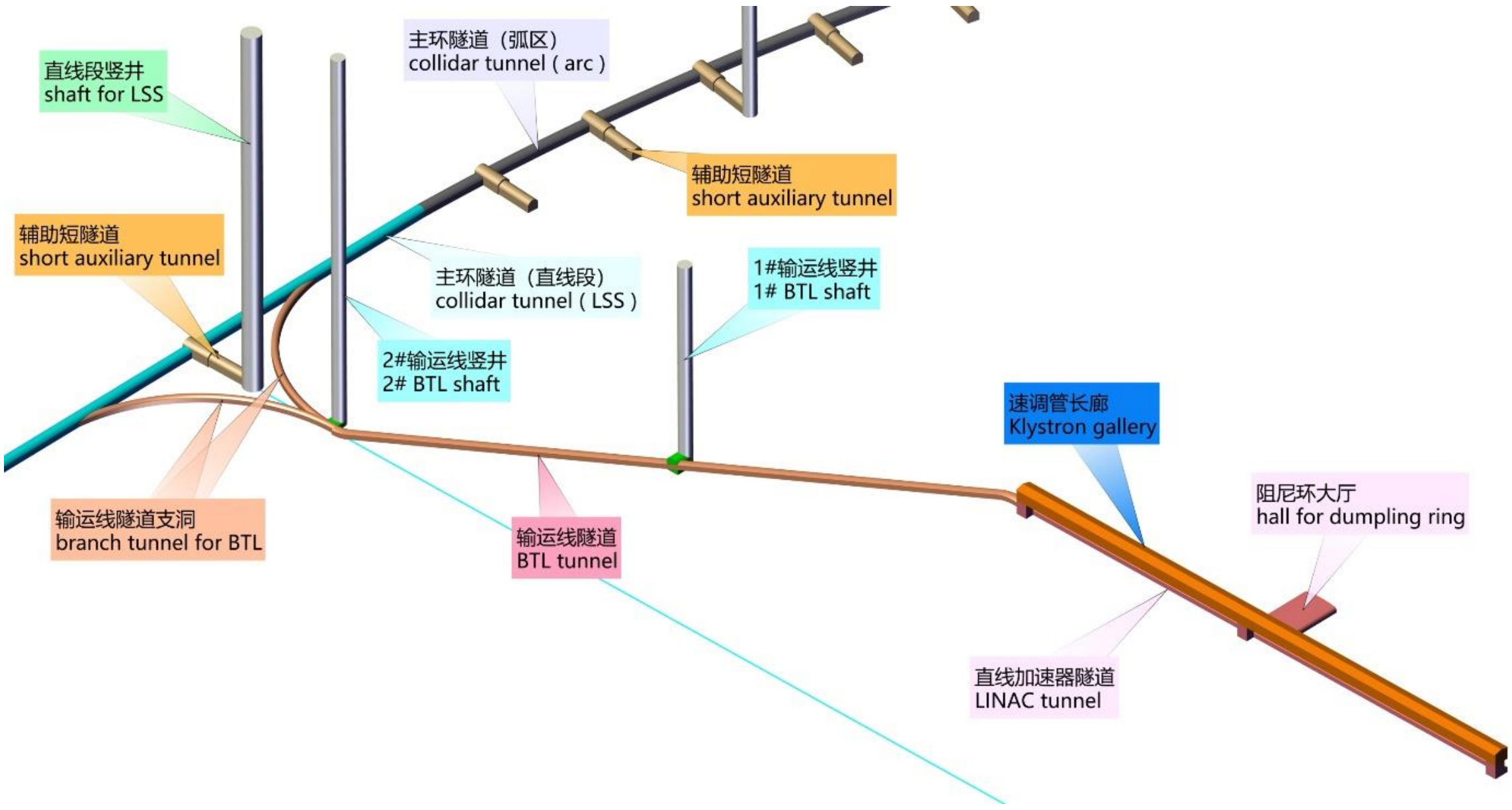

**Figure 9.2.10:** Layout plan of the Linac and BTL tunnels.

**Table 9.2.3.6:** Linac and BTL tunnels

| Item | | Unit | Qty | Remarks |
|---|---|---|---|---|
| Linear booster tunnel | Qty | Nos. | 1 | Local width 5.5 m |
| | length of individual tunnel | m | 1200 | |
| | Dimension (∩-shaped) | m | 3.50 × 3.50 | |
| Klystron gallery | Qty | Nos. | 1 | |
| | length of individual tunnel | m | 1200 | |
| | Dimension | m | 6.00 × 8.00 | |
| Damper ring cavern | Qty | Nos. | 1 | |
| | length of individual tunnel | m | 40 | |
| | Dimension (∩-shaped) | m | 20.00 × 3.50 | |
| BTL tunnel before branch tunnel | Qty | Nos. | 1 | |
| | length of individual tunnel | m | 1825 | |
| | Dimension (∩-shaped) | m | 4.50 × 3.50 | |
| Branch tunnel of BTL | Qty | Nos. | 2 | 2 branch lines |
| | length of individual tunnel | m | 459 | |
| | Dimension (∩-shaped) | m | 3.00 × 3.00 | |
| Transport shaft for LINAC | Qty | Nos. | 3 | Transport shaft for LINAC shows a square shape for connecting the LINAC tunnel to klystron gallery |
| | Dimension | m | 6.00 × 6.00 | |
| Shaft for BTL | Qty | Nos. | 2 | |
| | Diameter | m | 7.00 | |

*9.2.3.2.8 Gamma-ray Source Tunnel and Experiment Hall*

The gamma source tunnel is located at the counter-clockwise end of LSS2 and LSS4 and connects the collider ring tunnel and the gamma source equipment hall. The dimensions are in Table 9.2.3.7.

**Table 9.2.3.7:** Gamma source civil construction structures

| Item | | Unit | Qty | Remark |
|---|---|---|---|---|
| Gamma source tunnel | Qty | Nos. | 2 | Positioned at LSS2 and LSS4 |
| | length of individual tunnel | m | 300 | |
| | Dimension (∩-shaped) | m | 4.00 × 4.00 | |
| Equipment hall | Qty | Nos. | 2 | |
| | length | m | 15 | |
| | Dimension (∩-shaped) | m | 10.00 × 6.00 | |
| Shaft | Qty | Nos. | 2 | |
| | Diameter | m | 6.00 | |

*9.2.3.2.9 Access Tunnels*

Access tunnels are located at the inner side of IP1 and IP3 and outer side of IP2 and IP4, mainly for the transportation of personnel and small experiment equipment and consumables. The cross section is 5.00 m wide and 5.50 m high. See Table 9.2.3.8.

**Table 9.2.3.8:** Access tunnels

| Item | Unit | Qty |
|---|---|---|
| Qty | Nos. | 4 |
| Dimension (∩-shaped) | m | 5.00×5.50 |
| Longitudinal gradient | | 8% |

*9.2.3.2.10 Vehicle Shafts*

- Each IR includes a main cavern transport shaft with diameter 16.00 m, a service cavern transport shaft with diameter 9.00 m, a service cavern access shaft with diameter 6.00 m, and a bypass tunnel shaft with diameter 7.00 m.
- Each RF section is provided with one vertical shaft with 15.00 m. diameter for transport and two shafts with 6.00 m diameter for access.
- Each linear section of the collider ring is provided with an access and pipe shaft with diameter 10.00 m.
- A shaft for transportation and pipeline purpose with diameter 6.00 m is located at each gamma source equipment hall.
- Each curved section is provided with one access and pipe shaft with diameter 10.00 m and two ventilation shafts with diameter 7.00 m.
- Two access and pipe shafts with diameter 7.00 m are set at the end and midpoint of the slope section of the transfer tunnel.

These 42 shafts are summarized in Table 9.2.3.9.

**Table 9.2.3.9:** List of shafts

| Region | Item | Qty | Diameter(m) |
|---|---|---|---|
| IR | Transport shaft | 2 | 16.00 |
| | Bypass tunnel access shaft | 2 | 7.00 |
| | Auxiliary shaft | 2 | 9.00 |
| | Auxiliary access shaft | 2 | 6.00 |
| RF | Transport shaft | 2 | 15.00 |
| | Transport shaft | 4 | 6.00 |
| Linear tunnel | Access & pipe shaft | 4 | 10.00 |
| Curve sections | Access & pipe shaft | 8 | 10.00 |
| | Ventilation shaft | 16 | 7.00 |

*9.2.3.2.11 Design of the Underground Structures*

1) Tunnel shape:

Circular, inverted-U, and horseshoe shapes have all been considered for the tunnel cross section. If the TBM method is used, circular will be selected. See Fig. 9.2.11. If the drill-blast tunneling method is used, the dimensions will be determined according to construction and transportation requirements during construction, as well as equipment layout and accessibility requirements during installation and operation. The tunnel shape and construction method will be determined through comprehensive technical and economic comparisons. The inverted U-shape, shown in Fig. 9.2.12 is selected at this stage.

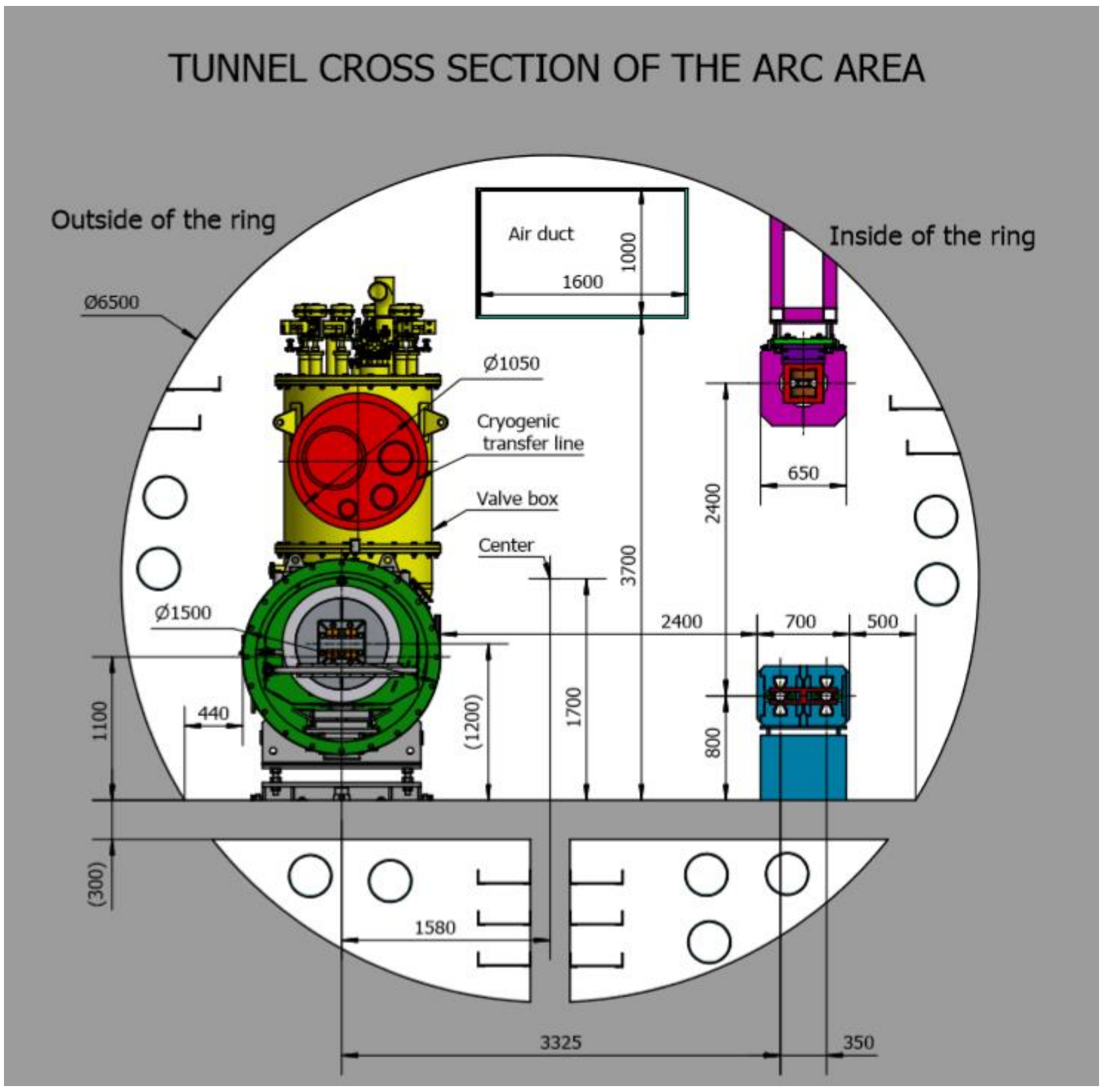

**Figure 9.2.11:** Circular cross section option in the Collider arc section.

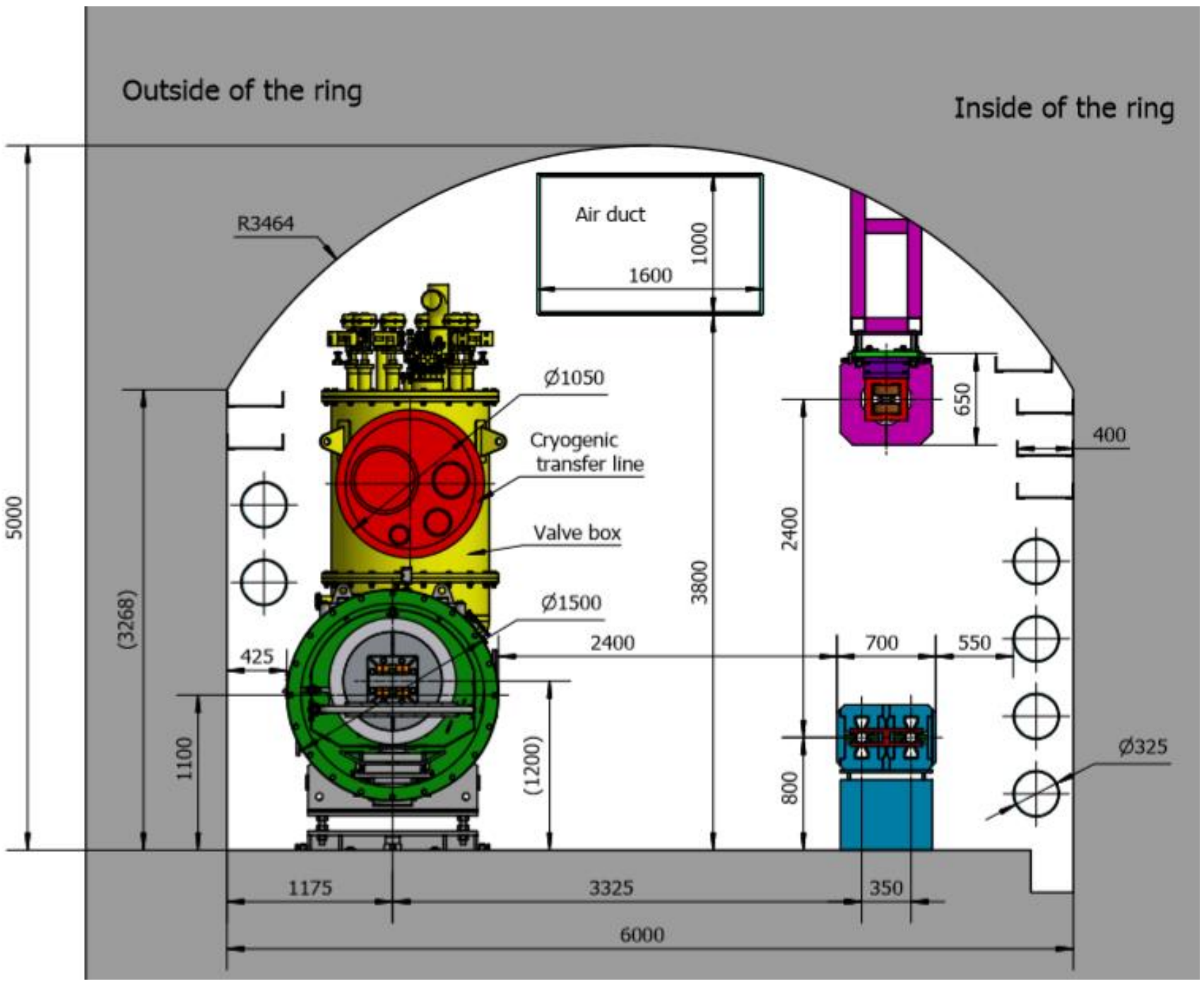

**Figure 9.2.12:** Inverted U-shape option in the Collider arc section

2) Tunnel lining and waterproofing:
   Waterproofing of the underground caverns is Grade I. Support and lining structures shall meet structural requirements and waterproof requirements. There are the following types of linings: bolt-shotcrete, reinforced concrete, steel fiber concrete and steel structure. Waterproof materials includes waterproof membrane, waterproof coating, rigid waterproof material and concrete admixture. Since the lining structure and waterproof material has a significant economic impact, the types of lining structure and waterproof material will be determined through comprehensive technical and economic comparison according to structural and waterproof requirements. At this stage, the following types are considered: drain holes + profiled steel sheets for the crown in Class II surrounding rocks, drain holes + profiled steel sheets for the crown + damp-proof decorative partitions in Class III surrounding rocks, waterproof membranes / boards / coating + 25~50 cm thick waterproof concrete lining in Class IV~V surrounding rocks.
3) Shaft structure
   Many shafts are distributed around the tunnels. Their size is determined by their functions. For example, the dimension of the transport shaft is determined by the size of the equipment to be transported, pipeline layout, evacuation passage and thickness required for support. The dimension of the shaft for construction and ventilation purposes is determined by construction ventilation requirements.
   Sprayed anchor + reinforced concrete lining is used for shaft support. The thickness of shotcrete and lining concrete is determined by shaft diameter and depth, surrounding rock type, groundwater and other factors.
4) Experiment halls
   The span is large; class I and II surrounding rocks are selected as much as possible for the cavern locations. The region of large geological tectonic belts, fault fracture zones, joint fissure development zones, high in-situ stress zones, goaf zones (where muck has been removed and the space filled with waste) and copious groundwater shall be avoided. The cavern depths should be determined by comprehensive analysis of the lithology, rock mass completeness, weathering unloading degree, in-situ stress magnitude, groundwater situation, construction conditions and experimental requirements and other factors. In general, the overburden thickness should not be less than twice the excavation width of the cavern.

A combination of flexible support and reinforced concrete lining is used due to the small depth of the experiment halls and the strict waterproof requirement. Flexible support is composed of one or several combinations of shotcrete, rock bolt, and anchor cable.

#### 9.2.3.3 ***Surface Structures***

All surface structures shall be as close to the access shafts as possible. In these buildings are located water cooling facilities, low-temperature facilities, ventilation systems, air compression systems, power transformer substations and electrical transmission and distribution and DC power supplies. The floor area of each surface structure is shown in Table 9.2.3.10.

**Table 9.2.3.10:** Floor Area of Surface Structures

| Surface Structure | | Location | | | | | | | | | | | | 合计 Total |
|---|---|---|---|---|---|---|---|---|---|---|---|---|---|---|
| | | P1 | P2 | P3 | P4 | P5 | P6 | P7 | P8 | C1-8 | V1-16 | LINAC | BT | |
| Control and duty rooms | $m^2$ | 1200 | 300 | 300 | 300 | 1200 | 300 | 300 | 300 | | | 400 | | 4600 |
| Magnet power source | $m^2$ | 2000 | 1500 | 1500 | 1500 | 2000 | 1500 | 1500 | 1500 | | | 500 | 400 | 13900 |
| High-frequency power source | $m^2$ | | | 4000 | | | | 4000 | | | | 8400 | | 16400 |
| 110kV substation | $m^2$ | 1500 | 1500 | 2500 | 1500 | 1500 | 1500 | 2500 | 1500 | | | | | 14000 |
| 10kV substation | $m^2$ | 600 | 500 | 1000 | 500 | 600 | 500 | 1000 | 500 | 2400 | 3200 | 400 | | 11200 |
| HVAC system | $m^2$ | 800 | 600 | 800 | 600 | 800 | 600 | 800 | 600 | 4800 | 3200 | 600 | 300 | 14500 |
| Low-temperature system (helium compression system) | $m^2$ | 1000 | | 4000 | | 1000 | | 4000 | | | | | | 10000 |
| Cooling water system | $m^2$ | 800 | 600 | 1500 | 600 | 800 | 600 | 1500 | 800 | 3200 | | 500 | 200 | 11100 |
| Experimental assembly hall | $m^2$ | 1500 | | | | 1500 | | | | | | | | 3000 |
| Magnet assembly hall | $m^2$ | 1500 | | | | 1500 | | | | | | | | 3000 |
| Transfer system | $m^2$ | 200 | 200 | 200 | 200 | 200 | 200 | 200 | 200 | 1600 | | 200 | 150 | 3550 |
| Air compression system | $m^2$ | 150 | 150 | 150 | 150 | 150 | 150 | 150 | 150 | | | 150 | | 1350 |
| Water cooling system | $m^2$ | 1500 | 1200 | 1200 | 1200 | 1500 | 1200 | 1200 | 1200 | 8000 | | 500 | | 18700 |
| Electronic room | $m^2$ | 1000 | 600 | 600 | 600 | 1000 | 600 | 600 | 600 | | | 450 | 100 | 6150 |
| Miscellaneous | $m^2$ | 500 | 500 | 500 | 500 | 500 | 500 | 500 | 500 | 2400 | 2400 | 200 | | 9000 |
| Total | $m^2$ | 15750 | 7650 | 18250 | 7650 | 12750 | 7650 | 18250 | 7850 | 22400 | 8800 | 12300 | 1150 | 140450 |

Note: C1-8 refer to surface structures corresponding to the ventilation & access shaft in the middle of each curved section, and V1-16 refer to surface structures corresponding to the ventilation shaft at each curved section.

## 9.2.4 Construction Planning

### 9.2.4.1 *Main Construction Conditions*

The Project is located 30.5 km from Qinhuangdao City. The project area is easily accessible as discussed in section 9.2.2. In addition, due to the support of national policies for poverty alleviation, there are village-to-village roads and household access to further improve the local traffic conditions.

The project area is located in a low mountain and hilly area, wide and open and favorable for a distributed arrangement of construction sites. It is also a suburban area with relative rapid socio-economic growth, and the arrangement of quarries/borrow area and spoil areas of the project will coincide with overall local planning, which may cause difficulties...

The project area is close to downtown Qinhuangdao. There are 110 kV and 220 kV substations in the Funing District and 35 kV substations in the townships. The power for each construction area will be by connection to these 35 kV substations. At a later stage in the construction a more permanent power supply arrangement will be considered. The project area passes twice through the Yanghe River. Since there is water in the river throughout the year and there are many rivers and reservoirs in the area, the project enjoys good water supply.

Part of the project area passes through residential areas with large population and numerous buildings, so there will be relatively severe disturbance from construction activities.

### 9.2.4.2 *Construction Scheme of the Main Structures*

#### *9.2.4.2.1 Collider Tunnel Construction*

Drill and blast and TBM tunnelling methods have been compared considering the layout as well as geological conditions, transportation, electric power availability, topography, transportation and construction duration. The drill and blast tunnelling method is chosen for construction of the experiment halls, auxiliary tunnels and access shafts. The 46 vertical shafts are for equipment transportation, ventilation and emergency egress.  They vary in diameter from 16 m to 7 m.

The method currently used in large and long tunnels mainly are the mining method and the tunnel boring machine (TBM) method, each of which has advantages and disadvantages. A comprehensive analysis will be done to choose which method to use for each of the various underground structures.  In choosing which method to use factors to consider are tunnel dimensions including depth, geology including rock compressive strength, structure and fragmentation degree, groundwater and environment conditions. The latter includes traffic capacity, water and power supply, layout of adits and the construction time line that best matches the production and installation of the CEPC technical components.

Drill and blast is conventional for medium and short tunnels, and has the advantages of flexible construction and ability to adapt to varying geology. When this method is used for the Collider ring, it uses various vertical shafts arranged around the circumference. There are 32 construction zones and 64 working faces. Each working face has a control length of about 1.6 km. The construction of the vertical shafts is carried out first. The collider ring is excavated by sectionalized full-face smooth blasting through drilling with self-made platform hand drills. The spoil is loaded to 4.0 $m^3$ mine carts through crawler loaders, transported to spoil bunkers at the bottom of the vertical shafts through electro mobile-traction mine carts, and then loaded to 4.0 $m^3$ buckets through crawler loaders. These buckets are lifted to the openings of vertical shafts with winches. The muck is directly loaded to 10 t dump trucks with chutes and then transported to the spoil area. The tunnel lining is done with formwork jumbo. Concrete is transported horizontally using a concrete mixer truck and delivered by a concrete pump.

Although the drill and blast method has been chosen for now, the reader of this CDR may wonder why TBMs are not preferred. China is making rapid development with this technology with among other things large high-speed rail systems and new subways. The TBM is a machine which excavates using mechanical energy to break the rocks. It avoids problems occurring in the traditional drill and blast such as impossible or extremely difficult location of construction adits, unreliable construction times due to constraints on the excavation surface, and difficulties in debris disposal and ventilation. On the other hand, a TBM has advantages under favorable geological conditions such as high excavation efficiency, small disturbance to surrounding rocks, small amount of over excavation, smooth excavation surface, good quality, a relatively safe working environment, small impact on the surrounding environment and minimal disturbance to the resident population. Suppose the construction of collider ring tunnel were to be carried out using TBMs. With adits around the circumference. 8 open-type TBMs would be installed, and the maximum control length of a single working face would be 17.2 km. The excavated spoil would be removed by belt conveyor with belt width 900 mm, and then transported with trucks. The transportation in the tunnel would be with locomotive-traction railcars.

##### *9.2.4.2.2 Shaft Construction*

The vertical shafts described above average a depth of about 129 m., but with considerable variation from a maximum of 245 m to a minimum of 50 m. They are all supported by bolt-shotcrete support + reinforced concrete lining.

The construction of the equipment transportation shafts for the experiment halls, diameters of 16 and 15 m, is carried out in three phases. In phase I, the construction of a 5 m. pilot shaft is done using the shaft-sinking method from the top down. After the pilot shaft is bottomed, enlargement in layers is carried out from the top down in phase II to the top elevation of the experiment hall, and then shaft wall concrete is placed using slip form from the bottom up in phase III.

The construction of other, smaller diameter shafts is carried out through drilling and masonry using the full-face shaft-sinking method from the top down.

The pilot shafts are excavated through full-face sectionalized drilling and blasting from the top down with the smooth blasting technique. Blast holes are drilled with an umbrella drilling rig and rock drill. The spoil is loaded to 4 $m^3$ hook type buckets with central rotary grab loaders, and then the buckets are lifted to the openings of vertical shafts using winches. The spoil is stirred automatically using hooks. The rock ballast is directly loaded to 10t dump trucks through chutes and then transported to the spoil area.

The enlargement of experiment hall shafts is carried out through bench blasting and excavation in layers, with protective layers reserved for pre-splitting. The bench height is 3 to 5 m. Presplit holes are drilled with down-the-hole drill and main blast holes are drilled with a hydraulic crawler drill. The rock ballast is stirred with a 1 $m^3$ hydraulic backhoe excavator and then transported from the shaft bottom through the pilot shaft, and finally loaded to 10t dump trucks through 2 $m^3$ side-dump loaders. Equipment and materials are lifted using a movable gantry crane at the shaft opening.

The concrete lining is done after excavation is complete, with concrete placed using slip form from the bottom up. The construction of concrete lining, excavation and support of other shafts is carried out through drilling and masonry, with concrete placed from the bottom up in a sectionalized manner using YJM type integrated steel forms moving downwards.

Concrete is produced by a surface concrete batching plant, horizontally transported by concrete mixer truck, vertically transported by vacuum chute, and manually vibrated and compacted by immersion vibrator.

##### *9.2.4.2.3 Experimental Hall Construction*

The dimensions and locations of the two experiment halls are described above. The burial depth of these large halls above the foundations is about 160 m. maximum and 80 m. minimum. The shafts and access tunnels serve for construction. The access tunnel is inverted U-shaped with dimension of 5 × 5.5 m and average slope gradient of 8%.

Construction of experiment halls is carried out in two phases: excavation of the chute shaft in phase I and excavation of the experiment hall in phase II.

For the chute shaft, the excavation diameter is 5.0 m and the construction method is the same as for the transport shafts. The excavation is carried out in layers from the top down, with the excavation height of 8 m for layer I and 5 to 7 m for the other layers.

Layer I has an excavation height of about 8 m and has 3 construction zones; its construction channels include the transport shaft of experiment halls and the chute shafts. The excavation of the pilot drift tunnel is carried out first, followed by adjacent excavation zones. The construction of each excavation zone is done through sectionalized full-face smooth blasting. Blast holes are drilled with a platform hand drill.

Each of the layers has one working zone divided into 4 parts for excavation and support. A 2.0 m thick protective layer is reserved close to the structural plane. Presplit holes are drilled with a 100B down-the-hole drill and main blast holes are drilled with a ROCD7 hydraulic crawler drill. Equipment and materials are lifted using a movable gantry crane at the shaft opening or transported using trucks through the access tunnels.

The spoil is removed in the same manner as in the shaft construction described above.

### 9.2.4.3 ***Construction Transportation and General Construction Layout***

##### *9.2.4.3.1 Construction Transportation*

Most sections along the route 30.5 km. from Qinhuangdao City are connected by simple roads which can be reconstructed and expanded to serve as on-site access roads. To meet the requirements for transportation of construction and experiment equipment, the site access roads in the areas of the experiment halls are of national standard Class III, each having a concrete or asphalt concrete pavement and a subgrade width of 8.5 m; the site access roads in high frequency zones are of national standard Class IV, each having a concrete or asphalt concrete pavement and a subgrade width of 6.5 m; the site access roads in the areas of the other access shafts and ventilation shafts are Class III single-lane roads for mines, each having a clay-bound macadam pavement and a subgrade width of 4.5 m.

The construction transportation shall be on local trunk roads as much as possible and for the shaft construction shall be on rural roads as much as possible. Transportation for the experiment hall construction shall be on permanent roads. We leave open the possibility of building some new roads.

##### *9.2.4.3.2 General Construction Layout*

Construction zones are of distributed close to the inlets of shafts and construction adits of all experiment halls, while temporary construction facilities such as workshops, warehouses and construction camps are in a centralized arrangement in each construction zone. Preliminary general construction layout planning specifies 32 centralized construction zones distributed at intervals of 3.2 km on the average.

Existing local resources shall be used. This includes roads, bridges, aggregate processing plants, concrete batching plants, production and living facilities, drainage facilities, and power transmission and communication lines.

The spoil area should be selected in cooperation with local cities and towns.

##### *9.2.4.3.3 Land Occupation*

The land permanently occupied is mainly where there are surface structures and permanent roads. The total area is about 1.28 million $m^2$. The land temporarily occupied is the material yard, the spoil area, temporary roads and construction sites, with a total area of about 2.70 million $m^2$, including 1.60 million $m^2$ temporarily occupied by the spoil area.

#### 9.2.4.4 ***General Construction Schedule***

##### *9.2.4.4.1 Comprehensive Indices*

Construction will occur in phases. First is a preparation phase. This includes land acquisition and resettlement, establishing supplies of water, power and compressed air, road connection and communications and site levelling. It is anticipated this will last for 6~8 months.

The construction of vertical shafts includes excavation, support and installation of lining. For a vertical shaft with diameter less than 10 m, the monthly advance is generally 20 to 40 m in depth through the upper overburden and then 60 to 80 m for the rock section. For a larger vertical shaft, say with a 15 m diameter, the shaft-sinking method is used and the advance is 30 to 50 m each month.

For tunnel excavation by drill and blast, the excavation progress is related to the length of the working face and the classification of the rock. Most rock in the collider ring tunnel are Class II and Class III rock. The working face length is 1.7 km, and the average advance is 80 to 100 m/month. Taking into account the auxiliary caverns, the excavation period is about 24 months. If, on the other hand, open-type TBMs are used, the average advance is 500 to 800 m/month, and the length of the working face is 17.2 km, so the tunnel excavation period is about 26 months. The excavation of auxiliary caverns should be carried out after completion of the main cavern excavation with multiple working faces and will take about 3 months.

Lining of ring tunnel: one concreting berth (12 m long) is finished every 2~3 days, and so the rate is 150 m/month. The length of the working face is 3.4 km in the tunnel lining phase. Waterproofing will be done concurrently. The period for tunnel lining and waterproofing is about 10 months in the case of a single working face. If the progress is too slow, the quantity of equipment or number of working faces can be increased.

Construction of experiment halls includes the pilot shaft and cavern excavation which will take about one month. The excavation and support of the top layer of the cavern is 4~5 months, and that of each of the remaining layers is about 2 months. The total excavation and support will take 10~13 months.

##### *9.2.4.4.2 Proposed Total Period of Construction with Drill-Blast Tunnelling Method*

The total construction period is 54 months, including 8 months for construction preparation, 43 months for construction of main structures and 3 months for completion.

The controlling project construction item is the collider ring tunnel from IP3 to LSS3. The critical path is as follows: construction preparation (8 months) → construction of vertical shafts (5 months) → tunnel excavation (24 months) → tunnel lining and waterproofing (10 months) → installation of ventilation equipment and access equipment of the shaft (4 months) → completion (3 months). The construction of surface structures is carried out as the project progresses, and is carried out concurrently with the underground work, so that will not lengthen the construction time line.

##### *9.2.4.4.3 Proposed Total Period for Construction with TBM Method (Using 8 Open-Type TBMs)*

If TBMs are used instead of drill and blast, then the length of the construction phases is somewhat different. Construction of the launching and arriving TBM shafts is combined with construction of the permanent vertical shafts. Designing and manufacturing of the TBMs would be done during the construction preparation phase. The TBMs will take 10 months for design and manufacturing, 2~3 months for transportation to the site and 2 months for installation and commissioning.

The total period of construction with the TBM method is 51 months, including 8 months for construction preparation, 40 months for construction of the main structures and 3 months for completion.

The critical path is as follows: Design, manufacture, and transportation of TBM equipment (13 months) → TBM assembly (2 months) → tunnel excavation with TBM (22 months) → collider ring tunnel short auxiliary tunnel as well as second time expansion excavation (3 months) → tunnel lining and waterproofing (8 months) → completion (3 months). The construction of surface structures is carried out concurrently.

The following measures should be taken to optimize the overall construction schedule for collider ring tunnel construction with the TBM method:

1) Speed up the design and manufacture of the TBMs;
2) Adopt advanced and reasonable TBM construction equipment to improve TBM boring speed;
3) Employ double-shield TBMs to enable tunnel excavation and lining be carried out at the same time, so as to reduce the time for lining the tunnel;
4) Use a combined approach of TBM boring and drill-blast tunneling methods. The drill-blast tunneling method should be used to deal with adverse geological segments, to reduce the time for the TBM to cope with adverse geological defects and thus improve the overall boring efficiency of the TBM;
5) Develop high-speed continuous vertical transport equipment for vertical shaft and vertical transport equipment for duct pieces. The large-diameter equipment transport vertical shaft in the hall will be used as the starting vertical shaft of the TBM and the vertical shaft for material transport. This reduces the amount of work required for the adits, reduces the TBM boring length and shortens the total period for construction with the TBM method.

Through combining the two methods, the advantages of TBM such as fast boring speed, low environmental impact, safe construction and flat and smooth excavation face can work together with the advantages of drill and blast such as flexible construction, adaptability to geological conditions, improve safety and lower the capital cost.

## 9.3 Electrical Engineering

### 9.3.1 Electrical System Design

#### 9.3.1.1 *Power Supply Range and Main Loads*

Electric power is primarily for the Collider complex and high-energy physics experiments but also for ventilation, air conditioning, lighting and elevators and other common needs.

Loads for the machine and experiments are generally composed of the first and second class loads. Loads for ventilation, air conditioning, lighting, elevator and other common facilities are second and third class loads. These are itemized in Table 9.3.1

**Table 9.3.1:** Main power loads for machine, experiments and general facilities

| | System for Higgs (30 MW /beam) | Location and Power Requirement (MW) | | | | | | Total (MW) |
|---|---|---|---|---|---|---|---|---|
| | | Collider | Booster | Linac | BTL | IR | Surface building | |
| 1 | RF Power Source | 103.8 | 0.15 | 5.8 | | | | 109.75 |
| 2 | Cryogenic System | 15.67 | 0.89 | | | 1.8 | | 18.36 |
| 3 | Vacuum System | 9.784 | 3.792 | 0.646 | | | | 14.22 |
| 4 | Magnet Power Supplies | 47.21 | 11.62 | 1.75 | 1.06 | 0.26 | | 61.9 |
| 5 | Instrumentation | 0.9 | 0.6 | 0.2 | | | | 1.7 |
| 6 | Radiation Protection | 0.25 | | 0.1 | | | | 0.35 |
| 7 | Control System | 1 | 0.6 | 0.2 | 0.005 | 0.005 | | 1.81 |
| 8 | Experimental Devices | | | | | 4 | | 4 |
| 9 | Utilities | 31.79 | 3.53 | 1.38 | 0.63 | 1.2 | | 38.53 |
| 10 | General Services | 7.2 | | 0.2 | 0.15 | 0.2 | 12 | 19.75 |
| | Total | 213.554 | 20.972 | 10.276 | 1.845 | 7.385 | 12 | 270.37 |

The total electrical load for physical experiments and general facilities is about 270 MW.

### 9.3.1.2 *Power Supplies*

It is proposed to use 220 kV for the project, and to have two 220 kV central substations (220kV/110kV/10kV) in the project area with two 210 MVA transformers in each substation.

A 110kV/10kV substation will be placed near the shaft ground exits of IR and RF (IP1-IP4) and collider ring tunnel straight sections (LSS1-LSS4), with 8 step-down substations in total. Two 50 MVA transformers and twenty-four (24) 10 kV outgoing lines will be provided for each of the substations, mainly to supply power for the shaft exits at ground level and the underground tunnels. Fig. 9.3.1 shows the electric distribution system network.

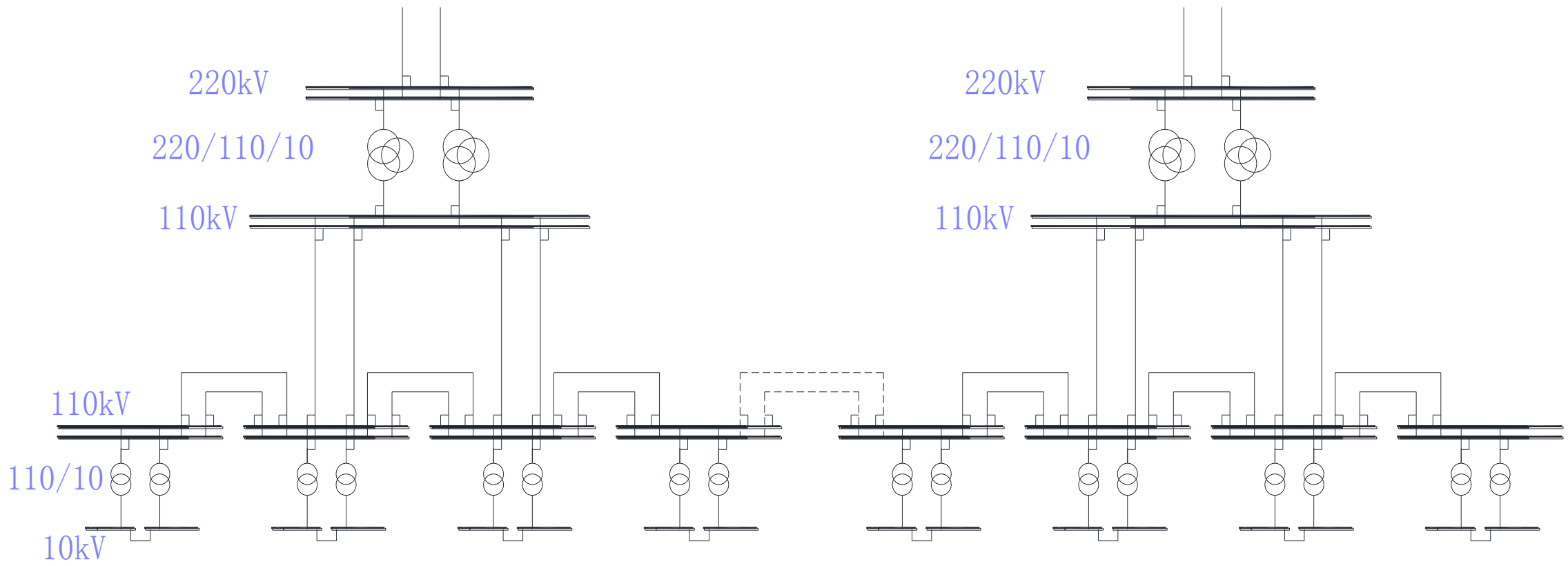

**Figure 9.3.1:** Schematic diagram for the 220 kV and 110 kV power supply network

Power for ancillary buildings at the exit of each ground shaft will be provided by 10kV/0.4 kV transformers at the ground exit of each shaft, with a total of 34 systems. The 10 kV power is from the 110 kV/10 kV step-down substation.

Underground 10 kV cables will go from the 110kV/10kV step-down substations to the underground tunnel via each shaft to form a 10 kV looped network in the tunnel. A 10kV/0.4kV transformer and distribution system will be near each load point. There will probably be 100 such systems.

For critical loads where a power failure could cause damage, diesel generators, EPS power supplies, or UPS will be installed.

### 9.3.1.3 *Additional Details on the 220 kV, 110 kV and 10 kV Systems*

#### *9.3.1.3.1 220 kV Power Supply System*

For each of the two 220 kV central substations there will be two 220 kV incoming lines and 4 110 kV outgoing lines. There will be 12 10kV outgoing lines for a total of 24 outgoing lines.

There will be double-bus wiring for 220 kV, double-sectionized double-bus wiring for 110 kV and sectionalized single-bus wiring for 10 kV.

These transformers will be three-phase three-volume natural oil circulating air-cooled on-load regulating. Outdoor GIS equipment is chosen for both the 220 kV and 110 kV systems. For the 10 kV system there will be withdrawable metal-enclosed high-pressure vacuum switch cabinets.

The outdoor 220 kV GIS equipment will be laid out in a single-row. The outdoor 110 kV system and the indoor 10 kV systems will be in a double-row layout.

The substations will be unattended. There will be a supervisory computer control system using micro-processor based protection relay and automatic safety devices, dual-configuration for protection of the 220 kV systems. DC systems will have a dual-changer, dual-storage configuration and sectionalized single-bus configuration with 110 V or 220 V. Communication with the electric power grid distribution station will be through optical fiber.

##### *9.3.1.3.2 110 kV Power Supply System*

One 110kV/10kV step-down substation will be installed in the IR and RF (IP1-IP4) and near the surface shaft exit in the RF cavity sections (RP1-RP4), with 8 substations in total.

In addition to the two 50 MVA transformers there're will 4 to 6 110 kV incoming lines and 24 10 kV outgoing lines.

There will be 110 kV double-bus wiring, 10 kV double-bus wiring and 10 kV sectionalized single-bus wiring.

The system will be three-phase with double-volume natural oil circulating air-cooled on-load regulating transformers. Outdoor GIS equipment is chosen for the 110 kV system. There will be withdrawable metal-enclosed high-pressure vacuum switch cabinets.

The main 110 kV transformer is indoors in single-row layout form and is provided with overhead incoming and outgoing lines; the 10kV indoor equipment is in double-row layout form and is provided with cable outlets.

The substation is unattended and the controls, distribution and communications with the substation are the same as for the 220 kV system described above.

##### *9.3.1.3.3 10 kV Power and distribution System*

A total of thirty-four (34) 10kV/0.4kV transformers and distribution systems will be installed.at ground level. In addition, there will be approximately one hundred 10kV/0.4kV transformers and distribution systems near each load point in the underground tunnel. All the 10 kV power supplies are fed from the 110 kV/10 kV step-down substations.

Sectionalized single-bus wiring is chosen for both the 10 kV and 0.4 kV systems.

It is necessary to use filtering devices to suppress the large number of higher harmonics produced by the large number of large-capacitance rectifying components. A centralized reactive compensation mode will be used.

Dry-type transformers are selected. There will be withdrawable metal-enclosed high voltage switch cabinets for the 10 kV systems and low voltage extraction type switch cabinets for the 0.4 kV systems. There will be active power filters.

The transformers are indoors in a double row and provided with cable outlets.

#### 9.3.1.4 ***Lighting System***

There are normal lighting and emergency lighting systems for both the ground facilities and the underground facilities. The emergency lighting system will insure good, visible lighting in important places for personnel excavation. The emergency lighting system may be powered with a diesel generator or EPS.

All lighting fixtures shall be energy-saving type and fluorescent lamps used throughout. Mining lamps shall be used in the experiment and assembly halls. Moisture-proof lamps shall be used in the tunnels.

### 9.3.2 **Automatic Monitoring System**

Objects to monitor include the power supply systems in shafts and underground tunnels, ventilation and air-conditioning systems and other common facilities. The monitoring of each 110 kV substation is accomplished by the substation integrated automation system included in the power supply system.

One ground monitoring center shall be established. It includes a data server, operator workstation, printer, network equipment and UPS. A local control unit will be provided near each controlled system to collect information. Redundant communication will be provided with industrial Ethernet and optical fiber.

### 9.3.3 **Communication System**

#### 9.3.3.1 ***Service Objects***

Internal communication includes voice, network communication and other communications at each experiment hall and equipment room in the shafts and tunnel. The system described below does not yet include the communication systems for the ground level buildings.

#### 9.3.3.2 ***Communication Mode***

Optical cables will be installed in cable trays in the shafts and tunnels. MSTP (Master/slave token passing) optical communication equipment shall be installed in each machine room to form a self-correcting and multi-service ring-type optical fiber communication network. The equipment bandwidth is10 Gb/s.

One soft switch system in each communication center provides service for each user. Telephone communication between each user and externally will be done with a software exchange system.

One leakage coaxial cable communication system around the tunnel provides wireless communication for personnel.

#### 9.3.3.3 ***Computer Network***

The computer network has a core layer, a convergence layer and an access layer. The core layer is in the ground-level control center and the convergence layer is in each experiment hall. The access Ethernet switch for each user is in the access layer. Redundancy is provided by the Ethernet switch and other equipment at the core and convergence layers. Fiber optics connects the layers and the transmission rate is several gigabits. The connection between the core layer equipment and external internet is done by the router and internet security equipment.

#### 9.3.3.4 ***Communication and UPS***

One 48 V high-frequency switching communication power supply (each equipped with two groups of 48 V storage batteries) is located in the machine room to supply power for optical communication equipment and the software exchange system. The UPS will supply power for at least three hours.

### 9.3.4 **Video Surveillance System**

The video surveillance system is composed of a network of HD cameras, transmission network and monitoring center equipment. These cameras will be installed in shafts, in the tunnel and experiment halls. The hard disk video and other storage devices, video servers and large-screen video monitoring devices shall be installed in the monitoring center.

## 9.4 Cooling Water System

### 9.4.1 **Overview**

Most of the electrical power consumed is absorbed by cooling water. In addition to its cooling function, it is quite critical for some subsystems where the operating temperature must be held constant.

The cooling water system consists of a low-conductivity water (LCW) closed-loop circuit, a cooling tower water (CTW) circuit, and a deionized water make-up system. The LCW system absorbs heat from the various devices, and this heat is transferred through heat exchangers to the cooling tower water circuits (CTW) and finally rejected into the atmosphere by cooling towers. A flow diagram of a typical cooling water system is shown in Fig. 9.4.1.

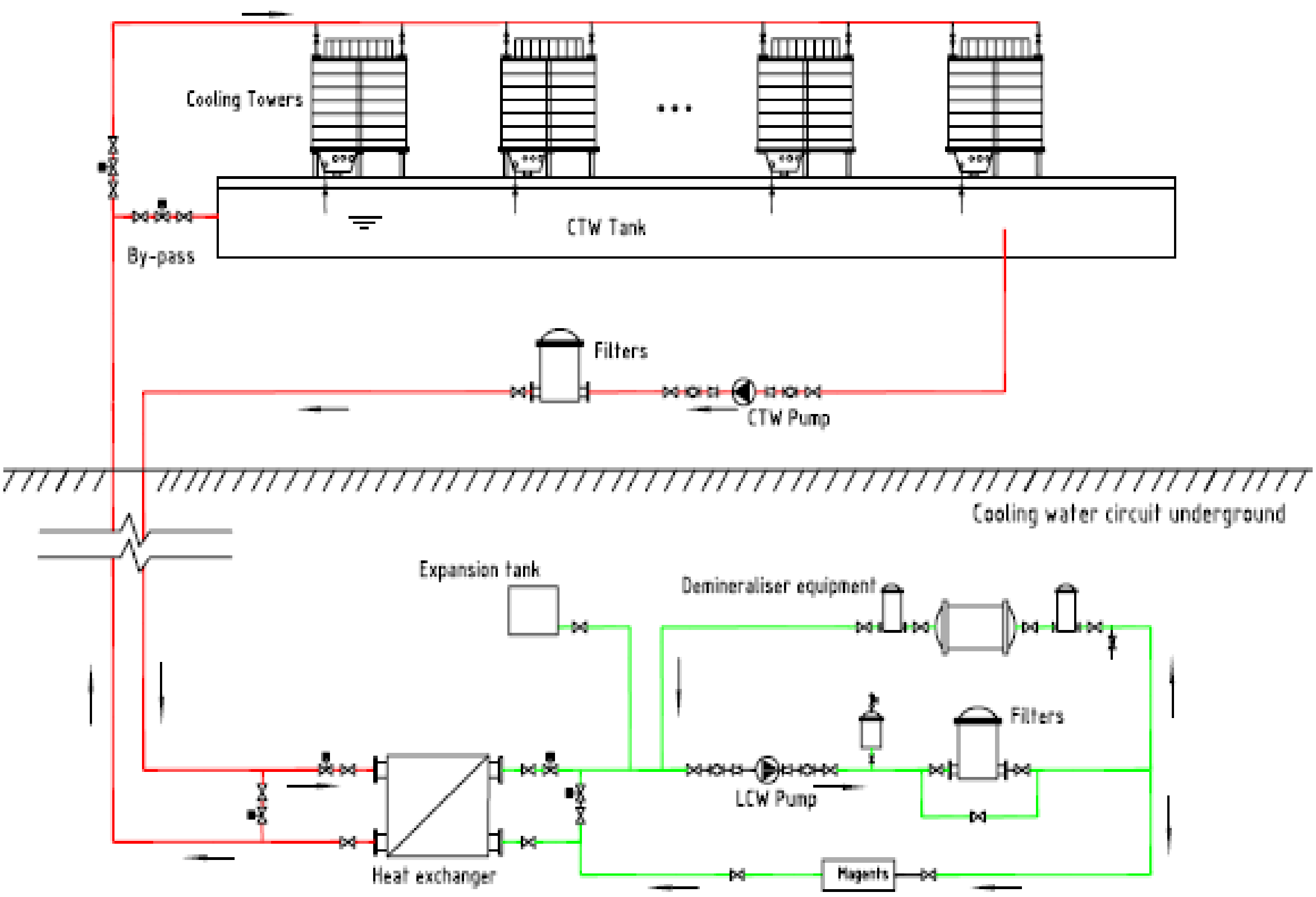

**Figure 9.4.1:** Flow diagram of a typical cooling water system

The major heat sources are RF power sources, magnets, vacuum chambers, cryogenic compressors and power converters (magnet power supplies). The total heat load dissipated is 212.387 MW during operation for Higgs physics. The estimated heat loads are summarized in Table 9.4.1.

**Table 9.4.1:** Estimated cooling water heat loads (at Higgs)

| System | Location and heat loads (MW) | | | | | |
|---|---|---|---|---|---|---|
| | Ring | Booster | Linac | BTL | IR | Total |
| Accelerating tube / Waveguide | | | 2.32 | | | 2.32 |
| Power source | 43.8 | 0.15 | 3.48 | | | 47.43 |
| Cryogenics | 11.62 | 0.68 | | | 1.8 | 14.1 |
| Experimental devices | | | | | 2.5 | 2.5 |
| Magnets | 33.763 | 7.604 | 1.367 | 0.838 | | 43.572 |
| Vacuum chamber of ring | 60 | | | | | 60 |
| Power for magnets | 4.721 | 1.162 | 0.175 | 0.093 | 0.026 | 6.177 |
| Condenser in stub tunnel | 18.169 | | | | | 18.169 |
| Pump | 16.787 | | 0.745 | 0.121 | 0.466 | 18.119 |
| Total | 188.86 | 9.596 | 8.087 | 1.052 | 4.792 | 212.387 |

There are 16 CTW pump stations at each point of the ring and an additional CTW system for the Linac. The CTW equipment will be installed close to the access shafts, and its LCW circuits will be located underground at machine level in order to reduce pipe pressure. There will be a deionized water plant at each point of the ring and at the Linac. These plants will supply low-conductivity makeup water for each of the LCW circuits in the area. The layout of cooling water circuits in the main ring is shown in Fig. 9.4.2.

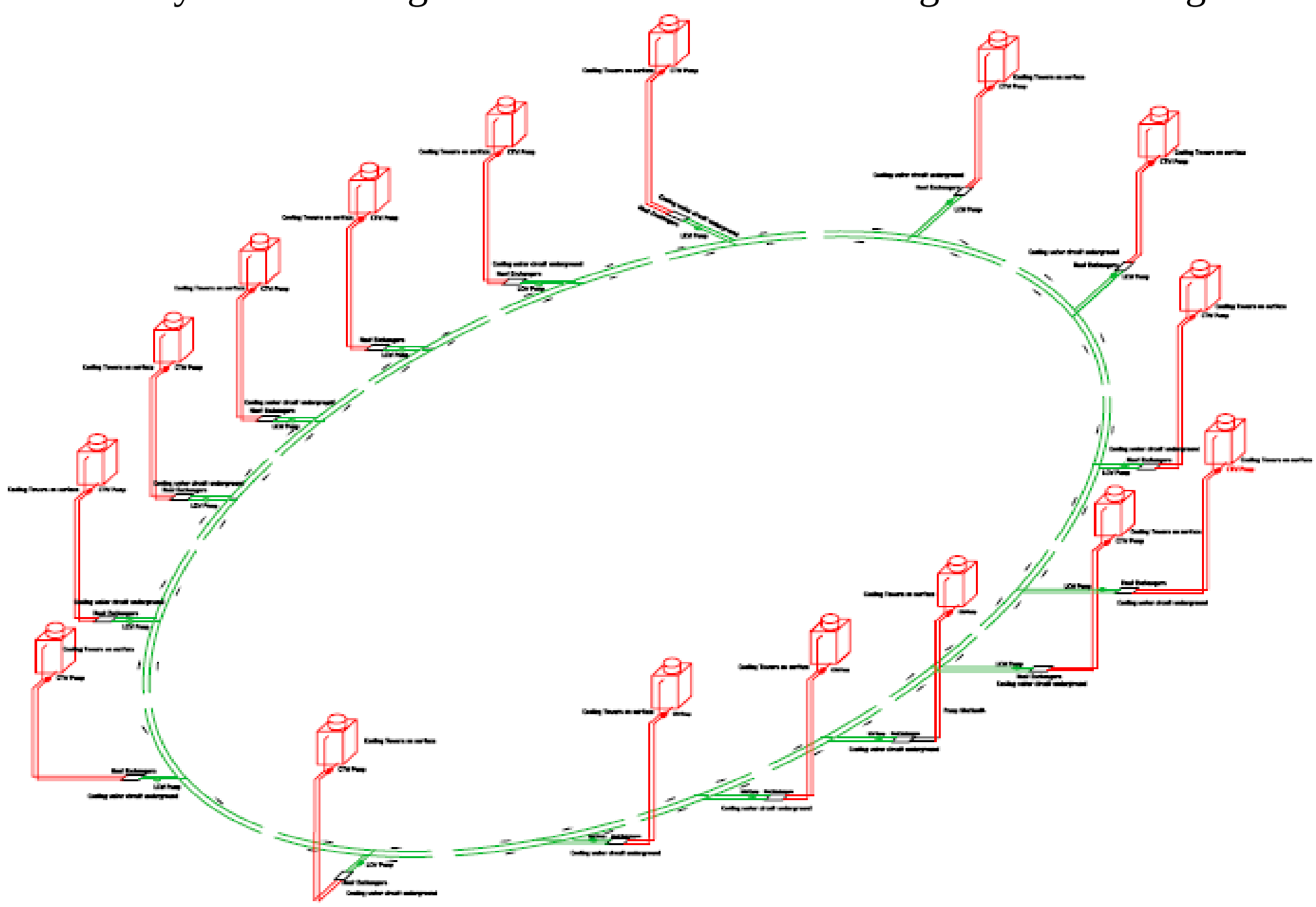

**Figure 9.4.2:** Layout of cooling water circuits in the ring

### 9.4.2 Cooling Tower Water Circuits

The cooling tower water system provides coolant for the heat exchangers in the LCW circuits. They will be installed at each point of ring and Linac area where service buildings will house pumps and filters and other components. Supply water temperature will be 29°C based on a wet-bulb air temperature of 26°C ambient. The main parameters of those circuits are shown in Table 9.4.2. Make-up water is introduced through an automatic valve which is connected to the raw water pipeline.

**Table 9.4.2:** Parameters of the cooling tower water system

| Parameters | Ring | Booster | Linac | BTL | IR | Total |
|---|---|---|---|---|---|---|
| Heat load (MW) | 188.86 | 9.596 | 8.087 | 1.052 | 4.792 | 212.387 |
| Supply water temperature | 29 ℃ | | | | | |
| Temperature rise | 5 ℃ | | | | | |
| Flow rate ($m^3$/h) | 32478 | 1650 | 1390 | 180 | 824 | 36522 |

### 9.4.3 Low Conductivity Water (LCW) Circuits

There are several closed-loop LCW subsystems in the LINAC, BTL, main ring, and experimental areas. They are defined by equipment characteristics, operational requirements, and location. They are:

- LINAC area: accelerating tube, waveguide circuit, klystron circuit, and the beam transfer line circuit.
- Main ring area: magnet circuits, vacuum circuits, RF circuits, and power resource circuits at each point. Each of magnet and vacuum circuits will serve two half adjacent sixteenth section of ring.
- Experimental area circuits.

The heat generated is removed by those LCW systems, which is bypassed with about 1% flowrate through a demineralizer to maintain the low conductivity water at over 1 MΩ.cm specific resistance. The main design parameters of those circuits are shown in Table 9.4.3. Supply water temperature for the ring will be 31°C based on the outlet temperature of the cooling tower water which may, however, be exceeded in the summer months. This relatively high temperature has been chosen in order to avoid the use of expensive refrigerators.

**Table 9.4.3:** Parameters of low-conductivity circuits

| system | location | Heat loads | Flow rate |
|---|---|---|---|
| | | MW | $M^3$/hr |
| Accelerating tubes / Waveguide circuits | Linac | 2.519 | 419 |
| Power source circuit （Linac） | | 3.692 | 314 |
| Power supplies for magnets （Linac） | | 1.035 | 148 |
| Magnet circuit （Linac） | | 1.483 | 244 |
| Power for BTL magnets | BTL | 0.119 | 18 |
| Magnet circuit （BTL） | | 0.931 | 151 |
| Magnet and condenser circuit | Ring | 4.427*16 | 715*16 |
| Vacuum chamber circuit | | 4.029*16 | 340*16 |
| Power source circuit （Ring） | | 23.289*2 | 1984*2 |
| Power supplies for ring magnets | | 0.322*8 | 44*8 |
| Circuit for experiment areas | IP | 1.380*2 | 226*2 |
| **Total** | | 196.989 | 22946 |

### 9.4.4 **DI Water System**

There will be a deionized water system at sixteen ring locations and at the Linac. They will supply low-conductivity makeup water for each of the LCW circuits in the area. The specific resistivity of its supply water is more than 10 MΩ.cm. The membrane process will be utilized for demineralizing.

## 9.5 Ventilation and Air-Conditioning System

### 9.5.1 **Indoor and Outdoor Air Design Parameters**

#### 9.5.1.1 ***Outdoor Air Parameters***

The following is based on the proposed Funing site. The outdoor parameters in this area are as follows:

Winter

- Dry-bulb temperature during winter heating is -9.6°C
- Temperature during winder ventilation is -4.8°C
- Temperature during winter air conditioning is -12°C
- Relative humidity during winter air conditioning is 51%
- Wind speed in winter is 2.5 m/s

Summer

- Dry-bulb temperature during summer air conditioning is 30.6°C
- Wet-bulb temperature during summer air conditioning is 25.9°C
- Temperature during summer ventilation is 27.5°C
- Relative humidity during summer ventilation is 55%
- Daily mean temperature during summer air conditioning is 27.7°C
- Wind speed in summer is 2.3 m/s

#### 9.5.1.2 *Indoor Design Parameters*

The temperature of the tunnel is controlled within 30-34°C and is kept below 35°C The relative humidity is controlled within 50%-60% and is smaller than 65%.

The temperature of the 4 experiment halls is controlled to about 26°C and the relative humidity is controlled within 50%-60% and is smaller than 65%.

### 9.5.2 **Tunnel Air-Conditioning System**

The air conditioning cold load of underground collider ring tunnel is about 6 MW.

During operation, the heat generated by electrical equipment is mainly removed by the low-conductivity cooling water system and the remaining heat may be removed by the tunnel ventilation and air-conditioning system. The ventilation and air-conditioning system also is important for providing fresh air for personnel, keeping equipment temperatures down, dehumidification to prevent condensation, discharging gas generated in accidents, discharging air in the tunnel before personnel enter and filtering the exhaust gas.

The tunnel is divided into 32 sections, with about 3 km spacing, and by shafts in the experiment halls, vent shafts, access shafts and shafts in the RF region. Each section is considered to be independent for the ventilation and air-conditioning system. Each shaft is used for air supply and exhaust.

The 32 ventilation and air-conditioning equipment rooms are at ground level for 32 vent shafts. 3 combined air conditioning units are in each room (with 2 for use and 1 for standby). The processed air is sent to chambers connected to the underground tunnel with air vents, and to the air pipe in the tunnel via a relay blower. The air is then sent to an exhaust pipe through relay air exhaust blowers at the bottom of the shafts. Finally, the air is sent to the inlet of a combined air conditioning unit or vented outdoors.

The operation of the cooling water unit stops when the outdoor air temperature is smaller than the air supply temperature of the air conditioner. Then the outdoor air is sent directly to an air conditioning area without undergoing cooling. All exhaust air is discharged out of the loop tunnel.

### 9.5.3 **Air-Conditioning System in Experimental Halls**

Independent ventilation and air-conditioning systems are installed in the IP1, IP3 experiment halls and adjacent assembly areas. Specialized smoke, argon and gas emission systems will be installed at the same time. The atmospheric pressure in enclosures requiring gas emission is always higher than that in other adjacent cavities due to the continuous operation of the gas emission system. Main air discharge is done by the upper air exhaust and collection chamber. The air conditioner is located in ground level buildings and connected to the air supply and discharge point via the vent pipe in the vent shaft.

In case of fire, the ventilation system will stop automatically. The smoke extraction system may be switched on manually after personnel are evacuated and the fire is under control. For this reason, manual instructions for the fire brigade will be provided both at ground and underground locations.

### 9.5.4 **Ventilation and Smoke Exhaustion System**

The emergency smoke control and exhaust system guarantees the safety of personnel working underground. It will be combined with the mechanical air exhaust system. Emergency smoke exhaust will be provided in both the tunnel and the experiment halls.

Depending on the final site selection it may be possible that the large amount of waste heat that will be generated could provide energy to help implementation of national rural revitalization in surrounding villages. It is also possible to utilize waste heat for biological applications.

## 9.6 Fire Protection and Drainage Design

### 9.6.1 **Fire Protection Design**

#### 9.6.1.1 ***Basis for Fire Protection Design***

The fire protection of this project is based on fire prevention and supplemented by fire fighting. Comprehensive fire protection measures are adopted in four steps: "prevention, cut-off, extinguishing, and exhaust." Fire prevention and exhaust systems, hydrant and fire extinguisher systems, and fire detection and fire alarm systems are combined with building fire prevention and evacuation, to minimize fire hazards.

Design is based on relevant codes: Design on Building Fire Protection and Prevention, Design Code for Scientific Experiment Buildings, Design Code for Office Building, and Design of Extinguisher Distribution in Buildings.

The tunnel and the experiment hall contain many computers and electromagnetic equipment. During operation, the interior is unmanned. Only professional personnel will enter for maintenance. And then, generally only a few personnel will enter and they are all familiar with evacuation procedures. There is a small possibility of spontaneous combustion of internal equipment. Explosion is not considered in case of an internal fire and at most one ignition point is considered at the same time.

The experiment hall is categorized as an underground equipment room. The ring tunnel is of a special nature and there is no available industry standard for it in China. The fire protection design refers to the urban traffic tunnel sections in the Standard on Fire Protection for City Underground Tunnels and the Code of Design on Building Fire Protection and Prevention as well as underground engineering specifications such as the Technical Code for Urban Utility Tunnel Engineering, the Code for Design of Prevention of Mine Fires in Coal Mines, and the Coal Mine Safety Code. Besides, special measures based on specific conditions such as internal equipment characteristics, and the personnel job descriptions can be considered in the design.

#### 9.6.1.2 ***Fire-fighting for Buildings***

1) Fire Resistance of Building Structures
   The fire protection rating of the ground level buildings is Grade I or Grade II. Non-combustible or refractory materials are employed.
   To reduce the fixed fire load in the tunnel or the experiment halls, the building and interior decoration materials for tunnel lining, experiment hall and other parts of the building should all be incombustible except for joint materials. The air ducts of the ventilation system should be made of incombustible materials, while the flexible joints can be made of refractory materials. The cables in tunnels should be flame-retardant cables or mineral-insulated cables, and the cable trays and conduits shall be fire-resistant. For vertical and inclined shafts and independent refuge room connected with the main tunnel and used for safety evacuation, emergency refuge and other purposes, the fire resistance limit of their load-bearing structure should not be lower than the fire limit requirement of the main structure of the tunnel.

2) Safety Escape
   - The experiment halls shall meet the requirements (Clause 12.1.10 of the Code for Fire Protection Design of Buildings). The maximum allowable building area of each fire zone for underground equipment shall not exceed 1500 $m^2$. The area of all experiment halls of the project satisfies this requirement. Each experiment hall is provided with two and more vertical access shafts which are directly connected with the surface outdoors, so as to meet the requirement of two safe exits in every fire zone. Meanwhile, fire cuts are arranged around every hall. The area of the fire cuts is based on the maximum number of people simultaneously in the fire zone and meets the requirement of 5 persons per $m^2$.
   - For the ring tunnel the evacuation vertical shaft is combined with the air shaft. There is one vertical shaft directly connected with the outside at about every 3000 m; the evacuation staircase leading directly to the outside is inside the air shaft; a smoke free cavern and fire cut are at the entrance of the evacuation staircase, so as to guarantee the safety of the vertical shaft, and also create space and time for persons who are unable to escape on their own to wait for rescue.

3) Arrangement of Fire Partitions
   - The distance for staff to go to the evacuation exit from the farthest point is 1500 m. From a series of simulation tests in Europe, the maximum distance for persons inside a tunnel to escape when the smoke density does not create an issue initially, is 250 m. The existing vertical access shafts can barely meet such a requirement, so other facilities need to be added. Fire-proof board is used as fire partitions at intervals. The locations where pipes cross fire-proof board is sealed with fire-proof material, in order to form relatively independent fire zones; and a Class A fire door is placed between every two fire zones. Every fire zone is able to use the fire zones on both sides as the emergency escape channels. Considering the maximum escape distance from the farthest point, fire partitions can be installed at 500 m intervals, so as to guarantee personnel safety to the greatest extent, protect the nearby equipment, and reduce the loss from fire as much as possible.
   - If it is difficult to arrange fire partitions, fire cuts shall be added inside as per the requirement of Clause 12.1.8 of the Code for Fire Protection Design of Building This clause specifies that a single tunnel shall be provided with refuge facilities such as personnel evacuation exits directly leading to the outside, or with independent shelter. There is a short auxiliary tunnel independent from the collider ring tunnel at about every 1000 m; adding independent fire cuts can be used to meet the personnel refuge requirement within the maximum evacuation distance.

#### 9.6.1.3 *Water-based Fire-fighting*

1) Necessary items are the following: indoor fire hydrants and extinguishers underground; municipal water supply for outdoor fire protection; gas fire-extinguishing system for the control room; seepage drain pump based on experience from similar projects and an estimate of leakage.
2) Fire hydrants are designed following the regulations of the Code for Fire Protection Design of Building (GB50016-2014). The spacing between hydrants shall not be more than 50 m. Therefore, there will be a total of 2100 sets of type SNJ65 pressure stabilizing and reducing hydrants. The hydrant mouth is perpendicular to the wall surface, and 1.10 m above the floor. A combined type fire hydrant box (04S202-P21) with extinguisher is chosen; the fire hydrant box is complete with one DN65 hydrant, one Ø19 water gun, one 25 m rubber lined hose, and one fire-fighting coiled hose. Building fire extinguishers are also provided.
   All fire pipes leading to the hydrant are connected to the fire cistern; the location of the fire cistern is combined with the air shaft and located on the surface. Water flows automatically into the tunnel through two main pipes; after decompression by a reducing valve, the water is connected to hydrants. There are a total of 32 fire cisterns; the volume of each is 200 $m^3$ and the water is from the municipal water supply. Each fire cistern serves about 3.2 km of collider ring; the main fire pipe is arranged in a ring which forms an independent system. The water supply volume of hydrant inside the tunnel is 20 L/s. For a fire duration of 2 hours, the water consumption is 144 $m^3$. The fire cistern can supply this amount.
   Each hydrant system consists of one fire cistern, 65 hydrants, about 6400 m of plastic coated steel pipe, and 15 valves. The entire ring tunnel and the experiment halls comprise a total of 32 such systems.
3) Extinguishers:
   The distribution of mobile extinguishers shall be in accordance with relevant requirements of the Code for Design of Extinguisher Distribution in Buildings (GB50140-2005).
   The tunnel and the halls house the accelerator and the experiments. There is a great deal of electromagnetic equipment. According to the Code for Design of Extinguisher Distribution in Buildings (GB50140-2005), if a fire occurs, it is a class D fire and a class E electric fire. The entire tunnel is considered as one fire zone and provided with a total of 6500 distribution points; every distribution point has 2 sets of Grade 3A MF/ABC6 ammonium phosphate dry powder extinguishers for a total of 13000 extinguishers. The two 1200 $m^2$ experiment halls, each provided with 4 MFT/ABC20 wheeled fire extinguishers.

#### 9.6.1.4 *Smoke Control*

The smoke control system will use a combination of a mechanical smoke extraction system and mechanical exhaust system. In case of fire, the mechanical smoke extraction system can promptly extract the smoke and provide fresh air, in order to guarantee safe personnel evacuation.

In the tunnel, experiment halls, and auxiliary caverns, smoke will be exhausted during the accident whereas in the accessory rooms of the auxiliary cavern smoke will be extracted after the accident.

The vertical shaft of the experiment hall, the air shaft, and the vertical shaft at the RF section basically divide the collider ring tunnel into 32 sections; the upper part of the collider ring tunnel is provided with an exhaust duct; and the bottom of each vertical shaft is provided with a smoke exhaust fan.

The collider ring tunnel is divided into several smoke control zones; smoke screen is installed between smoke control zones; and one smoke exhaust port is arranged in the middle of every smoke control zone.

In case of fire inside the collider ring tunnel, the smoke exhaust port of the smoke control zone where the fire is occurring will be immediately opened to exhaust smoke. At the same time, the smoke exhaust fan at the bottom of vertical shaft will be activated to exhaust the smoke. A fire damper is installed before the smoke exhaust fan; when the smoke exhaust temperature is up to 280°C, the temperature fuse will be activated, the fire damper will be closed, and the smoke exhaust fan will stop running.

In case of fire in the experiment main hall, the smoke exhaust fan will be started to exhaust smoke immediately, and the smoke will be exhausted into the atmosphere through the smoke exhaust system.

#### 9.6.1.5 *Electricity for Fire-fighting*

1) Fire-fighting equipment is powered by a second class load and by the emergency power supplies of the 0.4 kV distribution system. In the case of an accident, a spare emergency power supply is used (Diesel generator, EPS power supply or DC system power supply) to guarantee reliable electric power for fire-fighting.
2) Power distribution for fire-fighting will be provided by separate, independent circuits. Fire-resistant cables will be used.
   Fire-fighting equipment requiring power includes pumps, smoke-control and exhaust equipment, automatic fire alarms and extinguishers, emergency lighting, evacuation signs and fire-resistant rolling shutter doors.
   Emergency lighting and evacuation signs are powered by battery backup and hold for at least 30 minutes.
3) Fire accident lighting and safe evacuation signs will be installed to mark evacuation passages, staircases and safe exits.
   The minimum illumination intensity of evacuation accident lighting should not be less than 0.5 Lx and lights should be installed on the wall or the ceiling.
   Evacuation signs at safe exits should be above each exit. Evacuation signs in each passage and around corners should be installed on the wall 0.8 m above the ground and spaced not more than 20 m. There should be inflammable covers for accident lighting and evacuation signs.

#### 9.6.1.6 *Fire Auto-alarms and Fire-fighting Coordinated Control System*

1) There is a fire alarm and fire-fighting control system. Once there is a fire, the fire alarm system can automatically or manually send alarm signals and fire-fighting coordinated control directions, record, display and print fire and coordinated control information according to different signals given by the fire detection equipment.
2) The surveillance scope of the fire alarm system includes the ring tunnel, experiment halls, auxiliary ring tunnels, transportation passages, auxiliary chambers, auxiliary buildings at the surface of shaft openings, buildings housing substations and electric power distribution, the control room, electronics rooms, power supply hall, assembly hall, refrigerating machine room and ventilation fan room.

Control equipment includes audible and visual alarms, fire hydrants, heating, ventilation and air conditioning, smoke exhaust fans, smoke control valves and fire resistance rolling shutter doors.

3) The fire auto-alarm system is mainly powered by AC 220 V AC in the 0.4 kV firefighting power supply. It is equipped with a UPS and batteries for backup.
4) The signal and power transmission lines of the fire auto-alarm system are made of fire-retardant copper core shielded cables. All cables are laid out to go through metal or flexible metal conduits or metal sealed ducts.
5) The fire auto-alarm system is grounded by the common grounding network with a grounding resistance of 4Ω at most.

### 9.6.2 **Drainage Design**

Water seepage amount into sumps can be based on the experience of similar projects and on preliminary water leakage estimates. The volume of each sump is provisionally determined to be 1,000 $m^{3.}$ There would be 24 such water-collecting wells.

Each will have 4 horizontal multistage clean water pumps with synchronous discharge and suction. The sump pump chosen is a TPD200-280-43X7 pump, 288$m^3$/h, raises the water to a maximum height of 286 m., and uses 260 kW. At each sump there are three for use and one for standby. Thus, a total of 96 pumps are required along with valves and plumbing. The drain header is DN350 (for 350 mm pipe).and is chosen for economical flow velocity.

## 9.7 Permanent Transportation and Lifting Equipment

Installation of accelerator components and the massive detectors in the experimental halls will closely follow the completion of the civil construction. Required will be transportation and handling equipment such as cranes for components and personnel. Part of the equipment will be transported to the underground experiment hall for assembly by an overhead traveling crane after it is assembled into a larger unit in the assembly hall.

Later during operations personnel and materials or equipment necessary for experiment maintenance, overhaul and replacement are transported at a lower frequency.

There will be 46 access shafts from the surface to the underground experiment halls. These shafts vary in depth below the surface from 50.6 m. to 244.38 m. and average about 129.17 m. Eight access shafts serve each of the halls at IP1 and IP3. At each hall 2 access shafts are 16 m in diameter, 2 are 9 m in diameter and used for materials. 2 6-m shafts are used for personnel, and the other 2 shafts with diameter of 7 m lead down to the Booster bypass tunnel.

Over at the other two IPs, IP2 and IP4 where are the RF cavities, 2 6-m access shafts are used for personnel transportation, and also 1 access shaft with diameter of 15 m is for materials; One 10-m access shaft in LSS1, LSS2, LSS3 and LSS4 experiment halls, is mainly for personnel transportation.

There are 10-m shafts for AST2, AST4, AST7, AST9, AST11, AST13, AST16, AST1 (8 total) mainly used during construction and for personnel transportation during operation.

Vent shafts, 16 total and 7 m in diameter are at ASC1, ASC2, ASC3, ASC4, ASC5, ASC6, ASC7, ASC8, ASC9, ASC10, ASC11, ASC12, ASC13, ASC14, ASC15, ASC16 and are mainly used for construction and then during operations for ventilation.

Two access shafts with diameter 7 m will be constructed for the BTL tunnel. The design of emergency exits conforms to the provisions of GBl6423—2006 Safety Regulations for Metal and Non-metal Mines.

There are 1500 $m^2$ magnet assembly halls at IP1 and IP3. Gantry cranes are used for transportation and assembly. Overhead cranes are used in the underground main and service caverns. Elevators will transport personnel from the surface to underground.

There will be 20 electric station wagons in the tunnel, one stationed at each access shaft area and one additional truck at the IP1 and IP3 experimental halls.

Permanent lifting and transportation equipment includes 5 gantry cranes, 2 overhead cranes, 24 elevators and 20 electric trucks.

These transportation and lifting assets are itemized in Table 9.7.1.

**Table 9.7.1:** Quantities of Transportation Lifting

| Project section | Equipment and specs | QTY | Unit weight (t) | Total weight (t) |
|---|---|---|---|---|
| AST1~AST18 | Access shaft elevator | 18 sets | 6 | 108 |
| | Rail for Access shaft elevator | 18 sets | | 20+38+22+28+36+36+42+40+40+40+43+43+26+25+25+18+16+33 |
| ASU3、ASU6 | Equipment transportation gantry crane for shaft with diameter of 15m 30t crane | 2 sets | 60 | 120 |
| | Rail for shaft gantry crane | 2 sets | | 72+50 |
| ASSBB1~ASBB2 | Elevator for shaft with diameter of 7m | 2 sets | 6 | 12 |
| | Rail for shaft elevator | 2 sets | | 20+40 |
| ASG1~ASG2 | Elevator for shaft with diameter of 6m | 2 sets | 6 | 12 |
| | Rail for shaft elevator | 2 sets | 46 | 25+31 |
| IP1, IP3 ground assembly hall | Ground 1500 ㎡ experiment assembly hall 1500t gantry crane | 1 sets | 2000 | 2000 |
| | Ground 1500 ㎡ experiment assembly hall 1000t gantry crane | 1 sets | 1200 | 1200 |
| | Ground 1500 ㎡ experiment assembly hall 80t gantry crane | 2 sets | 120 | 240 |
| | Gantry crane rail QU70 | 2 sets | 12 | 24 |

| Project section | Equipment and specs | QTY | Unit weight (t) | Total weight (t) |
|---|---|---|---|---|
| ASU2、ASU5 | Elevator for equipment transportation gantry crane of shaft with diameter of 9m | 2 sets | 6 | 12 |
| | Equipment Transportation shaft elevator rail | 2 sets | | 20+40 |
| IP1, IP3 underground main cavern, service cavern | IP1 and IP3 underground main cavern overhead crane 20t overhead crane L=28m | 1 sets | 50 | 50 |
| | Main cavern overhead crane rail QU80 | 1 sets | 5 | 5 |
| | IP1 and IP3 underground service cavern overhead crane 10t overhead crane L=18.5m | 1 sets | 30 | 30 |
| | Service cavern overhead crane rail QU70 | 1sets | 3 | 3 |
| Underground vehicle | Trucks for both passenger and goods (Power driven) | 20 sets | 5 | 100 |
| Total | | | | 4785 |

## 9.8 Green Design

### 9.8.1 Energy Consumption

The total CEPC power consumption is about 300 MW, so during operation a great deal of energy will be consumed. A preliminary estimate is 2 billion kWh, equivalent to 720,000t of standard coal, and the emission of carbon dioxide is about 1,886,400t, so in the design of CEPC, energy conservation and consumption reduction, and sustainability must be considered. This is also a crucial social responsibility.

### 9.8.2 Green Design Philosophy

Green Design (also known as Ecological Design, Design for Environment, and Environment Conscious Design) refers to design methodology that treats environmental attributes as design objectives and not as constraints. It aims at incorporating those attributes without compromising performance, quality, or functionality and makes use of natural materials. Green Design focuses on the ecological balance between humans and nature. In decision-making during design, the environmental benefits are fully taken into consideration, to minimize damage to environment.

For the CEPC, the reduction of the consumption of energy and materials will be goals in the design stage, and environmental factors and pollution prevention measures will be incorporated. The core of Green Design is "3R", namely, Reduce, Recycle and Reuse. It not only aims to reduce the consumption of substances and energy and reduce the emission of harmful substances, but also fully consider recycling or reuse in the entire project.

We advocate for the CEPC "simplicity and practicability, energy saving and consumption reducing, cyclic utilization, and environmental friendliness.”

### 9.8.3 Green Design Implementation

#### 9.8.3.1 *Reduce – Reduce Environmental Pollution and Energy Consumption*

1) Optimize the design parameters of physical test facilities, improve the energy efficiency of physical test equipment and reduce the energy consumption of the equipment.
2) In the civil engineering and infrastructure, building energy-saving design will research planned building zoning, building cluster and single building, building orientation, spacing, solar radiation, wind direction and external space environment by following the basic methods of climate sensitive design and energy saving. Buildings will be designed with low energy consumption and with eco-environment protection. Environmental protection materials are used and green construction is performed to reduce pollution and damage caused by the project.

#### 9.8.3.2 *Reuse and Recycle – Recycling, Regeneration and Reuse*

1) The cooling system will generate a large amount of residual heat. Since the residual heat is at low temperature (30°C ~ 40°C), it is usually directly discharged into the atmosphere. If this residual heat can be recycled, energy will be effectively saved.
   According to the source and characteristics of the residual heat, heat pump technology can be used to upgrade the residual heat, and then an advanced residual heat utilization technology can be used. This includes incorporating various heat exchange technologies, thermal power conversion technology, and residual heat refrigeration technology to make effective use of CEPC residual heat resources. It is possible to envision domestic heating for science and technology parks, and providing heat for refrigeration and air conditioning, and for agricultural greenhouses.
2) (Recycling) Utilization of waste water by introducing the concept of "Sponge City." Collect and process rainwater and waste water, and establish a recycling system, to build a water-saving "sponge technology park".
3) Utilization of Renewable Energy will incorporate into the project wind power and solar photovoltaic power stations, to increase the utilization of renewable energy and reduce carbon emission. There is ample room for large solar panel arrays.

#### 9.8.3.3 *Advanced Energy Management System*

An important part of the CEPC information system will collect and process energy consumption data, electricity, fuel gas, and water, and analyse building energy consumption. Through energy planning, energy monitoring, energy statistics, energy consumption analysis, the management of key energy consumption equipment and energy metering equipment, and other means will conserve energy.

# 10 Environment, Health and Safety Considerations

Environmental, safety and health aspects are integrated into the design, construction and operation of CEPC. Experience at existing accelerators helps us identify the principal hazards and associated risks. Discussed in this chapter (or referred to in the chapter to other sections of the CDR for more details) are work planning and training (10.2), environmental impact (10.3), ionizing radiation (10.4), fire safety (10.5), cryogenic and oxygen deficiency hazard (10.6), electrical safety (10.7), non-ionizing radiation (10.8), and general safety issues (10.9).

The preparation of environmental impact and occupational hygiene assessment documents will be carried out and specific requirements for implementation of safety-related codes and standards will be defined and detailed.

There are natural hazards such as seismic disturbances or extreme weather events that require emergency planning procedures. These are not addressed in this CDR.

In addition, surveys of sites of historic and prehistoric periods, and preservation and mitigation of project impact on them, will be conducted and coordinated with the national historic preservation law.

## 10.1 General Policies and Responsibilities

Construction and operation will be based on prevention and safety first. Environmental protection management will be strengthened. A major project goal will be the avoidance of all accidents during construction and operation. All persons, staff and contractors will receive relevant training in order to cultivate a good sense of safety and a dedicated and responsible work attitude.

The basic principles and expectations are:

- no safety incidents
- no casualties
- no environmental pollution

The CEPC's environmental, safety, and health plans are fully compliant with all government laws and regulations. It is the responsibility of each individual employee to be safe. We firmly believe that through the unremitting persistence and efforts of everyone, CEPC's goals of environmental protection, safety and health are fully achievable.

## 10.2 Work Planning and Control

### 10.2.1 Planning and Review of Accelerator Facilities and Operation

Before permission is given to begin a physics run there will be a review of safety systems that prevent accidental entry into radiation areas and of the shielding that makes areas outside the machine safely accessible. Consideration of beam loss accidents will be part of the safety review.

### 10.2.2 Training Program

The CEPC project will implement the relevant national laws and regulations, carry out environment, safety, and health (hereinafter referred to as ES&H) training according to the operating environment, hazard factors and ES&H management requirements, continuously improve the ES&H awareness and skills of all employees, and ensure the safety and health of personnel. Training programs will consist of lectures, practical exercises where relevant, and an examination that must be passed. Some training will be one-time-only, e.g. for new employees. Other training will be required to be renewed on a periodic, often annual, basis.

ES&H training includes but is not limited to:

- **Radiation safety and protection training:** All personnel entering radiation areas will receive training in radiation safety and protection.
- **Occupational health training:** Management personnel shall receive occupational health training.
- **Special job training:** Personnel doing potentially hazardous work, e.g. entering confined spaces or working with hazardous materials, must receive special safety training before being allowed to do those jobs.
- **Special equipment operation training:** Personnel required to operate potentially hazardous equipment, such as fork lifts or overhead cranes need to participate in the special equipment operation training before being allowed to use such equipment.
- **Other suitable ES&H training:** There will be a number of courses, some voluntary, some required, depending on a person's job description such as emergency preparedness management, first aid and computer security.

### 10.2.3 Access Control, Work Permit and Notification

Controlled access areas are defined in Section 7.3.1.1. Personnel entering these areas are limited by their training experience, personal protection equipment, and personal accumulated radiation dose.

All work in controlled radiation areas must be planned and optimized including an estimate of the collective dose and of the individual effective doses to the participants. They must be informed of the hazards that could be incurred in their work including abnormal situations. They must receive training adapted to the work to be performed.

## 10.3 Environment Impact

### 10.3.1 Impact of Construction on the Environment

The buildings and structures, both surface as well as underground, have been enumerated in Chapter 9. A potential site is located in Funing District, which is some 30 km from the closest major city, Qinhuangdao. There are some villages nearby and no large-sized buildings or structures. There are local underground pipeline networks that need to be identified and avoided during excavation. Impacts of construction on the environment mainly include the impact of blasting vibration and noise generated by underground construction and the impact on water quality of domestic sewage and wastewater generated by construction.

The key environmental protection issues during construction are water and noise protection. Optimization of the sewage treatment process and noise protection measures protect the eco-environment and health of the area population.

#### 10.3.1.1 *Impact of Blasting Vibration on the Environment and Countermeasures*

The project is located in suburban areas; some of the facility will be in residential areas and some in non-residential areas. The impact is mainly possible damage to houses caused by blast shock waves and the tolerance of residents towards vibration frequency and intensity. Impact on the ground environment is small as the working faces are mostly deep underground. Blasting can be controlled by proper selection of blasting parameters during construction. The impact on residents will depend on whether underground construction is drill-blast or TBMs or a combination of both.

#### 10.3.1.2 *Impact of Noise on the Environment and Countermeasures*

The impact on the sound environment is mainly from excavation blasting, crushing of sand and gravel, mixing of concrete, construction transport and operation of heavy machinery. Low noise equipment and necessary work force protection need to be adopted during construction.

#### 10.3.1.3 *Analysis of Impact on the Water Environment*

Sewage and wastewater generated in the project construction area include construction wastewater and domestic sewage. Sewage being discharged directly into the watercourse nearby without treatment may have a large impact on water quality. Treatment of construction wastewater and domestic sewage must be carried out and discharges brought up to standard or recycling implemented.

#### 10.3.1.4 *Water and Soil Conservation*

Water loss and soil erosion may be generated by inadequate design of living quarters, construction roads and disposal areas. Due to the thick overburden, if inadequately designed, there will be a large impact on the surrounding surface water. Therefore, engineering and biological measures need to be taken to prevent scouring of rainfall runoff to the construction site and disposal area, so as to reduce water loss and soil erosion.

### 10.3.2 **Impact of Operation on the Environment**

The main ring will be 50 to 150 m underground, Except for access points, the surface area will be freely accessible to members of the public, and customary activities, such as agricultural and residential uses, could continue. In assessing the environmental impact, it is primarily the release of radiation and radioactivity to areas not under project control that is of concern.

#### 10.3.2.1 *Groundwater Activation and Cooling Water Release Protection*

Ground-water activation occurs only when the water comes in close proximity to the tunnel and then only where the rock is not waterproof. But, most of the ring is waterproof and dry. However, importantly, radioactivity may be produced in ground water outside the tunnel, or be leached out from activated rock or soil by the ground water. The total activity produced in the electron collider is much smaller than will be the case with the proton machine. Water in underground aquifers can travel long distance and contaminate off-site wells with H-3 (tritium). This needs to be monitored.

Radioactivity produced in the cooling water circuits is by high-energy radiation and from the high-energy tail of synchrotron radiation. The cooling circuits will be closed and no release is expected during normal operation. When the circuits are drained (during repair or by accident) the cooling water will be accumulated in the general ring-drainage system, from which it will be pumped and collected. Water will be periodically sampled and monitored for activity before release. If needed, release from the ring drains can be postponed for a considerable time. In case of flooding, water can be pumped out immediately, as dilution will reduce activity concentrations to negligible levels.

#### 10.3.2.2 ***Radioactivity and Noxious Gases Released into Air***

The radionuclides produced in the tunnel air, primarily N-13 and O-15, and the noxious gases produced, such as O-3 and NOx, are discussed in detail in Section 7.3.3.

Adequate shielding of the vacuum chamber reduces the amount of synchrotron radiation escaping into the air, and therefore reduces the production of these isotopes and noxious gases. The height of the air release point and the release velocity in the air and ventilation system (Section 9.5) reduces the concentration of these harmful components. Monitoring, both in the vent ports and at ground level should be done.

#### 10.3.2.3 ***Radioactive Waste Management***

Materials where activation is anticipated should be designed or chosen (e.g. tools or fixtures) to minimize volume and weight. National regulations specify the limit below which materials are classified as non-radioactive. Radioactive items will be handled as outlined in Section 7.3.6.

## 10.4 Ionization Radiation

Synchrotron radiation and radiation induced by lost beam are two kinds of ionizing radiation. The shielding design and radiation dose caused by these are calculated and listed in Chapter 4 and Section 7.3.

## 10.5 Fire Safety

Fire protection is based on fire prevention and supplemented by firefighting. Comprehensive measures are in four steps: "prevention, cutoff, extinguishing, and exhaust." Details are in the previous Chapter.

9.6.1.2 Fire-fighting for buildings
9.6.1.3 Water-based fire-fighting
9.6.1.4 Smoke control
9.5.4 Ventilation and smoke exhaustion system

**Electrical fire prevention measures:**

Power supplies for fire protection purposes shall be considered Grade II loads. Dual power supply automatic switch-over devices shall be arranged at the location where the power distribution at the last level in the distribution is located to provide electric power for electrical equipment for fire prevention. All electrical equipment for fire prevention is provided with an independent power supply circuit. All such circuits have fire-resistant cables.

Emergency lighting and signs shall be provided at each evacuation passage, staircase rooms and exits. Emergency lights have battery backup.

Additional details are in Section 9.6.1.6.

## 10.6 Cryogenic and Oxygen Deficiency Hazards

### 10.6.1 **Hazards**

Cryogens include liquid nitrogen and liquid helium. Cryogenic hazards include cold burns (frostbite), explosions, oxygen deficiency or asphyxiation. Oxygen deficiency hazard (ODH) can lead to asphyxiation when the oxygen content falls below 19.5%. Some of these same rules and procedures can apply to confined spaces even if there are no cryogens present.

### 10.6.2 **Safety, Environment and Health Measures**

Personnel Controls: Carry out targeted safety education and post technical training for staff according to various cryogenic and oxygen deficiency laws and regulations. Eye, hand, and body protection must be provided to prevent potential hazards. Adopt a two-person rule or a three-person rule in cryogenic and oxygen deficiency work areas. Personnel entering ODH areas must be certified medically and be trained. The ODH certification must be renewed periodically.

There must be adequate ventilation in such areas. Personnel can carry oxygen monitors in addition to the permanently installed monitors and warning signs.

### 10.6.3 **Emergency Controls**

Formulate accident emergency rescue plan in cryogenic and oxygen deficiency working areas. Organize emergency rescue drills.

## 10.7 Electrical Safety

Knowledge of safety and recommended practices is necessary to protect against electrical hazards (electrical shock or burns or other delayed effects). The electrical safety program also provides hazard awareness information to those who use electrical equipment. "Lock-out tag-out" training will ensure that personnel are knowledgeable and trained in the electrical tasks they are asked to perform.

## 10.8 Non-ionization Radiation

The subject of protection against non-ionizing radiation (NIR) has been under consideration since 1974 by the International Radiation Protection Association (IRPA). Consequently, in many countries protection against hazards arising from the use of NIRs has been added to the task of the radiation protection authorities.

There are a number of electromagnetic fields, microwaves, RF, lasers, and strong magnets, which potentially can cause health problems. Some of these are not well understood and are still controversial.

Radio frequency and microwave radiation is absorbed by body tissue, which causes a temperature rise. The acceleration of particles implies the use of very powerful radio frequency systems. High-power RF sources in the CEPC are klystrons. Since the klystrons are linked by a closed system to the RF cavities via waveguides, only leakage radiation is of concern. It is planned to carefully monitor RF power levels during startup. Experience has shown that the main hazard associated with RF systems arises from X-ray exposure rather than from microwaves.

Regulation "GBZ1-2002, Hygienic Standard for Industrial Enterprise Design," sets the contact limit of high frequency radio wave in working areas.

## 10.9 General Safety

### 10.9.1 **Personal Protective Equipment**

CEPC will provides staff with necessary, reliable, and appropriate personal protective equipment. Such equipment will be regularly checked and maintained. Personal protective equipment (PPE) includes, but is not limited to:

- Safety Helmets (hard hats)
- Safety shoes: puncture-proof boots, insulated boots, oil-proof boots acid-resistant boots, anti-static shoes
- Eye, face protection: goggles, protective covers
- Hearing protection: earplugs, earmuffs, anti-noise caps, ear muffs
- Respiratory protection: dust masks, gas masks and air packs for firefighters
- Hand protection equipment: chemical-resistant gloves, insulating gloves, handling gloves, fire-resistant gloves.
- Workwear: work clothes (long sleeves), special work clothes (such as: anti-static overalls, wading overalls, waterproof overalls). Easy-to-identify nighttime reflective vests.
- Radiation operations:
  Radiation workers wear personal dosimeters, personal dose alarms, and various types of lead protective equipment: lead aprons, lead glasses, lead gloves. And for those entering areas with high radioactivity disposal clothing.
- PPE of traffic and vehicle:
  This includes dust protection facilities of special transport operator and anti-noise equipment of special transport operator.

### 10.9.2 **Contractor Safety**

CEPC is a very large construction project. Besides a large staff there will be many contractors and sub-contractors. It is necessary to complete all tasks smoothly, safely and with minimum damage to the environment. Therefore, in the project implementation, all contractors and subcontractors should be familiar with the environmental protection, safety and health policies, plans, site safety regulations, emergency handling and evacuation procedures, and other relevant safety regulations. They must comply with the supervision and management of CEPC security personnel and possess the qualifications to undertake their part of the project. Policies and accountability will conform to the national environmental protection, safety and health regulations and standards. Safety management personnel will establish a responsibility system for all levels of personnel. They will ensure that all equipment brought to the site is safe and reliable and in good condition. Subcontractors shall be responsible for the safety and health of their employees and prevent unsafe work.

### 10.9.3 **Traffic and Vehicular Safety**

For personnel protection all vehicles will require seatbelts. Also required will be dust protection facilities and anti-noise equipment for special transport operators.

The traffic tools mainly include new energy cars and trams, lifting systems (such as shaft hoist winches), accident rescue vehicles, fire trucks, engineering and material handling vehicles, forklifts, crane trucks, earth-moving vehicles. Four aspects of traffic and vehicle safety are considered: station and hub safety design; vehicle safety control; traffic facilities and the traffic safety management.

A traffic safety control system will be set up. It includes a signal control system to realize the coordinated management of all kinds of vehicles, with similar or different road rights. A dispatch and command center can be used to realize the supervision, commanding, dispatching, and emergency management of all types of vehicles.

CEPC's route and station design implements the necessary safety functions: road surfaces; entrances and exits; safety exits; intersections; rights management, and emergency evacuation sites.

The transportation facilities guarantee the personal safety of vehicle operators and traffic participants by the installation of traffic signs, isolation facilities, access control management, on-board safety facilities, and tire chain facilities.

### 10.9.4 **Ergonomics**

CEPC activities that require work in a restricted space or with awkward or static postures, repetitive motions, pressure points, vibrating tools, or forceful exertions can lead to injuries and reduced worker effectiveness.

Workers and supervisors should be actively screening activities and workplace conditions with potential ergonomic risks and will be encouraged to engage their ES&H coordinator or contact the program manager for assistance, ranging from informal consultations to formal evaluations.

# 11 R&D Program

Chapters 1–10 of this CDR is the conceptual design of the Circular Electron Positron Collider (CEPC). We believe this design is sound and the costs are understood. Nevertheless, as the project moves from concept to reality, a comprehensive R&D program will improve many details. Calculations and simulations will refine some of the parameters. Models and prototypes will be built to verify and in some cases improve on designs. Various experiments will be carried out to better define construction and installation procedures for this future facility. These R&D activities are the subject of this chapter.

CEPC will be operated as a Higgs factory at 240 GeV and will also run at 160 GeV as a W factory and at 90 GeV as a Z factory. The accelerator chain consists of a 10 GeV normal conducting linear accelerator with a damping ring, followed by a 10 ~ 120 GeV Booster and then the 240 GeV Collider, The Booster and the Collider are housed in the same 100-km circumference tunnel. This complex consists of more than 13 systems, with a total of more than 400,000 individual sets of equipment. In order to reduce the CEPC cost, most equipment will be constructed domestically. Major goals of the R&D program described in this chapter are to improve reliability and efficiency as well as lower costs.

Whereas all of the systems can benefit from R&D the key components in the R&D program are the high performance superconducting RF cavities, and high efficiency klystrons, the low-field dipole magnets and the very large cryogenic system. The R&D program will also organize and train key personnel who will become the cadres as facility construction ramps up. Also important, during the R&D phase, will be industrialization and preparation for mass production of components.

Since 2015 there has been considerable support of the R&D program. This has come from multiple sources.

- Innovative funds of the Institute of High Energy Physics (2015-2019) for work on the high Q superconducting cavities, the high efficiency klystron and the digital BPM system.
- Since 2016, support has been provided by the national R&D plan of the Ministry of Science and Technology "Advanced research of large scientific facilities, R&D of the physics and key technologies related to high energy circular electron positron collider." This work (2016-2021) includes R&D on key CEPC technologies: accelerator physics, the high Q superconducting cavity and the injector.
- Since 2017, support from the "Scientist Studio" (2017-2022), was on high efficiency klystrons and other accelerator technologies.
- The project of "key technology R&D and verification of high energy circular electron positron collider" (2018-2023), approved by the national R&D plan of the Ministry of Science and Technology in 2018, includes the development of a high-precision low-field dipole magnet prototype, the development of the main ring vacuum chamber prototype and the development of a high-energy electrostatic separator prototype.
- In 2017, the "advanced light source technology research and development and testing platform" (2017-2020) was under construction with financial support from the Beijing municipal government. It will be put into operation by the end of 2019, including the R&D and test platform of various frequency superconducting cavies, high efficiency klystrons, a testing platform for the high Q superconducting cavity and testing of domestically manufactured large refrigerators.

Over the past decade, the development of domestically manufactured superconducting cavities has made great progress through the construction of BEPCII, SSRF and ADS facilities. However, the CEPC superconducting cavity has higher requirements than the ILC and ADS superconducting cavities. The CEPC cavity needs to run with high current (CW 460 mA for running at the Z pole), with high gradient (CW 20 MV/m during operation for the H), with high Q value ($1\sim2\times10^{10}$ at 2K), and with high input power (CW 300 kW). R&D on the cavities will be carried out in stages as these goals are achieved in the initial SRF Technology R&D stage (2017-2020) and R&D for the mass production stage (2021-2023).

Beam power is greater than 50 MW and long lifetime, high efficiency klystrons are needed to reduce power losses, construction cost and operation cost. At present, there is no klystron that meets those needs constructed in China. Although there are similar commercial products in the world, they have normal efficiency klystrons, and the import price is high. R&D on a long lifetime and high efficiency CW klystron includes the development of normal efficiency high power continuous wave klystrons, the development of the high efficiency klystron and then industrial mass production. In 2017, the new "high efficiency RF power source development cooperation group" was jointly established by IHEP, the Institute of Electronics of the Chinese Academy of Sciences and the Kunshan Guoli high power device Industrial Technology Research Institute. This cooperation group is committed to the development and mass production of high efficiency, high power CW klystrons. At present, under the framework of the cooperation group, the physical design of the general efficiency 650 MHz/800 kW CW klystron is completed and mechanical design is being carried out. The first domestic high-power CW klystron will be produced at the end of 2018, and the high power test carried out in early 2019. The goal is for the first long lifetime high efficiency CW klystron meeting the CEPC requirements to be completed in 2023.

Since the injection energy of the Linac is 10 GeV, the minimum field of the Booster dipole magnet is only about 30 Gs. Such magnets with the required field uniformity and operating at such low fields have never been designed nor built. There are magnet prototypes developed by CERN for LEPII and LHeC, but the lowest fields of these is larger than 120 Gs. The design, development and measurement of these low field dipoles is very challenging. Some of the approaches will be:

- To use longitudinal and transverse iron core dilution to improve the flux density in the core and reduce the effect of remnant field on the magnetic field performance;
- To use a hollow coil to eliminate the influence of the core material on the field;
- To use a closed core to shield against the earth's magnetic field.

Development work will begin in July 2018 and the plan is to complete it in 2023.

In CEPC we need to separate the positron and electron beams in the RF area. The development of high energy electrostatic separators will begin in July 2018 and will be completed in 2023.

R&D on the large refrigerator will be carried out by the Institute of Physical Chemistry. The goal is to build an 18 kW refrigerator that meets CEPC needs.

The CEPC accelerators require a large quantity of magnets, superconducting cavities, vacuum equipment, klystrons and large refrigerators, and many other components. The base for meeting those needs will be mass production by domestic industrial enterprises. At the end of 2017, the CEPC Industrial Promotion Consortium (CIPC) was established to coordinate R&D on key technologies, organize joint breakthroughs and promote technical development and industrialization,

After completion, the CEPC will be a large scientific research device for sustainable development of zero carbon dioxide emissions. This requires the minimization of energy consumption, the use of green energy as much as possible and the ability to recycle the heat of the water-cooling system.

By the end of May 2018, the CEPC accelerator R&D fund is about 285 million RMB from variable resources, including about 25 million RMB supported by the Ministry of science and technology under project of "frontier research on large scientific facilities" for key R&D program, about 210 million RMB supported by Beijing Municipal government for "Platform of Advanced Photon source technology research and development and testing." about 40 million RMB supported by Chinese Academy of Sciences for Scientist studio project etc. These funds cover the R&D of the CEPC accelerator design and several key technology developments, including the accelerator physics design, the development of high Q superconducting cavity technology, the development of high efficiency klystron, and the development of low field and high precision dipole magnet.

Sections 11.1 through 11.15 describe R&D to further develop CEPC and prepare for its construction.

Section 11.16 describes the program to develop high-field magnets that will be required for the future SPPC.

## 11.1 Superconducting RF

There are two large Superconducting RF (SRF) systems: 336 cavities operating at 650 MHz in 56 cryomodules for the Collider and 96 cavities operating at 1300 MHz in 12 cryomodules for the Booster. This would be one of the largest SRF installations in the world. The design, fabrication, commissioning and installation of such a system requires a very significant investment in R&D, infrastructure and personnel.

### 11.1.1 Initial SRF R&D (2017-2020)

During this phase the goals will be to develop in cooperation with industry the prototypes of all the components and demonstrate that they meet the required performance.

#### 11.1.1.1 *Initial Technology R&D*

1. Develop an SRF cavity of each type; order several prototypes from industry; perform a series of tests to optimize the cavity surface treatment; build vertical test stands and perform tests to demonstrate the cavity performance goals. Weld helium jackets on the cavities, re-test and demonstrate the performance goals.
2. Design fundamental RF power (or input) couplers (FPCs); order at least two couplers of each type from industry; build FPC test stands; test the FPCs and demonstrate that their performance meets the CEPC requirements.

3. Design HOM dampers; fabricate one or two prototypes of each design; design and build test set ups; test the HOM dampers.
4. Design and fabricate frequency tuners and a LLRF control system.
5. Design and build a short (one or two cavities) horizontal cryomodule for each cavity type; build a test stand; demonstrate performance of all components integrated together into a cryomodule.

#### 11.1.1.2 *Infrastructure and Personnel Development*

For the initial R&D, most of the infrastructure (clean rooms, HPR system, vertical and horizontal test stands) is available on-site or can be accommodated in the new SRF Lab in Huairou in a northern Beijing suburb, about 50 km from the city center. Some existing facilities will have to be upgraded; additional project-specific equipment will be purchased and some additional space is needed. These needs can be estimated as soon as a detailed R&D plan is developed.

At this stage, it is very important to begin the hiring and development of personnel. The core project personnel must be in place by the middle of this 2017-2020 phase of the project. It will take at least 4 years with two teams working in parallel: one working on the Collider SRF and the other on the Booster SRF. Each core team should consist of about 10 people (physicists, engineers, technicians). Support from other technical groups will be requested when necessary. Collaboration with other laboratories (BNL, DESY, Fermilab, JLab, KEK) will help shorten this stage of the project.

### 11.1.2 **Pre-production R&D (2021-2023)**

The goal during this pre-production phase is to demonstrate robustness of fabrication and assembly processes of the cryomodule and its components. We will establish procedures, quality control steps, test set ups, and assembly sequences. For the production phase.

#### 11.1.2.1 *Pre-production R&D*

During the pre-production phase, we plan to build and test two Booster cryomodules and three Collider cryomodules. To accomplish this, the following will be necessary:

1. Build and test twenty 1.3 GHz Booster cavities and thirty 650 MHz Collider cavities. This will allow pre-qualification of vendors for future cavity mass production, establish treatment processes and debug all the procedures. This will demonstrate that the cavity fabrication and treatment approaches are adequately robust to produce cavities meeting requirements with high acceptance (~90 %). Several cavities of each type should be chosen for horizontal testing. Two or three cavity fabrication and treatment vendors should be pre-qualified by the end of this phase.
2. Build and test twenty FPCs for the Booster cavities and twenty FPCs for the Collider cavities.
3. Build and test a sufficient number of tuners and other ancillary components.
4. Build and test cryomodules and demonstrate cryomodule performance. A cryomodule beam test is recommended especially for determining HOM damping and heat load performance of the Collider cryomodule.

#### 11.1.2.2 *Infrastructure and Personnel Development*

A large scale SRF R&D and production facility (at least 10,000 $m^2$) must be built on the CEPC site. Before this, a superconducting RF Laboratory of 4,000 $m^2$ will be built in 2017-2019 with facilities and assembly lines sufficient for pre-production. The CEPC SRF team should make site visits at the beginning of the initial R&D phase and study facilities used at JLab, FNAL, KEK and Euro-XFEL (DESY, Saclay, LAL) as well as industries for SRF system production and scale them as appropriate to the size of the CEPC SRF system.

The final facility (on the CEPC site and in industry) should include: cavity inspection and local repair facilities, RF laboratory and tuning set ups, BCP (buffer chemical processing) and EP (electro polishing) treatment facilities, annealing furnaces, 4 vertical test stands, clean rooms, HPR systems, FPC (fundamental power coupler) preparation and conditioning facilities, cryomodule assembly lines, 4 cryomodule horizontal test stations, high power RF equipment and a cryogenics plant.

To build and install the Booster SRF system in three years (2024-2026), the production facility should have the capacity to assemble about 1 cryomodule per two months. To build and install the Collider SRF system in four years (2024-2027), the assembly lines should manufacture about one cryomodule each month. To sustain this rate, the vertical test stands should be able to test 2 Booster cavities and 2 Collider cavities each week.

Commissioning and operation of the pre-production facility should begin during the last two years of the initial R&D phase (2018-2019). The pre-production stage will take 4 years, two of which will be for equipment installation and commissioning. The pre-production capacity of the off-site facility should be one fifth to one quarter of the eventual production facility.

Beginning in the last year of the initial R&D (2019), the core SRF teams should begin hiring and training more personnel (~200 FTEs, mostly engineers and technicians), who will then work first in pre-production and then in production.

## 11.2 650 MHz High Efficiency Klystron

### 11.2.1 **Introduction**

Since the CEPC beam power is more than 60 MW, high efficiency RF sources are required in order to reduce power losses and cost. The most popular RF source for an accelerator is a klystron, which has the advantage that it can be operated at high power with a reasonably high efficiency. A single 800 kW klystron will drive two Collider cavities through a magic tee and two appropriately rated circulators and loads. The choice of one klystron for two cavities is justified technically by better control of microphonic noise and minimum perturbation in the case of a klystron trip. Table 11.2.1 summarizes these requirements.

**Table 11.2.1:** Collider SRF system parameters

| | | |
|---|---|---|
| Frequency | MHz | 650 |
| Cavity type | | 2-Cell |
| Cavity No. | | 240 |
| Klystron No. | | 120 |
| Klystron power | kW | 800 |

### 11.2.2 **Design Considerations**

The klystron, with its gun and collector, will be more than 4 m. in length. It can be manufactured industrially in a partnership between IHEP and a Chinese company. Computer simulation tools are used to design the klystron including the electron gun, electromagnet, cavities and RF output structure [1]. However, design and simulation are not sufficient and there needs to be a series of follow up experiments. The first step is to set up a beam test stand to verify the gun and collector designs, then to connect the cavity components to form a classical klystron prototype, and finally to change to high efficiency ones to verify the klystron performance.

Increasing the efficiency of the RF power source is high priority. The majority of existing commercial high power pulsed klystrons operate with an efficiency from 40% to 55%. Only a few continuous wave and multi-beam klystrons available in the market are capable of being operated at 65% efficiency or higher. A new method to achieve 90% RF power conversion efficiency in a klystron amplifier is in Baikov et al [2] where 80% efficiency seems attainable. One klystron prototype will be manufactured with the goal of reaching this efficiency. The design parameters are in Table 11.2.2.

**Table 11.2.2:** Klystron design parameters

| | First step | Goal |
|---|---|---|
| Frequency(MHz) | 650 | 650 |
| Output Power(kW) | 800 | 800 |
| Beam Voltage(kV) | 82 | 110 |
| Beam Current(A) | 16 | 9.5 |
| Efficiency (%) | 65 | 80 |

#### 11.2.2.1 *Electron Gun*

Our initial design was an electron gun with modulating anode (MA). The design was done using DGUN software [3]. Uniform beam trajectories, with a beam perveance of 0.64 $\mu A/V^{3/2}$ has been achieved with this design. There is a Ba-dispenser cathode of radius 35 mm with a Φ10 hole at the center. Current density on the cathode less than 0.45 $A/cm^2$ is obtained. The beam trajectories were also simulated over the entire length with a magnetic field of 180 Gauss [4]. Design parameters in Table 11.2.3 and simulation results using DGUN are in Fig. 11.2.1. Left (a) shows the beam trajectory with maximum electric field on the focusing electrodes. Right (b) shows the current density on the cathode.

**Table 11.2.3:** Klystron gun parameters

| | | |
|---|---|---|
| Cathode voltage | kV | -81.5 |
| MA voltage | kV | -48.0 |
| Beam waste diameter | mm | 35.6 |
| Beam/Gun perveance | $\mu A/V^{3/2}$ | 0.64/1.45 |
| Average cathode density | $A/cm^2$ | 0.45 |
| Cathode uniformity | | 1.24 |

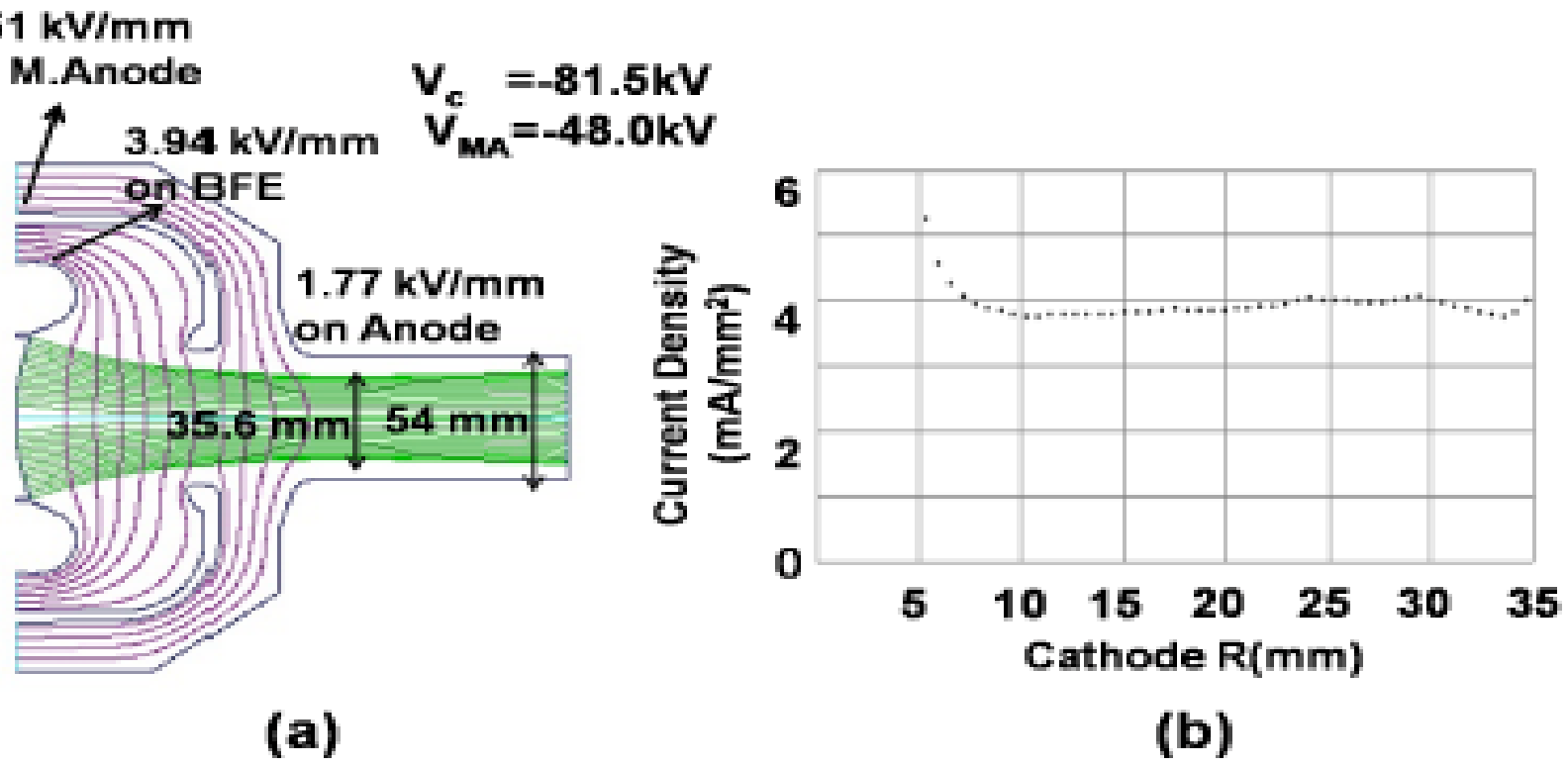

**Figure 11.2.1:** DGUN simulation results.

### 11.2.2.2 *RF Interaction and Cavities*

The klystron efficiency depends largely on the quality of electron bunching. High fundamental beam current and low velocity spread are prerequisites for obtaining high efficiency. After decades of development, the theory and technology of high efficiency klystrons have made great progress. Some techniques are mature such as perveance reduction, high order harmonic cavities and multi-beam klystrons, while some methods are prospective such as adiabatic bunching, COM (Core Oscillation Method), BAC (Bunch Align Collect) CSM (Core Stabilization Method) and depressed collector [5-10]. In order to obtain experience, initially the traditional and mature methods will be used to obtain moderate efficiency of 65%. Then the prospective methods will be applied to raise the efficiency up to 80%. The traditional design is shown in Fig. 11.2.2 and the parameters summarized in Table 11.2.4.

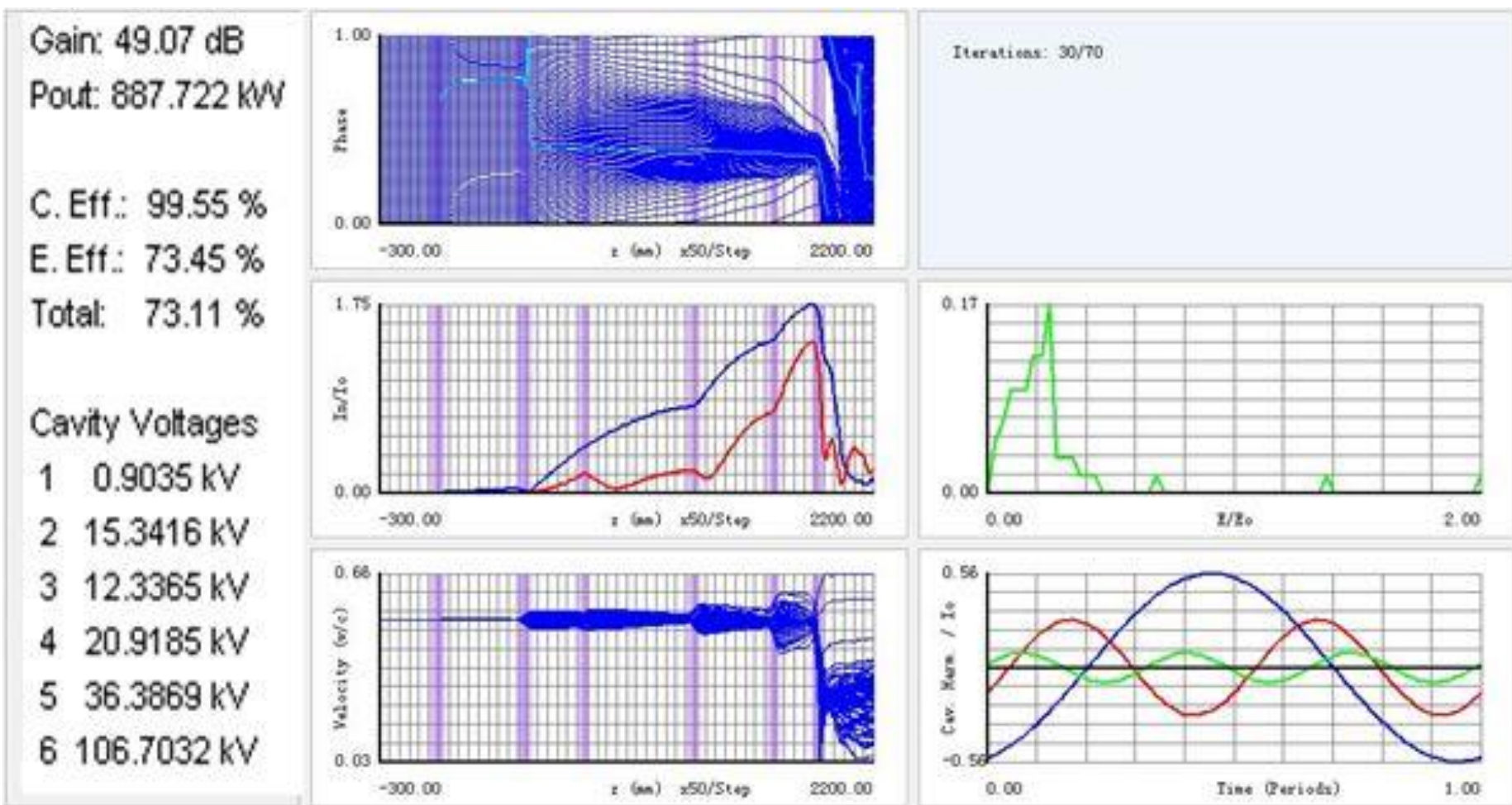

**Figure 11.2.2:** Traditional klystron design

**Table 11.2.4:** Summary of klystron parameters

| **Parameters** | **Traditional klystron** |
|---|---|
| Operating frequency | 650 MHz |
| Beam Voltage | 81.5 kV |
| Beam Current | 15.1 A |
| Beam Perveance | 0.65 $\mu A/V^{3/2}$ |
| Efficiency at rated Output Power | 73% |
| Saturation Gain | 49 dB |
| Output power | 887 kW |
| Brillouin field | 106 |
| Reduced Plasma Wavelength | 3.47 m |
| Number of Cavities | 6 |
| Normalized Drift Tube Radius | 0.63 |
| Normalized Beam Radius | 0.41 |
| Beam Fill Factor | 0.65 |
| Length | 2 m |

### 11.2.2.3 *Cavity Cooling Design*

The 650 MHz / 800 kW klystron consists of six re-entrant cavities, numbered 1 to 6, and shown in Fig. 11.2.3. From the beam dynamics study, the voltages should be 0.88 kV, 14.93 kV, 13.73 kV, 20.44 kV, 35.99 kV and 106.14 kV. The cavities are made of oxygen free copper with a high surface conductivity of $5.8 \times 10^7$ S/m. The specifications of this chain of cavities including the theoretical power loss and the heat load target are summarized in Table 11.2.5. The power dissipated in the output cavity is nearly 87% of the total power loss. Therefore, the cooling scheme of the output cavity plays an important role in the stability of high power klystron operation. The power loss in the output cavity is 3.88 kW. In order to leave a safe margin, power dissipation of 8 kW is in the simulation model as a thermal load.

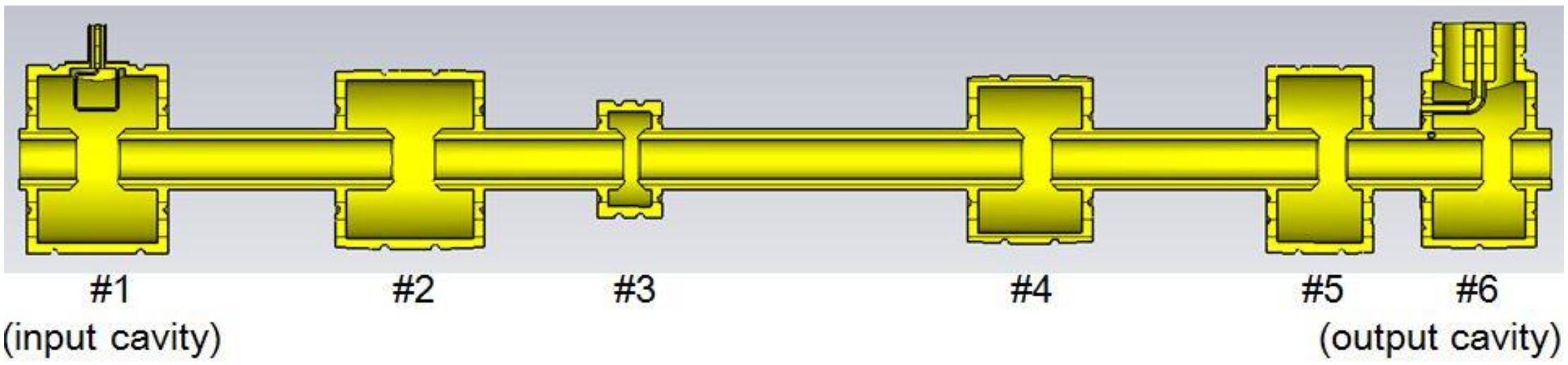

**Figure 11.2.3:** Schematic diagram of the cavity chain

**Table 11.2.5:** Specifications of the cavity chain

| **No.** | **Frequency [MHz]** | **Unload $Q_0$** | **Cavity voltage [kV]** | **Theoretical power loss[W]** | **Heat load target in simulation[W]** |
|---|---|---|---|---|---|
| 1 | 651.49 | 17159 | 0.8843 | 0.15 | 0.3 |
| 2 | 649.17 | 17380 | 14.9323 | 42.7 | 90.0 |
| 3 | 1293.10 | 11070 | 13.7366 | 113.0 | 250.0 |
| 4 | 668.80 | 15632 | 20.4474 | 114.0 | 250.0 |
| 5 | 667.92 | 17716 | 35.9962 | 332.0 | 700.0 |
| 6 | 649.50 | 14050 | 106.1440 | 3882.0 | 8000.0 |

The analysis, based on the CST multi-physics software package, consists of four steps. The analysis begins with RF simulation to calculate the resonant frequency, the unload Q factor and the electromagnetic field distribution. The surface magnetic field is used to calculate the heat flux. Then, we apply the heat flux on the cavity inner surface, along with the convection loads on the cooling pipes. The thermal simulation calculates the temperature distribution in the metallic part. Subsequently, the node temperature information is imported into the structural solver to evaluate the displacement and stress distributions caused by thermal expansion. Finally, the RF simulation of the deformed cavity is performed again to calculate the frequency in steady state operation.

Since, the power dissipation on the inner surface of the output cavity is 87% of the total loss of the cavities chain, the cooling scheme design is a key to the high power operation. Fig. 11.2.4 shows the electric field and the surface power loss surface density of the output cavity calculated by CST MWS. The maximum power loss density is 50.5 $W/m^2$ located at the coupling loop.

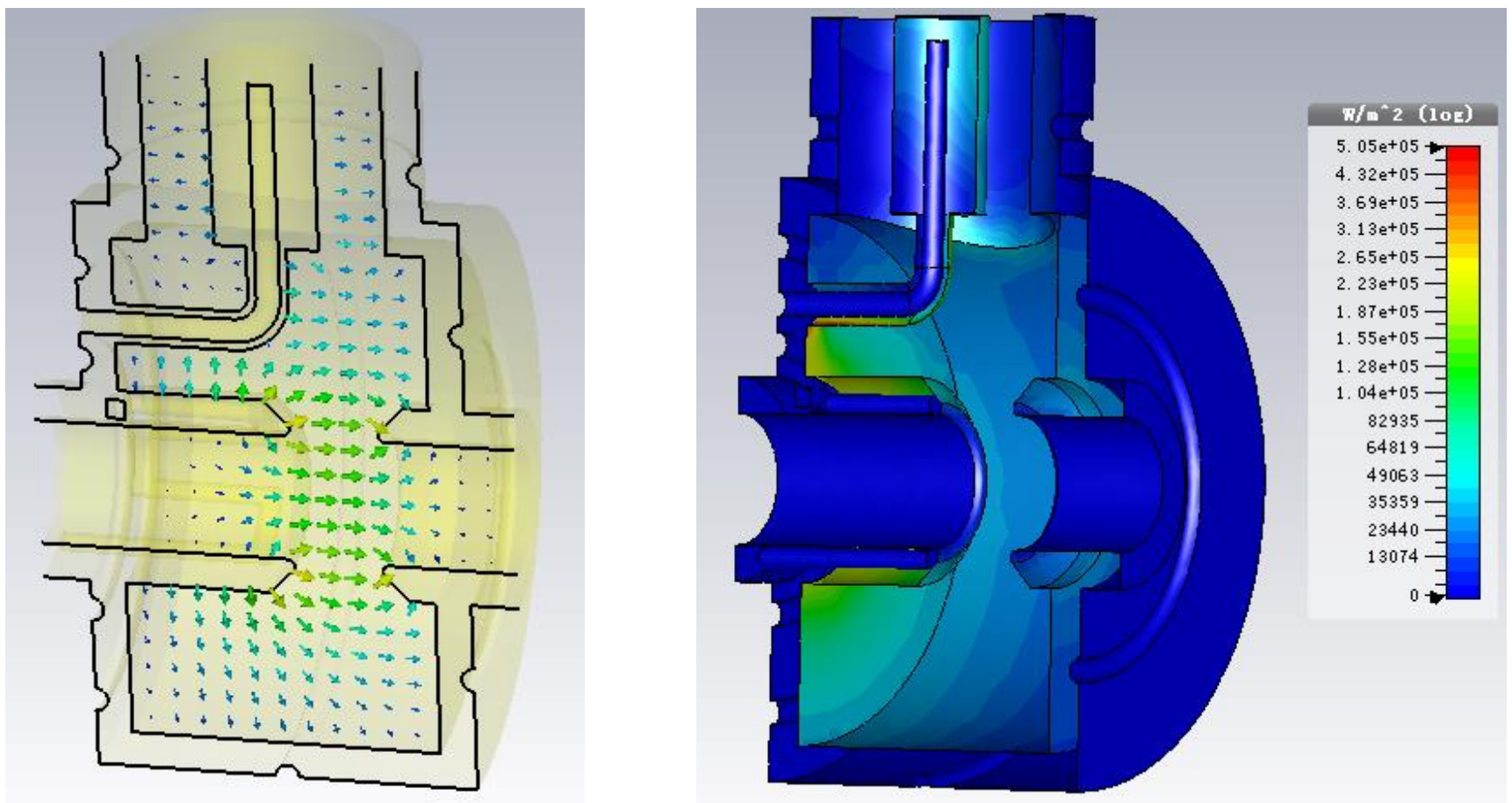

**Figure 11.2.4:** Electric field and surface power loss surface density of the output cavity calculated by CST MWS

After a series of thermal and structure simulations and optimizations, the final cooling scheme, composed of nose cooling, cavity cooling and coupling loop cooling is shown in Fig. 11.2.5. The fluid temperature is chosen to be 30 ℃ with a velocity of 2 m/s. The convection coefficients for these three cooling channels are 8598.8 $W/m^2$ ℃, 8223.7 $W/m^2$ ℃ and 8223.7 $W/m^2$ ℃, respectively. Fig. 11.2.6 shows the steady-state temperature distribution of the cavity. The maximum temperature is 49.9 ℃, which is located at the coupling loop.

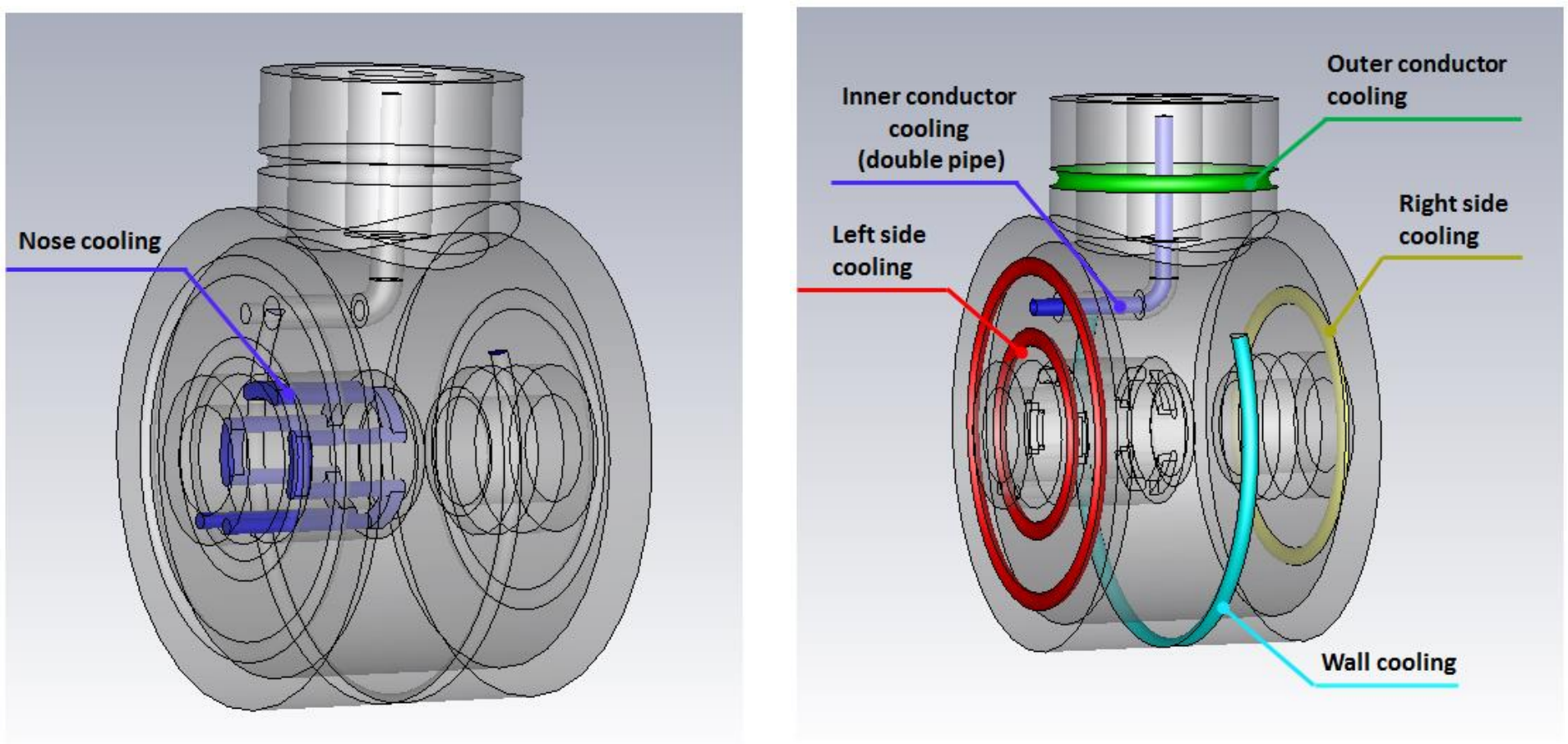

**Figure 11.2.5**：Cooling channels of the output cavity

The vacuum pressure rise and temperature rise lead to deformation of the cavity walls. A vacuum of -101.3 kPa is applied at the surfaces of the inner cavity walls. The temperature profile is imported into the structure analysis to calculate the displacement and the von Mises stresses. Fig. 11.2.7 shows the displacement distribution and the von Mises stress in steady-state operation. The maximum deformation is 131 µm. The maximum stress on the copper disc is calculated to be 16.4 MPa. The stresses are much lower than the yield strength limits of the copper (i.e. ~33 MPa). Failure is not a concern at these stress levels.

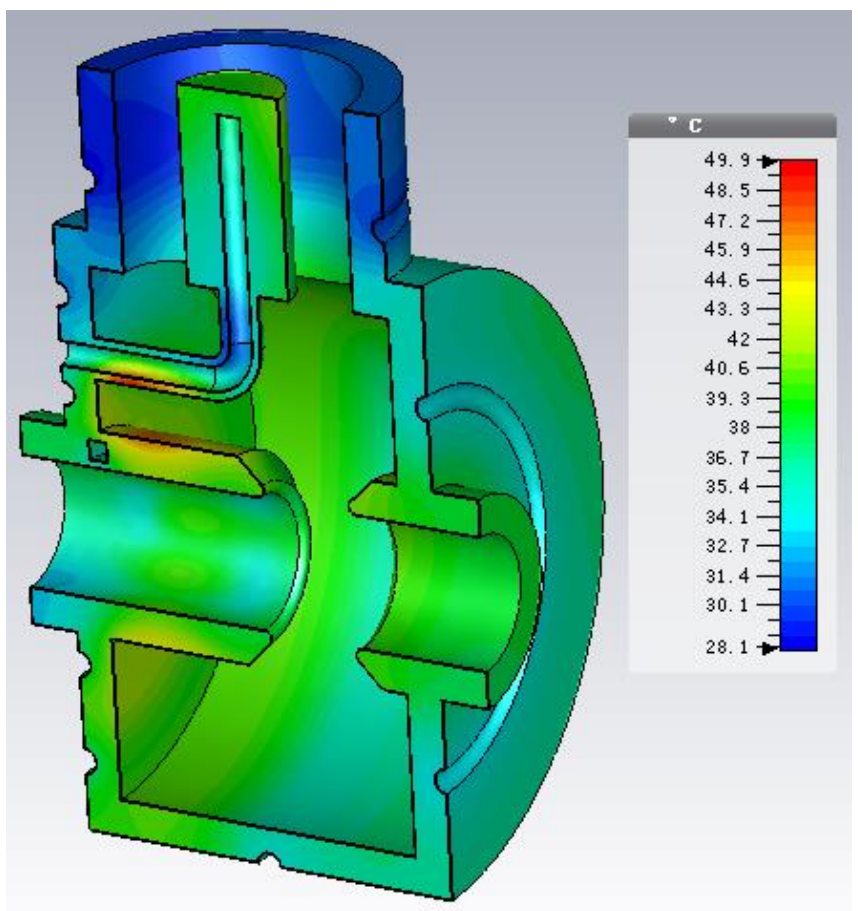

**Figure 11.2.6:** Temperature distribution in steady state operation

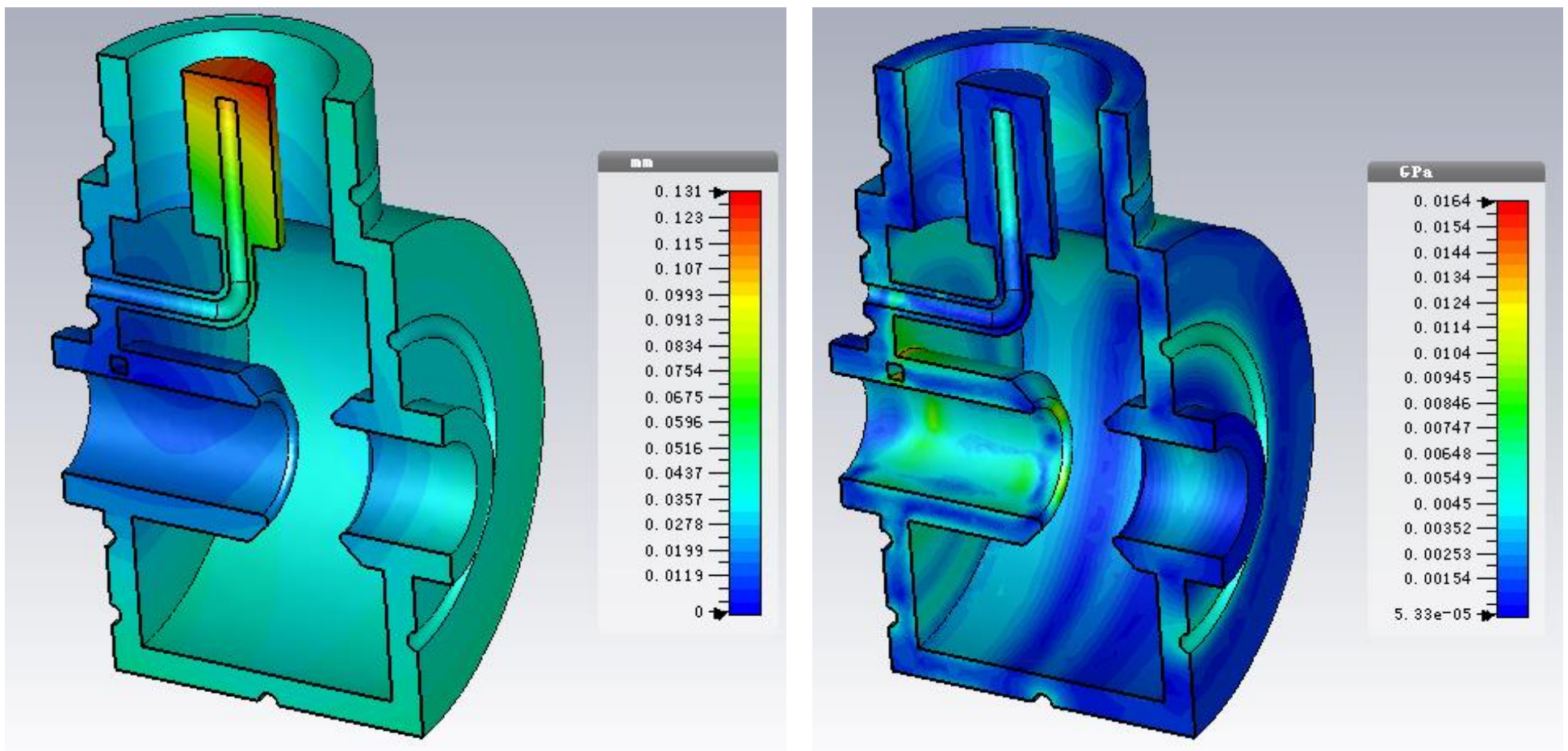

**Figure 11.2.7:** Displacement and von Mises stress distribution

The final RF analysis is performed to determine the frequency shift caused by cavity deformation. The result indicates that the steady-state frequency and the unload Q-factor are 642.049 MHz and 14023, respectively. The corresponding frequency shift is about 229.0 kHz, which should be correctible in the cavity tuning procedure.

The overall coupled RF, thermal and structural analysis of the output cavity is simulated and the cooling pipes designed and optimized to minimize the deformation and the stress. The von Mises stresses are much lower than the yield strength limit of the material. The results indicate that our cooling scheme can deal with the dissipated RF power source efficiently.

### 11.2.2.4 *Window Design*

The design of the coaxial output window is an important challenge in the development of high-power klystrons. For window design an average power capability and multipacting analysis for both fundamental and harmonic frequencies are important. Electromagnetic simulation of an output window was carried out using the CST Microwave Studio code. We have optimized the return loss, not only at the desired frequency but also over the entire range of the desired bandwidth. Fig. 11.2.8 shows the optimized reflection coefficient in the desired frequency range.

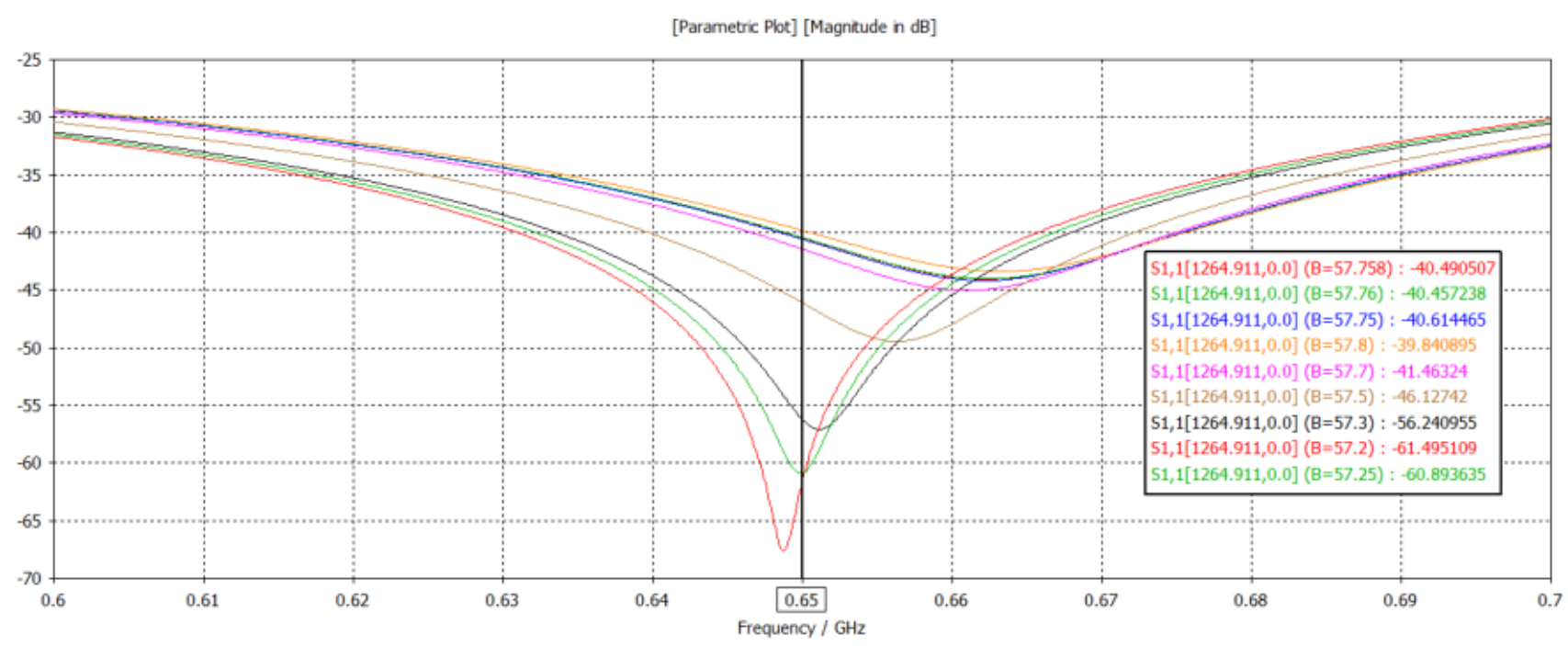

**Figure 11.2.8:** Parameter optimized reflection coefficient

The magnitude of the S-parameter for klystron operation with 800 kW of power at 650 MHz can reach -60 dB, while the S-parameter is well below -30 dB throughout the 600 MHz to 700 MHz range.

Multipacting absorbs RF energy, leading to significant power loss and wall heating. It may cause ceramic window breakdown in high power klystron operation. So multipacting simulations and experiments are needed. CST and Multipac codes are used to do the simulations. Results are shown and compared in Figs. 11.2.9 and 11.2.10.

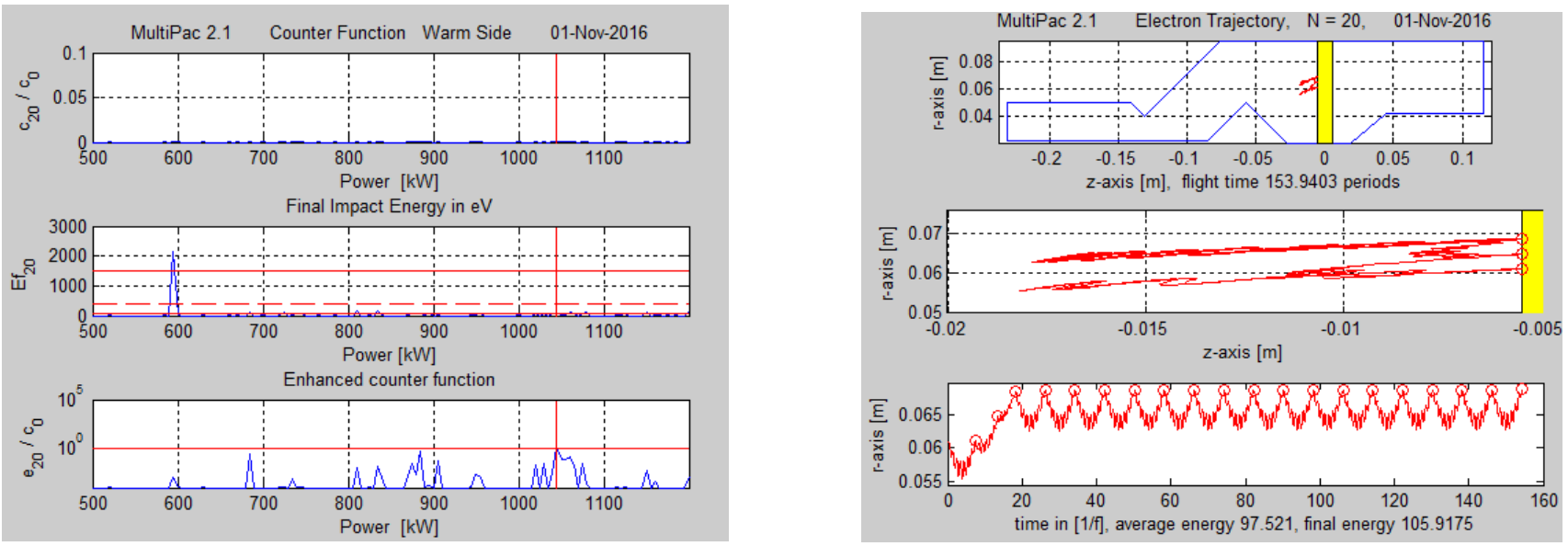

**Figure 11.2.9:** Multipacting simulation results using Multipac

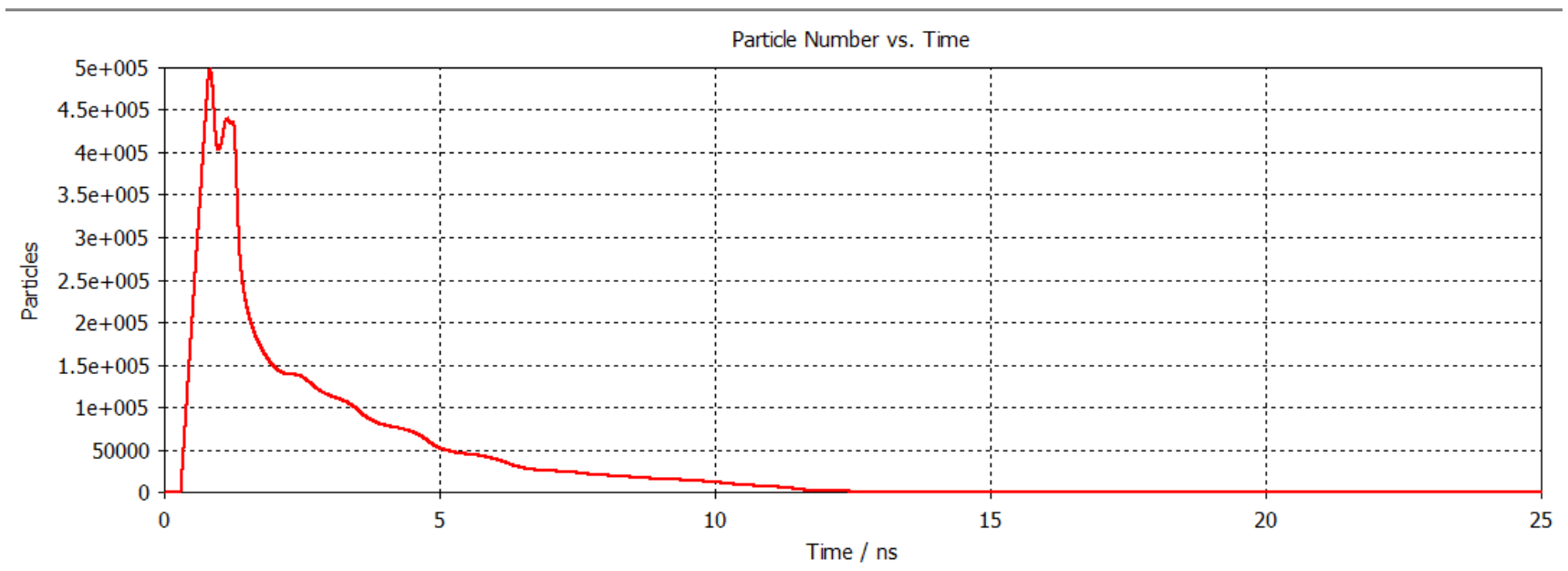

**Figure 11.2.10:** Multipacting simulation results using CST

From the top to the bottom of Fig. 11.2.9 (right), the electron counter function, the average impact energy of the last impact in eV and the enhanced electron counter are shown. The horizontal axis gives the average incident power in kW. The enhanced counter function e20/c0 is the ratio of the total number of secondary electrons after 20 impacts to the initial number of electrons, so if the relative enhanced counter function exceeds unity or the final impact energy of the electron after 20 impacts is in the range where the secondary emission coefficient is larger than 1, multipacting probably would occur. Even though the simulation results shown in the Fig. 11.2.9 (left) indicate that multipacting may exist, Fig. 11.2.9 (right) shows that the e20/c0 are much lower than 1 and the average energy is not in the range for the secondary emission coefficient to be larger than 1. So we conclude that multipacting will not occur. The CST simulation results, shown in Fig. 11.2.10, also strengthen this conclusion. It is seen that the number of particles rapidly decreases with increasing time.

### 11.2.2.5 *Solenoids*

Focusing magnets including the bucking coil were carefully designed so that a laminar beam was generated using DGUN, EGUN and Magic-2D codes. Beam ripple obtained is less than 5%. The Brillouin flow magnetic field is 107 Gauss and the focusing magnetic field was set to around 1.8 times that value to obtain semi-immersed flow. The focusing magnetic field is increased to 280 Gauss around the output cavity to avoid electrons from hitting the drift tube. Fig. 11.2.11 shows a profile of the solenoids in OPERA.

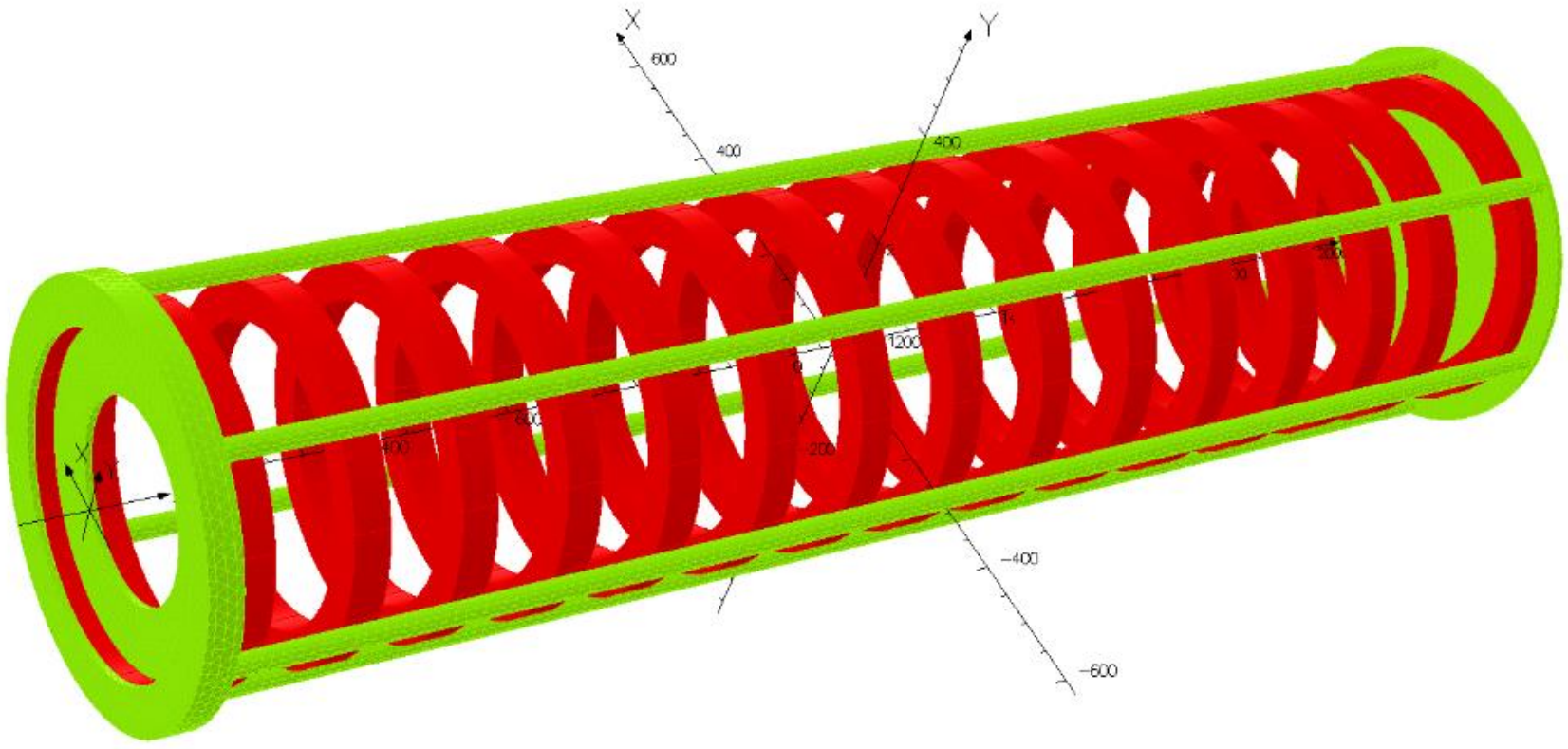

**Figure 11.2.11:** a profile of the solenoids in OPERA

### 11.2.2.6 ***Multipacting Suppression***

To suppress multipacting near the resonant cavity nose, periodic grooves are made on the nose along the circular direction. This is simpler and more practical than a coating of TiN on the whole inner surface of the cavity. There are 36 grooves on each nose along the circular direction. Every groove is at 5° to the central angle and the interval angle of adjacent grooves is also 5°. The angle between the corresponding groove units on the opposite noses is 5°, as shown in Fig. 11.2.12.

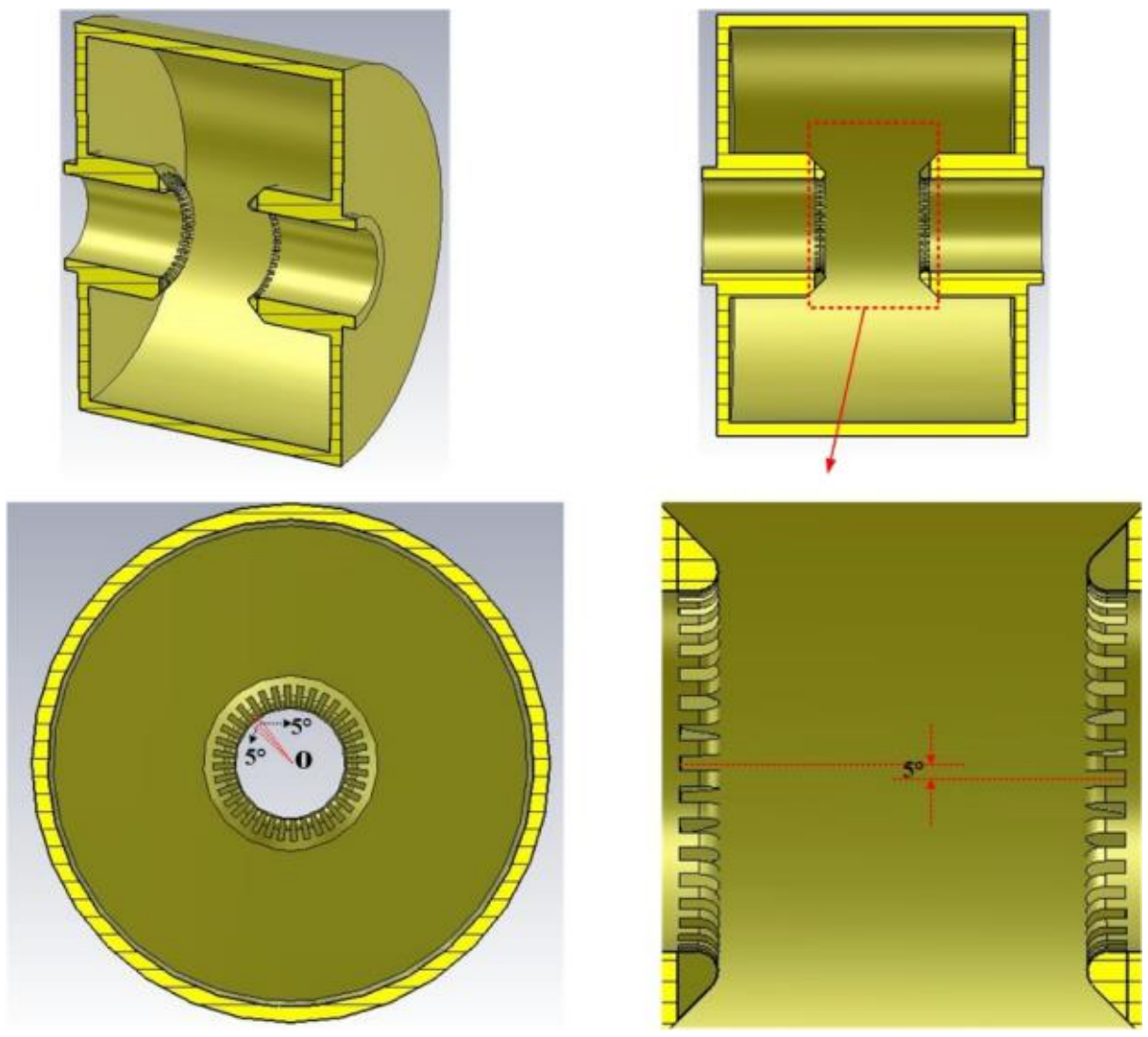

**Figure 11.2.12:** Grooves on the resonant cavity nose

The Secondary Electron Yield (SEY) simulation before and after the groove cutting is shown for a modulation voltage of 800 V (Fig. 11.2.13 left) or 900 V (Fig. 11.2.13 right).. The abscissa is the phase of the electromagnetic field at the nose, while the ordinate is the SEY. The SEY is effectively suppressed by the grooves.

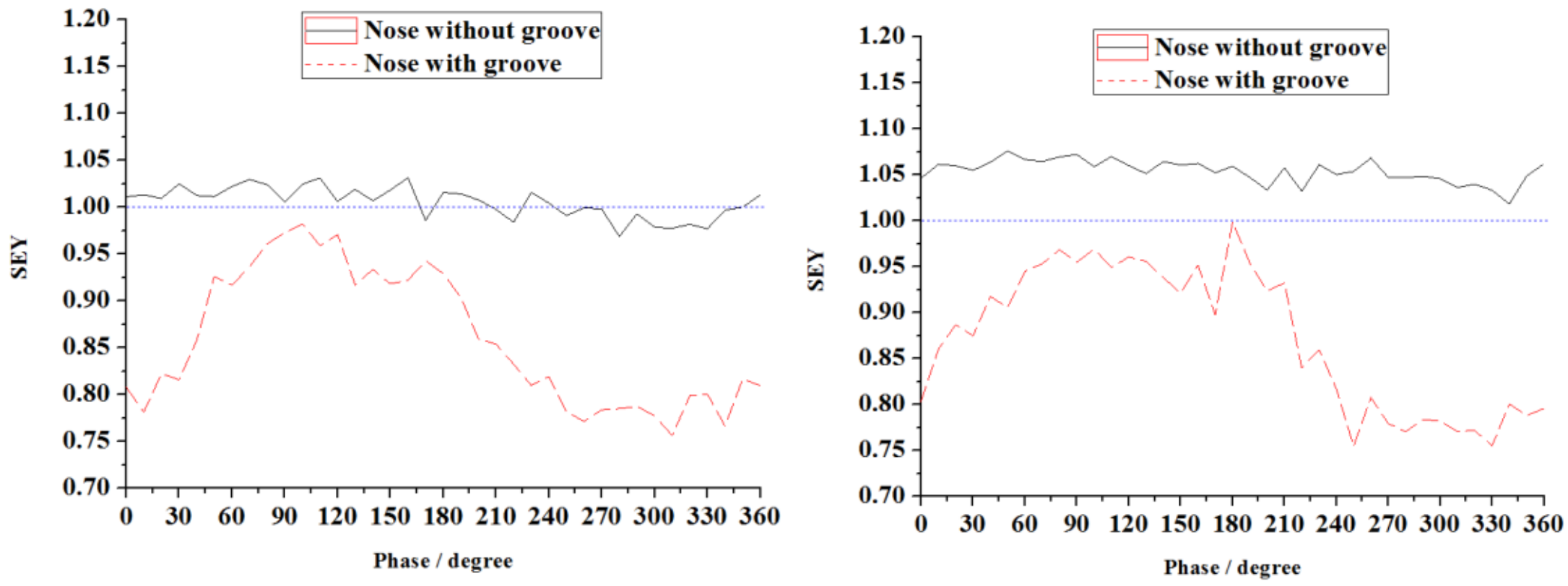

Figure 11.2.13: The comparison of SEY before and after grooves are cut at the modulation voltage of 800 V (left) and 900 V (right) of the resonant cavity

Curves of the maximum and average SEY (ASEY) versus the modulation voltage (from 50 V to 900 V) are shown in Fig. 11.2.14. The ASEY are averaged over 360 degrees of the electromagnetic field. Both, the maximum and average SEY are effectively suppressed by the grooves.

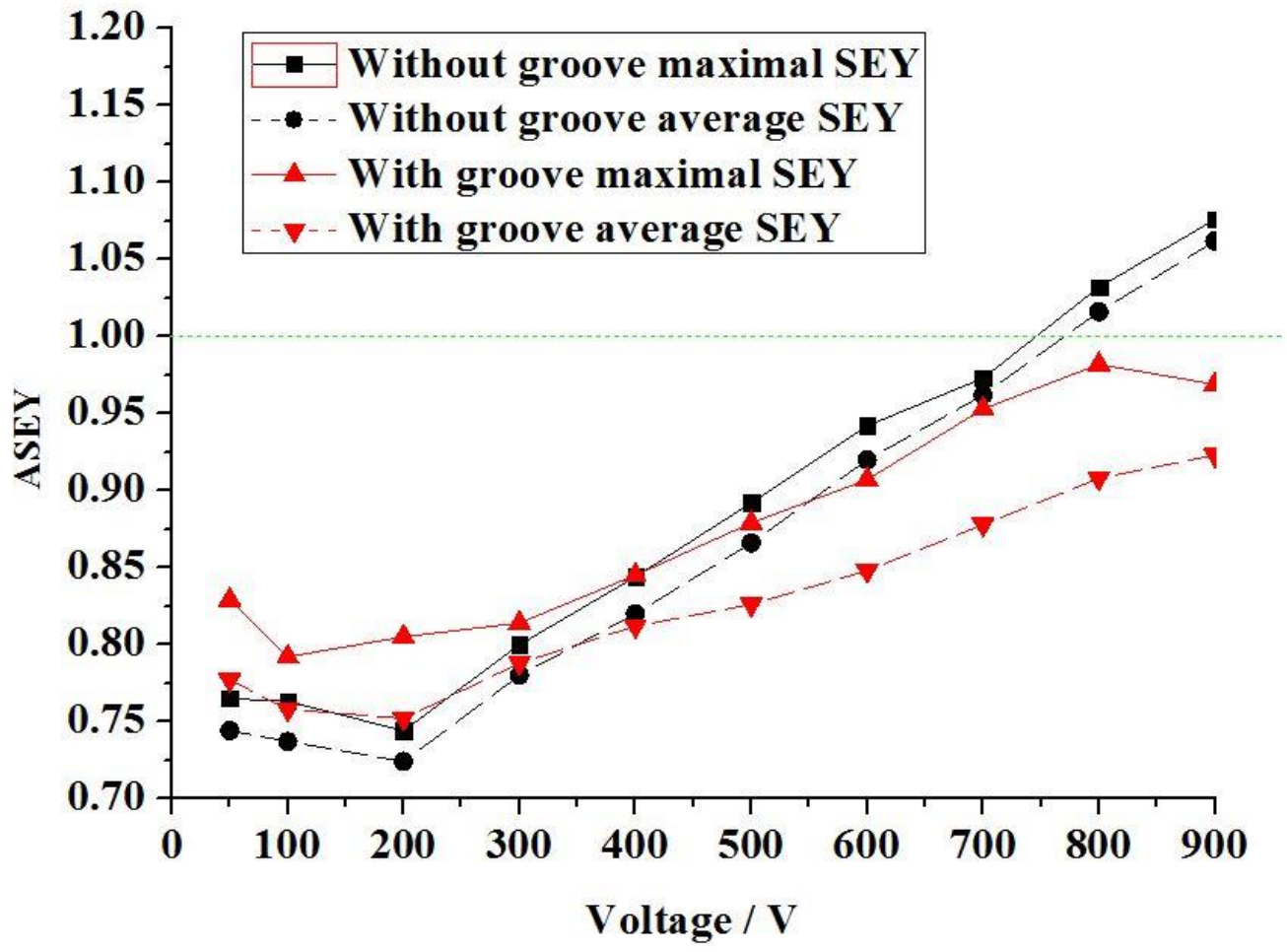

**Figure 11.2.14:** The maximum and average SEY at different modulation voltages of the resonant cavity

### 11.2.2.7 *Collector*

The capability of collector beam dissipation is a key issue. If all the beam power is dissipated without RF drive, it could reach 1.2 MW. The dissipated power is limited while klystron input power is switched on; it is less than 450 kW. Initially, water cooling was chosen to deal with this problem. We present two designs, a full beam power collector and a reduced size collector. A prototype employing a modulated anode (MA) gun will be manufactured and tested. The pulsed operation of the MA gun gives different pulsed power to the structure, and it is also important to evaluate the performance of a collector. Fig. 11.2.15 shows profiles of the collectors. The beam trajectory in the collector was cross-checked with the EGUN and MAGIC 2D codes.

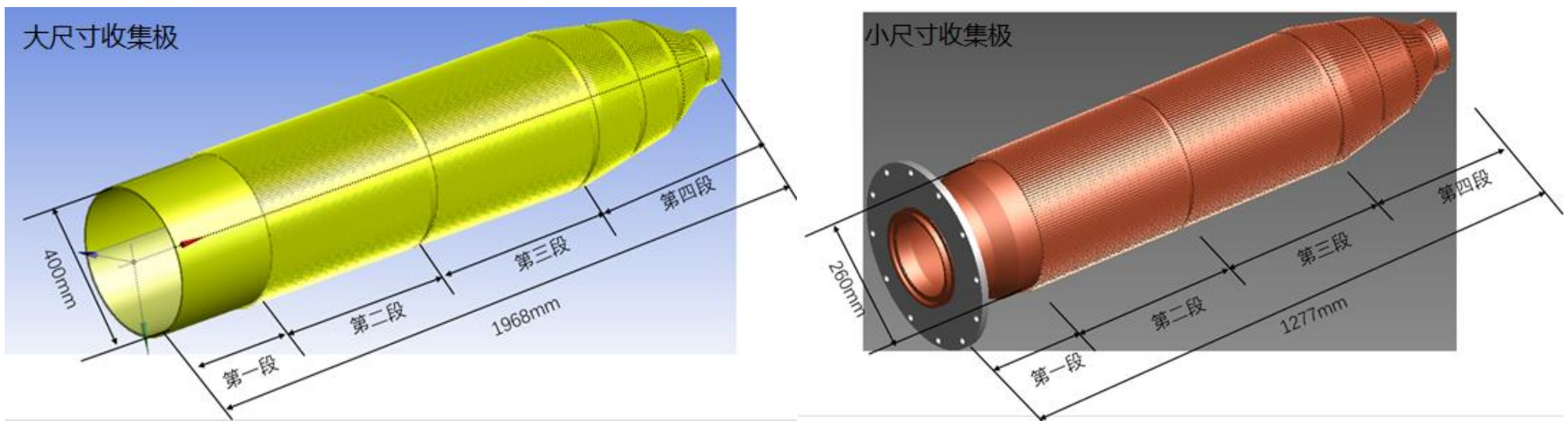

**Figure 11.2.15:** Full beam power collector (left) and reduced size collector (right)

The collector outer surface is grooved to enhance cooling efficiency. The number of grooves, the water flow rate and other parameters are optimized by a fluid flow and coupled heat transfer simulation. The maximum peak power dissipation density is 200 W/cm$^2$ for a full beam power collector and 500 W/cm$^2$ for a reduced size collector. Simulation results are shown in Fig. 11.2.16.

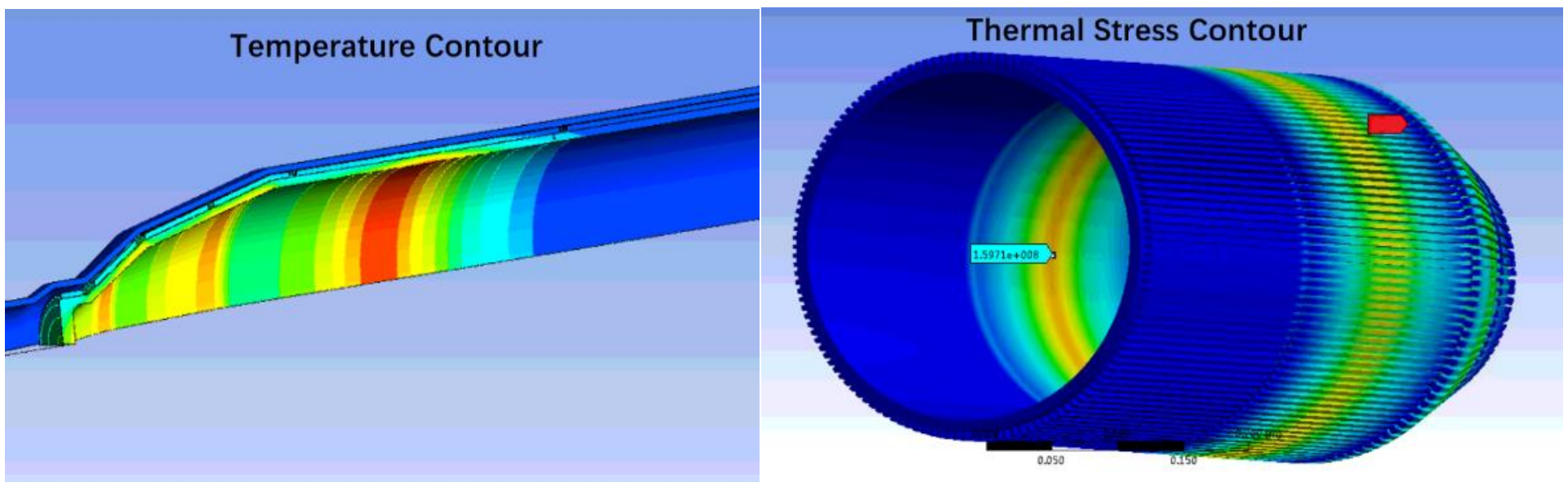

**Figure 11.2.16:** Collector thermal analysis and thermal stress simulation

## 11.2.3 **Summary**

A preliminary simulation shows that the klystron efficiency can be greater than 80%. The klystron will be manufactured by a Chinese company. The gun, cavities, collector and output window are designed to meet required specifications and capacities. The first classical designed klystron will be fabricated soon and the test stand will also be prepared in the near future.

## 11.3 Cryogenic System

The key technologies of the CEPC cryogenic system mainly include the large helium refrigerators and the study of the 2K superfluid helium. The R&D subjects are as follows:

(1) The design of a 4.5K refrigerator of 18 kW at 4.5K;
(2) Large helium turbine expander and its test bench;
(3) Research and prototype development of a large helium screw compressor;
(4) Research and design of a large helium centrifugal cold compressor with high pressure ratio and its test stand;
(5) Sub-cooled 2K Joule-Thomson counter flow heat exchanger.

### 11.3.1 **Large Helium Refrigerator**

The 18 kW at 4.5K refrigerators have been successfully used in LHC and all the components are available from industry. Design work on the large helium refrigerator for CEPC will be carried out by domestic company.

The refrigerator has the equivalent cooling capacity of 18 kW at 4.5K. It consists of the following subsystems: cold-box, main compressor station with oil removal system, vacuum pump group and gas management panel (GMP) with buffer tank and control system. There are five pressure levels in the cryoplant: 20 bara, 4 bara, 1.05 bara, 0.4 bara and 3 kPa. These are obtained with the high pressure screw compressor group, middle pressure screw compressor group, vacuum pumps and cold compressors.

### 11.3.2 **Turbine Expander**

The turbine expander is a key refrigerator component. There are different kinds of turbine expanders in large helium refrigerators. At high temperature, a large scale turbine expander is required to meet large flow rate. At low temperature, s small turbine expander which has a high speed with a small flow rate is needed. In the R&D program on the cryogenic system, analysis will be made of the dynamic properties of the high speed rotor bearing system. There will be an experimental study of the aerodynamic performance of the flow section, and structural design will improve the reliability of the cold quantity adjusting mechanism.

The detailed study includes:

(1) Design and development of a high stability radial bearing;
(2) Design and development of a high capacity thrust bearing;
(3) Development of a large refrigeration capacity helium turbine expander;
(4) Development of a test bench with large stable flow rate.

The helium turbine expander is the core of the system. Stability and good thermodynamic performance are important. The target is to guarantee the stability of the high-speed helium gas bearing of the turbine expander rotor bearing system, as well as to improve the thermodynamic efficiency.

### 11.3.3 **Screw Compressor**

The main study of the helium compressor group is the plan design, performance simulation and different pressure levels (1/4/20 bara) linkage regulation performance. The key design problem is to improve the efficiency of the host screw compressor, including optimization of rotor type line, reduction of the leakage triangle area, reduction of the contact line length, optimization of the meshing clearance. This will be accomplished using spray atomization cooling and high efficiency oil separation technologies.

The main items follow:

(1) Optimization of the rotor type line;
(2) Design and experimental study of the special seal structure of the power transmission shaft in the screw compressor;
(3) Design and test of the mechanical oil and gas separation system with the basic principle of mechanical centrifugal force and speed control;
(4) Construct a prototype of the oil injection type helium screw compressor.

Oil injection type helium screw compressor will be developed with domestic manufacturers. The dynamic characteristics of large screw compressor, helium screw rotor type line and assembly process will be studied. Then helium compression experiments and compressor performance tests will be carried out to improve the structure and processing technology.

The main technical parameters are as follows:

(1) Oil injection type helium screw compressor:
Volumetric efficiency ≥ 75% and Shaft seal leakage rate ≤ 10-3 Pa·m$^3$/s
(2) Precision oil separation system：Oil content ≤ 0.01ppm

### 11.3.4 **Centrifugal Cold Compressor**

The major R&D work on the centrifugal cold compressor is related to thermodynamic and mechanical performance as well as reliability. The adiabatic efficiency of the cold compressor will be 1% with 20 W heat leakage. Therefore, the length of the shaft will be increased and the overall sealing performance will be investigated. The stability of the cold compressor is affected by two aspects: the stable working area of the cold compressor, and the stability of the magnetic bearing.

The centrifugal cold compressor will be developed and an experimental study carried out. This will include determining thermodynamic characteristics, sealing insulation characteristics, internal flow instability characteristics, surge recovery features and many sets of a cold compressor cascade working characteristics. An electromagnetic bearing, high-speed motor performance test platform will be developed.

The main parameters of the centrifugal cold compressor are as follows:

(1) adiabatic efficiency: ≥ 60%
(2) compression ratio: ≥ 2
(3) leakage rate: $10^{-9}$ Pa·m$^3$/s
(4) high-speed motor output power: ≥ 1 kW
(5) high speed motor speed: ≥ 36k rpm

### 11.3.5 **2K Joule-Thomson Heat Exchanger**

The Joule-Thomson heat exchanger is one of the key components of the 2K cryogenic system. The 4.4K saturated helium is sub-cooled to about 2.2K in the Joule-Thomson counter flow heat exchanger and enters at 2.2K into the forward tubes of the cryomodules. The helium is expanded to 31 mbar via a JT valve, resulting in more than 80% helium-II liquid at 2K. The evaporation 2K helium with a pressure of 31 mbar is superheated to about 3.5K in the heat exchanger. The Joule-Thomson counter flow heat exchanger can improve the efficiency of the JT valve.

The sub-atmospheric pressure and low temperature heat exchanger has the characteristics of small volume and fluid resistance, and large heat transfer area. A finned tube is proposed to enhance the heat transfer and improve efficiency. In order to balance the heat transfer and pressure drop for the JT heat exchanger, the diameter of the tube and fin height will be optimized.

As a compact and efficient heat exchanger, Hampson type heat exchangers are widely used in the field of natural gas liquefaction and cryogenics. The design parameters of JT heat exchanger are as follows:

(1) temperature range: 2 K~4.5 K
(2) pressure drop: < 300 Pa
(3) efficiency: > 85%

## 11.4 Magnets

### 11.4.1 **Prototype Magnets for the Collider**

The field of the dual aperture dipole is about 140Gs at Z mode. The requirement of the field quality is hard to achieve at low field. To study the possible field distortions for Z running, a short prototype of dual aperture dipole will be developed and tested.

The dual aperture quadrupole uses multi-turn coils made of aluminum and a shielding plate to compensate for the non-systematic harmonics. The hollow aluminum conductor used is 11 × 11, Φ7, R1 [mm] and has not been used before. So it is necessary to make a real coil to verify feasibility. Simulation results show that the non-systematic harmonics induced by crosstalk are very sensitive to the thickness of the magnetic shielding plate placed in the middle of the two apertures. The transfer function of shielding thickness to the harmonics need to be tested. A dual aperture quadrupole prototype will be developed to study all these issues.

A permanent magnet has a great advantage in energy saving. The disadvantages are radiation resistance, difficulty of field adjustment and temperature stability. The radiation resistance of permanent magnet materials will be studied in the R&D program. A field adjustable dipole prototype will be developed to investigate this possibility.

### 11.4.2 **Prototype Magnets for the Booster**

In the R & D phase, one high precision low field prototype Booster dipole will be developed to study the technical issues of design and production. The specifications of the magnet are listed in Table 11.4.1.

**Table 11.4.1:** Booster prototype dipole magnet specifications

| | |
|---|---|
| Magnetic Length | 5 m |
| Gap | 63 mm |
| Working field | 29 to 338 Gs |
| Good field region | 55 mm |
| Field uniformity | $1 \times 10^{-3}$ |

The following describes the fabrication of these uniform field dipoles:

- The 5-m long core will have an H-type frame for better shielding of the earth's field;
- The core will be composed of stacks of 1 mm thick low-carbon steel laminations spaced by 1 mm thick aluminum laminations;
- The coil will be one turn per pole and will be solid aluminum bars without water cooling;
- By using supporters in the gap to compensate for the core weight and magnetic force the return yoke will be made as thin as possible;
- In the upper and lower poles of the laminations, 8 rectangular holes and 2 round holes will be stamped to reduce the core weight as well as to increase the field in the laminations;
- In order to install the vacuum chamber, the core is divided into upper and lower parts;
- Around the outside of each half-core, four long bars are used to weld the laminations of the half-core together. In the round hole of each pole, there is a long bar inserted to press the laminations together.

The cross section and flux distribution of the low field Booster prototype dipole is shown in Fig. 11.4.1.

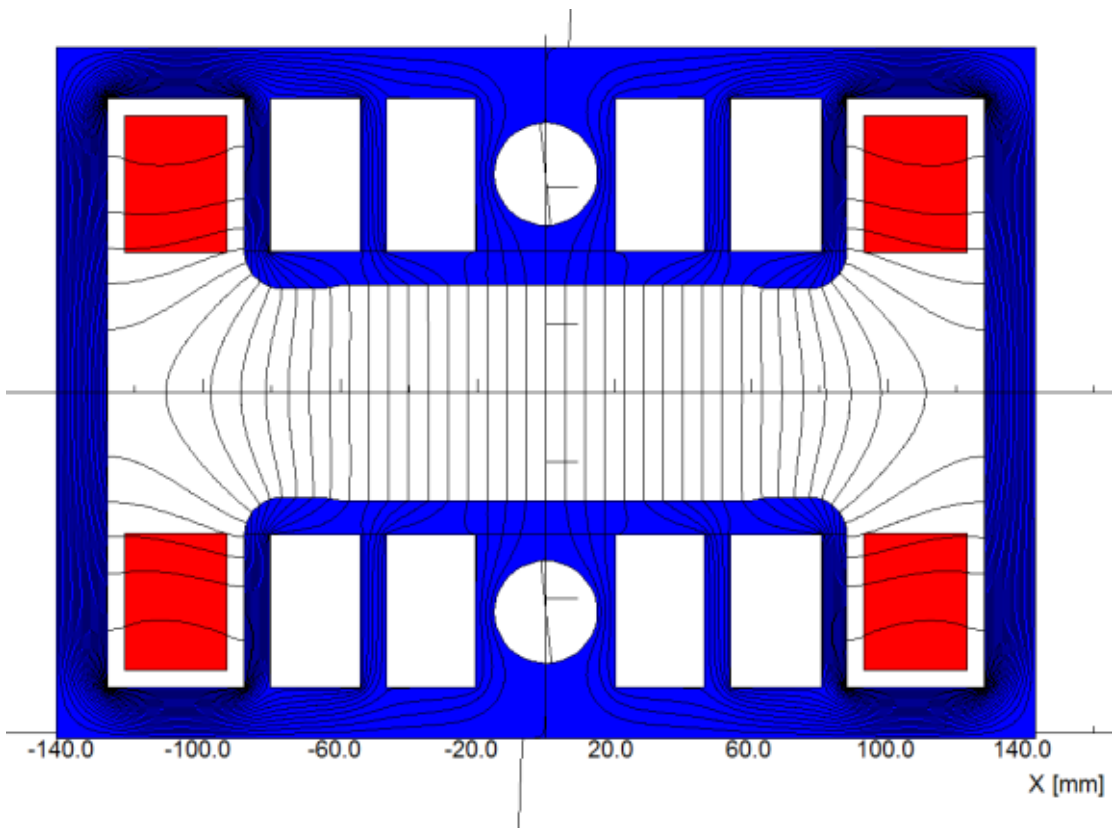

**Figure 11.4.1:** The cross section and flux distribution of the Booster dipole

Theoretically, a good design of dipole magnet without iron core can meet the high precision requirement at low field level of 29 Gs because the remnant field of the iron core is the key factor that destroys the field quality of the low field. However, the problem is how to improve the excitation efficiency at high field level of 338Gs. The cosθ and canted cosθ type coil of dipole magnet without iron cores will be tried to design the booster dipole magnet.

## 11.5 Magnet Power Supplies

All magnet power supplies will be produced domestically. There are two items requiring R&D.

### 11.5.1 **3000 A/10 V High Precision Power Supply**

Required for the superconducting magnets in the IPs are high current (~3kA) relatively low voltage (~10 V) supplies. In particular, DC stability is important. Design parameters are listed in Table 11.5.1.

**Table 11.5.1:** Parameters for the high current power supply

| | |
|---|---|
| Output Current – Imax (A) | 3000 |
| Output Voltage (V) | 10 |
| Stability(8h-10s)–referred to Imax (ppm) | 10 |
| stability (10s-0s)–referred to Imax (ppm) | 10 |
| Reproducibility - referred to Imax (ppm) | 100 |
| absolute accuracy - referred to Imax (ppm) | 100 |
| current ripple - referred to Imax (ppm, 50 Hz and greater) | 10 |

The power supply will use software switching techniques. The topology chosen is to split it into three modules:

- a diode rectifier connected to the AC mains with a damped L-C passive filter;
- a full-bridge zero voltage switching phase shift inverter;
- a high-frequency transformer, rectifier and output filter.

The converter will be split into n+1 modules, whose outputs are connected in parallel. N modules supply the nominal current, and one module will work in case of a trip. The modules can automatically switch between each other.

Radiation tolerance tests will be performed during the supply design, in case the supply is to be operated in a radiation environment.

### 11.5.2 **Digital Power Supply Control Module**

A DPSCM will be employed in all CEPC power supplies. The module will control the high precision current loop and communicate with the control room. The control module block diagram is shown in Fig. 11.5.1.

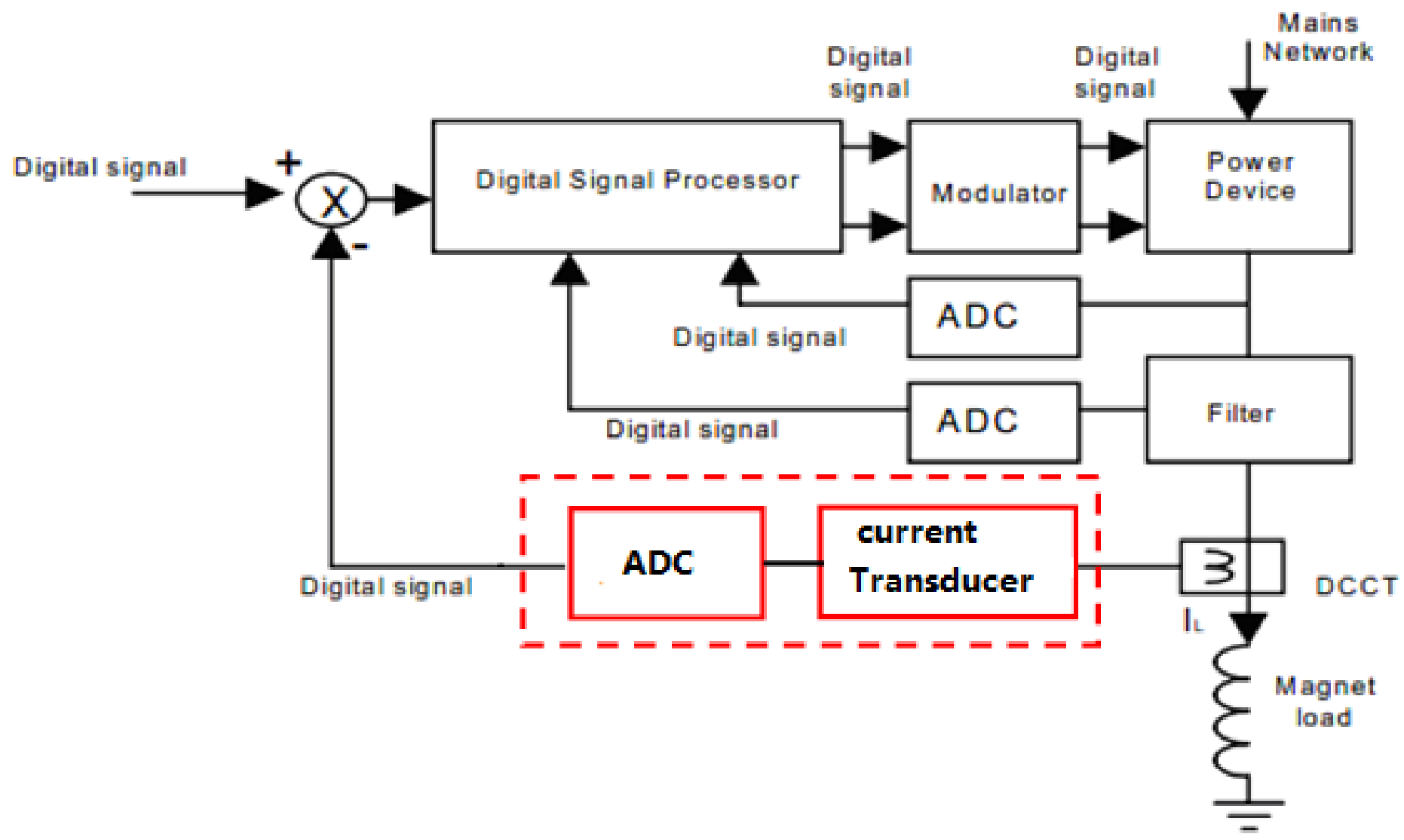

**Figure 11.5.1:** Digital controller for magnet power supplies.

To meet the high performance requirements, the following issues must be considered during the DPSCM development:

- Chip choice of digital signal processing based on the system-on-chip of FPGA (Altera). FPGA stands for Field Programmable Gate Array;
- ADC design: Low noise design on PCB; constant temperature protection for the ADC; anti-dithering circuit design.

Implementation of the digital control algorithm on FPGA: Embedded fuzzy logic and expert system into the digital control platform (DCP) for better diagnostics, faults analysis, auto-detection and self-calibration.

## 11.6 Electrostatic Separators

A set of electrostatic separators will deflect the e+ and e- bunches in the RF region. Each RF station is divided into two sections so that half of the cavities can be bypassed during operation in the W and Z modes as illustrated by Fig. 11.6.1. An electrostatic separator is combined with a dipole to avoid bending of the incoming beam. The gradient of the electrostatic separator is 1.8 MV/m and its total length 50 m. This is followed by a drift as long as 75 m to make the two-beam separation as large as 10 cm at the quadrupole entrance. In order to limit the beta functions, two triplets are used. Then the beam is further separated with dipoles.

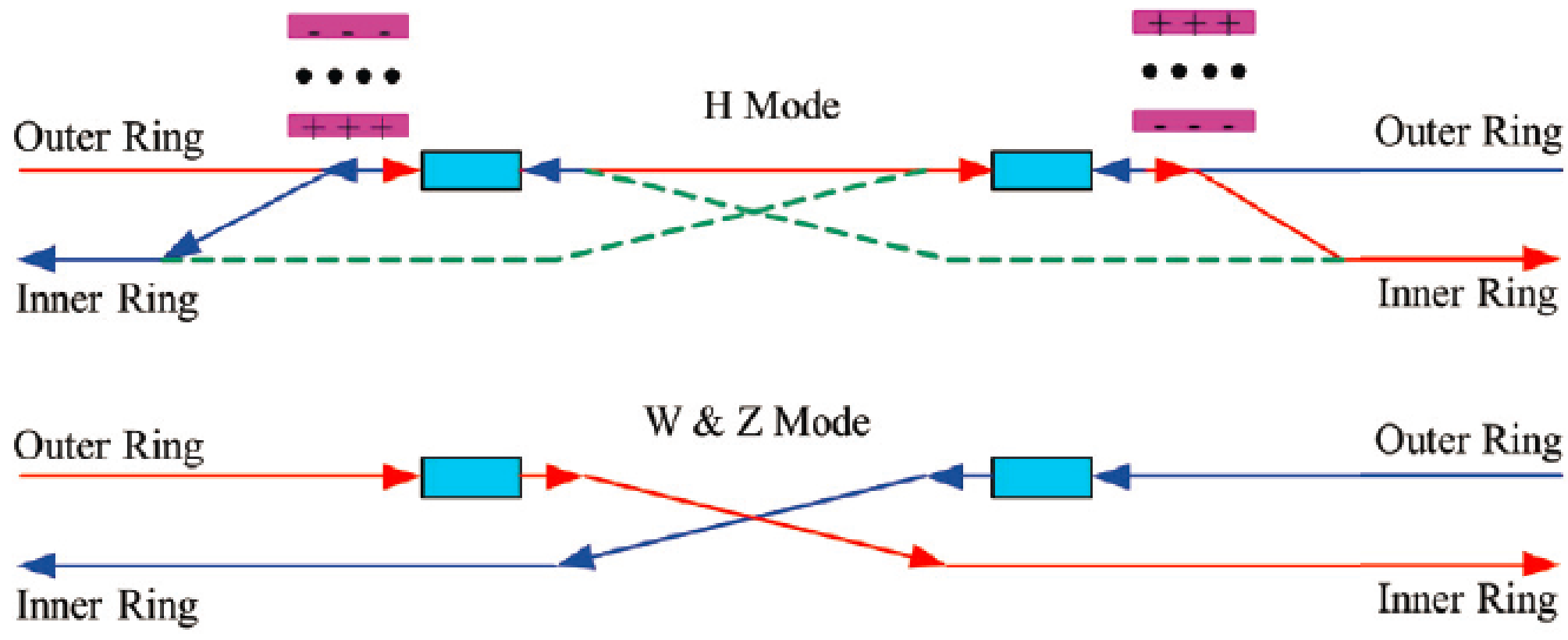

**Figure 11.6.1:** Layout of the RF region

Beams can be separated in either the horizontal or the vertical plane. With horizontal separation, the separation distance is larger since the beam size is larger vertically. With vertical separation, the separation distance is smaller, but separating in the vertical plane can easily induce large coupling between horizontal and vertical planes. As the coupling factor is limited to a small value to attain high luminosity, we have chosen horizontal separation.

Each RF region needs 24 electrostatic separators. Each one is 4.5 m. long and the inner diameter of the separator tank is 540 mm.

## 11.7 Vacuum System

### 11.7.1 **Vacuum Chamber**

As discussed in Section 4.3.6, the Collider will have an aluminum chamber for the electron beam and a copper chamber for the positron beam. The Booster vacuum chamber, described in Section 5.3.5 will be aluminum. The fabrication procedures for these chambers are described in those sections.

The aluminum chamber is similar to the LEP chamber. It has a beam channel, a cooling water channel, a pumping port used to install ion pump, and thick lead shielding blocks covering the outside.

The copper chamber has a beam channel and a cooling water channel. And, NEG coating will be used.

In the R&D program both types of chambers will be fabricated and tested.

### 11.7.2 **NEG Coating**

The NEG coating is a titanium, zirconium, vanadium alloy, deposited on the inner surface of the chamber through sputtering. The process is described in some detail in Section 4.3.6.4.1. As stated there, R&D is required so the sputtering process for NEG film deposition is optimized to avoid instability and lack of reproducibility. These problems can significantly change the gas sorption and surface properties (e.g. secondary electron yield, ion-induced gas desorption). During tests of the coating, all related parameters (plasma gas pressure, substrate temperature, plasma current, and magnetic field value) will be recorded and suitably adjusted to ensure stability of the deposition process. After coating, the chambers will be cooled down to room temperature, exposed to air and left to age for a couple of days before being visually inspected again. Aging is a recommended procedure, since it helps identify areas where the film adhesion is poor.

### 11.7.3 **RF Shielding Bellows Module**

Many narrow Be-Cu fingers slide along the inside of the beam passage as the bellows is compressed, as shown in Fig. 4.3.6.3 and Fig. 11.6.1 below. In the R&D program prototypes will be constructed and tested to ensure that they meet the requirements of maximum contraction and expansion.

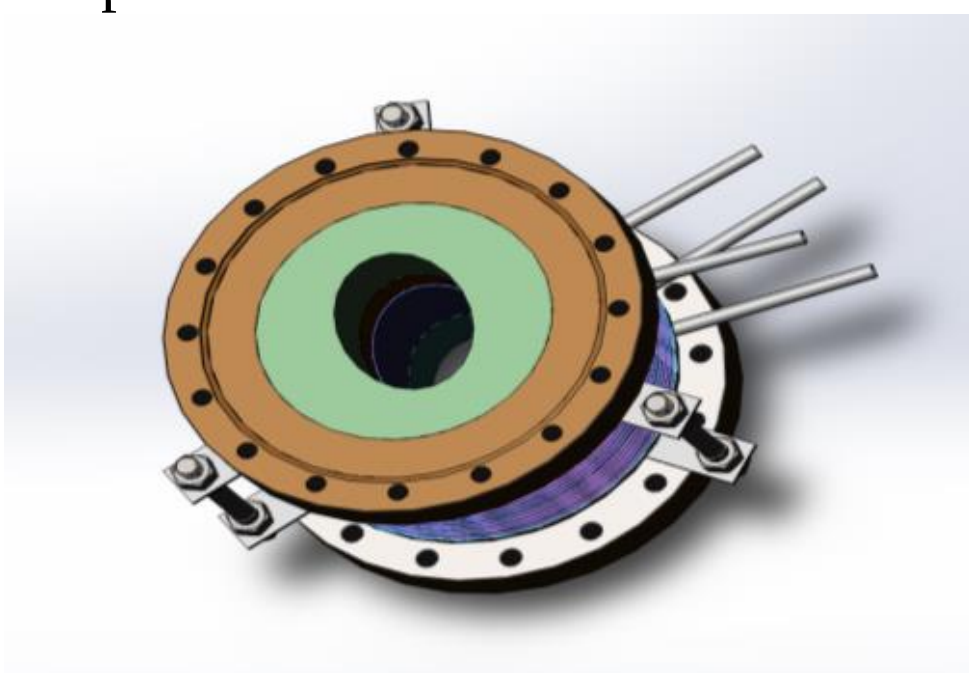

**Figure 11.6.2:** Bellows module with RF shielding

## 11.8 Instrumentation

The CEPC requires a total of 4,900 beam position monitor systems. These consist of a digital beam position monitor (DBPM), pick-ups and transmission cables. There are commercial products available for the DBPM from the I-Tech Corporation. However, they are expensive and the source code is not open, so development and upgrades are difficult and costly. In order to reduce cost and also to enable convenient maintenance, upgrades and personnel training, the DPBM will be developed in house. This R&D work benefits from seed money already granted and considerable progress has already been made. The DBPM chassis based on MTCA 4.0 was developed.

DBPM R&D is divided into three parts carried out in parallel. The DBPM architecture is shown in Fig. 11.8.1. The analog front-end electronics (AFE) is for signal filtering, attenuation and conditioning. Digital front-end electronics (DFE) is for analog to digital conversion, clock and data processing, frequency domain analysis and fast-Fourier conversion (FFT). The algorithm that is developed can be validated in commercial products before being switching to in-house fabrication.

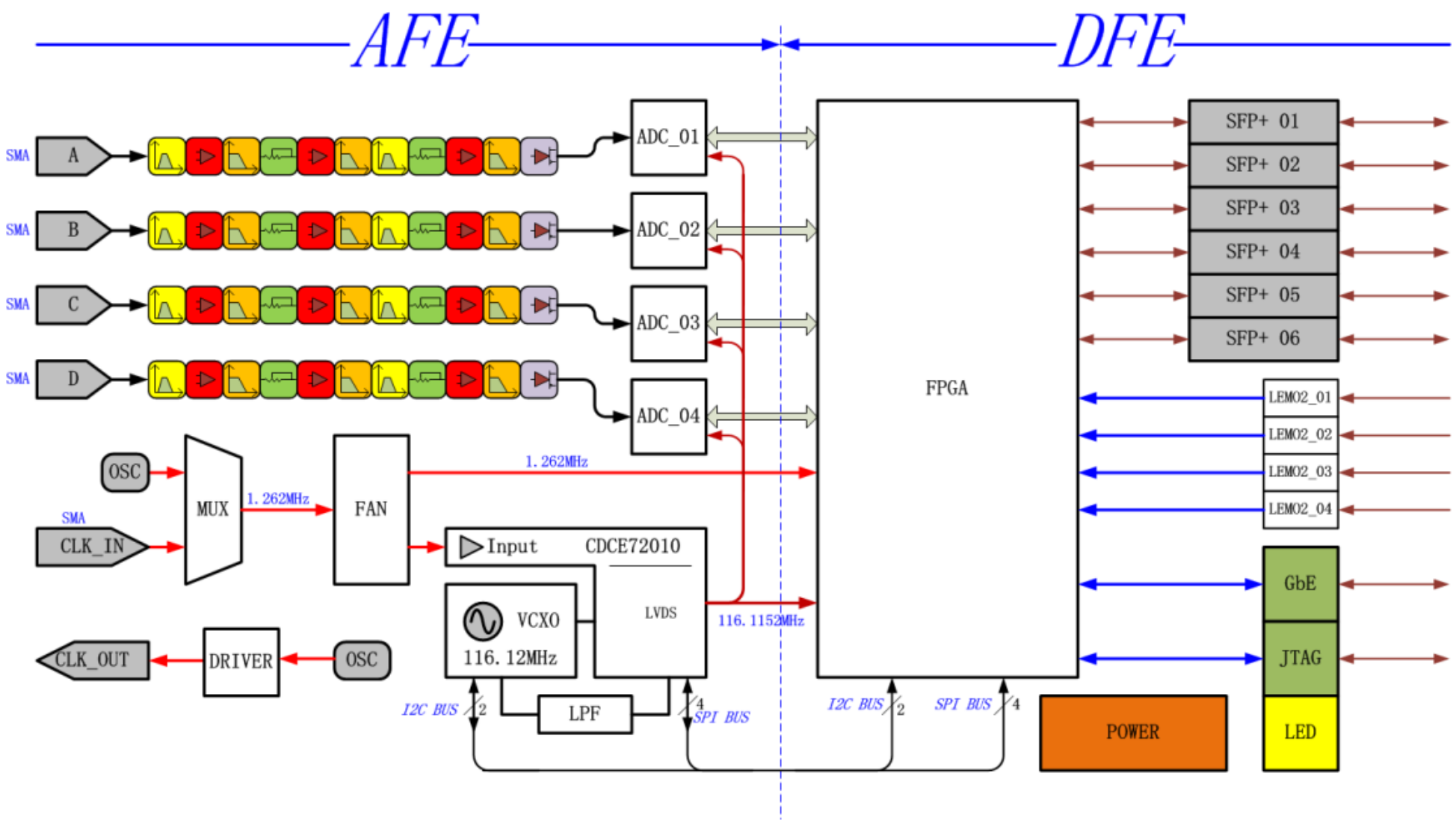

**Figure 11.8.1:** Architecture of the beam position monitor system

**AFE electronics**

The difficulty in the analog front-end electronics design lies in the crosstalk between the four channels. Four versions of the analog electronics have been done; the latest, version 4.0, is shown in Fig. 11.8.2. In this version, the cross-linked switch is cancelled, and the beam calibration method is used to ensure the consistency of the four channels. It was tested in the laboratory and the S-parameter (scattering parameter) results are shown in Fig. 11.8.3. Band pass smoothness and bandwidth meet the requirements. Crosstalk between channels is successfully solved and the resultant channel isolation is better than 65 dB.

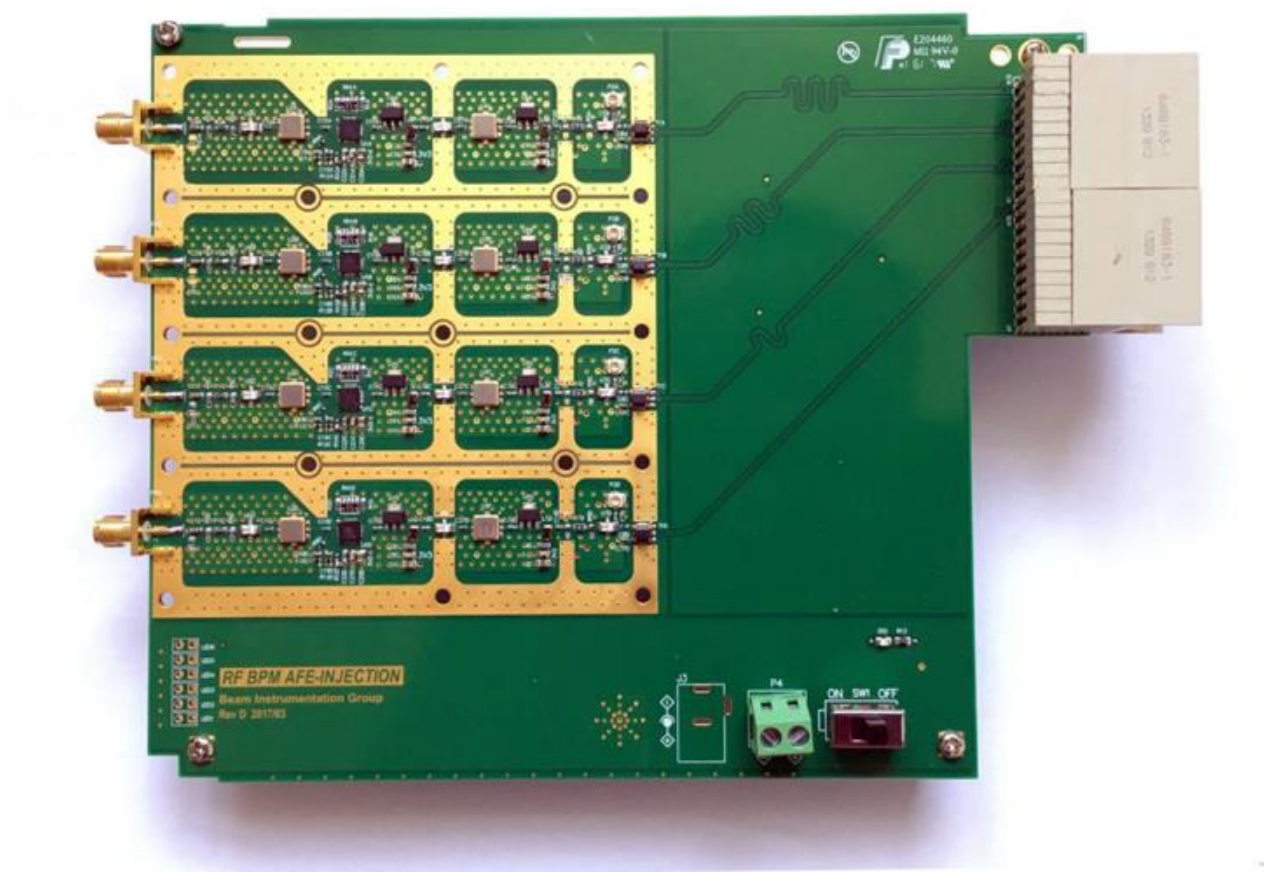

**Figure11.8.2:** Version 4.0 of the analog front-end electronics

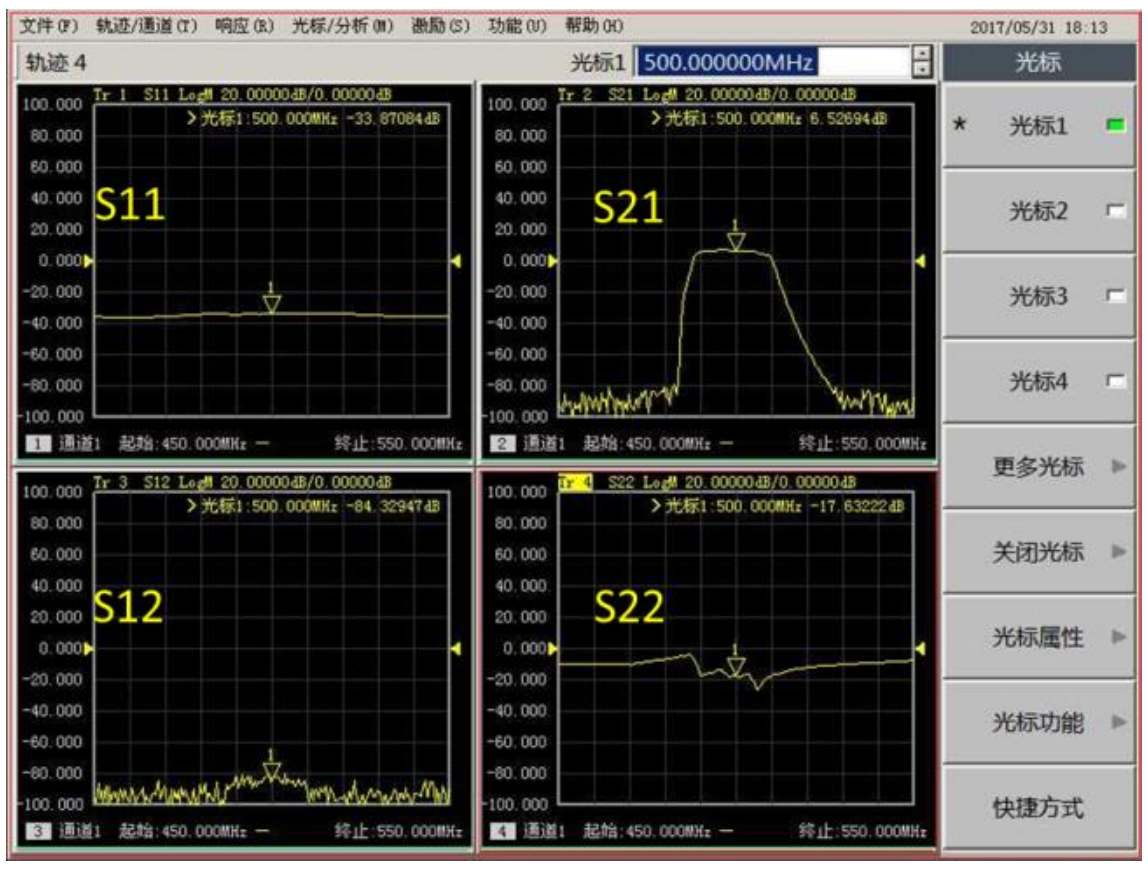

**Figure 11.8.3:** Receiver S-parameter characterization of the AFE

**DFE electronics**

R&D work on digital front-end electronics include hardware and firmware development. Two versions of the digital circuit board and optimized power modules, clock modules, DDR (double data rate) modules and FPGAs have been completed. The firmware is developed based on MATLAB results, first with simulation and then downloaded to FPGA for implementation. After hardware and firmware development, the AFE and DFE are connected together, as shown in Fig. 11.8.4, by an ADF connector for laboratory testing.

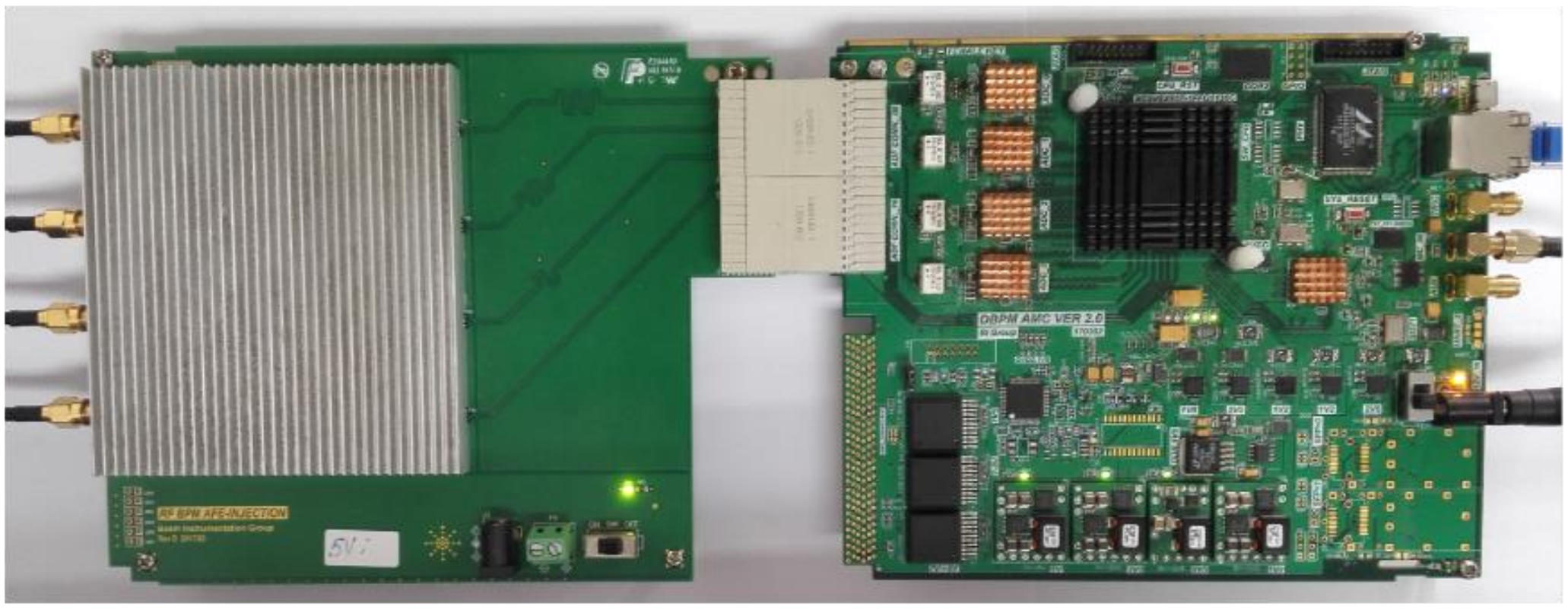

**Figure 11.8.4:** AFE and DFE joint laboratory test

**DBPM algorithm**

The algorithm is the core of the BPM and it and the associated hardware are developed in parallel. The algorithm is first verified on existing commercial products. RAW data of Libera (an Instrument Technology company specializing in accelerator applications) is used to check the data processing. Then, after checking, the algorithm is transplanted to the self-developed BPM electronics. Finally, after verification, the hardware and software was integrated and tested with beam in BEPCII. The test results obtained in July 2017, as seen in Fig. 11.8.5, show that the rms resolution is 1.229 μm for turn by turn data, 0.4 μm (3 s) for FA data, and 0.19 μm (10 s) for SA data. The result is shown in Fig. 11.8.5.

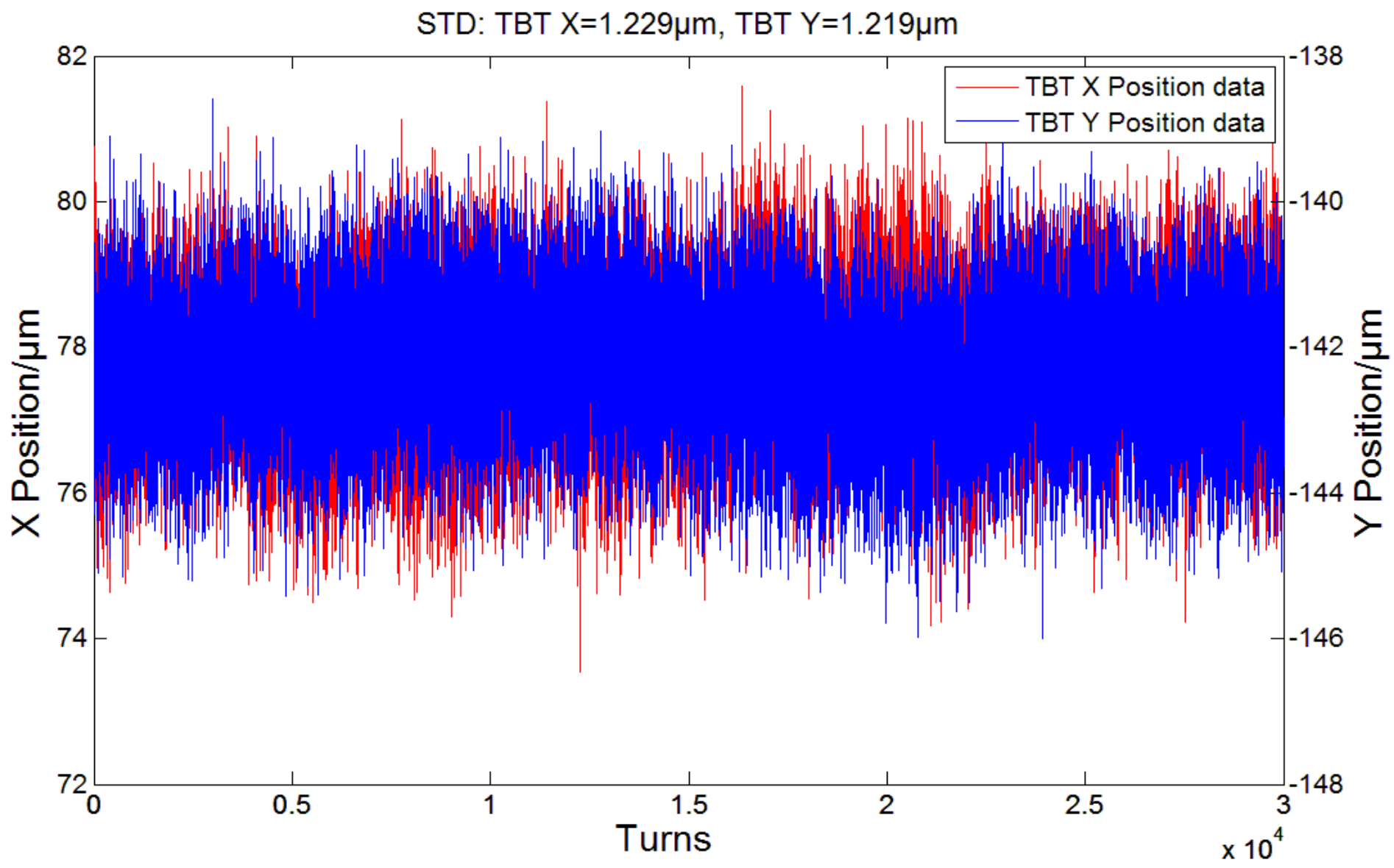

**Figure 11.8.5:** DBPM test results

There is still a lot of future work ahead. AFE and DFE must be transplanted to micro TCA (telecommunications architecture). Issues in future R&D are long-term stability of the integrated system, success rate of mass production, and future hardware upgrades.

## 11.9 Control System

Control system R&D is primarily concerned with the MPS (Machine Protection System). It steers the beam to the Beam Dump to protect devices from damage during the occurrence of faults. The beam revolution time is about 0.17 ms. So, the response time should be comparable.

For the CSNS (China Spallation Neutron Source) an MPS has been built successfully, and CEPC MPS R&D has been based on it. The signal type, signal transmission and FPGA chosen, need to be discussed.

The front board of the MPS is mother-daughter, where the mother board is responsible for signal processing and the daughter board is for signal collection and conditioning. These are shown schematically in Fig. 11.9.1, and pictorially in Fig. 11.9.2

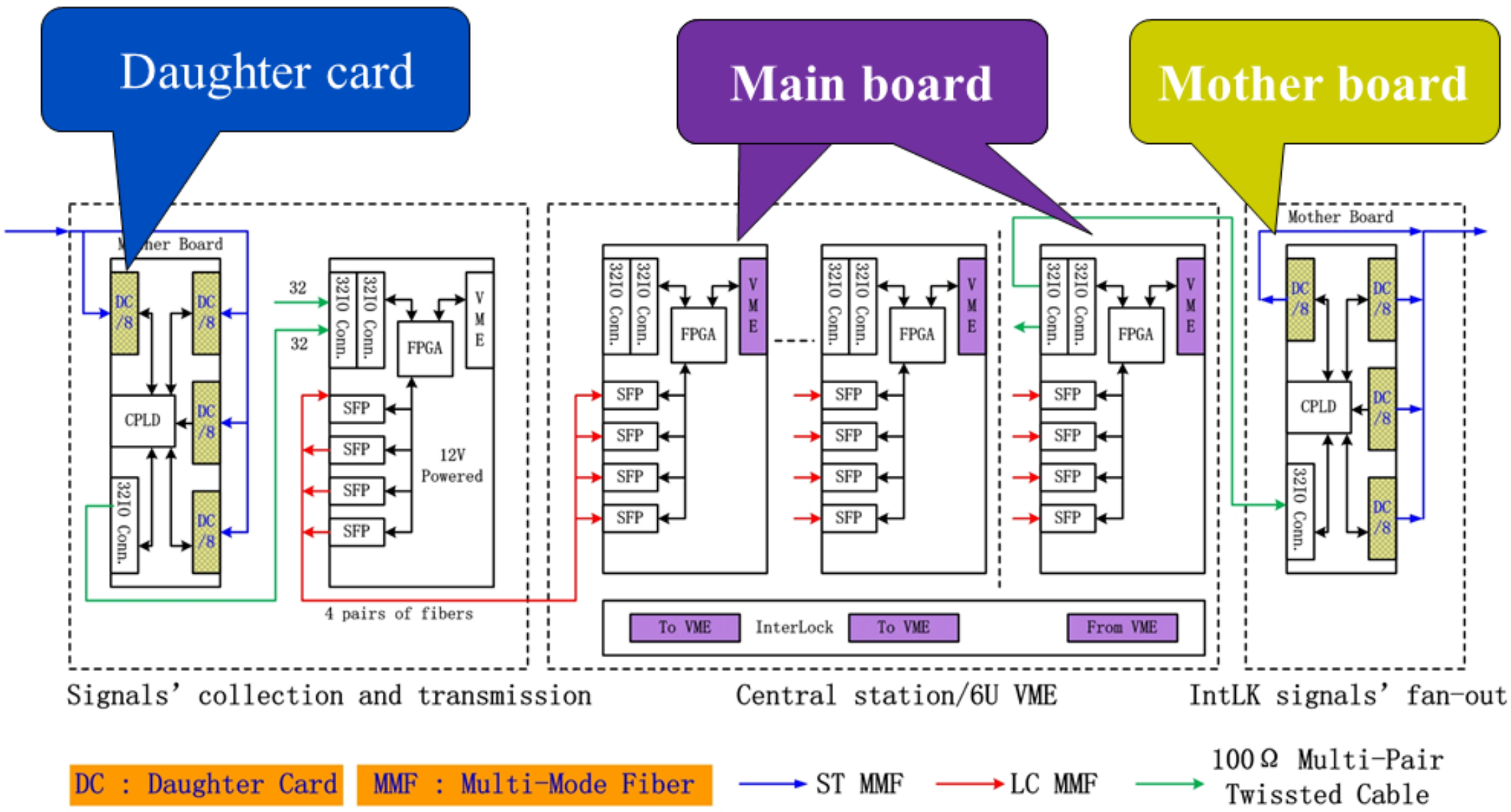

**Figure 11.9.1:** The mother-daughter structure of the MPS

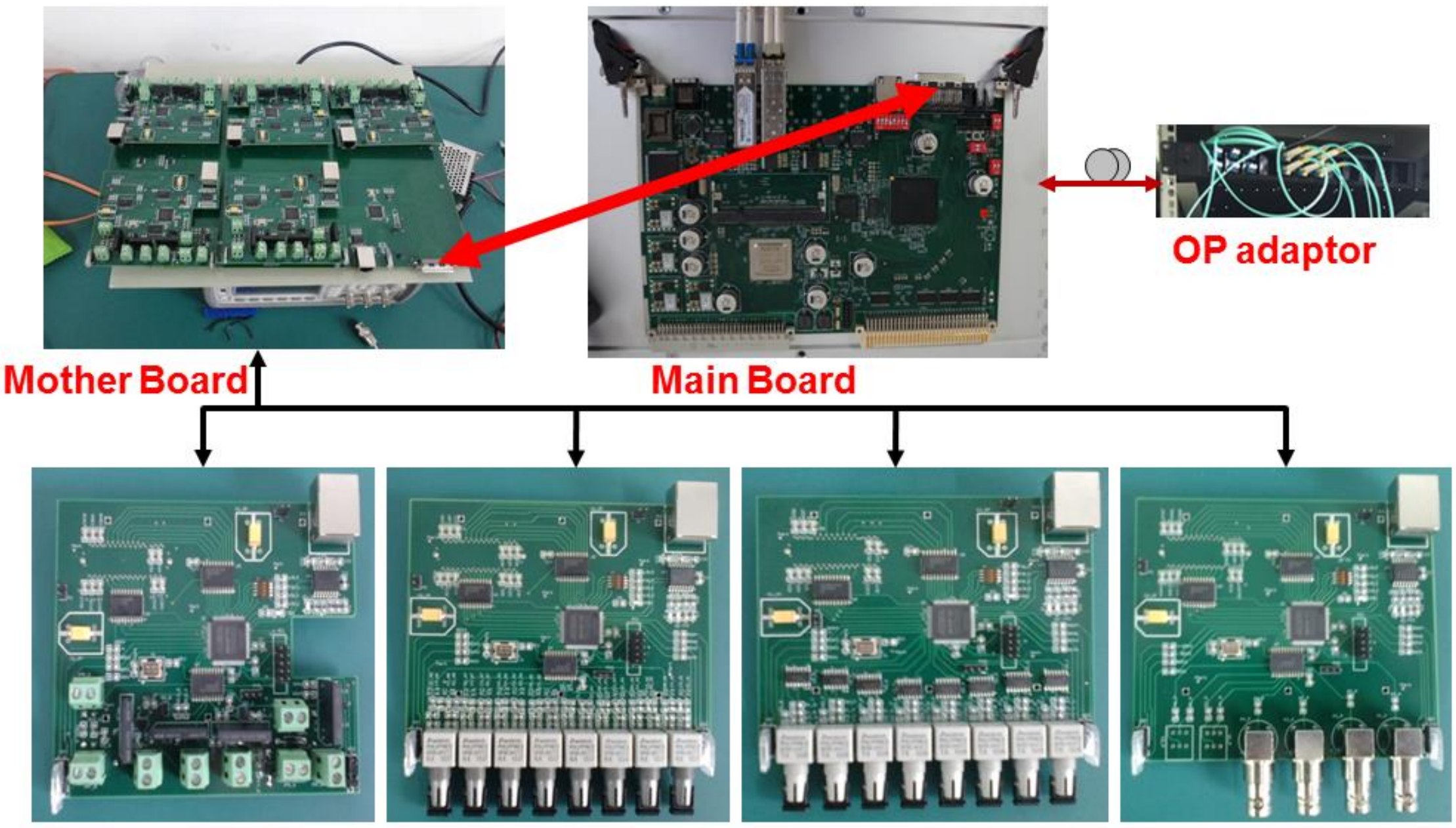

**Figure 11.9.2:** The MPS hardware

A MPS test bench has been built and the basic interlocking logic works well. To improve reliability and stability of the MPS, signal transmission and signal interference over long distances (up to 30 km) need to be carefully investigated. As part of future R&D, redundant MPS controllers will be considered.

## 11.10 Mechanical Systems

### 11.10.1 Development of the Collider Dipole Support System

The R&D goals are:

- Develop simple and reliable mechanics for safe mounting and easy alignment;
- Design the installation and replacement method for dipole magnets;
- Design the alignment method for dipole magnets;
- Develop necessary tools for installation and replacement of magnets;
- The systems must be stable with large time constants, avoiding creep and fatigue deformation;
- Reduce the cost through structural optimization and experiment.

The technical routes are:

- Design the structure of the supports;
- Optimize the structure and the position of the supports to reduce the deformation, stress and vibration using FEA;
- Design the installation and replacement methods of the supports. This job may require inventing specialized tools and fixtures;
- Design the alignment method based on the support and magnet structure;
- Develop a support prototype for one dipole magnet;
- Conduct experiments on installation, alignment and vibration;
- Summarize the results and validate the final design.

### 11.10.2 Development of the Booster Dipole Support System

The R&D goals and the technical routes for achieving them are the same as enumerated above for the Collider dipole with one addition. Since the Booster magnet is hung, special attention must be paid to the hanging location and support scheme. Topologic optimization will be used. The steel frame should be optimized for static and dynamic stability.

### 11.10.3 Development of a Tunnel Mockup

A mockup of the tunnel will be done after the creation of 3D models of individual components. It will be about 6,000 mm long (almost the length of one dipole magnet). The arc-section will be chosen for the mockup because it forms the majority of the tunnel.

For the mockup of the tunnel, the R&D goals are:

- Double-check the integration of 3D models into the tunnel and imitate the transportation, installation and alignment of the devices in the tunnel;
- Provide space to do tests such as magnet vibration;
- Provide a visual presentation of the tunnel;
- Design and develop the mockup of the arc-section and a support system to support all the prototypes needed to be installed in the mockup;
- Develop necessary tools for the mockup.

The technical routes are:

- Design and develop the mechanical structure of a 6-meter long tunnel prototype and the support system of the mockup;
- Design the transportation and installation methods of all the devices in the tunnel cross section;
- Develop the necessary tools and fixtures;
- Assemble all the devices in the tunnel (The dipole magnets and their supports are available; for other devices use wood or other materials).
- Test the transportation and installation methods;
- Test the alignment methods for the magnets;
- Summarize the results and validate the final design.

### 11.10.4 **Development of Movable Collimators**

Movable collimators will be designed during the project R&D phase; the R&D goals are:

- Design a movable collimator with low impedance and with simple and flexible mechanical structure;
- Build one prototype.

The technical routes are:

- Design the inner profile of the collimator based on the physical requirements. Optimize the profile to obtain low impedance;
- Calculate the thermal and mechanical stress using FEA. Design modification may be required based on the results;
- Design and develop a collimator prototype with a simple and flexible structure;
- Test the mechanical performance and measure the impedance. Summarize the experiments and validate the design.

## 11.11 Radiation Shielding

There are two aspects to radiation protection. Clearly it reduces or eliminates the risk of ironing radiation and protects staff and equipment. However, because of the enormous size and scope of the facility, it will be quite visible to the general public. They will read about the facility, its construction and its operations, its milestones and discoveries. And there will probably be controlled access to the site for visitors. Therefore, there is an important public relations aspect. For both these reasons the R&D program for radiation protection is important.

### 11.11.1 **Radiation Shielding Design Research**

MC simulation codes are widely used in the shielding design. To check the accuracy of the simulation results, we should design benchmark experiments and compare the results obtained with the simulation results to verify that the simulation codes are suitable for shielding design.

The highest electron beam energy for CEPC is 120 GeV (with a possible upgrade to 175 GeV), so the production of secondary particles should be researched, and then the radiation field around the tunnel assessed. Continuing research can be done on the synchrotron radiation and on the beam dump to make sure that the machine can be run successfully and safely for its entire scientific life.

### 11.11.2 **Dose Monitor Technology Research**

Radiation dose monitors are distributed throughout the facility to monitor all workplaces. The data acquisition system that keeps track of them will use various methods: IP/TCP, GPRS (General Packet Radio Services), wireless and GPS.

The dose rate induced by synchrotron radiation is very high and can damage sensitive components, so we should develop passive dose monitors and accumulated dose monitors.

In order to measure and assess the impact of CEPC to the environment, we also need to develop the methods for monitoring radiation and radioactivity outside the protected areas. Comparisons between simulation and measurements of samples is important. Samples can be measured by a low background HPGe (high-purity germanium) $\gamma$ spectrometer and liquid scintillation counter. Also we should establish an assessment method for air and water activation.

## 11.12 Survey and Alignment

### 11.12.1**Automatic Observation System**

Using the traditional survey method of a laser tracker, will require about 16,667 stations, 60 staff personnel and 100 working days to finish the ring survey. Therefore, it is necessary to research an automatic observation system to improve efficiency and decrease this heavy burden.

An automatic observation system will make use of high-precision photogrammetry. This system consists of a vision instrument, an automatic mobile platform, photogrammetric targets and a data processing center and is shown pictorially in Fig. 11.12.1.

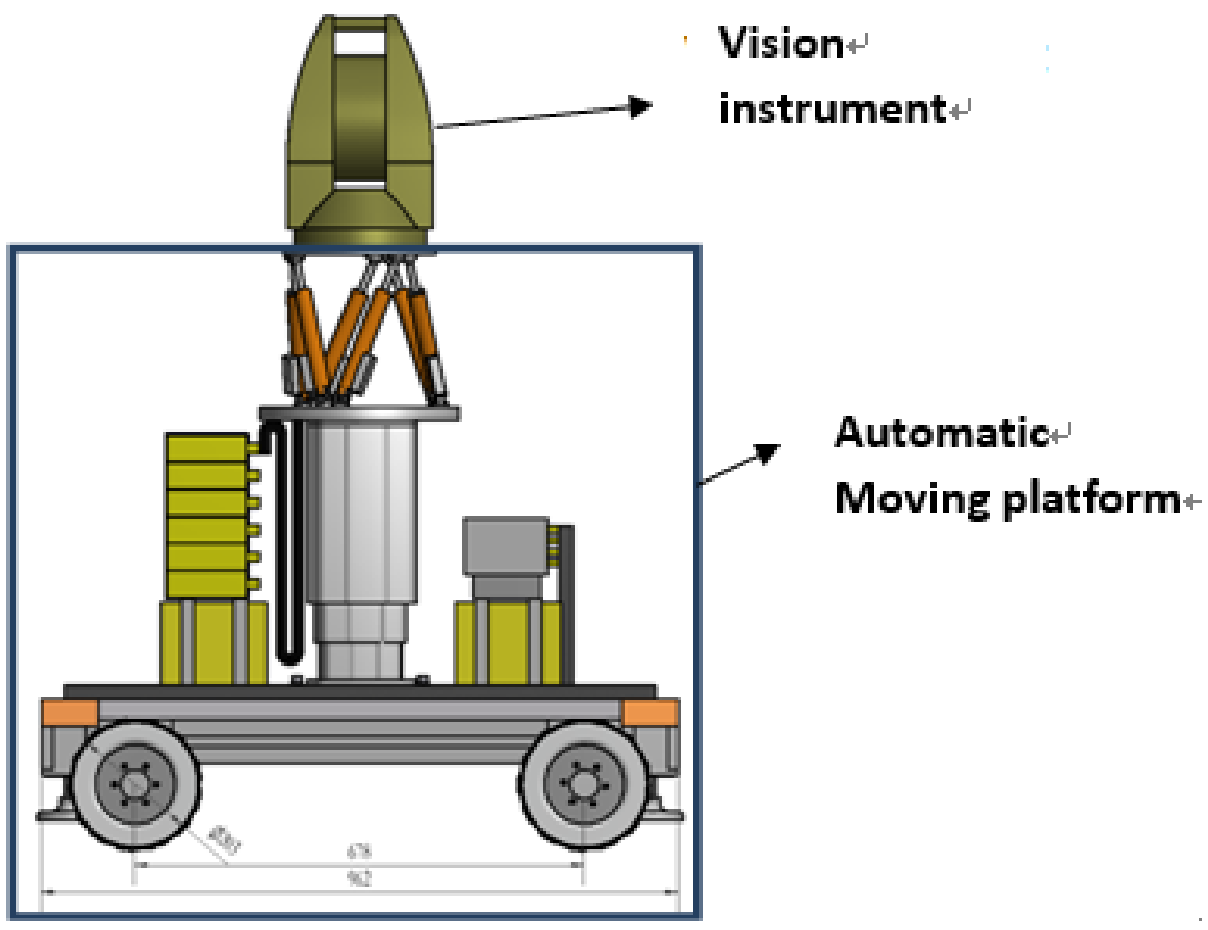

**Figure 11.12.1:** Automatic observation system

The vision instrument is installed on an automatic mobile platform. Before measurements begin a “walking” and observation program is input. The system will then travel along the indicated route and automatically do the survey work. Using wireless transmission, the observation data will be sent to the data processing center, processed there and final coordinates obtained. Below we outline some of the R&D that is required to develop this system and prepare it for use on the CEPC facility.

### 11.12.2 **Automatic Mobile Platform**

The main parts of the automatic mobile platform are the omnidirectional mobile module, the elevating module, the self-balancing module, the power supply module and control modules. It needs to have the following functions:

- omnidirectional smooth movement;
- automatic leveling;
- lift function, for observations at different heights;
- tracking, automatic obstacle avoidance.

#### 11.12.2.1 ***Omnidirectional Mobile Module***

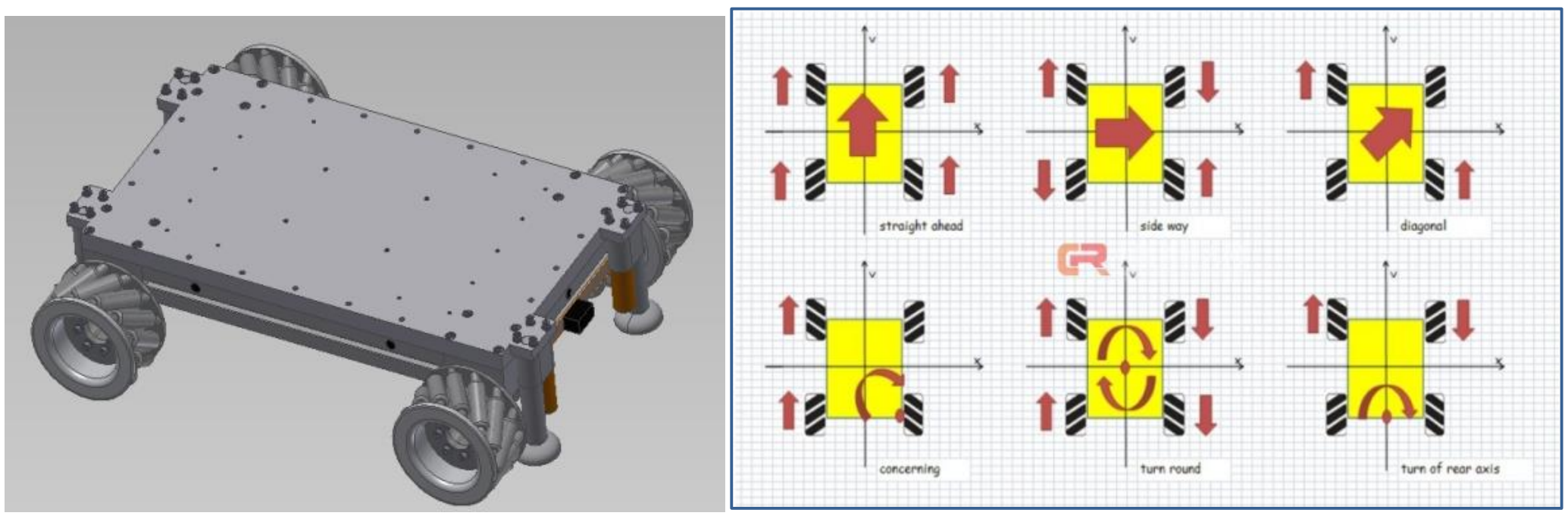

**Figure 11.12.1:** Mobile module

The omnidirectional mobile module shown in Fig. 11.12.1, mainly consists of a sports chassis, 4 McNam wheels and 4 supporting legs. The chassis size is 600 mm × 800 mm and can handle loads of over 150 kg. McNam wheels allow the platform to move in all directions. Speed is less than 5 m / s, and accuracy of movement is better than ± 10 mm. Supporting legs provide stability.

The platform movement is by program control. It can automatically travel along a guide stripe to accurately reach a specified position. After arriving there, the supporting legs will prop up the platform, lift the wheels off the ground and ensure platform stability. A total of 6 infrared obstacle avoidance sensors will be installed in the four directions of the moving module to detect and stop it before an obstacle at a distance of 20 cm.

#### 11.12.2.2 ***Lifting Module***

In order to automatically measure components at different heights, the lifting module will have a wide range of height adjustment. Considering the equipment height, convenience and stability, the module will use a third-order lift pillar with a 500 mm adjustment range. The module relies on a linear drive unit and has a load capacity of 200 kg. This design ensures stability and safety of the measuring instrument.

#### 11.12.2.3 ***Self-balancing Module***

The self-balancing module is placed on the lifting module and has pitch, yaw and rotation degrees of freedom. In order to improve rotation accuracy and still have adequate stiffness, 6-UPS (Universal joints, Prismatic pairs, Spherical joints) parallel mechanism and inclination sensors composed of closed-loop control systems will achieve high-precision automatic leveling. A Stewart-type six-degrees-of-freedom parallel mechanism will be used. The configuration is a six-branch 3-UPS (universal hinge - cylindrical deputy - ball joint), mechanism with accuracy ≤ 0.01 mm, level accuracy ≤ 0.001°. The range of angle adjustment is ≤ 10°. It is shown in Fig. 11.12.2.

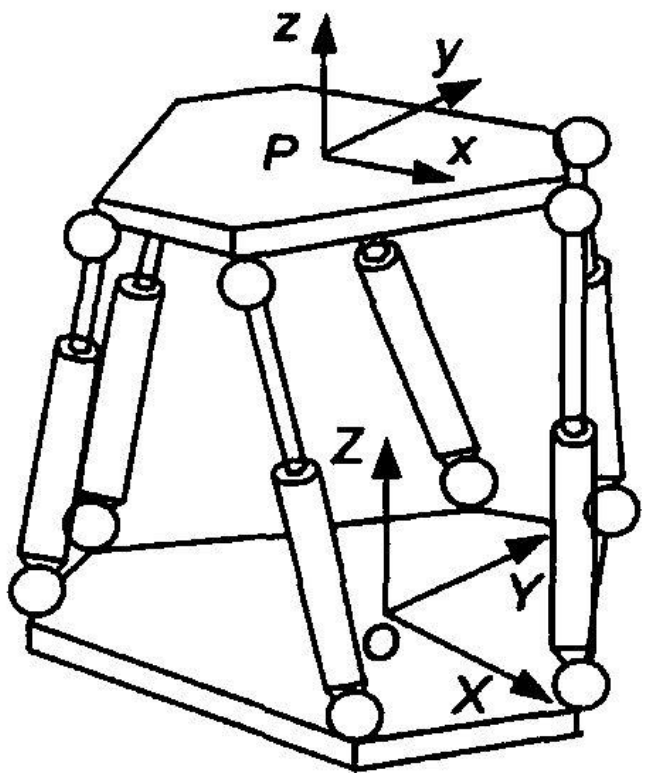

**Figure 11.12.2:** 6-UPS parallel mechanism

The displacement in the three directions is ±15 mm to achieve small stroke high-precision height adjustment. The whole structure uses electric drive and is powered by an automobile battery. It has the advantages of large driving force (second only to hydraulic power), high speed, high sensitivity, precise control, simple structure and easy maintenance. The disadvantage is complex control.

### 11.12.3 **Vision Instrument**

The vision instrument consists of frame, horizontal dial, vertical dial, lens, image sensor, distance sensor, level sensor, horizontal and vertical adjustment knobs, display screen and flash illumination. This instrument combines angle measurement technology, laser ranging technology, photogrammetry technology, and has non-contact, high precision and high efficiency characteristics.

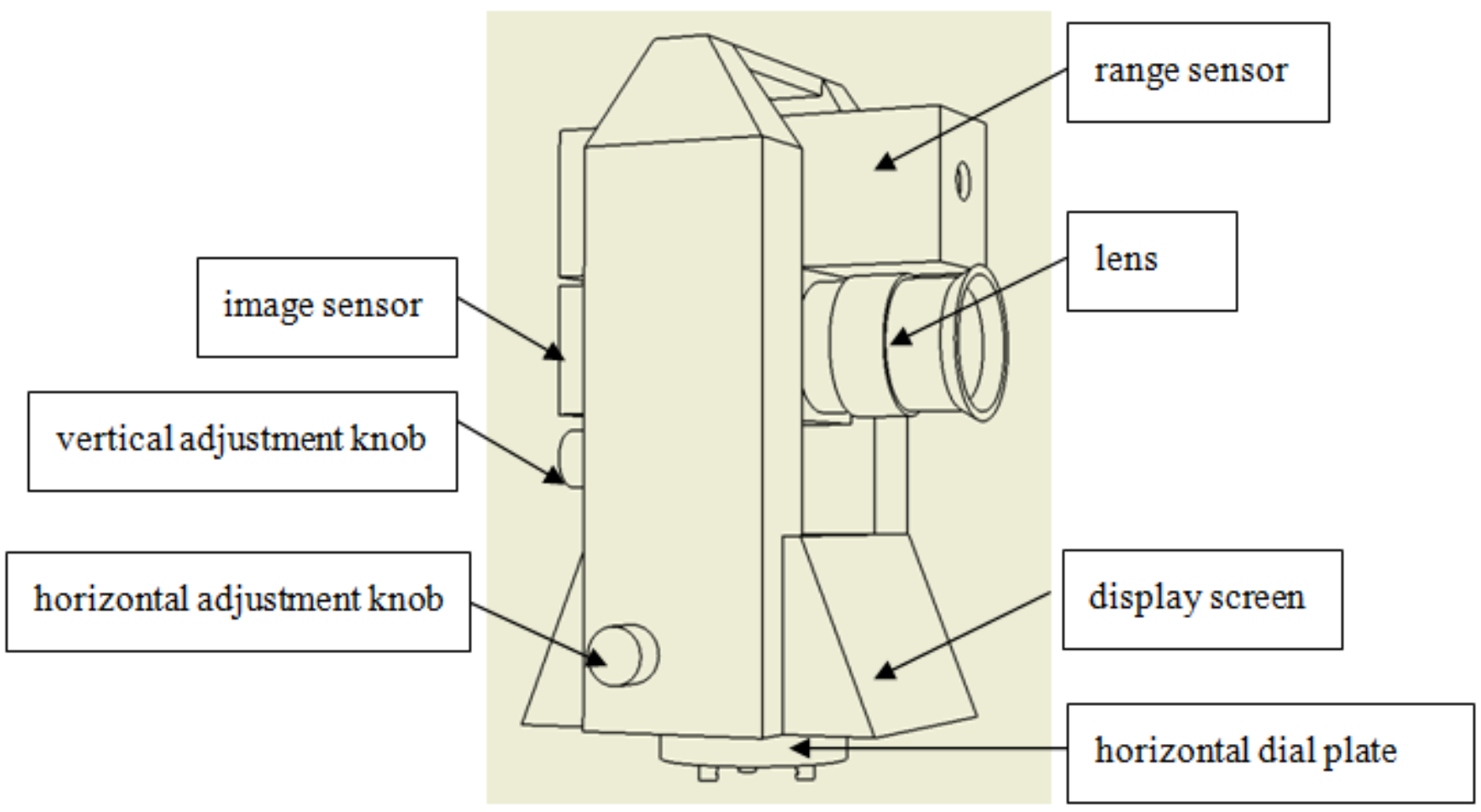

**Figure 11.12.3:** Vision instrument

The vision instrument, shown in Fig. 11.12.3 integrates a digital close-range photogrammetry system, using surface array measurement. The efficiency is very high, and the measurement target can be installed on the measured object permanently to meet the requirements of non-contact measurement. The digital close-range photogrammetry accuracy and external azimuth element totally relies on external target points. At the same time, using angle and range to obtain the external azimuth element solves the problem of limited capacity of the photographic measurement targets for the code matching. In addition, a horizontal sensor is installed on the vision instrument to provide a vertical reference, thus solving the problem of measuring the horizontal attitude and height difference of the object.

In order to meet the measurement requirements of the tunnel control network, the parameters in Tables 11.12.1 and 11.12.2 are necessary.

**Table 11.12.1:** Parameters of the image measurement

| No. | Items | Requirements |
|---|---|---|
| 1 | Focal length of lens | Focal length/60mm, Aperture/5 |
| 2 | Pixel resolution of image sensor | 7100 MP |
| 3 | Pixel size | 3.1 μm |

**Table 11.12.2:** Parameters of the angular measurement

| Items | Requirements |
|---|---|
| Axis wobble | elevation≤3″, azimuth≤5″ |
| Orthogonality between collimating axis and transverse axis | ≤3″ |
| Orthogonality between transverse axis and vertical axis | ≤7″ |
| Index error of vertical circle | ≤7″ |
| One measuring-process standard deviation in horizontal direction | ≤0.5″ |
| One measuring-process standard deviation in vertical direction | ≤0.5″ |
| Maximum rotary speed | 180°/s （30rpm） |
| Maximum accelerated speed | 360°/s2 |
| Load-bearing | ≥10 kg |
| Operating temperature | -20℃ to 50℃ |

### 11.12.4 **Laser Tracker**

The R&D on the laser tracker, shown in Fig. 11.12.4, has the following objectives: achieve 3 micron distance measurement precision, 2” angle measurement precision, and 2” leveling precision.

The critical areas to develop are:

1) Precise bearing system;
2) Develop error compensation and software;
3) Develop orientation adjustment methods and software.

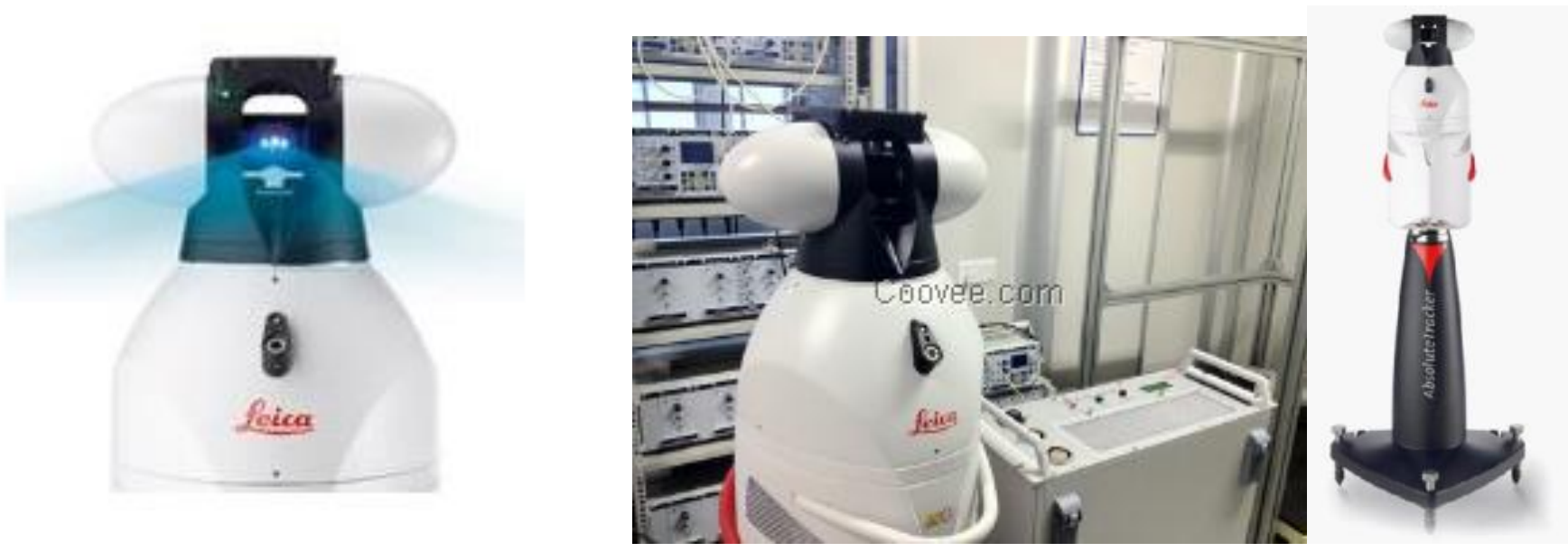

**Figure 11.12.4:** Laser tracker

### 11.12.5 **Accelerator Local Geoid Refinement**

The alignment challenge presented by the CEPC project requires us to look closely at the gravity field, because the earth can't be simply taken as an ellipsoid. We have two main datum plane schemes, but these surfaces are not accurate enough to take into account the anomalies from the vertical gravitational effects of mountains, lakes or geomagnetic variations. An equipotential gravity surface, called a geoid, to which the force of gravity is perpendicular everywhere must be defined. R&D work will be necessary to refine the geoid, shown in the sketch, Fig. 11.12.5, to 0.05 mm/100 m, or about 5 mm within the whole CEPC circle.

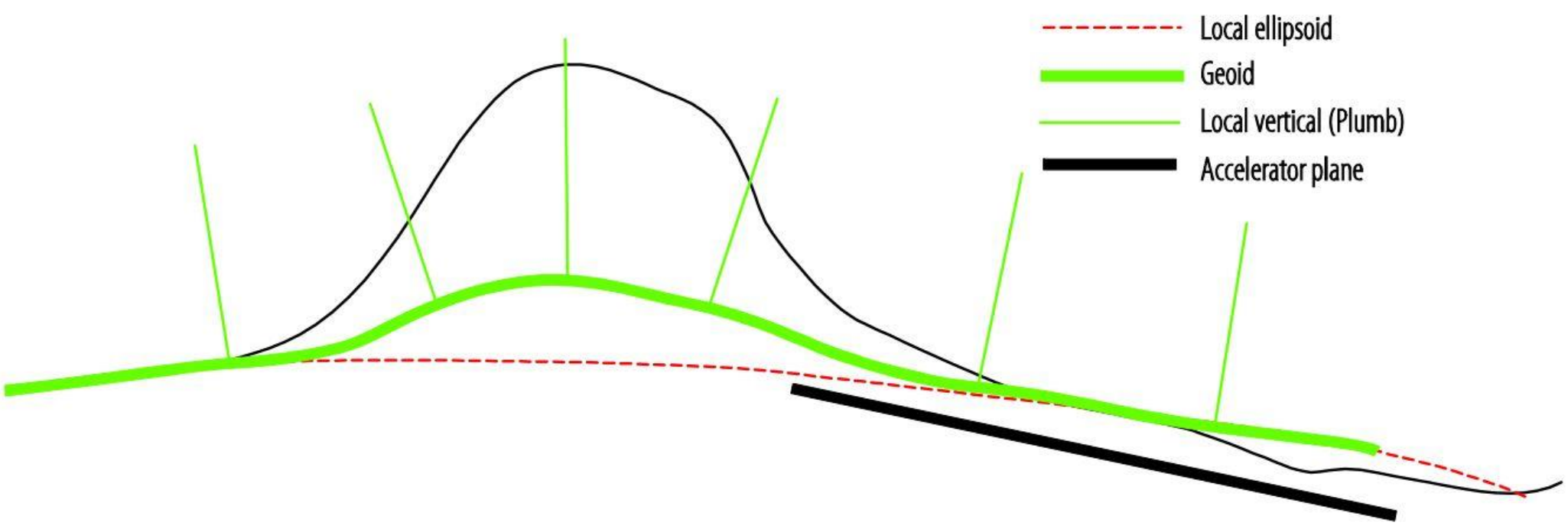

**Figure 11.12.5:** The geoid

## 11.13 $e^+$ and $e^-$ Sources

### 11.13.1 **Polarized Electron Gun**

For the future development of collisions between a polarized electron beam and an unpolarized positron beam, a photocathode dc gun type electron source using a specially prepared GaAs/GaAsP superlattice will be considered as an option. This polarized electron gun will enable the Linac to produce a high-intensity and low-emittance beam with high polarization. The proposed polarized electron beam routinely yields at least 85% polarization with a maximum QE (quantum efficiency) of ~1%. The high voltage between cathode and anode is 150 - 200 kV. This is a rather new technology and there is no experience to develop it domestically. So a R&D program on a polarized electron gun is necessary. Table 11.13.1 shows the design parameters of a polarized electron gun for the CEPC Linac.

**Table 11.13.1:** Design parameters of CEPC polarized electron source

| Design parameters of CEPC polarized electron source | |
|---|---|
| Gun type | Photocathode DC Gun |
| Cathode | Super-lattice GaAs/GaAsP photocathode |
| HV | 150-200kV |
| QE | 0.5% |
| Polarization | >80% |
| Electrons per bunch | $2\times10^{10}$ |
| Repetition rate | 50Hz |
| Drive laser | 790nm（±20nm），10μJ@1ns |

This challenging work involves advanced photocathode materials, a high intensity pulsed laser, high energy beam polarization measurements and an ultra-high vacuum. There are several key problems that will need to be solved:

1) Superlattice GaAs/GaAsP photocathode which is able to emit very high polarization (>80%) electron beam now is the most ideal photocathode. There is no experience on using this up to now, so international cooperation is certainly desirable and necessary;
2) Research on 790 nm wavelength high intensity pulsed laser. In order to obtain a high polarization, a drive laser with 790 nm wavelength has been selected. The single bunch population is $2\times10^{10}$, which requires the drive laser to have a 10 μJ pulse energy of 1 ns pulse length (peak power 10 kW);
3) Design and development of a Mott polarimeter and a Wien filter system for 200 keV electron beam;
4) The optimization design of 200 kV DC-Gun and its ultra-high vacuum system which is very important to maintain the QE and photocathode lifetime.

### 11.13.2 **High Intensity Positron Source**

The technology of conventional positron sources is mature and can satisfy the requirements for the CEPC Linac, but considering that the CEPC positron source requires a bunch charge of 3 nC, two orders of magnitude higher than BEPCII, R&D will be necessary and important to create a high intensity positron source. The R&D will be focused on the following aspects:

1) Use Geant4/FLUKA code to simulate the generation of a positron beam by a high energy electron beam incident on a converter target. Simulate the positron yield by optimizing target material, thickness and capture efficiency;
2) Use ANSYS to complete the thermal analysis of the converter target and determine its structure and cooling system;
3) Complete the design of a flux concentrator system which has a 6 Tesla peak magnetic field by using Opera code. Develop a 15 kV/15 kA/5 μs pulsed power supply and set up a flux concentrator test platform, and then finally achieve 6 Tesla peak magnetic field output through full HV conditioning. The machining of the flux concentrator is complicated and is part of the R&D program;

4) The parameters of the CEPC positron source is similar to the positron source of Super-KEKB in Japan, so an international collaboration between IHEP and KEK on high intensity positron source design and beam tests will be carried out.

## 11.14 Linac RF System

An S-band accelerating structure is selected for the Linac RF system. S-band, SLAC type, is a mature technology. However, the RF power feed is through a single coupling-hole which results in a field asymmetry. The time dependent multipole fields in the coupler induce a transverse kick along the bunch and cause an increase of beam emittance. An S-band accelerator structure adopting a dual-feed racetrack design instead of the single-feed couplers will be developed to minimize the multipole field effects and improve beam quality.

The Superfish program will be used to optimize the cavity shape. Rounding cavity shape instead of disc-loaded one can increase the Q value of the cavity more than 10%. The shape of the iris optimization shows that elliptical cross section can minimize the maximum value of the electric field. 3D program such as HFSS or CST will be used to design the dual-feed racetrack coupler.

A 3-meters long structure will be constructed after simulation. High power test will be carried out after cold measurement.

## 11.15 Superconducting Magnets for CEPC

In the R&D phase, superconducting prototype magnets for the CEPC interaction region will be developed in three consecutive steps:

1) Double aperture superconducting quadrupole prototype magnet QD0.
2) Short combined function superconducting prototype magnet including QD0 and the anti-solenoid.
3) Long combined function superconducting prototype magnet including QD0, QF1 and anti-solenoid.

The specification of the combined function superconducting prototype magnet is listed in Table 11.15.1.

**Table 11.15.1:** Specification of the combined function superconducting prototype magnet

<table>
<tr><th>Magnet</th><th>Central field (T/m or T)</th><th>Magnetic length (m)</th><th>Width of GFR (mm)</th><th>Field harmonics</th><th></th></tr>
<tr><td>QD0</td><td>136</td><td>2.0</td><td>19.51</td><td rowspan="2">Each multipole field content in each aperture ≤5.0×10<sup>-4</sup></td><td rowspan="2">Double aperture magnet; L* is 2.2 m; The angle between two aperture centerlines is 33 mard.</td></tr>
<tr><td>QF1</td><td>110</td><td>1.48</td><td>27.0</td></tr>
<tr><td>Anti-solenoid</td><td>7.2</td><td>4.8</td><td>---</td><td>---</td><td>Total field of the anti-solenoid and detector solenoid is nearly zero.</td></tr>
</table>

The key technical issues to be studied and solved in the R&D are listed below:

1) Magnetic and mechanical design of the superconducting quadrupole magnet and anti-solenoids with high field strength and limited space;
2) Fabrication technology of small size Rutherford cable with keystone angle;
3) Fabrication procedure of the twin aperture quadrupole coil with small diameter;
4) Fabrication procedure of the anti-solenoids with many sections and different diameters;
5) Assembly of the combined function coils including QD0, QF1 and anti-solenoids;
6) Development of the long cryostat for the combined function superconducting magnet;
7) Development of magnetic field measurement system for a small aperture long superconducting magnet;
8) Development of quench protection system for a combined function superconducting magnet;
9) Cryogenic test and field measurement of the small aperture long superconducting magnet.

## 11.16 Superconducting Magnets for SPPC

R&D on high field accelerator magnets is ongoing at IHEP. The development of these magnets is the key for the future Super Proton-Proton Collider (SPPC). A 12-T twin-aperture subscale magnet is under development as the first step. Four NbTi coils and two $Nb_3Sn$ coils will be fabricated to provide 12-T dipole fields in the two apertures with load line ratio of around 20% at 4.2 K. After that, HTS (IBS, ReBCO, and Bi-2212) coils will be inserted between the $Nb_3Sn$ coils to increase the field. ReBCO coils with block-type configuration will be inserted between the $Nb_3Sn$ coils with a common-coil configuration (Combined Common-coil and Block type configuration, CCB), to make the wide dimension of the tape conductor parallel with the magnetic flux and maximize its current carrying capacity. The first hybrid (NbTi + $Nb_3Sn$) magnet was tested in January 2018.

### 11.16.1 **Subscale Magnet R&D with $Nb_3Sn$ Technology**

#### 11.16.1.1 ***Magnetic Design***

The magnet will be first fabricated with low temperature superconductors, $Nb_3Sn$ and NbTi. The cross section is shown in Fig. 11.16.1 top. The red section in the figure is the iron yoke of the magnet, with outer diameter of 500 mm. The blue sections are the superconducting coils. There are two apertures in this dipole magnet with diameter 10 mm. The interaperture spacing is temporarily set to be 180 mm. For each layer of the coil, the minimum bending radius is 60 mm. The main parameters are in Table 11.16.1.

Fig. 11.16.1 bottom shows the coil cross section in the first quadrant. There are two layers of $Nb_3Sn$ coils and four layers of NbTi coils. The total width is 74.5 mm, and the height is 57.6 mm. For each $Nb_3Sn$ layer, there are 28 turns. For each NbTi layer in the middle two layers, there are 32 turns. And for the NbTi coils in the outer two layers, there are 33 turns for each double-pancake layer. Main parameters of the cables and strands are in Table 11.16.2. The $Nb_3Sn$ coils are wound with 20-strand IHEPW5 cable. The middle

NbTi coils are wound with 38-strand IHEPWN1 cable. The outer NbTi coils are wound with 24-strand IHEPWN2 cable. The required lengths for such a dipole is 4.5 km, of $Nb_3Sn$ strand and 16.1 km of NbTi strand.

The peak field in the $Nb_3Sn$ coils is 12.11 T, with an operating margin of about 20% at 4.2 K, corresponding to an operating current of 5910 A. The peak field in the NbTi coils is 6.65 T, with an operating margin of around 21% at 4.2 K. Fig. 11.16...2 shows the field distribution in the magnet.

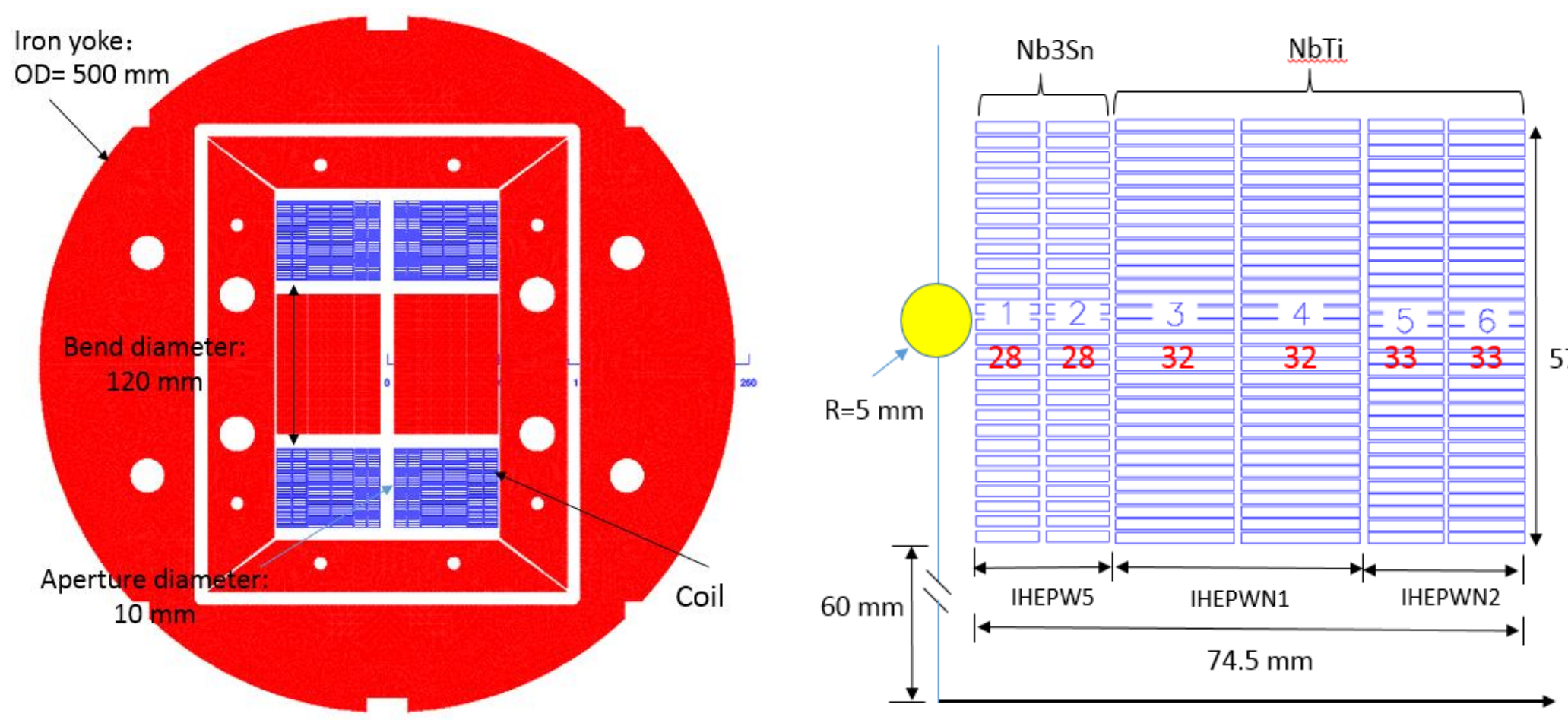

**Figure 11.16.1:** Left – cross section of the 12-T subscale magnet; Right – coil cross section in the first quadrant.

**Table 11.16.1:** Main parameters of the 12-T subscale magnet

| Parameter | Unit | Value |
|---|---|---|
| Number of apertures | - | 2 |
| Aperture diameter | mm | 10 |
| Inter-aperture spacing | mm | 180 |
| Operating current | A | 5910 |
| Operating temperature | K | 4.2 |
| Coil peak field | T | 12.11 |
| Margin along the load line | % | 20 |
| Stored energy | MJ/m | 0.48 |
| Inductance | mH/m | 33.8 |
| Number of $Nb_3Sn$ coils | - | 2 |
| Length of Nb3Sn coils | mm | 435.2 |
| Number of NbTi coils | - | 4 |
| Length of the middle NbTi coils | mm | 435.2 |
| Length of the outer NbTi coils | mm | 535.2 |
| Outer diameter of the iron | mm | 500 |
| Lorentz force Fx/Fy per aperture | MN/m | 4.26/0.49 |
| Peak stress in coils | MPa | 79 |
| Minimum bending radius | mm | 60 |

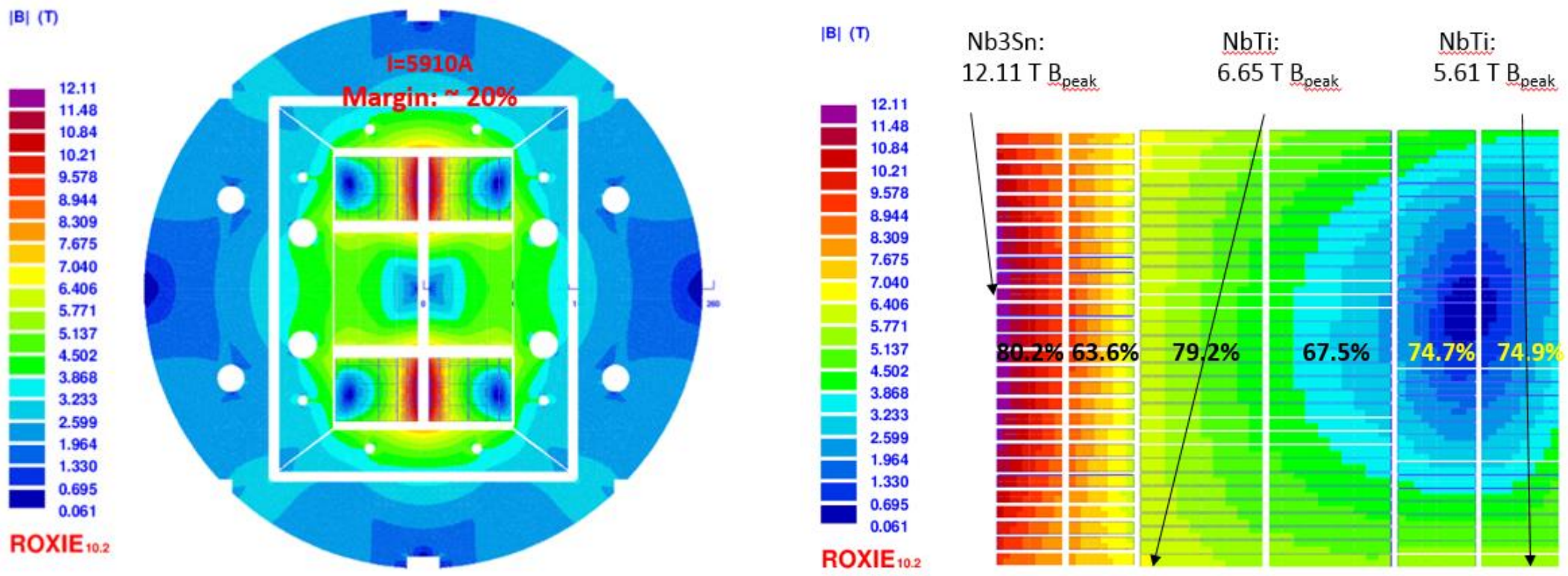

**Figure 11.16.2:** Left – Field distribution of the magnet; Right – Field distribution in coil in the 1st quadrant.

**Table 1.16.2:** Parameters of cables and strands

| Parameter | Unit | Inner | Middle | Outer |
|---|---|---|---|---|
| Superconductor types | - | Nb3Sn | NbTi | NbTi |
| Number of layers | - | 2 | 2 | 2 |
| Number of turns per layer | - | 28 | 32 | 33 |
| Cable width | mm | 8.5 | 16 | 10.2 |
| Cable inner height | mm | 1.45 | 1.5 | 1.5 |
| Cable outer height | mm | 1.45 | 1.5 | 1.5 |
| Number of strands | - | 20 | 38 | 24 |
| Insulation thickness | mm | 0.3/0.2 | 0.15 | 0.122 |
| Strands diameter | mm | 0.802 | 0.82 | 0.82 |
| Copper to superconductor ratio | - | 1 | 1 | 1 |
| Reference field | T | 12 | 5 | 5 |
| Jc @ reference field, 4.2K | A/mm2 | 2700 | 2613 | 2613 |
| dJc/dB | A/mm2 | 400 | 550 | 550 |
| Required strand length per 1m coil | Km | 4.48 | 9.73 | 6.34 |

The coil ends have been optimized using the ROXIE and OPERA codes. The results from the two codes are consistent, as shown in Fig. 11.16.3. All the coils are bent in the soft way and the bending radius is quite large, so this is easy on the $Nb_3Sn$ strand which is a stress-sensitive material. [9] To reduce the field enhancement in the ends, the iron yoke only covers the straight section of the coils in this design. [10] The half length of the iron yoke in the axial direction is 100 mm. If we set the straight length of all the coils to be the same, the field strength in the NbTi coil ends tends to increase compared with the 2D simulation results. The load line ratio rises from 75% to much higher than 80% for the outer two layers of NbTi coils. To solve this problem, we use different lengths of the straight section for different coils. If we set the length of the outer two layers of NbTi coils about 100 mm longer than the inner coils, then we could get the minimum value of all the load line ratios. After optimization, the length of the straight section is set to 200 mm for $Nb_3Sn$ coils and middle NbTi coils, while for the outer NbTi coils the length of the straight section is set to 300 mm.

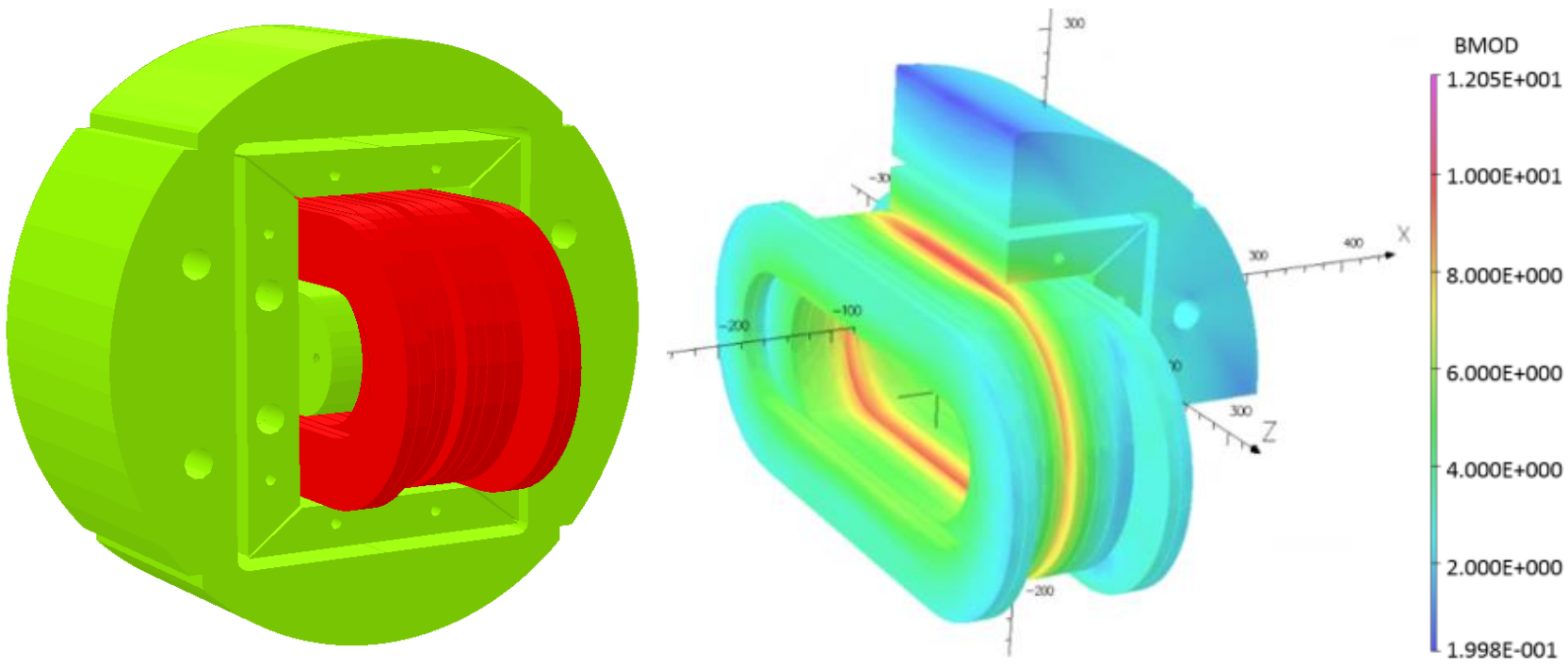

**Figure 11.16.3:** Left – 3D simulation model of the magnet with Opera; Right – field distribution in the magnet.

### 11.16.1.2 ***Mechanical Design***

The shell-based structure is adopted in our hybrid dipole magnet design with yoke diameter of 500 mm and the shell thickness of 60 mm. As shown in Fig. 11.16.4, to ensure that the horizontal preload can be efficiently applied to the coils, we have reduced the contact area between the H-pad and the coil pack (horseshoe, coil and island) by replacing the major part of the island with the no-touch iron center. The iron center can also help enhance the central magnetic field in the coil. It can be assembled into the island with shrink-fit technology. Iron is used for the yoke, H-pad, V-pad and iron center; aluminum alloy 6061 is used for the shell and the tie rod; stainless steel is used for the H-key and V-key and titanium alloy is used for the island and the horse shoe.

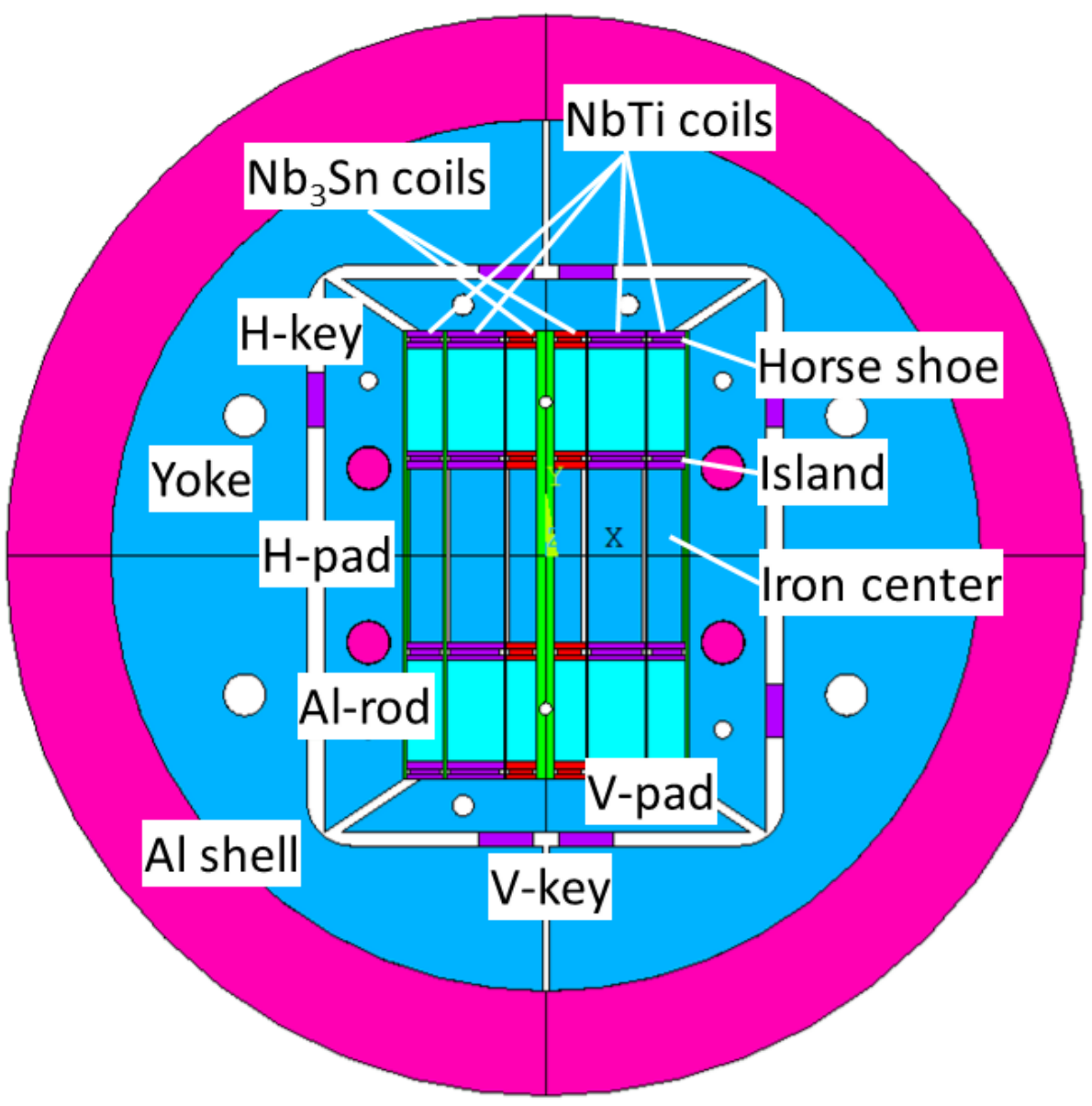

**Figure 11.16.4:** Cross section of a 12-T common-coil dipole magnet

ANSYS APDL is used to optimize the mechanical design of FECD1. The analysis steps are as follows:

a) Electromagnetic simulation. A 1/8 3-D FEA model containing the air volume is built to calculate the Lorentz force distribution in the magnet;

b) Interference analysis. This is used to simulate the horizontal preload applied by the bladder & key. During the analysis, all the air elements are deleted. Different types of contact pairs are built between adjacent components. The symmetric condition is applied to the three planes of $x$=0, $y$=0 and $z$=0;
c) Cool down. The temperature load of 4.2 K is applied to the whole magnet after setting both the uniform temperature and reference temperature as 300 K;
d) Excitation. The calculated Lorentz force at the first step is applied to the mechanical model.

The preload in three directions after cool down is when the interference between the H-key and the yoke is set to be 0.5 mm (corresponding to the water pressure of 67.7 MPa). The maximum shell stress after excitation is 225 MPa at the top inner of the shell. The stress level of the yoke after excitation is below 200 MPa except for some stress concentration areas. The tensile stress in the aluminum alloy tie rod is 141.5 MPa after excitation. The coil stress, the coil displacement and the factors that influence the cold shrinking force are discussed in detail below.

*11.16.1.2.1 Coil Stress and Coil Displacement*

Fig. 11.16.5 shows the magnetic force distribution in the coils at the xy plane (cross section at the straight part) and the xz plane (cross section at the end part). The total calculated magnetic force for the 1/8 3-D magnet model in the horizontal, vertical and axial direction is 760,217 N, 79,972 N and 41,769 N respectively. As described previously, the applied preload to the coils after cool down is sufficient to overcome the magnetic forces in different directions.

The stress contour in coils at different load steps is shown in Fig. 11.16.6. The maximum coil stress appears at the $Nb_3Sn$ inner coil is less than 114 MPa before and after excitation. There is no obvious stress change in the coils when comparing the results gained after cooldown and after excitation. The coil displacement is another important factor that guides the design. Large coil displacement during excitation may result in a premature quench. In addition, to ensure magnetic field uniformity, the required positional accuracy for superconducting coils must be maintained in the 20 μm range after excitation [15]. The coil displacement based on the present design is shown in Fig. 11.16.7. After assembly at room temperature, the coils are compressed in the horizontal and vertical directions, with a displacement of tens of micrometers. After cooldown, the coils shrink significantly in the three main directions with a magnitude of hundreds of micrometers. After excitation, the coil displacement keeps to almost the same value compared with that after cooldown. The change of the coil displacement in three main directions is less than 20 μm. The results of the coil stress and coil displacement indicate that the structure design of FECD1 is reasonable.

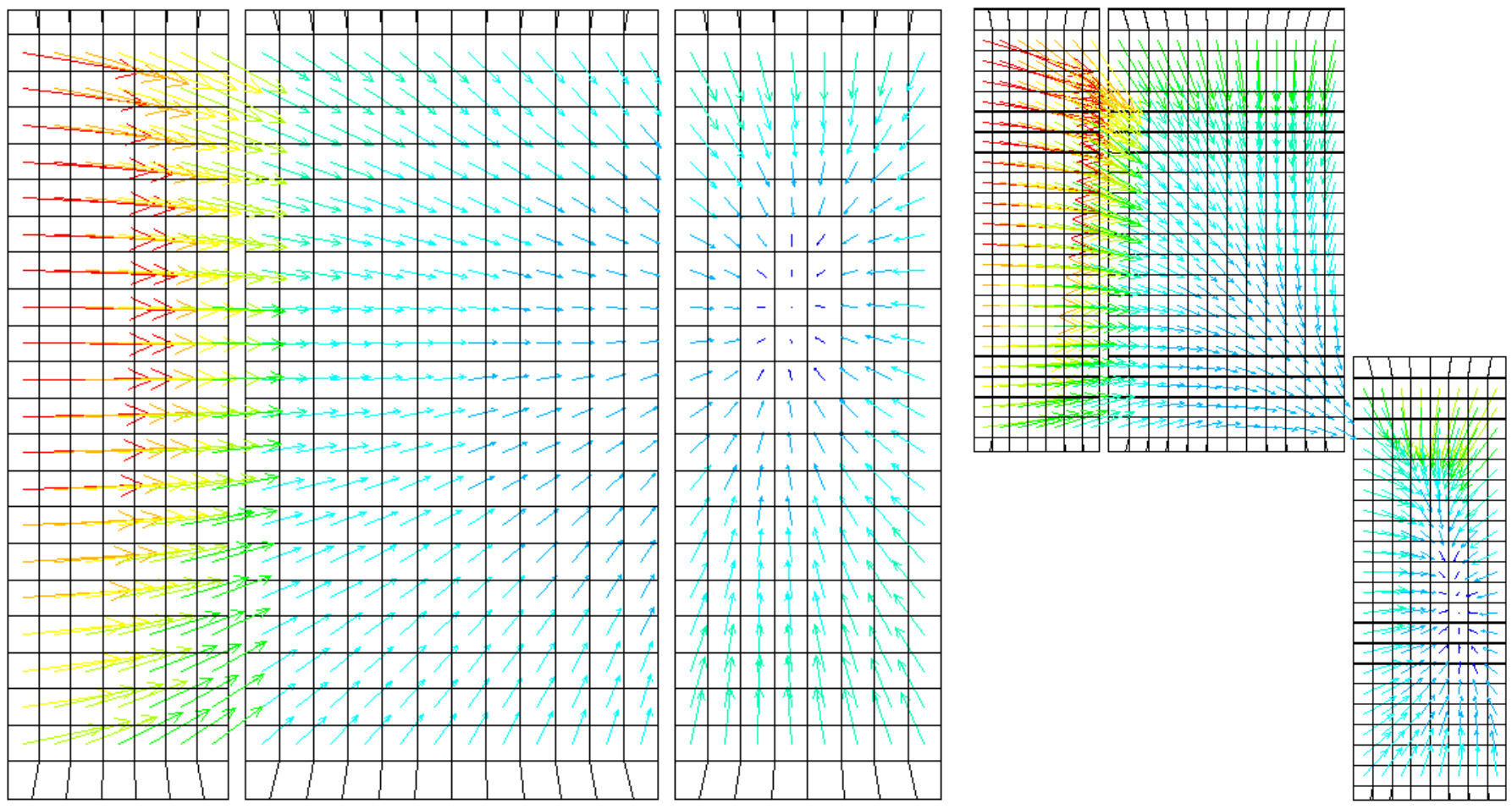

**Figure 11.16.5:** Magnetic force distribution in the coils at the xy plane (left) and xz plane (right).

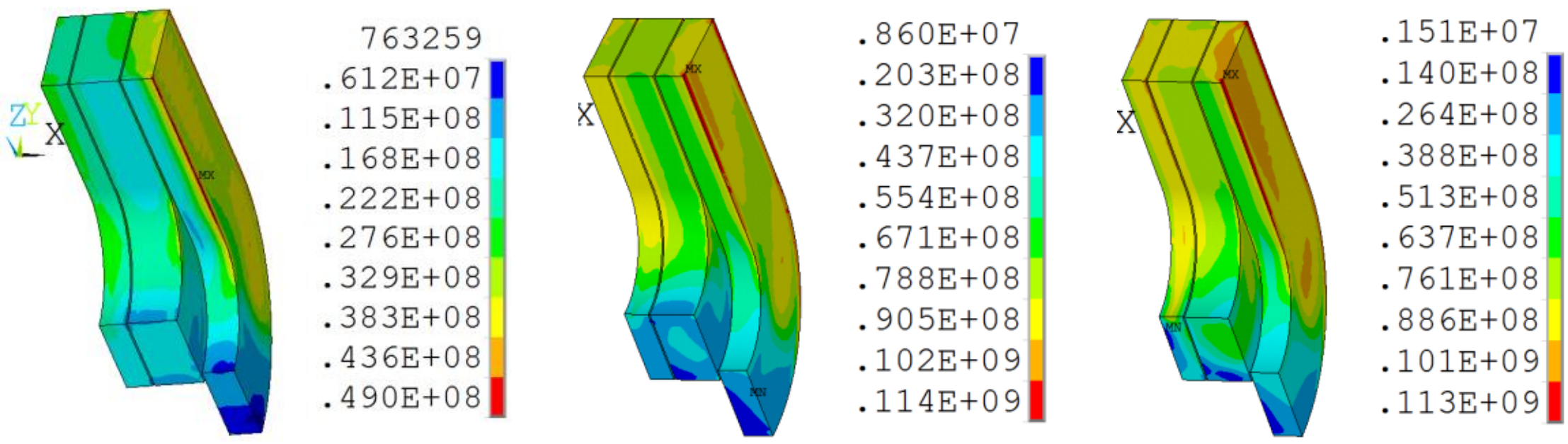

**Figure 11.16.6:** Von Mises stress in the coils at different load steps. From left to right: coil stress after the assembly, cool down and excitation.

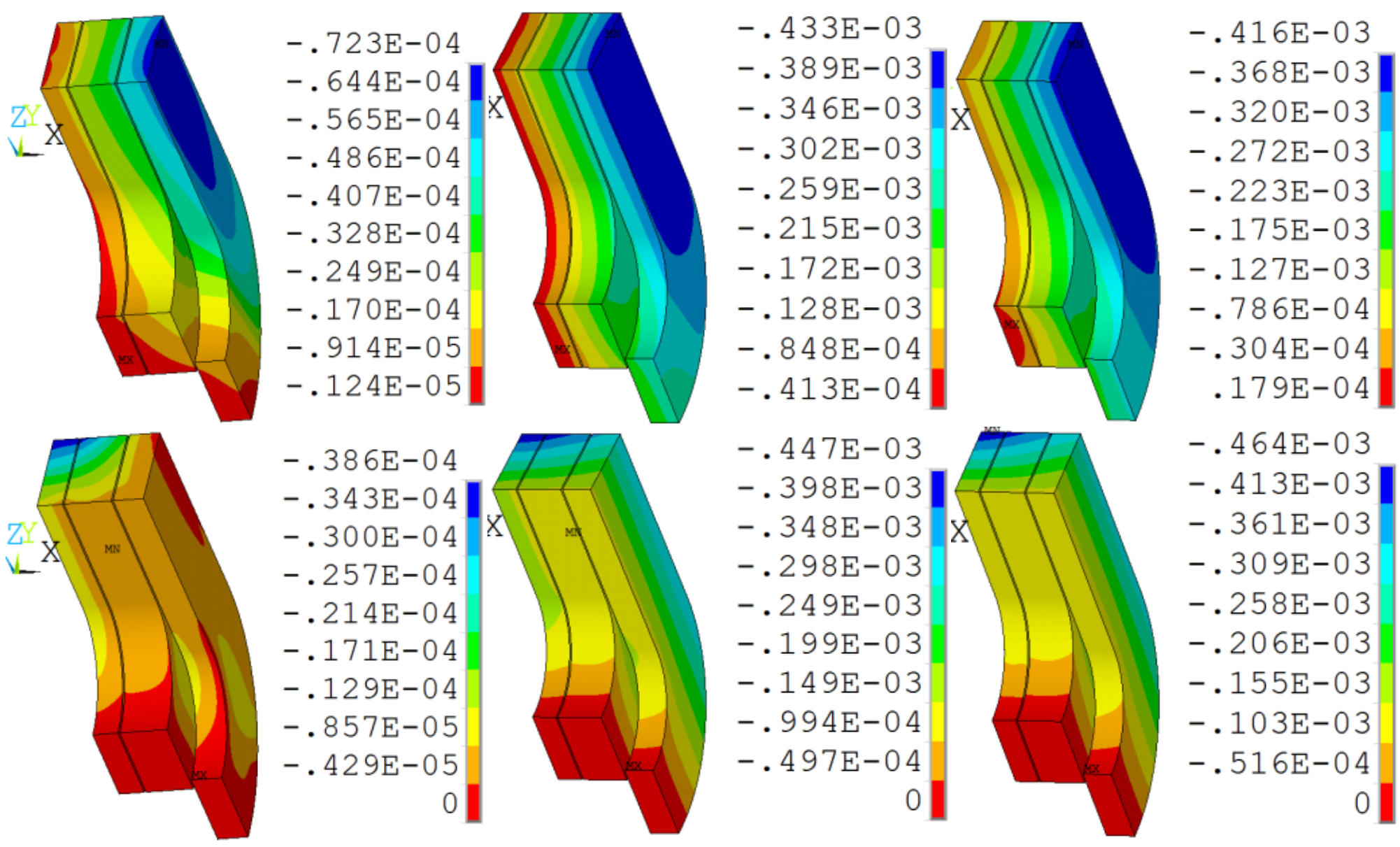

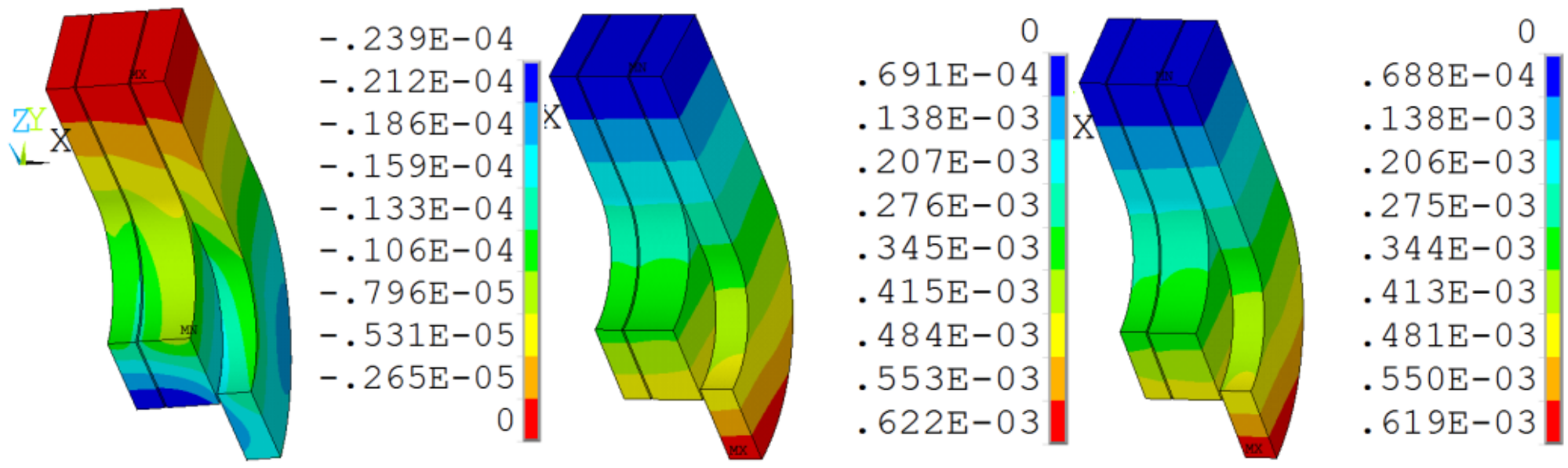

**Figure 11.16.7:** Coil displacement at different load steps. From left to right: coil displacement after the assembly, cool down and excitation; from the top to bottom: coil displacement in the horizontal, vertical and axial direction.

*11.16.1.2.2 Cold Shrinking Force*

The cold shrinking force applied to the coils after cooldown can be calculated from the mechanical structure. This helps us determine the required bladder pressure during the magnet assembly at room temperature. The present design assumes that the friction coefficient is 0.2 and the aluminum shell thickness is 60 mm. The total horizontal cold shrinking force is 1,038,100 N while the force shared by the coils is 400,745 N.

The influence of the key thickness and the friction coefficient on the cold shrinking force is shown in Fig. 11.16.8. The cold shrinking force rises from 977,400 N to 1,101,900 N when the Hkey thickness increases from 0.4 mm to 0.5 mm and the friction coefficient is set to be 0.2. The cold shrinking force decreases from 1,393,100 N to 1,038,100 N when the friction coefficient increases from 0 to 0.2 and the Hkey thickness is set to be 0.5 mm. To study the friction coefficient influence on the stress distribution in the whole magnet, we have collected the maximum hoop stress in the Al shell and the maximum Von Mises coil stress at different load steps. It can be concluded that the change of the friction coefficient value hardly influences the maximum hoop stress in the shell. However, the change of friction coefficient will influence the peak coil stress value after cool-down or excitation. The peak coil stress after excitation when the friction coefficient is set to be 0.2 is 17% larger than that after excitation when friction coefficient is set to be 0.

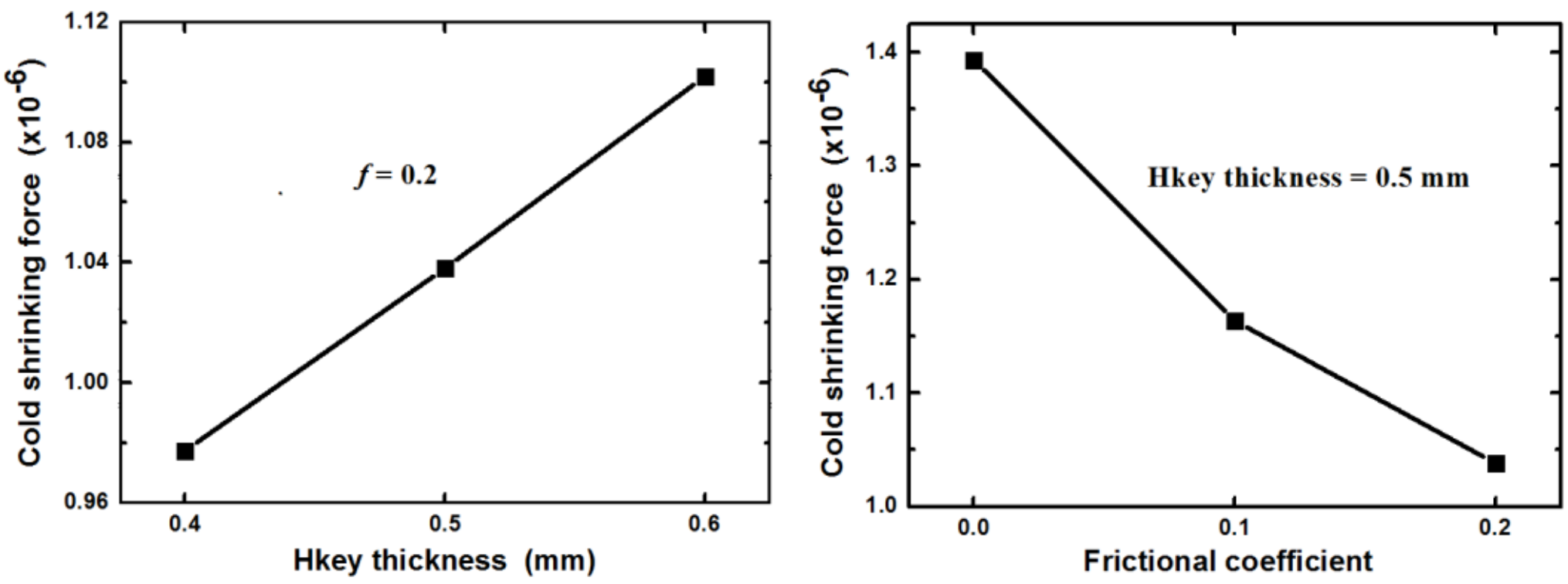

**Figure 11.16.8:** Left – relation between the cold shrinking force and shell thickness; Right – relation between the cold shrinking force and friction coefficient.

### 11.16.1.3 ***Fabrication of the Subscale Magnet***

#### *11.16.1.3.1 Cabling*

Toly Electric Co Ltd (Wuxi, China) has achieved great progress in making superconducting Rutherford cables for accelerator magnets since last year by collaborating with IHEP. By using the cabling machine shown in Fig. 11.16.12, they fabricated three different kinds of Rutherford cables for the 12-T NbTi/$Nb_3Sn$ subscale magnet. Fig. 11.16.13, from left to right, shows the 24 strands and Kapton insulated NbTi cable, the 38 strands and Kapton insulated NbTi cable and the 20 strands and S-glass (stiff glass) insulated $Nb_3Sn$ cable, respectively. The good news is that we find limited $I_c$ degradation from the extracted short samples from the cables.

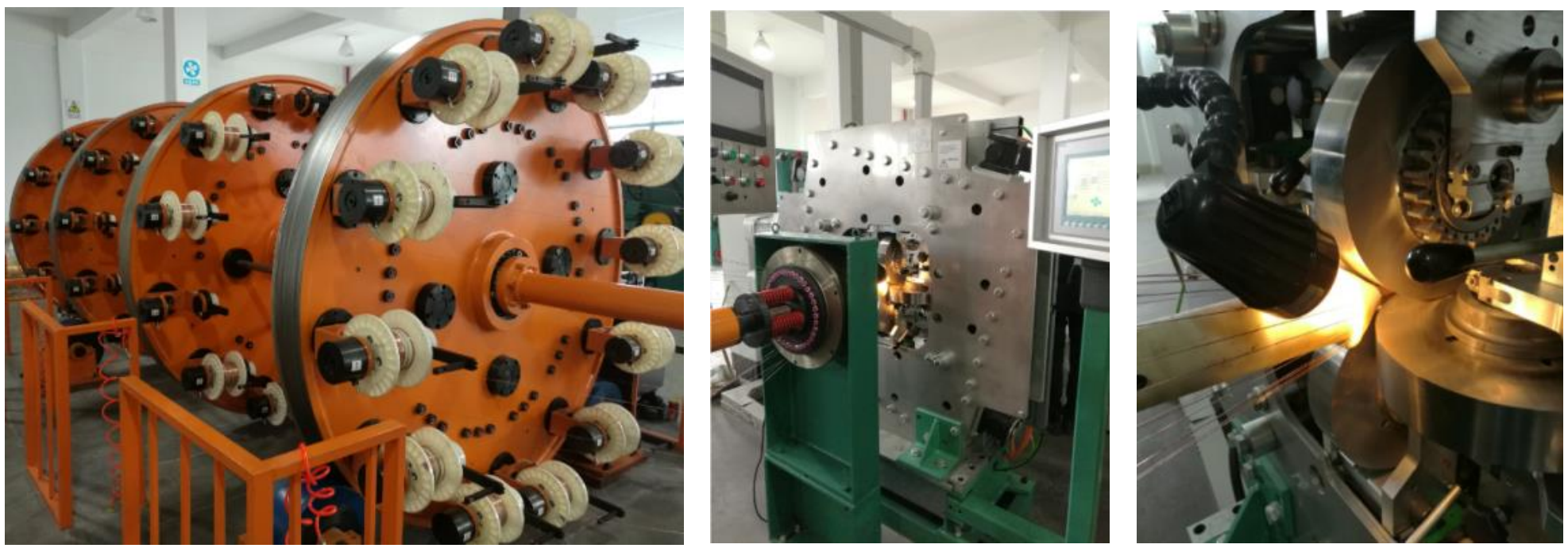

**Figure 11.16.12:** Superconducting Rutherford Cable winding machine

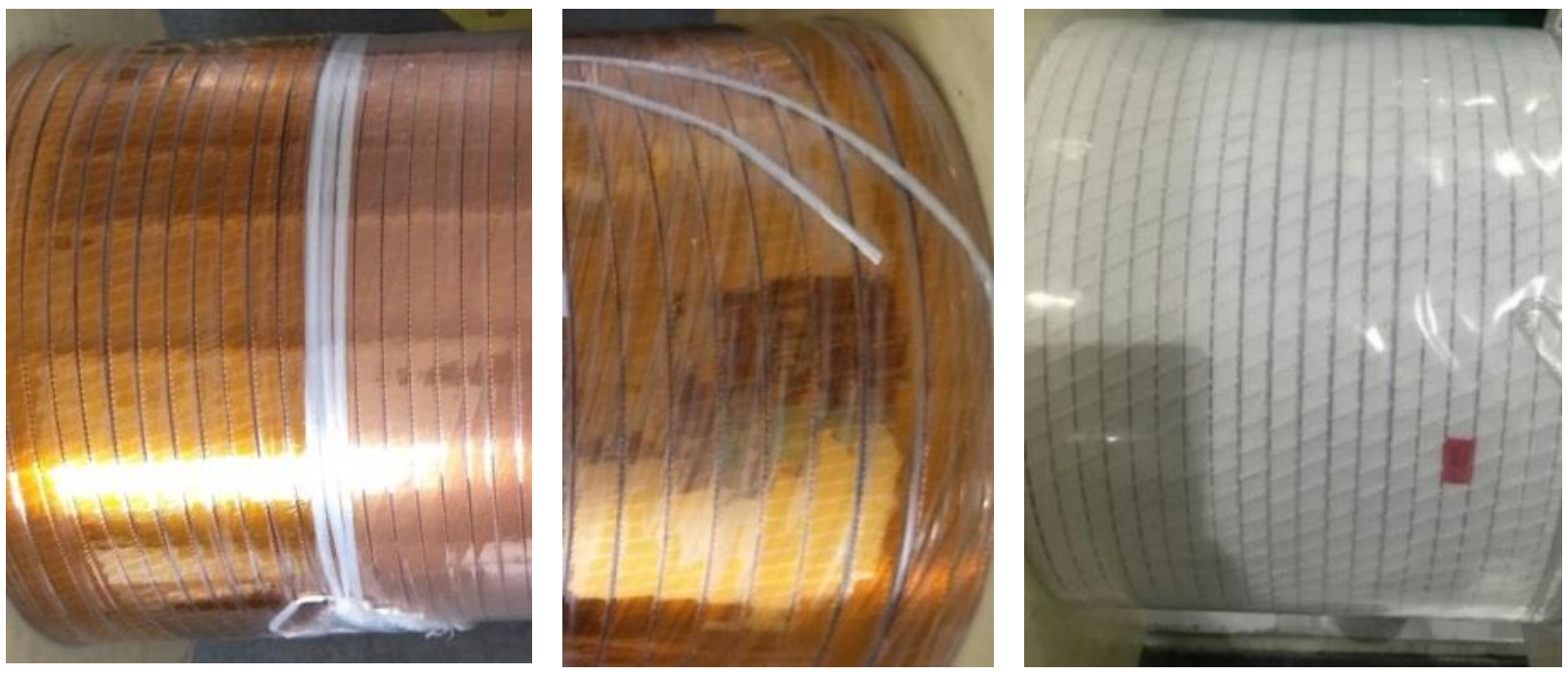

**Figure 11.16.13:** Rutherford Cables. Left – NbTi cables with 24 strands; Middle – NbTi cables with 38 strands; Right – $Nb_3Sn$ cable with 20 strands.

#### *11.16.1.3.2 Coil Winding*

Six racetrack coils were wound by using the multifunctional winding machine shown in Fig 11.16.14 (left). Fig 11.16.14 (right) shows the coil configuration of 4 NbTi coils after winding. Specifically, two outer NbTi coils were both wound with 2 layers and 33 turns with a straight length of 300 mm; two middle NbTi coils were both wound with 2 layers and 32 turns with a straight length of 200 mm; two inner $Nb_3Sn$ coils were both wound with 2 layers and 28 turns with a straight length of 200 mm. During the coil winding, glass fabric was added between the coil and the adjacent metal parts and between the two layers of the coil for insulation. Moreover, instead of using stainless steel horse shoe and end shoe for the NbTi coils, we used aluminum bronze counterparts for $Nb_3Sn$ coils. This was done to take into account consistency of the materials' CTE (coefficient of thermal expansion) during heat treatment.

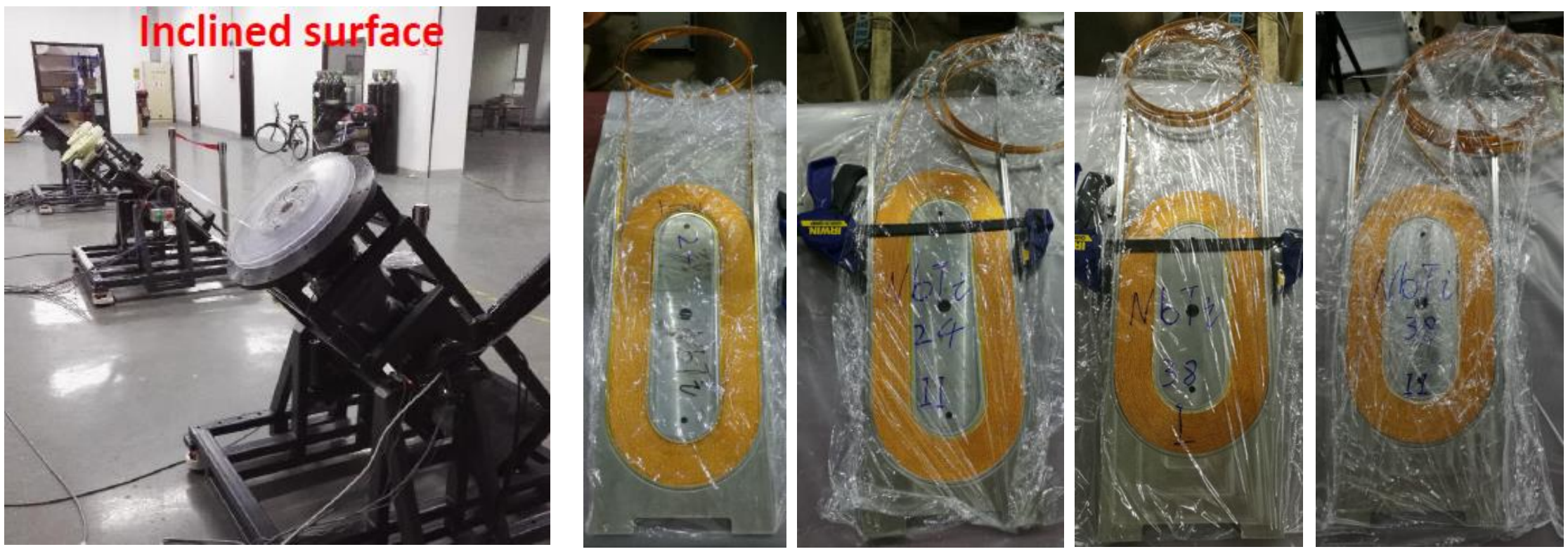

**Figure 11.16.14:** Left – Coil winding machine; Right – 4 NbTi coils (left two are outer coils and right two are middle coils)

*11.16.1.3.3 Heat Treatment*

After winding, the two $Nb_3Sn$ inner coils were constrained by pre-oxidized stainless steel plates and side bars. Importantly, a glass fabric was placed between the plate and the coil to avoid any contamination during reaction. The two coils were then placed into a vacuum furnace with high temperature control accuracy and heat treatment at 210 ºC for 48 hours, 400 ºC for 48 hours and 665 ºC for 50 hours. The coils were kept in the furnace during their cool down to room temperature. Fig 11.16.15 shows the $Nb_3Sn$ inner coil after winding and after heat treatment.

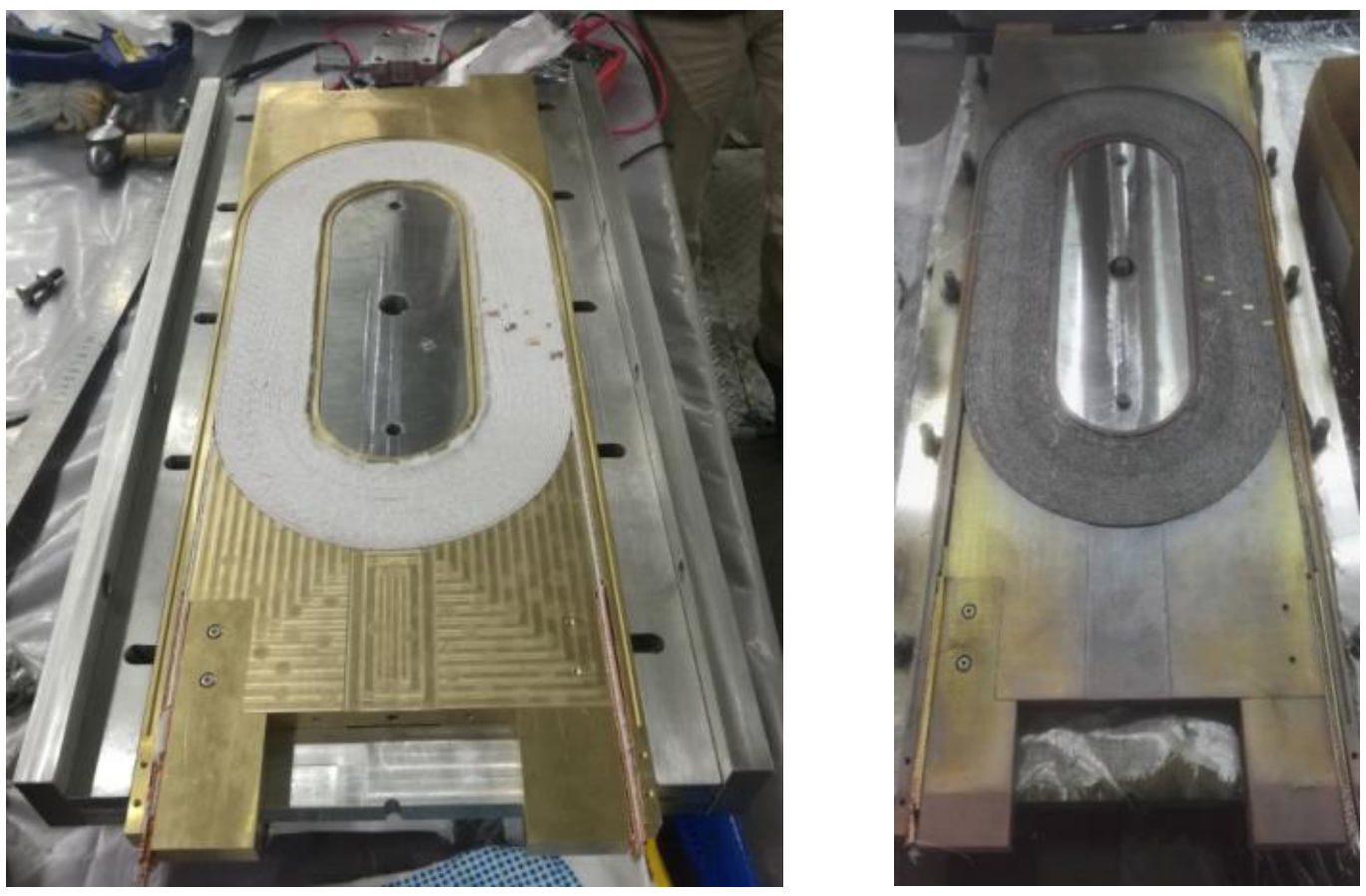

**Figure 11.16.15:** Left – $Nb_3Sn$ coils after winding; Right – $Nb_3Sn$ coils after heat treatment.

*11.16.1.3.4 NbTi/$Nb_3Sn$ Splices*

$Nb_3Sn$ cable is very brittle and stress-sensitive. We made the superconducting joints by soldering two NbTi cables onto each side of the $Nb_3Sn$ cable with commercial PbSn solder. As shown in Fig 11.16.16, the $Nb_3Sn$ coil was placed on an aluminum plate with its cable end clamped by a fixture. A heating block, together with a temperature controller, was used to heat the NbTi and $Nb_3Sn$ cables and make the NbTi/$Nb_3Sn$ splice. By using this method, we fabricated seven NbTi/NbTi splices and tested them at 4.2 K. Their resistance in nano-ohm are 0.495, 0.434, 0.334, 0.558, 0.500, 1.340, and 1.380, all below 2 nano-ohm.

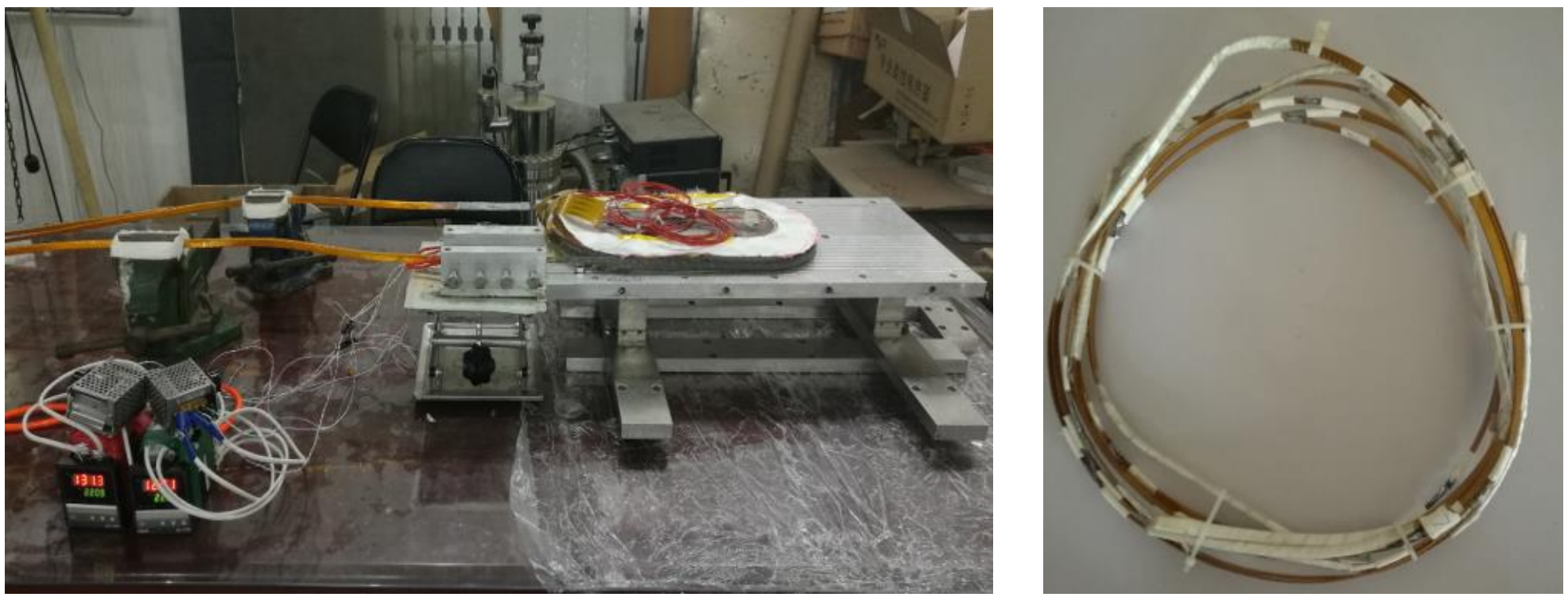

**Figure 11.16.16:** Left – Soldering tooling for making the NbTi/$Nb_3Sn$ splice; Right – Splice tested at 4.2 K

### *11.16.1.3.5 Vacuum Pressure Impregnation (VPI)*

Before VPI, all six coils were instrumented with multiple voltage taps for quench detection. For each side of a coil, as shown in Fig. 11.16.17 (a), the voltage taps were soldered onto a Kapton sheet which consists of multiple channels for tap connections and a quench heater for quench protection. After that, one piece of S-glass cloth was placed on the top of the Kapton sheet. The same operation was done on the other side of the coil. Finally, the coil was confined in aluminum alloy tooling made for VPI and placed into the vacuum vessel for impregnation.

Fig. 11.16.17 (b) shows the whole VPI system developed for the 12-T subscale magnet. The epoxy is expected to go through the coil seated in the vacuum vessel from a small tank after the epoxy is degassed and exposed to a 1 bar air environment. Here, CTD 101-K consisting of Part A (epoxy resin), Part B (hardener) and Part C (accelerator) is used for impregnating the 12-T subscale magnet. The epoxy composition was mixed with the recommended ratio and poured into a small tank. By using a magnetic stirrer, we kept pre-heating and degassing the epoxy composition at ~60 ºC until we couldn’t see any bubbles. This process took around half an hour. During epoxy impregnation, the coil was also pre-heated at around 60 ºC. Besides, the vacuum vessel is kept at an absolute pressure of around 100 Pa which is larger than the saturation vapor pressure of the epoxy at ~60 ºC to avoid re-degassing the epoxy inside the coil. After the coil was filled with epoxy, we closed the valve and left the coil overnight to let the epoxy fill into any of the small voids inside the coil and to check whether the liquid epoxy level went down or not. Finally, we cured the coil by heating it at 110 ºC for 5 hours and 135 ºC for 1.5 hours. Fig. 11.16.18 shows all six coils after epoxy impregnation.

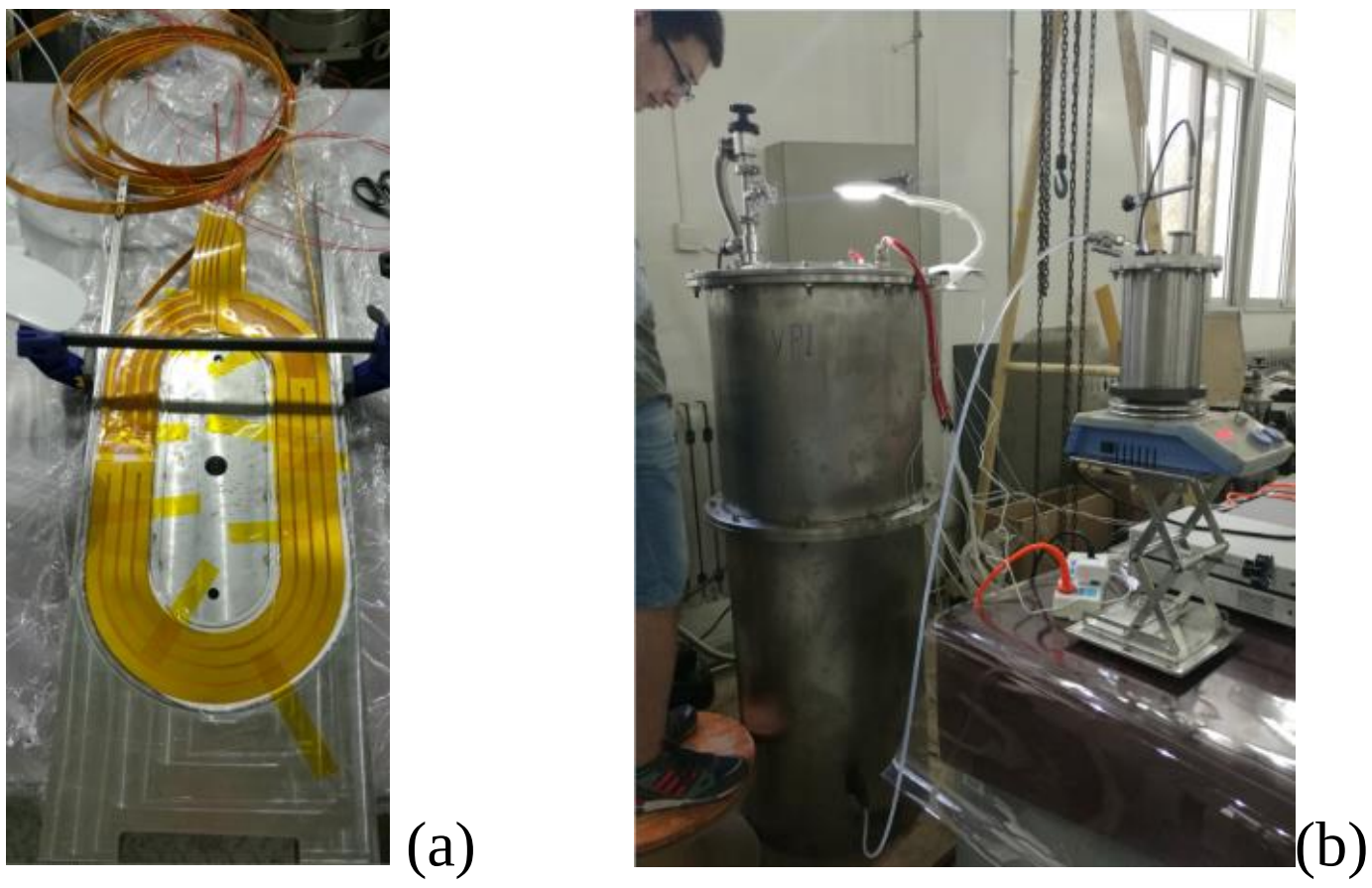

Figure 11.16.17: CTD-101k mixing procedure

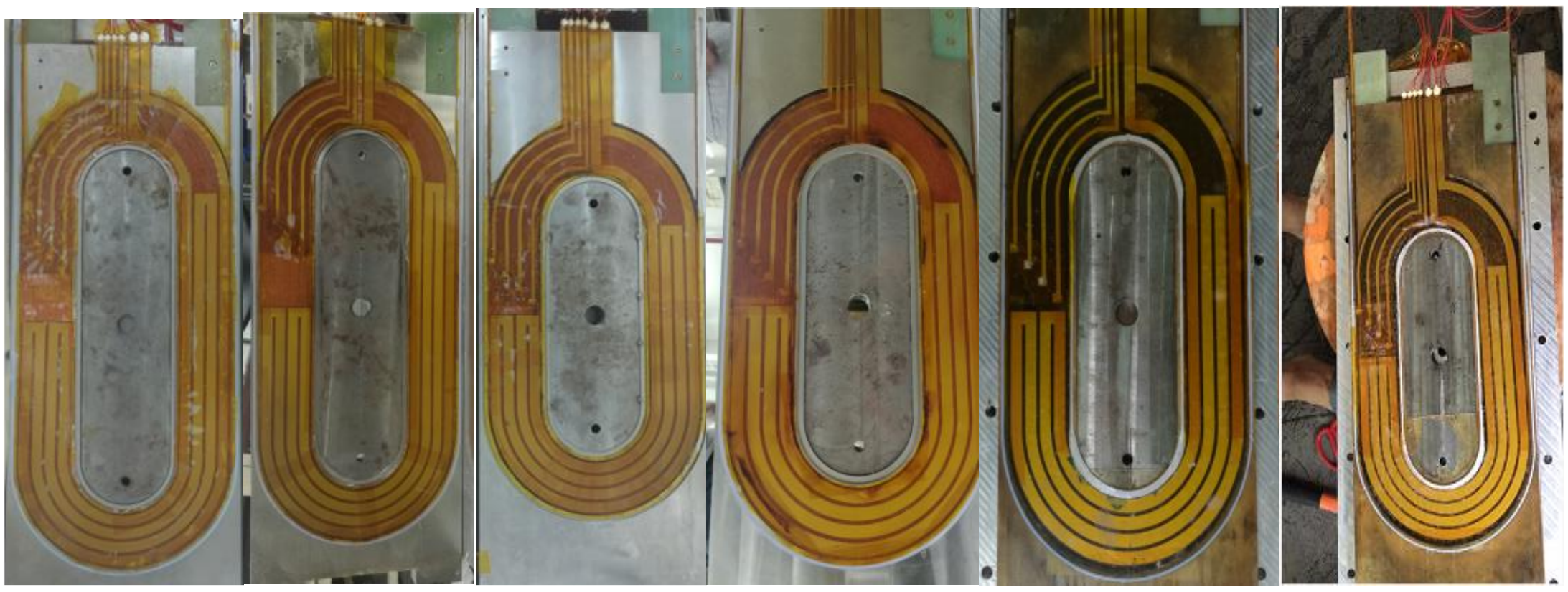

**Figure 11.16.18:** Coils after impregnation

*11.16.1.3.6 Magnet Assembly*

Fig. 11.16.19 shows the fabricated pads, shell, yoke, end plate and rods. By using the bladder & key technology, we tested the shell-based structure with G10 dummy coils successfully as shown in Fig. 11.16.20 (a). Both horizontal and vertical directions were pre-stressed and the 1/4 bridge based strain gauge measurement system worked well. As shown in Fig. 11.16.20 (b), we are now working on magnet assembly with real coils.

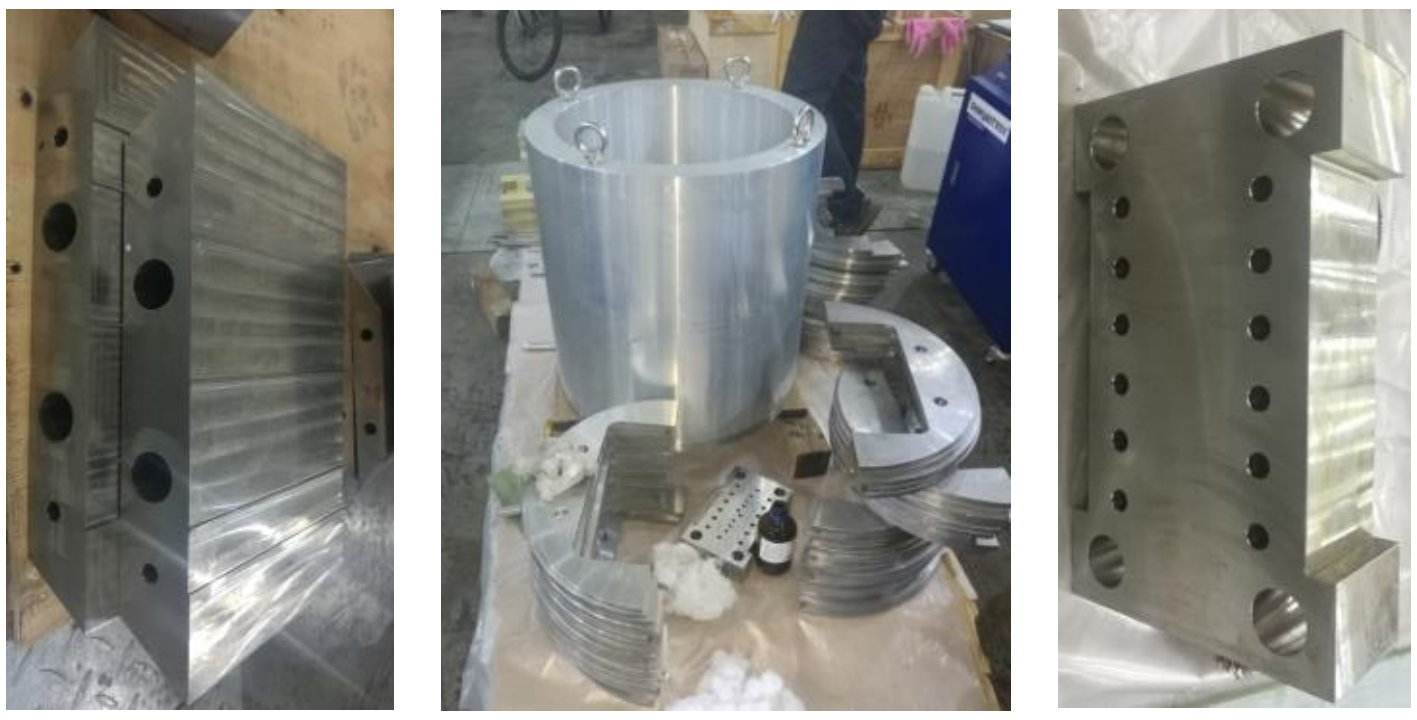

**Figure 11.16.19:** Main components of the magnet

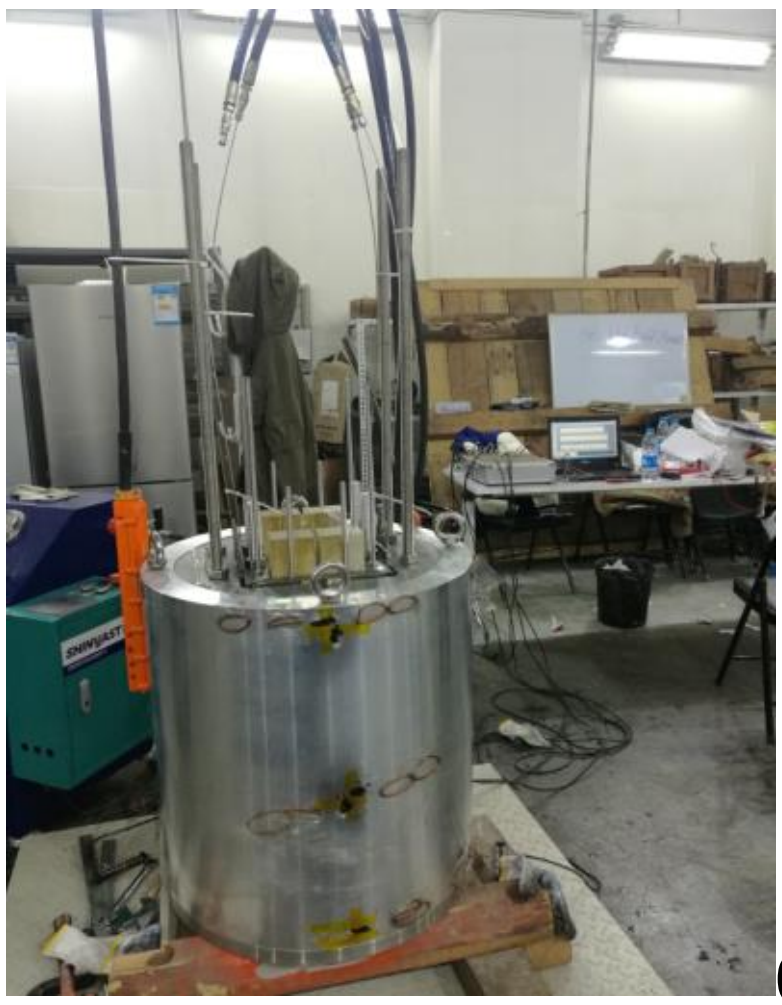
(a)
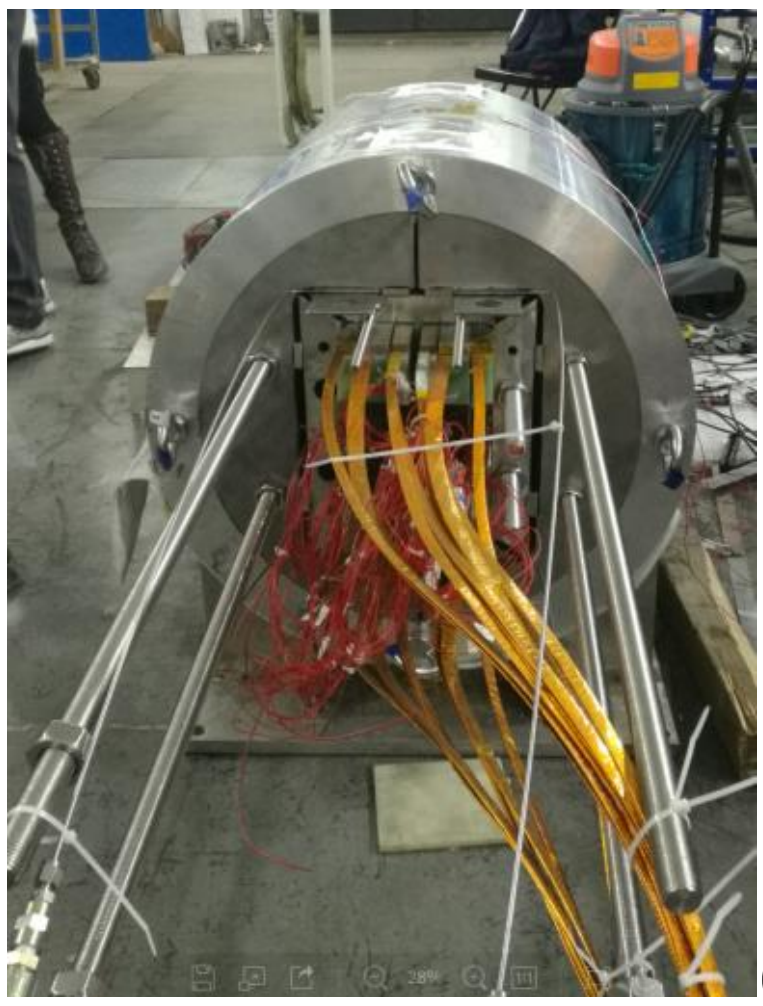
(b)

**Figure 11.16.20:** (a) Test with dummy coil; (b) Magnet assembly with real coils

11.16.2 **Subscale Magnet R&D with Hybrid ($Nb_3Sn$ & HTS) Technology**

A 12-T common-coil dipole magnet with two apertures will be fabricated with HTS (IBS, ReBCO, and Bi-2212) and $Nb_3Sn$ superconductors to test the field optimization method for HTS coils. The cross section of this magnet is shown in Fig. 11.16.9 top. There are two apertures in this dipole magnet. The clear bore diameter is temporarily set at 30 mm and the inter-aperture spacing is 180 mm. The outer diameter of the iron yoke is 500 mm. The coil cross section in the first quadrant is shown in Fig. 11.16.9 bottom. The whole coil width is 54.5 mm, and the height is 57.6 mm. The outer $Nb_3Sn$ coils are with the same parameters as the subscale magnet. The field quality has been optimized at the main field of 12 T. All the higher order multipoles are less than 1 unit. The YBCO insert coils are fabricated with 4-mm width and 0.2-mm thickness YBCO tape. This tape, of size 4*0.2 $mm^2$, reference temperature and field of 4.2K and 12 T, has Ic@BrTr and dIc/dB of 1200 and 40 respectively.

In this design the wide dimension of the ReBCO tape is set to be parallel with the magnetic flux to maximize its current-carrying capacity. The $I_c$ of ReBCO is three times higher than when the wide tape dimension is vertical to the magnetic flux. The required length of the IHEPWCJC strand ($Nb_3Sn$) is 11.65 km (4.48 km of IHEP W5, 7.17 km. of IHEP W6) and 0.52 km. of YBCO tape.

The main field with 100% load line ratio is 15.8 T at 4.2 K, corresponding to an operating current of 1100 A in the YBCO tapes and 13700 A in the $Nb_3Sn$ cables. The peak field is 17.09 T in the YBCO coil and 14.5 T in the $Nb_3Sn$ coils. Or we can get a main field of 12 T with an operating margin of 24% at 4.2 K, corresponding to an operating current of 825 A in the YBCO tapes and 9950 A in the $Nb_3Sn$ cables. The peak field is 13.61 T in the YBCO coil and 11.03 T in the $Nb_3Sn$ coils. Fig. 11.16.10 shows the field distribution in coils with an operating margin of 24% at 4.2 K.

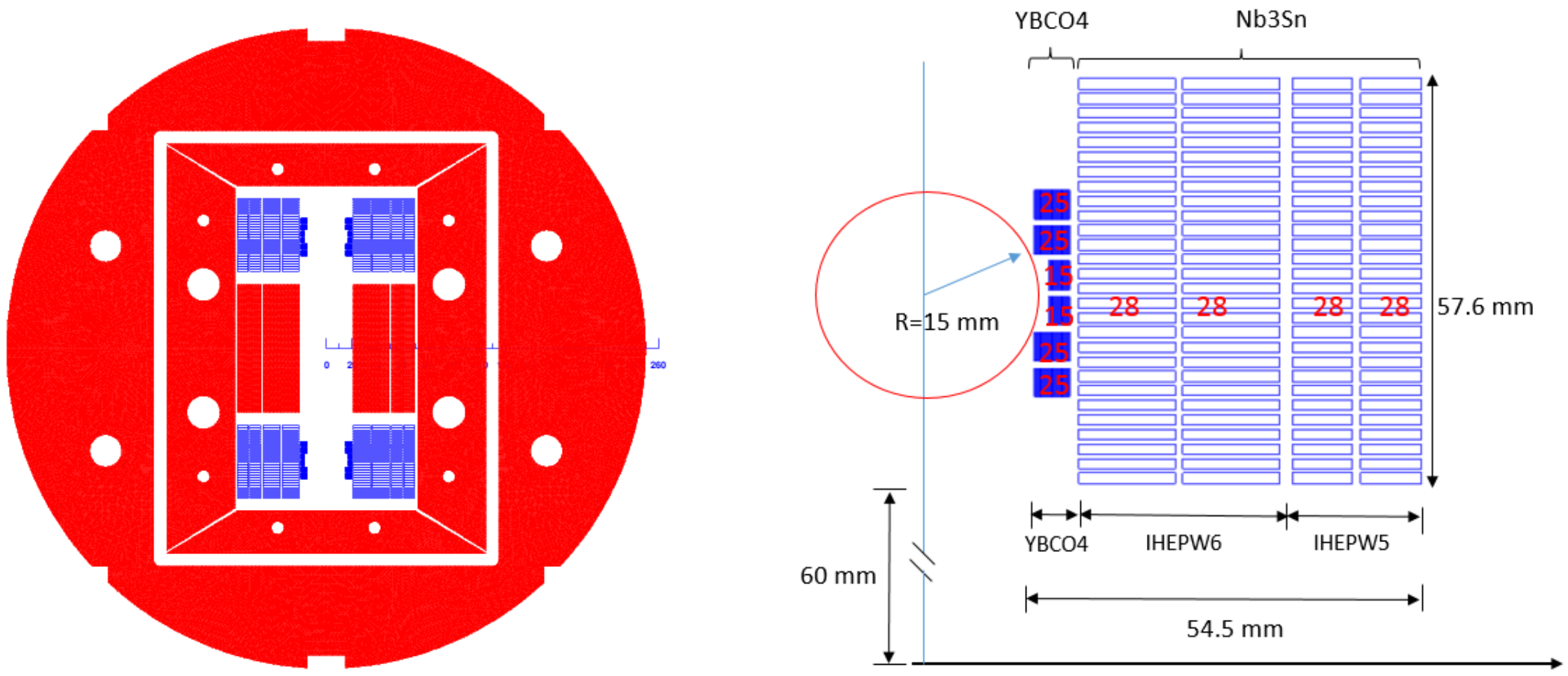

**Figure 11.16.9:** Left – Cross section of 12-T common-coil dipole; Right – Coil cross section 1st quadrant.

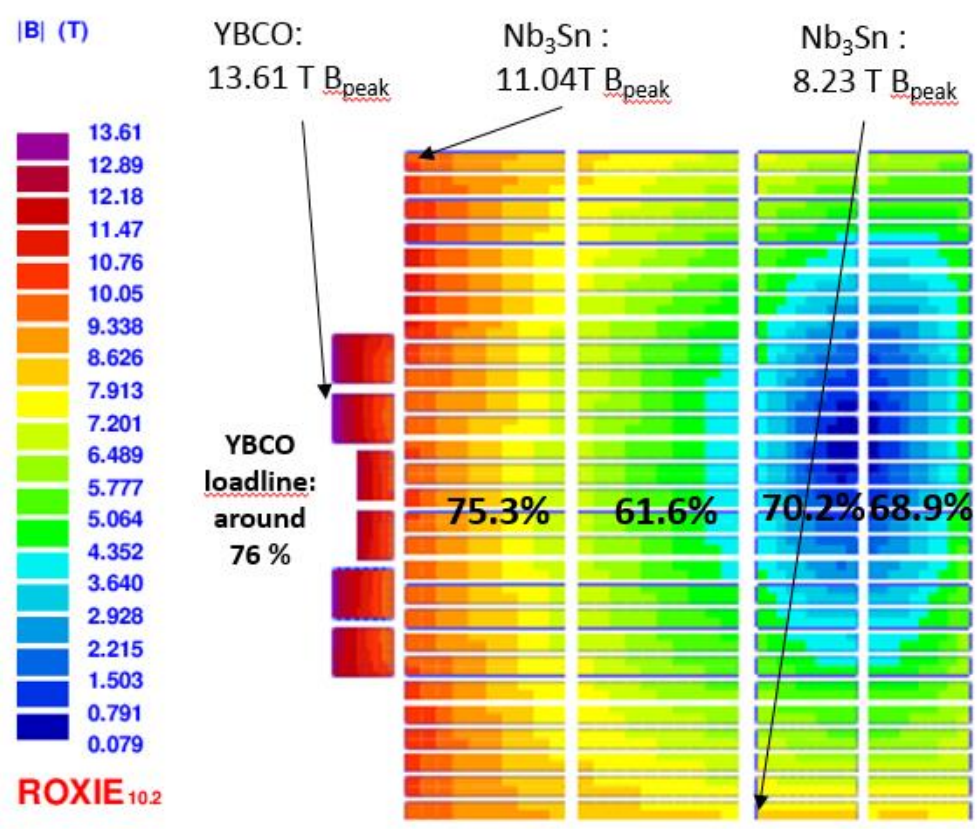

**Figure 11.16.10:** Field distribution in the coil (in first quadrant)

The current-carrying capacity $J_c$ of the ReBCO tape is highly anisotropic to the magnetic field direction [11]. After development of the 12-T NbTi & $Nb_3Sn$ magnet, the ReBCO coils with the block-type configuration will be inserted between the $Nb_3Sn$ coils with the common-coil configuration, to make the wide dimension of the ReBCO tape parallel with the magnetic flux and maximize its current-carrying capacity. We named this type of configuration as Combined Common-coil and Block type configuration (CCB).

By calculating the angular difference between the ReBCO coil block and the magnetic field, as shown in Table 11.16.3, we rotate the ReBCO block to minimize the angular deviation between the ReBCO tape and the magnetic field. With 4 mm ReBCO tape, we can get a maximum deviation of 2 degree between the ReBCO coil and the magnetic flux.

Since the deviation is mainly from the top and the bottom side of the block, if we shorten the width of the ReBCO tape from 4 mm to 2 mm, we can get a maximum deviation of 1.3 degree, as shown in Table 11.16.4. With this well aligned design, the critical current density of the ReBCO tape is increased from 750 A/mm2 to 2150 A/mm2.12-13 The ReBCO coils provide a magnetic field of 1.5 T, as show in Fig. 11.16.11.

**Table 11.16.3:** Angular deviation between coil blocks and the magnetic field with 4mm ReBCO tape

| Block number | Maximum (degree) | Minimum (degree) | Rotate angle | Deviation |
|---|---|---|---|---|
| 1 | 3.82 | 0.06 | 2 | 1.82 |
| 2 | 2.00 | -0.40 | 1 | 1.40 |
| 3 | 1.18 | -0.62 | 0 | 1.18 |
| 4 | 0.47 | -1.26 | 0 | 1.26 |
| 5 | 0.25 | -1.87 | -1 | 1.25 |
| [illegible] | [illegible] | [illegible] | [illegible] | [illegible] |

**Table 11.16.4:** Angular deviation between coil blocks and the magnetic field with 2mm ReBCO tape

| Block number | Maximum (degree) | Minimum (degree) | Rotate angle | Deviation |
|---|---|---|---|---|
| 1 | 3.00 | 0.41 | 1.7 | 1.30 |
| 2 | 1.87 | -0.05 | 1 | 1.05 |
| 3 | 1.12 | -0.17 | 0 | 1.12 |
| 4 | 0.66 | -0.39 | 0 | 1.05 |
| 5 | 0.26 | -0.60 | 0 | 0.86 |
| 6 | -0.14 | -1.31 | 0 | 1.31 |
| 7 | -0.06 | -1.87 | -1.2 | 1.26 |
| 8 | -0.41 | -2.92 | -1.6 | 1.32 |

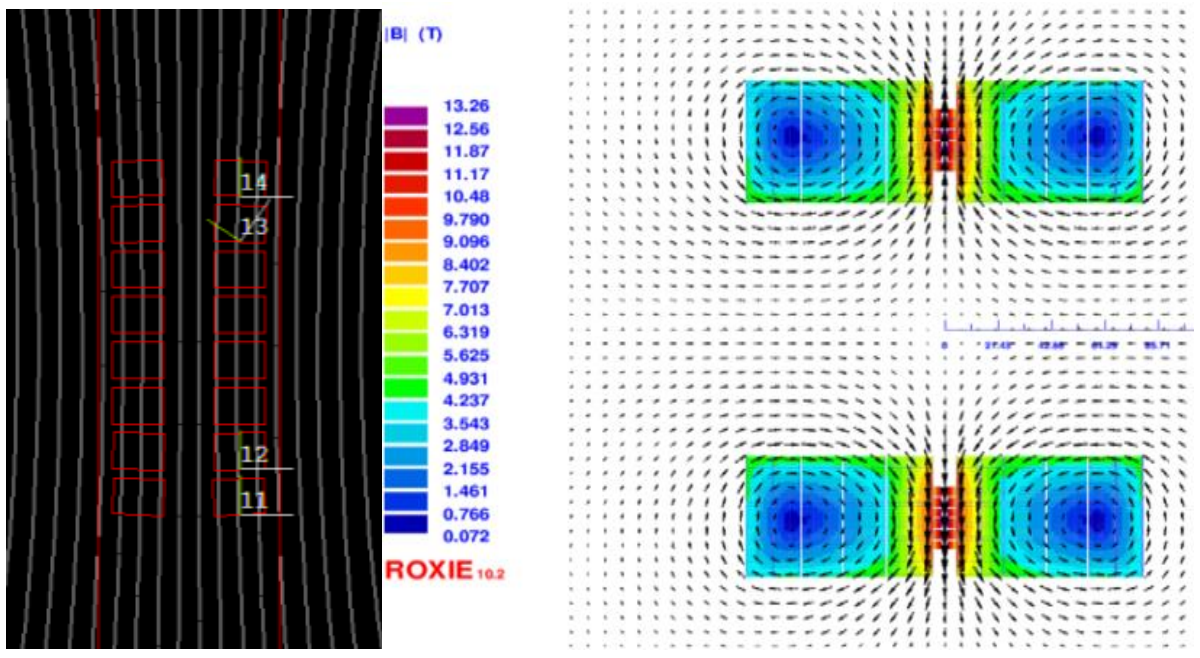

**Figure 11.16.11:** Flux lines and the potential vectors of the ReBCO insert coils

# 12 Project Cost, Schedule and Planning

## 12.1 Construction Cost Estimate

Compared to the Pre-CDR, there are three major design changes in the CDR: (1) the ring circumference is increased from 54 km to 100 km; (2) the Collider is changed from single ring to double ring; and (3) the synchrotron radiation power of the $e^+$ and $e^-$ beams is reduced from 50 MW to 30 MW each. These changes have a major impact on the cost.

As the beam power is reduced, the cost of the so-called "big three" technical systems – the superconducting RF (SRF), the RF power sources and the cryogenic system – is greatly reduced. On the other hand, as the ring circumference is increased and the number of beam pipes doubled, the cost of other systems is increased, including the magnets, magnet power supplies, vacuum system, instrumentation, control and mechanical systems. The civil construction cost is almost doubled.

The CEPC has two SRF systems:

- 650 MHz 2-cell cavities for the Collider, similar to the ADS and PIP-II, powered by klystrons;
- 1.3 GHz 9-cell cavities for the Booster, similar to the ILC, XFEL and LCLS-II, powered by solid state amplifiers (SSA).

This synergy makes it possible to make a reliable cost estimate based on the experience from these other accelerators,

Two references were particularly useful: the actual cost of LEP1 and LEP2, and the cost estimate of the LCLS-II 4 GeV SRF linac.

Some information on LEP1 costs is available [1, 2]. The total in 1986 prices was 1.3 billion Swiss francs (BCHF). LEP2 added 288 SRF systems in the 1990s for about 0.5 BCHF [3]. Taking into account inflation, the construction of LEP1 and LEP2 would cost roughly 2.6 BCHF today. As the CEPC is nearly four times as large as LEP, plus having a full-energy Booster as well as a new linac, the cost would be about 12 BCHF or higher were it to be built in Switzerland. But the cost in China is lower, not only the civil construction, but also technical systems. The cost saving is more than 50%.

To verify potential cost differences between the two countries, two cost estimate exercises were carried out at the IHEP: one by the magnet group, another by the vacuum group. Each group was given the LEP design and was asked to estimate the cost if the identical magnet or vacuum system was built in China. The result showed that the LEP magnet would cost 30% less if fabricated in China. But the saving on the vacuum system was smaller because China does not have the advanced aluminium extrusion technology.

The LCLS-II is another useful reference. Its 4 GeV linac uses 1.3 GHz 9-cell ILC type cavities and cryomodules. The cost is 2.7 million US dollars (USD) per module, or a total of 105 million USD for 38 modules. But this cost does not include non-superconducting RF components (klystron, modulator, RF distribution, etc.) [4]. The CEPC Booster needs 12 cryomodules (1.3 GHz), and the Collider 40 cryomodules (650 MHz). The LCLS-II figure was used as a cross check for the cost estimate of the CEPC cryomodules.

The actual CEPC cost estimate was done by using a bottom-up method. Namely, given the design parameters, each technical system manager made a cost estimate for each component, and then added them up. The result is presented in a Work Breakdown Structure (WBS) format, which is widely used for accelerator projects especially in the United States. Table 12.1 shows the WBS at Level 1. There are 12 items at this level as shown in the table.

In this CDR, the WBS of most systems was detailed down to Level 3 or 4, and in some cases reached Level 5. In the next stage for CEPC, the Technical Design Report (TDR), the complete WBS will reach Level 7.

**Table 12.1:** Level 1 of Work Breakdown Structure (WBS) of the CEPC

**CEPC Work Breakdown Structure (WBS)**

| black | |
|---|---|
| Level 1 | WBS Element Title |
| | **TOTAL** |
| 1 | Project Management |
| 2 | Accelerator Physics |
| 3 | Collider (Ch 4) |
| 4 | Booster (Ch 5) |
| 5 | Linac, Damping Ring and Sources (Ch 6) |
| 6 | Transport Lines (Ch 5.2.4, 6.2.2) |
| 7 | Systems Common to Accelerators (Ch 7) |
| 8 | Conventional Facilities (Ch 9) |
| 9 | Gamma Ray Source (Appendix 6) |
| 10 | Detector (Volume 2) |
| 11 | Research and Development (Ch 11) |
| 12 | Contingencies (10%) |

Using the WBS numbers, Fig. 12.1 shows the relative cost of accelerator, conventional facilities, detectors, gamma-ray sources, project management and contingency. Fig. 12.2 is the accelerator cost breakdown among the major components: Collider, Booster, Linac, Damping Ring and sources, and common systems. Fig. 12.3 is the cost breakdown among accelerator technical systems.

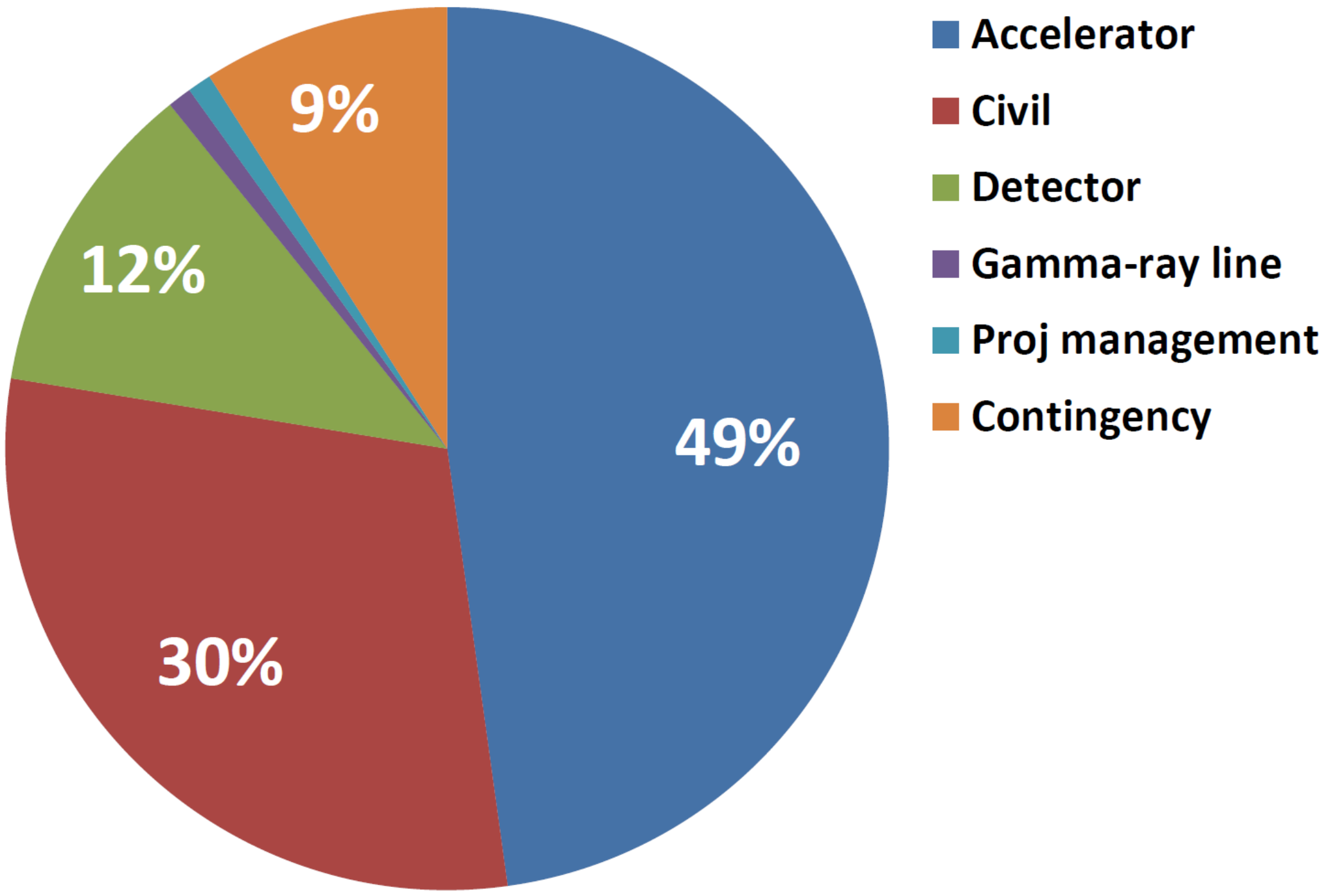

**Figure 12.1:** Relative cost of the CEPC project constituents.

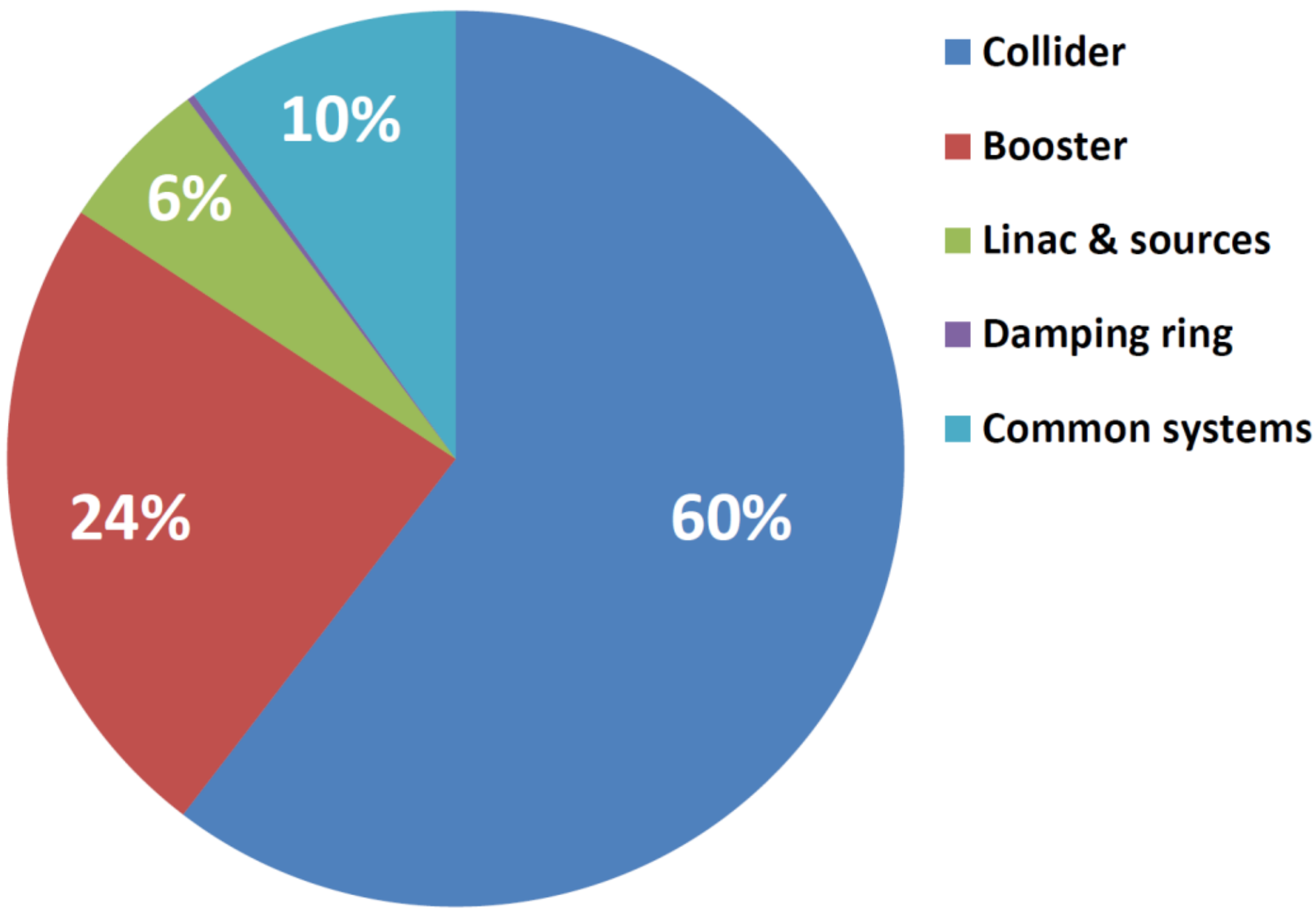

**Figure 12.2:** Cost breakdown of the CEPC major accelerator components.

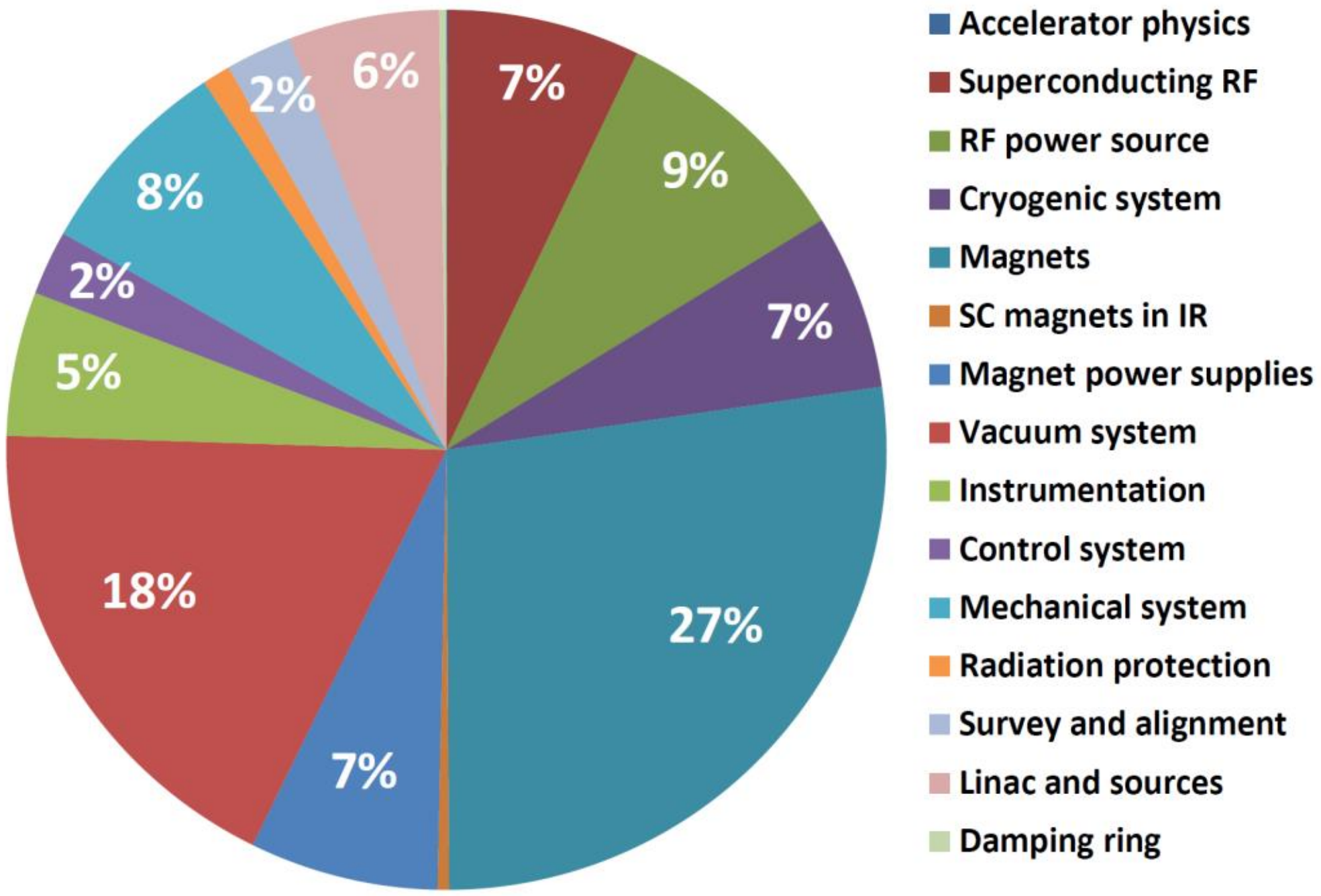

**Figure 12.3:** Cost breakdown of the CEPC accelerator technical systems.

The accelerators are the most expensive part of the project, representing half of the total construction cost. Civil construction is about 30% of the total.

In the accelerator complex, the Collider cost is 60%, followed by the Booster (24%), the common systems (10%) and the Linac and sources (6%).

Among the accelerator systems, the "big three" – SRF, RF power source and cryogenics – account for about 23% of the cost, whereas the cost of the magnets (27%) and the vacuum system (18%) are significant because of their length of hundreds of kilometres.

## 12.2 Operations Cost Estimate

In addition to the capital construction cost, the operations cost is another major consideration in the design. It is mainly determined by the power consumption to operate the CEPC. When the Tevatron was running, the average total power usage at Fermilab was 58 MW. When the LHC was running at 3.5 × 3.5 TeV, CERN used 183 MW (average during 2012). The consensus for operating a future circular Higgs factory is that the total power should not exceed 300 MW.

The RF and magnets are the two principal power consumers. To reduce power consumption, a number of measures were taken in the design of these systems:

- Limit synchrotron radiation power to 30 MW per beam
- Use superconducting RF cavities
- Use high efficiency klystrons
- Use a 2-in-1 structure for Collider dipoles and quadrupoles
- Combine dipole and sextupole functions in the Collider bending magnets
- Use a large coil cross section in the quadrupoles
- Use permanent magnets in the kilometer-long Linac to Booster transport lines

With these measures, the total facility power for operation as a Higgs factory is estimated to be 270 MW, below the 300 MW target. For operation in the Z and W mode, the power consumption is much lower as shown in Appendix 3. Fig. 12.4 shows the relative power consumption of each CEPC system.

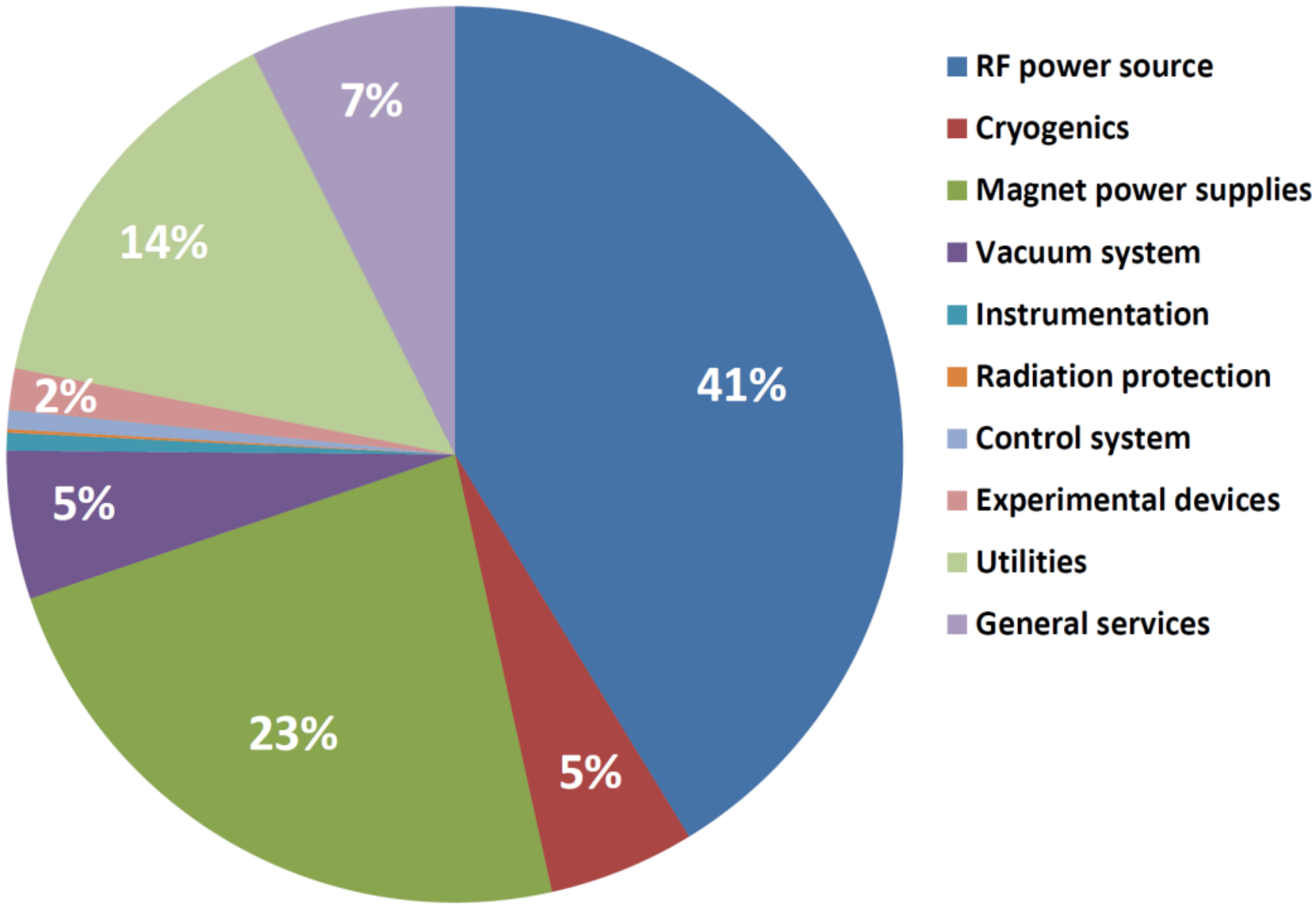

**Figure 12.4:** Relative power consumption of each CEPC system.

The energy efficiency of the CEPC, expressed as a ratio between the beam power (60 MW) over the total power from the grid (270 MW), is 22%, higher than other accelerator facilities. For example, the PSI cyclotron in Switzerland has an energy efficiency of 18%, which is the highest among all existing or previously existed accelerator facilities. The energy efficiency of the SNS (an SRF linac plus a storage ring) at Oak Ridge, USA is 8.6%, and that of the J-PARC (a linac and two synchrotrons) in Japan is 3%. Another comparison is with the ILC, a future linear collider, which has a design efficiency of 5% (beam power 5.28 MW, total facility power 117.3 MW).

We will continue to investigate effective ways for energy efficiency improvement, including possible reuse and recycling of waste power from the accelerator.

As stated in Chapter 3, the CEPC is planned to operate 6,000 hours each year. At 270 MW, the electricity usage will be $1.6 \times 10^9$ kW-hours a year, resulting in an electricity bill of RMB one billion (about USD 150 million).

## 12.3 Project Timeline

Fig. 12.5 shows our current concept of a timeline for the CEPC project. It consists of the following stages:

- The first stage is to complete a Preliminary Conceptual Design Report (Pre-CDR) in 2015 and a Conceptual Design Report (CDR) in 2018. With the publication of this report, these goals have been achieved.
- The next stage is a 5-year period from 2018 to 2022 for R&D and for completion of a Technical Design Report (TDR).

- Construction will start in 2022 in the government's 14th Five-Year Plan and continue in the 15th Five-Year Plan. Construction will be completed by 2030.
- Experiments can begin as early as 2030 when the 16th Five-Year Plan starts.
- The experiments will continue for about 10 years until 2040 as outlined in Chapter 3.
- After 2040, the superconducting magnets for the SPPC project are expected to be ready for installation, and the SPPC era will begin.

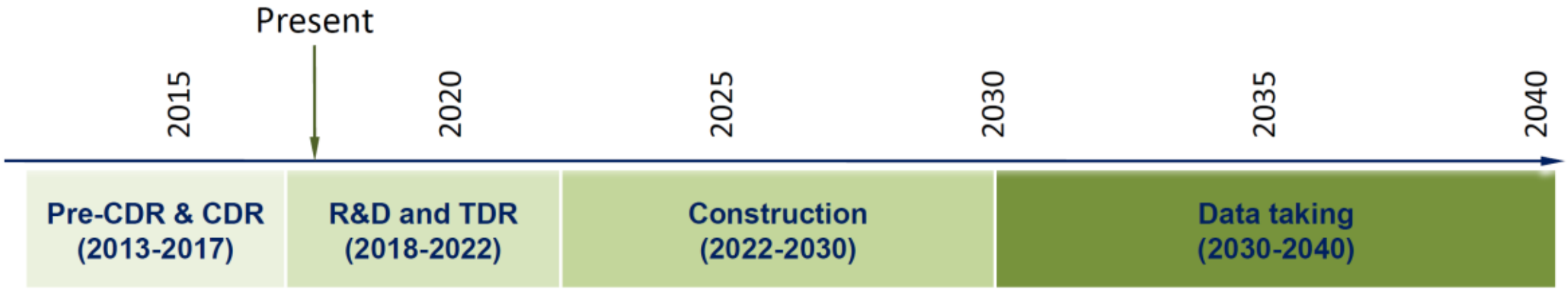

**Figure 12.5:** A possible timeline.

Of course, the realization of such an ideal timeline depends on many factors. Some are under our control, some are not. After completion of this CDR, the focus turns to the R&D.

There are several critical paths in the CEPC timeline:

- Successful R&D for the two SRF systems:
  - Collider: 650 MHz, 240 2-cell cavities in 40 cryomodules.
  - Booster: 1.3 GHz, 96 9-cell cavities in 12 cryomodules.
  - A large RF facility similar to those at JLab, Fermilab, KEK and DESY is currently under construction at Huairou, a city about 60 km north of Beijing. It will be used for cavity inspection and tuning set ups, RF laboratory, several vertical test stands, clean rooms, high pressure rinse (HPR) systems, fundamental power coupler (FPC) preparation and conditioning facility, cryomodule assembly lines, horizontal test stations, high power RF equipment, and will have a cryogenic plant.
- Successful R&D for high efficiency klystrons:
  - A collaboration between the IHEP and an industrial company in Kunshan city was formed. The goal is to design and prototype klystrons with a saturation efficiency of 80% or higher.
- During the construction period, some technical systems will require longer time than the others and this will be taken into account in the planning. For example:
  - Civil construction: 55 months
  - SRF: 3 years
  - RF sources: 2.5 years
  - Cryogenic system: 5 years
  - Magnets: 8 years, assuming 5 production lines each for the Collider and Booster magnets
  - Vacuum: 8 years for NEG coating of the 100-km long Cu pipes, assuming 10 or more companies working on this job
  - Installation and alignment: 3-5 years
  - MDI: from the experience at the SuperKEKB, MDI modules took 10 years from prototyping to completion

## 12.4 Project Planning

The CEPC has been identified as the No. 1 priority for the next high-energy physics (HEP) project in China by the HEP Division of the Chinese Physical Society. It has received strong support from HEP physicists both at home and from abroad.

A significant amount of R&D funds has been received from the Chinese government for the 5-year R&D plan as described in Chapter 11.

During the past several years, a core team of accelerator physicists and engineers has been formed. They are young, talented, motivated and able to steer the CEPC project from the design to R&D to the construction and to operation.

Site selection is a complicated process. There are currently several candidate sites as stated in Chapter 9. Preliminary geological studies were performed at these sites and reports were written. However, the actual selection will not start until the project receives government approval to proceed.

For a project of this size, collaboration is the key for its success. The IHEP has a long history in collaboration with other accelerator laboratories and universities in China (IMP, SINAP, USTC, Tsinghua University, Peking University, etc.). To encourage the participation of industrial companies, a CEPC Industrial Promotion Consortium (CIPC) was established with scores of companies. Its goal is to promote the construction of the CEPC and to prepare an industrial base for manufacturing the very large number of high-tech components required by the project.

Internationally, the IHEP has signed collaboration agreement on circular $e^+e^-$ colliders with a number of institutions, including KEK (Japan), BINP and MEPhI (Russia) and INFN (Italy). The US-China HEP Collaboration Panel holds annual meetings and provide a regular channel for communication between the two countries. A CEPC International Advisory Committee consisting of two dozen world renowned HEP physicists meets once a year in Beijing and gives valuable advice to the project. The collaboration with several other proposed future collider projects – the ILC, CLIC and FCC – includes visitor exchanges and jointly organized workshops (e.g., the ICFA *eeFACT* workshop series on high luminosity circular $e^+e^-$ colliders) and accelerator schools.

The Chinese government has announced an ambitious plan to establish several grand international scientific research centers in the coming decade in China. The CEPC is a strong and viable candidate to be one of them. It will be proposed to the government and compete for a place in the plan.

# Appendix 1: Parameter List

## A1: Collider

| Fundamental constants | Unit | H | W | Z | |
|---|---|---|---|---|---|
| | | | | 3T | 2T |
| electronic charge | C | 1.60E-19 | | | |
| speed of light | m/s | 3.00E08 | | | |
| $C_q$ | | 3.83E-13 | | | |
| fine structure constant | | 0.0073 | | | |
| classical radius of the electron [$r_e$] | m | 2.82E-15 | | | |
| Euler's constant [$\gamma_E$] | | 0.577 | | | |
| electron Compton wavelength [$\lambda_e$] | m | 3.86E-13 | | | |
| rest mass energy of the electron | MeV | 5.11E-01 | | | |
| **Accelerator Parameters** | | | | | |
| Beam energy [E] | GeV | 120 | 80 | 45.5 | |
| Circumference [C] | km | 100.0164 | | | |
| Luminosity [L] | $cm^{-2}s^{-1}$ | 3E34 | 1E35 | 1.7E35 | 3.2E35 |
| SR power/beam [P] | MW | 30 | 30 | 16.5 | |
| Bending radius [$\rho$] | m | 10700 | | | |
| N(IP) | | 2 | | | |
| No. of bunches | | 242 | 1524 | 12000 | |
| Filling factor [$\kappa$] | | 0.72 | | | |
| Lorentz factor [$\gamma$] | | 234834 | 156556 | 89041 | |
| Revolution period [$T_0$] | s | 3.33E-04 | | | |
| Revolution frequency [$f_0$] | Hz | 3003 | | | |
| Magnetic rigidity [$B\rho$] | T·m | 400.27 | 266.85 | 151.77 | |
| Momentum compaction factor [$\alpha_p$] | | 1.11E-05 | | | |
| Energy acceptance requirement [$\eta$] | % | 1.35 | 0.40 | 0.23 | |
| Cross-section for radiative Bhabha scattering [$\sigma_{ee}$] | $cm^2$ | 1.6E-25 | 1.8E-25 | 1.9E-25 | |
| Lifetime due to radiative Bhabha scattering [$\tau_L$] | min | 64.7 | 84.2 | 254.7 | 131.7 |
| build-up time of polarization [$\tau_p$] | hr | 2. | 16 | 271 | |
| **Beam Parameters** | | | | | |
| Beam current [I] | mA | 17.4 | 87.9 | 461.0 | |
| Bunch population [$N_e$] | $10^{10}$ | 15.0 | 12.0 | 8.0 | |
| emittance-horizontal [$\varepsilon_x$] | nm·rad | 1.21 | 0.54 | 0.18 | |
| emittance-vertical [$\varepsilon_y$] | pm·rad | 3.1 | 1.6 | 4.0 | 1.6 |
| coupling factor [$\kappa$] | | 0.00256 | 0.00296 | 0.0222 | 0.00888 |
| Bunch length SR [$\sigma_{s.SR}$] | mm | 2.72 | 2.98 | 2.42 | |
| Bunch length total [$\sigma_{s.tot}$] | mm | 4.4 | 5.9 | 8.5 | |
| **Interaction Region Parameters** | | | | | |
| Betatron function at IP-vertical [$\beta_y$] | m | 0.0015 | 0.0015 | 0.0015 | 0.001 |
| Betatron function at IP-horizontal [$\beta_x$] | m | 0.36 | 0.36 | 0.2 | |
| Transverse size [$\sigma_x$] | μm | 20.9 | 13.9 | 6.0 | |
| Transverse size [$\sigma_y$] | μm | 0.068 | 0.049 | 0.078 | 0.04 |
| Beam-beam parameter [$\xi_x$] | | 0.031 | 0.013 | 0.004 | |
| Beam-beam parameter [$\xi_y$] | | 0.109 | 0.106 | 0.056 | 0.072 |
| Hourglass factor | Fh | 0.89 | 0.94 | 0.99 | |
| Lifetime due to Beamstrahlung-Telnov [$\tau_{BS}$] | min | 60 | | | |
| Lifetime due to Beamstrahlung [simulation] | min | 100 | | | |
| | | | | | |

| | | | | |
|---|---|---|---|---|
| **RF Parameters** | | | | |
| RF voltage [Vrf] | GV | 2.17 | 0.47 | 0.10 |
| RF frequency [frf] | MHz | 650 | | |
| Harmonic number [h] | | 216816 | | |
| Synchrotron oscillation tune [$\nu_s$] | | 0.065 | 0.0395 | 0.028 |
| Energy acceptance RF [$\eta$] | % | 2.06 | 1.47 | 1.70 |
| **Synchrotron Radiation** | | | | |
| SR loss/turn [$U_0$] | GeV | 1.73 | 0.34 | 0.036 |
| Damping partition number [Jx] | | 1 | | |
| Damping partition number [Jy] | | 1 | | |
| Damping partition number [$J\varepsilon$] | | 2 | | |
| Energy spread SR [$\sigma_{\delta.SR}$] | % | 0.1 | 0.066 | 0.038 |
| Energy spread BS [$\sigma_{\delta.BS}$] | % | 0.089 | 0.0724 | 0.07 |
| Energy spread total [$\sigma_{\delta.tot}$] | % | 0.134 | 0.098 | 0.08 |
| Average number of photons emitted per electron during the collision [$n\gamma$] | | 0.29 | 0.35 | 0.55 |
| Transverse damping time [$\tau_x$] | ms | 46.5 | 156.4 | 849.5 |
| Longitudinal damping time [$\tau_\varepsilon$] | ms | 23.5 | 74.5 | 425.0 |
| **Ring Parameter** | | | | |
| Circumference [C] | km | 100.016 | | |
| Revolution period [$T_0$] | s | 3.34E-04 | | |
| Revolution frequency [$f_0$] | Hz | 3003 | | |
| Betatron tune [Q x/y] | | 363.10 / 365.22 | | |
| Damping time [$\tau$ x/y/s] | ms | 46.5/46.5/ 23.5 | 156.4/156.4/ 74.5 | 849.5/849. 5/425.0 |
| Number of arc regions | | 8 | | |
| Number of interaction regions | | 2 | | |

| | | |
|---|---|---|
| Number of straight section regions | | 4 |
| Number of RF regions | | 2 |
| Total number of dipoles | | 2546 |
| Total number of quadrupoles | | 3524 |
| Total number of sextupoles | | 1864 |
| Total number of horizontal correctors | | 5808 |
| **Regular lattice period parameters** | | |
| Lattice type | | FODO |
| Cell numbers in each period | | 5 |
| Phase advance/2π (x/y) | | 1.25/1.25 |
| Period length | m | 343 |
| Dipole type in regular lattice | | B0 |
| Maximum β value | m | 114 |
| Minimum β value | m | 21 |
| Maximum dispersion | m | 0.245 |
| Number of dipoles [B0] | | 10 |
| Dipole length | m | 5.737×5 |
| Strength of dipole | T | 0.0373 |
| Quadrupole type in regular lattice | | QF/QD |
| Number of quadrupoles [QF/QD] | | 10 |
| Quadrupole length | m | 2 |
| Strength of quadrupole | $Tm^{-1}$ | 8.421 |
| Sextupole type in regular lattice | | SF/SD |
| Number of sextupoles [SF/SD] | | 4 |
| Sextupole length | M | 0.7/1.4 |
| Strength of SF | $Tm^{-2}$ | 506.2 |
| Strength of SD | $Tm^{-2}$ | 500.7 |
| Correcting dipole type | | CH/CV |
| Number of correcting dipoles [CH/CV] | | 10 |
| Strength of correcting dipoles [CH/CV] [maximum] | T | 1371 |
| **Arc region** | | |
| Length | m | 10218 |
| Horizontal phase advance/2π | | 37.26 |
| Vertical phase advance/2π | | 37.25 |
| Maximum βx value | m | 113 |
| Maximum βy value | m | 116 |
| Maximum dispersion Dx | m | 0.245 |
| **Interaction region** | | |
| Length | m | 3320 |
| Horizontal phase advance/2π | | 10.25 |
| vertical phase advance/2π | | 10.75 |
| Maximum βx value | m | 373 |
| Maximum βy value | m | 3992 |
| Maximum dispersion Dx | m | 0.570 |
| **Straight section region** | | |
| Length | m | 1198 |
| Horizontal phase advance/2π | | 3.47 |
| Vertical phase advance/2π | | 3.62 |
| Maximum βx value | m | 599 |
| Maximum βy value | m | 392 |
| **RF region** | | |

| | | | | |
|---|---|---|---|---|
| Length | m | 3420 | | |
| Horizontal phase advance/2π | | 15.35 | | |
| Vertical phase advance/2π | | 15.59 | | |
| Maximum βx value | m | 449 | | |
| Maximum βy value | m | 506 | | |
| Maximum dispersion Dx | m | 0.291 | | |
| **Main RF parameter** | | | | |
| Frequency | GHz | 0.65 | | |
| Harmonic number | | 216816 | | |
| Cavity type | | 2-cell cavity | | |
| Cavity operating voltage | MV | 9 | 9 | 9 |
| Cavity operating gradient | MV/m | 19.7 | 9.5 | 3.6 |
| Number of cavities per cryomodule | | 6 | | |
| Cavity active length (two-cells) | m | 0.46 | | |
| Cryomodule length | m | 11 | | |
| Total number of cryomodules | | 40 | 36 | 20 |
| Total number of cavities | | 240 | 216 | 120 |

## A2: Booster

| Accelerator Parameters | Unit | Injection | Extraction | | |
|---|---|---|---|---|---|
| | | | Higgs | W | Z |
| Beam energy [E] | GeV | 10 | 120 | 80 | 45.5 |
| Circumference [C] | Km | 100.0164 | | | |
| Revolution frequency [$f_0$] | kHz | 2.99 | | | |
| Lorentz factor [γ] | | 19569 | 234830 | 156560 | 89041 |
| Magnetic rigidity [Bρ] | T.m | 33.35 | 400.28 | 266.85 | 151.77 |
| Bunch number [Nb] | | | 242 | 1524 | 6000 |
| Beam current [I] | mA | | 0.523 | 2.63 | 6.91 |
| Bunch charge [$N_e$] | nC | | 0.72 | 0.58 | 0.41 |
| emittance-horizontal [$\varepsilon_x$] | nm.rad | 0.025 | 3.58 | 1.59 | 0.51 |
| RF voltage [Vrf] | MV | 62.7 | 1.97E3 | 0.59E3 | 0.29E3 |
| RF frequency [frf] | GHz | 1.3 | | | |
| Harmonic number [h] | | 433633 | | | |
| SR loss / turn [U0] | GeV | 7.35E-5 | 1.52 | 0.3 | 0.032 |
| Transverse damping time [$\tau_x$] | s | 90.7 | 52E-3 | 177E-3 | 963E-3 |
| Betatron tune $\nu x$ | | 263.2 | | | |
| Betatron tune $\nu y$ | | 261.2 | | | |
| Momentum compaction factor | | 2.44E-5 | | | |
| Synchrotron oscillation tune [$\nu_s$] | | 0.1 | 0.13 | 0.1 | 0.1 |
| Energy acceptance RF [$\eta$] | % | 1.9 | 1.0 | 1.2 | 1.8 |
| Energy spread [$\sigma_\delta$] in equilibrium | % | 0.0078 | 0.094 | 0.062 | 0.036 |
| from Linac | % | 0.16 | | | |
| Bunch length [$\sigma_\delta$] in equilibrium | mm | 0.295 | 2.8 | 2.4 | 1.3 |
| from Linac | mm | 1.0 | | | |
| Number of arcs | | 8 | | | |
| Number of short straight sections | | 4 | | | |
| Number of long straight sections | | 2 | | | |
| Number of straight sections with RF | | 2 | | | |
| Number of bypass at IP | | 2 | | | |
| Total number of dipoles | | 16320 | | | |
| Total number of quadrupoles [QF/QD] | | 2036 | | | |
| Total number of sextupoles | | 448 | | | |

| | | | | |
|---|---|---|---|---|
| [SF/SD] | | | | |
| **Regular lattice period parameters** | | | | |
| Lattice type | | FODO | | |
| Phase advance (horizontal/vertical) | | 90°/90° | | |
| Cell length | m | 101.18 | | |
| Number of dipoles in a cell | | 20 | | |
| Dipole length | m | 4.64 | | |
| Defection angle of dipole | mrad | 0.397 | | |
| Magnetic field of the dipole at injection | T | 0.002817 | | |
| Magnetic field of the dipole at ejection | T | 0.0338 | 0.0225 | 0.0128 |
| Number of quadrupoles | | 2 | | |
| Quadrupole length | m | 1.0 | | |
| Strength of QF | $m^{-2}$ | 0.0281409 | | |
| Strength of QD | $m^{-2}$ | -0.028140 | | |
| Sextupole length | m | 0.4 | | |
| Strength of SF | $m^{-3}$ | 0.565 | | |
| Strength of SD | $m^{-3}$ | -1.1368 | | |
| Length of BH/BV | m | 0.3 | | |
| Maximum strength of BH | T | 0.02 | | |
| Maximum strength of BV | T | 0.02 | | |
| Maximum horizontal $\beta$ value | m | 171.1 | | |
| Minimum horizontal $\beta$ value | m | 29.9 | | |
| Maximum Vertical $\beta$ value | m | 171.1 | | |
| Minimum Vertical $\beta$ value | m | 29.9 | | |
| Maximum dispersion | m | 0.542 | | |
| **Dispersion suppressors** | | | | |
| Length | m | 202.36 | | |
| Horizontal phase advance/2π | | 0.5 | | |
| Vertical phase advance/2π | | 0.5 | | |
| Number of dipoles | | 20 | | |
| Dipole length | m | 4.64 | | |
| Strength of dipole at injection | T | 0.002817 | | |
| Strength of dipole at ejection | T | 0.0338 | 0.0225 | 0.0128 |
| Number of quadrupoles | | 4 | | |
| Quadrupole length | m | 1.0 | | |
| Strength of SF | $m^{-3}$ | 0.0 | | |
| Strength of SD | $m^{-3}$ | -0.0 | | |
| **Arcs** | **Unit** | **Value** | | |
| Length | m | 10219 | | |
| Number of cells per ARC | | 97 | | |
| Number of dispersion suppressors per arc | | 2 | | |
| Horizontal phase advance/2π | | 25.25 | | |
| Vertical phase advance/2π | | 25.25 | | |
| **Arc Magnet Type** | **Length [m]** | **Strength** | | |
| B | 4.64 | 0.002817~0.0338 [T] | | |
| QF/QD | 1.0 | 0.028 [$m^{-2}$] | | |
| SF | 0.4 | 0.565 [$m^{-3}$] | | |
| SD | 0.4 | -1.1368 [$m^{-3}$] | | |
| **Short Straight section** | **Unit** | **Value** | | |
| Length | m | 1201 | | |

| | | |
|---|---|---|
| Horizontal phase advance/2π | | 2.63 |
| Vertical phase advance/2π | | 2.62 |
| **SS Magnet Type** | **Length [m]** | **Strength** |
| QF/DM1 | 1 | 0.0232~0.0277 [$m^{-2}$] |
| **Long Straight section with RF** | **Unit** | **Value** |
| Length | m | 3421 |
| Horizontal phase advance/2π | | 16.82 |
| Vertical phase advance/2π | | 15.87 |
| **LS Magnet Type** | **Length [m]** | **Strength** |
| QF/DRF | 1.5 | 0.0608 [$m^{-2}$] |
| QF/DM4 | 1 | 0.023~ 0.0569 [$m^{-2}$] |
| **Bypasses at IP** | **Unit** | **Value** |
| Length | m | 3929 |
| Number of cells with dipoles | | 8 |
| Number of dispersion suppressors | | 0 |
| Horizontal phase advance/2π | | 10.02 |
| vertical phase advance/2π | | 10.00 |
| **Bypass Magnet Type** | **Length [m]** | **Strength** |
| B2 | 3.08 | 0.002817~0.0338 [T] |
| QF/DM2 | 1 | 0.0281~0.0296 [$m^{-2}$] |
| **The main RF system parameter** | | |
| Frequency | GHz | 1.3 |
| Harmonic number | | 433633 |

| | | | | |
|---|---|---|---|---|
| Cavity type | | 9-cell cavity | | |
| Cavity operating voltage (extraction) | MV | 20.5 | 9.1 | 9.0 |
| Cavity operating gradient | MV/m | 19.8 | 8.8 | 8.6 |
| Number of cavities per cryomodule | | 8 | | |
| Cavity active length (nice-cells) | m | 1.038 | | |
| Cryomodule length | m | 12 | | |
| Total number of cryomodules | | 12 | 8 | 4 |
| Total number of cavities | | 96 | 64 | 32 |

## A3: Linac, Damping Ring and Sources

| **Main parameter of Linac** | **Unit** | **Value** |
|---|---|---|
| Electron beam energy [$E_{e-}$] | GeV | 10 |
| Positron beam energy [$E_{e+}$] | GeV | 10 |
| Repetition rate [$f_{rep}$] | Hz | 100 |
| Electron bunch population [$N_{e-}$] | | $9.4\times10^9$ |
| Positron bunch population [$N_{e+}$] | | $9.4\times10^9$ |
| Energy spread ($e^+/e^-$) [$\sigma_E$] | | $< 2\times10^{-3}$ |
| Emittance ($e^-$) | nm | 120 |
| Emittance ($e^+$) | nm | 120 |
| **Electron Gun** | | |
| Gun type | Thermionic Triode Gun | |
| Cathode | Y796 (Eimac) Dispenser | |
| Beam Current (max.) | A | 15 |
| High Voltage of Anode | kV | 150-200 |
| Bias Voltage of Grid | V | 0 ~ -200 |
| Pulse duration （FWHM） | ns | 1 |
| Repetition Rate | Hz | 100 |
| Electron operation | nC | 3.3 |
| Positron operation | nC | 11 |
| **Positron source** | | |
| Electron beam energy on the target | GeV | 4 |
| Electron bunch charge on the target | nC | 10 |
| Target material | | W |
| Target thickness | mm | 15 |
| Flux Concentrator | T | 5.5 |
| Positron bunch charge after capture | nC | 3.0 |
| Positron energy after capture section | MeV | > 200 |
| **Damping Ring** | | |
| Energy | GeV | 1.1 |
| Circumference | m | 58.5 |
| Repetition frequency | Hz | 100 |
| Bending radius | m | 3.62 |
| Dipole strength $B_0$ | T | 1.01 |
| $U_0$ | keV | 35.8 |
| Damping time x/y/z | ms | 12/12/6 |
| $\delta_0$ | % | 0.05 |
| Nature $\varepsilon_0$ | mm.mrad | 287.4 |
| Nature $\sigma_z$ | mm | 7 (23 ps) |
| $\varepsilon_{inj}$ | mm.mrad | 2500 |
| $\varepsilon_{ext\ x/y}$ | mm.mrad | 704/471 |
| $\delta_{inj}$ /$\delta_{ext}$ | % | 0.3/0.06 |
| Energy acceptance by RF | % | 1.0 |
| $f_{RF}$ | MHz | 650 |
| $V_{RF}$ | MV | 1.8 |
| | | |

| **Accelerating structure** | | |
|---|---|---|
| Operation frequency | MHz | 2860 |
| Operation temperature | °C | 30.0 ± 0.1 |
| Number of cells | | 84 +2 coupler cells |
| Section length | mm | 3048 |
| Phase advance per cell | | 2π/3 - mode |
| Cell length | mm | 34.966 |
| Shunt impedance (r0) | MW/m | 60 ~ 69 |
| Q factor | | 15465~15370 |
| Group velocity (vg/c) | | 0.020 ~ 0.0080 |
| Filling time | ns | 850 |
| Attenuation factor | Neper | 0.50 |

# Appendix 2: Technical Component List

| | System | Name | Number | Typical parameters | remarks |
|---|---|---|---|---|---|
| **1** | **Magnet** | | | | |
| | Collider | B0I | 1160 | Dual aperture dipole, Gap 66mm, Field 550Gs, Length 28.7m. | |
| | | B1I | 32 | Dual aperture dipole, Gap 66mm, Field 275Gs, Length 28.7m. | |
| | | B0O | 1160 | Dual aperture dipole, Gap 66mm, Field 550Gs, Length 28.7m. | |
| | | B1O | 32 | Dual aperture dipole, Gap 66mm, Field 275Gs, Length 28.7m. | |
| | | BMV01IRD | 4 | Dipole, Gap 37mm, Field 150Gs, Length 61m. | |
| | | BMVIRD | 20 | Dipole, Gap 66mm, Field 710Gs, Length 44.2m. | |
| | | BMHIRD | 16 | Dipole, Gap 66mm, Field 310Gs, Length 28.5m. | |
| | | BGM1 | 32 | Dipole, Gap 66mm, Field 300Gs, Length 31.8m. | |
| | | BGM2 | 32 | Dipole, Gap 66mm, Field 220Gs, Length 9.7m. | |
| | | BMV01IRU | 4 | Dipole, Gap 37mm, Field 150Gs, Length 93.4m. | |
| | | BMVIRU | 20 | Dipole, Gap 66mm, Field 300Gs, Length 67m. | |
| | | BMHIRU | 16 | Dipole, Gap 66mm, Field 130Gs, Length 45m. | |
| | | BRF0 | 2 | Dipole, Gap 66mm, Field 180Gs, Length 50m. | |
| | | BRF | 16 | Dipole, Gap 66mm, Field 410Gs, Length 34.1m. | |
| | | QI | 1192 | Dual aperture quadrupole, Gap 75mm, Field 12T/m, Length 2.0m. | |
| | | QHI | 4 | Dual aperture quadrupole, Gap 75mm, Field 12T/m, Length 1.0m. | |
| | | QO | 1192 | Dual aperture quadrupole, Gap 75mm, Field 12T/m, Length 2.0m. | |
| | | QHO | 4 | Dual aperture quadrupole, Gap 75mm, Field 12T/m, Length 1.0m. | |
| | | Q3IRD | 4 | Quadrupole, Gap 37mm, Field 53T/m, Length 1.0m. | |
| | | Q4IRD | 4 | Quadrupole, Gap 37mm, Field 57T/m, Length 1.0m. | |
| | | Q5IRD | 4 | Quadrupole, Gap 37mm, Field 6T/m, Length 1.0m. | |
| | | QDVHIRD | 32 | Quadrupole, Gap 66mm, Field 18T/m, Length 0.5m. | |
| | | QFVIRD | 16 | Quadrupole, Gap 66mm, Field 18T/m, Length 1.0m. | |
| | | QFHHIRD | 32 | Quadrupole, Gap 66mm, Field 22T/m, Length 0.6m. | |
| | | QDHIRD | 16 | Quadrupole, Gap 66mm, Field 21T/m, Length 1.3m. | |
| | | QCMIRD | 24 | Quadrupole, Gap 66mm, Field 22T/m, Length 3.0m. | |
| | | QDCIRD | 16 | Quadrupole, Gap 66mm, Field 19T/m, Length 0.5m. | |
| | | QFCIRD | 8 | Quadrupole, Gap 66mm, Field 19T/m, Length 1.0m. | |
| | | QMIRD | 32 | Quadrupole, Gap 66mm, Field 23T/m, Length 3.5m. | |
| | | QMIRD | 16 | Quadrupole, Gap 66mm, Field 21T/m, Length 0.6m. | |
| | | QMIRD | 8 | Quadrupole, Gap 66mm, Field 21T/m, Length 1.3m. | |
| | | QDGMH | 64 | Quadrupole, Gap 66mm, Field 21T/m, Length 0.6m. | |
| | | QFGM | 32 | Quadrupole, Gap 66mm, Field 21T/m, Length 1.3m. | |
| | | Q3IRUJ2 | 4 | Quadrupole, Gap 37mm, Field 34T/m, Length 1.0m. | |
| | | Q4IRUJ2 | 4 | Quadrupole, Gap 37mm, Field 33T/m, Length 1.0m. | |
| | | QFVIRU | 16 | Quadrupole, Gap 66mm, Field 12T/m, Length 1.0m. | |
| | | QDVHIRU | 32 | Quadrupole, Gap 66mm, Field 12T/m, Length 0.5m. | |
| | | QDHIRU | 16 | Quadrupole, Gap 66mm, Field 17T/m, Length 1.0m. | |
| | | QFHHIRU | 32 | Quadrupole, Gap 66mm, Field 18T/m, Length 0.5m. | |
| | | QCMIRU | 24 | Quadrupole, Gap 66mm, Field 20T/m, Length 3.0m. | |
| | | QDCIRU | 16 | Quadrupole, Gap 66mm, Field 13T/m, Length 0.5m. | |
| | | QFCIRU | 8 | Quadrupole, Gap 66mm, Field 13T/m, Length 1.0m. | |
| | | QMIRU | 16 | Quadrupole, Gap 66mm, Field 17T/m, Length 1.0m. | |
| | | QMIRU | 8 | Quadrupole, Gap 66mm, Field 20T/m, Length 2.0m. | |
| | | QMIRU | 8 | Quadrupole, Gap 66mm, Field 13T/m, Length 2.0m. | |
| | | QMIRU | 16 | Quadrupole, Gap 66mm, Field 13T/m, Length 1.0m. | |
| | | QSEP | 20 | Quadrupole, Gap 66mm, Field 23T/m, Length 1.2m. | |
| | | QRF | 132 | Quadrupole, Gap 66mm, Field 23T/m, Length 2.3m. | |
| | | QFSTRH | 8 | Quadrupole, Gap 66mm, Field 10T/m, Length 1.2m. | |
| | | QFSTR | 12 | Quadrupole, Gap 66mm, Field 10T/m, Length 2.3m. | |
| | | QSEP | 12 | Quadrupole, Gap 75mm, Field 15T/m, Length 1.0m. | |
| | | QSTRH | 4 | Quadrupole, Gap 75mm, Field 11T/m, Length 1.0m. | |
| | | QSTR | 32 | Quadrupole, Gap 75mm, Field 11T/m, Length 2.0m. | |
| | | QAI | 40 | Quadrupole, Gap 75mm, Field 11T/m, Length 2.0m. | |
| | | QHAI | 4 | Quadrupole, Gap 75mm, Field 11T/m, Length 1.0m. | |
| | | QHAO | 4 | Quadrupole, Gap 75mm, Field 11T/m, Length 1.0m. | |
| | | QAO | 44 | Quadrupole, Gap 75mm, Field 13T/m, Length 2.0m. | |
| | | QSTR | 32 | Quadrupole, Gap 75mm, Field 11T/m, Length 2.0m. | |
| | | QSTRH | 4 | Quadrupole, Gap 75mm, Field 11T/m, Length 1.0m. | |
| | | QSEP | 12 | Quadrupole, Gap 75mm, Field 15T/m, Length 1.0m. | |
| | | QDI | 48 | Quadrupole, Gap 75mm, Field 13T/m, Length 2.0m. | |
| | | QSTRHI | 8 | Quadrupole, Gap 75mm, Field 12T/m, Length 1.0m. | |
| | | QINJI | 32 | Quadrupole, Gap 75mm, Field 5T/m, Length 2.0m. | |
| | | QFI | 44 | Quadrupole, Gap 75mm, Field 12T/m, Length 2.0m. | |
| | | QDO | 48 | Quadrupole, Gap 75mm, Field 13T/m, Length 2.0m. | |
| | | QFSTRHO | 8 | Quadrupole, Gap 75mm, Field 12T/m, Length 1.0m. | |
| | | QINJO | 32 | Quadrupole, Gap 75mm, Field 5T/m, Length 2.0m. | |
| | | QFO | 44 | Quadrupole, Gap 75mm, Field 12T/m, Length 2.0m. | |
| | | Q1IRD | 2 | Superconducting quadrupole, Field 136T/m, Length 2.0m. | |
| | | Q2IRD | 2 | Superconducting quadrupole, Field 110T/m, Length 1.48m. | |
| | | Q1IRUJ2 | 2 | Superconducting quadrupole, Field 136T/m, Length 2.0m. | |
| | | Q2IRUJ2 | 2 | Superconducting quadrupole, Field 110T/m, Length 1.48m. | |
| | | SFI | 448 | Sextupole, Gap 80mm, Field 740T/m^2, Length 0.7m. | |
| | | SDI | 448 | Sextupole, Gap 80mm, Field 740T/m^2, Length 1.4m. | |
| | | SFO | 448 | Sextupole, Gap 80mm, Field 740T/m^2, Length 0.7m. | |
| | | SDO | 448 | Sextupole, Gap 80mm, Field 740T/m^2, Length 1.4m. | |
| | | VSCIRD | 16 | Sextupole, Gap 66mm, Field 200T/m^2, Length 0.3m. | |
| | | HSCIRD | 16 | Sextupole, Gap 66mm, Field 590T/m^2, Length 0.3m. | |
| | | SC0IRD | 4 | Sextupole, Gap 66mm, Field 590T/m^2, Length 1m. | |
| | | VSCIRU | 16 | Sextupole, Gap 66mm, Field 200T/m^2, Length 0.3m. | |
| | | HSCIRU | 16 | Sextupole, Gap 66mm, Field 470T/m^2, Length 0.3m. | |
| | | SC0IRU | 4 | Sextupole, Gap 66mm, Field 600T/m^2, Length 1m. | |
| | | VSIRD | 8 | Superconducting sextupole, Gap 66mm, Field 1635T/m^2, Length 0.6m. | |
| | | HSIRD | 8 | Superconducting sextupole, Gap 66mm, Field 1882T/m^2, Length 0.8m. | |
| | | VSIRU | 8 | Superconducting sextupole, Gap 66mm, Field 1562T/m^2, Length 0.6m. | |
| | | HSIRU | 8 | Superconducting sextupole, Gap 66mm, Field 1999T/m^2, Length 0.6m. | |
| | | CH | 2904 | Horizontal corrector, Gap 66mm, Field 0.2T, Length 0.88m. | |
| | | CV | 2904 | Vertical corrector, Gap 85mm, Field 0.2T, Length 0.88m. | |
| | | Anti-solenoid | 4 | Max field 7.2T, total length 4.8m | |
| | | | | | |
| | Booster | BDISARC | 640 | Dipole, Gap 63mm, Field 338Gs, Length 2.35m. | |
| | | BARC | 15360 | Dipole, Gap 63mm, Field 338Gs, Length 4.71m. | |
| | | BIR | 320 | Dipole, Gap 63mm, Field 392Gs, Length 1.68m. | |
| | | QDM | 140 | Quadrupole, Gap 63mm, Field 11.8T/m, Length 1.0m. | |
| | | QFM | 106 | Quadrupole, Gap 63mm, Field 12.2T/m, Length 1.0m. | |
| | | QFMRF | 4 | Quadrupole, Gap 63mm, Field 12.4T/m, Length 1.5m. | |
| | | QDMRF | 4 | Quadrupole, Gap 63mm, Field 15.3T/m, Length 1.5m. | |
| | | QFARC | 848 | Quadrupole, Gap 63mm, Field 11.1T/m, Length 1.0m. | |
| | | QDARC | 816 | Quadrupole, Gap 63mm, Field 11.1T/m, Length 1.0m. | |
| | | QFRF | 60 | Quadrupole, Gap 63mm, Field 16.6T/m, Length 2.2m. | |
| | | QDRF | 58 | Quadrupole, Gap 63mm, Field 16.6T/m, Length 2.2m. | |
| | | SF | 224 | Sextupole, Gap 63mm, Field 217T/m^2, Length 0.4m. | |
| | | SD | 224 | Sextupole, Gap 63mm, Field 437T/m^2, Length 0.4m. | |
| | | CH | 48 | Horizontal corrector, Gap 63mm, Field 0.02T, Length 0.58m. | |
| | | CV | 302 | Vertical corrector, Gap 63mm, Field 0.02T, Length 0.58m. | |
| | | CHb | 856 | Horizontal corrector (attached to dipoles), Gap 66mm, Field 0.0025T, Length 4.71m. | |
| | | | | | |
| | Linac | B | 4 | Dipole, Gap 34mm, Field 1T, Length 2.356m. | |
| | | CB1 | 4 | Dipole, Gap 54mm, Field 0.5T, Length 0.279m. | |
| | | AM1 | 2 | Dipole, Gap 34mm, Field 0.3T, Length 0.262m. | |
| | | AM2 | 1 | Dipole, Gap 44mm, Field 0.8T, Length 5.236m. | |
| | | AM3 | 1 | Dipole, Gap 44mm, Field 1T, Length 5.847m. | |
| | | 100Q | 48 | Quadrupole, Gap 100mm, Field 10T/m, Length 0.3m. | |
| | | 60SQ | 6 | Quadrupole, Gap 60mm, Field 15T/m, Length 0.1m. | |
| | | 60LQ | 3 | Quadrupole, Gap 60mm, Field 15T/m, Length 0.2m. | |
| | | 40SQ | 88 | Quadrupole, Gap 40mm, Field 28T/m, Length 0.2m. | |
| | | 40LQ | 50 | Quadrupole, Gap 40mm, Field 28T/m, Length 0.4m. | |
| | | 32SQ | 38 | Quadrupole, Gap 32mm, Field 36T/m, Length 0.3m. | |
| | | 32LQ | 19 | Quadrupole, Gap 32mm, Field 36T/m, Length 0.6m. | |
| | | S1 | 1 | Solenoid,Aperture 100 mm, Field 0.1T, Max.Length 80mm | |
| | | S2 | 1 | Solenoid,Aperture 100 mm, Field 0.1T, Max.Length 120mm | |
| | | S3 | 20 | Solenoid,Aperture 100 mm, Field 0.1T, Max.Length 50mm | |
| | | S4 | 15 | Solenoid,Aperture 400 mm, Field 0.5T, Max.Length 1m | |
| | | FS1 | 4 | Solenoid,Aperture 90 mm, Field 0.06T, Max.Length 80mm | |
| | | 100C | 17 | Corrector x and y, Gap 100mm, Field 0.015T, Length 0.25m. | |
| | | L60C | 3 | Corrector x and y, Gap 60mm, Field 0.015T, Length 0.1m. | |
| | | L40C | 46 | Corrector x and y, Gap 40mm, Field 0.08T, Length 0.1m. | |
| | | L32C | 19 | Corrector x and y, Gap 32mm, Field 0.085T, Length 0.2m. | |
| | | | | | |
| | Damping Ring | BARC | 32 | Dipole, Gap 36mm, Field 1.015T, Length 0.71m. | |
| | | LTB44B | 24 | Dipole, Gap 44mm, Field 1T, Length 0.646m. | |
| | | LTBCB | 8 | Dipole, Gap 44mm, Field 1T, Length 1.614m. | |
| | | DRLAM | 2 | Lambertson,Gap 44mm, Field 1T, Length 0.646m. | |
| | | DRkicker | 2 | Kicker,Gap 44mm, Field 0.06T, Length 0.5m. | |
| | | DR36Q | 48 | Quadrupole, Gap 36mm, Field 20T/m, Length 0.2m. | |
| | | LTD54Q | 44 | Quadrupole, Gap 54mm, Field 20T/m, Length 0.2m. | |
| | | DR36S | 24 | Sextupole, Gap 36mm, Field 160T/m^2, Length 60mm. | |
| | | L36C | 24 | Corrector, Gap 36mm, Field 0.025T, Length 0.1m. | |
| | | L54C | 40 | Corrector, Gap 54mm, Field 0.025T, Length 0.1m. | |
| | | | | | |
| **2** | **SRF** | | | | |
| | Collider | 650 MHz 2-cell Nb cavity | 240 | $Q_0$ > 4E10 at 22 MV/m | |
| | | 650 MHz power coupler | 240 | > 300 kW CW | |
| | | 650 MHz HOM coupler | 480 | > 1 kW CW | |
| | | 650 MHz cavity tuner | 240 | slow tuning > 340 kHz, fast tuning > 1.5 kHz | |
| | | 650 MHz cavity HOM absorber | 80 | > 5 kW CW at room temperature | |
| | | 650 MHz cryomodule | 40 | 11 m | |
| | | 650 MHz cavity vacuum system | 40 | < 5E-8 Pa | |
| | | | | | |
| | Booster | 1.3 GHz 9-cell Nb cavity | 96 | $Q_0$ > 3E10 at 24 MV/m | |
| | | 1.3 GHz power coupler | 96 | > 30 kW peak, 4 kW average | |
| | | 1.3 GHz cavity tuner | 96 | slow tuning > 420 kHz, fast tuning > 1 kHz | |
| | | 1.3 GHz cavity HOM absorber | 12 | > 10 W CW at 80 K | |
| | | 1.3 GHz cryomodule | 12 | 12 m | |
| | | 1.3 GHz cavity vacuum system | 12 | < 5E-8 Pa | |
| | | | | | |
| | Damping Ring | 650 MHz 5-cell Cu cavity | 2 | 1 MV/m | |
| | | 650 MHz power coupler | 2 | 50 kW CW | |
| | | 650 MHz tuner | 4 | plunger range -20 mm to +40 mm | |
| | | cavity vacuum and cooling system | 2 | | |
| | | | | | |
| **3** | **RF Power Source** | | | | |
| | Collider | Klystron | 120 | 800kW | |
| | | PSM Power Supply | 120 | 110kV/15A | |
| | | Circulator | 120 | 800kW | |
| | | Load | 120 | 800kW | |
| | | Phase shift | 120 | 800kW | |
| | | Directional Coupler | 360 | 30dB | |
| | | Waveguide | 120 | 650MHz/800kW | |
| | | LLRF | 120 | 0.10% | |
| | | | | | |
| | Booster | SSA | 96 | 1.3GHz/25kW | |
| | | Waveguide | 96 | 1.3GHz/25kW | |
| | | LLRF | 96 | 0.10% | |
| | | | | | |
| | Linac | Klystron | 74 | 80MW | |
| | | Modulator | 74 | 400kV | |
| | | | | | |
| | Damping Ring | SSA | 2 | 650MHz/50kW | |
| | | Waveguide | 2 | 650MHz/50kW | |
| | | LLRF | 2 | 0.10% | |
| | | | | | |
| | Electron Source | Gun body | 1 | 150kV | |
| | | High Voltage Power Supply | 1 | 200kV | |
| | | Cathode grid | 1 | 10A | |
| | | Accessory equipment | 1 | 200kV | |
| | | | | | |
| **4** | **Survey and Alignment** | | | | |
| | | GPS | 16 | 0.5 mm/km | |
| | | measuring accessory | 8 | | |
| | | Zenith plummet | 8 | 1 / 200000 | |
| | | gyroscope | 4 | 3" | |
| | | Relative gravimeter | 2 | 1μgal | |
| | | laser tracker | 36 | 0.015mm/m | |
| | | reflector target | 424 | 0.5"-1.5" | |
| | | FARO arm | 8 | 0.025-0.036mm/2.4M | |
| | | optical level | 32 | 0.1mm/km | |
| | | digital level | 8 | 0.2mm/km | |
| | | transit square | 32 | 0.0254mm | |
| | | total station | 16 | 0.5" | |
| | | electronic gradienter | 32 | 0.001mrad | |
| | | vision instrument | 11 | 1μm+1ppm | |
| | | Automatic mobile platform | 9 | | |
| | | hydrostatic level | 200 | 0.03mm | |
| | | laser collimator | 4 | 0.1mm/100m | |
| | | indoor GPS | 4 | 0.1mm | |
| | | alignment telescope | 4 | CZW | |
| | | tool microscope | 4 | 0.01mm | |
| | | CMM | 4 | 0.005mm | |
| | | laser interferometer | 8 | 0.005mm | |
| | | baseline guide | 4 | 0.2mm/60m | |
| | | measuring tool | 4 | 0.005-0.02mm | |
| | | Marble reference parts | 16 | 0.005-0.02mm | |
| | | marble table | 16 | 1.6x0.1 | |
| | | marble table | 16 | 2.4x1.6 | |
| | | adjust software | 2 | MMA | |
| | | permanent control point | 16 | 0.1mm | |
| | | ground control point | 35872 | 0.1mm | |
| | | wall control point | 35872 | 0.1mm | |
| | | instrument stand | 80 | 0.5m-1.5m | |
| | | lifting stand | 32 | 0.3m-1.5m | |
| | | target stand | 32 | 0.7m | |
| | | All-around target | 1000 | 0.1mm | |
| | | target pillar | 96 | 2m | |
| | | reference parts | 52 | 0.01mm | |
| | | measuring equipment | 60 | | |
| | | alignment tools | 32 | | |
| | | Setting out tool | 16 | 0.5mm | |
| | | UPS power | 36 | 3KW | |
| | | Instrument trolley | 40 | TC-8A | |
| | | two way radio | 32 | 30km | |
| | | notebook PC | 36 | | |
| | | PC | 20 | | |
| | | work station | 12 | | |
| | | Epoxy glue | 35872 | Araldite | |
| | | Gasket | 251104 | | |
| | | Expansion bolt | 143488 | M8 | |
| | | Diamond drilling machine | 20 | Φ76 | |
| | | Impact drill | 20 | Φ16 | |
| | | Electric hammer | 10 | Φ24 | |
| | | pick-up truck | 4 | | |
| | | measurement hut | 16 | | |
| | | Instrument transportation | 300 | | |
| | | | | | |
| **5** | **Instrument-ation** | | | | |
| | Collider | | | | |
| | | BPM | 2900 | | |
| | | BLM | 5800 | | |
| | | DCCT | 2 | | |
| | | BCM | 2 | | |
| | | Transverse Feedback | 2 | | |
| | | Longitudinal Feedback | 2 | | |
| | | Synchrotron light monitor | 4 | | |
| | | Tune measurement | 2 | | |
| | | | | | |
| | Booster | | | | |
| | | BPM | 1808 | | |
| | | BLM | 3600 | | |
| | | DCCT | 2 | | |
| | | BCM | 2 | | |
| | | Transverse Feedback | 2 | | |
| | | Longitudinal Feedback | 2 | | |
| | | | | | |
| | Linac + 2 transport lines | | | | |
| | | BPM | 180 | | |
| | | Profile | 80 | | |
| | | ICT | 50 | | |
| | | Beam energy and spread | 3 | | |
| | | beam emittance | 4 | | |
| | | Bunch charge measurement | 4 | | |
| | | | | | |
| **6** | **Vacuum** | | | | |
| | Collider | | | | |
| | | vacuum chamber | 34000 | 75´56H*V | |
| | | Bellow | 24000 | RF shielding | |
| | | Pump | 35200 | 100L/s | |
| | | Gauge | 4320 | ccg | |
| | | Valve | 1040 | RF-CF100 | |
| | | | | | |
| | Booster | | | | |
| | | vacuum chamber | 17000 | 316L | |
| | | Bellow | 12000 | 316L | |
| | | Pump | 8400 | 50L/s | |
| | | Gauge | 2160 | CCG | |
| | | Valve | 520 | DN100 | |
| | | | | | |
| | Linac | | | | |
| | | vacuum chamber | 300 | 304/316L | |
| | | Bellow | 300 | 304/316L | |
| | | Pump | 1060 | 200L/s | |
| | | Gauge | 530 | CCG | |
| | | Valve | 30 | CF100 | |
| | | | | | |
| **7** | **Linac** | | | | |
| | Positron source | | | | |
| | | Positron conversion device | 1 | 6 Tesla | |
| | | Flux concentrator power supply | 1 | 15kA/15kV | |
| | | Positron focusing coil | 10 | 0.5 Tesla | |
| | | DC power supply | 6 | 600A/500V | |
| | | | | | |
| | RF system | | | | |
| | | sub-harmonic buncher | 2 | 142.8375MHz/571.35MHz | |
| | | buncher | 1 | $f_0$: 2856.75MHz | |
| | | Accelerating structure | 279 | Accelerating gradient：21MV/m | |
| | | Big-hole accelerating structure | 6 | Accelerating gradient：21MV/m@φ25mm | |
| | | RF Pulse compressor | 74 | Energy multiplication factor: 1.6 | |
| | | Waveguide system | 74 | VSWR⩽1.1 | |
| | | RF measurement system | 74 | VSWR⩽1.1 | |
| | | Low level RF | 74 | | |
| | | | | | |
| **8** | **Cryogenic System** | | | | |
| | for SC cavities | | | | |
| | | Refrigerator (18 kW@ 4.5 K) | 4 | 18kW@4.5K | |
| | | Main distribution valve box | 4 | heat loss: ≤30W, vacuum insulation leakage rate : ≤1E-8 Pa.m3/s | |
| | | Connection valve box | 52 | heat loss: ≤15W, vacuum insulation leakage rate: ≤1E-8 Pa.m3/s | |
| | | Cryogenic transfer line | 4000 | heat loss: ≤0.5 W/m, vacuum insulation leakage rate: ≤1E-8 Pa.m3/s | |
| | | Medium pressure Helium storage tank | 32 | working pressure: 2 MPa, volume: 100m3, helium purity:99.999% | |
| | | High pressure and high purity helium cylinder | 4 | working pressure: 20 MPa, volume: 25m3 | |
| | | High pressure impure helium cylinder | 4 | working pressure: 20 MPa, volume: 25m3 | |
| | | High pressure helium gas recovery compressor | 8 | flow rate:100m3/h, working pressure: 20MPa | |
| | | High pressure helium purifier | 4 | flow rate:200m3/h, working pressure: 20MPa | |
| | | | | | |
| | for SC magnets | | | | |
| | | Refrigerator (3 kW, 4.5 K) | 2 | 3kW@4.5K | |
| | | Liquid Helium Dewar | 2 | 5000L/tank | |
| | | Main distribution valve box | 2 | heat loss: ≤30W, vacuum insulation leakage: ≤1E-8 Pa.m3/s | |
| | | Connection valve box | 36 | heat loss: ≤15W, vacuum insulation leakage: ≤1E-8 Pa.m3/s | |
| | | Cryogenic transfer line | 4000 | heat loss: ≤0.5 W/m, vacuum insulation leakage rate: ≤1E-8 Pa.m3/s | |
| | | Medium pressure Helium tank | 8 | working pressure: 1.6 MPa, volume: 100m3, helium purity:99.999% | |
| | | High pressure and high purity helium cylinder | 2 | working pressure: 20 MPa, volume: 25m3 | |
| | | High pressure impure helium cylinder | 6 | working pressure: 20 MPa, volume: 25m3 | |
| | | High pressure helium gas recovery compressor | 4 | flow rate:100m3/h, working pressure: 20MPa | |
| | | High pressure helium purifier | 2 | flow rate:100m3/h, working pressure: 20MPa | |
| | | | | | |
| **9** | **Magnet Supports** | | | | |
| | Collider | SPT_D28686_C | 2384 | Range and accuracy of adjustment: | |
| | | SPT_D60976_C | 4 | X≥±20mm， | |
| | | SPT_D44200_C | 20 | Y≥±30mm， | |
| | | SPT_D28450_C | 16 | Z≥±20mm， | |
| | | SPT_D31843_C | 32 | θX≥±10mrad， | |
| | | SPT_D9667_C | 32 | θY≥±10mrad， | |
| | | SPT_D93378_C | 4 | θZ≥±10mrad， | |
| | | SPT_D68950_C | 20 | | |
| | | SPT_D44950_C | 16 | | |
| | | SPT_D50000_C | 2 | | |
| | | SPT_D34074_C | 16 | | |
| | | SPT_Q2000T_C | 2384 | | |
| | | SPT_Q1000T_C | 8 | | |
| | | SPT_Q1000_C | 172 | | |
| | | SPT_Q500_C | 128 | | |
| | | SPT_Q625_C | 112 | | |
| | | SPT_Q1250_C | 56 | | |
| | | SPT_Q3000_C | 48 | | |
| | | SPT_Q3500_C | 32 | | |
| | | SPT_Q2000_C | 412 | | |
| | | SPT_Q1150_C | 28 | | |
| | | SPT_Q2350_C | 144 | | |
| | | SPT_S700/1400_C | 448 | | |
| | | SPT_S300_C | 64 | | |
| | | SPT_S1000_C | 8 | | |
| | | SPT_C875_C | 5808 | | |
| | | SPT_Q2000SC_C | 4 | | |
| | | SPT_Q1480SC_C | 4 | | |
| | | SPT_S300SC_C | 32 | | |
| | | SPT_VacuumT_C | 34000 | | |
| | | SPT_valve_C | 1040 | | |
| | | SPT_BI_C | 2900 | | |
| | | | | | |
| | Booster | SPT_D4711_B | 15360 | | |
| | | SPT_D2356_B | 640 | | |
| | | SPT_D1683_B | 320 | | |
| | | SPT_D5000_LTB | 48 | | |
| | | SPT_D4000_LTB | 28 | | |
| | | SPT_D5000_BTC | 68 | | |
| | | SPT_Q1000_B | 1910 | | |
| | | SPT_Q1500_B | 8 | | |
| | | SPT_Q2200_B | 118 | | |
| | | SPT_Q900_LTB | 80 | | |
| | | SPT_Q2000_BTC | 40 | | |
| | | SPT_S400_B | 448 | | |
| | | SPT_C583_B | 350 | | |
| | | SPT_C200_LTB | 24 | | |
| | | SPT_C300_BTC | 30 | | |
| | | SPT_Sep1000_LTB | 4 | | |
| | | SPT_Sep1000_BTC | 140 | | |
| | | SPT_K500_LTB | 2 | | |
| | | SPT_K1000_BTC | 20 | | |
| | | SPT_VacuumT_B | 17000 | | |
| | | SPT_valve_B | 520 | | |
| | | SPT_BI_B | 1808 | | |
| | | | | | |
| | Linac | SPT_D2356_L | 4 | | |
| | | SPT_D279_L | 4 | | |
| | | SPT_D262_L | 2 | | |
| | | SPT_D5236_L | 1 | | |
| | | SPT_D5847_L | 1 | | |
| | | SPT_Q300_L | 48 | | |
| | | SPT_Q400_L | 6 | | |
| | | SPT_Qtri600_L | 3 | | |
| | | SPT_Qtri1200_L | 44 | | |
| | | SPT_Qtri1800_L | 19 | | |
| | | SPT_So80_L | 5 | | |
| | | SPT_So120_L | 1 | | |
| | | SPT_So50_L | 20 | | |
| | | SPT_So1000_L | 15 | | |
| | | SPT_C250_L | 17 | | |
| | | SPT_C100_L | 49 | | |
| | | SPT_C200_L | 19 | | |
| | | SPT_AccT3000_L | 277 | | |
| | | SPT_AccT2000_L | 6 | | |
| | | SPT_VacuumT_L | 500 | | |
| | | SPT_valve_L | 30 | | |
| | | SPT_BI_L | 86 | | |
| | | | | | |
| | Damping Ring | SPT_D710_DR | 32 | | |
| | | SPT_D646_DR | 24 | | |
| | | SPT_D1614_DR | 8 | | |
| | | SPT_Q200_DR | 92 | | |
| | | SPT_S60_DR | 24 | | |
| | | SPT_C100_DR | 64 | | |
| | | SPT_Lam646_DR | 2 | | |
| | | SPT_Kicker500_DR | 2 | | |
| | | SPT_AccT3000_DR | 2 | | |
| | | SPT_VacuumT_DR | 60 | | |
| | | SPT_valve_DR | 4 | | |
| | | SPT_BI_DR | 40 | | |
| | | | | | |
| **10** | **Collimators** | | | | |
| | | Movable collimator | 20 | | |
| | | Fixed collimator | 80 | | |
| | | | | | |
| **11** | **Radiation Protection** | | | | |
| | | PLC | 38 | | PPS |
| | | Interlock key | 39 | | PPS |
| | | Access control system | 1 | | PPS |
| | | LED | 92 | | PPS |
| | | Video surveillance | 92 | | PPS |
| | | Sheilding door | 37 | | PPS |
| | | Interlock door | 37 | | PPS |
| | | Interlock system research platform | 1 | | PPS |
| | | Areal monitor detector | 300 | | Dose detection |
| | | Environmental monitor detector | 21 | | Dose detection |
| | | OSL Personal Dose Detector System | 3 | | Dose detection |
| | | CR-39 Personal Dose Detector System | 2 | | Dose detection |
| | | Cooling Water Monitoring Detector | 16 | | Dose detection |
| | | Air Ventilation Monitoring Detector | 16 | | Dose detection |
| | | Portable Monitoring Detectors | 21 | | Dose detection |
| | | Low background gamma spectrum | 2 | | Dose detection |
| | | Liquid scintillation counter | 2 | | Dose detection |
| | | Beam dump | 3 | | Dose detection |
| | | | | | |
| **12** | **Power Supply** | | | | |
| | Collider | | | | |
| | | | | **Stability /8hours** | **Output Rating** |
| | | D-aperture Dipole | 8 | 100ppm | 1170A/1200V |
| | | S-aperture Dipole | 162 | 100ppm | 180A/50V |
| | | D-aperture Quadrupole | 192 | 100ppm | 180A/750V |
| | | S-aperture Quadrupole1 | 920 | 100ppm | 180A/70V |
| | | S-aperture Quadrupole2 | 80 | 100ppm | 180A/240V |
| | | S-aperture Quadrupole3 | 22 | 100ppm | 180A/850V |
| | | Sextupole1 | 896 | 100ppm | 180A/70V |
| | | Sextupole2 | 72 | 100ppm | 40A/75V |
| | | Corrector1 | 2904 | 500ppm | 40A/75V |
| | | Corrector2 | 2904 | 500ppm | 40A/110V |
| | | SCQ | 8 | 50ppm | 2700A/11V |
| | | Anti-Solenoid | 4 | 50ppm | 1100A/16V |
| | | SC-Corrector1 | 8 | 100ppm | 140A/10V |
| | | SC-Corrector2 | 16 | 100ppm | 55A/13V |
| | | | | | |
| | | | | **Electrode length/width** | **Nominal gap** |
| | | Electrostatic separator | 48 | 4.0m/260mm | 110mm |
| | | | | | |
| | Booster | | | | |
| | | | | **Stability /tracking error** | **Output Rating** |
| | | Dipole | 16 | 500ppm / 0.1% | 940A/820V |
| | | Quadrupole | 32 | 500ppm / 0.1% | 320A/2100V |
| | | Sext.D | 16 | 1000ppm / 0.1% | 140A/650V |
| | | Sext.F | 16 | 1000ppm / 0.1% | 140A/650V |
| | | Corrector | 350 | 1000ppm | 25A /20V |
| | | | | | |
| | Transport Line | | | | |
| | | | | **Stability /8hours** | **Output Rating** |
| | | Dipole1 | 60 | 500ppm | 340A/50V |
| | | Dipole2 | 4 | 500ppm | 340A/200V |
| | | Quadrupole | 120 | 500ppm | 20A/15V |
| | | Corrector | 54 | 500ppm | 25A/5V |
| | | | | | |
| | Linac | | | | |
| | | Dipole | 11 | 500ppm | 240A/200V |
| | | Quadrupole | 177 | 500ppm | 150A/40V |
| | | Solenoid-1 | 22 | 500ppm | 11A/30V |
| | | Solenoid-2 | 5 | 500ppm | 400A/460V |
| | | Corrector | 110 | 500ppm | 40A/7V |
| | | | | | |
| | Damping Ring | | | | |
| | | | | **Stability /tracking error** | **Output Rating** |
| | | Dipole1 | 32 | 500ppm | 240A/25V |
| | | Dipole2 | 6 | 500ppm | 300A/35V |
| | | Quadrupole1 | 52 | 500ppm | 140A/11V |
| | | Quadrupole2 | 13 | 500ppm | 160A/30V |
| | | Sextupole | 2 | 500ppm | 10A/7V |

# Appendix 3: Electric Power Requirement

The main power consumption is by the RF power source, magnet power supplies, cryogenics system, heat removal devices.

The wall-plug power consumption for RF is estimated in Table A3.1 for the Collider and in Table A3.2 for the Booster. The Linac klystrons consume considerable power and that is estimated in Table A3.3.

**Table A3.1:** Collider RF Wall Plug Power Efficiency

| | | |
|---|---|---|
| Wall to PSM power supply/modulator | 95% | |
| Modulator to klystron | 96% | |
| Klystron to waveguide | 70% | |
| Waveguide to coupler | 95% | |
| Coupler to cavity | ~100% | |
| Cavity to beam | ~100% | |
| Overall efficiency | ~60.6% | |
| LLRF control | 5% more power | |

Required wall plug power for Collider RF in Higgs and W modes = beam power × 1.05 / 0.606 = 103.8 MW (beam power = 60 MW for operation in Higgs and W modes). Required wall plug power for Collider RF in Z mode = beam power × 1.05 / 0.606 = 57.1 MW (beam power = 33 MW for operation in Z mode).

**Table A3.2:** Booster RF Wall Plug Power Efficiency

| | | |
|---|---|---|
| Wall to SSA power supply | 95% | |
| SSA number | 96 | |
| Operation pulsed power | 18.2 kW | Higgs mode |
| Duty factor | 3.8% | Higgs mode |
| SSA to waveguide | 50% | |
| Waveguide to coupler | 98% | |
| Coupler to cavity | ~100% | |
| Cavity to beam | ~100% | |
| LLRF control | 5% more power | |

Required wall plug power for Booster RF in Higgs mode =1.05 × 96 × 18.2 × 3.8% /50% /95% /98% = 0.15 MW.

**Table A3.3:** Linac Wall Plug Power Efficiency

| | | |
|---|---|---|
| Wall to Modulator | 95% | |
| Modulator to klystron | 70% | |
| Klystron to accelerator structure | 40% | |
| Klystron operation pulsed power | 60 MW | |
| Duty factor | 0.04% | |
| Klystron number | 64 | |

Required wall plug power for Linac RF = 64 × 60 × 0.04% /40% /70% /95%= 5.8 MW.

Power consumption by magnet power supplies for three different operating modes is summarized in tables A3.4, A3.5 and A3.6.

Magnet power consumption is about 70% of the total loss. The magnet power supplies use switched mode. The estimated efficiency is about 90% for full load, and 80% for low load. The power loss of power supplies can be calculated as follows: (cable loss + magnet loss) /9.

Cable loss: Magnets in two adjacent half-arcs will be connected in series and powered by one or two power supplies, housed in ground level halls. Power supplies for single magnet loads will be installed in auxiliary stub tunnels around the main tunnel. Cables are copper. Cable current density is less than 2A/mm2 and the cable loss is about 20% of the total loss.

**Table A3.4:** Power consumption of magnet power supplies in H mode

| Power consumption for H (30 MW /beam) | | | | |
|---|---|---|---|---|
| Location | Magnet | Power supply | Magnet cable | Total |
| Collider | 33.7631 | 4.7214 | 8.7297 | 47.21 |
| Booster | 7.6037 | 1.1621 | 2.8547 | 11.62 |
| Transport Line | 0.8382 | 0.1056 | 0.1124 | 1.06 |
| Linac | 1.1747 | 0.1454 | 0.1337 | 1.45 |
| Damping Ring | 0.1925 | 0.0300 | 0.0772 | 0.30 |
| IR | 0.0000 | 0.0255 | 0.2299 | 0.26 |
| Total | 43.5723 | 6.1900 | 12.1376 | 61.90 |

**Table A3.5:** Power consumption of magnet power supplies in W mode

| Power consumption for W (30 MW /beam) | | | | |
|---|---|---|---|---|
| Location | Magnet | Power supply | Magnet cable | Total |
| Collider | 15.9454 | 3.2267 | 4.1772 | 23.35 |
| Booster | 3.4236 | 0.7861 | 1.3096 | 5.52 |
| Transport Line | 0.3780 | 0.0715 | 0.0527 | 0.50 |
| Linac | 1.1747 | 0.1454 | 0.1337 | 1.45 |
| Damping Ring | 0.1925 | 0.0300 | 0.0772 | 0.30 |
| IR | 0.0000 | 0.0172 | 0.1035 | 0.12 |
| Total | 21.1143 | 4.2769 | 5.8538 | 31.24 |

**Table A3.6:** Power consumption of magnet power supplies in Z mode

| Power consumption for Z (16.5 MW /beam) | | | | |
|---|---|---|---|---|
| Location | Magnet | Power supply | Magnet cable | Total |
| Collider | 5.8165 | 2.1405 | 1.5583 | 9.52 |
| Booster | 1.1435 | 0.5259 | 0.4667 | 2.14 |
| Transport Line | 0.1269 | 0.0478 | 0.0201 | 0.19 |
| Linac | 1.1747 | 0.1454 | 0.1337 | 1.45 |
| Damping Ring | 0.1925 | 0.0300 | 0.0772 | 0.30 |
| IR | 0.0000 | 0.0115 | 0.0345 | 0.05 |
| Total | 8.4542 | 2.9011 | 2.2905 | 13.65 |

The cryogenic system for the SC cavities supplies 2K superfluid helium for the cryomodules in both the Collider and the Booster. When running in the Higgs mode, the total 4.5K equivalent heat load of Booster and Collider is 56.2 kW. With a 20% margin, 4 individual 18 kW at 4.5K refrigerators will be employed. The total cryogenic capacities are equivalent to 72 kW at 4.5K. The LHC large refrigerator, has a COP (Coefficient of Performance) at 4.5K of about 218.8 W/W and the required installed power is 15.75 MW. Using these figures, the CEPC installed power for Collider and Booster is 14.88 MW and 0.87 MW respectively. The IR consumption is 1.8 MW giving a total of 17.35 MW.

The cryogenics for SC Magnets supply 4.5K helium for the cryomodules of the IR magnets. The total 4.5K equivalent heat load of Booster and Collider is 5.73 kW. Two 3 kW at 4.5K refrigerators will be employed. The total cryogenic capacities are equivalent to 6 kW at 4.5K. Referred to the ADS refrigerator, the COP at 4.5K is about 300 W/W. The required installed power is 1.8MW.

The electrical consumption during operation for the working modes of H, W and Z, the calculated 4.5K equivalent heat load of Booster and Collider is shown in tables A3.7 and A3.8. COP is 218.8W/W. For the magnets of the IR regions, the calculated 4.5K equivalent heat load is 5.73 kW. The electrical consumption is 1.72 MW.

**Table A3.7:** Electrical consumption during operation for Cryogenics

| Cryogenics | Location and electrical consumption (MW) | | | | | Total (MW) |
|---|---|---|---|---|---|---|
| | Collider | Booster | LINAC | BTL | IR | |
| H | 11.62 | 0.68 | | | 1.72 | 14.02 |
| W | 5.94 | 0.41 | | | 1.72 | 8.07 |
| Z | 2.91 | 0.31 | | | 1.72 | 4.94 |

**Table A3.8:** The calculated 4.5K equiv. heat load of Booster and Collider

| | Collider | | | Booster | | |
|---|---|---|---|---|---|---|
| | H | W | Z | H | W | Z |
| Equiv. heat load at 4.5 K (kW) | 53.08 | 27.12 | 13.3 | 3.12 | 1.88 | 1.41 |

The power requirement for heat removal devices (water cooling, ventilation and air conditioning) is estimated to be about 15% of the total electric power, which is a figure we should strive to achieve with a careful design.

We add in other power consumption for vacuum, instrumentation, control, radiation protection, detectors (“experimental devices”) and general services to calculate the total power consumption for CEPC running in the H, W, Z modes and summarize it below in Tables A3.9, A3.10, and A3.11:

**Table A3.9:** Total facility power consumption in H mode

| | System for Higgs (30 MW /beam) | Location and Power Requirement (MW) | | | | | | Total (MW) |
|---|---|---|---|---|---|---|---|---|
| | | Collider | Booster | Linac | BTL | IR | Surface building | |
| 1 | RF Power Source | 103.8 | 0.15 | 5.8 | | | | 109.75 |
| 2 | Cryogenic System | 15.67 | 0.89 | | | 1.9 | | 18.36 |
| 3 | Vacuum System | 9.784 | 3.792 | 0.646 | | | | 14.22 |
| 4 | Magnet Power Supplies | 47.21 | 11.62 | 1.75 | 1.06 | 0.26 | | 61.9 |
| 5 | Instrumentation | 0.9 | 0.6 | 0.2 | | | | 1.7 |
| 6 | Radiation Protection | 0.25 | | 0.1 | | | | 0.35 |
| 7 | Control System | 1 | 0.6 | 0.2 | 0.005 | 0.005 | | 1.81 |
| 8 | Experimental Devices | | | | | 4 | | 4 |
| 9 | Utilities | 31.79 | 3.53 | 1.38 | 0.63 | 1.2 | | 38.53 |
| 10 | General Services | 7.2 | | 0.2 | 0.15 | 0.2 | 12 | 19.75 |
| | Total | 213.554 | 20.972 | 10.276 | 1.845 | 7.385 | 12 | 270.37 |

**Table A3.10:** Total facility power consumption in W mode

| | System for W (30 MW /beam) | Location and Power Requirement (MW) | | | | | | Total (MW) |
|---|---|---|---|---|---|---|---|---|
| | | Collider | Booster | Linac | BTL | IR | Surface building | |
| 1 | RF Power Source | 103.8 | 0.15 | 5.8 | | | | 109.75 |
| 2 | Cryogenic System | 5.94 | 0.41 | | | 1.72 | | 8.07 |
| 3 | Vacuum System | 9.784 | 3.792 | 0.646 | | | | 14.22 |
| 4 | Magnet Power Supplies | 23.35 | 5.52 | 1.75 | 0.5 | 0.12 | | 31.24 |
| 5 | Instrumentation | 0.9 | 0.6 | 0.2 | | | | 1.7 |
| 6 | Radiation Protection | 0.25 | | 0.1 | | | | 0.35 |
| 7 | Control System | 1 | 0.6 | 0.2 | 0.005 | 0.005 | | 1.81 |
| 8 | Experimental Devices | | | | | 4 | | 4 |
| 9 | Utilities | 26.24 | 2.92 | 1.38 | 0.58 | 1.2 | | 32.32 |
| 10 | General Services | 7.2 | | 0.2 | 0.15 | 0.2 | 12 | 19.75 |
| | Total | 178.464 | 13.992 | 10.276 | 1.235 | 7.245 | 12 | 223.21 |

**Table A3.11:** Total facility power consumption in Z mode

| | System for Z (16.5 MW /beam) | Location and Power Requirement (MW) | | | | | | Total (MW) |
|---|---|---|---|---|---|---|---|---|
| | | Collider | Booster | Linac | BTL | IR | Surface building | |
| 1 | RF Power Source | 57.1 | 0.15 | 5.8 | | | | 63.05 |
| 2 | Cryogenic System | 2.91 | 0.31 | | | 1.72 | | 4.94 |
| 3 | Vacuum System | 9.784 | 3.792 | 0.646 | | | | 14.22 |
| 4 | Magnet Power Supplies | 9.52 | 2.14 | 1.75 | 0.19 | 0.05 | | 13.65 |
| 5 | Instrumentation | 0.9 | 0.6 | 0.2 | | | | 1.7 |
| 6 | Radiation Protection | 0.25 | | 0.1 | | | | 0.35 |
| 7 | Control System | 1 | 0.6 | 0.2 | 0.005 | 0.005 | | 1.81 |
| 8 | Experimental Devices | | | | | 4 | | 4 |
| 9 | Utilities | 19.95 | 2.22 | 1.38 | 0.55 | 1.2 | | 25.3 |
| 10 | General Services | 7.2 | | 0.2 | 0.15 | 0.2 | 12 | 19.75 |
| | Total | 108.614 | 9.812 | 10.276 | 0.895 | 7.175 | 12 | 148.77 |

# Appendix 4: Advanced Partial Double Ring Scheme

## A4.1: Introduction

To reduce the CEPC cost and still have adequate performance, a Partial Double Ring (PDR) scheme has been proposed [1-2]. The energy saw-tooth effects in the PDR scheme can be solved by making correction in the partial-double-ring region and by using more RF stations [3]. However, it is found that beam loading effect is the main limitation for reaching the desired luminosities both for running as a Higgs factory and at the Z-pole. An Advanced Partial Double Ring (APDR) scheme has been proposed to solve this problem [4-5] and is defined as CEPC Alternative Scheme by CEPC steering committee on January 14th of 2017. The main issues in the APDR scheme, the energy saw-tooth and beam loading effects, will be addressed in this appendix.

The luminosity goals for the Higgs and Z modes are respectively $2 \times 10^{34} cm^{-2} s^{-1}$ and $1 \times 10^{34} cm^{-2} s^{-1}$ per interaction point. To achieve the goal luminosity for the Z mode, the bunch number should be larger than 3400. The detector requires the bunch spacing to be larger than 15 ns. Considering beam loading effects [6], the number of double ring regions should be large than 8. To achieve a smaller saw-tooth energy deviation, the RF stations are distributed uniformly between each of the two double rings. Thus, the number of RF stations should be larger than the number of the double ring regions. More RF stations raises the total cost. Finally, the length of each double ring is around 4 km. Fig. A4.1 shows the layout of the CEPC ARDR scheme.

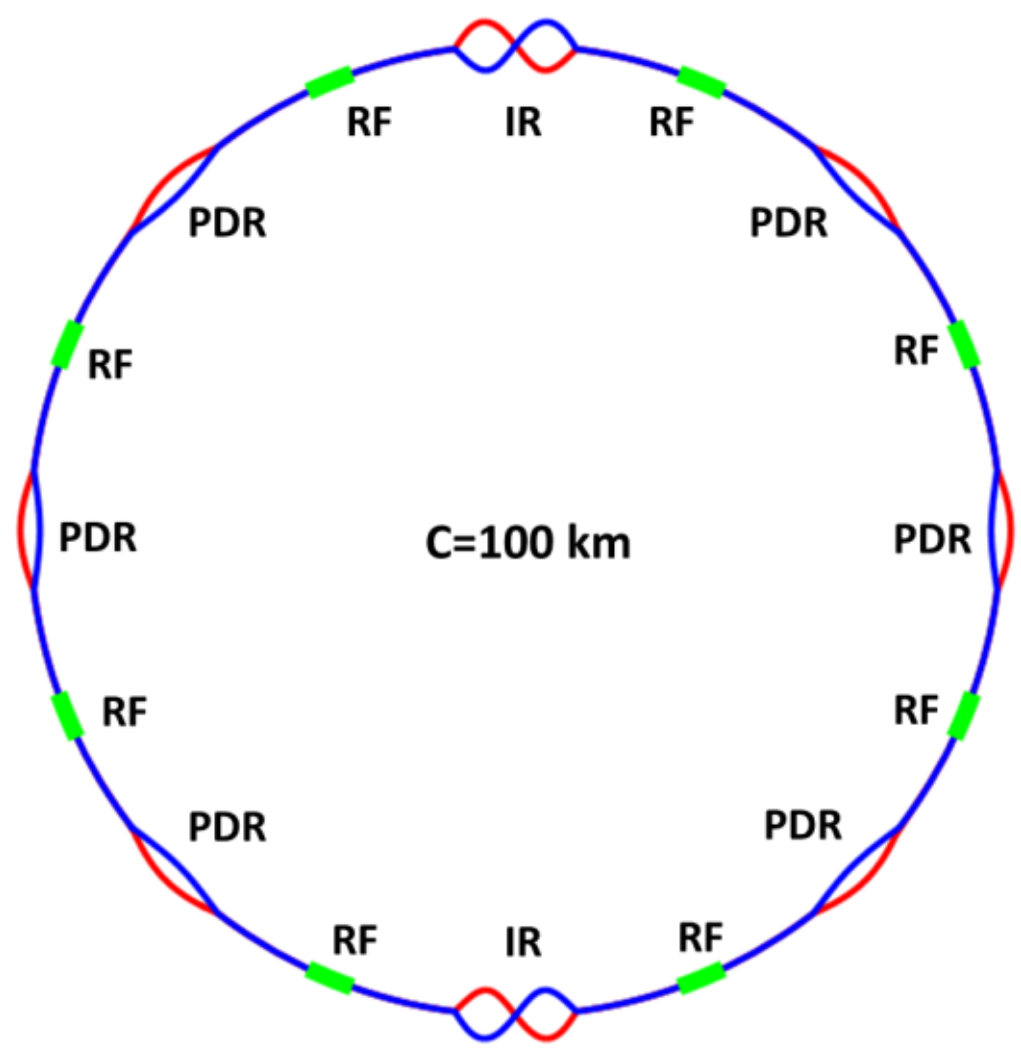

**Figure A4.1:** Layout of the CEPC advanced partial double ring.

## A4.2: Main Parameters

Same as the design of double ring scheme, one has to consider the key beam physics limitations, such as beam-beam effects, crab waist enhancement effect, beamstrahlung effect, and also the economical and technical limitations, such as synchrotron radiation power, RF phase shift induced by beam loading and high order mode power in each superconducting RF cavity. CEPC parameters for the APDR scheme have been listed in Table A.4.1.

The luminosities of W and Z are mainly limited by the beam loading effect. The maximum phase shift of RF system is controlled at 10 degrees. The design luminosity of H is $2.0\times10^{34}$ $cm^{-2}s^{-1}$ and the luminosity at Z pole is $1.02\times10^{34}$ $cm^{-2}s^{-1}$ which can reach the minimum design goal of CEPC.

The luminosities listed in Table A.4.1 include the bunch lengthening effect according to the impedance budget assuming the same material and size of beam pipe as the baseline design. The bunch lengthening is about 4% for H, 24% for W and 44% for Z.

**Table A4.1:** Main parameters of the APDR.

| | **Higgs** | **W** | **Z** |
|---|---|---|---|
| Number of IPs | 2 | 2 | 2 |
| Energy (GeV) | 120 | 80 | 45.5 |
| Circumference (km) | 100 | 100 | 100 |
| SR loss/turn (GeV) | 1.61 | 0.32 | 0.033 |
| Half crossing angle (mrad) | 16.5 | 16.5 | 16.5 |
| Piwinski angle | 2.28 | 4.4 | 8.83 |
| $N_e$/bunch ($10^{10}$) | 9.68 | 6.0 | 2.6 |
| Bunch number | 420 | 900 | 3400 |
| Beam current (mA) | 19.5 | 26.0 | 42.5 |
| SR power /beam (MW) | 31.4 | 8.3 | 1.41 |
| Bending radius (km) | 11.4 | 11.4 | 11.4 |
| Momentum compaction ($10^{-5}$) | 1.15 | 1.15 | 1.15 |
| $\beta_{IP}$ x/y (m) | 0.36/0.002 | 0.36/0.002 | 0.36/0.002 |
| Emittance x/y (nm) | 1.18/0.0036 | 0.52/0.0016 | 0.17/0.0029 |
| Transverse $\sigma_{IP}$ (um) | 20.6/0.085 | 13.7/0.056 | 7.85/0.076 |
| $\xi_x/\xi_y$/IP | 0.025/0.085 | 0.016/0.098 | 0.0097/0.049 |
| RF Phase (degree) | 128 | 135 | 151 |
| $V_{RF}$ (GV) | 2.03 | 0.45 | 0.069 |
| $f_{RF}$ (MHz) (harmonic) | 650 | 650 | 650 |
| Nature $\sigma_z$ (mm) | 2.75 | 2.96 | 2.92 |
| Total $\sigma_z$ (mm) | 2.85 | 3.68 | 4.2 |
| HOM power/cavity (kw) | 0.42 (2cell) | 0.16 (2cell) | 0.1(2cell) |
| Energy spread (%) | 0.096 | 0.064 | 0.036 |
| Energy acceptance (%) | 1.1 | | |
| Energy acceptance by RF (%) | 1.98 | 1.48 | 1.2 |
| $n_\gamma$ | 0.19 | 0.18 | 0.13 |
| Life time due to beamstrahlung_cal (minute) | 63 | | |
| $F$ (hour glass) | 0.93 | 0.963 | 0.987 |
| $L_{max}$/IP ($10^{34}$ $cm^{-2}s^{-1}$) | 2.0 | 2.12 | 1.02 |

## A4.3: Optics Design

The optics of APDR scheme is designed taking into account the Higgs mode running parameters and the beam loading effects for running at the Z pole. For the W and Z modes, the optics is obtained from the H mode by scaling down the magnet strength with energy.

***Interaction Region***

The design of the interaction region is basically the same as with the full double ring. To allow as many as 3400 bunches for each beam, the length of the interaction region is extended to be 4 km. Figs. A4.2 show the optics of the IR of APDR scheme.

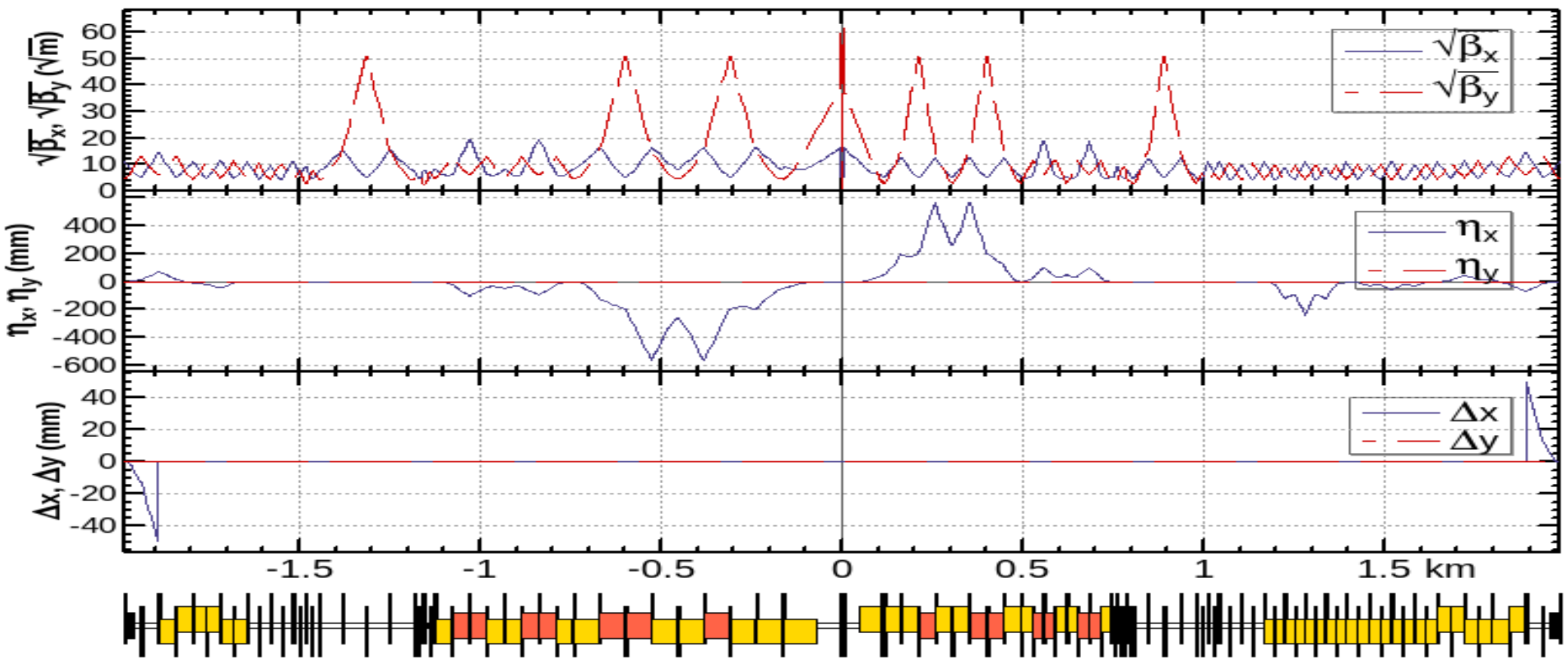

**Figure A4.2:** Optics of the interaction region.

***Arc Region***

In the arcs, the optics structure, phase advance and configuration are the same as with the double ring scheme. The dispersion suppressor at the ends of arc region was designed with half-bending-angle FODO structure. The optics of the arc region is shown in Fig. A4.3

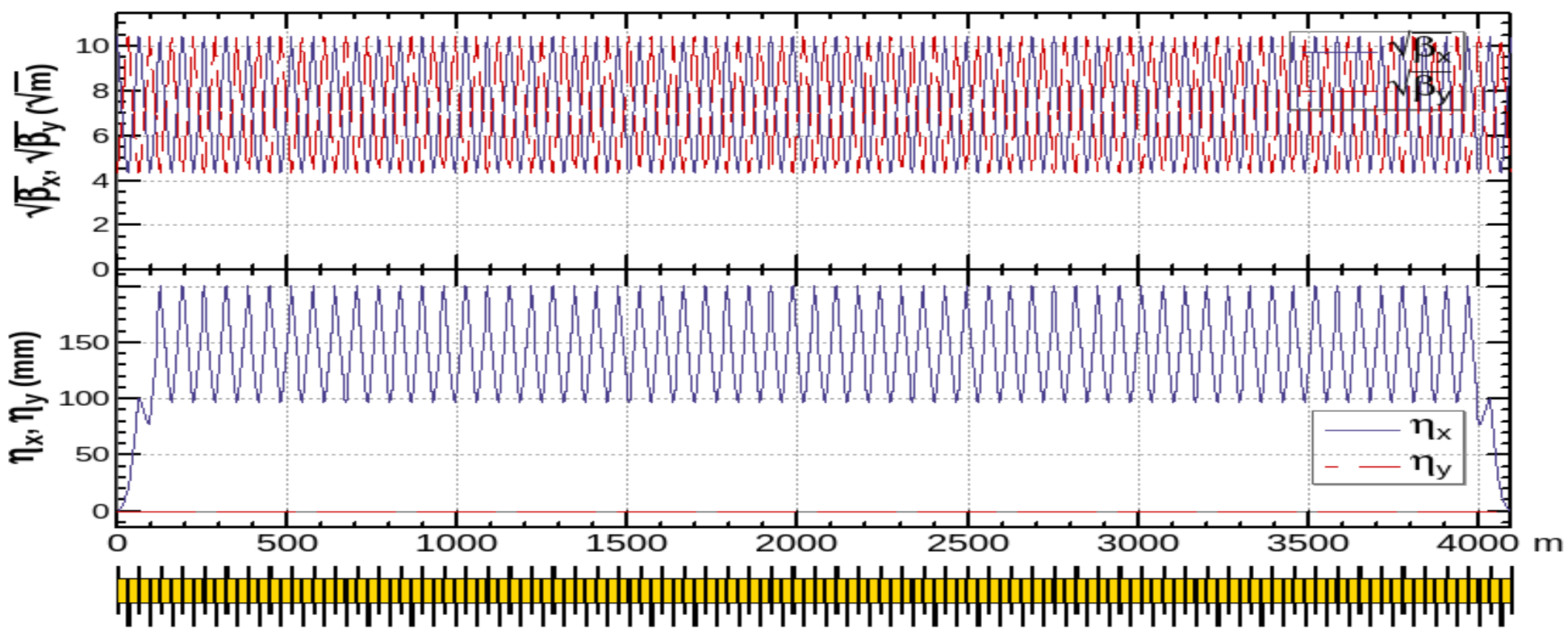

**Figure A4.3:** Optics of the arc region

***RF Region***

In the RF region, the RF cavities are shared by the two beams. Small average beta functions are favored to reduce the multi-bunch instability caused by the RF cavities. Phase advance of 90/90 degrees is chosen and a quadrupole distance of 13.7 m allows room for the chosen cryomodule. Fig. A4.4 shows the optics of RF region.

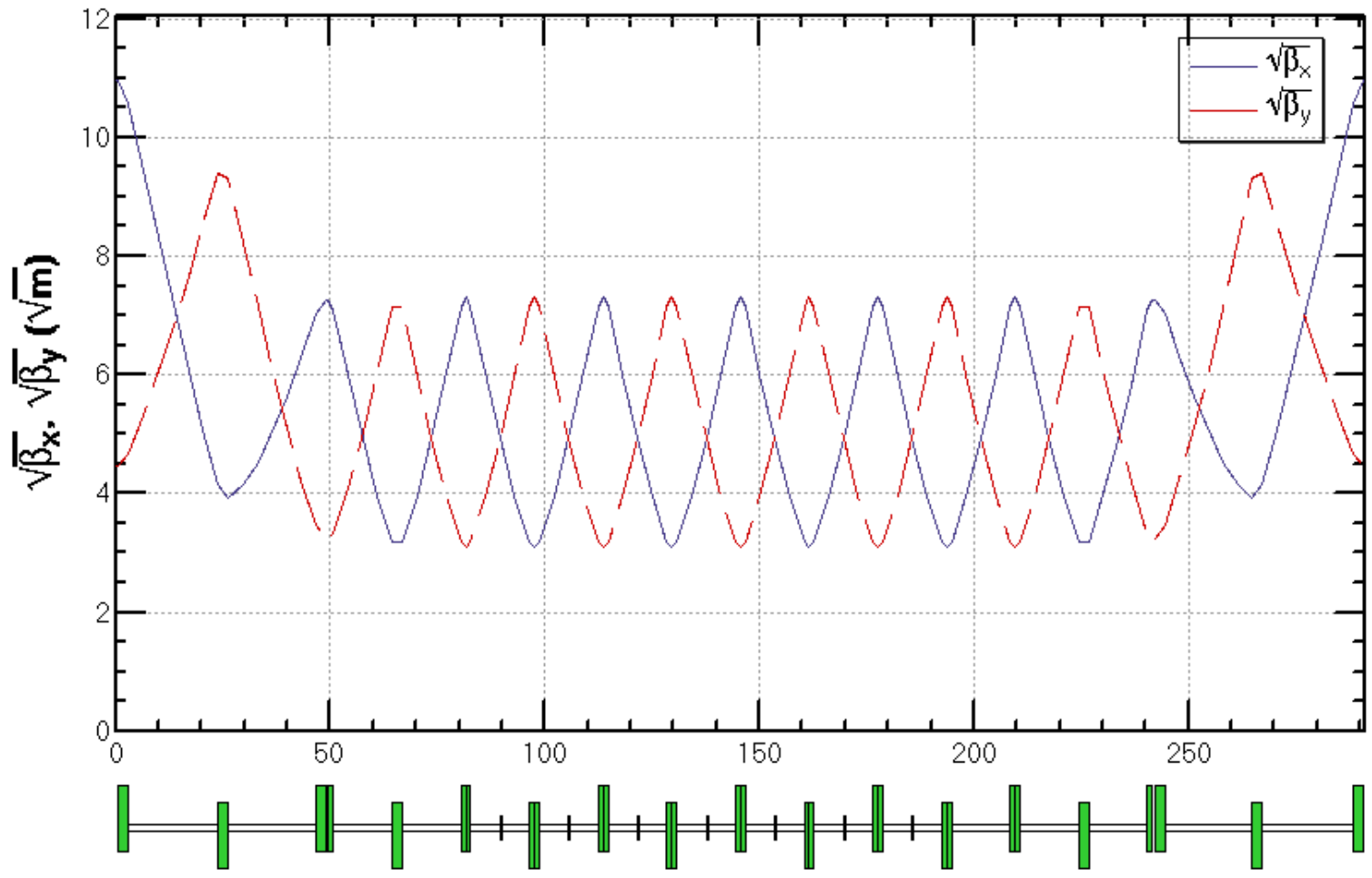

**Figure A4.4:** Optics of the RF region.

***Double Ring Region***

The length of double ring region is 4 km and there are a total of 6 double ring regions. There are two kinds of double ring regions. The four DRs near IP1 and IP3 include beam separation, arc and straight sections. The two DRs at IP2 and IP4 include beam separation and long straight section. The straight sections are designed for the compatibility of SPPC. A4.5 show the optics of the two kinds of double ring region.

The two beams are separated by 6 electrostatic separators and further separated by a quadrupole. The gradient of the electrostatic separators is 2.0 MV/m. After the quadruple, there is a drift space to make the two beam separation as large as 10 cm at the entrance of the independent quadrupoles. The deviation of the two beams in the double ring region is 0.35 m.

Same as the full double ring, twin-aperture dipoles and quadrupoles are used in the arc parts of the double ring region to reduce the power. The distance of 0.35 m between the two beams is chosen.

The optics of whole ring is shown in A4.6. The non-zero closed orbits indicate the separation region.

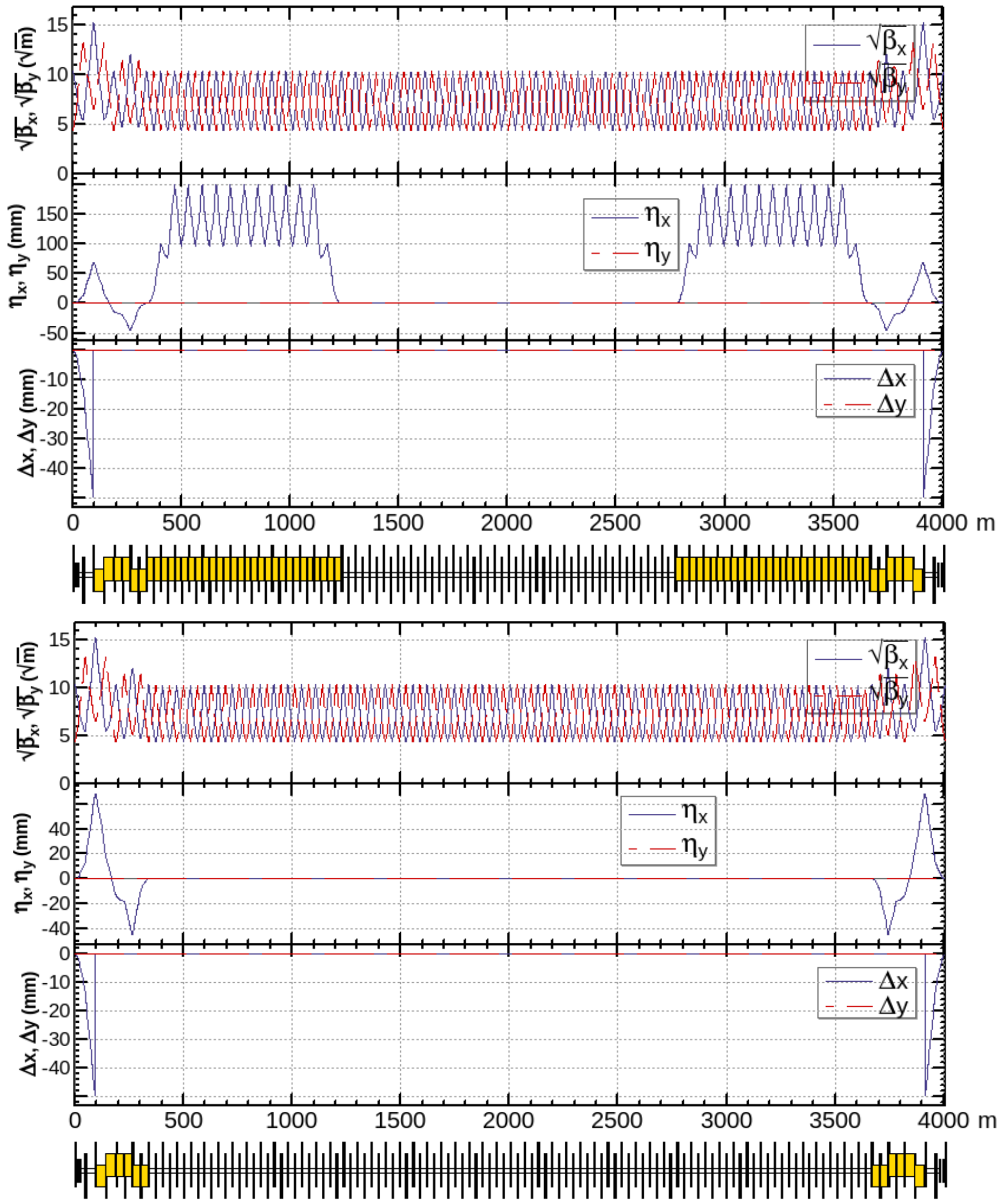

**Figure A4.5:** The optics of the double ring regions (outer rings).

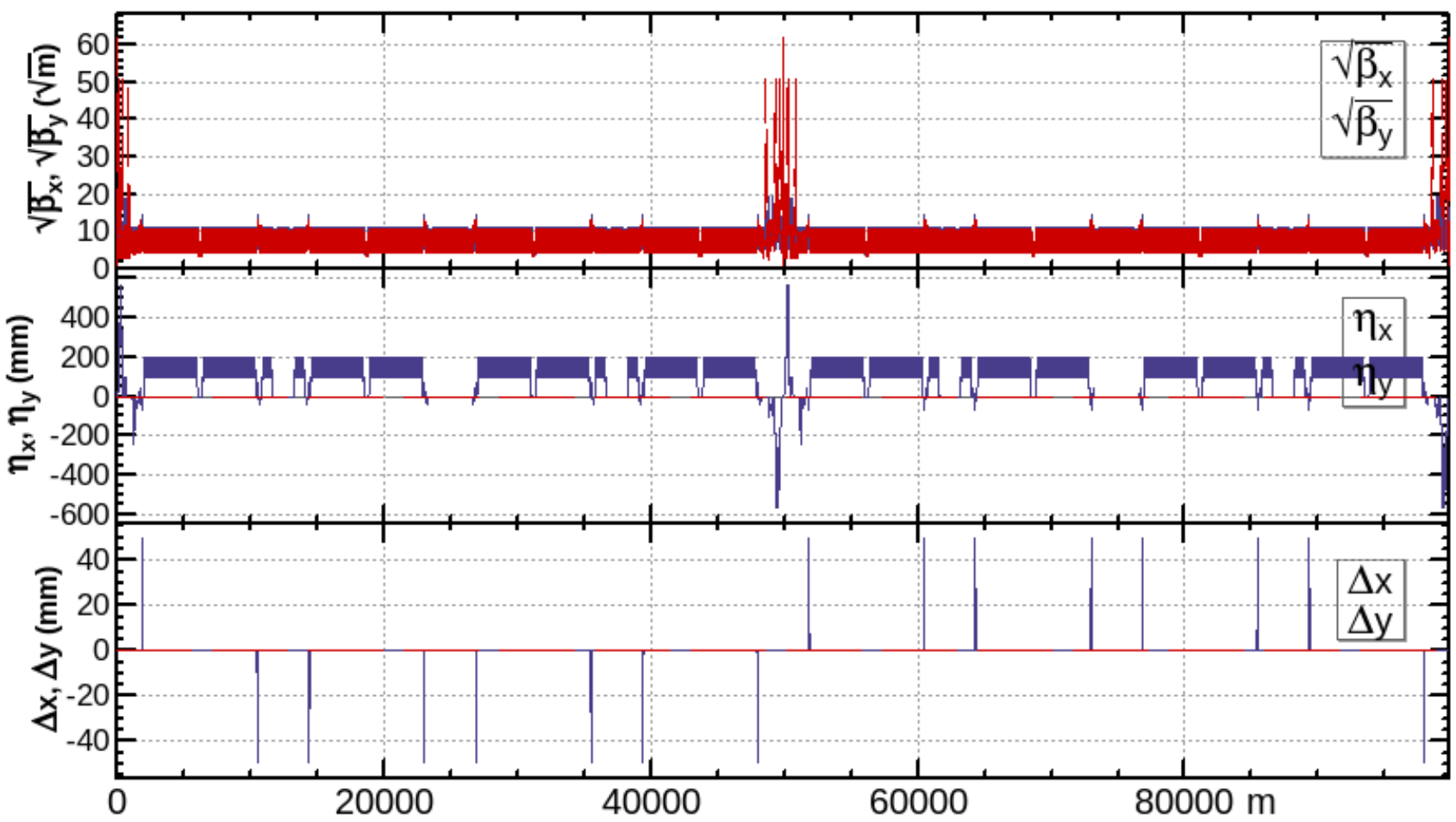

**Figure A4.6:** Main Ring optics

## A4.4: Energy Saw-tooth Effect

The beam energy loss in the dipoles and their gain in the cavities leads to a saw-tooth energy profile. This phenomenon will be significant with a beam energy as high as 120 GeV and will lead to significant closed orbit and optics functions distortion, and thus an increased emittance and a reduced dynamic aperture. For the double ring scheme, the effects of the energy saw-tooth can be corrected by tapering the magnet strength with beam energy along the beam line independently in the two rings [7].

For the APDR, the effects of energy saw-tooth can't be completely corrected as most of the circumference of the APDR consists of a single ring. Fortunately, the mitigation of the saw-tooth effect with the APDR scheme is possible. With tapering, orbit and optics correction in the double ring region, the orbit and optics distortion only occur in the arc and RF region. Then the emittance growth is negligible and the dynamic aperture reduction is small.

Figs. A4.8 and A4.9 show the energy saw-tooth, closed orbit and optical function distortions without and with tapering, respectively. Without tapering, the relative energy deviation is around 0.1% and the distortion of the horizontal closed orbit is around 1 mm. With tapering, orbit and optics correction in the double ring region, the distortion of the horizontal closed orbit is around 1 um; the relative distortion of the beta functions is less than 1% and the emittance growth is less than 1%. Re-optimization of the dynamic aperture with arc sextupoles is done by constraining symmetry for the two interaction points and the two beams. Fig. A4.10 shows the dynamic aperture result before and after optimization [8].

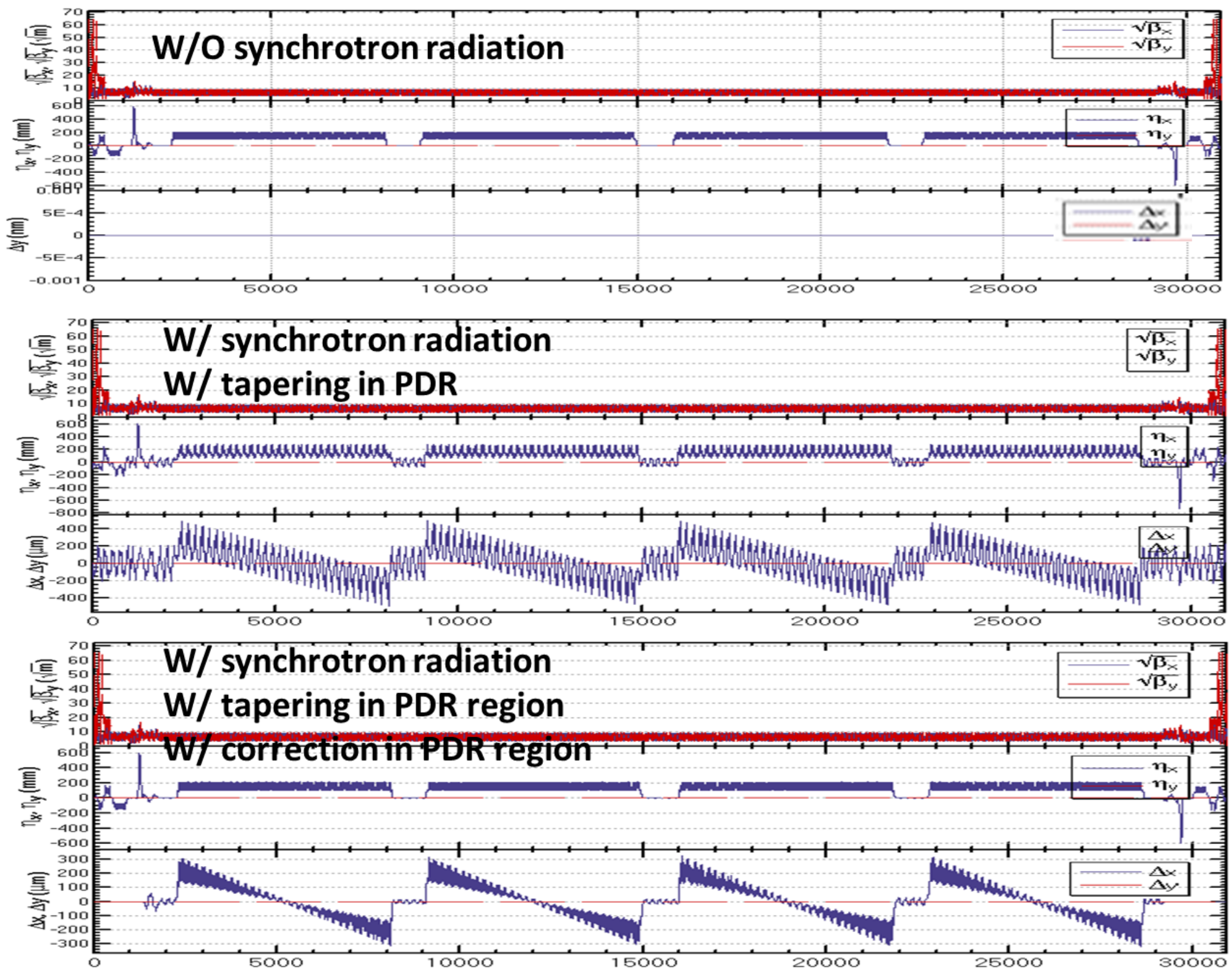

**Figure A4.8:** The closed orbit and optics functions for the whole ring.

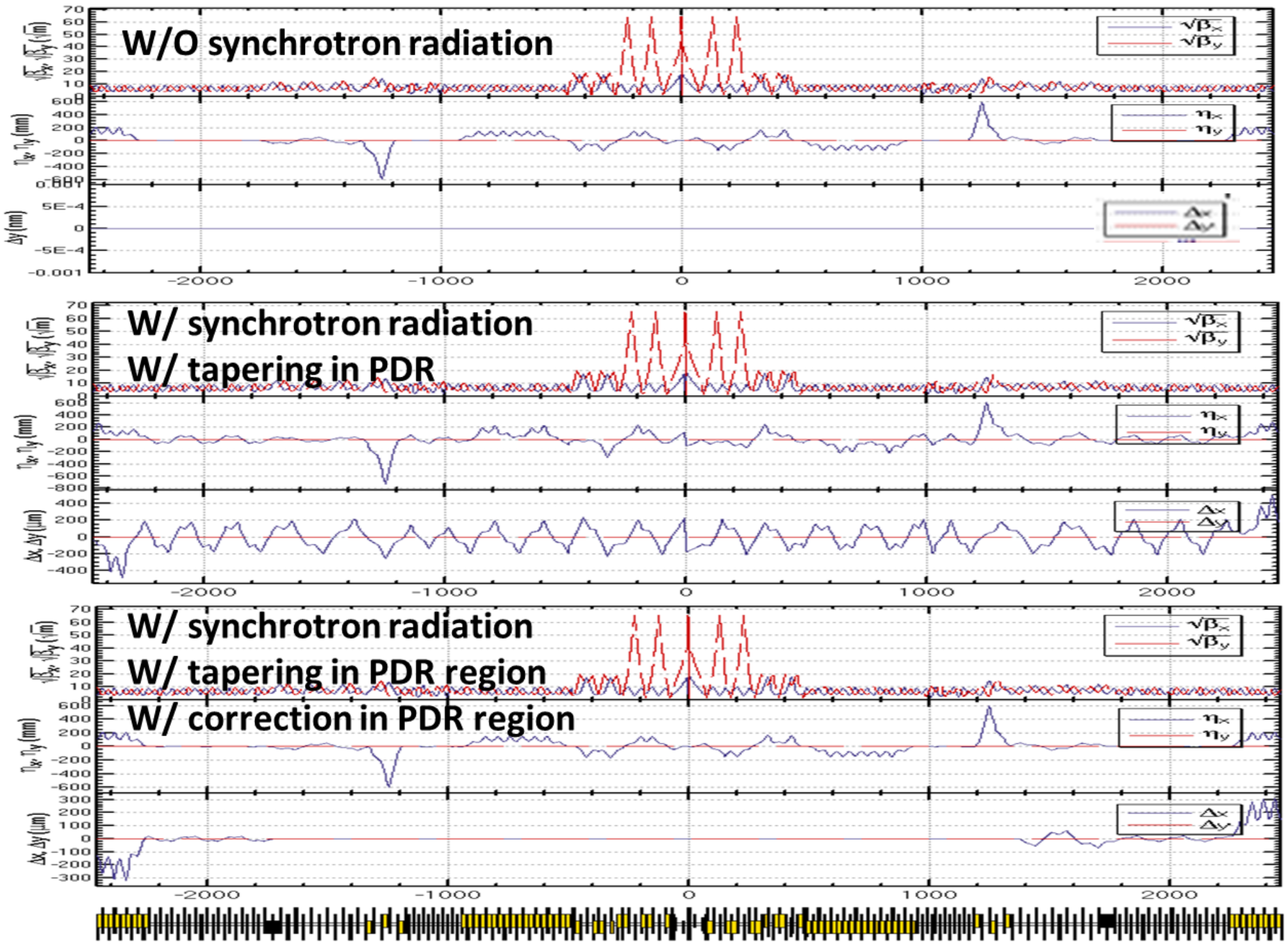

**Figure A4.9:** The closed orbit and optics functions for the double ring region.

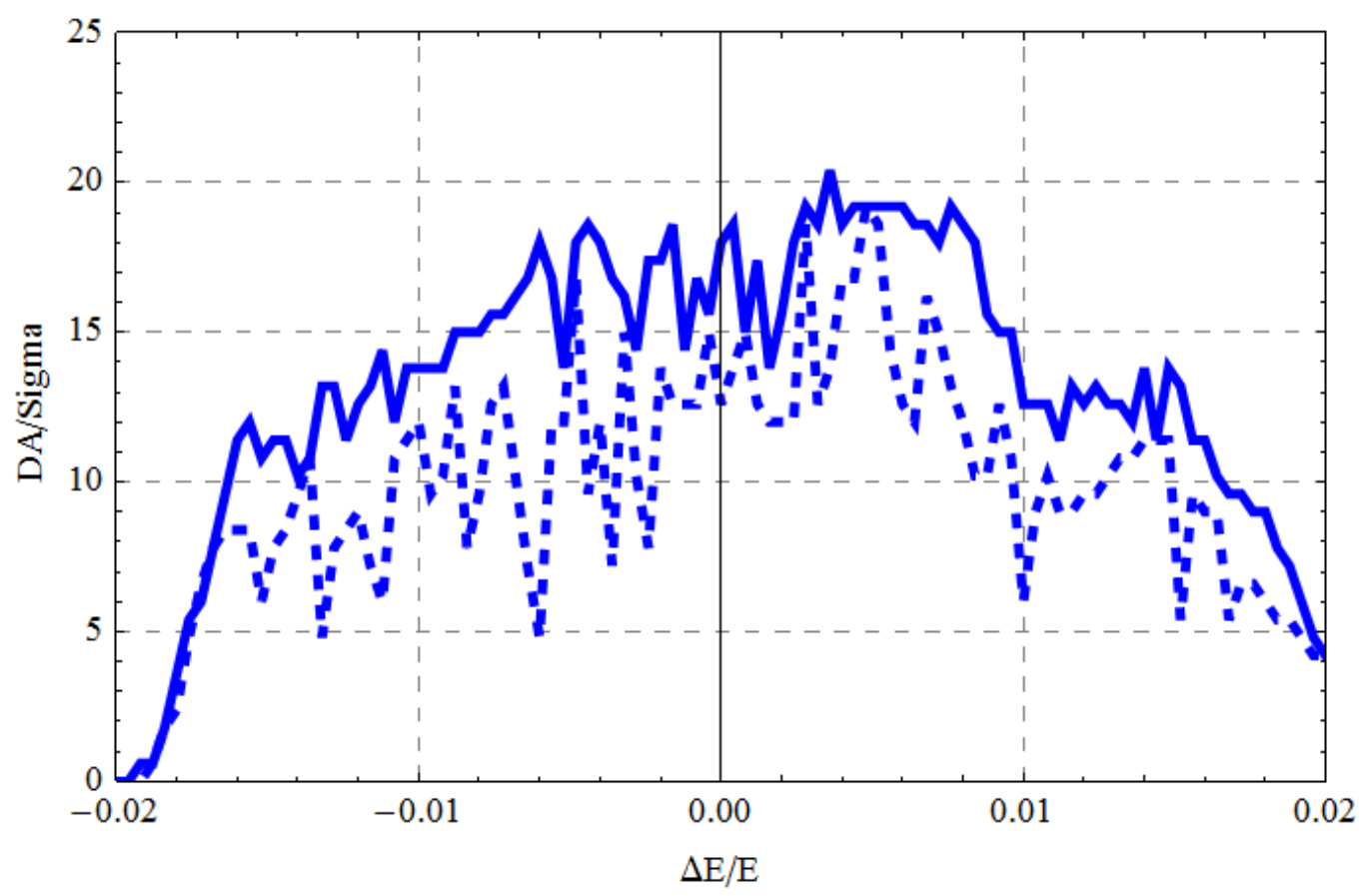

**Figure A4.10:** Dynamic aperture.

## A4.5: Beam Loading Effect

The APDR saves capital costs but present a large challenge for the SRF system. In the Collider, the electron and position bunches are compressed into 4 short partial double rings respectively, so both the beam gaps and the pulse currents are large. Besides, the electron and position beams share the same RF system. The RF system of APDR suffers from a serious beam loading effect, especially in operation at the Z-pole.

Eight 4 km-long double rings proposed for the APDR are equally spaced around the circumference. They contain 4 electron bunch trains and 4 position trains. 8 superconducting RF stations are equally spaced along the Collider; 336 650 MHz 2-cell

SRF cavities are located at the RF stations. The time structure of the beam is shown in the Fig. A4.11.

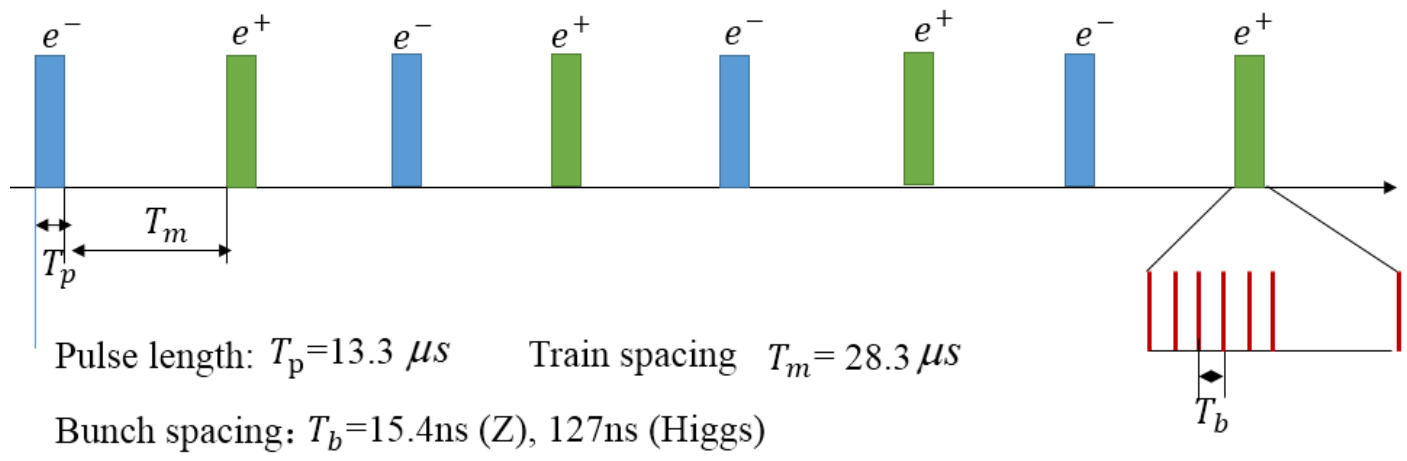

**Figure A4.11:** Time structure in the APDR [6]

The main machine parameters are nearly the same as for the full DR. The RF parameters for the APDR are listed in Table A4.2.

**Table A4.2:** APDR RF parameters

| **Parameter** | **Unit** | **Higgs** | **W** | **Z** |
|---|---|---|---|---|
| Beam Energy | GeV | 120 | 80 | 45.5 |
| Circumference | km | 100 | 100 | 100 |
| SR loss/ turn | GeV | 1.61 | 0.32 | 0.033 |
| Beam current | mA | 19.5 | 25.9 | 42.4 |
| SR power/beam | MW | 31.4 | 8.3 | 1.4 |
| Bunch number | | 420 | 900 | 3400 |
| Bunch number/ train | | 105 | 225 | 850 |
| Bunch charge | nC | 15.5 | 9.6 | 4.2 |
| RF frequency | MHz | 650 | 650 | 650 |
| RF voltage | GeV | 2.03 | 0.45 | 0.069 |
| Cavity number in use | | 336 | 64 | 12 |
| Synchrotron phase | deg | 37.5 | 44.7 | 61.4 |
| Cavity voltage | MV | 6.04 | 7.34 | 5.75 |
| Input power/ cavity | kW | 187 | 259 | 233 |
| Loaded Q ($10^5$) | | 9.5 | 9.5 | 6.9 |
| Optimal detuning | kHz | 0.26 | 0.33 | 0.87 |
| Cavity bandwidth | kHz | 0.7 | 0.7 | 0.9 |
| Cavity stored energy | J | 43 | 64 | 39 |
| Luminosity ($10^{34}$) | $cm^{-2}s^{-1}$ | 2.0 | 2.1 | 1.03 |
| Max voltage decrease | | 7.6% | 8.4% | 17.5% |
| Max phase shift | deg | 6.7 | 6.8 | 11.2 |

The voltage decrease and phase shift are calculated by K. Bane's formula [9].

The simulation program based on the beam transfer function [10] is also used for the phase shift calculation. The result is same as obtained with the analytical estimation formula.

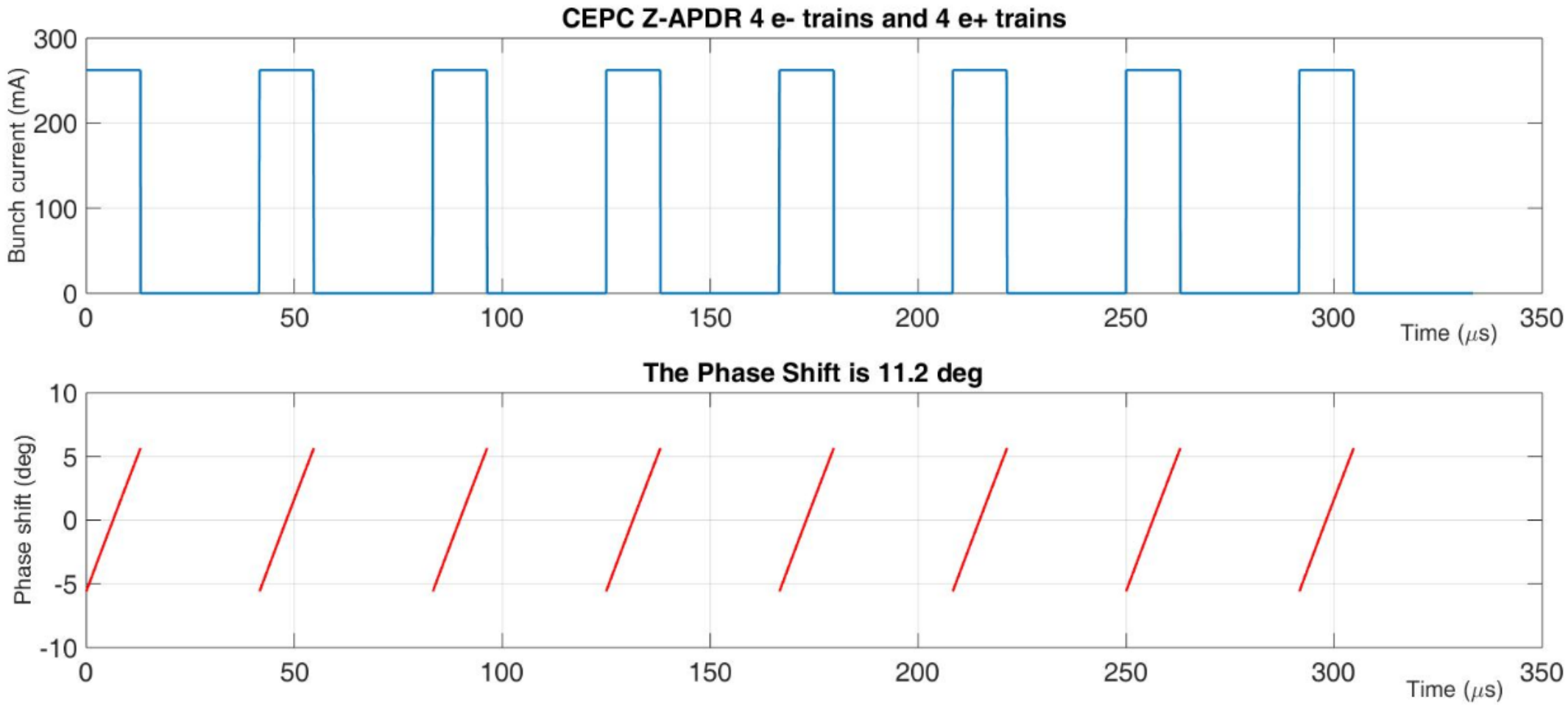

**Figure A4.12:** Phase shift of CEPC APDR Z by simulation code [6]

As shown in Fig. A4.12 for operation of the APDR at the Z-pole, the phase shift is 11.2 degrees, and will cause a maximum 12% bunch length increase, 12% synchrotron frequency decrease, and 13% RF acceptance drop.

## A4.6: References

# Appendix 5: CEPC Injector based on a Plasma Wakefield Accelerator

## A5.1: Introduction

### A.5.1.1 Plasma-based Wakefield Acceleration (PWA)

Currently the energy frontier of particle physics is more than several TeV, and the colliders that are needed to reach this high energy in the center of mass are gigantic machines that are difficult and expensive to build. Since the 1980s, many novel acceleration concepts have been proposed and tested to explore the possibility of high gradient acceleration, with a long term goal of significantly reducing the scale and cost of these machines. Among these schemes, plasma based wakefield acceleration [1-2] has made impressive progress in recent years and has achieved several key milestones for producing high quality and efficiency electron and positron acceleration.

In a PWA, a driver (either an ultra-short intense laser pulse or a charged particle beam) propagating through a low density plasma can produce a plasma wave (wakefield) with a phase velocity very close to the speed of light and a large longitudinal accelerating field, which can be utilized to accelerate a charged particle beam to high energy. With a laser driver, such a machine is called a laser wakefield accelerator (LWFA) [1], and if a particle beam is the driver, it is called a plasma wakefield accelerator (PWFA) [2]. Since the plasma is an ionized medium that is generated on every beam cycle, this avoids the problem of breakdown. Furthermore, such a wave is capable of supporting very large coherent fields on the order of 10-100 GeV/m, more than 2-3 orders of magnitude larger than is used in conventional accelerators. Therefore, 10 - 100 GeV level acceleration can be achieved in principle in 1 - 10 meter long plasma structures. These wakefields typically have very small structure sizes, around $\sim 10 - 100\ \mu\mathrm{m}$, which naturally leads to the acceleration of beams with high peak current and low emittance.

### A5.1.2: A Plasma-based High Energy Injector for CEPC

In the CEPC, a 100 km long main ring is used for colliding electrons and positrons at full energy. To inject the electron and positron beams into the main ring, a high energy injector with sufficient energy (45 GeV or more) is preferred to avoid the difficulty of low injection magnetic fields. A straightforward way is to use a linac several km long as an injector, but the cost of such a linac is very high.

Here we discuss the requirements for a high energy injector based on PWA with significantly higher energy than the planned 10 GeV linac. Based on the requirements of CEPC injector, a repetition rate around 100 Hz is needed and the normalized emittance for the injected beams is below a few mrad. Both conditions are relatively easy to achieve, giving significant flexibility on what accelerator schemes to use, including advanced high gradient concepts like PWA.

At the current stage of development, a linear collider based on plasma technology is far from ready, and may need decades of further development. Nevertheless, it is reasonable to think about utilizing current plasma technology to significantly reduce the size and cost of a 45 - 120 GeV high energy injector for the CEPC Collider. Due to the much relaxed requirements on beam parameters, the plasma based injector concept not only seems attractive, but also to be feasible. In the next section, a preliminary conceptual design for a 45 GeV level injector based on a high transformer ratio (TR) electron beam driven plasma wakefield accelerator (PWA) is presented. Its feasibility is mainly due to the availability of high average power driver technology (10 kW or more).

### A5.1.3: References

## A5.2: Preliminary Design

### A5.2.1: Overall Conceptual Design based on a Single-Stage HTR PWFA

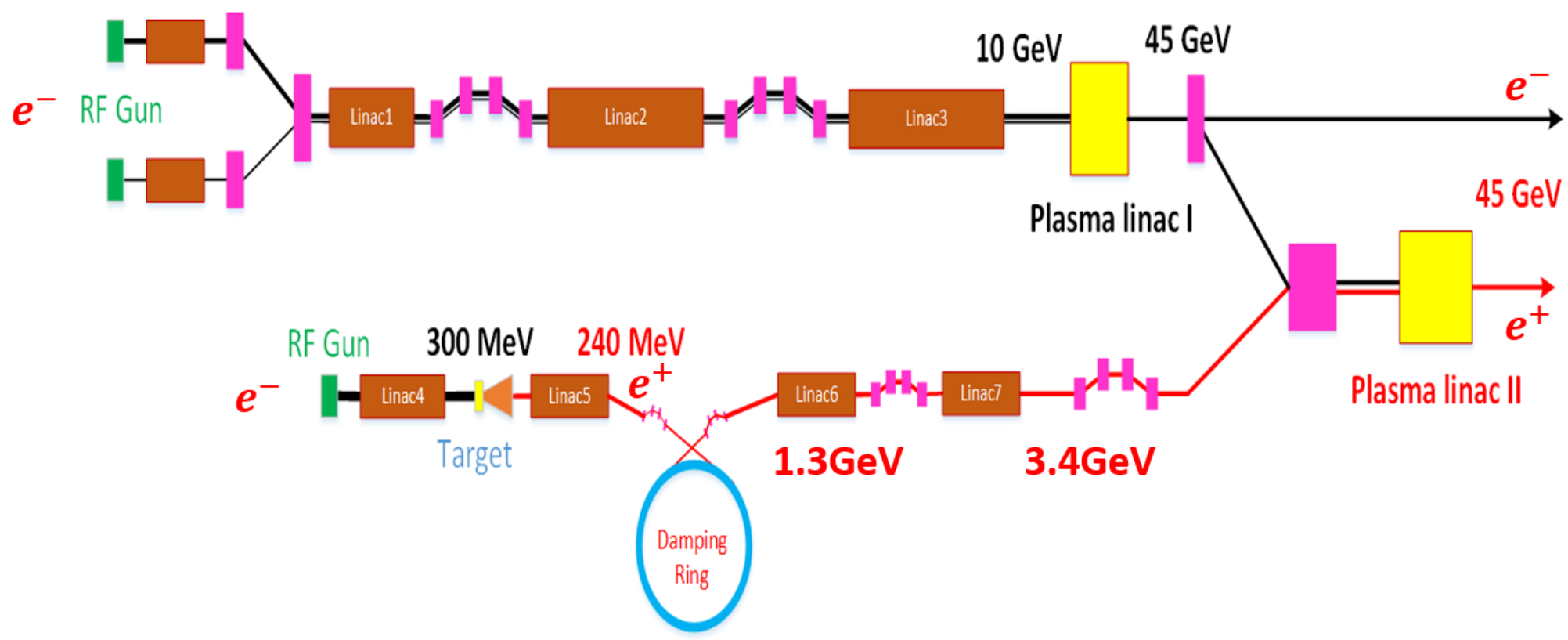

**Figure A5.1**: layout of a PWFA injector

In the layout in Fig. A5.1, two RF guns are used to generate two electron bunches with proper timing delay and bunch shapes. Combined with bunch compression and acceleration, these two beams are accelerated to 10 GeV energy together. After passing through a meter scale plasma acceleration module, the trailing bunch will be accelerated to about 45 GeV (with TR = 3.5). For the positron beam, a beam line including a positron target, a damping ring and bunch compression chicane is used to generate a short positron beam (50 fs) of a few GeV. This positron beam then is accelerated by a PWFA driven by the 45 GeV electron beam to about 45 GeV.

To generate a high transformer ratio in the PWFA, the electron beam driver needs to have a current profile with slow rise and fast decay (triangle shape). Such profiles can be produced by combing the RF gun with bunch compression. Fig. A5.2 shows an example (7 nC) from beam line simulation with an S-band photoinjector. A transformer ratio TR = 3.5 can be generated using this driver in a PWFA, as shown in Fig. A5.3.

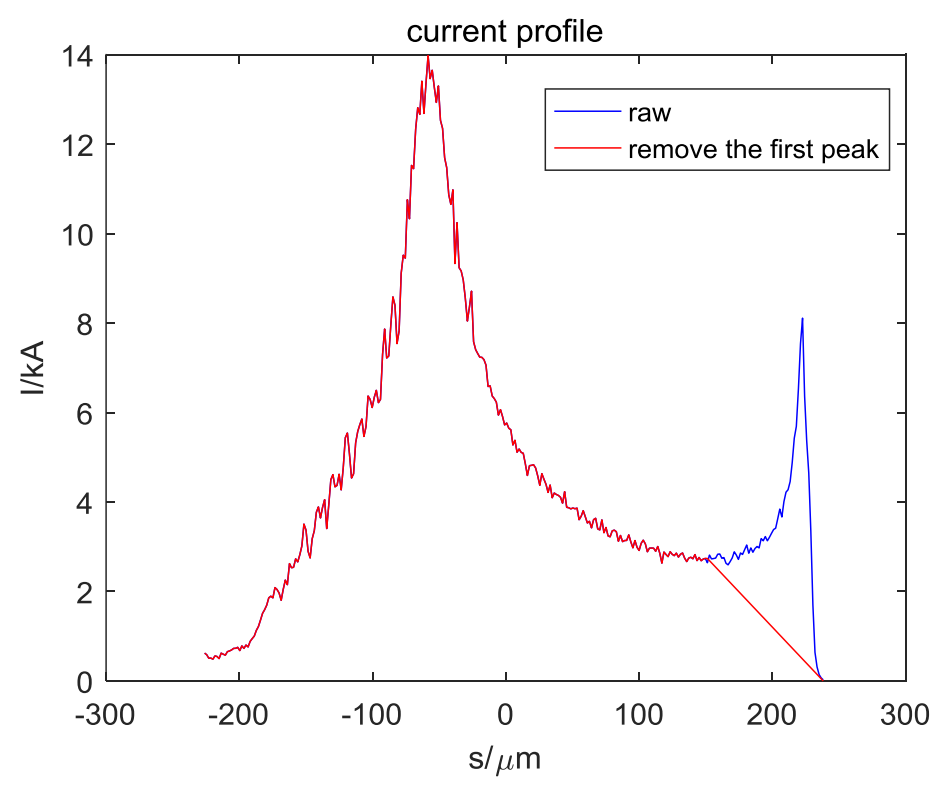

**Figure A5.2**: S-band based nearly trianglular shape current pulse

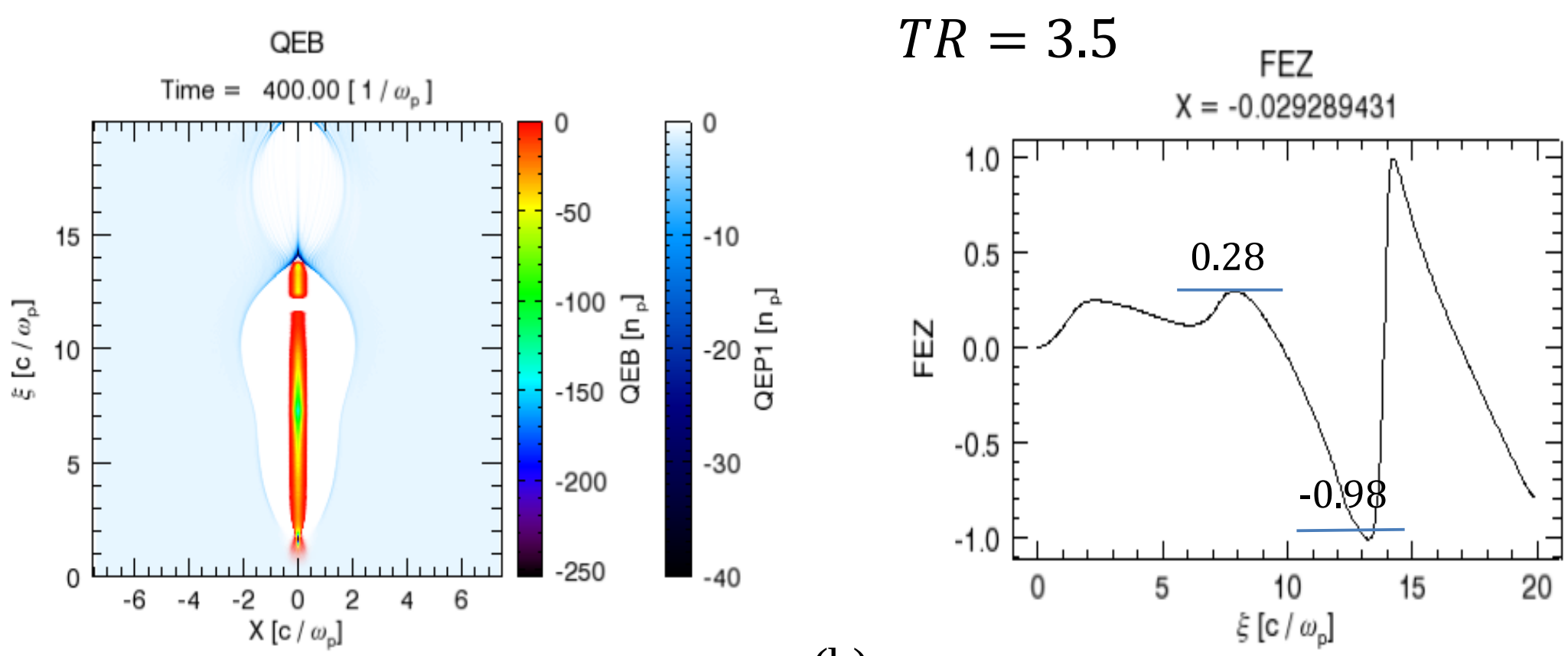

(a) (b)

**Figure A5.3**: (a) plasma wake generated by electron bunch with current profile in Figure A6.2.1.2; (b) the $E_z$ field with a TR=3.5

### A5.2.2: Parameters for the HTR PWFA (Electron Acceleration)

In Table A5.1, the beam and plasma parameters for the electron acceleration stage are listed. The beam current profiles and the wake structure in the plasma are shown in Fig. A5.4.

**Table A5.1**: Input parameters for HTR PWFA

| | |
|---|---|
| Plasma density $n_0(cm^{-3})$ | $5.15 \times 10^{16}$ |
| Driver charge $Q_d(nC)$ | 6.47 |
| Driver energy $E_d(GeV)$ | 10 |
| Driver length $L_d(\mu m)$ | 285 |
| Driver RMS size $\sigma_d(\mu m)$ | 10 |
| Driver normalized emittance $\epsilon_{nd}(mm\ mrad)$ | 10 |
| Trailer charge $Q_t(nC)$ | 1.25 |
| Trailer energy $E_t(GeV)$ | 10 |
| Trailer length $L_t(\mu m)$ | 35 |
| Trailer RMS size $\sigma_t(\mu m)$ | 5 |
| Trailer normalized emittance $\epsilon_{nt}(mm\ mrad)$ | 100 |

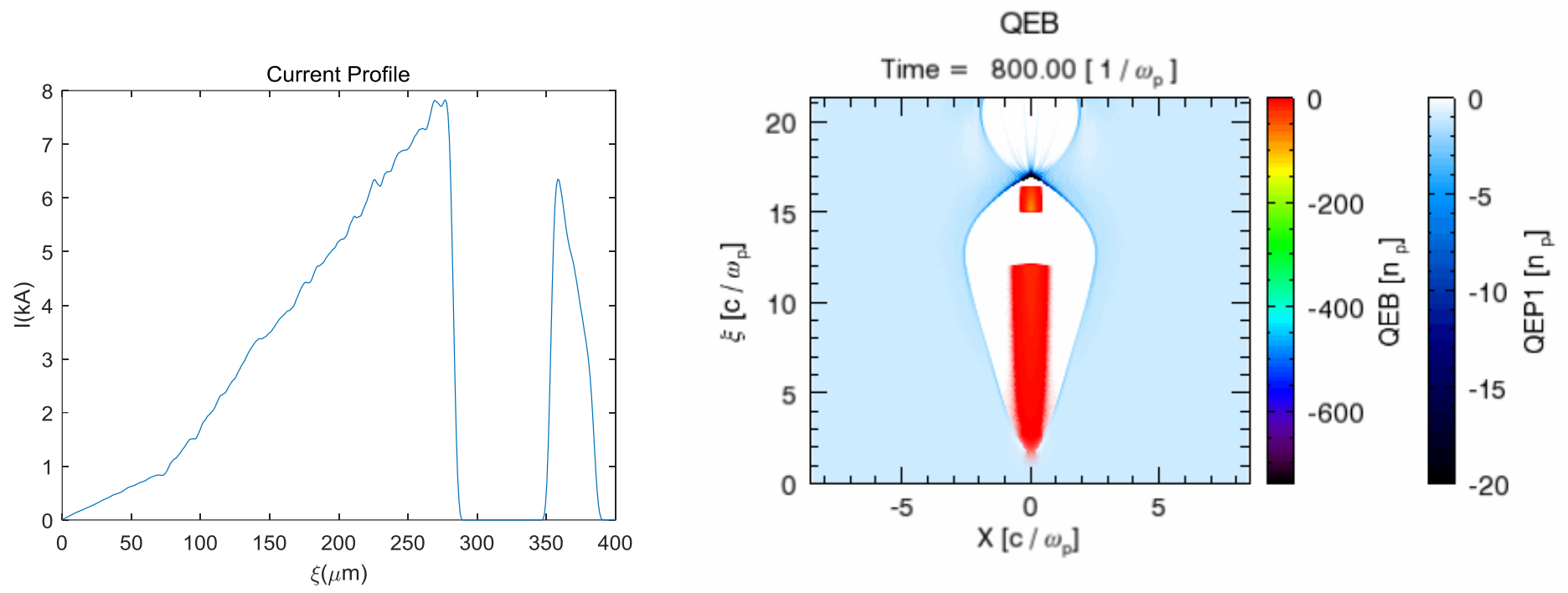

**Figure A5.4**: Beam current profiles and wake structure in the plasma

Table A5.2 lists the output parameters of the HTR PWFA stage (1.9 m.) using 3D PIC simulations.

**Table A5.2**: Output parameters for HTR PWFA

| | |
|---|---|
| Trailer energy $E_t(GeV)$ | 45.5 |
| Trailer normalized emittance $\epsilon_{nt}(mm\ mrad)$ | 98.9 |
| TR | 3.55 |
| Energy spread $\delta_E(\%)$ | 0.7 |
| Efficiency (driver -> trailor) | 68.6% |

### A5.2.3: Parameters for the Positron Acceleration Stage

In this preliminary design, positron acceleration is implemented using a hollow channel plasma wakefield accelerator. In this scheme, the electron and positron beams are combined together before sending them through the hollow channel plasma. When the parameters are properly chosen, the positron beam can gain energy uniformly with high efficiency.

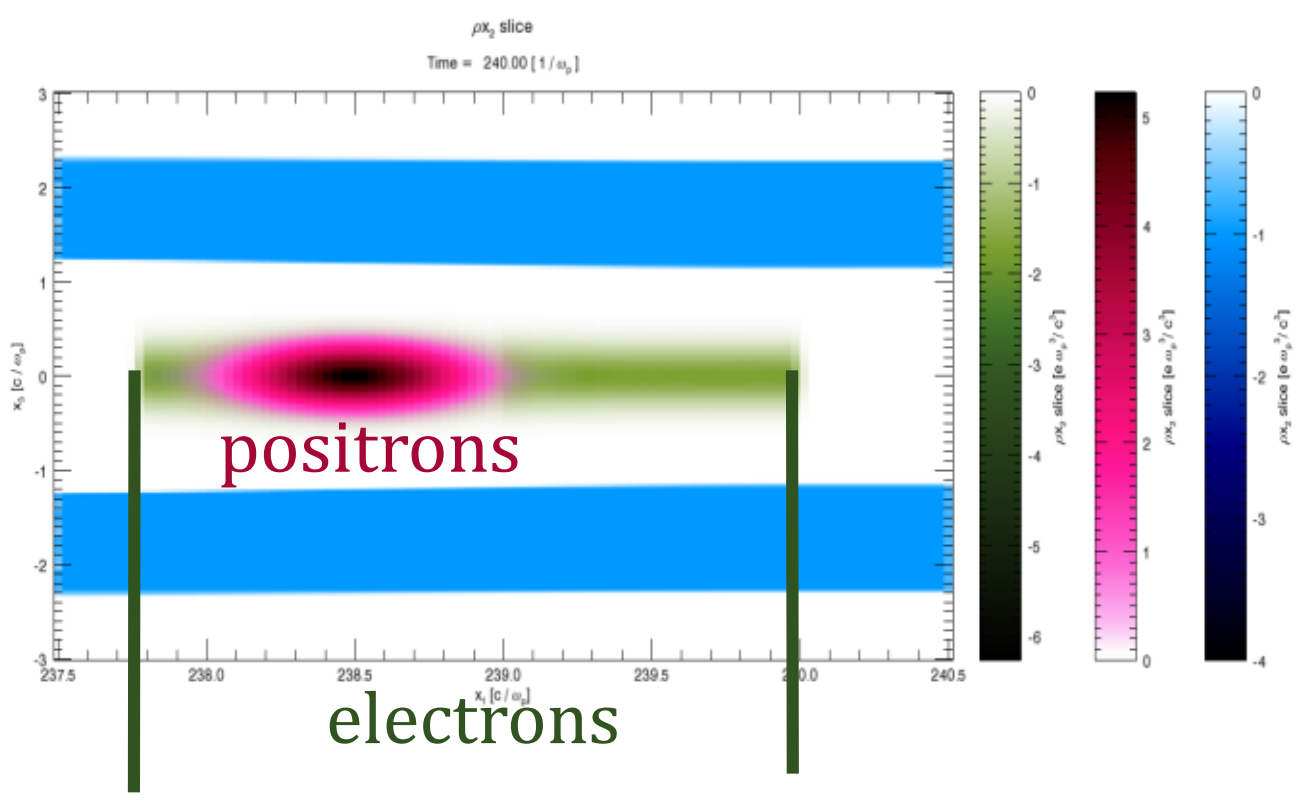

**Figure A5.5**: beam and plasma densities for positron acceleration

In Fig. A5.5, the layout of the electron and positron beams is shown on top of the plasma channel. Since the current profile of the positron beam is mainly determined by the damping ring (assuming as a Gaussian profile), the electron beam profile needs to be properly designed to assure uniform acceleration for the positron beam. Fig. A5.6 is an example of uniform high efficiency positron acceleration in a hollow plasma channel.

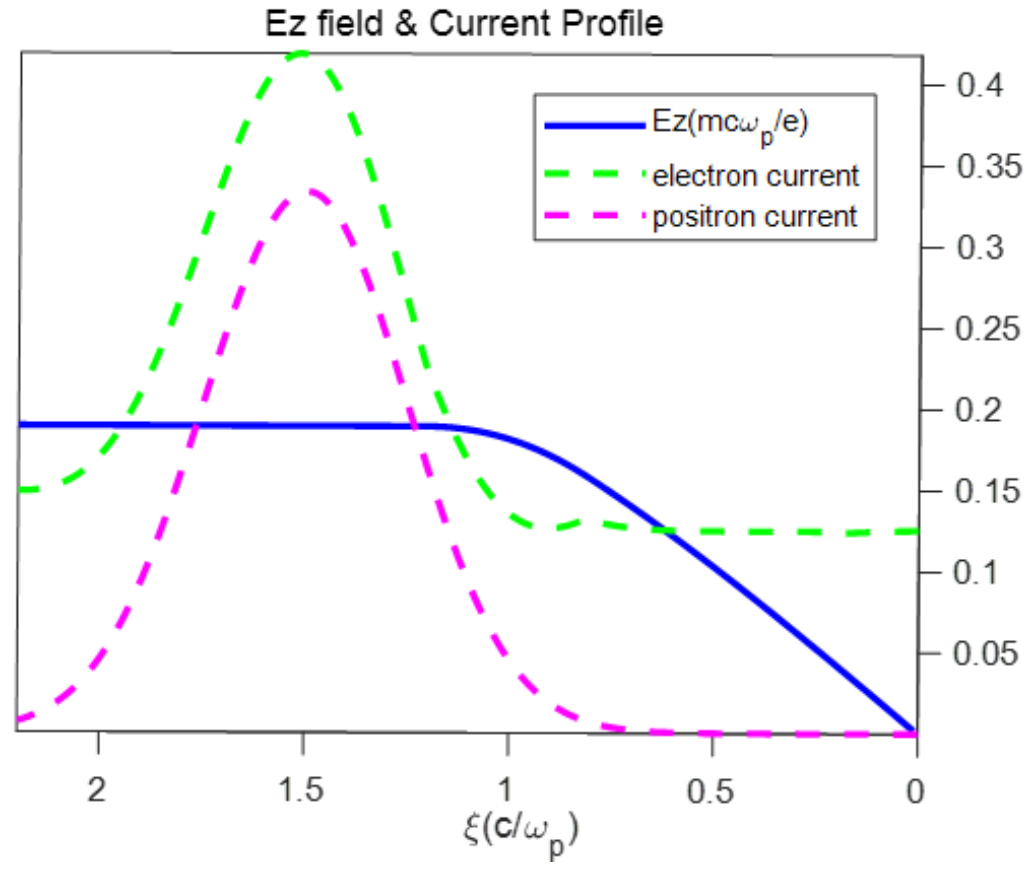

**Figure A5.6**: electron and positron current profiles and $E_z$ field in the hollow channel

In Table A5.3, the beam and plasma parameters for the positron acceleration stage are listed. In Table A5.4, the output of the positron acceleration stage using 3D PIC simulations is shown. We note that the energy spread of the beam can be further reduced down to 0.2% by parameter optimization or post-processing.

**Table A5.3**: input parameters for positron acceleration

| | |
|---|---|
| Plasma density $n_0(cm^{-3})$ | $1.77 \times 10^{16}$ |
| Channel inner radius $r_1(\mu m)$ | 45 |
| Channel out radius $r_2(\mu m)$ | 90 |
| Bunches | |
| Driver charge $Q_d(nC)$ | 2.13 |
| Driver energy $E_d(GeV)$ | 45.5 |
| Driver length $L_d(\mu m)$ | 88 |
| Driver RMS size $\sigma_d(\mu m)$ | 10 |
| Driver normalized emittance $\epsilon_{nd}(mm\, mrad)$ | 100 |
| Trailer charge $Q_t(nC)$ | 1 |
| Trailer energy $E_t(GeV)$ | 3.4 |
| Trailer length $L_{t-rms}(\mu m)$ | 10 |
| Trailer RMS size $\sigma_t(\mu m)$ | 10 |
| Trailer normalized emittance $\epsilon_{nt}(mm\, mrad)$ | 100 |

**Table A5.4**: output parameters for positron acceleration

| | |
|---|---|
| Trailer energy $E_t(GeV)$ | 45.5 |
| Trailer normalized emittance $\epsilon_{nt}(mm\, mrad)$ | 100 |
| TR | 1 |
| Energy spread $\delta_E(\%)$ | 0.3 |
| Efficiency (driver -> trailer) | 46.9% |

### A5.2.4: Parameters for the Electron/Positron Energy Dechirper

Due to the very short wavelength of a plasma accelerator, the acceleration phase interval occupied by trailing beam is much larger than that in a traditional accelerator, which leads to relatively large energy chirp. However, this chirp is typically much larger than the intrinsic slice energy spread of the beam. Here, we use a low density uniform plasma or hollow channel plasma as a near-ideal dechirper for both electrons and positrons. This scheme can reduce the projected energy spread down to 0.2%, satisfying the requirement of CEPC.

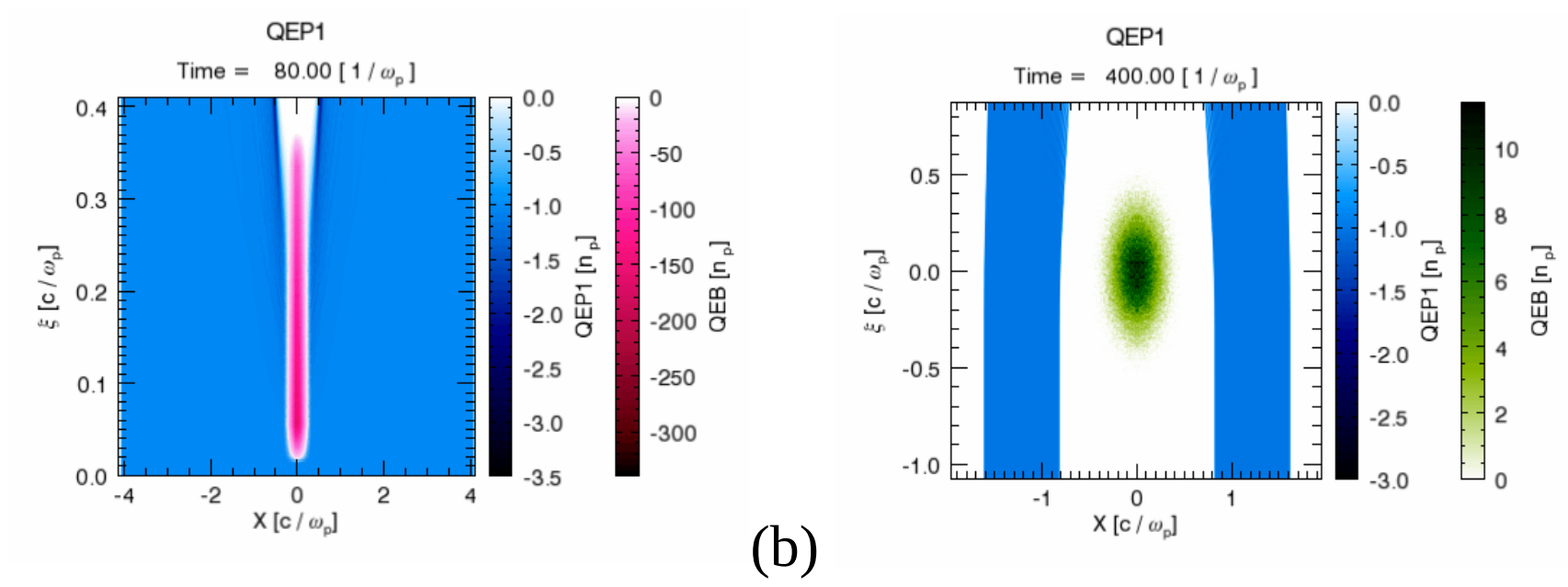

(a) (b)

**Figure A5.7**: Beam and plasma densities in dechirper regime for (a) electrons in uniform plasma and (b) positrons in hollow channel plasma

In Fig. A5.7, we illustrate the dechirper for both electrons and positrons. For positrons, hollow channel plasma is implemented to preserve its emittance. In this dechirper, beam bunch length is very short compared to the plasma wavelength, so the longitudinal wakefield is approximately linear in the beam region. Thus the linear energy chirp, which is the main part of the projected energy spread for the beam, can be removed. In Table A5.5 and A5.6, the output parameters for electron and positron beam after the dechirper are shown respectively. One can see that after the dechirper, the beam energy spreads are within the 0.2% requirement of the CEPC injector.

**Table A5.5**: Output parameters for electron dechirper

| | |
|---|---|
| Trailer energy $E_t(GeV)$ | 45.3 |
| Trailer normalized emittance $\epsilon_{nt}(mm\,mrad)$ | 169.9 |
| Energy spread $\delta_E(\%)$ | 0.2 |
| Efficiency | 86.6% |

**Table A5.6**: Output parameters for positron dechirper

| | |
|---|---|
| Trailer energy $E_t(GeV)$ | 45.2 |
| Trailer normalized emittance $\epsilon_{nt}(mm\,mrad)$ | 100 |
| Energy spread $\delta_E(\%)$ | 0.14 |
| Efficiency | 97.0% |

## A5.3: Summary

In this design, a high transformer ratio PWFA with TR = 3.5 is used to accelerate the electron beam to 45 GeV in a single meter scale plasma stage. This 45 GeV electron beam can also be used to accelerate the positron beam in a hollow channel plasma to a similar energy. A low-density plasma dechirper is used after acceleration to reduce the energy spread down to 0.2% or even less. 3D PIC simulations are used to verify these basic designs. Based on these simulations, we conclude that a plasma based high energy injector at 45GeV level is feasible using available plasma technology.

# Appendix 6: Operation as a High Intensity γ-ray Source

Very high-quality synchrotron radiation can be obtained from the CEPC Collider electron beam and used to carry out important research and application experiments.

Nuclear astrophysics, nuclear physics and the radiation effects on space probes have led, in recent decades, to important advances in these areas. However, there are still numerous problems and challenges. It is important to develop non-destructive testing methods to investigate the internal structure of large samples. A synchrotron radiation γ-source provided as a CEPC operating option would be a powerful research tool for important applications.

In this appendix, we will elaborate on the parameters of a CEPC synchrotron radiation γ-ray source, and discuss its application and experiment techniques.

## A6.1: Parameter of a CEPC Synchrotron Radiation γ-ray Source

As shown in Fig. A6.1 one of the two injection points can be used for the synchrotron radiation γ-ray source.

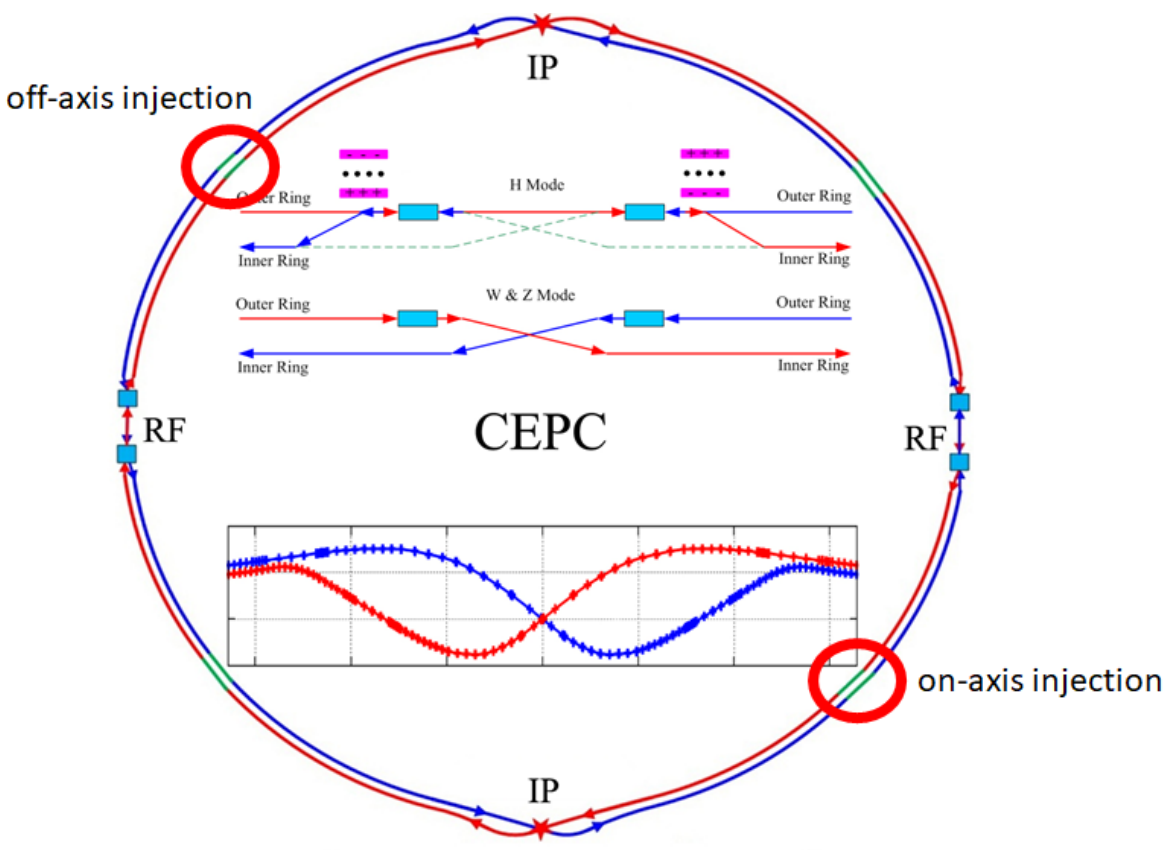

**Figure A6.1:** Possible positions for the γ-source at CEPC.

### A6.1.1: Production of the γ Beam from Different Insertion Devices

Fig. A6.2 shows the flux, brilliance and power density for different possible insertion devices. The calculations are based on running in the H mode. (Same calculations can be done for the W and Z mode.) For emission from a bending magnet, the characteristic energy is about 358.2 keV, the flux is more than $5 \times 10^{12}$ phs/s/0.1%B.W., with angular acceptance of $0.2 \times 0.2$ mrad.

Using an undulator of total length 10 m, period length 140 mm, period number 71, minimum gap 82 mm. and maximum magnetic field 0.21 T, the maximum K (undulator strength parameter) is 2.74. The energy will be several MeV, the flux is about $10^{12}$ phs/s/0.1%B.W. the brilliance about $10^{18}$ phs/s/mm$^2$/mrad$^2$/0.1%B.W., and the maximum power is 69 kW.

Using a wiggler of total length 32 m., period length 320 mm, period number 100, minimum gap 82 mm. and maximum magnetic field 2.0 T, the maximum K is 59.7. The energy will be in the tens of MeV, the flux about $10^{15}$ phs/s/0.1%B.W.; the critical energy is 19.2 MeV. The maximum power is 20 MW and the maximum power density is 428 MW/mrad$^2$.

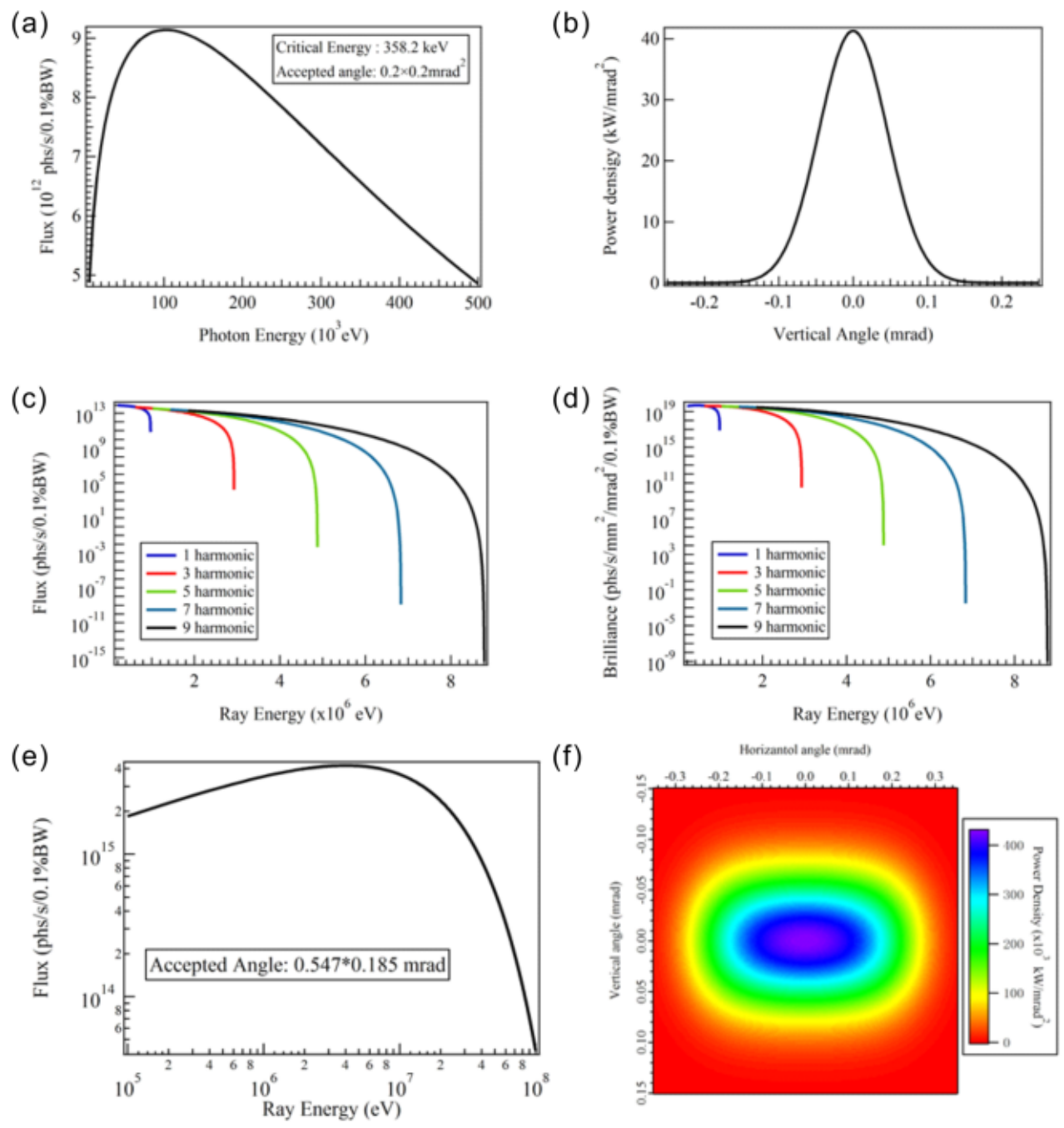

(g)

| Source | Energy (MeV) | Horizontal emittance (mrad) | Vertical emittance (mrad) | Flux (ph/s/1‰bw) |
|---|---|---|---|---|
| Beading Magnet | 0.05 | 0.2 | 0.2 | 8.58E+12 |
| | 0.1 | | | 9.06E+12 |
| | 0.2 | | | 8.44E+12 |
| | 0.3 | | | 7.21E+12 |
| | 0.4 | | | 5.96E+12 |
| Undulator | 1 | 0.25 | 0.2 | 8.90E+13 |
| | 5 | | | 3.00E+13 |
| | 10 | | | 2.50E+13 |
| | 15 | | | 2.49E+11 |
| Wiggle | 0.5 | 0.547 | 0.185 | 2.97E+15 |
| | 1 | | | 3.52E+15 |
| | 5 | | | 4.16E+15 |
| | 10 | | | 3.64E+15 |
| | 50 | | | 5.40E+14 |

**Figure A6.2:** Different insertion devices to generate the γ-ray beam; (a, b) Flux and power density distribution from a bending magnet; (c, d) flux and brilliance from an undulator; (e, f) flux and power density distribution from a wiggler; (g) emittance and flux at a specific energy for different insertion devices.

### A6.1.2: Comparison with Other γ-ray Source

Today's state-of-the-art γ-ray source is based on laser-electron Thomson scattering. The performance of several important such sources are listed in Table A6.1 and compared with what would be possible from CEPC. The flux of CEPC gamma rays is much higher than the flux of all other laser gamma sources in the world.

**Table A6.1:** Performance comparison between a CEPC gamma source and the main laser gamma sources in the world.

| Source | CEPC BM | CEPC Undulator | CEPC Wiggler | SSRF (China) | TUNL-HIGS (USA) | TERAS (Japan) | ALBL (Spain) |
|---|---|---|---|---|---|---|---|
| Gamma energy rang (*MeV*) | 0.1~5 | 0.1~10 | 0.1~100 | 0.4-20<br>330-550 | 2-100 | 1-40 | 0.5-16<br>16-110<br>250-530 |
| Energy resolution (*ΔE/E*) | continuous | ~1% | continuous | 5% | 0.8~10% | | |
| Flux (*phs/s*) | ＞$10^{12}$ @0.1% | ＞$10^{13}$ @0.1% | ＞$10^{16}$ @0.1% | $10^6$ | $10^8$ | $10^4$~$10^5$ | $10^5$~$10^7$ |

## A6.2: Applications of a High Intensity CEPC-SR γ-ray Source

The brightness from a CEPC synchrotron radiation source can reach $10^{17}$ photons/s/mm2/mrad2/0.1%b.w. at an energy of several hundreds of keV, several orders of magnitude greater than existing synchrotron radiation sources. High brightness MeV X-ray or γ-ray sources are especially important for experiments in photon-nuclear physics, nuclear astrophysics and quantum electrondynamics (QED) phenomena, in additions to applications in the life sciences and cultural heritage.

### A6.2.1: Nuclear Astrophysics Application I

An application to the nuclear astrophysics is the reaction $^{12}C(\alpha,\gamma)^{16}O$ at the Gamow peak ($E_0 \approx 300 \pm 80$ keV). With the gamma beam being developed at LCS/IHEP, this key reaction can be studied by the time-reversal reaction $^{16}O(\gamma,\alpha)^{12}C$ through the well-known detailed-balance principle. The $^{13}C(\alpha,n)^{16}O$ background is not expected in this kind of photodissociation experiment. In addition, one can measure detailed angular distributions, thus obtaining accurate values of the E2/E1 ratio that are crucial for an accurate extrapolation to stellar energies. The E1 cross section is shown in Fig. A6.3.

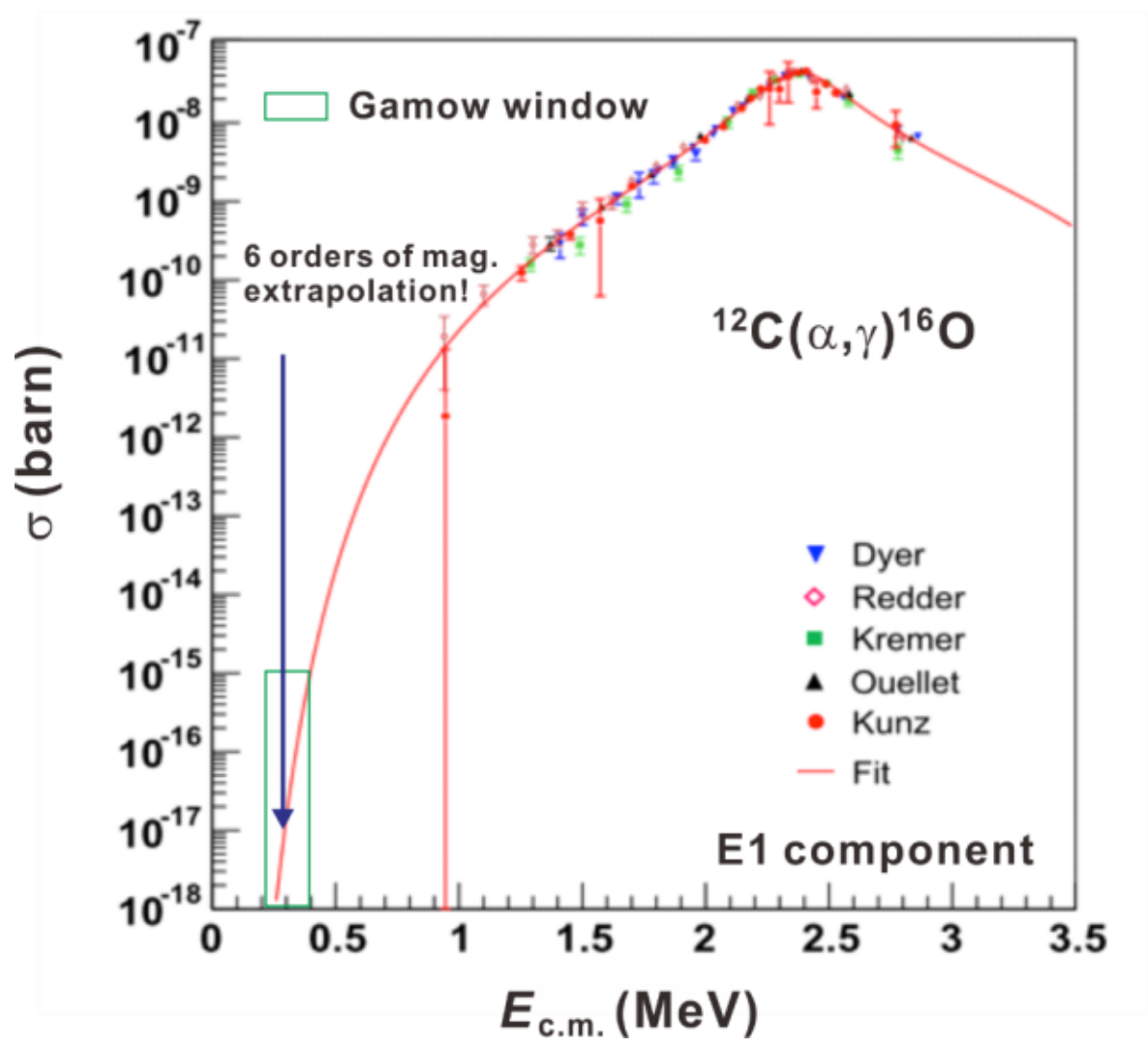

**Figure A6.3:** Reaction cross section of the E1 component

This "Holy-grail" reaction, $^{12}C(\alpha,\gamma)^{16}O$ at the Gamow peak ($E_0 \approx 300 \pm 80$ keV), would be the most important application of a CEPC synchrotron radiation beam from a wiggler magnetic field. The central energy of 7.16±0.3MeV is designed for this reaction.

### A6.2.2: Nuclear Astrophysics Application II

Another application to the nuclear astrophysics is the photo-disintegration reaction measurement, a probe of the p-process. The p-process has been proposed to explain the synthesis of the stable neutron-deficient nuclides heavier than iron observed in the solar system. These nuclides cannot be produced by slow or rapid neutron captures (s- or r-process), but can be formed from seed nuclides formed in the s- or r-process through photodisintegration reactions or capture reactions. To calculate the p-process, more than 2000 nuclides and 20000 reactions have to be taken into account. The Hauser-Feshbach statistical model is used. More experimental data on photodisintegration reactions are required for reference and restrictions on that model. A CEPC-SR **γ**-ray source can supply high intensity **γ**-rays from 1 MeV to 20 MeV to produce the required photodisintegration reactions.

### A6.2.3: Nuclear Science and Technology Application

An application to the nuclear science and technology is the giant resonance and "mini" giant dipole resonance of nuclear reactions. The CEPC synchrotron radiation from a wiggler can cover the energy from 1 MeV to 300 MeV. The gamma-rays with energy of 7 – 40 MeV can be used to study the giant resonance, the giant dipole resonance of nucleus and the "mini" giant dipole resonance.

### A6.2.4: Interfacial Structure and Physics application

Angular distribution measurements in photoemission have been established as a powerful tool in the determination of the electronic structure of bulk solids and surfaces as well as in research on thin epitaxial layer systems. These latter systems are of interest for understanding semiconductor hetero-structures or thin magnetic films. For example, photoelectron diffraction as a technique has been successfully applied to the study of the epitaxial growth of metal and semiconductor layers and this has made important contributions in our understanding of the growth mechanism of these systems.

## A6.3: Detection Methods of High-Intensity CEPC-SR γ-ray

In the energy range, 1 keV to 2 MeV, Si-PIN, CdZnTe and SiC detectors and prototypes are employed.

For the Si-PIN detector, the detection energy range is 5 - 30 keV and the energy resolution can attain 3 keV at 59.5 keV (-5° C).

The CdZnTe detector can achieve 6 keV - 2 MeV energy coverage, an energy resolution of <6% at 662 keV and a counting rate of 2000 - 20000 Hz.

In order to meet the needs of high temperature and radiation-resistant extreme environments, SiC Schottky barrier diode (SBD) detectors are fabricated using a 4H-SiC epitaxial wafer for α and γ-ray detection, and their performance changes are studied in high temperature and radiation environment. The results show that the detector has an energy resolution of 9.49% at 59.5 keV and can work normally from 25 to 125°C. After 1 Mrad $^{60}$Co γ-ray irradiation, the performance of the detector remains basically unchanged.

The energy range we need to cover is from 0.1 MeV to 300 MeV. There is no single detector that can cover such a large range with acceptable energy resolution. We consider three parts: 0.1 to 10 MeV, 10 to 20 MeV, and 20 to 300 MeV.

HPGe detectors are designed to detect the gamma rays in the 0.1 to 10 MeV range. Their resolution is 1.8 keV at 1332 keV.

In the 10 to 20 MeV range, LaBr3(Ce) detectors can be used. Their energy resolution is about 2-3% and is smaller than that of a NaI(Tl) detector (about 6-7%). HPGe or LaBr3(Ce), with a BGO anti-Compton enclosure should be used to decrease the background from Compton scattering.

For higher energy gamma rays, we choose a CsI detector as used in the BES electromagnetic calorimeter (BESEMC). This has been calibrated from 300 MeV to 2 GeV. The important region is below 500 MeV. The energy resolution is about $2.5\%/\sqrt{E}$ (GeV). Another possible solution is Cerenkov detection.

# Appendix 7: Operation for *e-p, e-A* and Heavy Ion Collision

## A7.1: Introduction

The SPPC would explore the energy frontier accessible with accelerators, whereas the CEPC will provide clean and much needed precision for the study of the Higgs to shed light on mass generation and the mystery behind spontaneous symmetry breaking. But, neither the SPPC nor the CEPC is the natural facility for exploring the precise internal structure of nucleons and nuclei, although both facilities can create hadronic matter from the energy of the collisions. The construction of CEPC and SPPC in a common accelerator complex provides a great opportunity to realize collisions of ultra high energy protons or ions with very high energy electrons or positrons (*e-p* or *e-A* where *e* stands for either *e*- or *e*+). Such an *e-p* or *e-A* collider can reach a center-of-mass (c.m.) energy up to 4.2 TeV and ultimately 6 TeV after an energy upgrade of SPPC, far beyond the energy of any proposed future lepton-hadron colliders including the Electron-Ion Collider (EIC) in the United States (JLEIC and eRHIC, up to 0.14 TeV) and the Large Hadron-electron Collider (LHeC) at CERN (1.3 TeV). The following table lists the c.m energy range for various lepton-hadron colliders presently envisioned or under study.

**Table A7.1:** Center-of-mass. energy of possible future *e-p* or *e-A* colliders

| | JLEIC | eRHIC | LHeC | FCC-*he* | CEPC-SPPC *e-p* |
|---|---|---|---|---|---|
| c.m. energy (TeV) | 0.02 – 0.065 | 0.02 – 0.14 | 1.3 | 3.5 | 4.2 – 6 |

With precise measurements of the scattered leptons, such a lepton-hadron facility exploring the ultra-deep inelastic scattering (DIS) region will provide a clean and fully controlled probe of the inner structure and quantum fluctuations of the dynamics of a proton down to the unprecedented distance of $10^{-4}$ fm (or one ten-thousandth of the proton size). Such a study could be sensitive to the dynamics that might restore the spontaneously broken symmetries of the standard model, and the quantum fluctuations caused by physics beyond the standard model. These measurements of a proton with a momentum transfer of over one TeV lead to the finest tomographic images of the spatial distributions of quarks and gluons of momenta ranging from the one tenth to the one thousandth of the proton's momentum. This information is sensitive to the color confinement of QCD.

The SPPC with proton beams replaced by heavy ion beams will produce the hottest quark-gluon plasma ever. Such extreme conditions could only have existed in the first few microseconds of our universe. With the option of colliding leptons with heavy ions in an *e-A* collider, the heavy ions with various atomic weight could act as the smallest vertex detectors in the world to map out the dynamics of color neutralization and probe the emergence of hadrons, a necessary phase in the evolution of our universe. With the CEPC and the SPPC in a common accelerator complex, we could have a unique facility in the world able to explore the fundamental structure of matter, and its birth and evolution

It is relatively straight forward to bring a lepton beam from CEPC and a hadron beam from SPPC into collision at one or multiple interaction points (IP). The estimated luminosity in the *e-p* or *e-A* collisions can reach several times $10^{33}$/cm$^2$/s at each detector. The challenge is to design an interaction region (IR) to optimize performance while managing the facility resources including the site power while running both lepton and hadron facilities at the same time. Furthermore, an additional challenge is to bypass the non-colliding beams near the detectors if the *e-p* or *e-A* collisions are to be run simultaneously with the SPPC programs.

In this appendix, we summarize a preliminary design concept of an *e-p* or *e-A* collider based on CEPC and SPPC. It should be understood that the positron beam can also be used for *e-p* or *e-A* collisions and there is virtually no change of the design parameters for positron-proton or positron-ion collisions.

## A7.2: *e-p* or *e-A* Accelerator Design Considerations

An assumption is that no major upgrade will be required in either the CEPC or SPPC facilities for the realization of *e-p* or *e-A* collisions. The *e-p* or *e-A* collider conceptual design follows the operational limits of CEPC and SPPC such as the synchrotron radiation power budget as well as fundamental beam effects. Within these boundaries, we are free to alter some of the beam or machine parameters for achieving optimized *e-p* or *e-A* collider performance.

Since *e-p* or *e-A* collisions are proposed as an additional capability of the CEPC-SPPC facility, naturally there is a question about whether *simultaneous* operation of *e-p* or *e-A* collisions with *e+e-* or *pp* collisions is feasible while still being able to deliver performance acceptable to the multiple physics programs.

The two major differences between the beams from CEPC and SPPC are bunch structure and beam emittance. The electron beam in CEPC has only 242 bunches in the current baseline design for a Higgs factory program since its current is limited by the synchrotron radiation power budget. On the other hand, the proton or ion beams in SPPC have 10080 bunches. Clearly it would be very inefficient to maintain the original CEPC and SPPC beam bunch structures since the majority of the proton or ion bunches will not collide with electron or positron bunches. This effectively excludes the option of simultaneous operations of *e+e-* and *e-p/A* collisions in the complex. Here it is assumed the lepton energy for *e-p* or *e-A* collisions is 120 GeV. However, for physics at a lower lepton energy (such as 45.5 GeV) the conclusion is quite different; and simultaneous operations e+e- and *e-p* or *e-A* may be possible. Nevertheless, in this appendix, we only focus on high lepton energy e+e- and *e-p* collisions, i.e. with CEPC running in H mode.

Without the constraint of running *e+e-* and *e-p* or *e-A* collisions simultaneously, the electron beam is no longer limited to 242 bunches; thus it can be altered to match the bunch numbers of a proton or ion beam from SPPC. In addition, since only one lepton beam is required in the CEPC ring while running *e-p* or *e-A* collisions, the electron beam current can be doubled (to 34.8 mA) under the same limit of synchrotron radiation power (30 MW per beam), an advantage which could double the *e-p* or *e-A* luminosities.

Selection of the beam focusing parameters is also driven by considerations of interaction region designs. The proton beam in SPPC is basically a round beam; the electron beam can also be made round by utilizing transverse optical coupling. This should greatly simplify matching of the beam spot sizes, therefore resulting in a significant increase in *e-p* or *e-A* luminosity.

Since the proton beam energy in SPPC is 37.5 TeV and ultimately could be 75 TeV, synchrotron radiation and its effect on the beam emittance is no longer negligible. The damping time of a proton or heavy ion beam is similar or even shorter than the time of a beam store. As a consequence, the proton or ion beam emittance will approach an equilibrium value (a balance of synchrotron radiation damping and quantum excitation, and intra-beam scatterings) during the store. This will affect the peak luminosity as well as the integrated luminosity. This will be discussed further in section A7.4.

This *e-p* or *e-A* collider based on CEPC-SPPC is highly asymmetric, with an energy ratio of the two colliding beams of 312.5 (or 625) with a 37.5 TeV (or 75 TeV) proton energy and 120 GeV electron beam energy. This would be much higher than any other lepton-hadron colliders ever built, designed or studied. Particles from collisions are highly concentrated around zero scattering angle relative to the hadron beam direction. It is expected that forward detection of particles with extremely small scattering angles will be a critical requirement of the detector design.

## A7.3: *e-p* Collisions

Table A7.2 presents the nominal parameters for *e-p* collisions for 120 GeV electrons colliding with 37.5 TeV protons (or about 14 TeV per nucleon for completely stripped lead ions). The electron beam current is 34.8 mA and is still under the operational limit of 60 MW synchrotron radiation power. To achieve luminosity optimization and also a better interaction region design, the electron beam emittance will be made round by introducing transverse optical coupling. The *e-p* luminosities at other energies can be estimated following a similar design approach as well as similar parameter limits.

**Table A7.2:** CEPC-SPPC *e-p* and *e-A* Design Parameters

| Particle | | Proton | Electron | Lead ($^{208}Pb^{82+}$) | Electron |
|---|---|---|---|---|---|
| Beam energy | TeV | 37.5 | 0.12 | 14.8 | 0.12 |
| CM energy | TeV | 4.2 | | 2.7 | |
| Beam current | mA | 730 | 34.8 | 730 | 34.8 |
| Particles per bunch | $10^{10}$ | 15 | 0.72 | 0.18 | 0.72 |
| Number of bunch | | 10080 | | 10080 | |
| Bunch filling factor | | 0.756 | | 0.756 | |
| Bunch spacing | ns | 25 | 25 | 25 | 25 |
| Bunch repetition rate | MHz | 40 | 40 | 40 | 40 |
| Norm. emittance, (x/y) | μm rad | 2.35 | 282 | 0.22 | 282 |
| Bunch length, RMS | Cm | 7 | 0.5 | 7 | 0.5 |
| Beta-star (x/y) | Cm | 75 | 3.7 | 75 | 0.88 |
| Beam spot size at IP (c/y) | Mm | 6.6 | 6.6 | 3.25 | 3.25 |
| Beam-beam per IP(x/y) | | 0.0004 | 0.12 | 0.0016 | 0.12 |
| Crossing angle | mrad | ~0.95 | | ~0.95 | |
| Hour-glass (HG) reduction | | 0.77 | | 0.34 | |
| Luminosity/nuclei per IP, with HG reduction | $10^{33}/cm^2/s$ | | | 1.0 | |
| Luminosity/nucleon per IP, with HG reduction | $10^{33}/cm^2/s$ | 4.5 | | 23.6 | |

In Table A7.2, the proton beam parameters are identical to these of the SPPC design presented in Chapter 8. The bunch numbers for both electron and proton beams are 10080, maintaining a 40 MHz repetition rate (thus 25 ns bunch spacing) as well as a gap (or multiple gaps) of 24.4 km in the beam bunch trains. The final focusing of the proton beam is also identical to that for *pp* collisions; however, the electron $\beta^*$ is increased to 3.67 cm in order to match the beam spot sizes at the collision point. As a comparison, the CEPC e+e- vertical β* is only 1.5 mm.

The geometric correction factors to the *e-p* collision luminosity are the crab crossing and hour-glass effects. Due to small bunch spacing, a crossing angle is introduced to enable rapid beam separation near the interaction point to alleviate the parasitic beam-beam effect. The required minimum crossing angle is 0.95 mrad which provides a horizontal separation of 5 ($\sigma_e$+$\sigma_p$) ≈ 3.6 mm. at the first parasitic collision point (3.75 m. from it); $\sigma_e$ and $\sigma_p$ are the rms sizes of the electron and proton beams at that location. We propose to utilize SRF crab cavities on both sides of a collision point to restore head-on collisions; otherwise the luminosity loss due to a crab crossing angle is enormous. The required transverse kick voltages are estimated to be ~63 MV and ~1 MV for the proton and electron beams respectively, assuming a 650 MHz RF frequency and modest values for the betatron functions (400 m. and 200 m.) at the location of the crab cavities. We assume there is no luminosity reduction from crab collisions after compensation. It can be shown that due to a significant increase of the electron β*, the luminosity reduction factor due to the hour-glass effect is 77%.

## A7.4: *e-A* Collisions

The conceptual design of *e-A* collisions at the CEPC-SPPC facility follows the same design principles as for *e-p* collisions; nevertheless, the synchrotron radiation damping effect on the heavy ion beams is much stronger than that on the proton beams, thus requiring some additional considerations and beam parameter adjustments.

It can be shown by scaling that the synchrotron radiation damping time of an ion beam is a factor of $A^4/Z^5$ shorter than that of a proton beam in a storage ring, assuming both beams have the same magnetic rigidity. A is the atomic number and Z is the number of stripped electrons from the ion. Taking a fully stripped lead ion ($^{208}Pb^{82+}$) as an example, the above damping time reduction factor is about 0.5. Thus damping of the lead ion beam is twice as fast as that of a proton beam. The equilibrium emittance of an ion beam has an even higher reduction factor, $Z^3/A^4$, which equals 0.0003 for fully stripped lead ions. This means the lead ion beam equilibrium emittance (in a balance of synchrotron radiation damping and quantum excitations) is four orders of magnitude smaller than that of the proton beam. This is clearly a non-physical result since intra-beam scattering which can cause emittance growth has not been taken into account. In fact, after a couple of damping times, the ion beam emittance will reach an equilibrium value in a balance of radiation damping, quantum excitations and intra-beam scatterings induced heating. It has been estimated that this true equilibrium emittance of a fully stripped lead ion beam is about 0.22 μm rad in the SPPC, approximately 10 times smaller than the proton emittance. We use this value to estimate the luminosity of *e-A* collisions shown in the fifth and sixth columns of Table A7.3.1. An accurate estimate of the ion beam emittance depends on the lattice design of the SPPC ring. Other parameters in Table A.3.1, such as the bunch length and number and final focusing, are identical to the *e-p* collision design shown in the third and fourth columns. The luminosity reaches $1.0\times10^{32}/cm^2/s$ per nuclei and $2.1\times10^{34}/cm^2/s$ per nucleon at each detector.

## A7.5: Additional Comments

Interaction region design is critical for studies of an *e-p* or *e-A* collider. Forward particle detection will be a requirement for the detector. A large detector space is required for both lepton and hadron beams. In the case of the proton or ion beams, this requirement will be similar to that of SPPC. On the other hand, the CEPC detector space is relatively small. A specialized *e-p* or *e-A* experimental hall and dedicated excavation will be required.

Another critical issue is to provide sufficient separation of the colliding beams at the locations of the final focusing magnets. The separation due to the crab crossing angle is merely a few cm if the detector space is ±25 m. and smaller than the physical size of warm or superconducting magnets. Additional schemes must be implemented to avoid interference between the beam transport and these magnets.

In the CEPC-SPPC design, the lepton rings are placed on the inner side of the tunnel and the hadron rings on the outer side, separated by a few meters. This arrangement causes a circumference difference (up to 20 m.) between the two rings, which must be corrected in order to maintain beam synchronization at multiple interaction points.

Estimating beam and luminosity lifetime, and evaluating and mitigating various effects that limit these lifetimes is critical and yet to be carried out. Nonlinear and collective beam dynamics, particularly the beam-beam effect, must be thoroughly studied.

# Appendix 8: Opportunities for Polarization in the CEPC

## A8.1: Introduction

One of the future experiments at CEPC can be a precise measurement of the mass of the Z using resonant depolarization [1]. To achieve this goal one needs a method for obtaining polarized electron and positron beams. In this appendix we consider the major issues for obtaining the radiative self-polarization of particles with the current CEPC design parameters at 45 GeVand at 80 GeV.

## A8.2: Special Wigglers to Speed up Polarization

The well-known Sokolov-Ternov mechanism of radiative self-polarization of particles in an ideal storage ring is characterized by the time $\tau_p$ of polarization build-up to the extent $P_0 = 0.92$ [2]:

$$\frac{1}{\tau_p} = \frac{5\sqrt{3}}{8} \frac{r_e \Lambda_e c \gamma^5}{R^3} \left\langle K^3 \right\rangle, \tag{1}$$

where $r_e$, $\Lambda_e$, $\gamma$ are the electron radius, Compton wave length and relativistic factor respectively; $K$ is the orbit curvature in units of the inverse machine radius $R$. It is averaged over the storage ring azimuth ($\vartheta$). The time for the buildup of this radiative polarization in the 100 km CEPC is huge: 260 hours at 45 GeV! At 80 GeV, this time falls as $(45/80)^5$ to 16 hrs. To speed up the polarization process, it is possible to apply a method [2] based on the use of $N_w$ special wiggler magnets (the so-called shifters) with distribution of the vertical field along the orbit such that $\int B_w d\vartheta = 0$ and $\int B_w^3 d\vartheta \neq 0$.

Let every shifter consist of three bending magnets. The field of the edge magnets ($B_-$) is much smaller in magnitude than the field of the central one ($B_+$) and opposite in sign. The field of the latter is in the same direction as the bending field in the arcs. Since $|B_+|^3 >> |B_-|^3$, the equilibrium degree of polarization in the ideal case is close to the maximum $P_0$. The shifters decrease the polarization time as shown in the equation below ($L_-$ and $L_+$ are the corresponding magnet length):

$$\tau_p^w = \tau_p \left[ 1 + N_w \frac{B_+^3 L_+ + 2|B_-|^3 L_-}{2\pi R \langle B_0 \rangle B_0^2} \right]^{-1}. \tag{2}$$

The fraction of radiation energy loss enhancement is

$$u = N_w \frac{B_+^2 L_+ + 2B_-^2 L_-}{2\pi R \langle B_0 \rangle B_0}. \tag{3}$$

The harmful effect of the shifters is to increase the beam energy spread:

$$\frac{\Delta E_w}{\Delta E} = \left[ \frac{\tau_p}{\tau_p^w} \cdot \frac{1}{1+u} \right]^{1/2}. \tag{4}$$

The effectiveness of this system as applied to CEPC can be judged by its parameters in Table A8.1. The equations (2-4) are written in the isomagnetic approximation (the characteristic field in the CEPC magnets is $B_0 \approx 0.013\,\mathrm{T}$ at 45.6 GeV; the averaged-over-azimuth field $< B_0 > \approx 0.01$ T).

**Table A8.1**: Parameters of the special wiggler system (45.6 GeV)

| $N_w$ | $B_+$ [T] | $L_+$ [m] | $B_-$ [T] | $L_-$ [m] | $\tau_p / \tau_p^w$ | $u$ | $\Delta E_w / \Delta E$ |
|---|---|---|---|---|---|---|---|
| 10 | 0.5 | 1 | 0.125 | 2 | 8.3 | 0.20 | 2.6 |
| 10 | 0.6 | 1 | 0.15 | 2 | 13.6 | 0.29 | 3.3 |

The reduction of $\tau_p$ by an order of magnitude, down to 20 or 30 hours, means that it becomes possible to polarize the beams up to 10% in a few hours. This degree of polarization is sufficient for observation by a laser polarimeter using the resonant depolarization technique for determining the particle energy [3]. As the analysis below shows, a further increase in the wiggler field and their number leads to an undesirable increase of depolarizing effects due to the large energy spread.

## A8.3: Depolarizing Effects of Quantum Fluctuations

Quantum fluctuations lead to the scattering of particle trajectories in the beam relative to the equilibrium orbit. In turn, this causes diffusion of the vertical projection of the spins in the presence of imperfections in the guide field. The corresponding depolarizing effect is characterized by the depolarization time $\tau_d$. As a result, an actual equilibrium polarization degree $P < P_0$ is established with a relaxation time $\tau_{rel} = (1/\tau_p + 1/\tau_d)^{-1} < \tau_p$. The depolarization factor

$$G = P / P_0 = \tau_{rel} / \tau_p \tag{5}$$

depends on the distributed radial magnetic and vertical electric fields which perturb the trajectories of particles in the vertical plane. The strongest depolarizing effect is produced by the sources that cause vertical distortions $y_0 = y_0(\vartheta)$ of the closed orbit. Their influence increases in the vicinity of the integer spin resonances $\nu = k$. Here $\nu = \gamma a$ is the spin tune, a number of spin precession cycles per one turn ($\gamma$ and $a$ are the Lorentz factor and the anomalous part of the gyromagnetic ratio of electron, respectively.) Because of synchrotron oscillations with frequency $\nu_\gamma$, the spin tune is modulated by $\tilde{\nu} = \nu + \Delta \cdot \cos\psi_\gamma$. $\Delta$ is the amplitude related to the amplitude of energy oscillation. The distribution function of $\Delta$ is

$$f(\Delta) = \frac{\Delta}{\sigma_\nu^2} \exp\left( -\frac{\Delta^2}{2\sigma_\nu^2} \right), \tag{6}$$

$\sigma_\nu = \nu\sigma_\gamma = \overline{(\Delta^2/2)}^{1/2}$ is the spin tune spread due to the beam energy spread $\sigma_\gamma$.

Modulation leads to the dependence of the factor $G$ on the detuning from the modulation resonances $\nu = k \pm m\nu_\gamma$ ($m$ is integer).

To estimate the degree of polarization that is actually achievable, and taking into account synchrotron modulation, one can use the well-known formula [4]

$$G \approx \left\{1+\frac{11\nu^2}{18}\sum_{k,m}\frac{|w_k|^2 I_m(\sigma_\nu^2/\nu_\gamma^2)\exp(-\sigma_\nu^2/\nu_\gamma^2)}{\left[(|\nu-k|-m\nu_\gamma)^2-\nu_\gamma^2\right]^2}\right\}^{-1}, \quad (7)$$

Here $I_m(x)$ is the modified Bessel function; $w_k$ is the $k$-th azimuthal Fourier harmonic amplitude of the spin perturbations: $w_k \approx \left\langle \frac{\nu}{R}\frac{d^2 y_0}{d\vartheta^2}\exp(-ik\vartheta)\right\rangle$. In (7) the following expansion in a series of Bessel functions is used:

$$w_k \exp\left(-i\frac{\Delta}{\nu_\gamma}\sin\psi_\gamma\right) = w_k\sum_{l=-\infty}^{l=\infty} J_l\left(\frac{\Delta}{\nu_\gamma}\right)\exp\left(-il\psi_\gamma\right), \quad (8)$$

averaged over the ensemble of beam particles. In a strict sense, the expansion (8) and then equation (7) for $G$ are valid if [5]

$$\sigma_\nu^2\Lambda_\gamma << \nu_\gamma^3, \quad (9)$$

$\Lambda_\gamma$ is the radiation decrement of synchrotron oscillations in units of inverse turns.

The parameter

$$\Gamma = \frac{11\nu^2}{18\nu_\gamma^3\tau_p f_0} = \frac{\sigma_\nu^2\Lambda_\gamma}{2\nu_\gamma^3} \quad (10)$$

characterizes the precession phase increment due to diffusion per the period of synchrotron oscillations $(\nu_\gamma f_0)^{-1}$ ( $f_0$ is the revolution frequency). Condition (9) is almost equivalent to the condition $\Gamma << 1$, which means that such an increment is negligibly small. Under the condition $\Gamma \geq 1$, the spectrum of spin perturbations becomes blurry, i.e. it must differ from a strictly linear spectrum (8).

In Table A8.2, data for three colliders at the Z-pole energy are presented. On can compare the parameter $\Gamma$. In the case of LEP, as well as of CEPC and FCCee without the use of wigglers, this parameter is small and the expansion (8) is quite applicable.

## A8.4: Polarization Calculation Example

In Fig. A8.1, the polarization at LEP and CEPC is calculated from (7) for the magnitude of the resonance harmonic of random perturbations of a closed orbit $|w_k| = 2\cdot 10^{-3}$ for its dependence on beam energy and spin tune. Only the two nearest integer spin resonances with $k = 103$ and $k = 104$ are taken into account. Because of the chaotic nature of the field imperfection sources we put $|w_k| = |w_{k+1}|$. The red curves are the case without consideration of the modulation when (in the approximation $|\nu - k| >> \max(\sigma_\nu, \nu_\gamma)$):

$$G \approx \left[1+\frac{11\nu^2}{18}\sum_k\frac{|w_k|^2}{(\nu-k)^4}\right]. \quad (11)$$

The estimate for LEP for the harmonic amplitude $|w_k| = 2\cdot 10^{-3}$ is compared with real data. The polarization at LEP reached 40% at 44.7 GeV) [6]. This approximately corresponds to the calculation at $\nu = 103.46$. In the LEP polarization simulations [7], performed long before LEP began operation, the same polarization degree was obtained for the 50 μm random vertical displacements of the LEP quads.

**Table A8.2**: comparison of three colliders in terms of the precession phase diffusion increment at 45.6 GeV ($\nu = 103.5$). A star indicates the cases with special wigglers

| | $\sigma_\nu$ | $\nu_\gamma$ | $f_0$ [kHz] | $\tau_p$ [hr] | $\Gamma$ |
|---|---|---|---|---|---|
| LEP | 0.061 | 0.083 | 11 | 5 | 0.054 |
| CEPC | 0.039 | 0.028 | 3 | 268 | 0.103 |
| | 0.128 | 0.028 | 3 | 19.7* | 1.40 |
| | 0.103 | 0.056 | 3 | 32.3* | 0.107 |
| FCCee | 0.038 | 0.025 | 3 | 150 | 0.244 |
| | 0.113 | 0.025 | 3 | 15* | 2.44 |

The graphs for CEPC in Fig. A8.1 relate to two cases. The bottom left graph is that of the current CEPC design. It is characterized by $\nu_\gamma = 0.028$ and moderate spreads of beam energy and spin tune. The maximal equilibrium polarization degree is close to 50%. The modulation resonances can be neglected. The time to build up polarization is too large: $\tau_{rel} \approx 0.5\tau_p = 134$ hrs. From that point of view, obtaining polarization with this mode is only relevant if one injects a polarized beam from the Booster. In this case, the relaxation of the polarization occurs to a level of 50%, which guarantees the preservation of a high degree of polarization for the whole life time of the injected beam.

Another case is an example of CEPC using special wigglers to speed up polarization more than 8 times. Refer to the bottom right graph in Fig. A8.1. The energy and spin tune spreads are increased approximately 3 times. This reduces approximately by a factor of two the degree of polarization at the maximum.

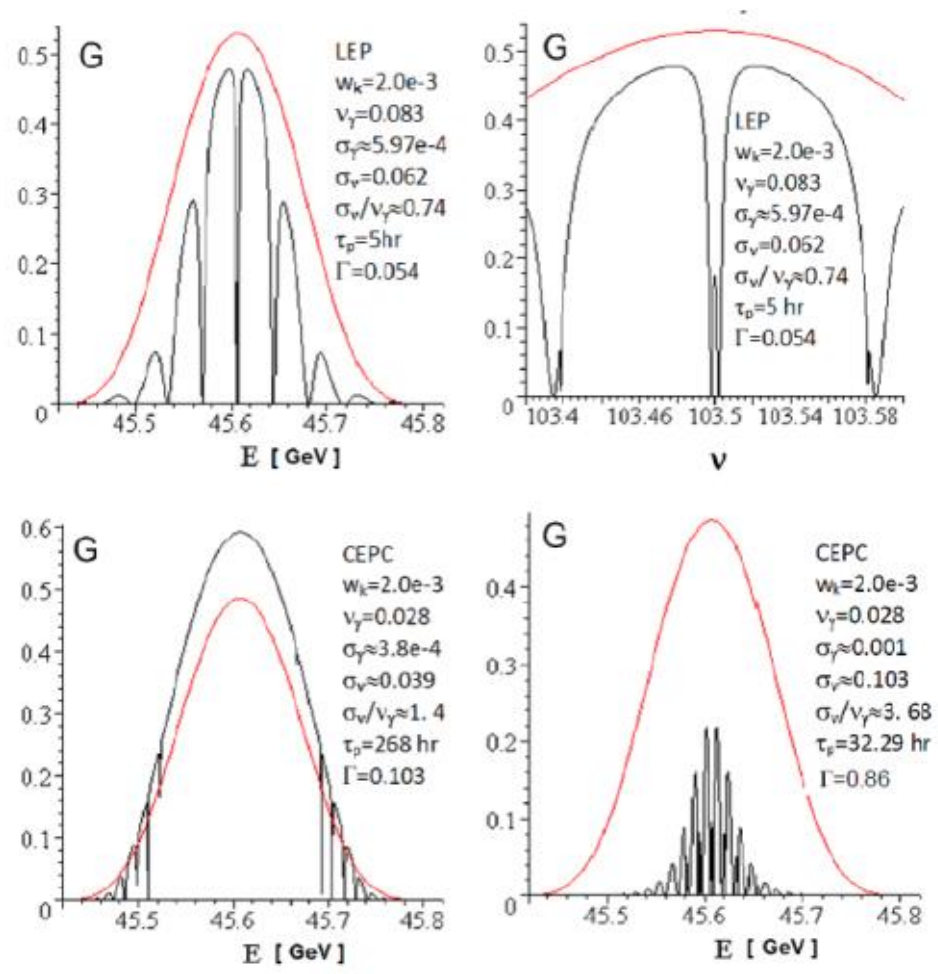

**Figure A8.1**: the polarization at LEP and CEPC vs. energy and spin tune. The red curve corresponds to the case without consideration of synchrotron modulation (11).

In order to increase the equilibrium degree of radiation polarization in the wiggler mode, it is necessary to decrease the harmonic amplitudes of the two nearest integer spin resonances (in this case, $k = 103$ and $k = 104$) due to special correction of the vertical closed orbit. Estimates show that in the case of CEPC, one should reduce the harmonic to a level of $|w_k| = 10^{-3}$ or even less. It is also necessary to limit the wiggler fields, since the increase in the spin tune spread caused by them leads to a drop in the degree of polarization. As can be seen from Fig. A8.2, a wiggler field of 0.5 T is preferable in comparison to 0.6 T; this gives 7-fold greater polarization.

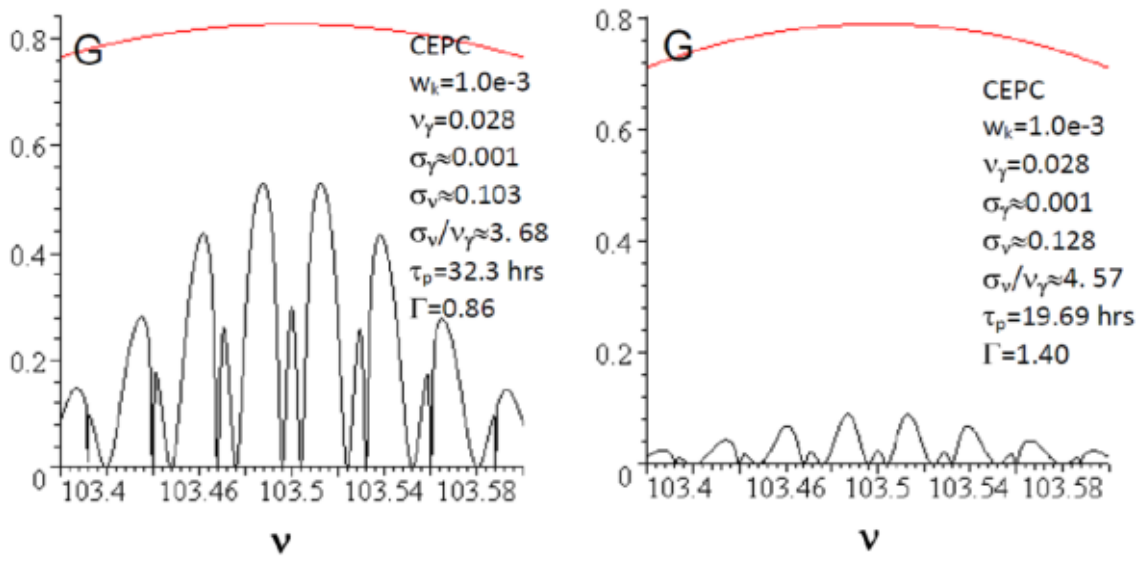

**Figure A8.2:** the polarization at 45 GeV CEPC vs. the spin tune in two cases of the special wiggler parameters from Table A8.2.1: $B_+ = 0.5\,\mathrm{T}$ (left) and $B_+ = 0.6\,\mathrm{T}$ (right).

With increasing beam energy, the depolarizing effects of guide field imperfections intensify. At the threshold energy of W pair production, in order to obtain polarization, an even more thorough correction of the spin harmonics associated with the distortions of the vertical closed orbit is needed. But, as shown below, the wigglers will not be needed to speed up the polarization process. The dependence of the polarization on the energy/spin tune in the region of the W-threshold is plotted in Fig. A8.3 for the harmonic amplitude $|w_k| = 5 \cdot 10^{-4}$ .

## A8.5: Time to Reach 10% Polarization

The polarization increases in time according to the law $P(t) = G \cdot P_0 \left[1 - \exp\left(-t/\tau_{rel}\right)\right]$. Let us define the operation time spent to reach the polarization degree of $\eta = P(t = t_\eta)$ percentages in the conditions under consideration: $t_\eta = -\tau_{rel} \ln\left(1 - \frac{\eta}{92G}\right)$.

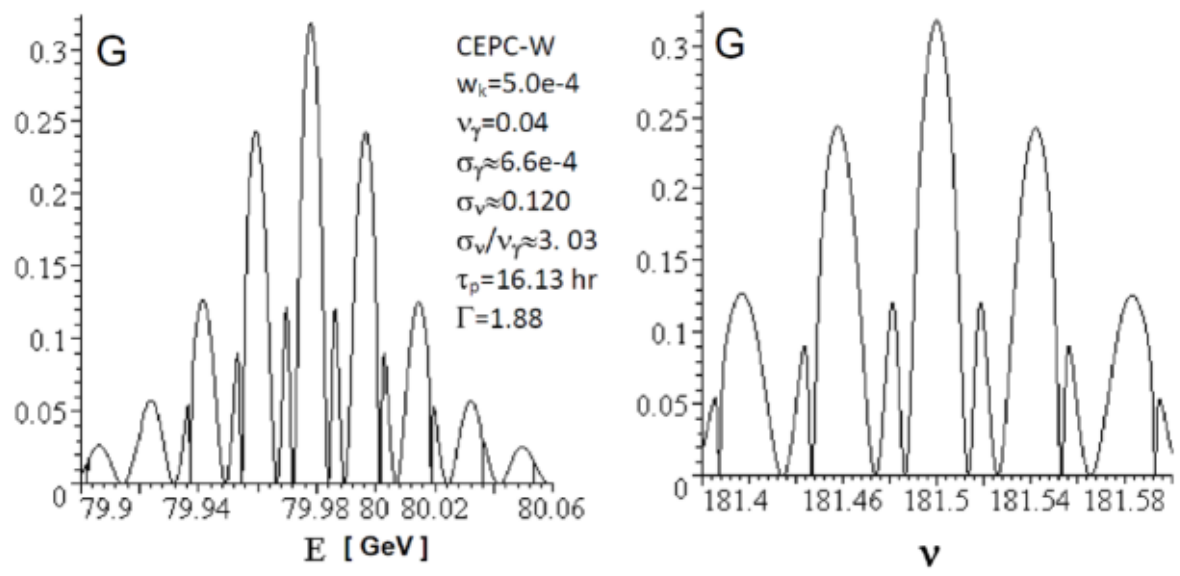

**Figure A8.3:** CEPC polarization near the W pair production threshold vs. the beam energy (left) and the spin tune (right).

Here, $\tau_{rel} = G\tau_p^w$ or $\tau_{rel} = G\tau_p$ depending on whether special wigglers are used or not. Based on the maximum values of the factor $G$ in Figs. A8.2 and A8.3, a table of parameters for obtaining polarization at Z-pole and W pair production threshold is compiled, including the time $t_\eta$, which in the considered cases ranges from 2 to 4 hours (see Table A8.3).

## A8.6: Polarization Scenario

About 100 pilot electron/positron bunches of relatively small total current $I_p$ (of the order of 1% of the main train current) are stored to be partially polarized up to 10% in a few hours using 10 shifter magnets with field about 0.5 T. The SR power from each 0.6 T shifter magnet (Table A8.1) is 2 kW at $I_p$=2 mA. When the polarization process ends the shifter magnets are turned off. Then the main bunch train is stored. The pilot bunches are not in collision. Their lifetime is about $10^5$ seconds due to scattering of particles on thermal radiation photons [8]. The polarized bunches are used one by one for the resonant depolarization calibration of beam energy every 15 minutes. So, a single polarized bunch train is spent per day while data taking occurs.

**Table A8.3**: Parameters for obtaining polarization at CEPC; * and ** indicate the cases with the use of special wigglers of $B_+ = 0.5$ T and $B_+ = 0.6$ T, respectively

| $E$, GeV | $\lvert w_k \rvert$ | $G_{\max}$ | $\nu_\gamma$ | $\tau_{rel}$, hr | $\eta$, % | $t_\eta$,hr |
|---|---|---|---|---|---|---|
| 45.602 | $10^{-3}$ | 0.53 | 0.028 | 17.1* | 10 | 3.93 |
| 45.602 | $10^{-3}$ | 0.09 | 0.028 | 1.8** | 6 | 2.28 |
| 79.978 | .0005 | 0.32 | 0.040 | 4.8 | 10 | 2.14 |

## A8.7: Summary

Particle polarization of at least 10% is needed to apply the resonant depolarization technique in precise measurement of the Z mass. Because of the excessively long time for radiative polarization, it becomes necessary to add to the Collider strong non-uniform wiggler magnets to speed up the polarization process. The wigglers cause a multiple increase in the spread of the spin precession frequency. This, in turn, leads to an intensification of the depolarizing effect of quantum fluctuations. The calculations of the depolarizing factor of vertical closed orbit distortions were performed taking into account the synchrotron modulation of the spin tune. Obtaining the required degree of polarization in CEPC at 45 and 80 GeV for an acceptable time is tentatively possible provided that the resonance spin harmonics of vertical closed orbit distortions are corrected to the levels indicated in the corresponding numerical examples. This conclusion is indirectly supported by similar estimates for LEP and their achieved polarization.

## A8.8: References